\documentclass{aastex63}

\usepackage{newtxtext,newtxmath,float,xspace,threeparttable,bm}

\newcommand{\todo}{\ifmmode \text{\color{red}\Huge{\(\bullet\)}} \else {\color{red}{\Huge$\bullet$}}\fi}
\newcommand{\tido}{\ifmmode {{\color{red}\bullet}} \else {\color{red}$\bullet$}\fi}

\newcommand{\Halpha}{\ifmmode {\rm H}\alpha \else H$\alpha$\fi}
\newcommand{\halpha}{\Halpha}

\newcommand{\Hbeta}{\ifmmode {\rm H}\beta \else H$\beta$\fi}
\newcommand{\hbeta}{\Hbeta}

\newcommand{\oiii}{\ifmmode \left[{\rm O}\,\textsc{iii}\right] \else [O\,{\sc iii}]\fi}
\newcommand{\OIII}{\ifmmode \left[{\rm O}\,\textsc{iii}\right]\,\lambda5007 \else [O\,{\sc iii}]\,$\lambda5007$\fi}
\newcommand{\nii}{\ifmmode \left[{\rm N}\,\textsc{ii}\right]  \else [N\,\textsc{ii}]\fi}
\newcommand{\NII}{\ifmmode \left[{\rm N}\,\textsc{ii}\right]\,\lambda6584 \else [N\,\textsc{ii}]\,$\lambda6584$\fi}

\newcommand {\nh}{\ifmmode N_{\rm H} \else $N_{\rm H}$\fi}

\newcommand {\Lsoftint} {\ifmmode L^{\rm in}_{\mathrm{2-10\ keV}} \else $L^{\rm in}_{\mathrm{2-10\ keV}}$\fi}

\newcommand {\ciao}{{\sc CIAO}\xspace}
\newcommand {\sherpa}{{\sc SHERPA}\xspace}

\newcommand {\ergpersec} {\ifmmode {\rm erg~s}^{-1} \else erg~s$^{-1}$ \fi}

\newcommand {\nhunit} {cm$^{-2}$\xspace}

\def\micron{{\mbox{$\mu{\rm m}$}}}
\def\arcsec{{\mbox{$^{\prime \prime}$}}}

\def \swift {{\em Swift\ }}

\def \swiftbat {{\em Swift}-BAT\ }
\def \swiftbatsh {{\em Swift}-BAT}

\def \wise{{\em WISE\ }}
\def \wisesh{{\em WISE}}
\def \herschelsh{{\em Herschel}}
\def \herschel{{\em Herschel\ }}
\def \iras{{\em IRAS\ }}
\def \irassh{{\em IRAS}}

\newcommand{\akari}{\emph{Akari\ }}
\newcommand{\akarish}{\emph{Akari}}

\def \hst{{\em HST\ }}

\def\arcsec{{\mbox{$^{\prime \prime}$}}}
\def\cm{{\rm\thinspace cm}}

\def\erg{{\rm\thinspace erg}}

\def\g{{\rm\thinspace g}}

\def\K{{\rm\thinspace K}}

\def\km{{\rm\thinspace km}}

\def\Lsun{\hbox{$\rm\thinspace L_{\odot}$}}
\def\m{{\rm\thinspace m}}

\newcommand{\Msun}{\ifmmode M_{\odot} \else $M_{\odot}$\fi}

\def\s{{\rm\thinspace s}}

\def\ergps{\hbox{$\erg\s^{-1}\,$}}

\def\kmps{\hbox{$\km\s^{-1}\,$}}

\def\psqcm{\hbox{$\cm^{-2}\,$}}

\def\apjl{\hbox{ApJL}}
\def\apjs{\hbox{ApJS}}

\def\aap{\hbox{AAP}}
\def\micron{{\mbox{$\mu{\rm m}$}}}
\def\arcsec{{\mbox{$^{\prime \prime}$}}}

\newcommand{\lledd}{\ifmmode L/L_{\rm Edd} \else $L/L_{\rm Edd}$\fi}
\newcommand{\mbh}{\ifmmode M_{\rm BH} \else $M_{\rm BH}$\fi}

\newcommand{\kms}{\ifmmode {\rm km\,s}^{-1} \else ${\rm km\,s}^{-1}$\fi}

\newcommand{\nuvr}{\ifmmode {\rm NUV}-r \else NUV-$r$\fi}
\newcommand{\mh}{\ifmmode M_{\rm H_2} \else $M_{\rm H_2}$\fi}
\newcommand{\mhi}{\ifmmode M_{\rm HI} \else $M_{\rm HI}$\fi}
\newcommand{\mstar}{\ifmmode M_{\ast} \else $M_{\ast}$\fi} 
\newcommand{\must}{\ifmmode \mu_{\ast} \else $\mu_{\ast}$\fi}
\newcommand{\hmol}{\ifmmode H_2 \else $H_2$\fi}
\newcommand{\rmol}{\ifmmode R_{\rm mol} \else $R_{\rm mol}$\fi}
\newcommand{\tdep}{\ifmmode t_{\rm dep}({\rm H_2}) \else $t_{\rm dep}({\rm H_2})$\fi}
\newcommand{\tdepHI}{\ifmmode t_{\rm dep}({\rm HI}) \else $t_{\rm dep}({\rm HI})$\fi}
\newcommand{\fgas}{\ifmmode f_{\rm H_2} \else $f_{\rm H_2}$\fi}
\newcommand{\fhi}{\ifmmode f_{\rm HI} \else $f_{\rm HI}$\fi}
\newcommand{\xco}{\ifmmode \alpha_{\rm CO} \else $\alpha_{\rm CO}$\fi}

\newcommand{\deltams}{$\Delta$(MS)}
\newcommand{\deltasfr}{$\Delta$(SFR)}

\newcommand{\Nobs}{213} 
\newcommand{\Nunobserve}{135}
\newcommand{\perobs}{61\%}
\newcommand{\perunobs}{39\%}
\newcommand{\Napex}{165} 
\newcommand{\Njcmt}{35} 
\newcommand{\Nreduced}{200}
\newcommand{\Nanalyzed}{207} 
\newcommand{\Narchival}{13}

\newcommand{\Nlowzall}{348}
\newcommand{\Nxcoldgassinactive}{344}
\newcommand{\ntarosfr}{171}
\newcommand{\ntaroco}{129}

\usepackage[T1]{fontenc}
\usepackage{ae,aecompl}

\usepackage{graphicx}	
\usepackage{amsmath}	
\usepackage{amssymb}	
\received{}
\revised{}
\accepted{}

\submitjournal{ApJS}

\shorttitle{BASS -- XX: Molecular Gas In AGN Galaxies}
\shortauthors{Koss et al.}

\begin{document}

\title{BAT AGN Spectroscopic Survey -- XX: Molecular Gas in Nearby Hard X-ray Selected AGN Galaxies}

\correspondingauthor{Michael Koss}
\email{mike.koss@eurekasci.com}

\author[0000-0002-7998-9581]{Michael J. Koss}
\affiliation{Eureka Scientific, 2452 Delmer Street Suite 100, Oakland, CA 94602-3017, USA}

\author[0000-0002-7606-3679]{Benjamin Strittmatter}
\affiliation{Institute for Particle Physics and Astrophysics, ETH Z{\"u}rich, Wolfgang-Pauli-Strasse 27, CH-8093 Z{\"u}rich, Switzerland}

\author[0000-0003-3336-5498]{Isabella Lamperti}
\affiliation{Department of Physics and Astronomy, University College London, Gower Street, London WC1E 6BT, UK}

\author[0000-0002-2125-4670]{Taro Shimizu}
\affiliation{Max-Planck-Institut f{\"u}r extraterrestrische Physik, Postfach 1312, D-85741 Garching, Germany}

\author[0000-0002-3683-7297]{Benny Trakhtenbrot}
\affiliation{School of Physics and Astronomy, Tel Aviv University, Tel Aviv 69978, Israel}

\author[0000-0003-4357-3450]{Amelie Saintonge}
\affiliation{Department of Physics and Astronomy, University College London, Gower Street, London WC1E 6BT, UK}

\author[0000-0001-7568-6412]{Ezequiel Treister}
\affiliation{Instituto de Astrof{\'i}sica, Facultad de F{\'i}sica, Pontificia Universidad Cat{\'o}lica de Chile, Casilla 306, Santiago 22, Chile}

\author[0000-0003-0522-6941]{Claudia Cicone}
\affiliation{Institute of Theoretical Astrophysics, University of Oslo, Postboks 1029, Blindern 0315 Oslo, Norway}

\author[0000-0002-7962-5446]{Richard Mushotzky}
\affiliation{Department of Astronomy, University of Maryland, College Park, MD 20742, USA}

\author[0000-0002-5037-951X]{Kyuseok Oh}
\affiliation{Korea Astronomy \& Space Science institute, 776, Daedeokdae-ro, Yuseong-gu, Daejeon 34055, Republic of Korea}
\affiliation{Department of Astronomy, Kyoto University, Kitashirakawa-Oiwake-cho, Sakyo-ku, Kyoto 606-8502, Japan}
\affiliation{JSPS Fellow}

\author[0000-0001-5231-2645]{Claudio Ricci}
\affiliation{N\'ucleo de Astronom\'ia de la Facultad de Ingenier\'ia, Universidad Diego Portales, Av. Ej\'ercito Libertador 441, Santiago 22, Chile}
\affiliation{Kavli Institute for Astronomy and Astrophysics, Peking University, Beijing 100871, People's Republic of China}
 \affiliation{George Mason University, Department of Physics \& Astronomy, MS 3F3, 4400 University Drive, Fairfax, VA 22030, USA}
 
 \author[0000-0003-2686-9241]{Daniel Stern}
\affiliation{Jet Propulsion Laboratory, California Institute of Technology, 4800 Oak Grove Drive, MS 169-224, Pasadena, CA 91109, USA}

\author[0000-0001-8211-3807]{Tonima Tasnim Ananna}
\affiliation{Department of Physics \& Astronomy, Dartmouth College, 6127 Wilder Laboratory, Hanover, NH 03755, USA}

\author[0000-0002-8686-8737]{Franz E. Bauer}
\affiliation{Instituto de Astrof\'{\i}sica  and Centro de Astroingenier{\'{\i}}a, Facultad de F\'{i}sica, Pontificia Universidad Cat\'{o}lica de Chile, Casilla 306, Santiago 22, Chile}
\affiliation{Millennium Institute of Astrophysics (MAS), Nuncio Monse{\~{n}}or S{\'{o}}tero Sanz 100, Providencia, Santiago, Chile}
\affiliation{Space Science Institute, 4750 Walnut Street, Suite 205, Boulder, Colorado 80301, USA}

\author[0000-0003-3474-1125]{George C. Privon}
\affiliation{Department of Astronomy, University of Florida, 211 Bryant Space Science Center, Gainesville, FL 32611, USA}
\affiliation{National Radio Astronomy Observatory, 520 Edgemont Rd,
Charlottesville, VA 22903, USA}

\author[0000-0001-5481-8607]{Rudolf E. B\"{a}r}
\affiliation{Institute for Particle Physics and Astrophysics, ETH Z{\"u}rich, Wolfgang-Pauli-Strasse 27, CH-8093 Z{\"u}rich, Switzerland}

\author[0000-0002-6637-3315]{Carlos De Breuck}
\affiliation{European Southern Observatory, Karl Schwarzschild Str 2, 85748 Garching, Germany}

\author{Fiona Harrison}
\affiliation{Cahill Center for Astronomy and Astrophysics, California Institute of Technology, Pasadena, CA 91125, USA}

\author[0000-0002-4377-903X]{Kohei Ichikawa}
\affiliation{Frontier Research Institute for Interdisciplinary Sciences, Tohoku University, Sendai 980-8578, Japan}

\author[0000-0003-2284-8603]{Meredith C. Powell}
\affiliation{Institute of Particle Astrophysics and Cosmology, Stanford University, 452 Lomita Mall, Stanford, CA 94305, USA}

\author[0000-0002-0001-3587]{David Rosario}
\affiliation{Centre for Extragalactic Astronomy, Department of Physics, Durham University, South Road, DH1 3LE Durham, UK}

\author{David B. Sanders}
\affiliation{Institute for Astronomy, 2680 Woodlawn Drive, University of Hawaii, Honolulu, HI 96822, USA}

\author[0000-0001-5464-0888]{Kevin Schawinski}
\affiliation{Modulos AG, Technoparkstrasse 1, CH-8005 Zurich, Switzerland}

\author[0000-0003-2015-777X]{Li Shao}
\affiliation{National Astronomical Observatories, Chinese Academy of Sciences, 20A Datun Road, Chaoyang District, Beijing, China, 100012}

\author[0000-0002-0745-9792]{C. Megan Urry}
\affiliation{Yale Center for Astronomy \& Astrophysics, Physics Department, New Haven, CT 06520, USA}

\author[0000-0002-3158-6820]{Sylvain Veilleux}
\affiliation{Department of Astronomy, University of Maryland, College Park, MD 20742, USA}



\begin{abstract}
  We present the host galaxy molecular gas properties of a sample of \Nobs\ nearby ($0.01 < z < 0.05$) hard X-ray selected AGN galaxies, drawn from the 70-month catalog of \swiftbatsh, with 200 new CO(2--1) line measurements obtained with the JCMT and APEX telescopes. We find that AGN in massive galaxies ($\log(\mstar/\Msun){>}10.5$) tend to have more molecular gas, and higher gas fractions, than inactive galaxies matched in stellar mass. {When matched in star formation, we find AGN galaxies show no difference from inactive galaxies with no evidence of AGN feedback affecting the molecular gas.}    The higher molecular gas content is related to AGN galaxies hosting a population of gas-rich early types with an order of magnitude more molecular gas and a smaller fraction of quenched, passive galaxies ($\sim$5\% vs. 49\%) compared to inactive galaxies.  The likelihood of a given galaxy hosting an AGN (L$_{\rm bol}{>}10^{44}$\ergps) increases by $\sim$10-100 between a molecular gas mass of $10^{8.7}$\Msun\ and $10^{10.2}$\Msun.
 Higher Eddington ratio ($\log$(\lledd)$>$$-$1.3) AGN galaxies tend to have higher molecular gas masses and gas fractions.    {Higher column density AGN galaxies ($\log(\nh/\psqcm)>23.4$) are associated with lower depletion timescales and may prefer hosts with more gas centrally concentrated in the bulge that may be more prone to quenching than galaxy wide molecular gas.}  The significant {average} link of host galaxy molecular gas supply to SMBH growth may naturally lead to the general correlations found between SMBHs and their host galaxies, such as the correlations between SMBH mass and bulge properties and the redshift evolution of star formation and SMBH growth. 
\end{abstract}

\keywords{catalogs --- surveys}



\section{Introduction} \label{sec:intro}
The study of cold molecular gas and its properties in galaxies hosting Active Galactic Nuclei (AGN) is crucial to understanding the growth of supermassive black holes (SMBHs) and how they affect their host galaxies.   There appears to be an evolutionary link between the merger of gas-rich spirals, ultra-luminous infrared (IR) galaxies, and luminous quasars \citep[e.g.,][]{Sanders:1988:L35,Hopkins:2008:356,Veilleux:2009:587}.  However, our understanding of this evolution and AGN fueling is biased by the IR pre-selection, as most CO surveys have focused on far-IR (FIR) bright AGN \citep{Rigopoulou:1997:493,Monje:2011:23,Kirkpatrick:2019:41} to ensure a large number of detections.    

 Early studies surveyed optically selected Seyferts in the accessible $^{12}$CO J=1--0 transition\footnote{Hereafter, we simply refer to $^{12}$CO as CO.} \citep{Heckman:1989:735,Maiolino:1995:95}.  Some surveys have focused on \hbox{X-ray} selected FIR bright AGN \citep{Yamada:1994:L27,Monje:2011:23}.   Some of these studies found that obscured AGN that lack broad Balmer lines (i.e., Seyfert 2 AGN) have increased infrared emission from their disk, suggesting the nuclear toroidal structure hypothesized to obscure the broad line region in type 2 Seyferts \citep[e.g.,][]{Antonucci:1993:473}.  However, other studies have found no difference \citep{Maiolino:1997:552}.  
 

Several observational studies have found that AGN host galaxies appear to lie in the ``green valley'' on the color-magnitude diagram, in between actively star-forming galaxies in the blue cloud and passively evolving galaxies on the red sequence \citep{Silverman:2008:1025,Georgakakis:2009:623, Schawinski:2015:2517}.  The lower star formation rates (SFRs) associated with the ``green valley''  AGN galaxies are attributed to AGN feedback where any reservoir of gas is either heated and/or expelled or simply not accreted \citep{Schawinski:2009:1672,Povic:2012:A118,Cimatti:2013:L13,Mahoro:2017:3226}.  Recent interferometric observations, particularly with ALMA \citep[see e.g., ][for a review]{Combes:2018:5}, have found many examples of molecular outflows in both low and high redshift AGN galaxies  \citep[e.g.,][]{Feruglio:2010:L155,Dasyra:2012:L7,Cicone:2018:143,Brusa:2018:A29,Alonso-Herrero:2019:A65,Bischetti:2019:A118,Fluetsch:2019:4586}, though there are large uncertainties in the gas mass and other properties due to the assumptions and the limited samples considered.  Thus, studying molecular gas in AGN hosts can provide a direct observational test for the presence or destruction of the molecular gas reservoir.  

With the current sensitivity of large, single-dish mm-wavelength telescopes, larger ($N>50$) surveys directed toward unbiased samples of star-forming galaxies and AGN galaxies (i.e. not far-infrared biased) are now possible.  For nearby galaxies with extended structure, single-dish observations are better able to recover total extended CO emission than interferometers.    Such surveys have included low-mass galaxies \citep[ALLSMOG;][]{Bothwell:2014:2599,Cicone:2017:A53} and late-type galaxies belonging to the \herschel Reference Survey \citep[HRS;][]{Boselli:2014:A65}. The IRAM 30m CO Legacy Database for the GALEX Arecibo SDSS Survey \cite[COLD GASS;][]{Saintonge:2011:61,Saintonge:2011:32, Kauffmann:2012:997} with 350 galaxies was one of the first large CO studies in the nearby Universe whose sample was purely stellar mass-limited ($\log$ (\mstar/\Msun) $>$10 ).  The survey was further extend to 532 galaxies in xCOLD GASS \citep{Saintonge:2017:22}.  The survey has been used to trace the abundance of molecular gas in galaxies as a function of their mass, color, and morphology to understand the mechanisms that contribute to galaxy growth and quenching.  The COLD GASS surveys were based on IRAM-30m measurements and complementary APEX CO(2--1) observations, including both CO(1--0) and CO(2--1) line fluxes and offset pointings.  This provides a robust CO luminosity function for inactive galaxies, well-calibrated CO(2--1)/CO(1--0) excitation corrections, and scaling relations between gas fraction (\fgas), depletion timescales [\tdep$\equiv M_{H2}/{\rm SFR}$], specific star formation rates (${\rm sSFR}\equiv {\rm SFR}/M_*$), and global galaxy properties. Thus xCOLD GASS serves as an ideal reference comparison sample for studies of AGN host galaxies.  

There are a handful of Seyferts in the xCOLD GASS sample, many of which appear to have large molecular gas reservoirs, but the statistics are too poor to understand the link to AGN host galaxies in detail. A study of (molecular and atomic) gas based on dust masses inferred from the FIR-sub-mm spectral energy distribution (SED) fitting of stacked samples of AGN galaxies at $z<1$ found both higher gas mass and higher gas fractions in AGN hosts \citep{Vito:2014:1059}.  Conversely, several small studies of powerful, distant AGN (N=3--15, $z{=}1$--3) hosts have found lower molecular gas fractions and depletion timescales \citep{Fiore:2017:A143,Kakkad:2017:4205,Perna:2018:A90}.  A dedicated AGN galaxy survey, with comparatively large numbers ($N>200$) to xCOLD GASS, is therefore needed to understand the average molecular gas and depletion timescales in nearby AGN galaxies. 

The Burst Alert Telescope (BAT; \citealp{Barthelmy:2005:143}) on board the {\it Neil Gehrels Swift Observatory} \citep{Gehrels:2004:1005} has opened up a fundamentally new way to identify and study accreting SMBHs, as ultra-hard X-rays ($>$15\,keV) can penetrate gas and dust to directly probe the central engine of the AGN. Ultra-hard X-rays are therefore less biased by host galaxy properties, which can strongly affect optical and IR techniques for finding AGN \citep[e.g.,][]{Mateos:2012:3271,Ichikawa:2017:74}.  In addition, the ultra-hard X-rays are less affected by obscuration of the AGN, provided the material is thin to Compton scattering [i.e., $\log(\nh/\psqcm)\lesssim 24$; e.g., \citealp{Ricci:2015:L13,Koss:2016:85}].  The AGN detected with BAT are mainly nearby ($\langle z \rangle{=}0.035$), which allows observers to obtain high signal-to-noise (S/N) ancillary data for a large subset of these systems. 
Thus, studying BAT AGN galaxies allows for a high level of completeness, avoiding previously present biases related to FIR selection.  

Previous studies have found that the morphologies and star formation characteristics of the host galaxies of ultra-hard X-ray selected AGN galaxies differ significantly from inactive galaxies, and from optically-selected AGN galaxies.  
Ultra-hard X-ray selected AGN appear to be predominantly hosted by rare, massive spirals or mergers \citep{Koss:2010:L125,Koss:2011:57}.  
At high stellar masses ($\log(M_*/\Msun)>10.5$) these AGN are $\sim$10--100 times more likely to be in spiral hosts, compared with inactive galaxies or optically selected AGN, which are hosted almost exclusively in red ellipticals.  BAT AGN galaxies are significantly more FIR luminous than inactive galaxies or optical Seyferts at the same redshift ($0.01{<}z{<}0.05$), as the majority of BAT AGN (54\%) galaxies are detected with \akari\ at 90 \micron, compared to only 4\% of optically-selected AGN galaxies (i.e., emission-line selection) and only 5\% of inactive galaxies in the SDSS \citep{Koss:2011:57}.  
Further studies based on SED modeling, constructed from \textit{WISE}, \herschel PACS and SPIRE data, have shown that most of the FIR emission ($>$50\%) at 70 \micron\ originates from star formation for most of the sources  \citep[e.g., 76\%,][]{Shimizu:2017:3161}.  The link with star formation also indirectly suggests that AGN activity may be driven by the stochastic accretion of cold gas that should be more prominent among late-type systems \citep{Hopkins:2006:1}.  However, a more direct test would be to measure the amount of molecular gas in these systems. This sample is therefore excellent for understanding the link between molecular gas and AGN fueling.

In this study, we focus on a sample of nearby BAT AGN galaxies ($0.01{<}z{<}0.05$) to test whether the total host galaxy molecular gas is more prominent among luminous AGN and whether this is a key part of their triggering mechanism.  High-quality multi-wavelength data and associated products for the BAT AGN and their hosts are now available through the BAT Spectroscopic Survey \citep[BASS;][]{Koss:2017:74},\footnote{http://www.bass-survey.com} including black hole mass measurements, X-ray modelling \citep{Ricci:2017:17} to measure the intrinsic AGN bolometric luminosity and line-of-sight obscuration, and extensive continuum modeling of the FIR emission from \herschel\citep{Mushotzky:2014:L34,Melendez:2014:152,Shimizu:2016:3335,Shimizu:2017:3161}.  In this paper, we present the CO(2--1)  observations of \Nobs\ BAT AGN galaxies to study the relationship between black hole growth and host galaxy molecular gas. Throughout this work, we adopt $\Omega_{\rm M} = 0.3$, $\Omega_\Lambda = 0.7$, and $H_0 = 70 \,\kms\,{\rm Mpc}^{-1}$.

\section{Survey Description, Samples, and Data Reduction}

\begin{figure*} 
\centering
\includegraphics[width=16cm]{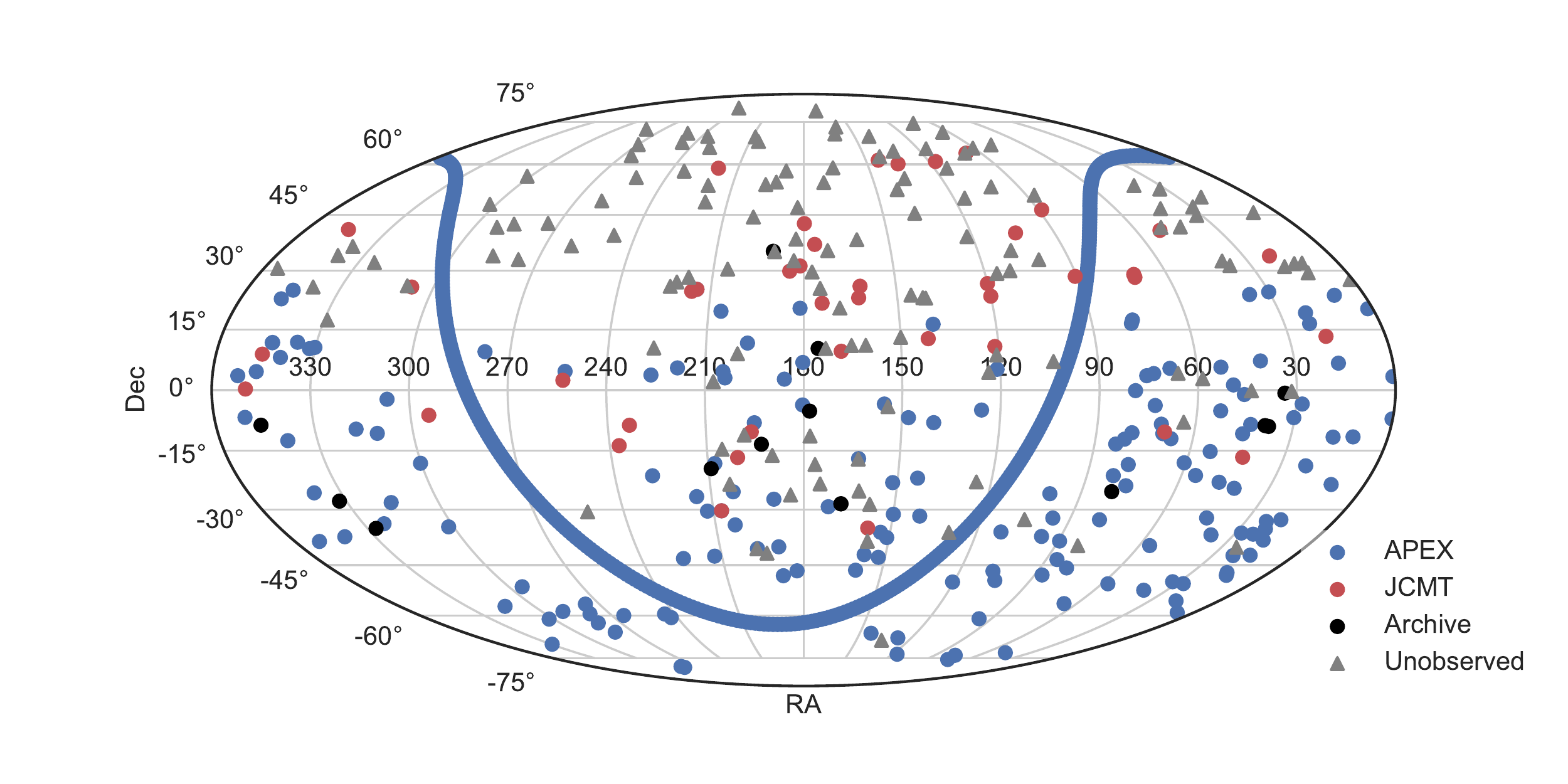}
\caption{An overview of the AGN galaxies drawn from the 70-month \swiftbat all-sky catalog relevant for our molecular gas observations (shown in equatorial coordinates and a Mollweide projection).  
Sources with CO(2--1) observations available either through our APEX or JCMT campaigns, or from the literature, are shown with blue, red, and black circles, respectively.  
Unobserved sources, mainly in the northern hemisphere, are shown in grey triangles.  
The Galactic plane is indicated by the light line.
}
\label{skycoverage}
\end{figure*}

\subsection{AGN Sample}

Our AGN parent sample consists of 836 ultra-hard X-ray selected (14--195 keV) AGN included in the 70-month \swiftbat all sky catalog \citep{Baumgartner:2013:19}. Of these we focused on \Nlowzall\ un-beamed AGN with low Galactic extinction [$E(B{-}V)<0.5$] and in a redshift range of $0.01{<}z{<}0.05$. Sources outside of the Galactic plane ($b > 10^{\circ}$) were given higher priority, but in 7 cases, sources with low Galactic extinction within the Galactic plane were observed. The lower redshift bound was chosen so that a single pointing would recover most of the CO flux from the host galaxy (see relevant angular scales below), while the higher redshift bound was chosen to guarantee a high detection rate within a reasonable integration time.
The results for the complementary sample of $z<0.01$ BAT AGN galaxies are presented in \cite{Rosario:2018:5658}. In total, \Nreduced\ AGN galaxies were newly observed --- \Napex\  with the Atacama Pathfinder Experiment (APEX) and 
\Njcmt\ with the James Clerk Maxwell Telescope (JCMT) --- 
while \Narchival\ were obtained from the literature.  
Figure \ref{skycoverage} gives an overview of the sky distribution of the \Nobs/\Nlowzall\ (\perobs) observed and the \Nunobserve/\Nlowzall\ (\perunobs) unobserved AGN galaxies between $0.01{<}z{<}0.05$.  A further study of BAT AGN galaxies in the northern hemisphere ($\delta > 0^{\circ}$), based on the IRAM 30m telescope, will appear in a future publication (Shimizu et al., in prep.).  An ALMA campaign on 32 BAT AGN galaxies in CO (2-1) has also recently been completed (Izumi et al., in prep).

	


\subsubsection{APEX Observations}
APEX 12 m antenna observations were taken for \Napex\ AGN galaxies, comprised of multiple 400-second-long integration frames, with a 100\arcsec\ throw between the ON and OFF-beam. The program totalled 254 hours with 2288 400-second-long scans, taken over 67 days between 2016 March and 2017 September.  Observing conditions had Precipitable Water Vapour (PWV) ranging from 0.4--4.5 mm with a median of 1.5 mm.  The observing programs involved were mainly an ESO Large program (PI M.\ Koss, $\sim$150 hrs), a follow-up ESO program (PI B.\ Trakthenbrot, $\sim$50 hrs), and Chilean time (PI E.\ Treister, $\sim$75 hrs).
In addition to our own programs, we also reduced data from archival programs for 6 BAT AGN galaxies, of which only the spectrum of IC 5063 was already published \citep{Morganti:2013:L4}.

We observed the CO(2--1) transition ($\nu_{\rm rest}$${=}$ 230.538 GHz) using the Swedish Heterodyne Faciliy Instrument  \citep[SHFI;][]{Vassilev:2008:1157} with the eXtended Fast Fourier Transform Spectrometer (XFFTS) backend (213--275\,GHz), tuned to the source-specific redshifted CO line frequencies.  At the observed frequencies ($\sim$220--229\,GHz), the APEX effective primary beam size is 26.3--27.5\arcsec half power beam width (HPBW), corresponding to scales between 6--27\,kpc at the distances of our sources.  Observations were terminated either after (1) a 5-$\sigma$ line detection was reached, (2) $\sim$3 hours of observations (including overheads; i.e., 40 minutes on-source), or (3) a sensitivity of 1\,mK rms {in the antenna temperature scale ($T_A^*$)} (corrected for atmospheric attenuation, the forward efficiency, and signal band gain) in a 50\,\kmps channel  was reached (whichever criterion came first).

The data were reduced using CLASS, a package of the GILDAS\footnote{http://www.iram.fr/IRAMFR/GILDAS} programming language. {Each 400-second-long scan} was examined by hand, and  were rejected if they had baseline ripples that clearly do not originate from astronomical sources; or strong artifacts at the edge of the spectral range ($\pm$1000 \kms).  On average, about 10\% of the individual scans of each target were discarded, and in extreme cases up to 50\% of the frames. 
Flux densities were derived assuming a constant conversion factor of 39 Jy beam$^{-1}$ K$^{-1}$ for the APEX telescope.\footnote{http://www.apex-telescope.org/telescope/efficiency/}

\subsubsection{JCMT Observations}
JCMT observations of the CO(2--1) molecular line were taken for \Njcmt\ AGN galaxies, between 2011 February and 2013 April.  Archival data for one additional galaxy was also reduced (NGC 6240). 
We used the A3 (211--279 GHz) receiver with the ACSIS spectrometer with a beam size of 20.4$\arcsec$ HPBW.  Each galaxy was initially observed for 30 minutes. For weak detections, additional observations were obtained, but totalling no more than 2 hours per target.  The individual scans for each target were first-order baseline-subtracted and then co-added. The narrow spectral coverage of the ACSIS ($\sim1300$ \kmps) does not cover enough line-free regions to enable an estimation of the spectral noise, and therefore we added a 10\% error to the uncertainty on the line fluxes to account for any spectral baselining uncertainties \citep[e.g.,][]{Rosario:2018:5658}. The RxA3 observations took place in weather band 5 ($\tau_{225\,\text{GHz}} = 0.20$--0.32).  Flux densities were derived assuming a constant conversion factor of 28 Jy beam$^{-1}$ K$^{-1}$ (i.e., an aperture efficiency of 0.55).

\subsubsection{Literature Measurements}

We incorporate \Narchival\ literature CO line measurements of AGN galaxies from our parent BAT AGN sample, and which we did not re-observe (see Table \ref{COtab}). 
This includes two CO(2--1) observations from the 15-m Swedish ESO Sub-millimetre Telescope (SEST; HPBW=23\arcsec\ at 230 GHz) published in \cite{Strong:2004:1151}, two from the JCMT published in \cite{Papadopoulos:2012:2601}, and five from the Caltech Submillimeter Observatory (CSO; HPBW=32\arcsec) published in \cite{Monje:2011:23}. Finally, we also include four AGN galaxies from IRAM which were previously published \citep{Bertram:2007:571}, but we use the CO(1--0) line (HPBW=22\arcsec) to avoid the smaller beamwidth associated with the CO(2--1) line.


\section{Measurements}
\label{sec:measurements}

The principal analysis steps following the data reduction involve integrated measurements of the CO(2--1) line for each AGN galaxy in our sample, as well as line measurements using simple (Gaussian) profile fitting, to derive CO line properties.  These spectral measurements are performed consistently for all \Nreduced\ AGN galaxy observed with either APEX or JCMT.
For the \Narchival\ literature measurements, we have only applied beam corrections to the integrated measurements (see \S\ref{subsec:beam_corr}).
The new CO measurements are complemented with the rich collection of ancillary data available for our BAT AGN through BASS (\S\ref{othermeas}).
We finally describe the sample of inactive galaxies we use for comparison (\S\ref{subsec:comp_sample}).

\subsection{Integrated Measurements}
\label{subsec:int_measure}

\begin{figure*}
\centering
\raggedright
\includegraphics[height=2.2in]{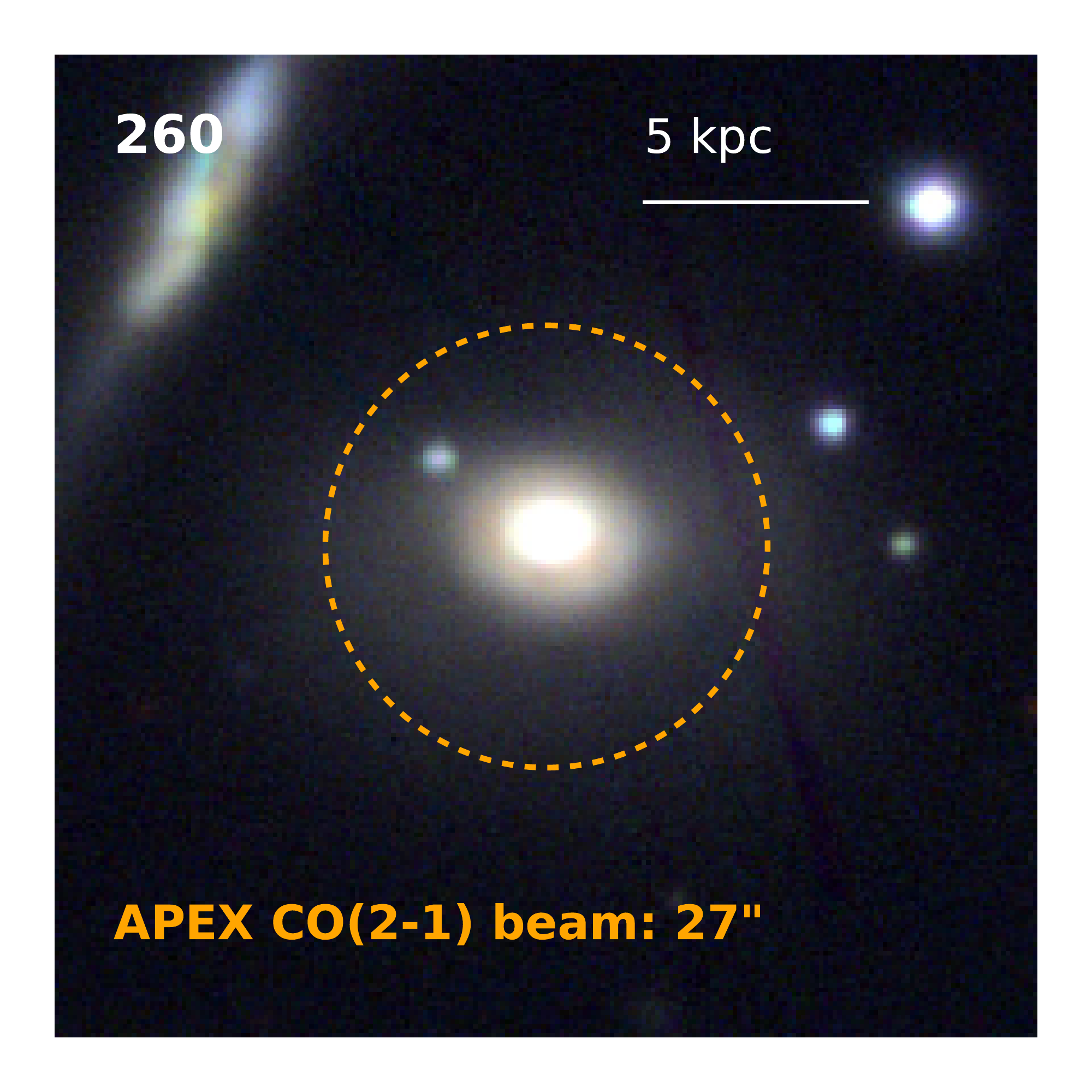}\includegraphics[height=2.2in]{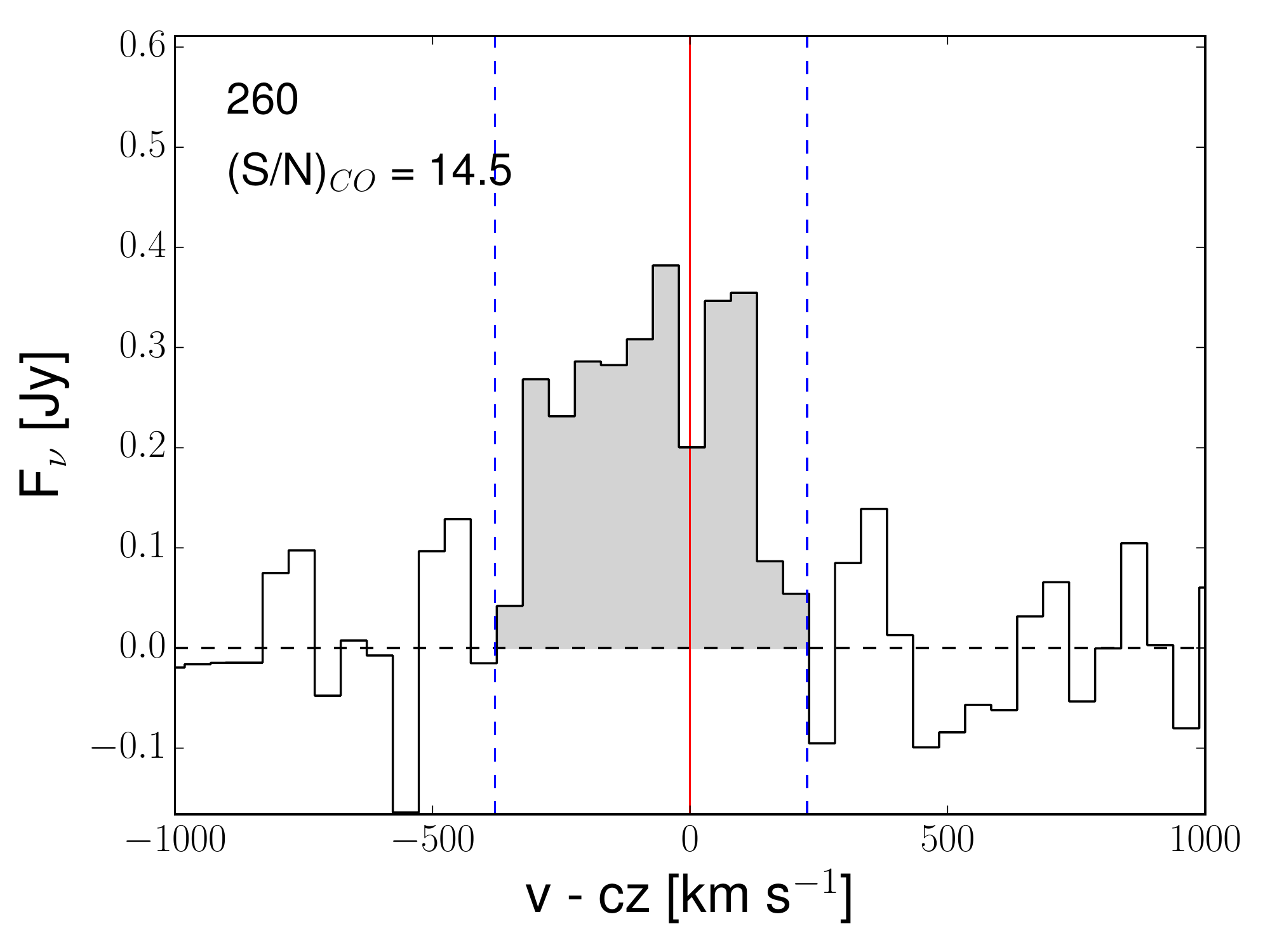}
\includegraphics[height=2.2in]{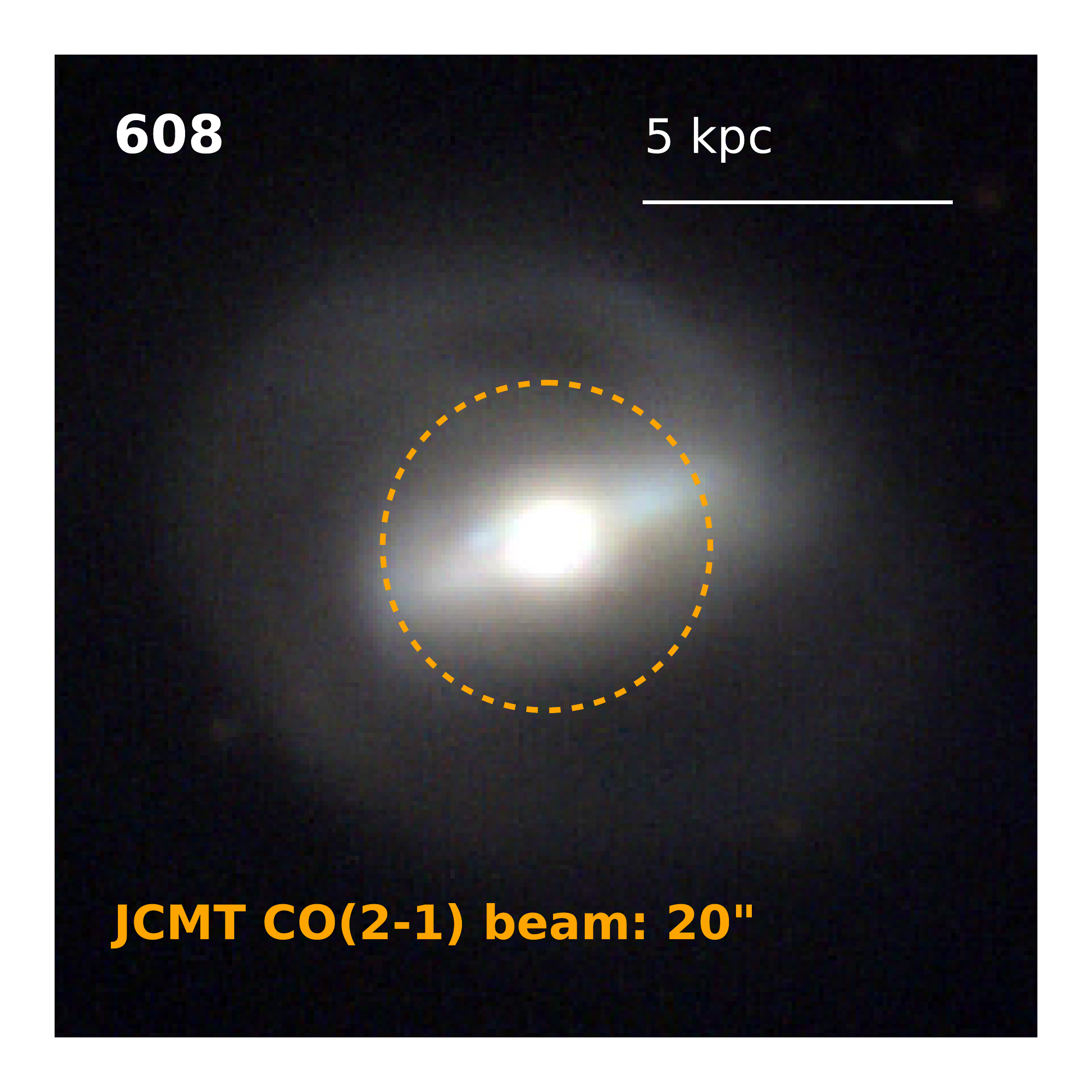}
\includegraphics[height=2.2in]{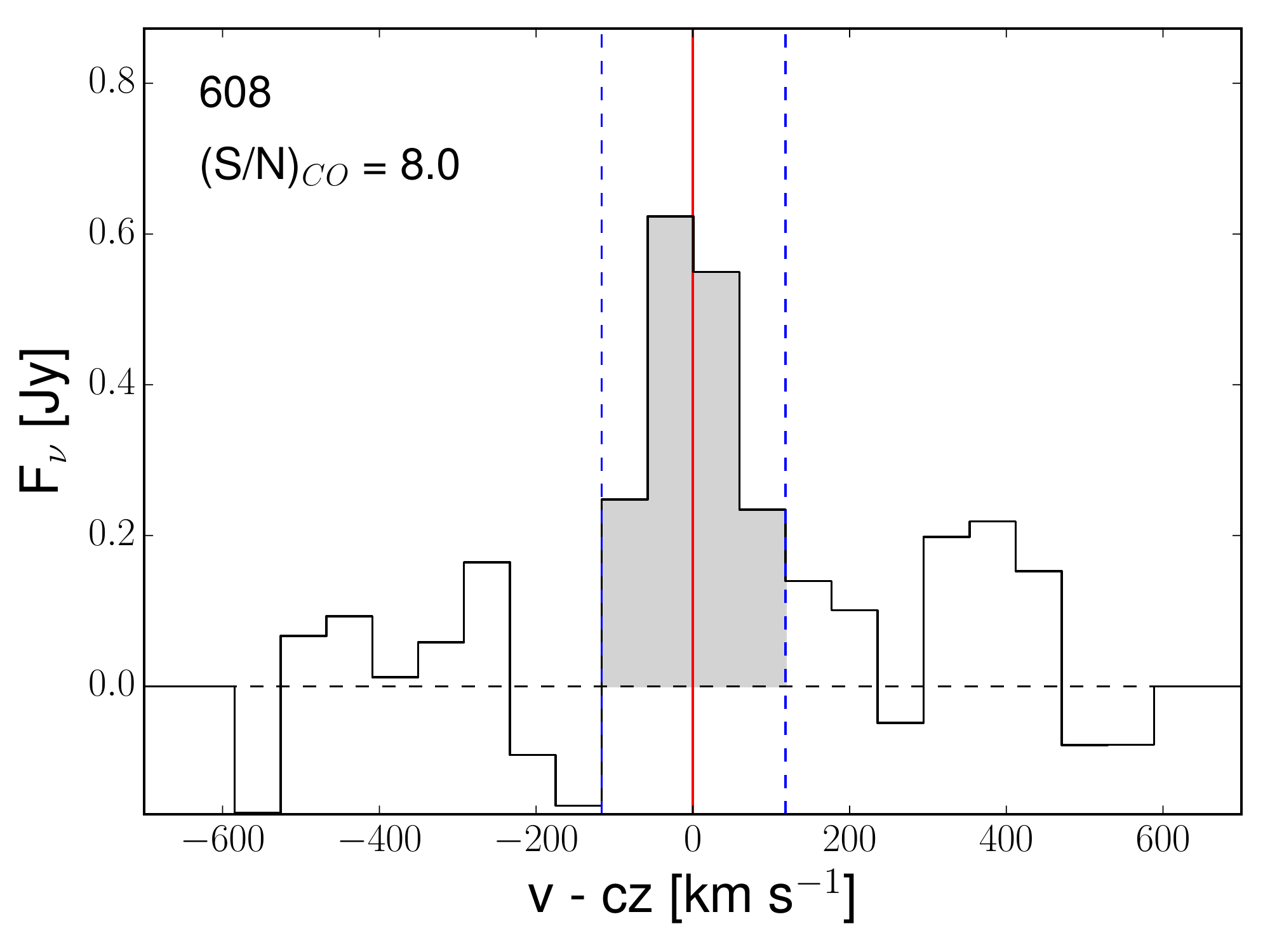}\\
\includegraphics[height=2.2in]{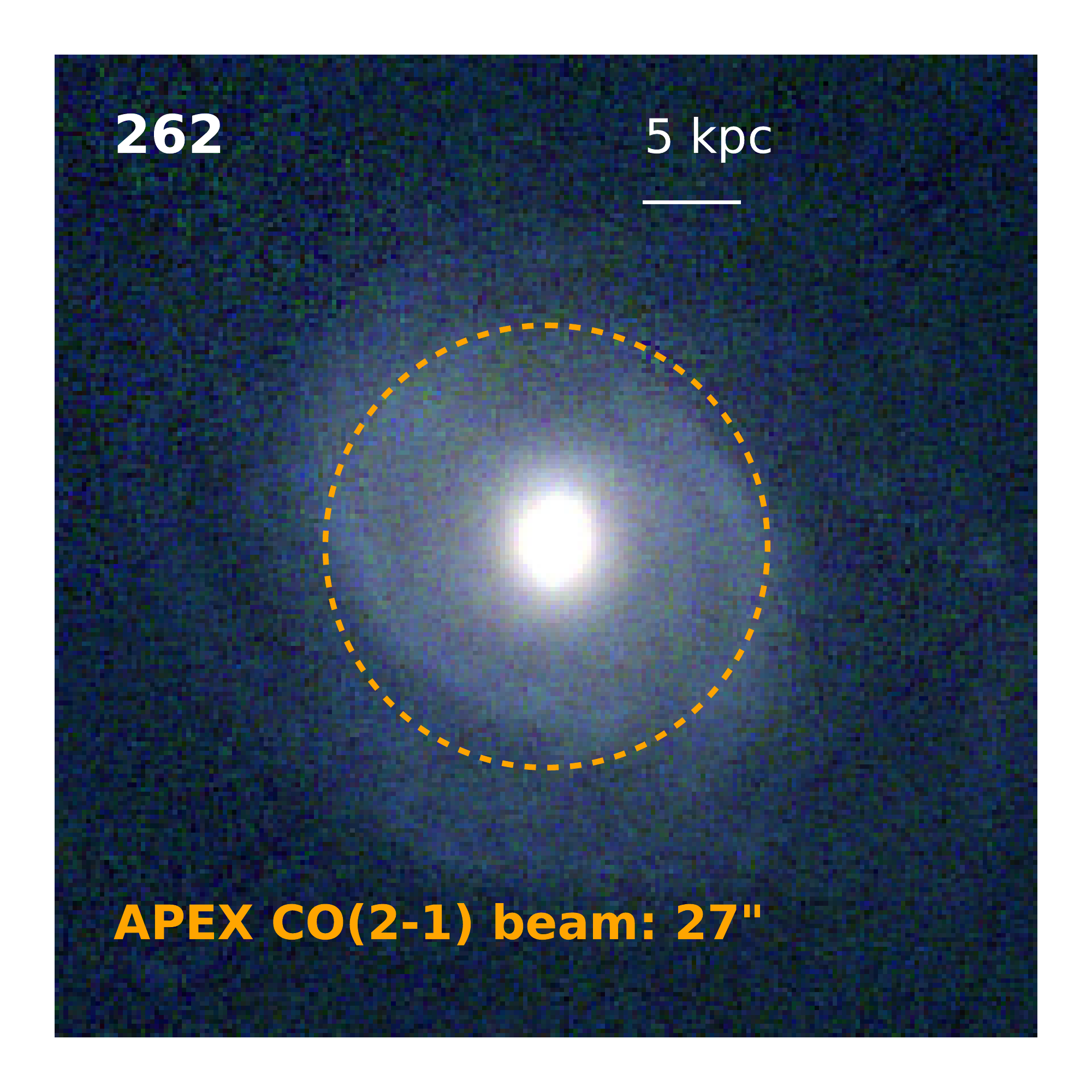}\includegraphics[height=2.2in]{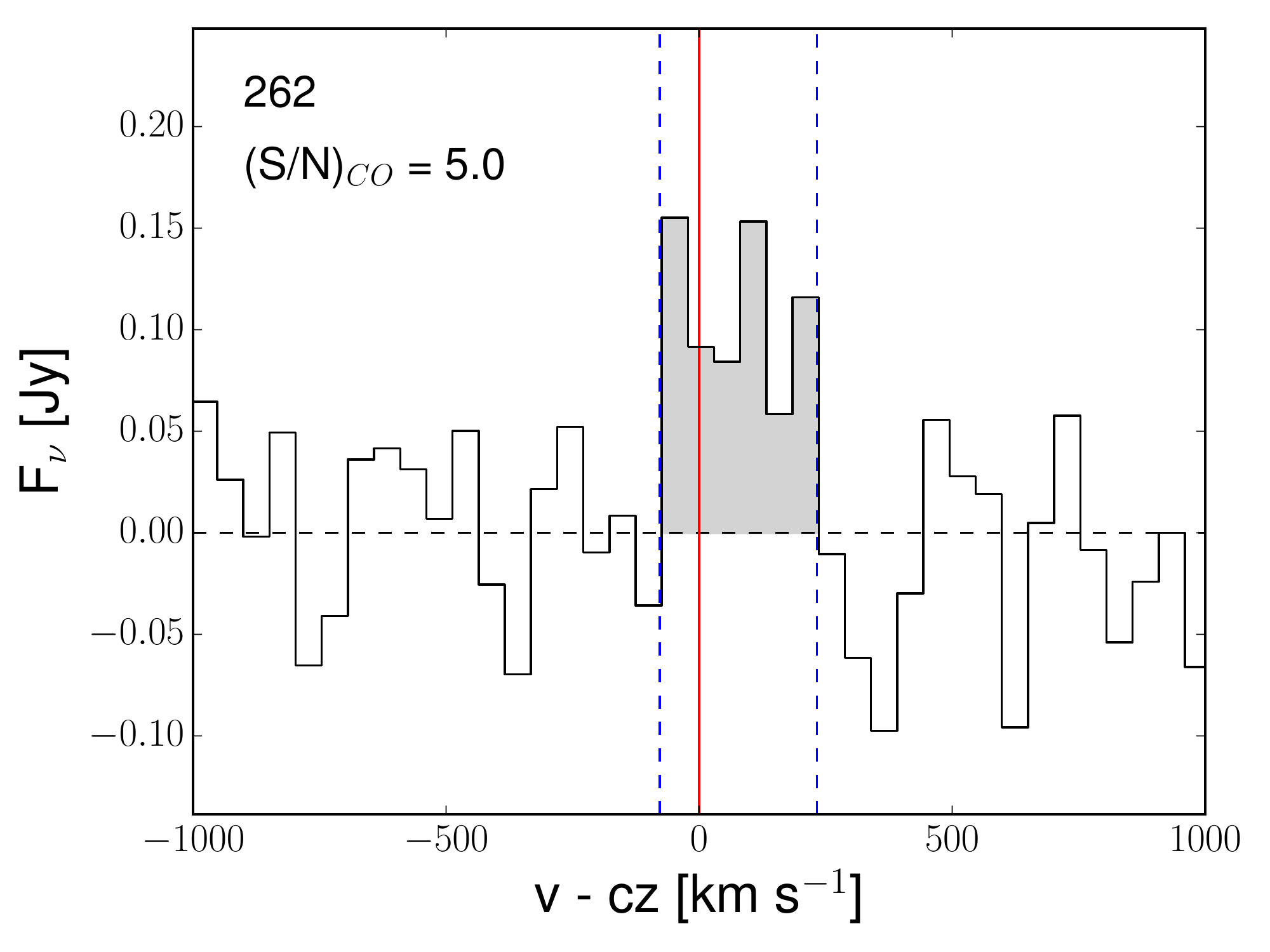}

\caption{Examples of CO(2--1) detections for three of our BAT AGN galaxies.
\textit{Left}: Pan-STARRS $1'\times1'$ $gri$ color cutouts of the three AGN galaxies, with the beamsize of the CO observations marked in orange.  
\textit{Right}: CO(2--1) spectra of the same BAT AGN galaxy.  
The spectra are centered at the position of the CO(2--1) line. Solid red lines mark the central velocity of the optical redshift of each AGN or HI of the host galaxy which was set at the central velocity ($cz$), while the dashed blue lines indicate the velocity range within which we have integrated the CO(2--1) line fluxes.  
The three AGN galaxies shown here illustrate CO signal-to-noise ratios typical of the top quartile, median, and bottom quartile of the sample of 151 detections (from top to bottom; $S/N=14.5, 8.0$, and $5.0$, respectively). 
The figure set including all detections and non-detections (\Nreduced\ images and all associated CO(2--1) spectra) is available in the online version of this paper.
} 
\label{fig:CO21_spectra}
\end{figure*}

\begin{figure*}
\centering
\raggedright
\includegraphics[height=2.2in]{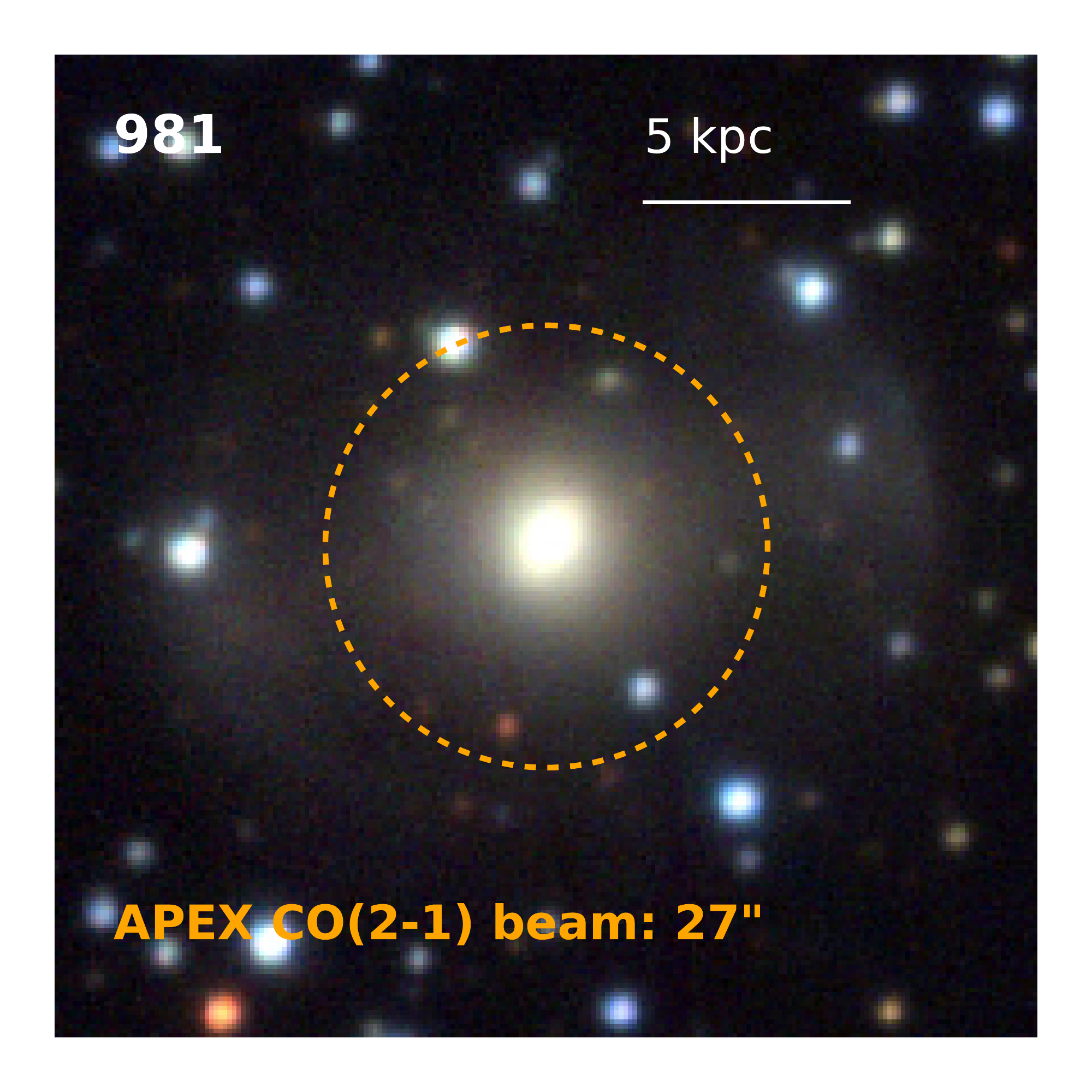}\includegraphics[height=2.2in]{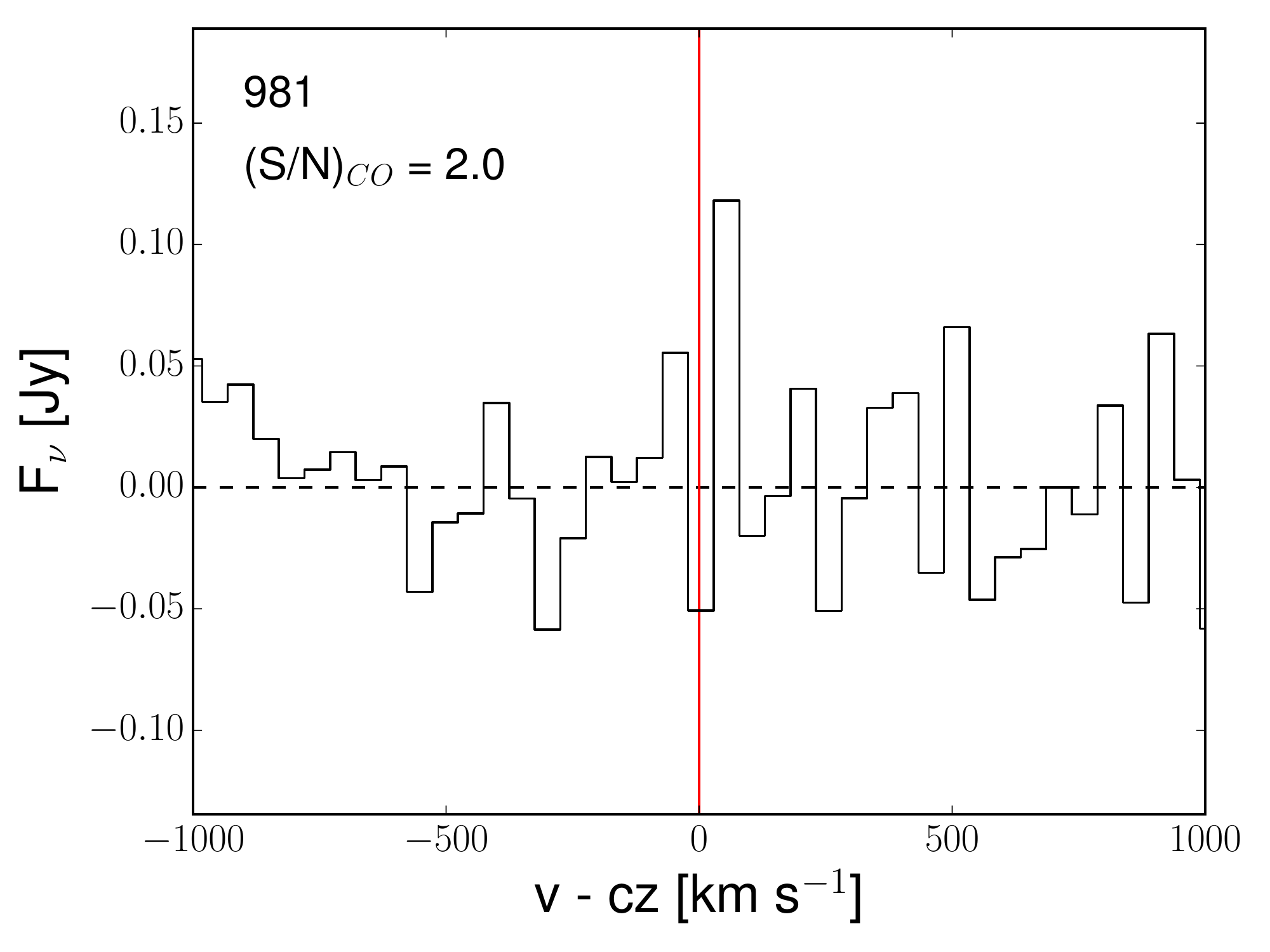}
\includegraphics[height=2.2in]{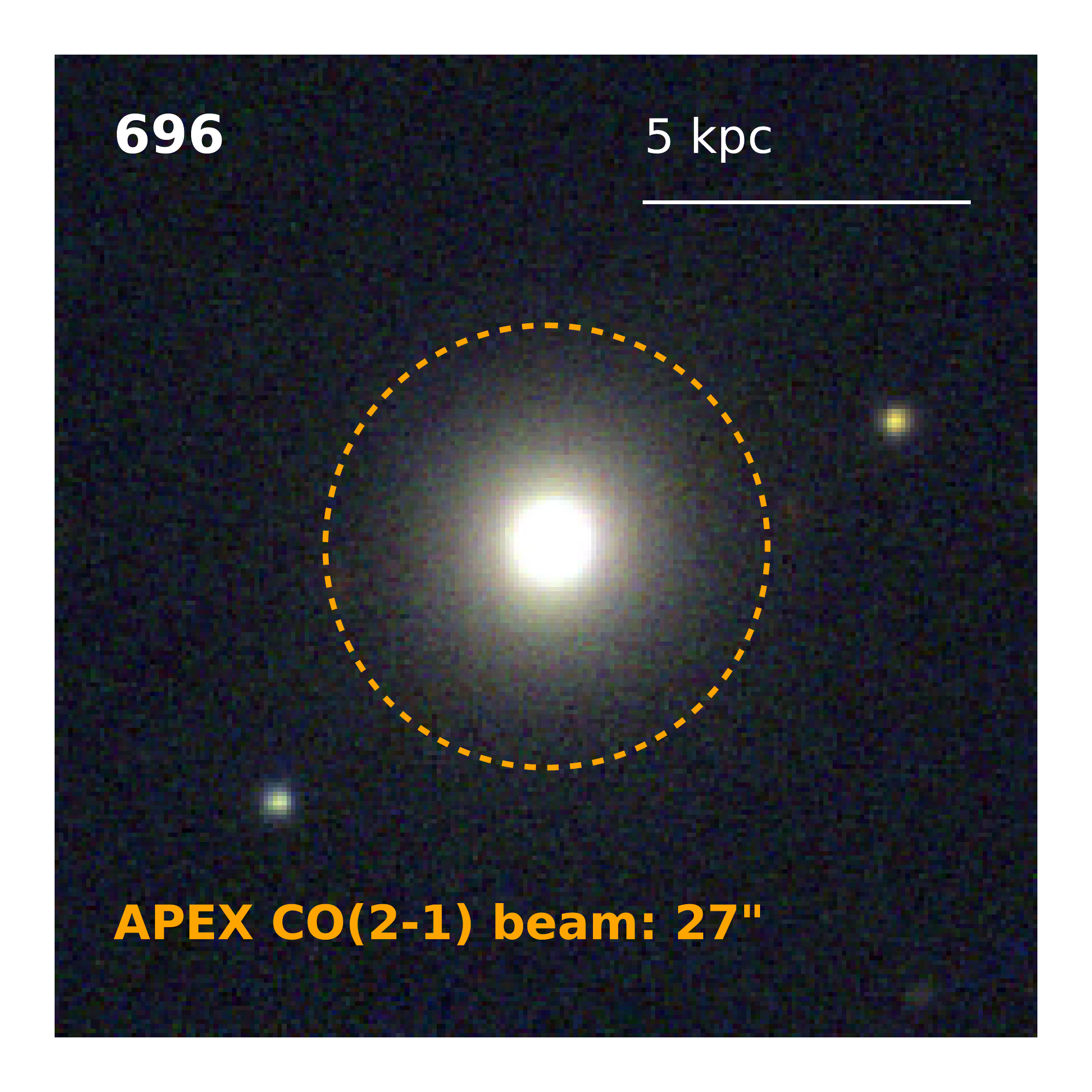}\includegraphics[height=2.2in]{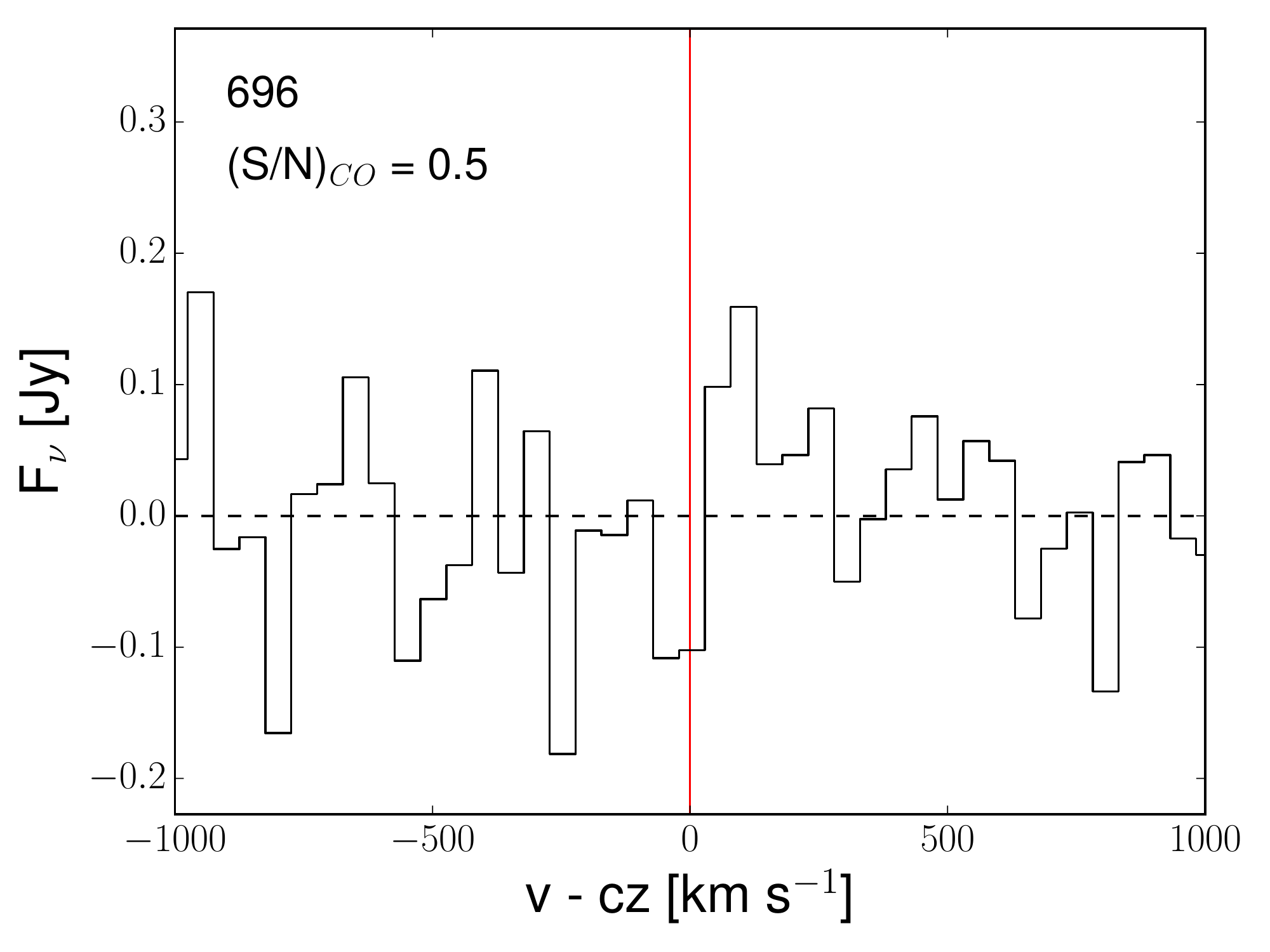}
\includegraphics[height=2.2in]{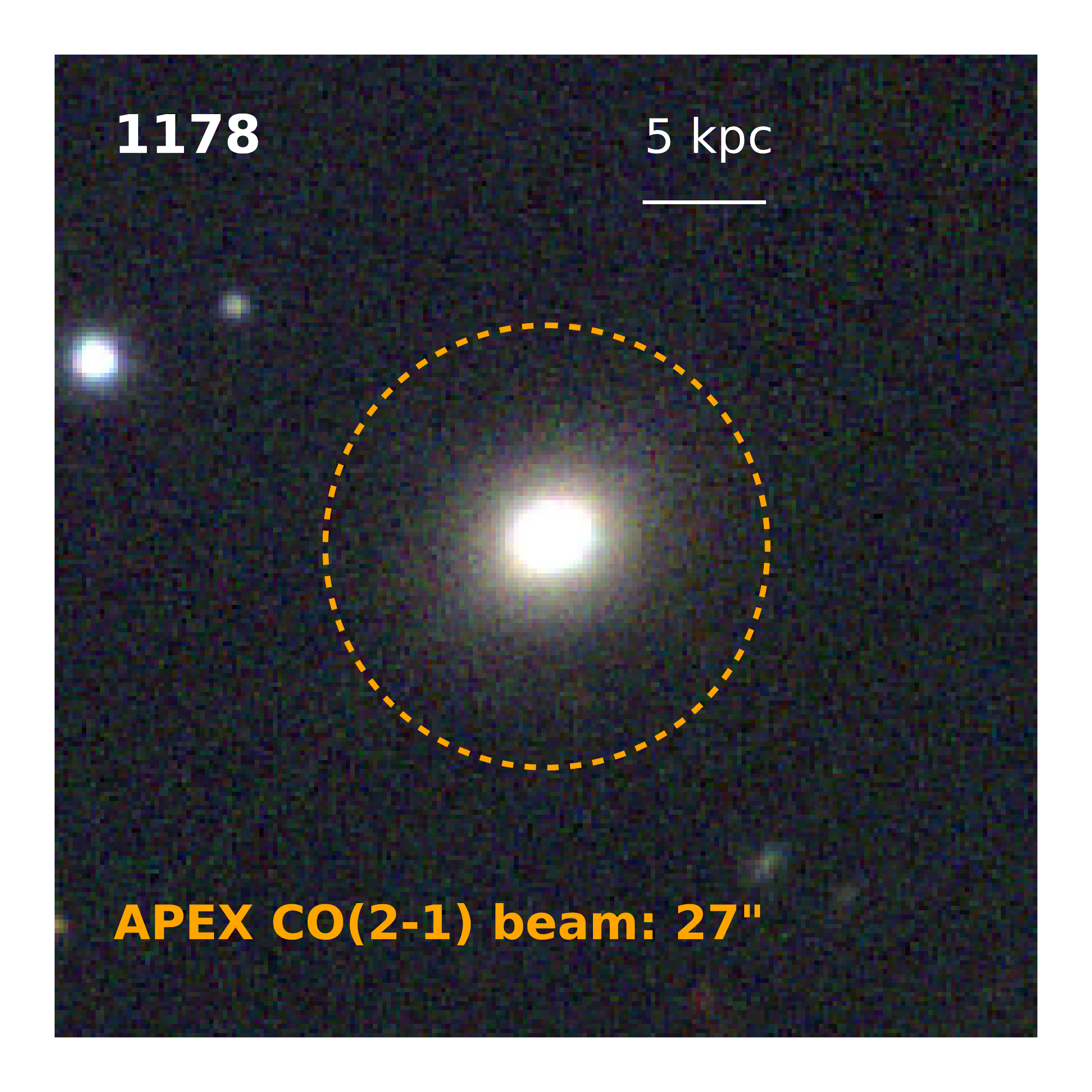}\includegraphics[height=2.2in]{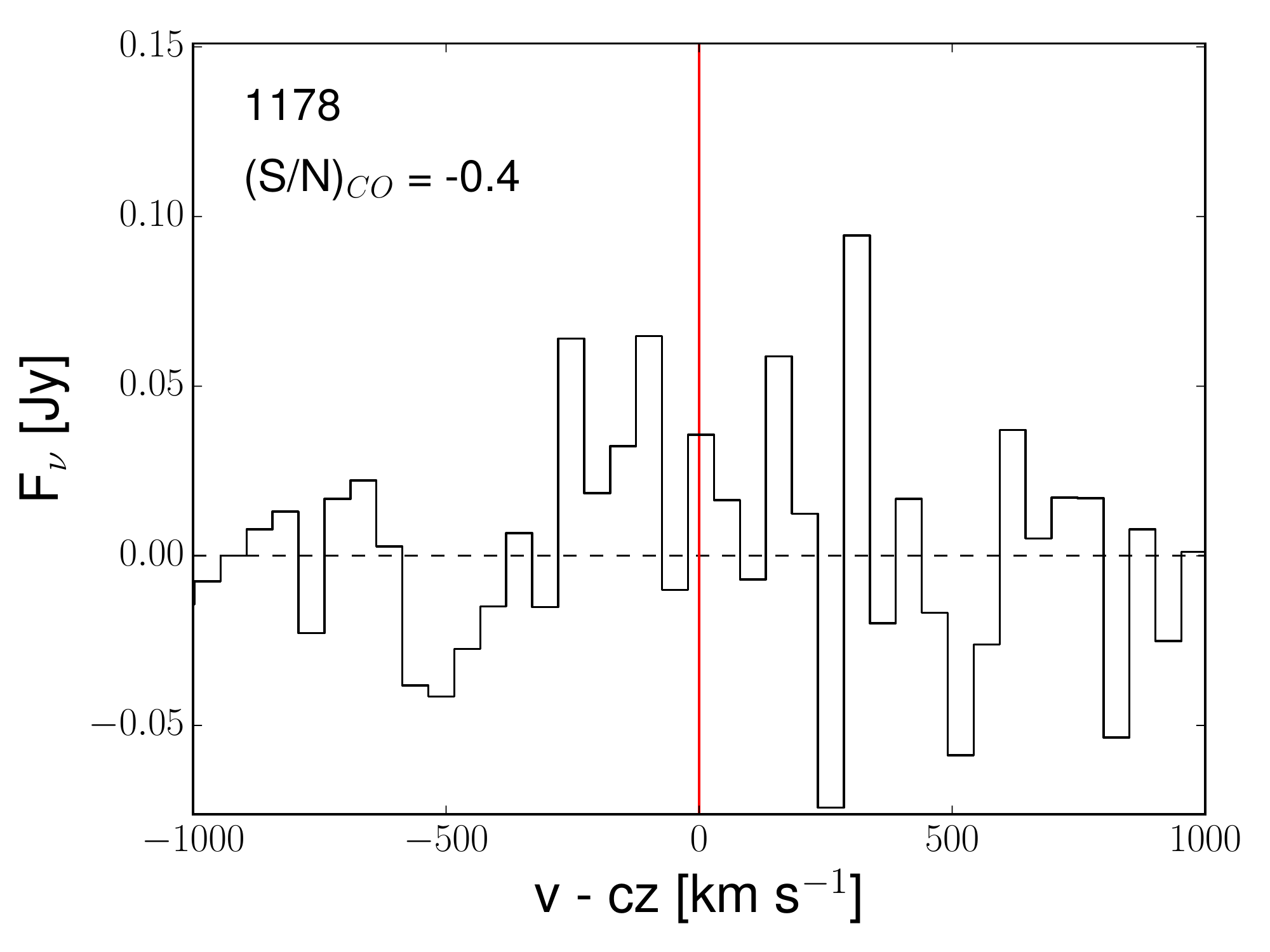}
\caption{Similar to Figure \ref{fig:CO21_spectra}, but for {\it non}-detections of the CO(2--1) line, for three of our BAT AGN galaxies.
Here, too, the examples illustrate the typical $S/N$ in the top quartile, median, and bottom quartile among the 63 AGN galaxies with CO non-detections ($S/N=2$, $0.5$, and $-0.4$, respectively).
} 
\label{fig:CO21_spectra_undetet}
\end{figure*}

We measure the velocity-integrated line flux $S_{\rm CO}$ (in units of Jy km s$^{-1}$) from the reduced spectra by summing the signal within a spectral range defined by hand, following the technique in xCOLD GASS \citep{Saintonge:2017:22}.  If the CO(2--1) line is undetected or very weak, the window is set to a width of 300 \kmps to estimate uniform upper limits.  
The CO line luminosities are then calculated following the relation
\begin{equation}
L'_{\text{CO(2--1)}}= 3.25 \times 10^7  S_{\text{CO(1--0)}} \nu_{obs}^{-2} D_L^2 (1+z)^{-3},
\end{equation}
as given in \cite{Solomon:1997:144}, where the line luminosity $L'_{\rm CO}$ (in K km s$^{-1}$) is derived from the velocity integrated line flux, $S_{\rm CO} \Delta v$ (in Jy km s$^{-1}$) and the luminosity distance, $D_{L}$ (in Mpc). 
The error on the line flux is defined as the rms noise achieved around the CO(2--1) line, in spectral channels with width $\Delta w_{\rm ch}^{-1} = 50$ \kms, and $W_{\rm CO}$ is the width of the CO(2--1) line in \kms. 
The detection threshold is set to $S/N{=}3$.  
We provide three examples of CO line detections and non-detections in Figures \ref{fig:CO21_spectra} and \ref{fig:CO21_spectra_undetet}, respectively.  {For 15 sources characterized by particularly narrow CO emission, we produced a higher resolution spectrum to optimize the profile fit using either 5, 10, or 20 km/s binning.  The complete sample of CO line measurements is found in Appendix \ref{allspec_appen}.}

\subsection{Line Fitting}
\label{subsec:line_fitting}
We have additionally measured the line widths (\texttt{W50$_{\rm CO21}$}) and systemic velocities using Gaussian profile fitting for the detected sources, following the same approach as used in the xCOLD GASS studies \citep{Tiley:2016:3494}. 
Each reduced spectrum is first fitted with a single Gaussian through Monte Carlo minimization, using the \sherpa Python package in \ciao 4.11.  
As discussed in \citet{Tiley:2016:3494}, a symmetric Gaussian Double Peak function may be more appropriate for higher $S/N$ spectra, as it minimises potential biases as a function of $S/N$, inclination (apparent width) and rotation velocity (intrinsic width). We therefore also fit each spectrum with a Gaussian Double Peak function (a parabolic function surrounded by two equidistant and identical half-Gaussians forming the low and high velocity edges of the profile),  which reduces to a single Gaussian in the limit where the half-width of the central parabola is zero.  We report the results obtained from the fitting method which provides the lowest $\left| \chi^2_\nu \right|$ (where $\chi^2_\nu$ is the $\chi^2$ of the fit divided by the degrees of freedom).  
We show three examples of line fits in Figure \ref{fig:CO21_profile}.  We find good agreement between profile fitting and integrated measures (see Appendix \ref{profile_fitting_appen}).

\begin{figure*}
\centering
\raggedright
\includegraphics[height=1.7in]{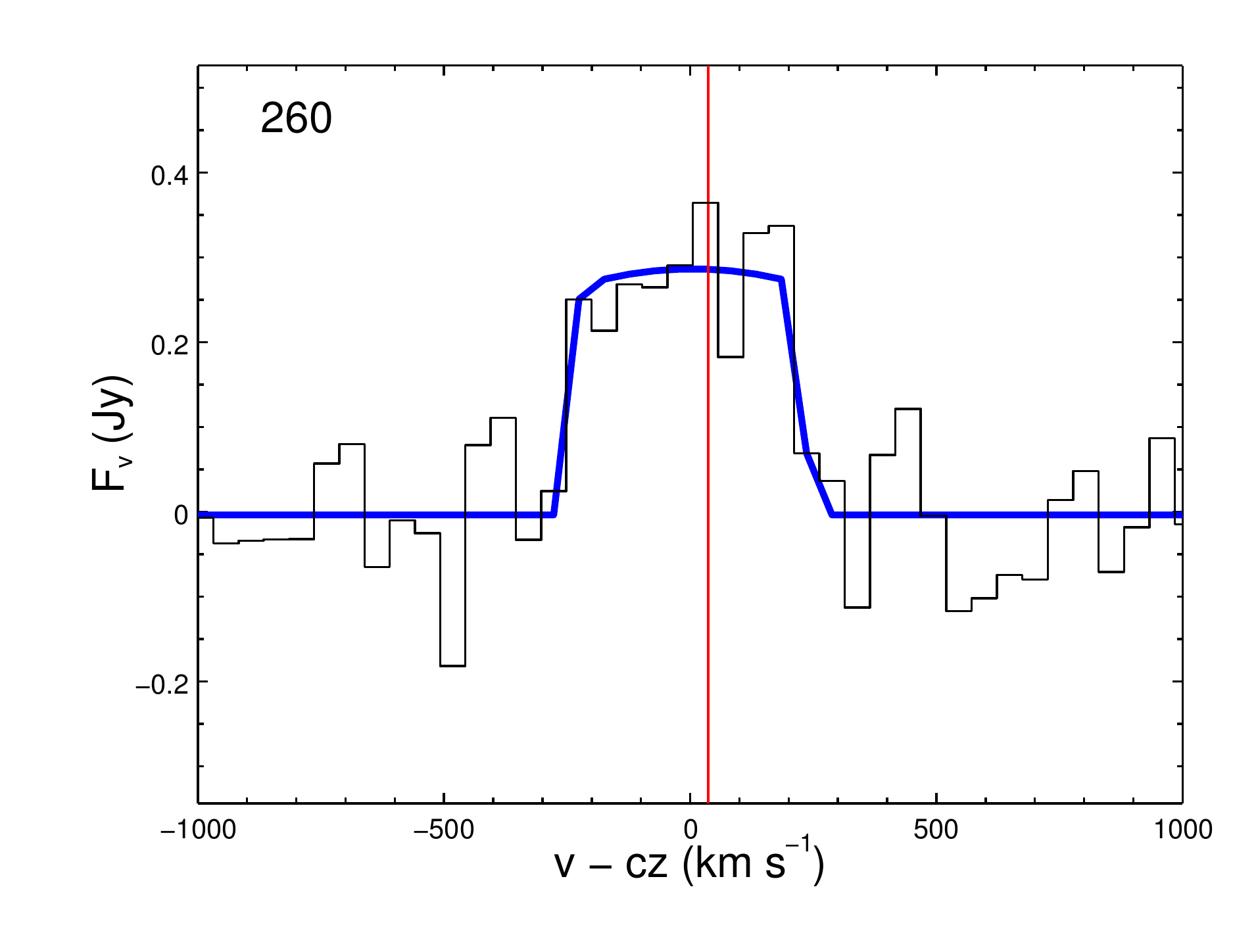}
\includegraphics[height=1.7in]{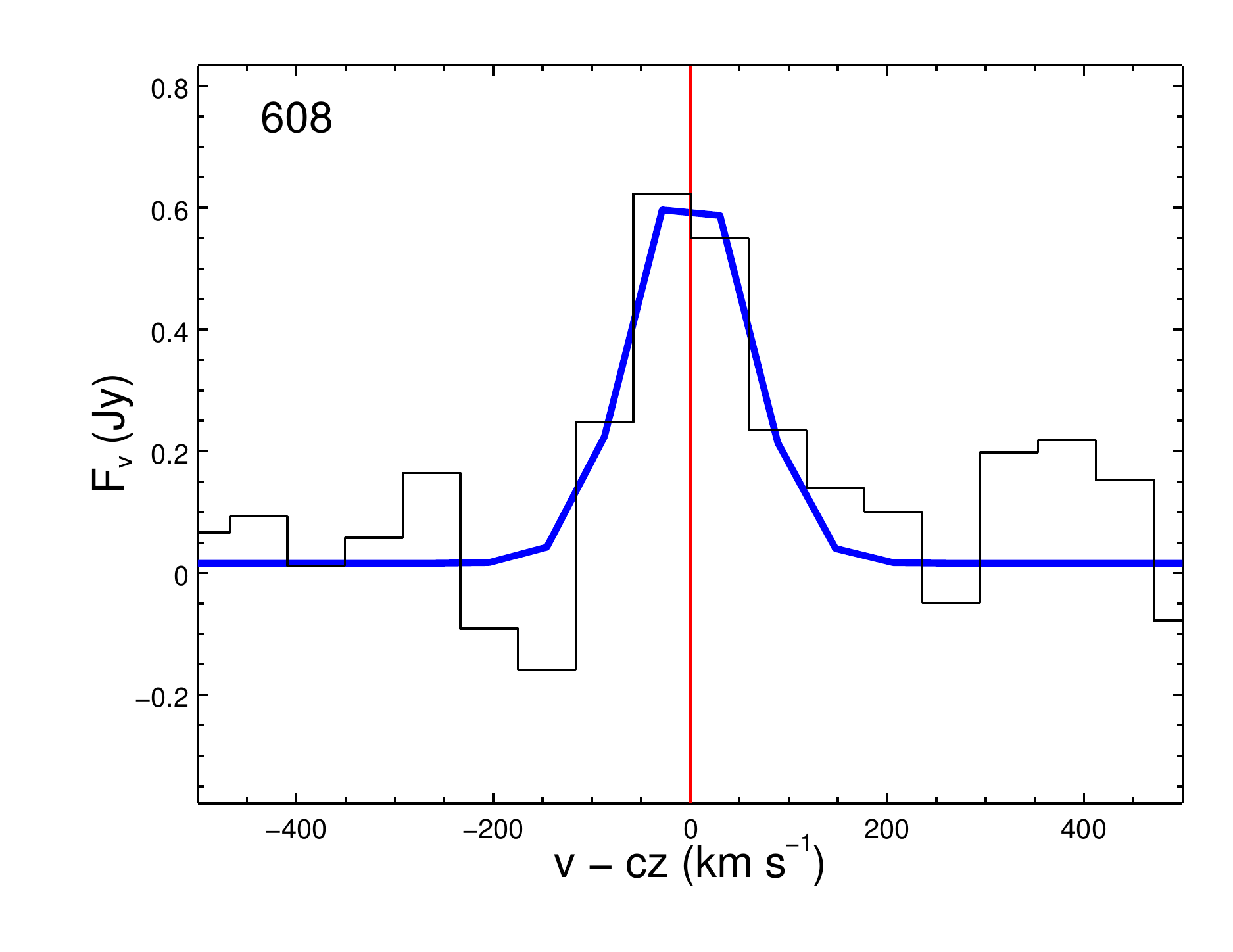}
\includegraphics[height=1.7in]{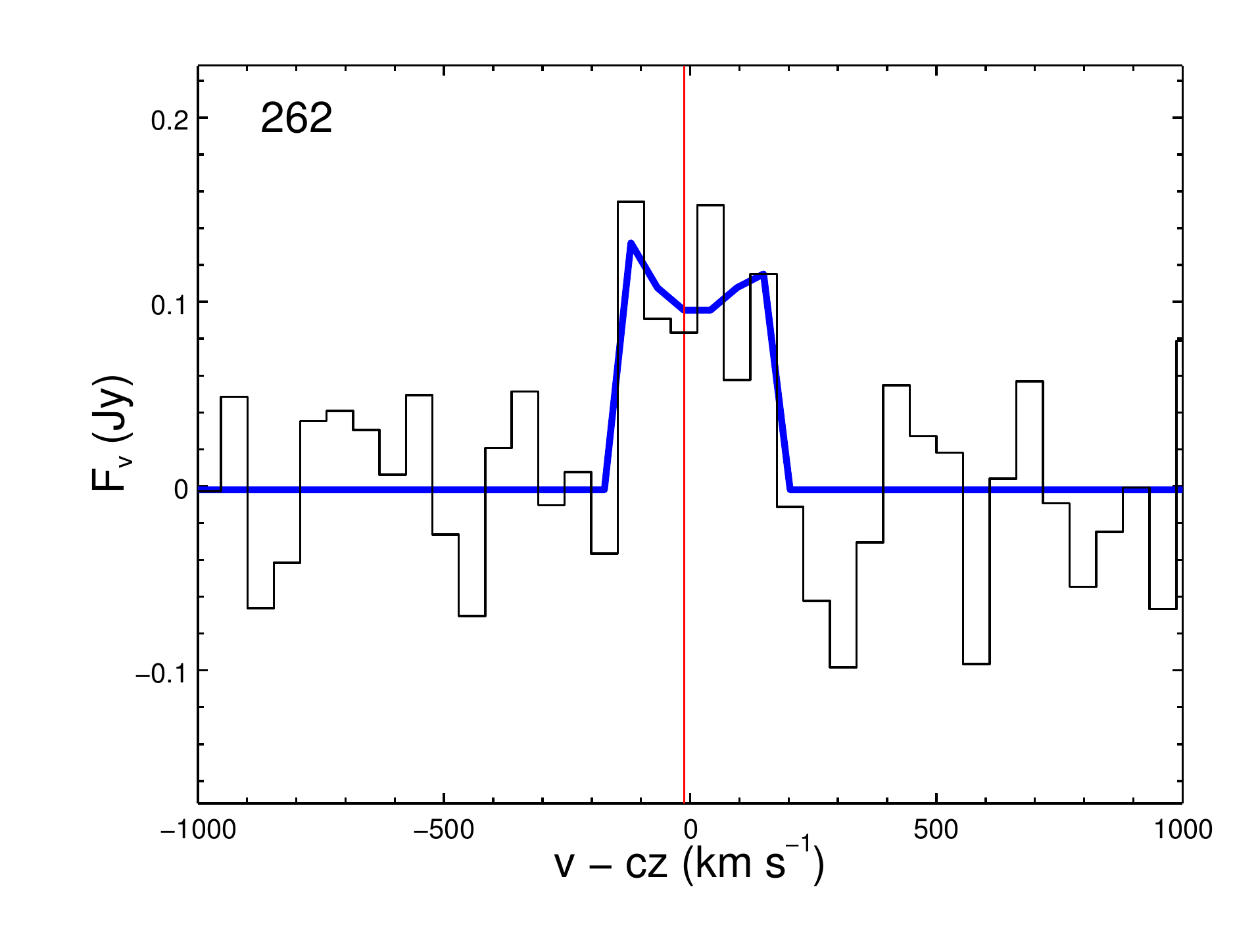}
\caption{CO(2--1) line profile fits to detected BAT AGN galaxies for the same sources shown in Figure \ref{fig:CO21_spectra},  representing the top quartile, median, and bottom quartile of $S/N$ among our detected sample. BAT AGN galaxy IDs 260 (left) and 262 (right) are best fit with a Gaussian Double Peak function (a parabolic function surrounded by two equidistant and identical half-Gaussians forming the low and high velocity edges of the profile), while BAT AGN galaxy ID~608 (center) is best fit with a single Gaussian.  
The vertical red lines indicate the central velocity of the CO(2--1) line.
} 
\label{fig:CO21_profile}
\end{figure*}

\subsection{Ancillary AGN and host galaxy properties from BASS}
\label{othermeas}

We complement the new CO measurements with various galaxy and AGN physical properties, drawn from BASS \citep{Koss:2017:74} --- a large effort to collect optical and X-ray spectra \citep{Ricci:2017:17} and various other measurements for \swiftbat AGN, in order to assess black hole masses, accretion rates, and bolometric luminosity estimates for an unbiased local AGN sample.  
Optical spectra can be accessed through the BASS website.\footnote{http://www.bass-survey.com} We incorporate products from both Data Releases 1 (DR1) and 2 (DR2); DR1 is currently publicly available, while DR2 will be published in the near future (Koss et al., in prep). 
For all the BAT AGN galaxies in our sample, we use the following key parameters (see Appendix \ref{parameters_appen} for a further discussion of the uncertainties in parameter estimation), listed in Table \ref{sampleparams} or in associated BASS publications:

\begin{enumerate}

\item \texttt{BAT ID:} Catalog ID in the BAT survey.\footnote{https://swift.gsfc.nasa.gov/results/bs70mon/}

\item \texttt{$\alpha_{\rm J2000}$} and \texttt{$\delta_{\rm J2000}$}: Right ascension and declination of the optical/IR counterpart of the BAT AGN, in decimal degrees, based on WISE positions.

\item \texttt{AGN type} and \texttt{Redshift}: Measurements from BASS/DR1 \citep{Koss:2017:74} and DR2 (Oh et al., in prep), when available.  The redshift is based on optical emission-line fitting (mainly of \OIII).  While the emission line redshifts may have an offset from the systemic host velocities \citep{Rojas:2020:5867}, they provide a rough estimate of the likely CO centroid.

\item \texttt{rk20fe}: The isophotal radius at a surface brightness of 20 mag arcsec$^{-2}$ in the $K$-band, taken from the 2MASS extended source catalog \citep{Jarrett:2000:2498}.

\item \texttt{Inclination angle}:  Galaxy inclination angle based on $r$-band optical imaging from the SDSS, Kitt Peak \citep{Koss:2011:57},  or from the  2MASS extended source catalog \citep{Jarrett:2000:2498}.  
In order to determine the inclination from the observed ratio of the semi-minor to semi-major axes of the galaxy, we use the prescription of \cite{Davis:2011:968} for both early and late-type galaxies.  

\item \texttt{Morphology}: Morphological classification from Galaxy Zoo \citep{Lintott:2008:1179}, when available, followed by the Third Reference Catalog of Bright Galaxies \citep[RC3,][]{deVaucouleurs:1995}, and finally visual inspection following the approach of \citet{Koss:2011:57}, which is based on optical imaging.  
Values can be `elliptical', `spiral', or `uncertain'. See Appendix \ref{morph_appen} for a further discussion of uncertainty in morphology measures.

\item \texttt{Stellar mass}: Stellar masses of the BAT AGN host galaxies. We combined near-IR data from 2MASS, which is more sensitive to stellar emission, with mid-IR data from the AllWISE catalog \citep{Wright:2010:1868}, which is more sensitive to AGN emission (i.e., re-processed by the circumnuclear dusty gas). 
See \citet{Powell:2018:110} for a further description of these measurements.

\item \texttt{Star formation rate}: The majority (80\%, \ntarosfr/\Nobs) of SFR estimates were derived from the \herschelsh-based study of BAT AGN galaxies by \cite{Shimizu:2017:3161}, which measured the total IR (8--1000 \micron) luminosities by decomposing the IR SEDs into AGN- and galaxy-related components using \wise (3.4, 4.6, 12, and 22 \micron), \herschel PACS (70 and 160 \micron) and SPIRE (250, 350, and 500 \micron) imaging.  A subset of 19\% (40/\Nobs) AGN galaxies were not part of this \herschel imaging publication. For these we adopt the results of \cite{Ichikawa:2019:31}, who performed a similar IR SED decomposition but included any available \wisesh, \akarish, \irassh, and \herschel measurements. While both studies have similar values, we preferentially use the \herschelsh-based measurements of \cite{Shimizu:2017:3161} to avoid the much larger PSF of the \iras data compared to the CO. Ultimately, 15/\Nobs\  (7\%) of the SED estimates are only upper limits because no FIR detection ($>$60 \micron) was found.  Additionally, 3/\Nobs\ (1\%) sources lie within 10$^{\circ}$ of the Galactic plane and were not part of either study.  {See Appendix \ref{parameters_appen} for a further discussion about uncertainties and offsets in SFR measurements.}

\item \texttt{Offset from the Main Sequence of star-formation}:  A correlation between the stellar mass and SFR has been found for galaxies in the local and distant universe \citep[see review in e.g.,][]{Tacconi:2020:arXiv:2003.06245}.  The offset, \deltams, is defined as the difference on a logarithmic scale between the observed SFR of a galaxy and it's expected SFR in relation to the MS from \citet{Renzini:2015:L29}.  A value of \deltams$=1$ would indicate a galaxy is elevated above the expected value of SFR on the MS by a factor of 10, while \deltams$=-1$ indicates a factor of 10 lower than expected.


\end{enumerate}

\begin{table}
\centering
\small{}
\caption{Multiwavelength properties of the BAT AGN Galaxy sample.}
\begin{tabular}{llccccccccccc}
\hline
ID & Galaxy&$\alpha_{\rm J2000}$&$\delta_{\rm J2000}$&AGN Type&$z_{spec}$&rk20fe&Incl&Morph&log M$_*$&SFR&\deltams\\   && [deg] & [deg] &&  &  [\arcsec]& [deg]&  &[log M$_{\odot}$]& [M$_{\odot}$ yr$^{-1}$]& \\
  \hline \hline
3&NGC7811&0.6101&3.35191&Sy1.5&0.033&14.8&41&spiral&10.64&4.68&0.45\\
17&ESO112-6&7.68262&-59.00721&Sy2&0.0325&15.5&61&spiral&10.56&7.17&0.66\\
28&NGC235A&10.72004&-23.54104&Sy1.9&0.0158&17.5&38&spiral&10.9&8.05&0.61\\
44&ESO195-IG021&15.14568&-47.86772&Sy2&0.037&10.3&66&uncertain&10.95&9.48&0.68\\
50&ESO243-G026&16.40833&-47.07168&Sy2&0.0168&34.8&78&spiral&10.39&2.21&0.21\\
58&NGC424&17.86516&-38.08345&Sy1.9&0.0163&35.1&83&spiral&10.49&4.72&0.5\\
62&IC1657&18.52921&-32.65088&Sy2&0.0262&46.6&80&spiral&10.62&3.1&0.27\\
63&NGC454E&18.60391&-55.39704&Sy2&0.0164&35.2&77&uncertain&10.45&1.57&0.04\\
72&NGC526A&20.97654&-35.06544&Sy1.9&0.0169&13.9&57&elliptical&10.48&2.23&0.18\\
77&Mrk359&21.88548&19.17883&Sy1.5&0.0136&16.6&52&spiral&10.5&3.23&0.33\\
79&CGCG459-058&22.10183&16.45932&Sy2&0.0179&13.2&77&spiral&10.83&11.6&0.79\\
81&ESO244-IG030&22.46323&-42.32645&Sy2&0.0294&21.4&78&uncertain&10.66&6.25&0.57\\
83&ESO353-G009&22.96001&-33.11929&Sy2&0.0173&24.3&57&uncertain&10.68&6.13&0.55\\
95&ESO354-G004&27.92441&-36.18782&Sy1&0.0202&16.5&40&spiral&10.94&3.45&0.24\\
96&MCG-01-05-047&28.20432&-3.44684&Sy2&0.0249&66.7&89&spiral&10.88&11.65&0.78\\
101&UGC01479&30.07939&24.47383&Sy2&0.0129&39.1&80&spiral&10.6&3.22&0.3\\
102&NGC788&30.27691&-6.81588&Sy2&0.0127&37.6&42&elliptical&10.89&1.74&-0.04\\
114&ESO197-G027&32.71889&-49.69852&Sy2&0.0267&18.5&82&uncertain&11.04&15.65&0.88\\
116&Mrk 590&33.63984&-0.76672&Sy1.5&0.0289&27.3&41&uncertain&11.2&6.08&0.45\\
128&MCG+4-6-43&36.72751&23.79966&Sy1&0.0333&13.2&44&spiral&10.8&<1.35&<-0.12\\
\hline
\end{tabular}
\begin{tablenotes}
\item The full version of the table is available in its entirety for the \Nobs\ BAT AGN galaxies in a machine-readable form in the online journal. A portion is shown here for guidance regarding its form and content. A detailed description of this table's contents is given in Section \ref{othermeas}.
\end{tablenotes}
\label{sampleparams}
\end{table}

\subsection{Beam Corrections}
\label{subsec:beam_corr}
 Beam corrections are needed because the lowest redshift galaxies ($z{\sim}0.01$--0.02) with the largest angular sizes may have more flux extending outside of the beam that goes unmeasured compared to more distant sources ($z{\sim}0.04$--0.05) with smaller angular sizes, where most or all of the CO(2--1) flux is likely recovered. 

For each galaxy we estimate an aperture correction separately using a model of the gas distribution and either \herschelsh/PACS imaging \citep{Lamperti:2019:arXiv:1912.01026} or the galaxy diameter from the $K$-band effective radius \cite[see][for more details on this technique]{Saintonge:2011:32, Saintonge:2017:22}.  For the majority of the sample (81\%, 179/\Nobs), we use \herschelsh/PACS imaging.  We assume the CO luminosity is traced by the PACS 160 \micron\ \herschel emission (FWHM=12$\arcsec$) from \citet{Melendez:2014:152}.  We use the PACS 160 \micron\ images because the longer wavelength is less likely to be contaminated by AGN emission which can still contribute to a significant fraction of the 70 \micron\ emission \citep[e.g.,][]{Shimizu:2017:3161,Ichikawa:2019:31}. 

We first measure the total infrared 160 \micron\ emission of the galaxy within a radius big enough to include the entire galaxy.  Then we take the ratio between the flux from the map multiplied by the CO(2--1) Gaussian beam sensitivity function and the total infrared flux and we use this value to extrapolate the total CO(2--1) flux.

The images that we are using to trace the distribution of the FIR emission are not maps of the `true' distribution, but instead are maps of the `true' distribution convolved with the PSF of PACS (12\arcsec).  Therefore an additional correction is needed.

To correct for the effect of the PACS PSF smearing, we use a simulated galaxy gas profile, following the procedure described in \cite{Saintonge:2011:32}.  For each galaxy, we create a model galaxy simulating a molecular gas disk following an exponential profile, with a scale length equivalent to its half-light radius. Then the profile is tilted according to the inclination of the galaxy and we measure the amount of flux that would be observed from this model galaxy, using an aperture corresponding to the size of the beam.  We multiply the infrared image by a 2D Gaussian centered on the galaxy center and with FWHM equal to the beam size, to mimic the effect of the beam sensitivity of the telescope that took the CO observations. By taking the ratio of these two measurements, we estimate how much the flux changes due to the effect of the FIR PSF smearing. We perform this 160 \micron-based correction on 69\% (148/213) of images.  For an additional 7\% (15/213), we use the 70 \micron\ surface brightness, since the 160 \micron\ was a non-detection.

In cases where the PACS imaging resulted in a non-detection, or where no PACS imaging exist (51/\Nobs, or 24\%), we directly use the simulated galaxy gas profile discussed above as described in the COLD GASS and xCOLD GASS papers \citep{Saintonge:2011:32,Saintonge:2017:22} and solely use the $K$-band galaxy diameter ($D=2\times r_{k20fe}$) and inclination to estimate the correction factor. Further discussion of the beam corrections is in Appendix \ref{beamcorr_appen}.

\subsection{Inactive Galaxy Comparison Sample}
\label{subsec:comp_sample}

As the properties of (molecular) gas in galaxies are known to be related to stellar mass and morphology \citep[e.g.,][]{Saintonge:2011:32,Young:2011:940}, it is critical that our analysis of the BAT AGN galaxies is done in comparison to a large, unbiased control sample of {\it inactive} galaxies spanning a similar range in stellar mass.  

The IRAM 30m xCOLD GASS sample \citep{Saintonge:2017:22} is comprised of 532 galaxies from two large IRAM programs, with measurements and analyses of molecular gas across a large range of stellar masses [i.e. $9{<} \log(M_*/\Msun) {<} 11.5$], and over the same redshift range as our sample of BAT AGN galaxies (i.e., $0.01{<}z{<}0.05$).

The xCOLD GASS sample is primarily composed of IRAM measurements, meaning that CO(1--0), with a HPBW of $\approx$22\arcsec\, is a good match to our APEX (HPBW=27\arcsec) and JCMT (HPBW=20\arcsec) CO(2--1) data, even though a different line is used to probe the molecular gas. The xCOLD GASS sample also contains 28 galaxies with APEX(2--1) measurements, and found that the APEX CO(2--1) to IRAM CO(1--0) luminosity ratio for velocity-integrated measurements is $r_{21}=0.79\pm0.03$, based on robust measurements of both lines, with no systematic variations across the sample properties (e.g. SFR, sSFR, or \fgas).  The range typically varies between $\sim$0.6-1.0 in nearby galaxies \citep{Leroy:2009:4670}.  We adopt this $r_{21}$ conversion factor to convert our AGN galaxy measurements from to CO(2--1) to CO(1--0) before converting to molecular gas.  Further analysis of the line ratio will be discussed in Shimizu et al. (in prep.), which re-observed 56 BAT AGN galaxies from our APEX and JCMT sample using IRAM in both CO(1--0) and CO(2--1). 

 For the purposes of our study, we excluded 164 galaxies with low stellar masses [$\log(M_*/\Msun) {<} 10$], which are only rarely present in our BAT AGN galaxy sample.  We further ensured that the xCOLD GASS comparison sample does not include any AGN. 
First, no BAT AGN galaxies were present in xCOLD GASS, likely due to their low space density.  
Second, we used the SDSS nebular emission line measurements from the OSSY database \citep{Oh:2011:13} to test for AGN activity based on strong line diagnostics for the entire xCOLD GASS sample.  We identified 14 optically selected AGN based on the \oiii/\hbeta\ vs. \nii/\halpha\ diagnostics of \cite{Kewley:2006:961}.  
We then also checked against the SDSS-based catalog of broad \halpha\ line emitters compiled by \cite{Oh:2015:1}, but no galaxies were found to be part of this catalog.  
We finally checked the Veron-Cetty Catalog of Quasars and AGN \citep{Veron-Cetty:2010:A10} and found three xCOLD GASS sources were reported as hosting AGN.  We also removed four radio loud AGN galaxies 
\citep{Best:2012:1569}.  In the process of reviewing the xCOLD GASS sample, we also noticed three galaxies where the SDSS spectroscopic pipeline inaccurately targeted a spiral arm or other feature and the nucleus of the galaxy is more than 5\arcsec\ away, we have excluded these as well due to their unreliable photometry and measurements (ID=20790, 36169, and 31592). The final xCOLD GASS inactive galaxy comparison sample thus holds \Nxcoldgassinactive\ inactive galaxies.

Throughout this analysis, and for both samples, we adopt a Milky way-like conversion factor from CO luminosity to H$_2$ mass of \xco = 4.3 \mstar (K \kmps pc$^2$)$^{-1}$ \cite[see e.g., ][for review]{Bolatto:2013:450}.  Studies of nearby galaxies with kpc-scale-resolutions have found that \xco\ is generally flat with galactocentric radius, with average values of 3.1$\pm$0.3 for nearby galaxies.  While ULIRG type galaxies are thought to have lower \xco\ values \citep[e.g., $\alpha_{CO}\sim 1$ \Msun /(K km s$^{-1}$ pc$^2$);][]{Narayanan:2012:3127}, there are none of these sources in either sample.  Given that the stellar mass-metallicity relation is rather flat over the average stellar mass that we study \citep[$\log$ \mstar/\Msun$>$10.0,][]{Yates:2012:215} and the difficulty of measuring metallicity in AGN hosts using emission lines because of contamination, we prefer this simple approach rather than invoke metallicity dependent gradients and conversion factors \citep[e.g.,][]{Accurso:2017:4750}.   

The median beam correction for all the BAT AGN galaxies is 1.33, which is higher, by 16\%, than the xCOLD GASS survey (1.15), mainly driven by the lower redshift (${<}z{>}{=}0.026$ vs. ${<}z{>}{=}0.036$), and higher stellar masses of the galaxies in the AGN galaxy sample.  This is somewhat by design as the redshift range  ($z{=}0.026$--0.05) was selected to ensure the whole galaxy was in the IRAM beam for most massive galaxies ($\log$ \mstar/\Msun$>$10.0) in xCOLD GASS.  We verified that the results presented throughout the rest of the present study are not significantly affected by inadequate aperture corrections, by confirming that the distributions of key quantities and/or relations between them do {\it not} depend on galaxy redshift (or distance).

\subsection{CO Catalog Description}
\label{cocatsection}

\begin{figure*} 
\centering
\includegraphics[width=0.495\textwidth]{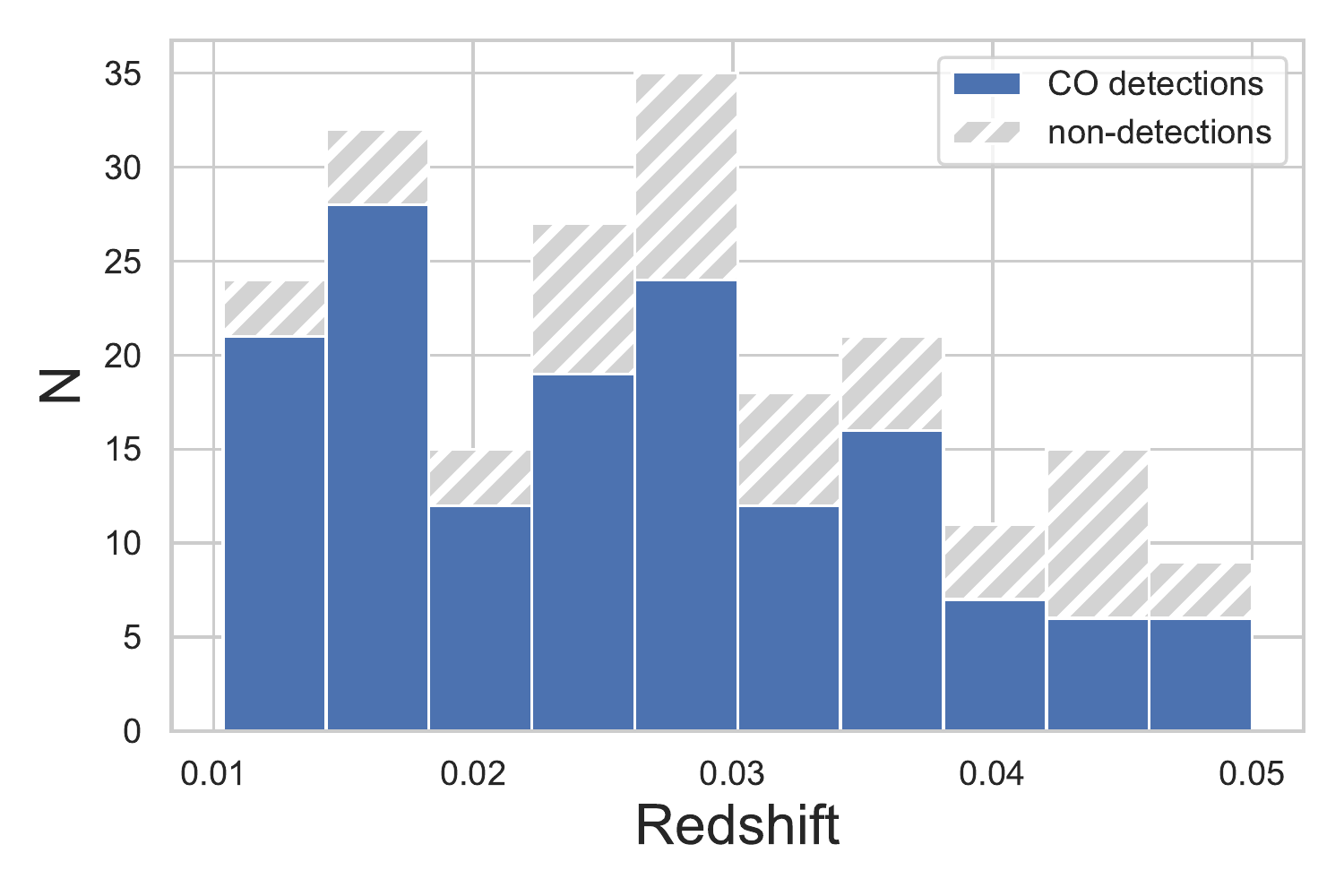}
\includegraphics[width=0.495\textwidth]{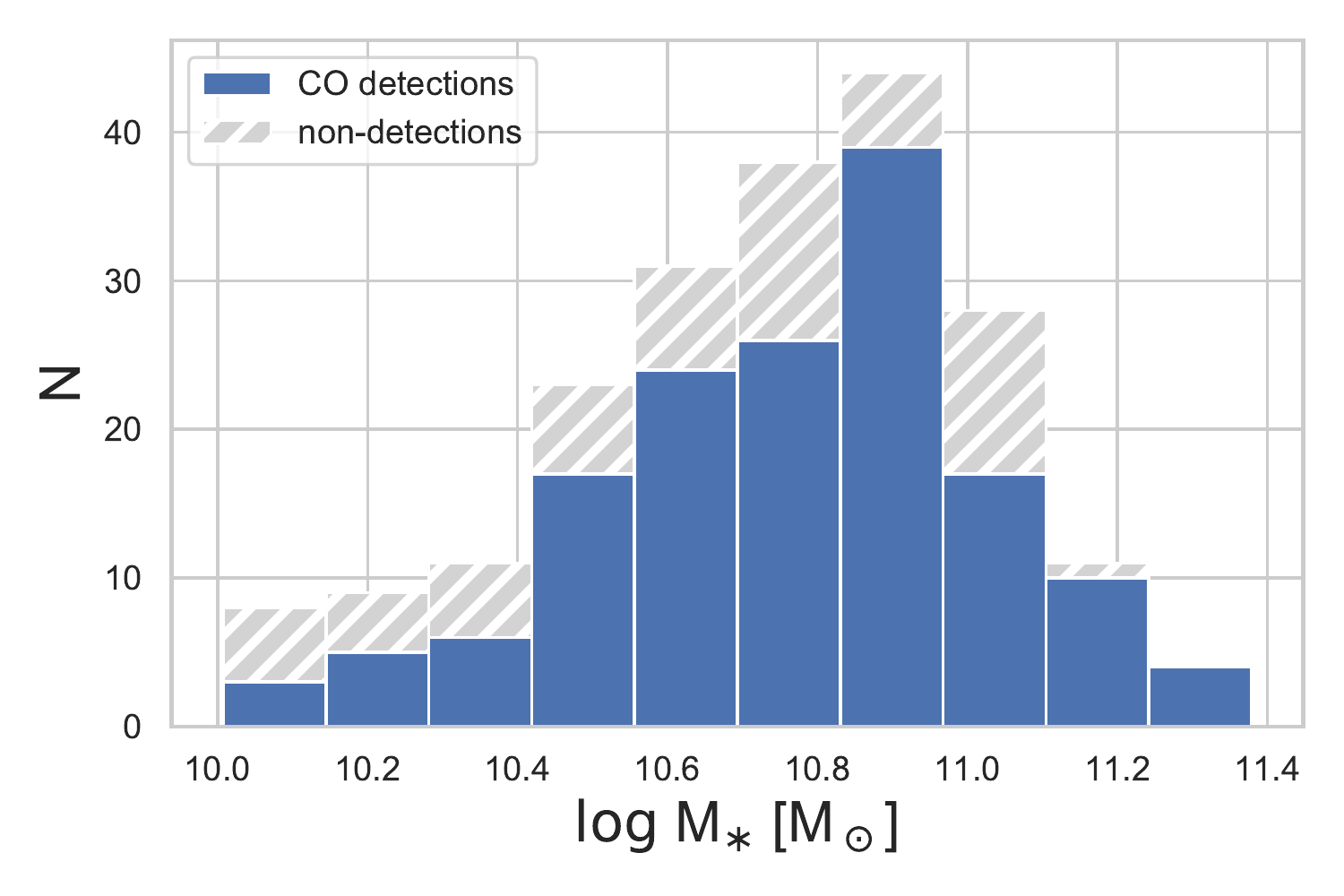}
\includegraphics[width=0.495\textwidth]{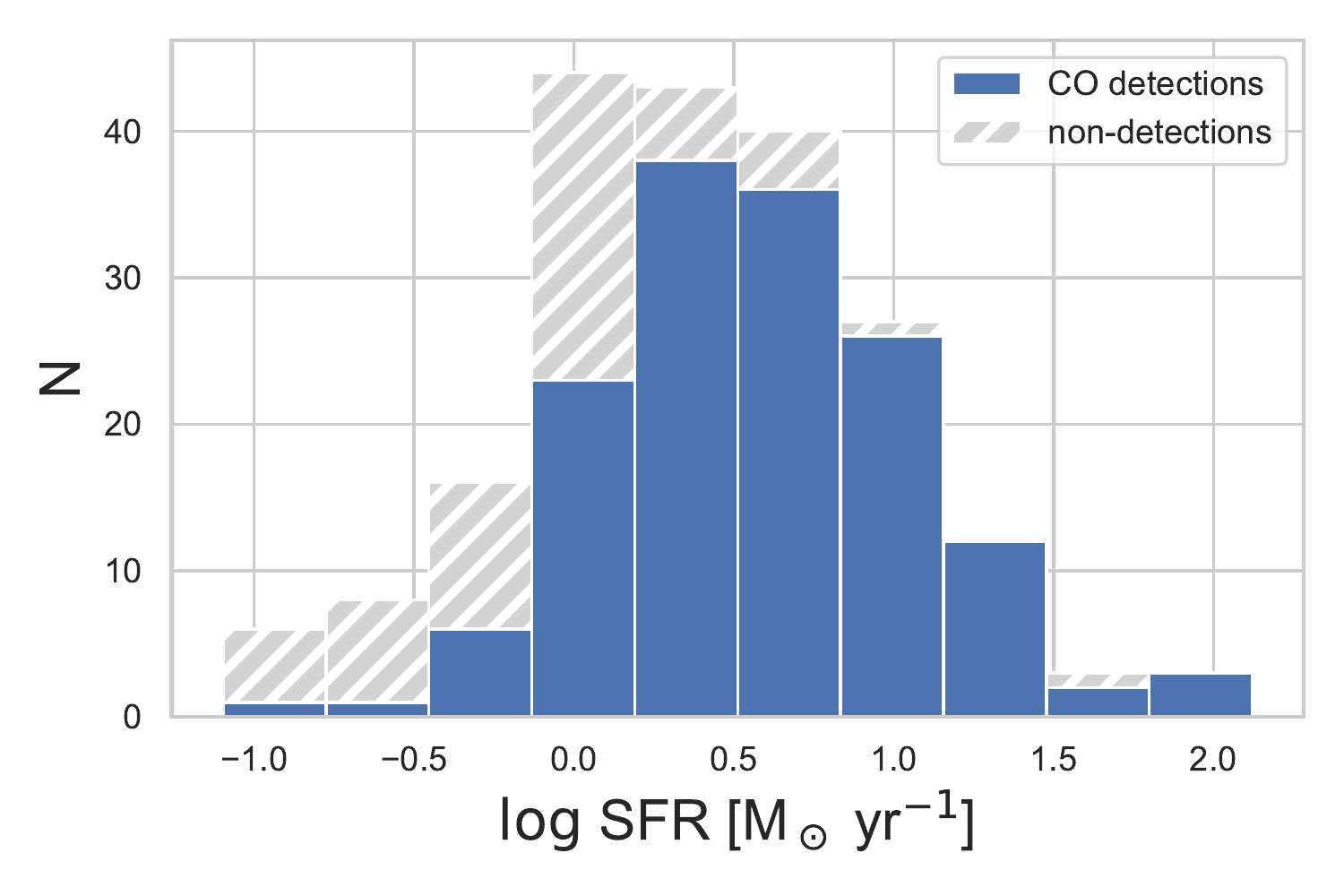}
\includegraphics[width=0.495\textwidth]{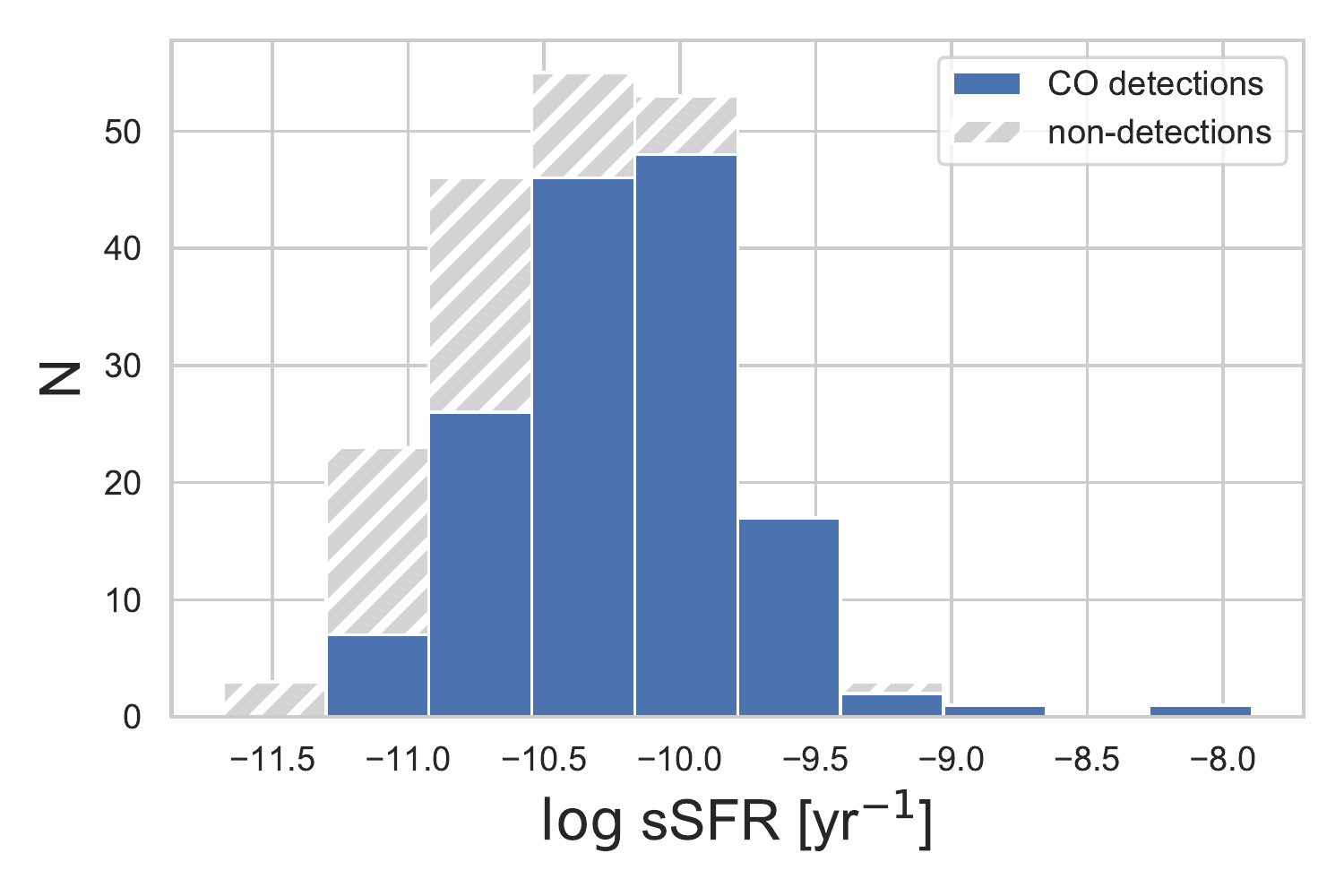}
\includegraphics[width=0.495\textwidth]{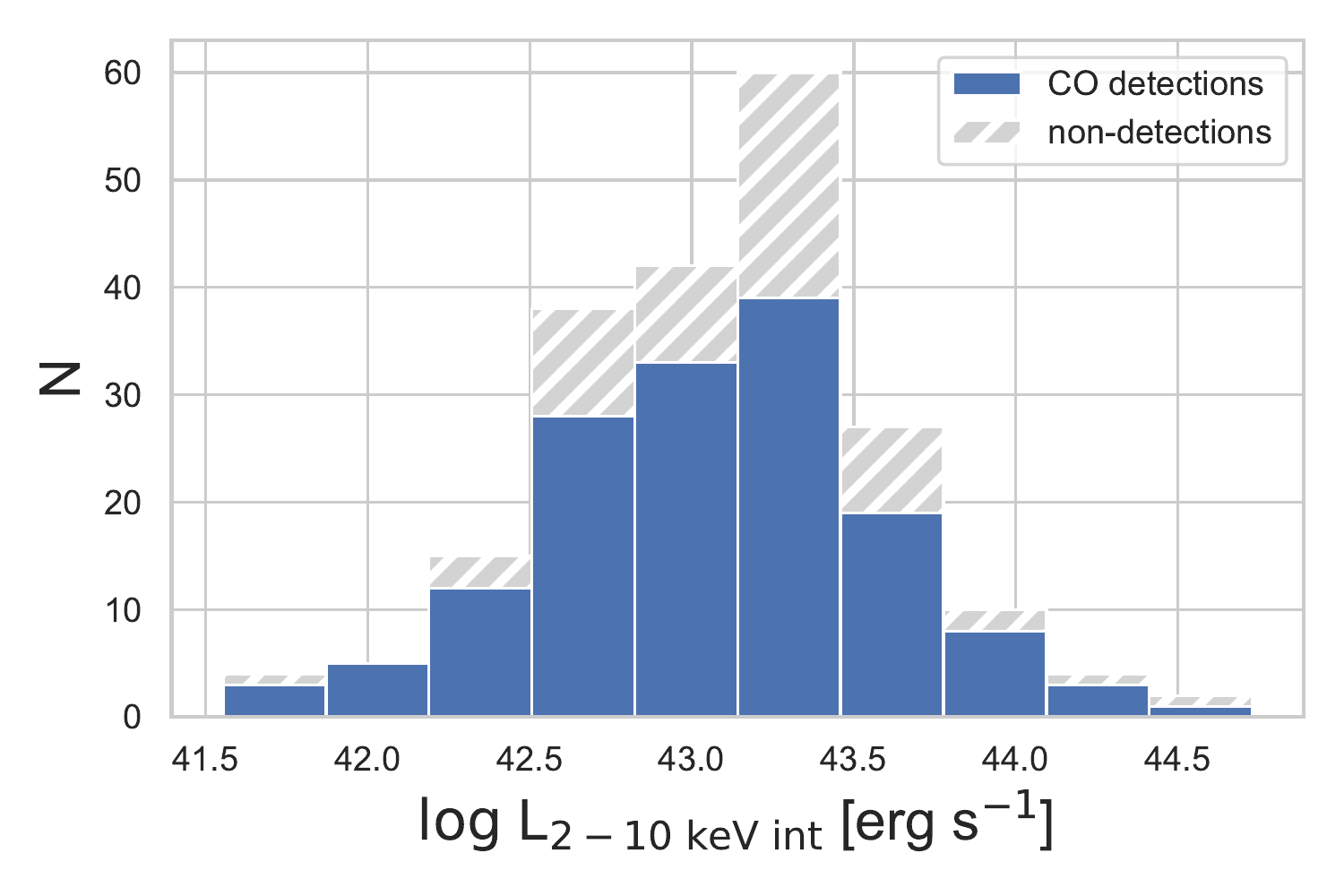}
\includegraphics[width=0.495\textwidth]{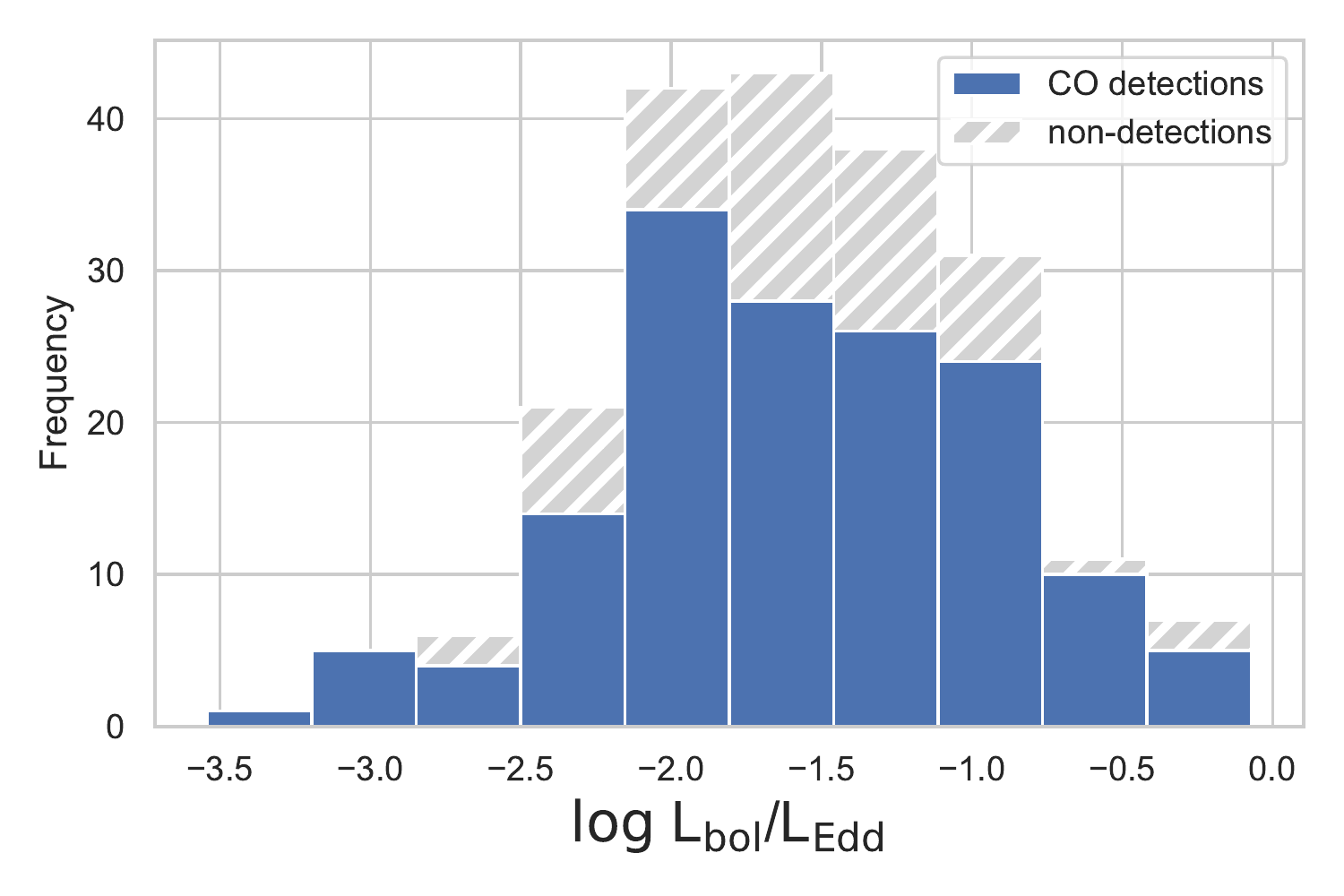}
\caption{Stacked bar chart distributions of the sample of BAT AGN galaxies, with CO detections (blue) and non-detections (grey), as functions of redshift, host galaxy stellar mass, SFR, sSFR, intrinsic X-ray luminosity, and Eddington ratio.  In nearly all cases, except at low star formation rates (SFR$<$1 \Msun yr$^{-1}$), the majority of the sample is detected.}
\label{detectfrac}
\end{figure*}

We provide the full catalog of CO(2--1) measurements and key derived quantities for all \Nobs\ BAT AGN galaxies in our sample.
A summary of possible trends between the CO detection fraction and various AGN galaxy properties is provided in Figure \ref{detectfrac}. 

The catalog includes the following quantities, included in Table \ref{COtab}:

\begin{enumerate}
\item \texttt{BAT ID}: Unique ID in the \swift-BAT 70 month catalog.\footnote{https://swift.gsfc.nasa.gov/results/bs70mon/}

\item \texttt{Telescope} and \texttt{HPBW}: The telescope used for the CO observation -- either JCMT or APEX for our own observations, or for the archival data, along with the associated beam size.

\item \texttt{FlagCO21}: A detection flag for the CO(2--1) line, with `1' indicating a detection, and `0' indicating a non-detection. In the cases where the line is not detected, the tabulated line luminosities and molecular gas masses (see below) are $3\sigma$ upper limits. 

\item $\mathtt{\sigma_{CO21}}$: {RMS noise achieved around the CO(1-0) line, in spectral channels with width $\Delta w_{ch}=50$ \kms.}  {The quoted RMS is at 50 \kms\ even for spectra with finer binning.}

\item \texttt{Beam correction factor}:  Beam aperture correction, derived either through simulations following the approach developed for the COLD GASS sample \citep{Saintonge:2011:32}, or by assuming the CO luminosity is traced by the 160 \micron\ emission observed by \herschelsh/PACS (FWHM=12$\arcsec$), as reported by \cite{Melendez:2014:152}.

\item $\mathtt{L^{\prime}_{\rm CO21, corr}}$: Total CO(2--1) line luminosity, calculated from $L^{\prime}_{\rm CO21, obs}$ and the aperture correction.  The error includes the measurement uncertainty, a 10\% flux calibration error, and the 15\% uncertainty on the aperture correction \citep{Saintonge:2011:32} -- all combined in quadrature. If the source is undetected in CO (FlagCO21$=$0 in Col. 3), then the tabulated value corresponds to a $3\sigma$ upper limit.

\item \texttt{$\log M_{\rm H2}$}: Total molecular gas mass, including the Helium contribution, calculated from the integrated CO(2--1) line luminosity assuming a conversion of $0.79$ from CO(2--1) to CO(1--0), and assuming a Milky Way-like conversion factor of \xco = 4.3\mstar\ (K \kmps pc$^2$)$^{-1}$ \citep{Bolatto:2013:450}. If the source is undetected in CO (FlagCO21$=$0 in Col. 3), then the tabulated value corresponds to a $3\sigma$ upper limit.

\item \texttt{\tdep}:
The gas depletion timescale (\tdep$\equiv M_{\rm H2}$/SFR).  A small number (8\%, 16/\Nanalyzed) of AGN galaxies have no measurements (denoted as `...'), since they lack robust SFR measurements.  Most of these (14/16) were not observed within the \herschel program, and thus have less sensitive SFR upper limits, or lie close to the Galactic plane where no measurements were performed (12\%, 2/16). Most are also undetected in molecular gas (75\%, 12/16).  As they have upper limits on SFR and molecular gas, a gas depletion timescale, which is the ratio of the two, cannot be estimated. 

\item \texttt{Pro$_{\rm type}$}:
 CO line measured with either a single or double peak Gaussian profile (as indicated by \texttt{W50$_{type}$};\citealt{Tiley:2016:3494}). 
 
\item $L^{\prime}_{\rm pro,\,corr}$:
Total CO(2--1) line luminosity, calculated from the profile fit $L^{\prime}_{\rm pro,\,corr}$ with an aperture correction.

\item \texttt{W50}, \texttt{$z_{\rm CO}$}:
FWHM of the line from the profile fit in \kms, the associated error, and redshift from the central velocity.  The tabulated errors correspond to $1\sigma$.

\end{enumerate}

\begin{table}
\begin{rotatetable}
\begin{center}
\caption{Catalog of CO(2--1) measurements}
\footnotesize{}
\label{COtab}
\begin{tabular}{llccccccccccccc}
\hline
ID  &   
Tele. & 
HPBW & 
Flag & $\sigma$ &
$C_{\rm aper}$& 
$S/N_{\rm inte}$ & 
$L^{\prime}_{\rm inte,\,corr}$ & 
$\log M_{\rm H2}$ & 
\tdep & 
$Pro_{\rm type}$ & 
$L^{\prime}_{\rm pro,\,corr}$&
$S/N_{\rm pro}$&
$W50$ & 
$z_{\rm CO}$ \\
& & (\arcsec) && (mK) & Factor &  &  
($10^8$ K \kmps pc$^2$) & 
($\log \Msun$) & 
($\log$ yr)&
 &
($10^8$ K \kmps pc$^2$)&
& 
(\kmps) &(\kmps) \\
\hline \hline
3&APEX&27&1&1.8&1.05&5.5&3&9.11&8.44&1&3&3.2&74$\pm$21&7659$\pm$10\\
17&APEX&27&1&0.9&1.04&11&7.7&9.52&8.66&2&8.1&10.4&561$\pm$158&8614$\pm$55\\
28&APEX&27&1&1.7&1.31&14.4&11.7&9.7&8.79&2&11.4&12.5&461$\pm$94&6604$\pm$42\\
44&APEX&27&1&1.3&1.21&7.4&25.2&10.03&9.05&2&23&9.2&469$\pm$123&14363$\pm$56\\
50&APEX&27&1&1.1&1.16&7&2.9&9.1&8.75&2&4.8&11.9&559$\pm$160&5737$\pm$51\\
58&APEX&27&1&0.8&1.51&6&0.7&8.5&7.83&1&0.6&6.2&499$\pm$237&3408$\pm$79\\
62&APEX&27&1&2&1.78&16.3&5.7&9.39&8.9&2&4.7&7.8&304$\pm$59&3572$\pm$26\\
63&APEX&27&1&0.8&1.17&7&0.9&8.61&8.41&1&0.5&5.7&158$\pm$135&3634$\pm$18\\
72&APEX&27&1&0.7&1.19&4.3&1&8.63&8.28&1&1.7&8&601$\pm$252&5748$\pm$83\\
77&APEX&27&1&2.1&1.3&8.4&4&9.24&8.73&1&4&5.8&86$\pm$23&5075$\pm$7\\
79&APEX&27&1&1.1&1.02&10.7&14.6&9.79&8.73&2&14.9&10.2&522$\pm$121&11416$\pm$52\\
81&APEX&27&1&1.3&1.35&13.2&9.6&9.61&8.82&2&10.2&9.9&381$\pm$120&7707$\pm$42\\
83&APEX&27&1&4.4&1.11&12.8&10.3&9.64&8.86&1&11.1&9.9&268$\pm$56&4934$\pm$23\\
95&APEX&27&1&1&1.06&6.8&5.9&9.41&8.87&2&6.2&8.4&409$\pm$166&10033$\pm$54\\
96&APEX&27&1&2.9&2.19&8.4&12.8&9.74&8.67&1&10.5&7.9&281$\pm$83&5006$\pm$32\\
101&APEX&27&1&2.1&1.21&9.2&4.8&9.32&8.81&1&5.3&10.2&323$\pm$62&4903$\pm$31\\
102&APEX&27&1&0.5&2.88&5.9&1&8.66&8.41&1&0.4&2.2&42$\pm$35&4170$\pm$9\\
114&APEX&27&1&0.9&1.42&12.4&26.4&10.05&8.86&2&24.5&8.8&533$\pm$115&14295$\pm$60\\
116&APEX&27&1&1.3&1.52&11.2&8.4&9.56&8.77&2&10.1&8.4&252$\pm$44&7926$\pm$19\\
128&APEX&27&1&1.2&1.04&4.7&3.4&9.16&...&1&3&3&48$\pm$54&10069$\pm$7\\
\hline\\
\end{tabular}
\begin{tablenotes}
\item The full version of this table is available in its entirety, for the \Nobs\ BAT AGN galaxies we study, in a machine-readable form in the online journal. A portion is shown here for guidance regarding its form and content. A detailed description of this table's contents is given in Section \ref{cocatsection}.)
\end{tablenotes}
\end{center}
\end{rotatetable}
\end{table}
\movetabledown=4in

The full BAT AGN galaxy CO catalog, including all CO spectra, SDSS and PS1 images, and galaxy and ancillary AGN observations from BASS will be available at the BASS website (\url{http://www.bass-survey.com}).

\begin{figure*}
\centering
\includegraphics[width=0.495\textwidth]{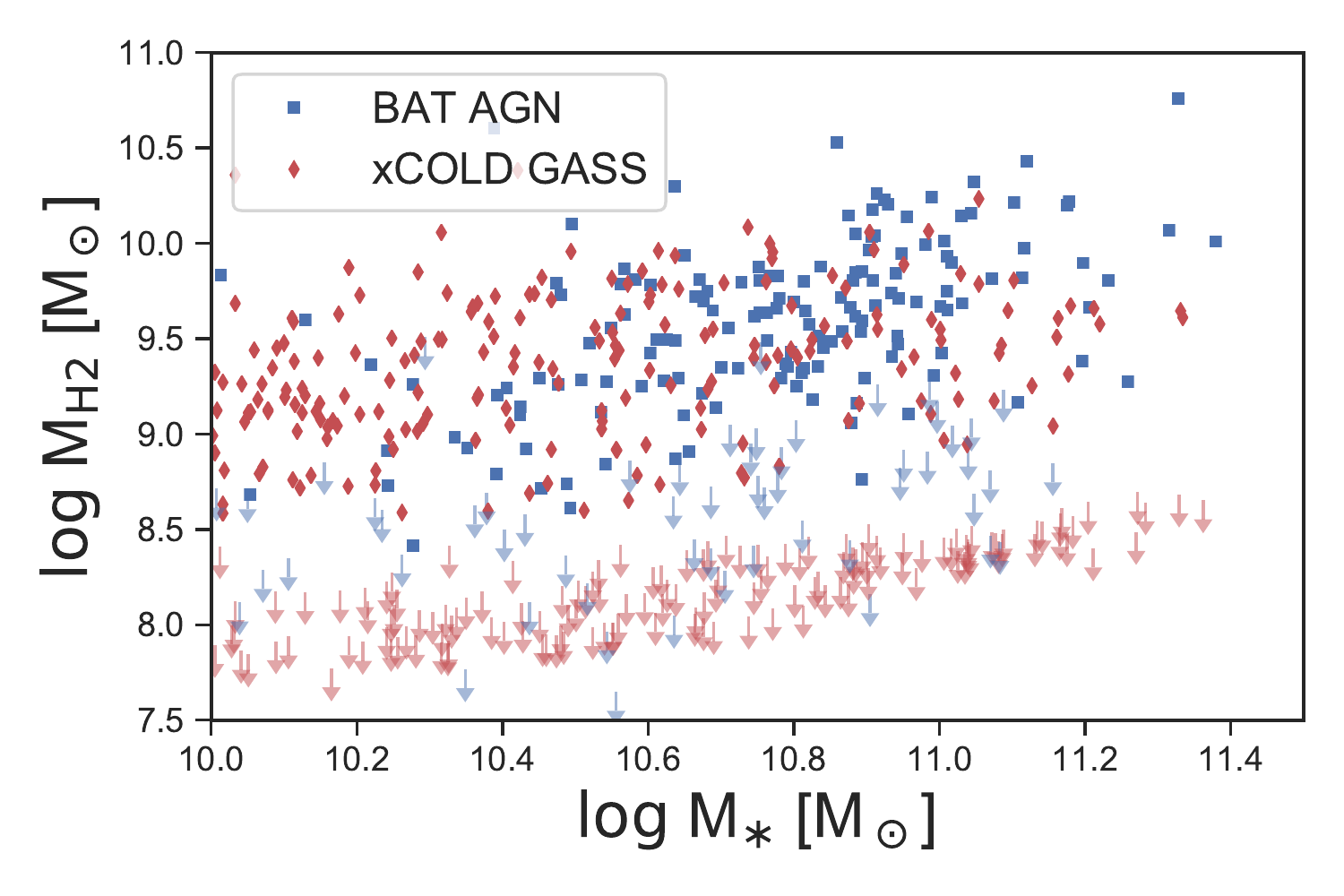}
\includegraphics[width=0.495\textwidth]{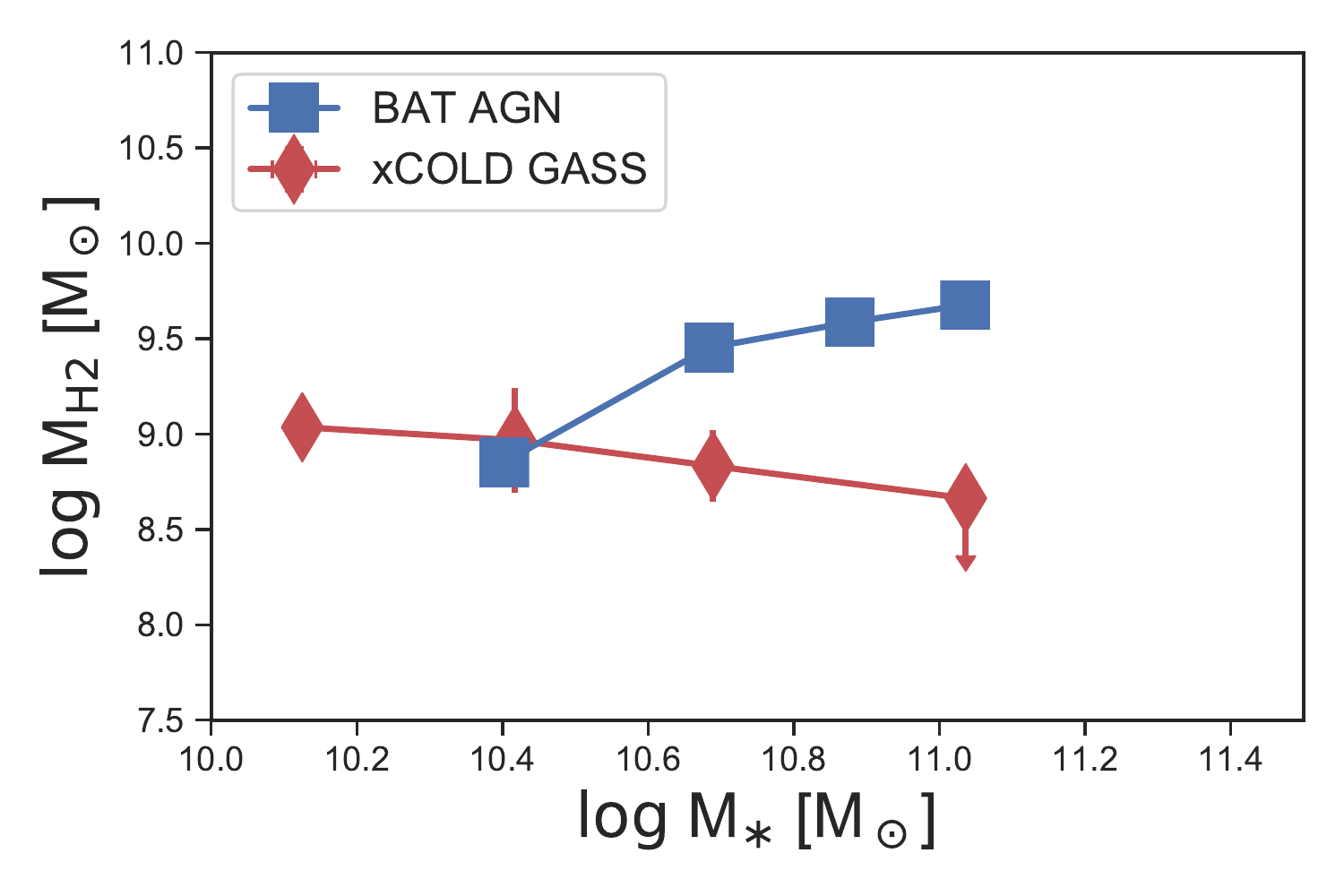}
\includegraphics[width=0.495\textwidth]{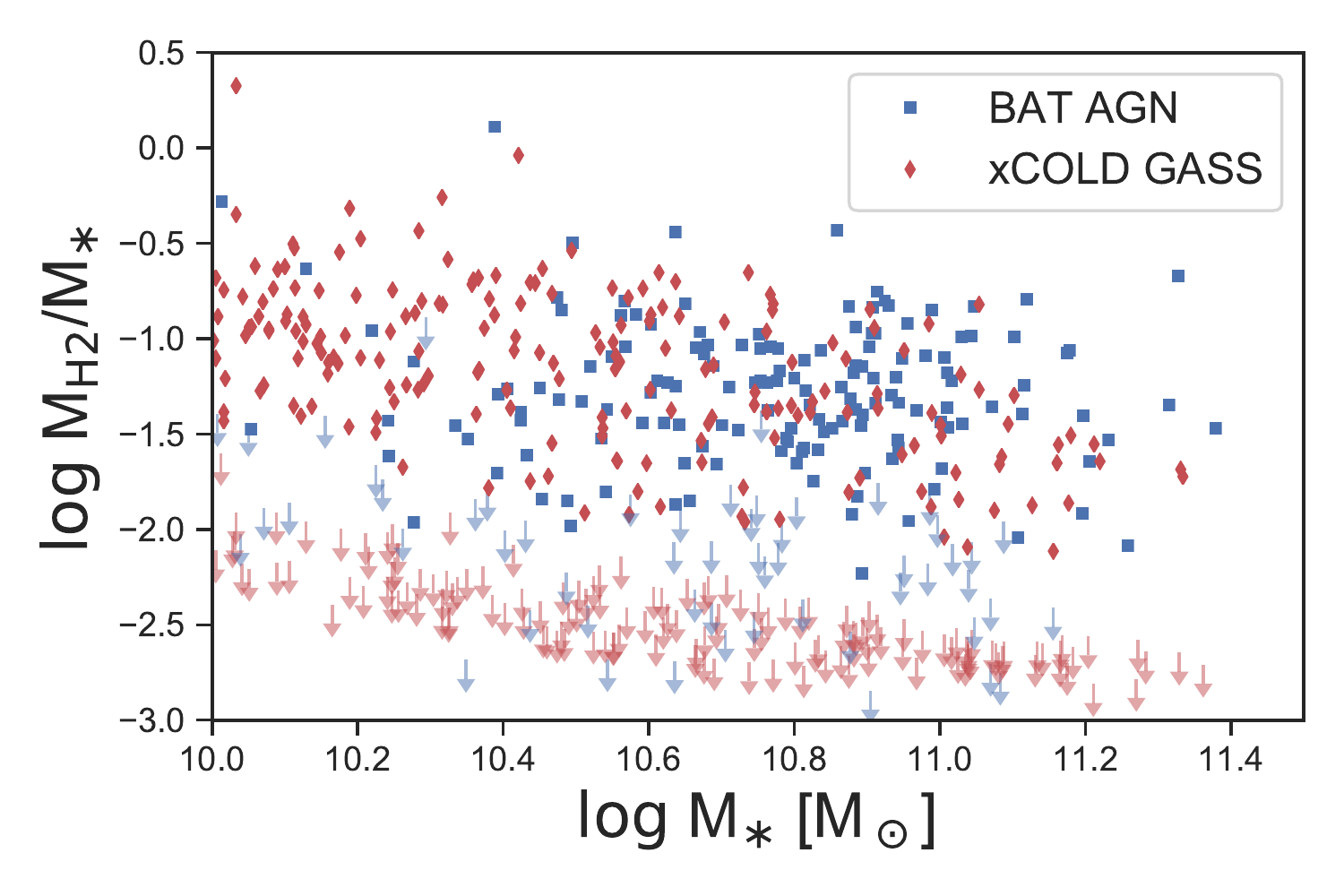}
\includegraphics[width=0.495\textwidth]{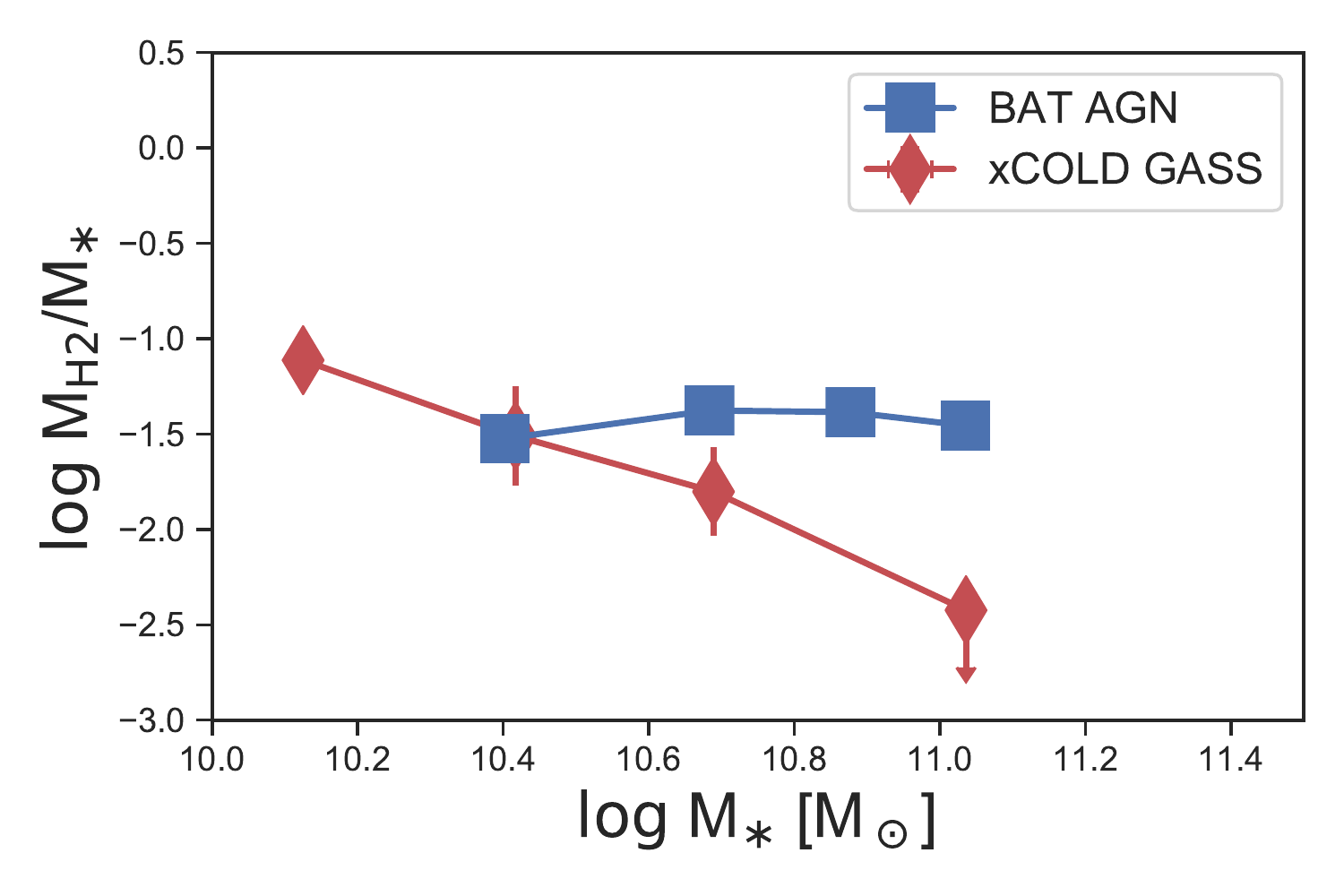}
\includegraphics[width=0.495\textwidth]{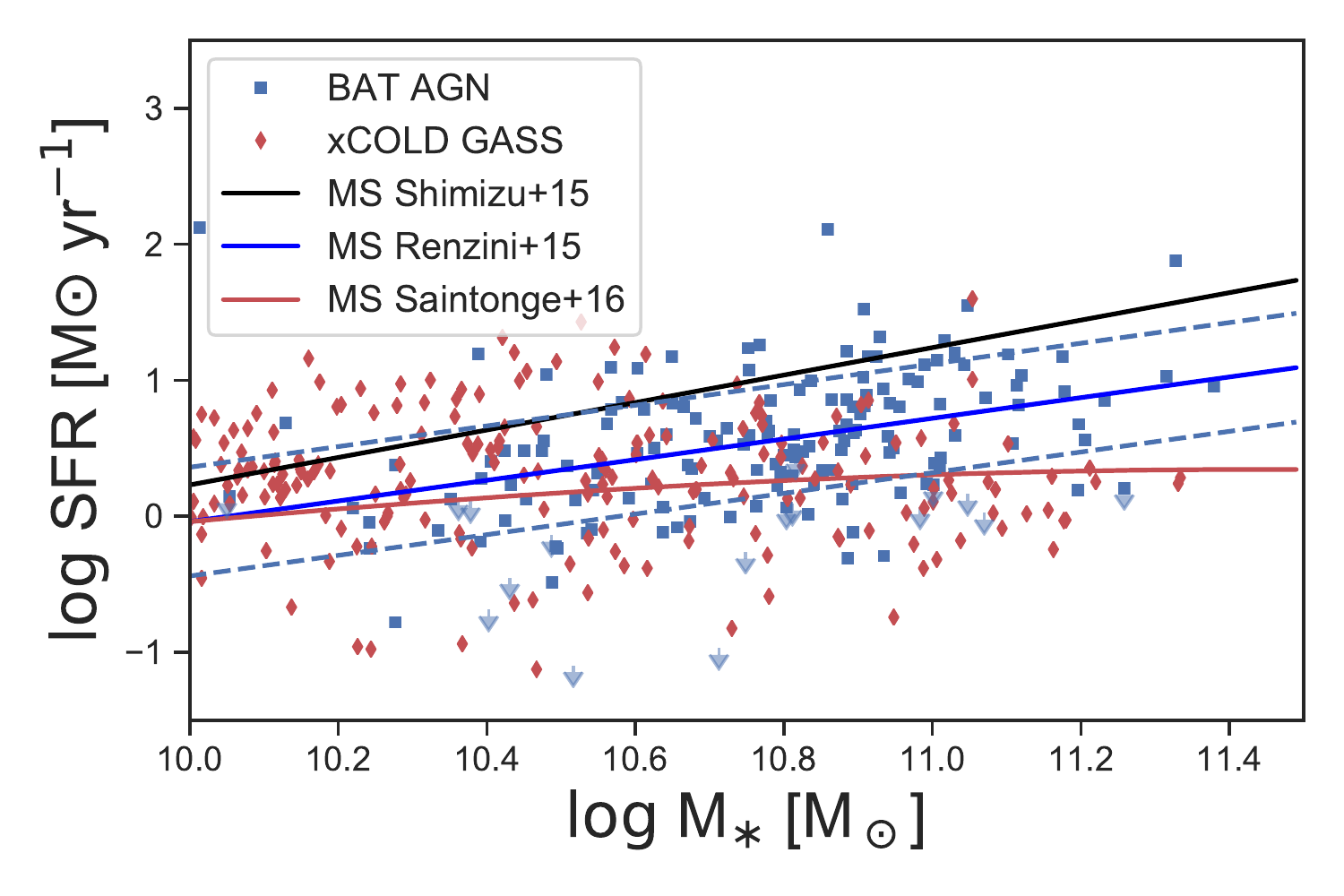}
\includegraphics[width=0.495\textwidth]{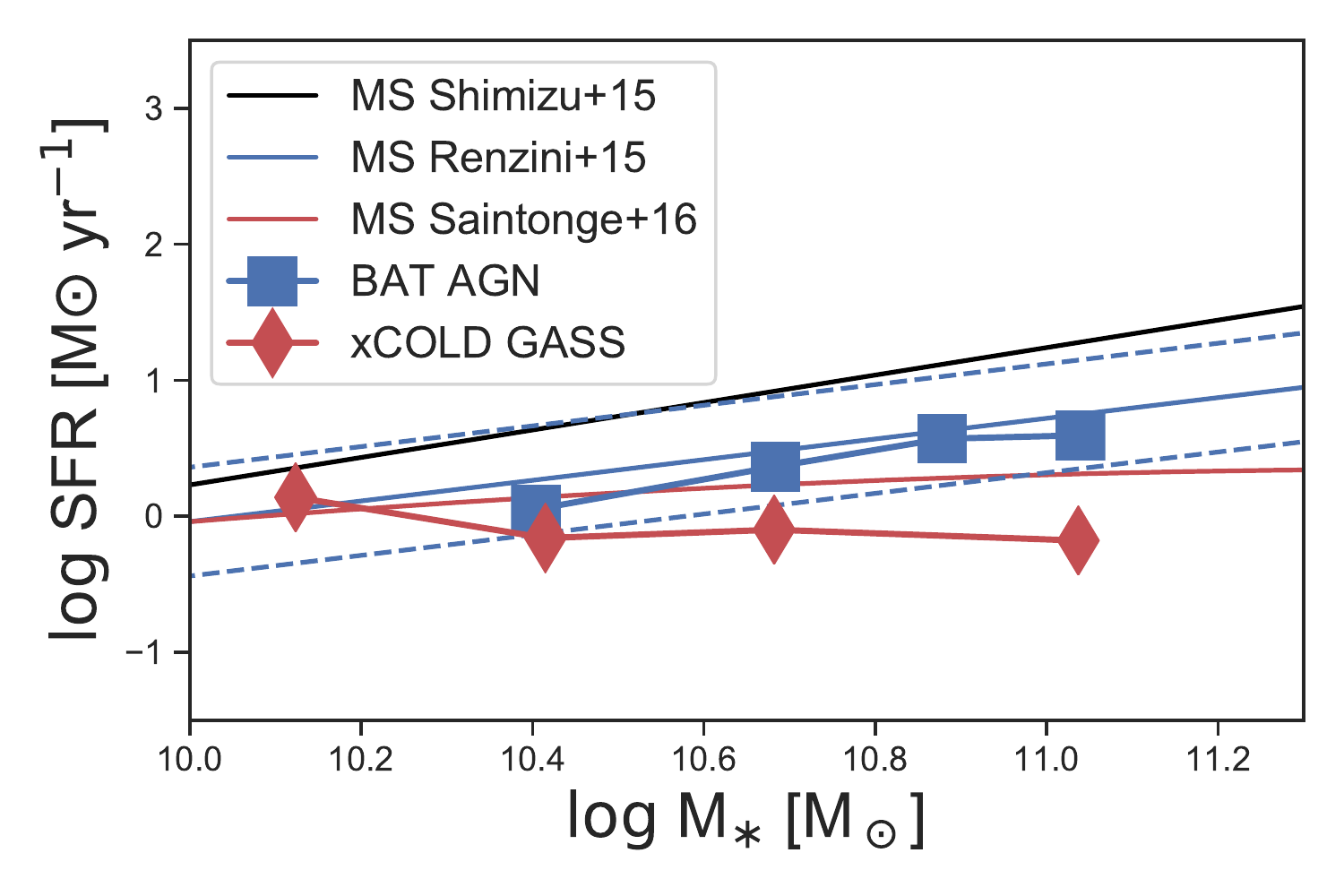}
\caption{Comparison of molecular gas content, SFR, and stellar mass between BAT AGN hosts and xCOLD GASS inactive galaxies.  Top panels show molecular gas masses vs. stellar masses while the middle panels show molecular gas mass fractions ($M_{\rm H2}/M_*$) vs. stellar masses.  {The bottom panels show the SFR vs. stellar masses. The solid line represents the Main Sequence (MS) of star forming galaxies as determined by \citet{Renzini:2015:L29} while the dashed lines indicate the 0.4$\sigma$ scatter around this line.  Due to the significant uncertainty at high stellar masses ($\log(M_*/\Msun)>10.5$) we have plotted additional MS lines \citep{Shimizu:2015:1841,Saintonge:2016:1749}.  Left panels show the individual data points, while right panels show the data binned in stellar mass.  
For presentation purposes, in the left panels individual upper limits are shown at the $1\sigma$ level, while detections are limited to those with ${>}3\sigma$. 
In the right panels, the bin sizes were constructed to have equal numbers of sources in each bin. Error bars on the plotted median values are equivalent to 1$\sigma$ and calculated based on a bootstrap procedure with 100 realizations.  
Upper limits on binned data are shown when more than half of the individual data points within that bin are themselves upper limits.  }
}
\label{co_all}
\end{figure*}

\section{Results}
Here we compare the molecular gas properties we measure in our sample of BAT AGN galaxies with those of the inactive galaxies studied within the xCOLD GASS survey.  
We further search for links between the molecular gas content of our BAT AGN galaxies and their basic AGN-related properties, such as their intrinsic X-ray luminosity (\Lsoftint) and Eddington ratio ($\lledd \equiv L_{\rm bol} / L_{\rm Edd}$).  
For the purposes of this analysis, we have excluded six BAT AGN which are hosted in particularly low stellar mass galaxies, compared with the overall sample [$\log (M_*/\Msun) < 10.0$; see Figure~\ref{detectfrac}].  
A comparison of low stellar mass BAT AGN galaxies with inactive galaxies, at $z<0.01$, can be found in \citet{Rosario:2018:5658}.  
Our final BAT AGN galaxy sample thus totals \Nanalyzed\ AGN galaxies with $\log(M_*/\Msun) > 10.0$.

As our sample includes a significant number of CO non-detections, and thus upper limits on a few key derived quantities (CO luminosities and molecular gas mass), our analyses of distributions and correlations rely on survival analysis. 
{ Specifically, we have utilized the Kaplan-Meier (KM) distribution to perform survival analysis using the software \texttt{lifelines} Python package version 0.25.4 \citep{cameron_davidson_pilon_2020_4002777}.    
In what follows, we test for statistically significant differences using the KM estimator, which is similar to the standard Kolmogorov-Smirnov test but allows for left-censoring (i.e. upper limits).  We adopt a 5 percent threshold probability when evaluating the significance of a difference from the Null Hypothesis between samples (i.e., $p<0.05$) and use this test to compare all sample differences.  In order to ensure the sample tests are not sensitive to the specific upper limit level, we reran significance tests assuming the upper limits were larger or smaller by 30\%, and noted any differences when they occurred.}

\begin{figure*} 
\centering
\includegraphics[width=7.5cm]{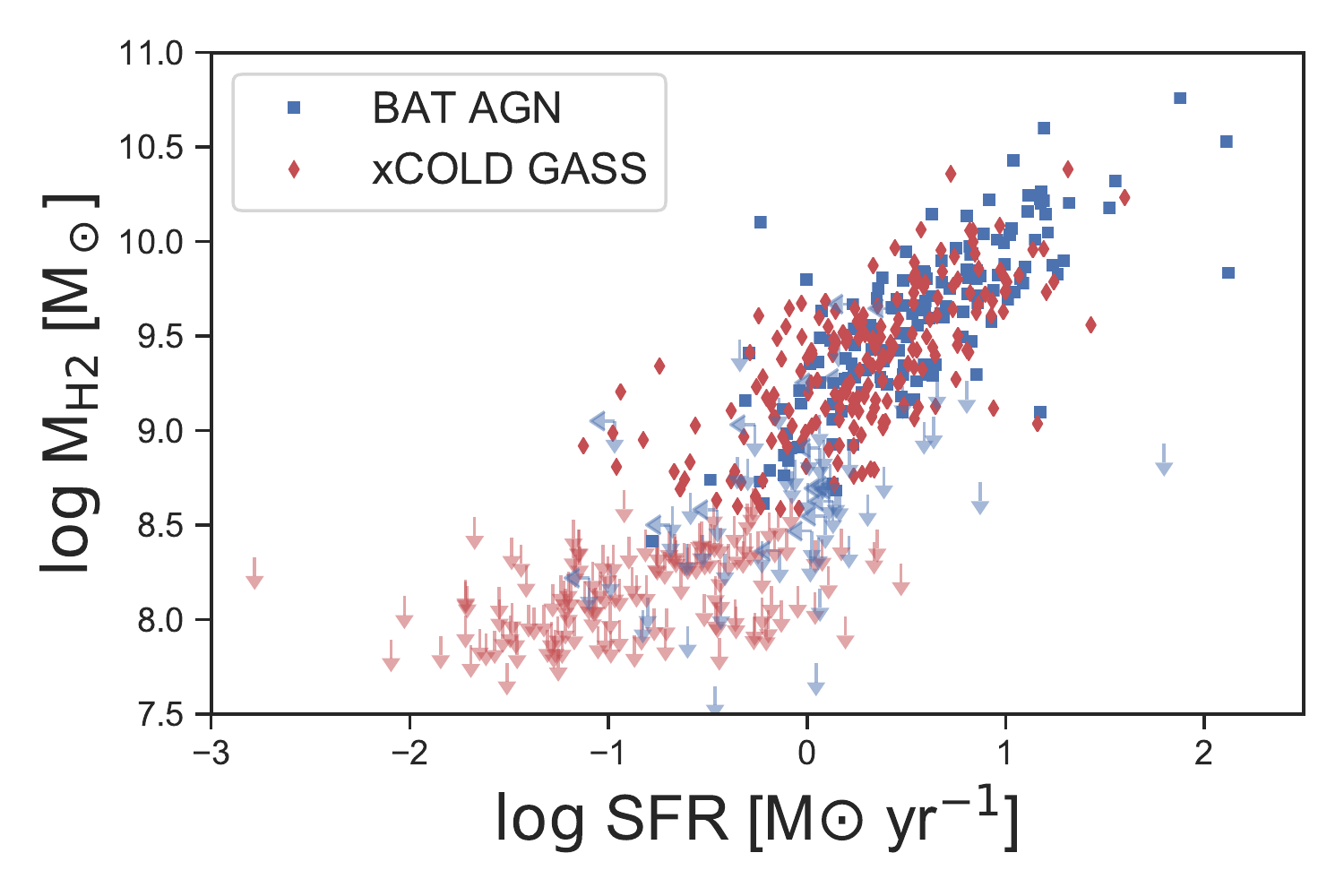}
\includegraphics[width=7.5cm]{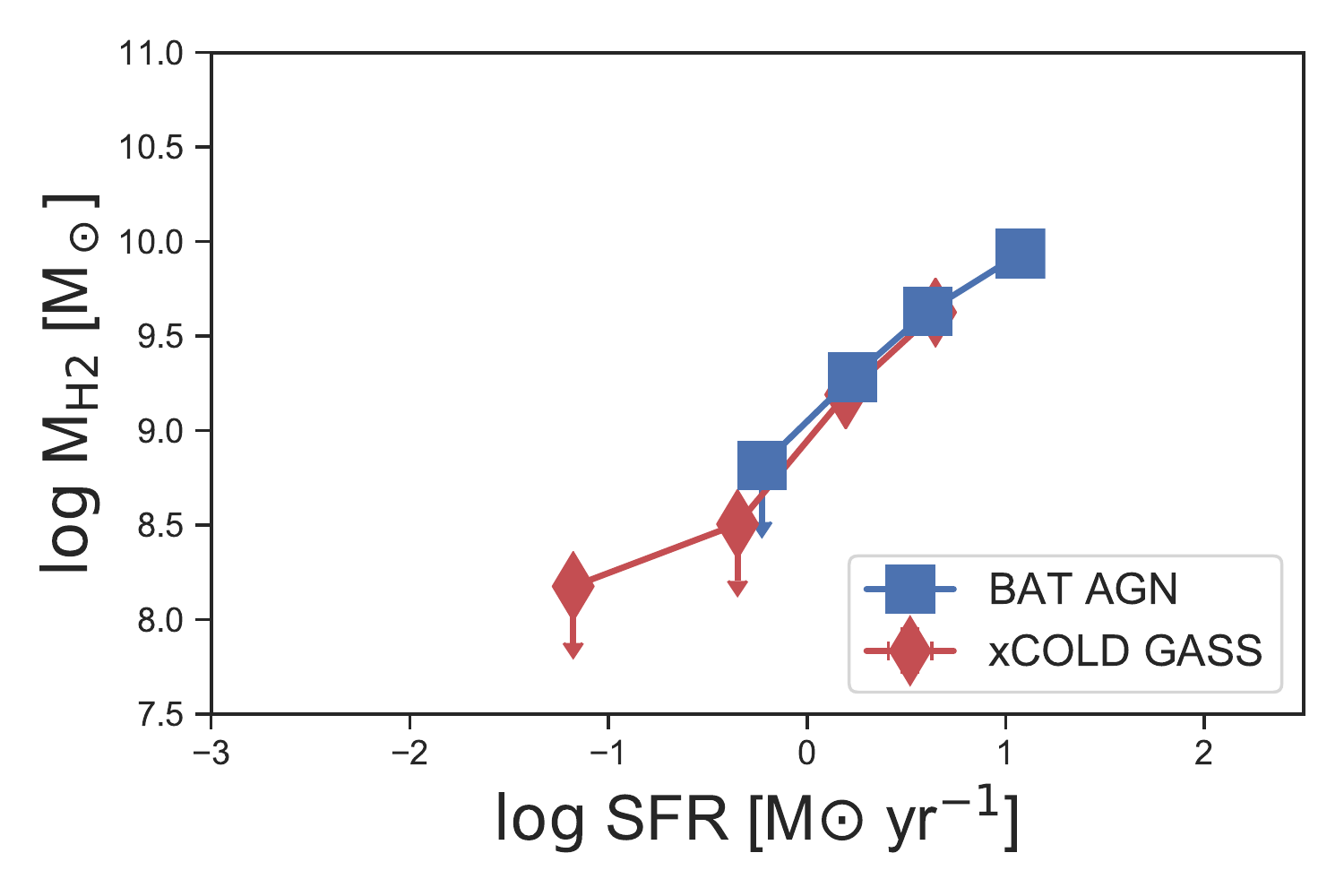}
\includegraphics[width=7.5cm]{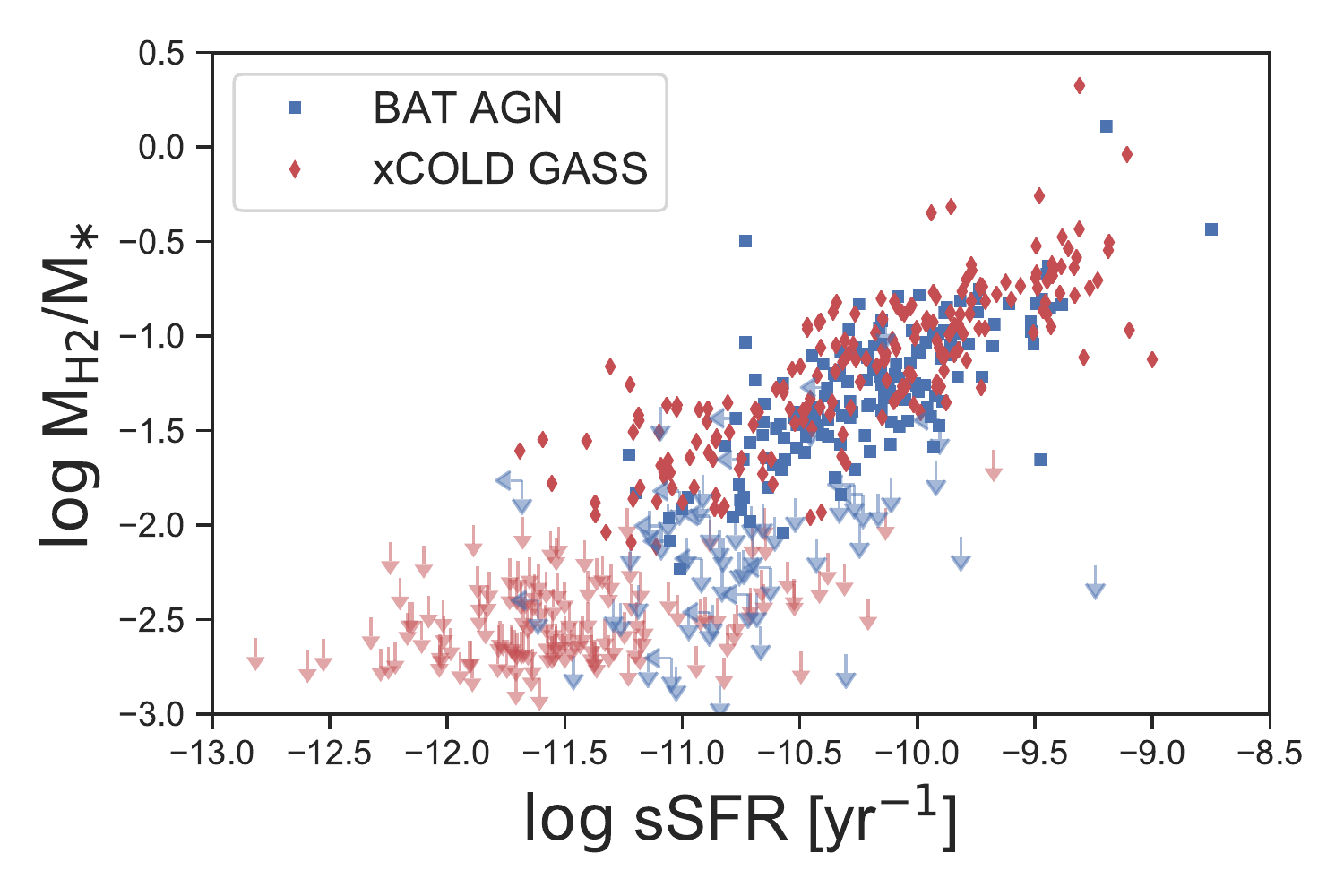}
\includegraphics[width=7.5cm]{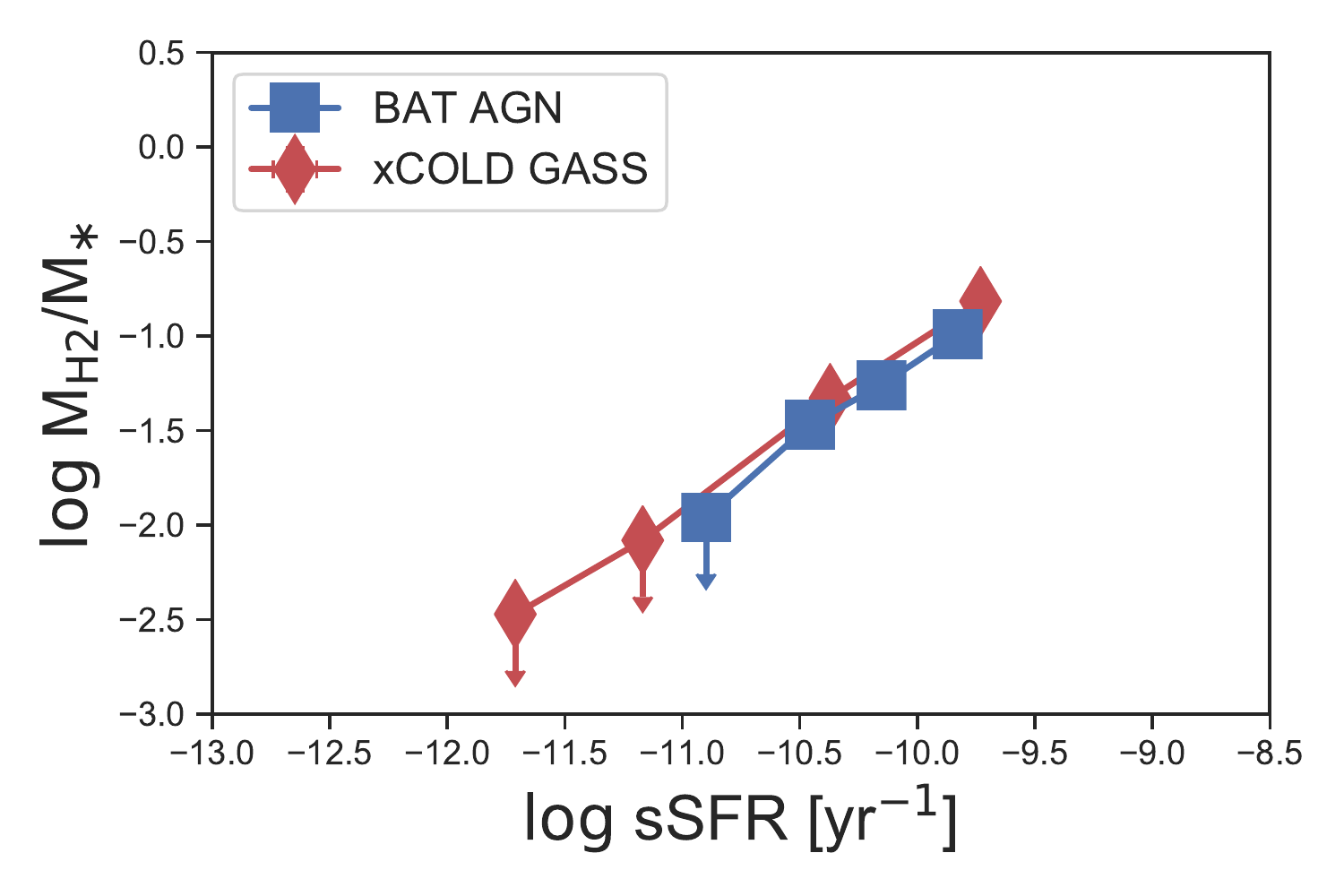}
\includegraphics[width=7.5cm]{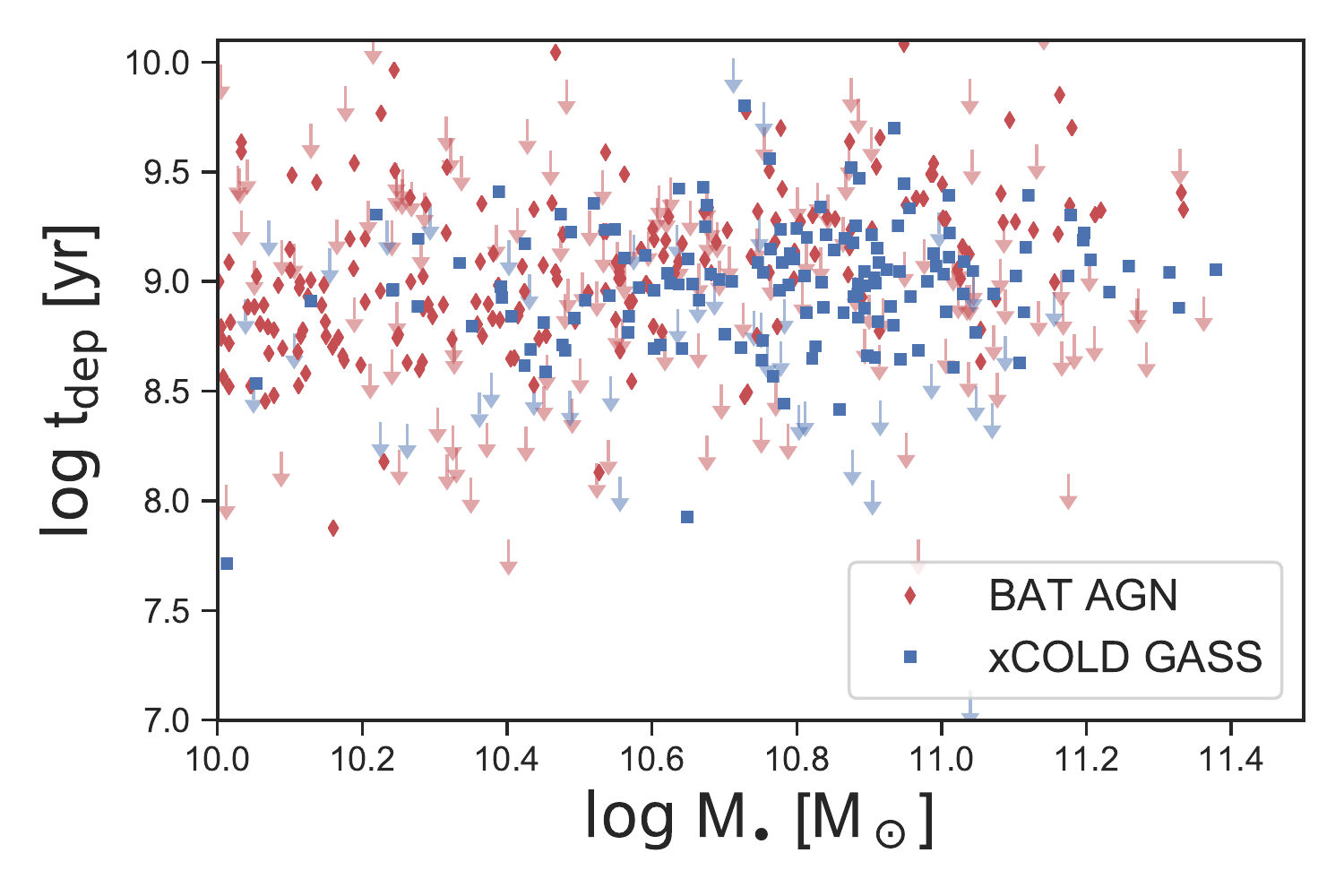}
\includegraphics[width=7.5cm]{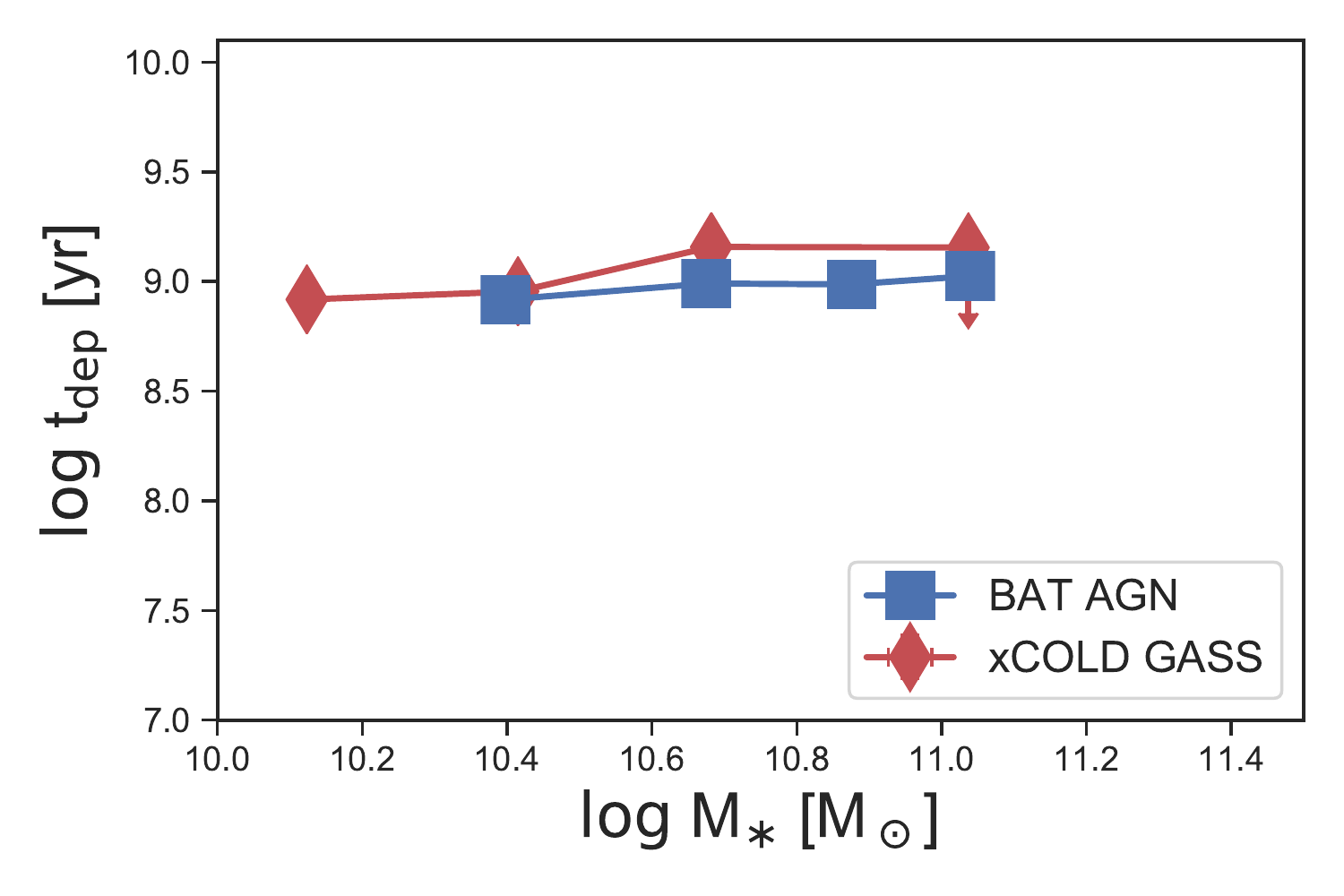}
\includegraphics[width=7.5cm]{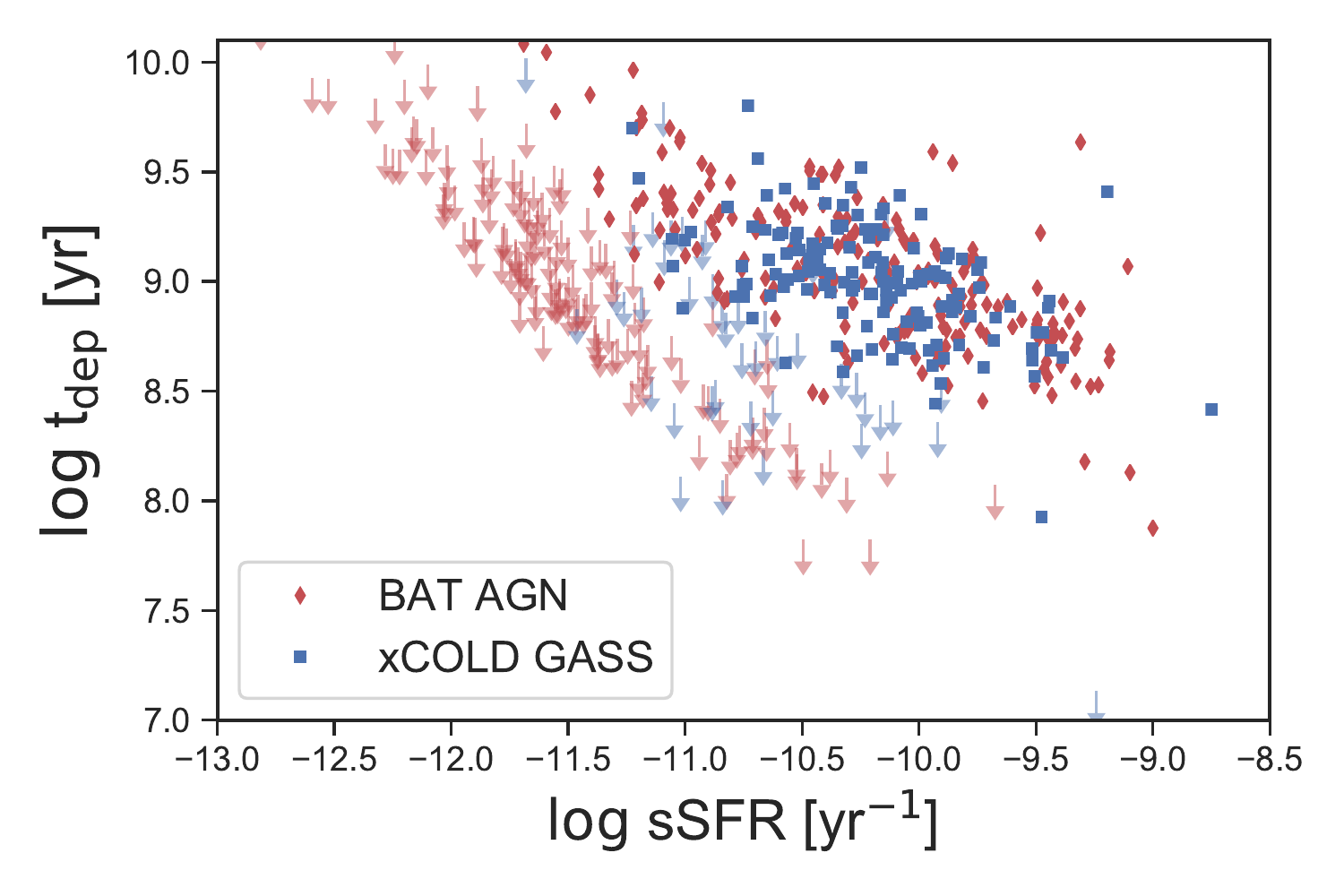}
\includegraphics[width=7.5cm]{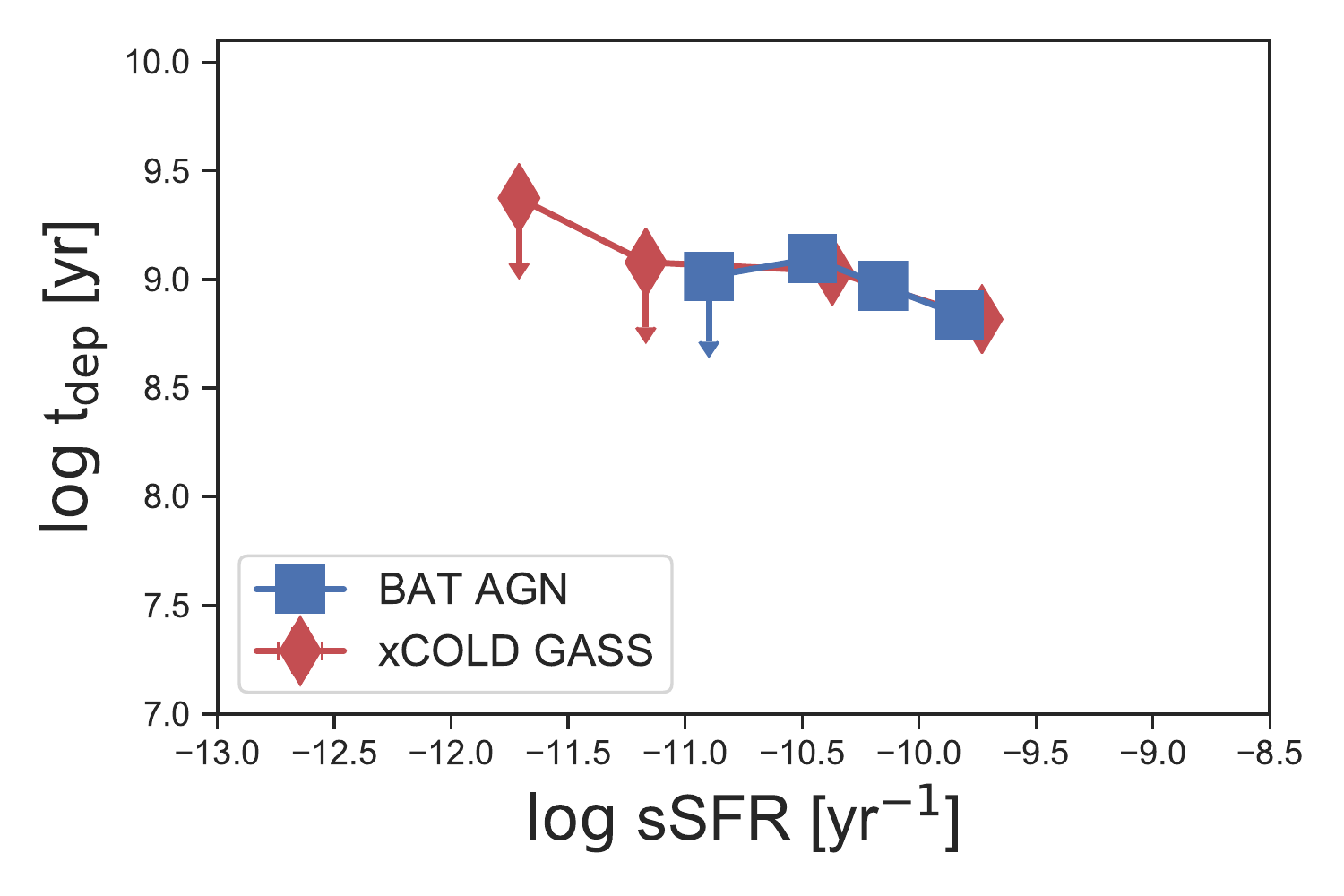}
\caption{
Links between molecular gas content and star formation in BAT AGN galaxies and xCOLD GASS inactive galaxies.
The symbols are identical to those of Figure \ref{co_all}, as is the binning procedure used to produce the right column of panels.
From top to bottom, we present: 
(1) molecular gas masses vs. SFRs;
(2) molecular gas {\it fractions} vs. specific SFRs (${\rm sSFR}\equiv {\rm SFR}/M_*$);
(3) molecular gas depletion timescales ($\tdep \equiv M_{\rm H_2}/{\rm SFR}$) vs. stellar mass; 
and 
(4) depletion timescales vs. sSFRs.
}
\label{fig:sfrdist}
\end{figure*}

\begin{figure*} 
\centering

\includegraphics[width=7.5cm]{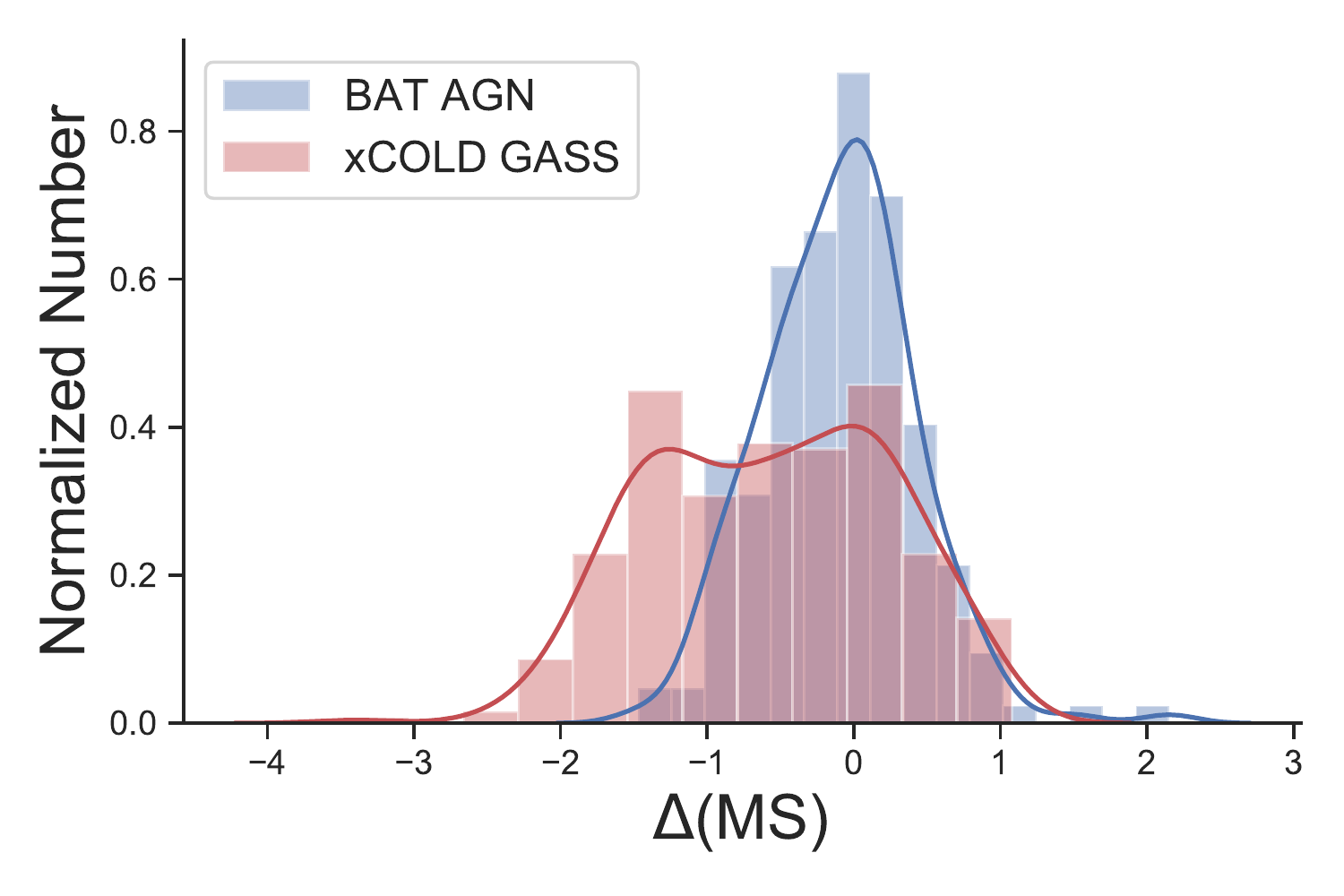}
\includegraphics[width=7.5cm]{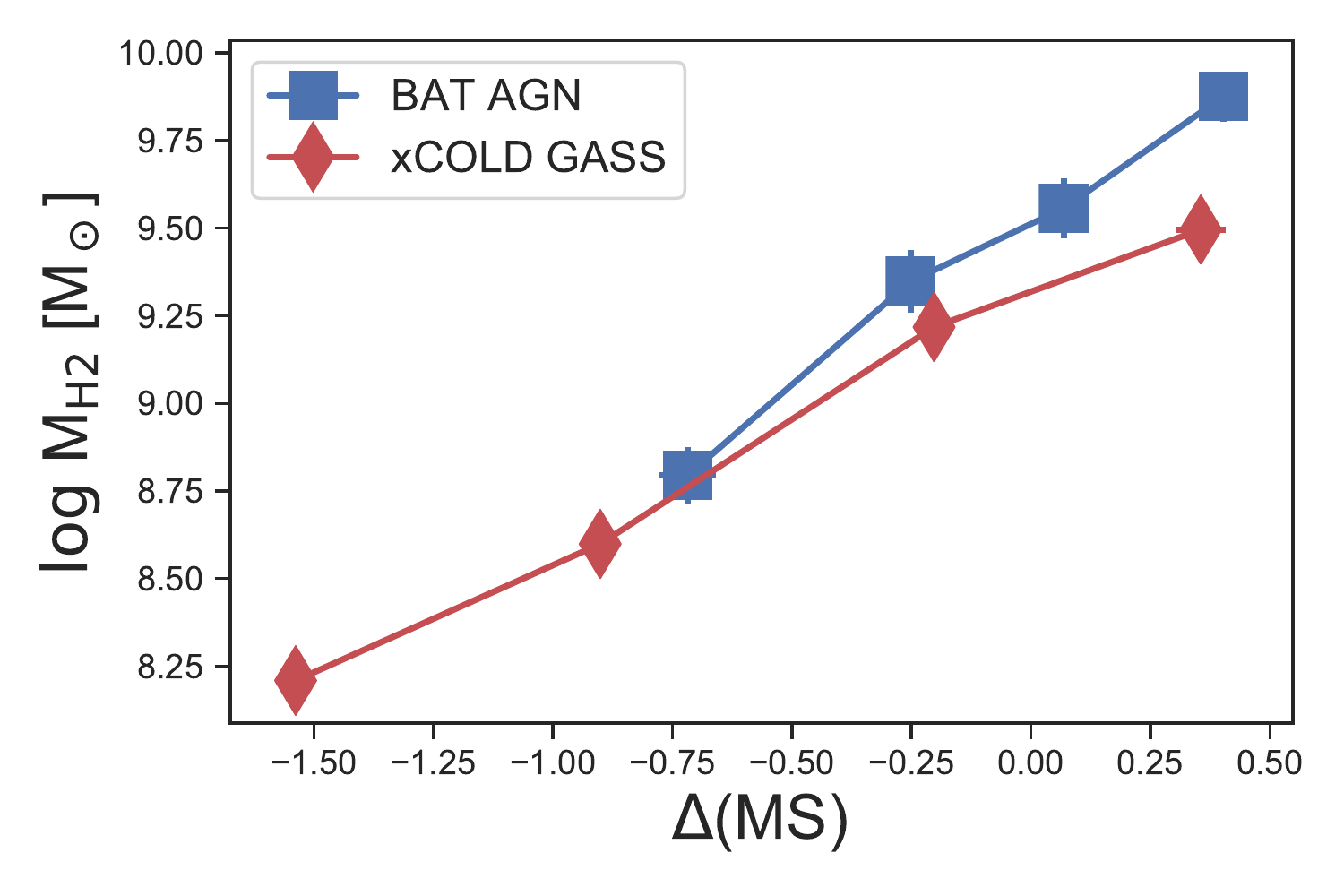}
\includegraphics[width=7.5cm]{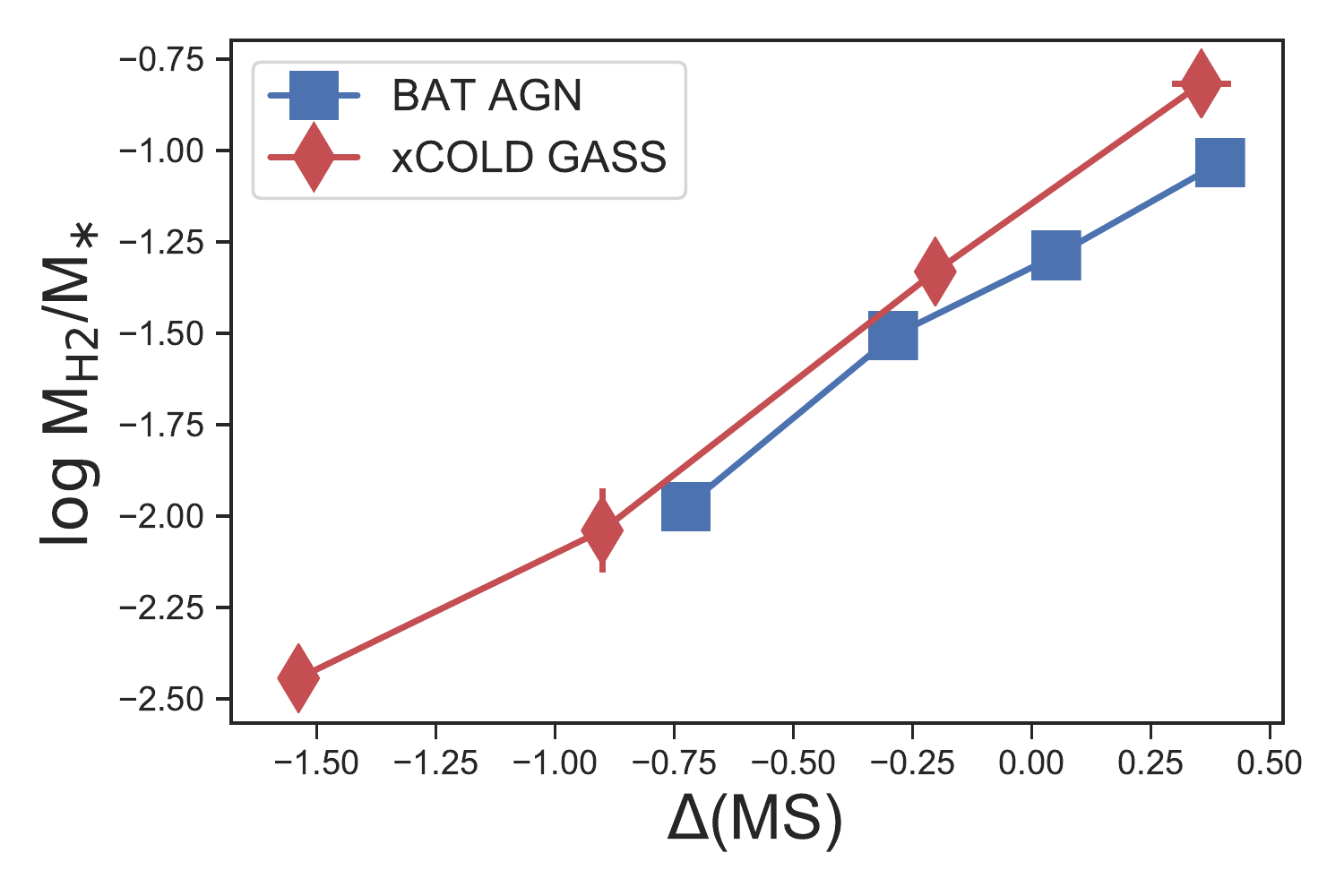}
\includegraphics[width=7.5cm]{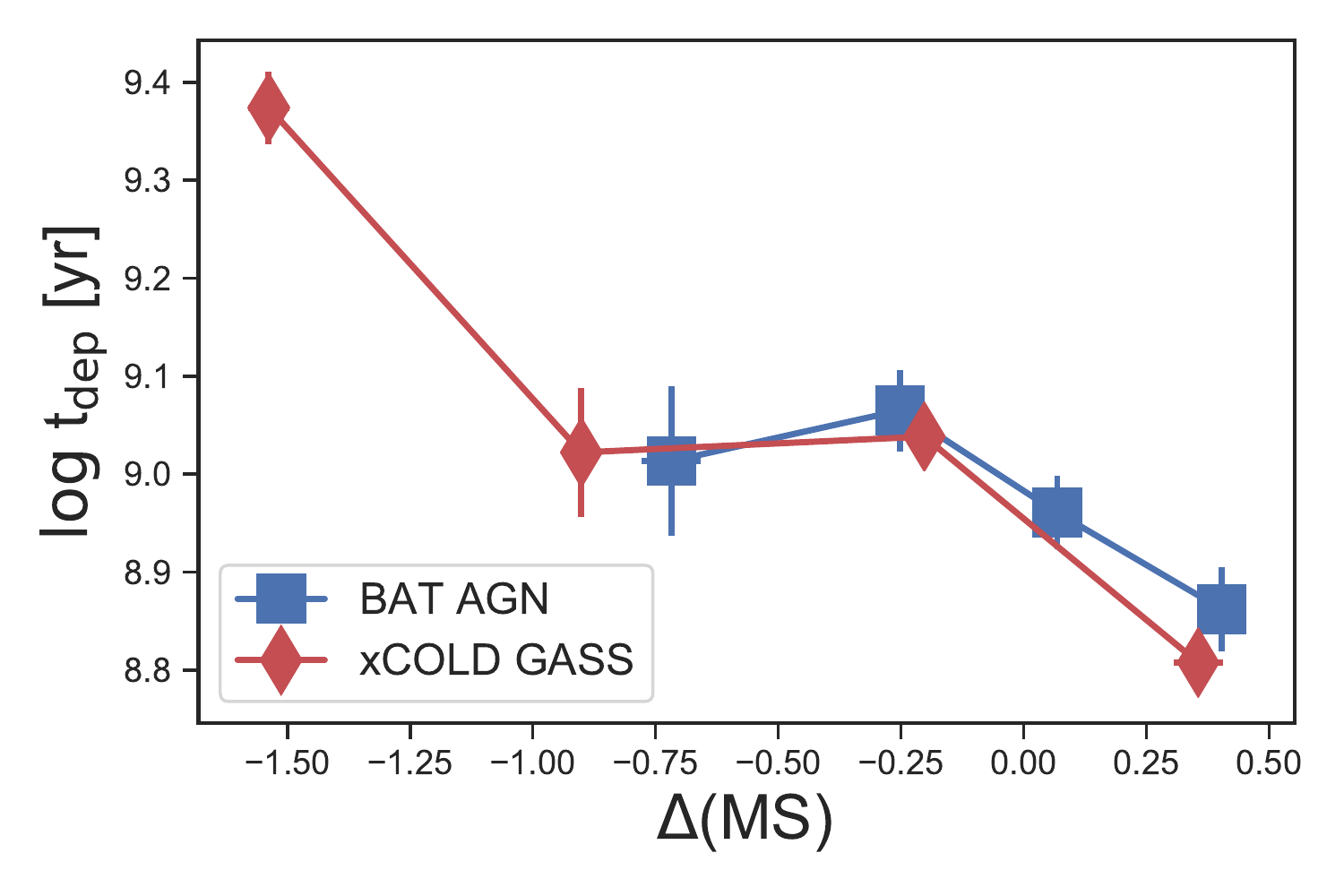}
\caption{
{Distribution of offsets from the MS for AGN hosts and inactive galaxies (upper left).  The offset, \deltams, is defined as the difference on a logarithmic scale between the observed SFR of a galaxy and it's expected SFR in relation to the MS from \citet{Renzini:2015:L29}.   Links between molecular gas (upper right), gas fraction (bottom left),  and depletion timescales (bottom right) in BAT AGN galaxies and xCOLD GASS inactive galaxies as compared to offset from the MS.   The symbols and binning procedure are identical to those in the right column of Figure \ref{co_all}.}}
\label{fig:msdist}
\end{figure*}

\subsection{Molecular gas content -- comparison to inactive galaxies}

As shown in Figure \ref{co_all}, our sample of BAT AGN galaxies strongly skews toward higher stellar masses compared to the xCOLD GASS sample of inactive galaxies   [$\log(M_*/\Msun){=}10.75{\pm}0.02$ vs. $\log(M_*/\Msun){=}10.58{\pm}0.02$, respectively].
Importantly, considering the regime where most BAT AGN galaxies reside ($\log(M_*/\Msun){>}10.5$), the median molecular gas mass and molecular gas fraction are systematically higher in BAT AGN galaxies than in the inactive galaxies, at fixed stellar mass ($p<0.001$; right panels of Figure \ref{co_all}).  Additionally, the BAT AGN galaxies typically lie on the MS of star forming galaxies compared to inactive galaxies in xCOLDGASS which lie below it for the majority of the sample ($p<0.001$, $\log(M_*/\Msun){>}10.2$).



In Figure \ref{fig:sfrdist} we present the molecular gas content of the BAT AGN galaxies and the xCOLD GASS inactive galaxies in the context of other star formation properties. The AGN galaxies clearly occupy a region of elevated SF (both in terms of SFR and sSFR).
In particular, passive galaxies --- i.e., those with $\log({\rm sSFR/yr}^{-1}){<}{-}11$ --- are a distinct minority among our BAT AGN galaxies (2--7\%, due to upper limits), whereas they comprise roughly half of the inactive galaxies (49\%).   
Over the limited range of SFRs where the two samples overlap ($0.2 {<} \log({\rm  SFR}/ \Msun\,{\rm yr}^{-1}) {<} 0.7$), the amount of molecular gas is not significantly different between the inactive galaxies and AGN galaxies.  The gas fractions, at a given sSFR, are however lower for AGN galaxies than for inactive galaxies ($p=0.004$).  
The gas depletion timescales [\tdep$\equiv M_{H2}$/SFR] for AGN galaxies are not significantly different.  

In Figure \ref{fig:msdist}, we show the molecular gas, gas fraction, and depletion timescale scales with offset from the MS as defined by \citep{Renzini:2015:L29}.  The BAT AGN tend to largely lie on this MS.  The xCOLDGASS inactive galaxies  have a much larger fraction of galaxies significantly below the MS consistent with their lower average SFR.  Over the limited range the two samples overlap in offset from the MS ($-0.5<$\deltams$<0.5$), the AGN galaxies have lower gas fractions ($p<0.01$), but not significantly different molecular gas masses or depletion timescales.

\subsubsection{Galaxy molecular gas luminosity function}

Another way to study how important molecular gas is to AGN activity is to compare the number of AGN galaxies with a given molecular gas mass for a given volume to the number of inactive galaxies.  {We do this by comparing the CO luminosity function from xCOLD GASS \citep[see Figure 2,][]{Fletcher:2020:arXiv:2002.04959}, which was constructed to measure the molecular gas of all galaxies of a given stellar mass ($\log(\mstar/\Msun)>9.0$.  The mass function was then created by normalizing to match the stellar mass function of local galaxies, and thus be representative of local galaxies.} The BAT AGN galaxy sample is then a subset of this local galaxy sample for all galaxies hosting an AGN above a certain X-ray luminosity across the sky.  {As the BAT sample is flux limited, we use the $V_{\rm max}$ correction method following \citet{Ananna:2020:17} to estimate the correction factor for each BAT AGN galaxy based on the observability of its intrinsic X-ray flux within the volume studied ($0.01<z<0.05$) and the sensitivity of BAT across the sky \citep{Baumgartner:2013:19}.  We focus on the 144 more luminous AGN galaxies (\Lsoftint${>}10^{42.8}$\ergps or L$_{\rm bol}{>}10^{44}$\ergps) in the sample where the survey sensitivity can detect sources in the majority of the volume surveyed and the $V_{\rm max}$ corrections are less uncertain.  We also correct for the unobserved BAT AGN galaxies (\perunobs) and the volume near the Galactic plane that was excluded ($b < 10^{\circ}$). } 

\begin{figure*} 
\centering
\includegraphics[width=14cm]{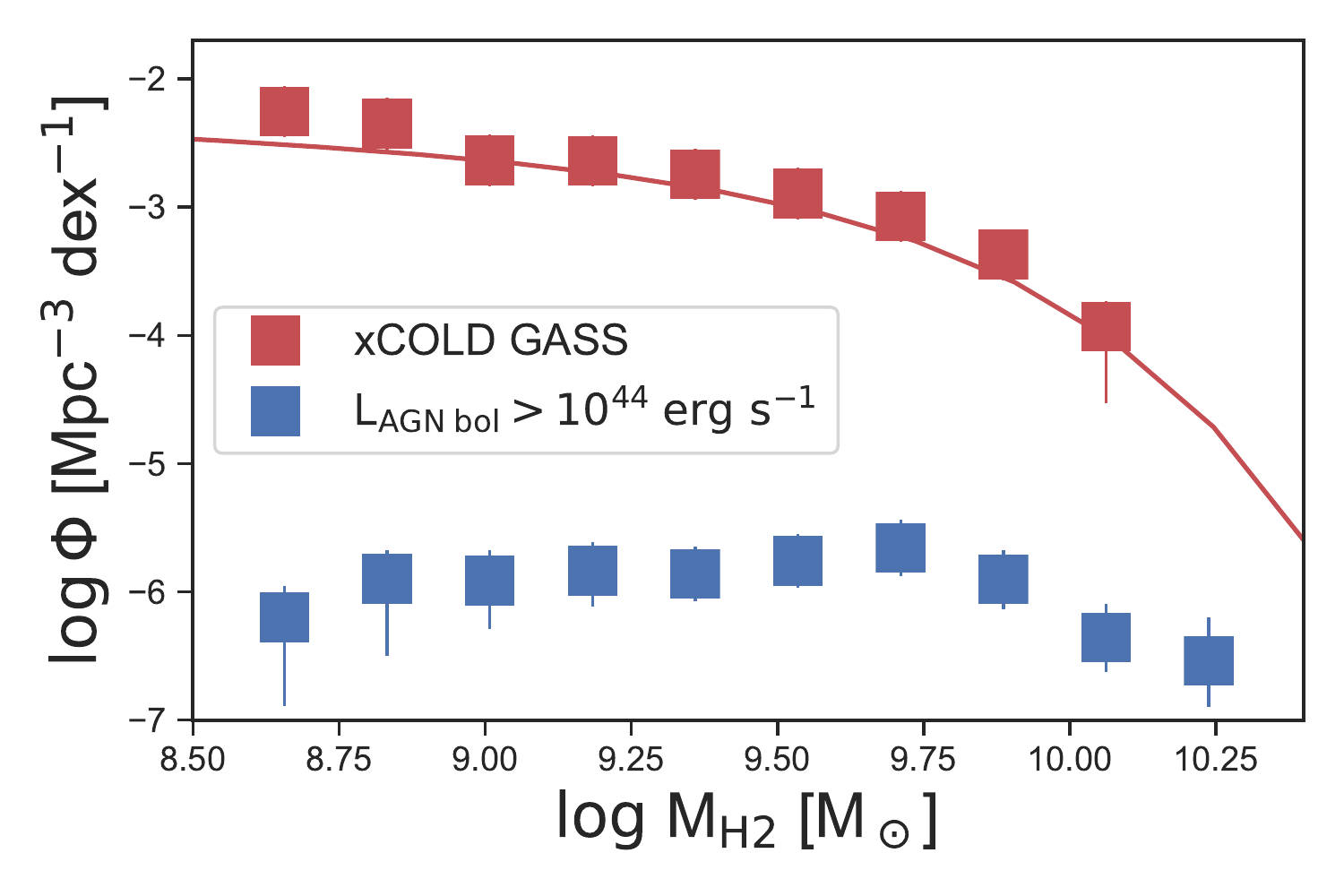}
\includegraphics[width=14cm]{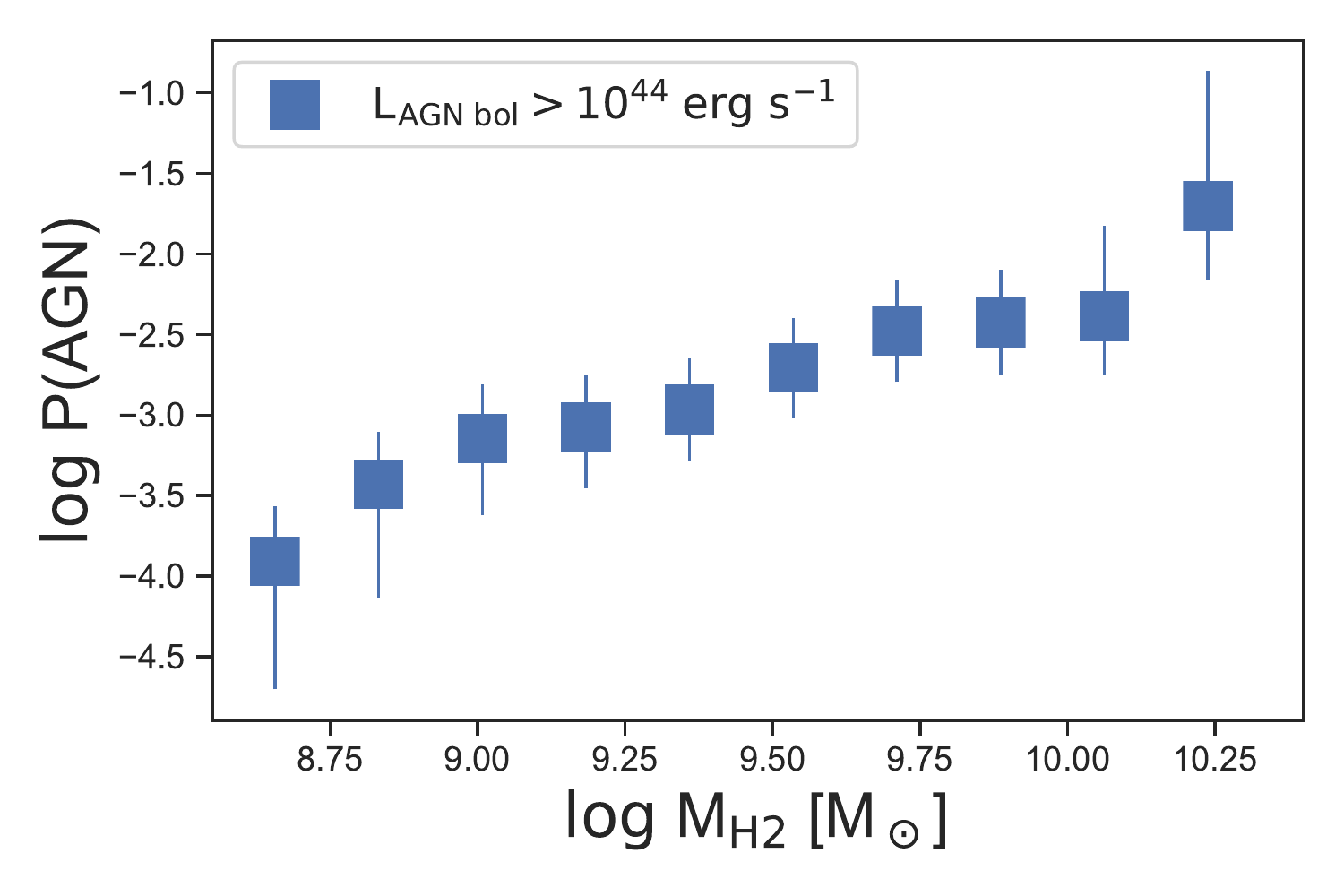}
\caption{Top: The H$_2$ mass function for xCOLD GASS \citep[see Figure 2,][]{Fletcher:2020:arXiv:2002.04959} compared to galaxies hosting a BAT AGN (\Lsoftint${>}10^{41.8}$\ergps or L$_{\rm bol}{>}10^{44}$\ergps).  Error bars are based on Poisson statistics and include the number of non-detections in the lower limits.  Only intervals with at least 5 sources per bin and with the majority detected are plotted.  A red line indicates the best fit Schechter function for xCOLD GASS.  Bottom: the likelihood of a galaxy of a given molecular mass hosting an AGN, based on dividing the H$_2$ mass functions.  The highest H$_2$ mass point (at $\log(\mh/\Msun)=10.23$) is based on an extrapolation of the best fit Schechter function because of the limited numbers of sources in xCOLD GASS.}
\label{fig:agn_likelihood}
\end{figure*}

The likelihood of a given galaxy hosting a luminous AGN clearly increases with gas mass (Figure \ref{fig:agn_likelihood}), but the absolute fraction maxes out at 1-10\% even among the most gas rich sources. Presumably key factors like the gas distribution, stochasticity, and ability to lose angular momentum play essential roles.

\subsubsection{Morphological Comparison}
We next look into the molecular gas properties of the BAT AGN hosts and the xCOLD GASS inactive galaxies as a function of their morphologies (spiral, elliptical, and `uncertain'), as shown in Figure \ref{morph}. For reference, we provide example images of elliptical, uncertain, and spiral host morphologies in Appendix \ref{morph_appen}.

Among spiral galaxies, the molecular gas masses and gas fractions are similar for both AGN galaxies and inactive galaxies, except at the highest stellar mass bins ($\log(M_*/\Msun){>}10.8$), where AGN hosts have significantly more molecular gas ($p=0.001$).  
For galaxies of uncertain morphologies, the difference between AGN  and inactive galaxies extends across all stellar masses ($p<0.001$) though the offset is not as large as the ellipticals. On the other hand, the difference in molecular gas content of BAT AGN ellipticals and xCOLD GASS ellipticals is profound across all stellar masses studied ($p<0.001$).  Roughly half (51\%, 21/41) of the BAT AGN galaxies with elliptical morphology hosts contain significant molecular gas mass reservoirs [$\log(\mh/\Msun)>9$], compared to almost no gas-rich ellipticals in the xCOLD GASS inactive sample (4\%, 3/81).

%
Both the median molecular gas mass [$\log(\mh/\Msun)=8.91^{+0.44}_{0.17}$ vs. $\log(\mh/\Msun)<8.4$] and the 
gas fraction [$\log \fgas = {-}1.71^{+0.24}_{0.38}$ vs. $\log\fgas = <2.71$] are significantly higher for elliptical hosts of AGN than for inactive ellipticals ($p<0.001$). While the gas depletion timescales [$\log(\tdep/{\rm yr}){=}8.88^{+0.15}_{0.12}$ vs. $\log(\tdep/{\rm yr}){<}8.5$] appear to be lower for inactive galaxies ($p=0.001$), though we find this result is not significant if the upper limits are increased by 30\%.  

{Finally, we note that if a stricter criteria were used for `uncertain', the general trends would remain with small differences.  The offset in molecular gas properties would be more significant for AGN galaxies in spirals compared to inactive galaxies.  Conversely, the size of the offset would be somewhat less for AGN galaxies in ellipticals.}


\begin{figure*} 
\centering
\includegraphics[width=8cm]{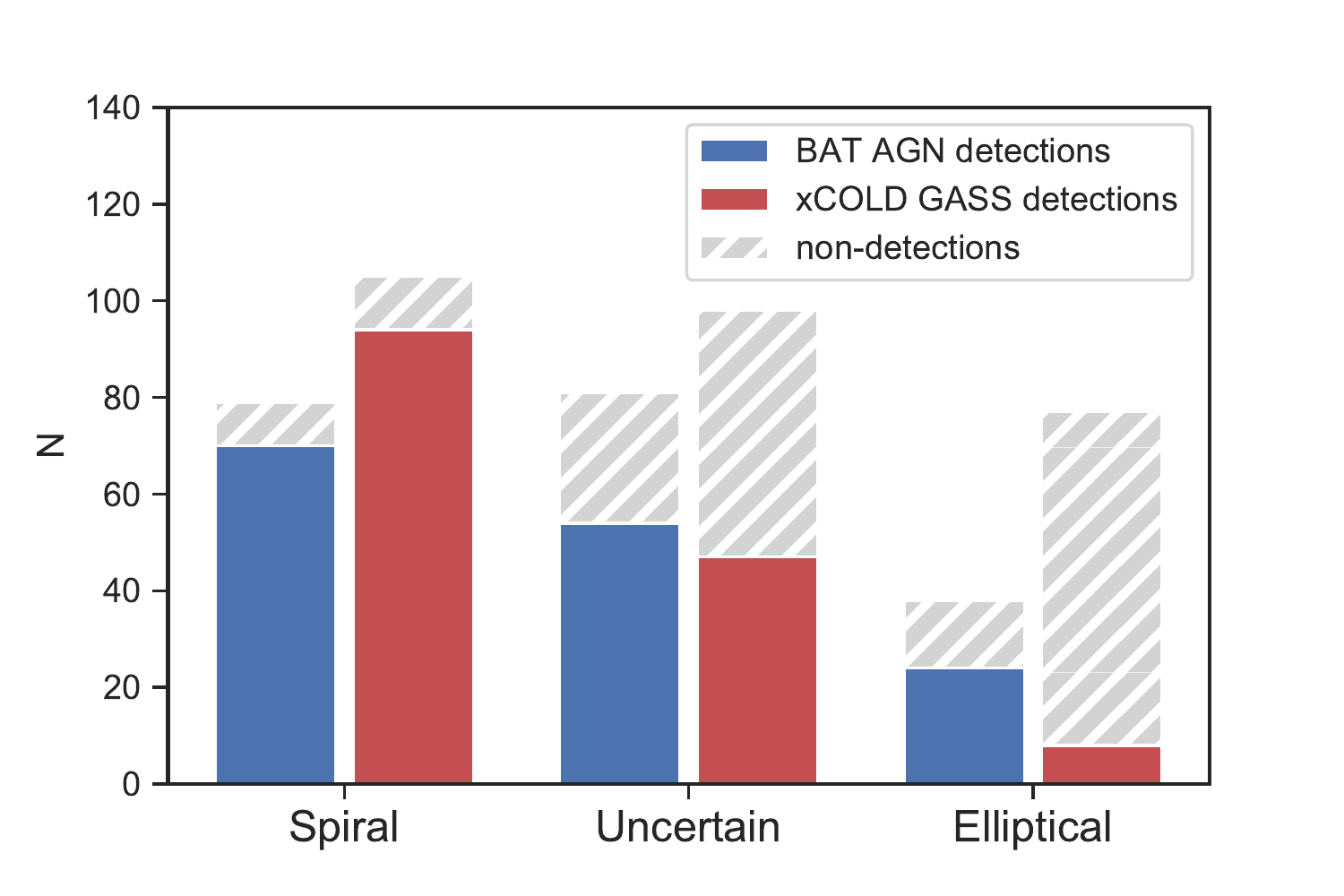}\\
\includegraphics[width=7.3cm]{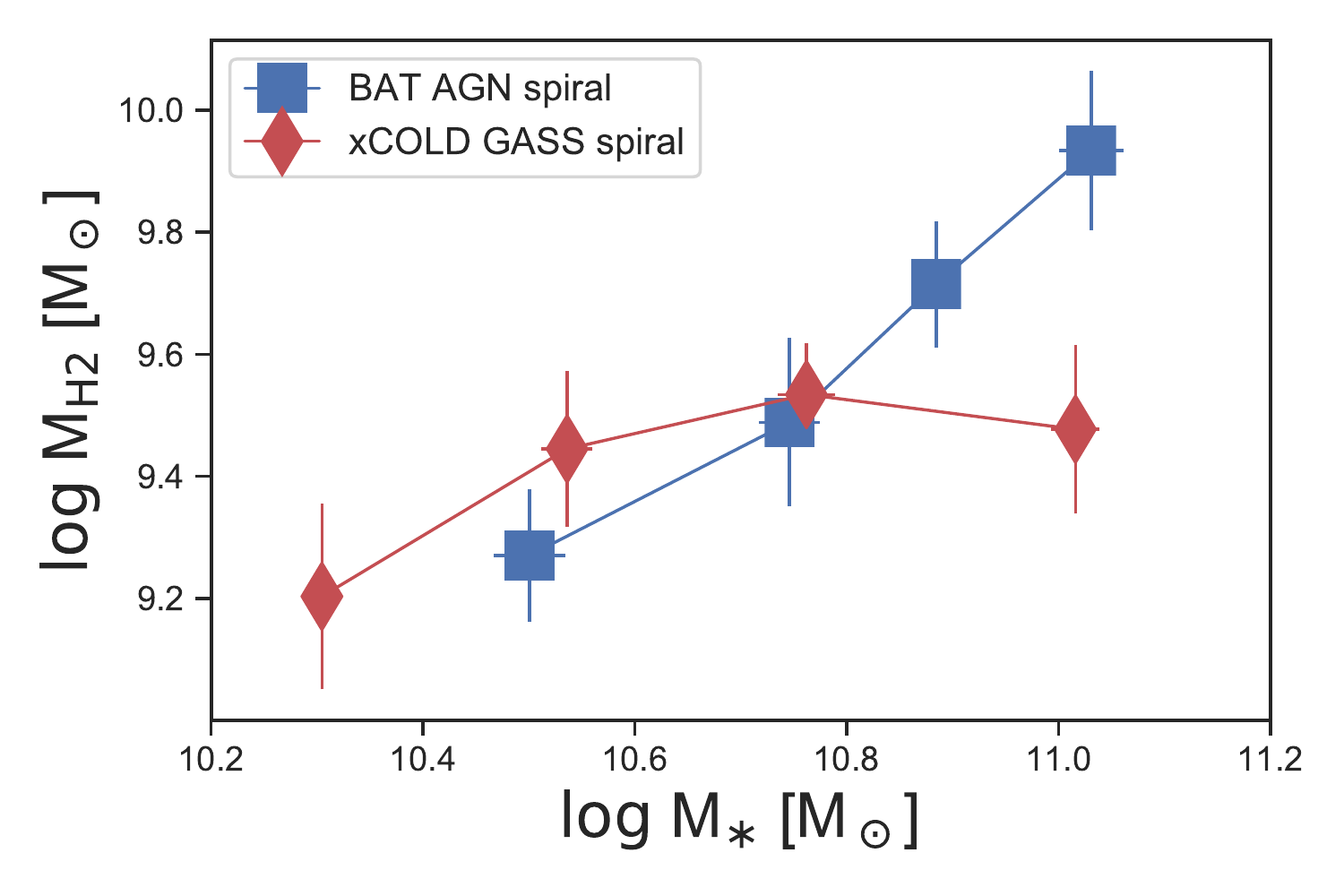}
\includegraphics[width=7.3cm]{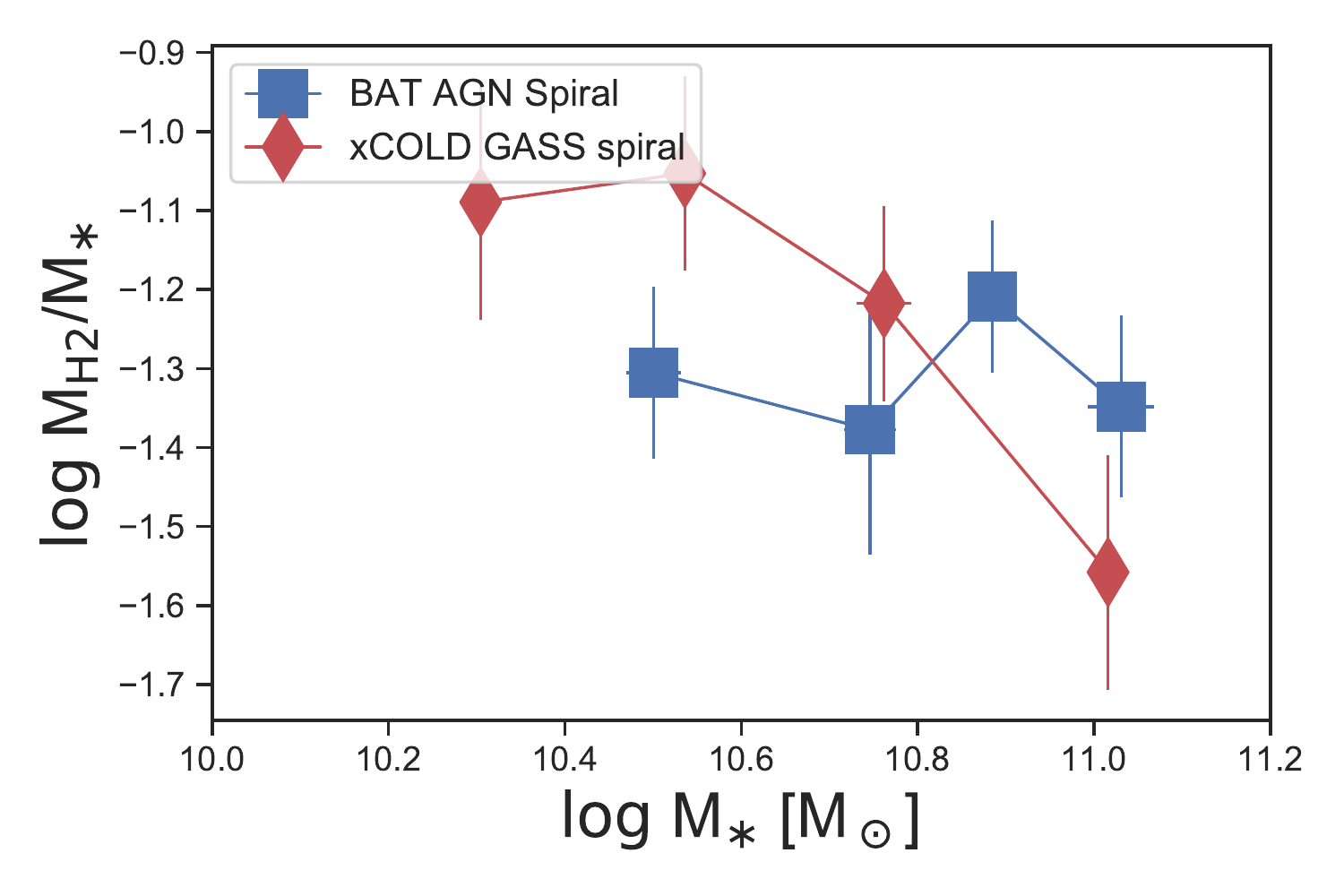}
\includegraphics[width=7.3cm]{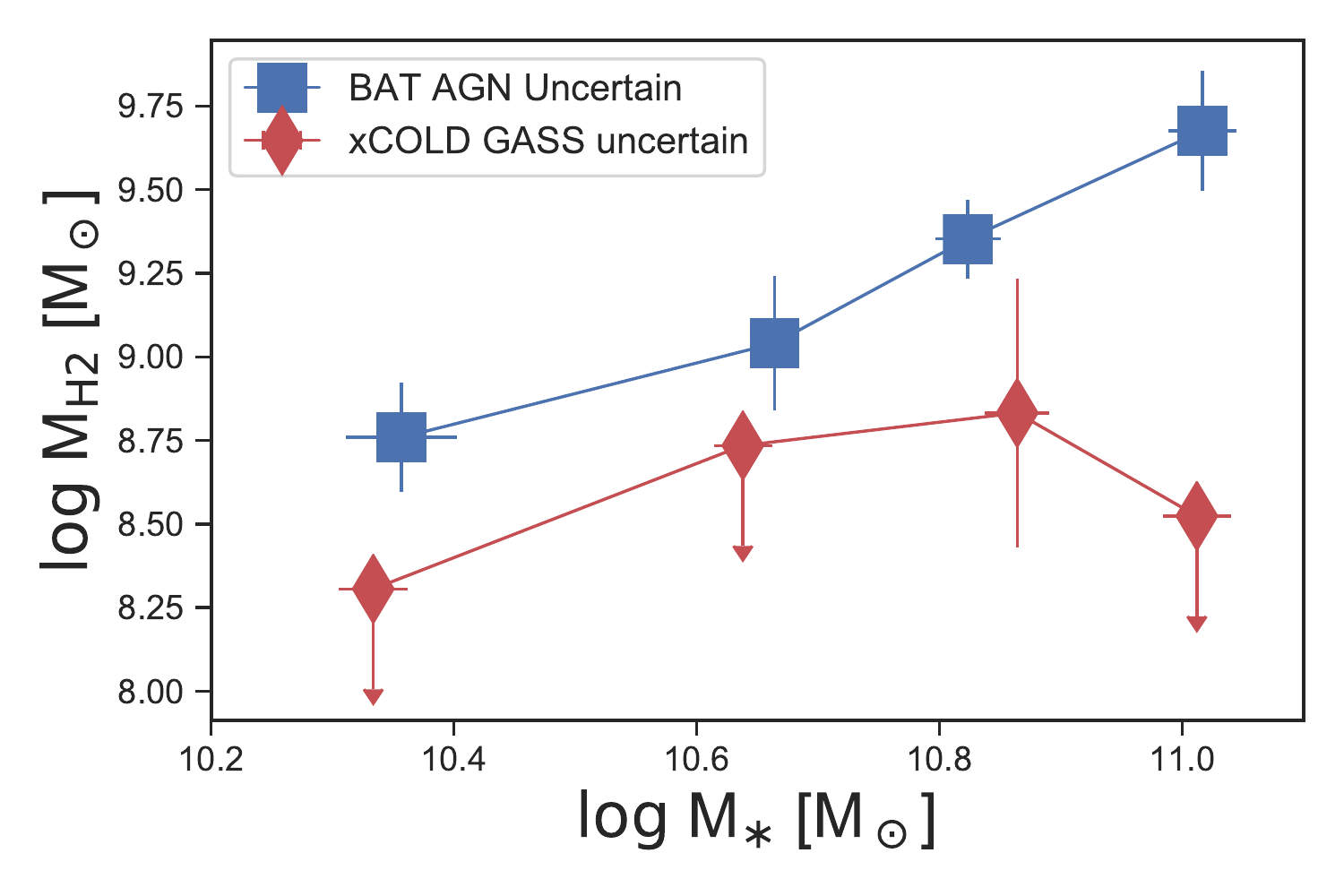}
\includegraphics[width=7.3cm]{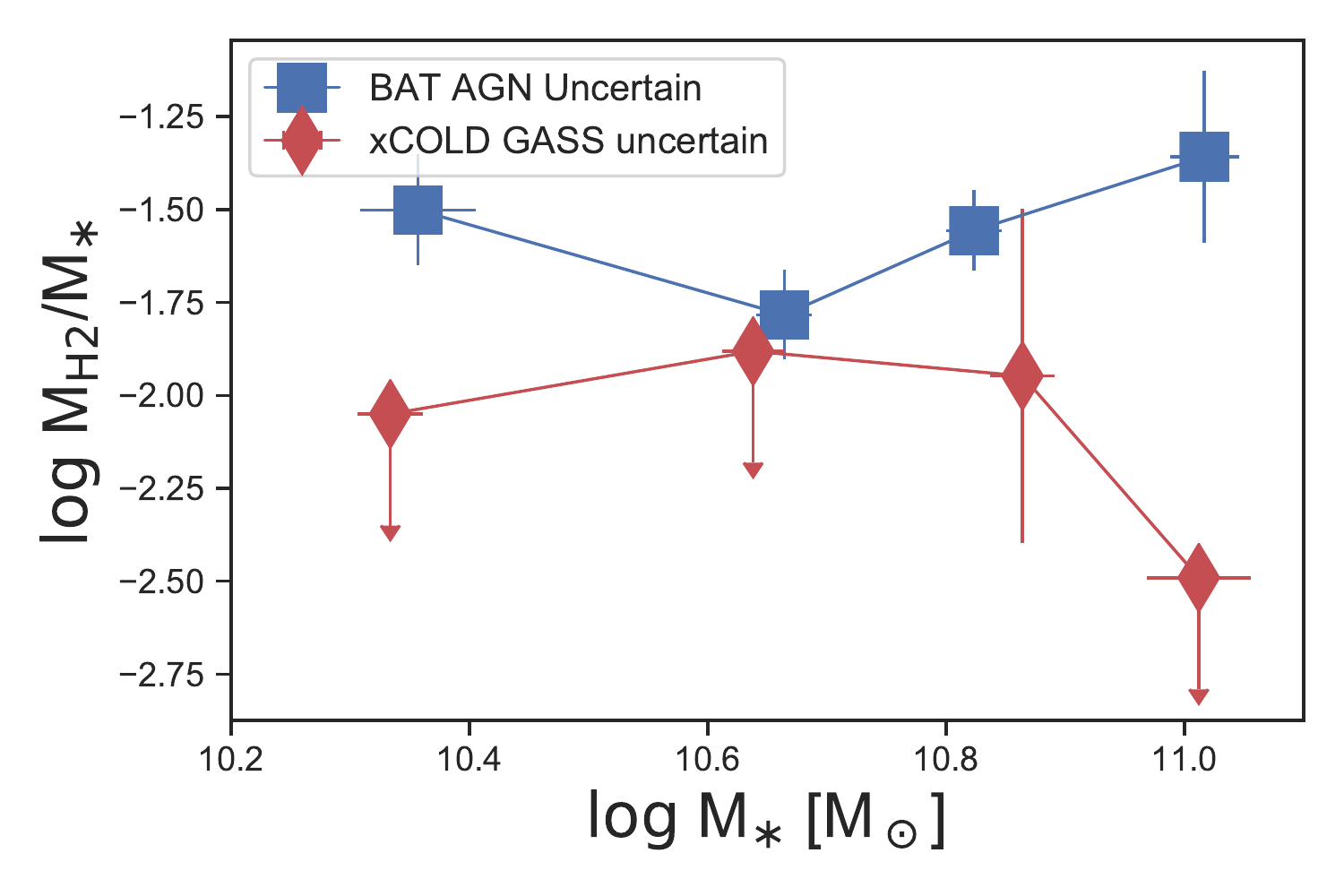}
\includegraphics[width=7.3cm]{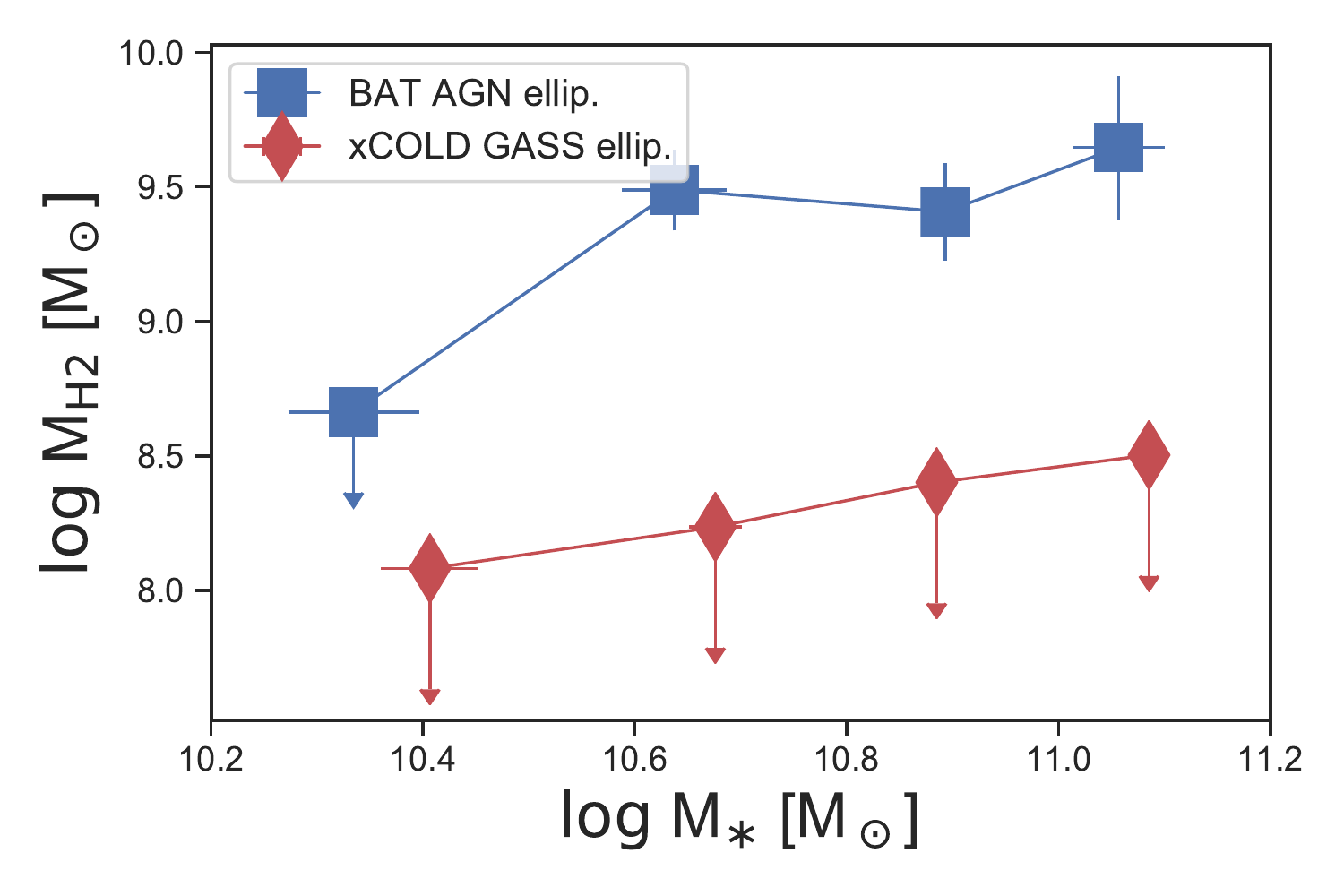}
\includegraphics[width=7.3cm]{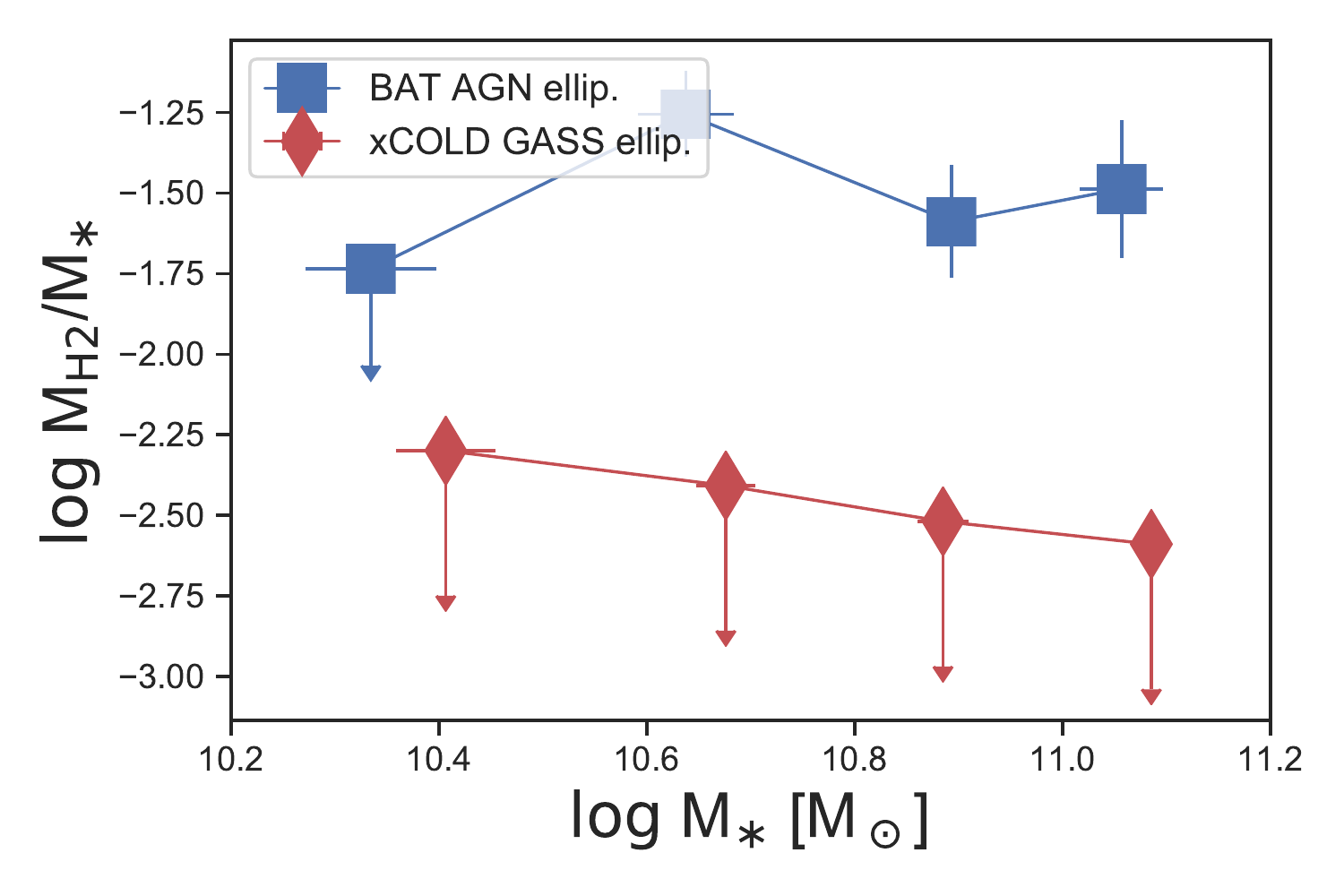}
\caption{Links between molecular gas content and galaxy morphology, for the BAT AGN galaxies and the xCOLD GASS inactive galaxy sample.
Top left panel: distributions of galaxy morphology.  
For each sample and morphology class, grey bars indicate sources with CO(2--1) upper limits.  
The other panels show the median molecular gas mass and gas fraction vs. stellar mass, for both AGN hosts and xCOLD GASS inactive galaxies, with either spiral, uncertain, or elliptical morphologies (second, third, and bottom row, respectively).
Upper limits are shown when more than half the sample is upper limits. Note that more than half of all inactive galaxies with elliptical or uncertain morphologies have no robust CO detections.}
\label{morph}
\end{figure*}

\subsection{Molecular gas content -- links to AGN galaxy properties}

Here we examine the molecular gas properties of our BAT AGN galaxy sample as a function of various AGN- and SMBH-related properties, such as optical AGN classification, intrinsic X-ray luminosity, black hole mass, Eddington ratio, and line-of-sight column density.  A summary of the median parameters and confidence limits from survival analysis for these parameters can be found in table \ref{samplemedians}.

\subsubsection{Optical AGN Classification}

We use the public BASS/DR1 and proprietary DR2 optical spectroscopy available for our AGN to determine their optical, spectroscopic AGN classification.
We classify sources that have both broad \halpha\ and \hbeta\ emission lines as Seyfert 1s. 
This includes sources that are often referred to as Type 1.0, 1.5, 1.8 etc. 
Sources with broad \halpha\ but only narrow \hbeta\ are classified as Seyfert 1.9s, and finally Seyfert 2s have only narrow Balmer lines.  
Our BAT AGN sample breaks down as 40\% (83/\Nanalyzed) Seyfert 1s, 16\% (34/\Nanalyzed) Seyfert 1.9s, and 43\% (90/\Nanalyzed) Seyfert 2s.

\begin{figure*} 
\centering
\includegraphics[width=16cm]{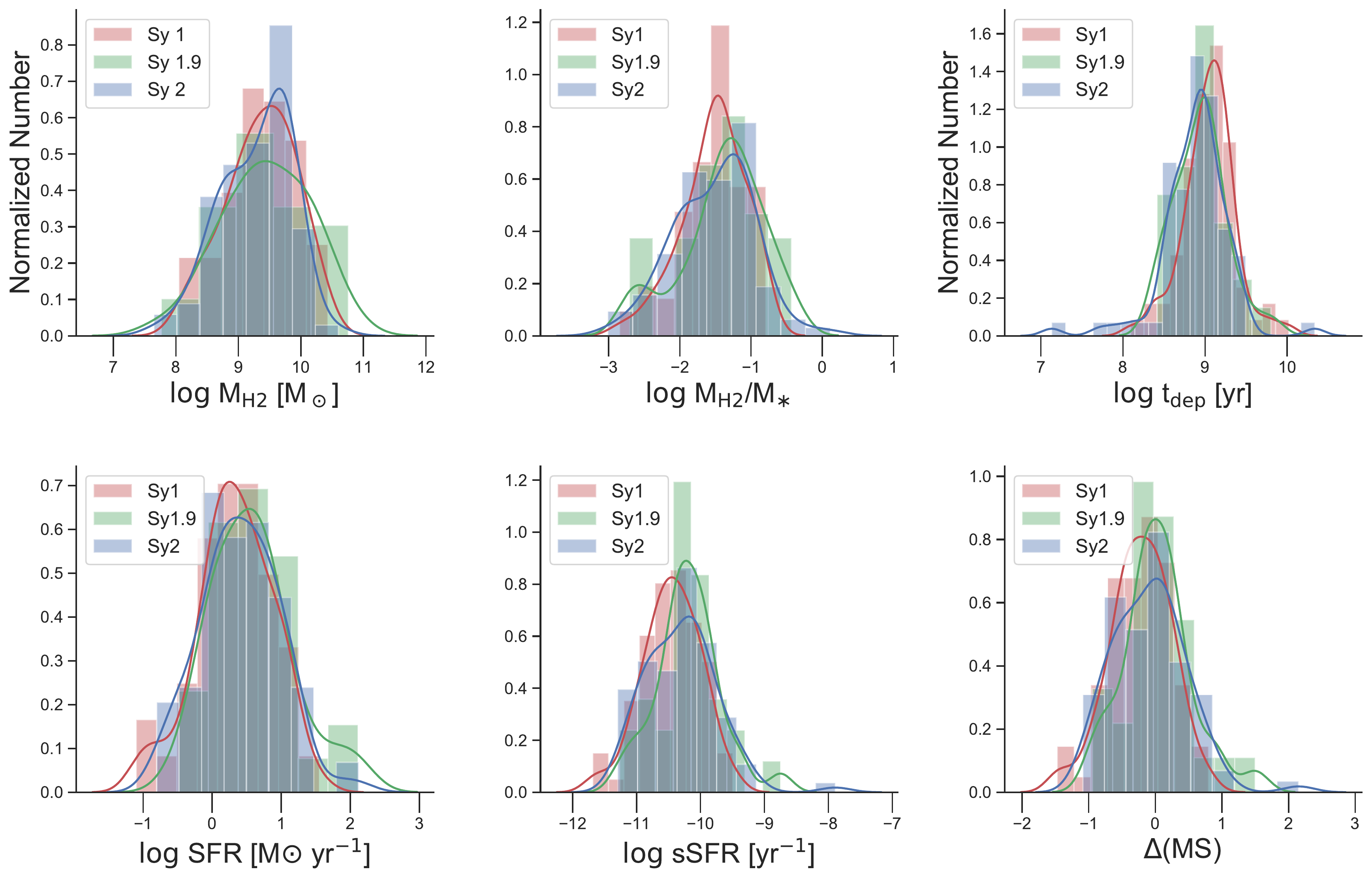}
\caption{Molecular gas properties for different optical AGN galaxy sub-classes.
Distributions of molecular gas mass (upper left), gas fraction (upper right), depletion timescale (lower left), and sSFR (lower right), colour-coded by the optical AGN classification. 
Sources that have both broad  \halpha\ and \hbeta\ (red) are considered `Sy 1' (e.g. type 1.0, 1.5, 1.8 etc.); 
sources with broad \halpha\ but only narrow \hbeta\ are `Sy 1.9' (green); 
and finally `Sy 2' have only narrow Balmer lines. {Lines represent Kernel Density Estimates for the distributions. }
Seyfert 1.9s are found to have higher gas fractions,
otherwise the distributions are statistically indistinguishable.}
\label{fig:agntype}
\end{figure*}

Figure \ref{fig:agntype} shows the distributions of molecular gas mass, molecular gas mass fraction, the associated depletion timescales, SFRs, sSFRs, and offset from the MS.  Overall the distributions are similar.  However, we do find the depletion timescales of Sy 2 galaxies are on average lower than Sy 1 galaxies ($p=0.002$).  

\subsubsection{Line-of-sight column density and molecular gas}

We next investigate the possible links between two completely independent probes of the (cold) gas content in our BAT AGN galaxies, which in principle originate from very different scales: the line-of-sight column density, \nh\ -- as determined from detailed spectral modeling of the X-ray SEDs \citep{Ricci:2017:17}, and the new molecular gas measurements.

\begin{figure*} 
\centering
\includegraphics[width=8cm]{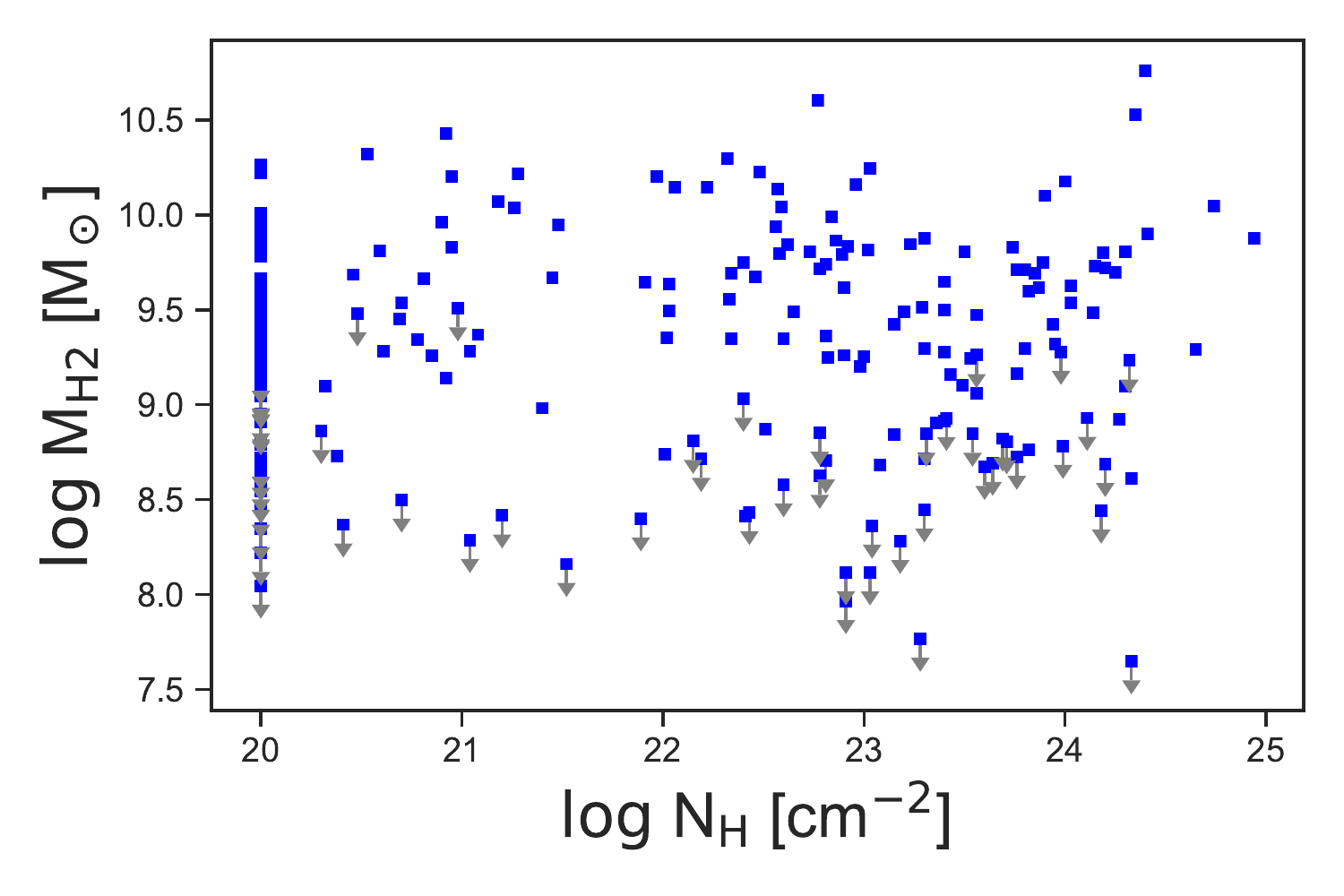}
\includegraphics[width=8cm]{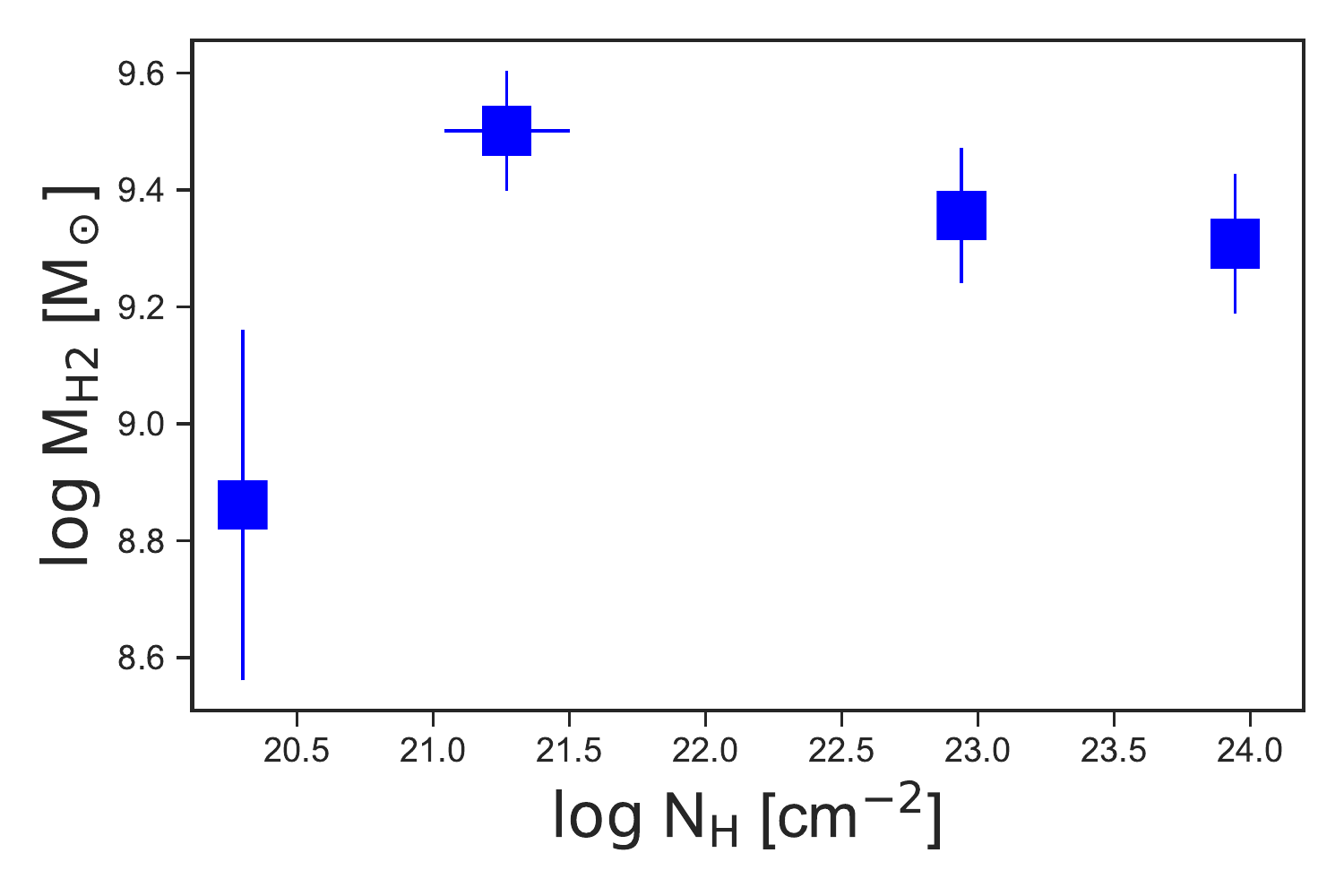}
\includegraphics[width=8cm]{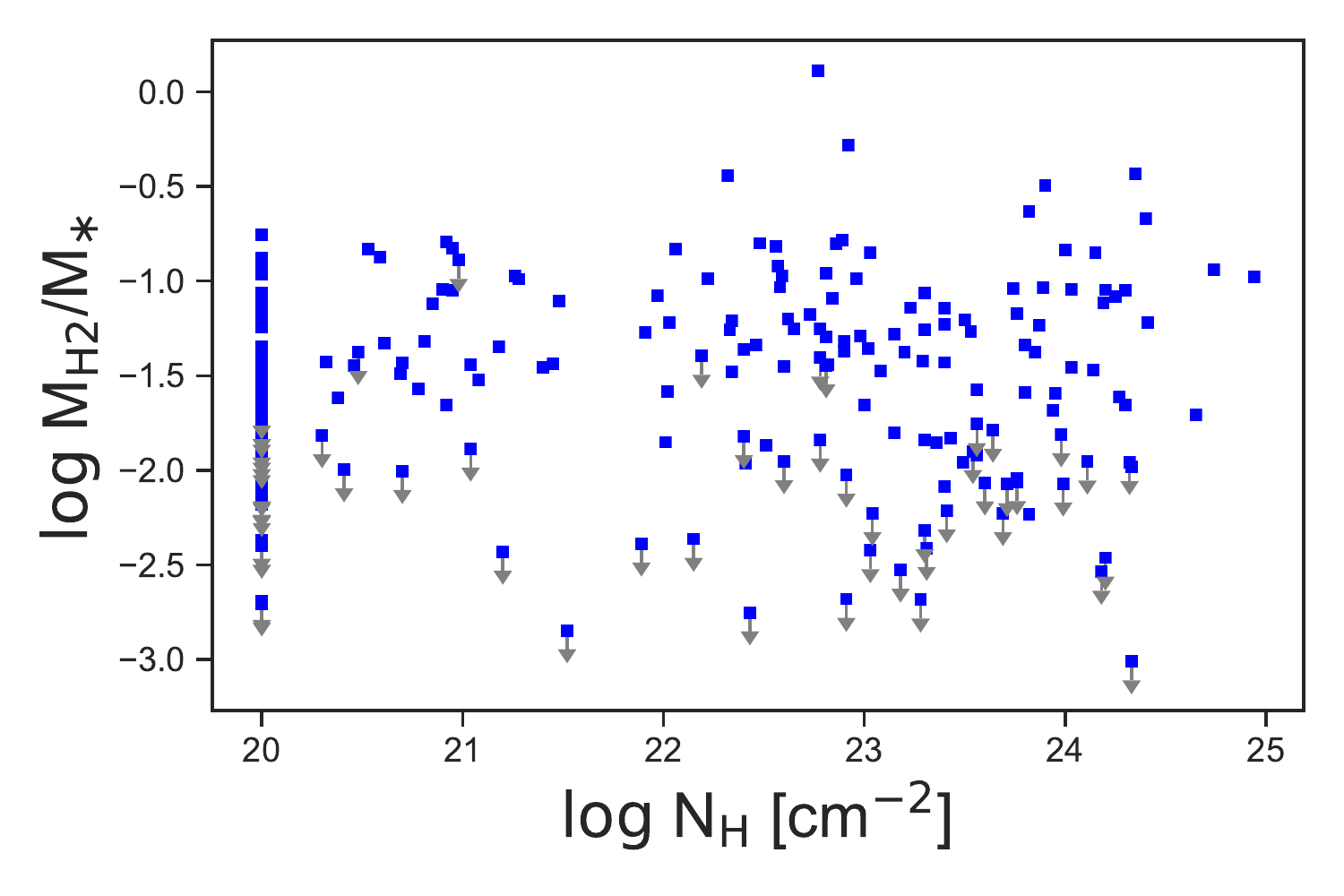}
\includegraphics[width=8cm]{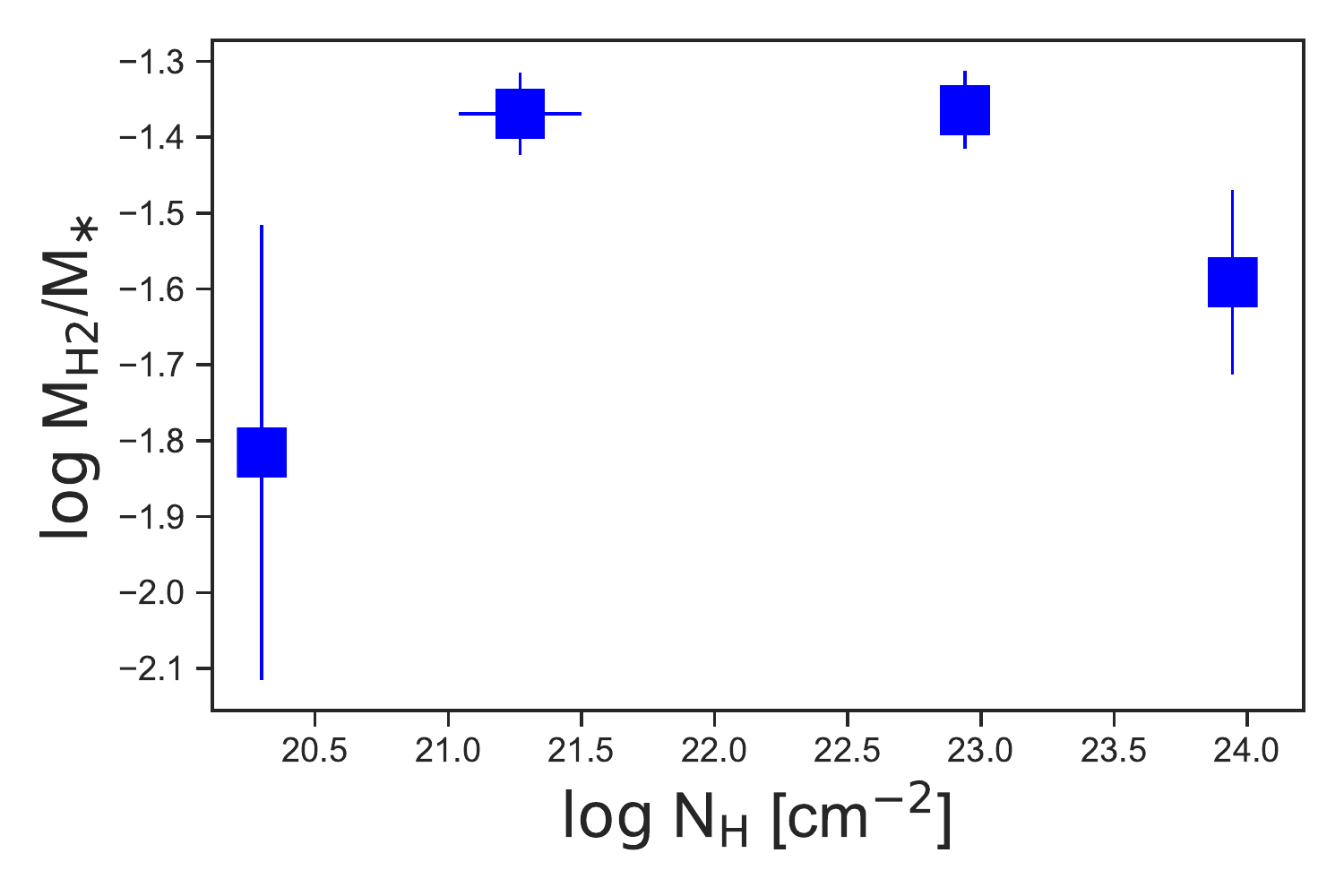}
\includegraphics[width=8cm]{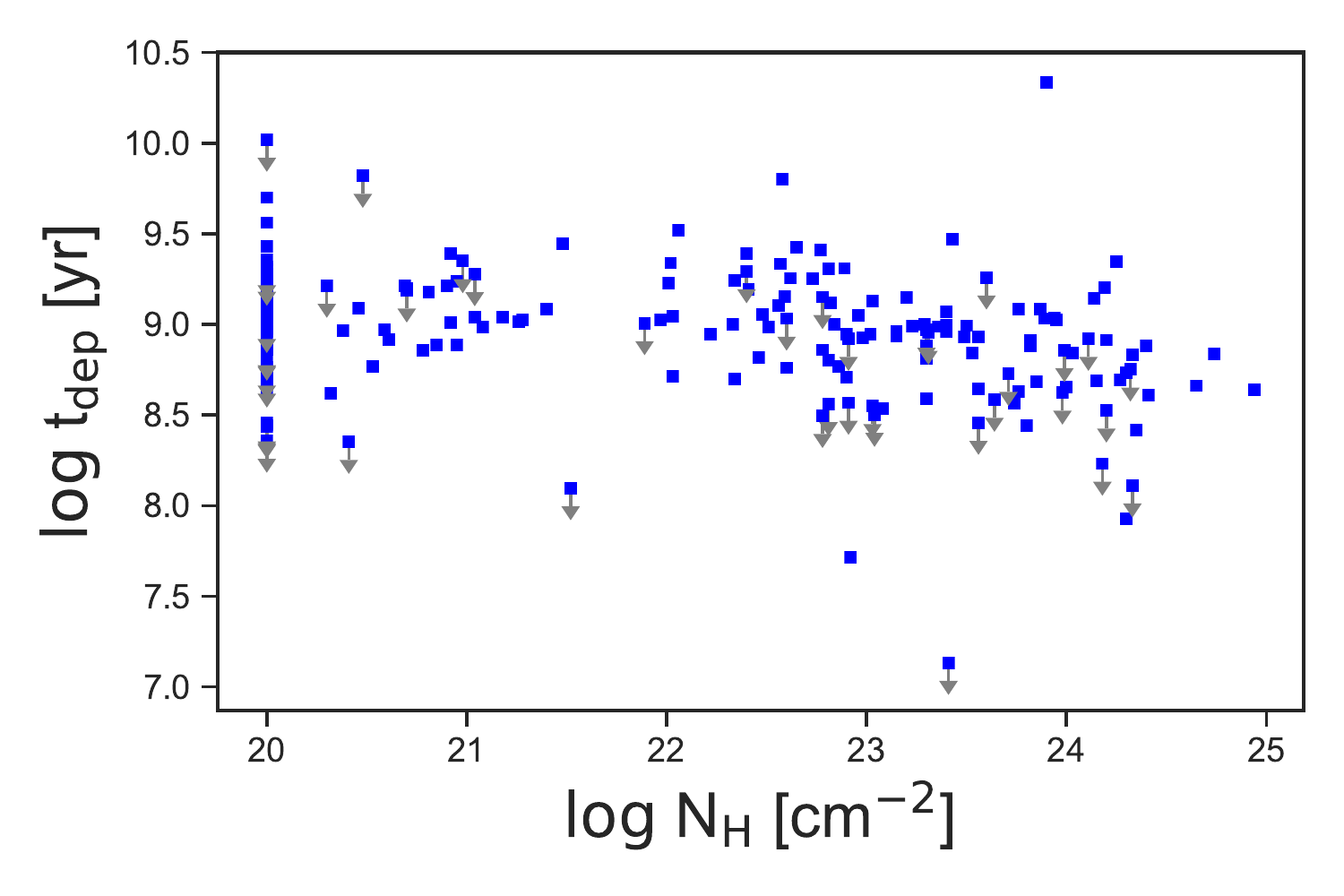}
\includegraphics[width=8cm]{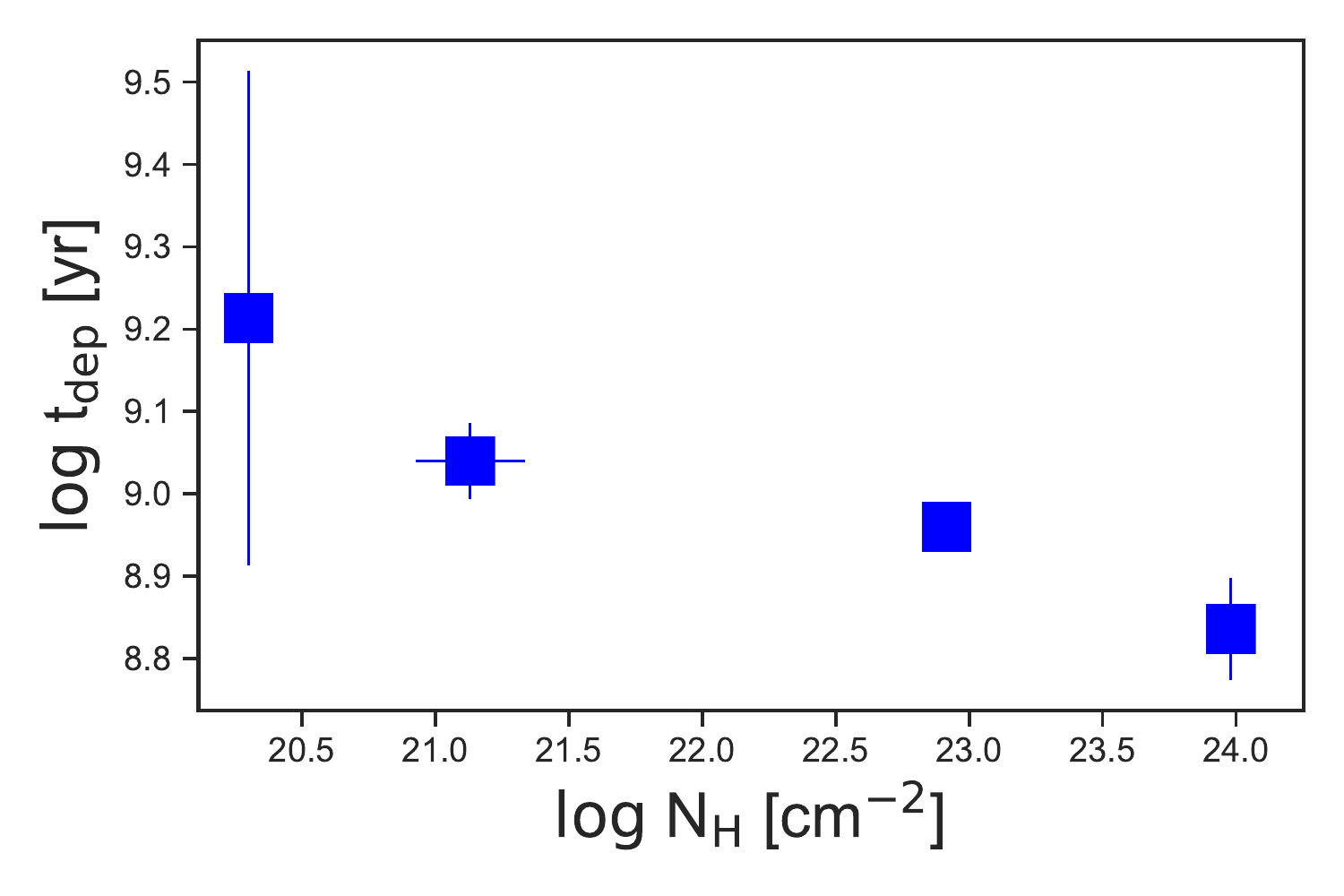}
\caption{Links between molecular gas properties and the X-ray line-of-sight column density. The different panels show the molecular gas mass (top), molecular gas fraction (center), and depletion timescale (bottom).  {52 AGN have a column density consistent with a lower limit (\nh$=10^{20}$ \nhunit)}. 
The scales of the median plots (right column) are smaller than those of the individual measurements (left column), to better highlight the subtle trends seen (or not seen).}
\label{tab:NHmolec}
\end{figure*}

In Figure \ref{tab:NHmolec} we plot \nh\ vs. molecular gas mass, gas fraction, and depletion timescales, both for individual sources (left panels) and binned (right panels). 

There are no significant trends between \nh\ and molecular gas mass. 
However, similar to the Seyfert 2 sample, a lower depletion timescale is found for the absorbed systems ($p=0.001$), which is most pronounced in the most absorbed systems ($\log(\nh/\psqcm)>23.4$).

\begin{table}
\centering
\small{}
\caption{Summary of median parameters for molecular gas and host galaxy parameters of AGN galaxies.}
\begin{tabular}{lccccccc}
\hline
&$\log M_{H2}$&$\log M_{\rm H2}/\Msun$&$\log$ \tdep&$\log$ SFR&$\log$ sSFR&\deltams&$\log$ M$_{\star}$\\
&$\log$ (\Msun)&&$log$ (yr)&$\log$ (M$_\odot$ yr$^{-1}$)&$\log$(yr$^{-1}$)&&$\log$(M$_\odot$)\\
\hline\hline&&&&&&&\\
\bf AGN Type&&&&&&&\\
Sy 1&$9.35^{+0.18}_{0.08}$&$-1.47^{+0.04}_{0.07}$&$9.02^{+0.06}_{0.05}$&$0.33^{+0.14}_{0.11}$&$-10.45^{+0.10}_{0.07}$&$-0.19^{+0.09}_{0.12}$&$10.80^{+0.07}_{0.03}$\\
Sy 1.9&$9.37^{+0.43}_{0.07}$&$-1.26^{+0.06}_{0.22}$&$8.98^{+0.04}_{0.15}$&$0.54^{+0.25}_{0.15}$&$-10.25^{+0.22}_{0.10}$&$0.00^{+0.16}_{0.18}$&$10.79^{+0.09}_{0.08}$\\
Sy 2&$9.36^{+0.13}_{0.12}$&$-1.42^{+0.08}_{0.23}$&$8.88^{+0.05}_{0.07}$&$0.40^{+0.11}_{0.16}$&$-10.32^{+0.12}_{0.13}$&$-0.08^{+0.10}_{0.19}$&$10.75^{+0.03}_{0.07}$\\
$p<0.05$&&&\bf $p=0.002$&&&&\\
\hline  $\bm{\log(\nh/\psqcm)}$ &&&&&&&\\
$<20.3$&$9.31^{+0.21}_{0.13}$&$-1.53^{+0.07}_{0.11}$&$9.05^{+0.06}_{0.09}$&$0.30^{+0.16}_{0.11}$&$-10.47^{+0.09}_{0.11}$&$-0.24^{+0.12}_{0.10}$&$10.80^{+0.11}_{0.04}$\\
$20.3-22.5$&$9.50^{+0.17}_{0.15}$&$-1.43^{+0.11}_{0.03}$&$9.02^{+0.02}_{0.04}$&$0.39^{+0.10}_{0.14}$&$-10.39^{+0.15}_{0.06}$&$-0.15^{+0.15}_{0.13}$&$10.79^{+0.08}_{0.04}$\\
$22.5-23.4$&$9.35^{+0.16}_{0.10}$&$-1.37^{+0.11}_{0.07}$&$8.94^{+0.05}_{0.09}$&$0.34^{+0.21}_{0.21}$&$-10.21^{+0.09}_{0.12}$&$-0.07^{+0.17}_{0.11}$&$10.64^{+0.07}_{0.05}$\\
$>23.4$&$9.32^{+0.28}_{0.16}$&$-1.59^{+0.25}_{0.12}$&$8.73^{+0.11}_{0.07}$&$0.53^{+0.16}_{0.18}$&$-10.27^{+0.15}_{0.25}$&$-0.06^{+0.20}_{0.21}$&$10.81^{+0.07}_{0.04}$\\
$p<0.05$&&&\bf $p<0.001$&&\bf $p=0.034$&&\\
\hline $\bm{\log(\Lsoftint/\ergpersec)}$&&&&&&&\\
$<42.8$&$9.28^{+0.15}_{0.07}$&$-1.44^{+0.13}_{0.08}$&$8.99^{+0.05}_{0.05}$&$0.28^{+0.19}_{0.13}$&$-10.32^{+0.12}_{0.14}$&$-0.18^{+0.20}_{0.10}$&$10.60^{+0.08}_{0.06}$\\
$42.8-43.2$&$9.34^{+0.19}_{0.20}$&$-1.49^{+0.11}_{0.08}$&$8.93^{+0.05}_{0.09}$&$0.34^{+0.13}_{0.21}$&$-10.40^{+0.12}_{0.16}$&$-0.19^{+0.10}_{0.17}$&$10.75^{+0.07}_{0.06}$\\
$43.2-43.4$&$9.50^{+0.20}_{0.15}$&$-1.34^{+0.13}_{0.23}$&$8.96^{+0.19}_{0.12}$&$0.55^{+0.15}_{0.19}$&$-10.18^{+0.07}_{0.17}$&$-0.03^{+0.16}_{0.10}$&$10.78^{+0.11}_{0.02}$\\
$>43.4$&$9.48^{+0.18}_{0.21}$&$-1.47^{+0.04}_{0.12}$&$8.88^{+0.14}_{0.22}$&$0.43^{+0.20}_{0.12}$&$-10.44^{+0.15}_{0.08}$&$-0.18^{+0.17}_{0.12}$&$10.88^{+0.11}_{0.05}$\\
$p<0.05$&&&&&&&\bf $p=0.001$\\
\hline  $\bm{\log(\lledd)}$&&&&&&&\\
$<-2$&$9.29^{+0.13}_{0.05}$&$-1.52^{+0.19}_{0.11}$&$8.99^{+0.01}_{0.05}$&$0.40^{+0.11}_{0.27}$&$-10.35^{+0.16}_{0.22}$&$-0.14^{+0.20}_{0.19}$&$10.78^{+0.09}_{0.08}$\\
$-2$ to $-1.6$&$9.35^{+0.13}_{0.10}$&$-1.48^{+0.04}_{0.20}$&$8.94^{+0.10}_{0.03}$&$0.37^{+0.12}_{0.17}$&$-10.42^{+0.13}_{0.09}$&$-0.20^{+0.11}_{0.11}$&$10.80^{+0.08}_{0.05}$\\
$-1.6$ to $-1.1$&$9.50^{+0.20}_{0.23}$&$-1.40^{+0.15}_{0.13}$&$9.02^{+0.06}_{0.14}$&$0.35^{+0.18}_{0.11}$&$-10.35^{+0.13}_{0.17}$&$-0.12^{+0.07}_{0.18}$&$10.81^{+0.07}_{0.07}$\\
$>-1.1$&$9.56^{+0.10}_{0.22}$&$-1.43^{+0.11}_{0.02}$&$8.88^{+0.07}_{0.08}$&$0.50^{+0.18}_{0.16}$&$-10.17^{+0.12}_{0.16}$&$0.07^{+0.09}_{0.22}$&$10.74^{+0.04}_{0.14}$\\
$p<0.05$&\bf $p=0.037$&\bf $p=0.042$&&&&&\\
\hline $\bm{\log (\mbh/\Lsun)}$&&&&&&&\\
$<7.4$&$9.29^{+0.24}_{0.18}$&$-1.43^{+0.14}_{0.04}$&$8.91^{+0.05}_{0.11}$&$0.33^{+0.16}_{0.09}$&$-10.23^{+0.12}_{0.15}$&$-0.09^{+0.24}_{0.10}$&$10.55^{+0.12}_{0.10}$\\
$7.4-7.8$&$9.45^{+0.18}_{0.20}$&$-1.37^{+0.14}_{0.10}$&$9.00^{+0.08}_{0.03}$&$0.24^{+0.27}_{0.09}$&$-10.30^{+0.15}_{0.10}$&$-0.09^{+0.14}_{0.15}$&$10.68^{+0.08}_{0.03}$\\
$7.8-8.1$&$9.35^{+0.16}_{0.05}$&$-1.53^{+0.08}_{0.30}$&$8.98^{+0.05}_{0.14}$&$0.34^{+0.25}_{0.12}$&$-10.52^{+0.12}_{0.15}$&$-0.28^{+0.13}_{0.12}$&$10.88^{+0.02}_{0.08}$\\
$>8.1$&$9.42^{+0.25}_{0.15}$&$-1.47^{+0.13}_{0.21}$&$8.93^{+0.06}_{0.12}$&$0.55^{+0.08}_{0.15}$&$-10.35^{+0.15}_{0.22}$&$-0.06^{+0.11}_{0.21}$&$10.91^{+0.04}_{0.03}$\\
$p<0.05$&&&&&&&\bf $p<0.001$\\
\hline
\end{tabular}
\begin{tablenotes}
\item {Summary of median values for AGN host galaxies and the 1$\sigma$ confidence limits based on a survival analysis using the KM estimator and the {\sc lifelines} survival analysis package.  A two sample survival analysis in time test is run between the two populations split in half between the highest and lowest values.}
\end{tablenotes}
\label{samplemedians}
\end{table}

\subsubsection{Intrinsic X-ray luminosity, black hole mass, and Eddington ratio}

We next examine possible links between the molecular gas content of the BAT AGN hosts and key properties of the SMBHs that power their central engines.
For this, we use measurements of intrinsic X-ray luminosities \cite[\Lsoftint; see][]{Ricci:2017:17}, BH mass (\mbh) and Eddington ratio \cite[\lledd; both from][]{Koss:2017:74}, available through the BASS project.

\begin{figure*} 
\centering
\includegraphics[width=8cm]{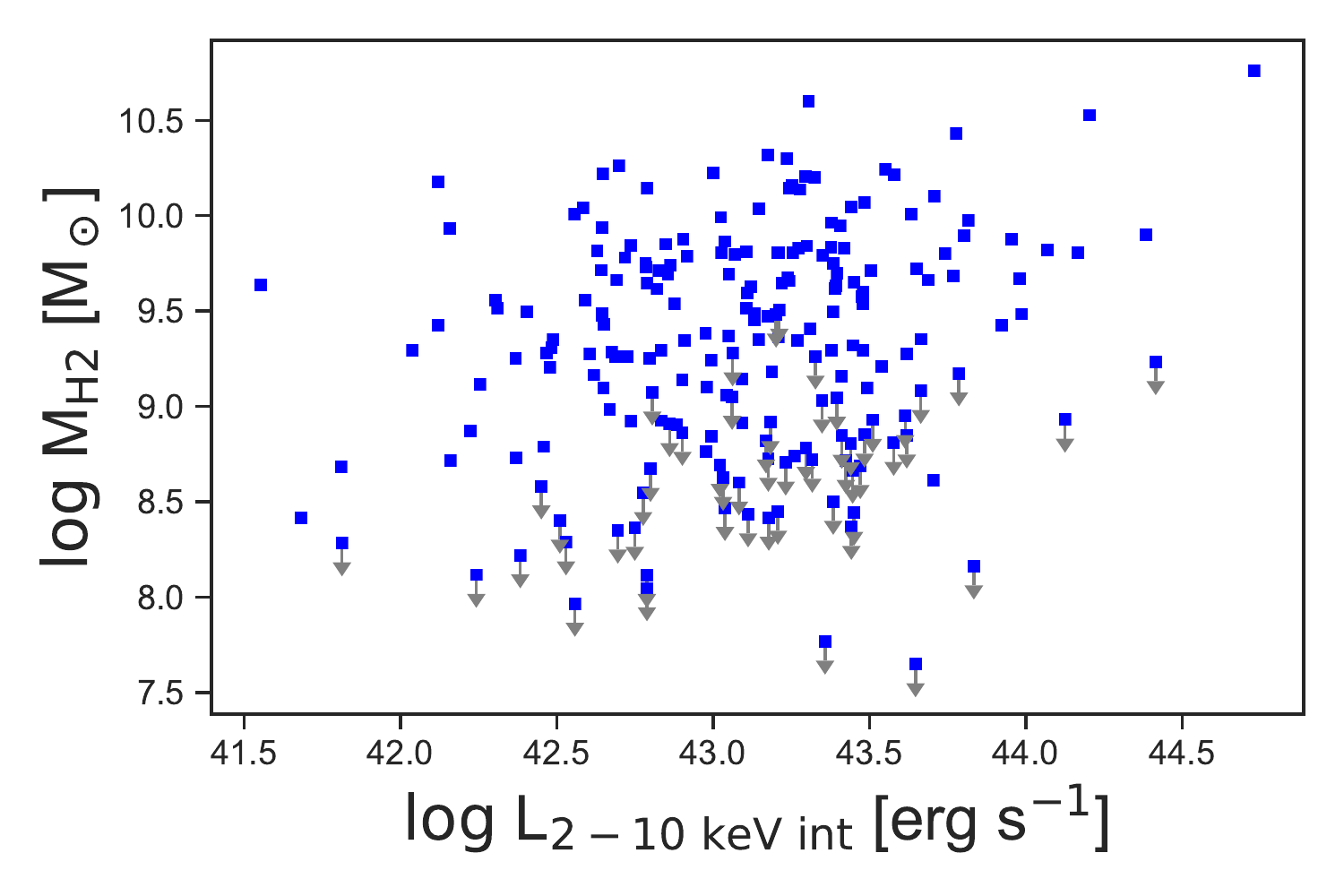}
\includegraphics[width=8cm]{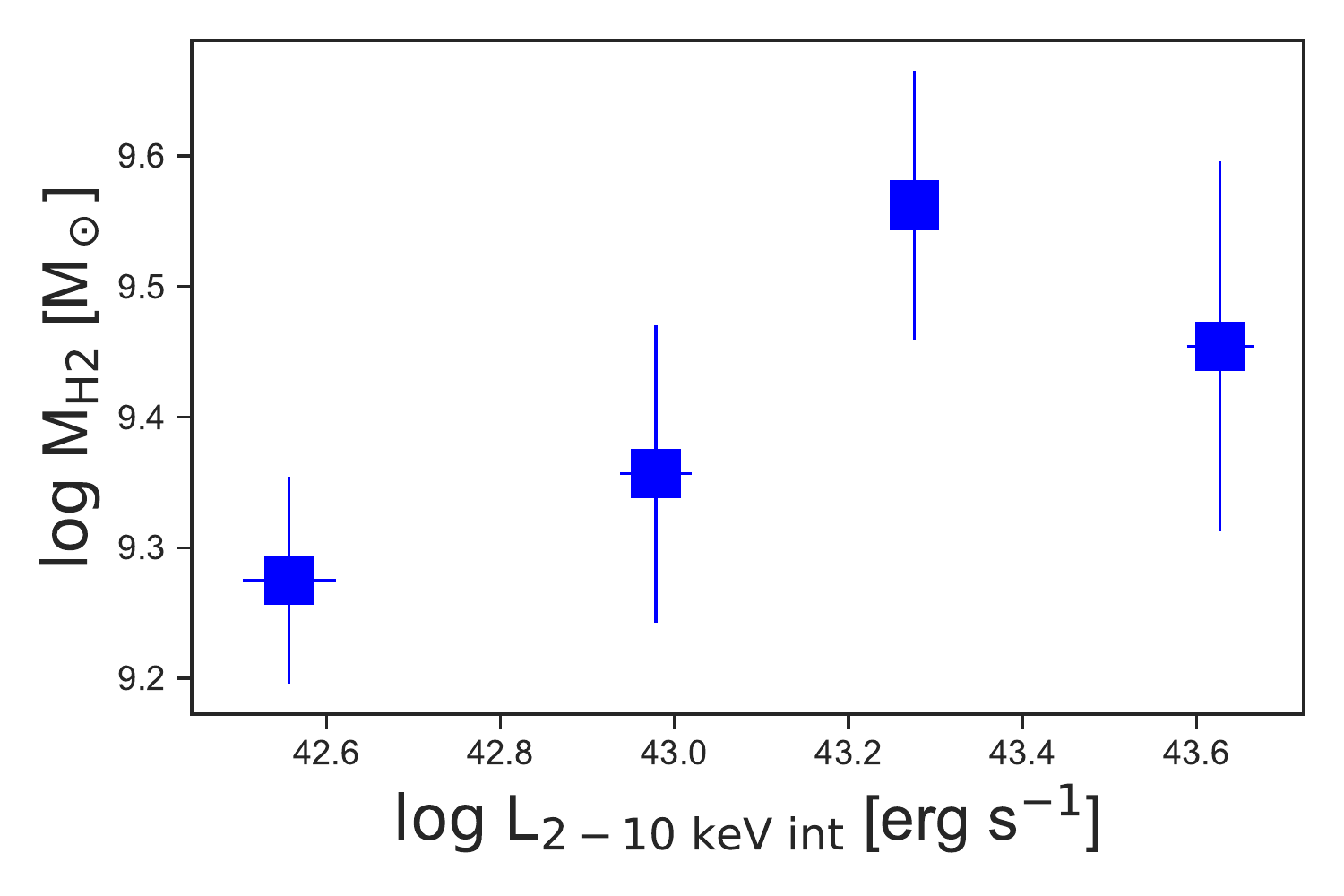}
\includegraphics[width=8cm]{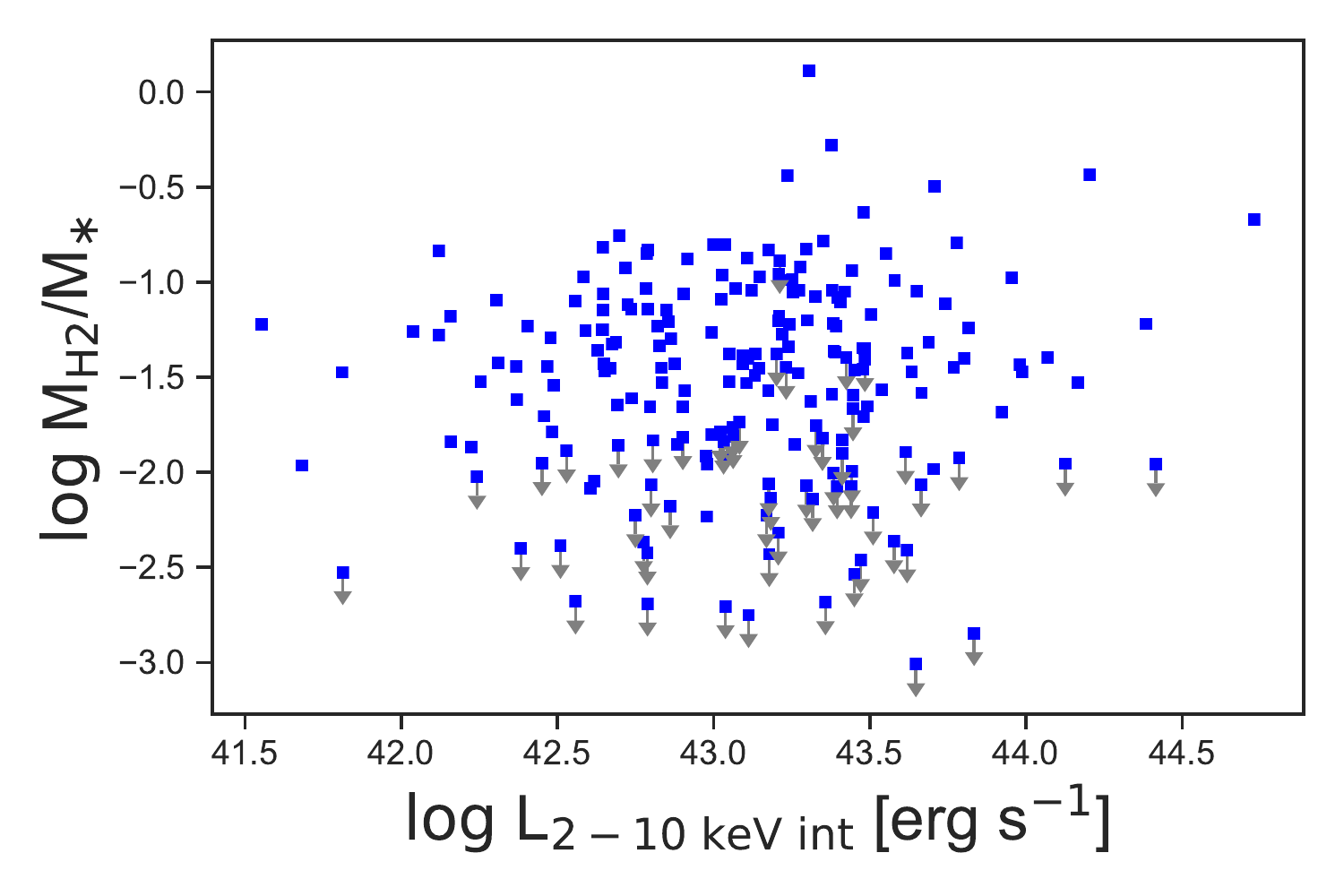}
\includegraphics[width=8cm]{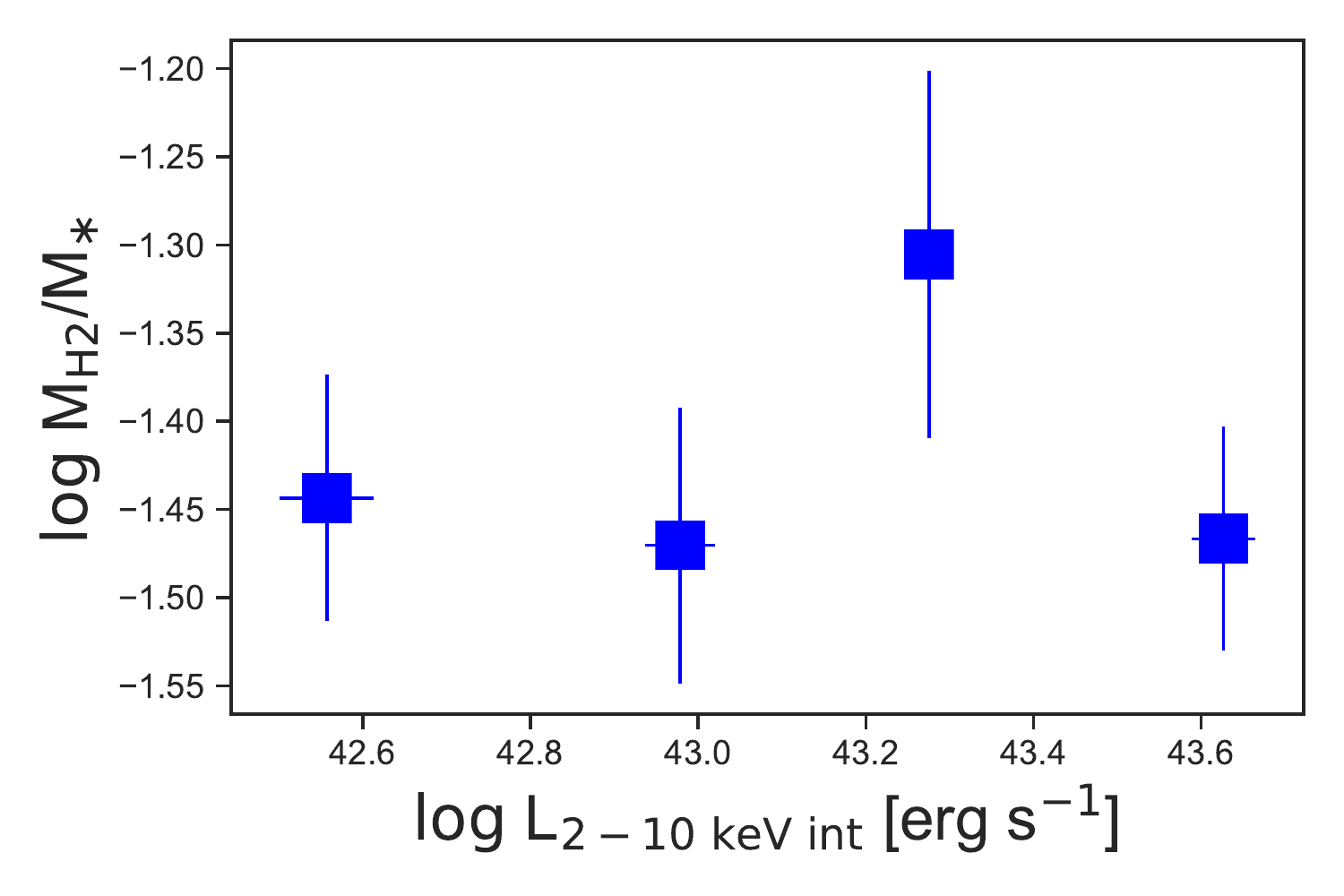}
\includegraphics[width=8cm]{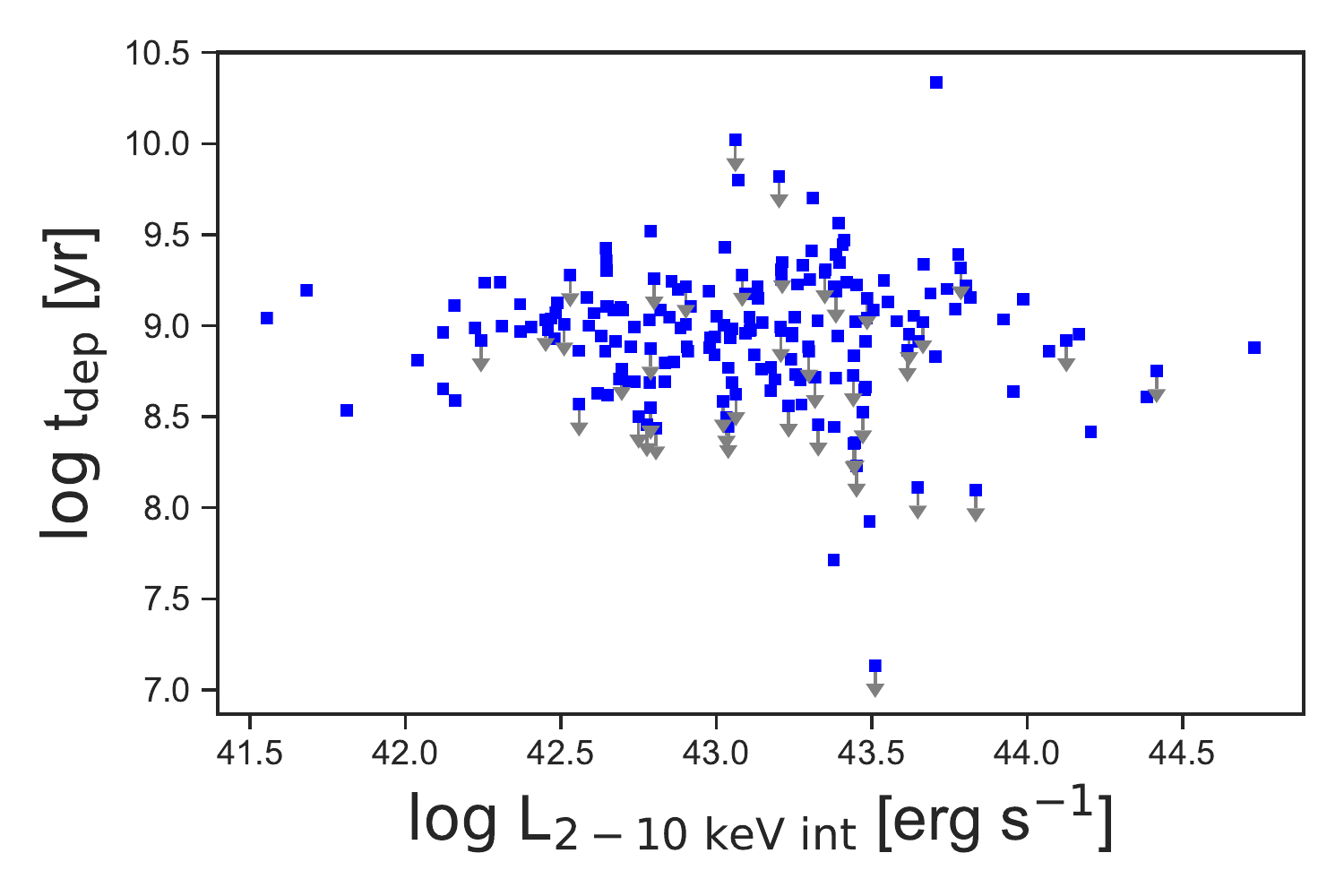}
\includegraphics[width=8cm]{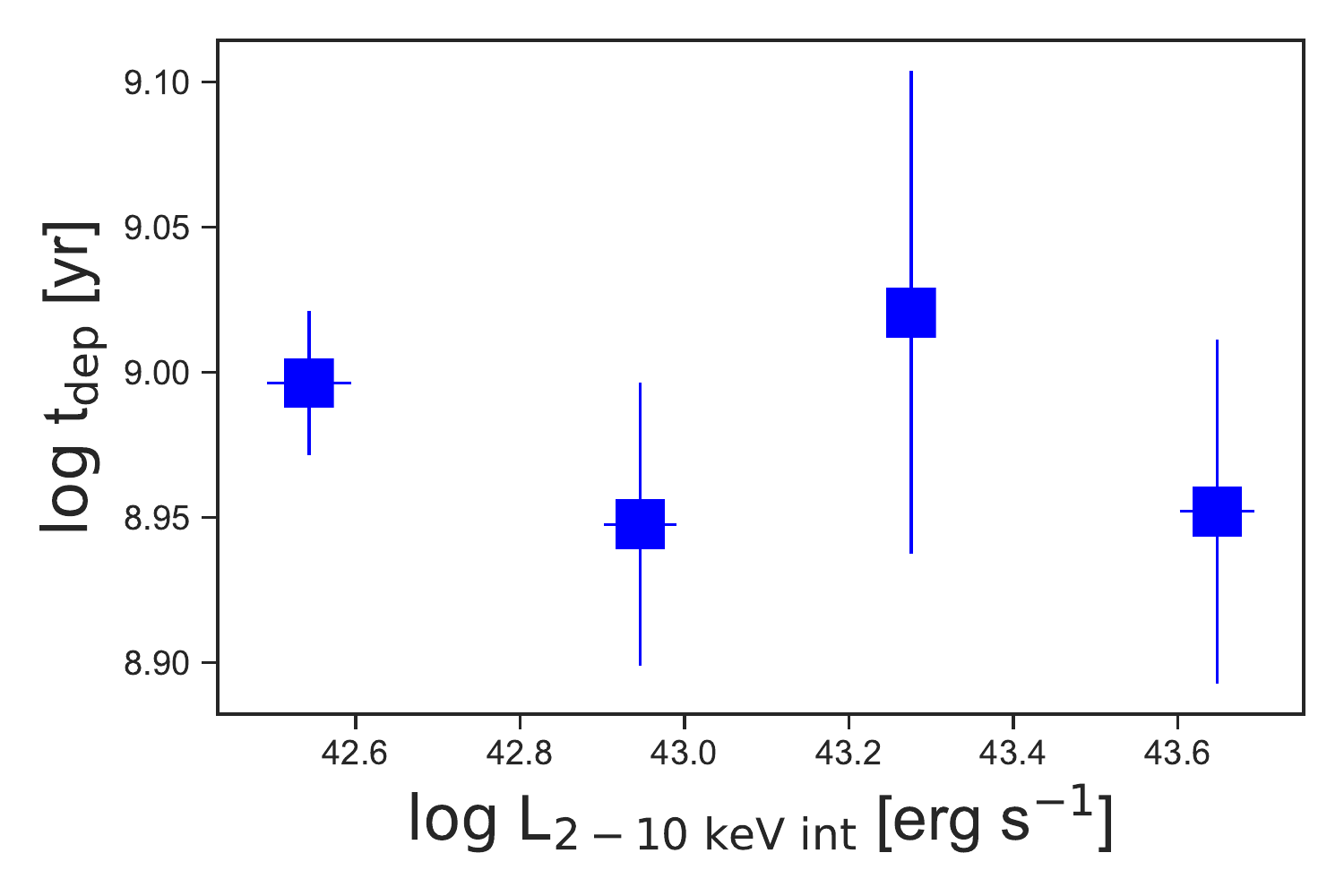}
\caption{Molecular gas mass, gas fraction, and depletion timescale compared to the intrinsic X-ray luminosity (top, middle, and bottom panels, respectively).}
\label{tab:Xraymolec}
\end{figure*}

In Figure \ref{tab:Xraymolec} we plot \Lsoftint\ vs. the molecular gas mass, gas fraction, and depletion timescale.  While there appears to be a slight increase in total molecular gas mass with X-ray luminosity, it is not statistically significant ($p=0.10$) when comparing the higher-luminosity half of our AGN to the lower-luminosity half (above and below $\log(\Lsoftint/\ergpersec)=43.2$, respectively).  One area where there is a significant correlation is between the stellar mass and increased \Lsoftint\ ($p=0.001$, see table \ref{samplemedians}).

\begin{figure*} 
\centering
\includegraphics[width=8cm]{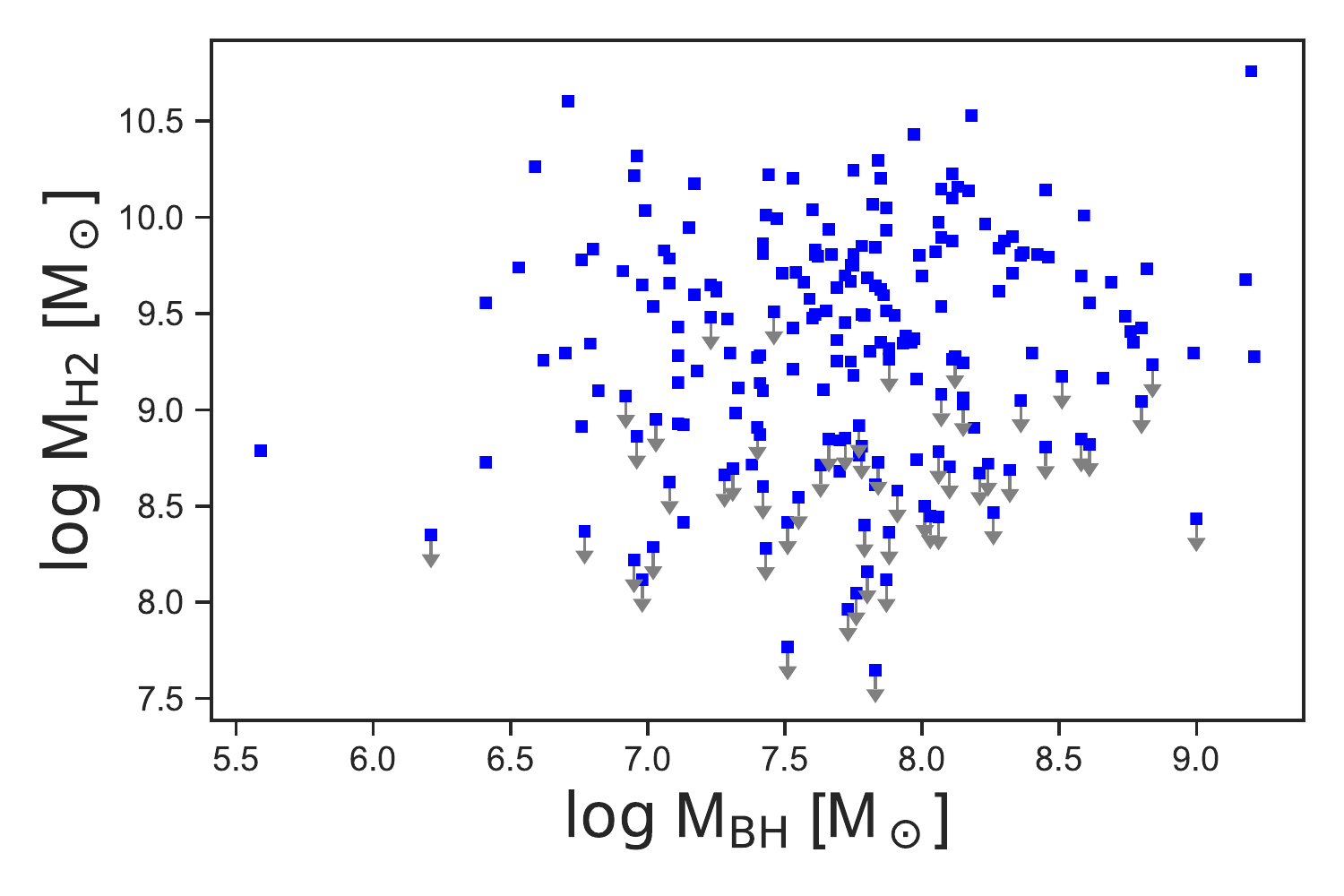}
\includegraphics[width=8cm]{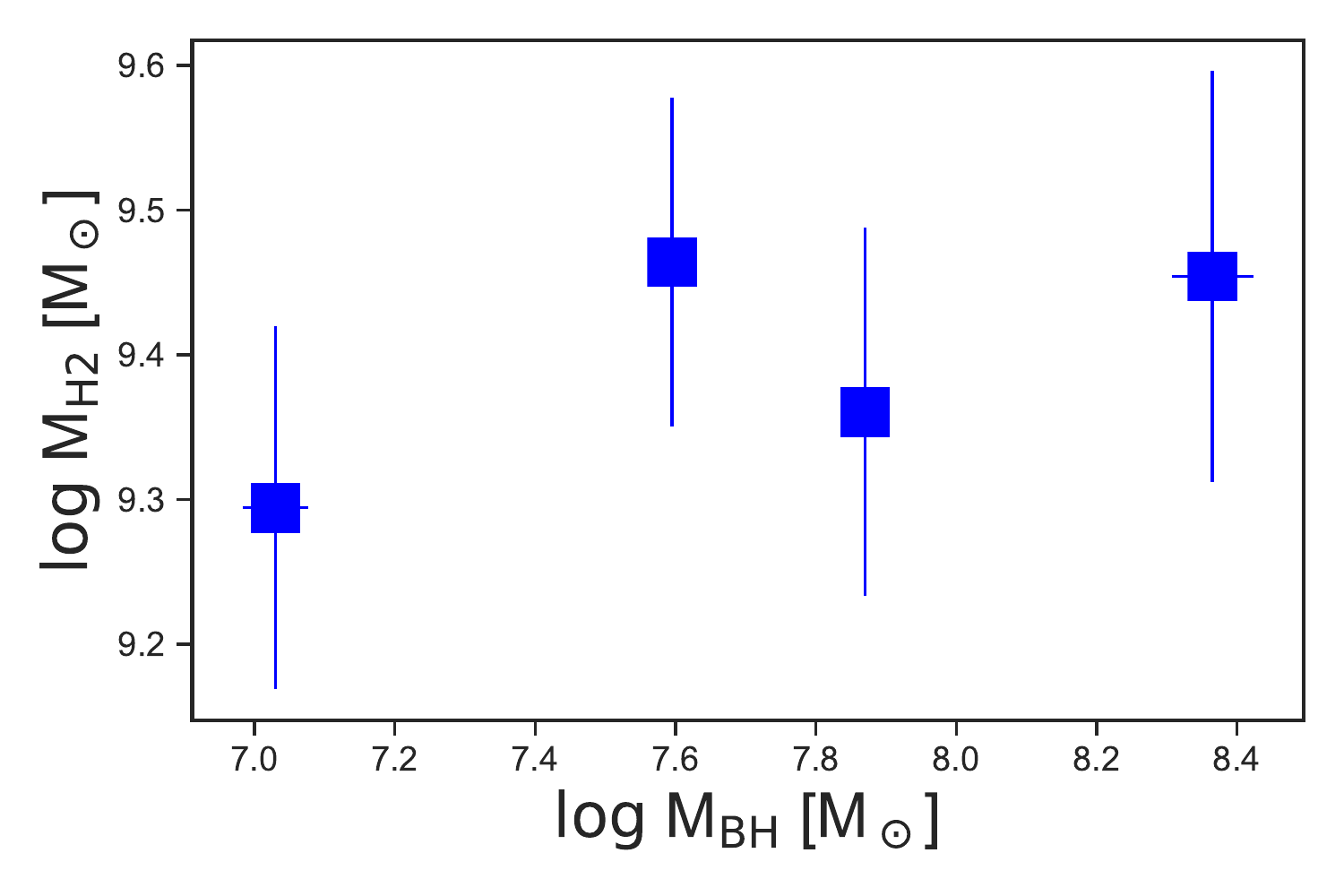}
\includegraphics[width=8cm]{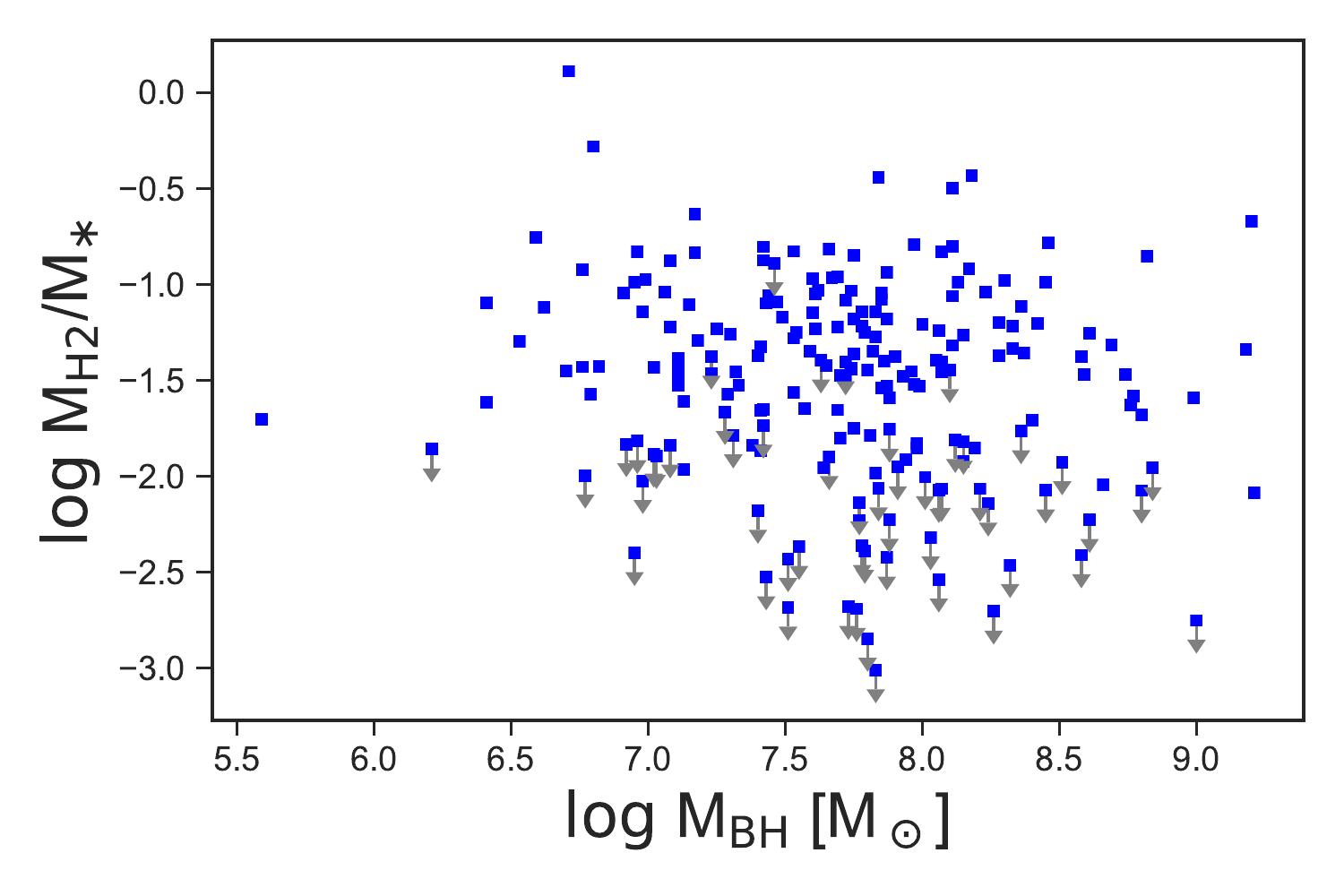}
\includegraphics[width=8cm]{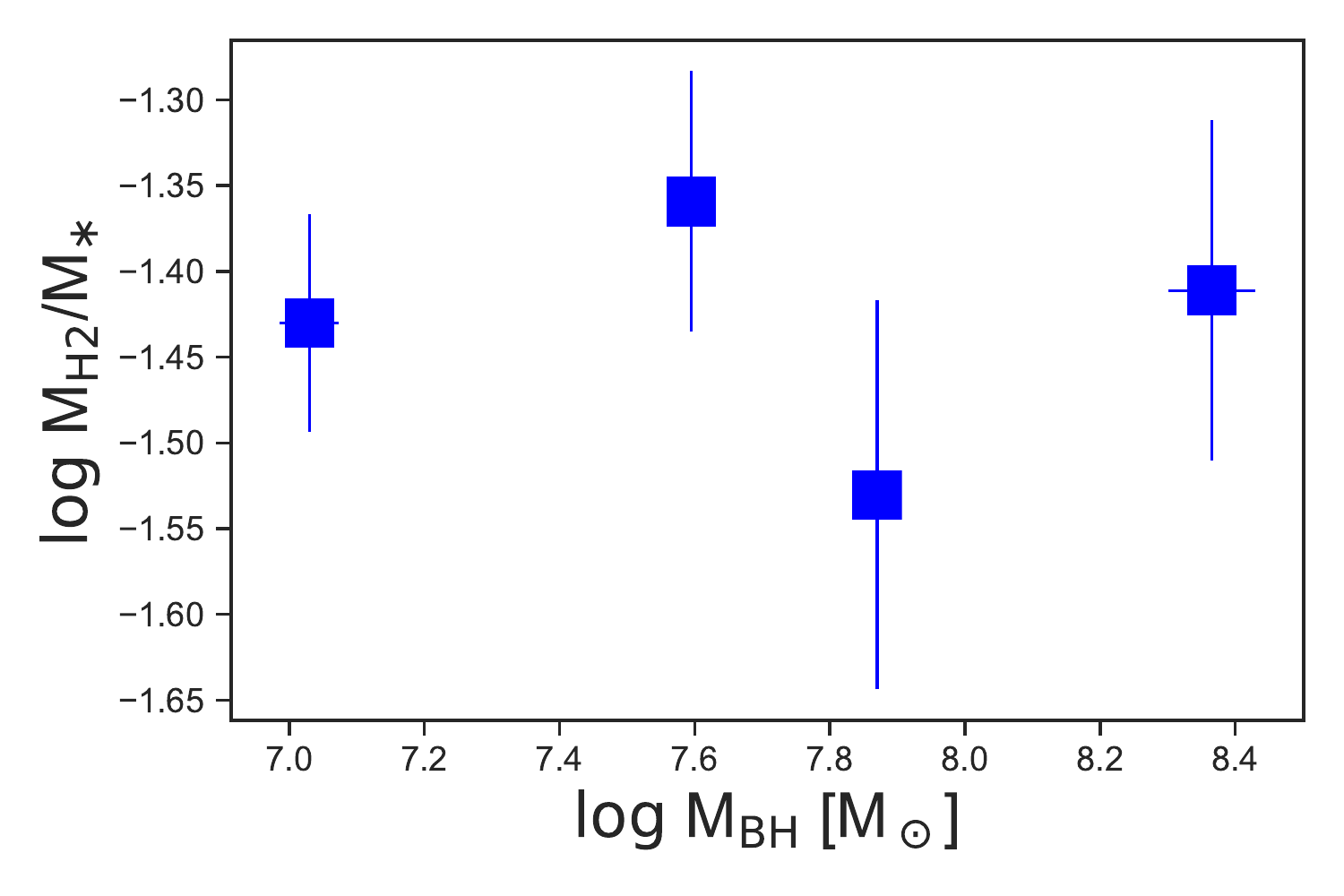}
\includegraphics[width=8cm]{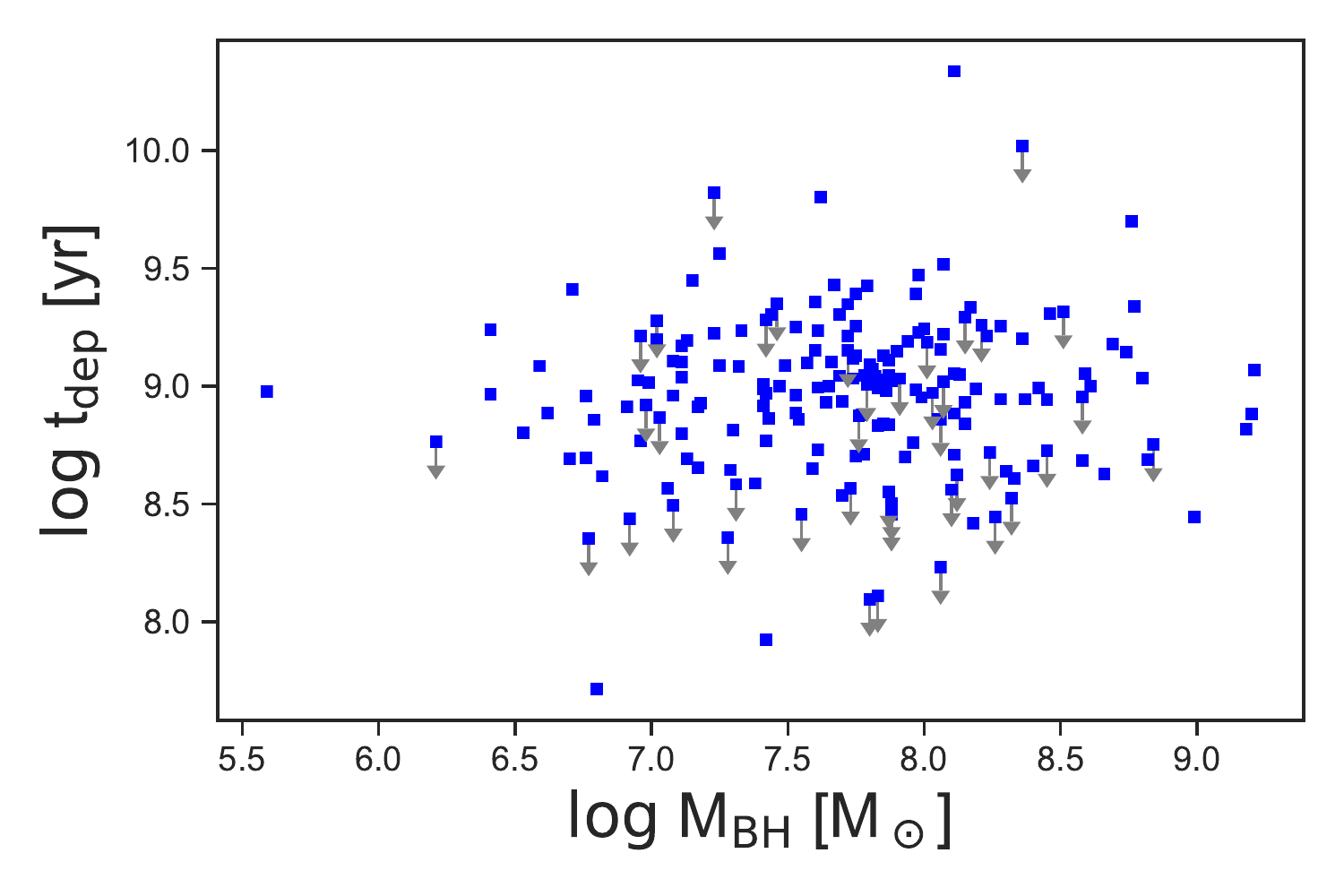}
\includegraphics[width=8cm]{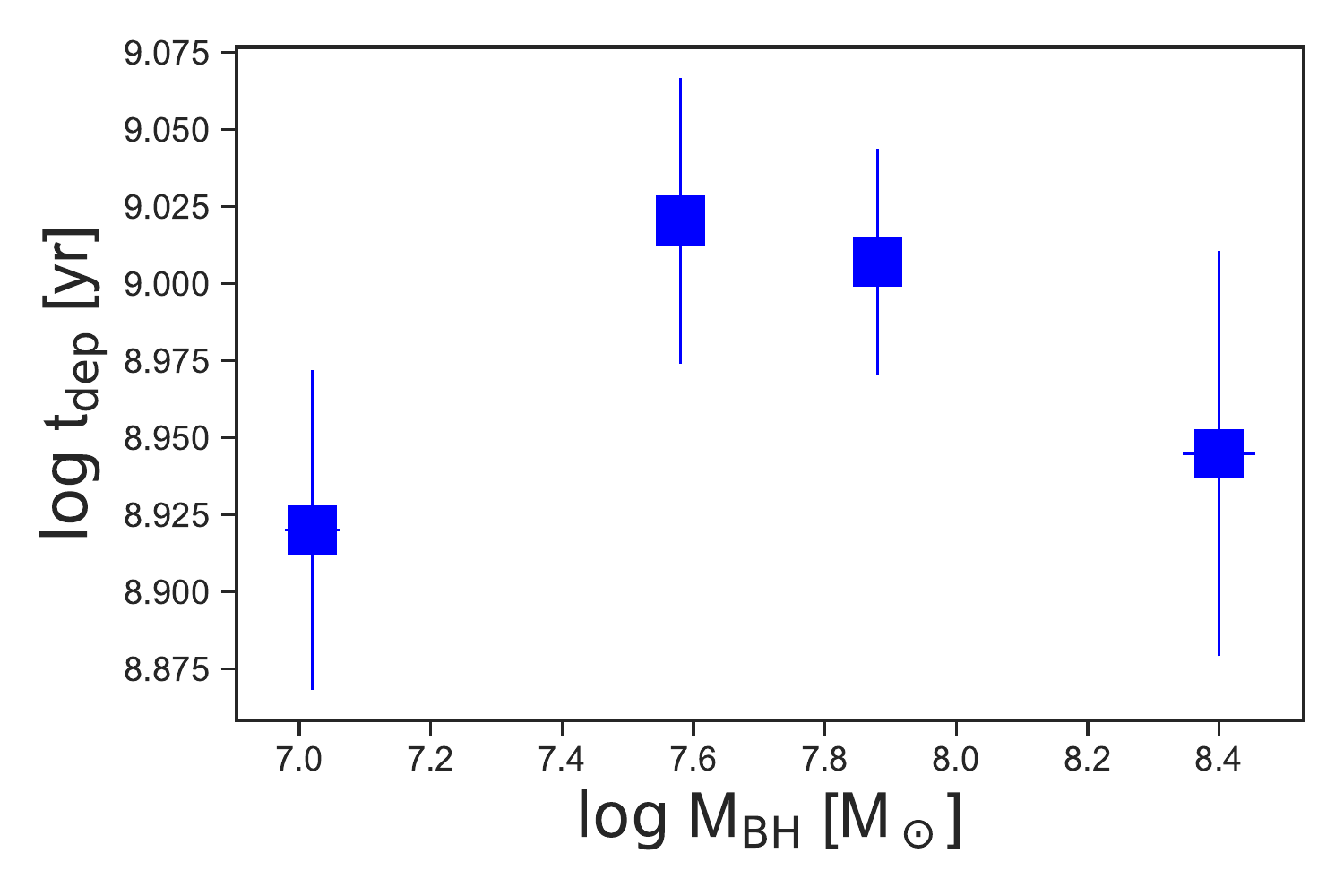}
\caption{Molecular gas mass, gas fraction, and depletion timescale compared to the black hole mass (top, middle, and bottom panel, respectively). 
As in other figures, we show both individual measurements (left column of panels) and medians within bins (right panels).
For presentation purposes the scales of the median plots (right column) are smaller than left column to show sample differences.}
\label{tab:mbhmolec}
\end{figure*}

Figure \ref{tab:mbhmolec} shows the molecular gas properties compared to \mbh. No significant trends are found in any of the properties (molecular gas mass, gas fraction, or depletion timescale, SFR, sSFR, \deltams).  There is a significant trend between stellar mass and increased black hole mass as expected ($p<0.001$).

\begin{figure*} 
\centering
\includegraphics[width=8cm]{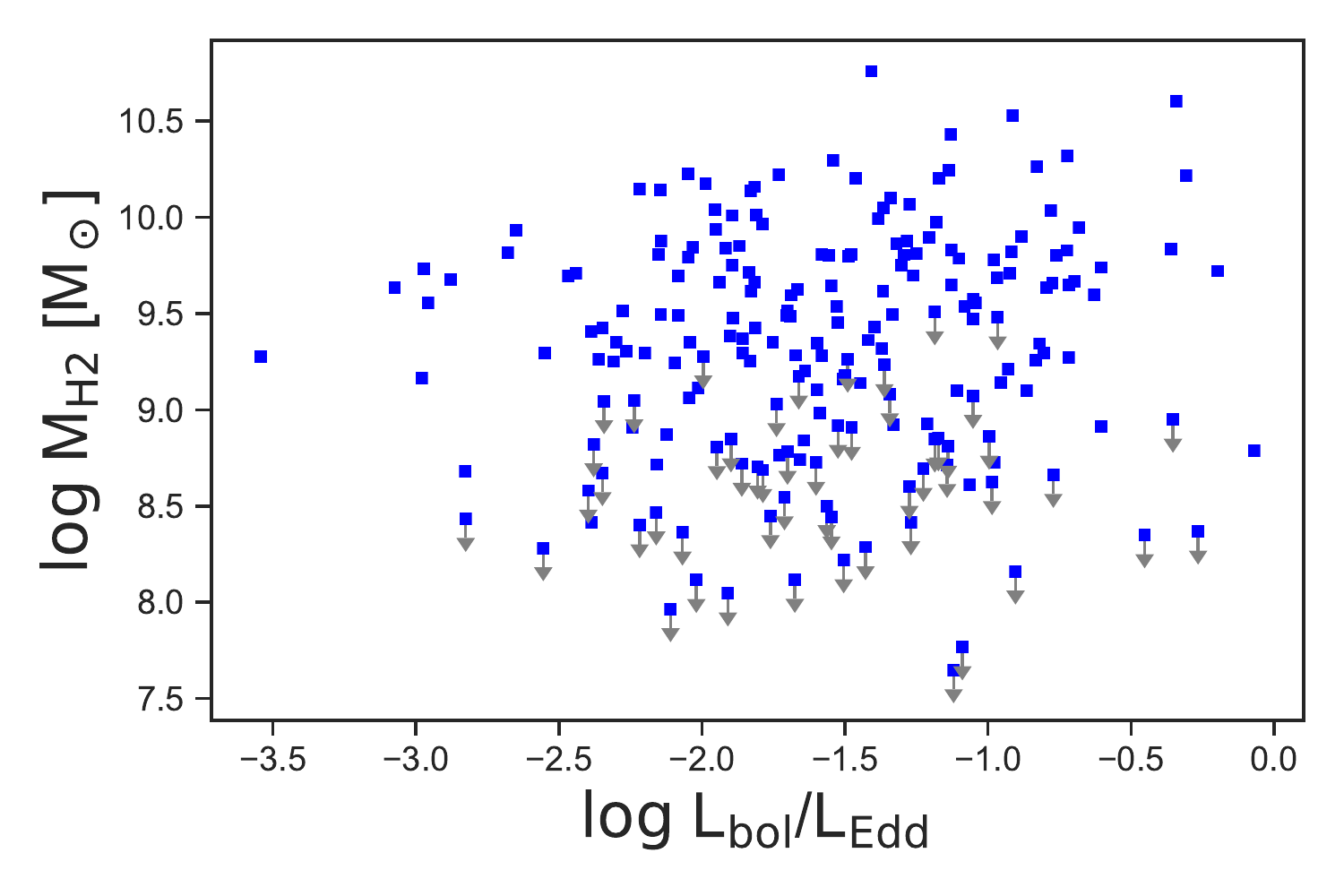}
\includegraphics[width=8cm]{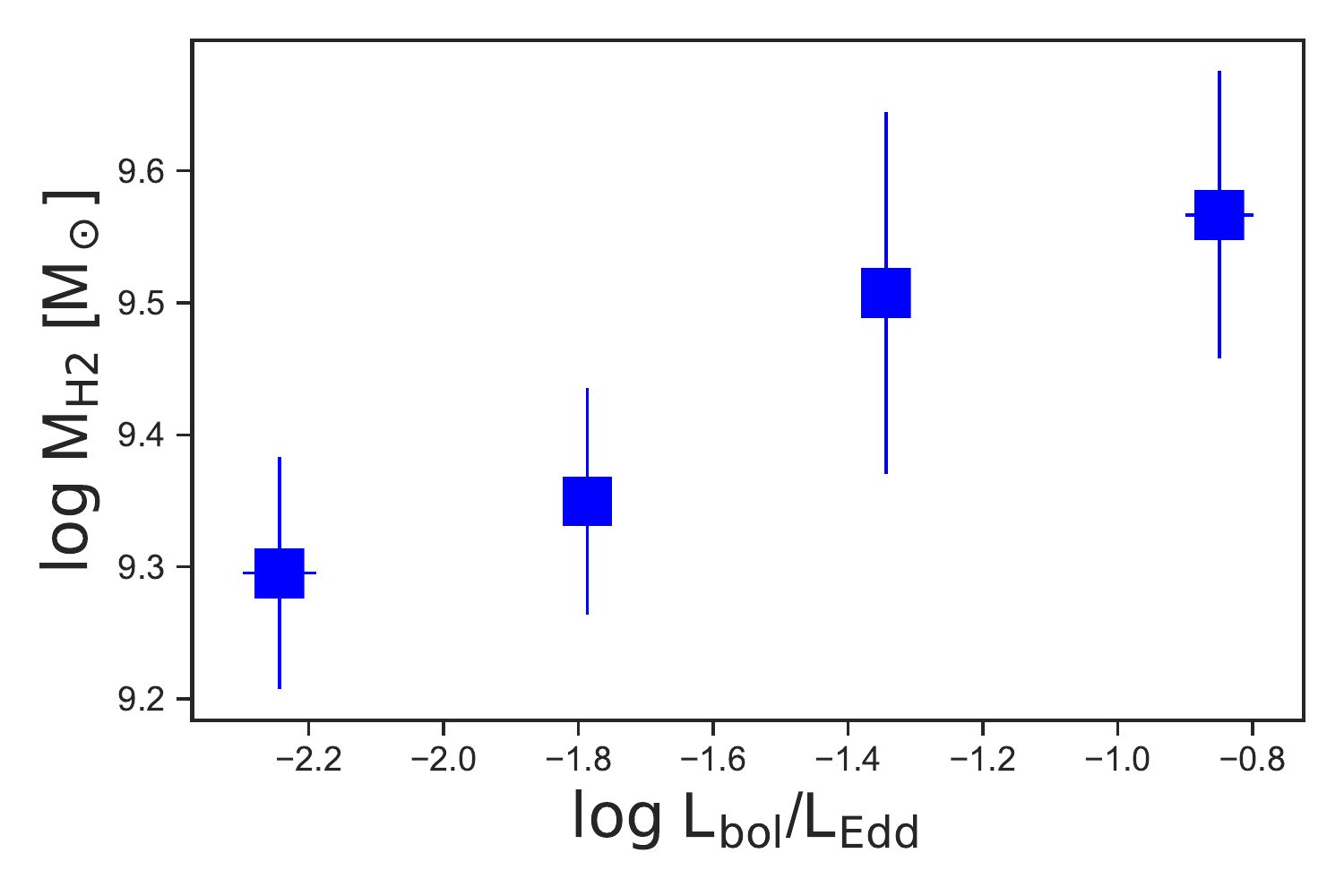}
\includegraphics[width=8cm]{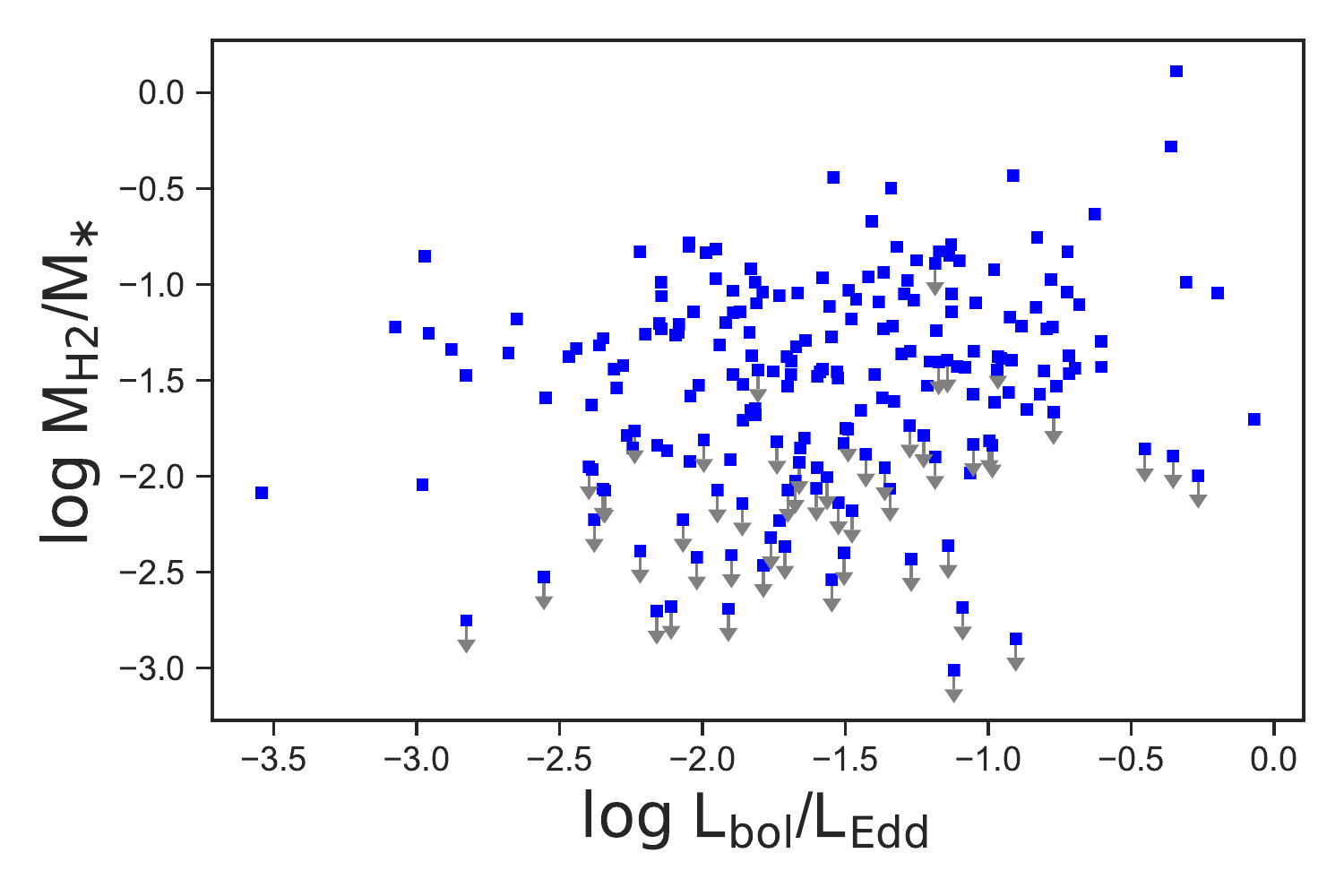}
\includegraphics[width=8cm]{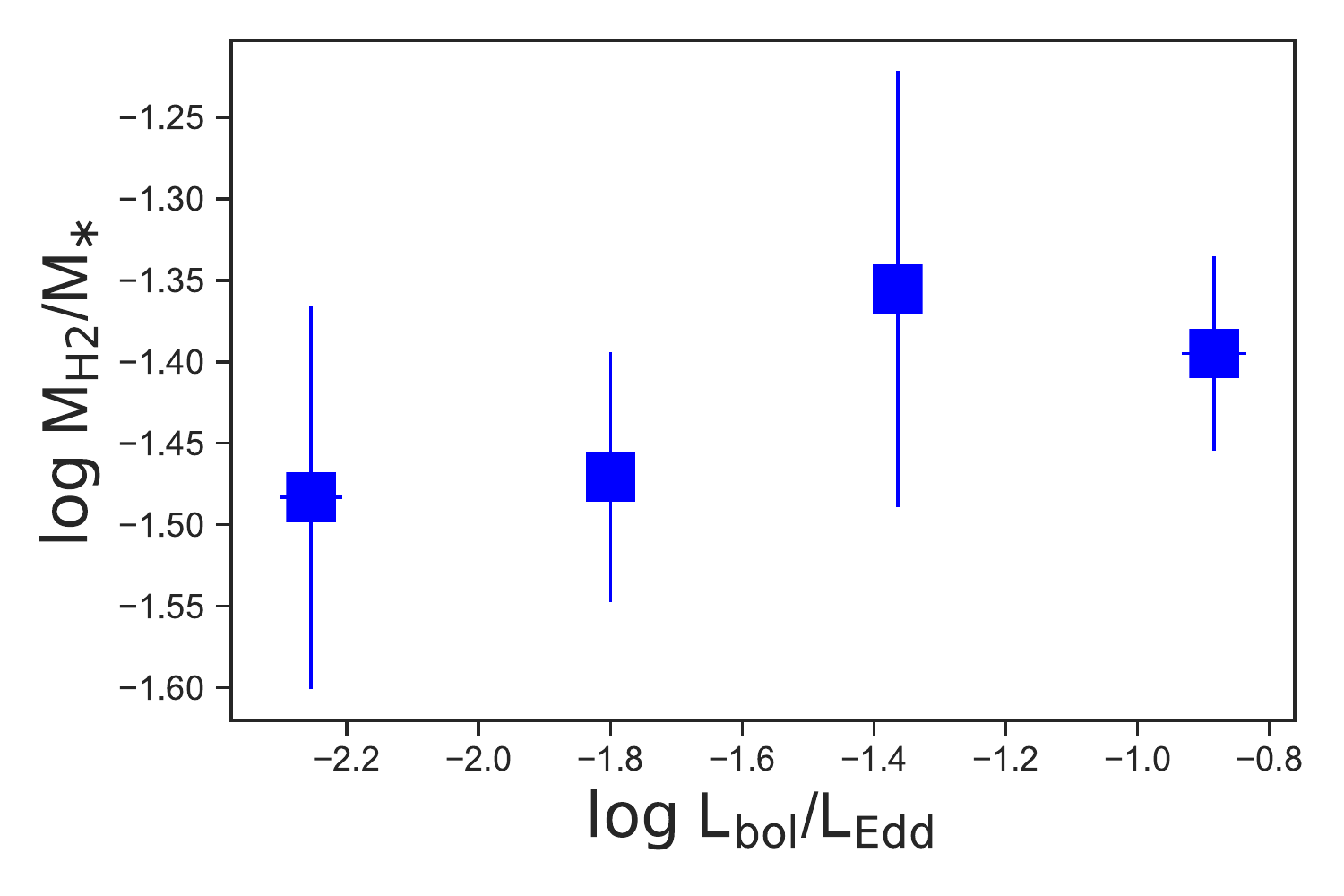}
\includegraphics[width=8cm]{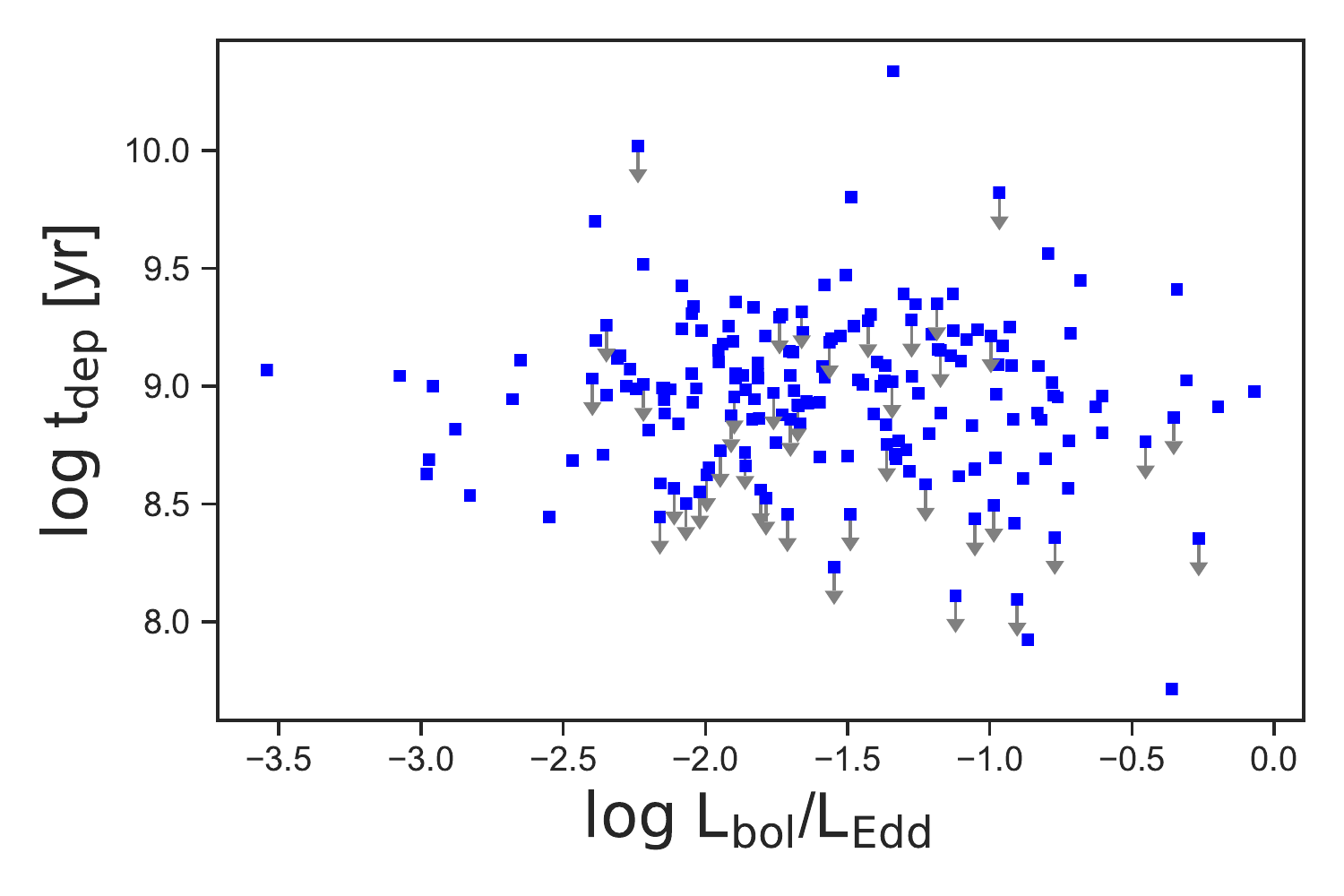}
\includegraphics[width=8cm]{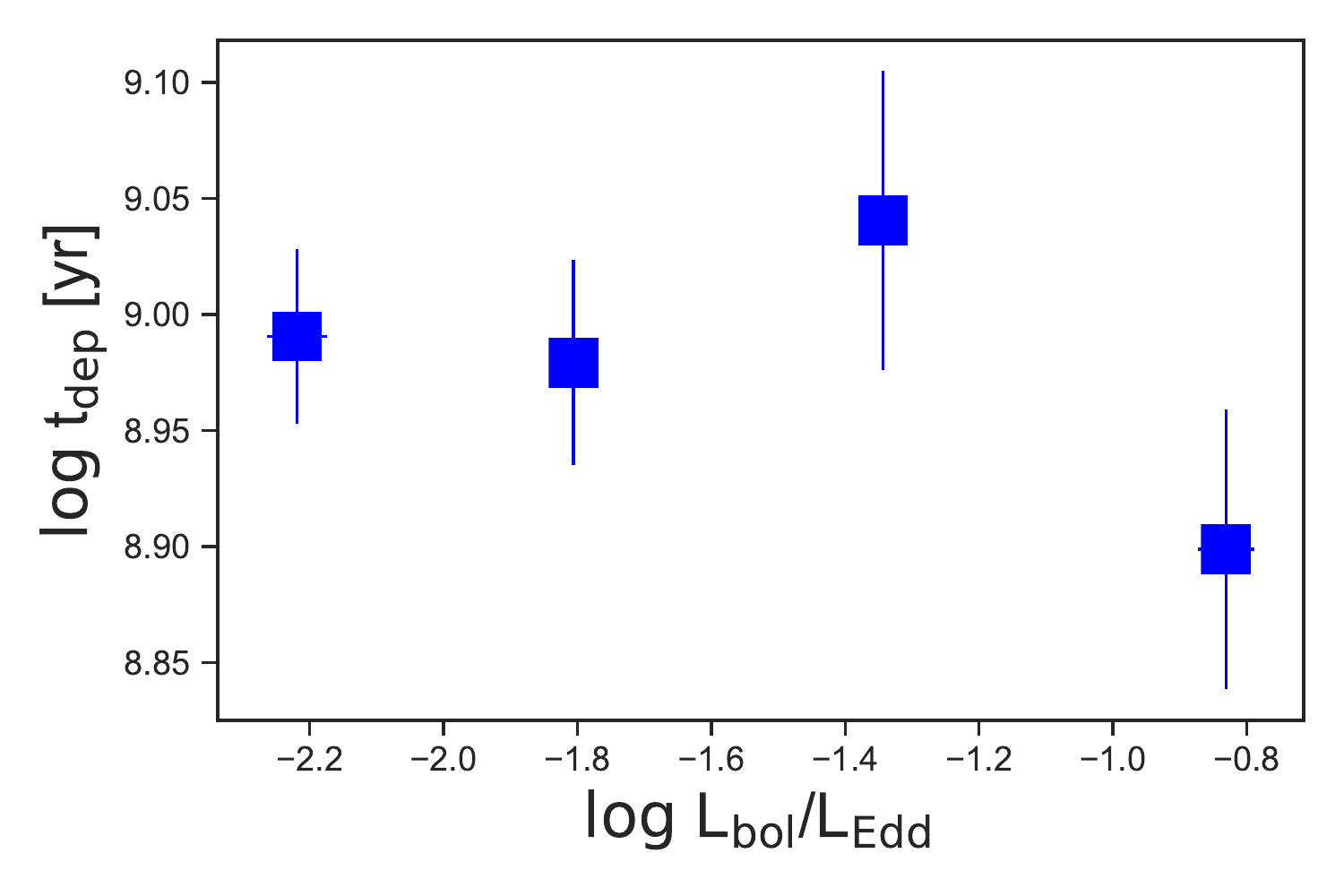}
\caption{Molecular gas mass, gas fraction, and depletion timescale compared to the Eddington ratio (top, middle, and bottom panel respectively).  
As in other figures, we show both individual measurements (left column of panels) and medians within bins (right panels).
There is a general increase of molecular gas mass and gas fraction with increasing \lledd, while there is no trend with gas depletion timescale.  
For presentation purposes the scales of the median plots (right column) are smaller than left column to show sample differences.  }
\label{tab:edd_molecular}
\end{figure*}

Finally, Figure \ref{tab:edd_molecular} shows the molecular gas properties compared to \lledd. The hosts of the highest Eddington ratio AGN ($\log\lledd > -0.16$) have molecular gas masses that are significantly higher ($p=0.037$) than hosts with the lowest Eddington ratio AGN, based on survival analysis.   Similarly, the highest Eddington ratio AGN have significantly higher ($p=0.042$) gas fractions than the lowest \lledd\ ones.  The gas depletion timescales, SFR, sSFR, \deltams, and stellar mass show no statistically significant trends with Eddington ratio.

\section{Discussion}

Our results generally support the idea that rapidly growing, radiatively efficient AGN are fueled by a significant supply of cold gas, but this difference is concentrated in early and uncertain morphological types and more pronounced at higher stellar masses. These AGN host galaxies are starkly different with higher gas fractions and SFRs than the population of quenched massive elliptical galaxies with undetectable star formation and molecular gas ($-12< \log({\rm sSFR}/{\rm yr}^{-1})<-11$; \citealt{Saintonge:2017:22}).    There are also trends that support the idea that higher gas fractions increase the average normalized black hole growth rate (\lledd).\footnote{Or, alternatively, that higher gas fractions are linked to shorter SMBH growth timescales, which scale inversely with \lledd.}  

\subsection{Links to coevolution and stochastic accretion}
The global Eddington ratio and gas fraction correlation, as well as the increased likelihood of a galaxy with high molecular gas host a luminous AGN, may be tracing an important part of SMBH and host galaxy co-evolution. High gas content may lead naturally to general correlations found in the SMBH and host \citep[e.g., the correlations between SMBH mass and bulge properties;][]{Kormendy:2013:511} and the similarity between the redshift evolution of star formation and SMBH growth \citep[see the review of][]{Heckman:2014:589}.   

The correlations found between the Eddington ratio and both the total molecular gas mass and gas fraction (over the entire host galaxy)  suggest that substantial cold gas may be concentrated in the nucleus of the galaxy.  There is some evidence for this from \herschel FIR imaging of AGN hosts \citep{Mushotzky:2014:L34,Lutz:2018:A9}, which show that a significant fraction of \herschelsh/PACS emission originates from compact sources, with typical scales of $<$2 kpc. However, high sensitivity, interferometric observations of the central cores of AGN host galaxies are critical to confirm this scenario. Several ALMA campaigns are currently trying to pursue this challenging goal, including our own BASS-focused effort (Izumi et al., in prep), but the samples are still rather small ($<$40 AGN hosts). {These high resolution studies, which can overcome some of the huge difference in scales between the entire AGN galaxy studied here,  and the small scales of ultimate accretion are critical.}

While statistically significant global trends and/or correlations ($p=0.001-0.04$) are found in the large sample studied here ($N=\Nanalyzed$), it is crucial to stress that there is significant variation between individual AGN hosts, most likely driven by the stochasticity of both SMBH fuelling and the emergent AGN emission, and the galaxy-scale SF processes \cite[e.g.,][]{Hickox:2014:9,Schawinski:2015:2517,Caplar:2019:3845,Wang:2019:arXiv:1912.06523}.  This is similar to the result that when averaging over the full star-forming galaxy population, the average SMBH growth is correlated with SFR \citep[e.g.,][]{Mullaney:2012:L30,Chen:2013:3,Jones:2019:110} as it is in the host galaxies of BAT AGN \citep{Shimizu:2017:3161}, though this correlation may be mainly linked to the host-galaxy stellar mass relation rather than star formation \citep{Yang:2017:72,Yang:2018:1022}.    

{ The ability to assess differences in star formation properties and offset from the MS complicated by systematic biases in measuring the SFR using different techniques in high stellar mass passive galaxies and AGN (and see Appendix B) and how "passive galaxies" are defined.  However, the BAT AGN do have on average a higher fraction of massive spirals, bluer colors, more mergers, and more LIRGs \citep[e.g.,][]{Koss:2011:57,Koss:2013:L26} than inactive galaxies, consistent with their higher molecular gas masses than inactive galaxies and typically higher SFRs than xCOLD GASS.   We note that a past study found that the BAT AGN predominantly lie below the MS of star forming galaxies \citep[but still well above passive galaxies;][]{Shimizu:2015:1841} based on the \herschel observations, though we show the BAT AGN galaxies {are typically within the scatter of the} main sequence. This is because there is significant difficulty establishing the low reshift main sequence at high stellar masses ($\log(\mstar/\Msun){>}10.3$), such as studied here, in that there is potentially a flattening as well as increased scatter at high stellar masses \citep[e.g.,][]{Popesso:2019:3213,Popesso:2019:5285}. Furthermore, the rarity of high mass galaxies ($\log(\mstar/\Msun){>}11$), also contributes to this difficulty.    This issue will be further explored in a subsequent publication (Shimizu et al., in prep).}


\subsection{AGN feedback}

Given the significant connection between gas-rich AGN host galaxies and black hole growth, the next question is to what extent AGN are able to suppress star formation activity throughout the host galaxy, or the infall of more (cold) gas onto the host.  There is no significant evidence that the most luminous X-ray AGN galaxies have depletion timescales that are significantly shorter than lower luminosity AGN galaxies.  {This suggests that these AGN are not the primary drivers of quenching or evolution in the \mstar-SFR plane}.  This could be because the measured `instantaneous' luminosities are weakly related due to the much longer timescales on which average stochastic accretion occurs.  Large samples with hundreds of sources, such as the one studied here, combined with studies of the AGN history on longer timescales \citep[e. g.,][]{Sartori:2018:L34} are crucial to overcome AGN variability and understand the role of AGN outbursts and feedback.  {Finally, high sensitivity, interferometric observations of the central cores of AGN host galaxies would also test whether signs of AGN feedback are present.}






\subsection{Gas-rich ellipticals}

The large amounts of molecular gas we find among BAT AGN hosted in elliptical galaxies is quite surprising and significant.  
Examples of some of the most gas-rich ellipticals in the BAT AGN galaxy sample are shown in Figure \ref{fig:ellip_pic}.  We caution that some of these galaxies may be better classified as early-types, as it is difficult to distinguish between true ellipticals and face-on lenticular hosts without deeper, higher-quality imaging (e.g., from \hst; see Appendix A) -- as has been noted with Galaxy Zoo classifications \citep[e.g. S0-SAs,][]{Lintott:2008:1179}. 
Importantly, ellipticals with significant molecular gas mass ($\log(\mh/\Msun)>9$) are quite rare among inactive galaxies. 
Among the 81 inactive massive ($\log(\mstar/\Msun)>10.2$) ellipticals from xCOLD GASS, only 4\% (3/81) have significant amounts of molecular gas.  Similarly, the fraction of early-type galaxies with ($\log(\mh/\Msun>9)$) within the ATLAS3D sample is even lower, at about 1\% \cite[3/260;][]{Young:2011:940}.   

In contrast, the {\it majority} (51\%, or 21/41) of the elliptical hosts of BAT AGN are comparatively gas-rich (i.e., meeting the $\log(\mh/\Msun)>9$ criterion).  Interestingly, the ATLAS3D survey found a significant trend between the molecular gas content and stellar specific angular momentum. Combined with our finding of a high fraction of gas-rich ellipticals among BAT AGN hosts, this may suggest that the amount of galaxy angular momentum may be coupled to SMBH growth via stochastic accretion.    

Given the lack of gas-rich early types among inactive galaxies, our results suggest that the many gas-rich early-type galaxies host luminous AGN. This is further supported by the most gas rich early-type galaxy in xCOLD GASS (ID 3819), being excluded from the study due to hosting an AGN.  Additionally, there is strong evidence that at {least} 2/3 early-type gas rich galaxies in ATLAS3D host AGN based on radio data \citep{Nyland:2016:2221}.  Further study is however, required to test this because of the difference in volumes surveyed, and the rarity of both gas-rich early type galaxies and of luminous AGN.\footnote{The BAT AGN survey probed a volume of about $2\times10^{7}\,{\rm Mpc}^{3}$, which is about 173 times larger than the ATLAS3D volume \citep{Cappellari:2011:813}.}

One obvious source of the extra gas in the massive galaxies is from mergers which have already been shown to be significantly more common in BAT AGN hosts \citep[e.g.,][]
{Koss:2010:L125,Koss:2011:L42,Koss:2018:214a}. One might also expect an increase in the molecular-to-atomic gas ratio as measured from HI during the merger. The relatively high gas fractions of BAT AGN galaxies may explain the enhanced occurrence rate of dual AGN \citep{Koss:2012:L22,Koss:2018:214a}, which exceeds predictions from (low-redshift) simulations \citep{Rosas-Guevara:2016:190}.  
As a specific example, looking at eight gas-rich BAT AGN galaxies with uncertain morphologies (Figure \ref{fig:uncertain_pic}), 38\% (3/8) are close mergers with separations between galaxy nuclei of less than 3.5 kpc \citep{Koss:2018:214a}, whereas the rate among inactive galaxies for similarly close mergers is less than 1\%. 

These observations may be able to measure the importance of externally acquired gas (e.g., from mergers and cold accretion). This was already clearly demonstrated in some inactive early-type galaxies (e.g., in the ATLAS3D sample; \citealt{Davis:2011:968}), as the specific angular momentum of the gas is dramatically different from that of the stars. Early efforts studying dozens of the nearest BAT AGN galaxies are currently underway with ALMA (Izumi et al., in prep), in HI with the Australia Telescope Compact Array (Wong et al., in prep.), and for 100s of BAT AGN galaxies with the Jansky Very Large Array (Smith et. al, in press).

\subsection{AGN Obscuration}
{The finding that higher column density AGN galaxies                                                                 ($\log(\nh/\psqcm)>23.4$) and Seyfert 2 AGN galaxies are associated with lower depletion timescales and suggests a connection to the gaseous environment of the nucleus.  The higher column density AGN are also associated with higher sSFR.  Obscured AGN may prefer hosts that with more molecular gas centrally concentrated in the bulge associated with the torus that may be more prone to quenching than galaxy wide molecular gas.} The relationship between X-ray line-of-sight column density, the amount of molecular gas, stellar mass, and the inclination angle of the galaxies will likely allow one to further understand the galaxy-scale distribution of dusty, molecular gas, star formation, and the associated galaxy-scale obscuration (Strittmatter et al., in prep).  
When also combined with higher resolution data, the closer (lower-redshift) subset of the BAT AGN galaxies sample may provide a unique ability to measure the amount of extended ($0.1-1$ kpc) molecular gas. This, in turn, will provide a template for interpreting the typical contribution of galaxy-scale contribution to the X-ray obscuration in distant AGN, observed in deep surveys. At high redshift, the gas fractions are generally higher \citep{Vito:2014:1059}, and galaxy-scale obscuration is likely much more important for these surveys.  \


%

\section{Summary and conclusions}

We reported observations of the $^{12}$CO(2-1) emission line for \Nobs\ ultra-hard X-ray selected AGN galaxies drawn from the {\it Swift}/BAT 70-month catalog. 
These data, obtained through large allocations of APEX and JCMT observing time, allowed us to reliably measure several key properties of the molecular gas in the hosts of these powerful AGN, including molecular gas mass and mass fraction, and depletion timescales.
These were compared to similar measurements for a large comparison sample of inactive galaxies, drawn from the xCOLD GASS survey.
Our analysis demonstrates how large ($N>200$) samples of AGN galaxies with high quality, multi-wavelength data and well defined control samples are essential to understand the possible links between significant black hole growth and the host galaxy scale molecular gas content. 

In summary, our main findings are:
\begin{itemize}
    \item 
    AGN in massive galaxies ($\log(\mstar/\Msun)>10.5$) tend to have more molecular gas and higher gas mass fractions than inactive galaxies matched in stellar mass ($p<0.001$).  The higher molecular gas content is related to AGN galaxies hosting a population of gas-rich early types with an order of magnitude more molecular gas, a smaller fraction of quenched, passive galaxies ($\sim$5\% vs. 49\%), and a {larger fraction of AGN galaxies on the Main Sequence} of star-formation compared to inactive galaxies.
    
    \item {When matched in star formation, specific star formation, or offset from the Main Sequence of star formation, AGN galaxies are similar to inactive galaxies with respect to molecular gas and depletion timescales. There is no evidence of AGN feedback affecting the host galaxy cold molecular gas.}
    
    \item
    The probability of hosting an AGN (\Lsoftint$>10^{41.8}$\ergps or L$_{\rm bol}>10^{44}$\ergps) is correlated with the total molecular gas and increases by a factor of $\sim$10-100 between a molecular gas mass of $10^{8.7}$\Msun\ and $10^{10.2}$\Msun.
    
    \item The hosts of AGN accreting at higher Eddington ratios have, on average, higher molecular gas masses ($p=0.037$) and gas fractions ($p=0.042$).
    
    \item When matched by stellar mass and morphology, early-type AGN hosts show much higher molecular gas mass and gas fractions -- by more than an order of magnitude, compared to inactive galaxies (e.g., $\log(\mh/\Msun)=9.58\pm0.07$ for AGN galaxies vs. $8.36\pm0.08$ for inactive galaxies).  
    This is related to the fact that the AGN galaxy sample lacks a significant number of massive and passive galaxies ($\log({\rm sSFR}/{\rm yr}^{-1})<-11$), which are common among the inactive ellipticals.  {AGN host galaxies are also more likely to be on the MS.}  Among spirals and uncertain morphologies the difference in molecular gas content is significant for only the highest stellar masses ($\log(\mstar/\Msun)>10.5$).
    
    \item {Higher column density AGN galaxies ($\log(\nh/\psqcm)>23.4$) and Seyfert 2 AGN galaxies are associated with lower depletion timescales.  Obscured AGN may prefer hosts that with more molecular gas centrally concentrated in the bulge that may be more prone to quenching than galaxy wide molecular gas.}

\end{itemize}

These results suggest that molecular gas has a critical role to play in black hole growth. Ultimately, a combination of interferometric CO observations (e.g., with ALMA or NOEMA), HI observations, and measurements of stellar dynamics from optical/NIR integral-field spectroscopy at scales of 100s of pc are critical to further interpret the excess of molecular gas in powerful AGN galaxies, and understand its role in fuelling SMBHs, and perhaps the way this SMBH growth may affect the host.


\acknowledgments
We acknowledge support from NASA through ADAP award NNH16CT03C and 80NSSC19K0749 (MK);  
the Israel Science Foundation through grant number 1849/19 (BT);
the Royal Society through the award of a University Research Fellowship (AS); 
ANID grants PIA ACT172033 (ET), Basal-CATA PFB-06/2007 and AFB170002 grants (ET, FEB), FONDECYT Regular 1160999, 1190818 (ET, FEB), and 1200495 (ET,FEB),
and Millennium Science Initiative ICN12\_009 (FEB);
the National Research Foundation of Korea (NRF-2020R1C1C1005462) (KO), 
the Japan Society for the Promotion of Science JSPS ID:17321 (KO); 
the ANID+PAI Convocatoria Nacional subvencion a instalacion en la academia convocatoria a\~{n}o 2017 PAI77170080 (CR);
the UK Science and Technology Facilities Council through grantsST/P000541/1 and ST/T000244/1 (DR); 
and the Jet Propulsion Laboratory, California Institute of Technology, under a contract with NASA (DS). 
We acknowledge the work that \swiftbat team has done to make this project possible. We acknowledge the help of Rozenn Boissay-Malaquin.

This publication is based on data acquired with the Atacama Pathfinder Experiment (APEX). APEX is a collaboration between the Max-Planck-Institut fur Radioastronomie, the European Southern Observatory, and the Onsala Space Observatory.  Based on observations collected at the European Organisation for Astronomical Research in the Southern Hemisphere under ESO programmes 198.A-0708(A), 0100.A-0384(A), 097.B-0757(A), 098.B-0152(A), 081.F-9405(A), and 091.F-9313(A) as well as Chilean programs C-097.F-9705A-2016, C-098.F-9700-2016, and C-0100.F-9715. The James Clerk Maxwell Telescope is operated by the East Asian Observatory on behalf of The National Astronomical Observatory of Japan; Academia Sinica Institute of Astronomy and Astrophysics; the Korea Astronomy and Space Science Institute; Center for Astronomical Mega-Science (as well as the National Key R\&D Program of China with No. 2017YFA0402700). Additional funding support is provided by the Science and Technology Facilities Council of the United Kingdom and participating universities in the United Kingdom and Canada. This project involved JCMT programs M11AH42C, M11BH35C, M12AH35C, M12BH03E, and M09BH34B. The {\tt Starlink} software \citep{Currie:2014:391} is currently supported by the East Asian Observatory.     This research made use of: the NASA/IPAC Extragalactic Database (NED), which is operated by the Jet Propulsion Laboratory, California Institute of Technology, under contract with the National Aeronautics and Space Administration and the SIMBAD database, operated at CDS, Strasbourg, France \citep{Wenger:2000:9}.

\facilities{APEX, JCMT, PS1, Swift (BAT)}

\software{astropy \citep{Collaboration:2013:A33},  
          Matplotlib \citep{Hunter:2007:90}, 
          Numpy \citep{vanderWalt:2011:22},
          APLpy \citep{Robitaille:2012:1208.017}
          }





%

\appendix
\section{Morphological Classification}
\label{morph_appen}
\begin{figure*}
\centering
\includegraphics[width=0.18\textwidth]{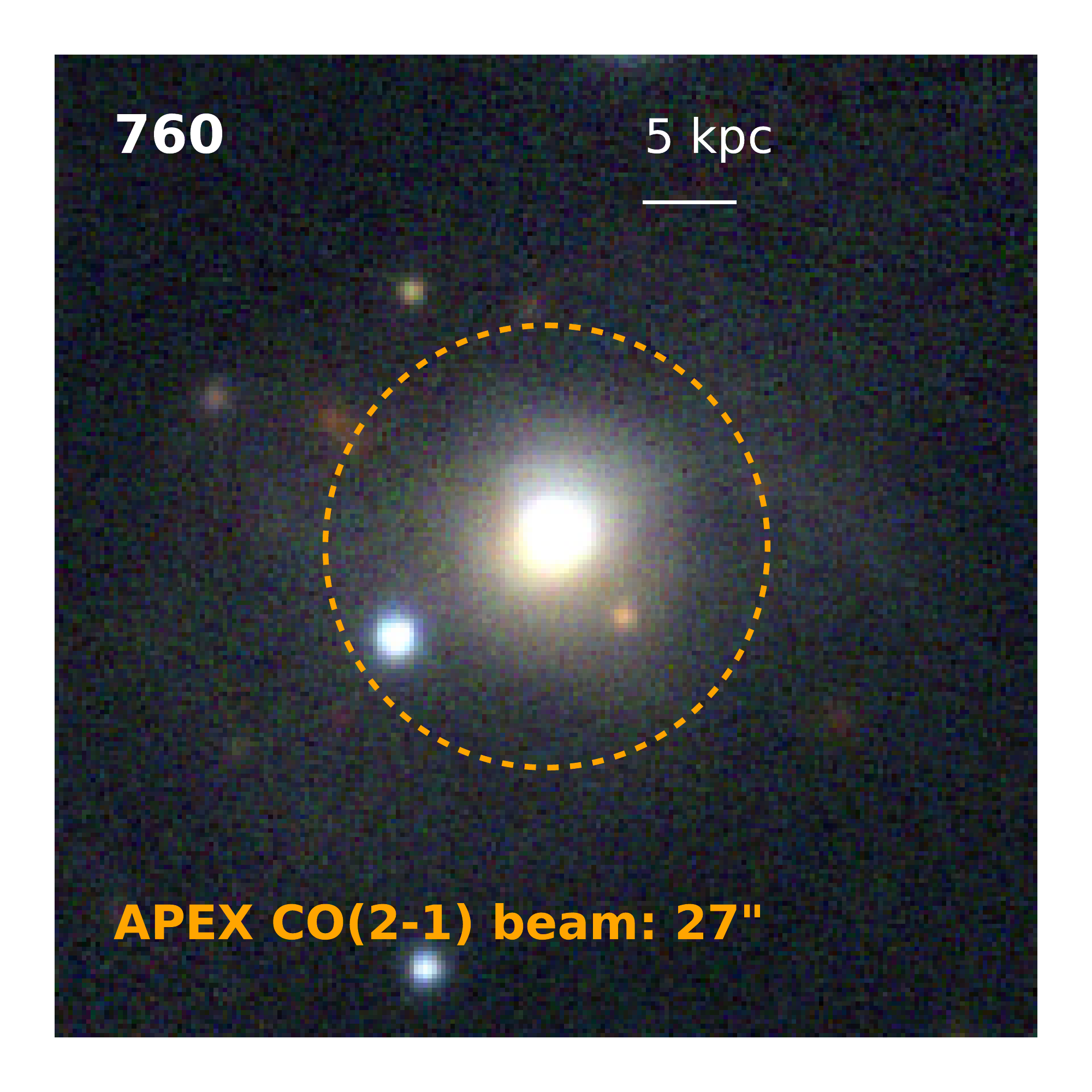}\includegraphics[width=0.26\textwidth]{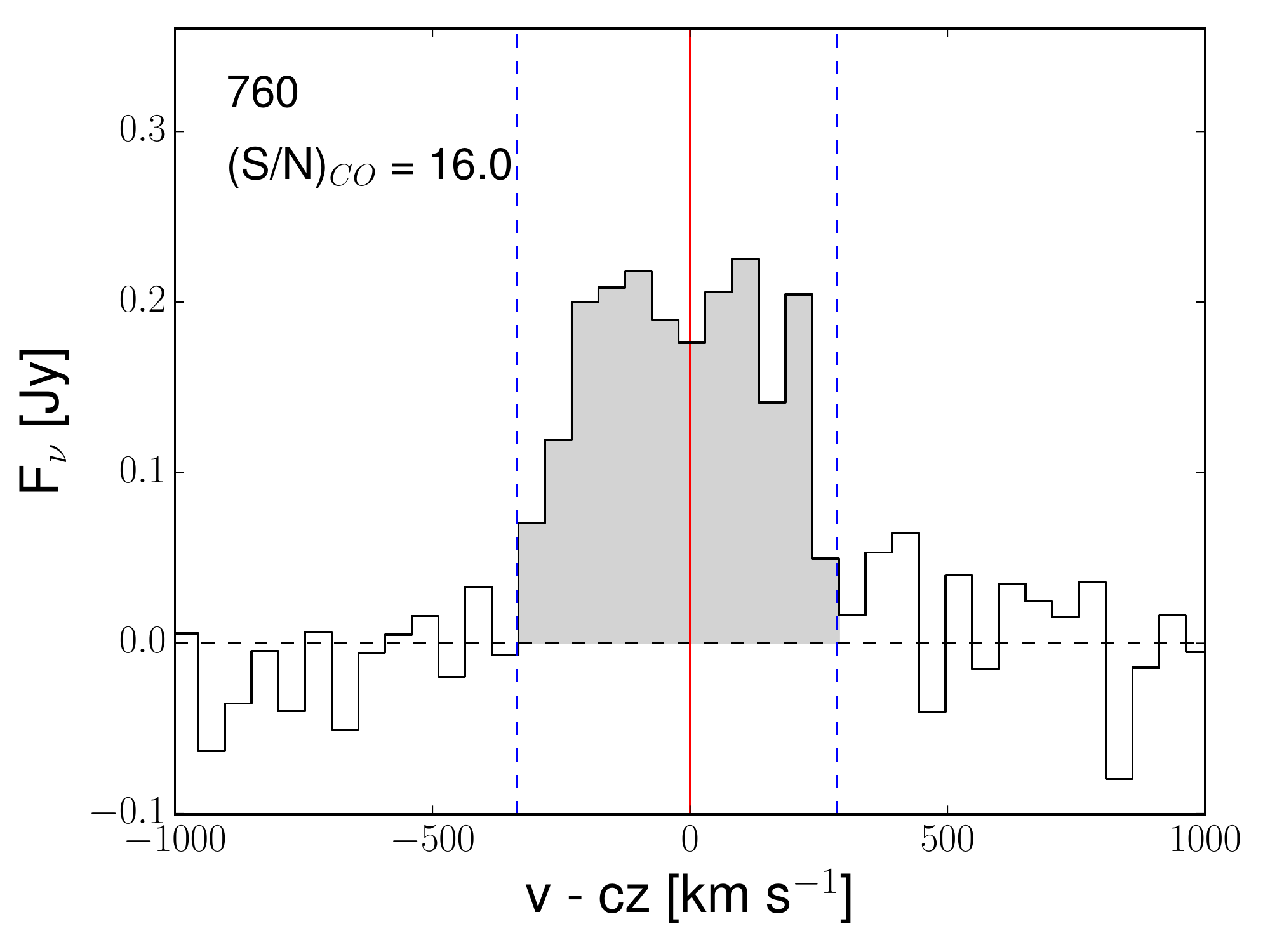}
\includegraphics[width=0.18\textwidth]{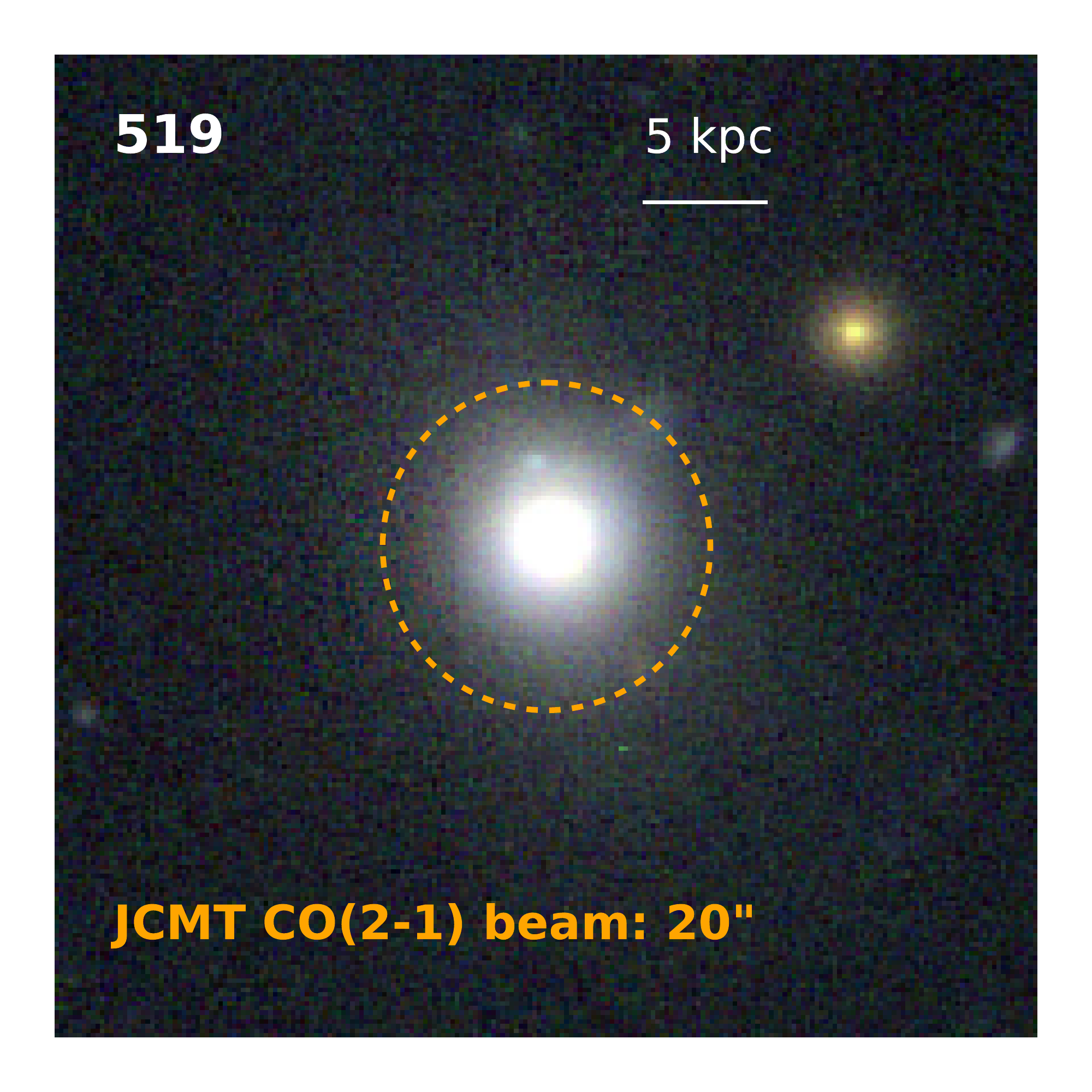}\includegraphics[width=0.26\textwidth]{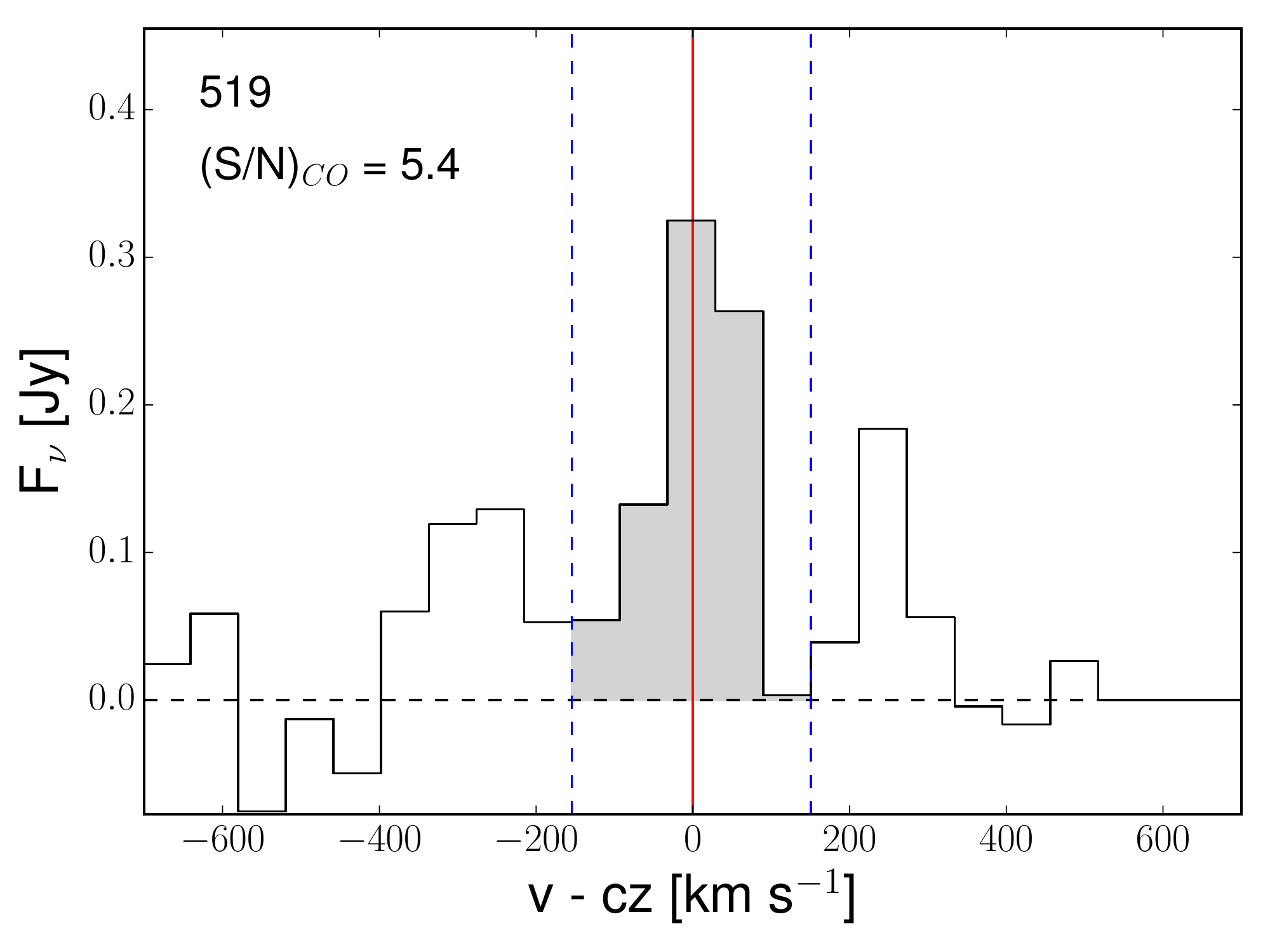}
\includegraphics[width=0.18\textwidth]{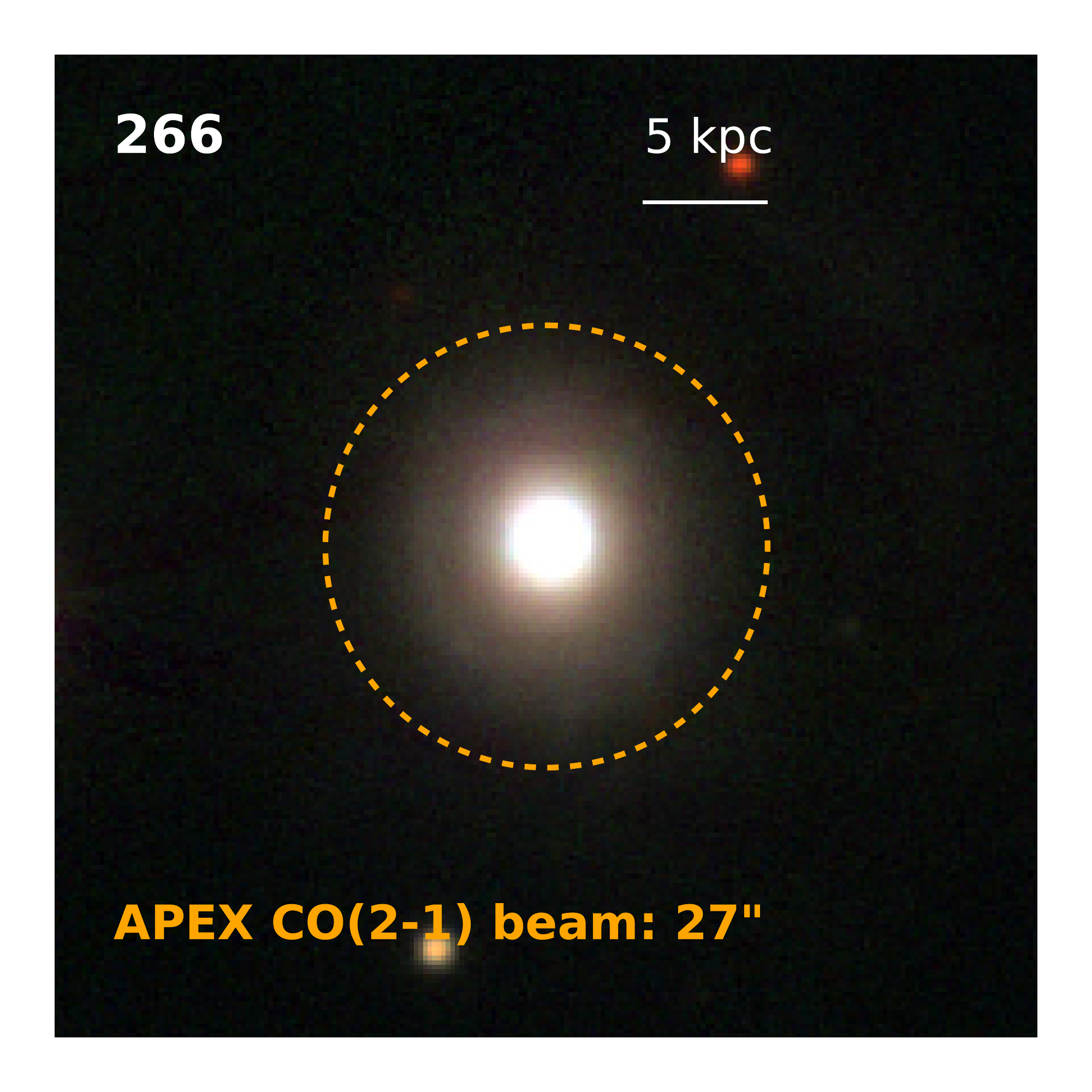}\includegraphics[width=0.26\textwidth]{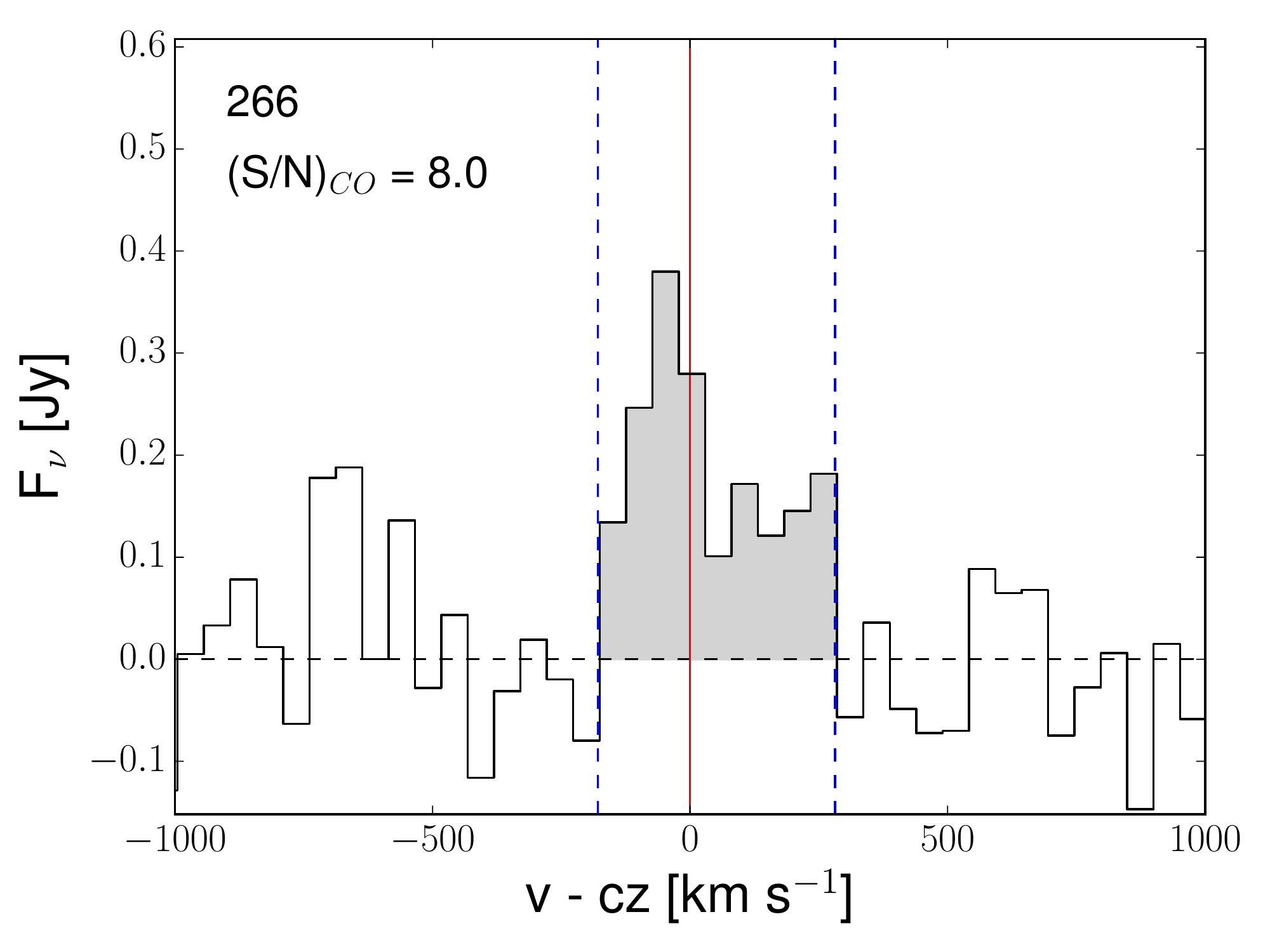}
\includegraphics[width=0.18\textwidth]{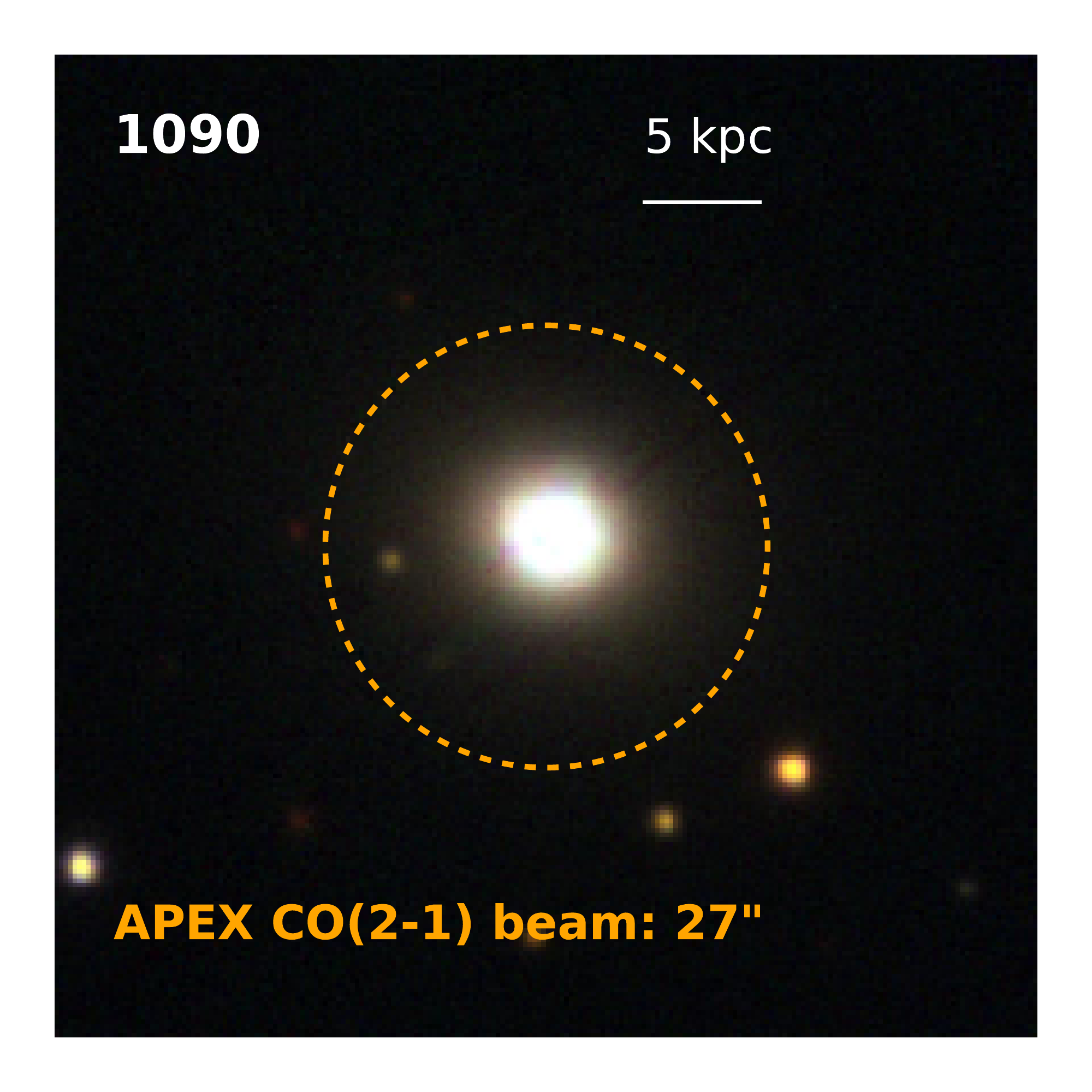}\includegraphics[width=0.26\textwidth]{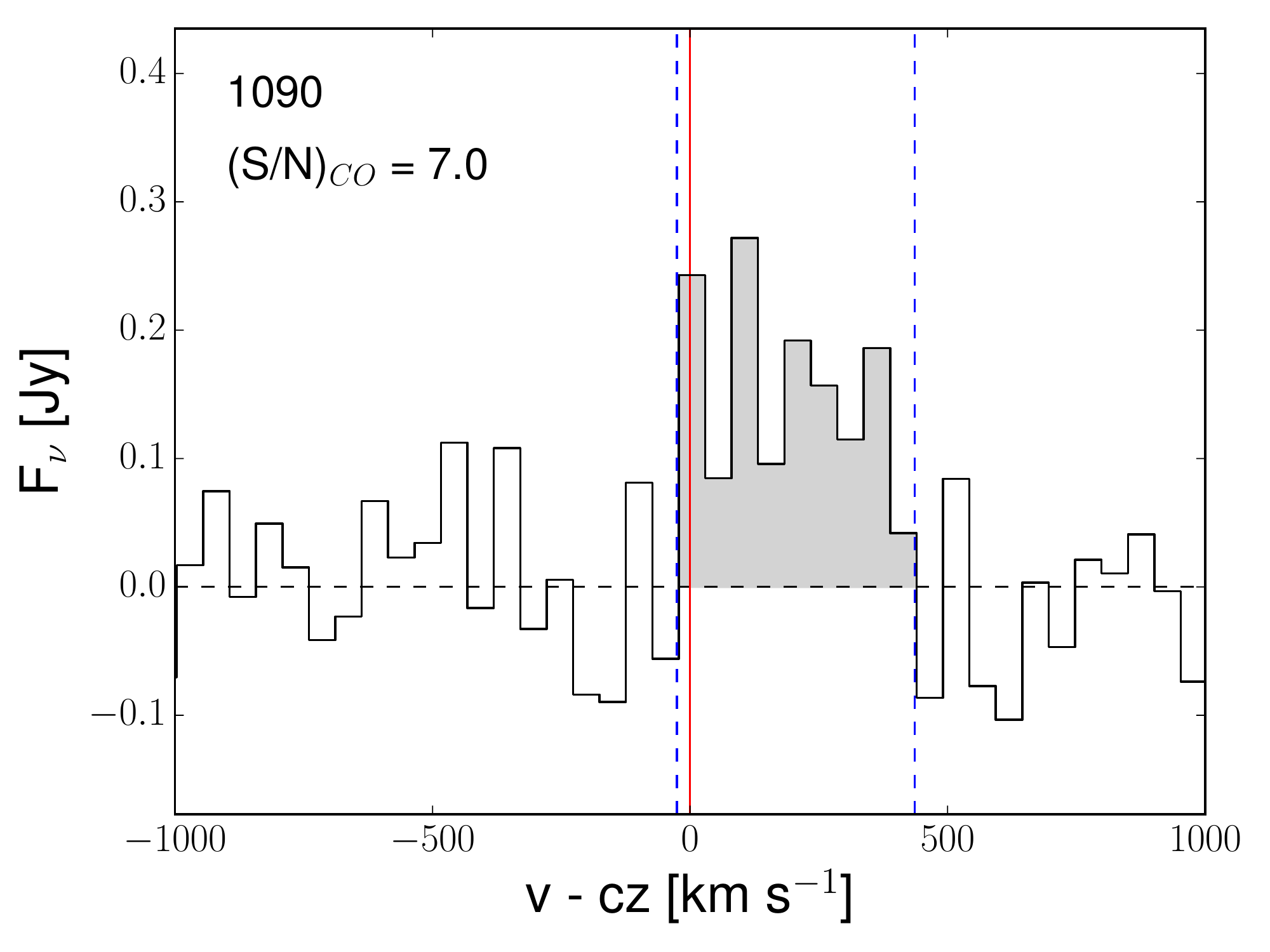}
\includegraphics[width=0.18\textwidth]{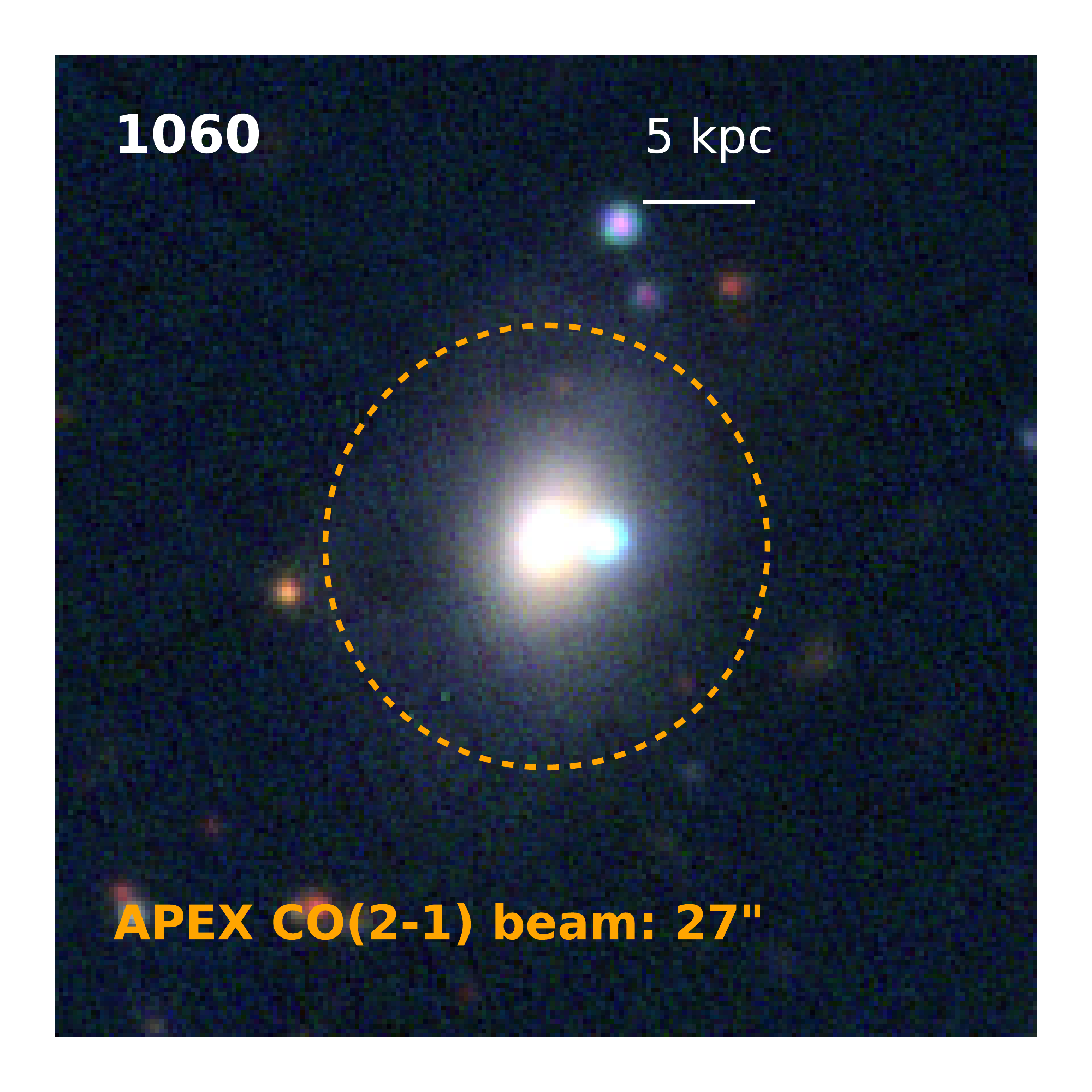}\includegraphics[width=0.26\textwidth]{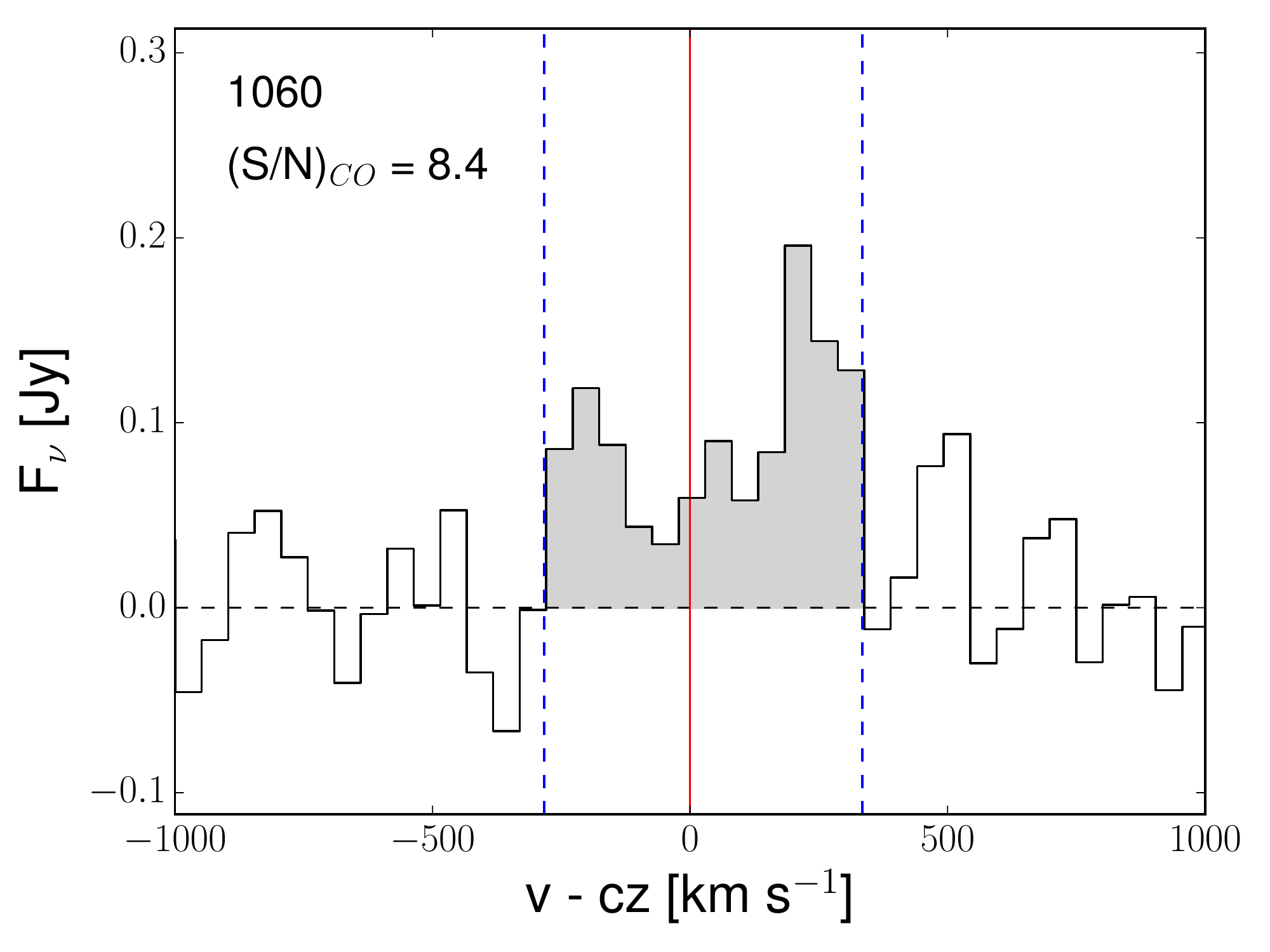}
\includegraphics[width=0.18\textwidth]{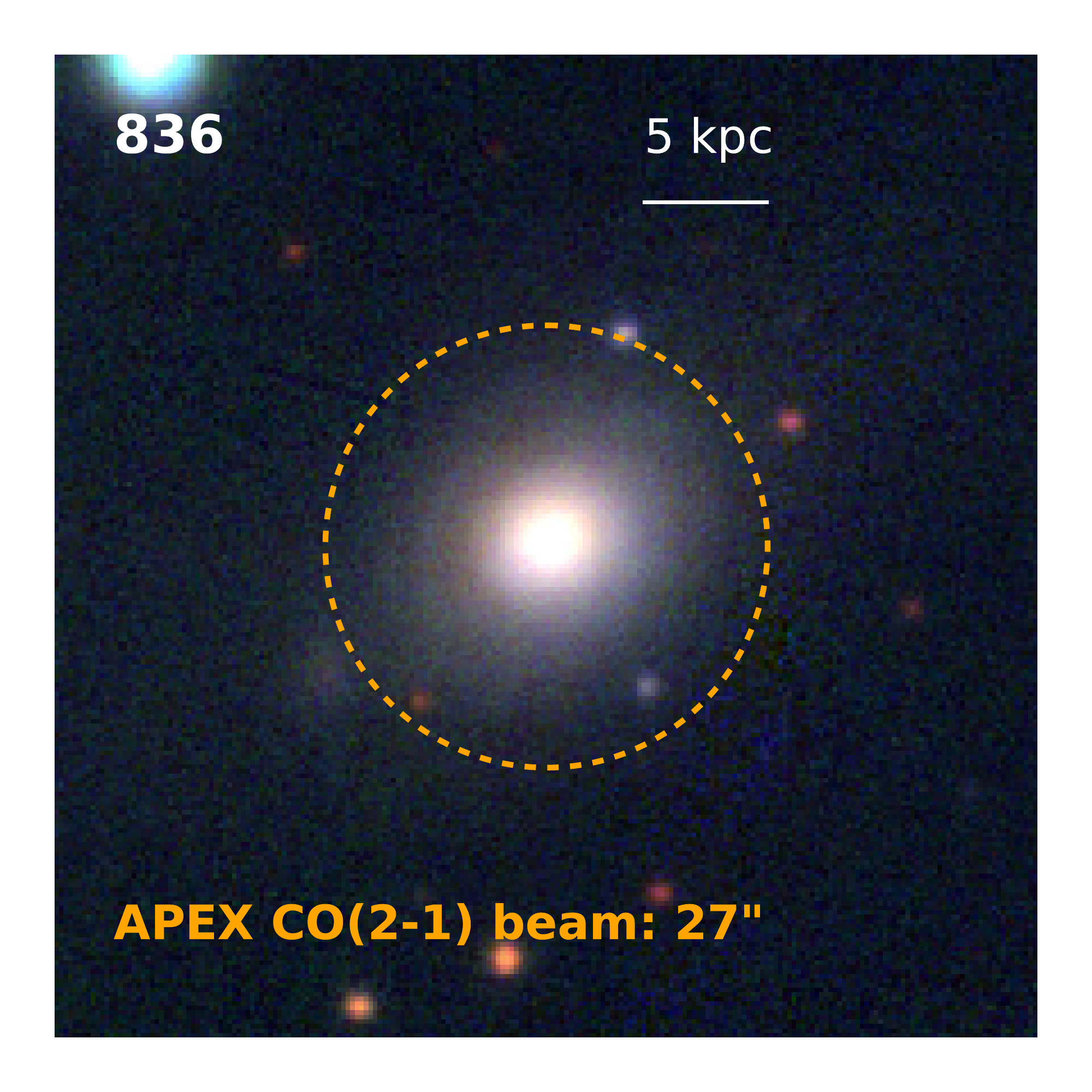}\includegraphics[width=0.26\textwidth]{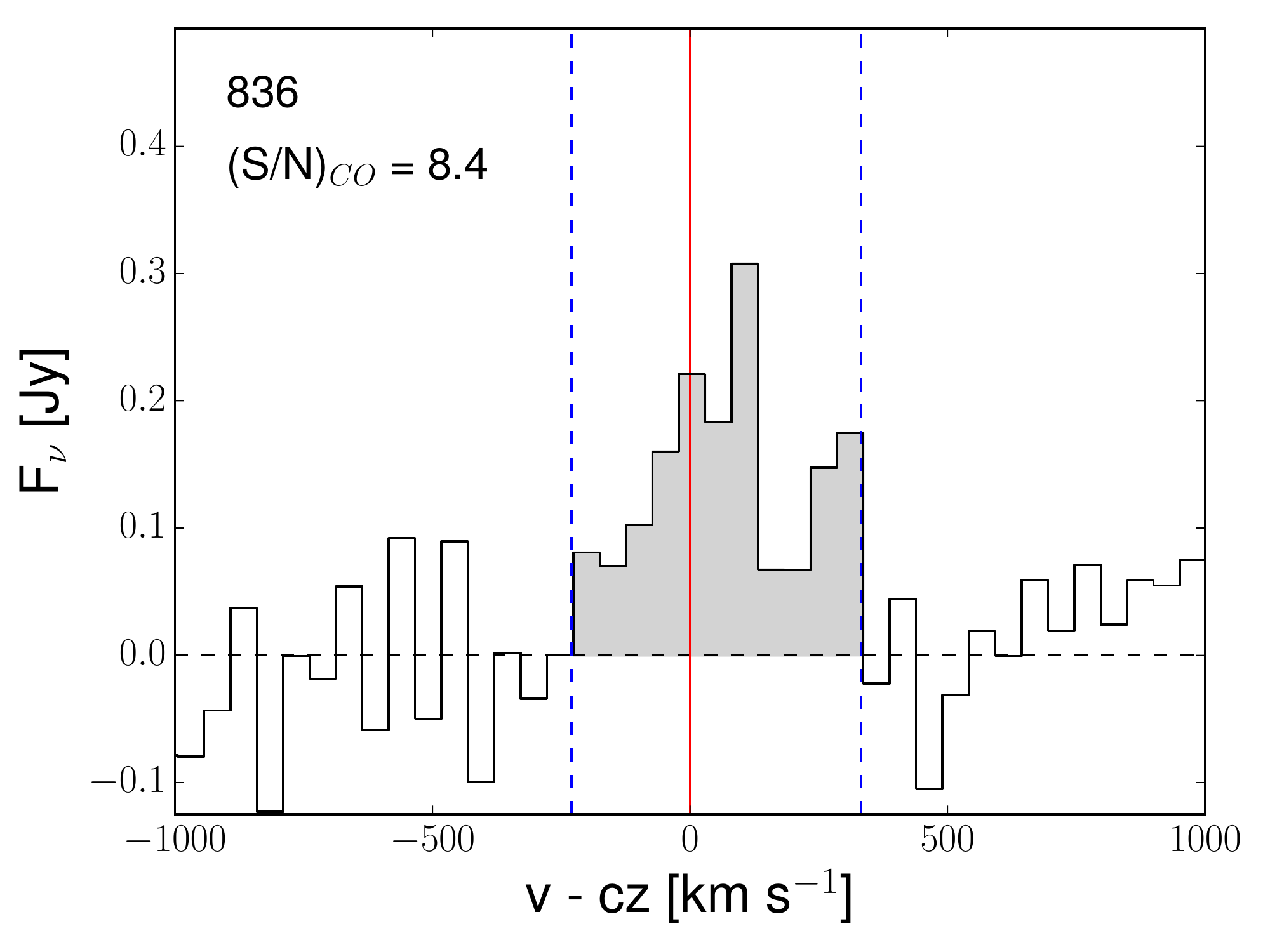}
\includegraphics[width=0.18\textwidth]{BAT260_PS1_image_size1024_pix_60arcsec_dpi30.pdf}\includegraphics[width=0.26\textwidth]{BAT260_CO21_spectrum_for_paper.pdf}
\includegraphics[width=0.18\textwidth]{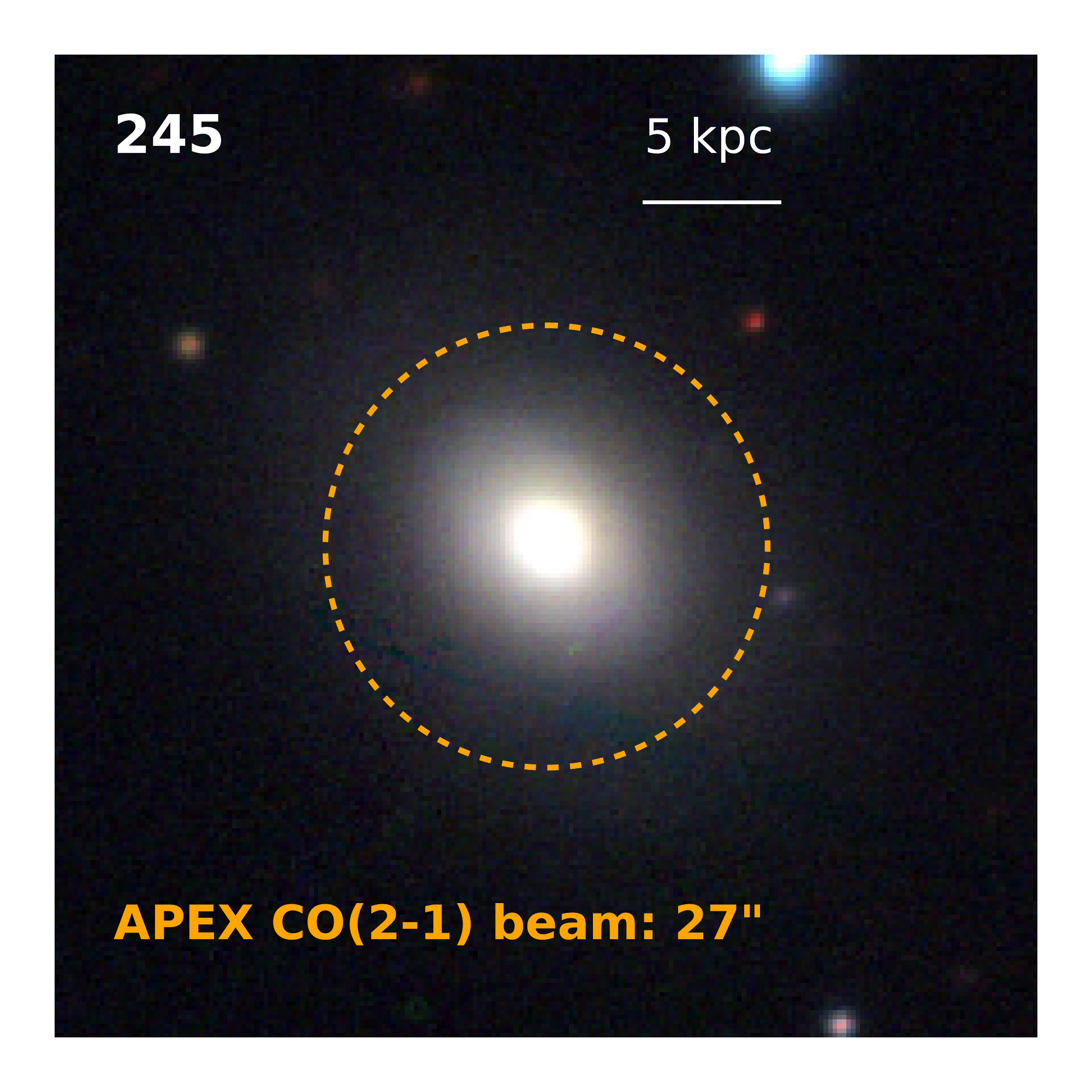}\includegraphics[width=0.26\textwidth]{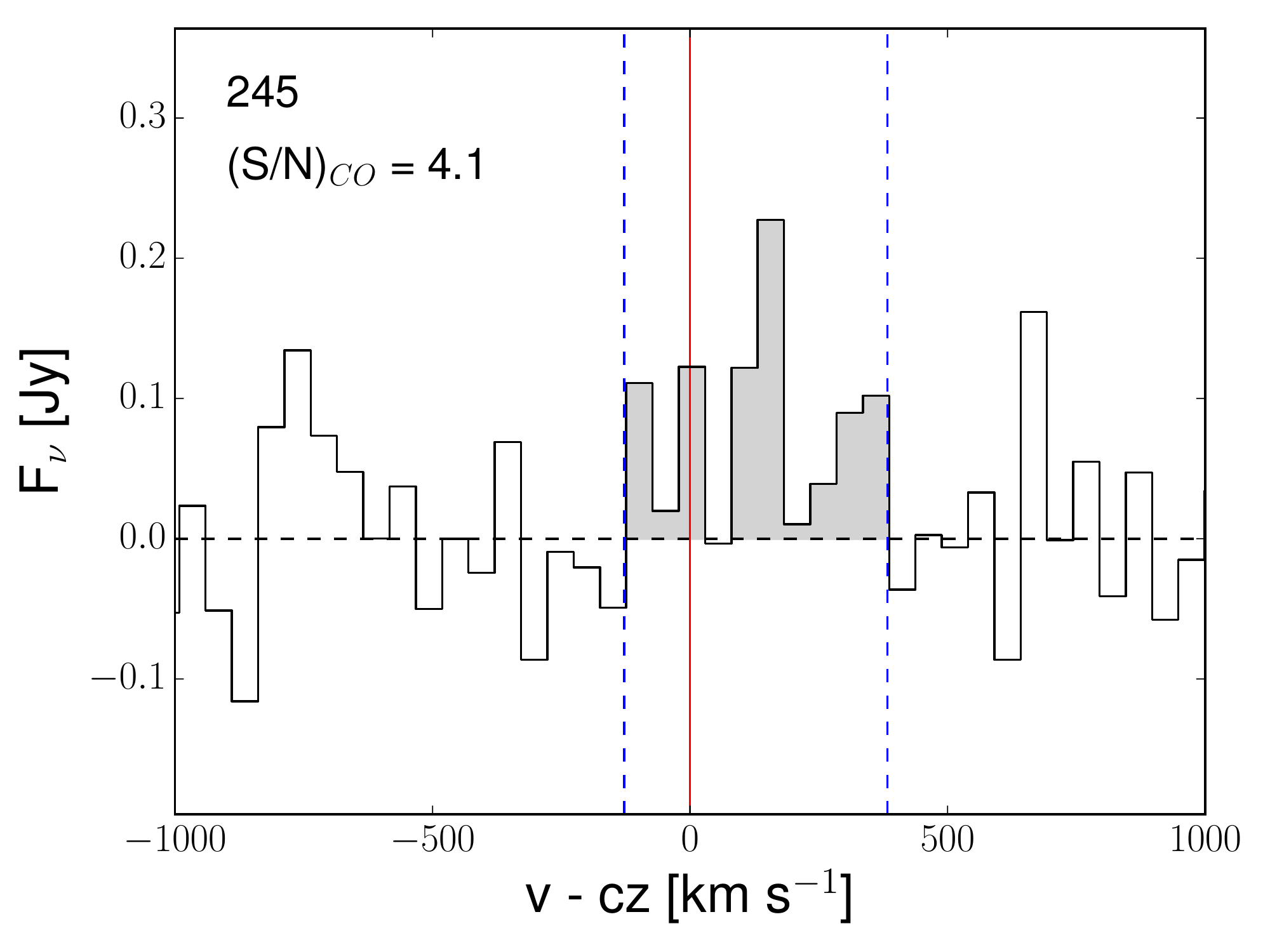}
\caption{Pan-STARRS $1'\times1'$ example $gri$ images of the BAT AGN sample for galaxies classified as elliptical with some of the highest molecular gas masses.  The galaxies are sorted from largest (upper left, ID=760, $\log M_{H2}=10.27$ \Msun) to smallest (bottom right, ID=102, $\log M_{H2}=9.53$ \Msun) in molecular gas content going from left to right and top to bottom.  All of these galaxies have more molecular gas than any of the 81 inactive massive ($\log$ \mstar$>10.2$ \Msun) ellipticals from xCOLD GASS. 
} 
\label{fig:ellip_pic}
\end{figure*}

\begin{figure*}
\centering
\includegraphics[width=0.18\textwidth]{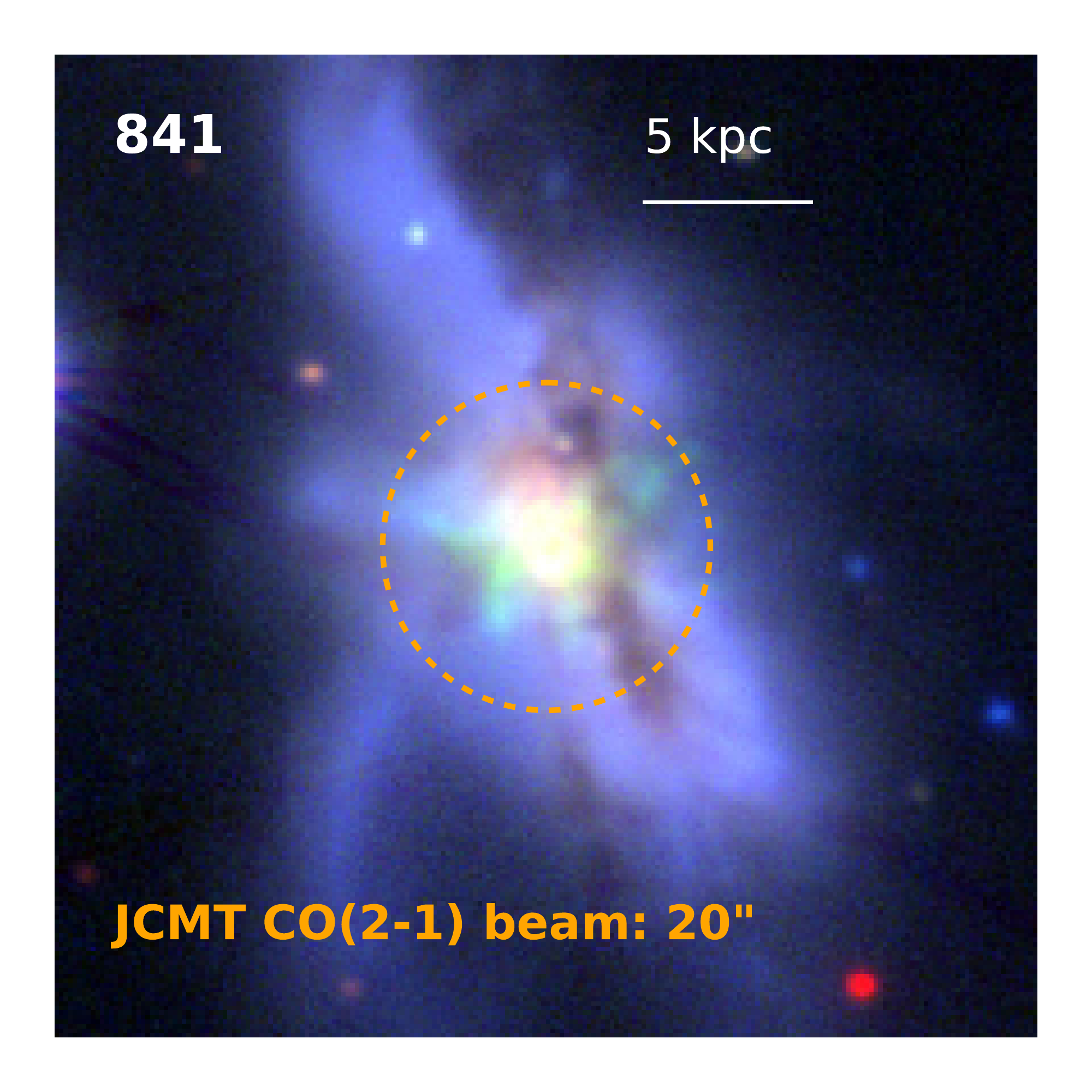}\includegraphics[width=0.26\textwidth]{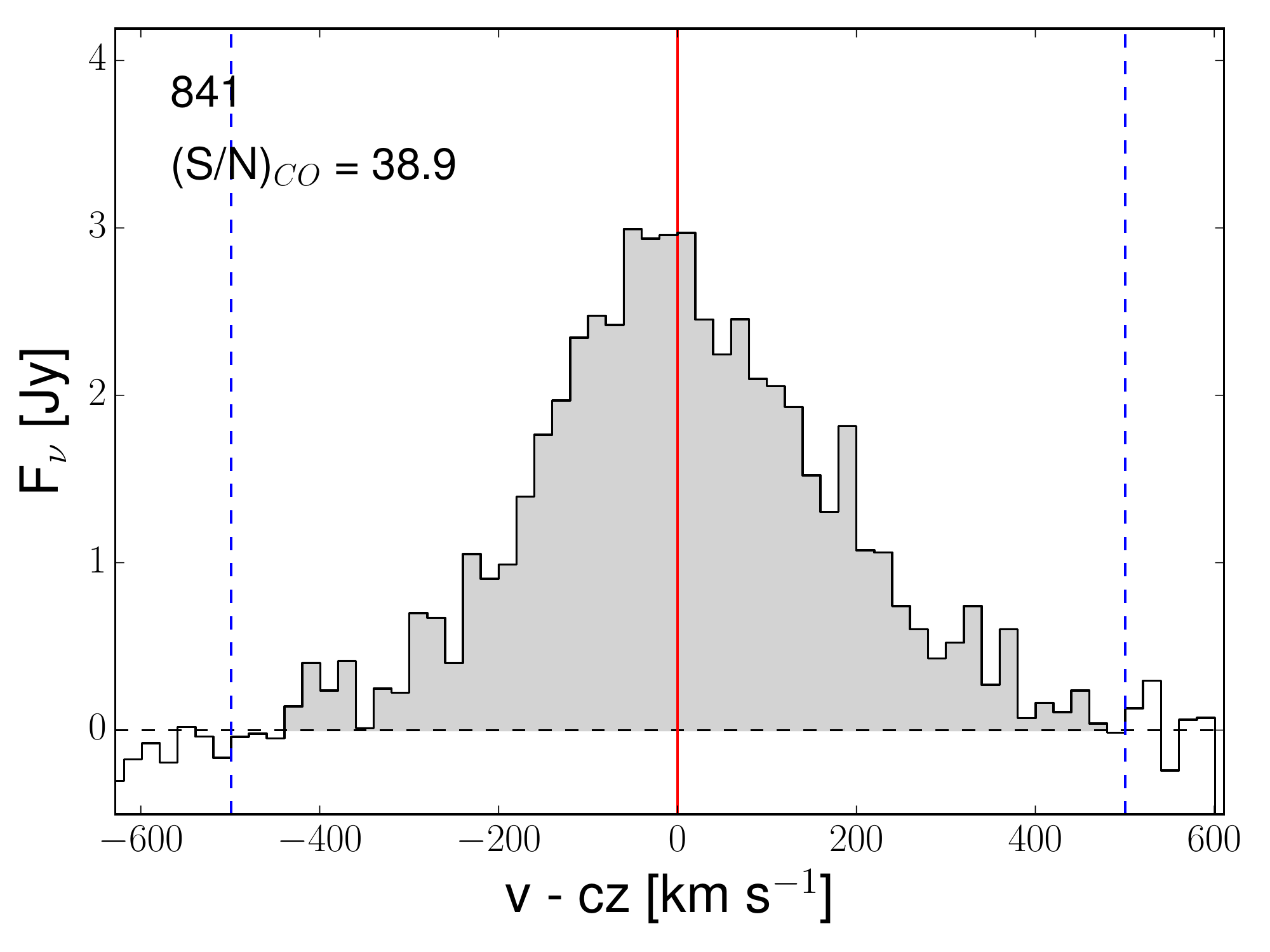}
\includegraphics[width=0.18\textwidth]{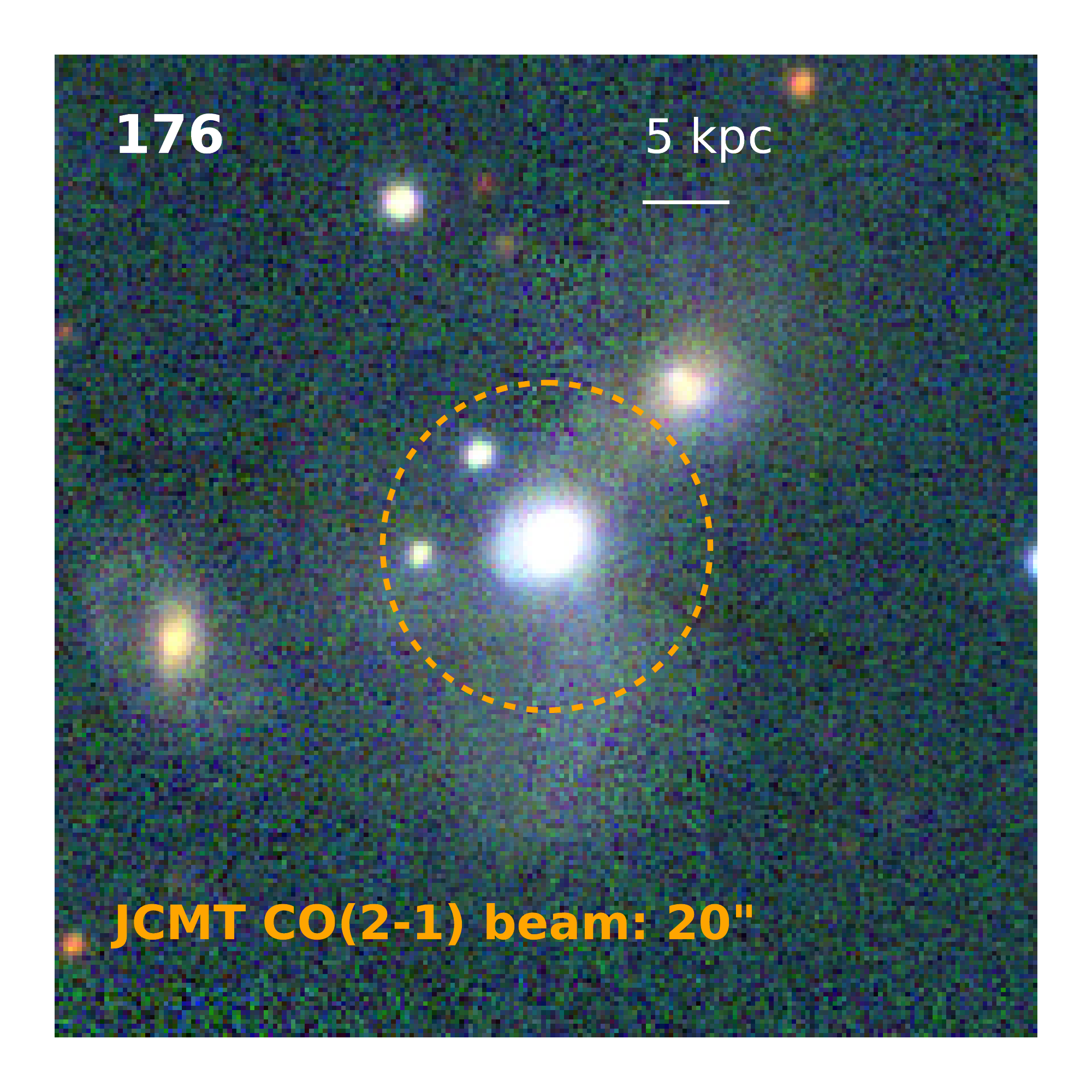}\includegraphics[width=0.26\textwidth]{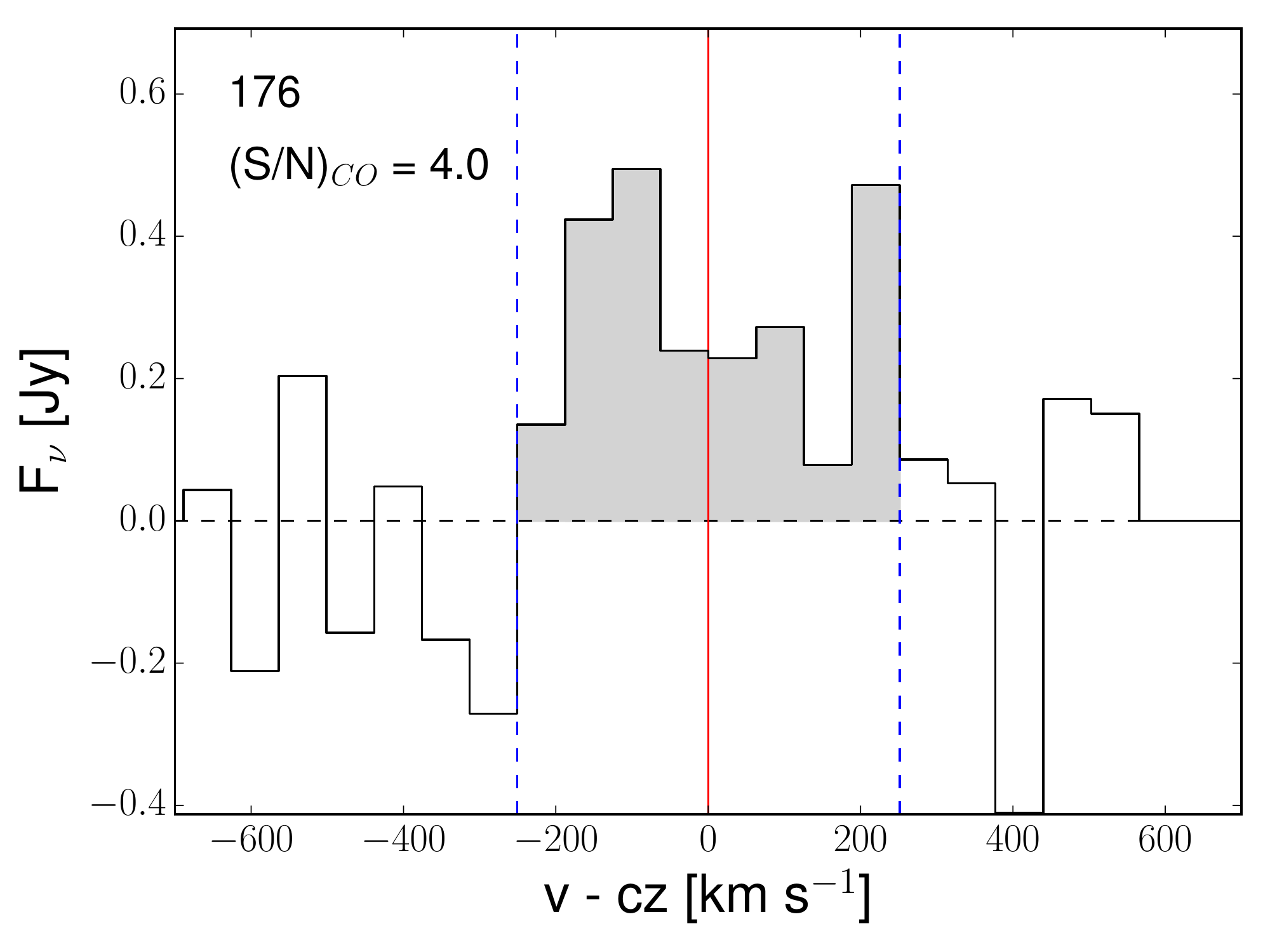}
\includegraphics[width=0.18\textwidth]{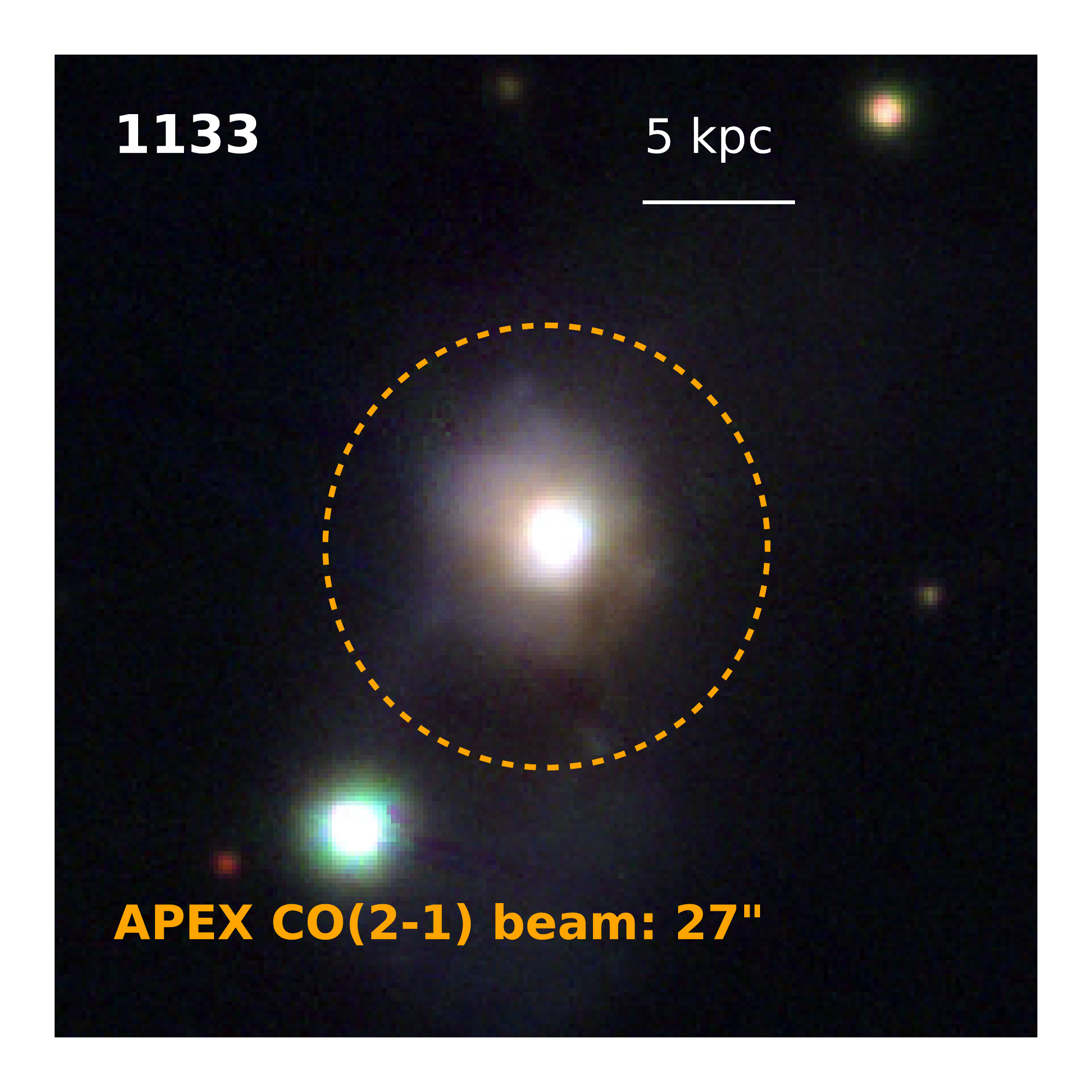}\includegraphics[width=0.26\textwidth]{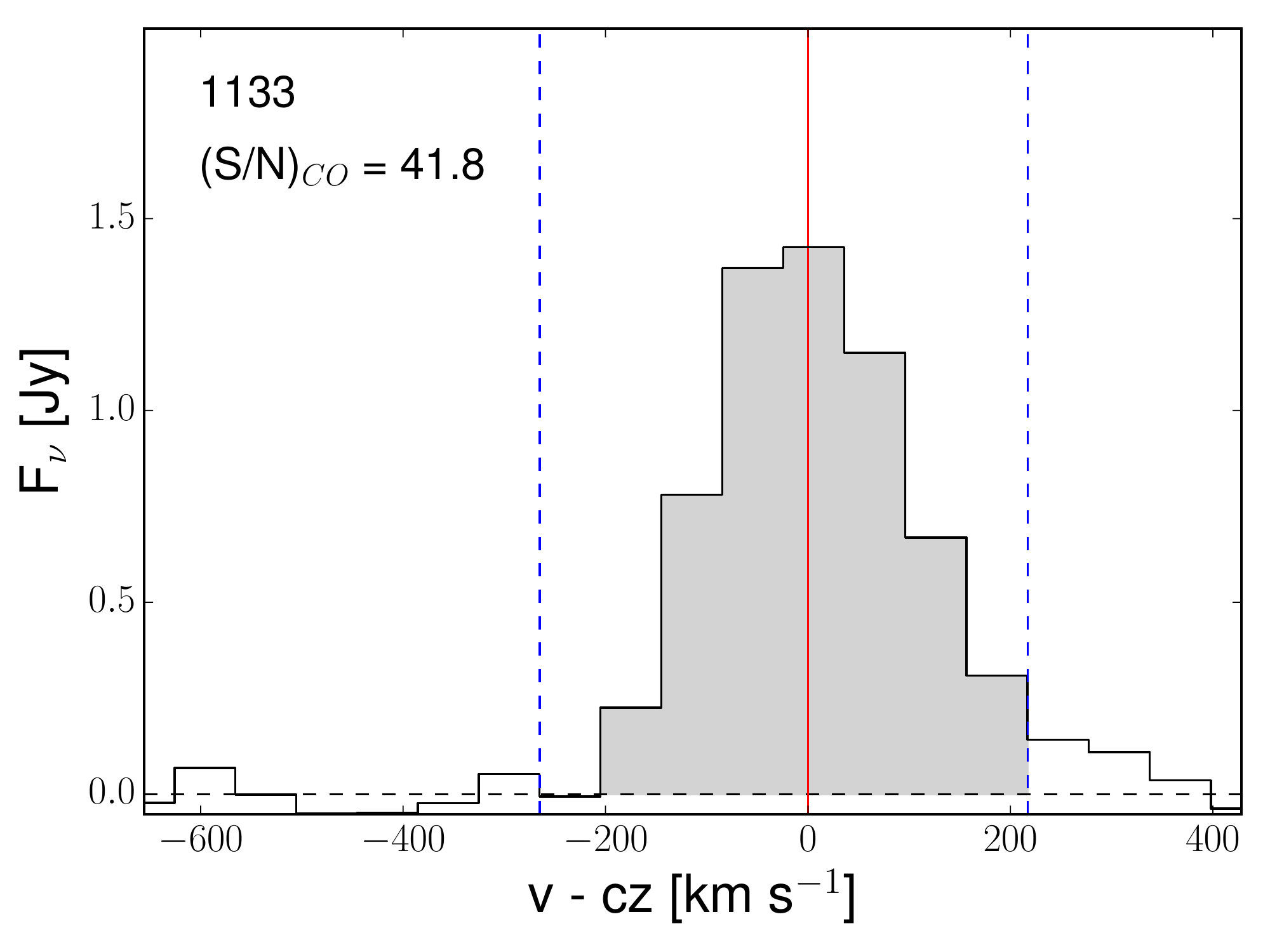}
\includegraphics[width=0.18\textwidth]{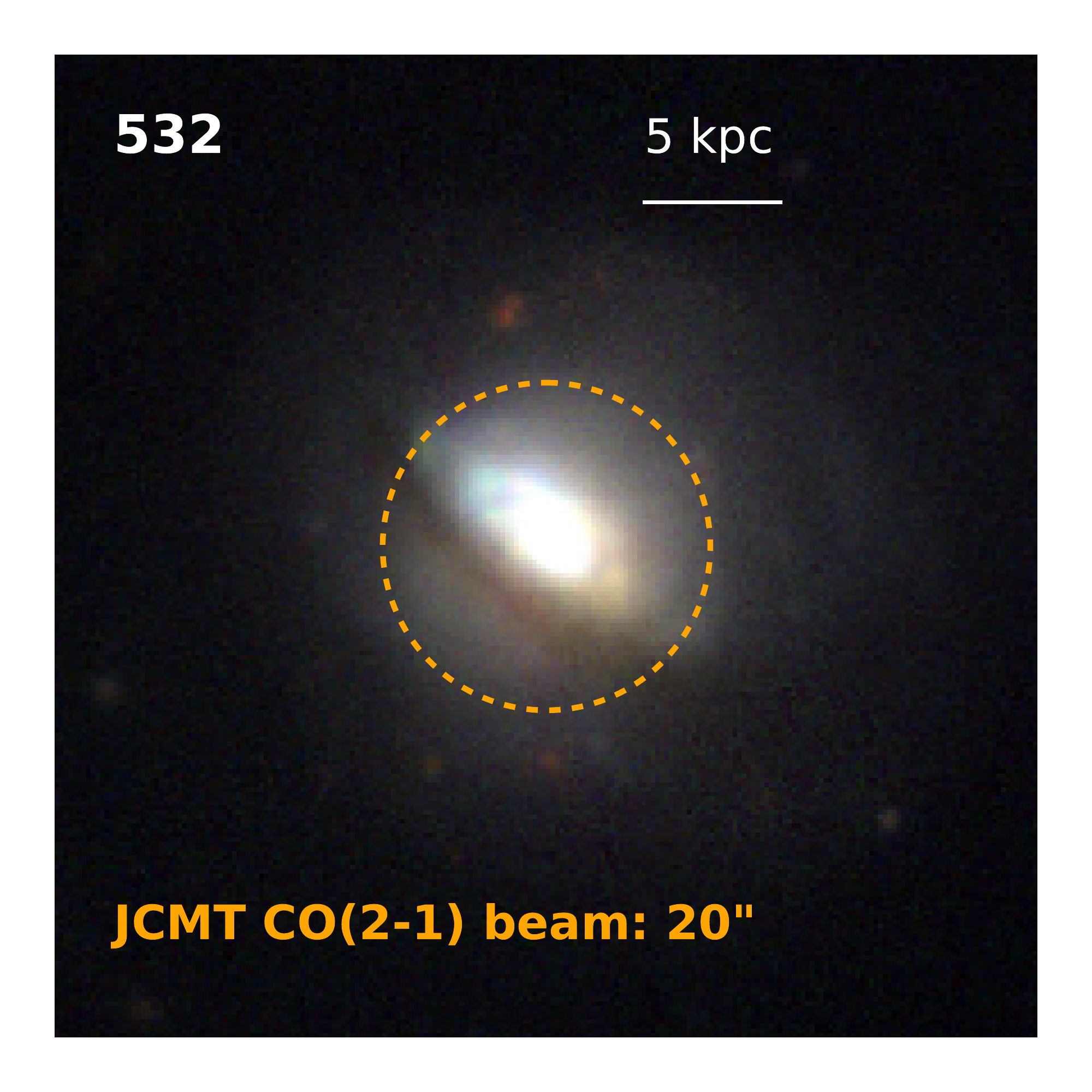}\includegraphics[width=0.26\textwidth]{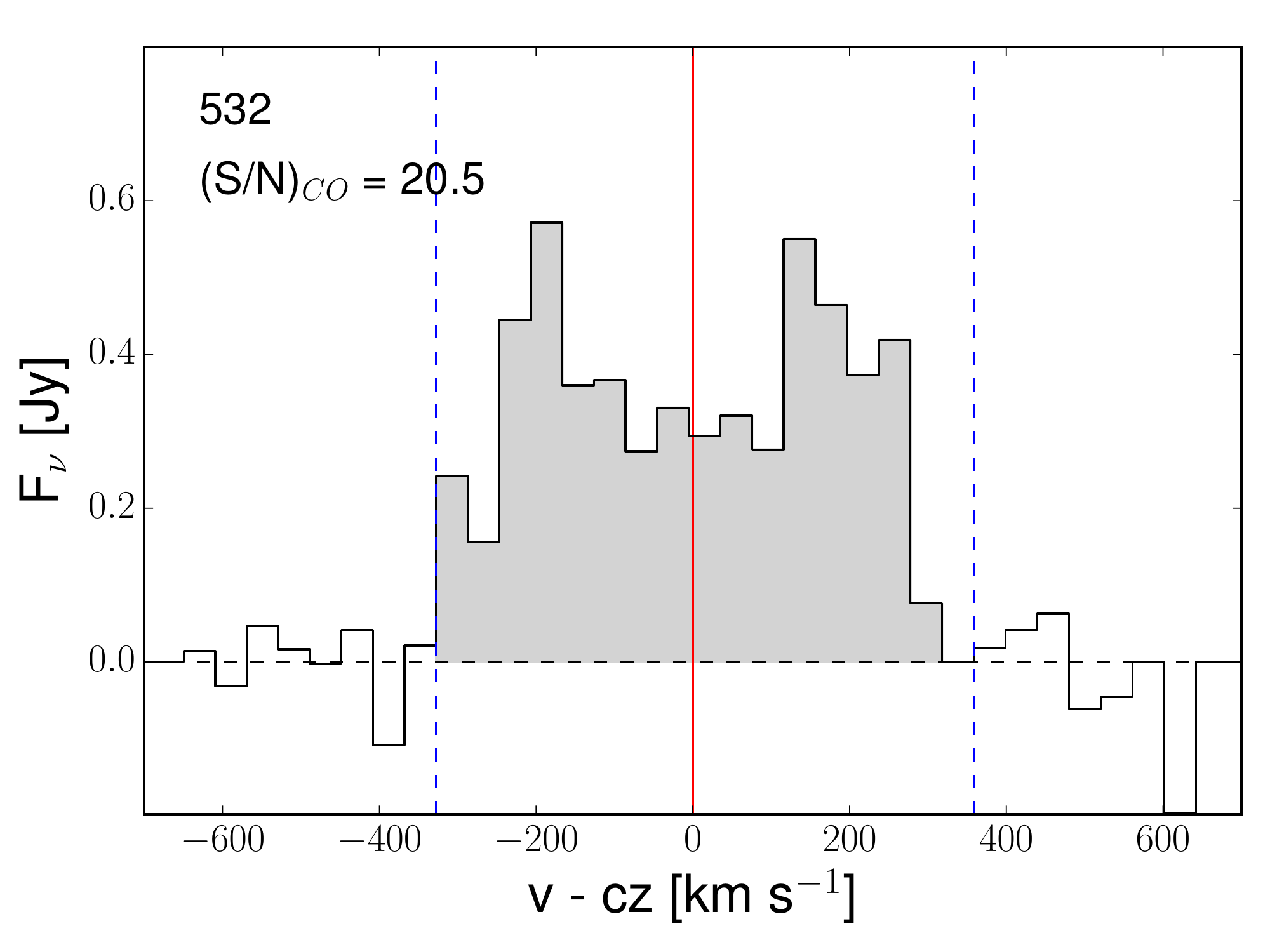}
\includegraphics[width=0.18\textwidth]{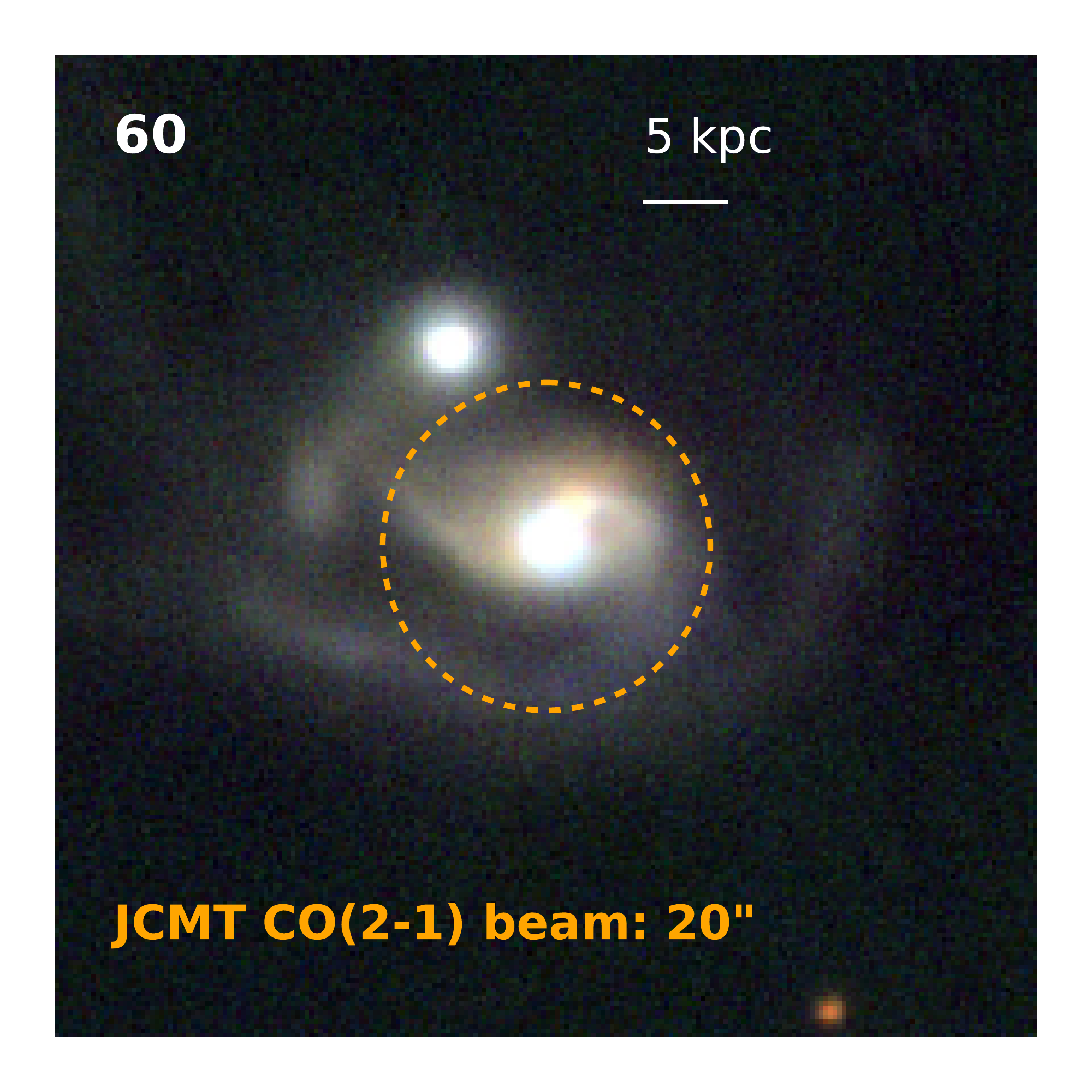}\includegraphics[width=0.26\textwidth]{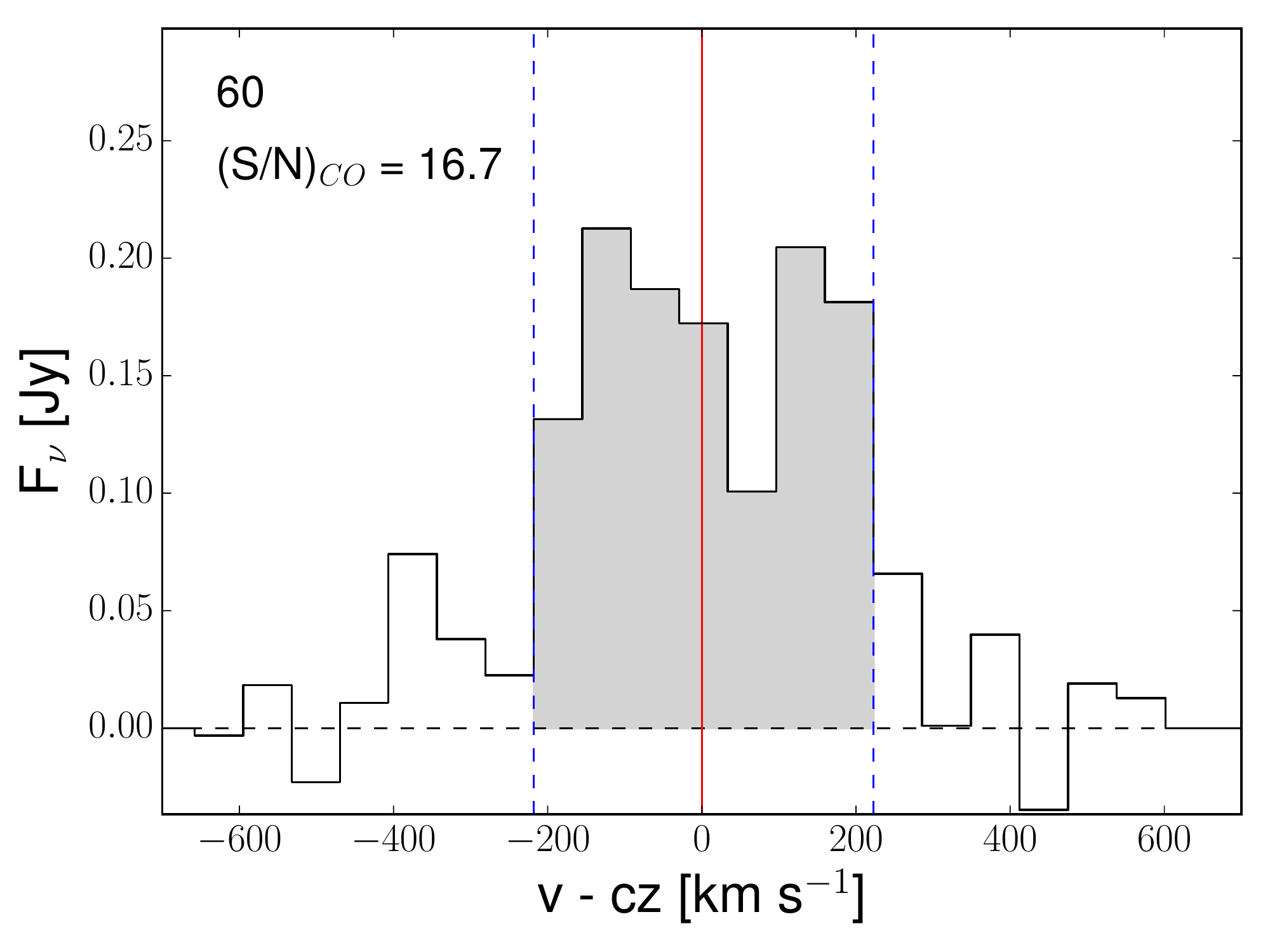}
\includegraphics[width=0.18\textwidth]{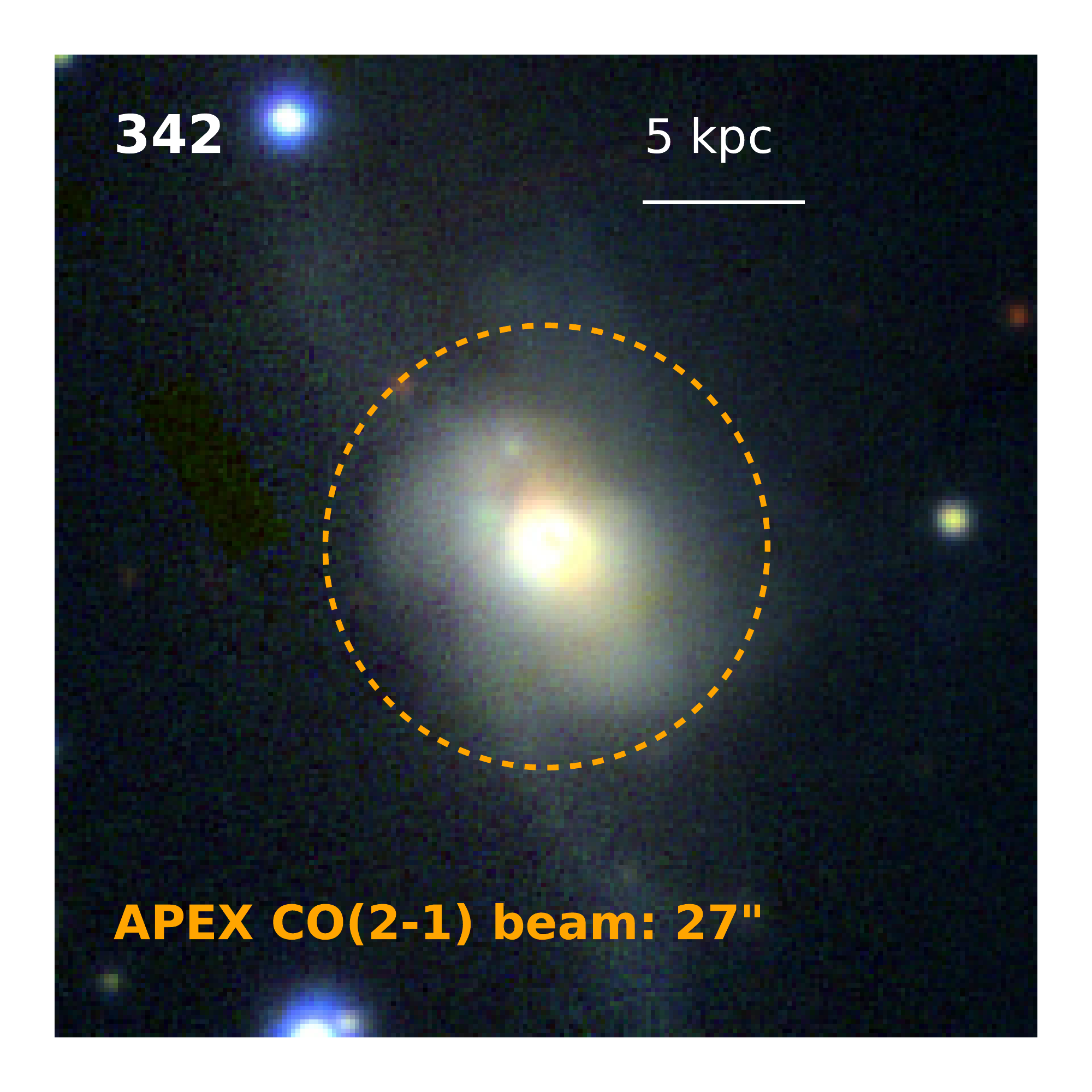}\includegraphics[width=0.26\textwidth]{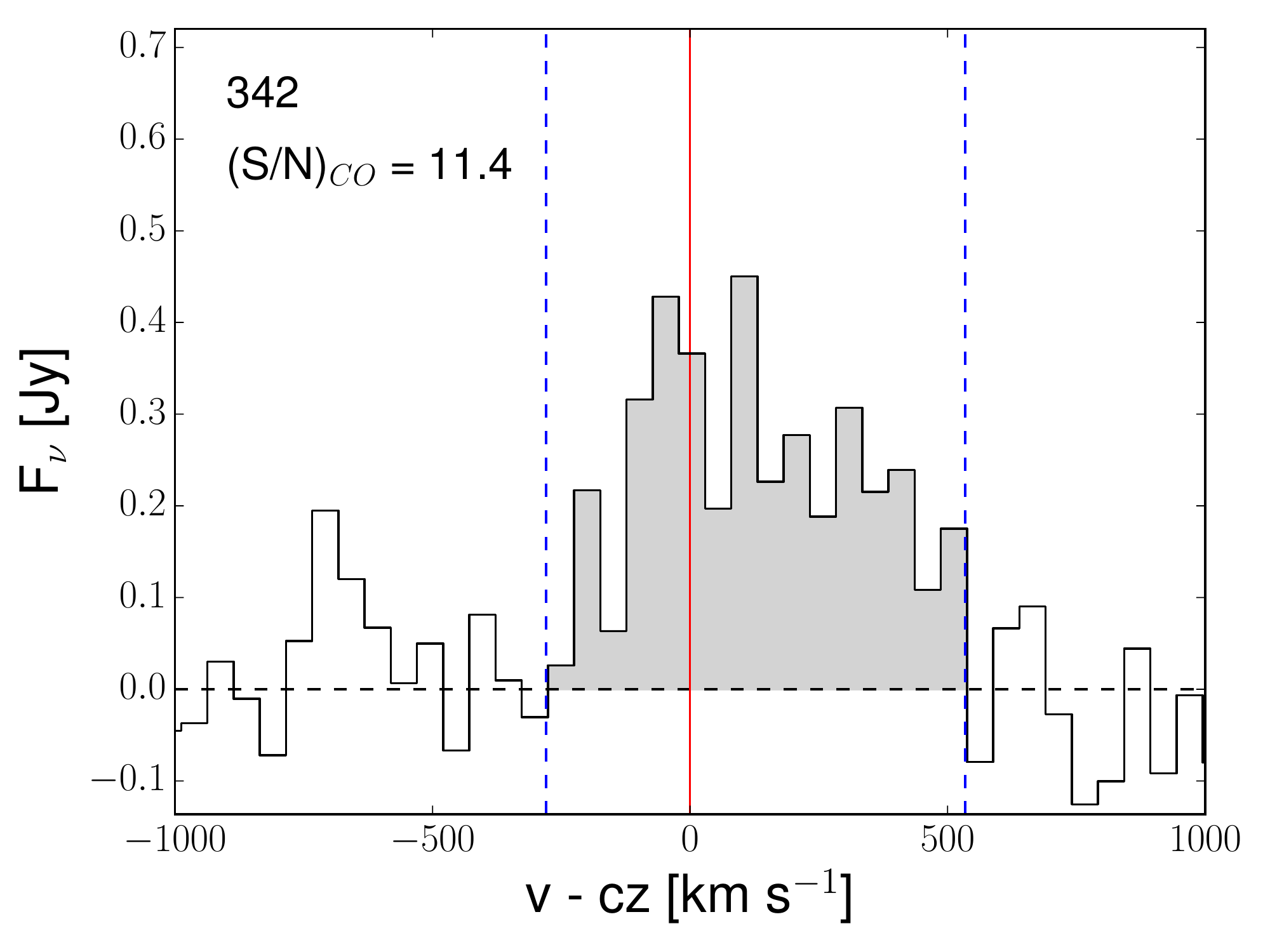}
\caption{Pan-STARRS $1'\times1'$ example $gri$ images of the BAT AGN sample for galaxies classified as uncertain with some of the highest molecular gas masses.  The galaxies are sorted from largest (upper left, ID=841, $\log (M_{H2}/\Msun)=10.76$ ) to smallest (bottom right, ID=342, $\log (M_{H2}/\Msun)=10.01$) in molecular gas content going from left to right and top to bottom.  All galaxies on the top row have more molecular gas than any of the 105 inactive massive ($\log (M_{H2}/\Msun)>10.2$) uncertain galaxy morphologies from xCOLD GASS. 
} 
\label{fig:uncertain_pic}
\end{figure*}

\begin{figure*}
\centering
\includegraphics[width=0.18\textwidth]{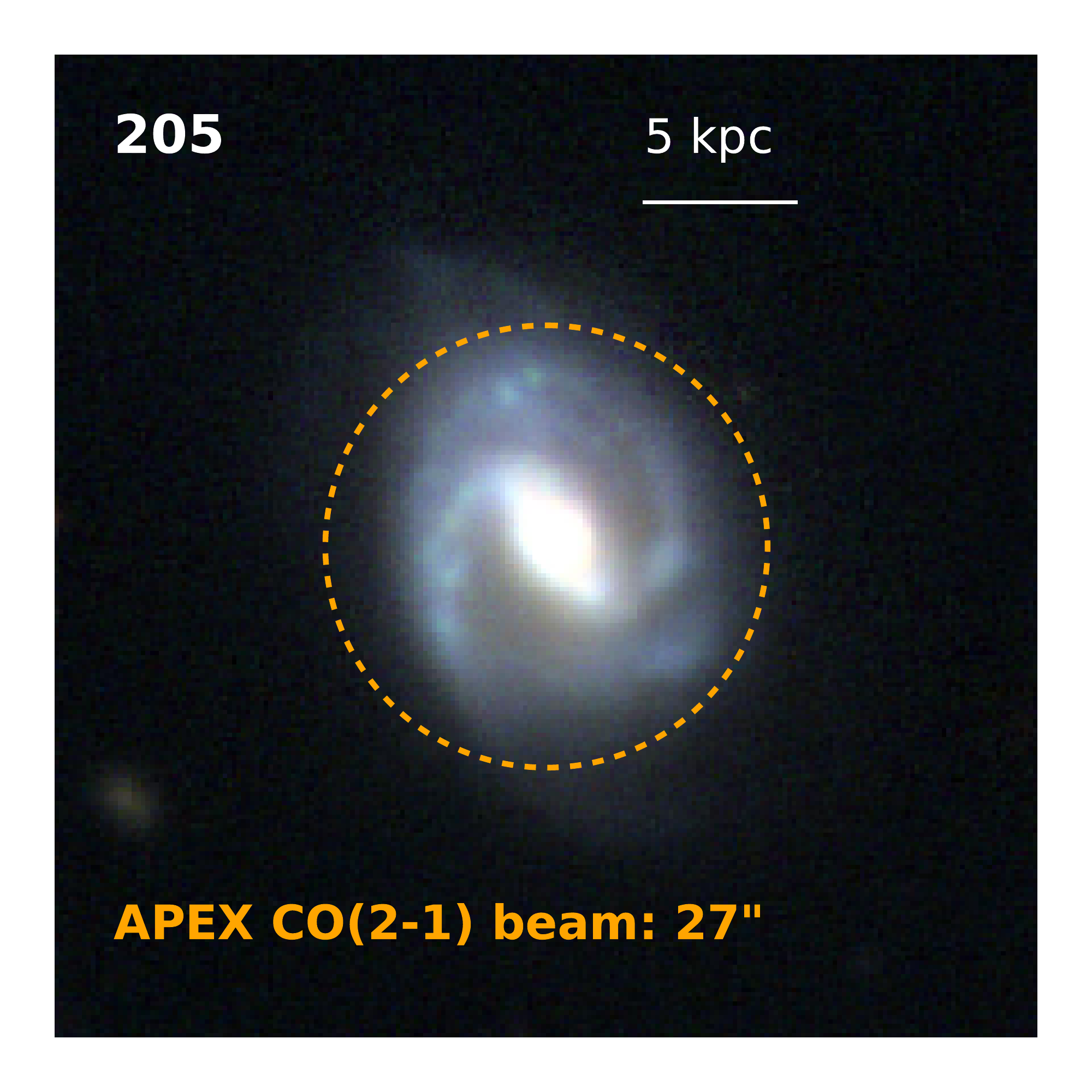}\includegraphics[width=0.26\textwidth]{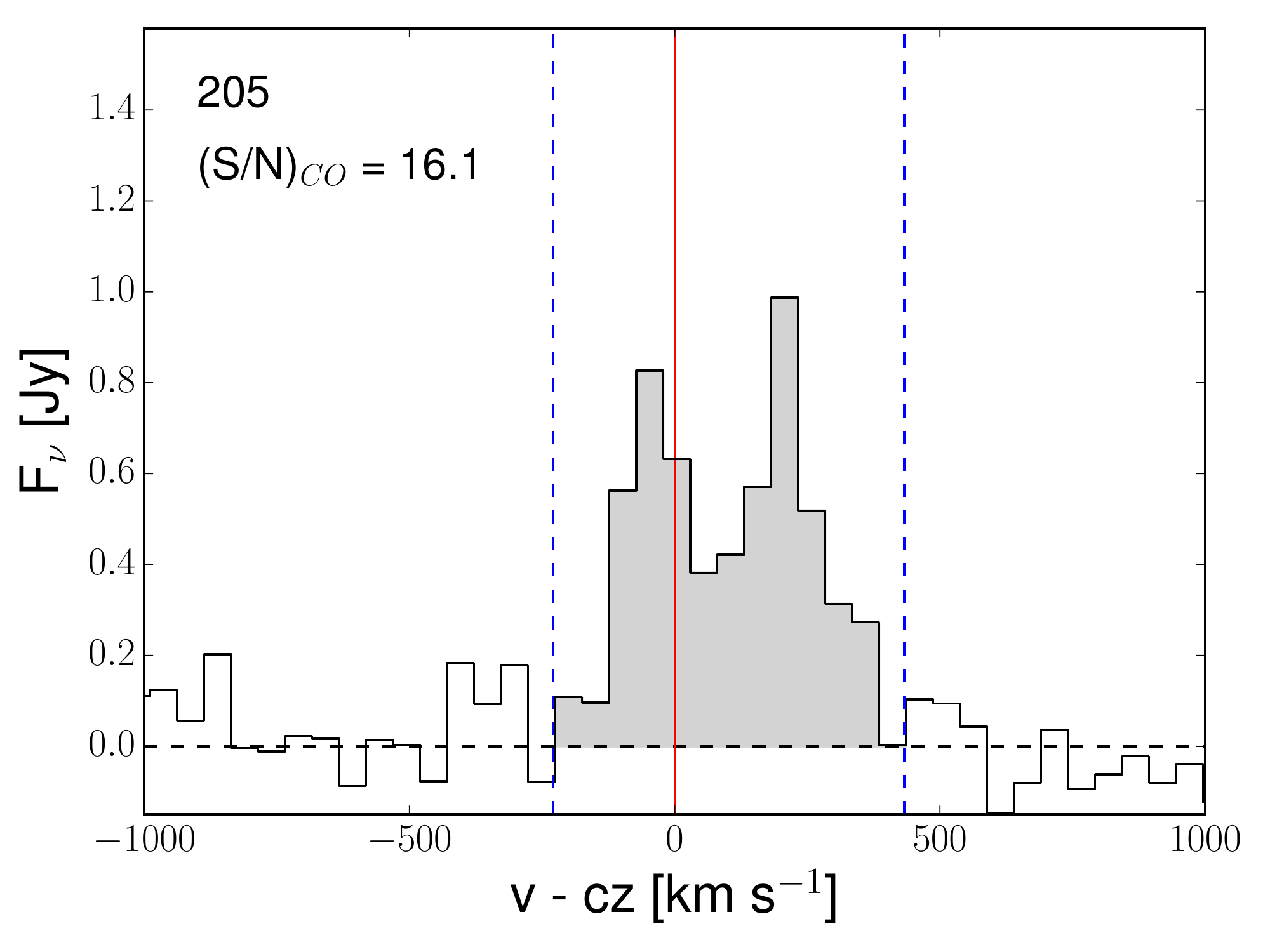}
\includegraphics[width=0.18\textwidth]{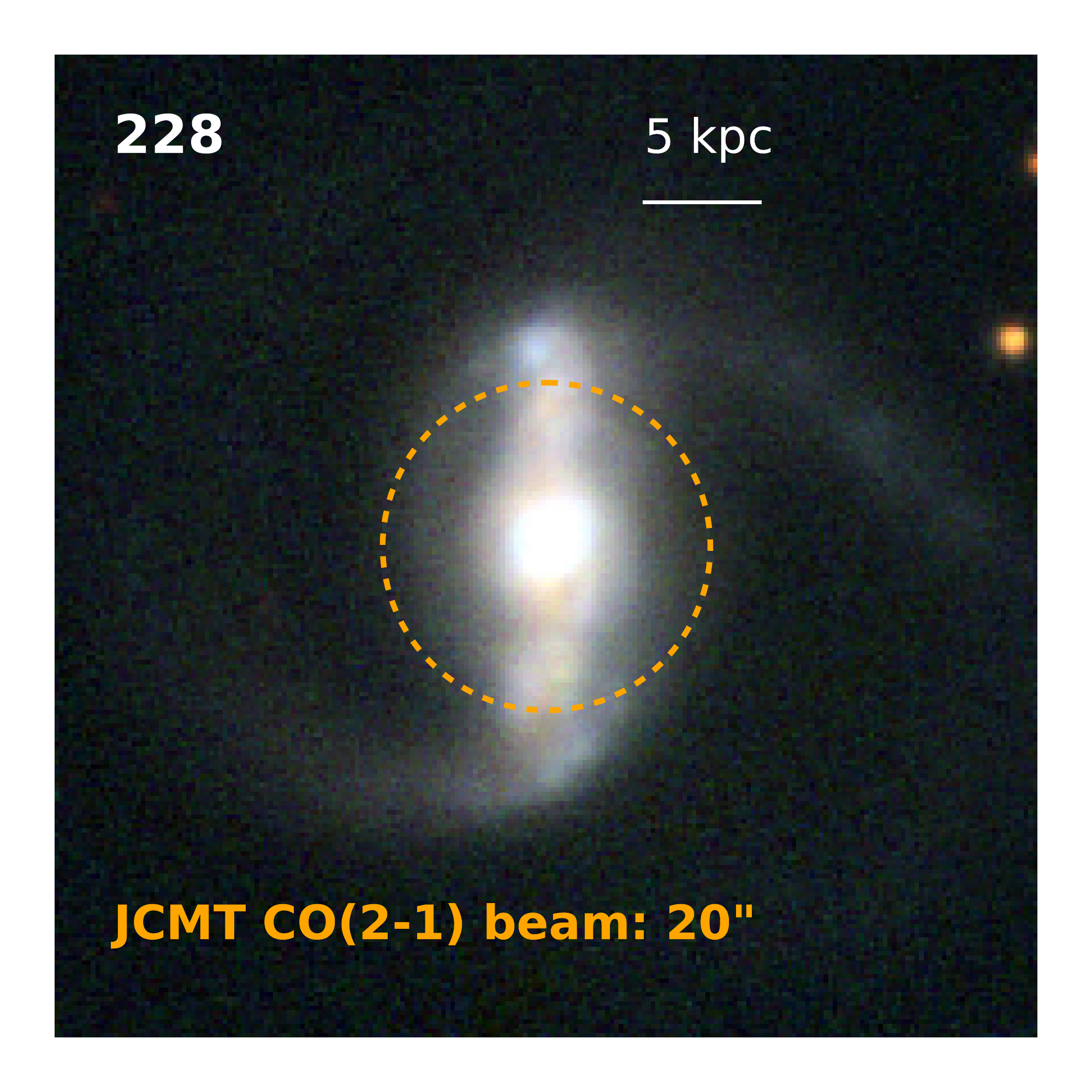}\includegraphics[width=0.26\textwidth]{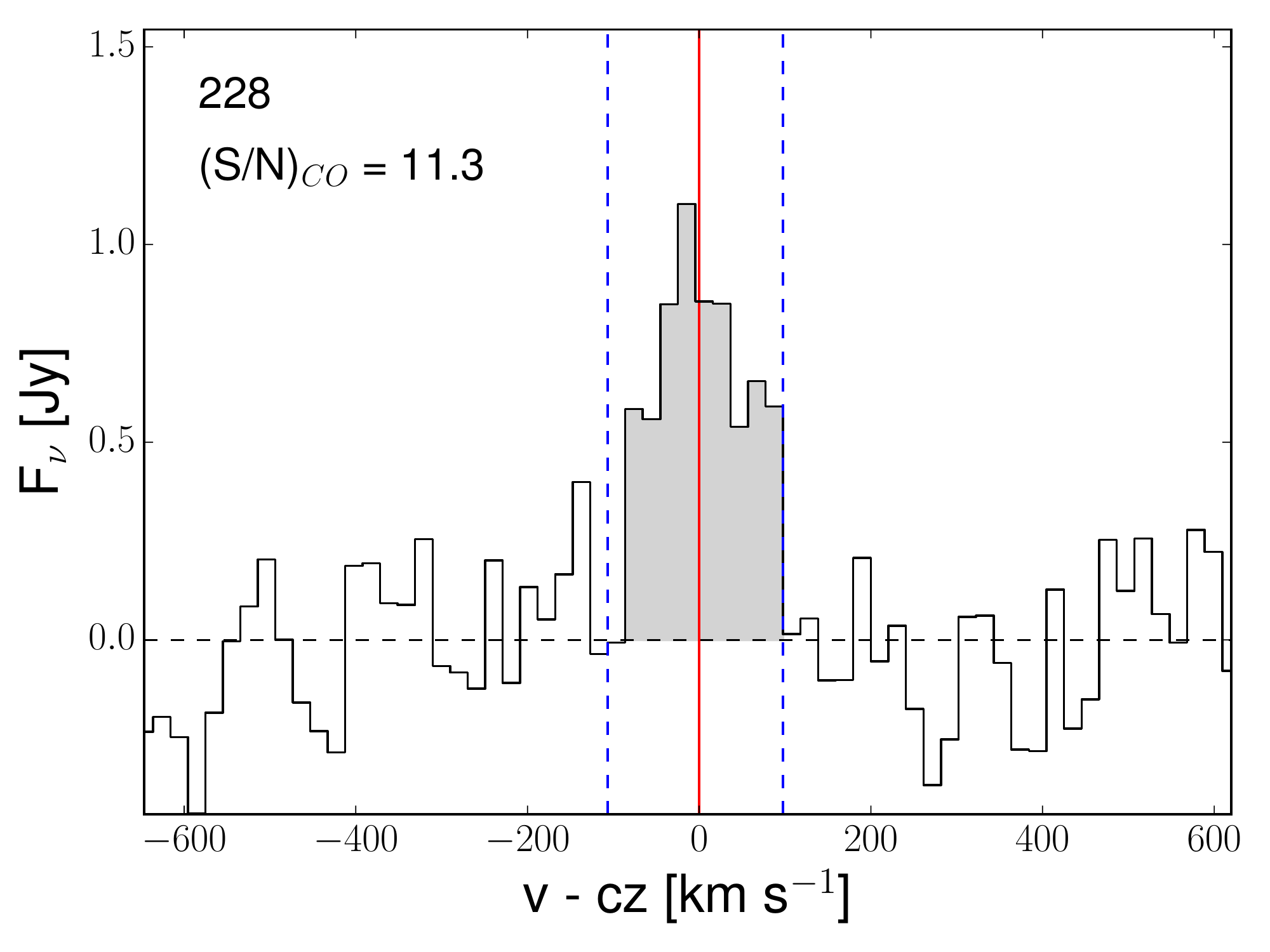}
\includegraphics[width=0.18\textwidth]{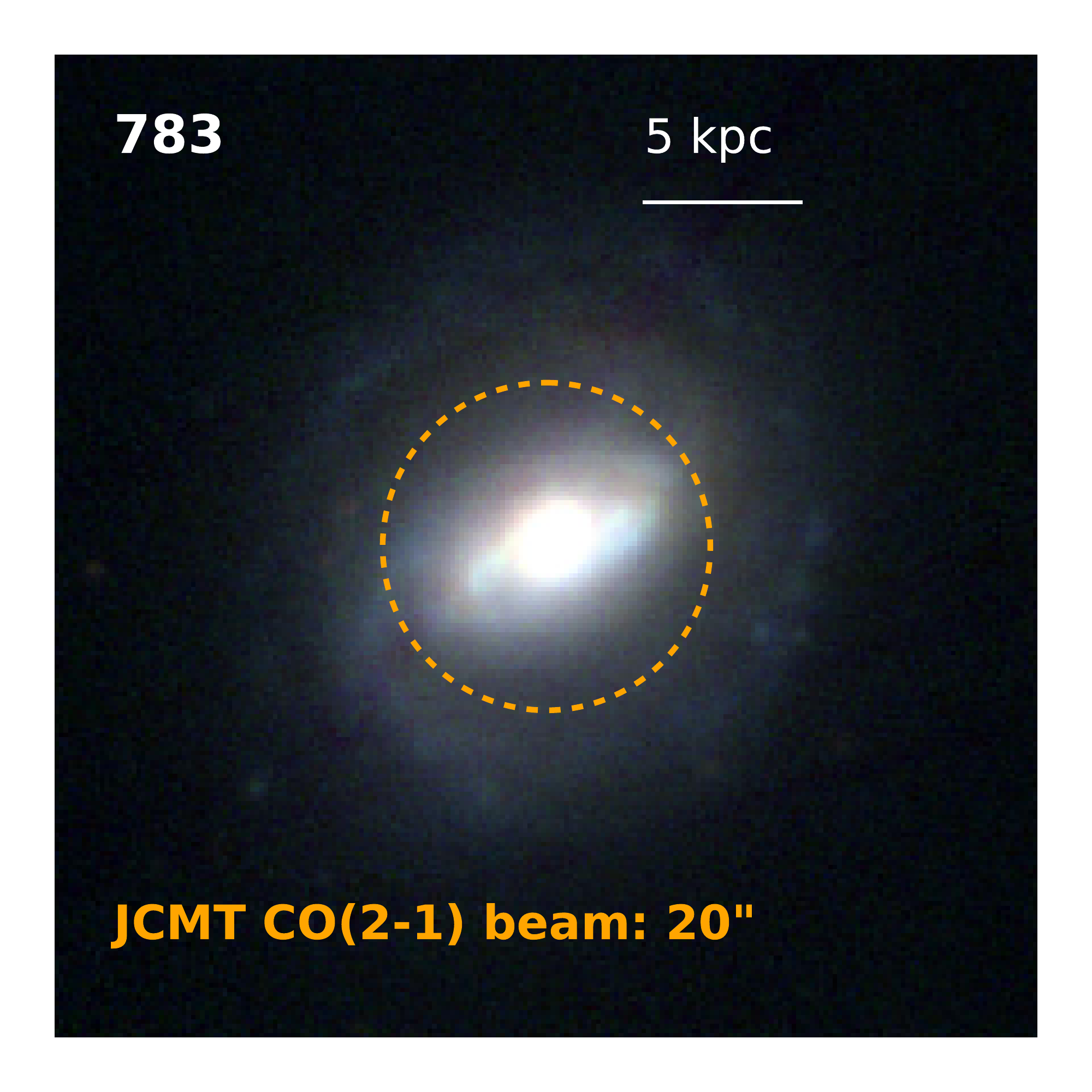}\includegraphics[width=0.26\textwidth]{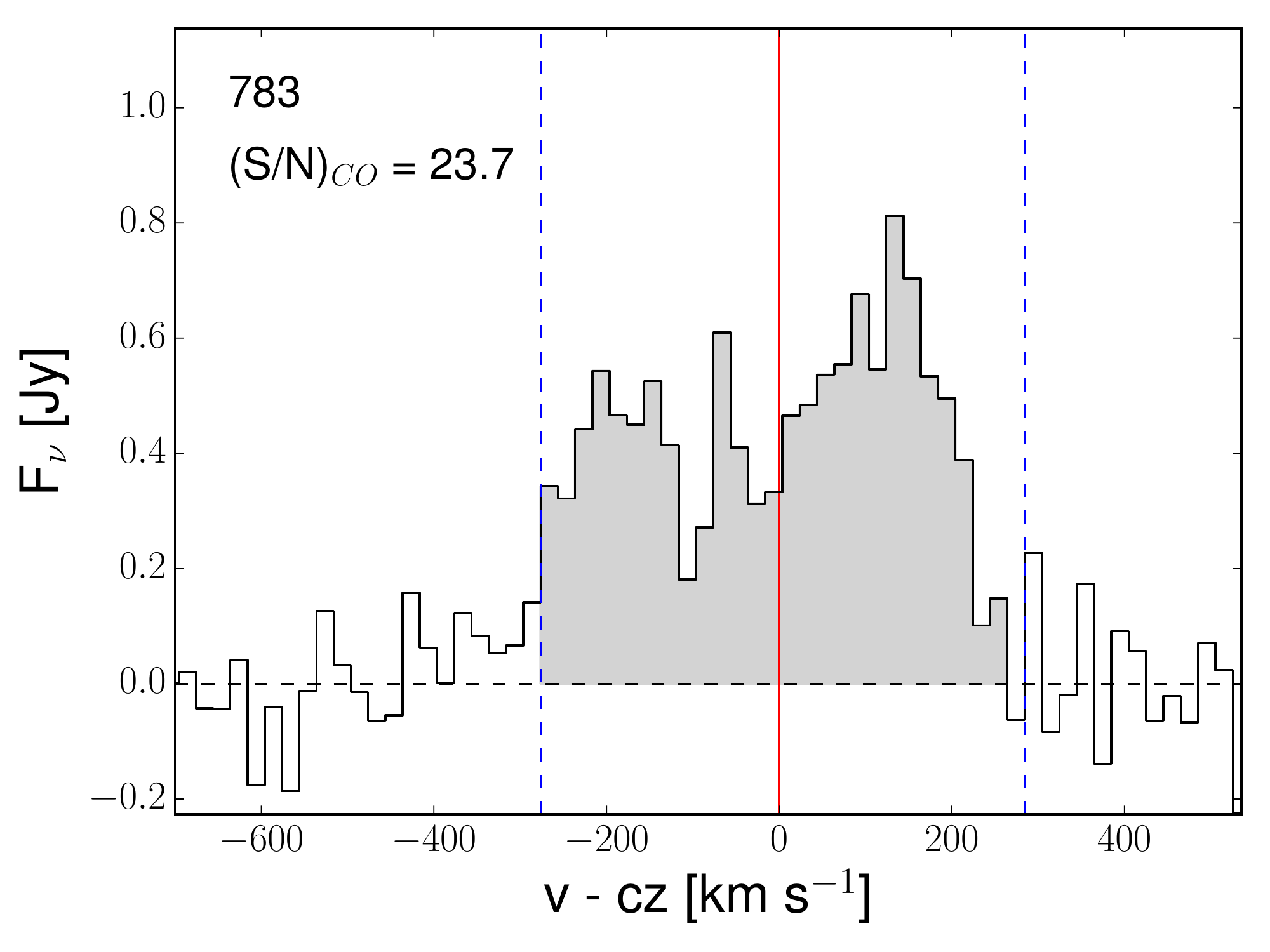}
\includegraphics[width=0.18\textwidth]{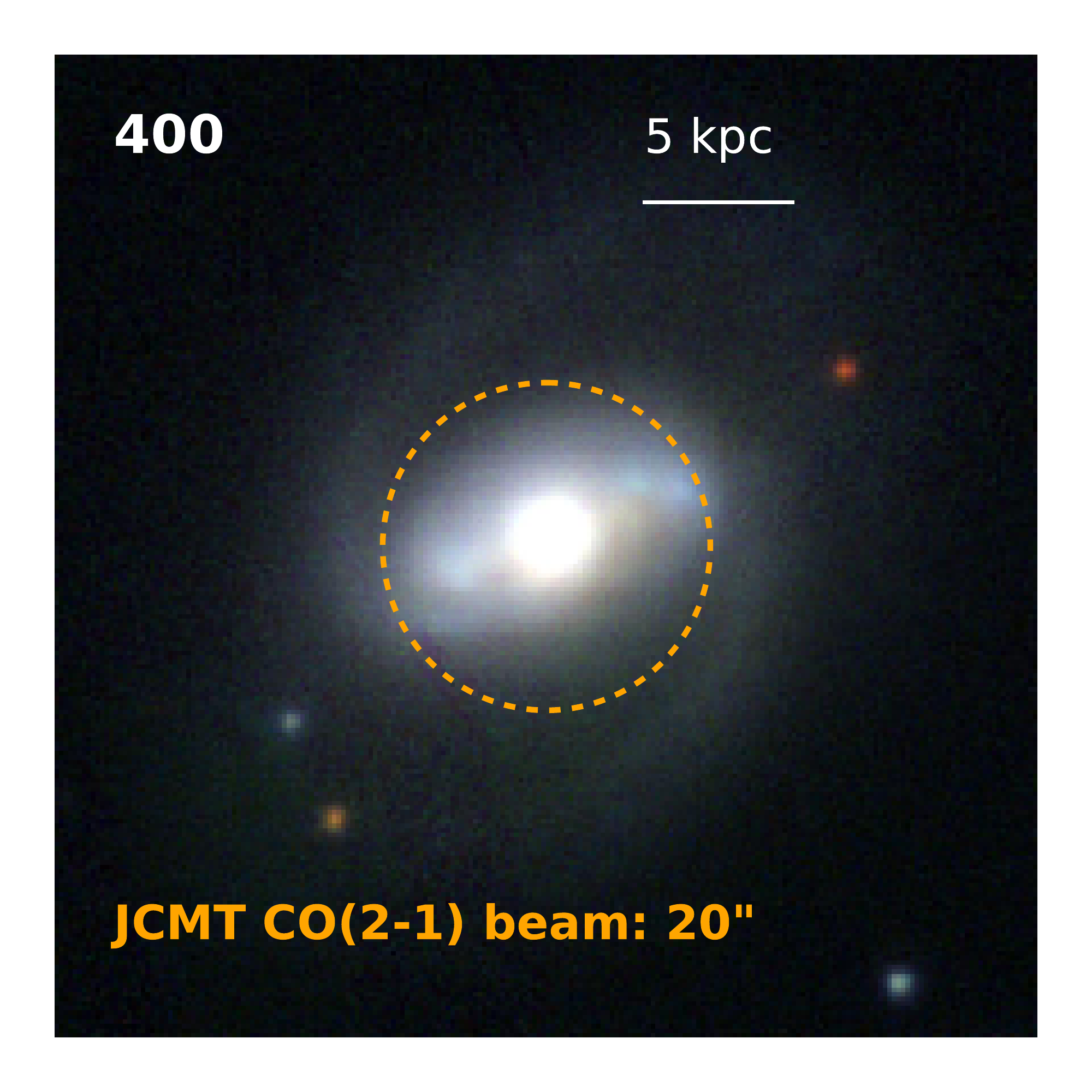}\includegraphics[width=0.26\textwidth]{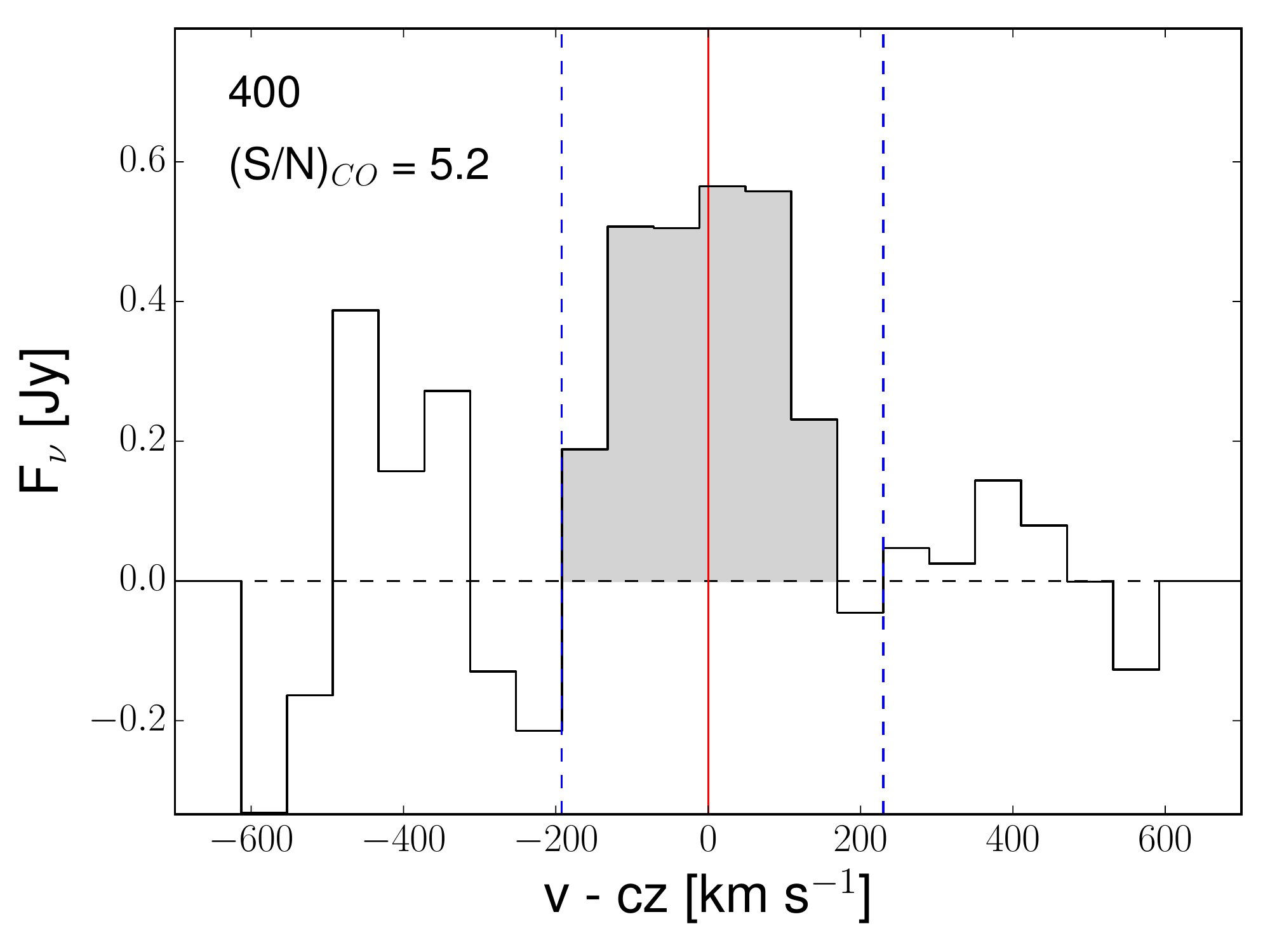}
\includegraphics[width=0.18\textwidth]{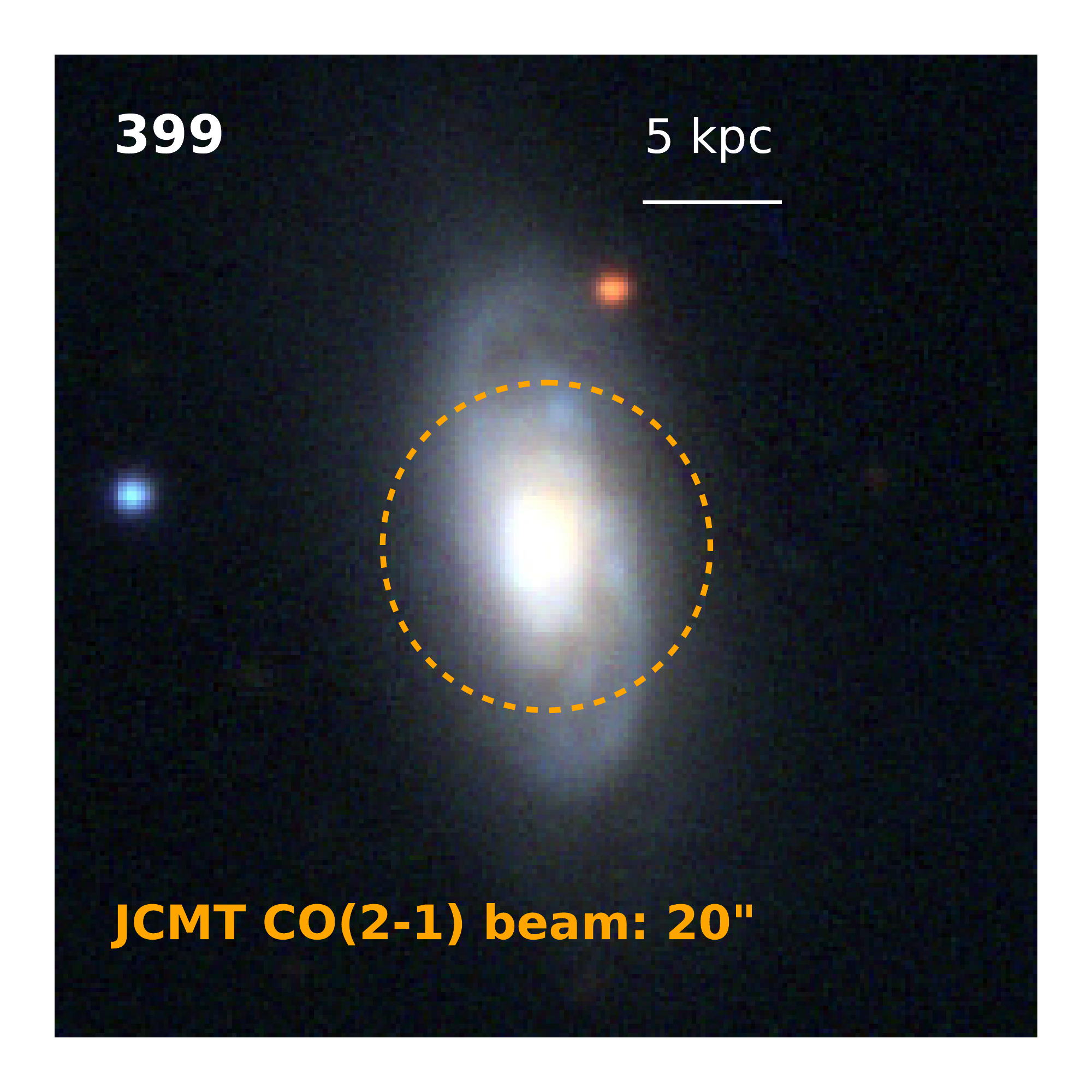}\includegraphics[width=0.26\textwidth]{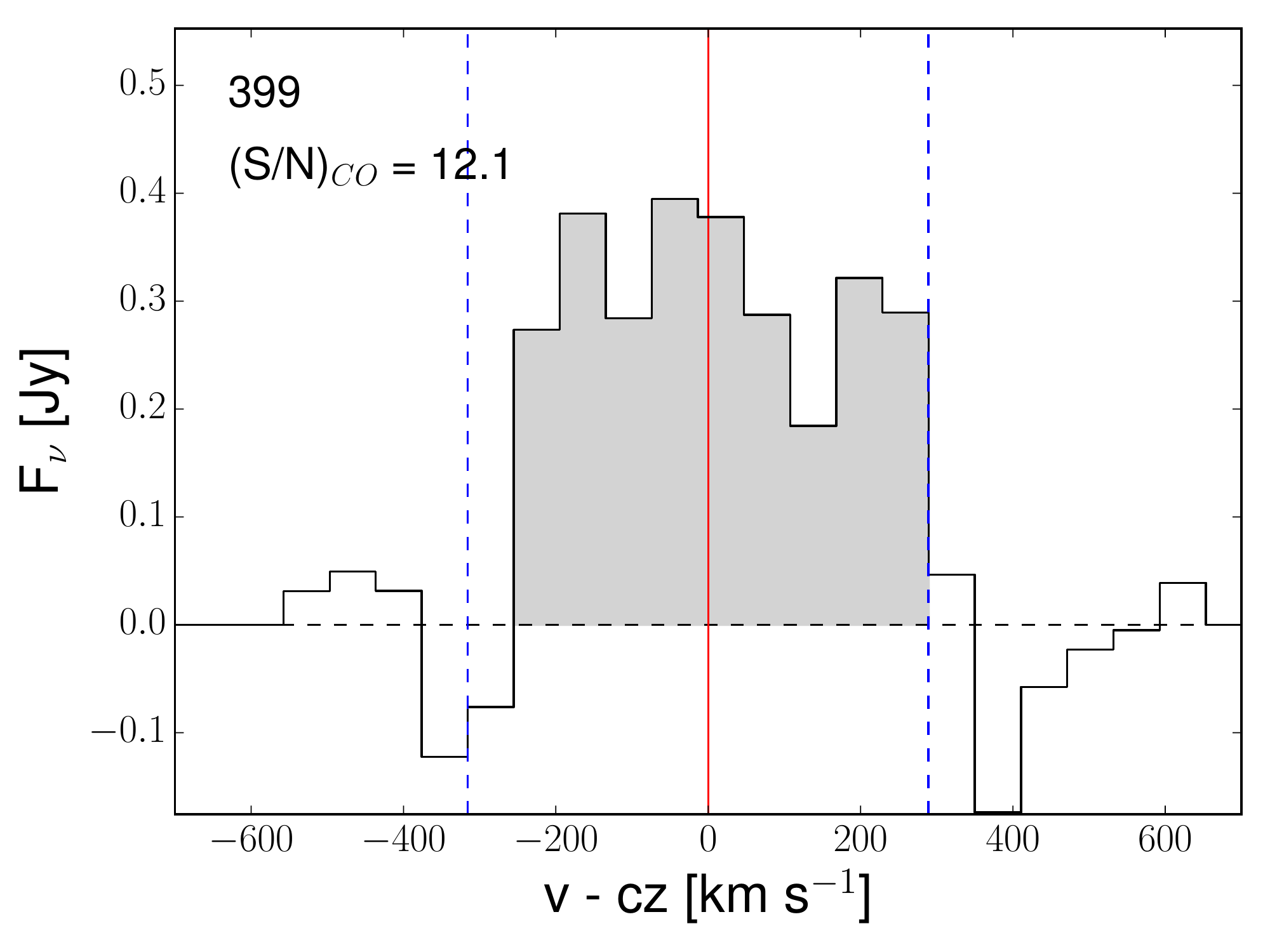}
\includegraphics[width=0.18\textwidth]{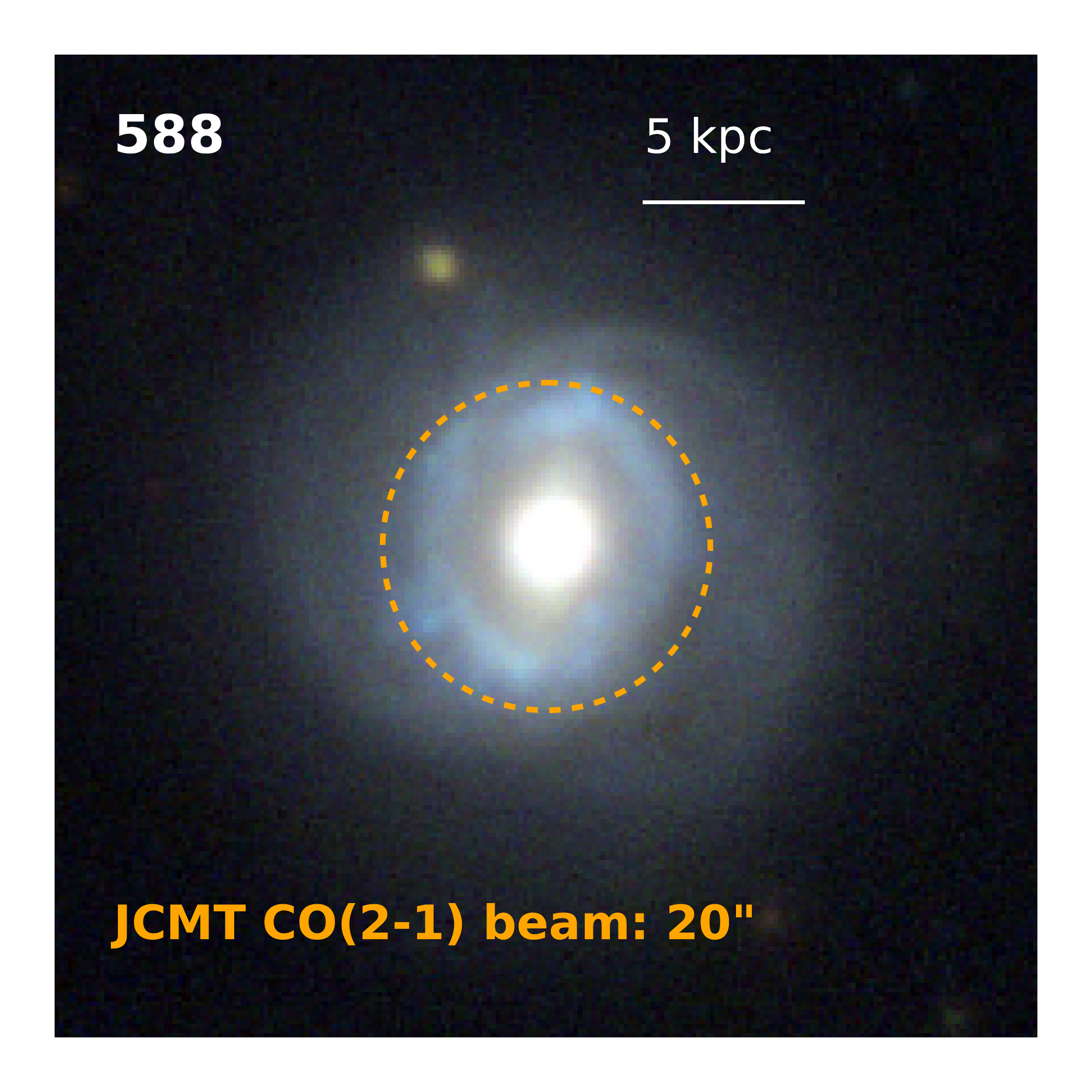}\includegraphics[width=0.26\textwidth]{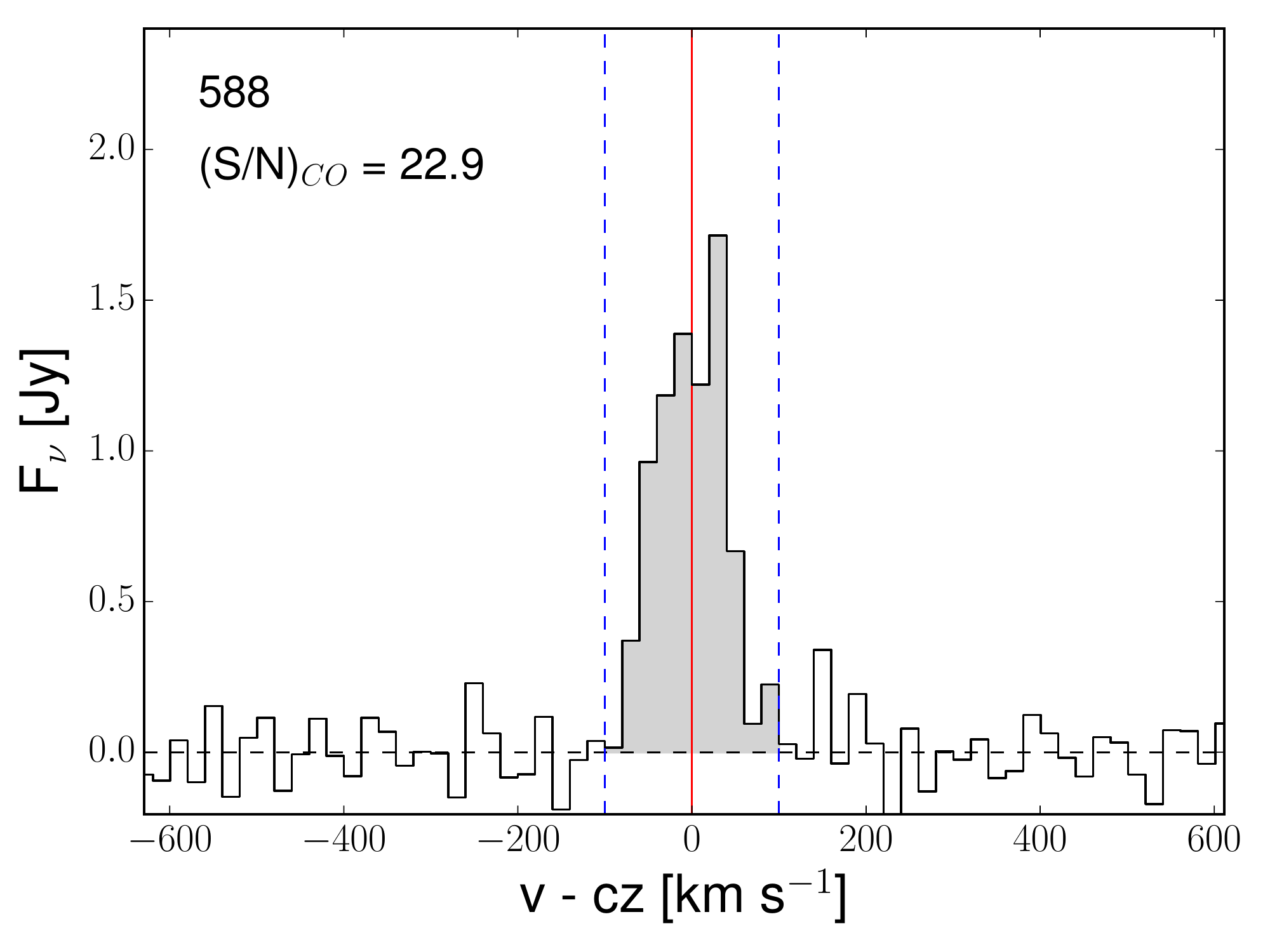}
\caption{Pan-STARRS $1'\times1'$ example $gri$ images of the BAT AGN sample for galaxies classified as spiral with some of the highest molecular gas masses.  The galaxies are sorted from largest (upper left, ID=205, $\log (M_{H2}/\Msun)=10.11$ ) to smallest (bottom right, ID=588, $\log (M_{H2}/\Msun)=9.93$) in molecular gas content going from left to right and top to bottom. 
} 
\label{fig:spiral_pic}
\end{figure*}

Here we discuss issues related to morphological classification using ground-based data even for nearby galaxies.  Examples of galaxy morphologies from the sample for elliptical morphologies can be found in Fig.~\ref{fig:ellip_pic}, for uncertain morphologies in Fig.~\ref{fig:uncertain_pic}, and for spiral morphologies in Fig.~\ref{fig:spiral_pic}.

We primarly investigate the elliptical classification since the excess of molecular gas among AGN galaxies compared to inactives was most significant. Galaxy Zoo morphologies, which are used in both samples, are known to include some of the S0-Sa continuum \citep{Lintott:2008:1179}.  Their is also a bias in comparing between samples that the tendency is that a fainter, higher redshift galaxy will be more likely to be classified as an elliptical due to the difficulty identifying faint spiral arms.

Our study contains 34\% (73/213)  of galaxies with only DSS images in the southern hemisphere ($\delta < -30^{\circ}$) which may, due to their poorer quality imaging have some bias to elliptical morphologies.  Of these 19\% (14/73)  have elliptical morphologies.  However, we do not find that these galaxies with DSS imaging have a significantly higher fraction of ellipticals than the majority with SDSS or PS1 imaging where the fraction is 20\% (28/140) suggesting this is not a significant issue.  

We then focus our study on the 13 BAT AGN galaxies with the most molecular gas ($\log (M_{H2}/\Msun)>$9.5), which is higher than any of the elliptical galaxies either from xCOLD GASS or the ATLAS sample.  First, we use Hyperleda to study whether morphological classifications from other publications are consistent with ours.  We find that of the 13 BAT AGN galaxies, nine are classified as elliptical, three as S0, and one as SA.  

Lower surface brightness features such as spiral disks can be found with higher quality imaging such as from \hst  \citep[e.g.,][]{Ellis:2000:10} as well as features such as nuclear spirals \citep[e.g.,][]{Martini:1999:2646,Martini:2003:353,Martini:2003:774} which are only resolved with high-spatial-resolution imaging.  We found that 4/13 sources have optical \hst imaging.  Of these, one was classified as an elliptical (Mrk 509), two as an SA (Fairall 49 and ESO 323-77), and finally one as a peculiar/SB system due to a nuclear spiral (Ark 120) based on past studies \citep{Malkan:1998:25,Bentz:2009:160}.  

Among the xCOLD GASS inactive elliptical sample, there are 8 detections with much lower molecular gas masses ($\log(\mh/\Msun)=8.5-9.3$). Four are reported as elliptical in Hyperleda, and four as S0-SAs. No optical \hst images exist for the sample. 

In summary, we find that the elliptical sample with gas detections includes a significant fraction of lenticular systems in both BAT AGN galaxies and the inactive galaxies.  Further large studies of morphology with \hst would be necessary to investigate how common weak spiral arms or nuclear spirals are in these systems.

\section{Uncertainty in derived parameters for inactive galaxies and AGN galaxies}
\label{parameters_appen}
The derived parameters (stellar mass, gas depletion timescales, etc.)\;for inactive and AGN host galaxies require a few assumptions because some of the parameters are derived differently because of the difficulty dealing with AGN contamination of the multiwavelength emission.  The stellar masses are both computed using SED fitting, but their is some uncertainty in the AGN emission.   In general the results in this paper are more significant than the likely scatter or possible offsets, however. 

First, for the stellar masses measurements, 18 BAT AGN overlap with the Max Planck Institute for Astrophysics/Johns Hopkins University (MPA/JHU) catalog which was used for xCOLD GASS stellar mass measurements.  The agreement in stellar masses for these AGN is good with the BAT AGN stellar masses somewhat lower than those in the catalog ($\log(\mstar/\Msun)=-0.07\pm0.15$).

Second, we have assumed the CO (2-1) can be reliably converted to the CO(1-0) assuming a conversion of $0.79\pm0.03$ which was derived from xCOLD GASS.  For the CO(2-1), \citet{Lamperti:2017:540} looked into the CO(3-2) and CO(1-0) for these AGN galaxies compared to xCOLD GASS inactive galaxies and found no significant offset so we would not expect an offset in CO(2-1).  \citet{Shangguan:2019:arXiv:1912.00085} looked at a sample of 15 Palomar-Green quasars and also found no offset in  CO (2-1) to CO (1-0).

One area where there may be systematic offsets is the derivation of SFR and hence depletion timescales which are derived from the SED fitting of the IR \citep{Shimizu:2017:3161,Ichikawa:2019:31}.  There is also some uncertainty in the level to which the AGN contaminates the FIR emission that was used to estimate the SFR after SED fitting to remove the AGN contribution \citep[e.g.,][]{Shimizu:2017:3161,Ichikawa:2019:31}, particularly in studies of luminous systems, which might affect depletion timescales.    In xCOLD GASS the SFR estimates are from an average of NUV, optical, and MIR fitting \citep{Janowiecki:2017:4795} as well as from the MPA/JHU catalog from optical measurements.  For the purpose of this study we used the SED fitting SFR measurements from xCOLD GASS to better match the \herschel based SED fits.  For emission line galaxies, the star formation rates of xCOLD GASS where found be in very good agreement with H-alpha (0.2 dex scatter with 0.01 offset across 3 orders of mag) which has been shown to be in good agreement with the FIR emission, but with some scatter consistent with past studies \citep[see e.g., ][for review]{Kennicutt:2012:531}.

\begin{figure*} 
\centering
\includegraphics[width=5cm]{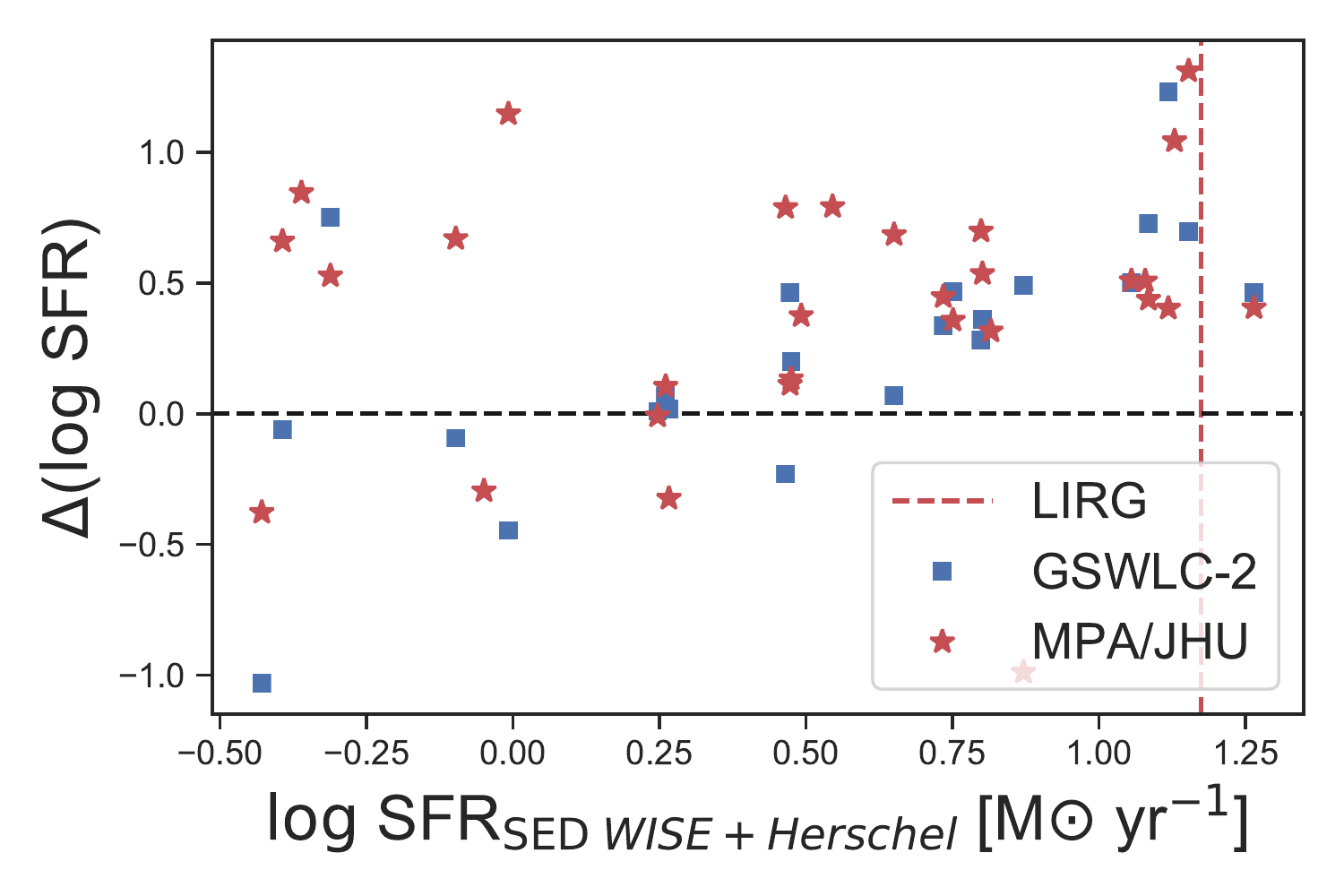}
\includegraphics[width=5cm]{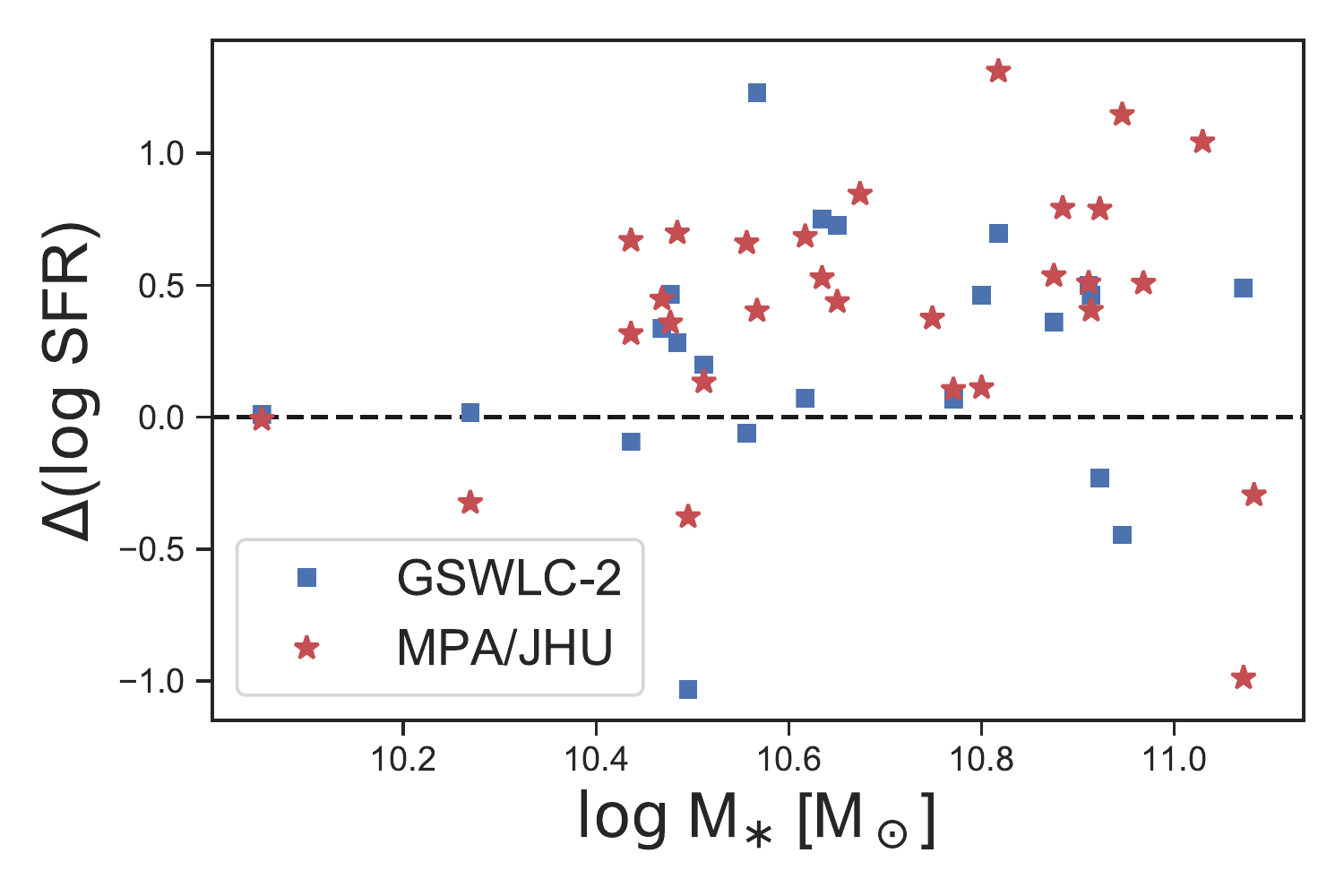}
\includegraphics[width=5cm]{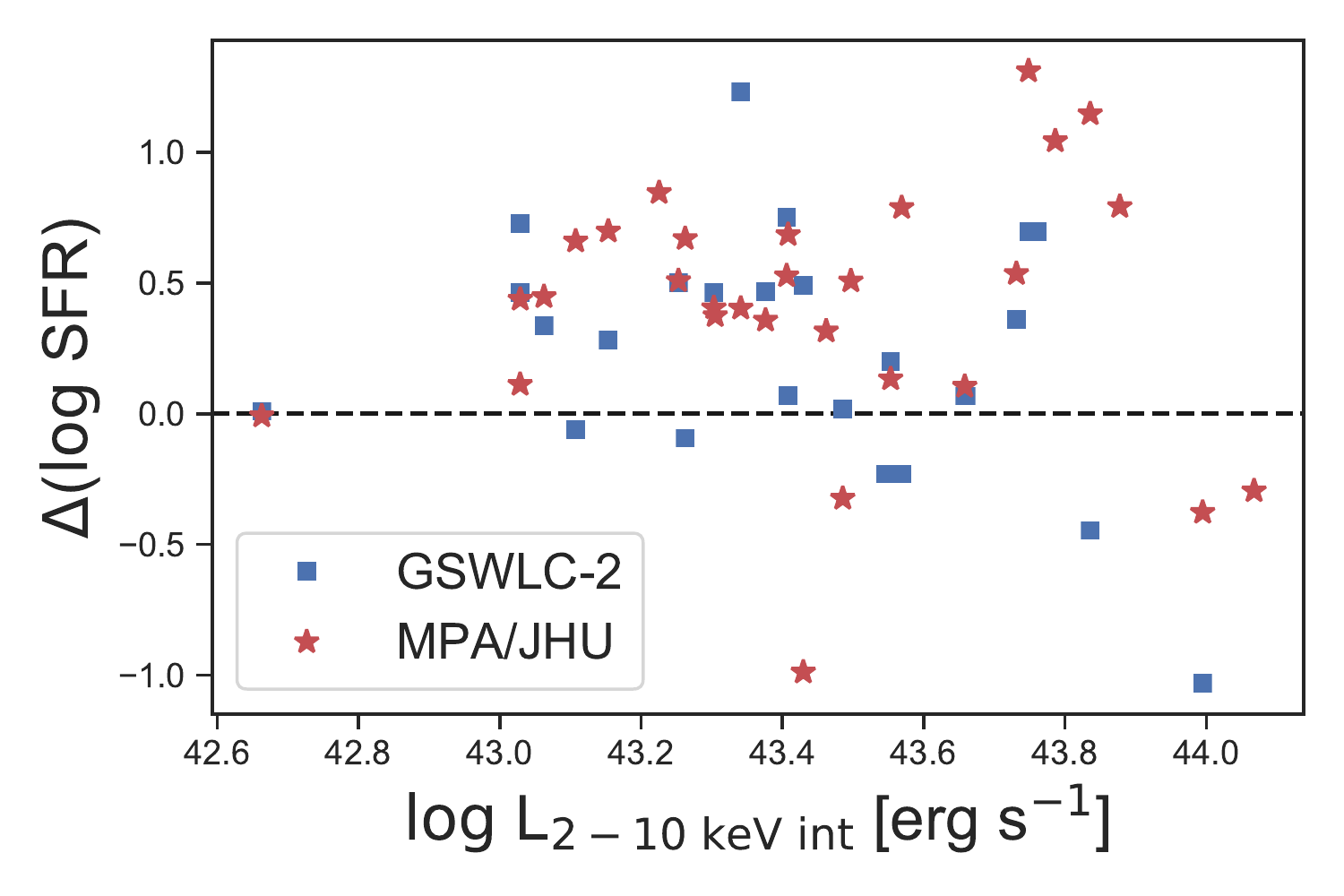}
\caption{Comparison between the SFR derived measurements for the same BAT AGN galaxies using different studies.  The offset, \deltasfr, is defined as the difference on a logarithmic scale between the measured SFR between studies.  The left panel shows the \deltasfr\ as compared to the SFR with a dashed line indicating the boundary with LIRGs.  The middle panel shows the offset with stellar mass.  Finally, the right panel shows the offset with AGN X-ray luminosity.}
\label{fig:sfrcomp}
\end{figure*}

{In order to better understand the offsets, we crossmatched the BAT AGN galaxies with measurements from either the MPA/JHU catalog or $GALEX$-SDSS-\wise Legacy Catalog version 2 (GSWLC-2) \citep{Salim:2018:11}.  GSWLC-2 determines SFRs from joint UV+optical+mid-IR SED fitting along with \texttt{CIGALE} \citep{Noll:2009:1793}.  Because only a small number of BAT AGN galaxies are in the SDSS footprint, we used the full sample of BAT AGN galaxies (e.g., including those unobserved as shown in Figure 1), limited to BAT AGN where the SFR was detected rather than a limit.  This left us with a sample of 32 BAT AGN galaxies with overlapping SFR measurements for the same AGN galaxies (Figure \ref{fig:sfrcomp}).  }

{We found the \herschel derived SED SFRs in our studies were typically significantly higher with the median offset and 1$\sigma$ confidence as $0.28^{+0.18}_{0.22}$ dex for GSWLC-2 and $0.45^{+0.08}_{0.07}$ dex for the MPA/JHU, respectively.  The SFR offsets are consistent with previous studies of AGN and massive galaxies when measured with \herschel \citep[e.g., see Figure 12,][]{Rosario:2016:2703}.   The AGN galaxies with the highest offsets tended to be those near the LIRG boundary, with some of the highest IR luminosities in the local universe in the sky indicating some of this offset might be in systems with lots of obscured star formation.  If the evolved stars are significantly contributing to the dust heating measured in the FIR, causing an overestimate of the IR based SFR, we might expect a correlation with stellar mass, which we do not find.  Finally, we do not see any correlation in the offsets with AGN X-ray luminosity, which we might expect if the AGN contribution to the IR is being incorrectly subtracted. }

 {In xCOLD GASS the SFR estimates which we directly compare to are from \citep{Janowiecki:2017:4795}, which itself is somewhat above the MPA/JHU (\deltasfr=0.28) and GSWLC-2 (\deltasfr=0.21 dex), so it's possible some of these offsets between AGN galaxies and xCOLD GASS cancel each other.  In summary, systematic offsets in SFR measurements between AGN and inactive galaxies and FIR vs. UV-optical-MIR may limit our ability to measure small differences ($\approx$0.3 dex) in gas depletion times, star formation properties, and offset from the MS.  But in the current sample the evidence suggests the \herschel based measurements may be better capturing obscured star formation.}

These issues will be further explored in a subsequent paper studying the combined results from this paper and a large IRAM program (N=\ntaroco\ BAT AGN galaxies) which has the benefit of both a larger sample size as well a overlap for 56 of the AGN galaxies from the APEX and JCMT data presented here (Shimizu et al., in prep) and will perform a full UV-optical-MIR fitting.

\section{Beam correction methods}
\label{beamcorr_appen}
We have used somewhat different beam corrections methods for the AGN and inactive galaxies.  Since the galaxies have not been mapped in CO beyond the primary beam, the critical issue is what other imaging wavelength best represents the CO scale lengths to provide the most accurate correction.  As discussed in the text, the xCOLD GASS sample used a combination of simulations and the $g$-band surface brightness.  The BAT AGN galaxy sample primarily used PACS 160 \micron\ surface brightness, but when not available used the $K_\prime$-band surface brightness. 


Understanding the effect of using different imaging for correction of the CO beam is still somewhat unclear, though the literature does suggest the use of FIR (e.g. 70 \micron\ or 160 \micron) tracers is most effective.  \citet{Leroy:2008:2782} found that in spirals the star formation efficiency (SFE=SFR/\mh) based on H$_2$ is nearly constant at 800 pc resolutions, compared to 3.6 \micron\ data.  However, when the interstellar medium is predominantly HI, the SFE drops with a smaller scale length.  A further study by \citet{Schruba:2011:37} used IRAM to study CO emission in 33 nearby spirals down to very low intensities.  They found a tight linear relationship between 24 \micron\ and 70 \micron that does not show any notable break between regions that are dominated by molecular gas or HI.  The correlation was much weaker and the scatter much larger for the FUV and H$\alpha$ emission.  \citet{Corbelli:2012:A32} found that the CO (1-0) flux correlates tightly and linearly with FIR fluxes observed by \herschelsh, especially with the emission at 160, 250 and 350 \micron. \citet{Bourne:2013:479} find that the correlation between CO (3-2) is strongest at 100 \micron, compared to longer or shorter \herschel wavelengths, but that CO (2-1) may be more closely related to colder dust at longer wavelengths.  \citet{Casasola:2017:A18} studied 18 face-on spiral galaxies and fitted the surface-brightness profiles, finding that the scale lengths at 160 \micron\ are somewhat smaller than at 3.6 \micron\ (0.9) and the 70 \micron\ smaller still than the 3.6 \micron\ (0.84). The total H$_2$ scale length from CO is found on somewhat smaller scales than both the 70 \micron\ (0.85), 160 \micron\ (0.77), and 0.5 \micron\ which is similar to the $g$-band central wavelength used for xCOLD GASS (0.65).

The median beam correction for all the BAT AGN galaxies is 1.33, which is higher, by 16\%, than the xCOLD GASS survey (1.15), though the BAT AGN galaxies are at lower redshifts and higher stellar masses so some offset is naturally expected.  The average $K$-band effective radius ($R_{k20}$) for the BAT AGN galaxy sample is 21.7\arcsec\ and for xCOLD GASS is 15.7\arcsec\, indicating the BAT AGN host galaxies are on average 38\% larger on the sky than the xCOLD GASS comparison sample.  The ratio of effective radius to HPBW is on average 0.86 for BAT AGN host galaxies and 0.71 for xCOLD GASS inactives, indicating the BAT AGN galaxies are on average 20\% larger compared to the beam size.  We then compare the ratio of the beam correction between BAT AGN and inactive galaxies based on the average ratio of the beam correction to $K$-band effective radius ($R_{k20}$), since presumably a larger galaxy would have a larger beam correction.  We find the BAT AGN galaxy beam correction per arcsecond effective radius is on average slightly lower ($0.0698\pm0.002$) than in inactive galaxies ($0.0724\pm0.001$).   

\section{Measuring CO using integrated emission vs. profile fitting}
\label{profile_fitting_appen}
In the catalog we have provided measurements by integrating the line flux, so that we can robustly compare with xCOLD GASS using the exact same measurement techniques.  For upper limits in the AGN galaxy sample, we have also assumed the exact same width (300 \kmps) as xCOLDGASS {which matches well the median width of the detected BAT sample (\texttt{W50}$=295$ \kms)}.   

\begin{figure*} 
\centering
\includegraphics[width=12cm]{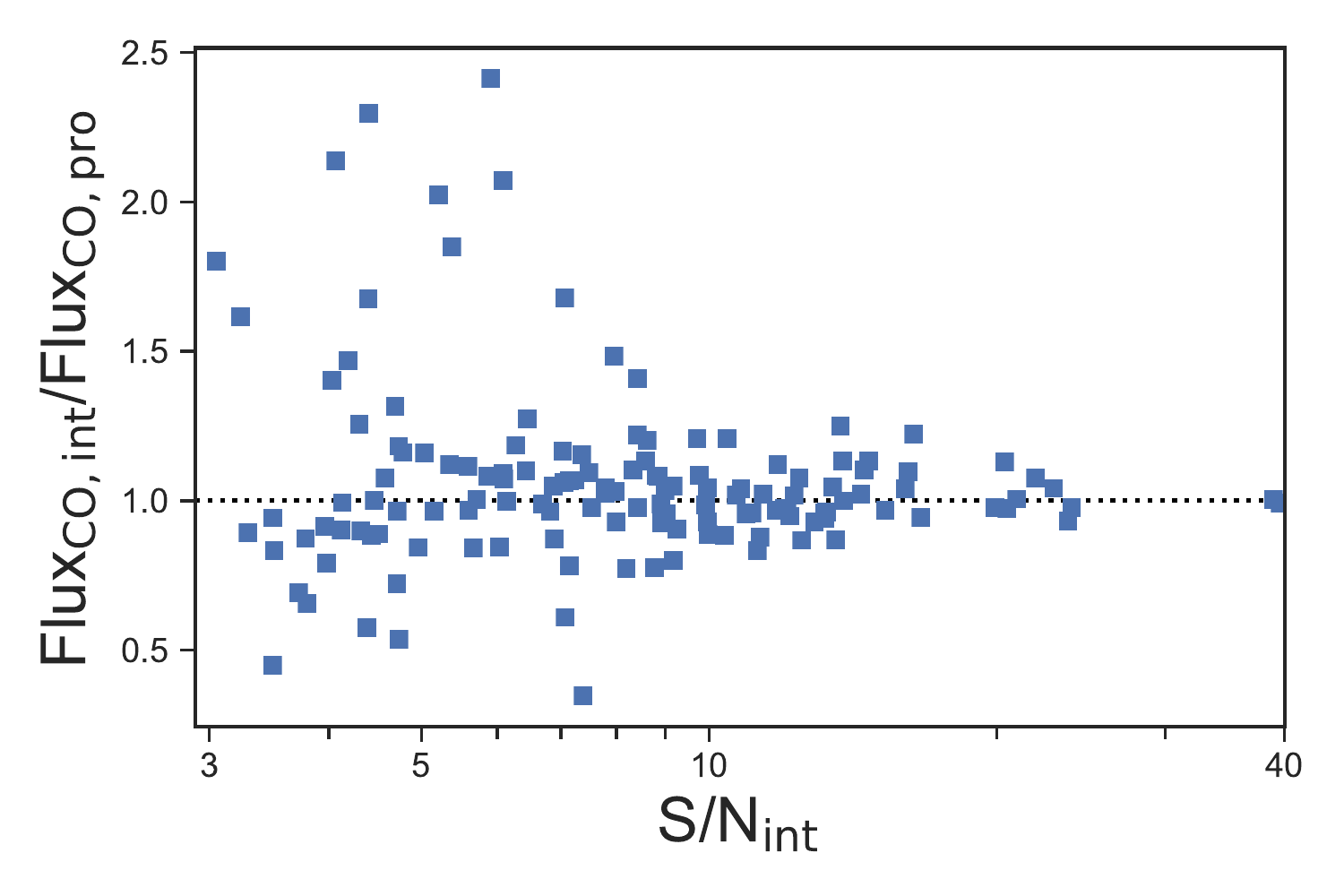}
\caption{Comparison between the fluxes from integrating the total CO emission vs. profile line fitting with Gaussians.  Other than at low S/N ($3<$SN$<5$), there is good agreement between the methods.}
\label{fig:fluxcomparison}
\end{figure*}

We have also provided catalog measurements using profile fitting with Gaussians, though they are not used in the sample analysis.  In Figure \ref{fig:fluxcomparison} we provide a comparison of the flux of the integrated fitting method compared to the line fitting methods for detected sources.  There is good agreement with the mean ratio between the fluxes of $1.05\pm0.10$ of the integrated flux to the flux from profile fitting with Gaussians.  

There are 11 sources that are detected using integrated measurements with SNR between 3.1 and 5.9, but are non-detections using a profile fit (SNR$<3$).  One source, ID=586, is at SNR=1.6 with a profile fit and the remaining sources are between SNR=2.2-2.7.  Of these, 6/11 have been reobserved with IRAM (Shimizu et al., in prep) and all six were confirmed detections (SNR=3.4-18.1). 

The line profiles and a comparison with IRAM observations will be further studied in Shimizu et al. (in prep), with a large number of northern hemisphere BAT AGN galaxies, ensuring the higher sensitivity required for this type of analysis.

\section{Spectra for all detected sources}
\label{allspec_appen}
Here we provide optical images and spectra for all detected sources not shown in the paper (Figure 21-37).

\begin{figure*}
\centering
\raggedright
\includegraphics[width=0.18\textwidth]{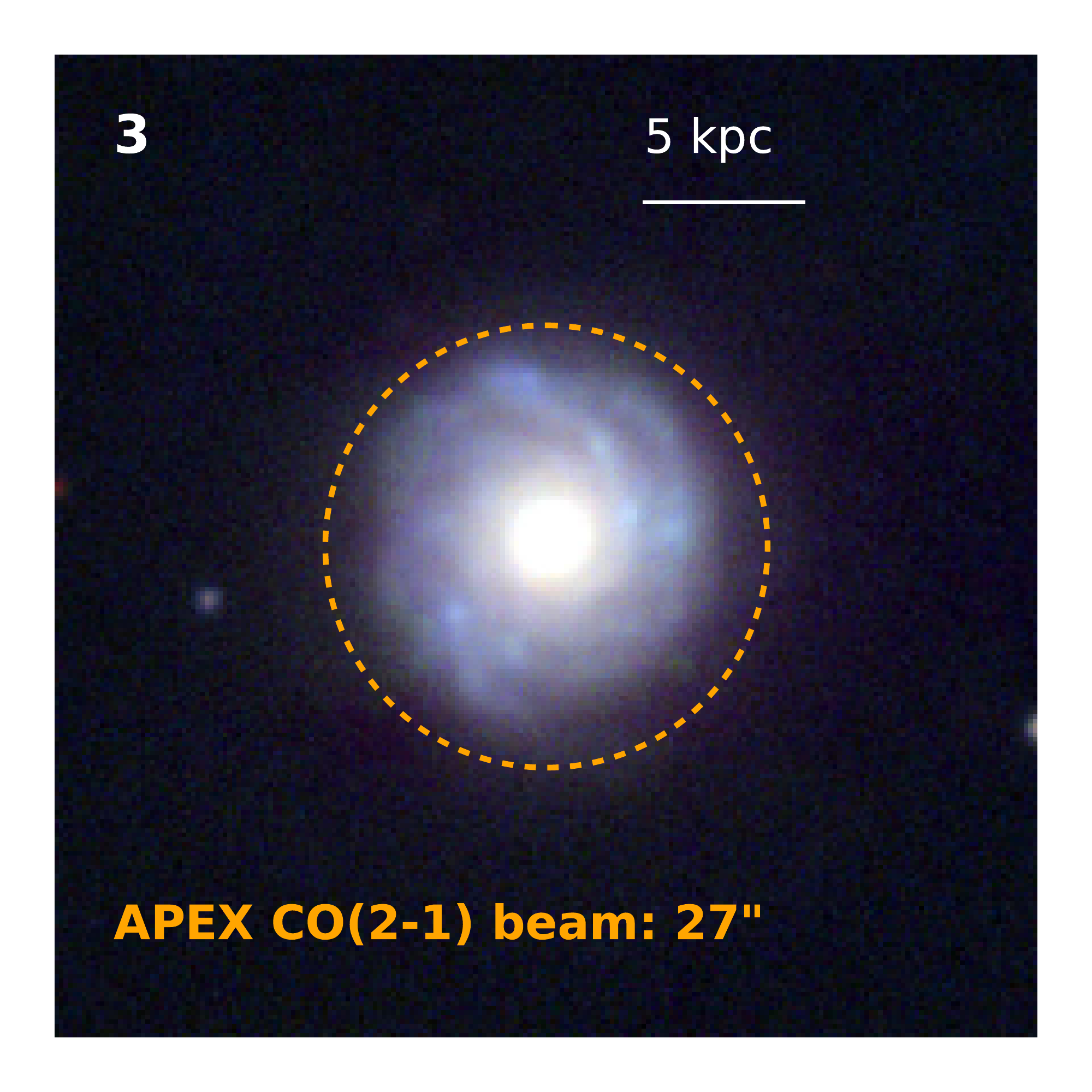}\includegraphics[width=0.26\textwidth]{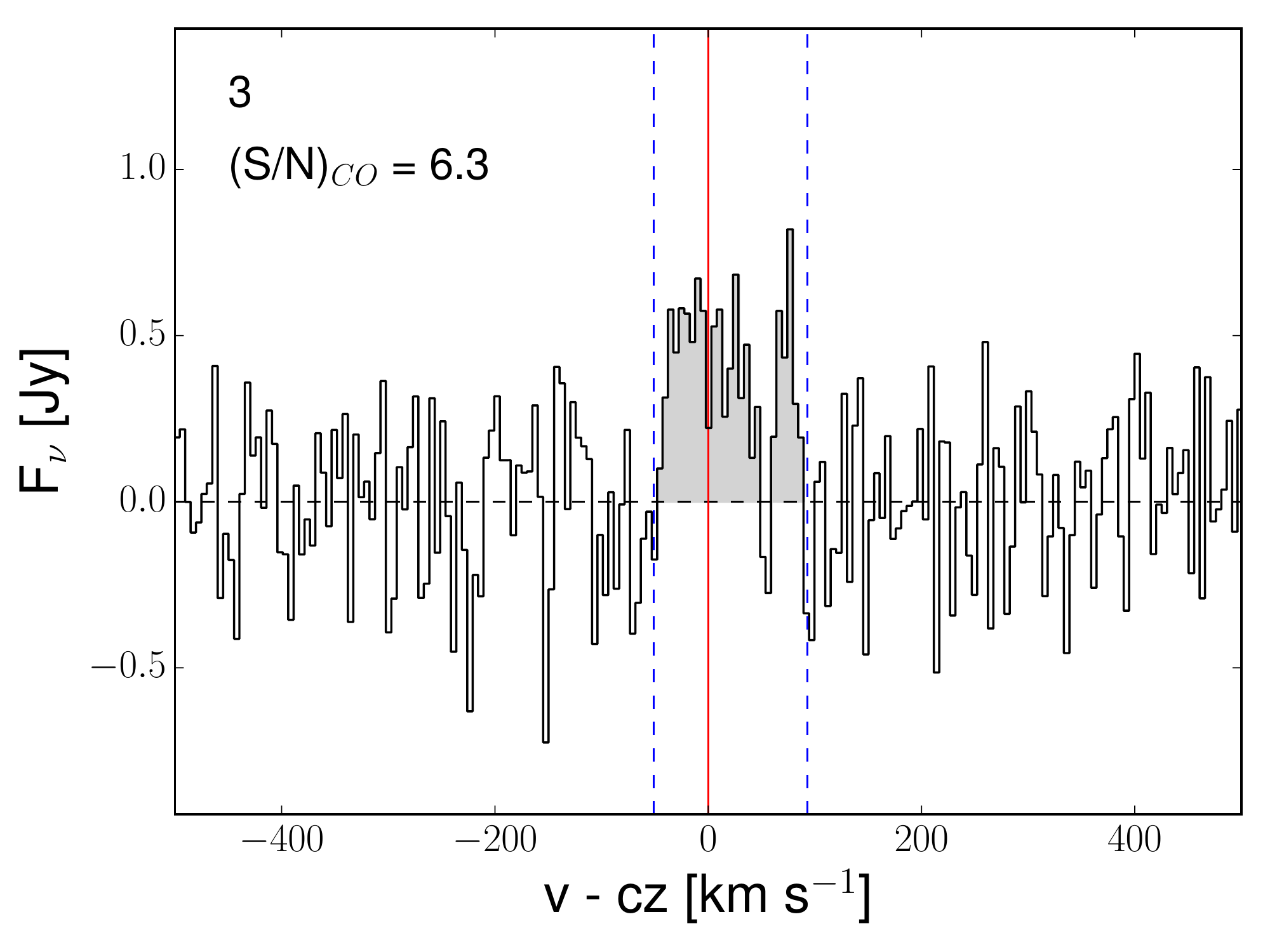}
\includegraphics[width=0.18\textwidth]{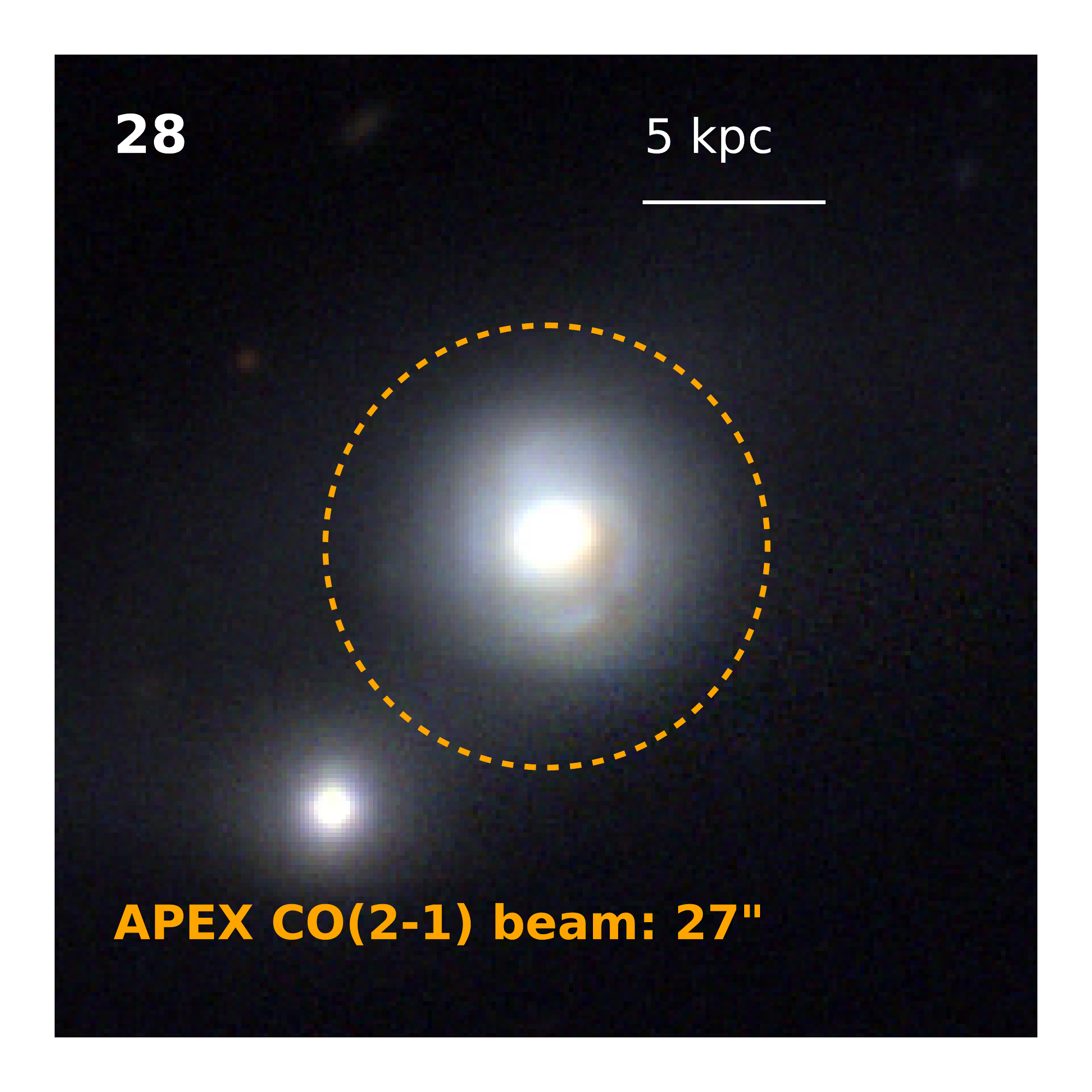}\includegraphics[width=0.26\textwidth]{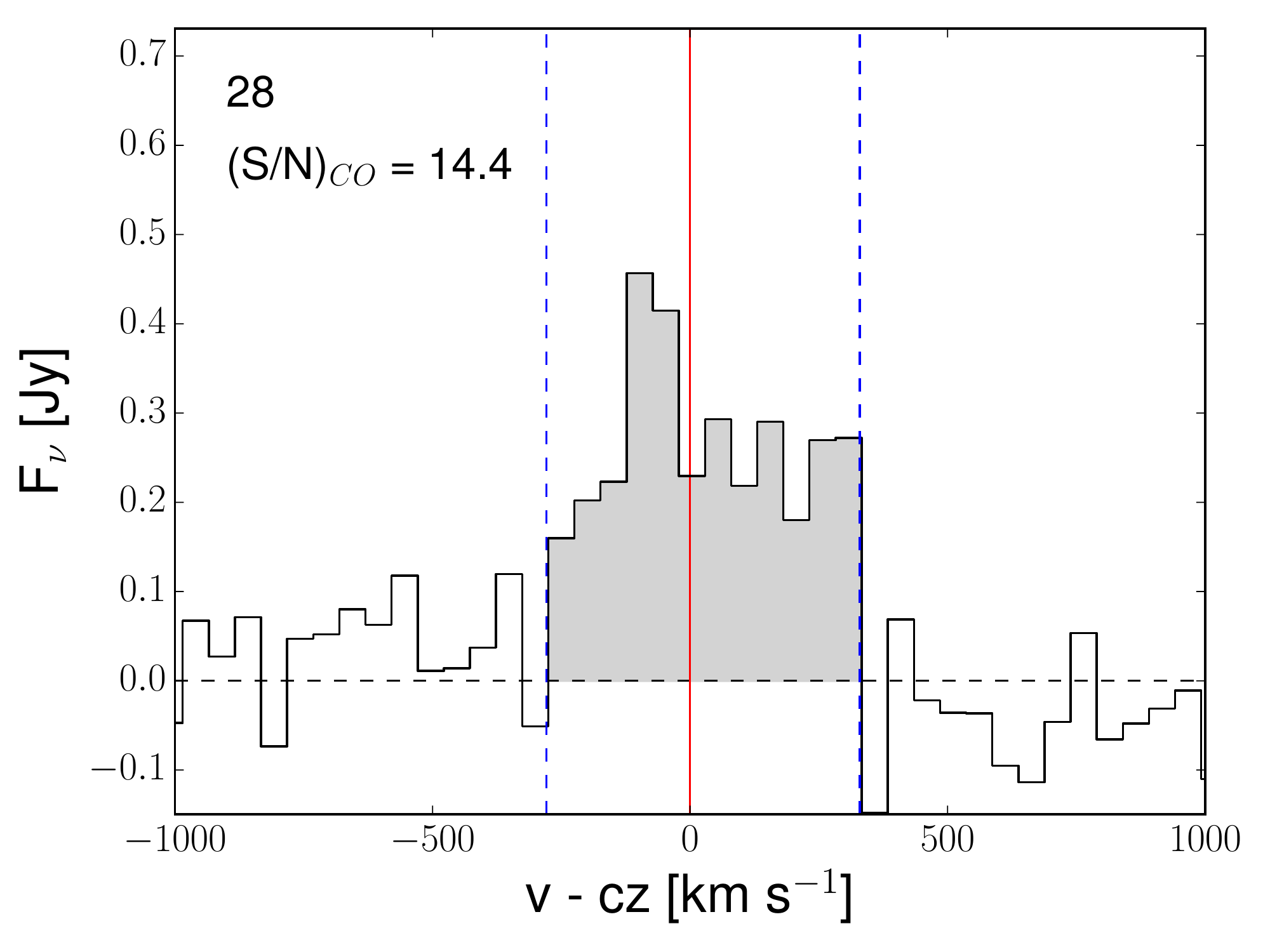}
\includegraphics[width=0.18\textwidth]{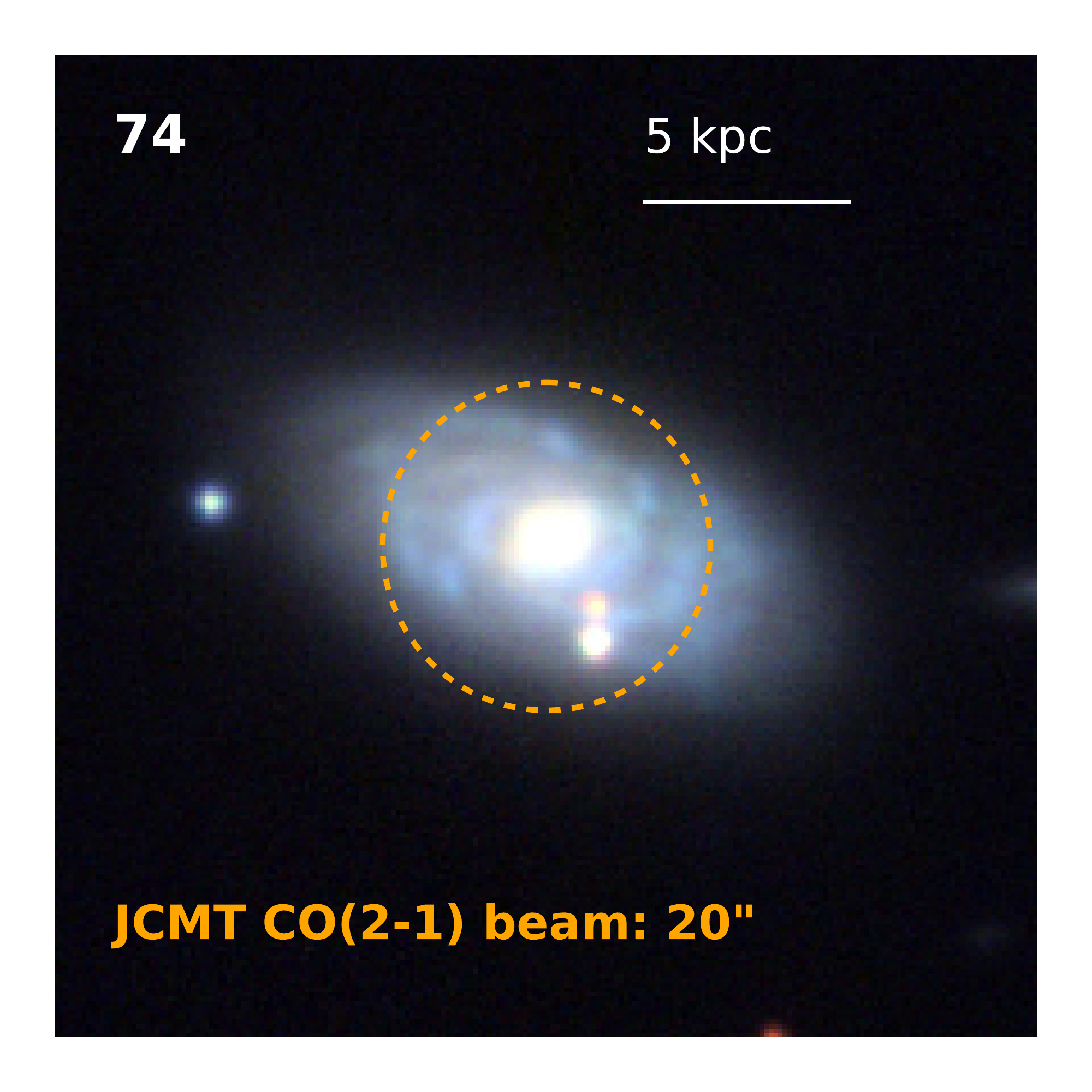}\includegraphics[width=0.26\textwidth]{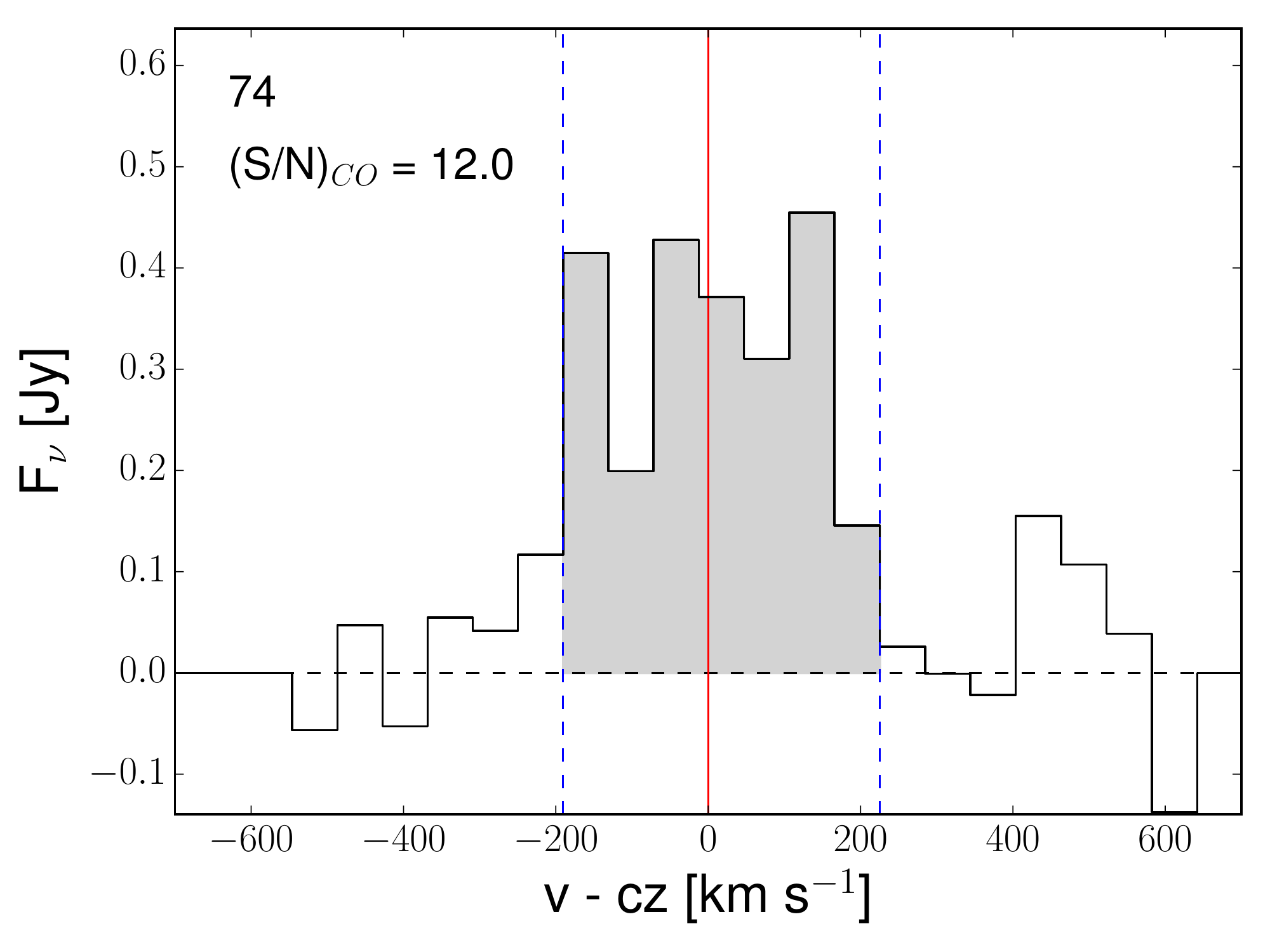}
\includegraphics[width=0.18\textwidth]{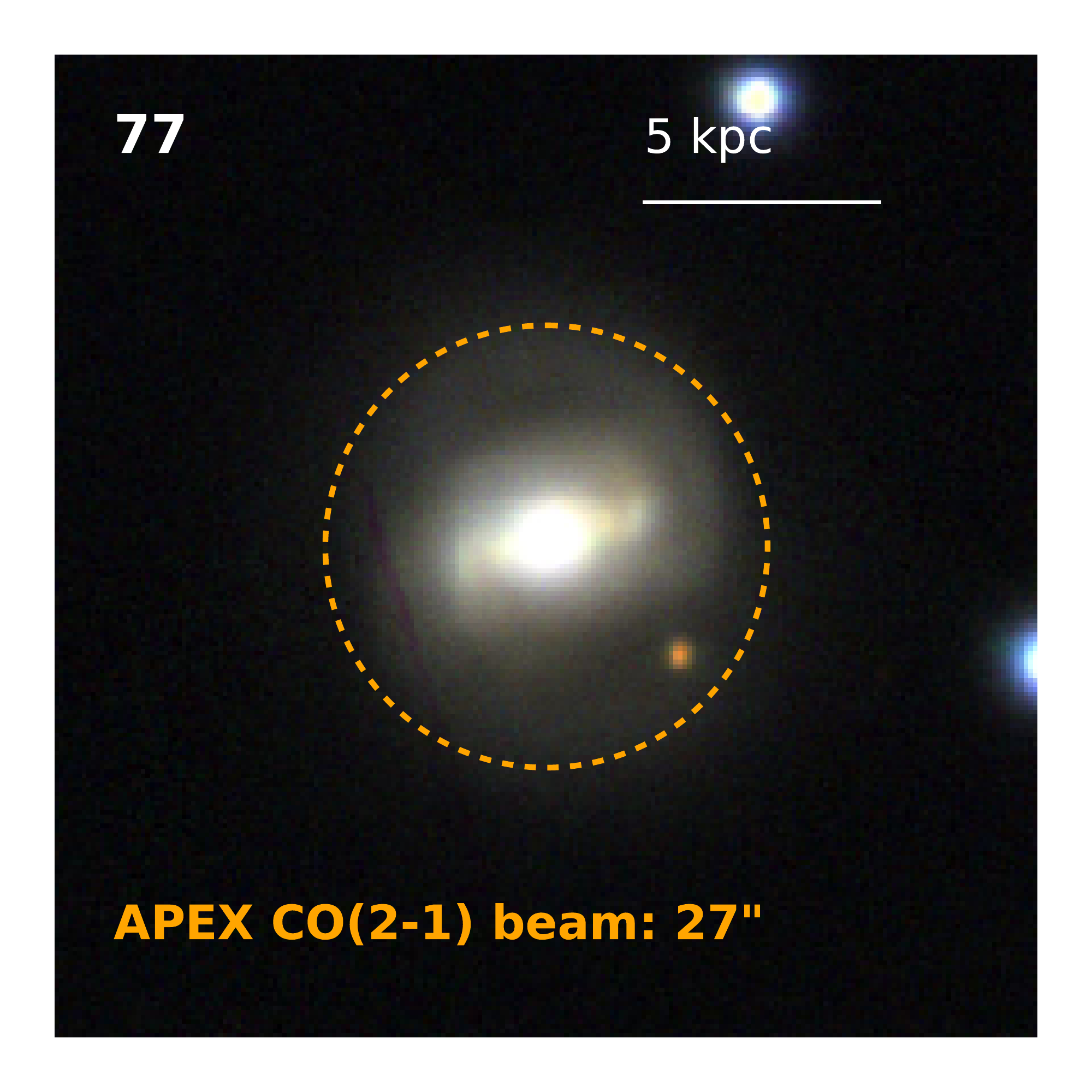}\includegraphics[width=0.26\textwidth]{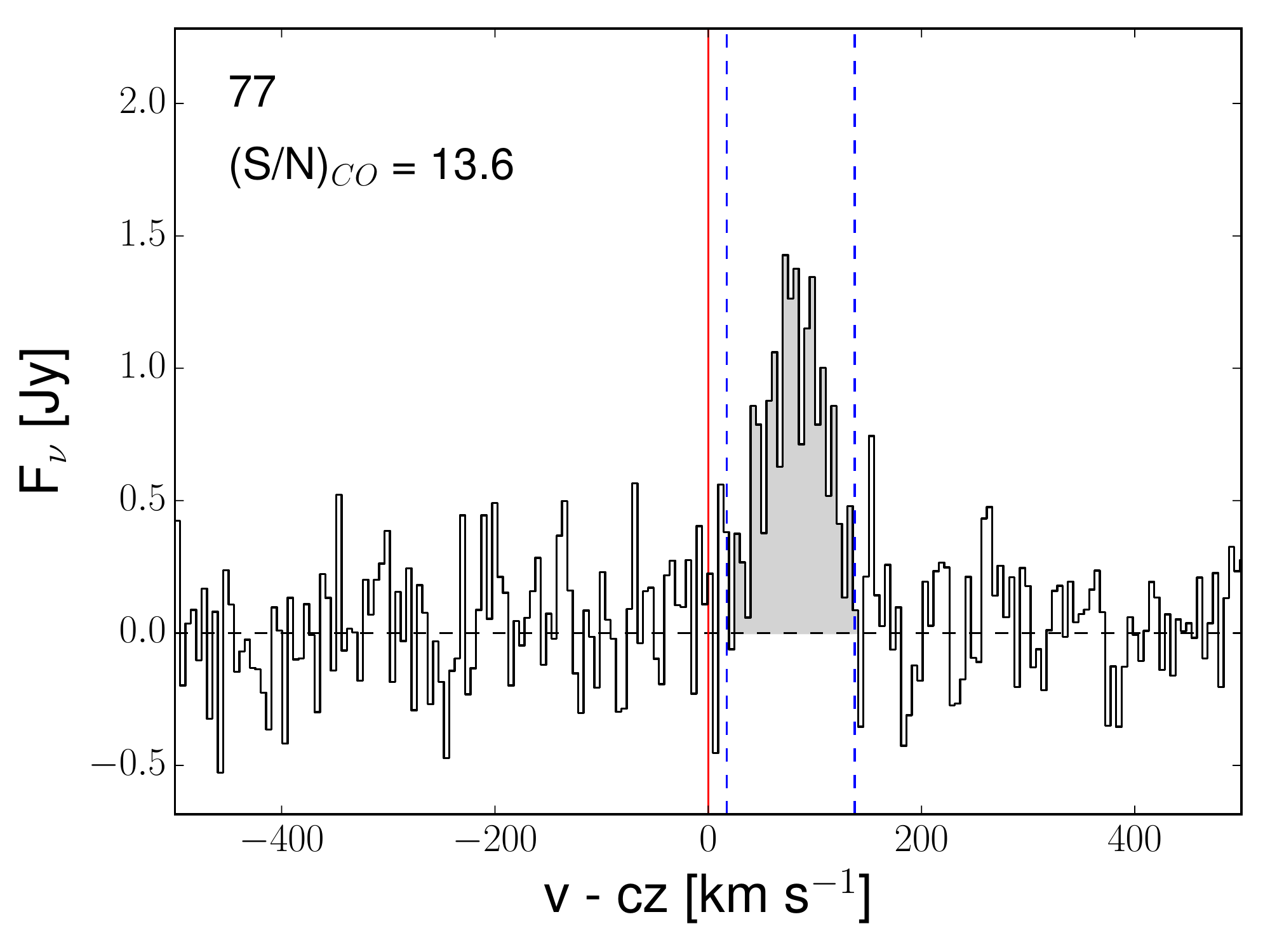}
\includegraphics[width=0.18\textwidth]{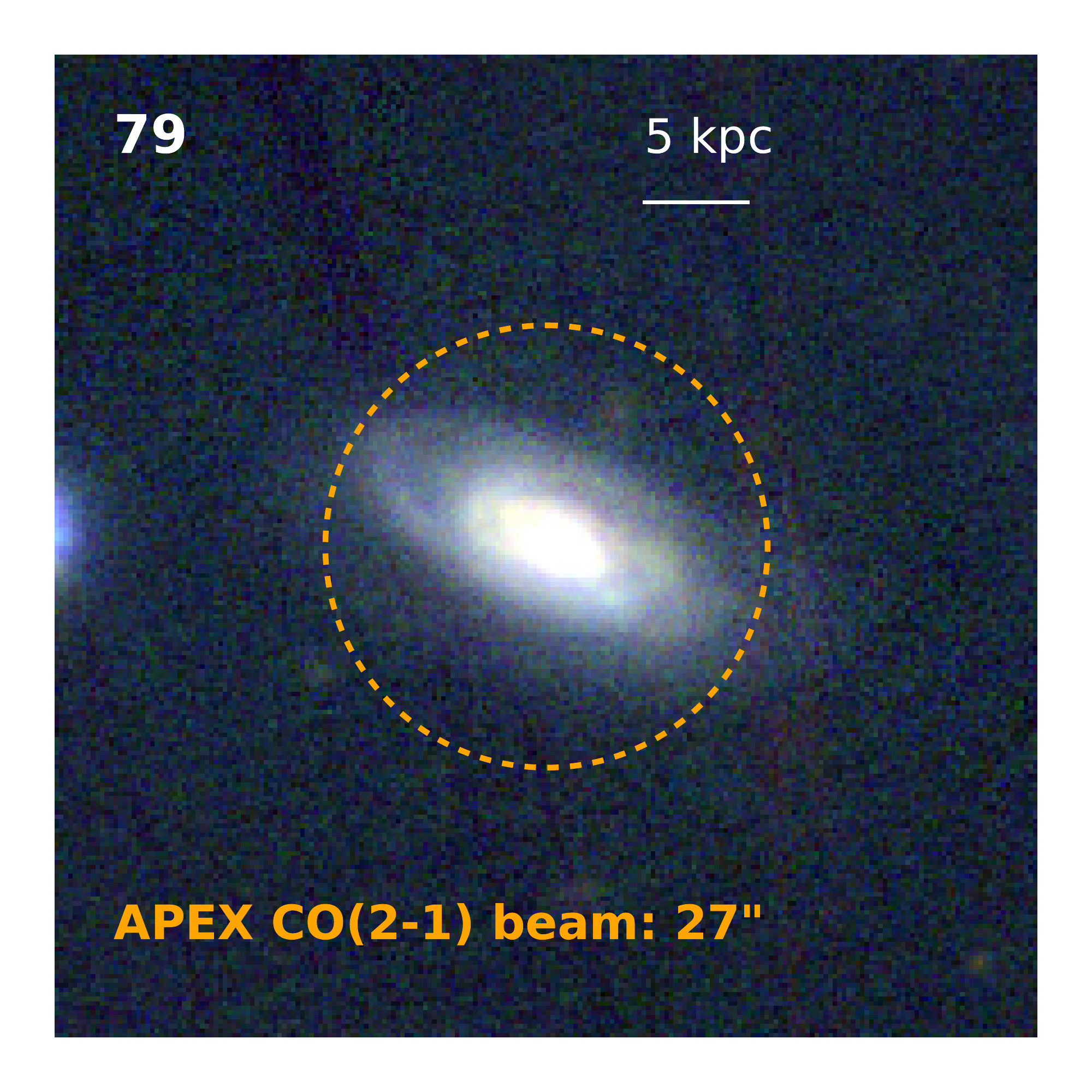}\includegraphics[width=0.26\textwidth]{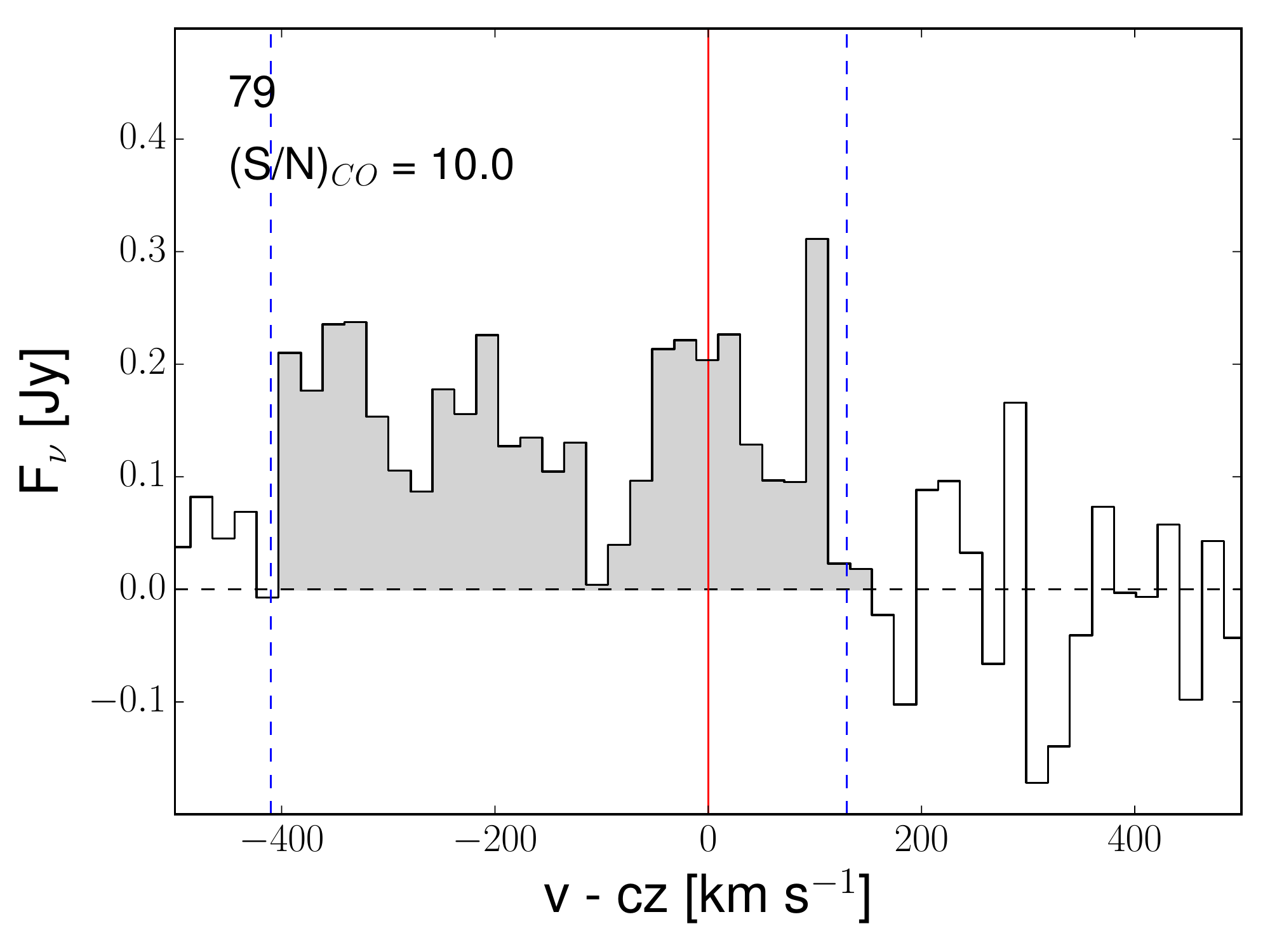}
\includegraphics[width=0.18\textwidth]{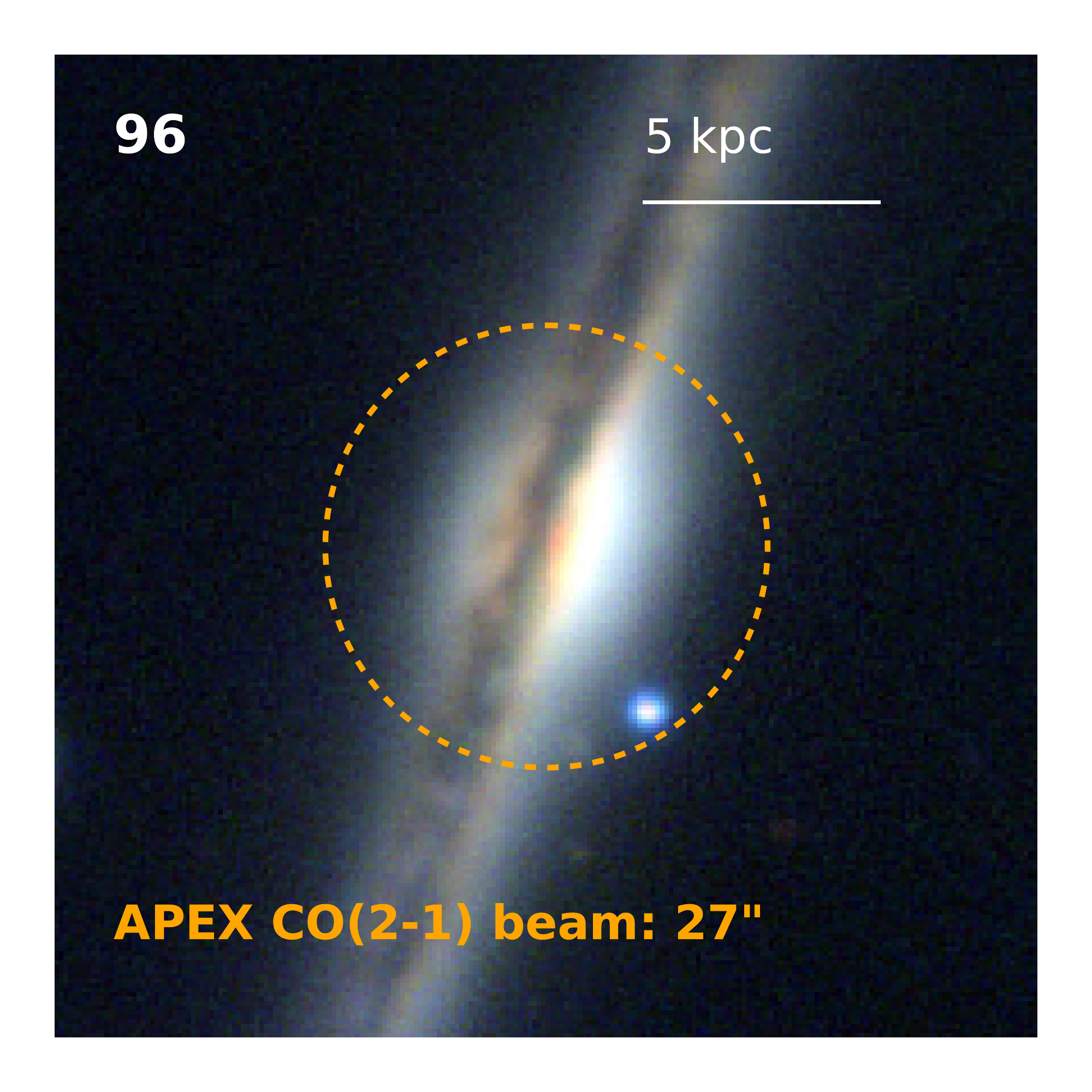}\includegraphics[width=0.26\textwidth]{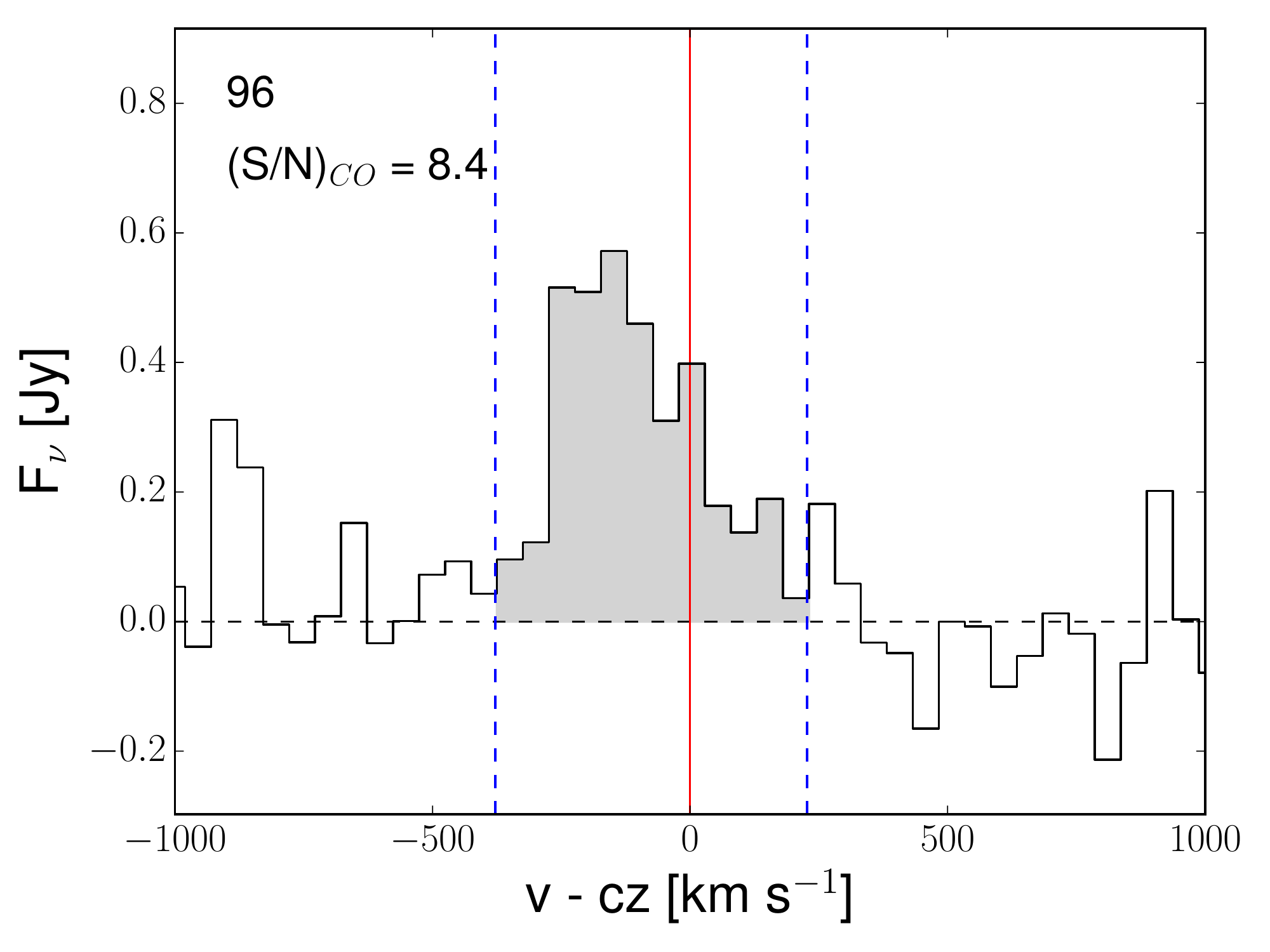}
\includegraphics[width=0.18\textwidth]{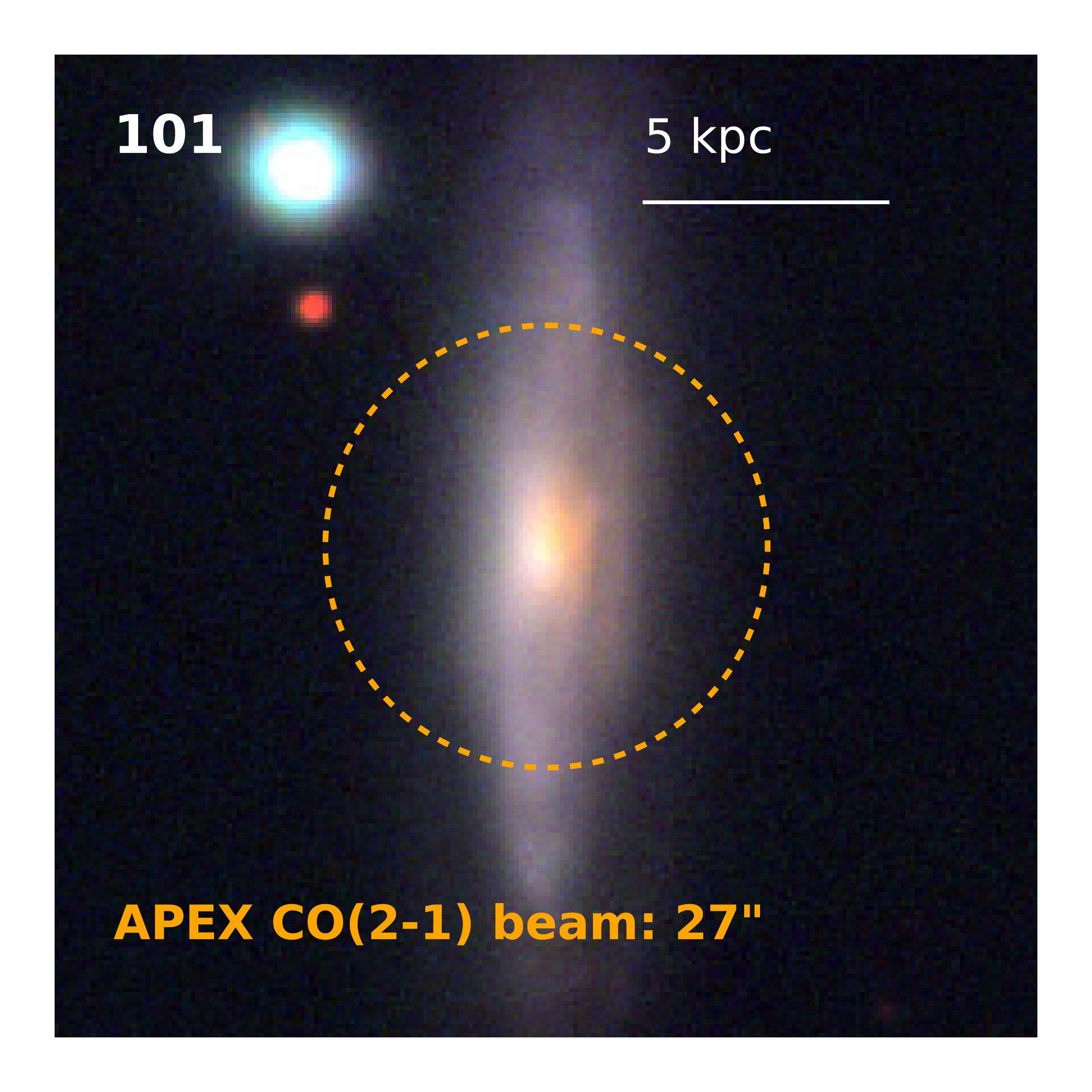}\includegraphics[width=0.26\textwidth]{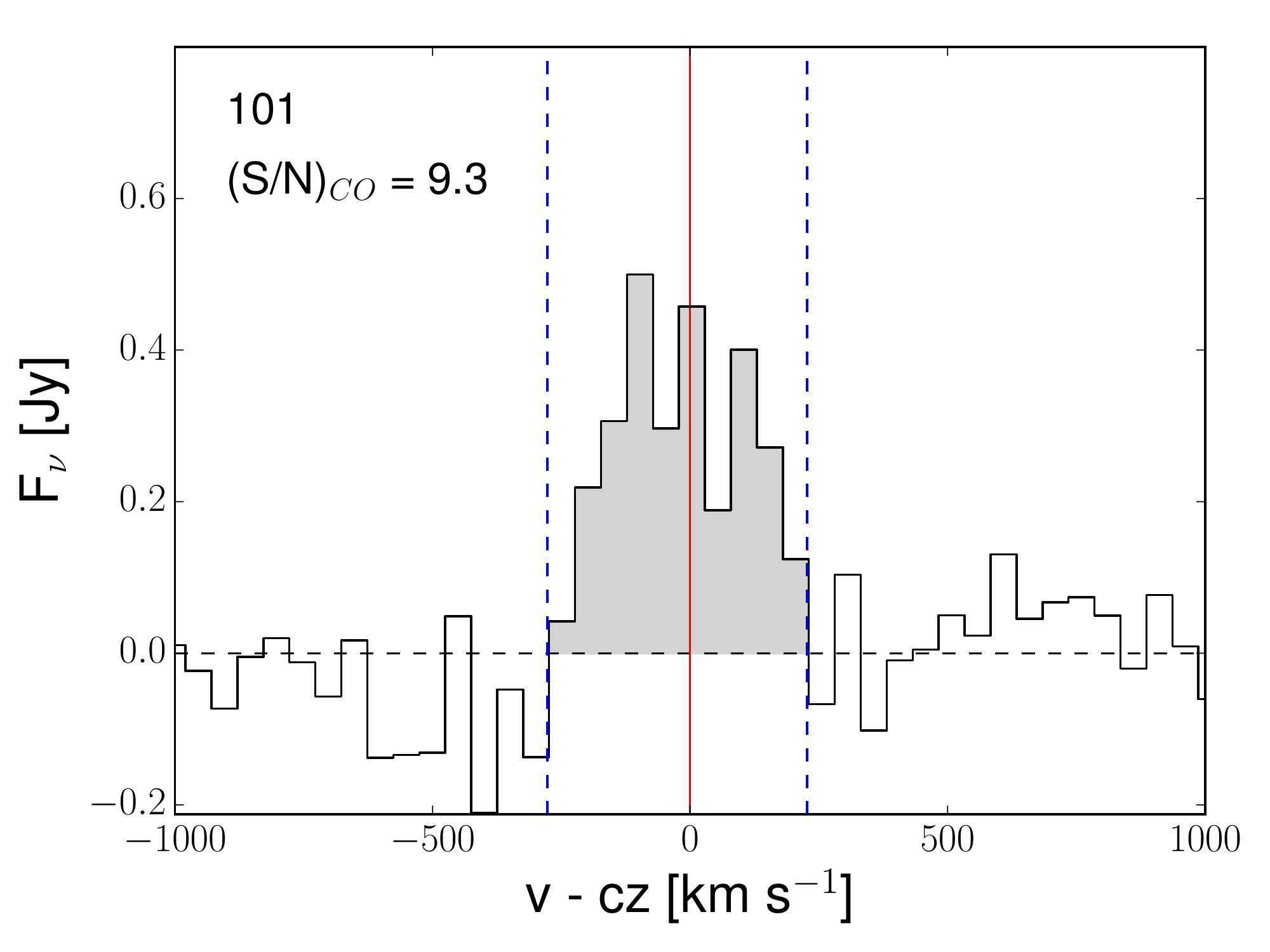}
\includegraphics[width=0.18\textwidth]{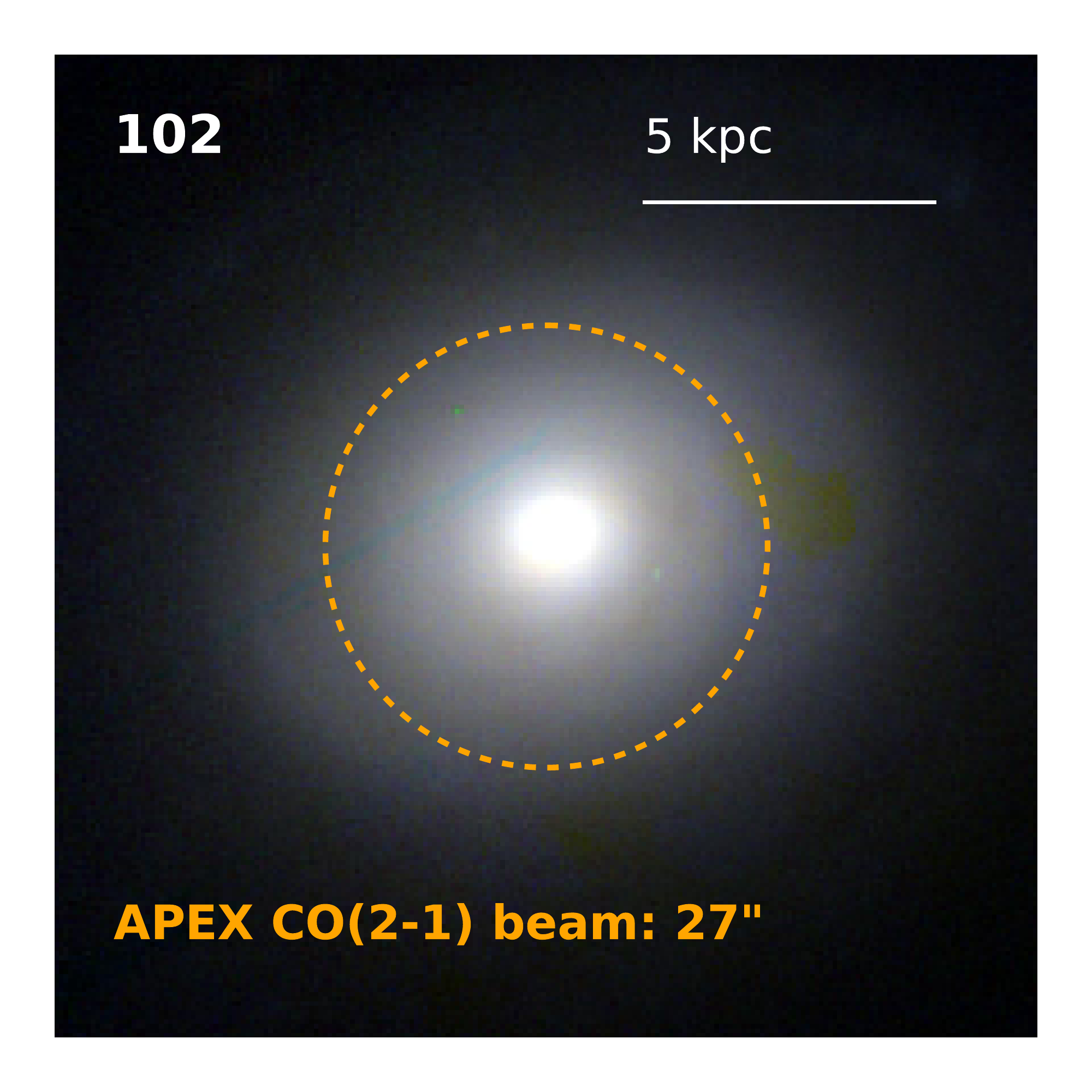}\includegraphics[width=0.26\textwidth]{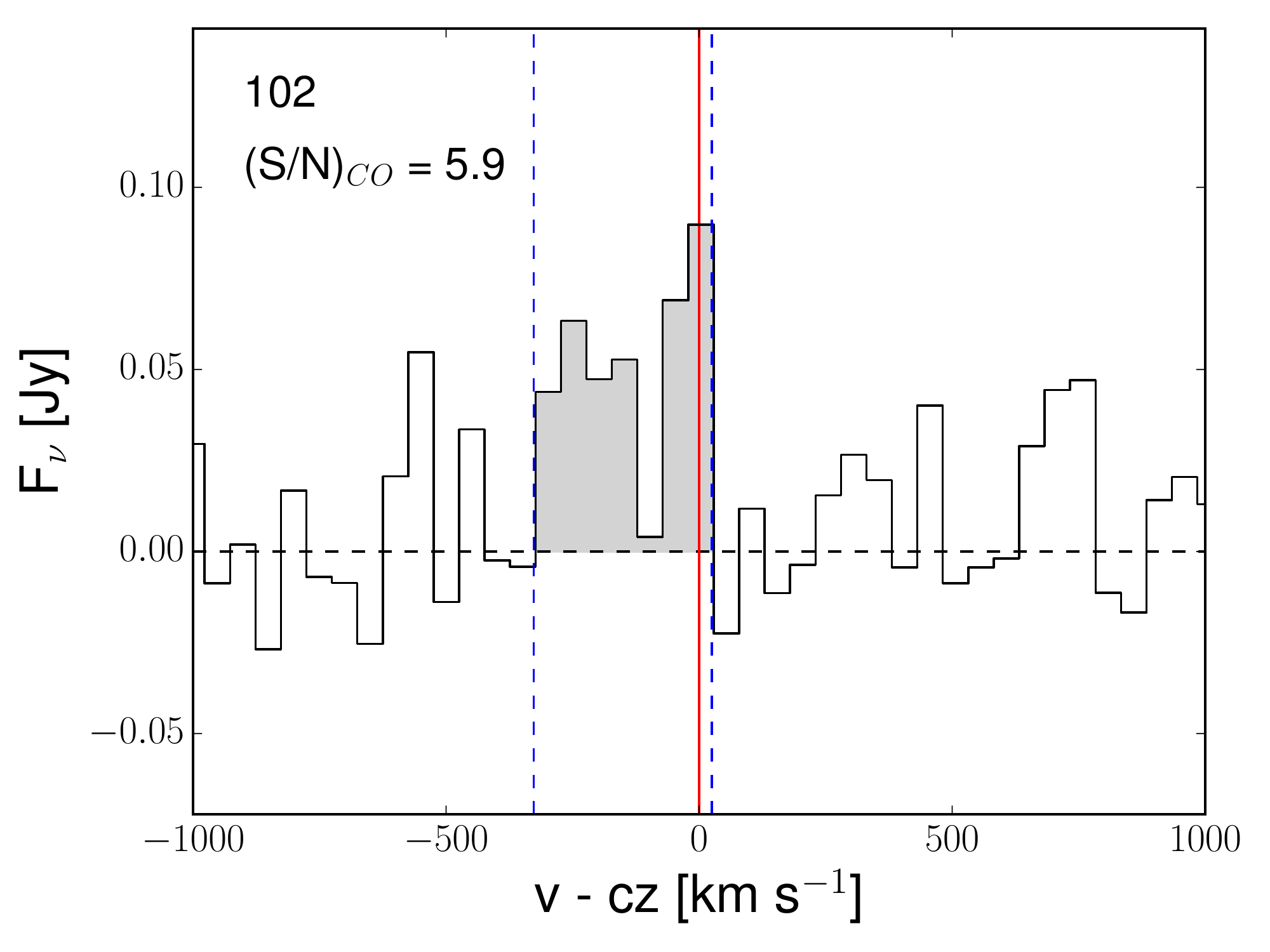}
\includegraphics[width=0.18\textwidth]{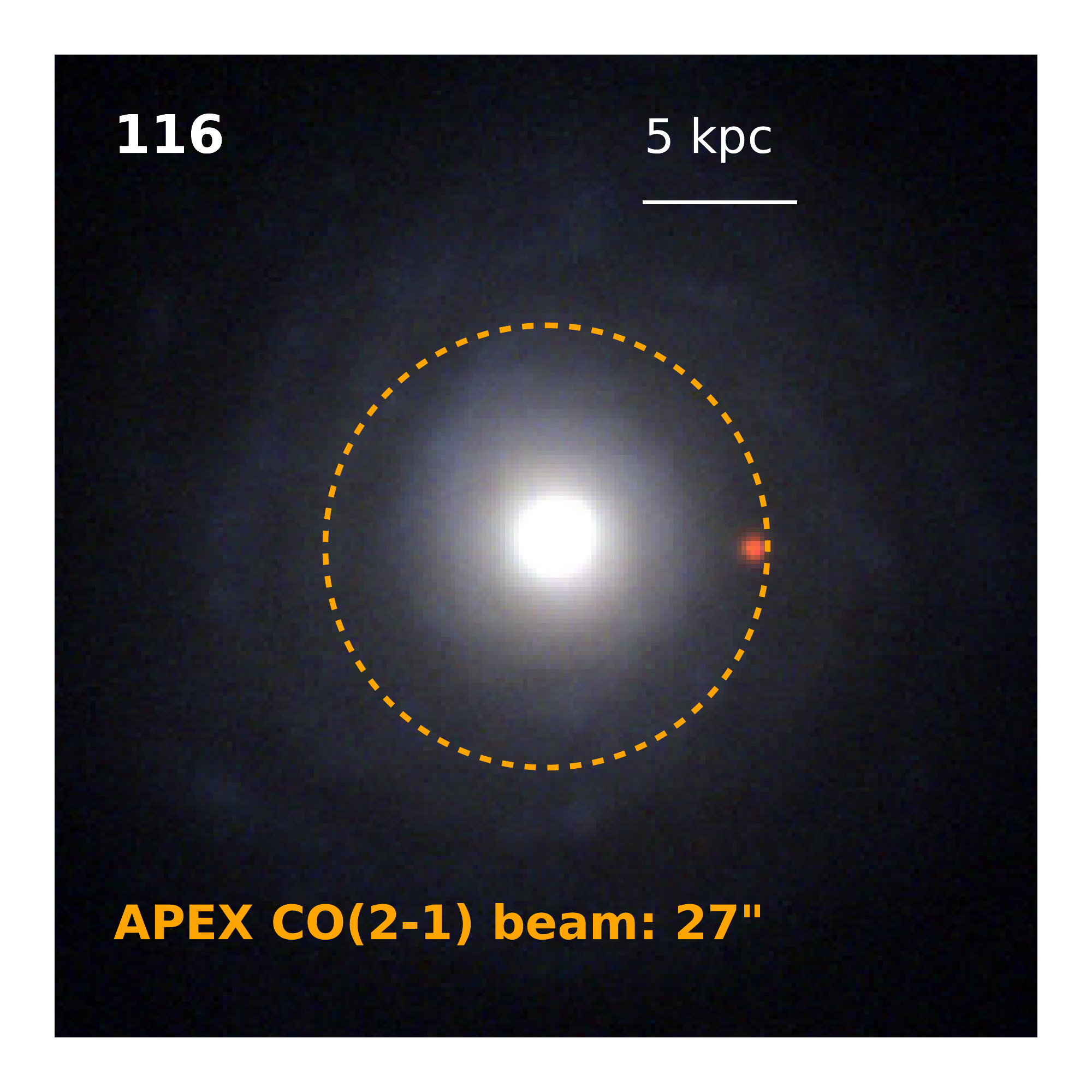}\includegraphics[width=0.26\textwidth]{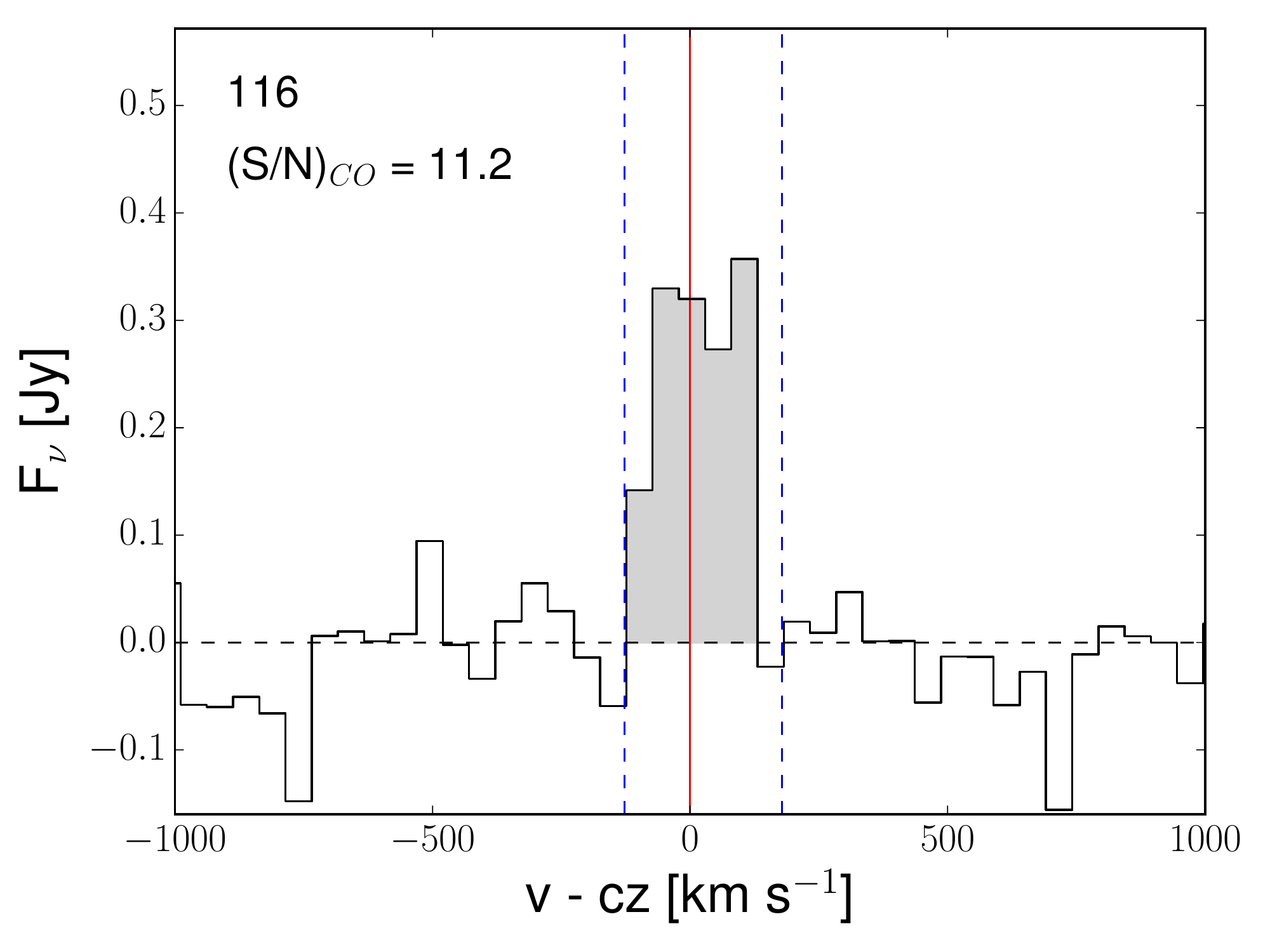}
\includegraphics[width=0.18\textwidth]{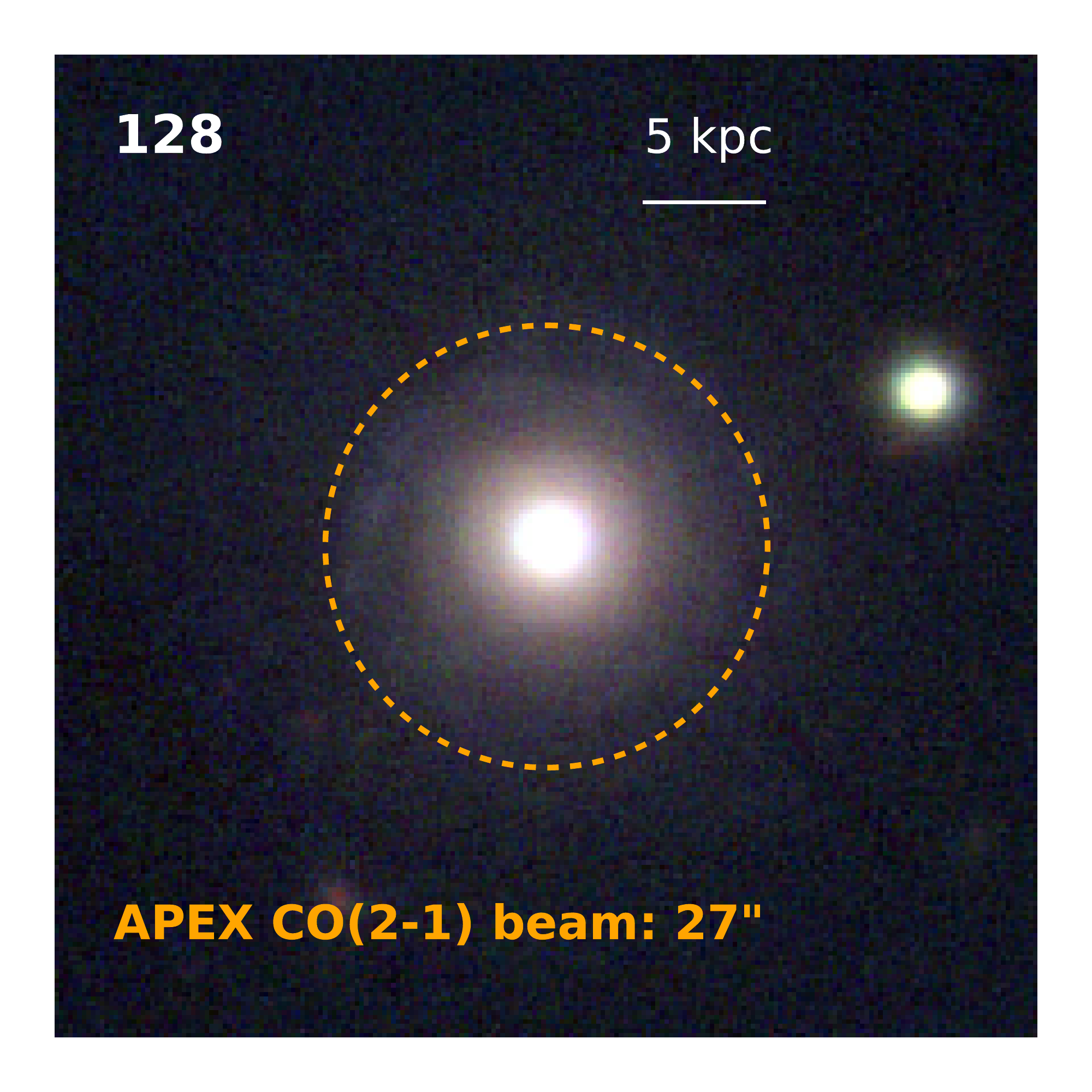}\includegraphics[width=0.26\textwidth]{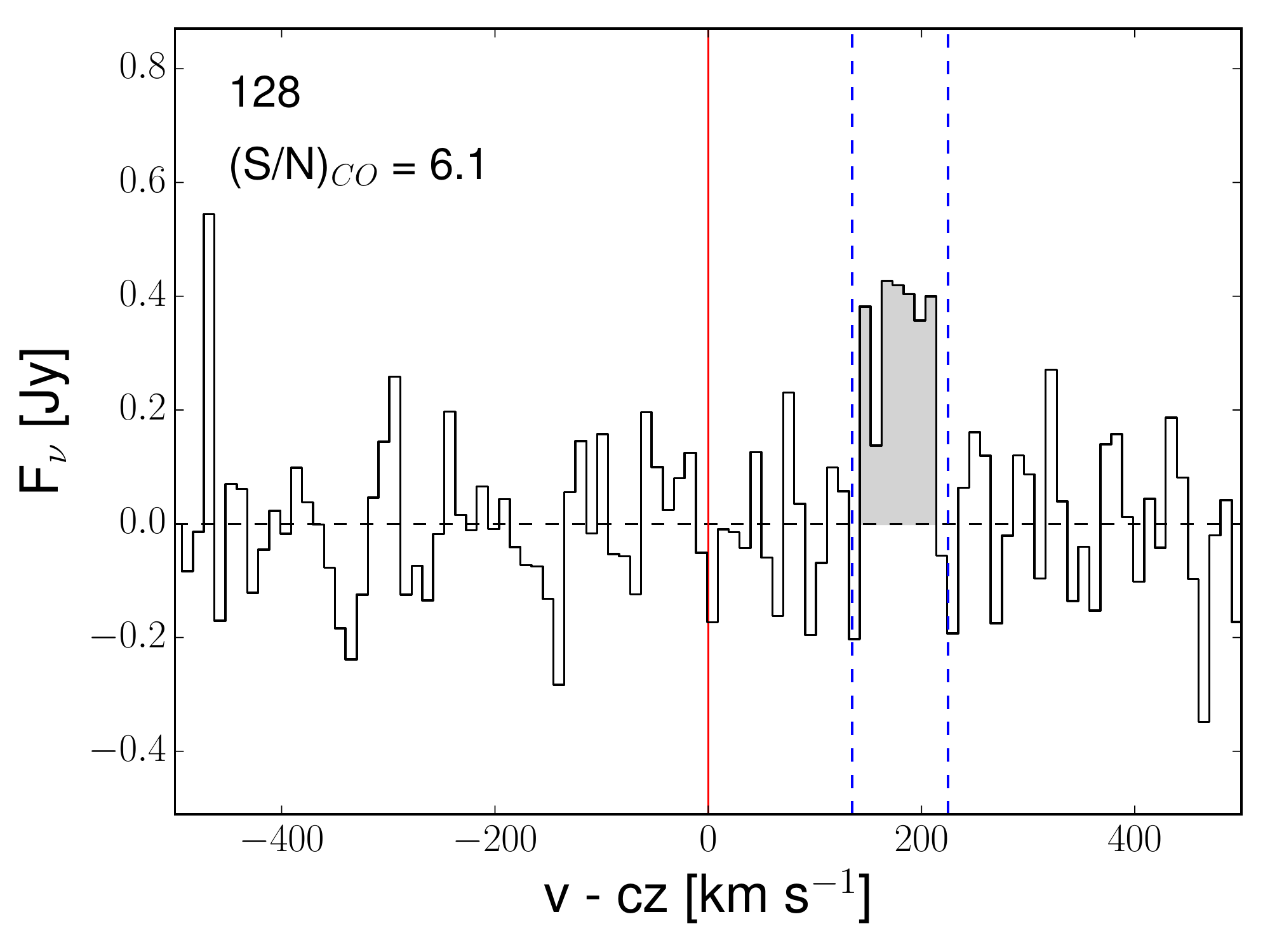}
\includegraphics[width=0.18\textwidth]{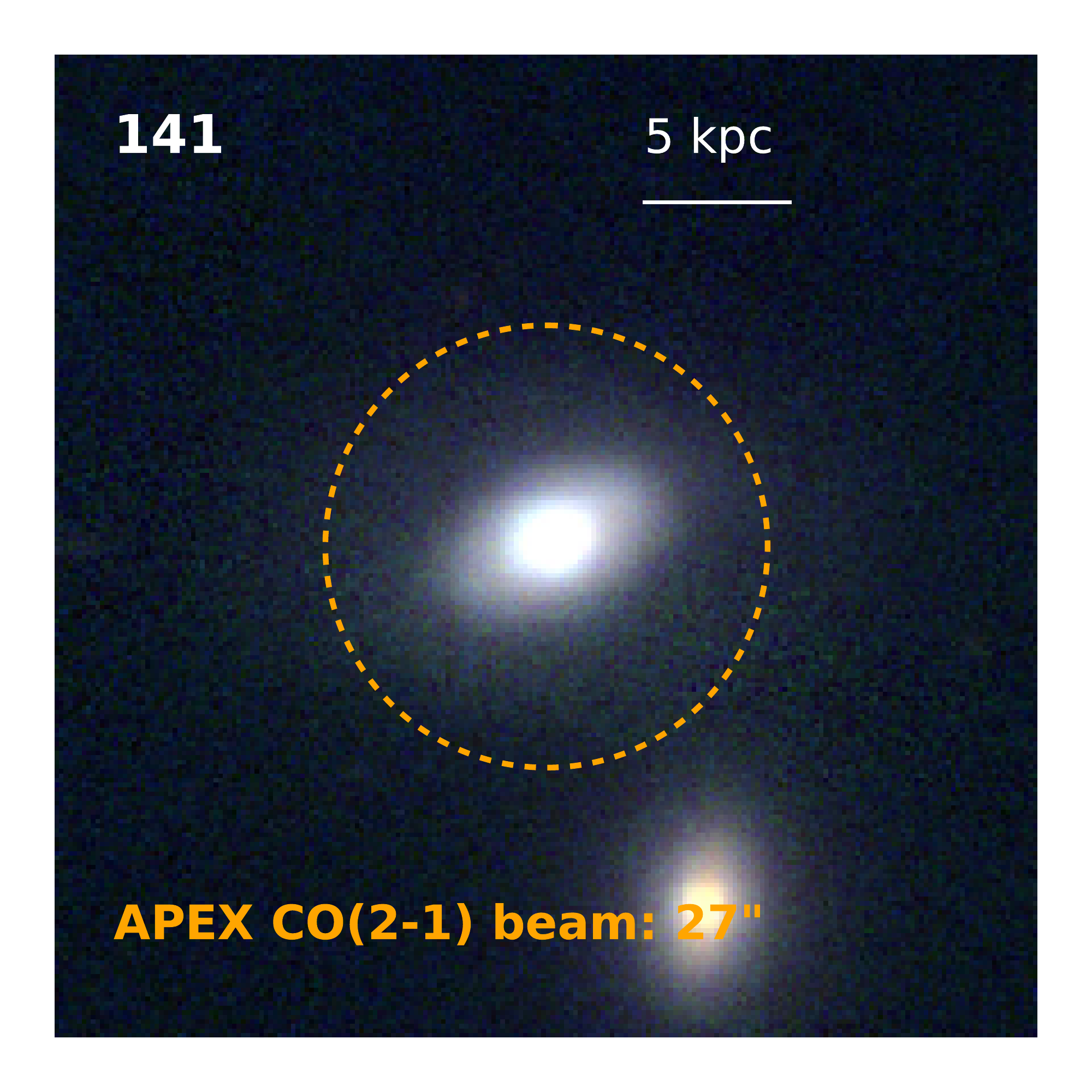}\includegraphics[width=0.26\textwidth]{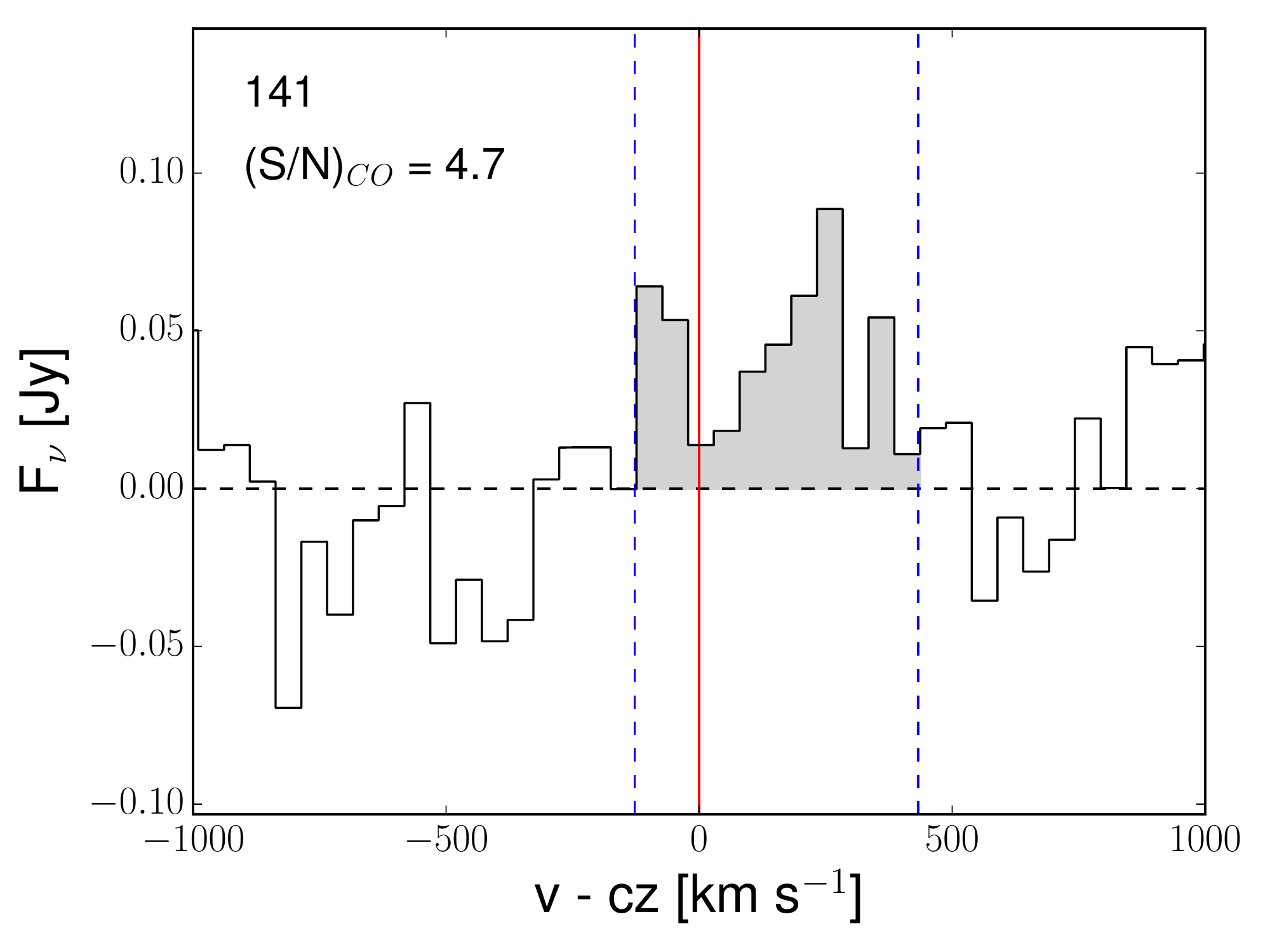}
\includegraphics[width=0.18\textwidth]{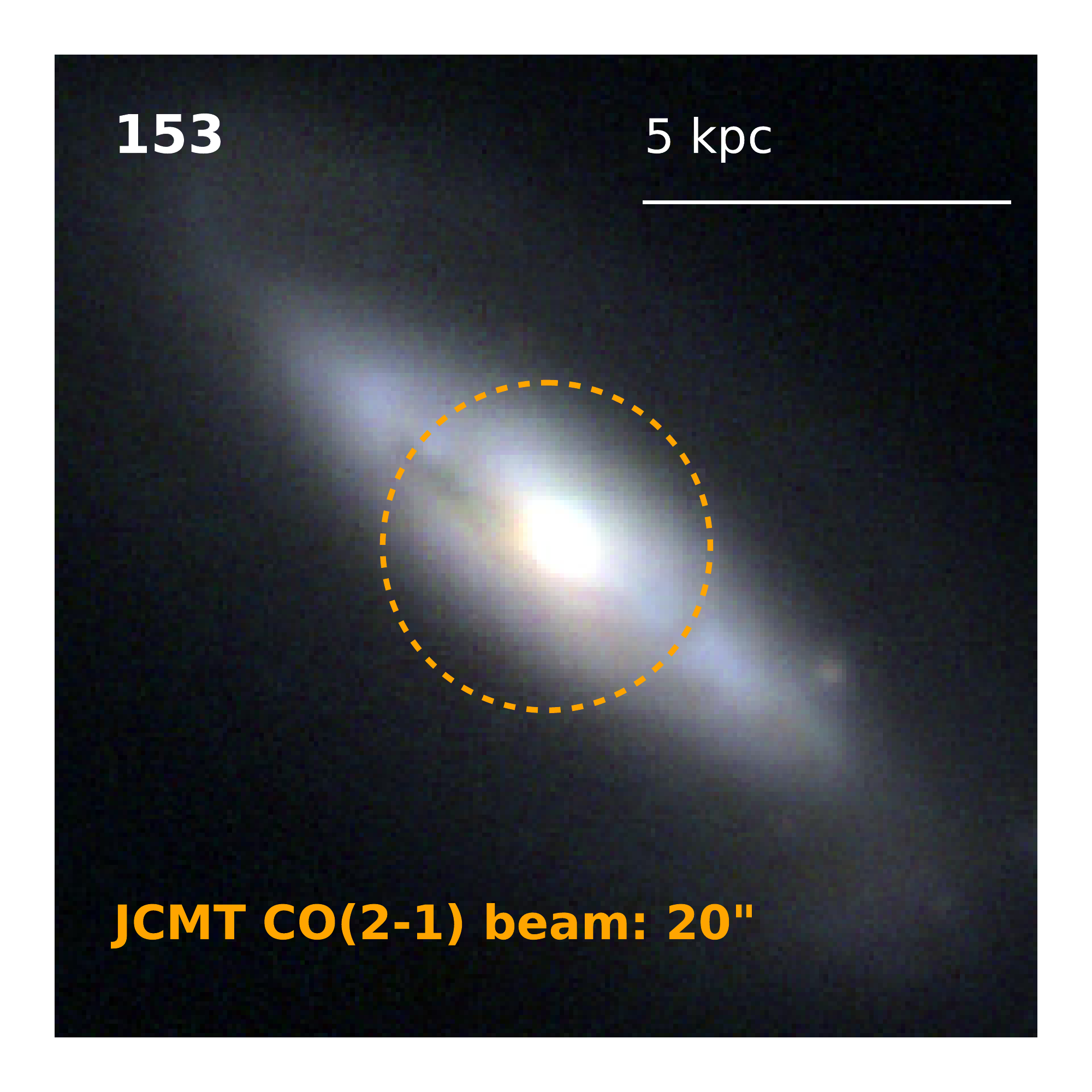}\includegraphics[width=0.26\textwidth]{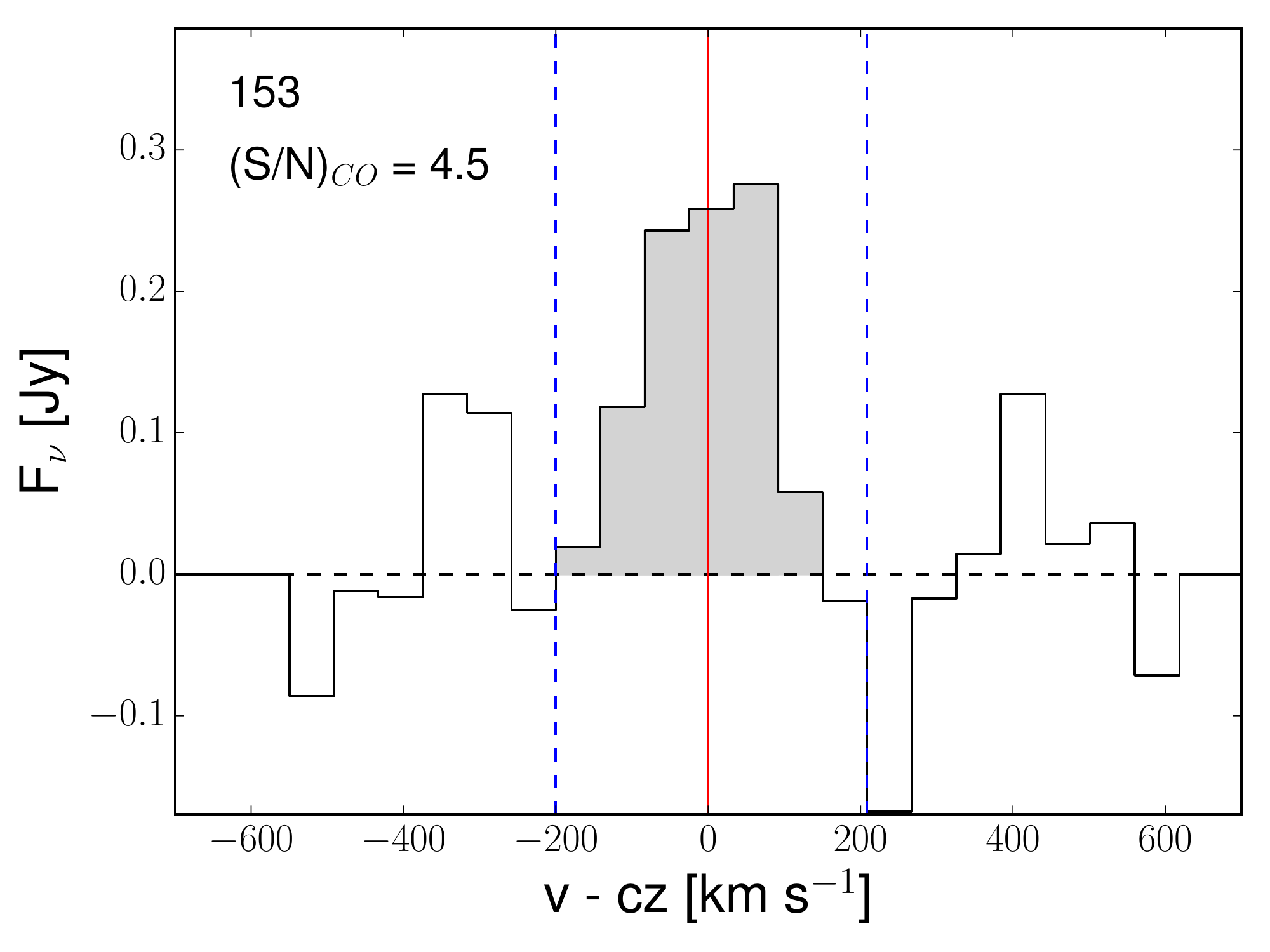}
\caption{Similar to Fig.~\ref{fig:CO21_spectra}, for the remaining detected sources not already shown in the text.\
} 
\label{fig:CO21_spectra_all_1}
\end{figure*}

\begin{figure*}
\centering
\raggedright
\includegraphics[width=0.18\textwidth]{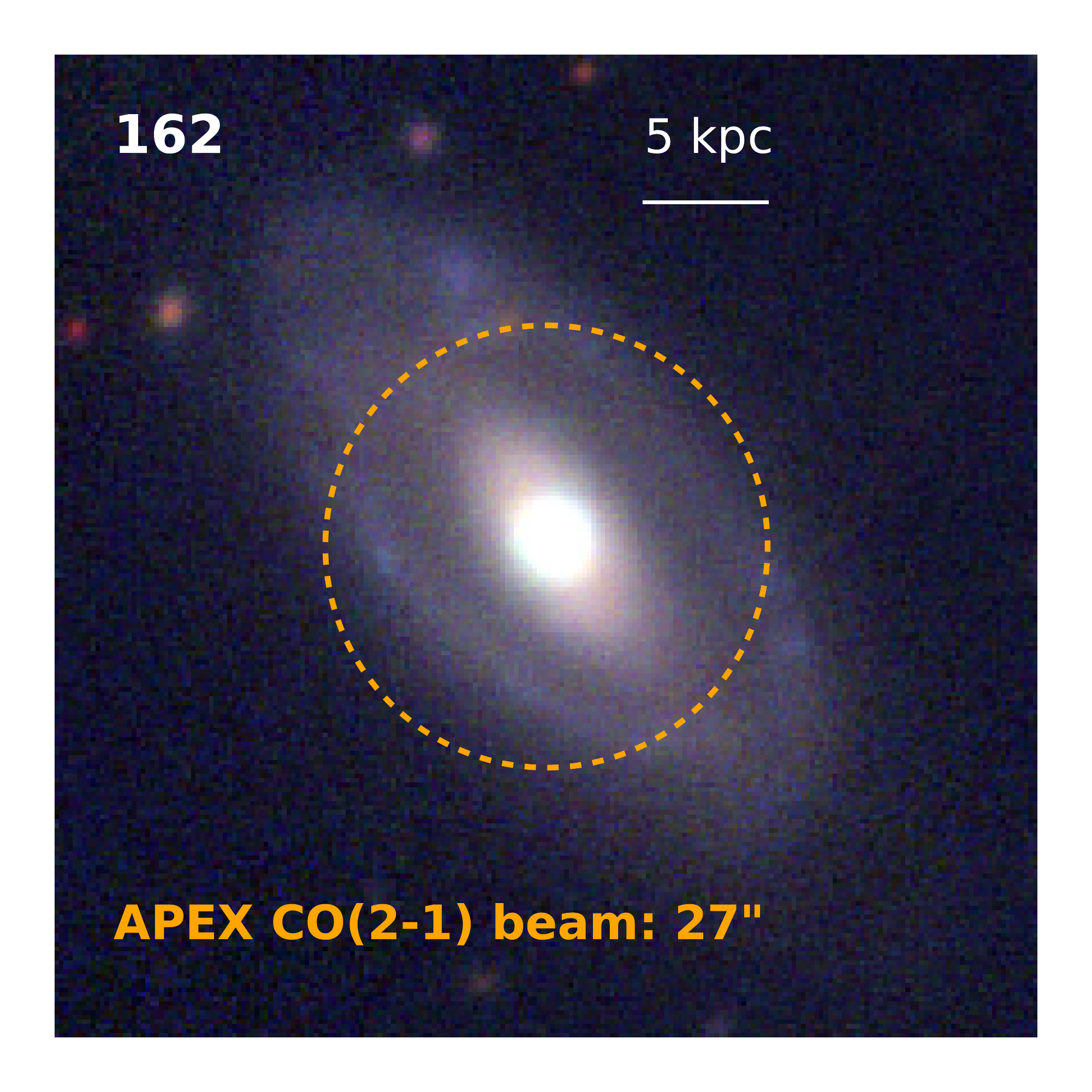}\includegraphics[width=0.26\textwidth]{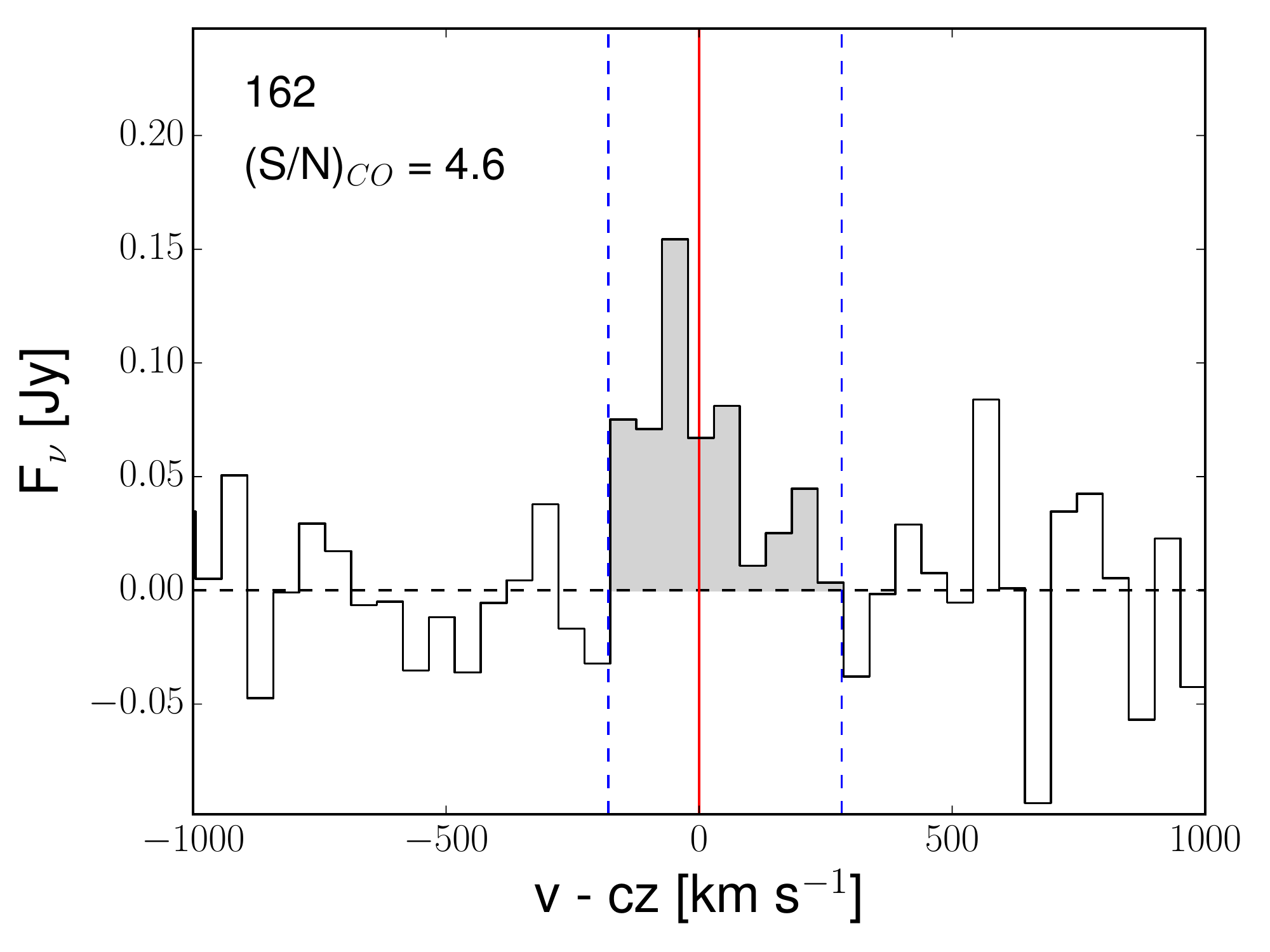}
\includegraphics[width=0.18\textwidth]{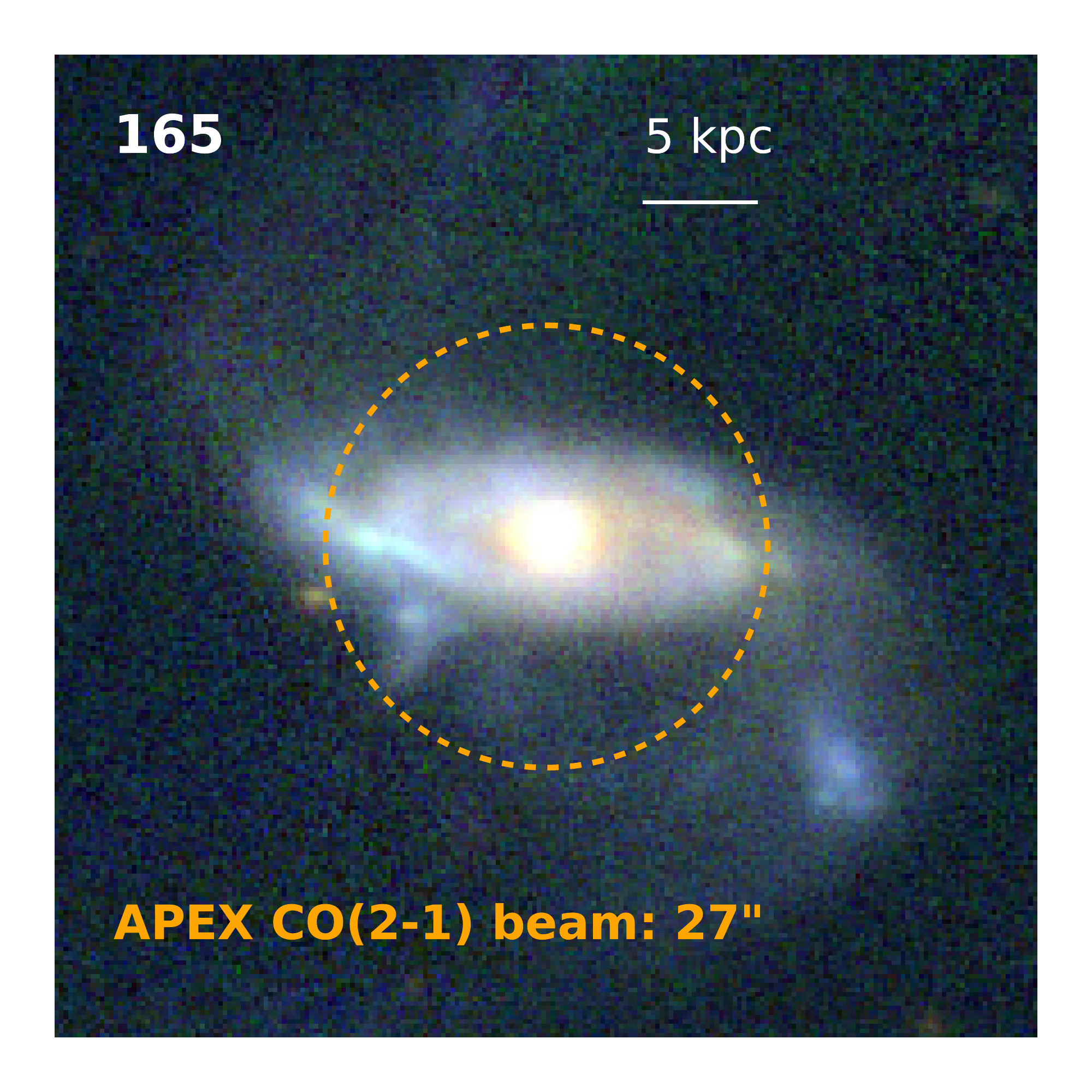}\includegraphics[width=0.26\textwidth]{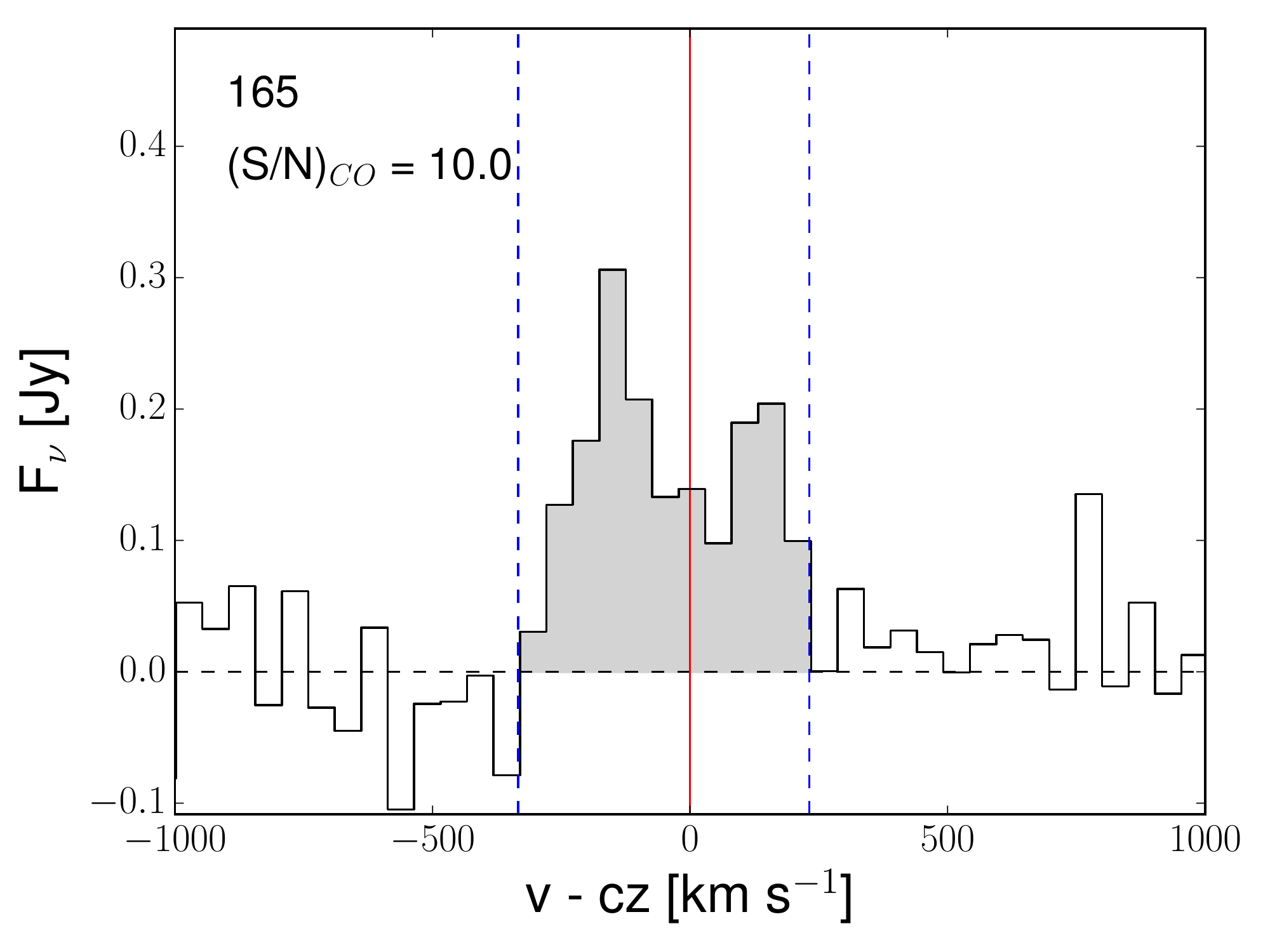}
\includegraphics[width=0.18\textwidth]{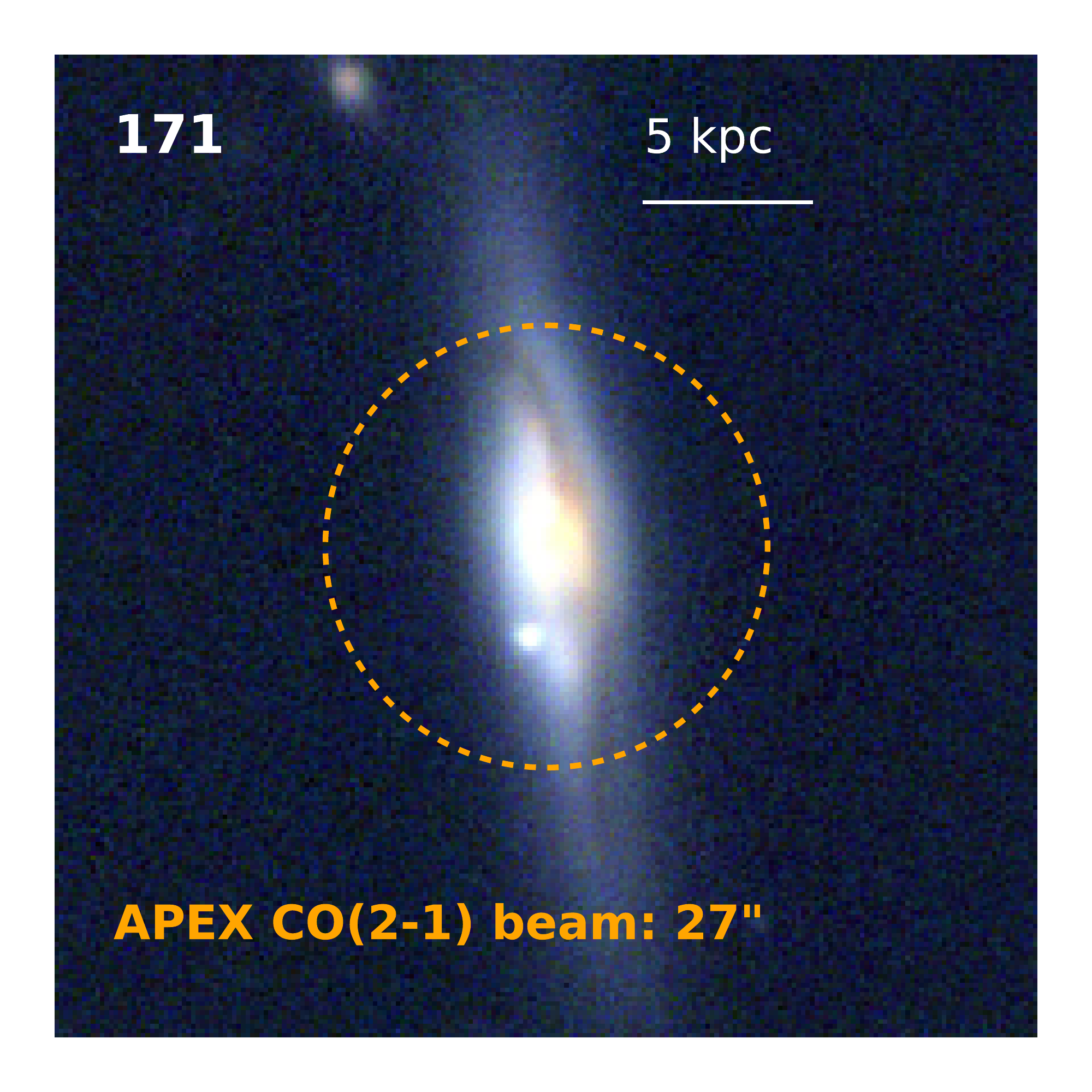}\includegraphics[width=0.26\textwidth]{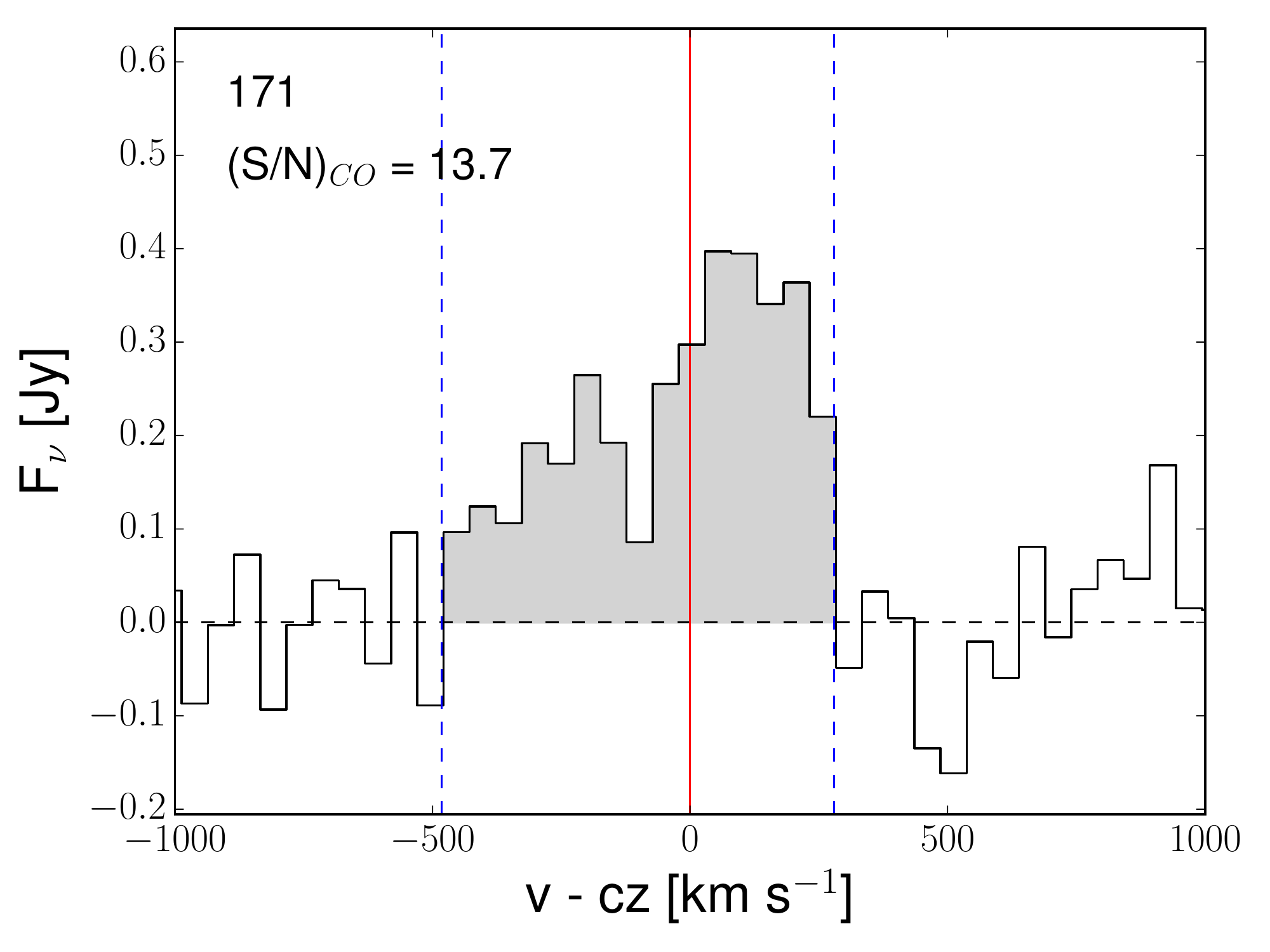}
\includegraphics[width=0.18\textwidth]{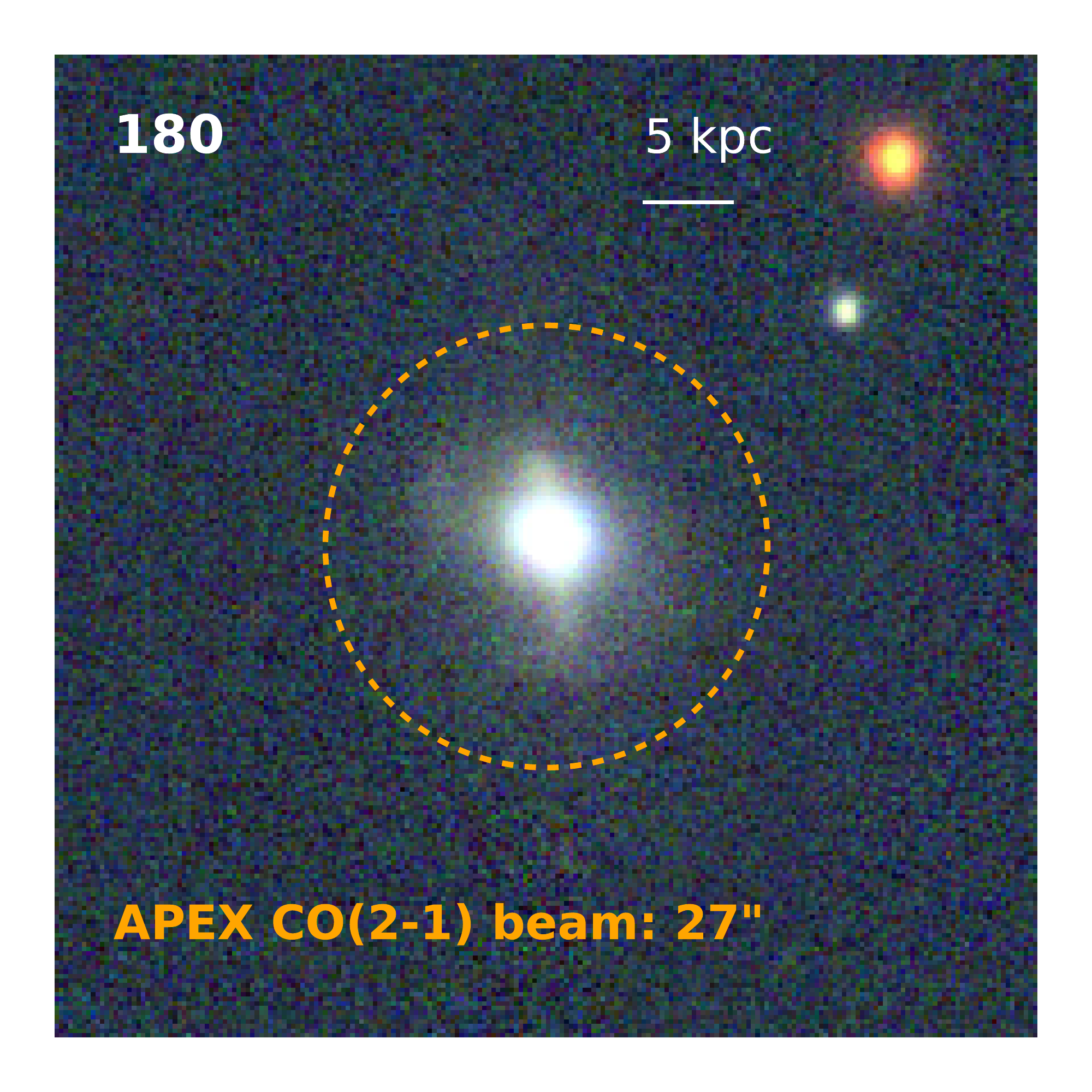}\includegraphics[width=0.26\textwidth]{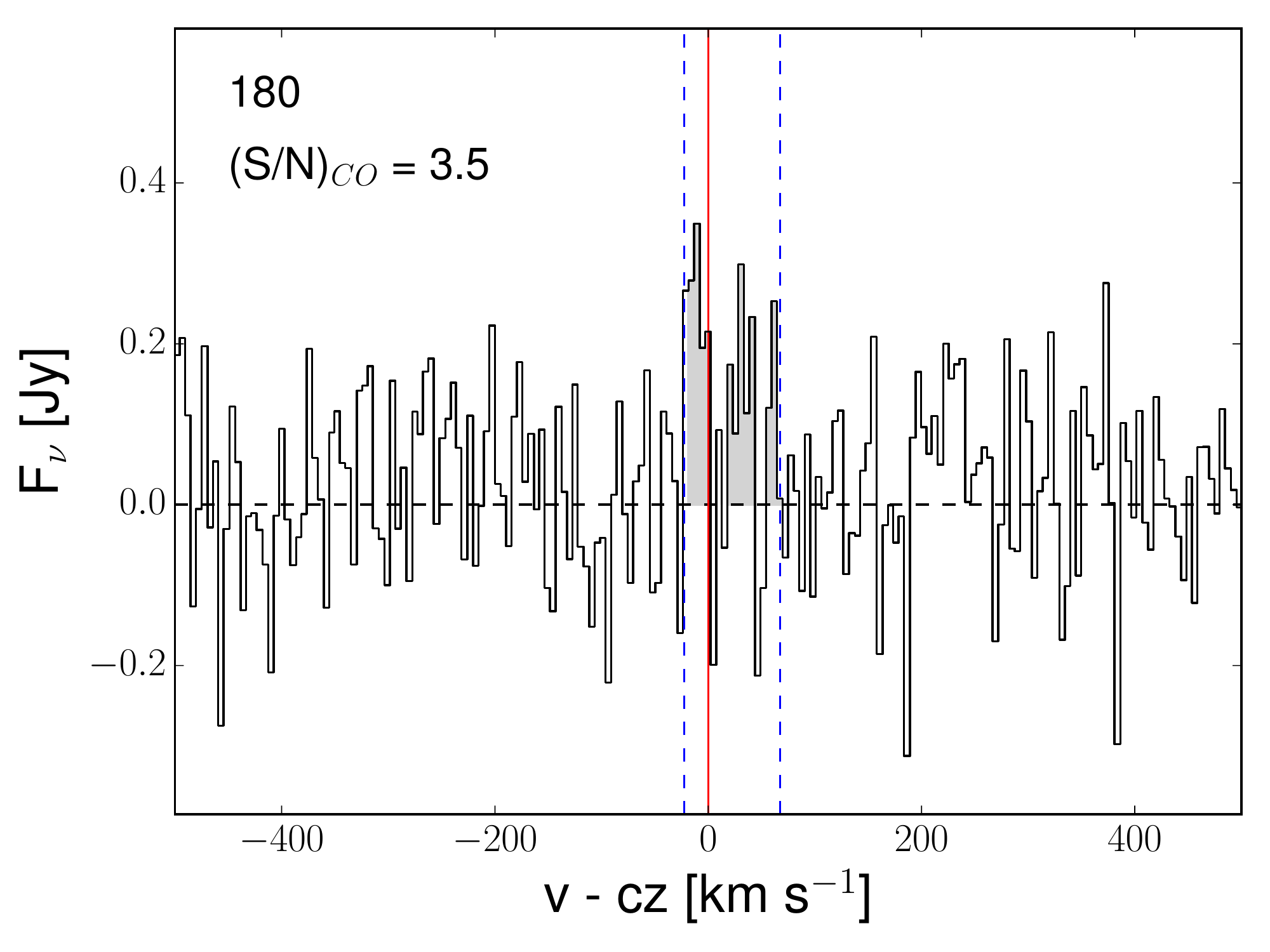}
\includegraphics[width=0.18\textwidth]{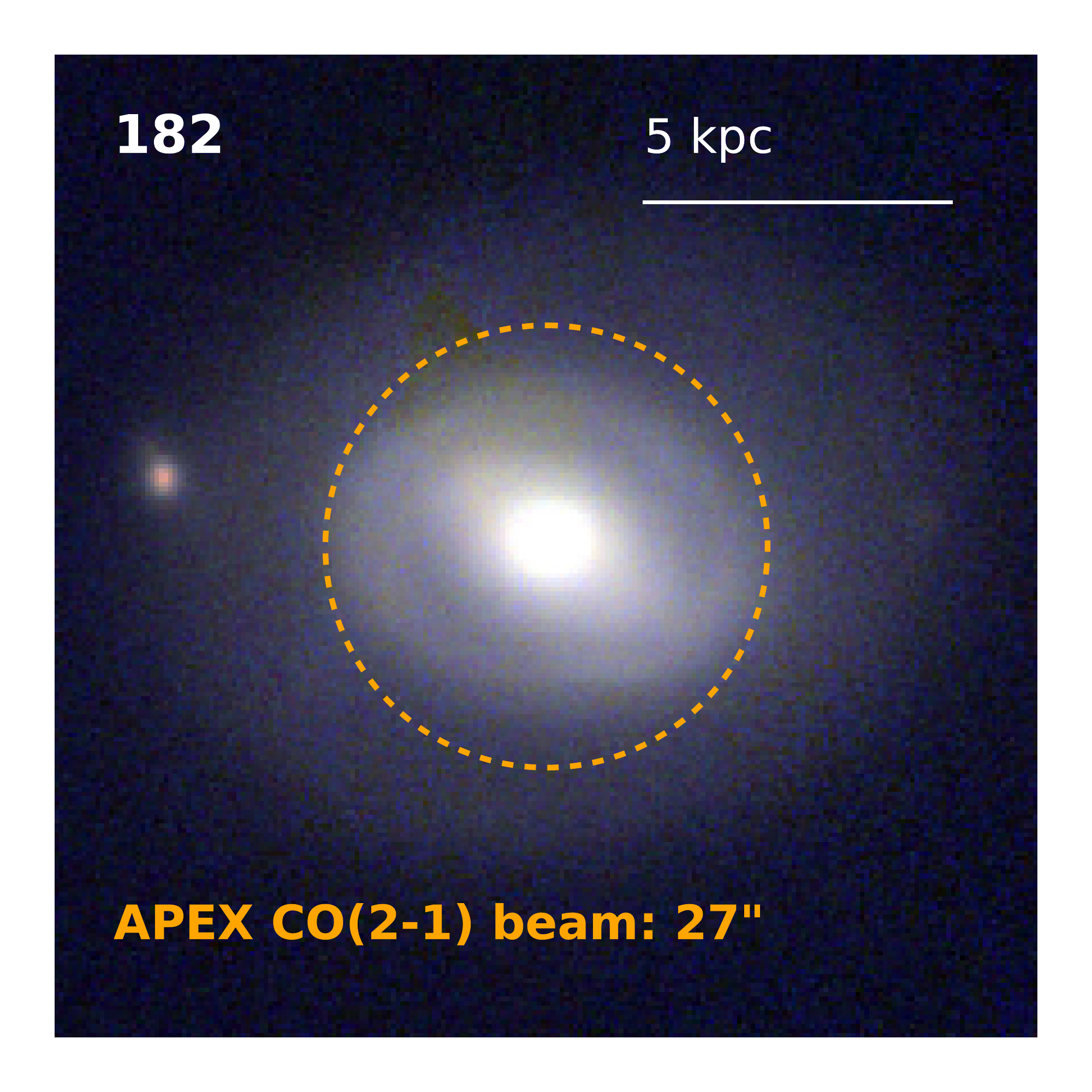}\includegraphics[width=0.26\textwidth]{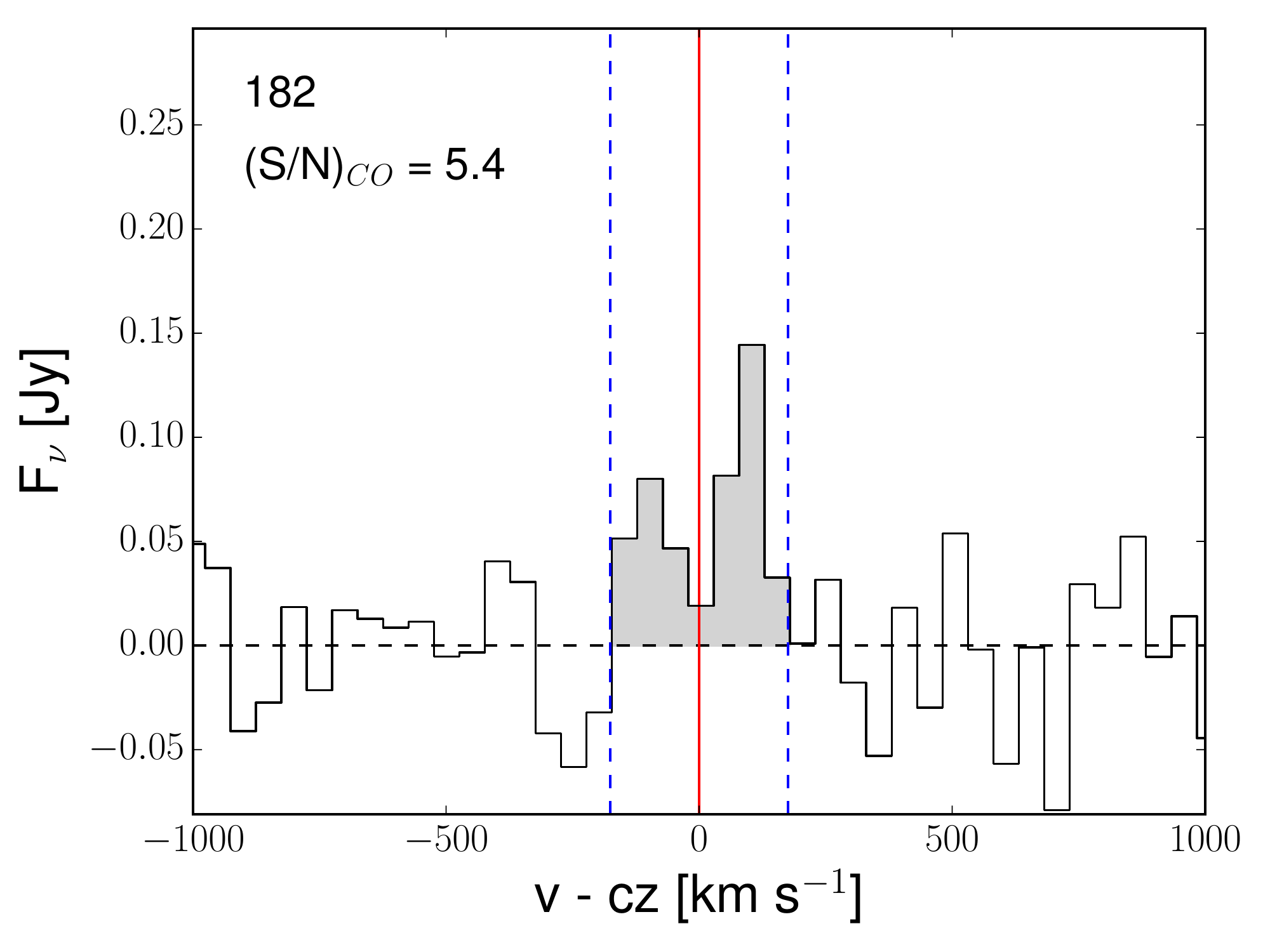}
\includegraphics[width=0.18\textwidth]{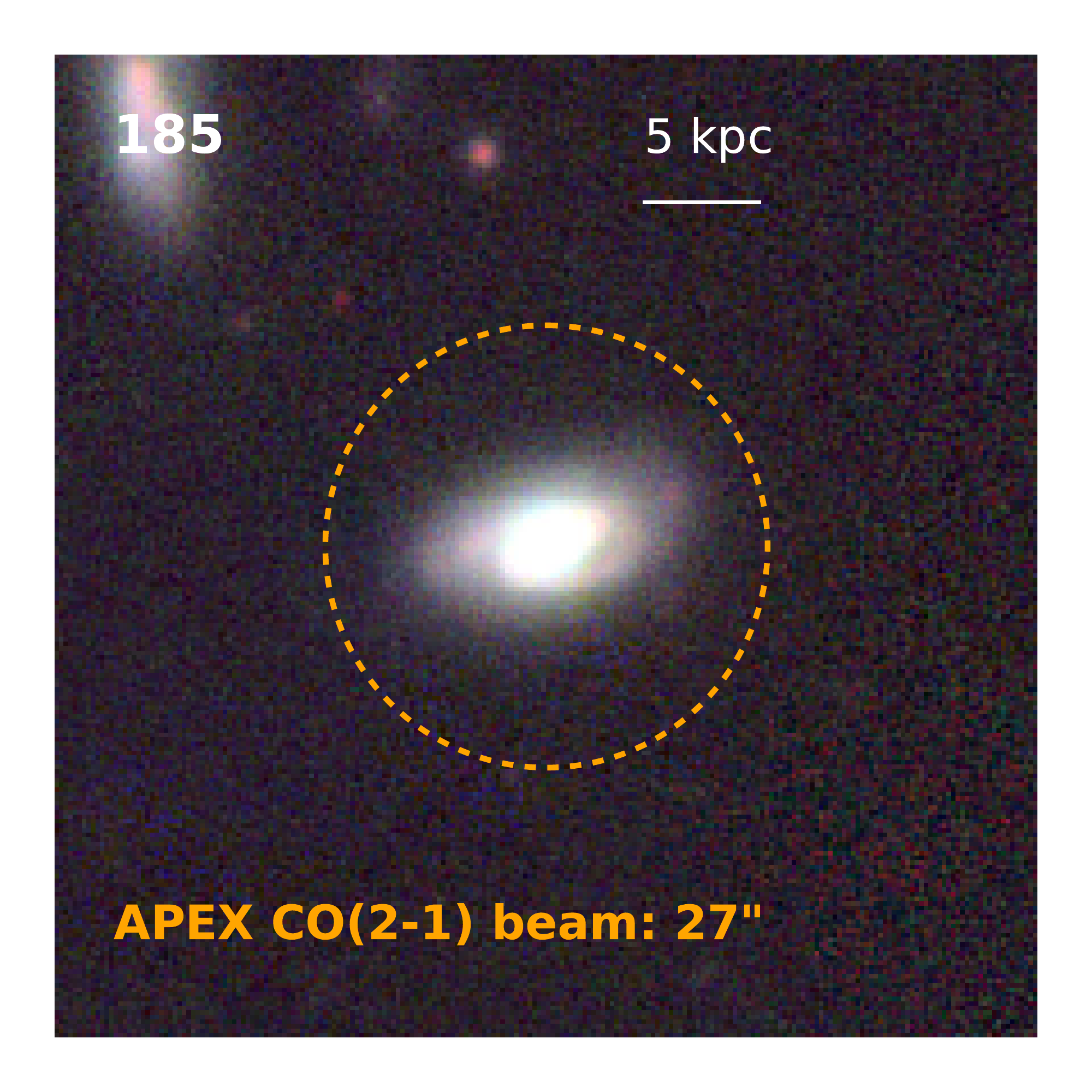}\includegraphics[width=0.26\textwidth]{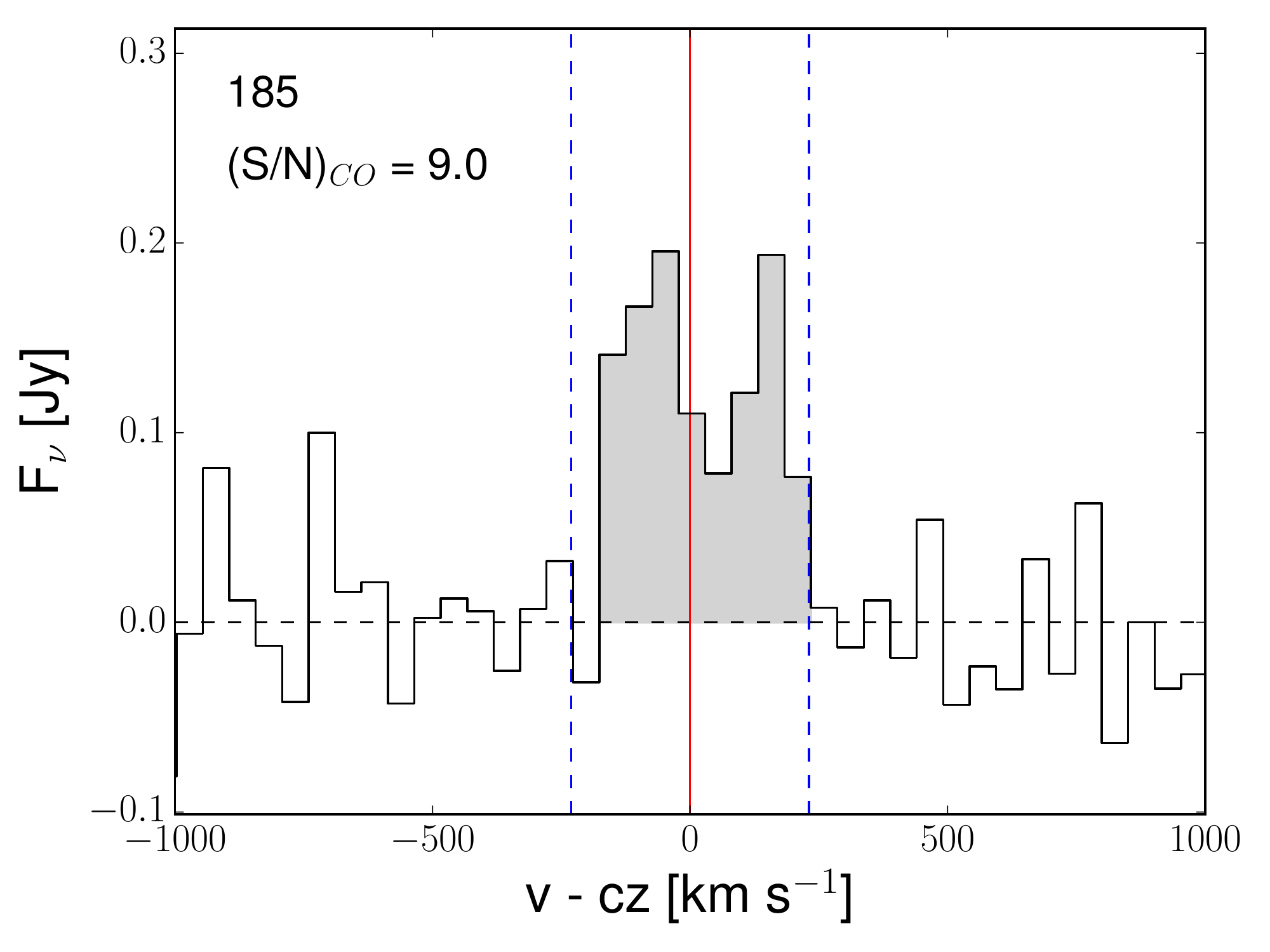}
\includegraphics[width=0.18\textwidth]{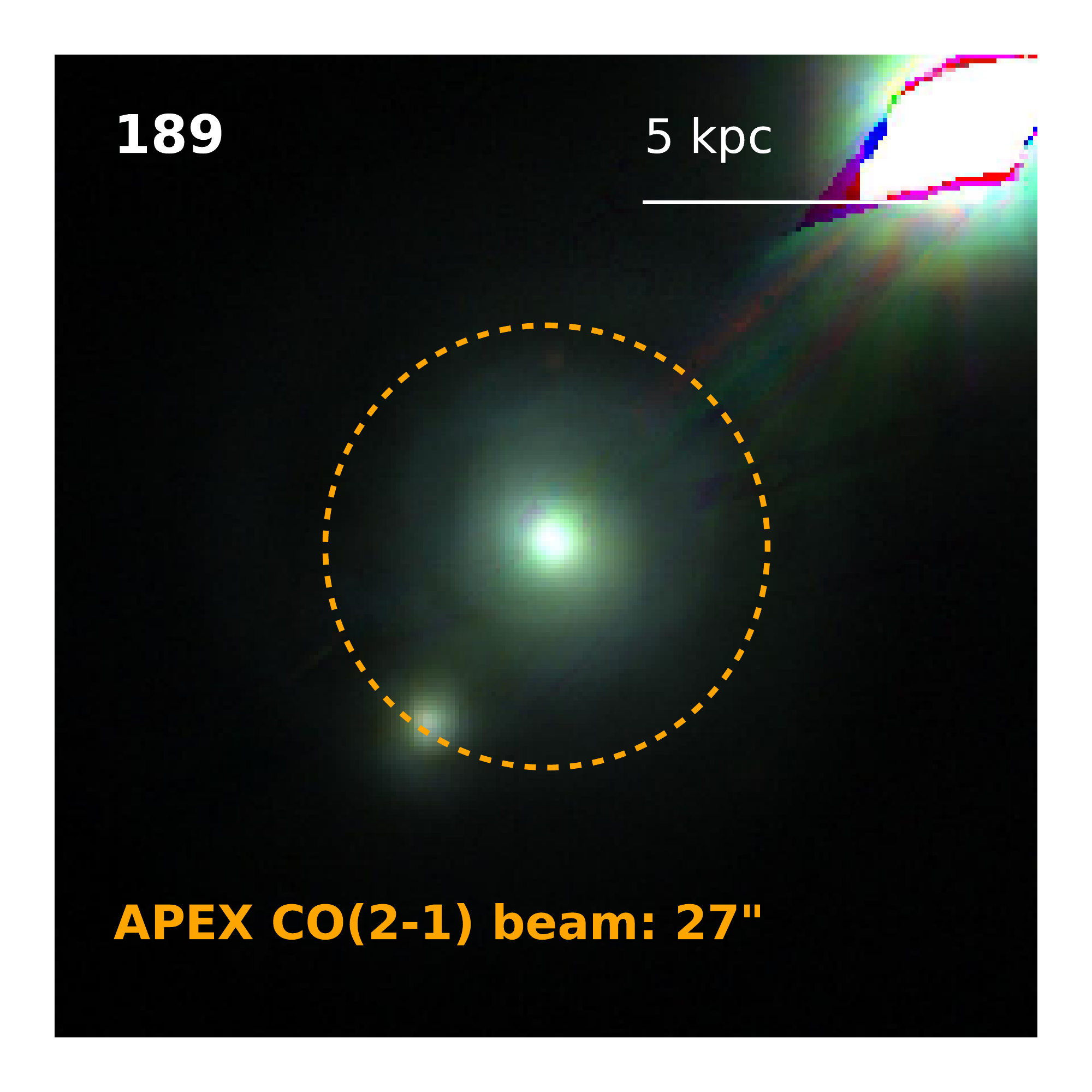}\includegraphics[width=0.26\textwidth]{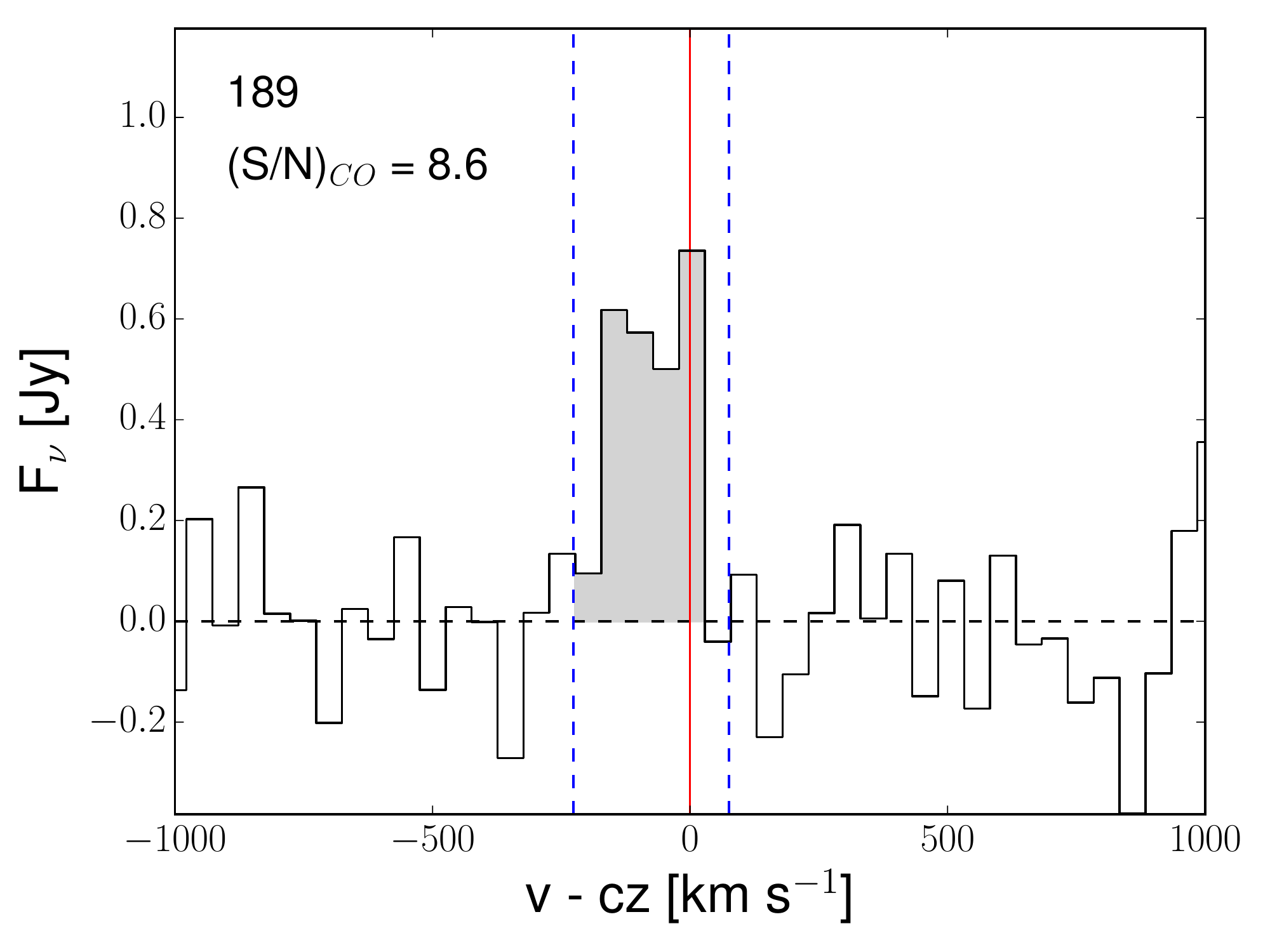}
\includegraphics[width=0.18\textwidth]{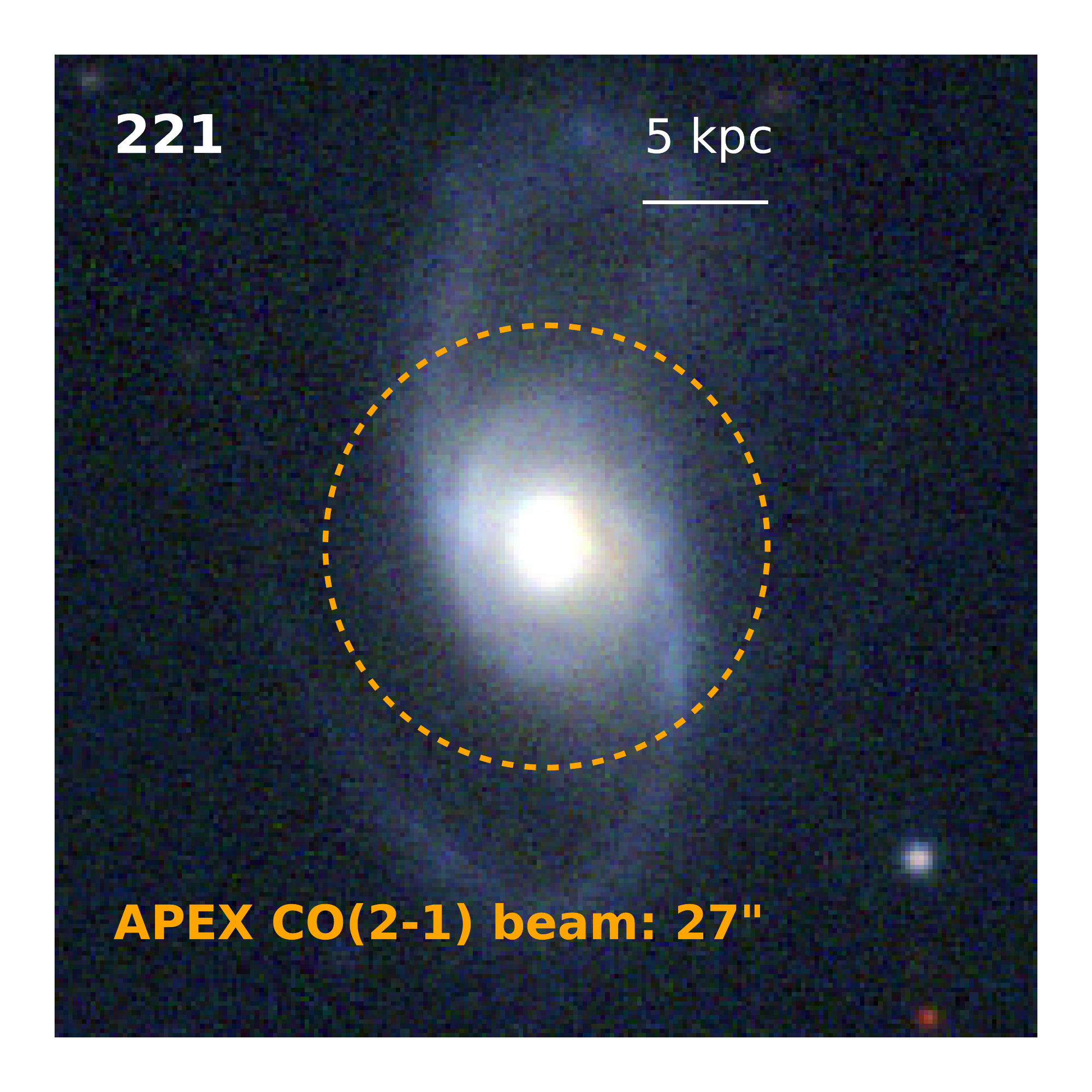}\includegraphics[width=0.26\textwidth]{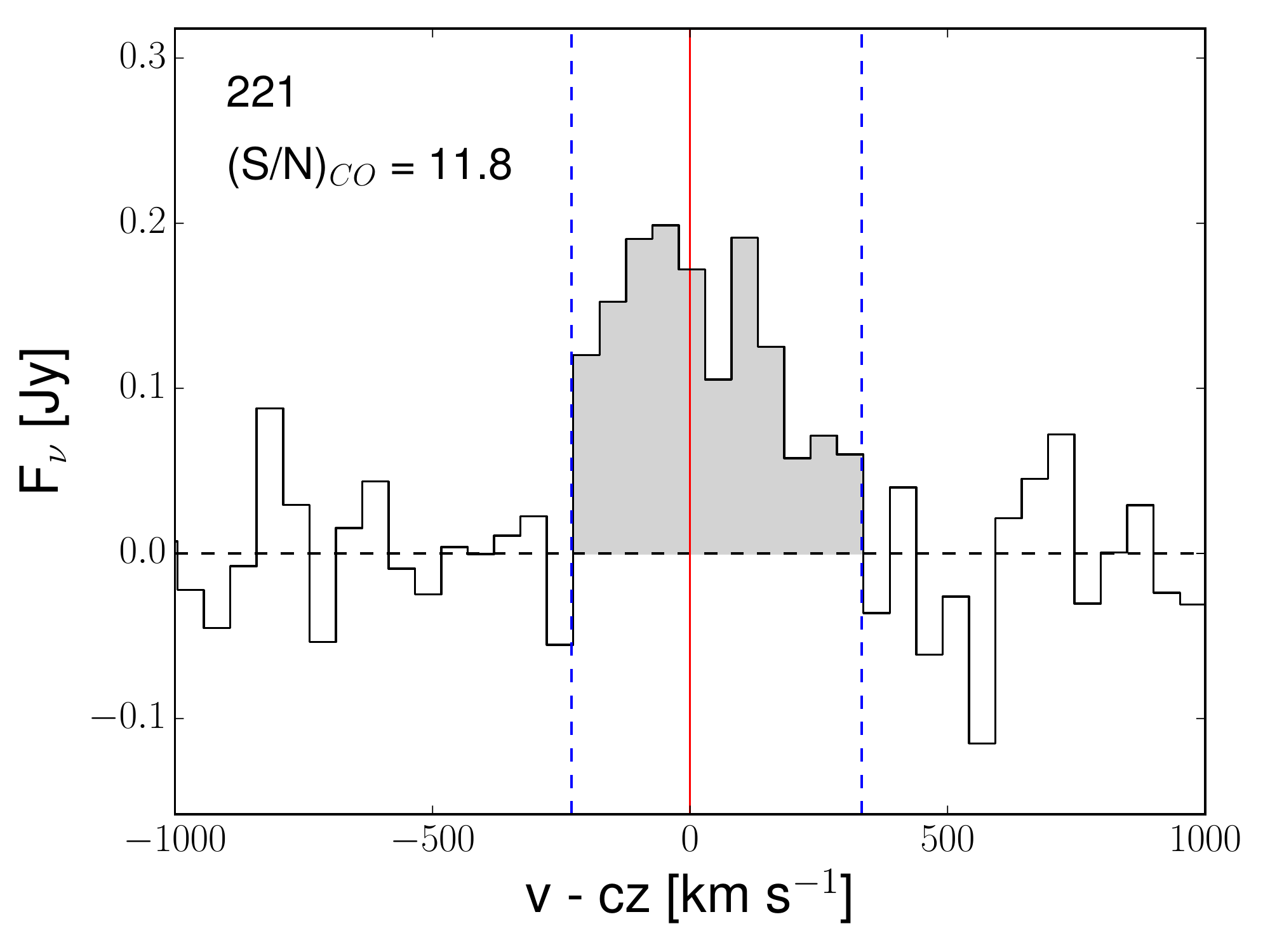}
\includegraphics[width=0.18\textwidth]{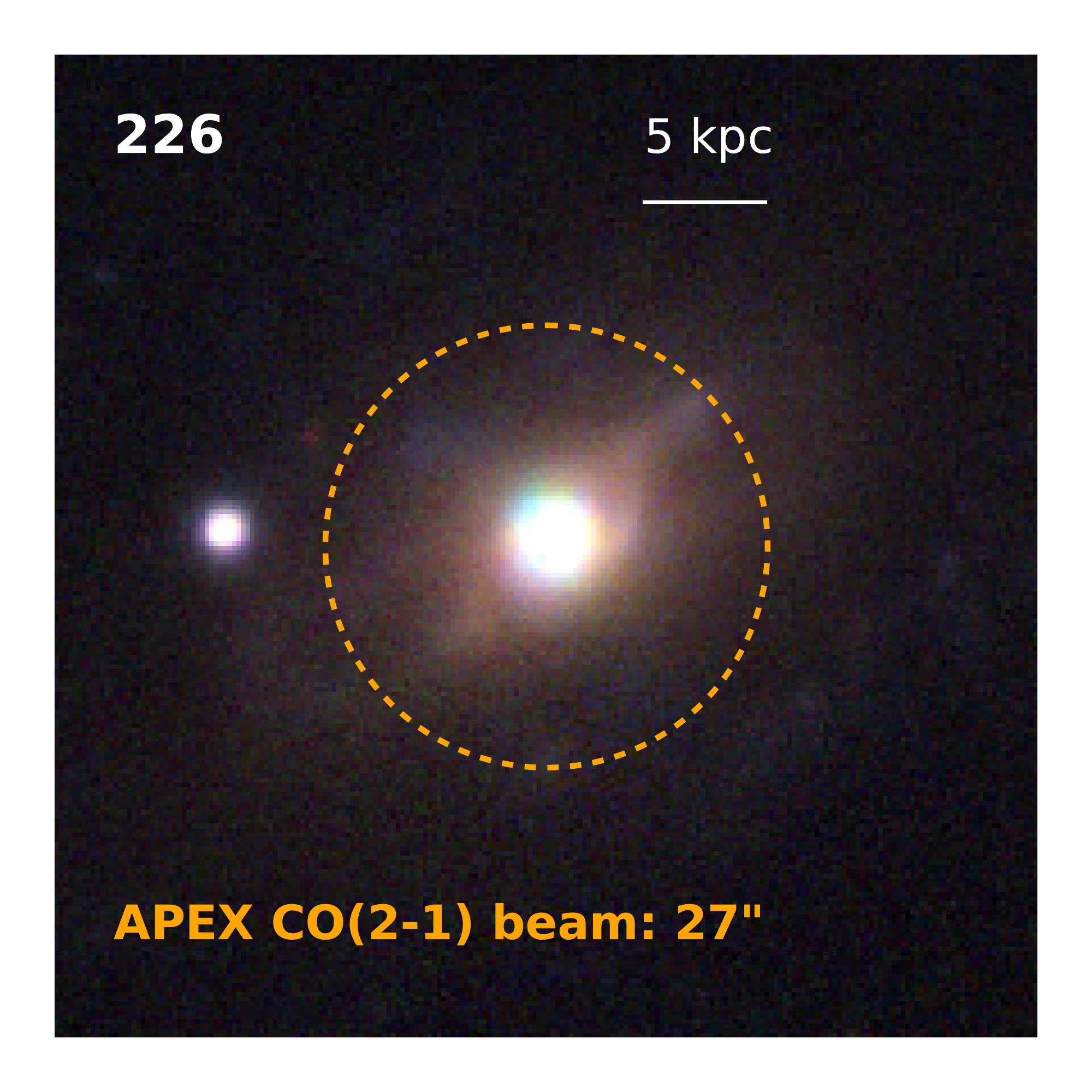}\includegraphics[width=0.26\textwidth]{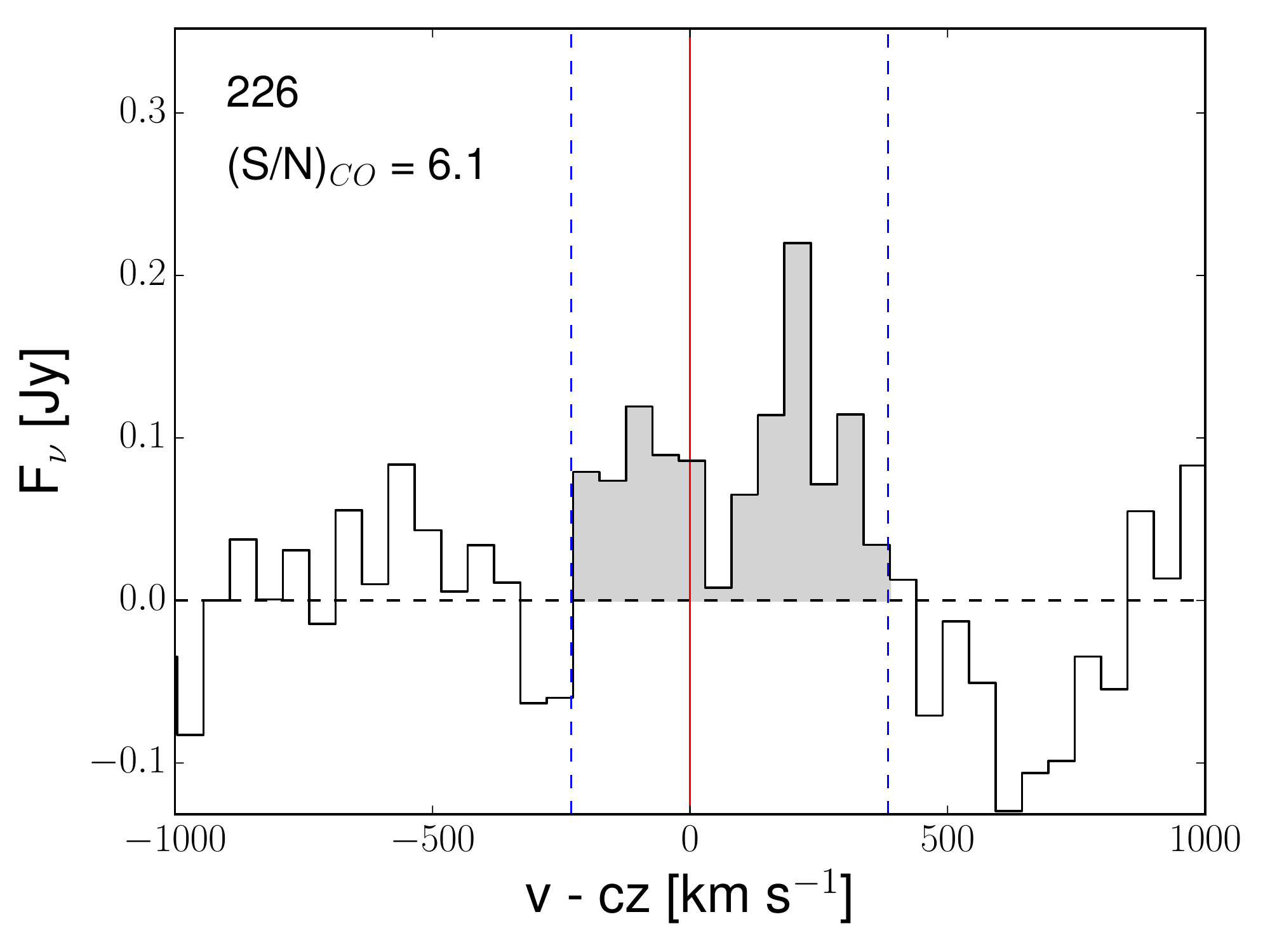}
\includegraphics[width=0.18\textwidth]{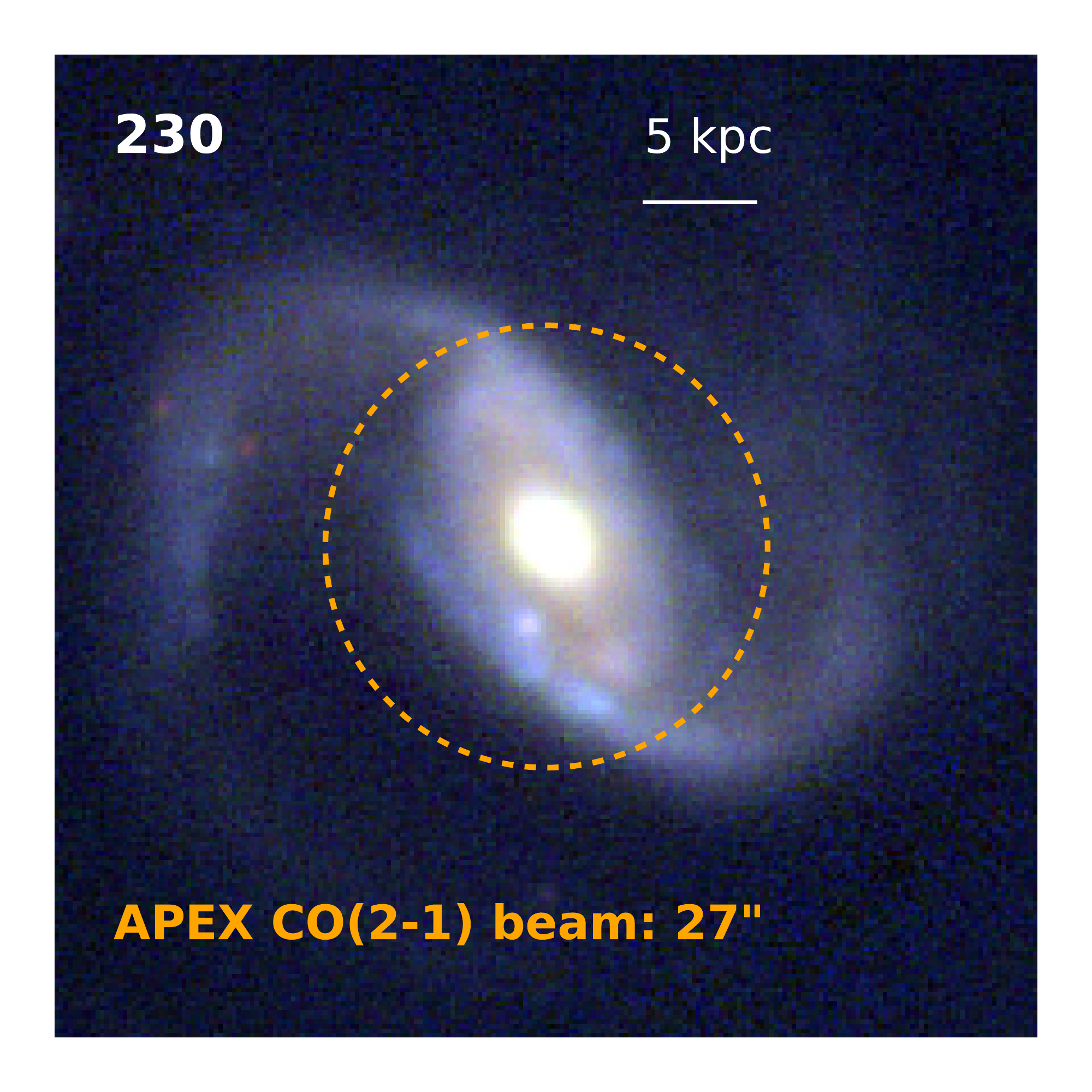}\includegraphics[width=0.26\textwidth]{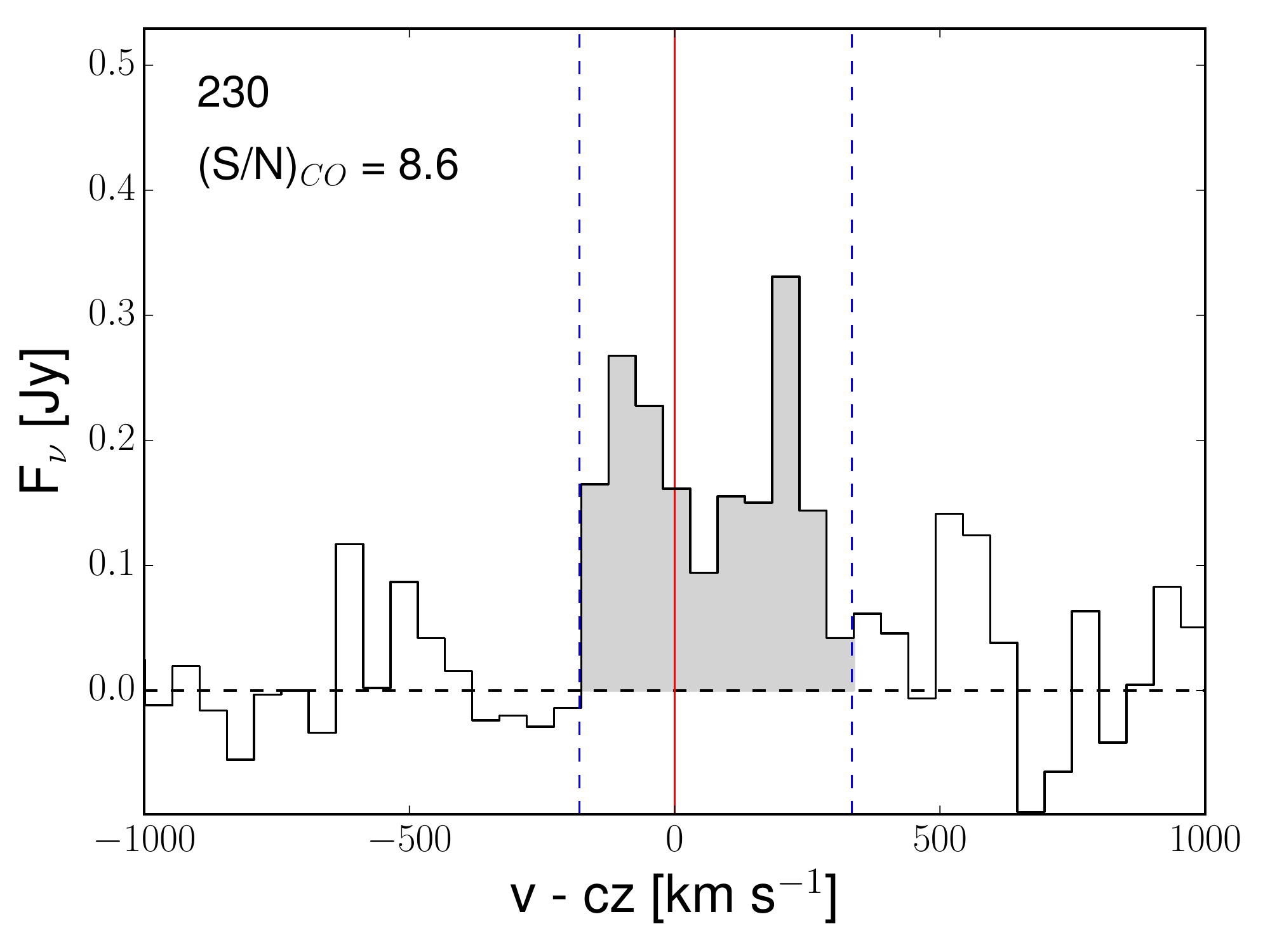}
\includegraphics[width=0.18\textwidth]{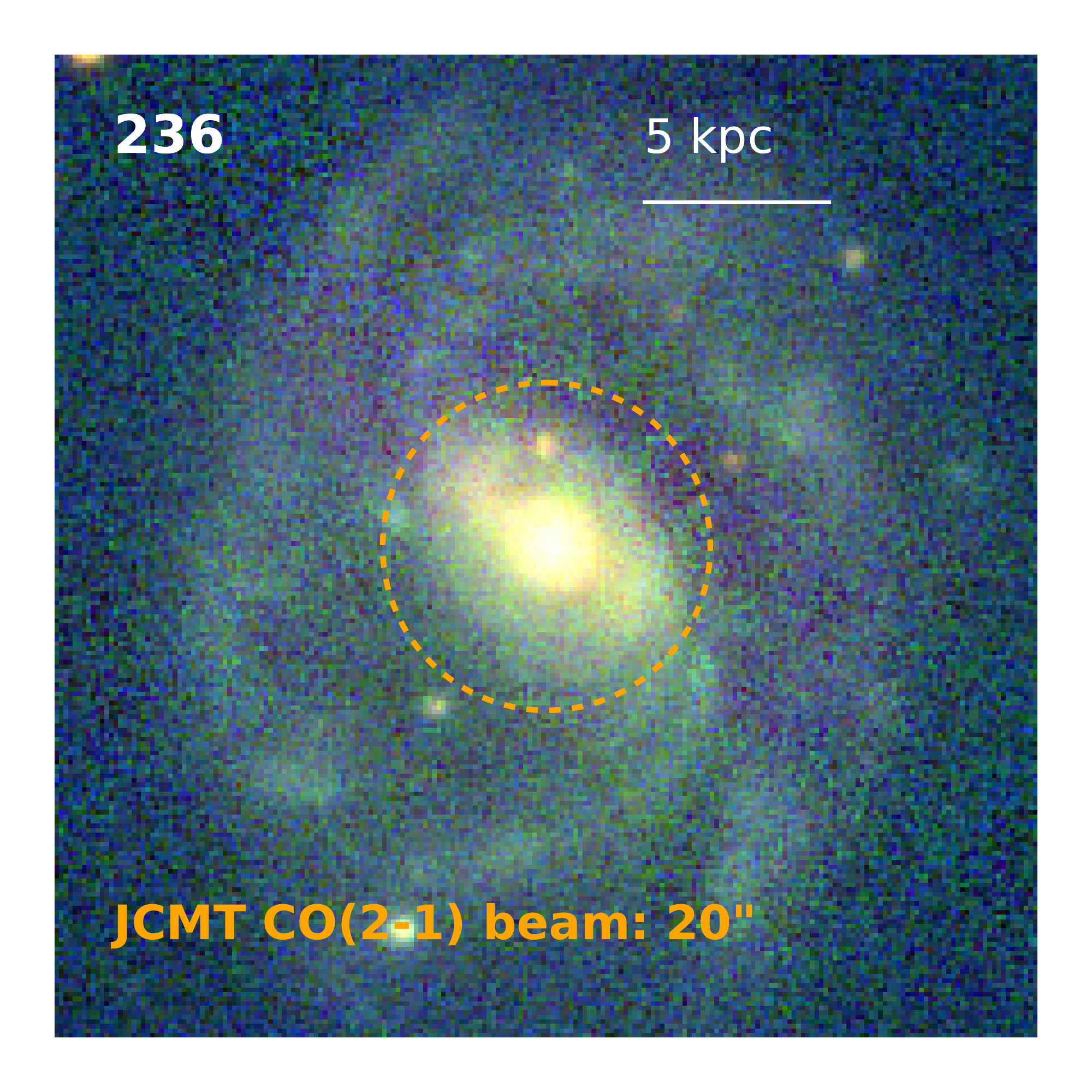}\includegraphics[width=0.26\textwidth]{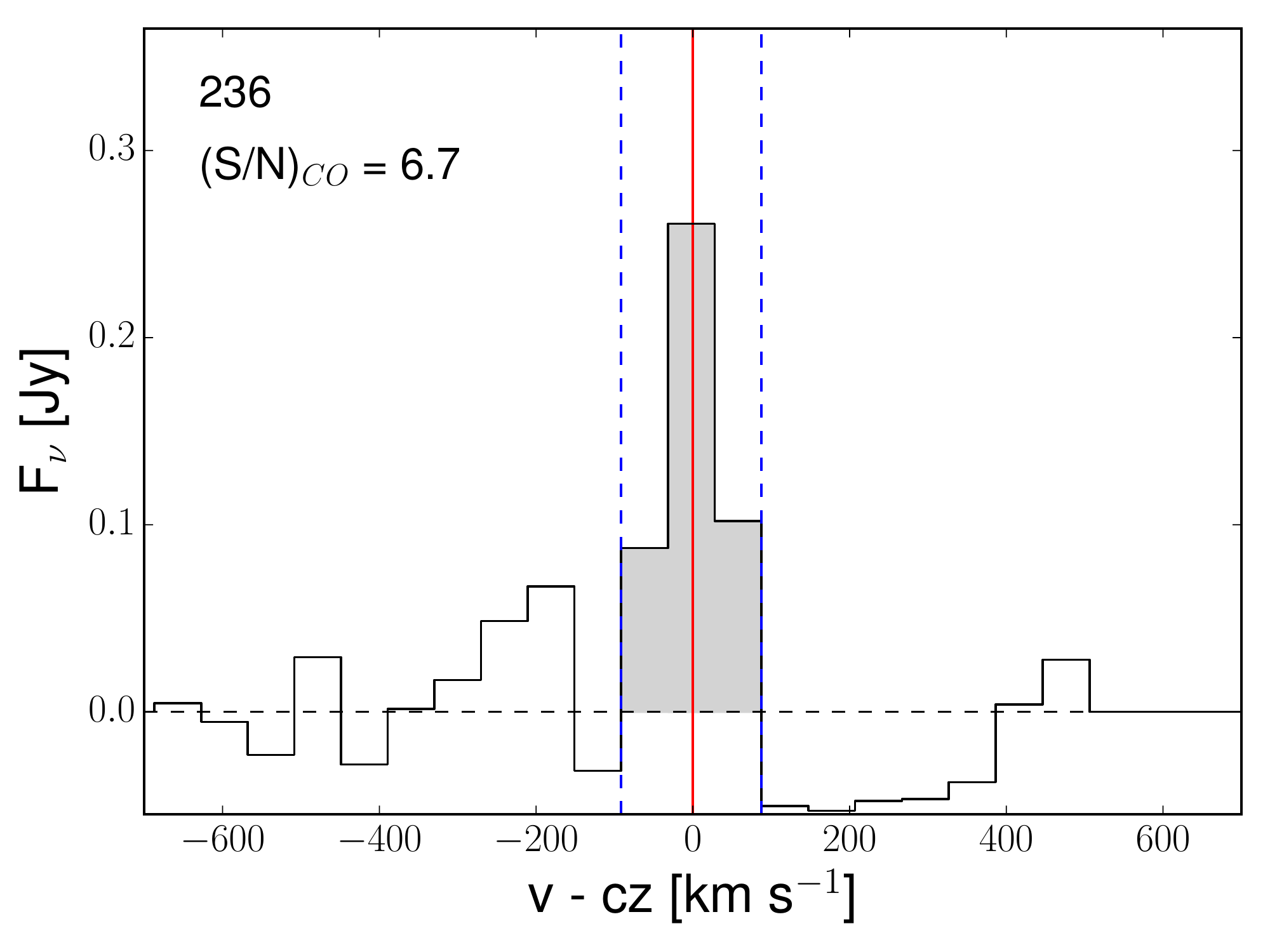}
\includegraphics[width=0.18\textwidth]{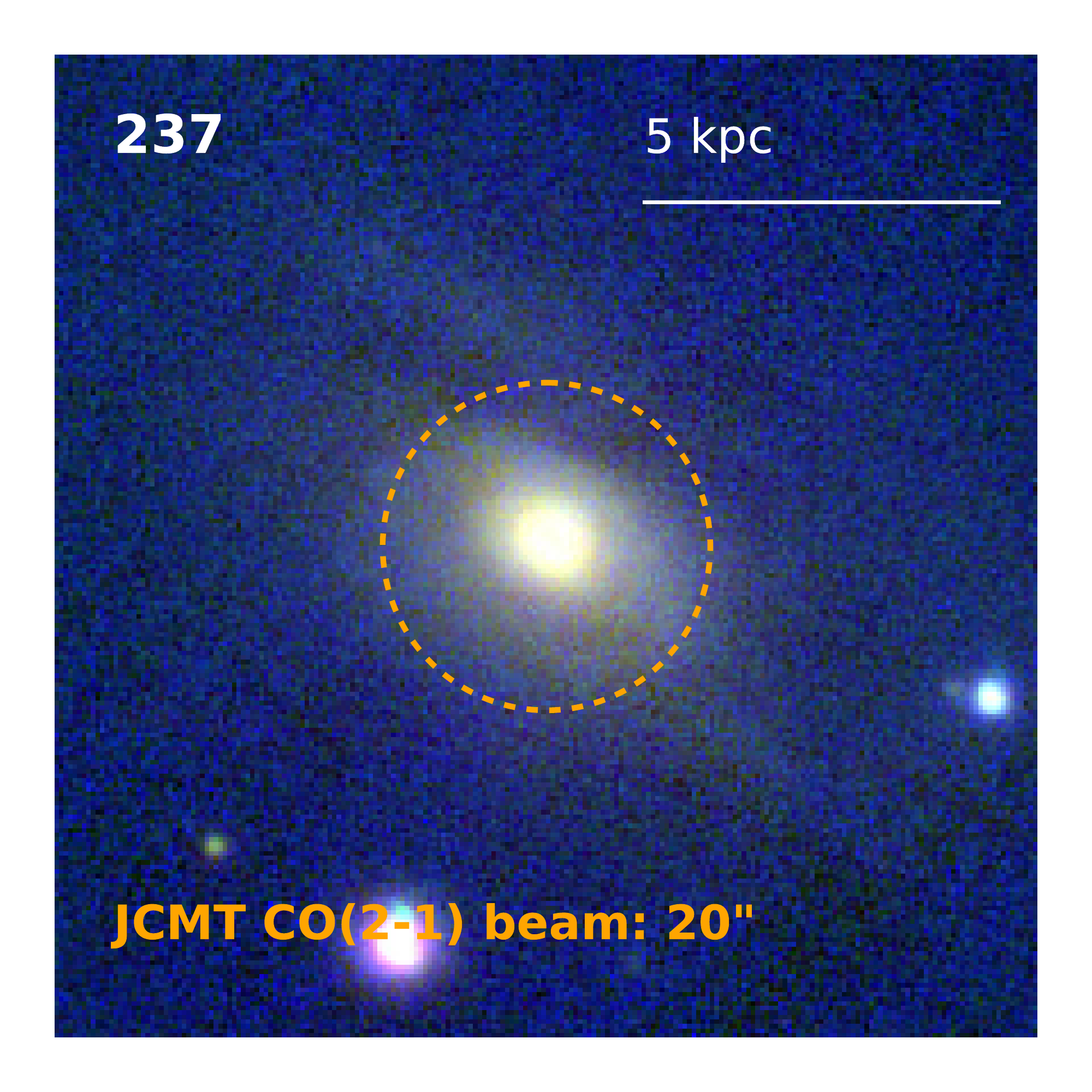}\includegraphics[width=0.26\textwidth]{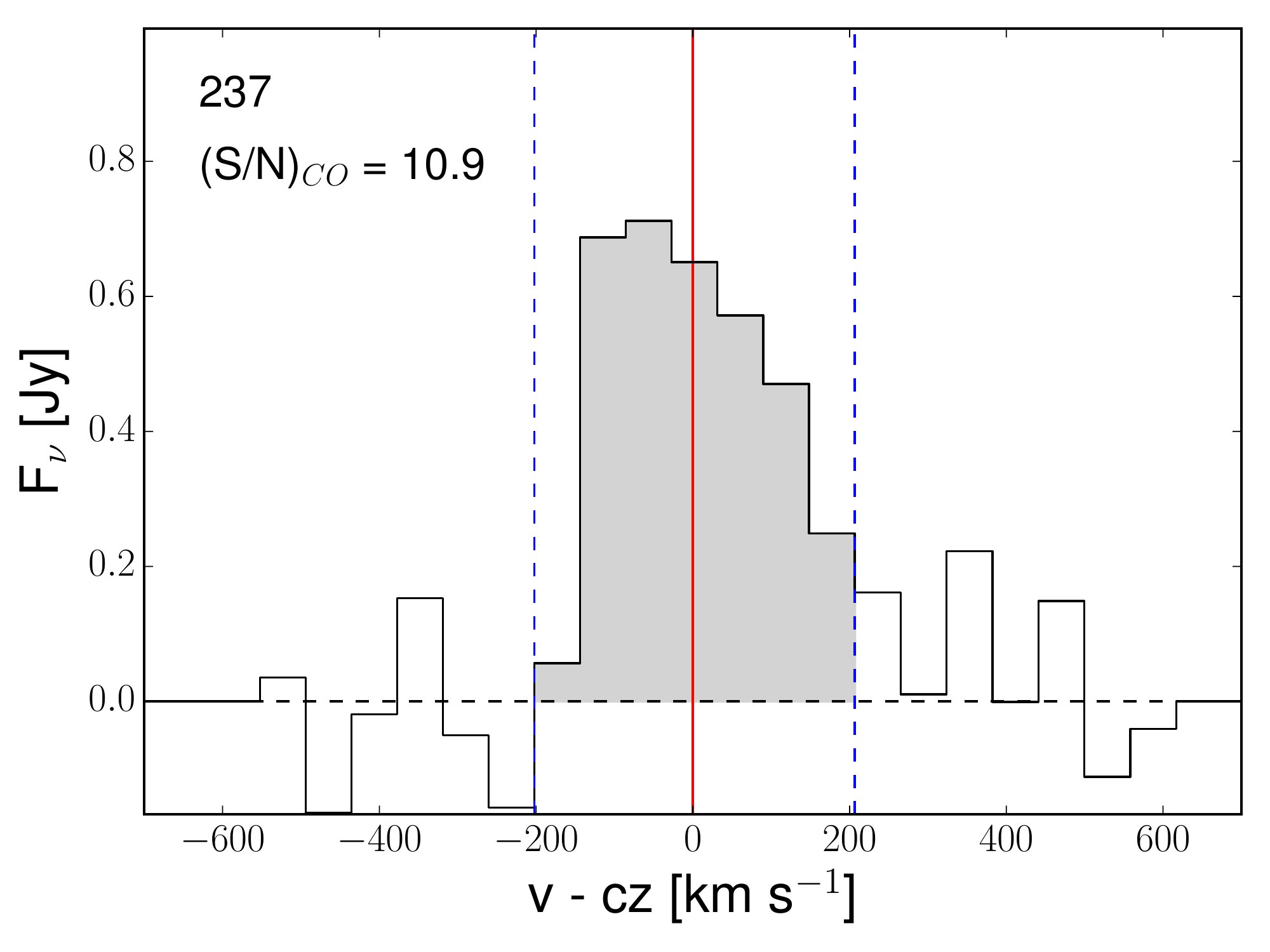}
\caption{continued from Fig.~\ref{fig:CO21_spectra_all_1}
} 
\label{fig:CO21_spectra_all_2}
\end{figure*}

\begin{figure*}
\centering
\raggedright
\includegraphics[width=0.18\textwidth]{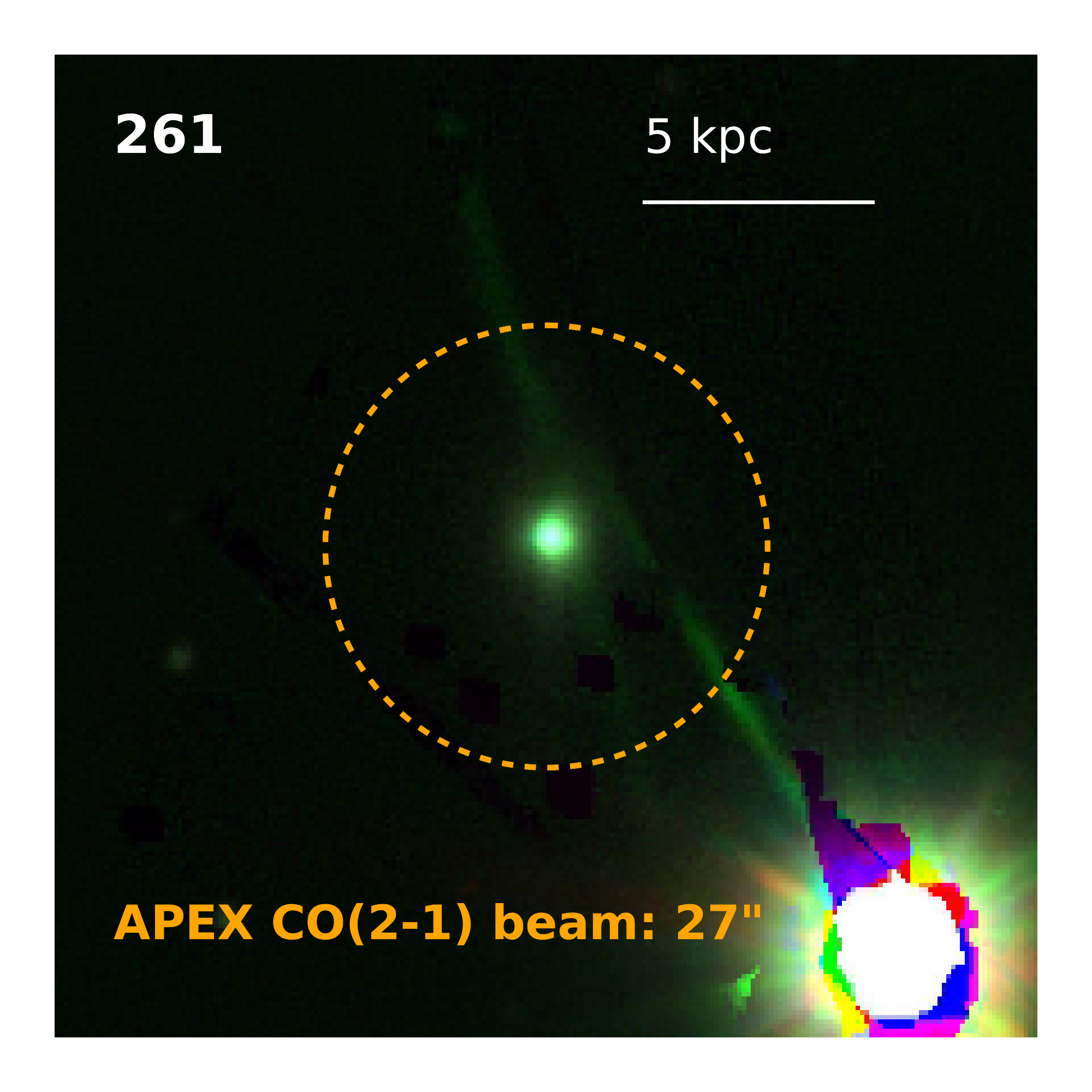}\includegraphics[width=0.26\textwidth]{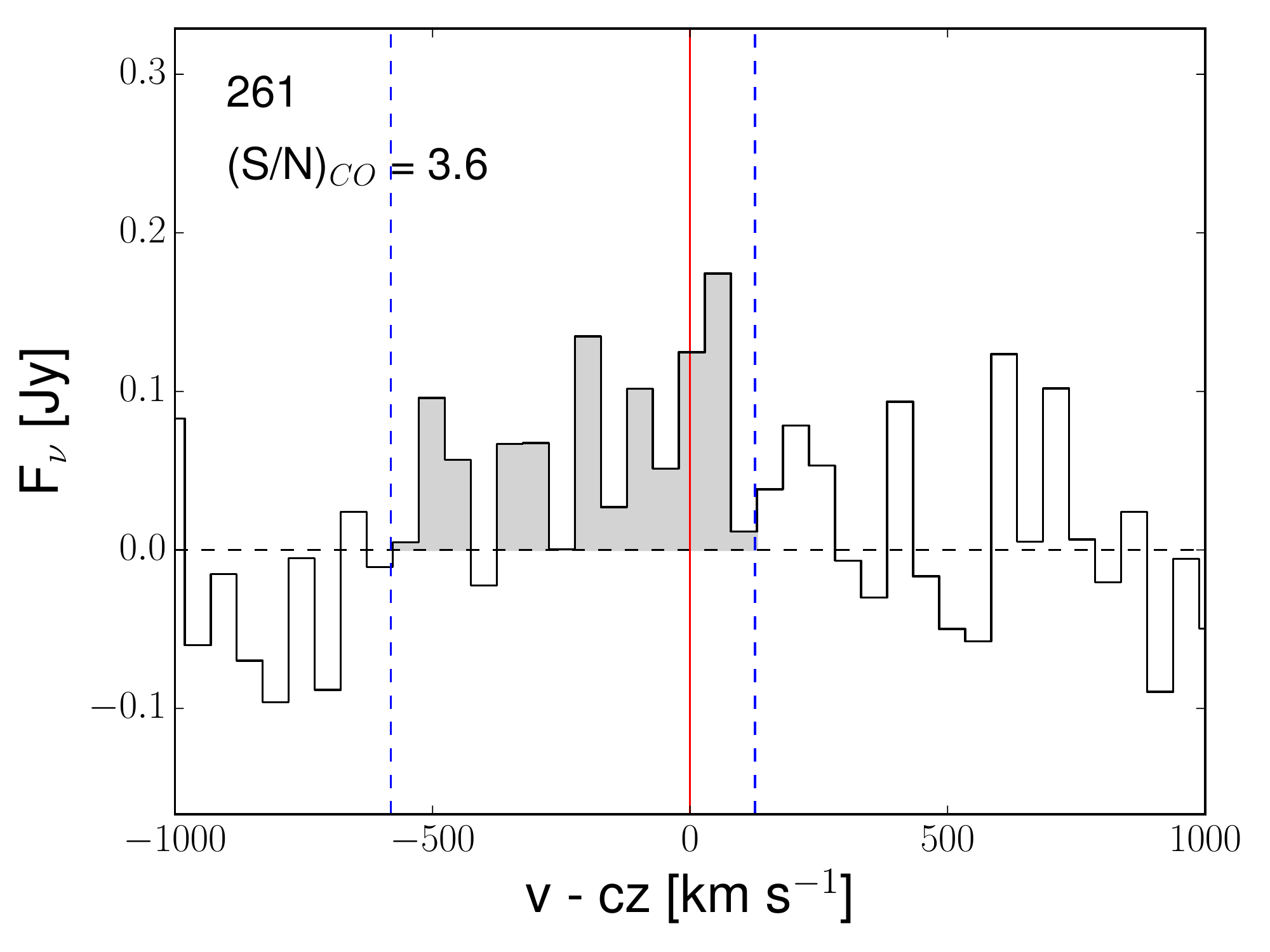}
\includegraphics[width=0.18\textwidth]{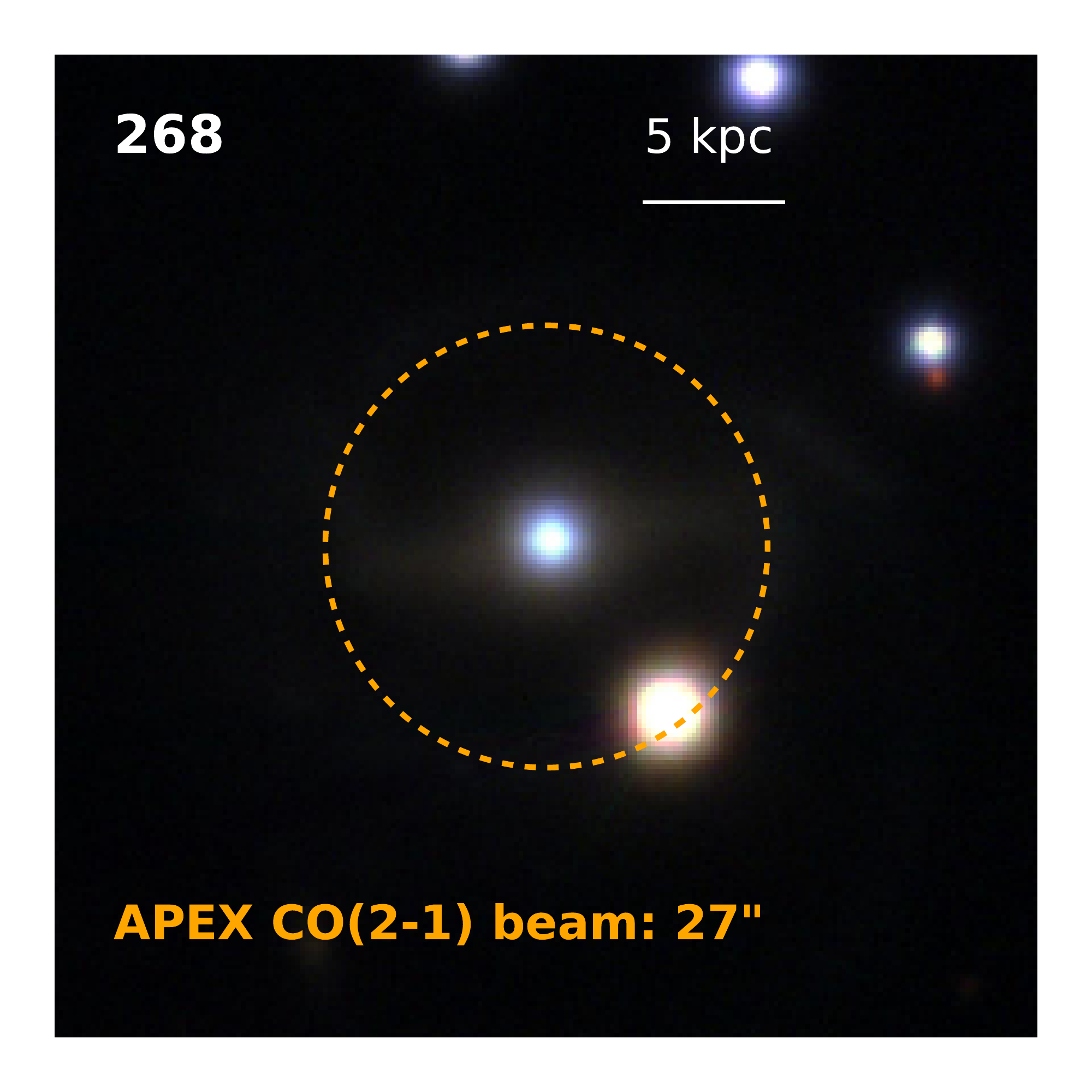}\includegraphics[width=0.26\textwidth]{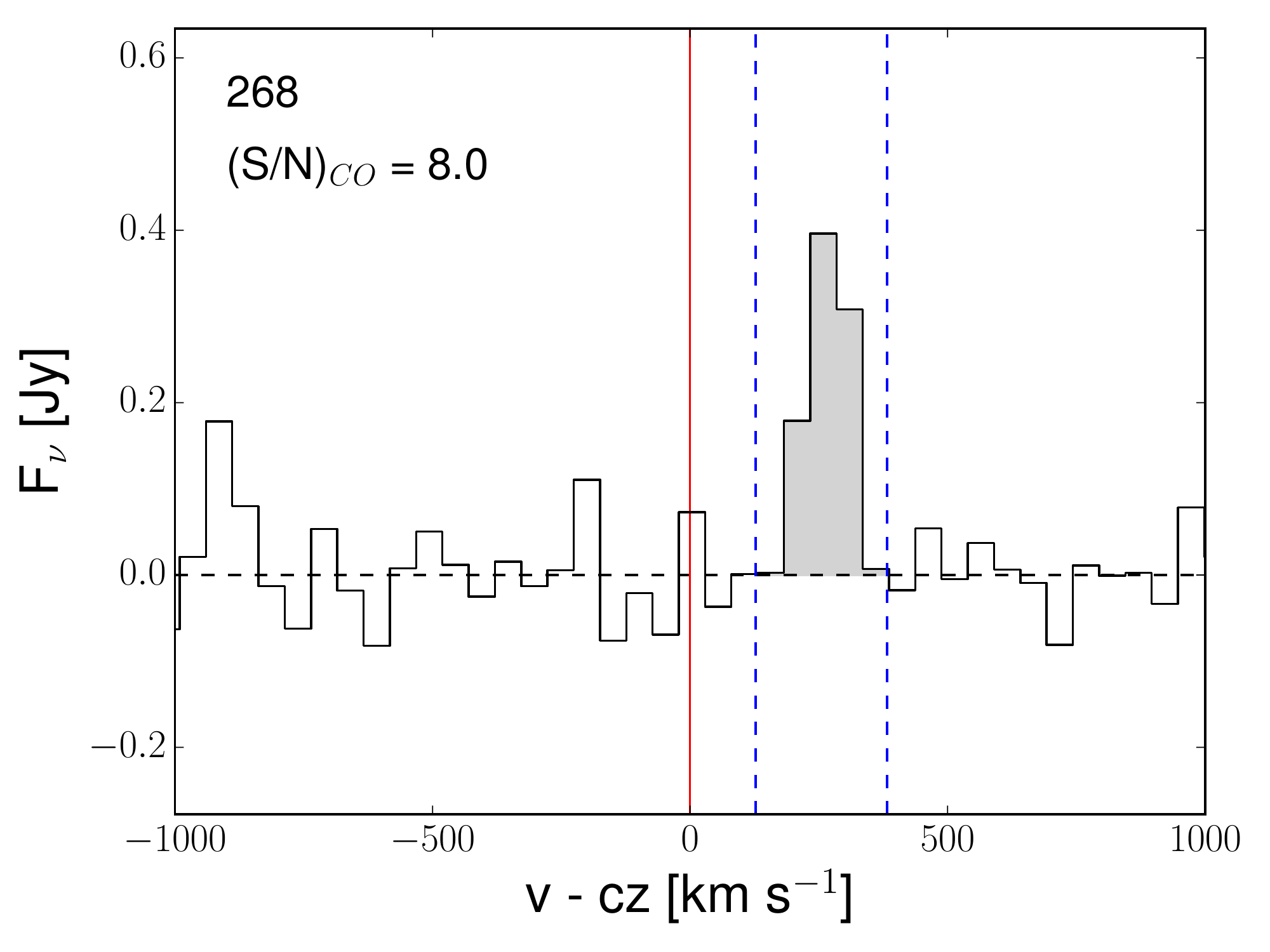}
\includegraphics[width=0.18\textwidth]{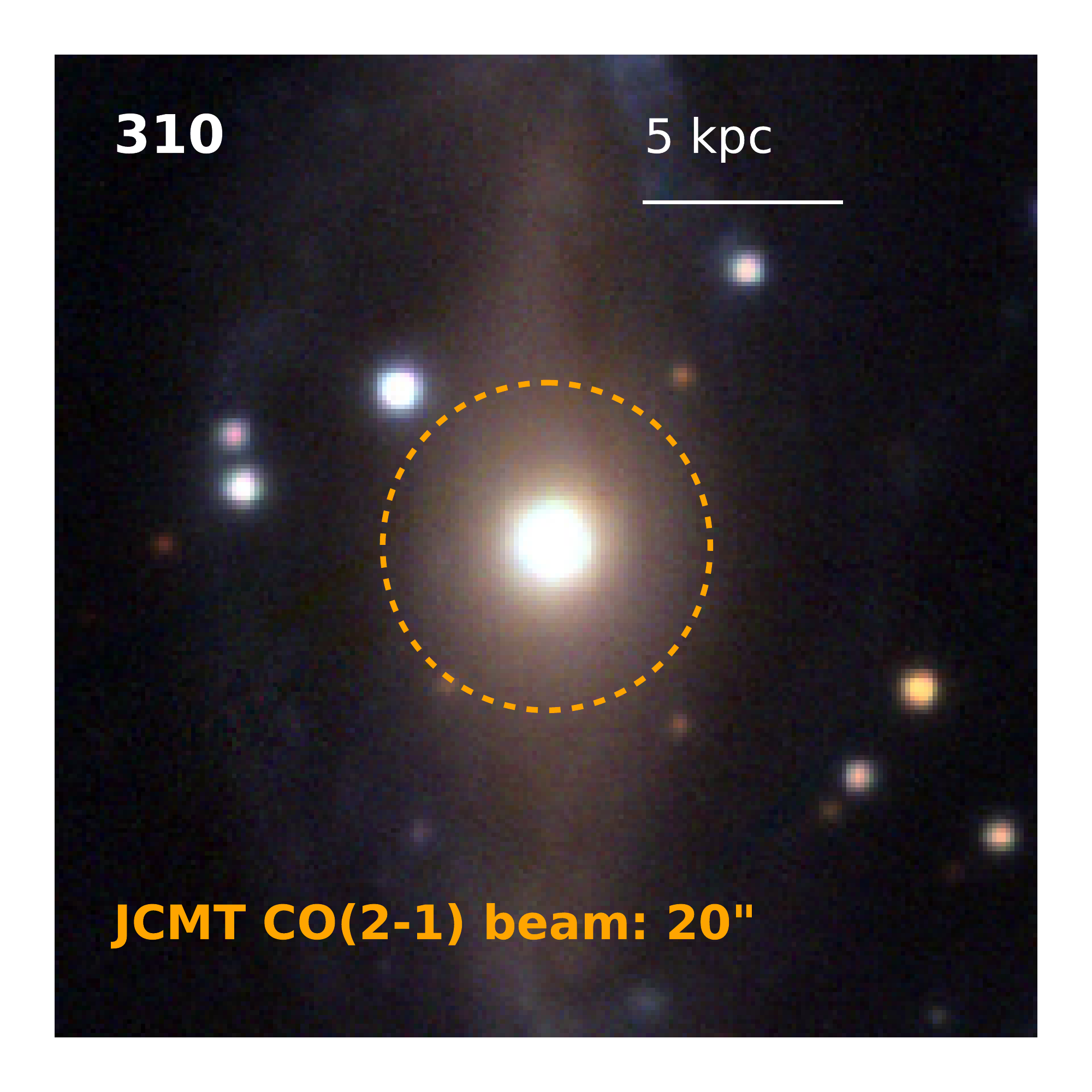}\includegraphics[width=0.26\textwidth]{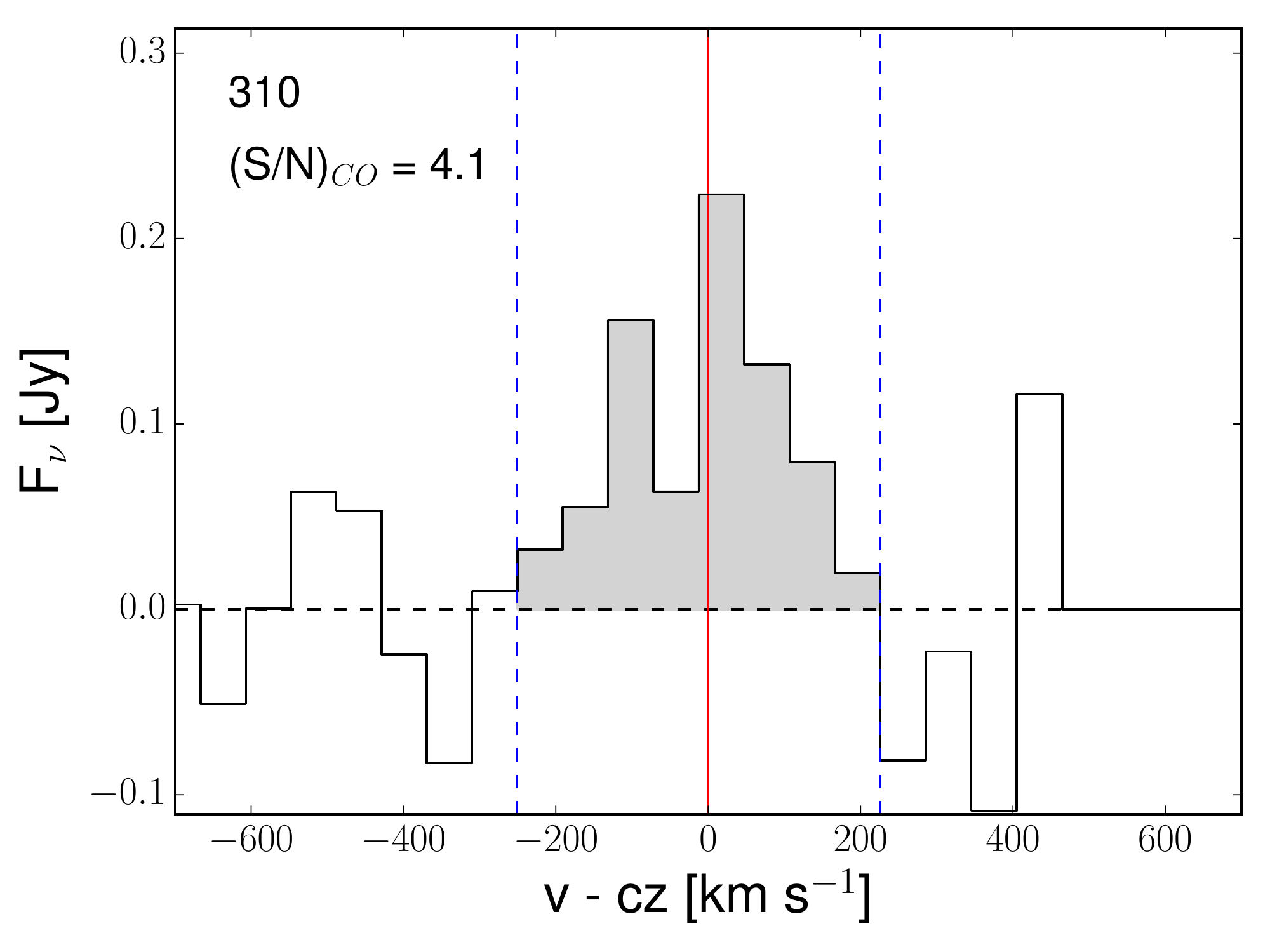}
\includegraphics[width=0.18\textwidth]{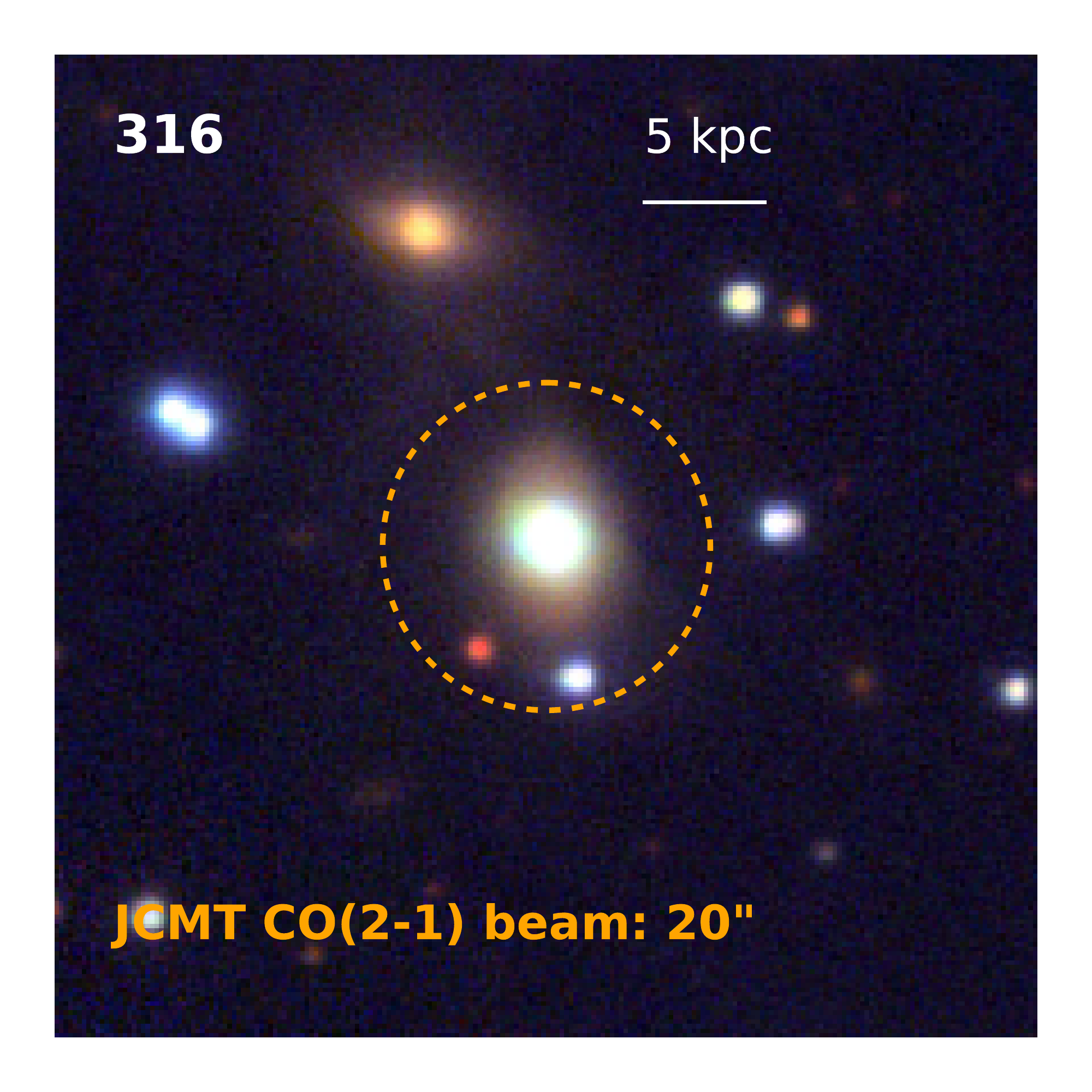}\includegraphics[width=0.26\textwidth]{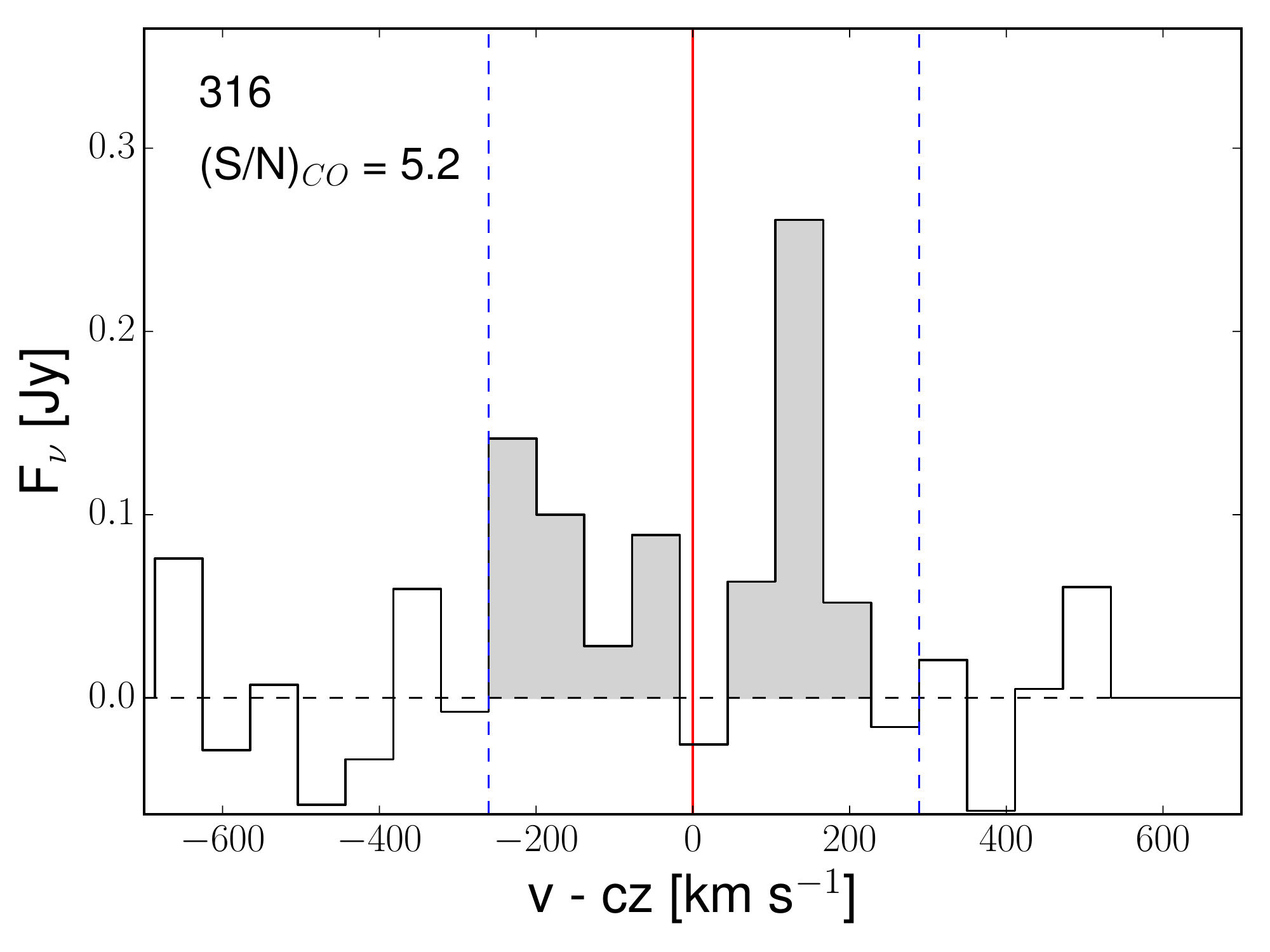}
\includegraphics[width=0.18\textwidth]{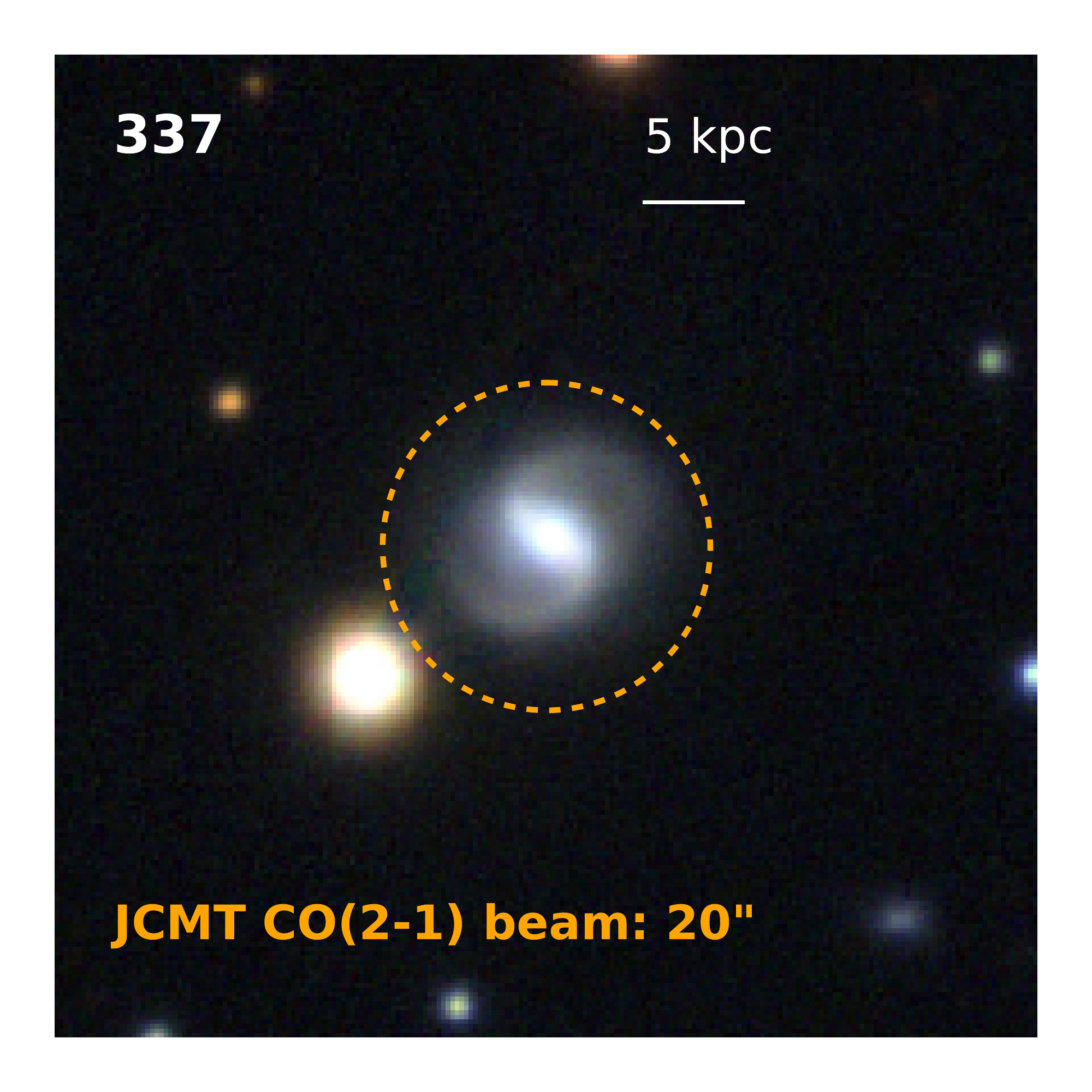}\includegraphics[width=0.26\textwidth]{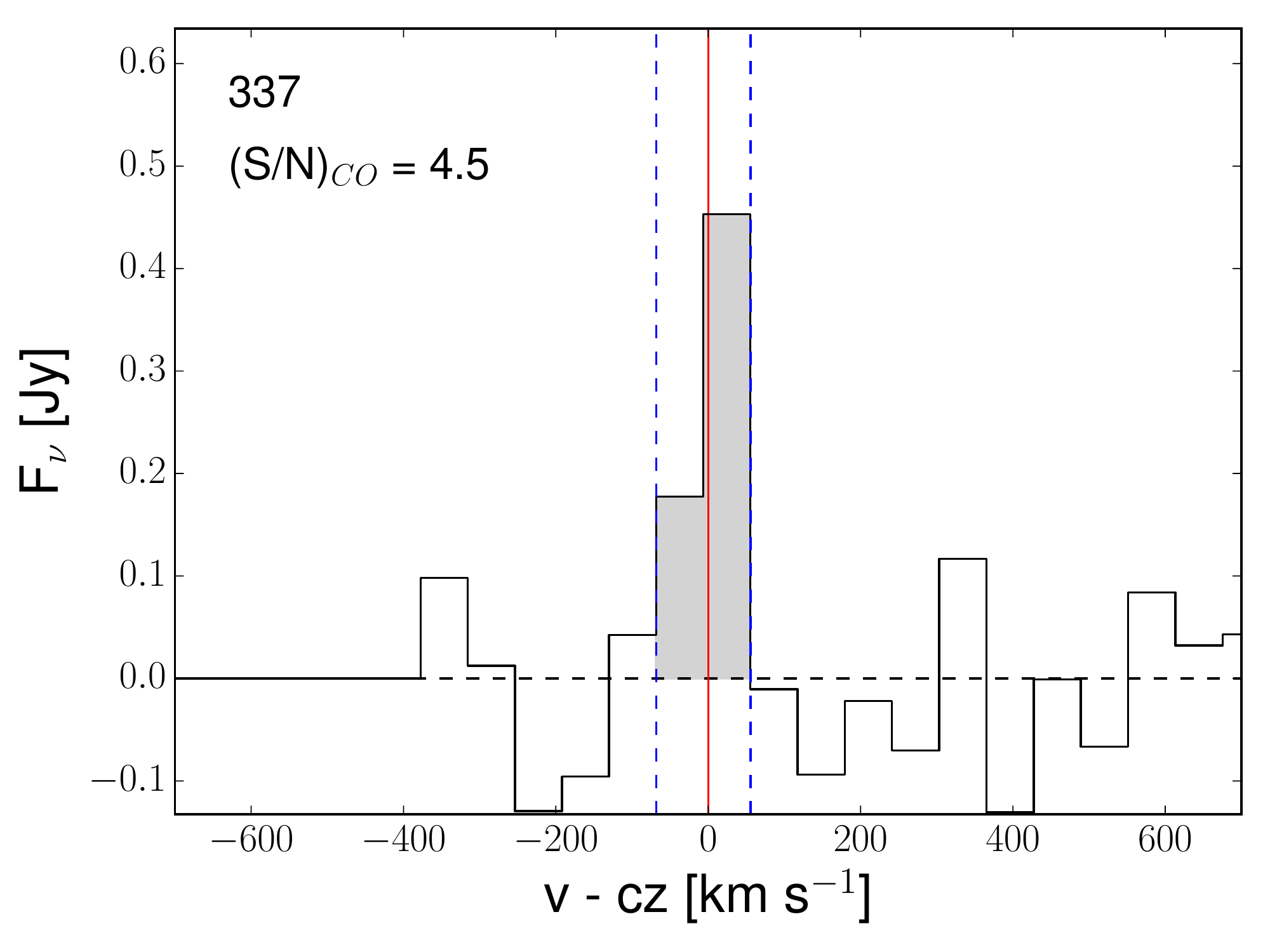}
\includegraphics[width=0.18\textwidth]{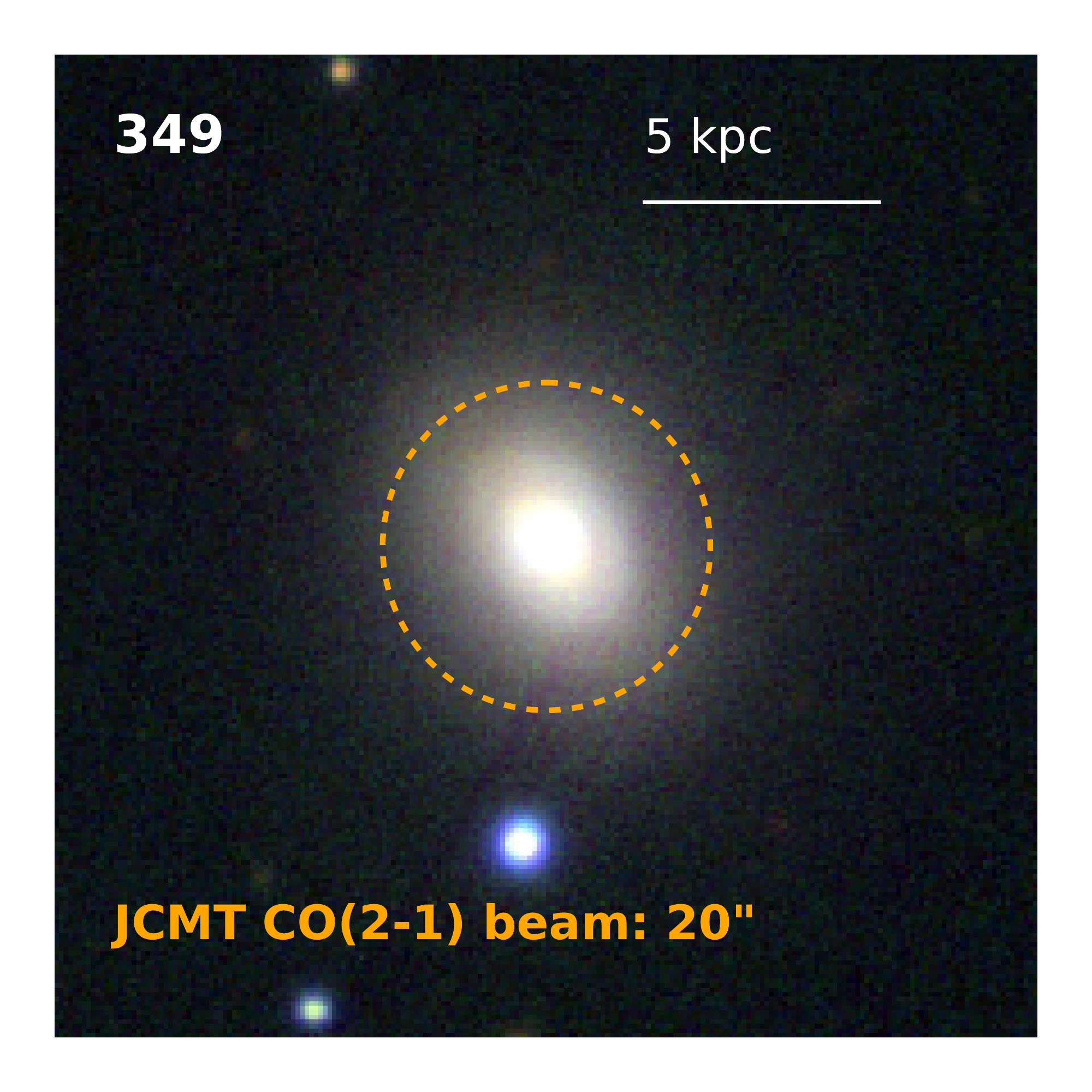}\includegraphics[width=0.26\textwidth]{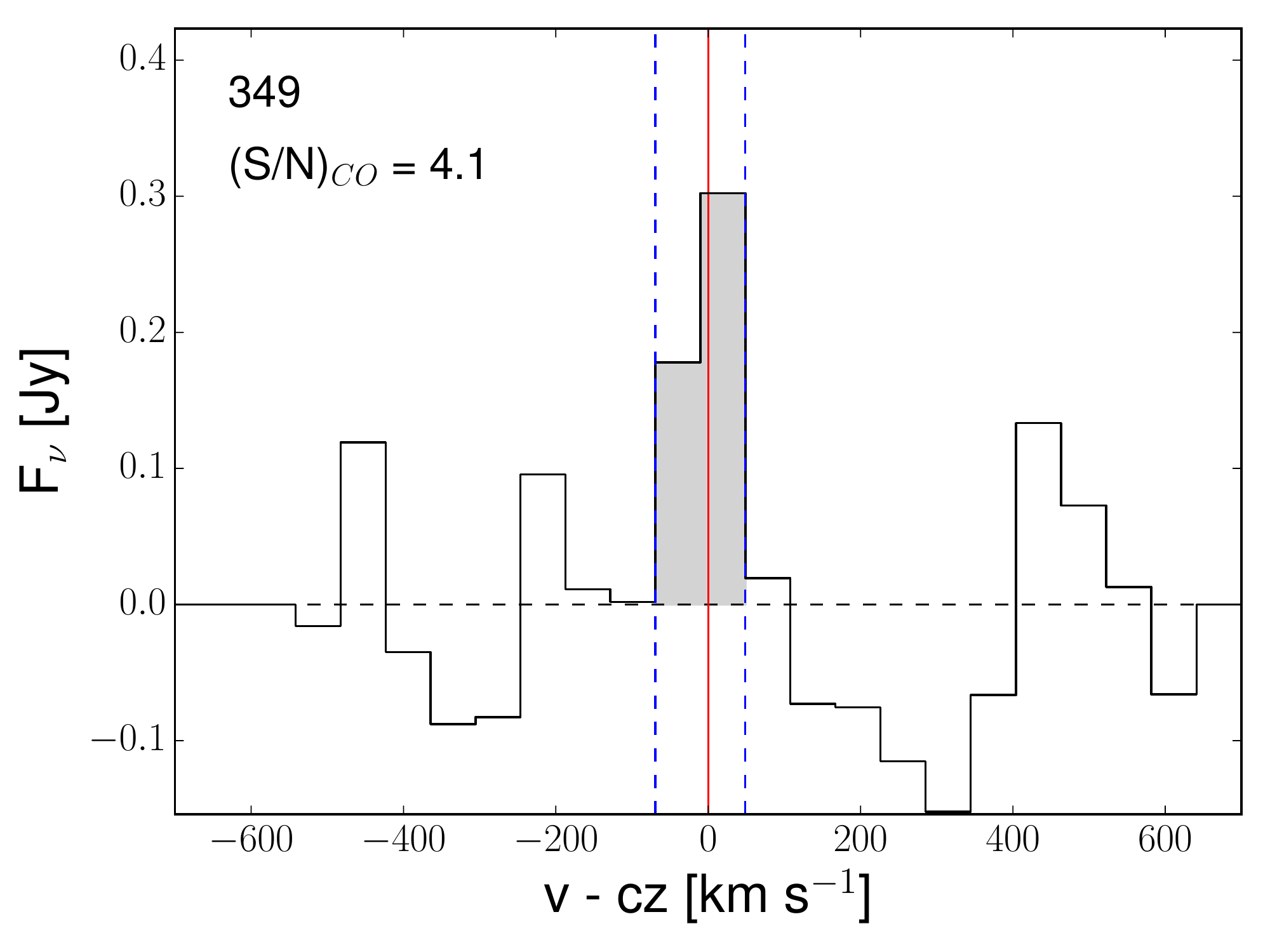}
\includegraphics[width=0.18\textwidth]{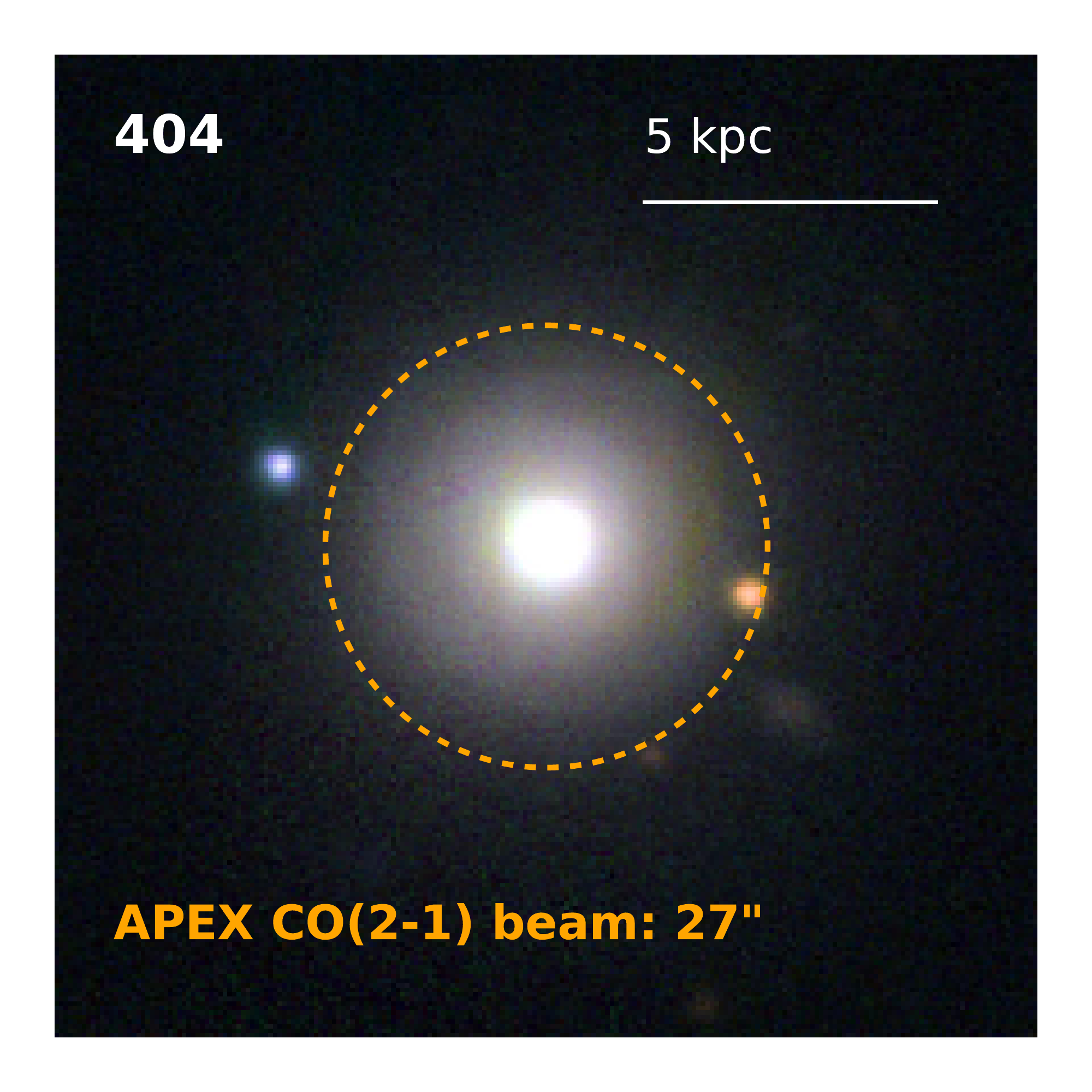}\includegraphics[width=0.26\textwidth]{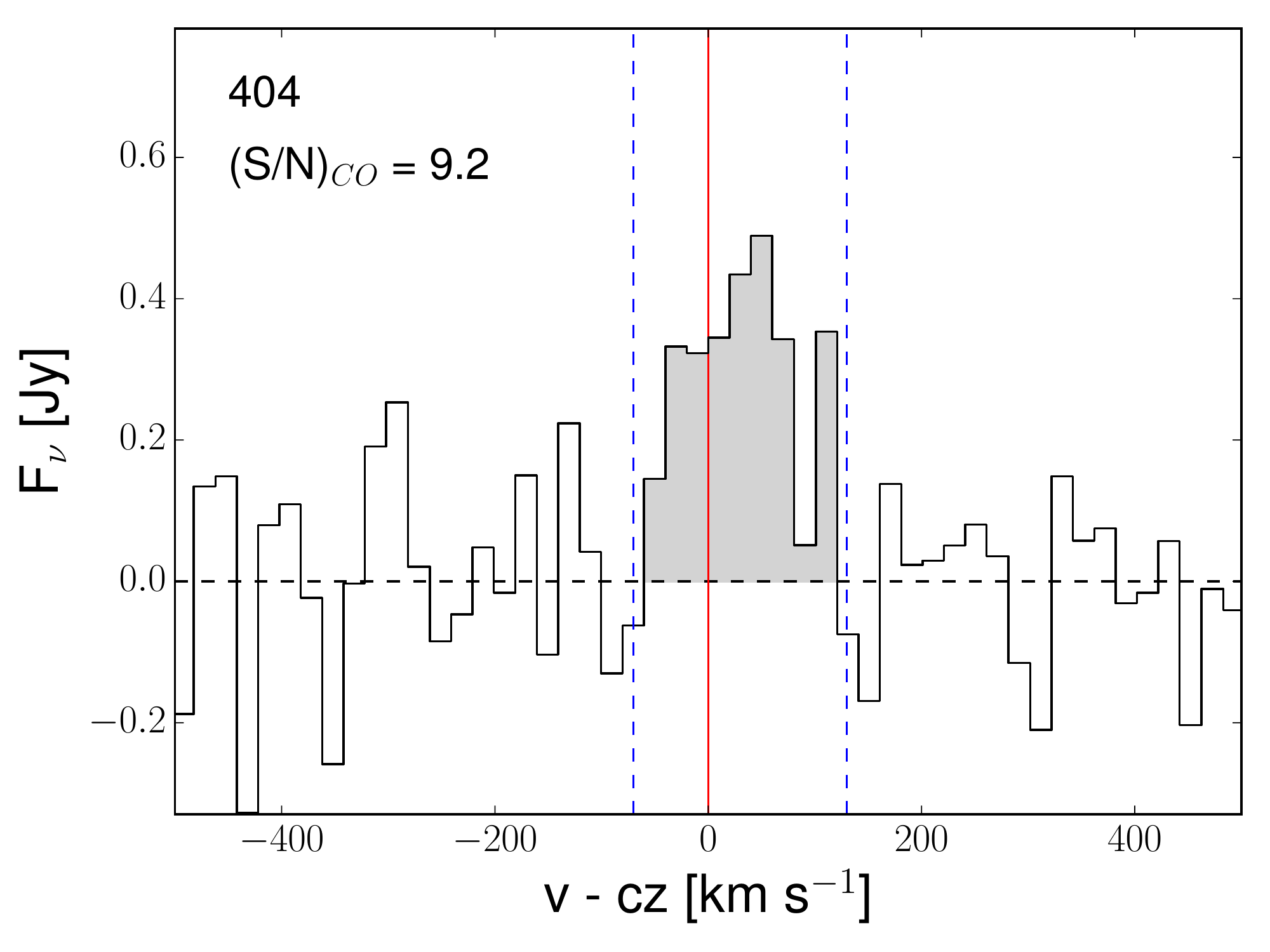}
\includegraphics[width=0.18\textwidth]{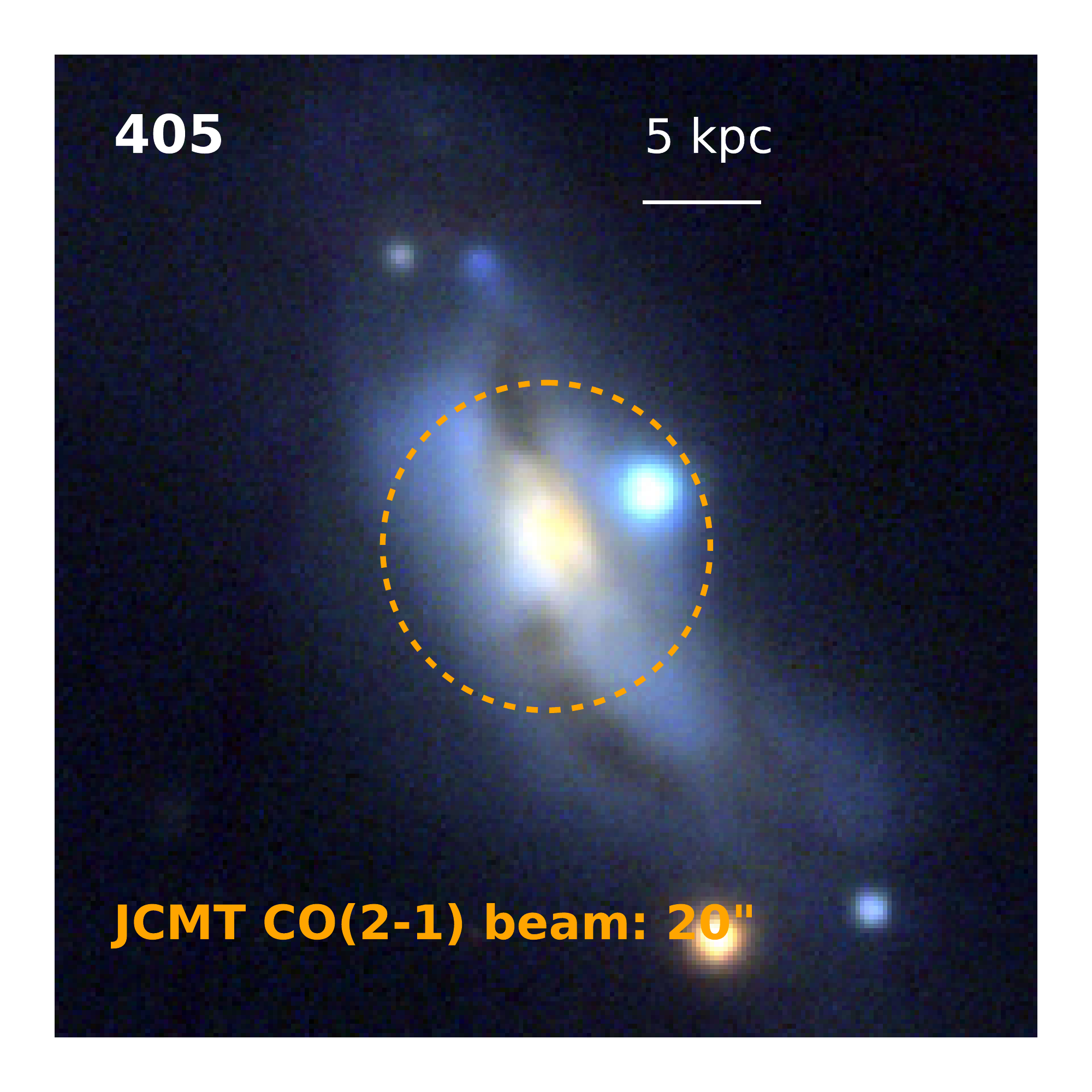}\includegraphics[width=0.26\textwidth]{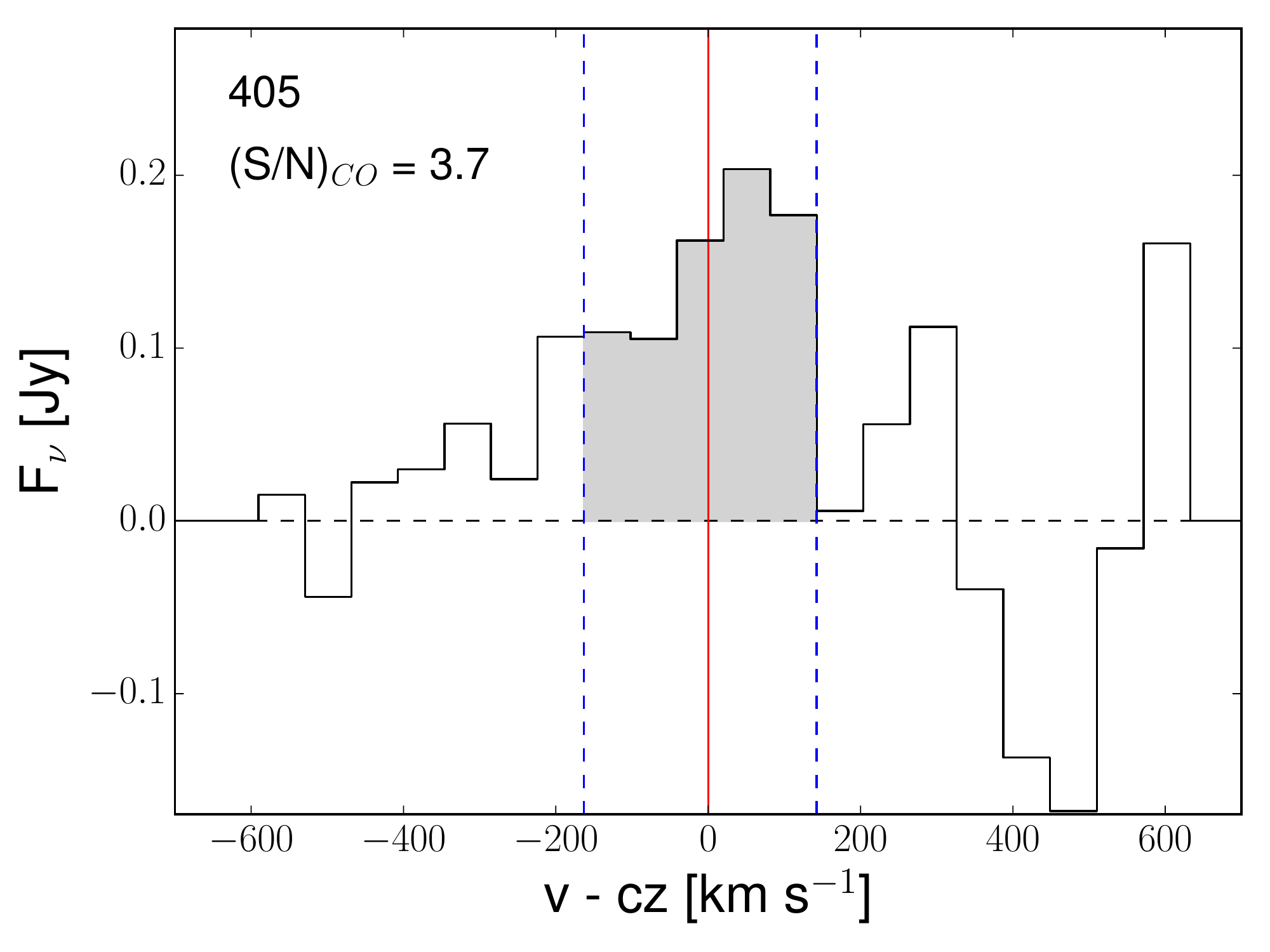}
\includegraphics[width=0.18\textwidth]{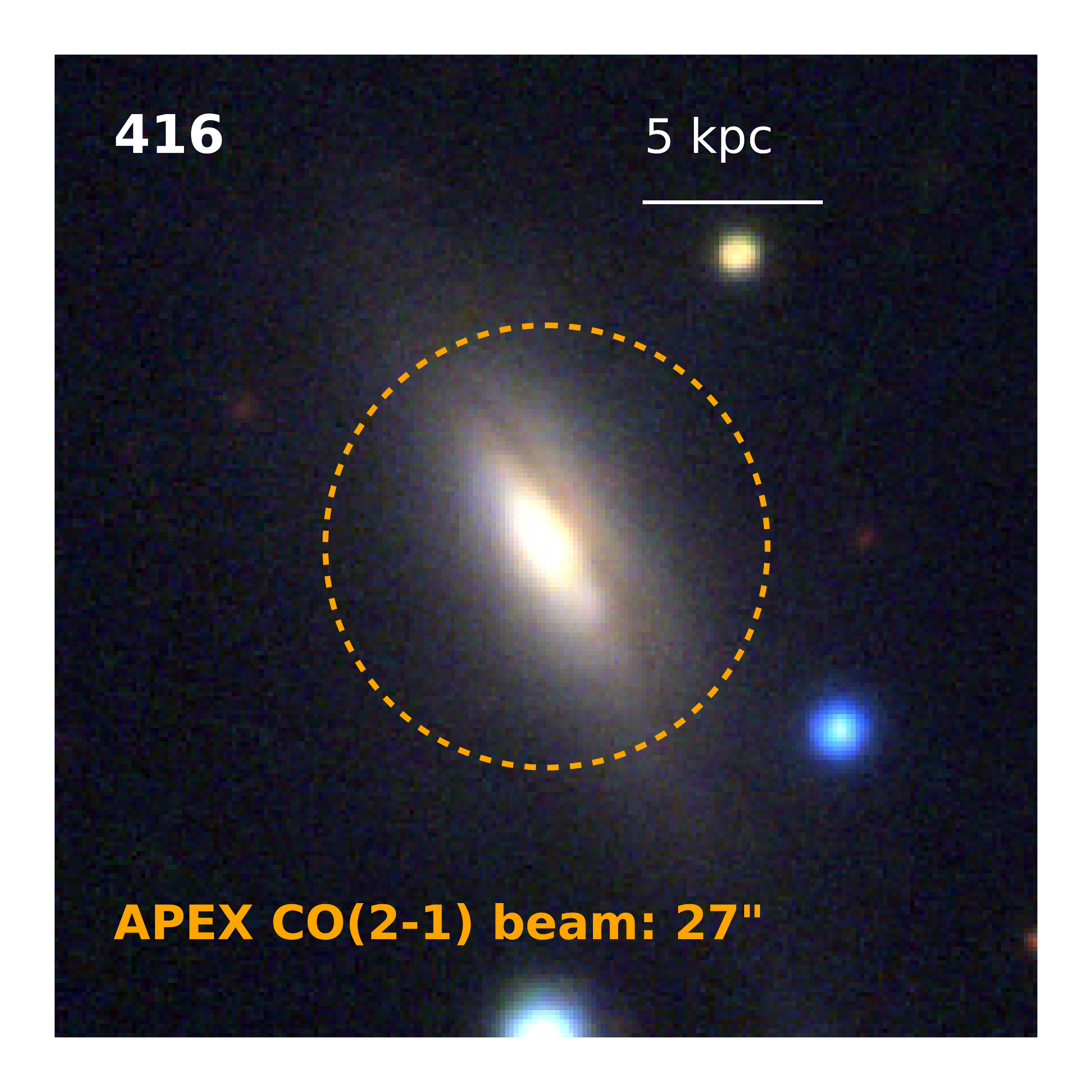}\includegraphics[width=0.26\textwidth]{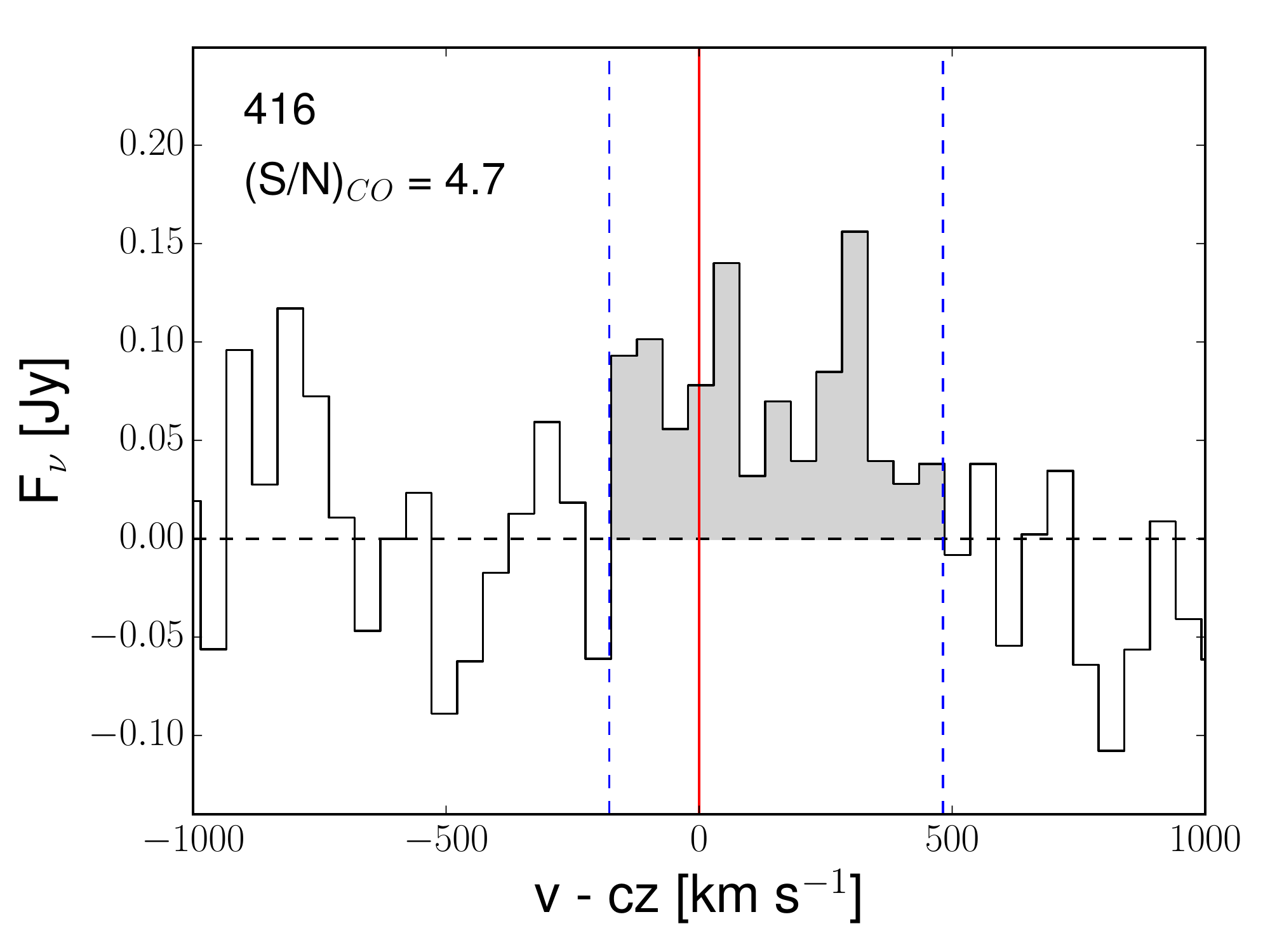}
\includegraphics[width=0.18\textwidth]{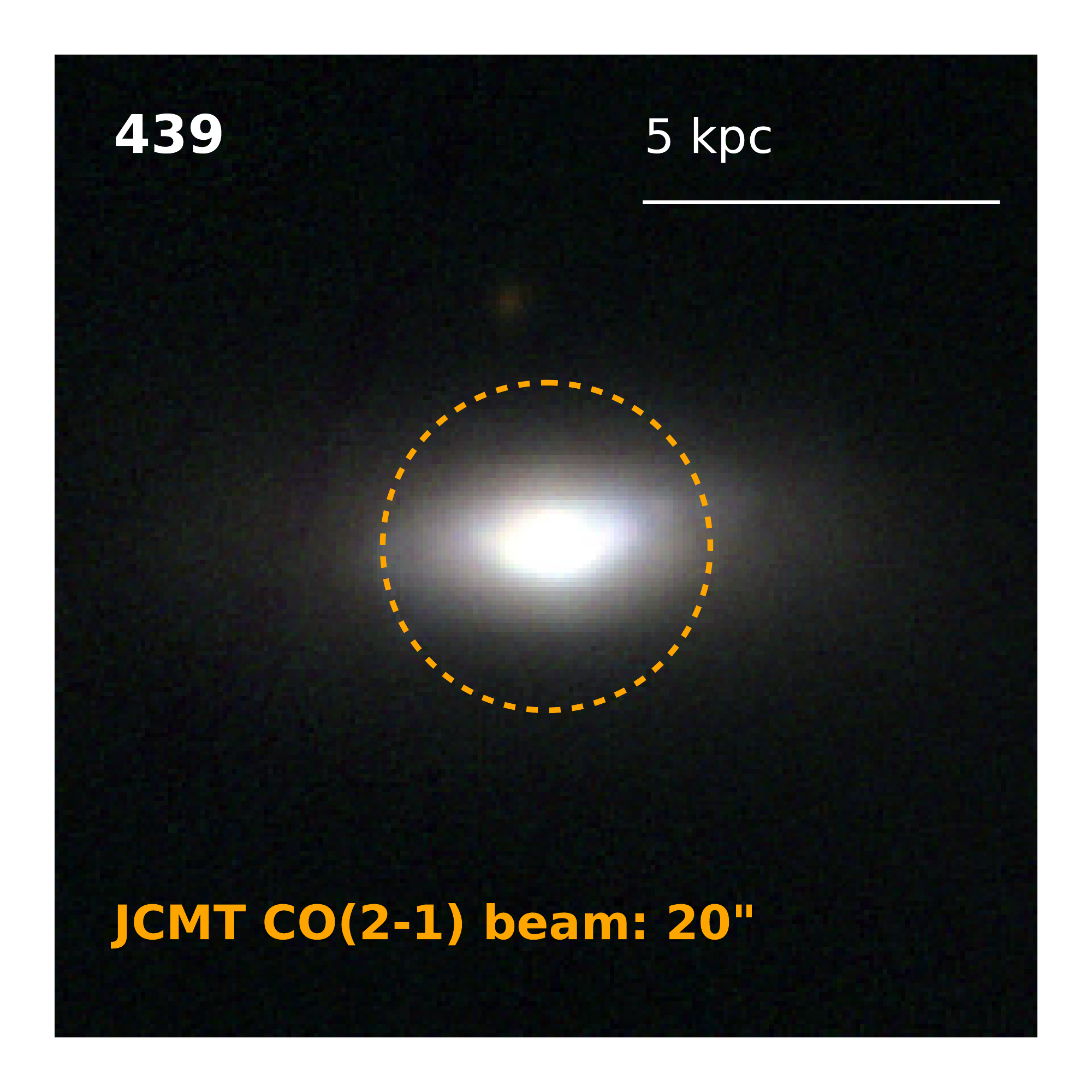}\includegraphics[width=0.26\textwidth]{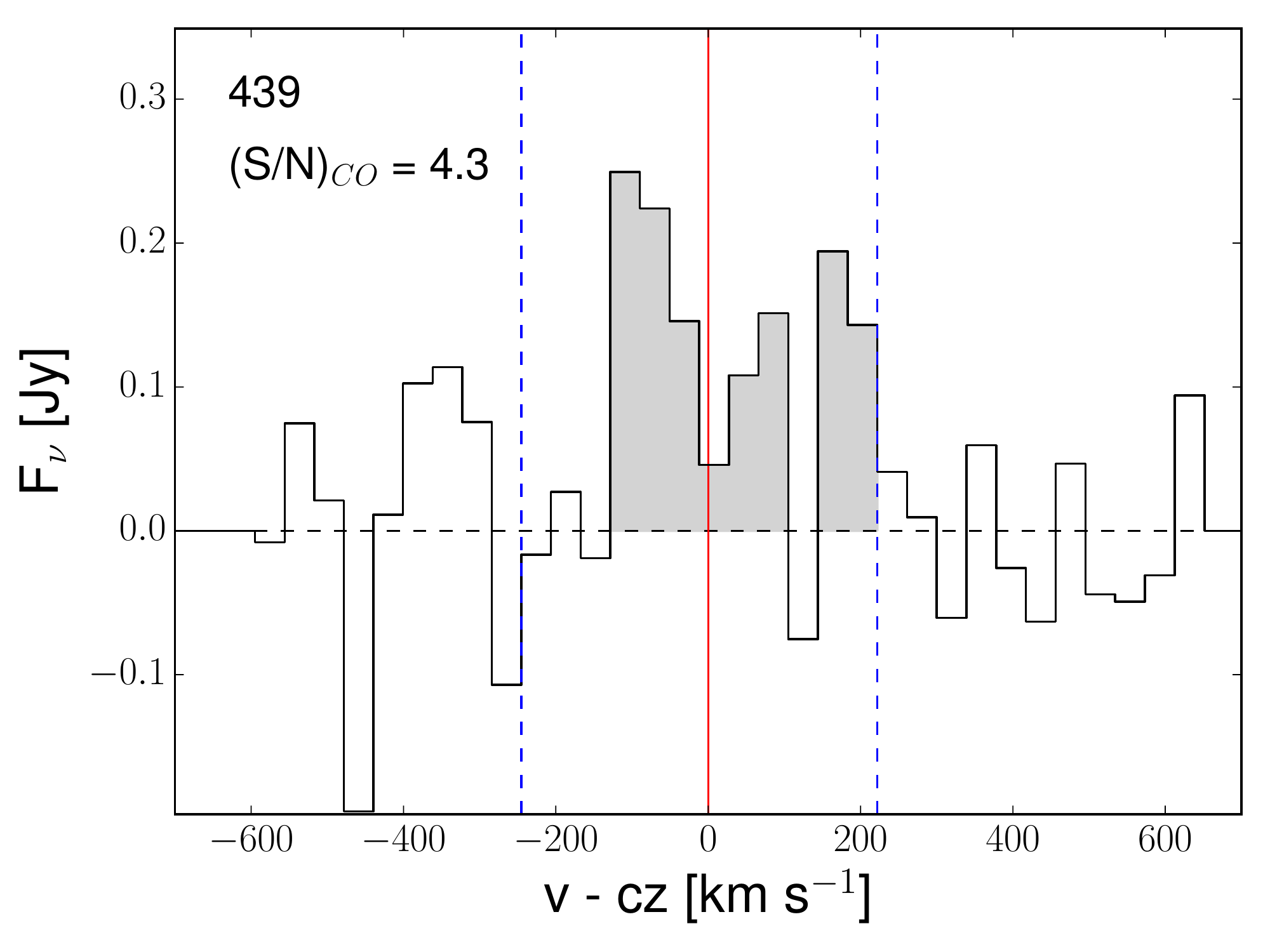}
\includegraphics[width=0.18\textwidth]{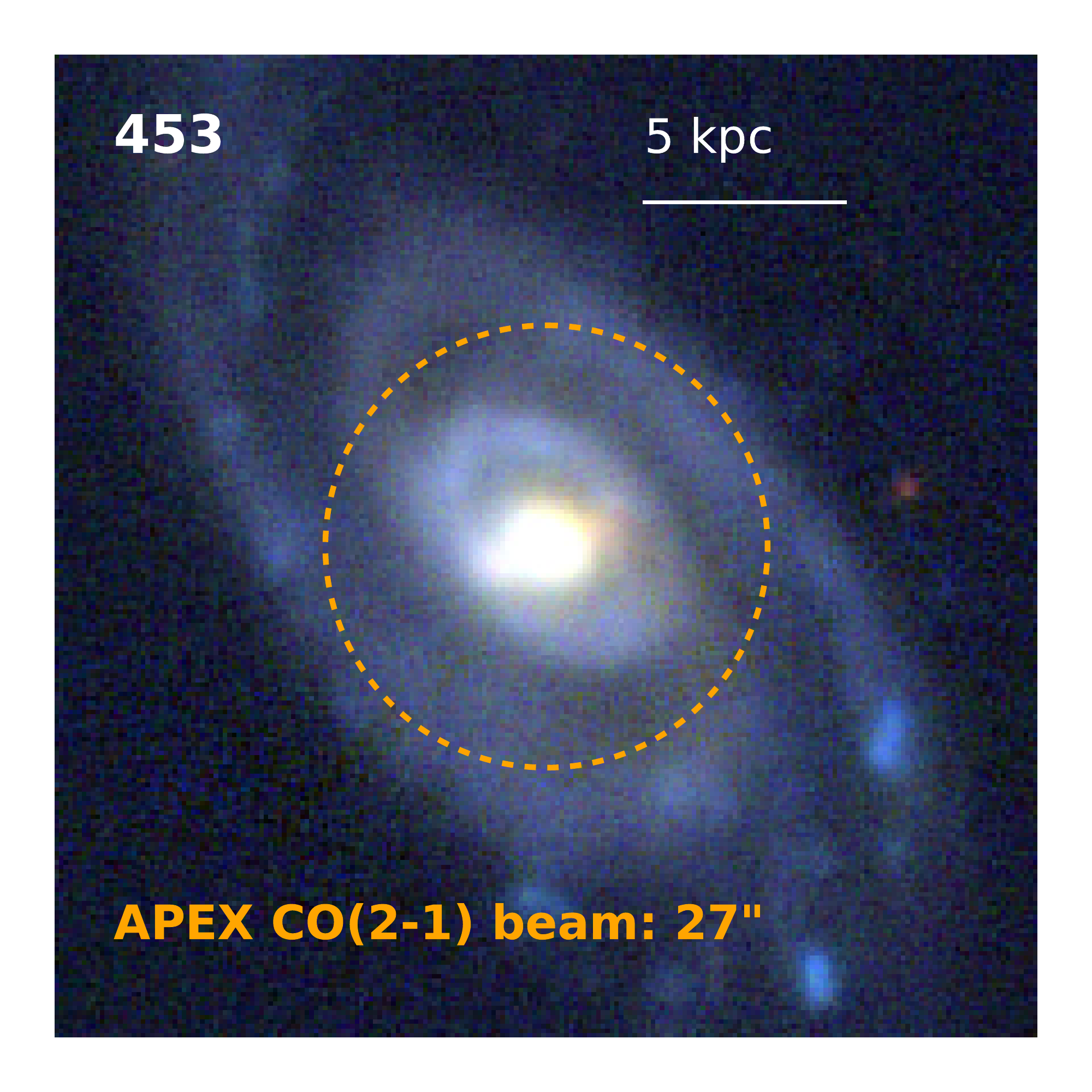}\includegraphics[width=0.26\textwidth]{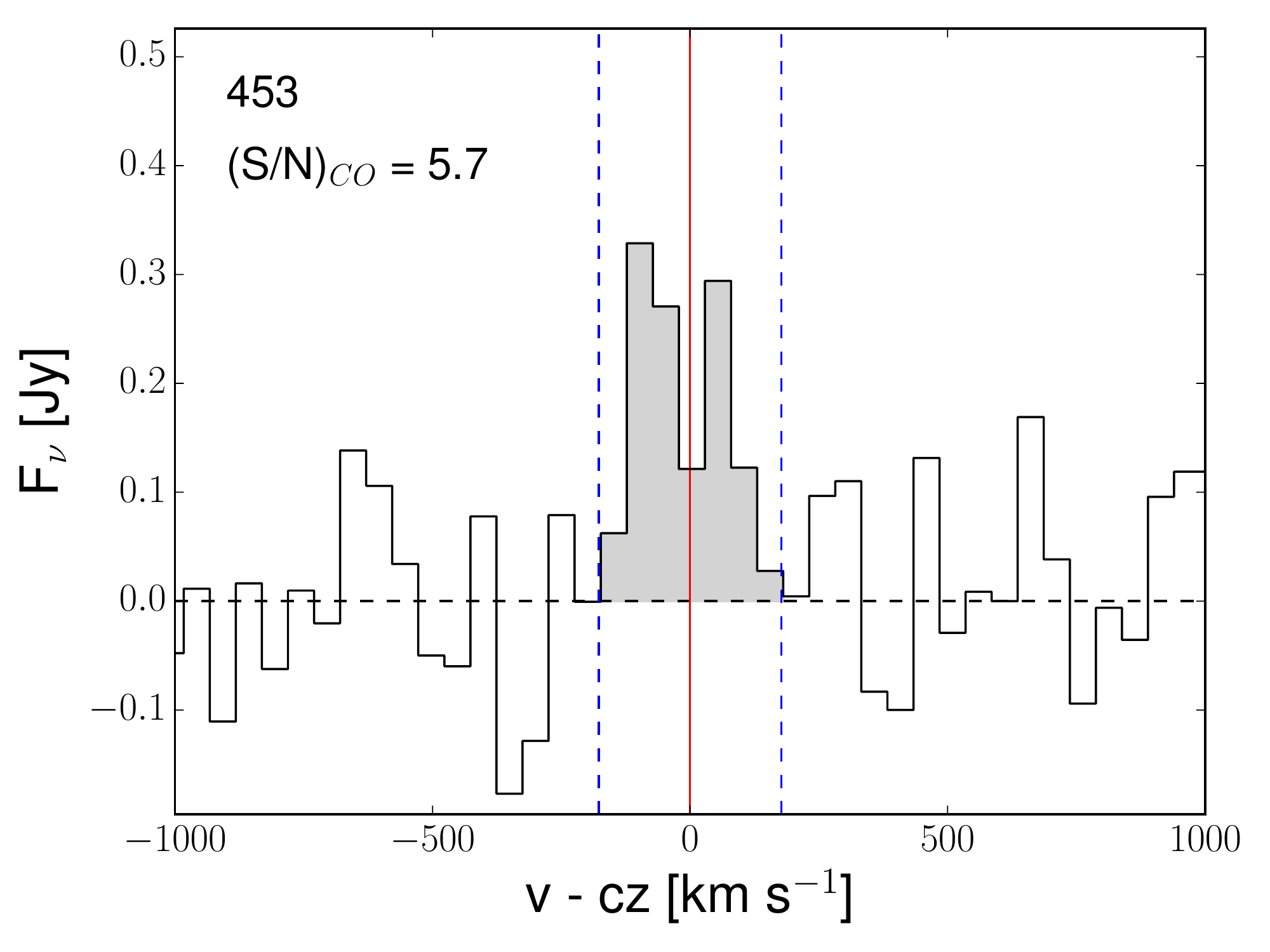}
\includegraphics[width=0.18\textwidth]{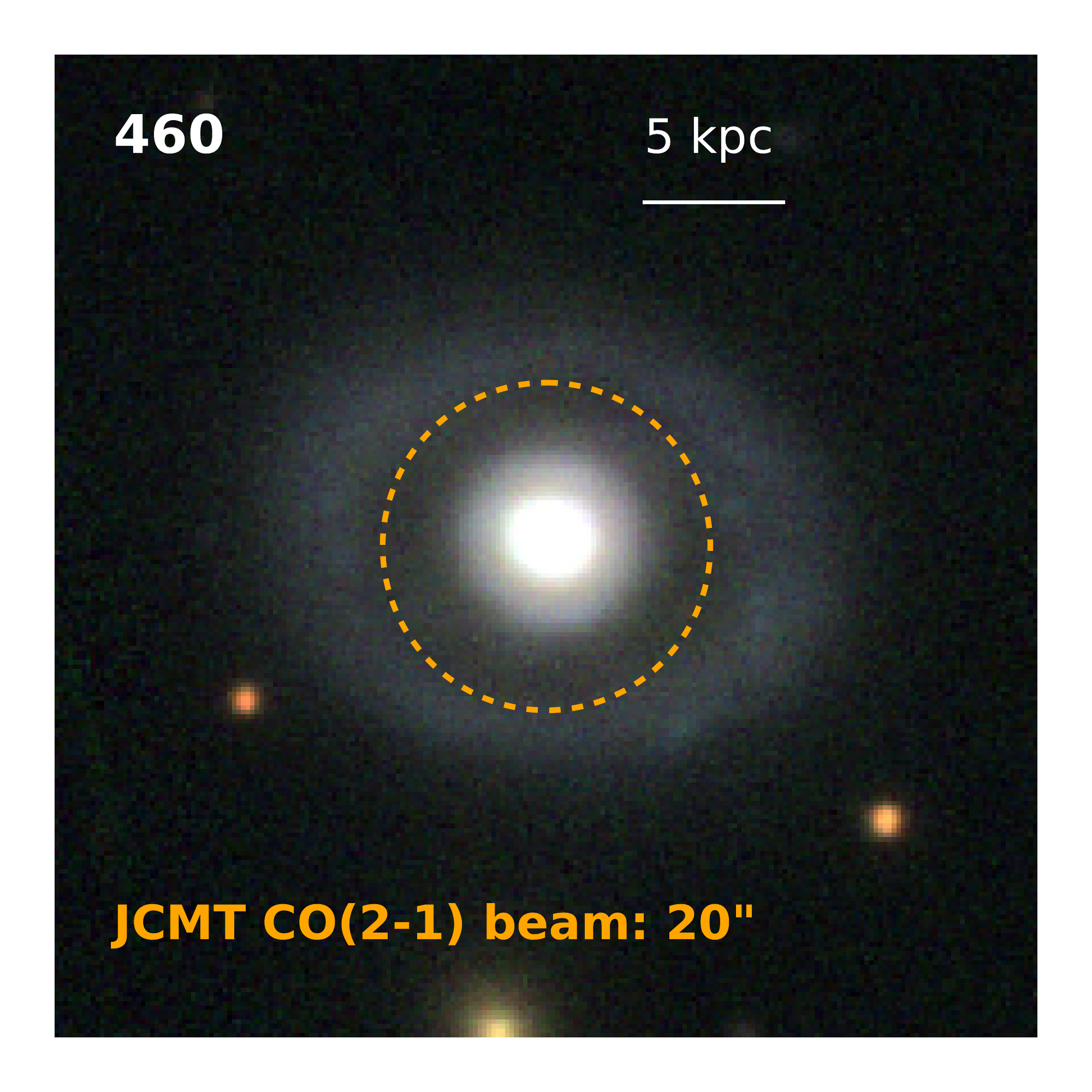}\includegraphics[width=0.26\textwidth]{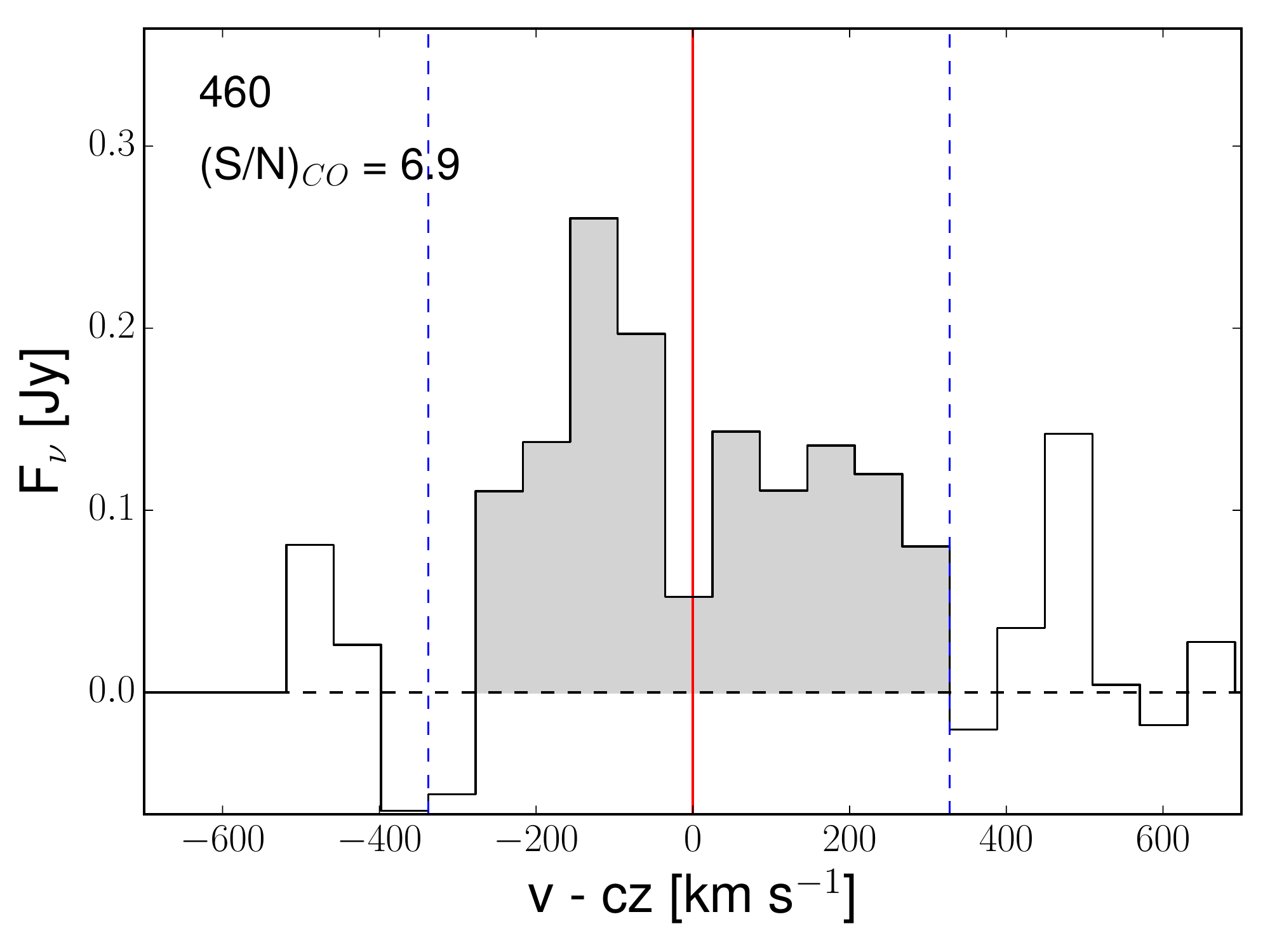}
\caption{continued from Fig.~\ref{fig:CO21_spectra_all_1}
} 
\label{fig:CO21_spectra_all_3}
\end{figure*}

\begin{figure*}
\centering
\raggedright
\includegraphics[width=0.18\textwidth]{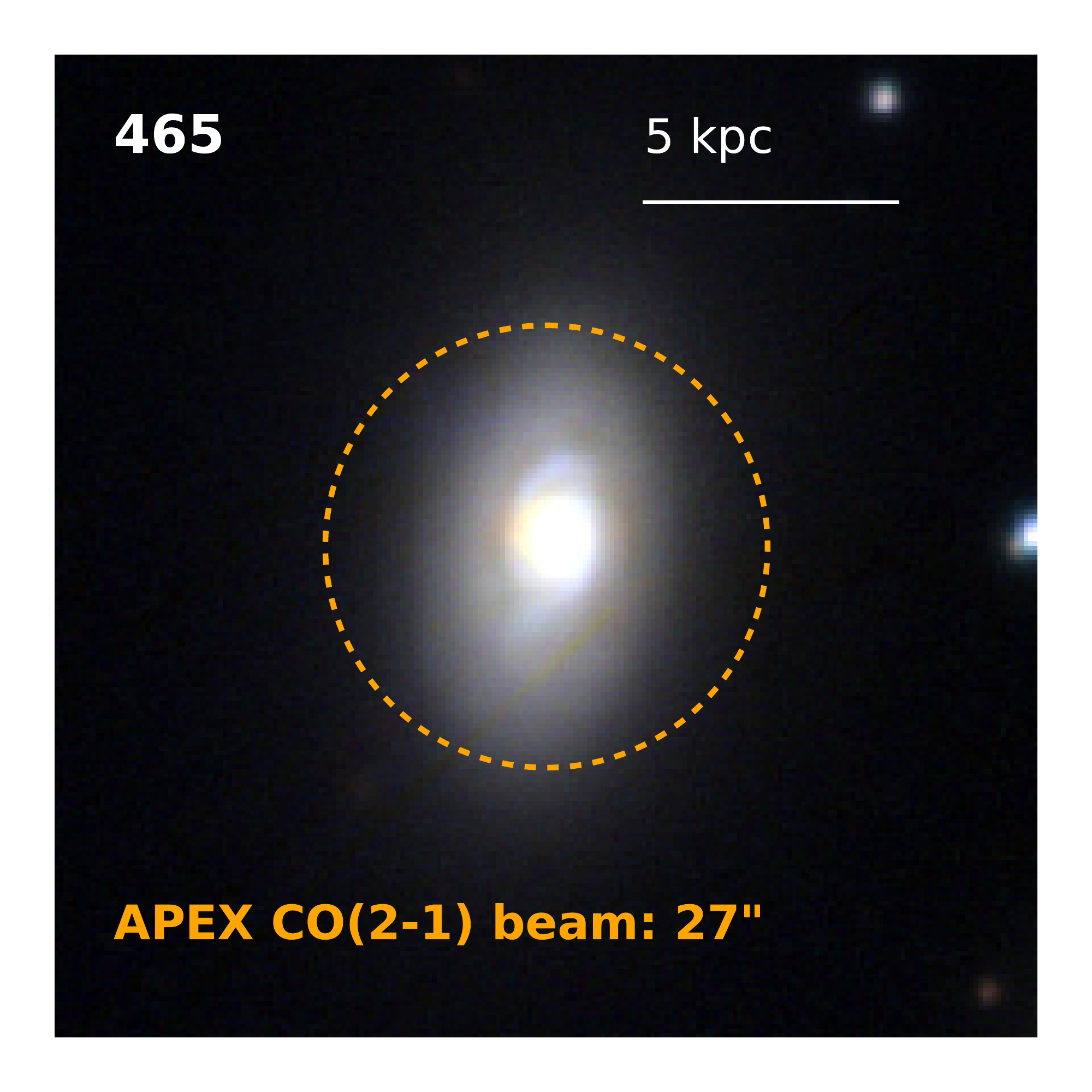}\includegraphics[width=0.26\textwidth]{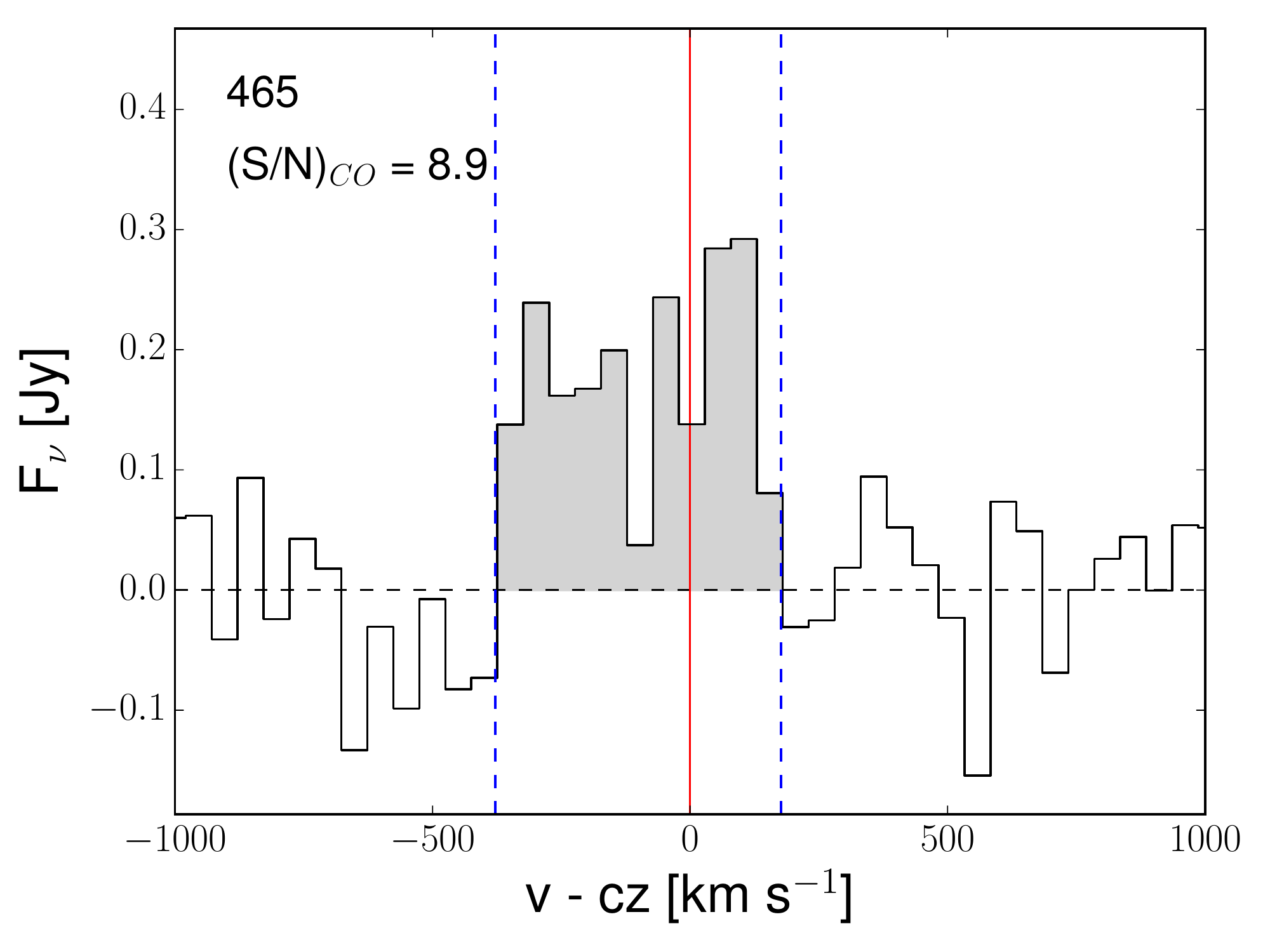}
\includegraphics[width=0.18\textwidth]{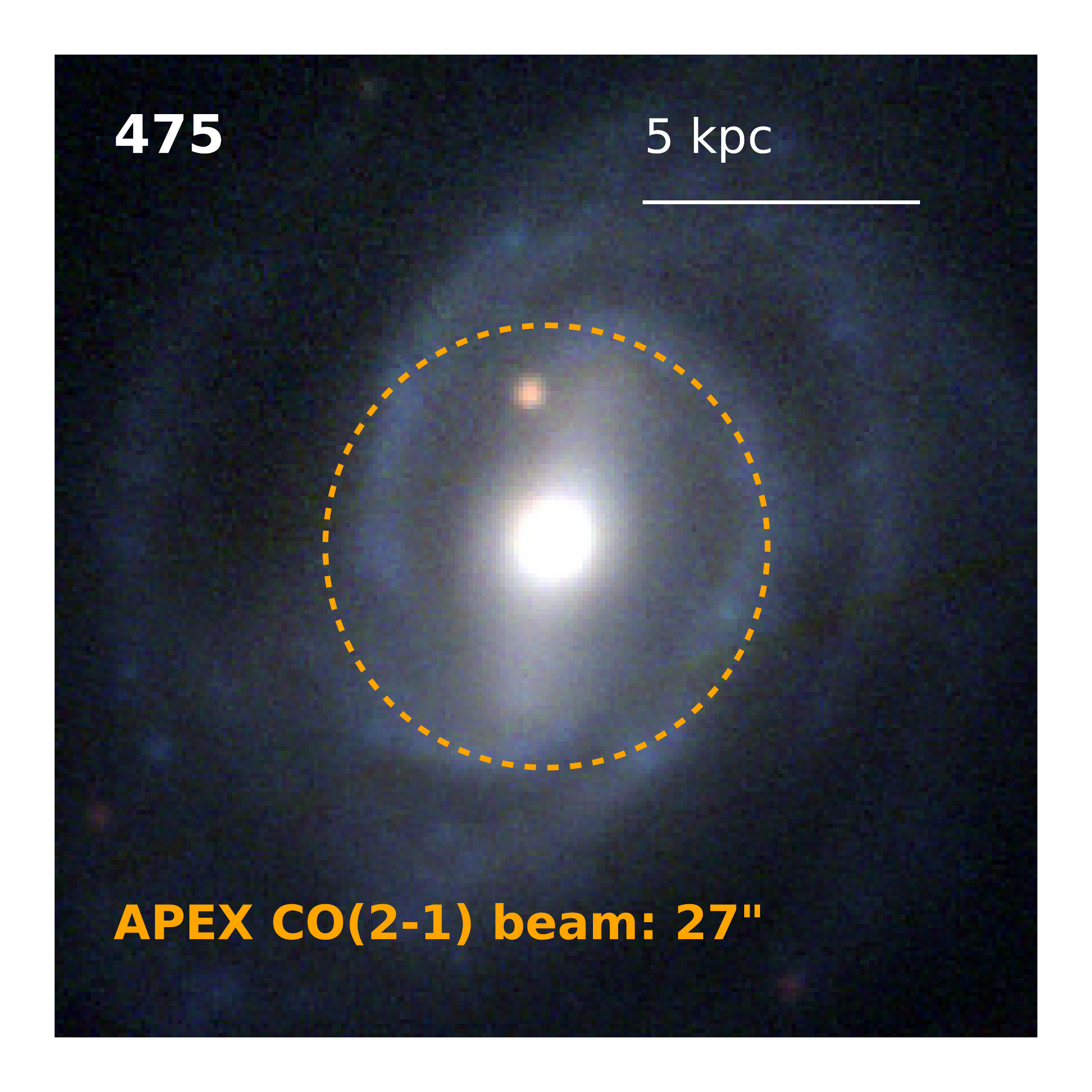}\includegraphics[width=0.26\textwidth]{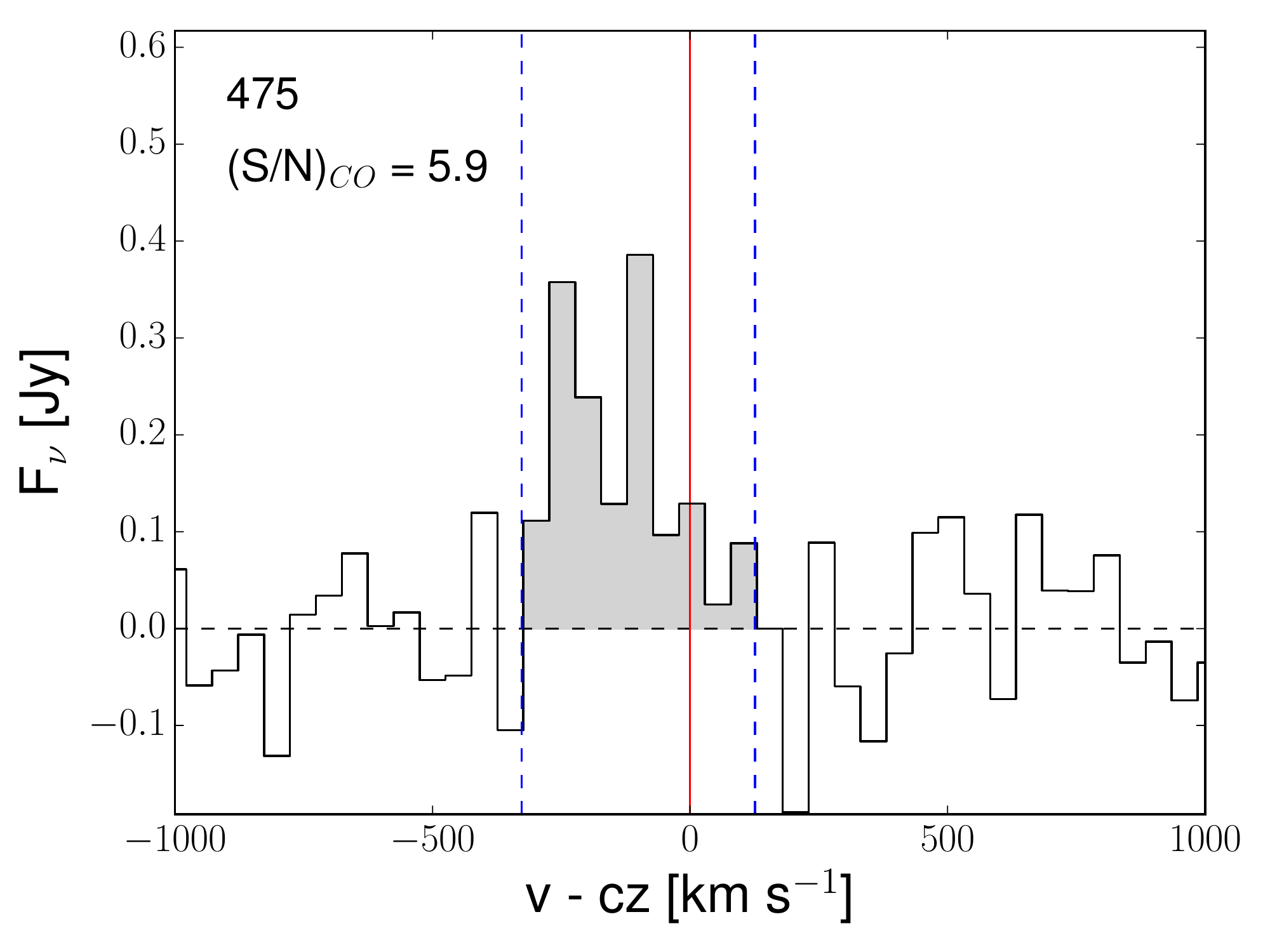}
\includegraphics[width=0.18\textwidth]{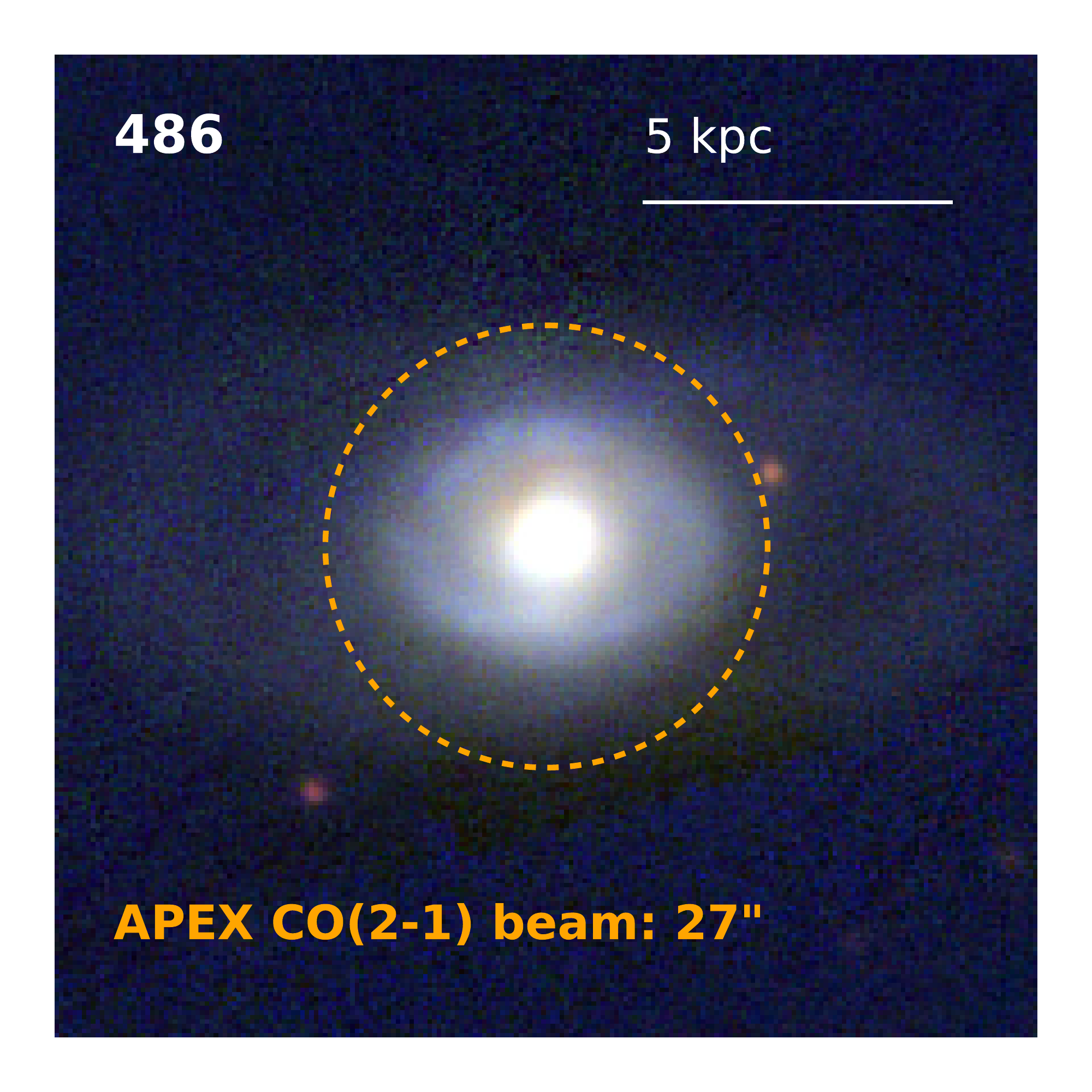}\includegraphics[width=0.26\textwidth]{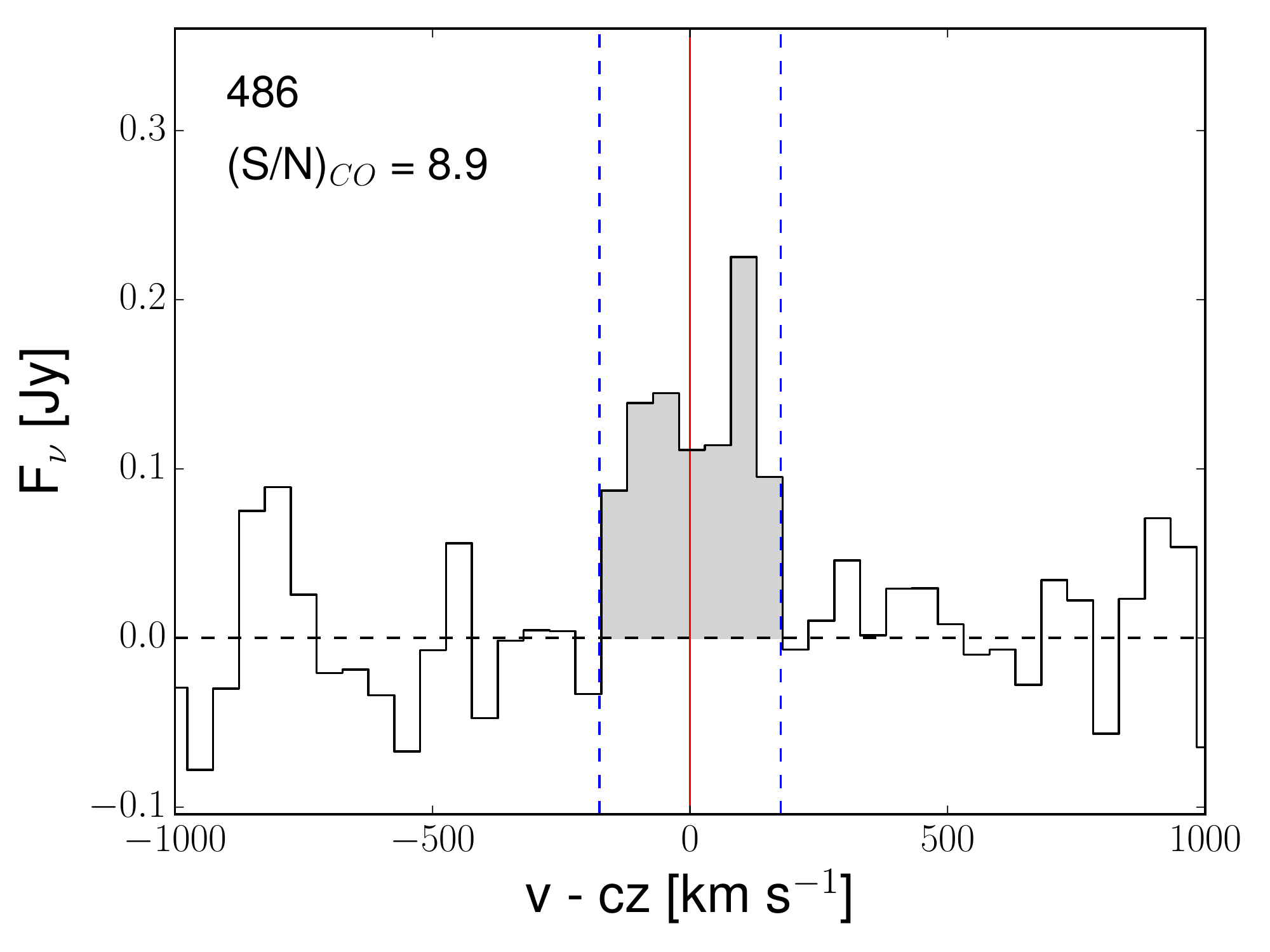}
\includegraphics[width=0.18\textwidth]{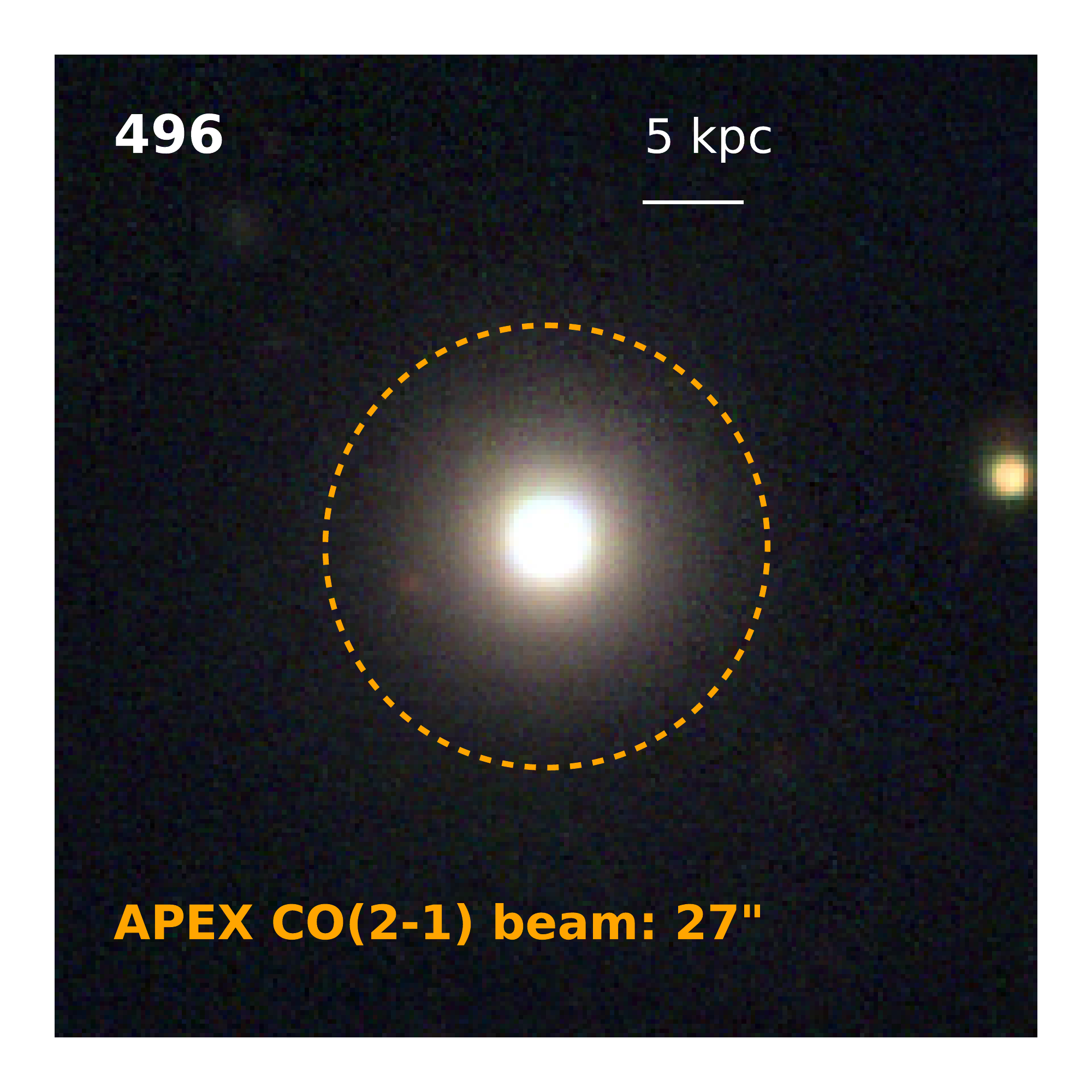}\includegraphics[width=0.26\textwidth]{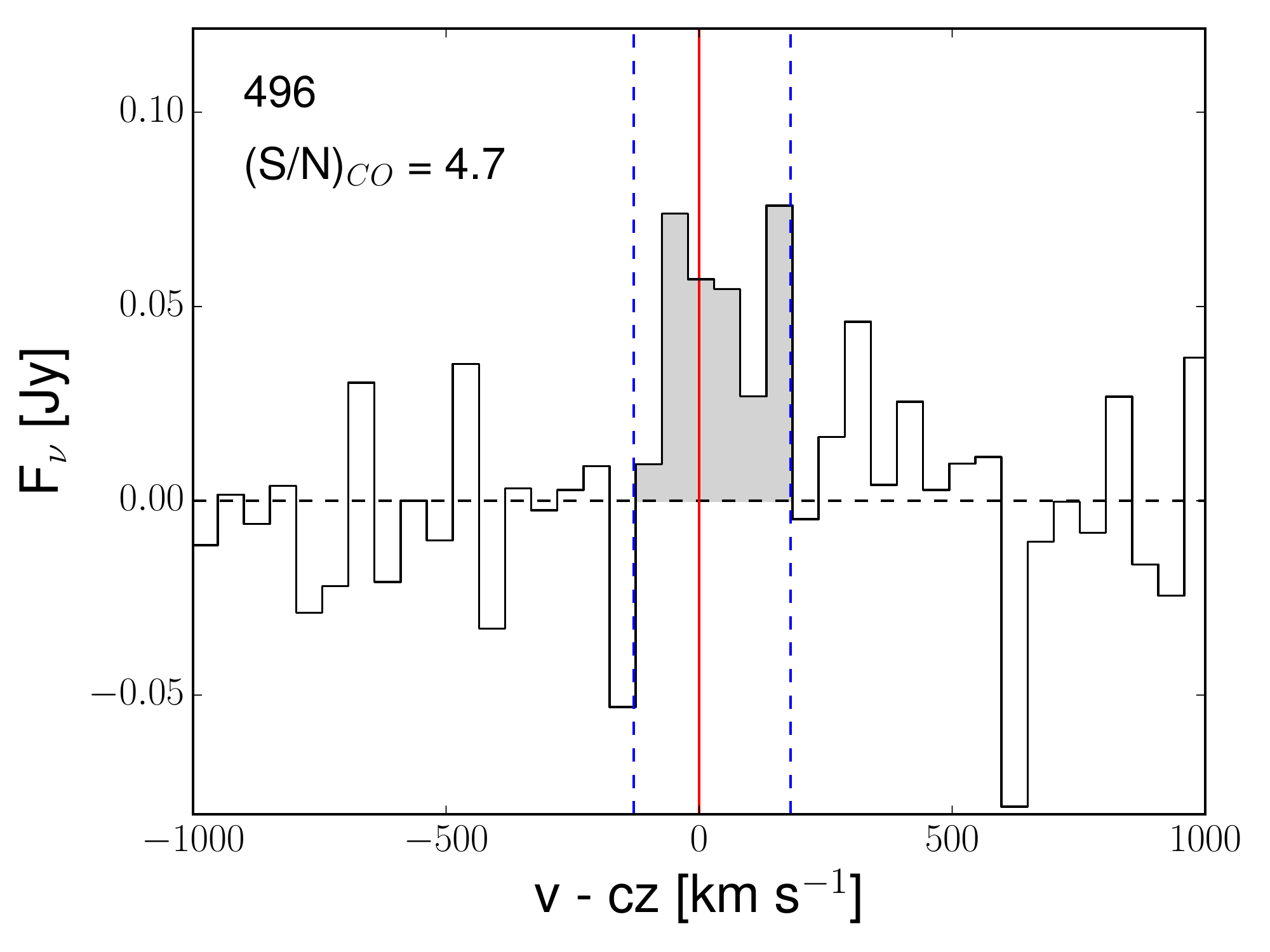}
\includegraphics[width=0.18\textwidth]{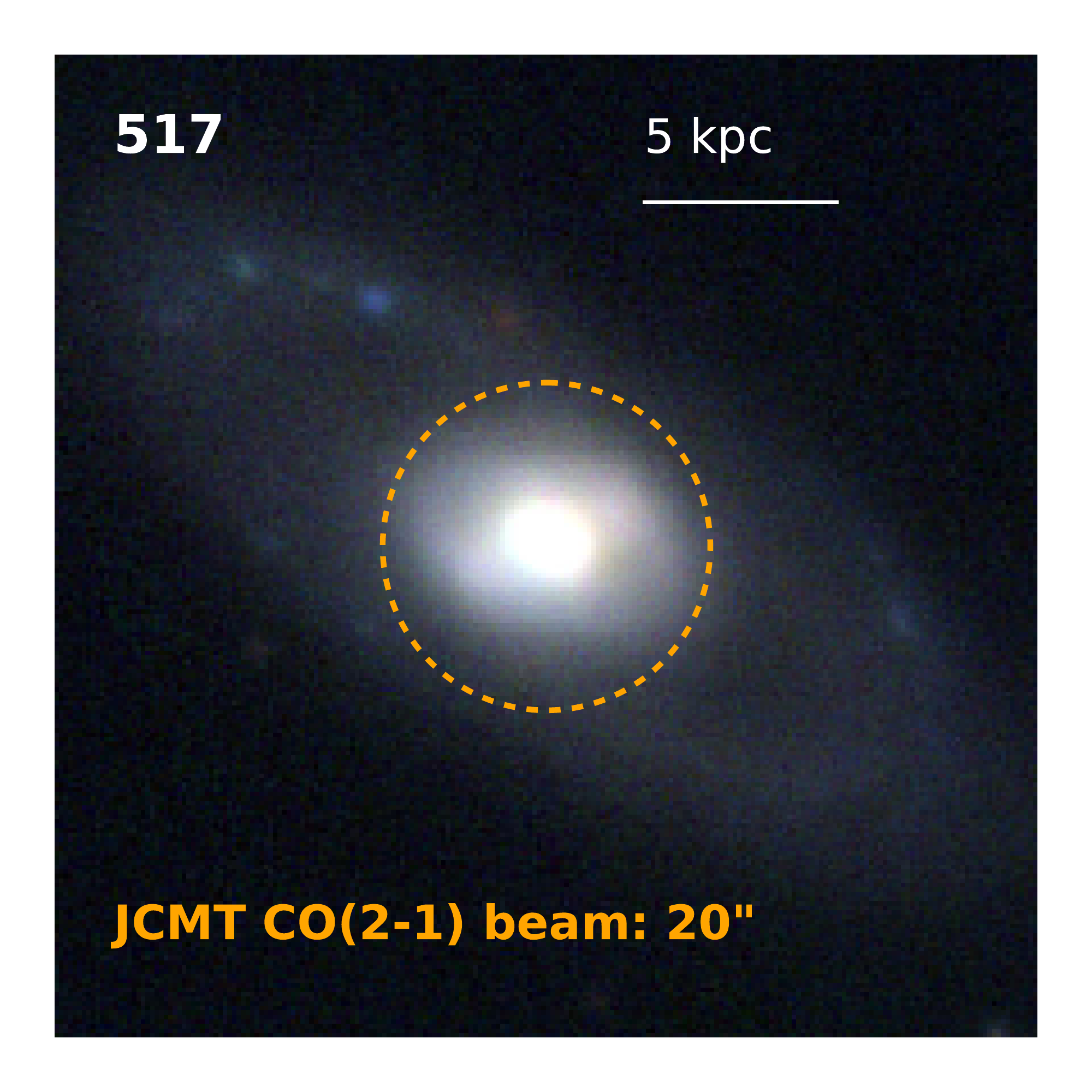}\includegraphics[width=0.26\textwidth]{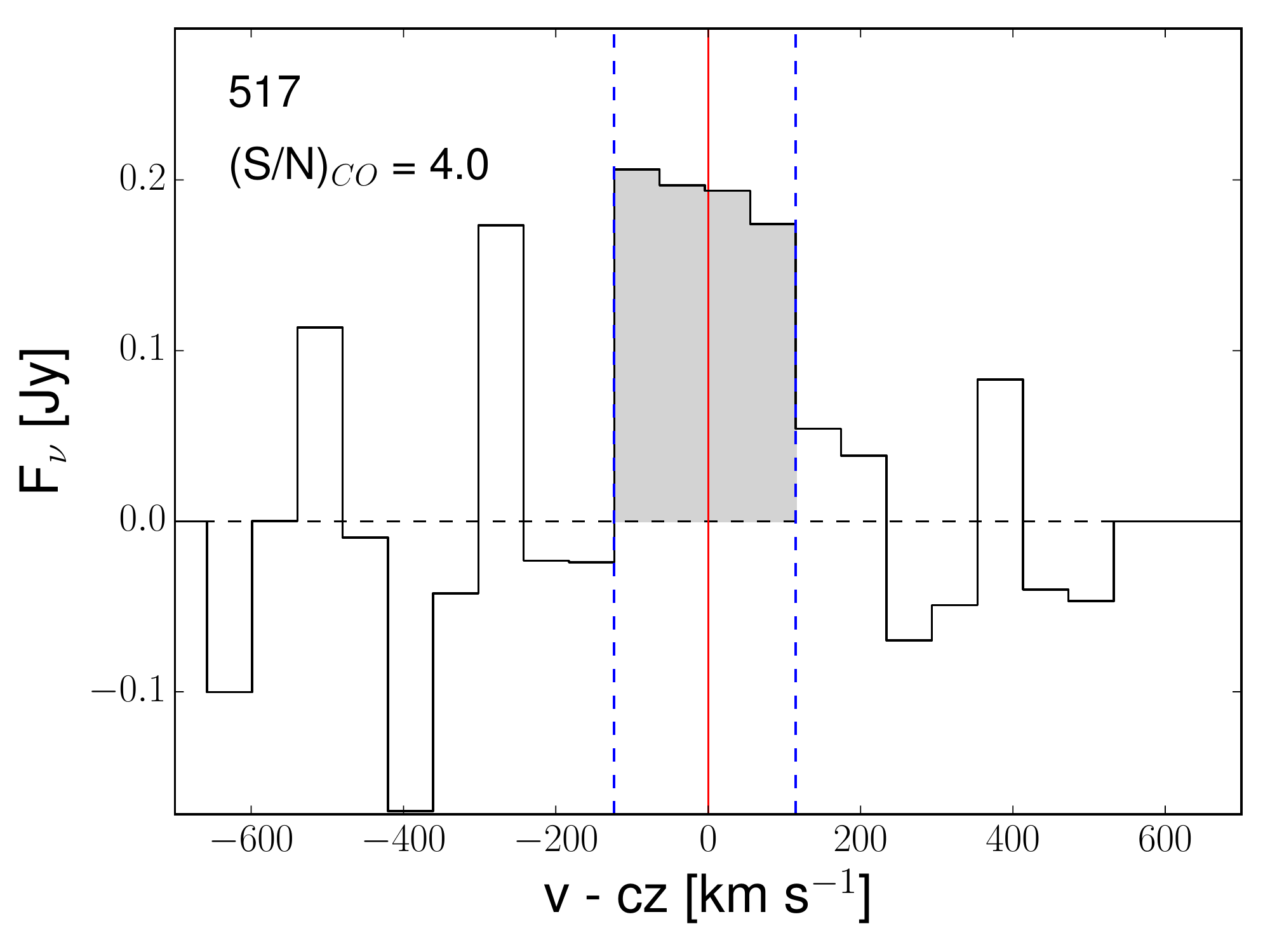}
\includegraphics[width=0.18\textwidth]{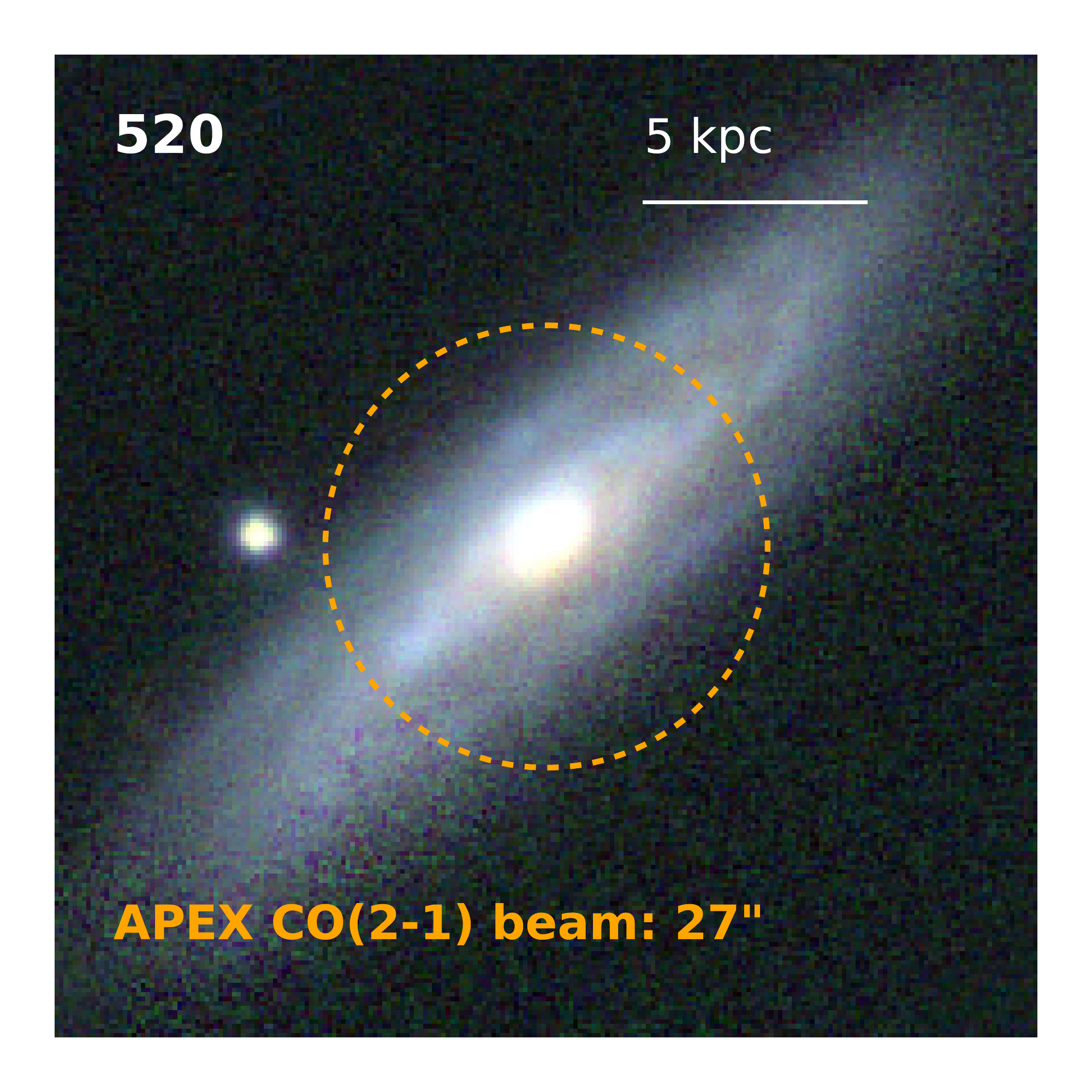}\includegraphics[width=0.26\textwidth]{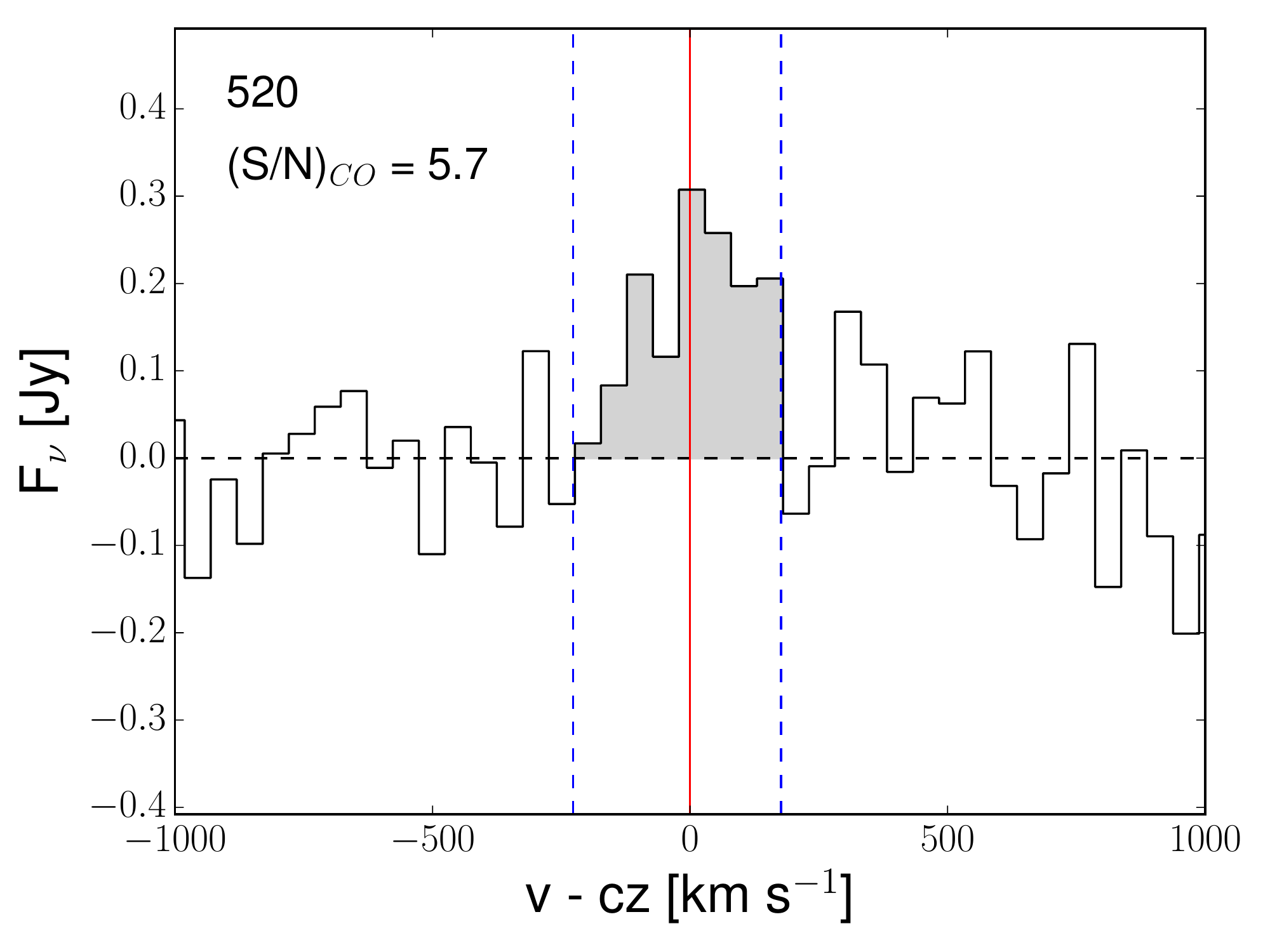}
\includegraphics[width=0.18\textwidth]{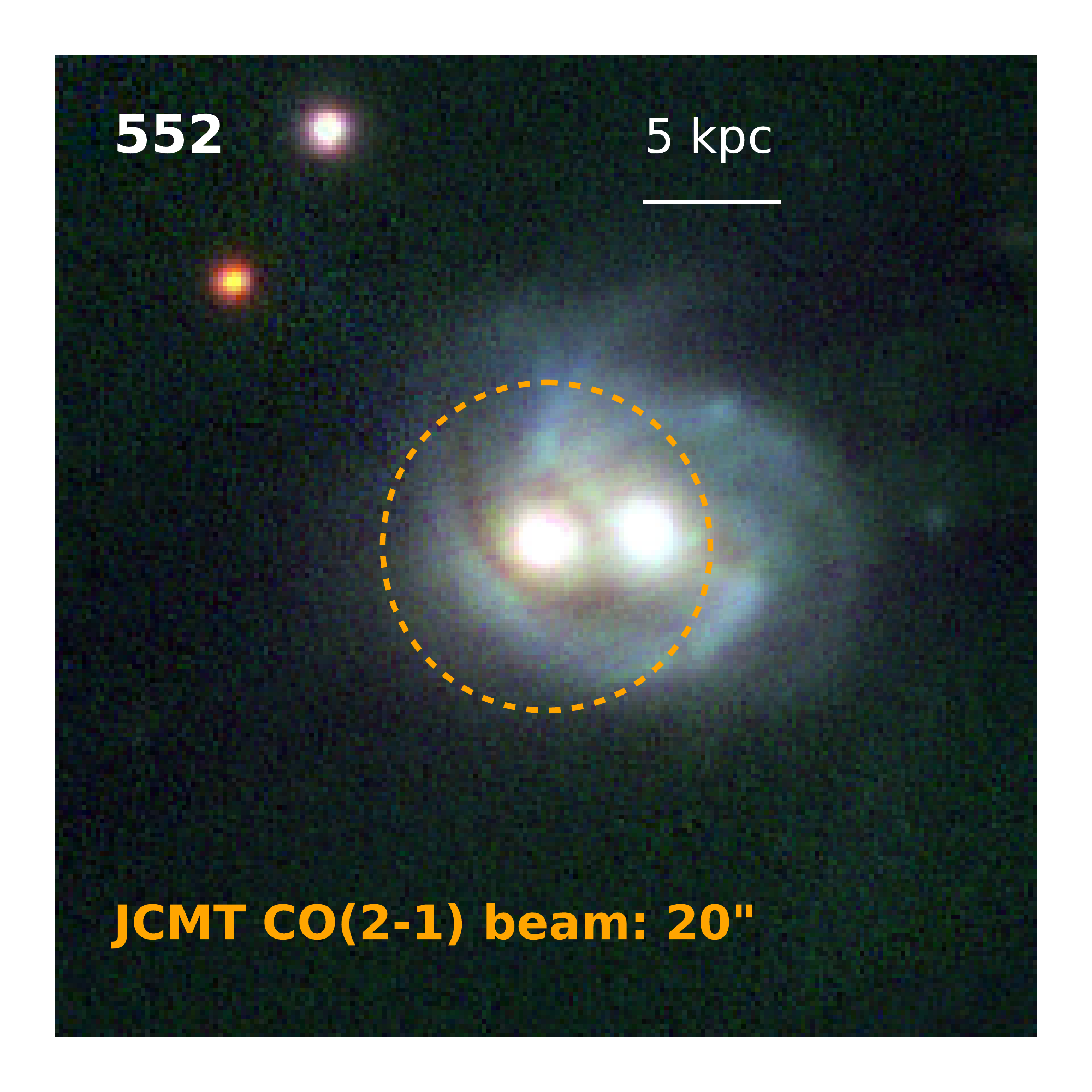}\includegraphics[width=0.26\textwidth]{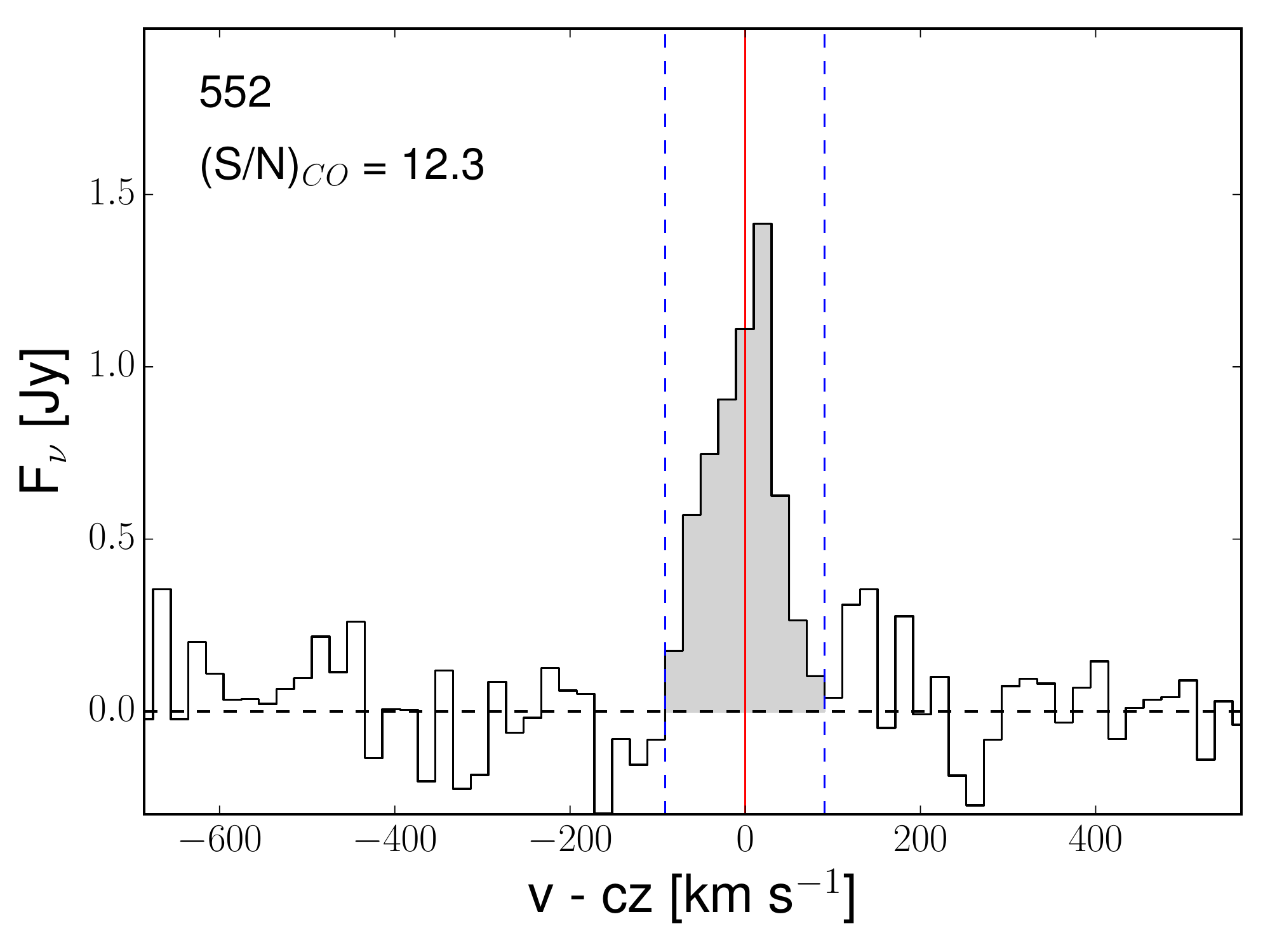}
\includegraphics[width=0.18\textwidth]{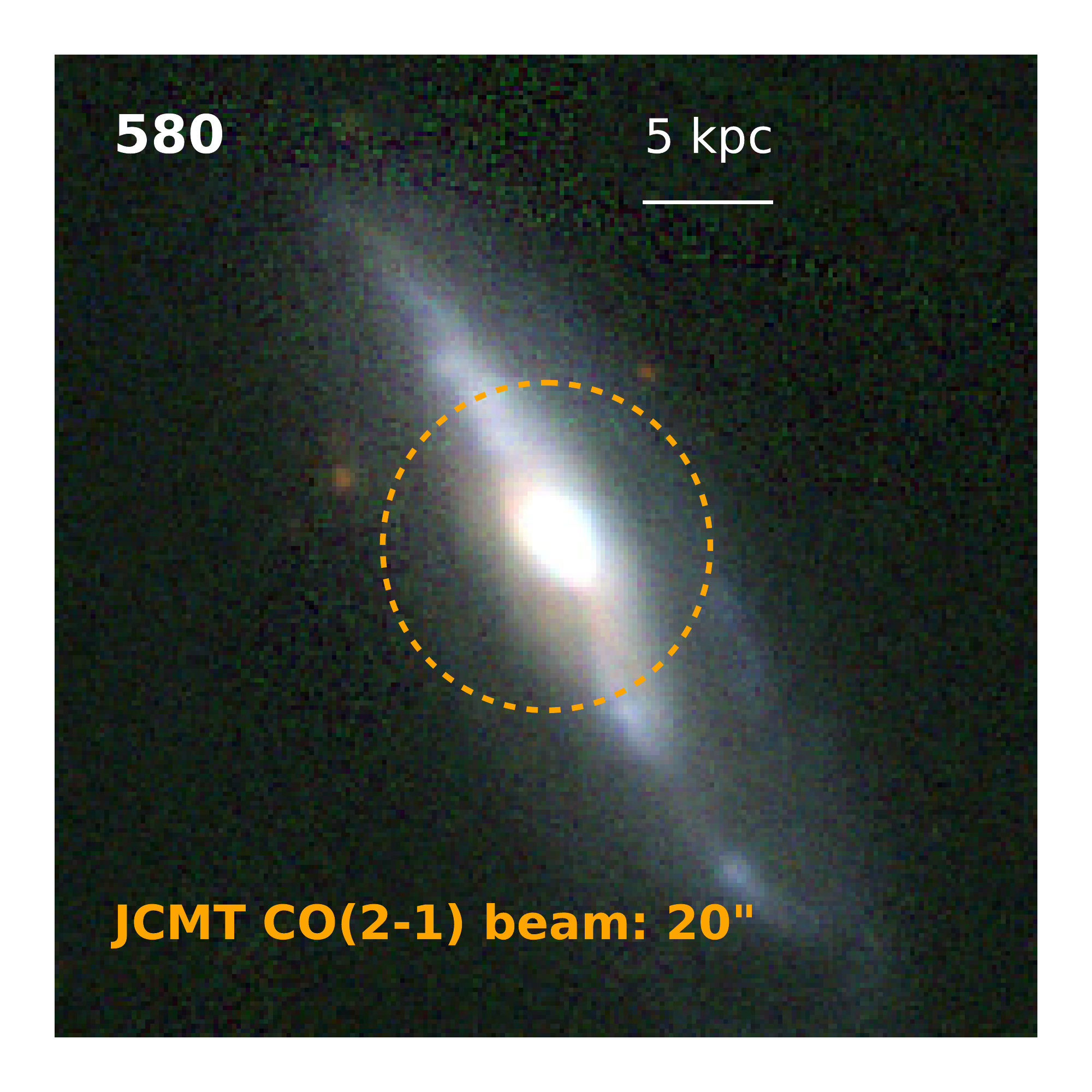}\includegraphics[width=0.26\textwidth]{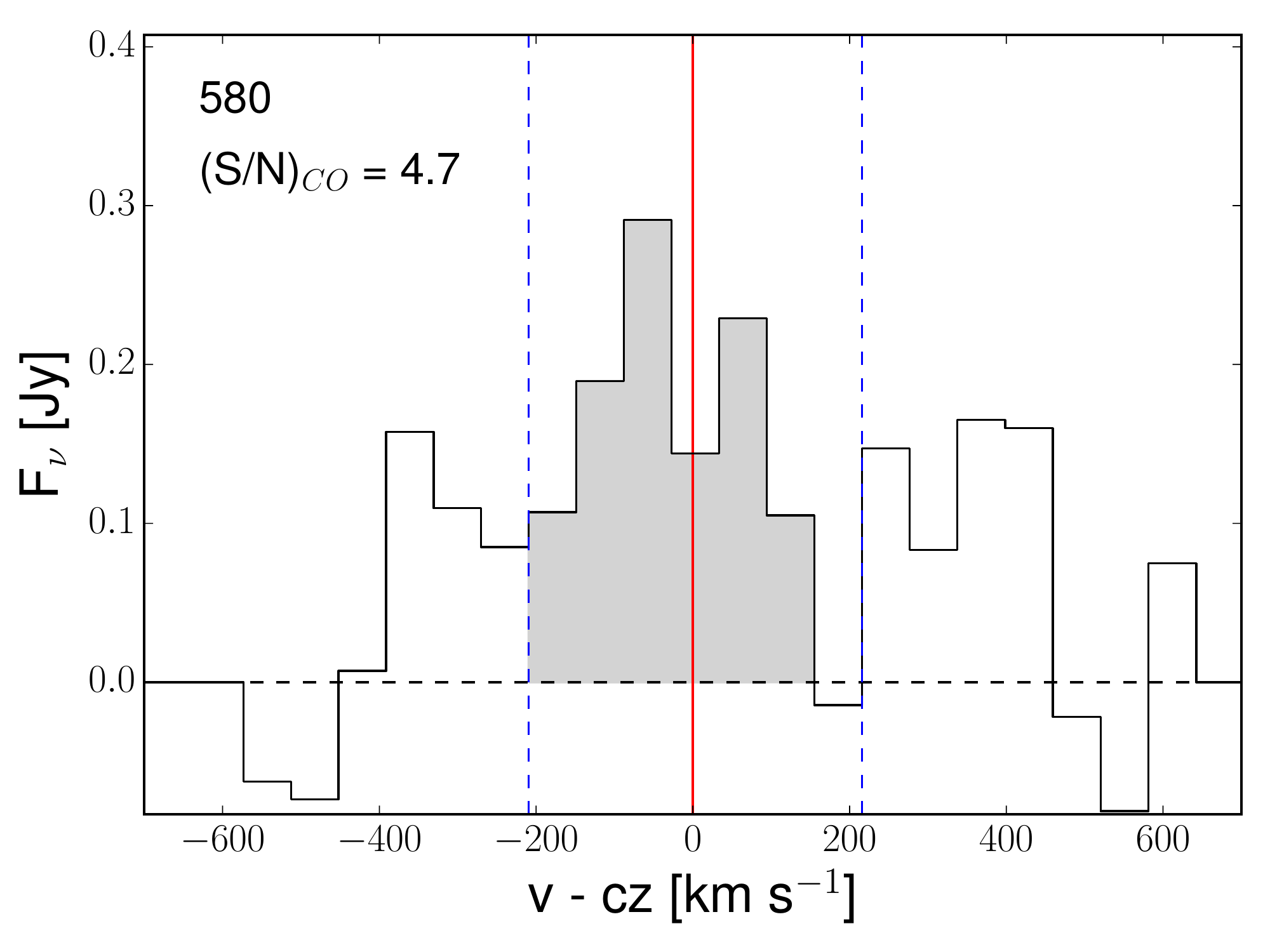}
\includegraphics[width=0.18\textwidth]{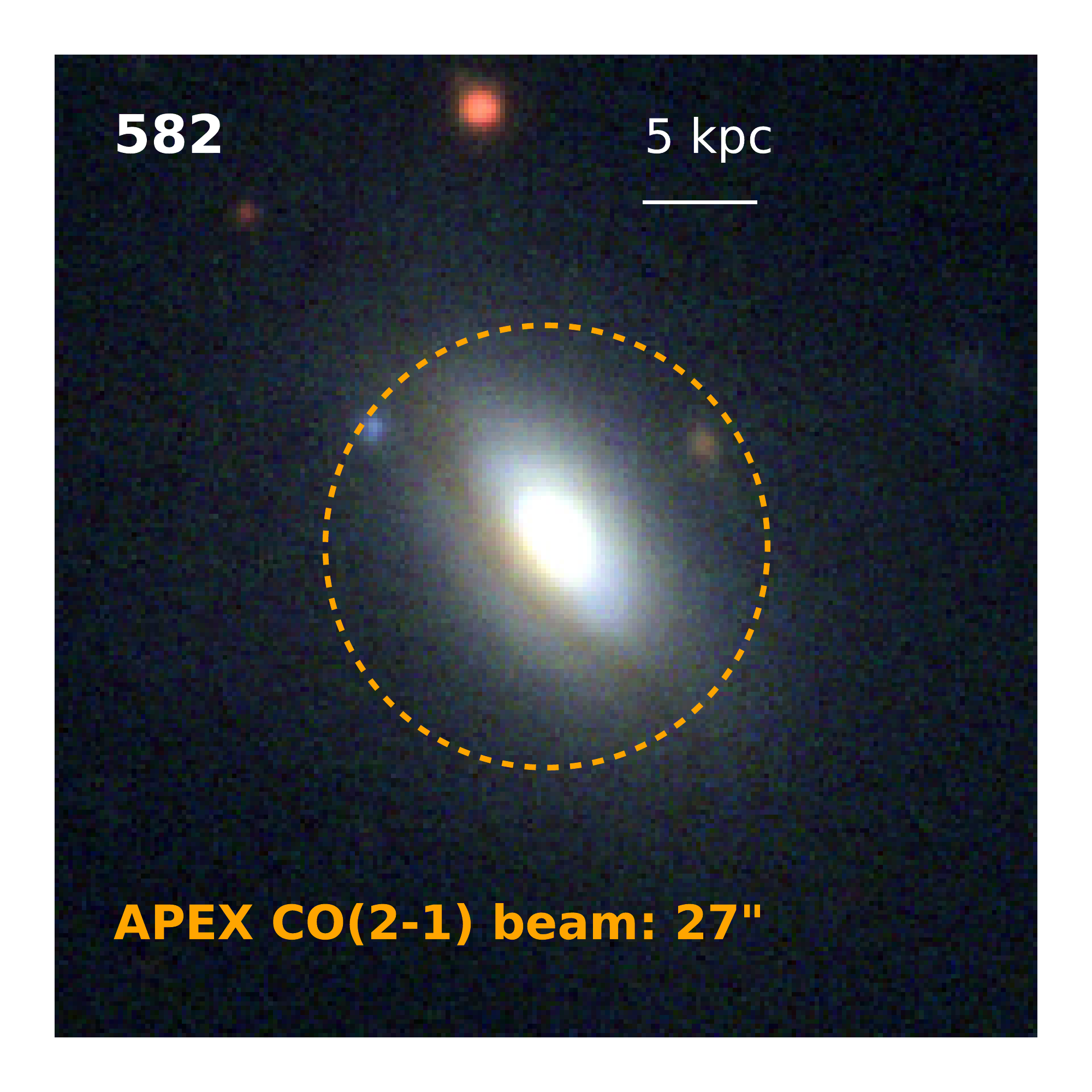}\includegraphics[width=0.26\textwidth]{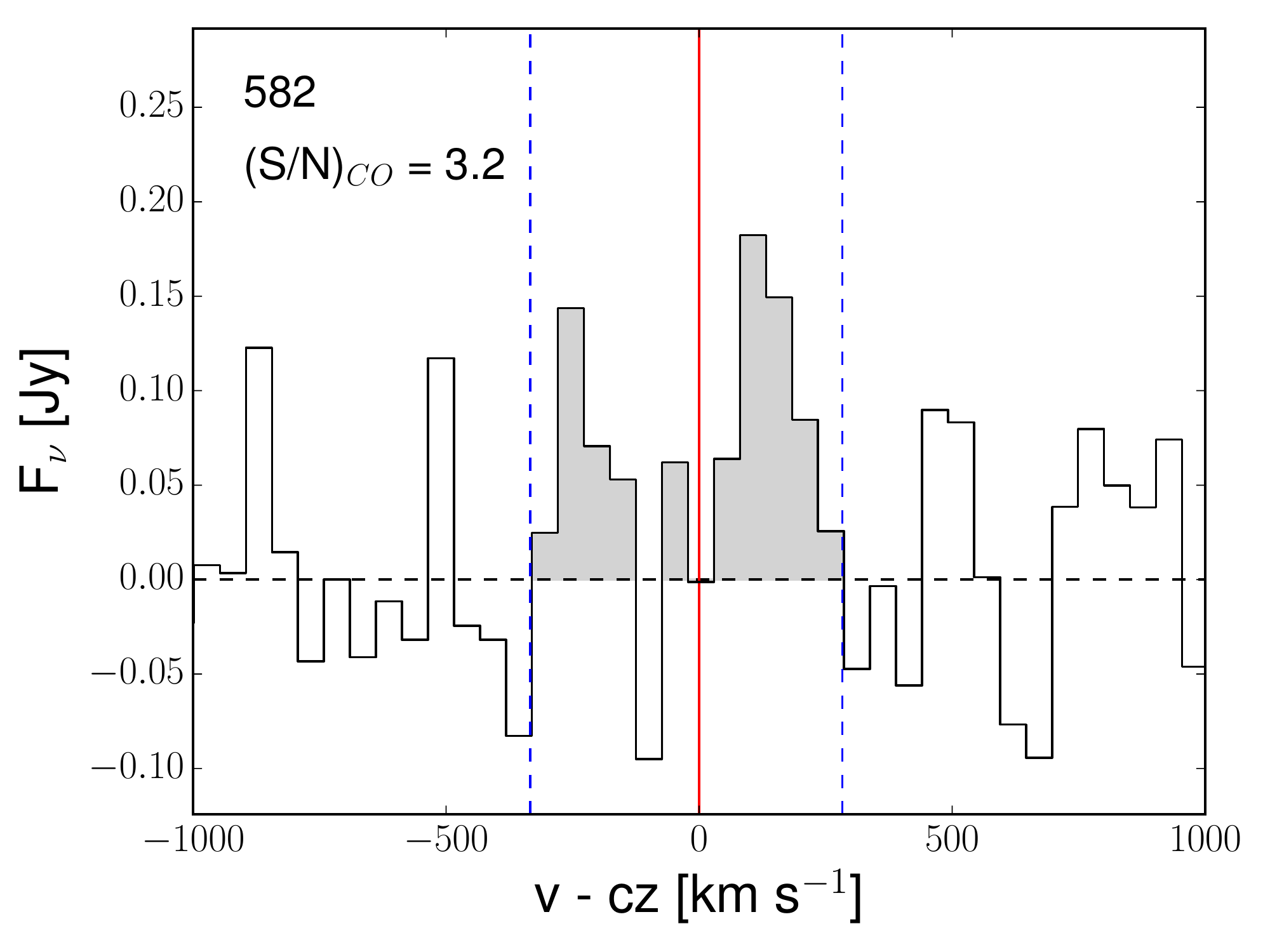}
\includegraphics[width=0.18\textwidth]{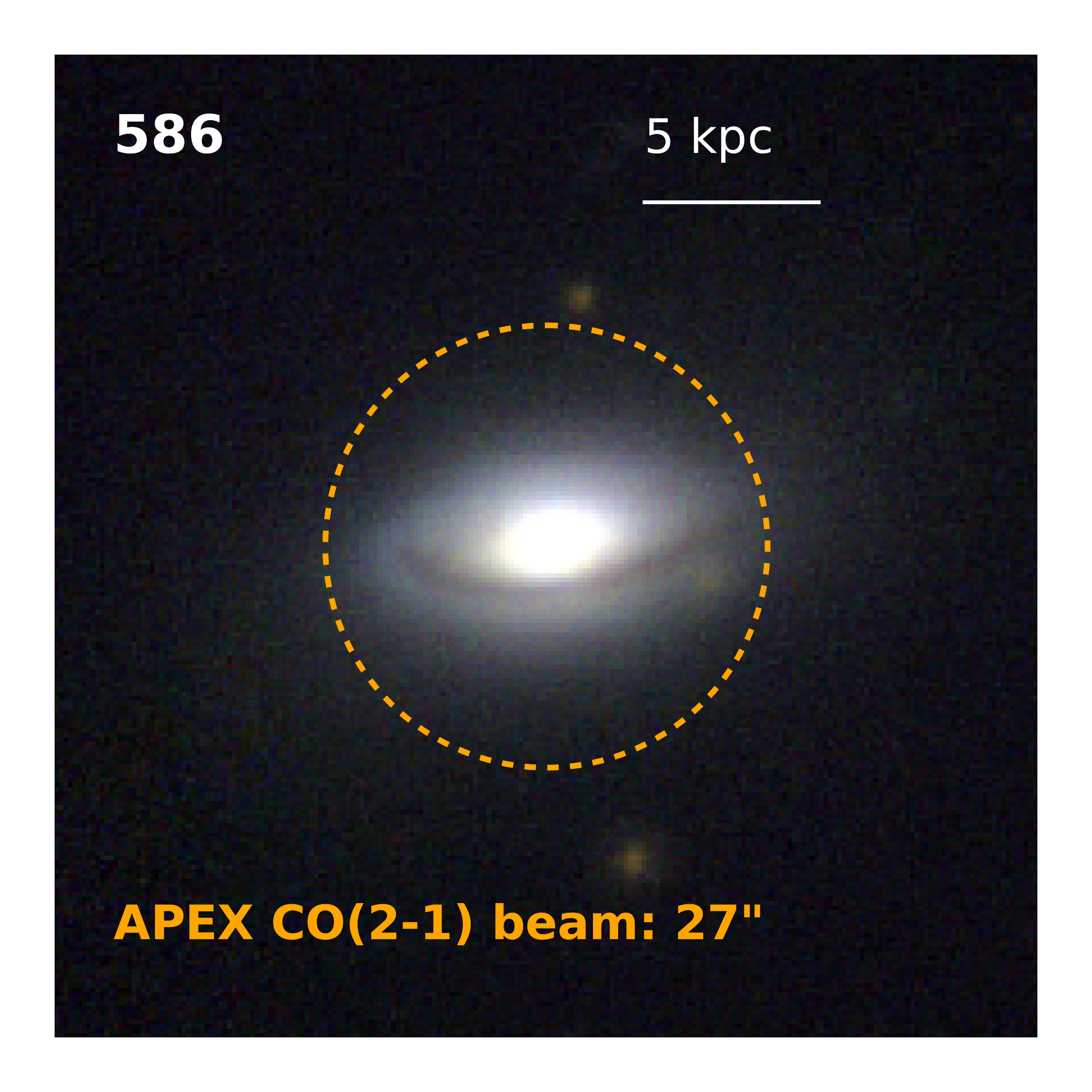}\includegraphics[width=0.26\textwidth]{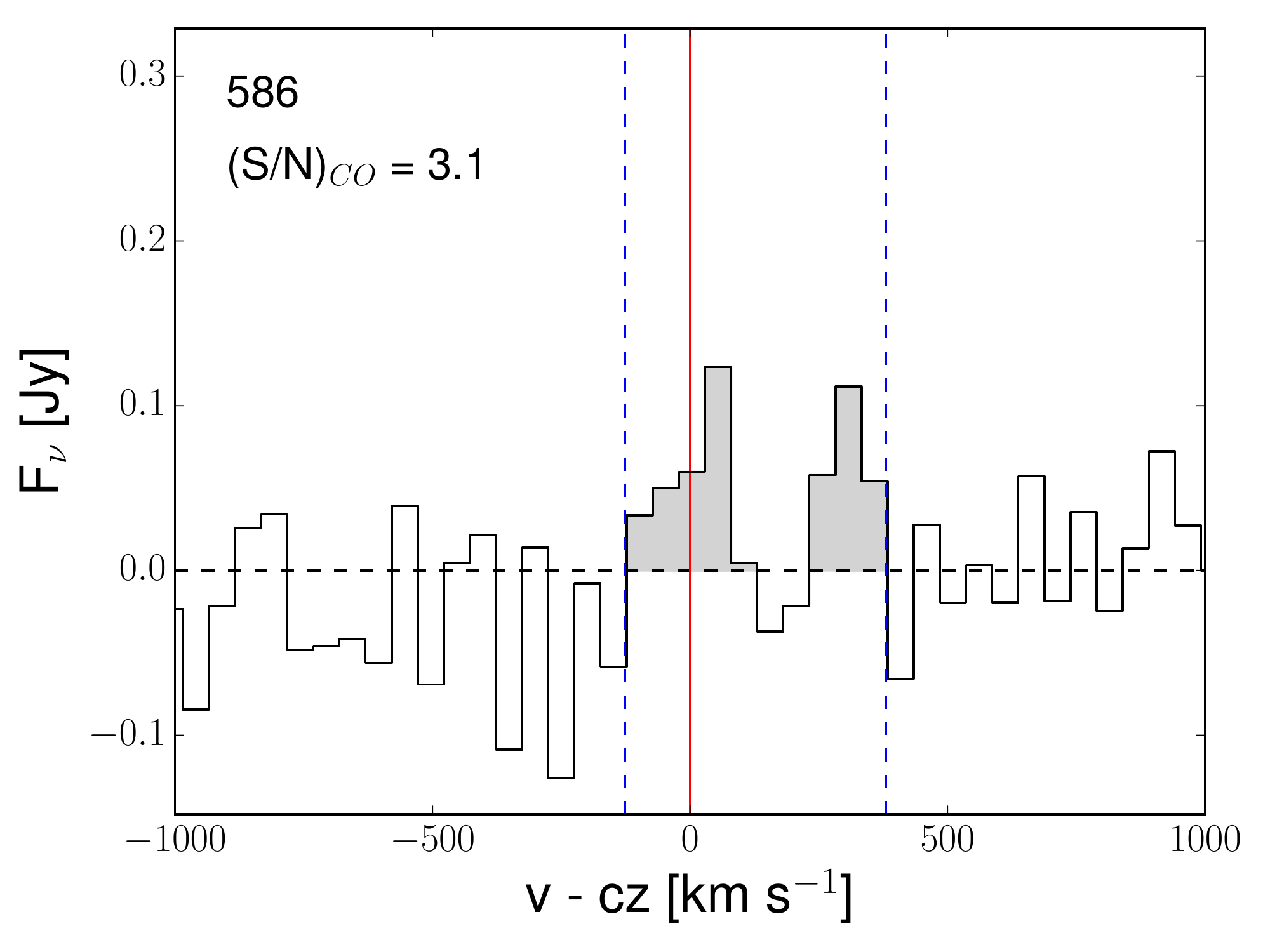}
\includegraphics[width=0.18\textwidth]{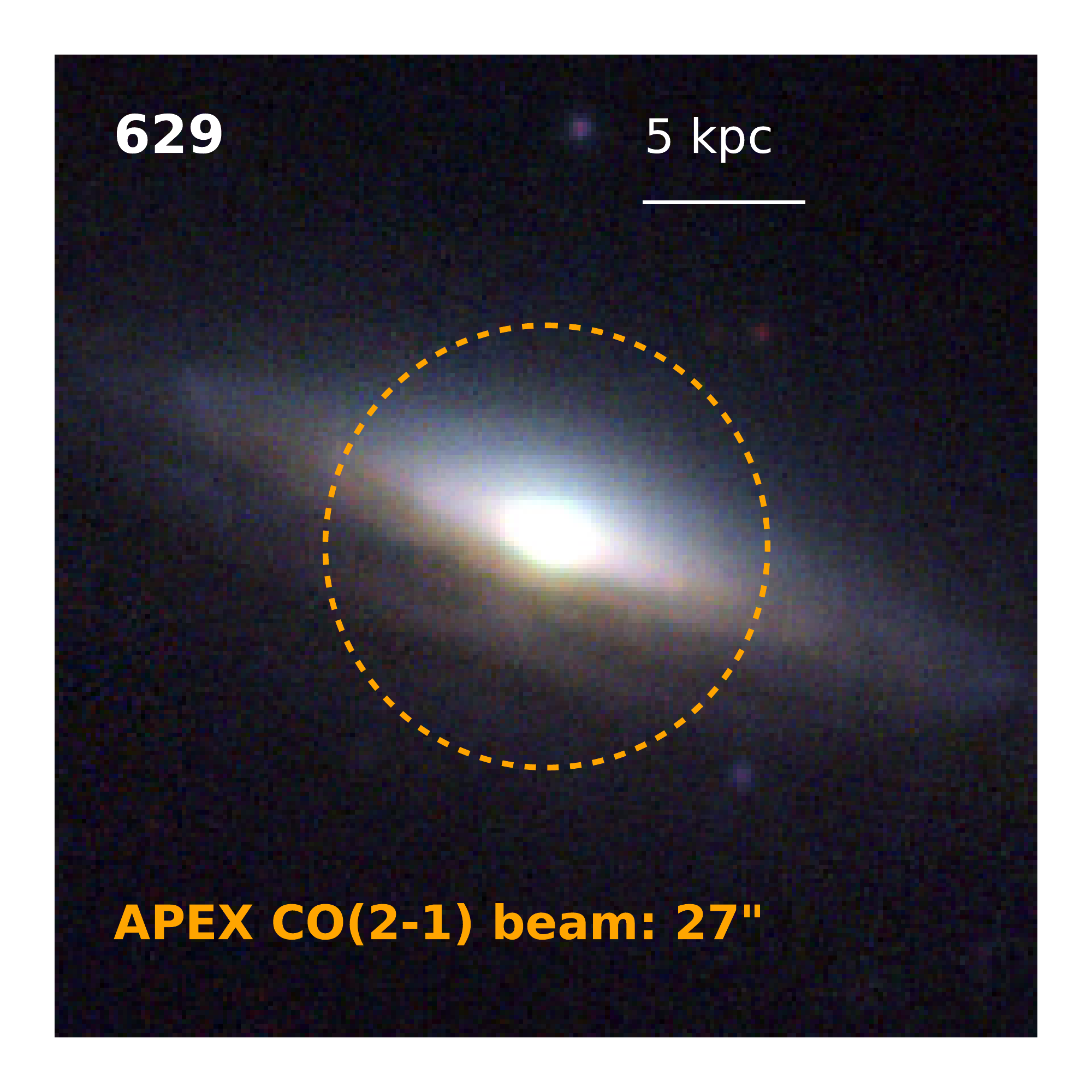}\includegraphics[width=0.26\textwidth]{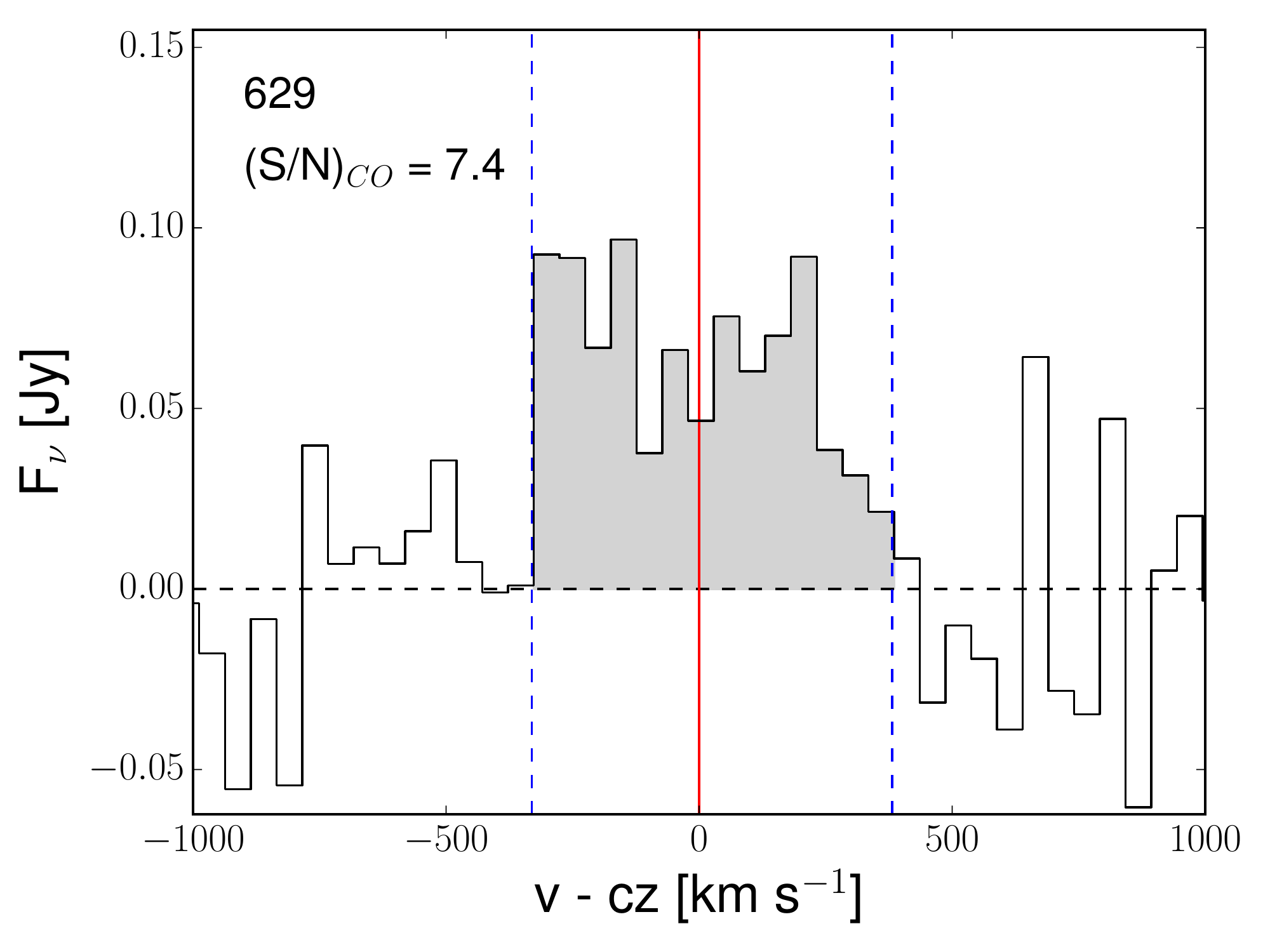}
\includegraphics[width=0.18\textwidth]{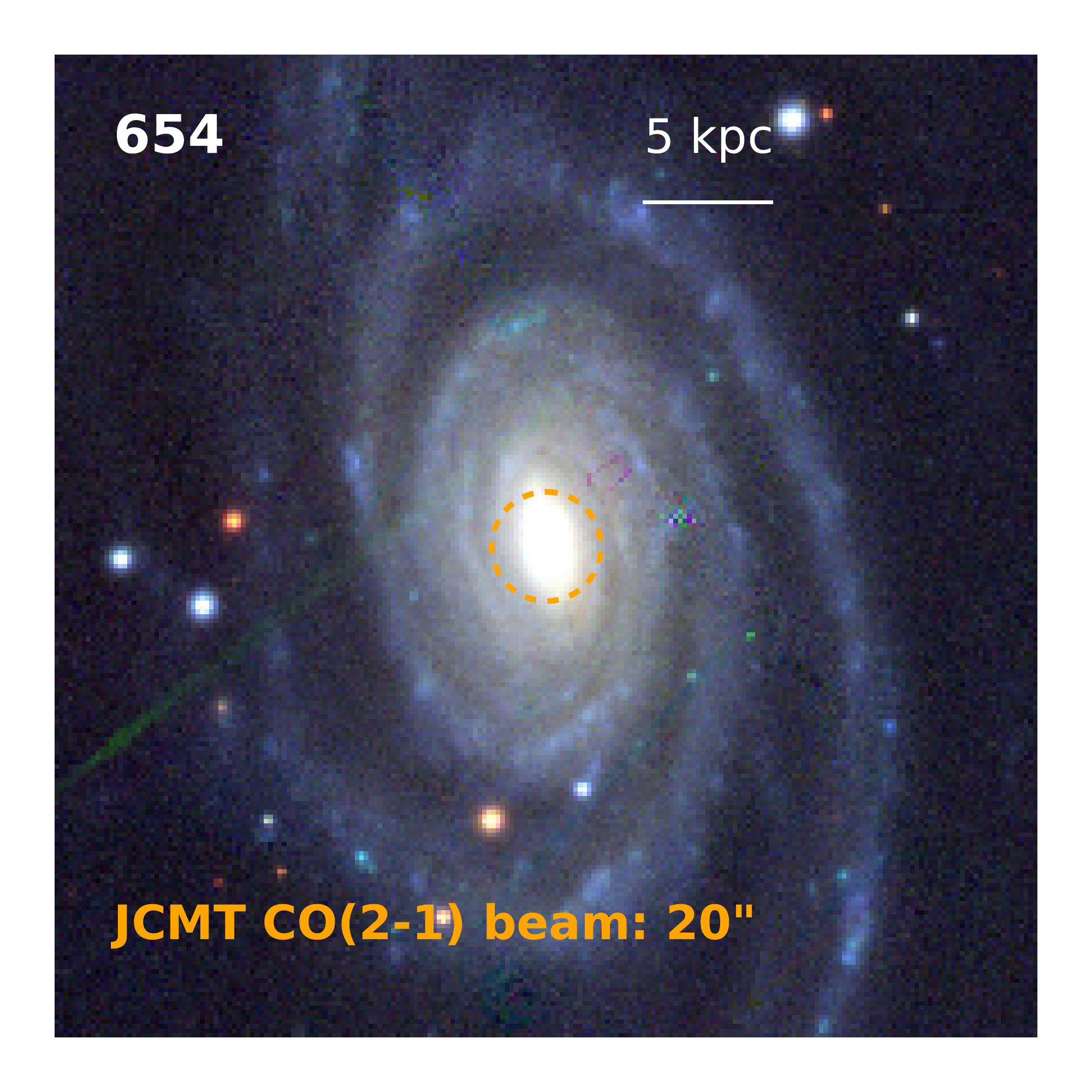}\includegraphics[width=0.26\textwidth]{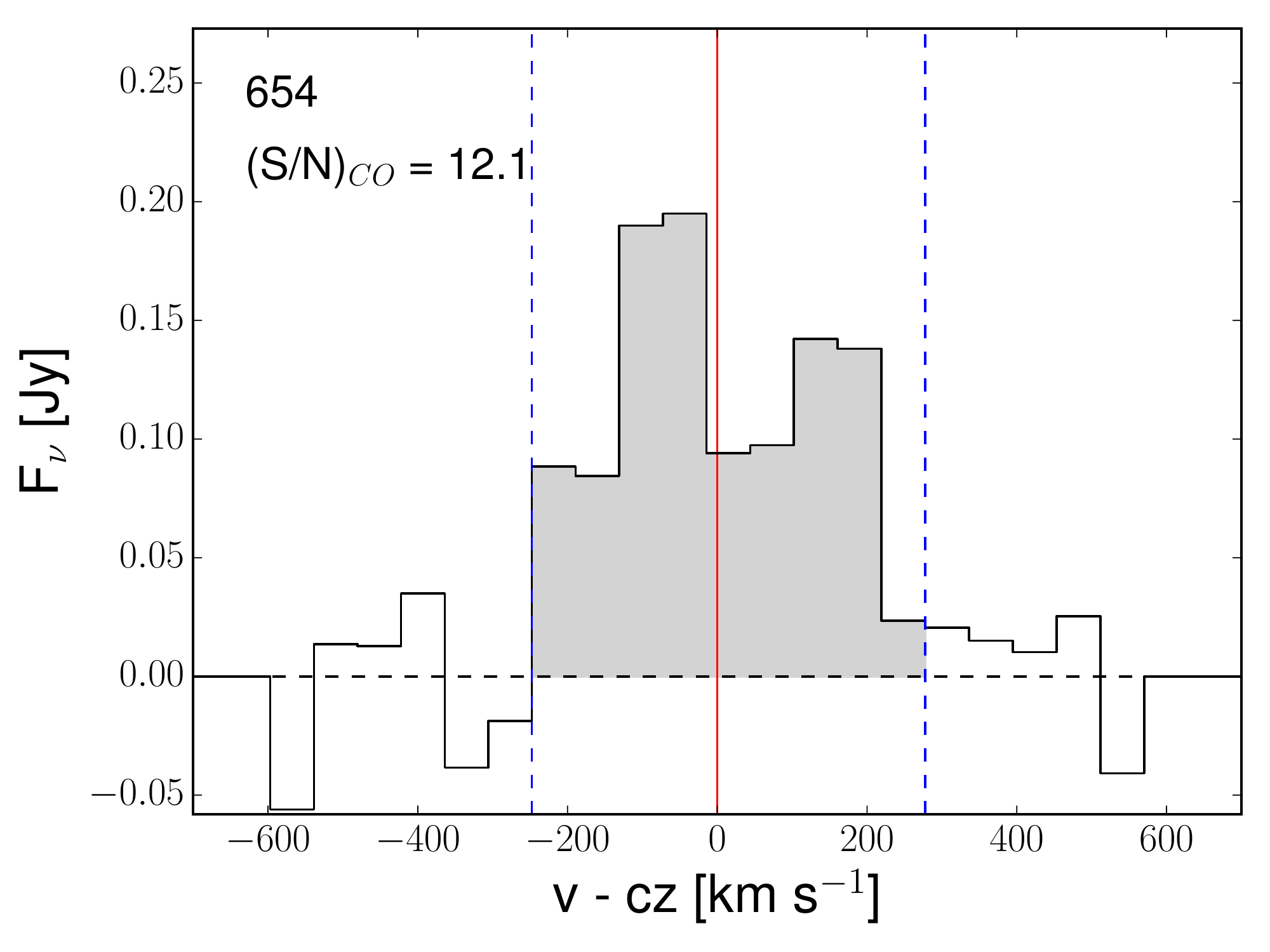}
\caption{continued from Fig.~\ref{fig:CO21_spectra_all_1}
} 
\label{fig:CO21_spectra_all_4}
\end{figure*}

\begin{figure*}
\centering
\raggedright
\includegraphics[width=0.18\textwidth]{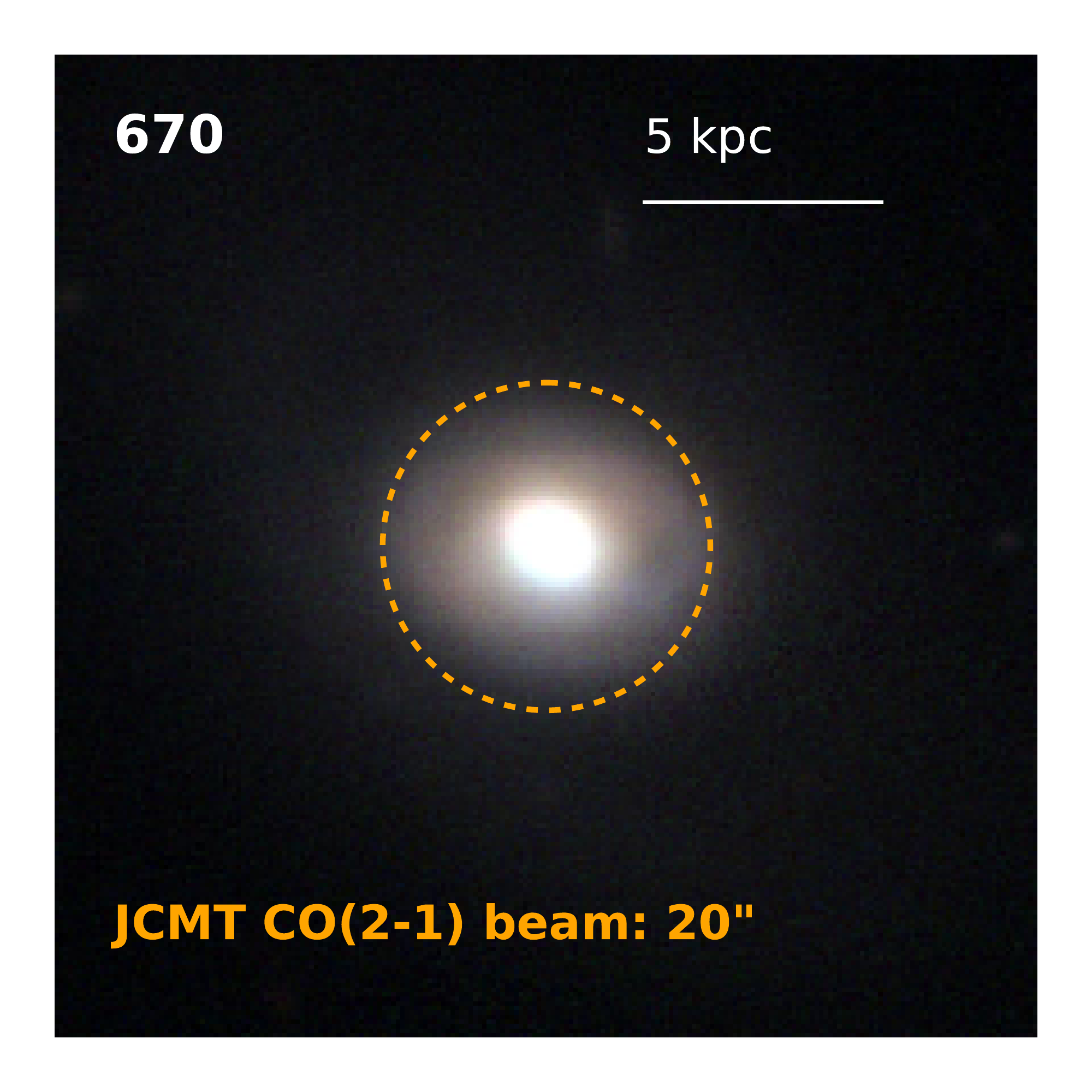}\includegraphics[width=0.26\textwidth]{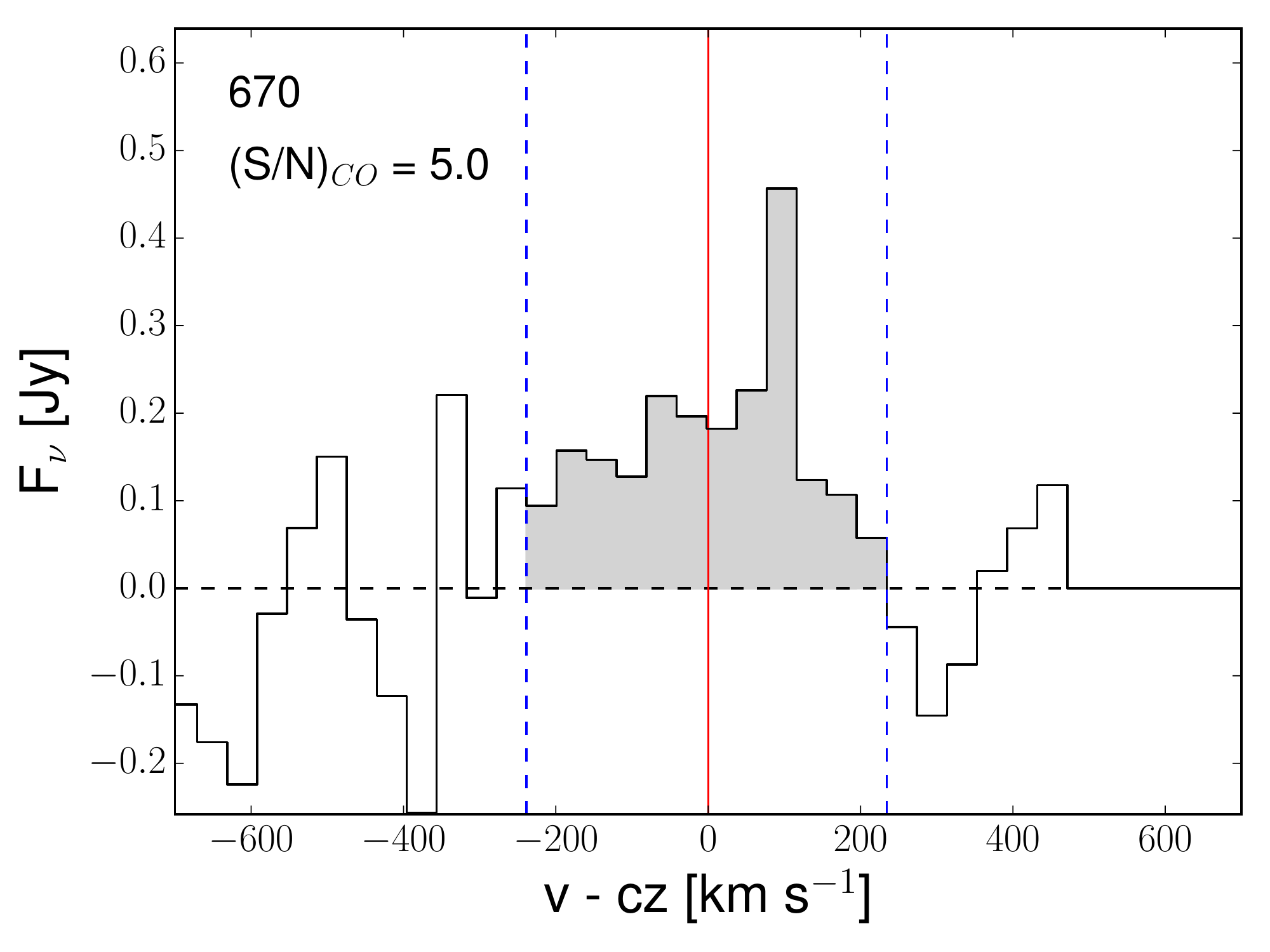}
\includegraphics[width=0.18\textwidth]{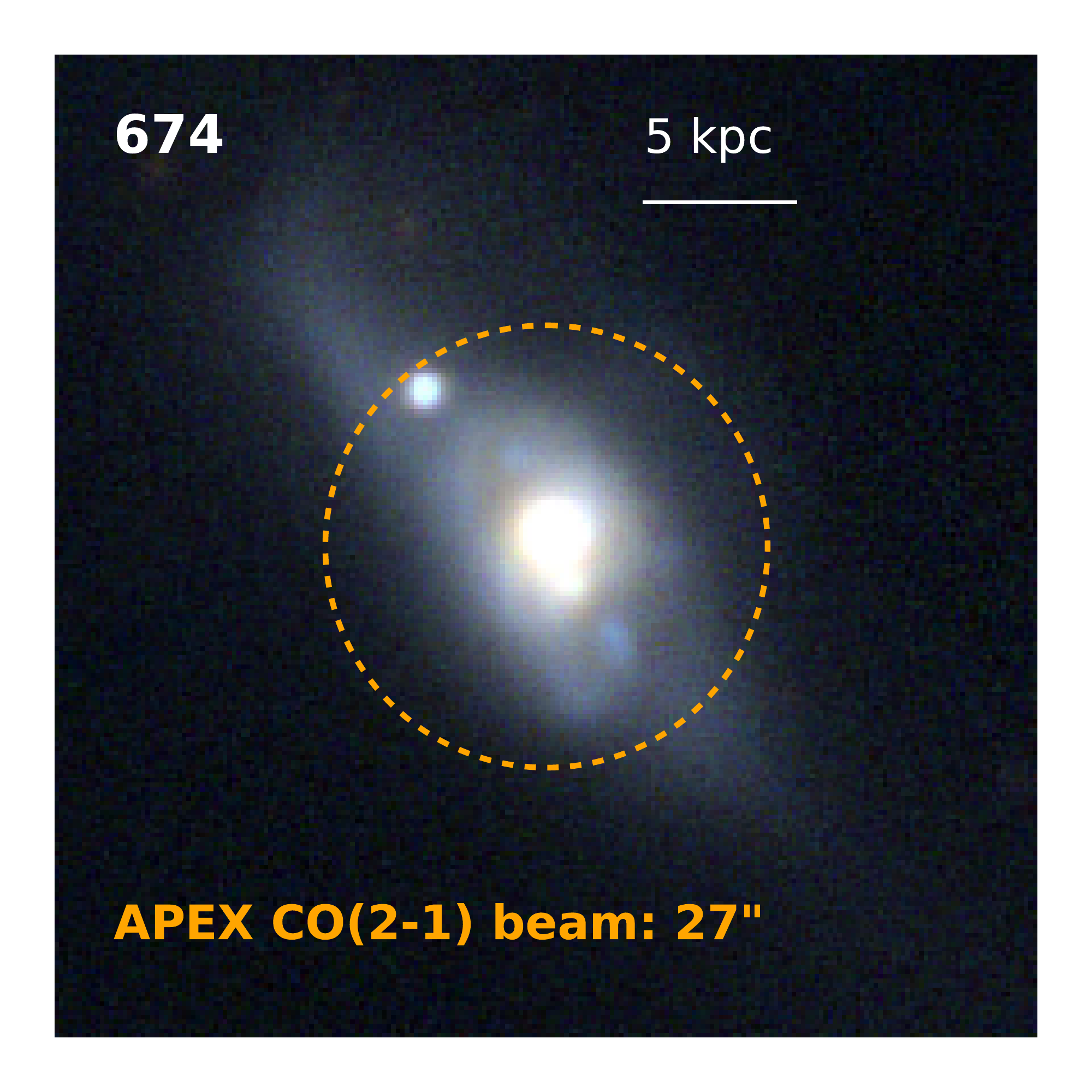}\includegraphics[width=0.26\textwidth]{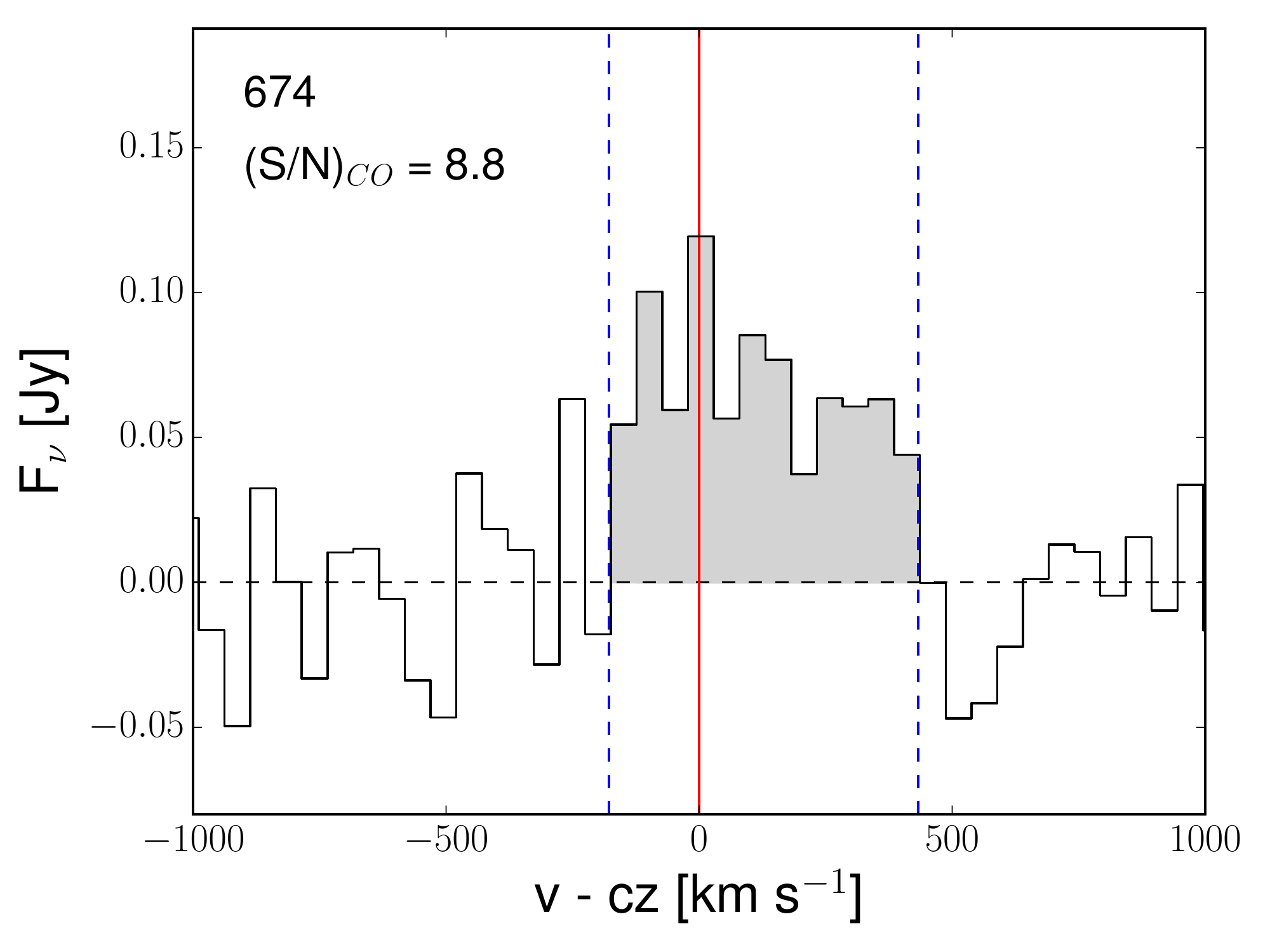}
\includegraphics[width=0.18\textwidth]{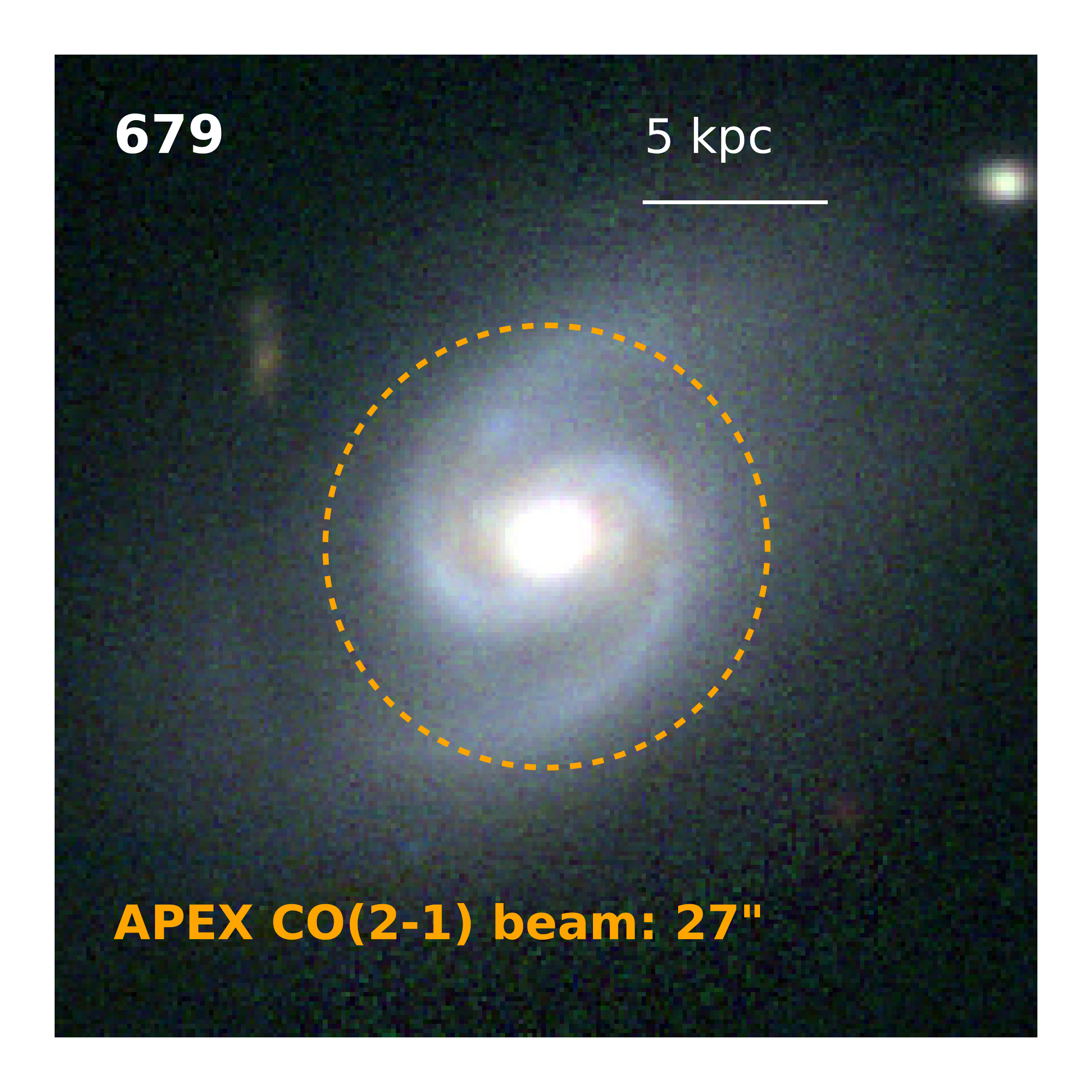}\includegraphics[width=0.26\textwidth]{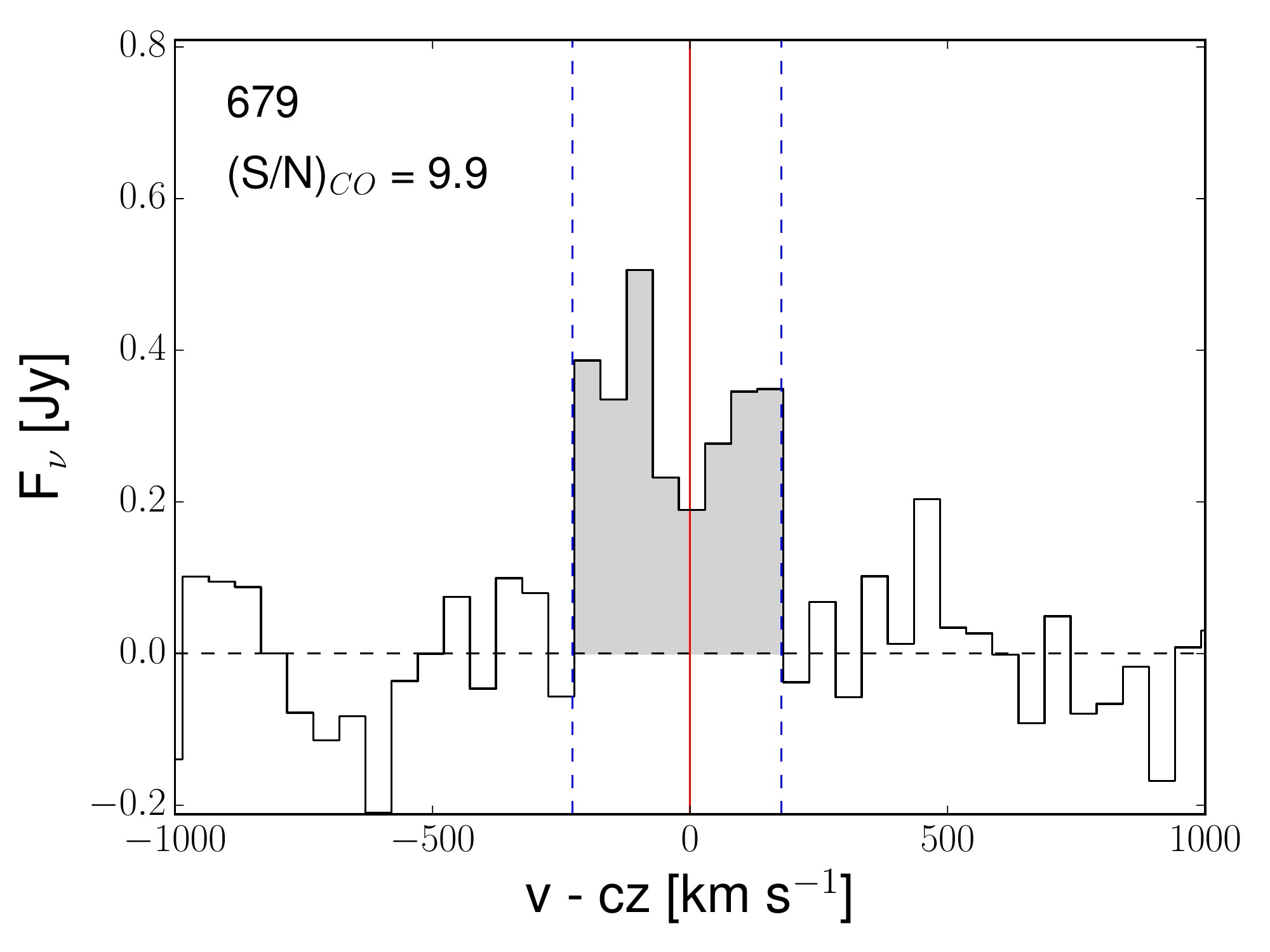}
\includegraphics[width=0.18\textwidth]{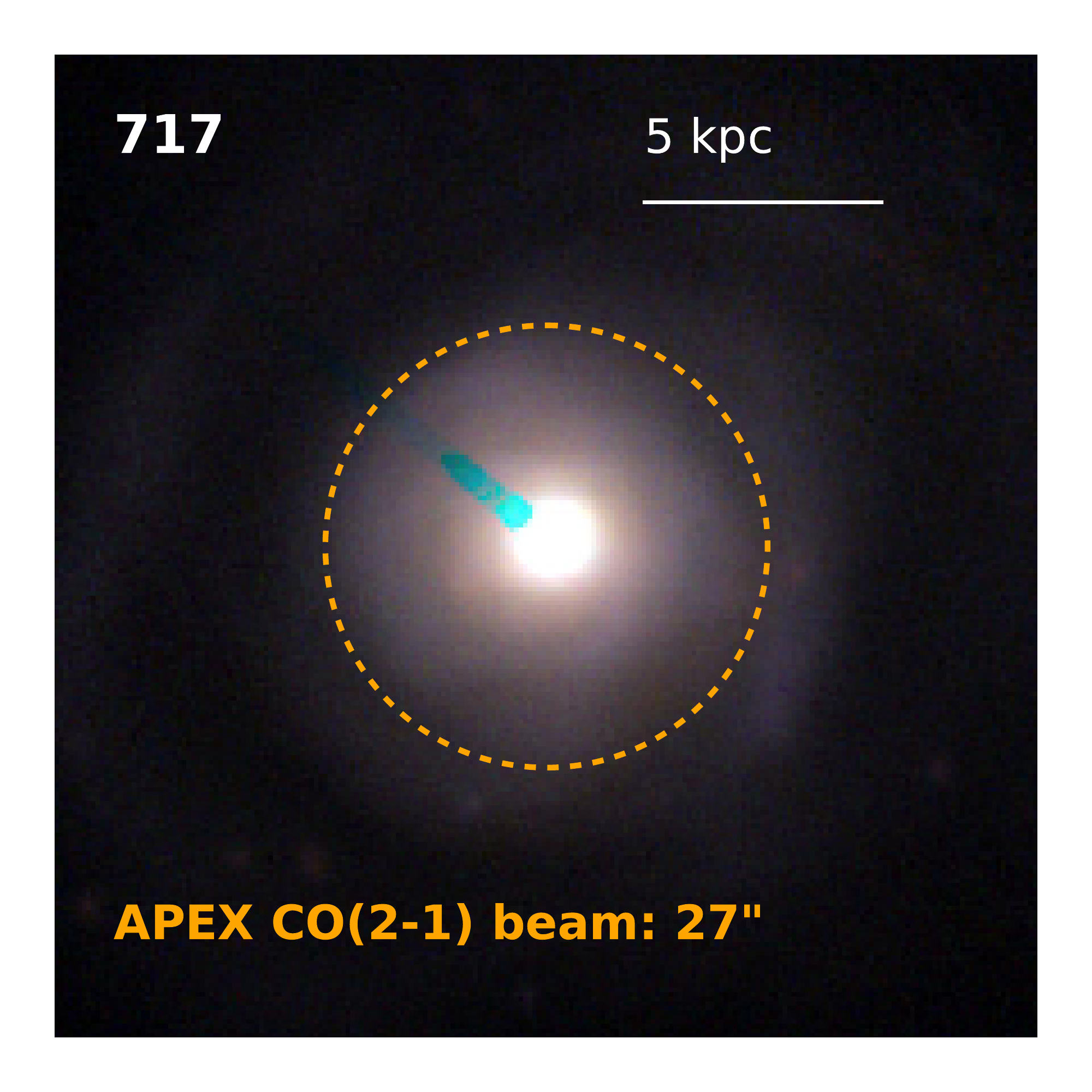}\includegraphics[width=0.26\textwidth]{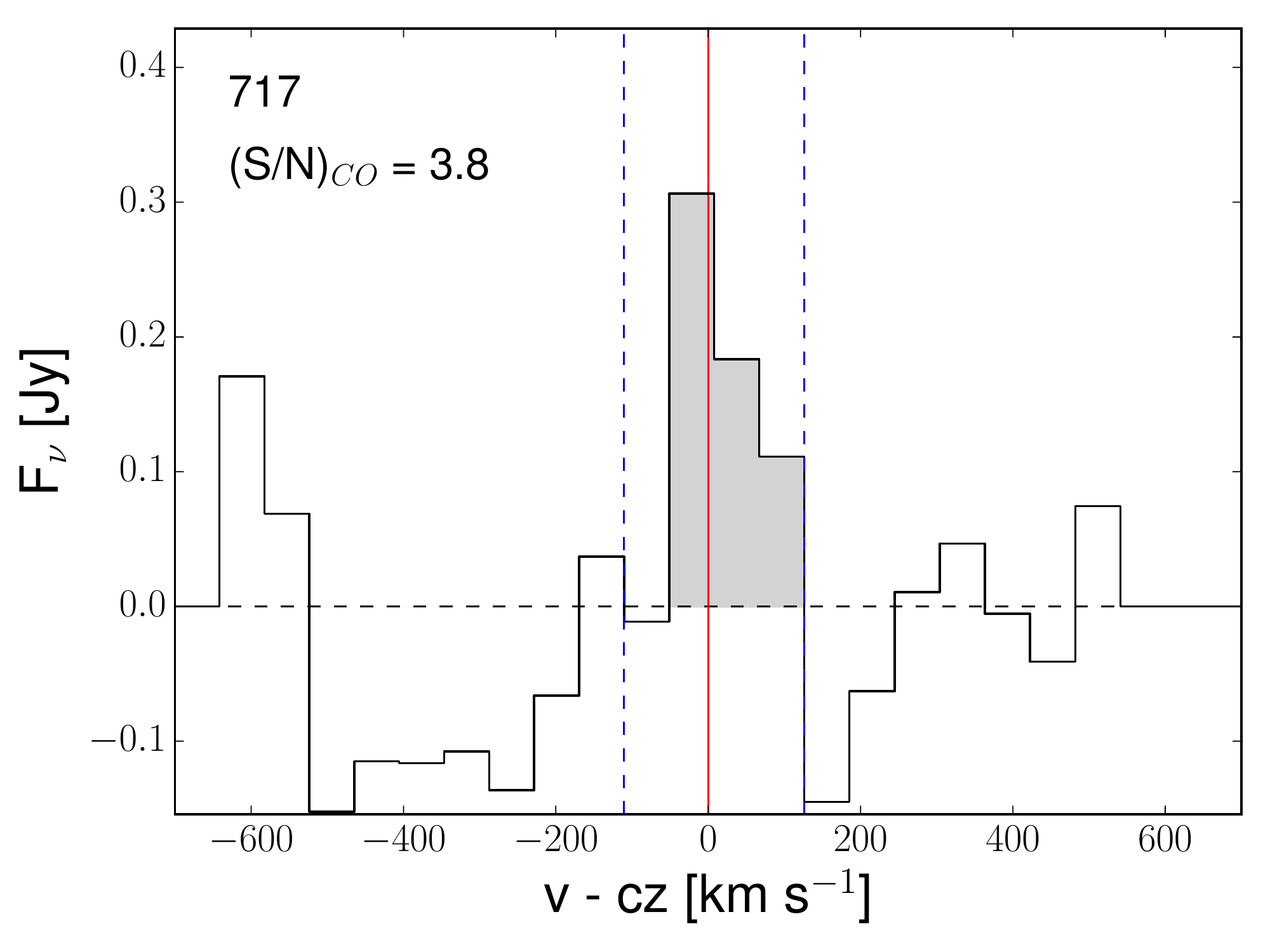}
\includegraphics[width=0.18\textwidth]{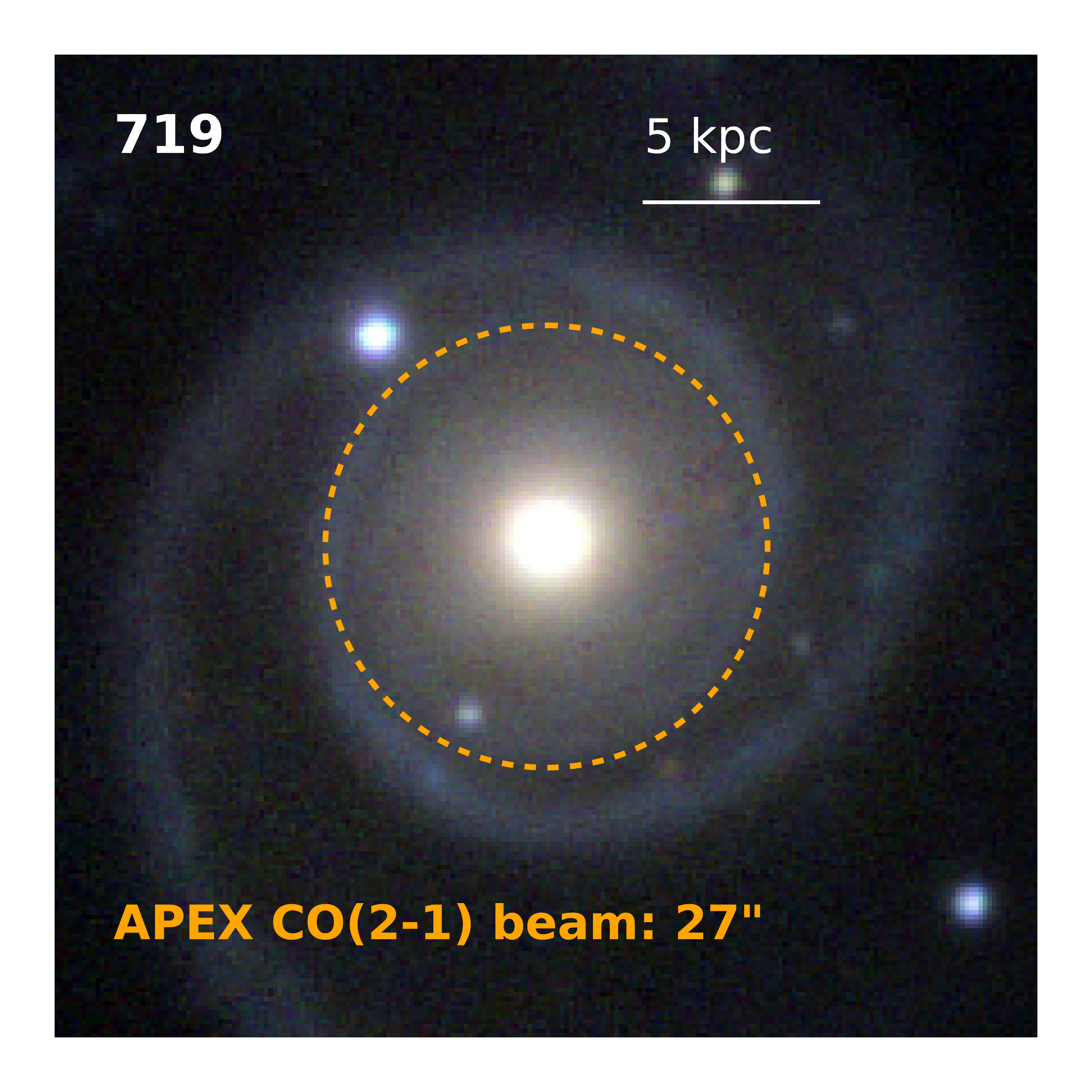}\includegraphics[width=0.26\textwidth]{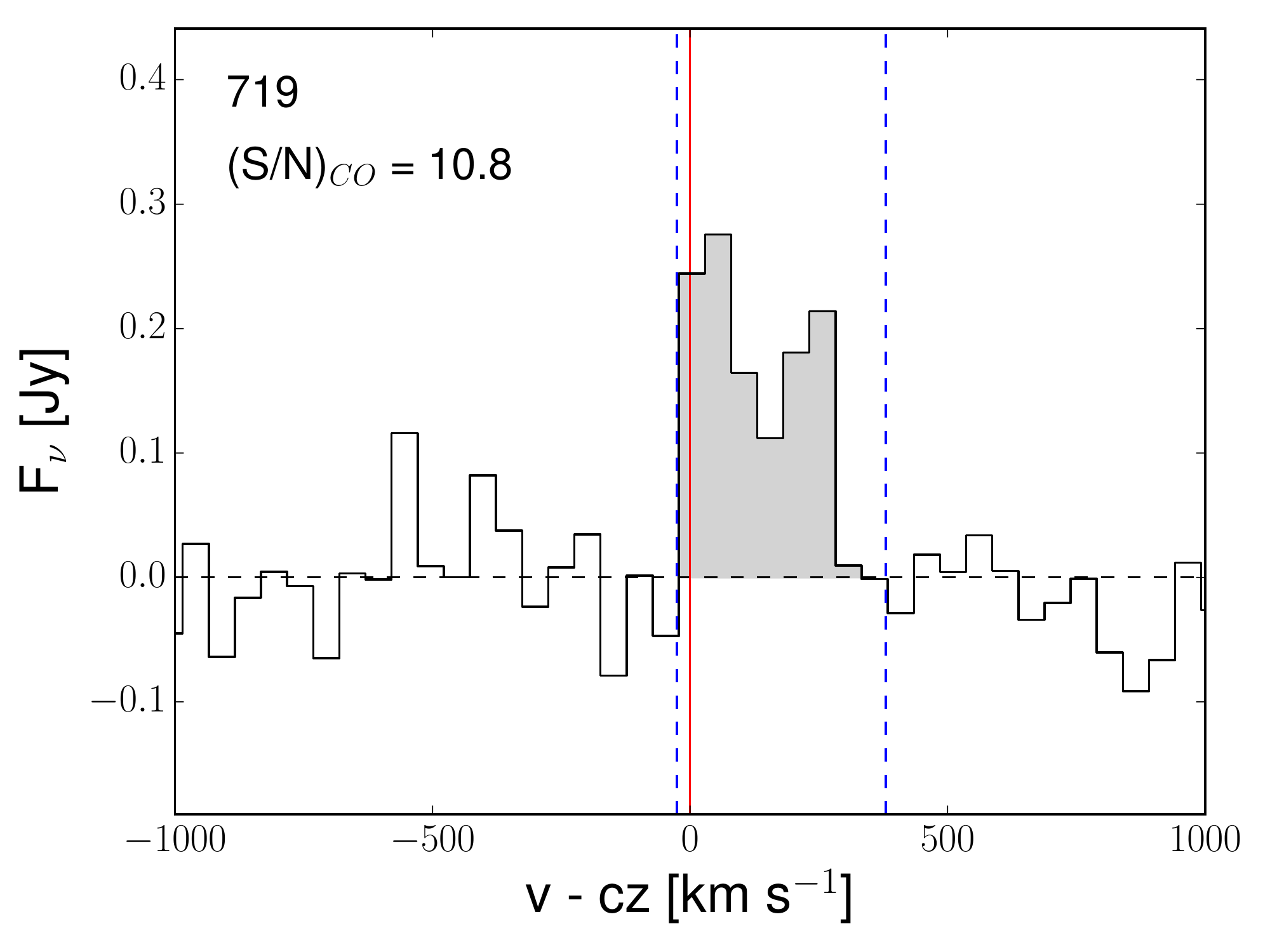}
\includegraphics[width=0.18\textwidth]{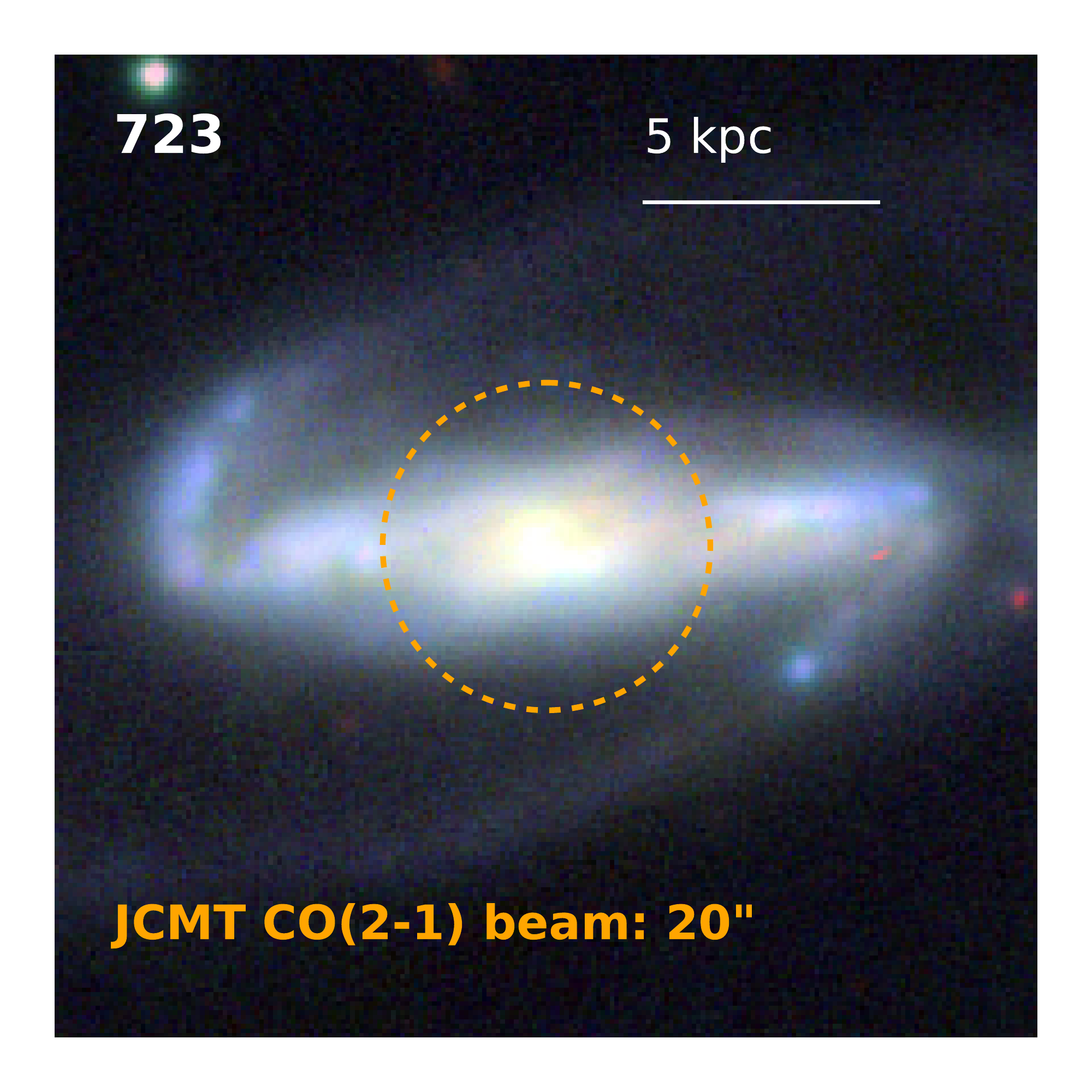}\includegraphics[width=0.26\textwidth]{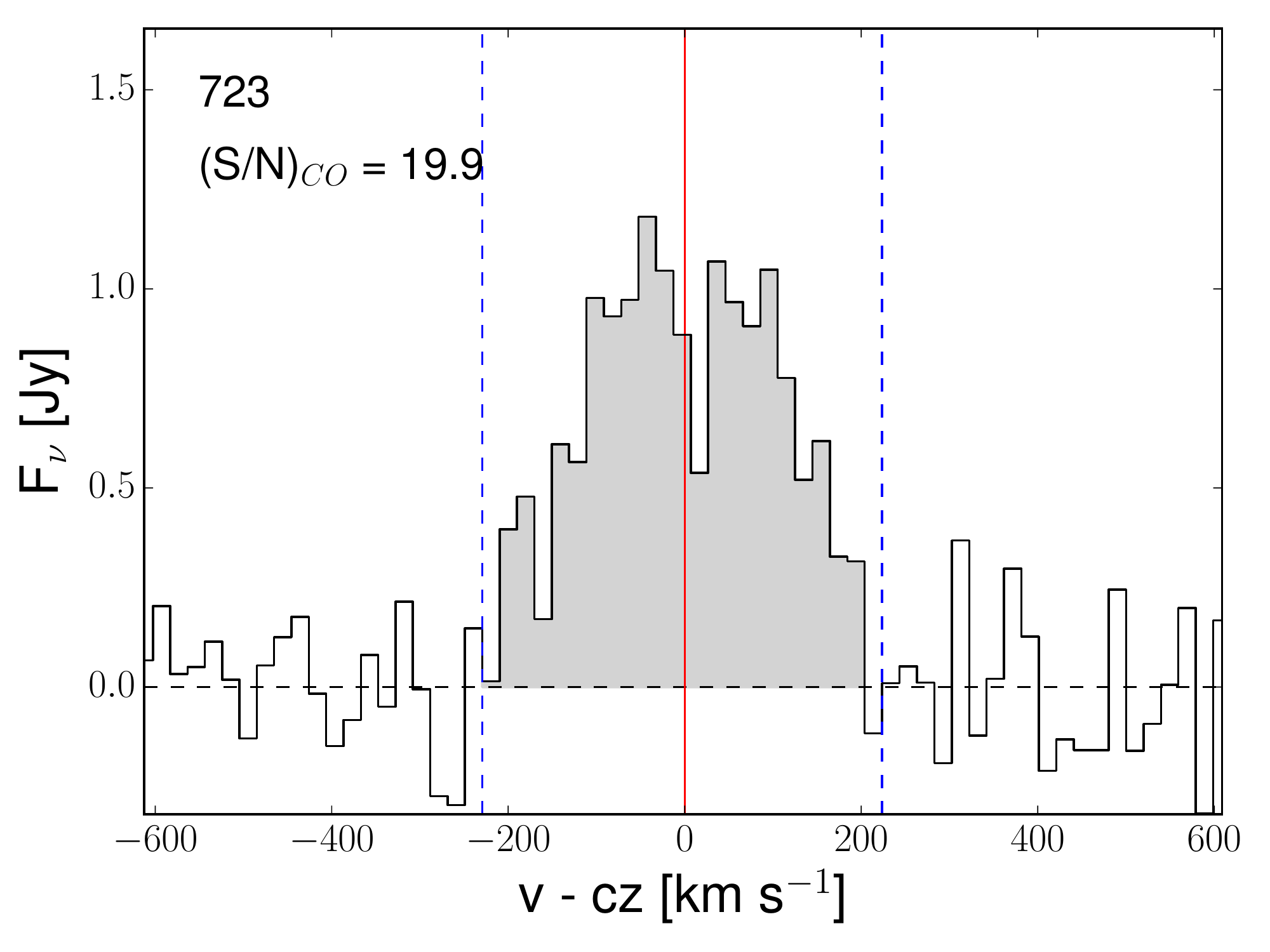}
\includegraphics[width=0.18\textwidth]{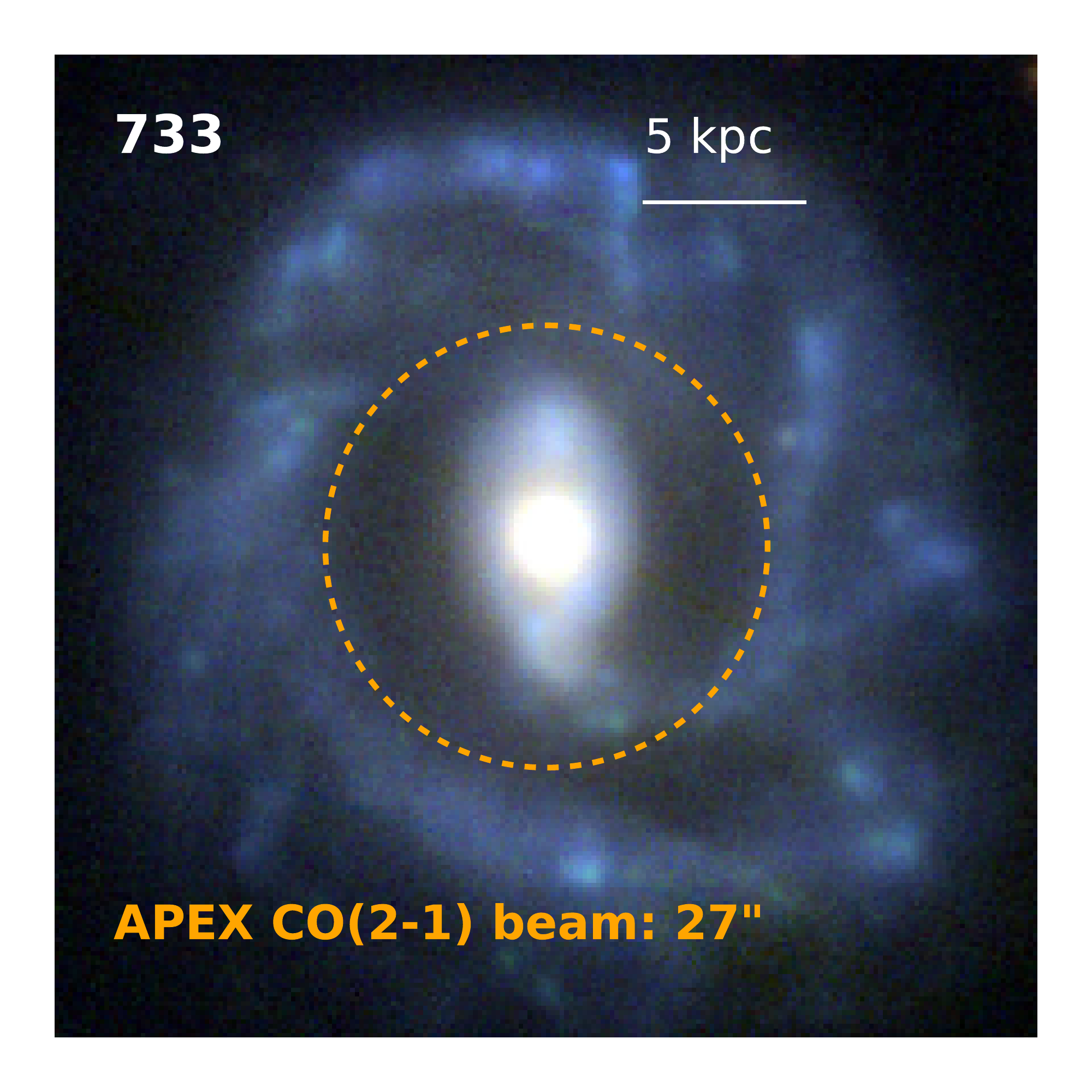}\includegraphics[width=0.26\textwidth]{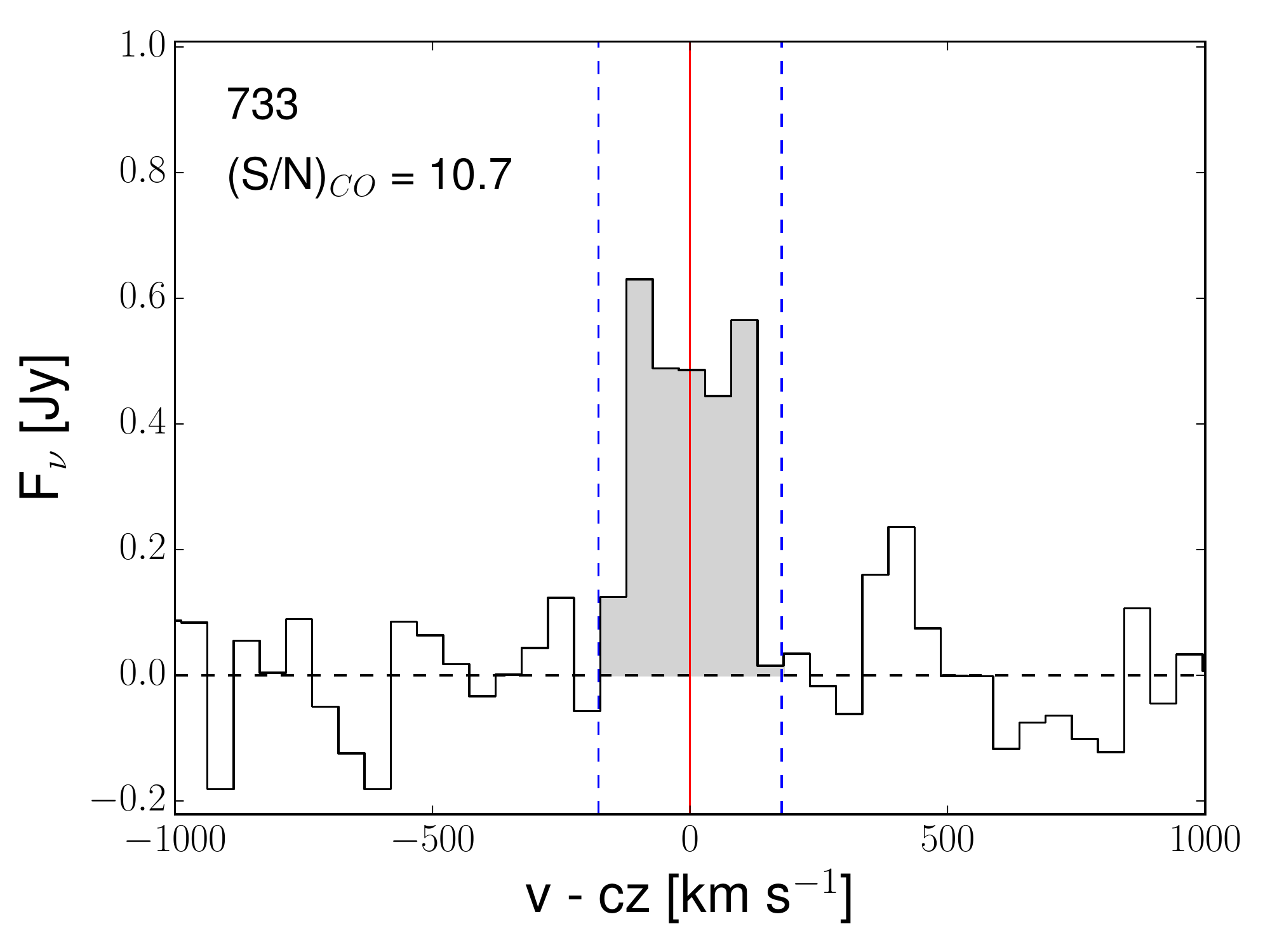}
\includegraphics[width=0.18\textwidth]{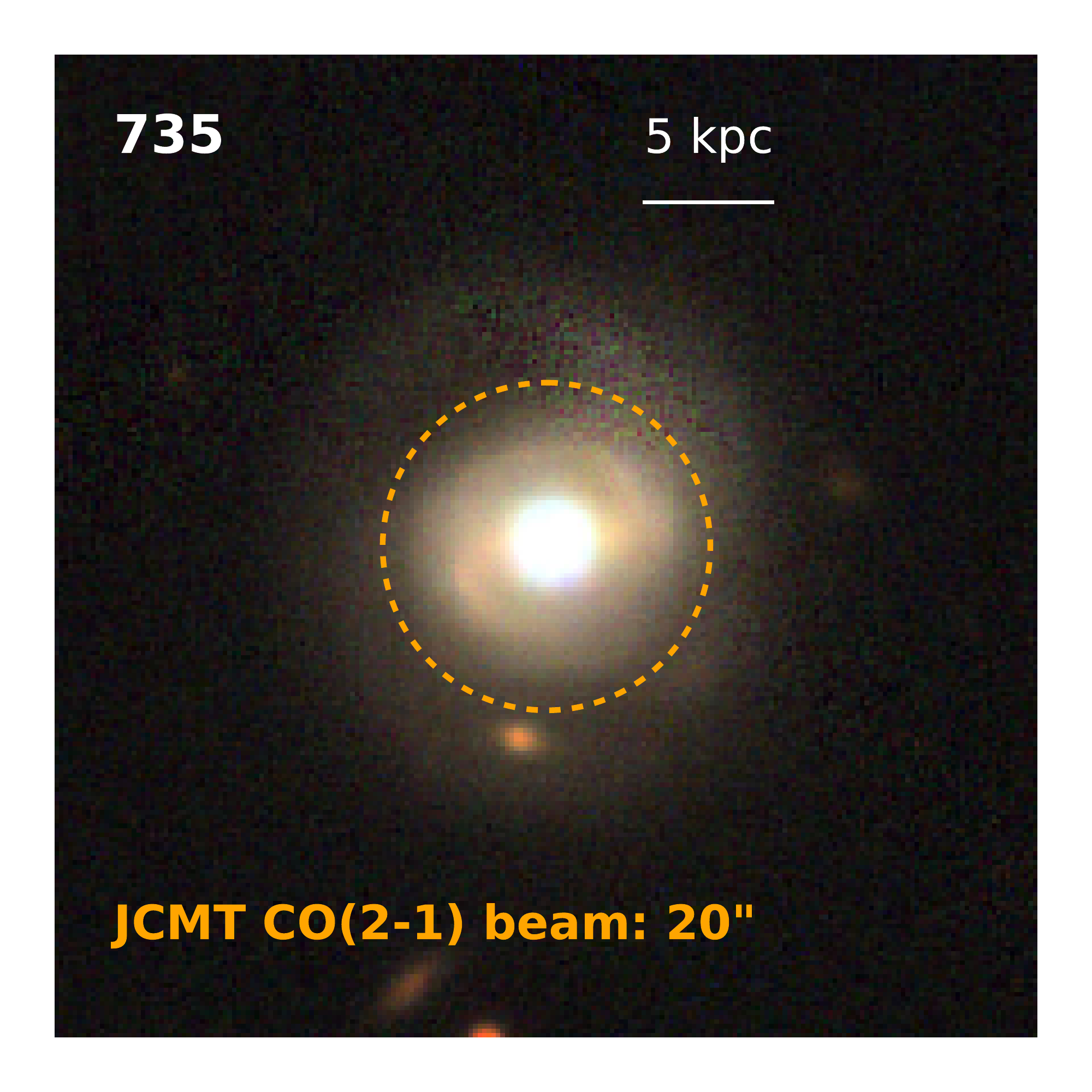}\includegraphics[width=0.26\textwidth]{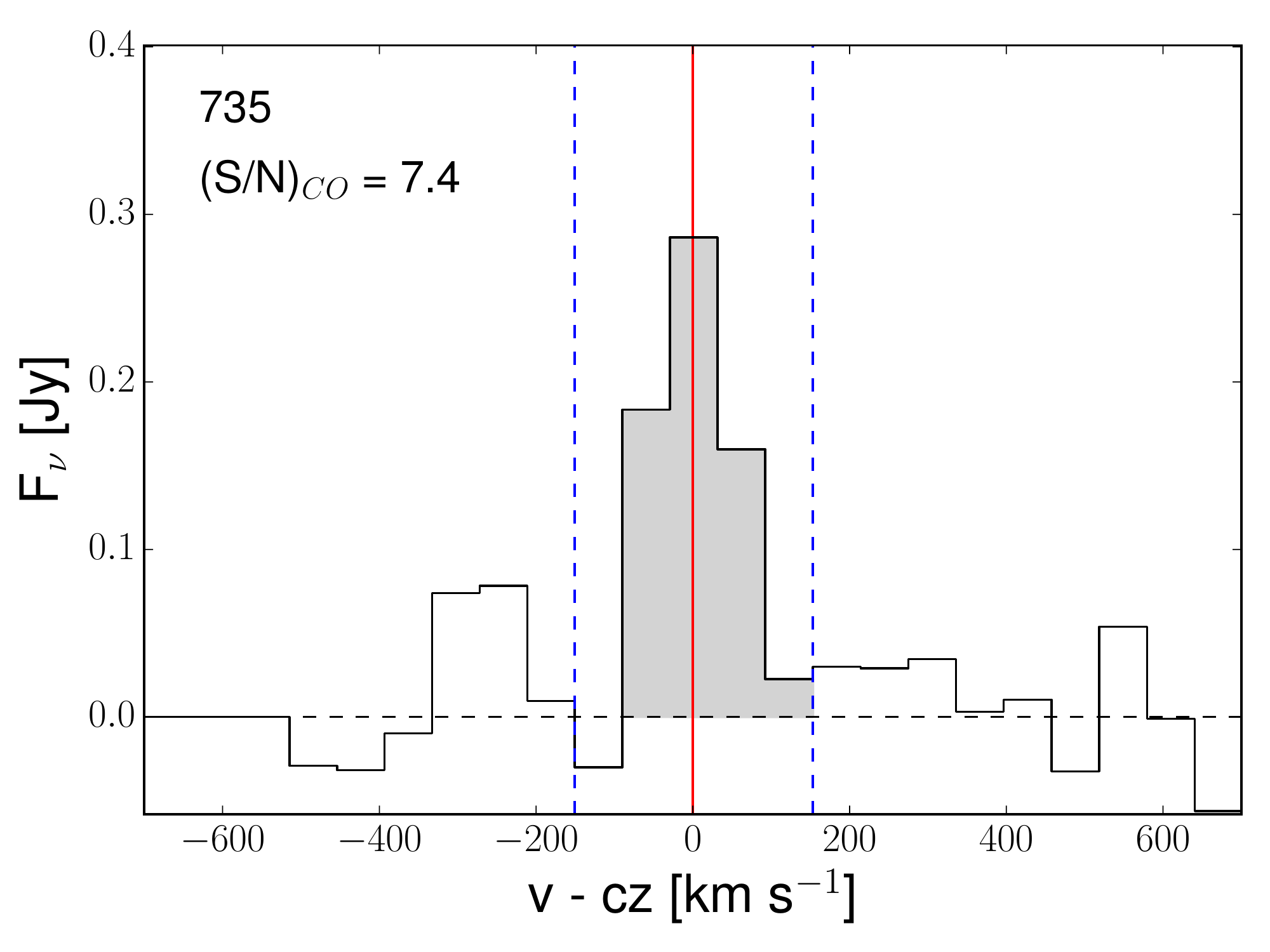}
\includegraphics[width=0.18\textwidth]{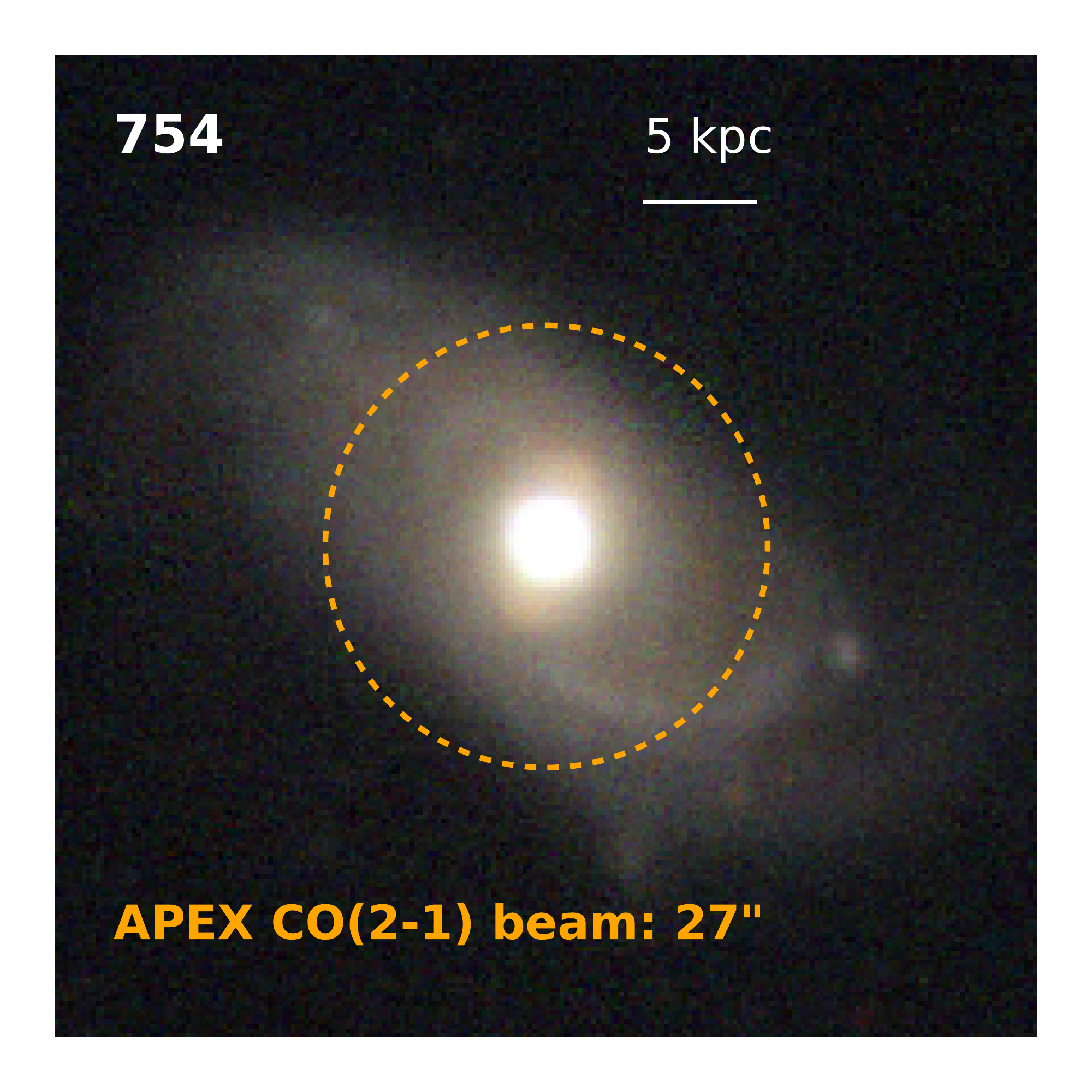}\includegraphics[width=0.26\textwidth]{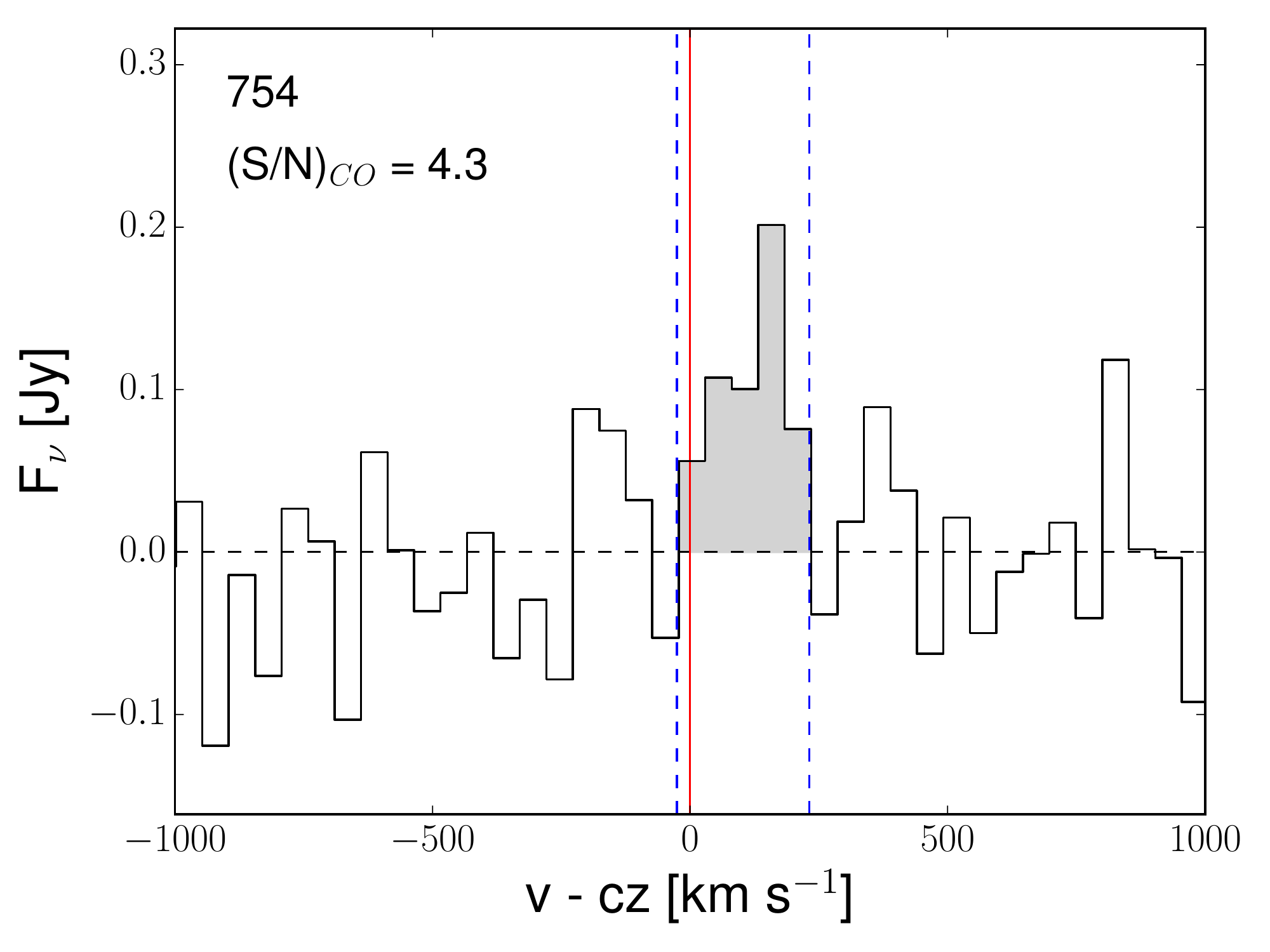}
\includegraphics[width=0.18\textwidth]{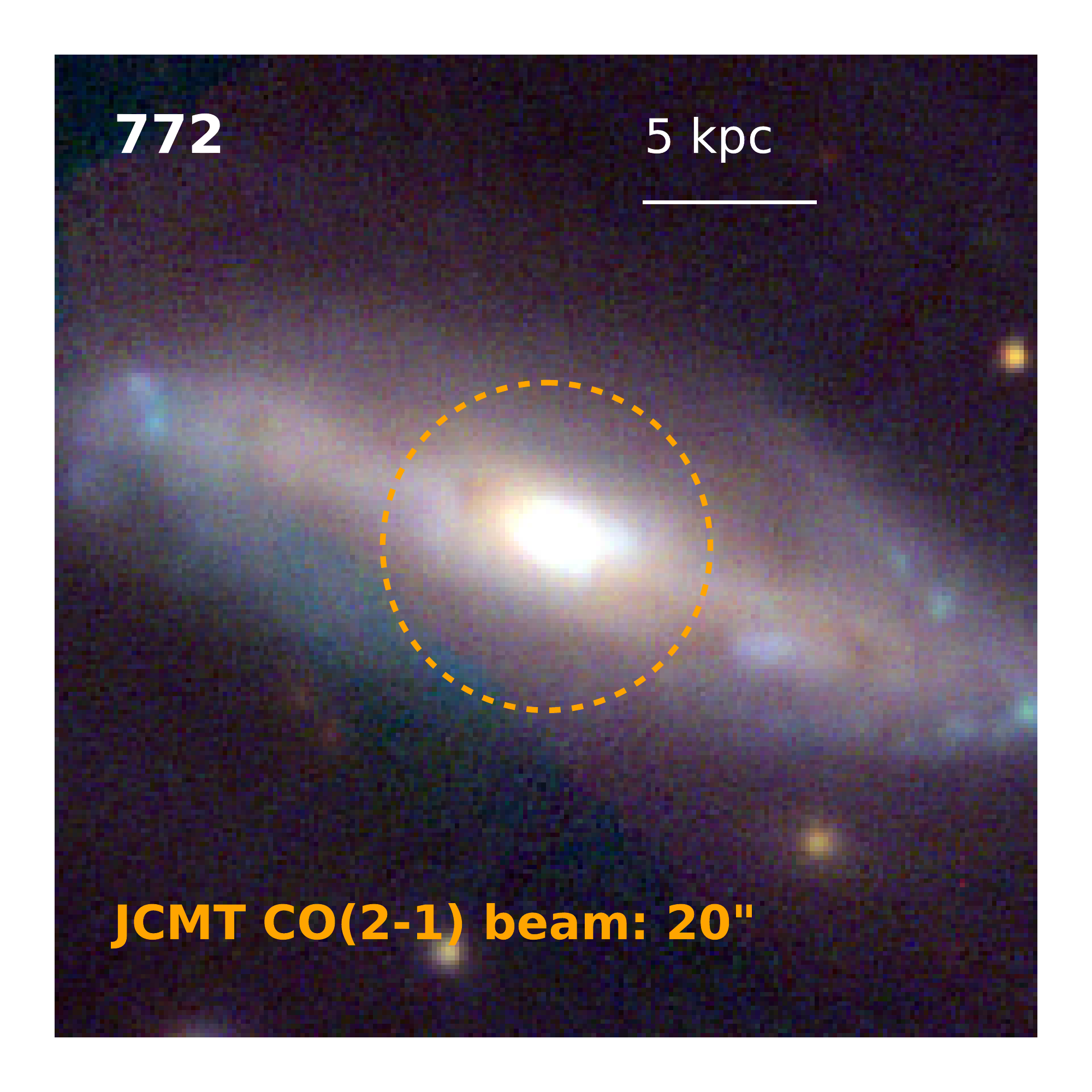}\includegraphics[width=0.26\textwidth]{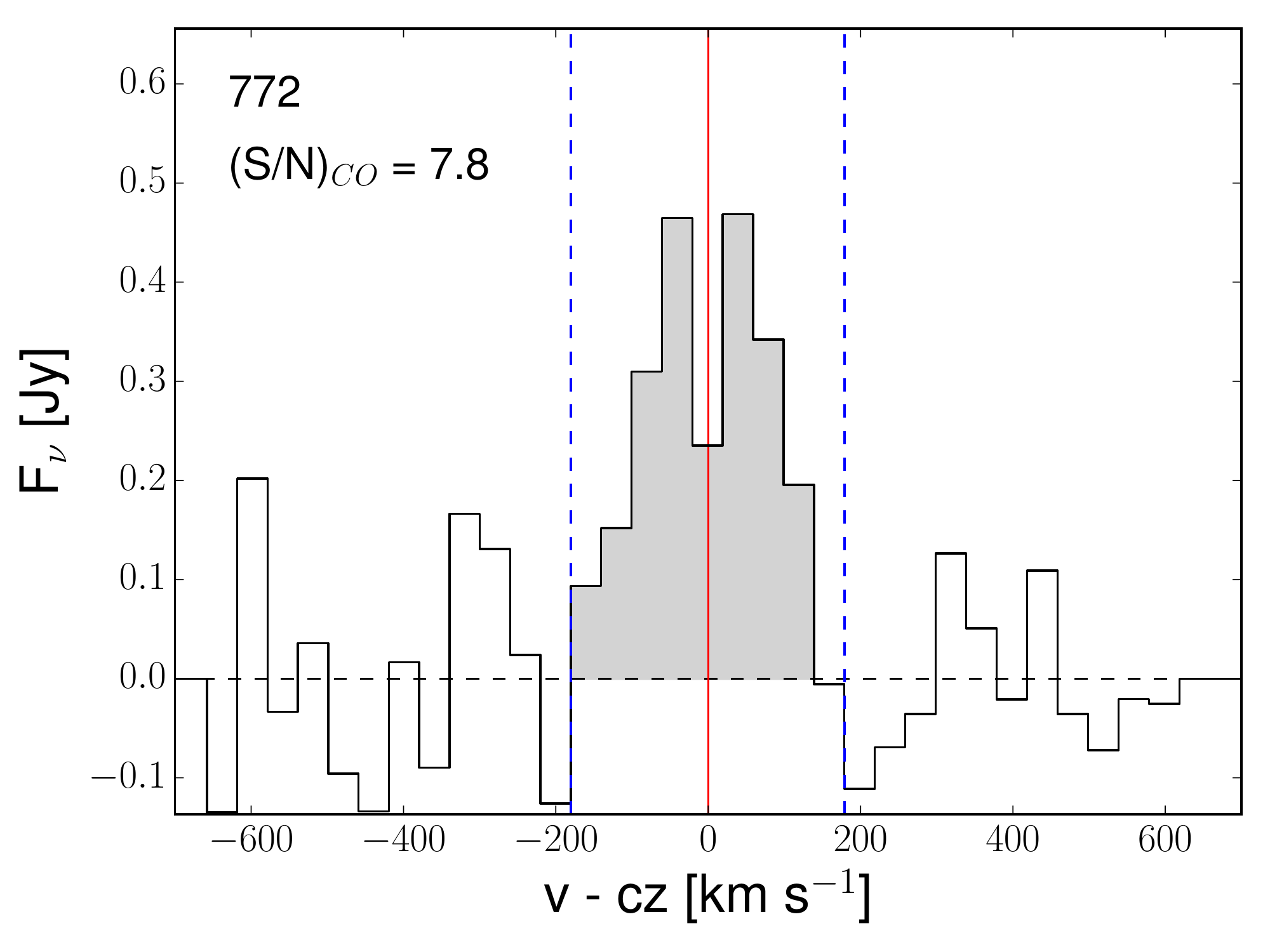}
\includegraphics[width=0.18\textwidth]{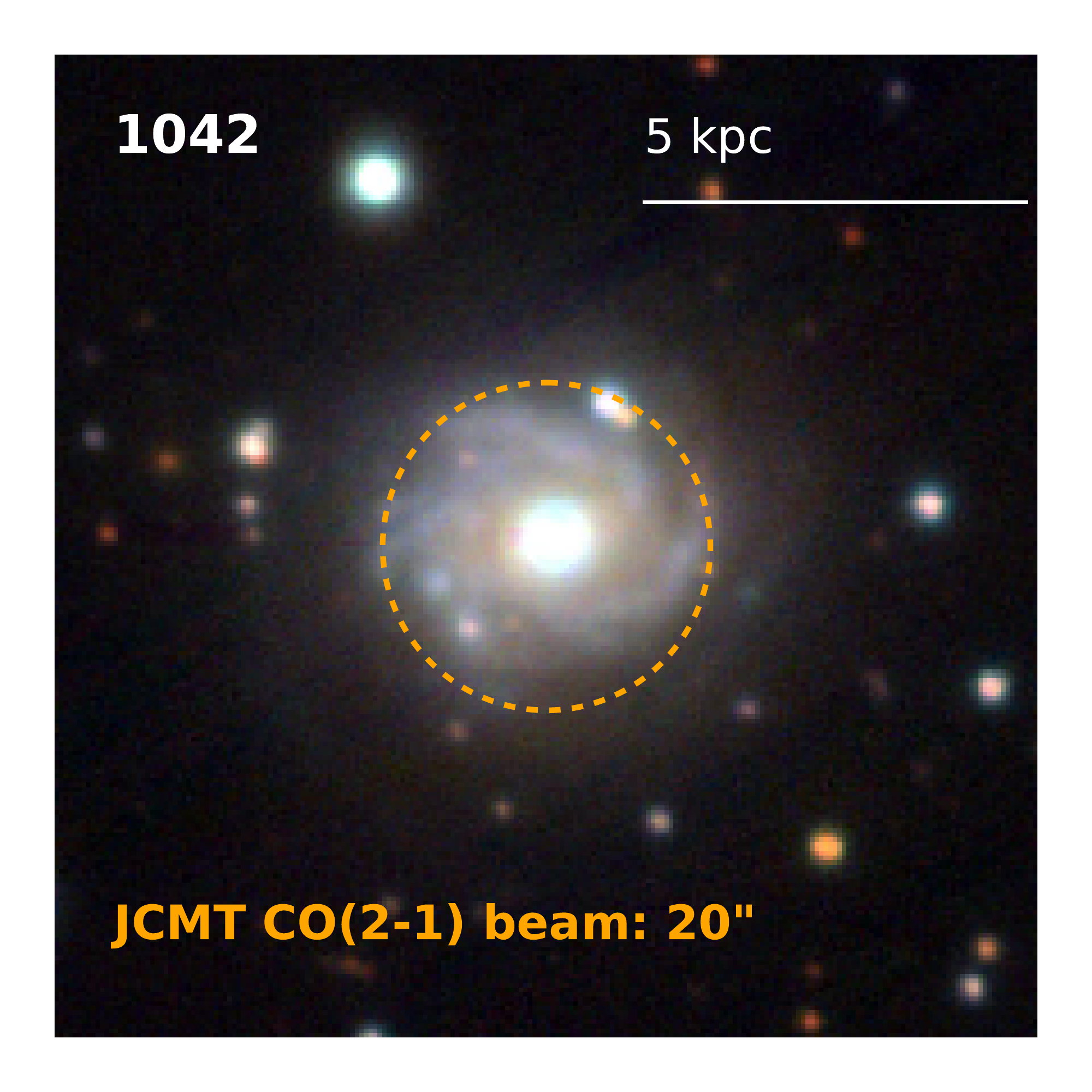}\includegraphics[width=0.26\textwidth]{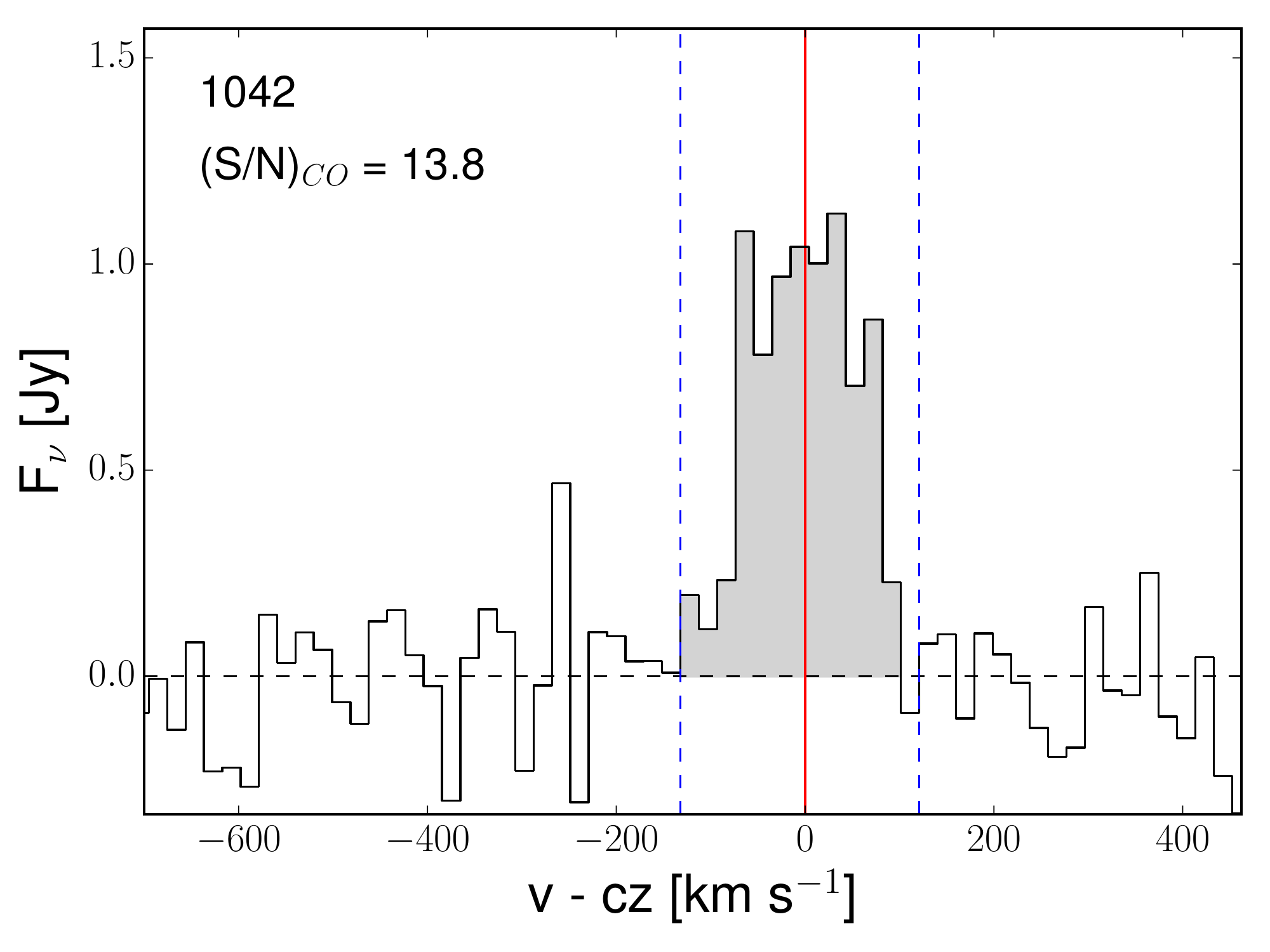}
\includegraphics[width=0.18\textwidth]{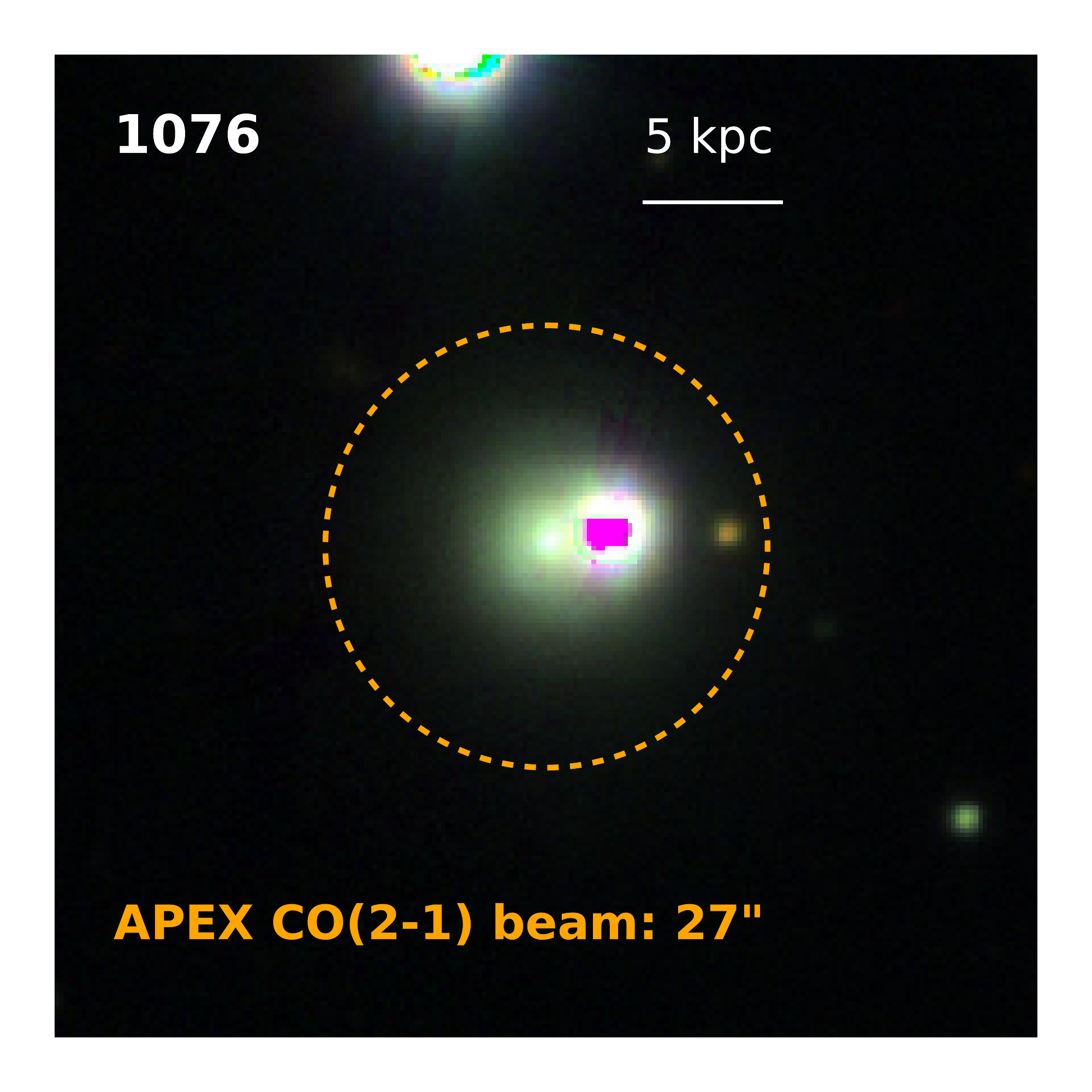}\includegraphics[width=0.26\textwidth]{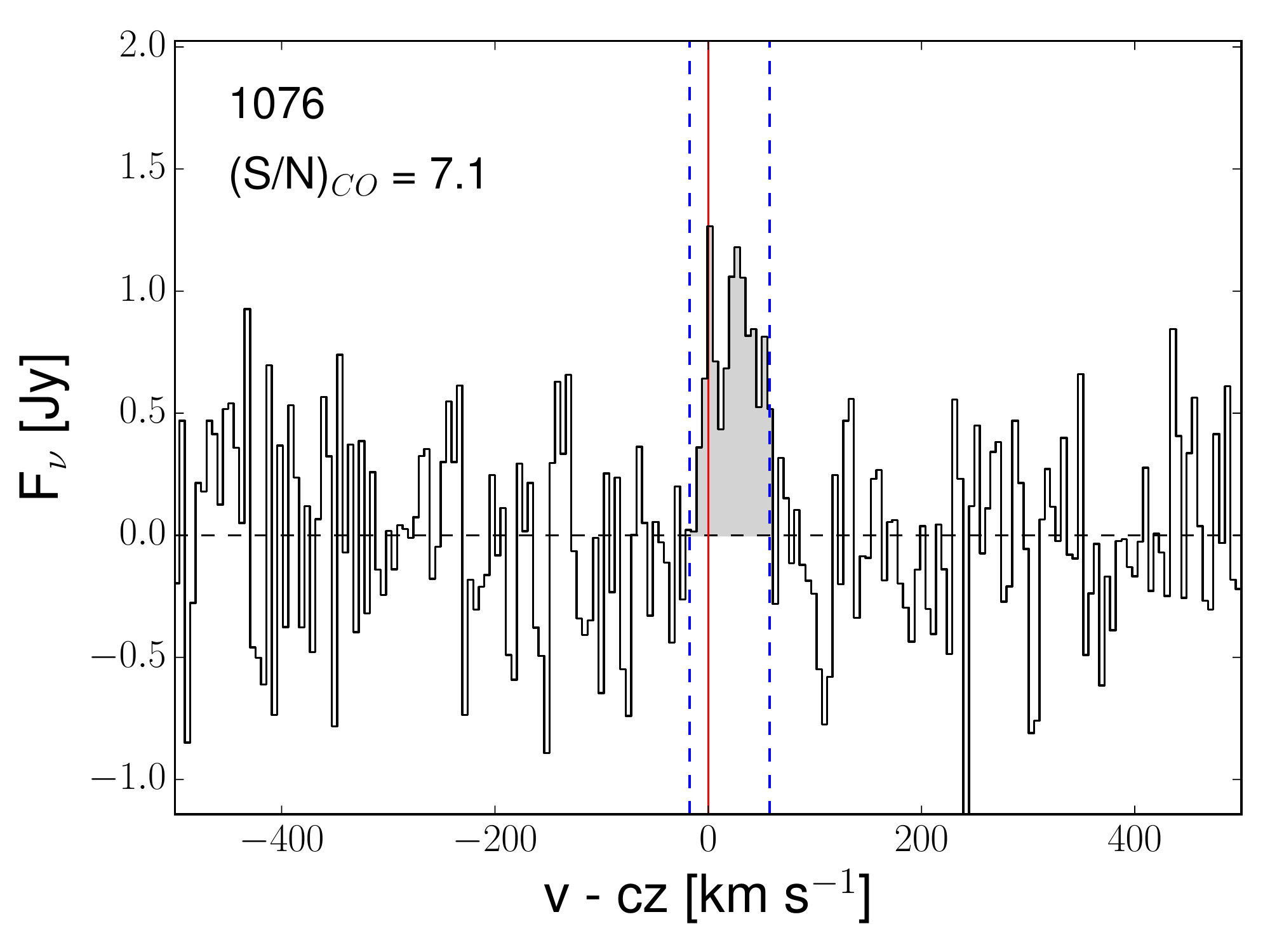}
\caption{continued from Fig.~\ref{fig:CO21_spectra_all_1}
} 
\label{fig:CO21_spectra_all_5}
\end{figure*}

\begin{figure*}
\centering
\raggedright
\includegraphics[width=0.18\textwidth]{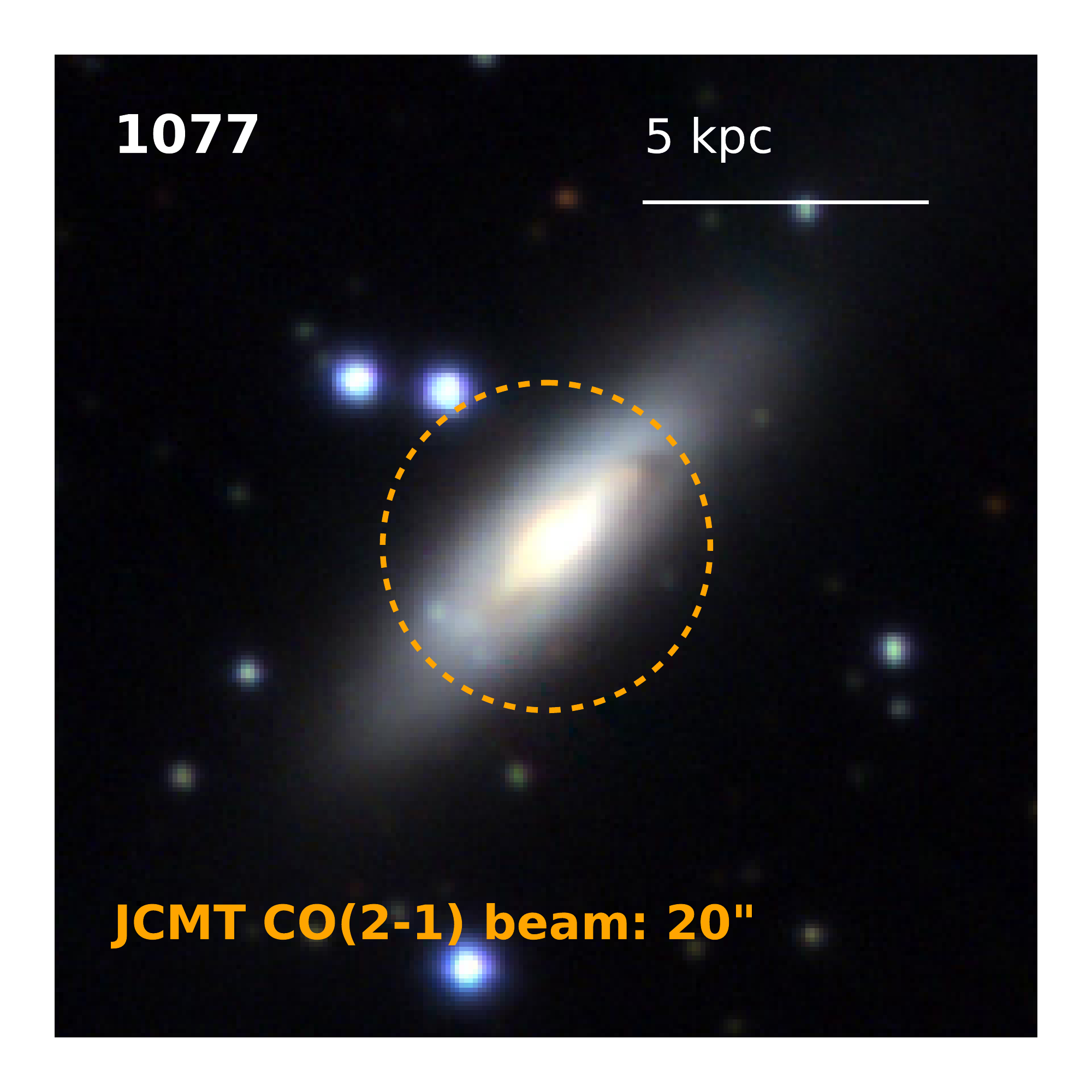}\includegraphics[width=0.26\textwidth]{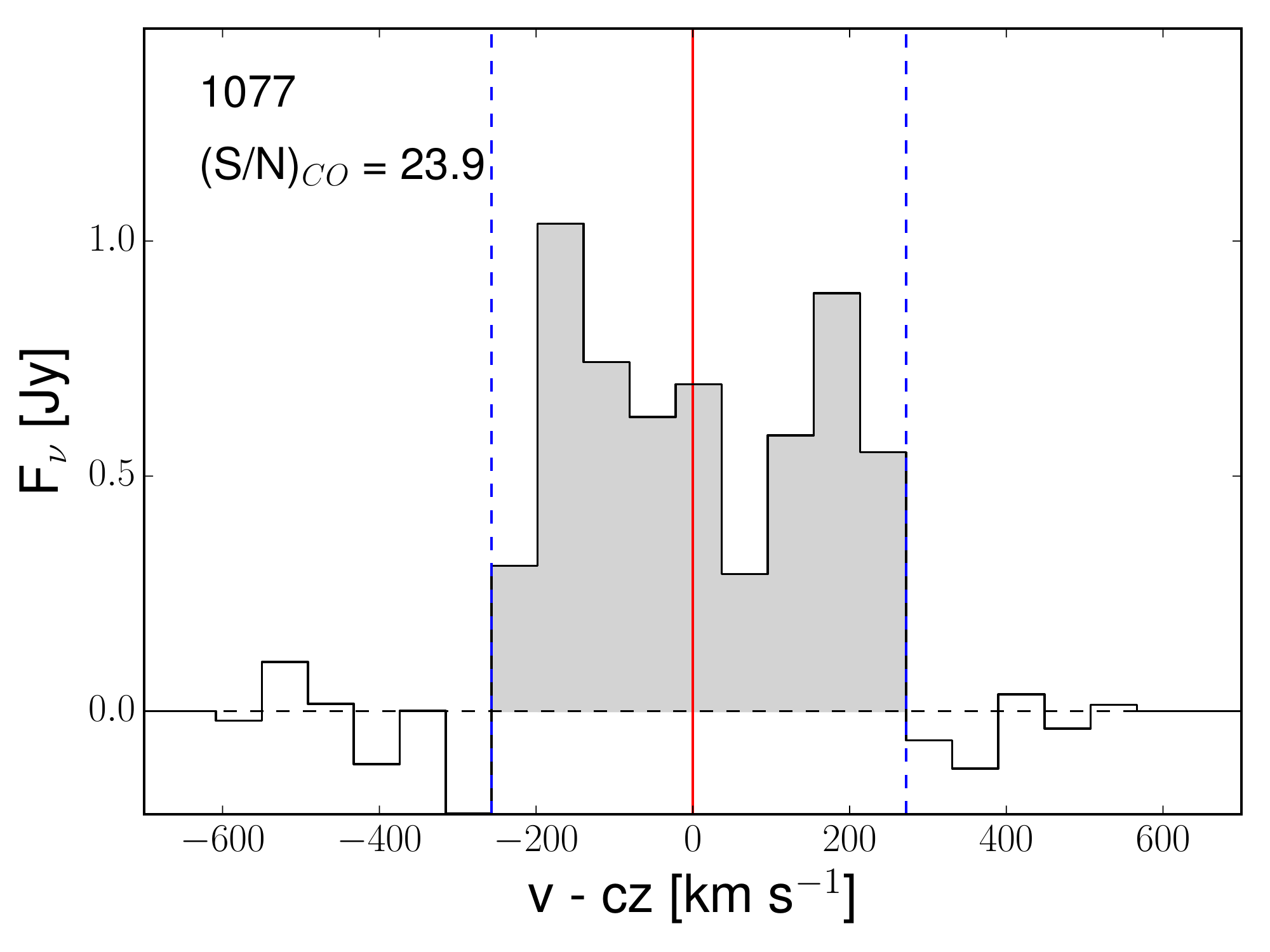}
\includegraphics[width=0.18\textwidth]{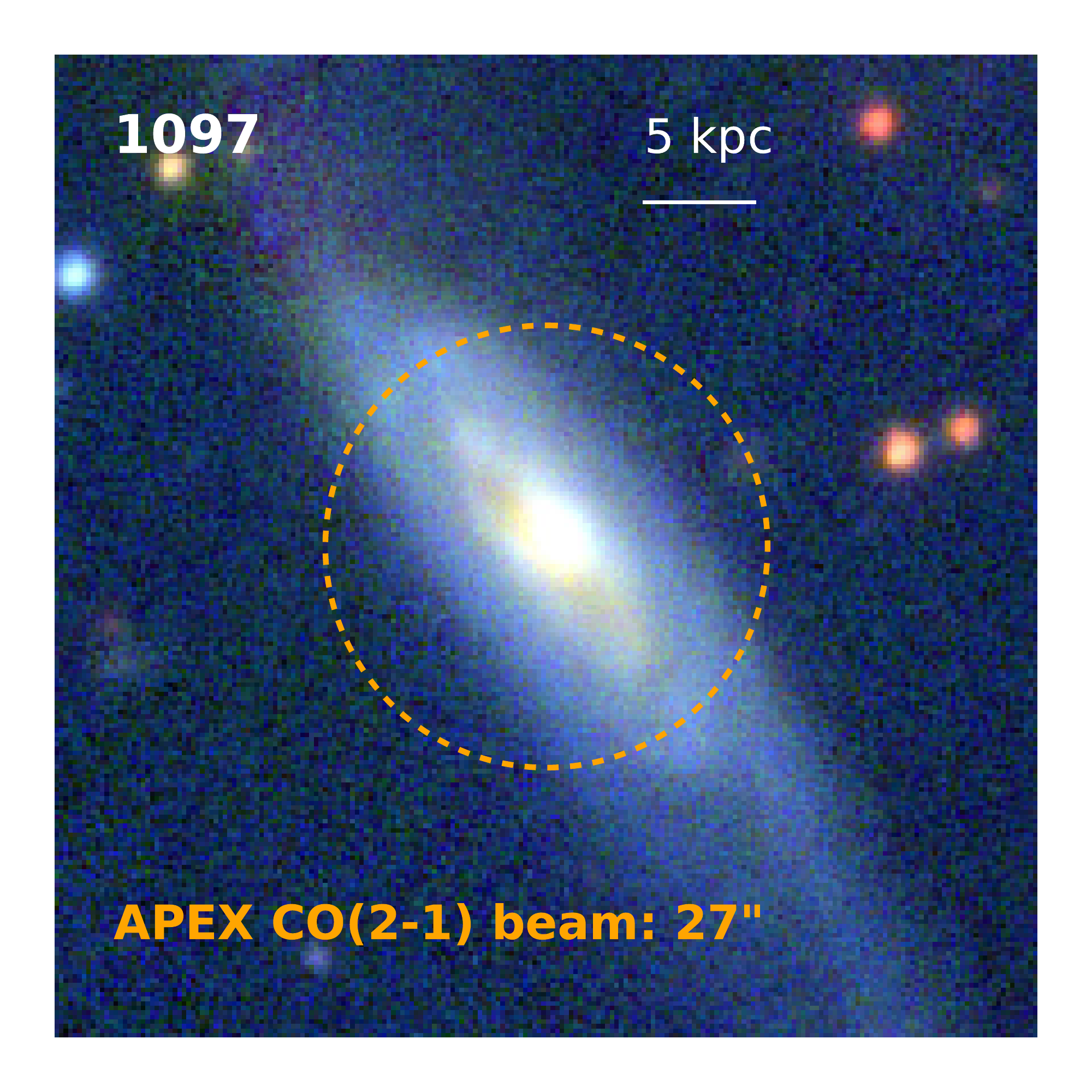}\includegraphics[width=0.26\textwidth]{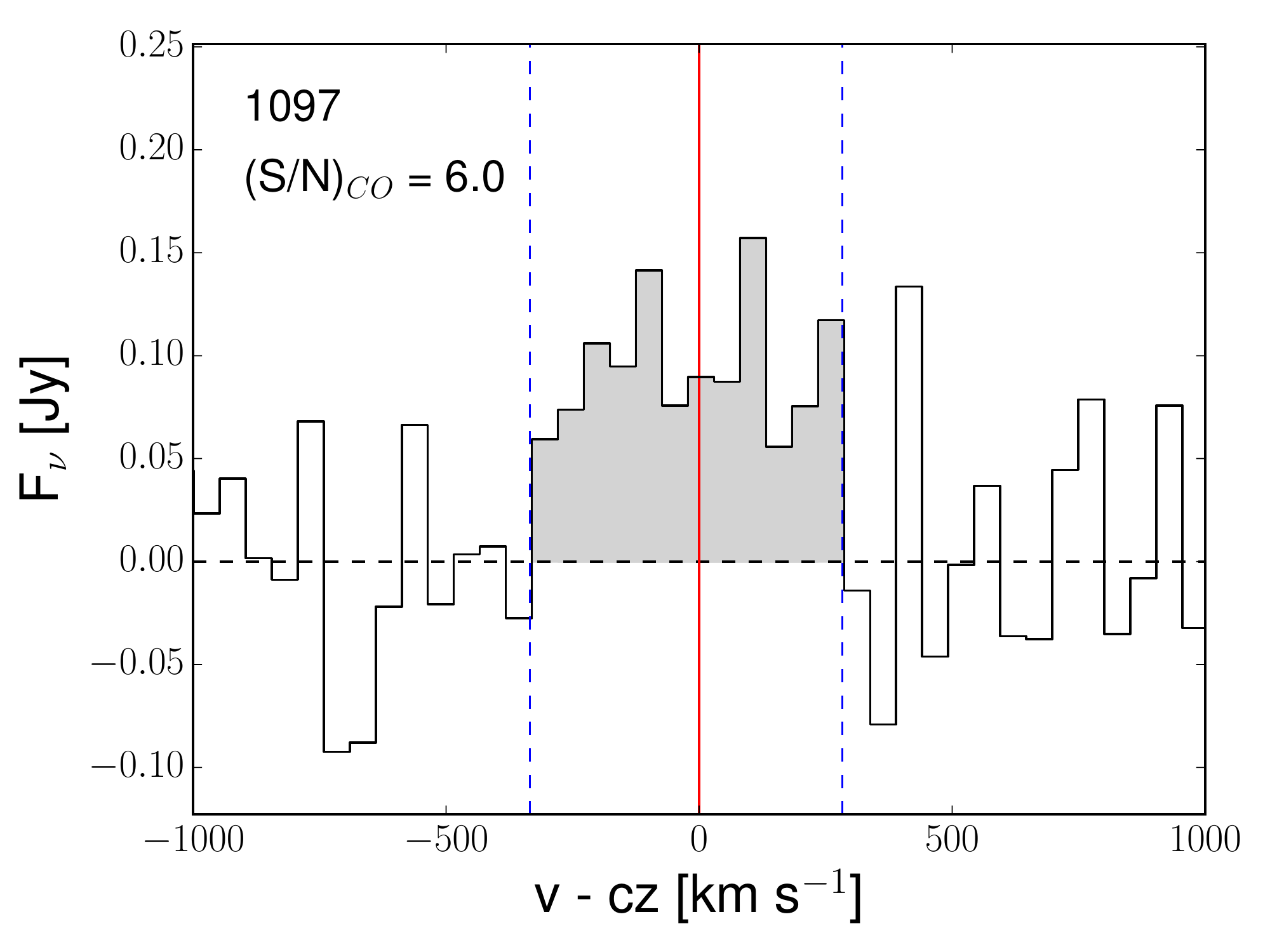}
\includegraphics[width=0.18\textwidth]{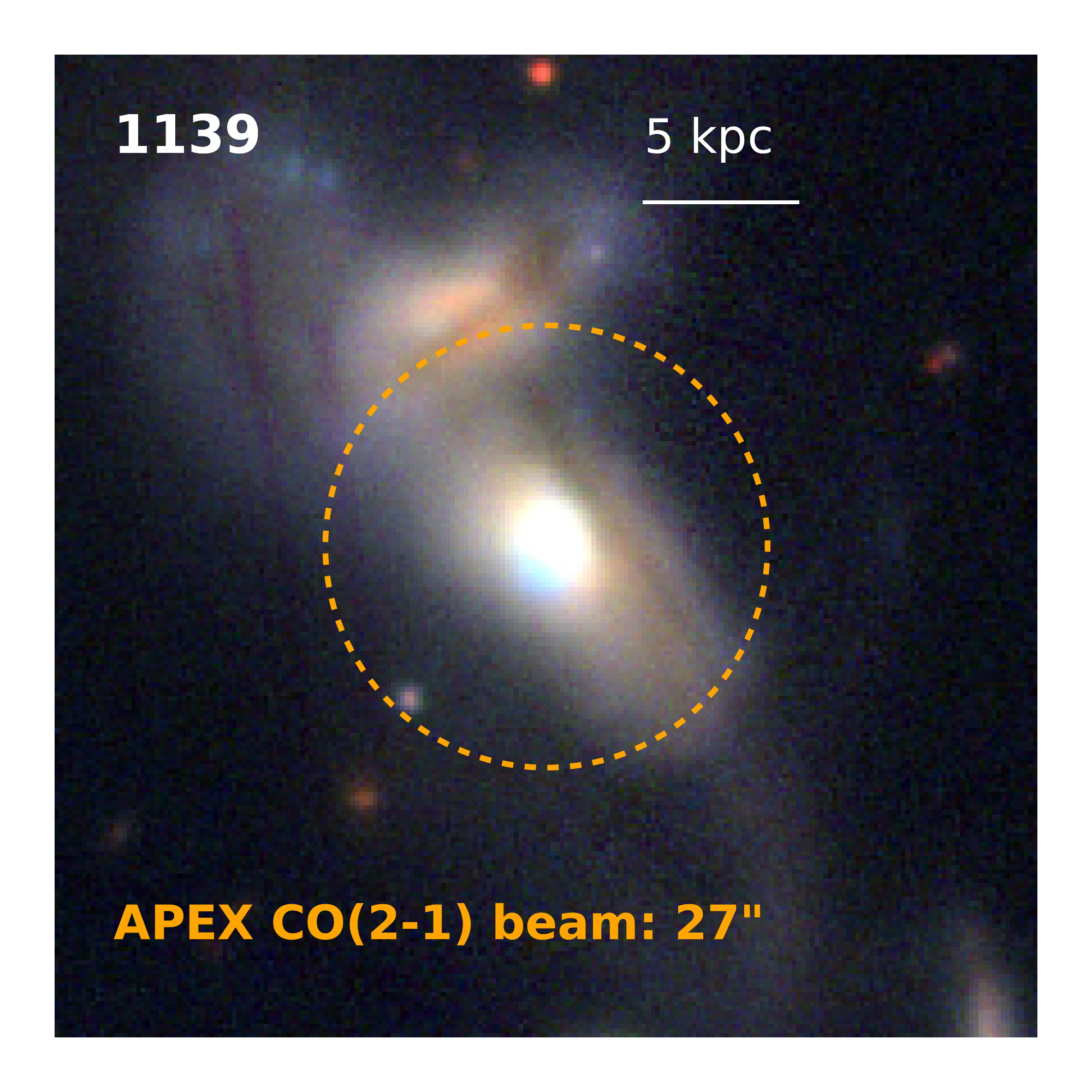}\includegraphics[width=0.26\textwidth]{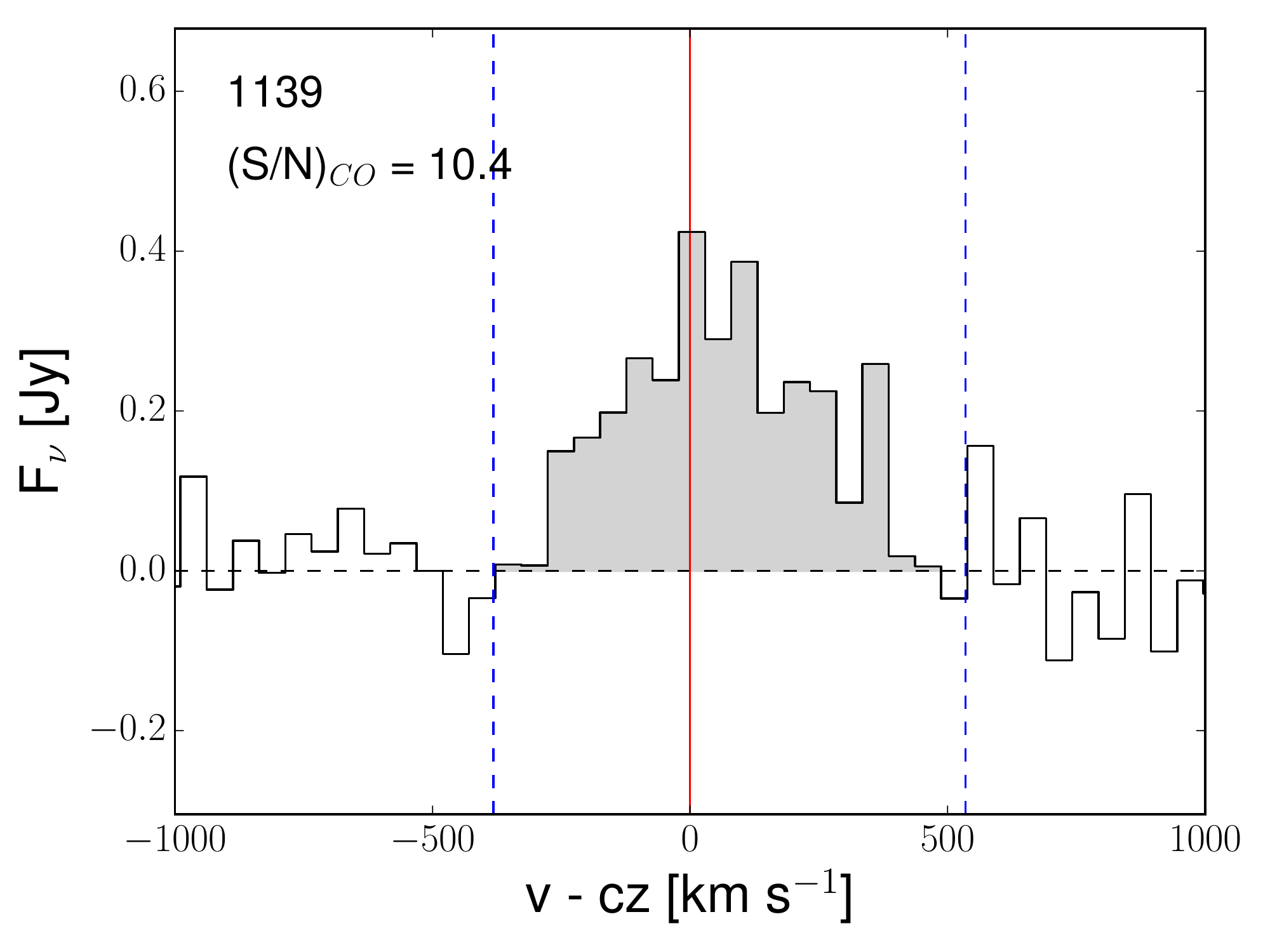}
\includegraphics[width=0.18\textwidth]{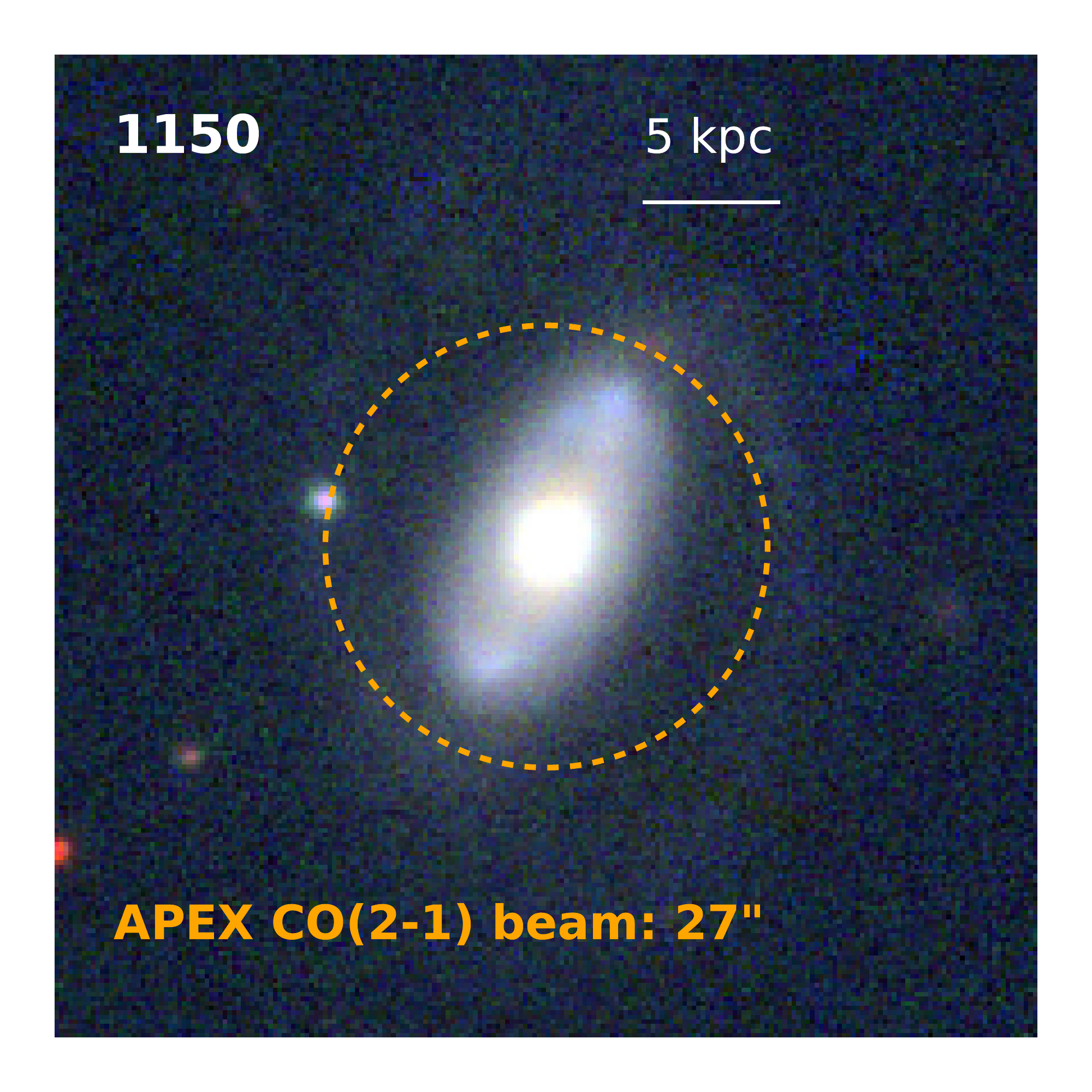}\includegraphics[width=0.26\textwidth]{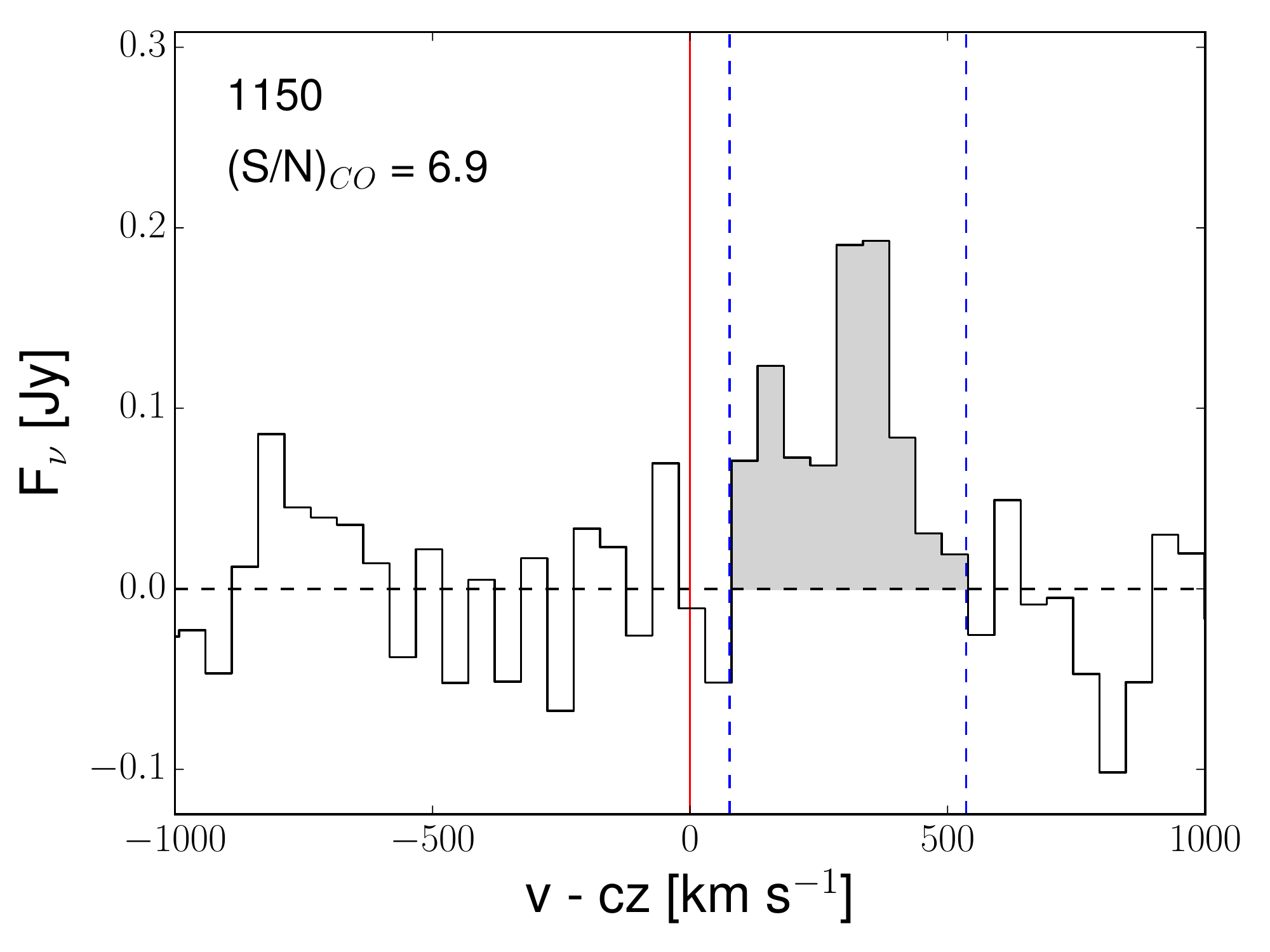}
\includegraphics[width=0.18\textwidth]{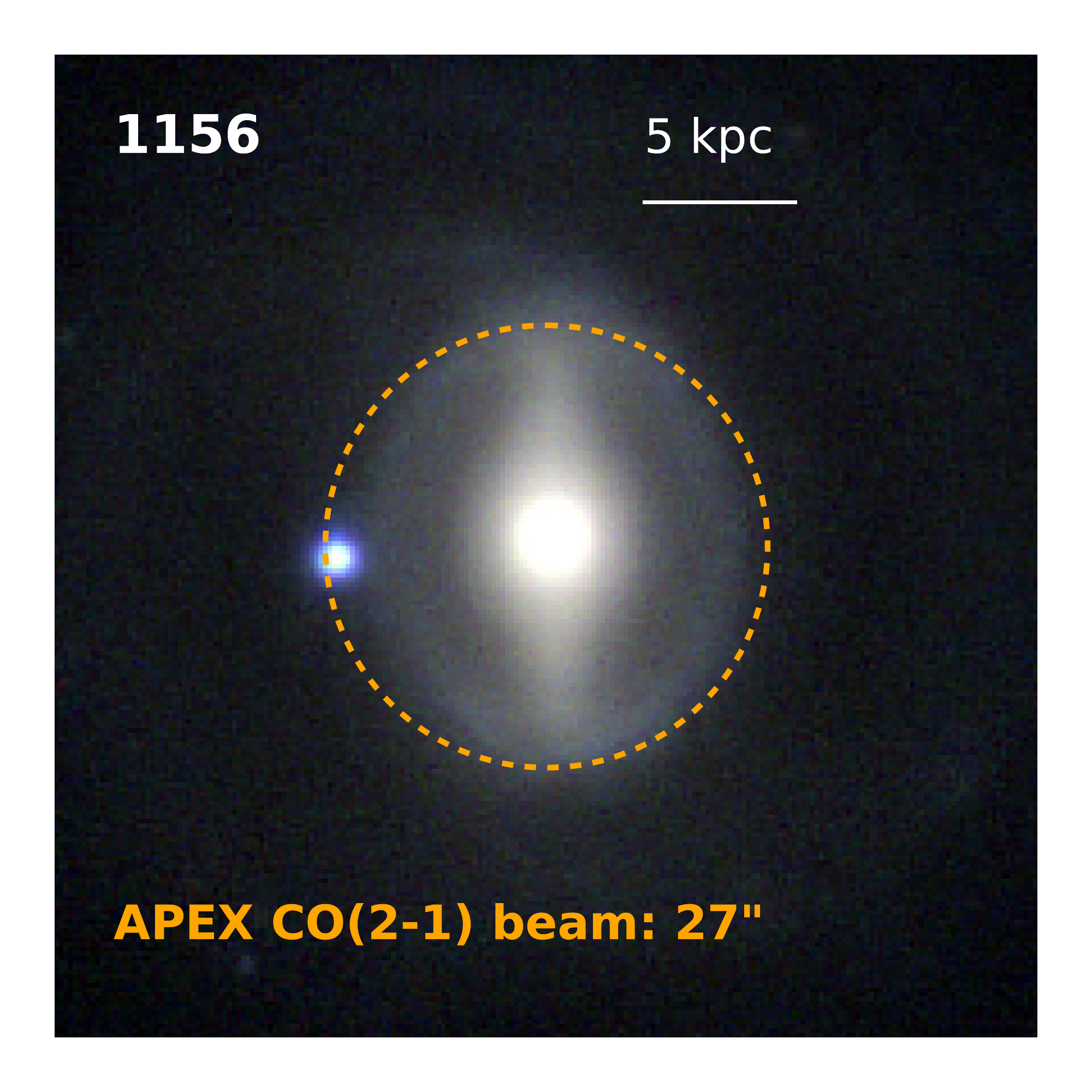}\includegraphics[width=0.26\textwidth]{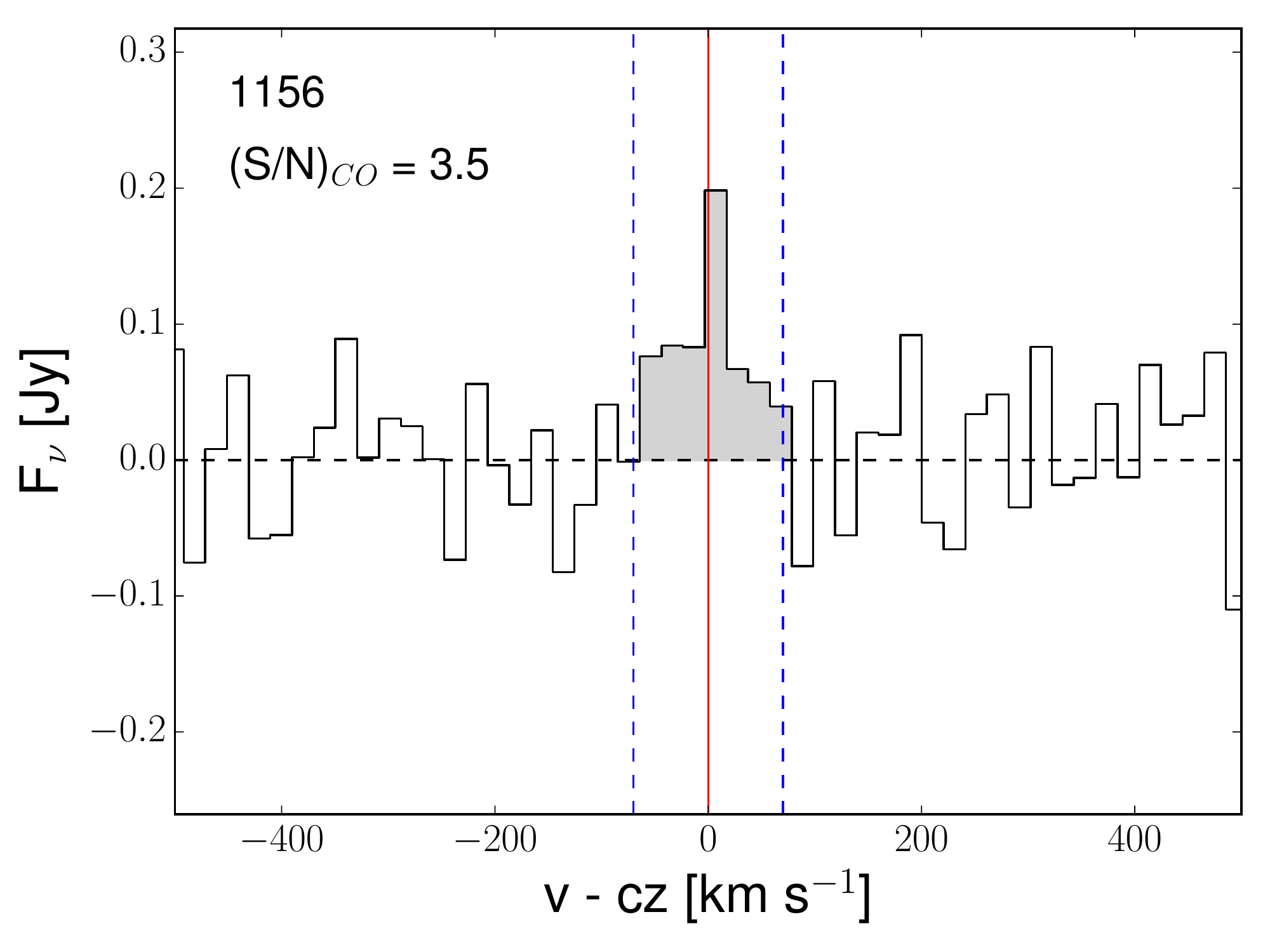}
\includegraphics[width=0.18\textwidth]{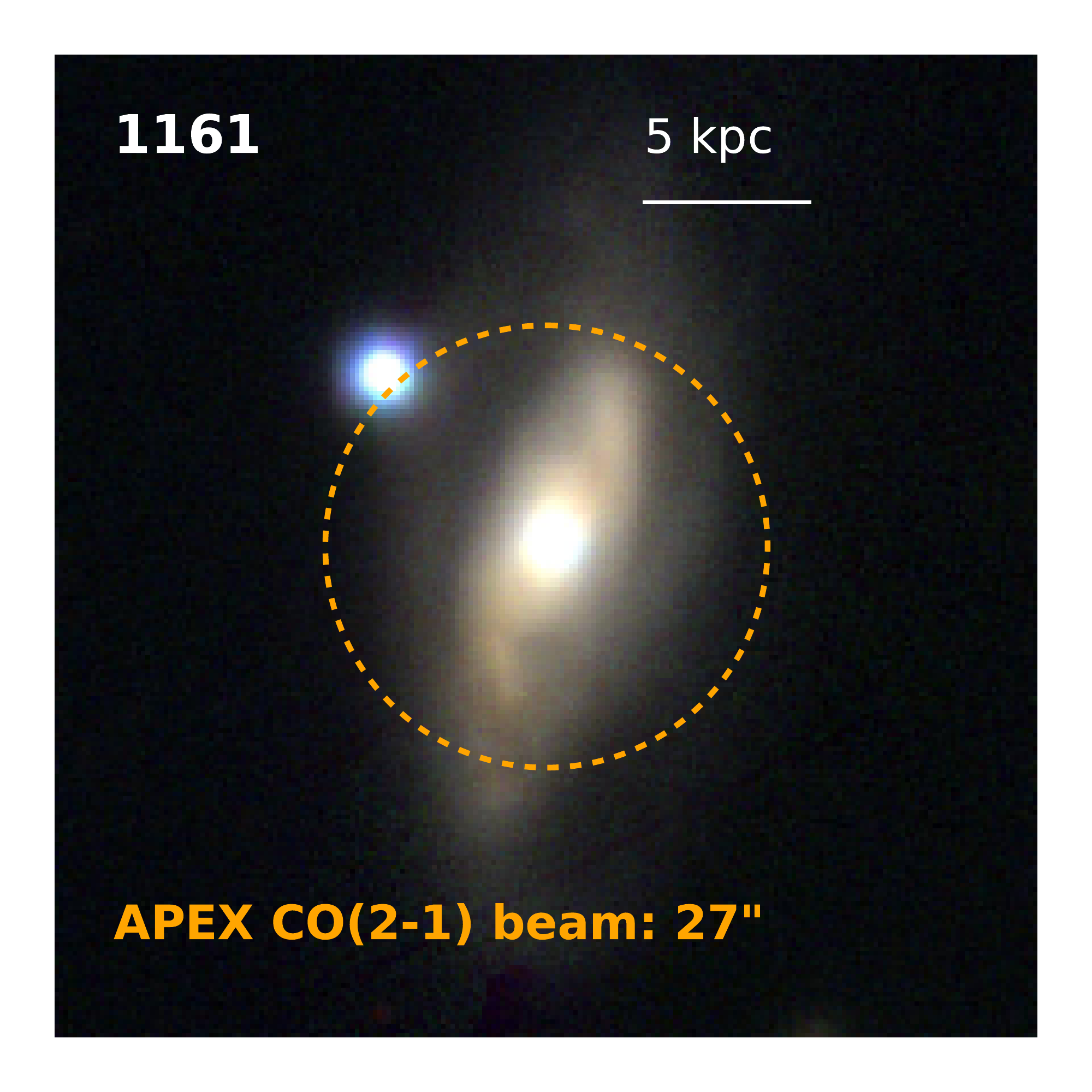}\includegraphics[width=0.26\textwidth]{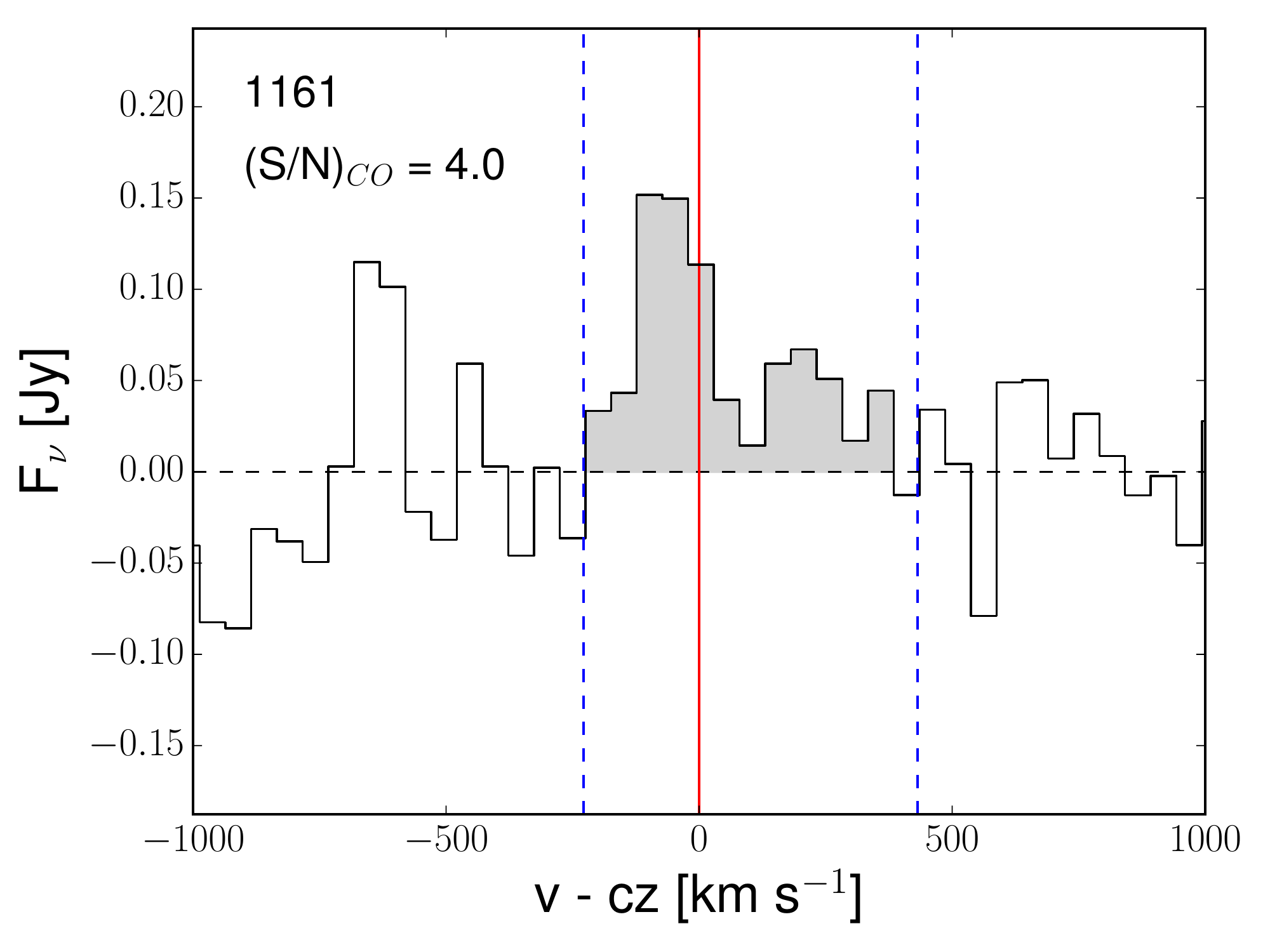}
\includegraphics[width=0.18\textwidth]{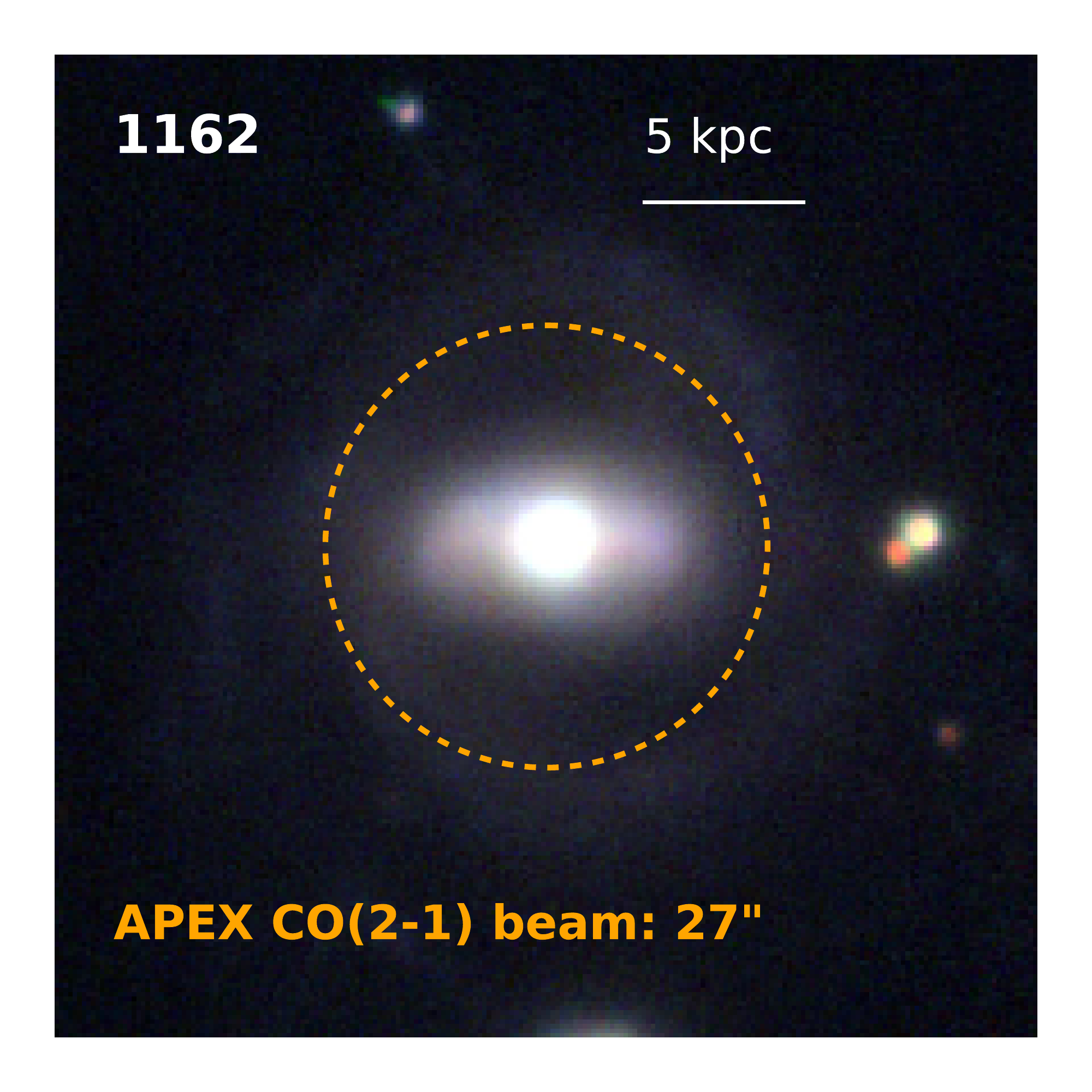}\includegraphics[width=0.26\textwidth]{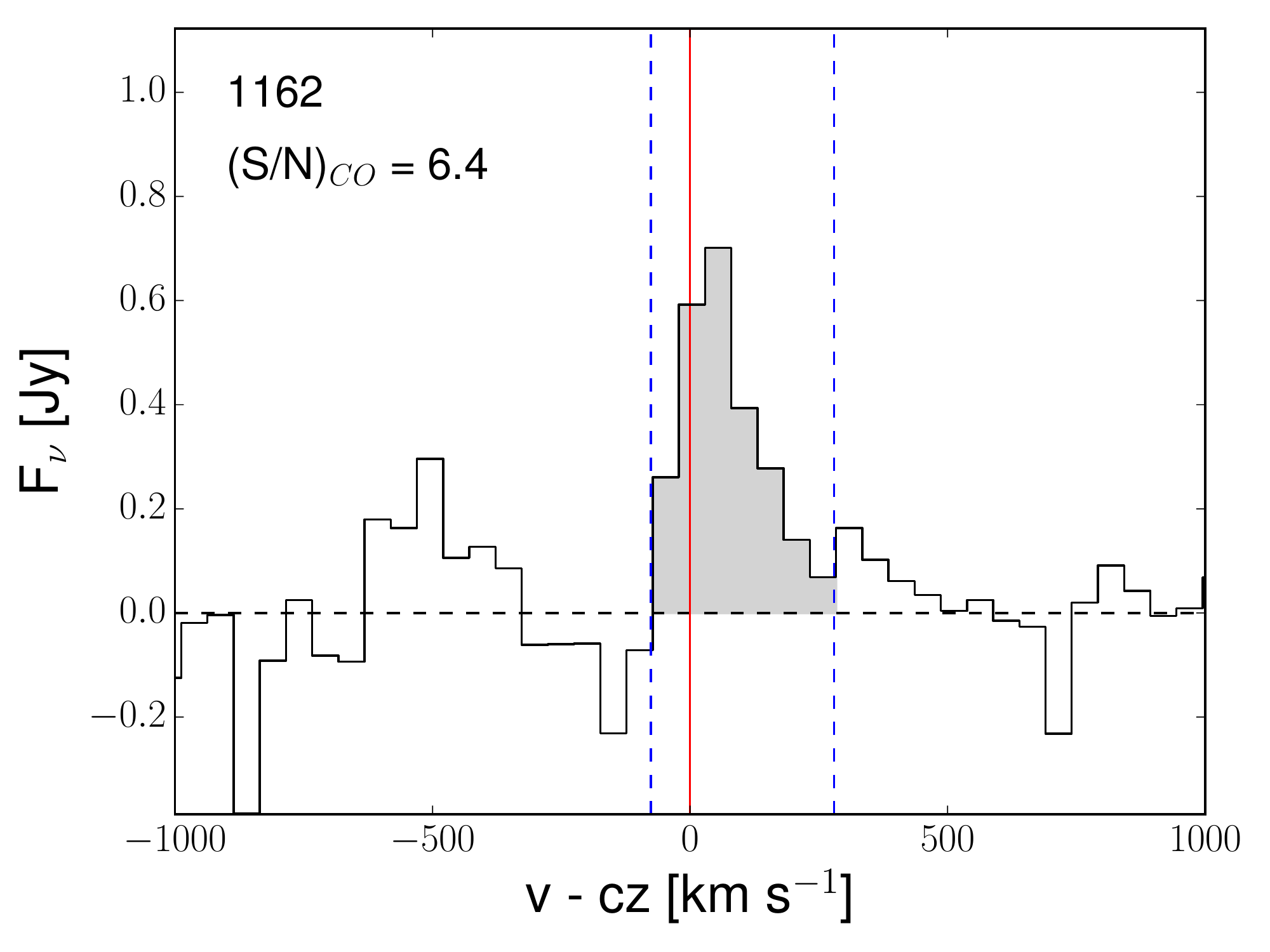}
\includegraphics[width=0.18\textwidth]{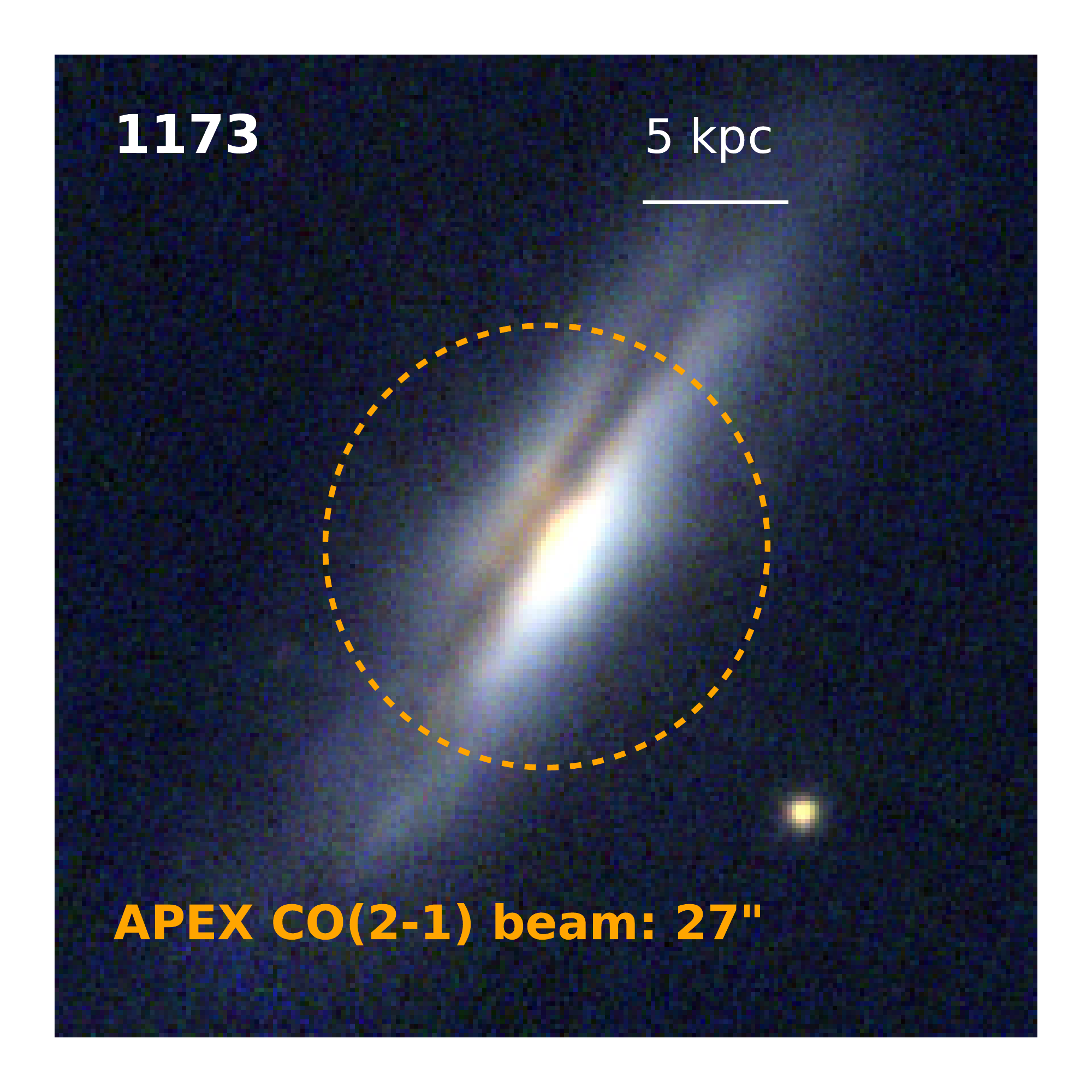}\includegraphics[width=0.26\textwidth]{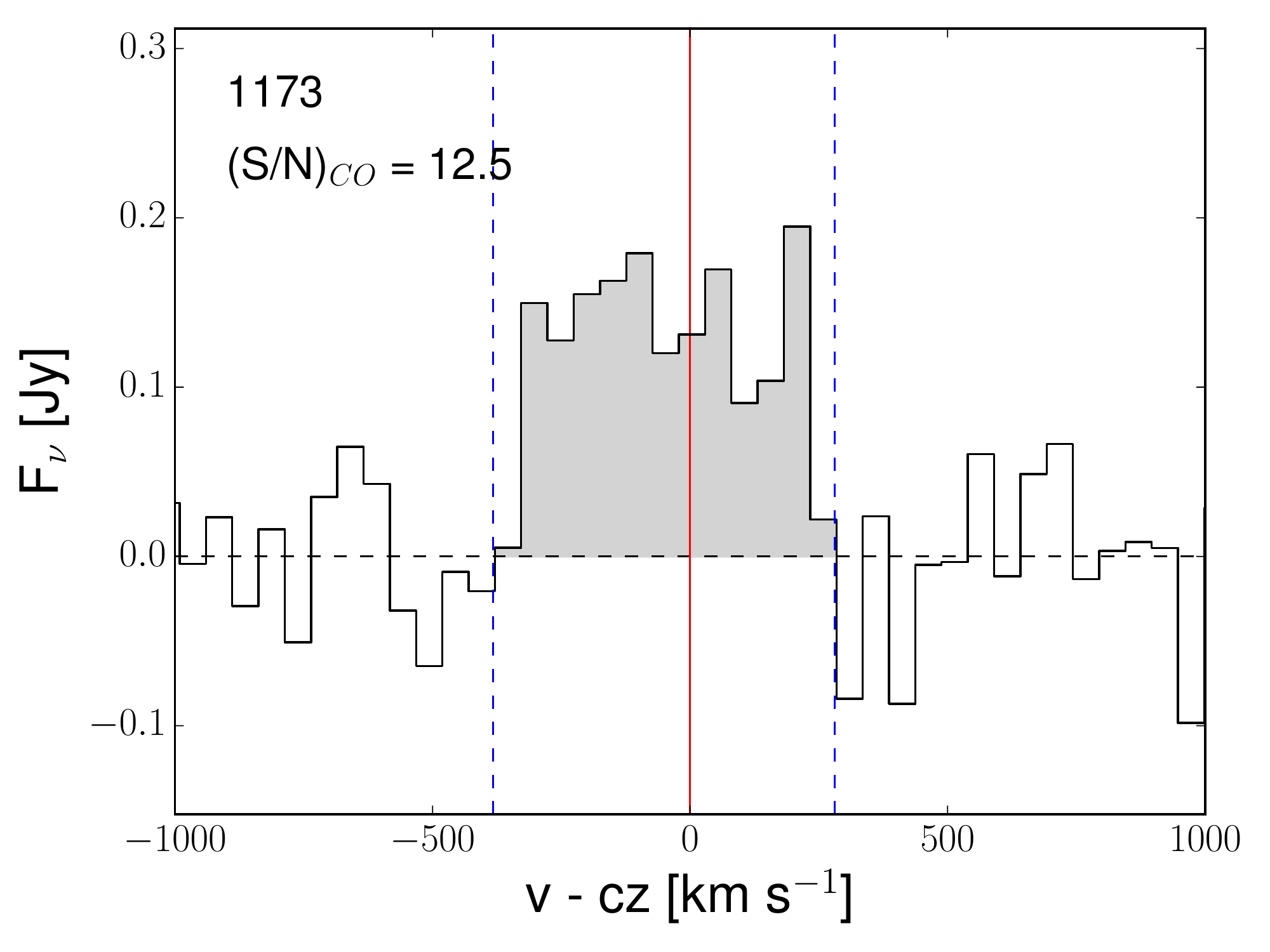}
\includegraphics[width=0.18\textwidth]{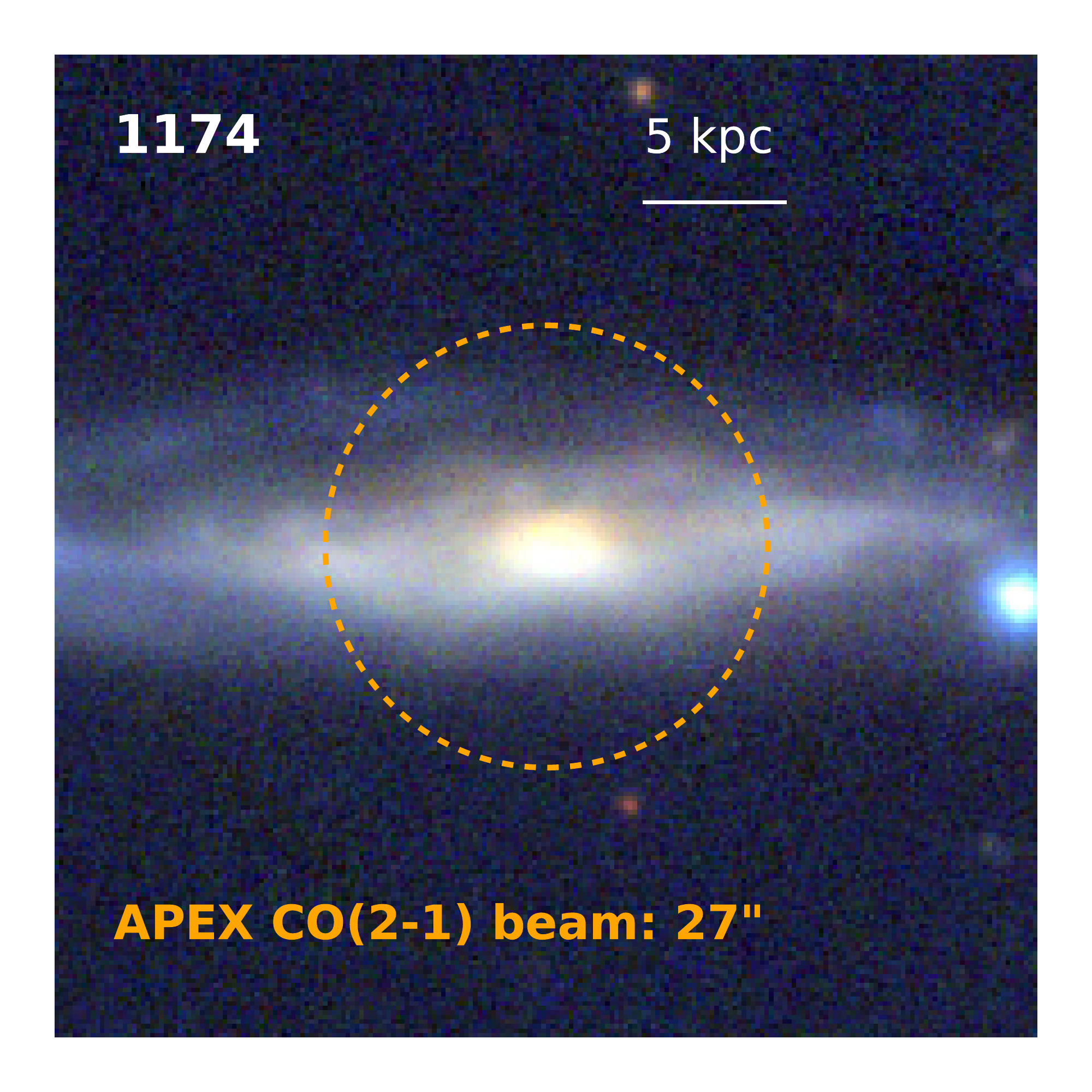}\includegraphics[width=0.26\textwidth]{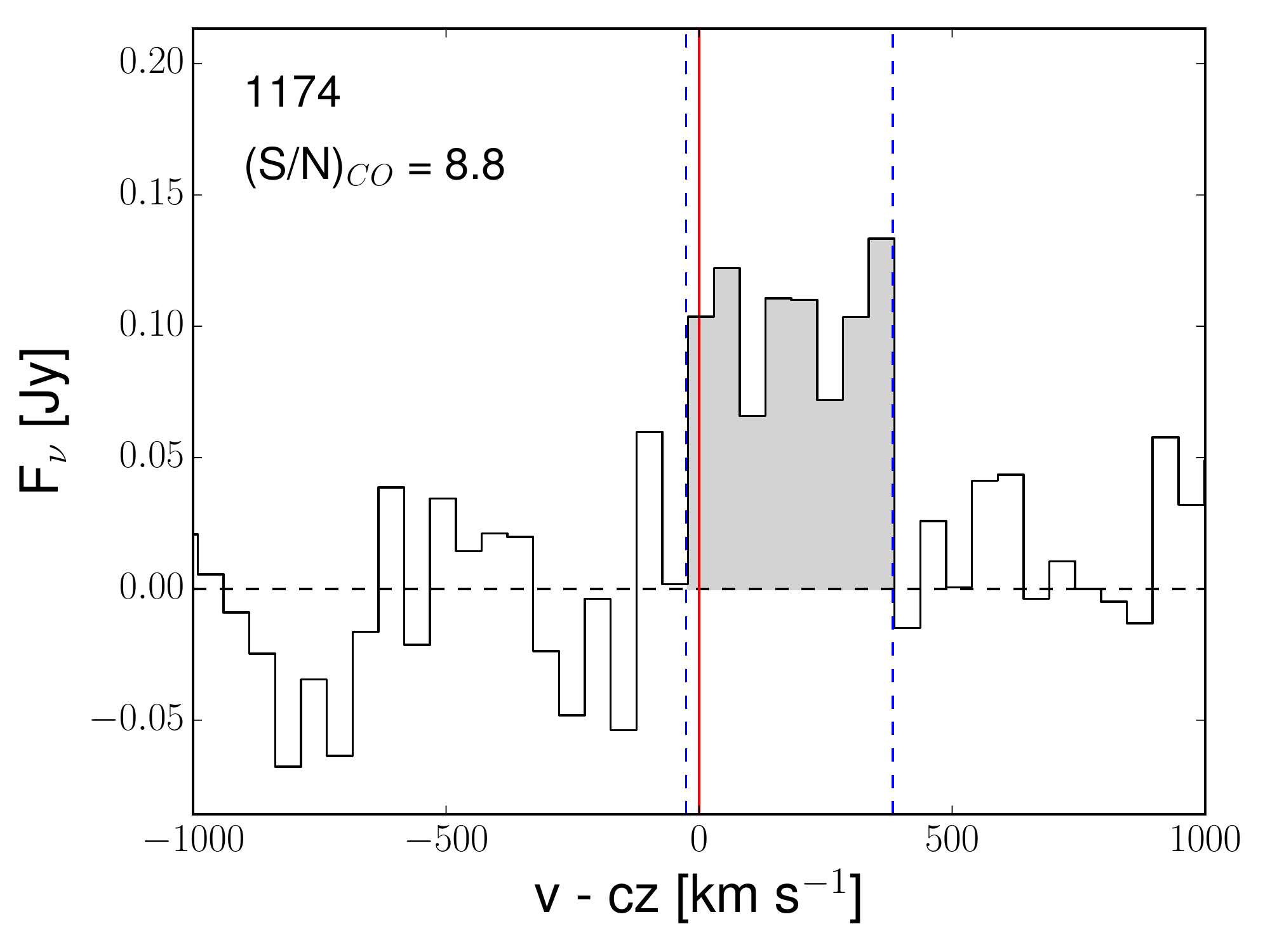}
\includegraphics[width=0.18\textwidth]{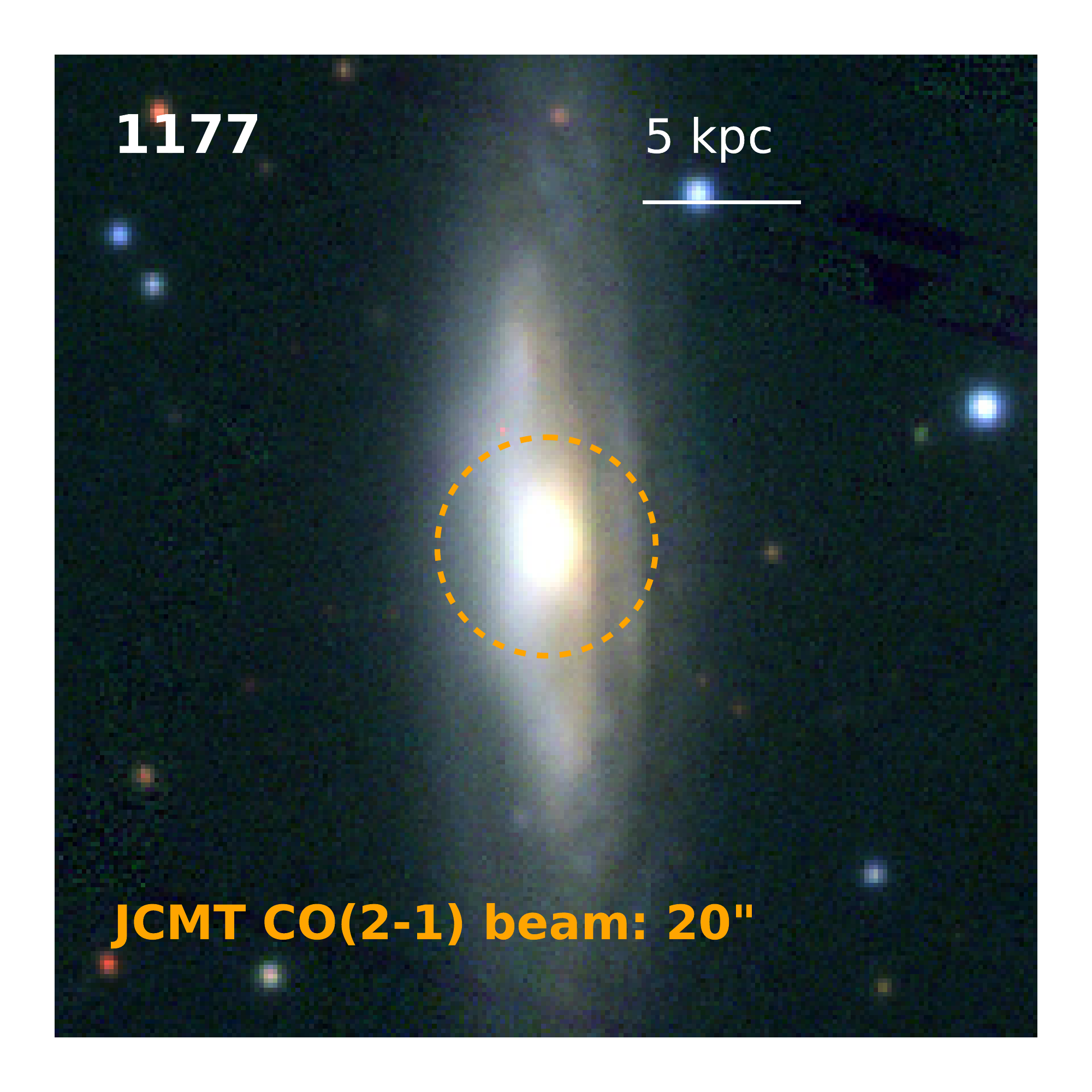}\includegraphics[width=0.26\textwidth]{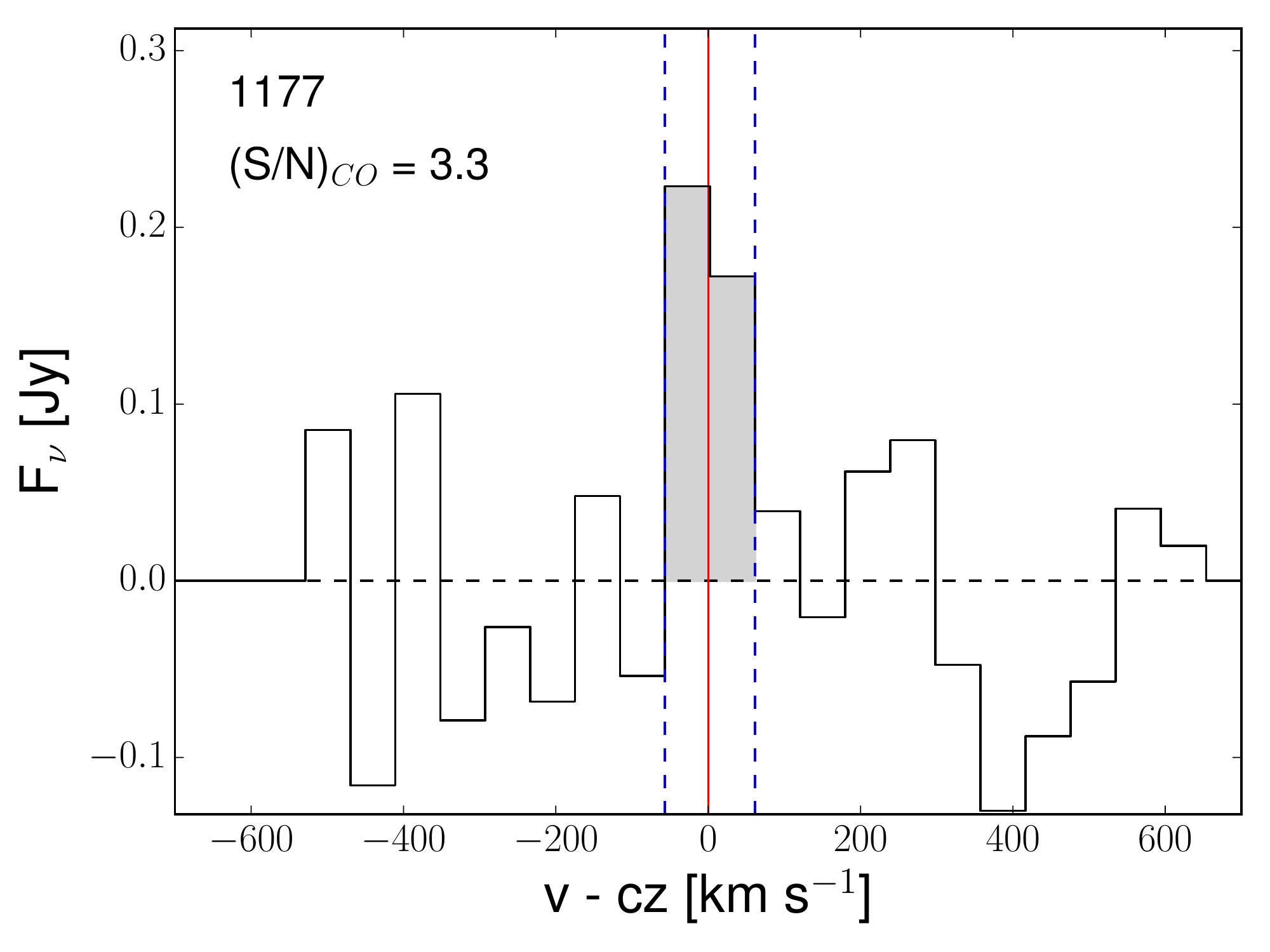}
\includegraphics[width=0.18\textwidth]{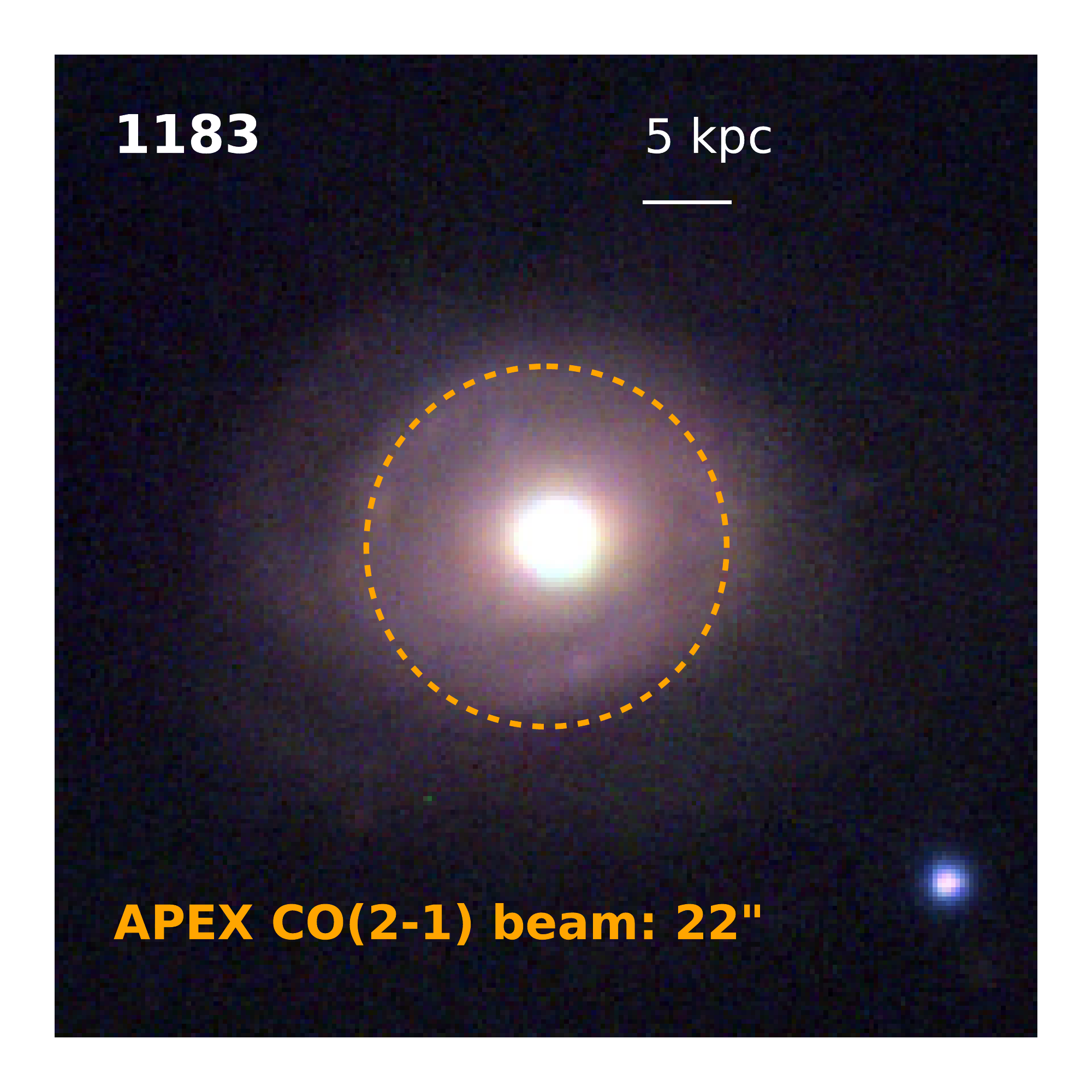}\includegraphics[width=0.26\textwidth]{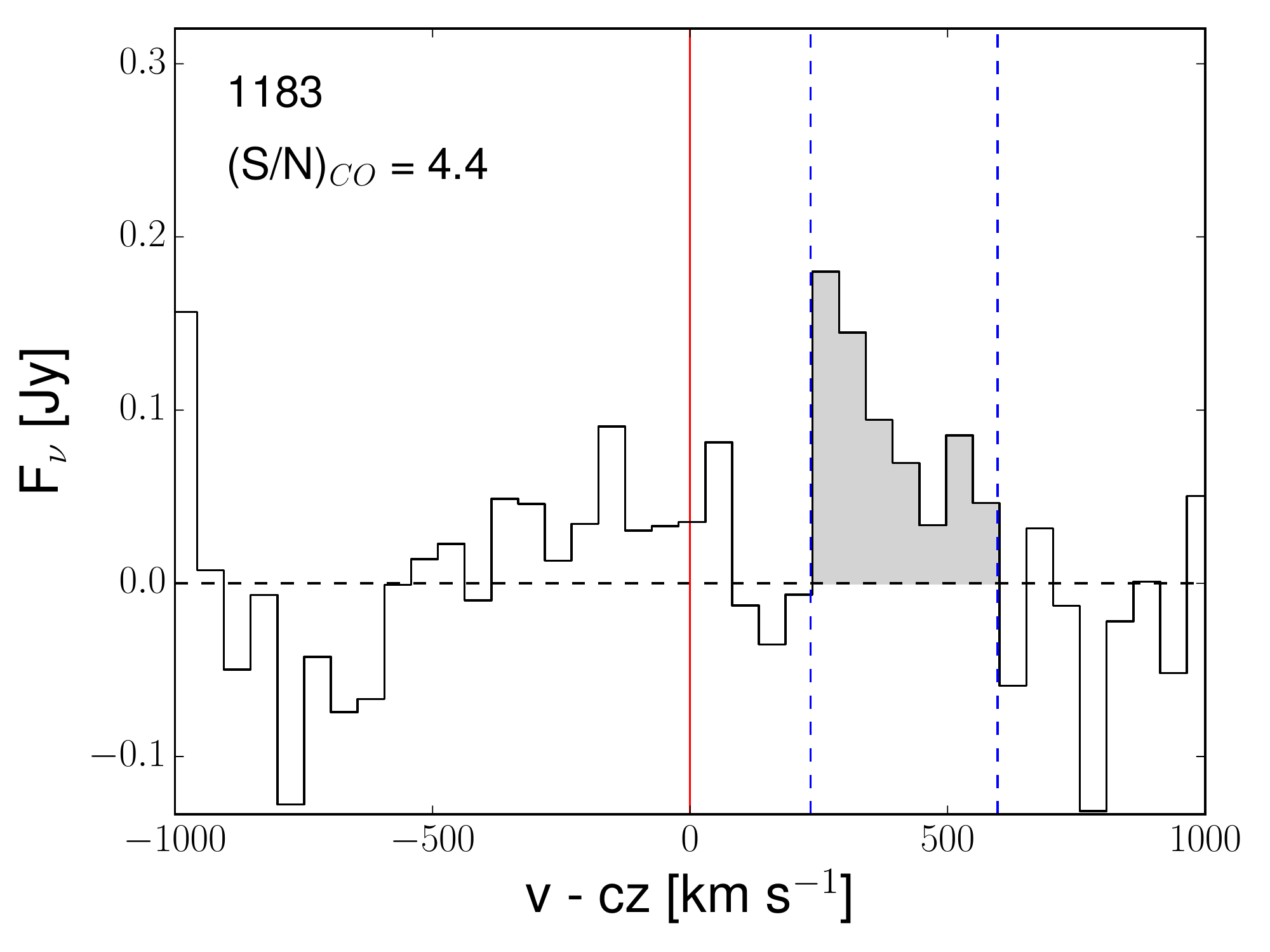}
\includegraphics[width=0.18\textwidth]{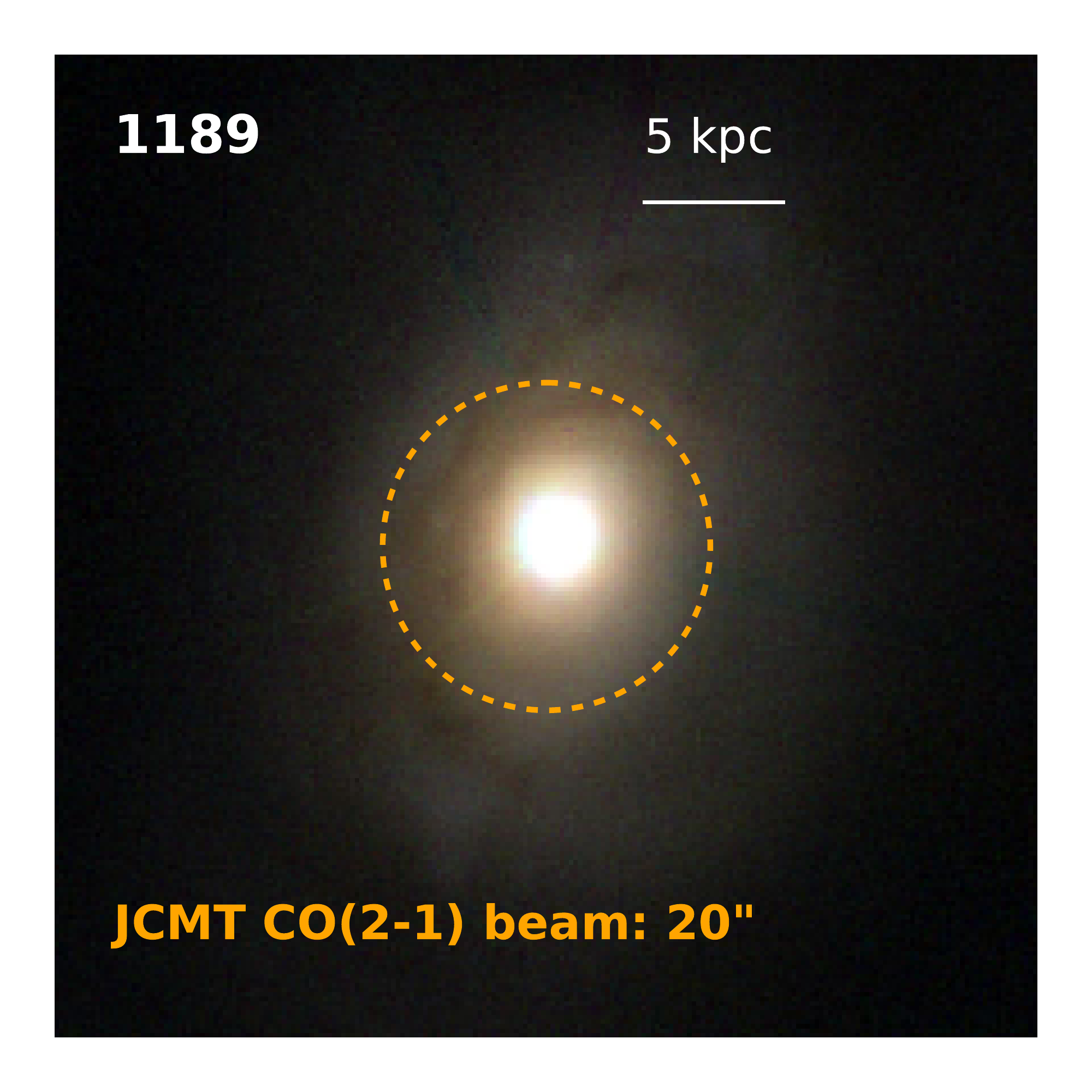}\includegraphics[width=0.26\textwidth]{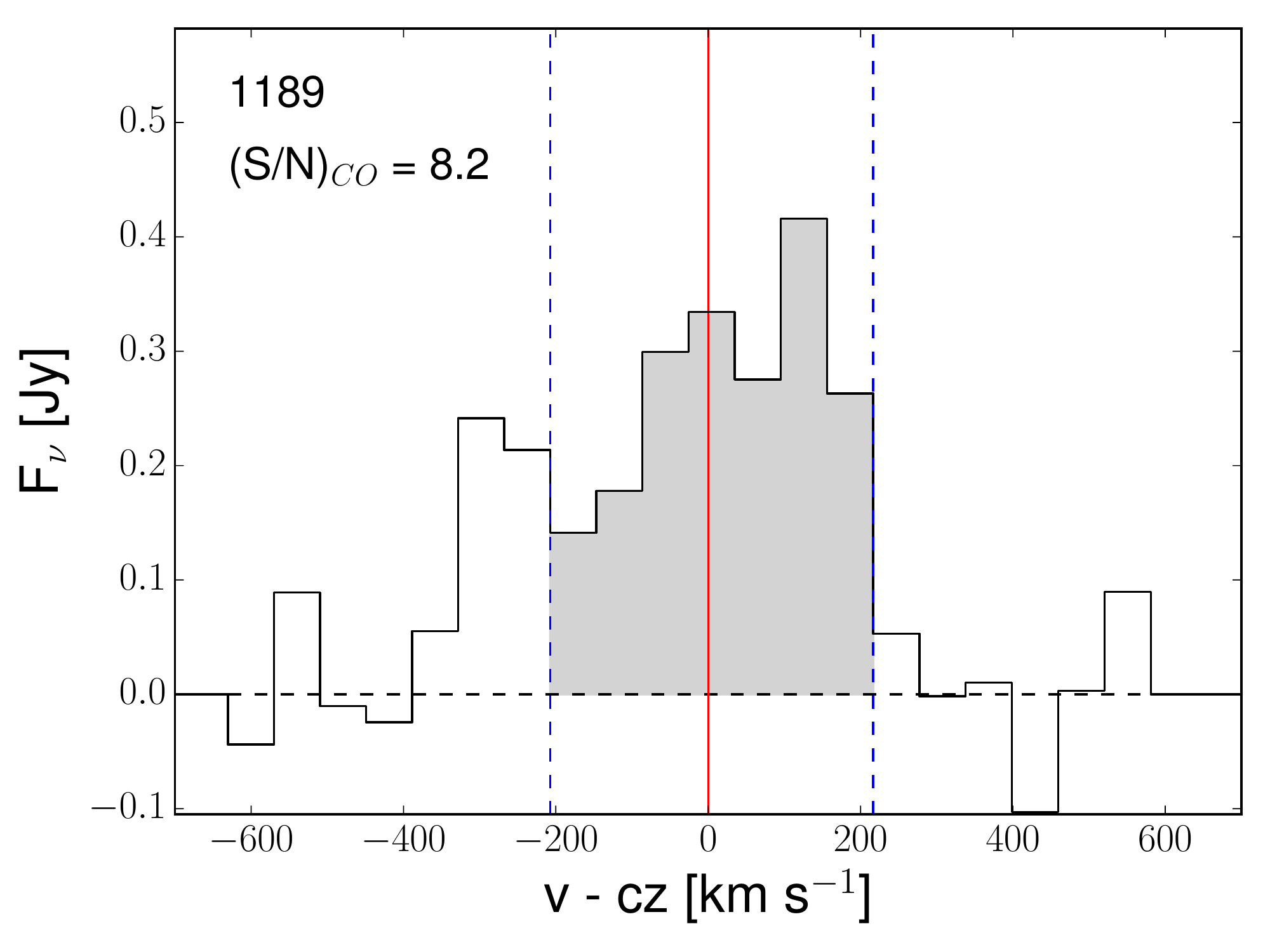}
\caption{continued from Fig.~\ref{fig:CO21_spectra_all_1}
} 
\label{fig:CO21_spectra_all_6}
\end{figure*}

\begin{figure*}
\centering
\raggedright
\includegraphics[width=0.18\textwidth]{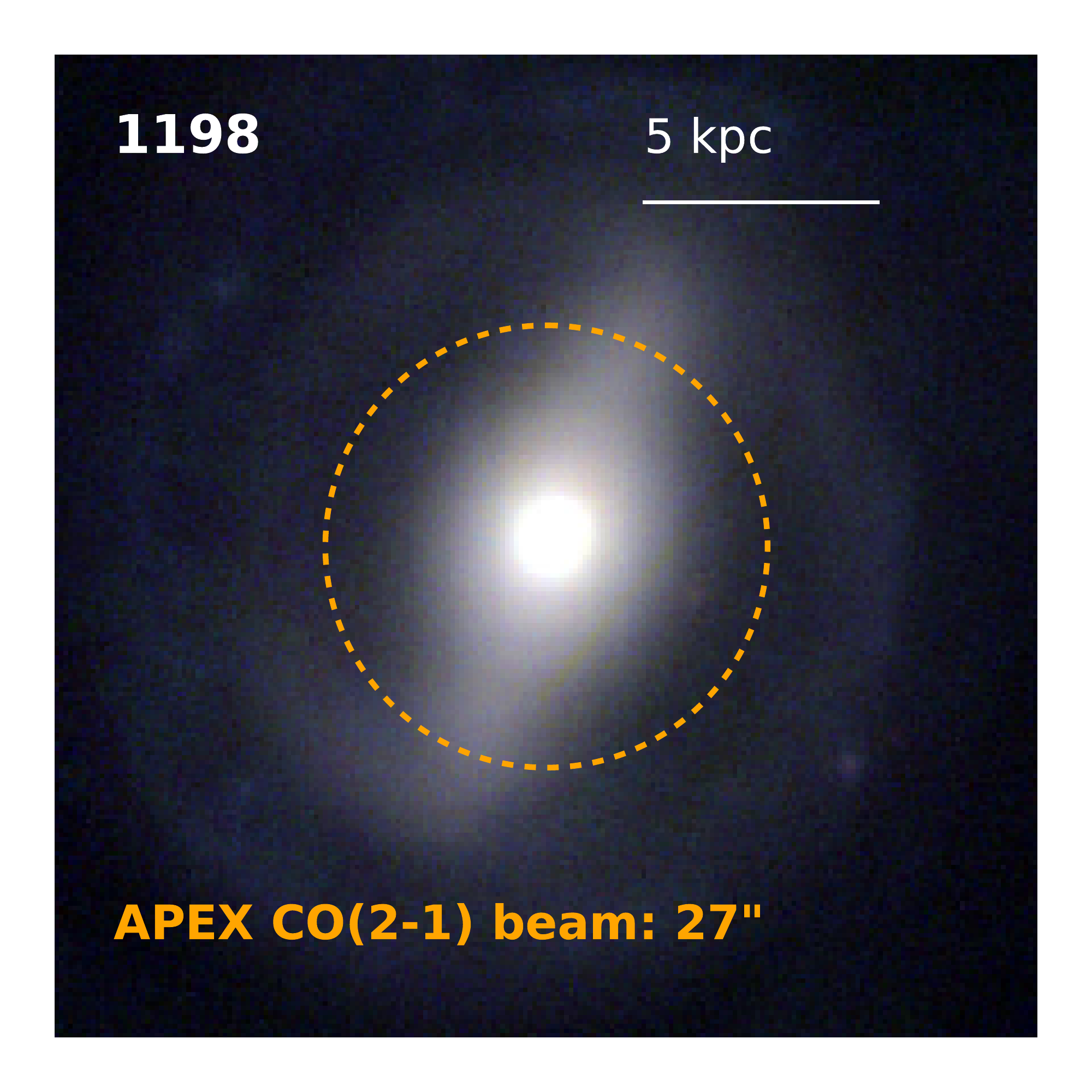}\includegraphics[width=0.26\textwidth]{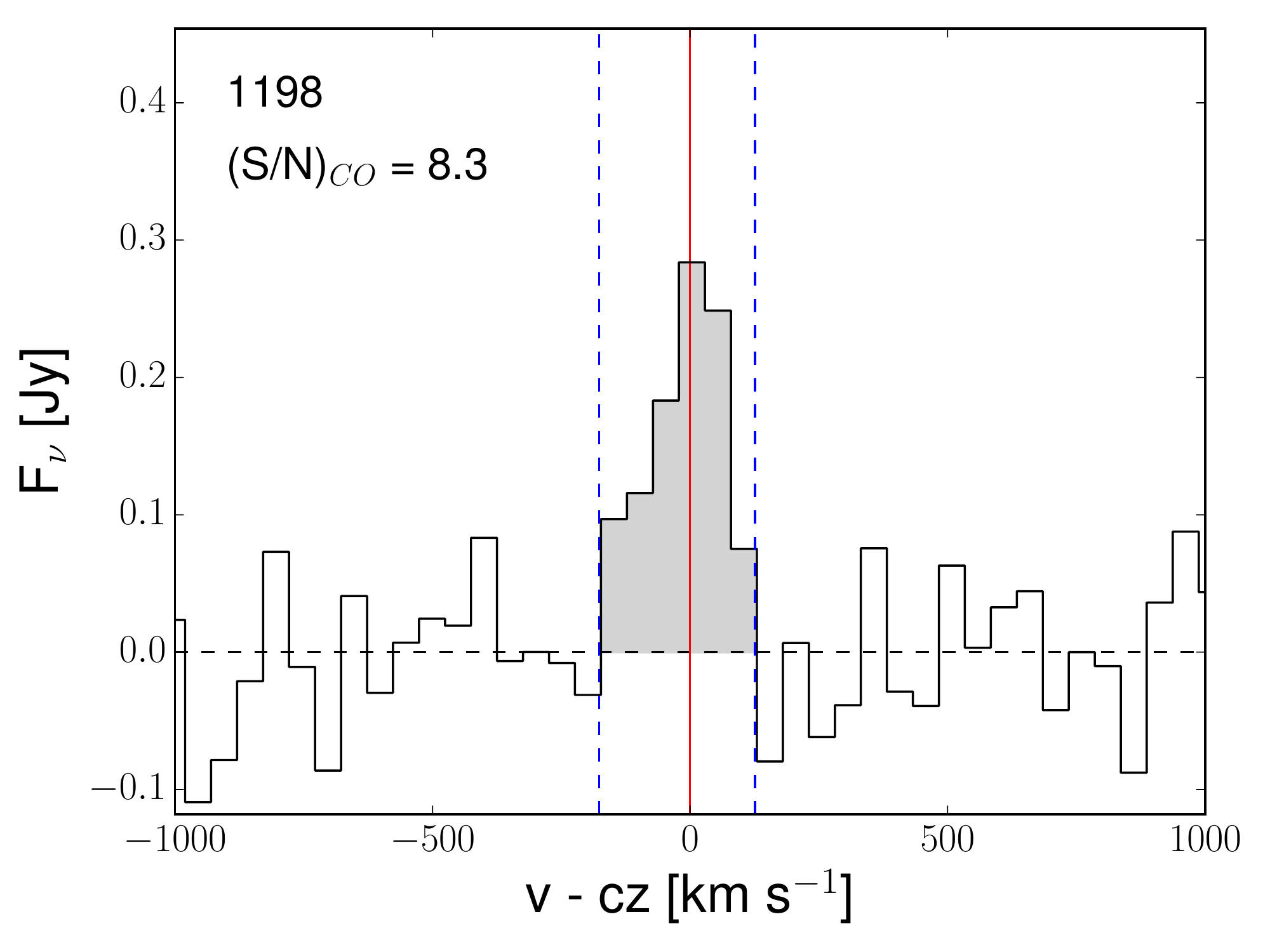}
\includegraphics[width=0.18\textwidth]{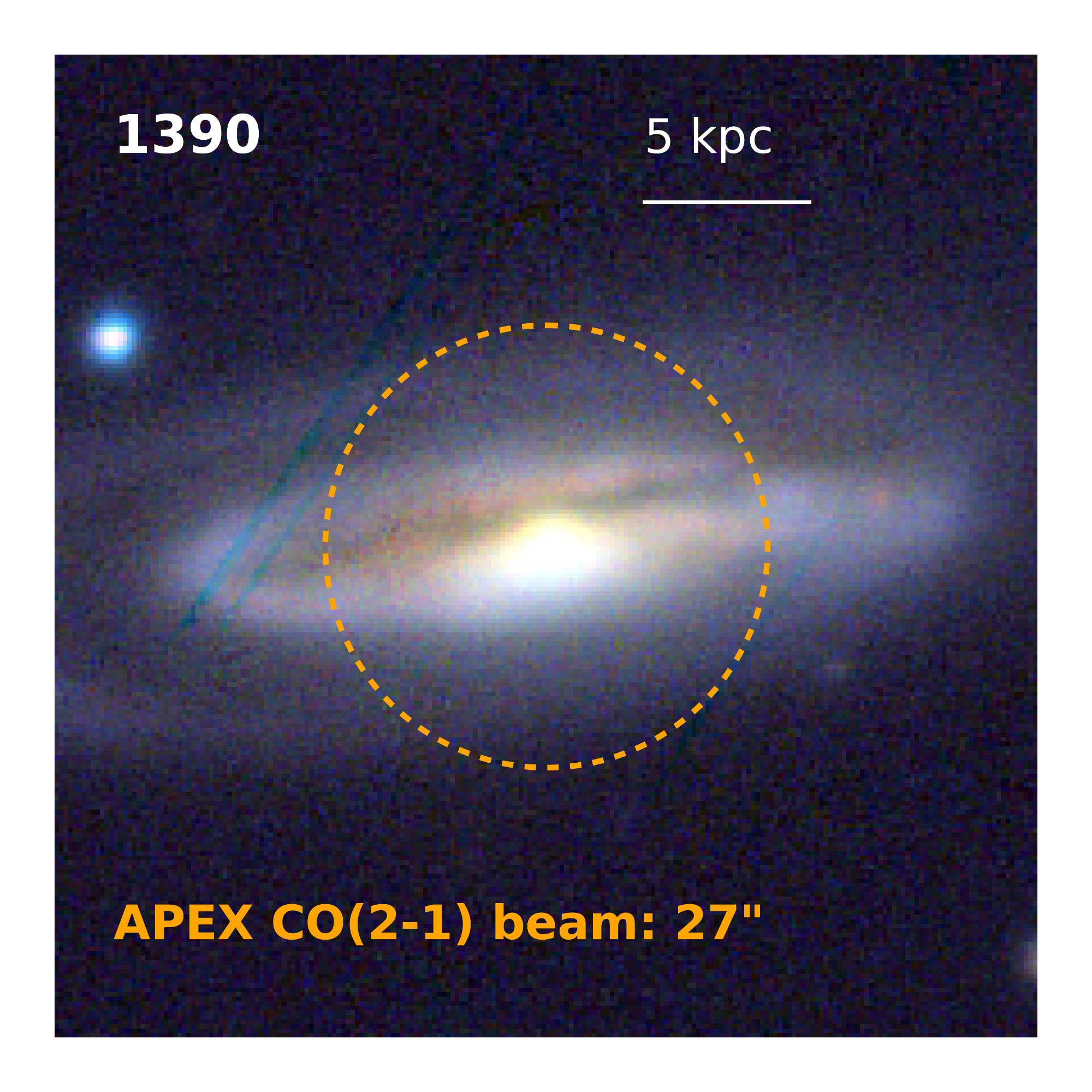}\includegraphics[width=0.26\textwidth]{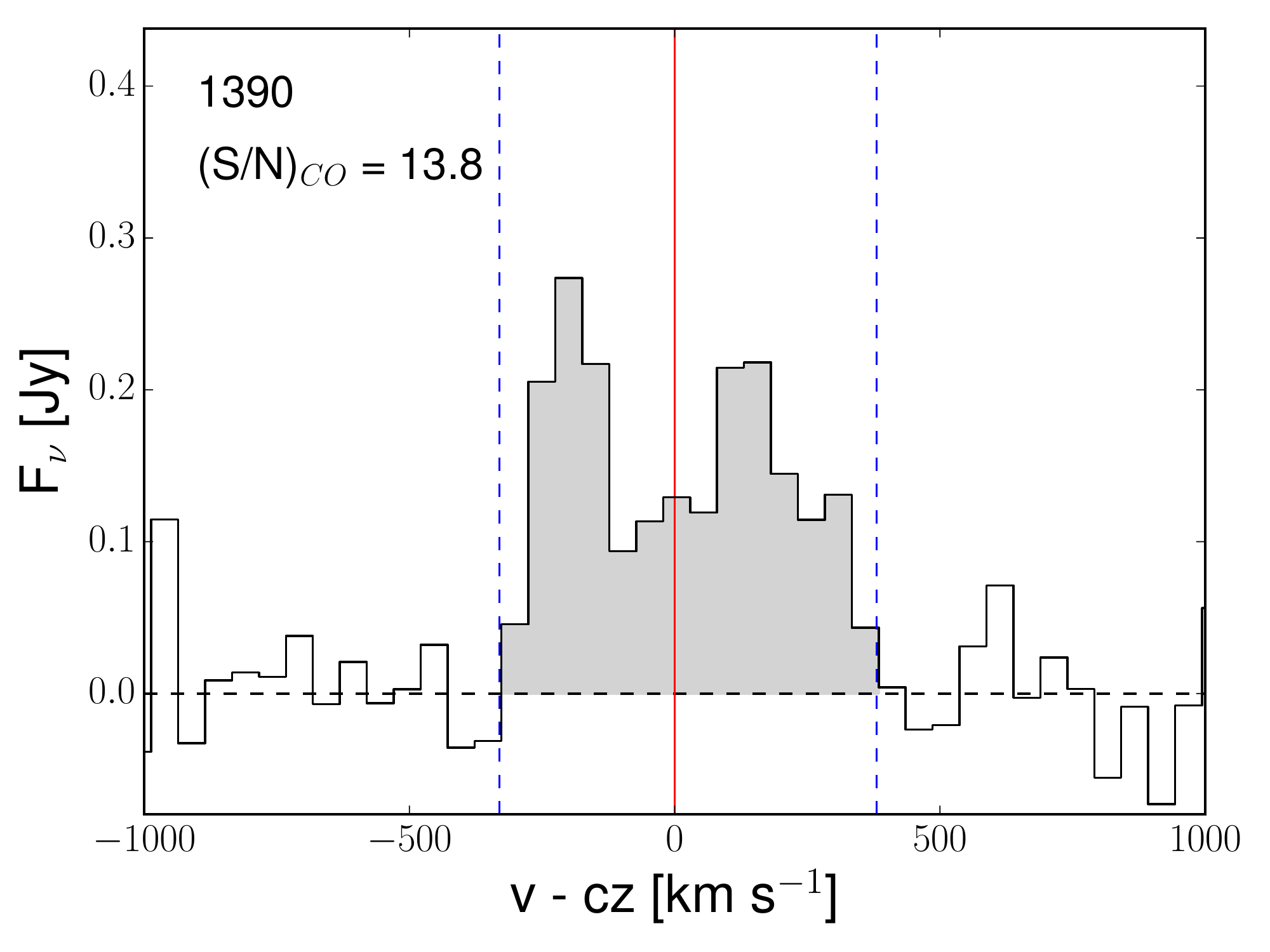}
\includegraphics[width=0.18\textwidth]{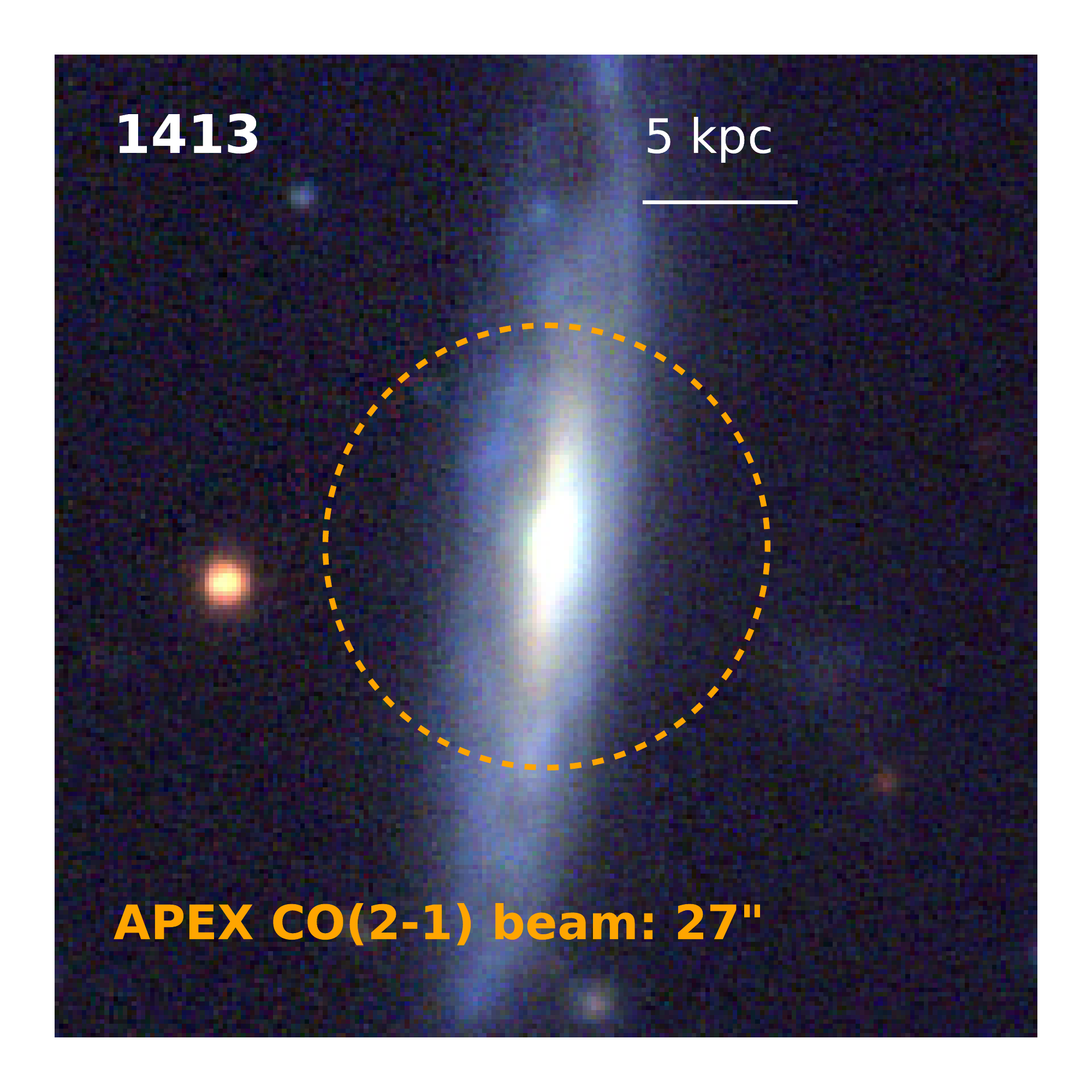}\includegraphics[width=0.26\textwidth]{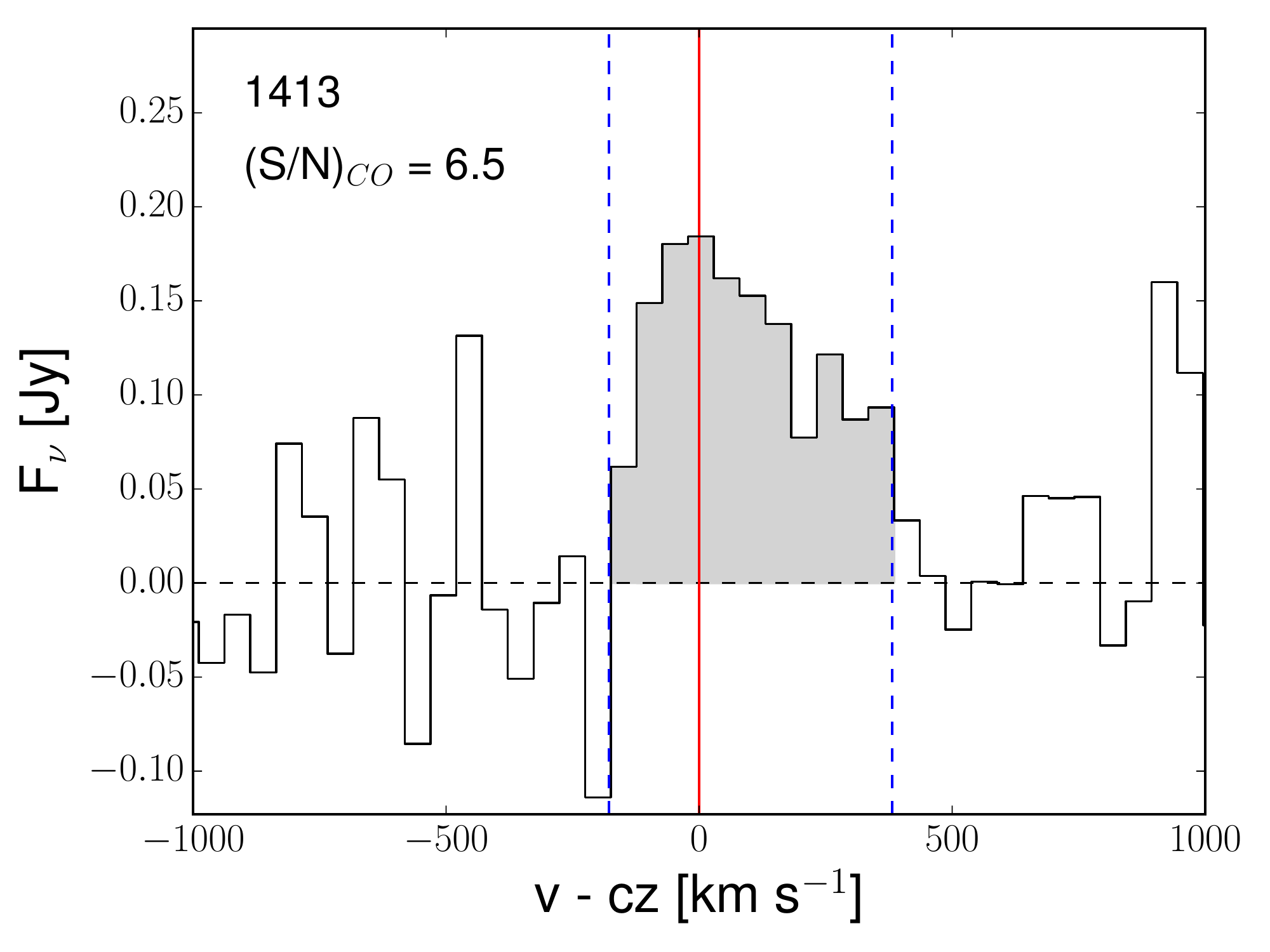}
\caption{continued from Fig.~\ref{fig:CO21_spectra_all_1}
} 
\label{fig:CO21_spectra_all_7}
\end{figure*}

\begin{figure*}
\centering
\raggedright
\includegraphics[width=0.18\textwidth]{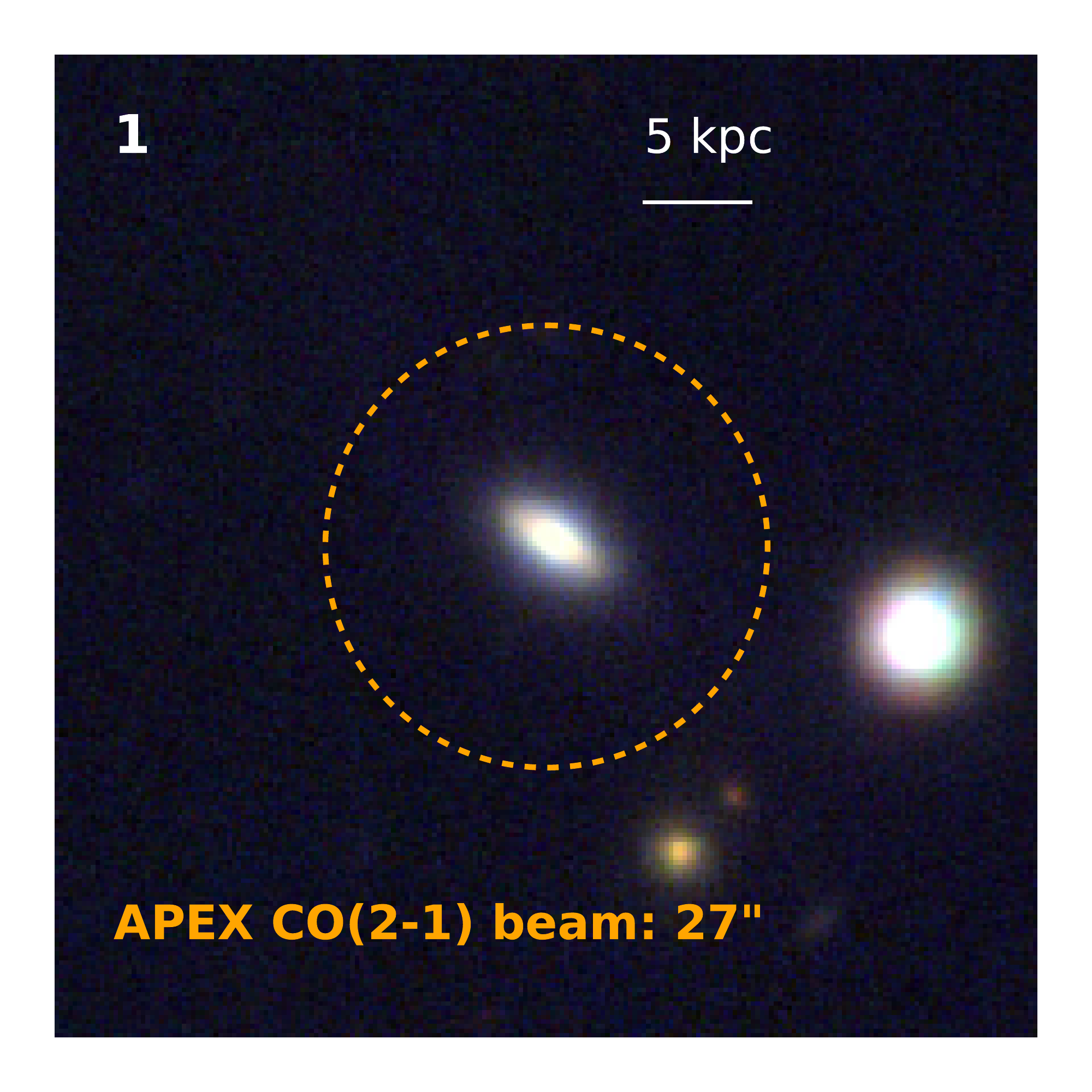}\includegraphics[width=0.26\textwidth]{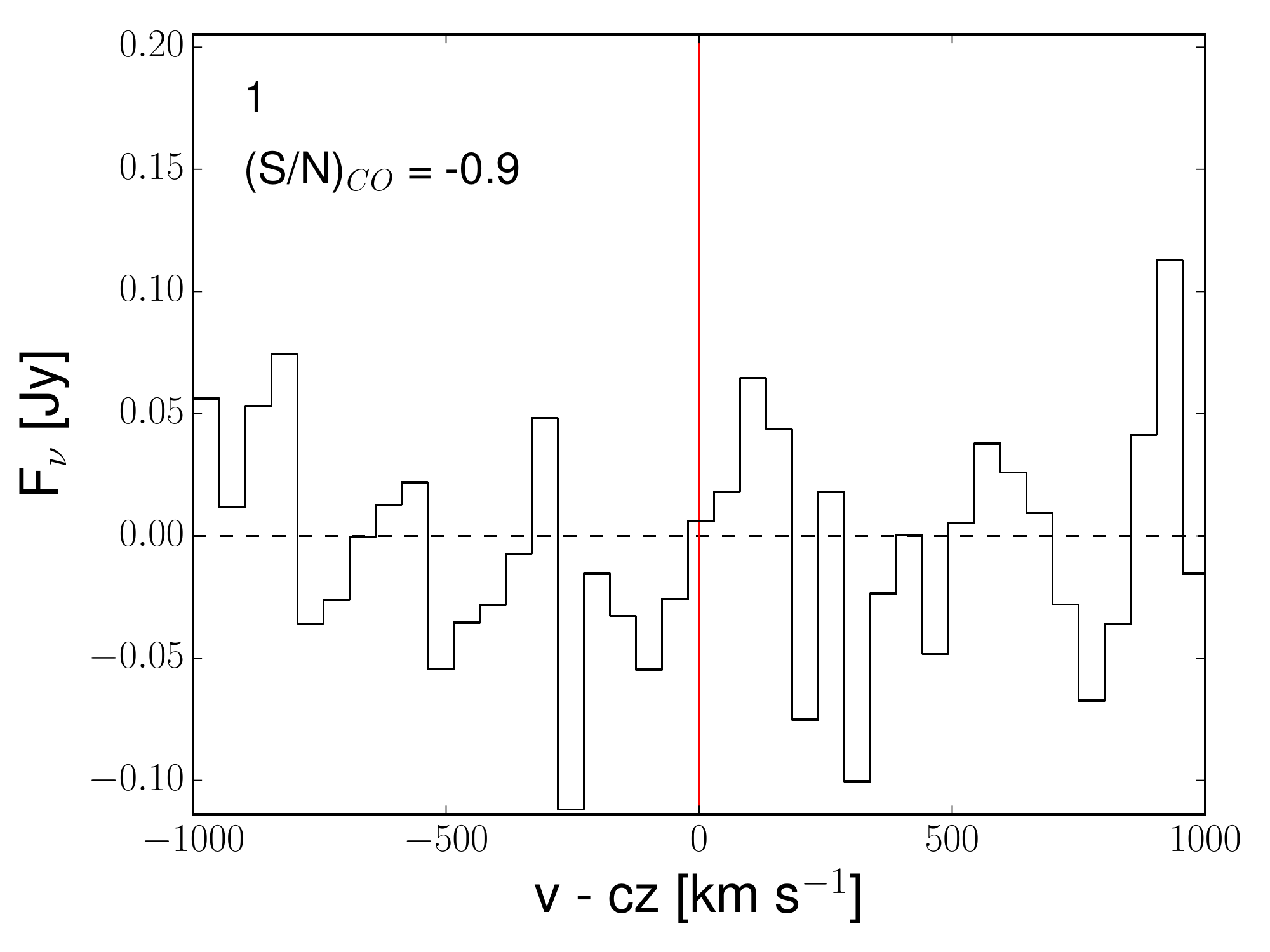}
\includegraphics[width=0.18\textwidth]{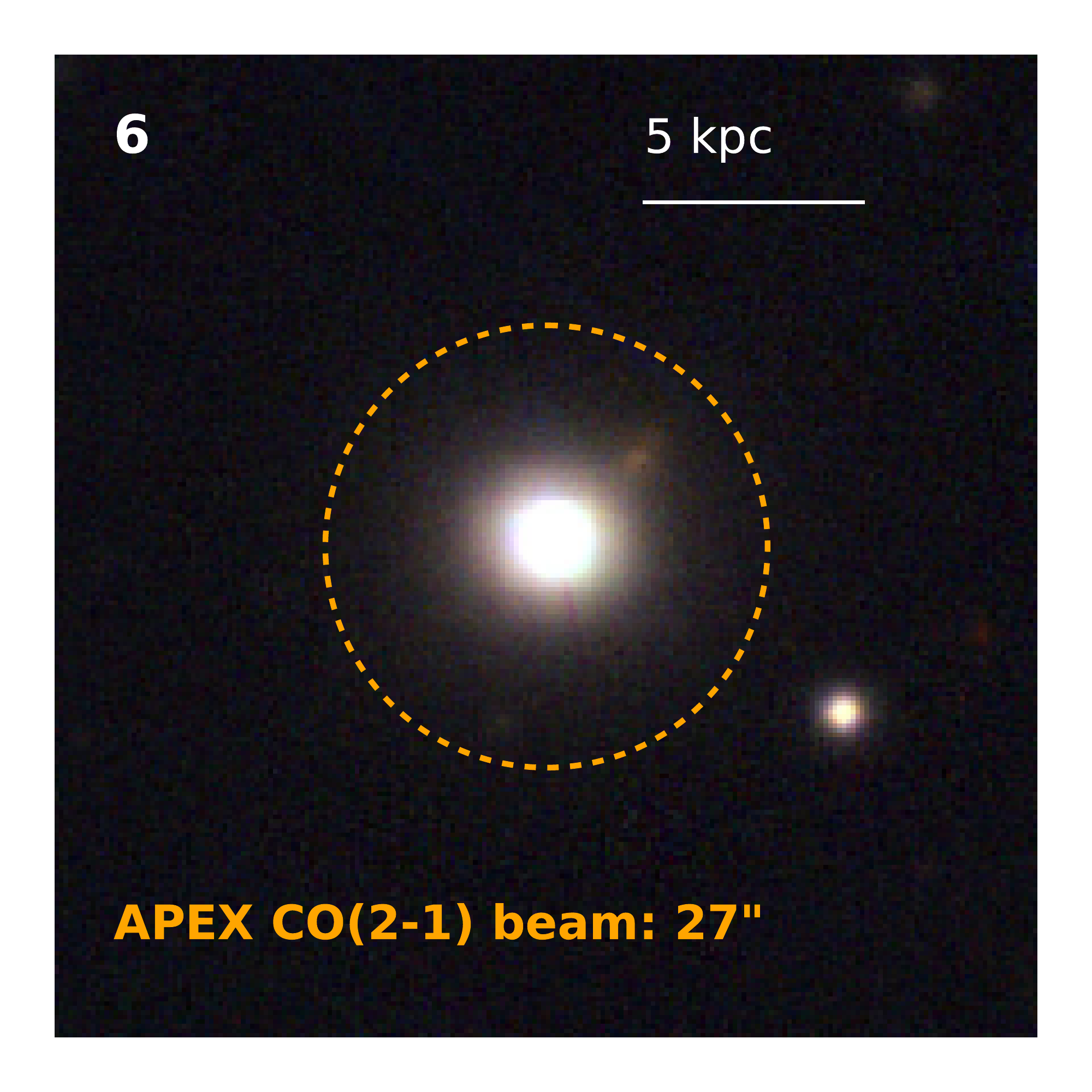}\includegraphics[width=0.26\textwidth]{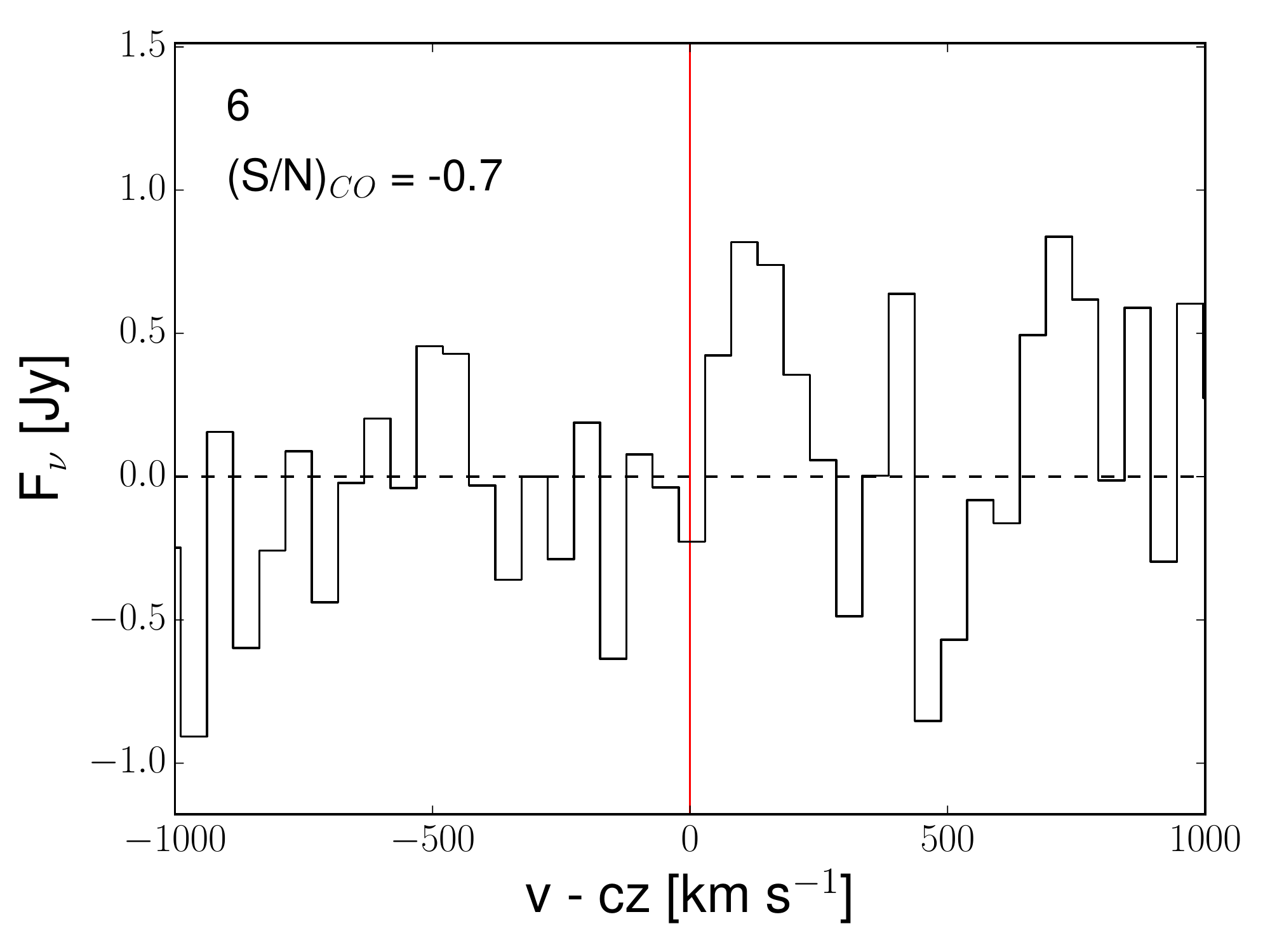}
\includegraphics[width=0.18\textwidth]{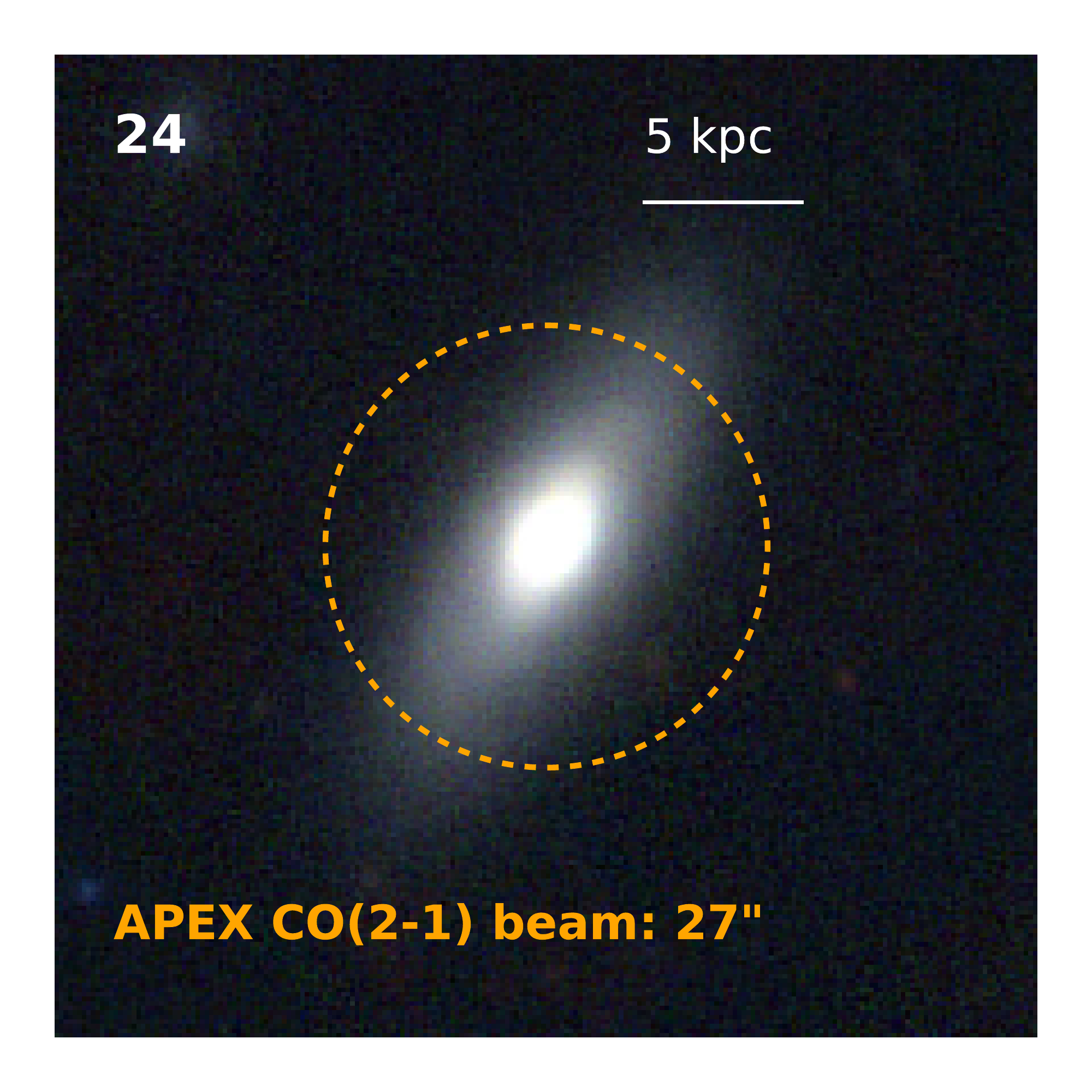}\includegraphics[width=0.26\textwidth]{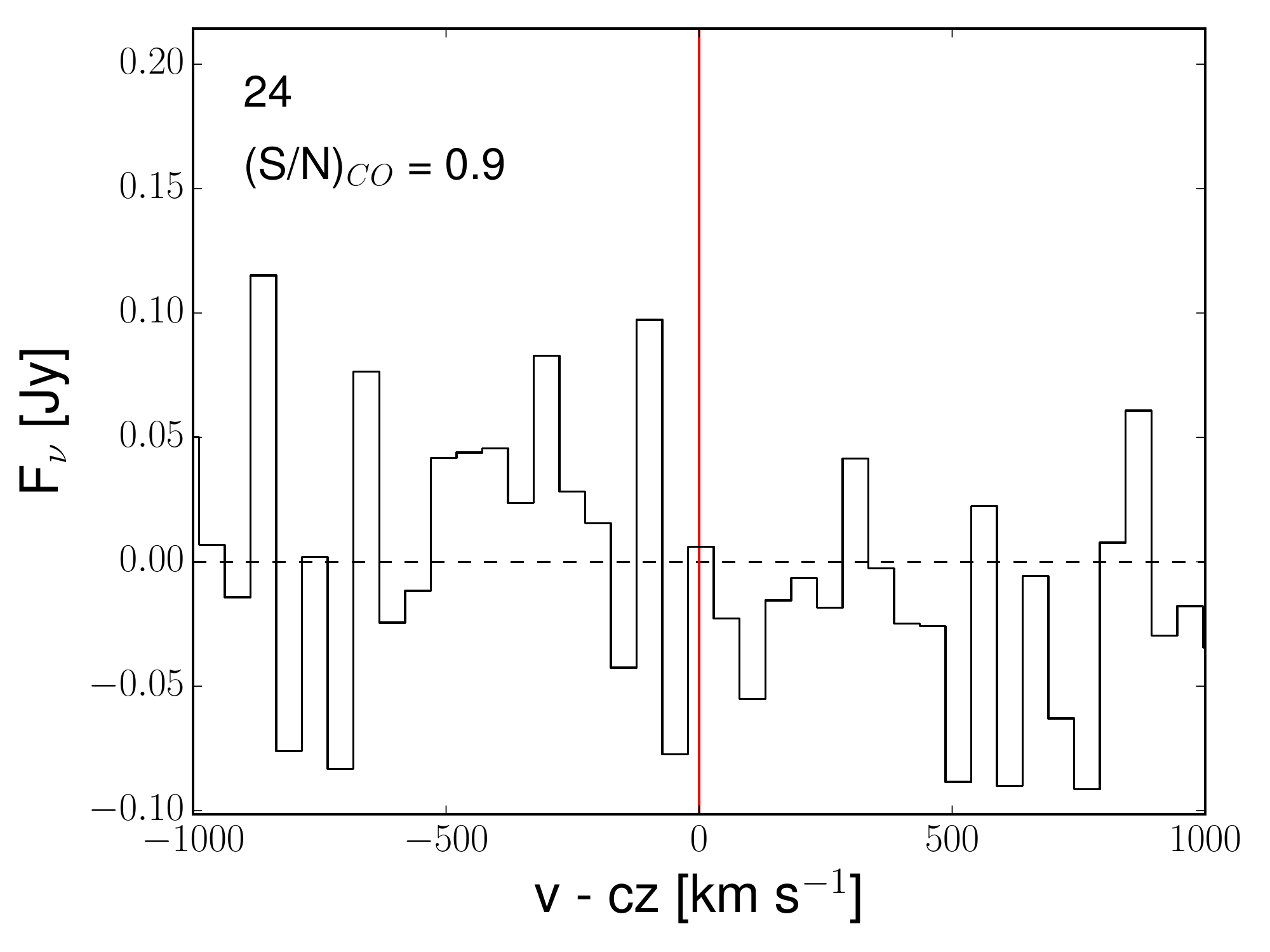}
\includegraphics[width=0.18\textwidth]{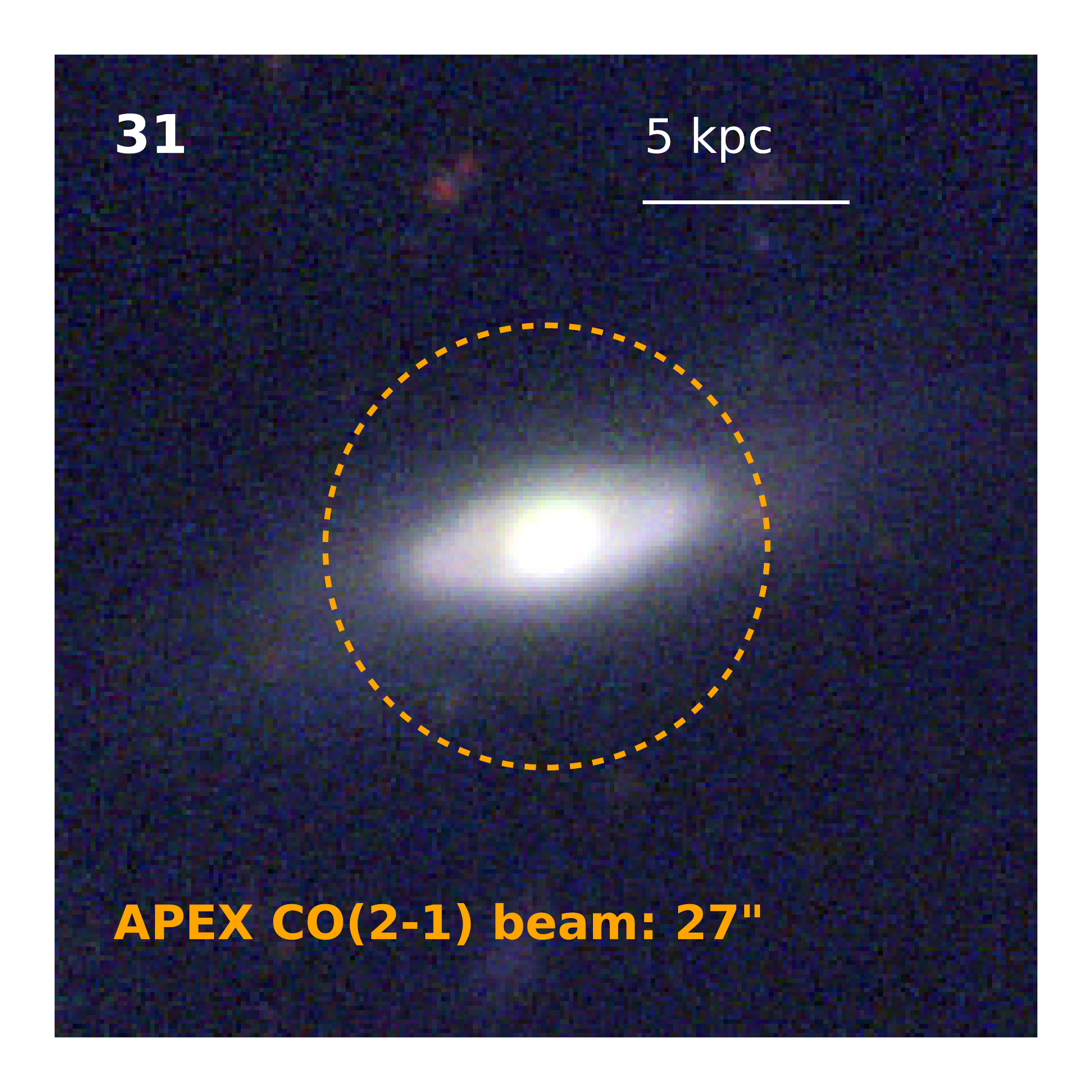}\includegraphics[width=0.26\textwidth]{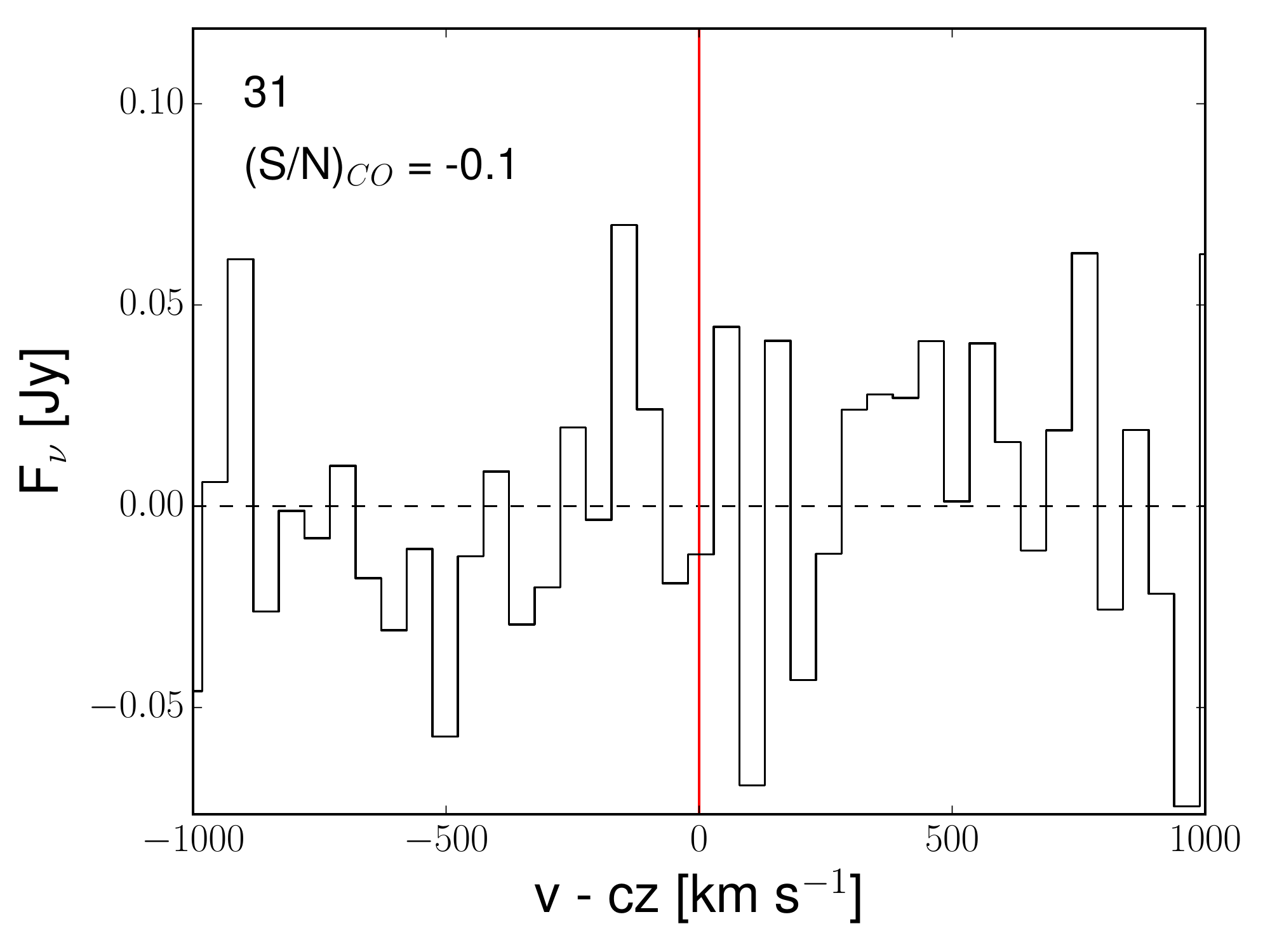}
\includegraphics[width=0.18\textwidth]{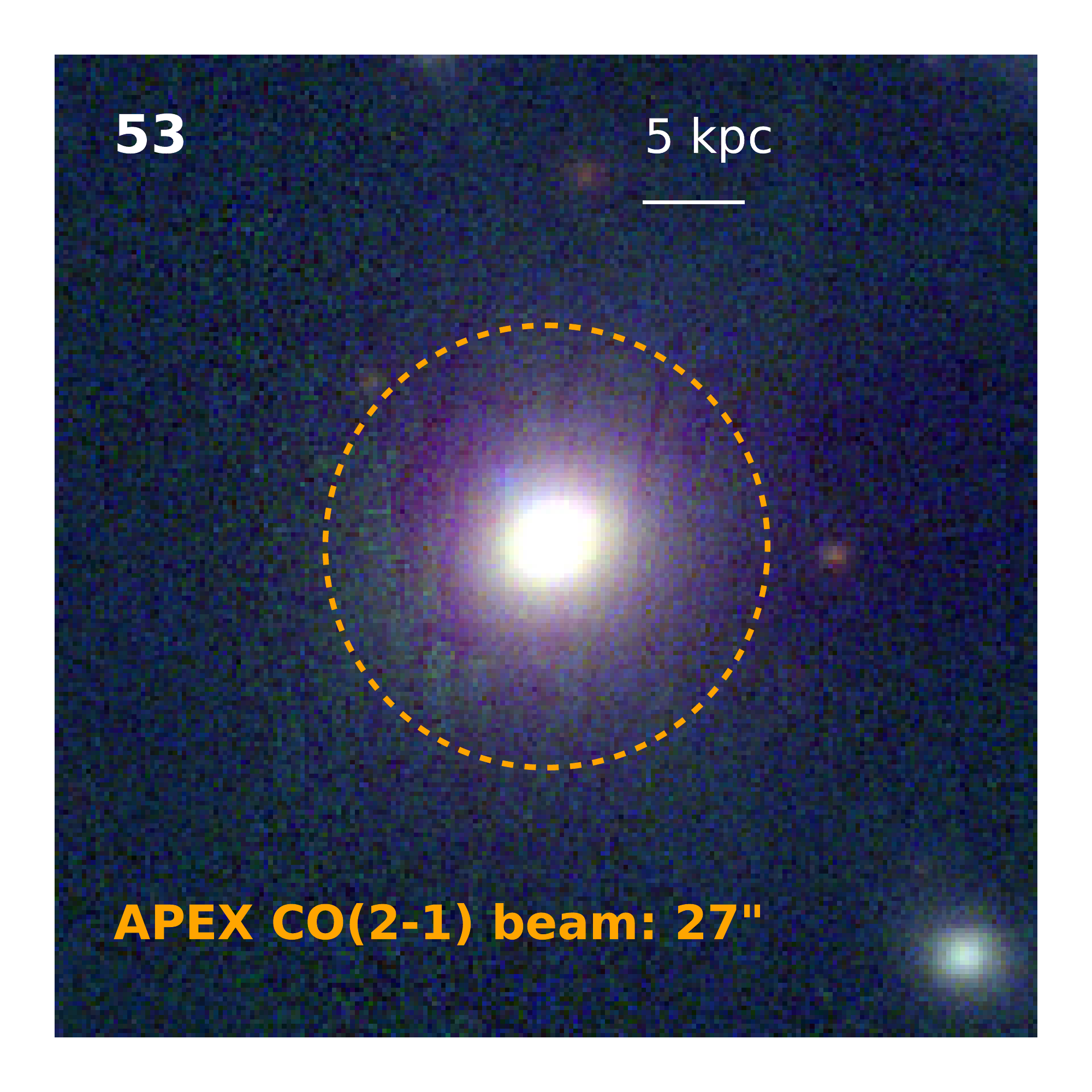}\includegraphics[width=0.26\textwidth]{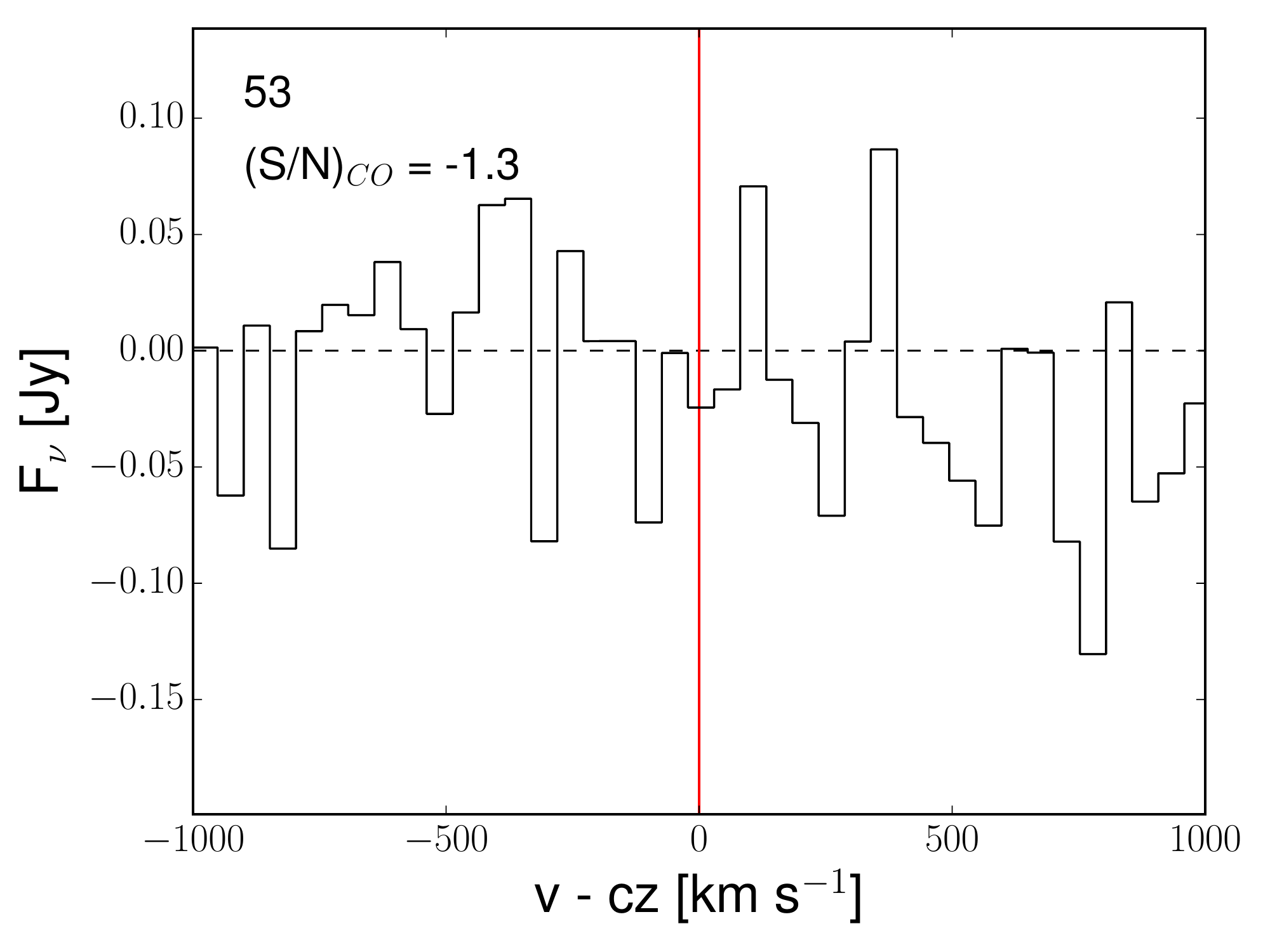}
\includegraphics[width=0.18\textwidth]{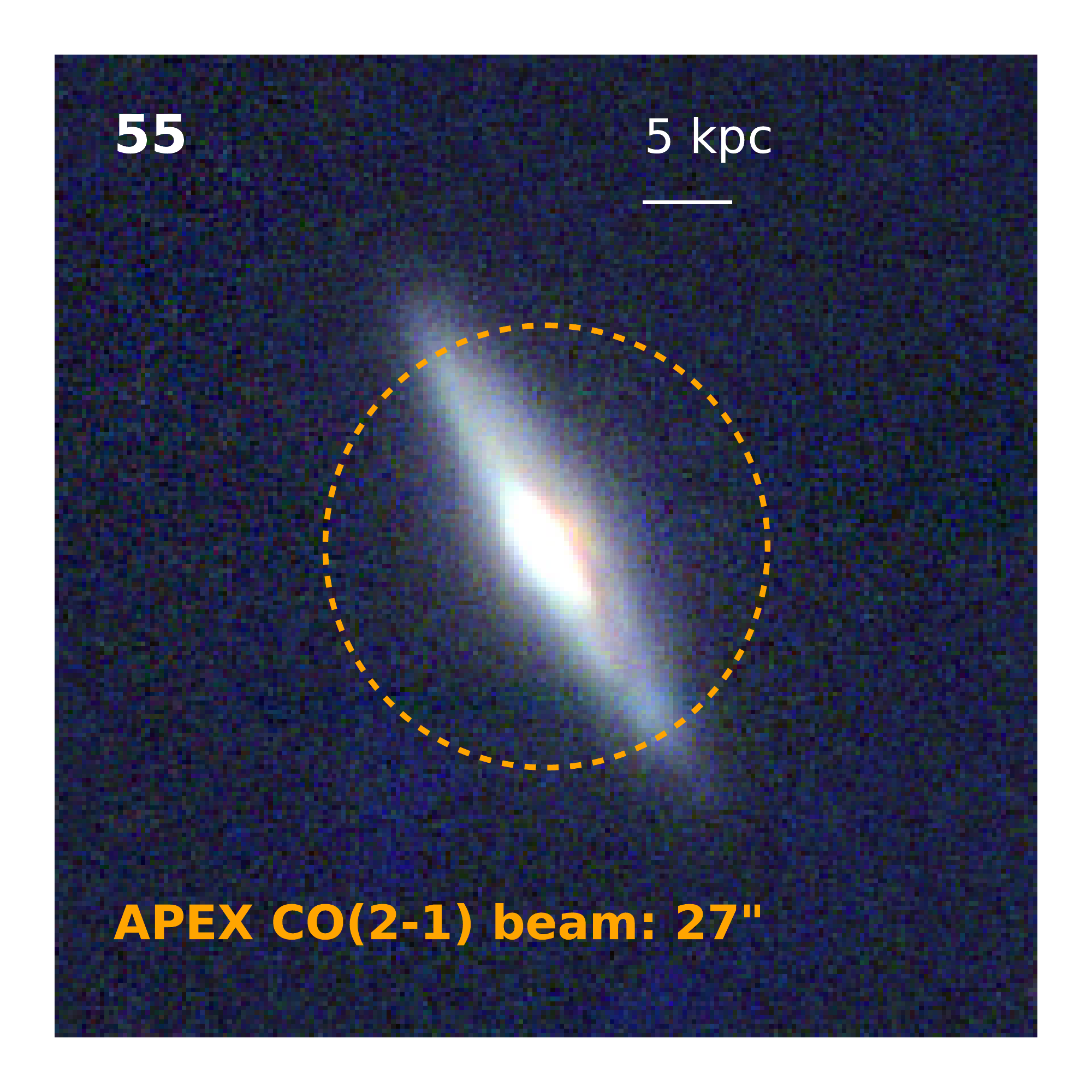}\includegraphics[width=0.26\textwidth]{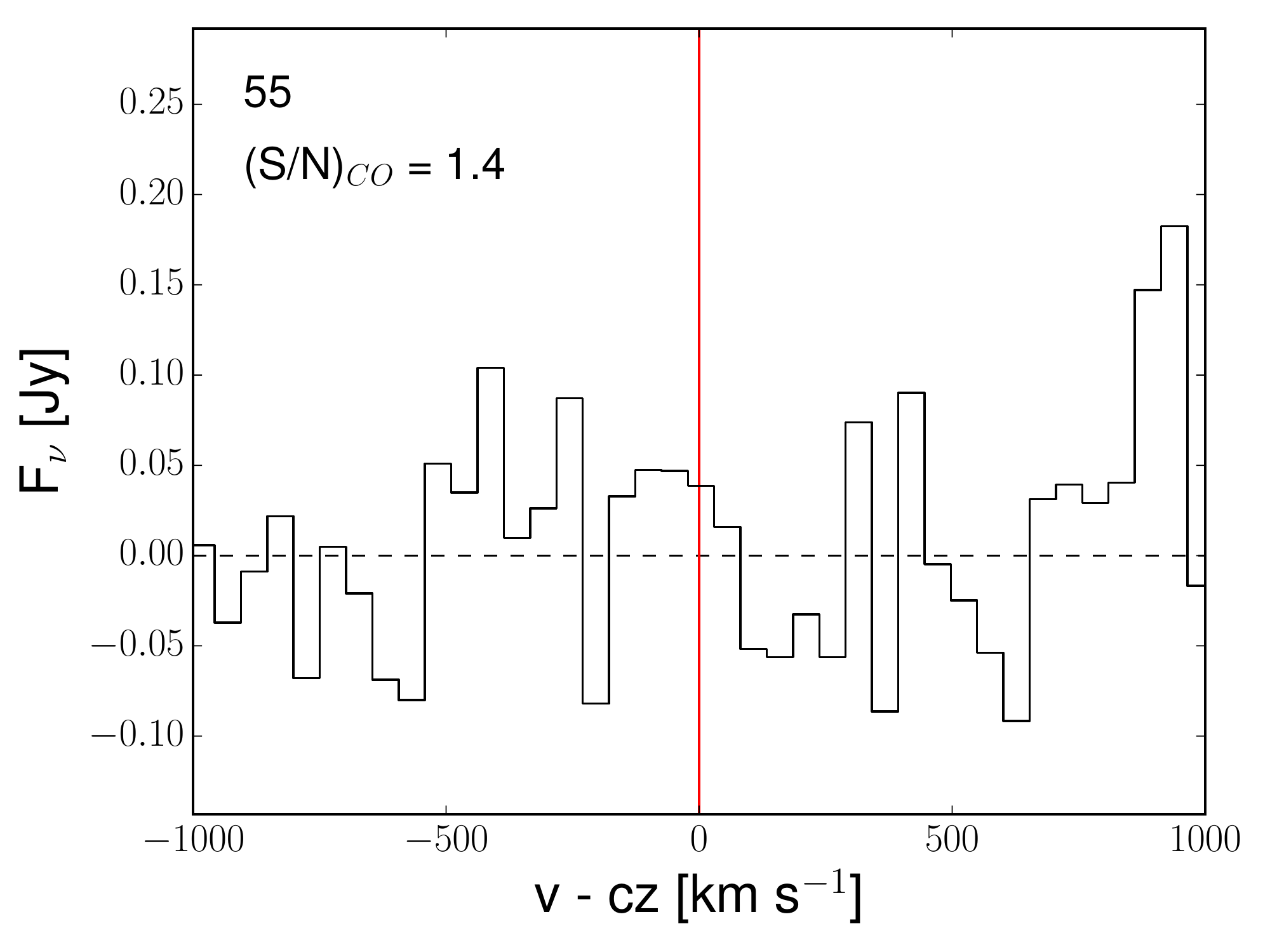}
\includegraphics[width=0.18\textwidth]{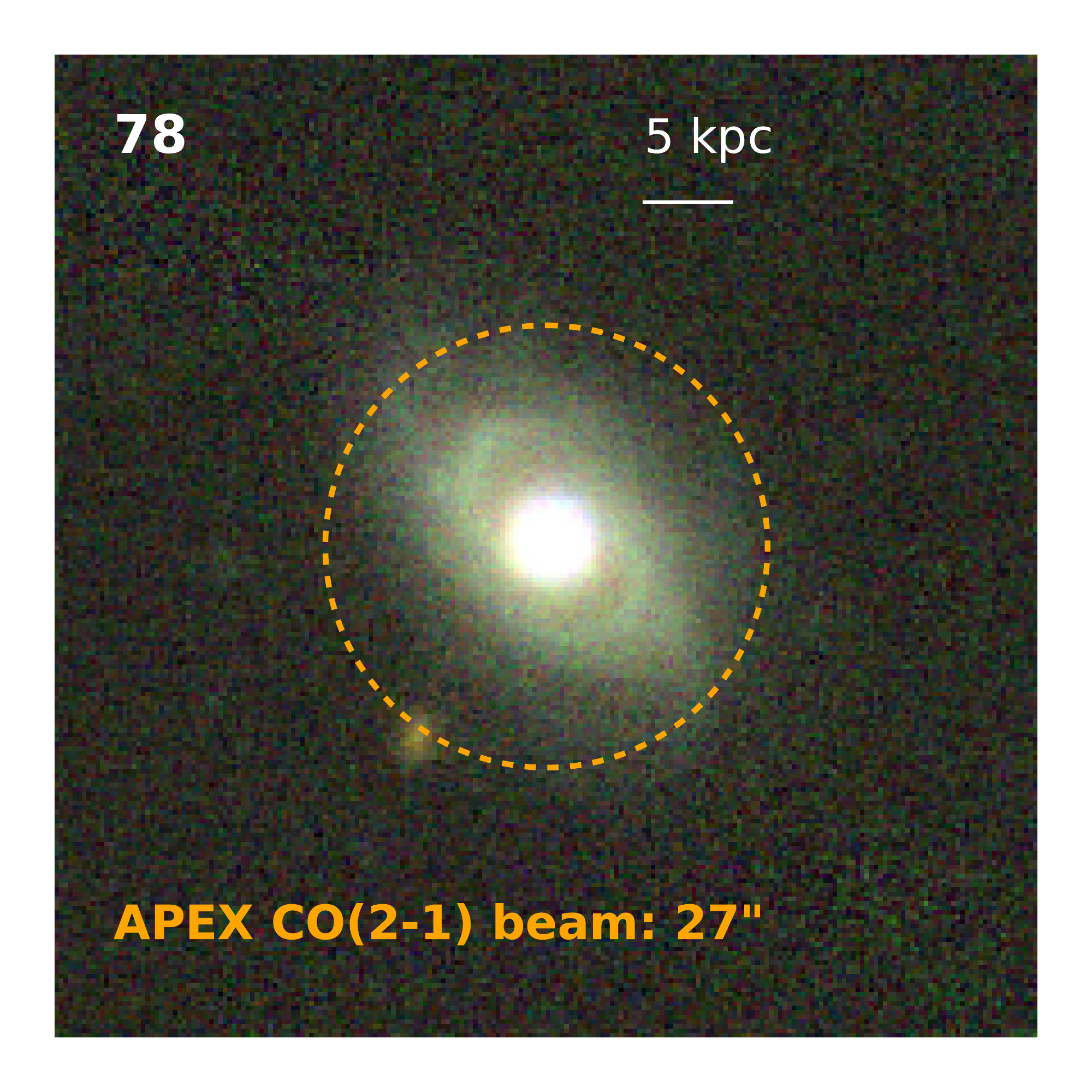}\includegraphics[width=0.26\textwidth]{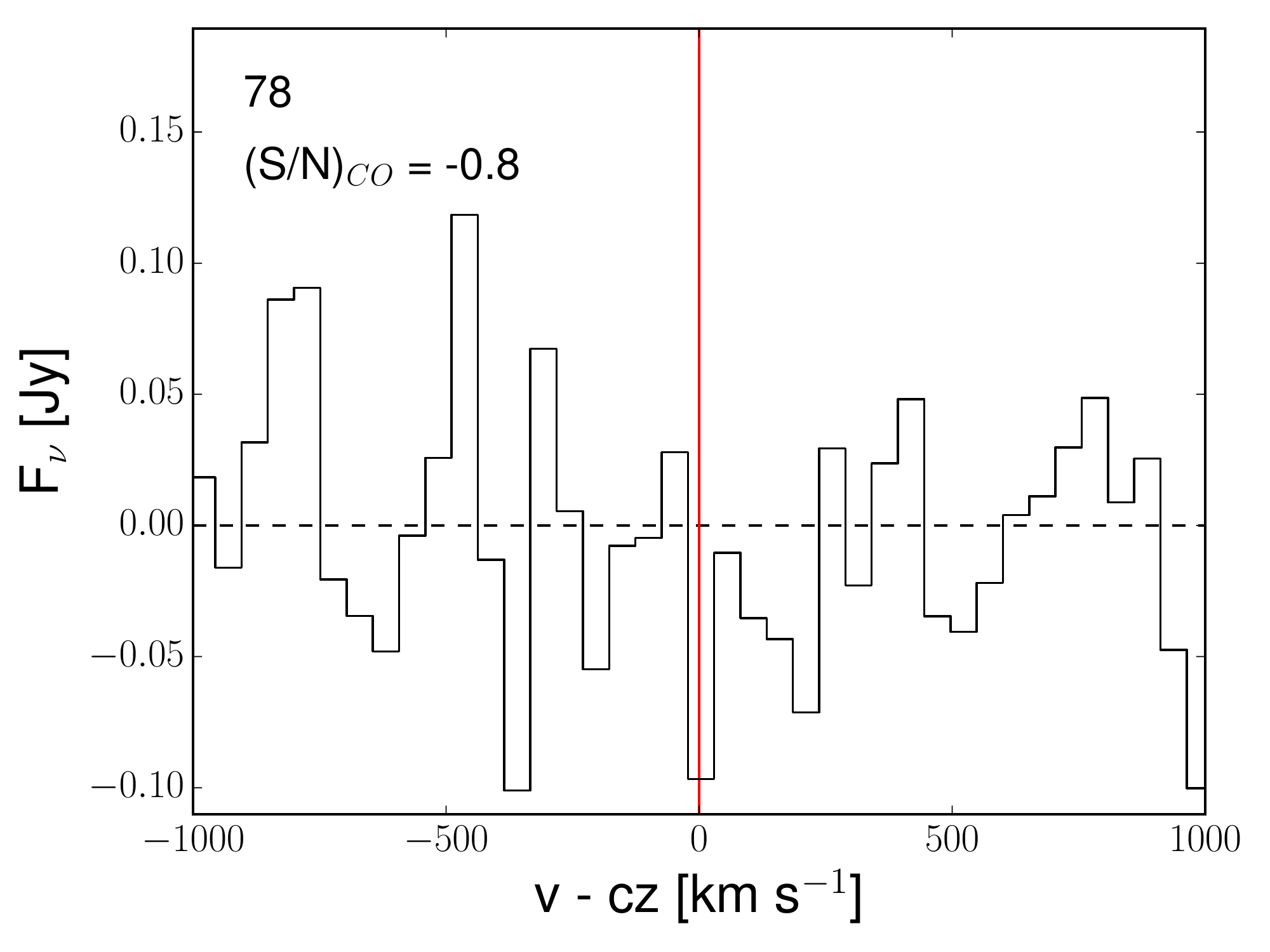}
\includegraphics[width=0.18\textwidth]{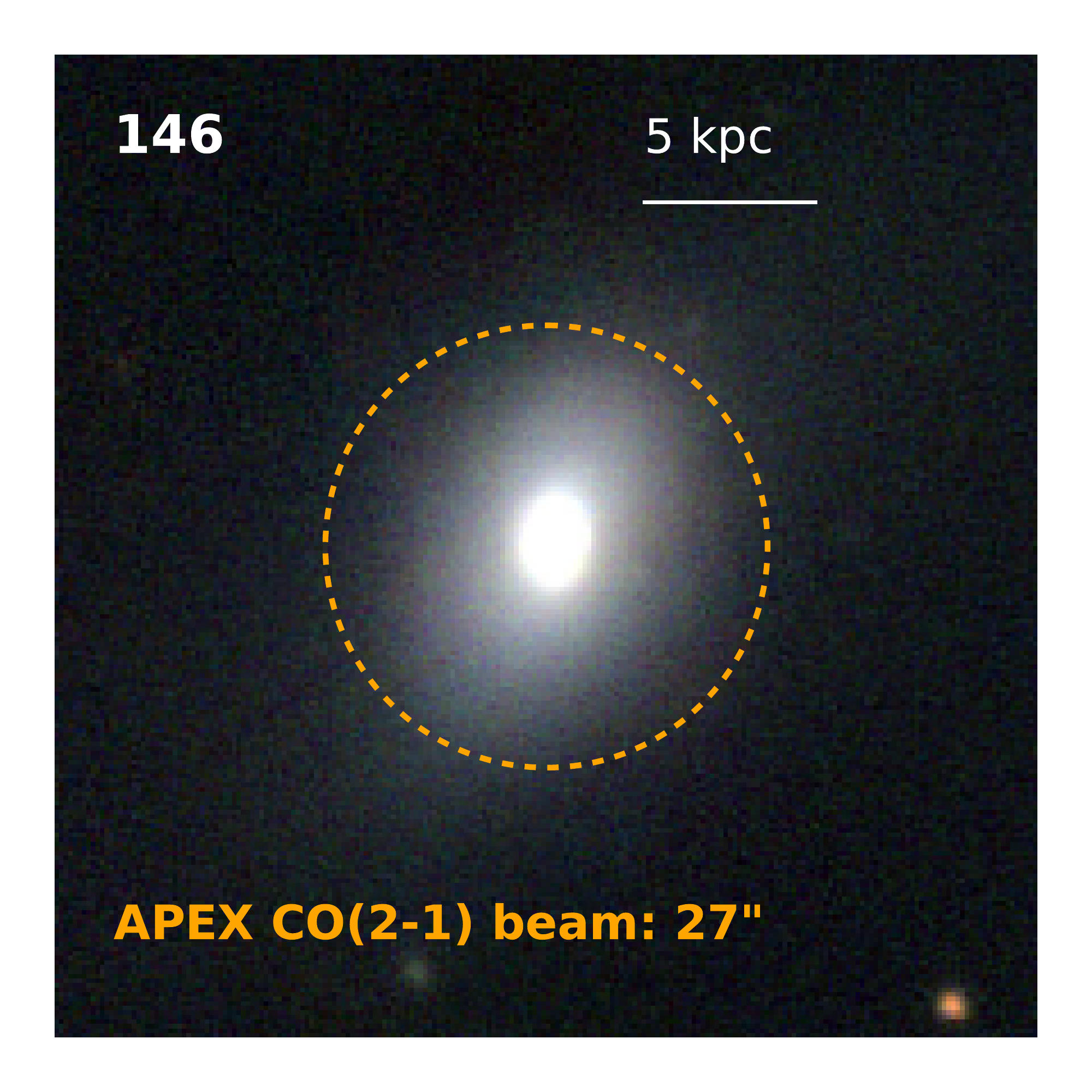}\includegraphics[width=0.26\textwidth]{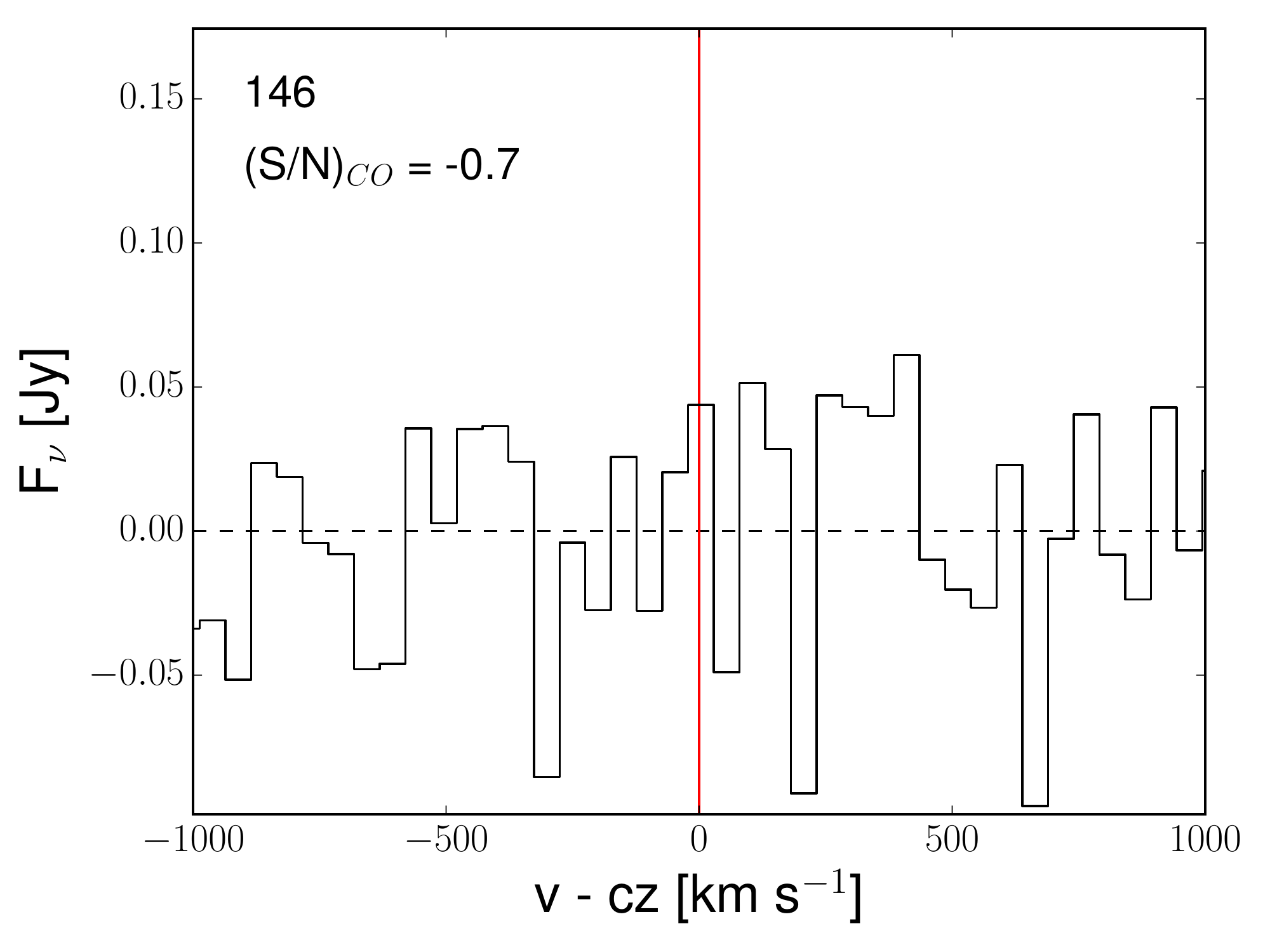}
\includegraphics[width=0.18\textwidth]{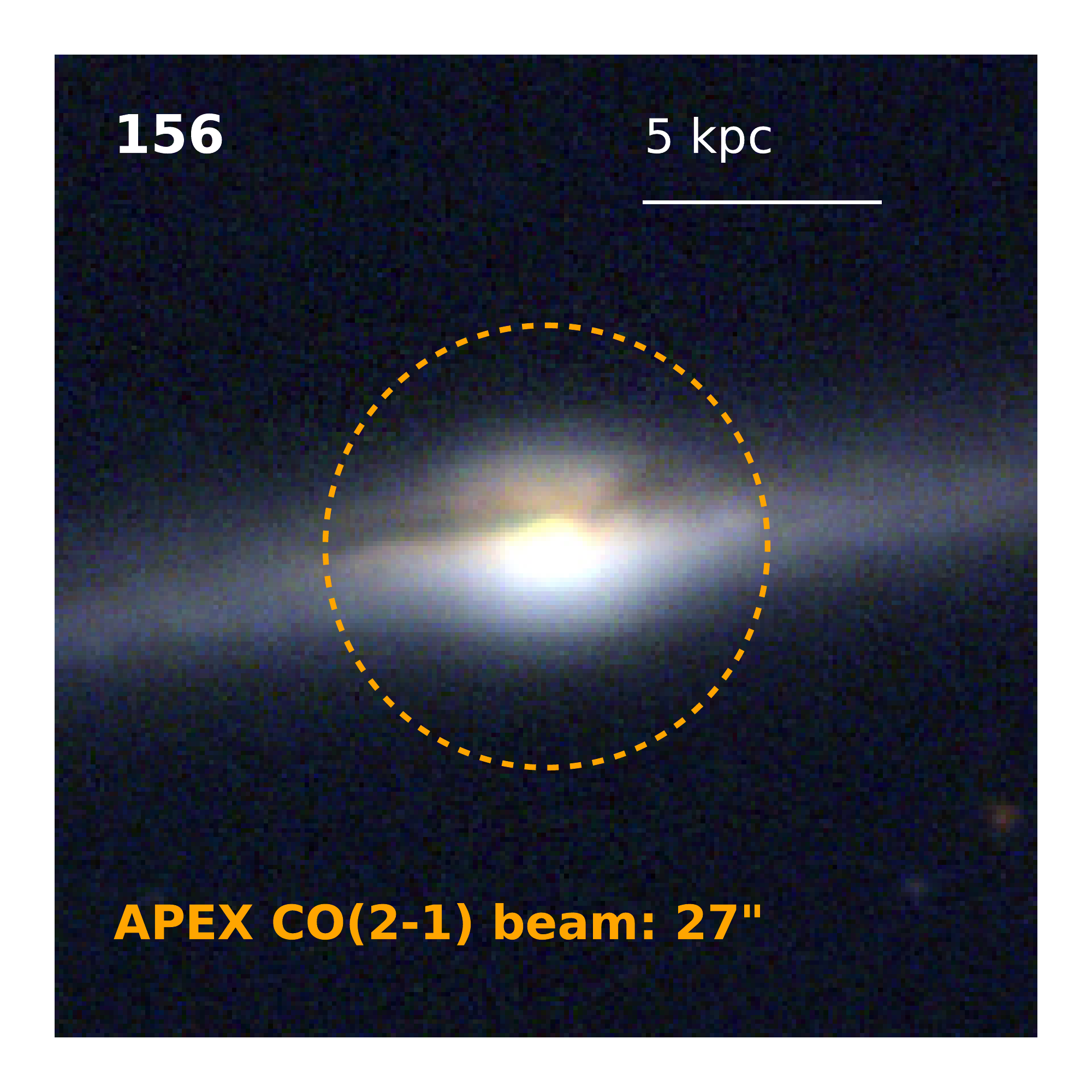}\includegraphics[width=0.26\textwidth]{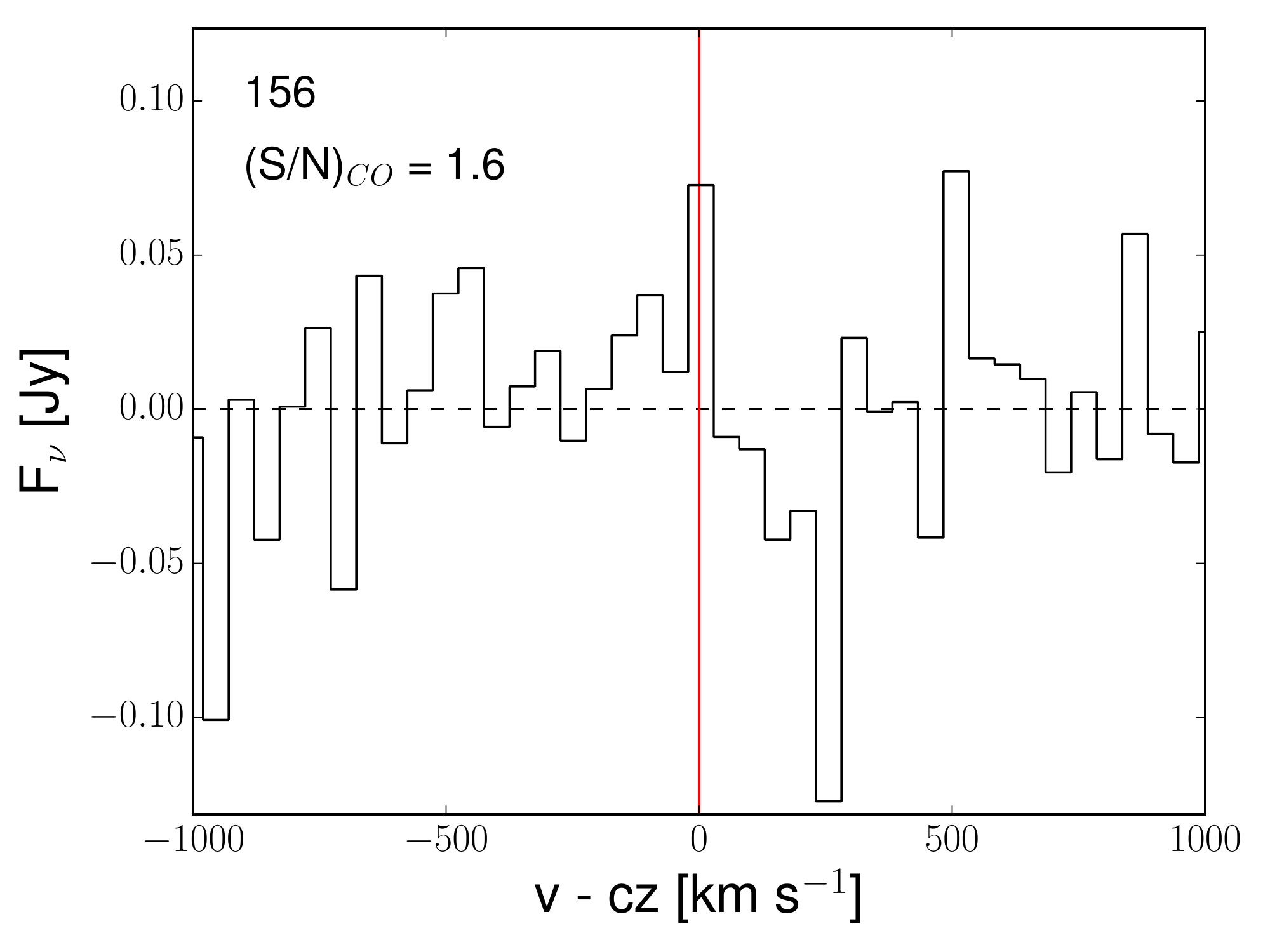}
\includegraphics[width=0.18\textwidth]{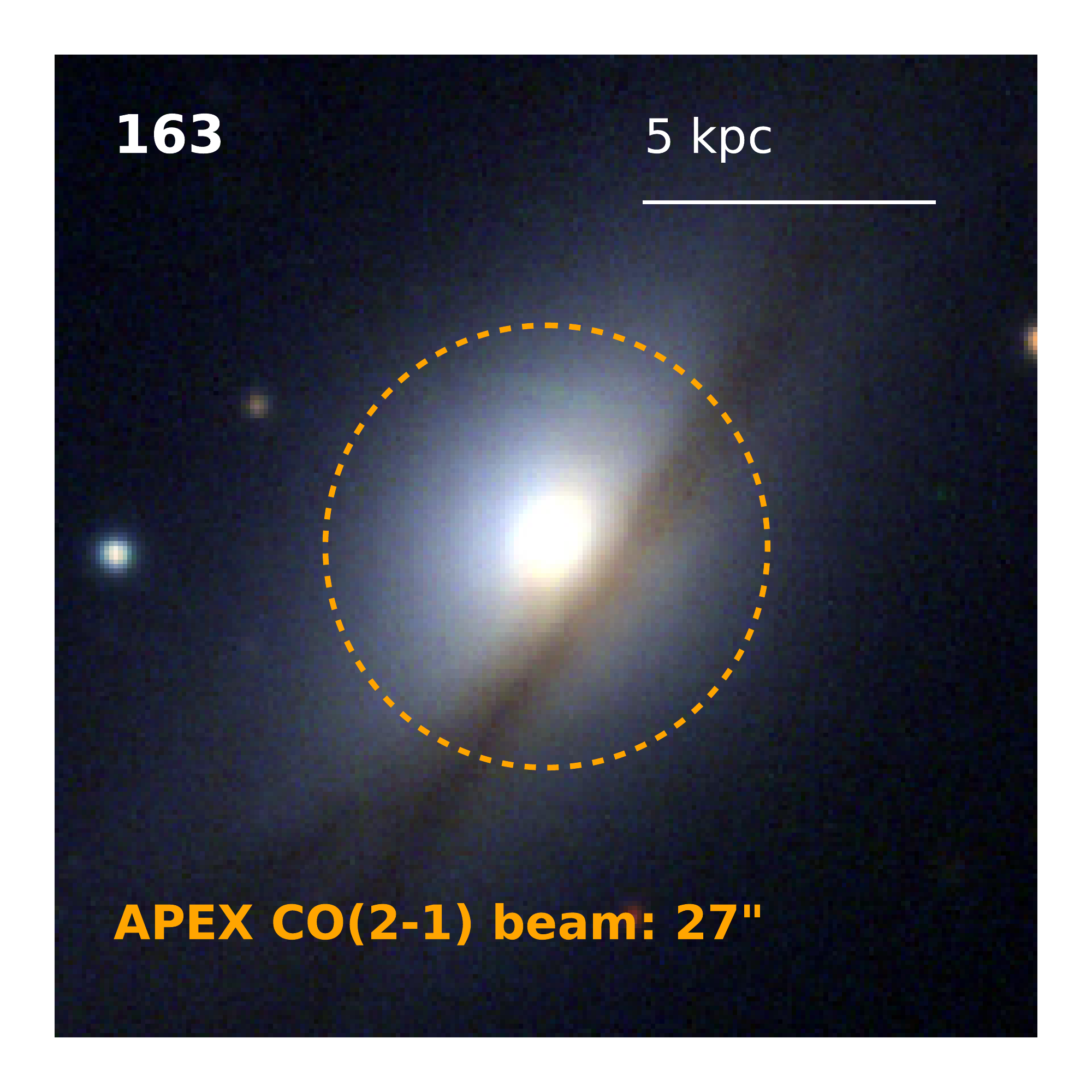}\includegraphics[width=0.26\textwidth]{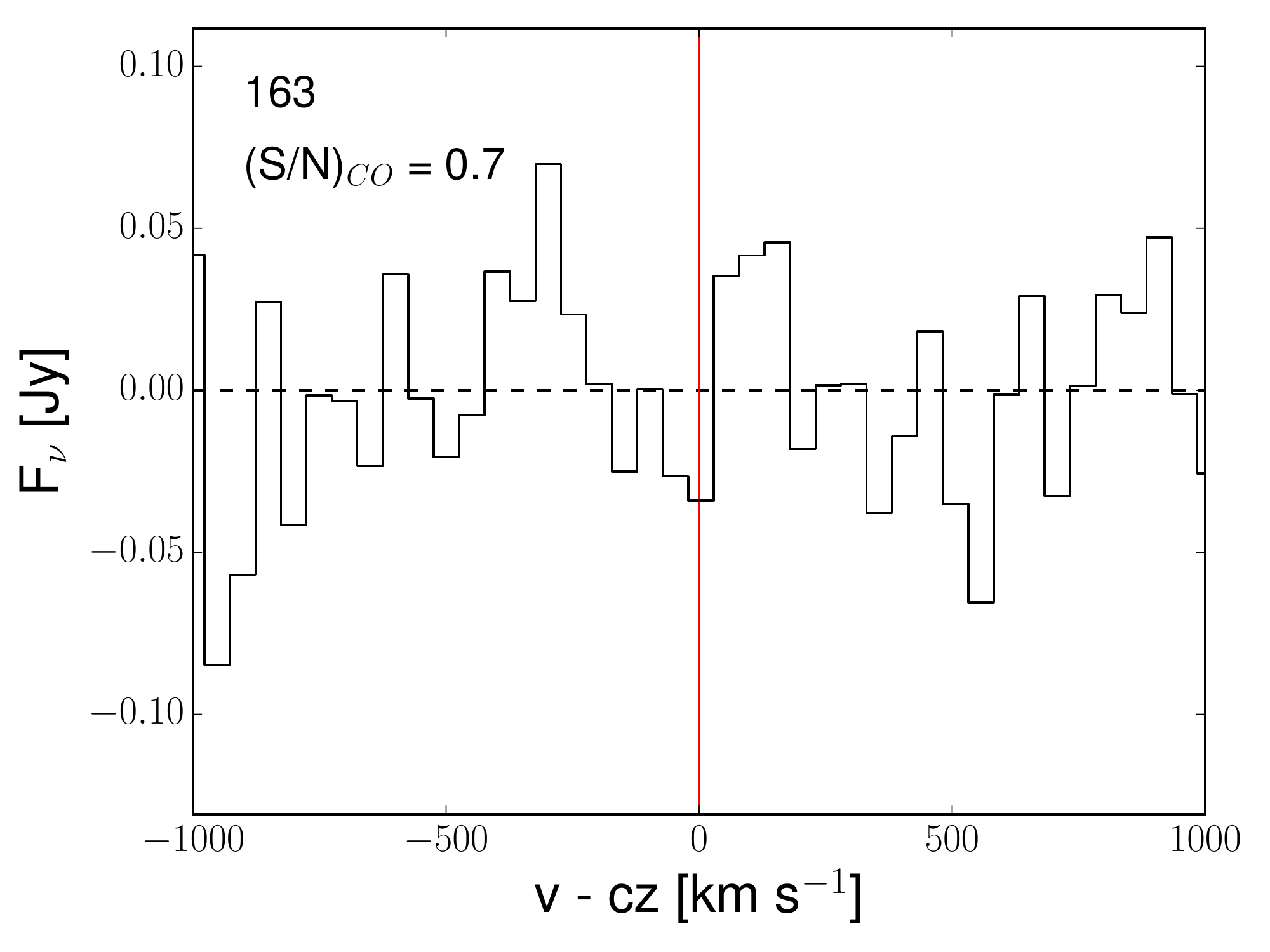}
\includegraphics[width=0.18\textwidth]{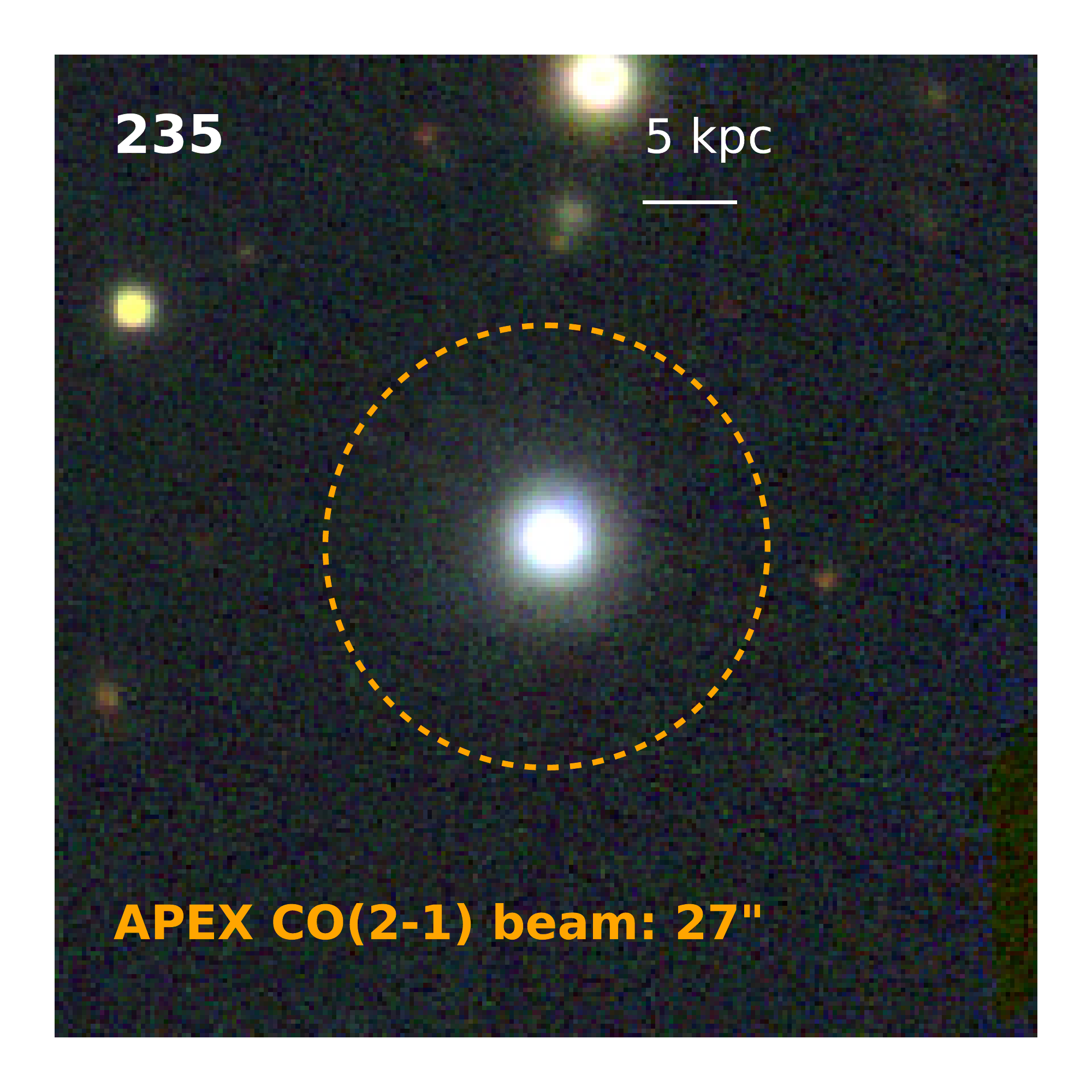}\includegraphics[width=0.26\textwidth]{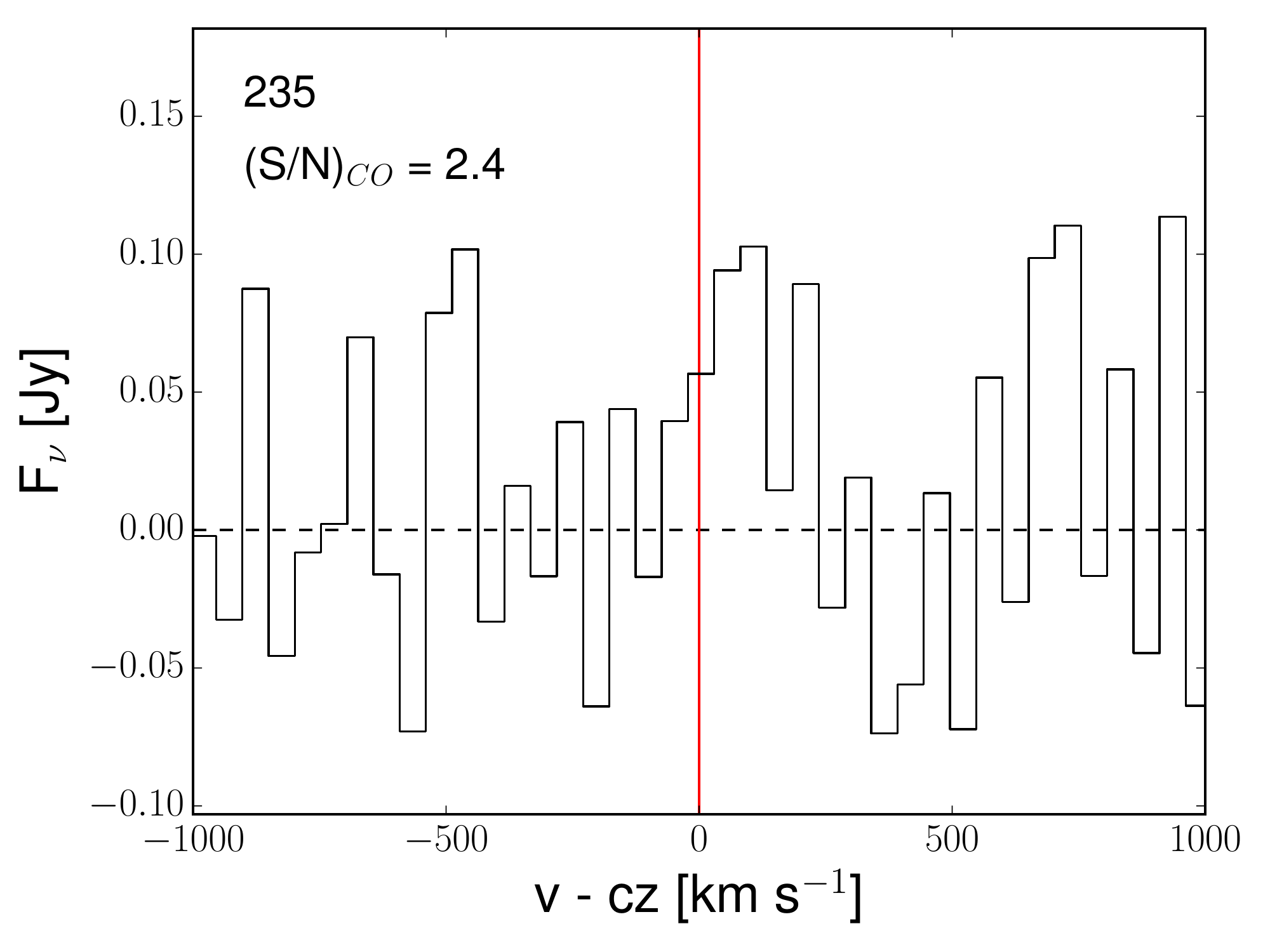}
\includegraphics[width=0.18\textwidth]{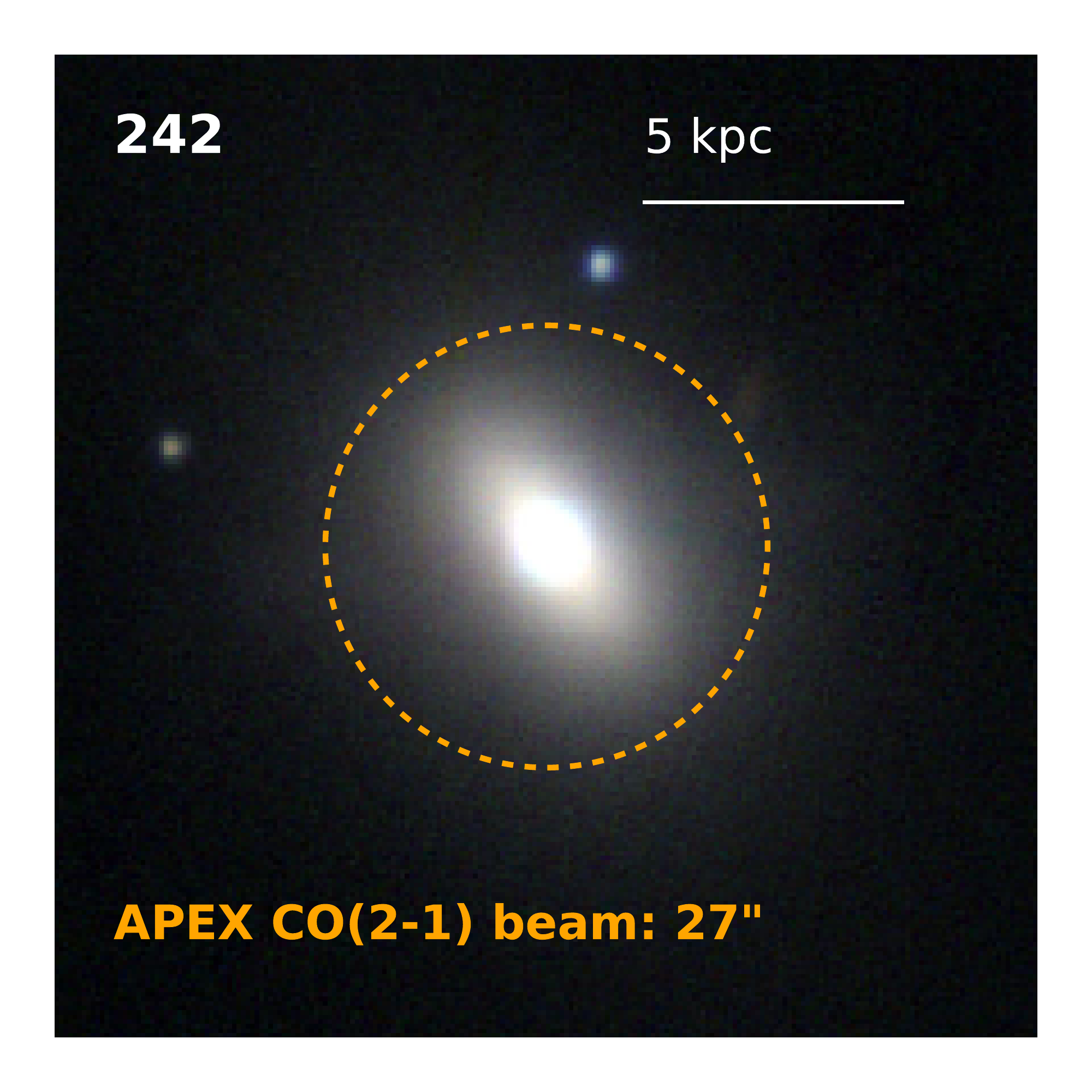}\includegraphics[width=0.26\textwidth]{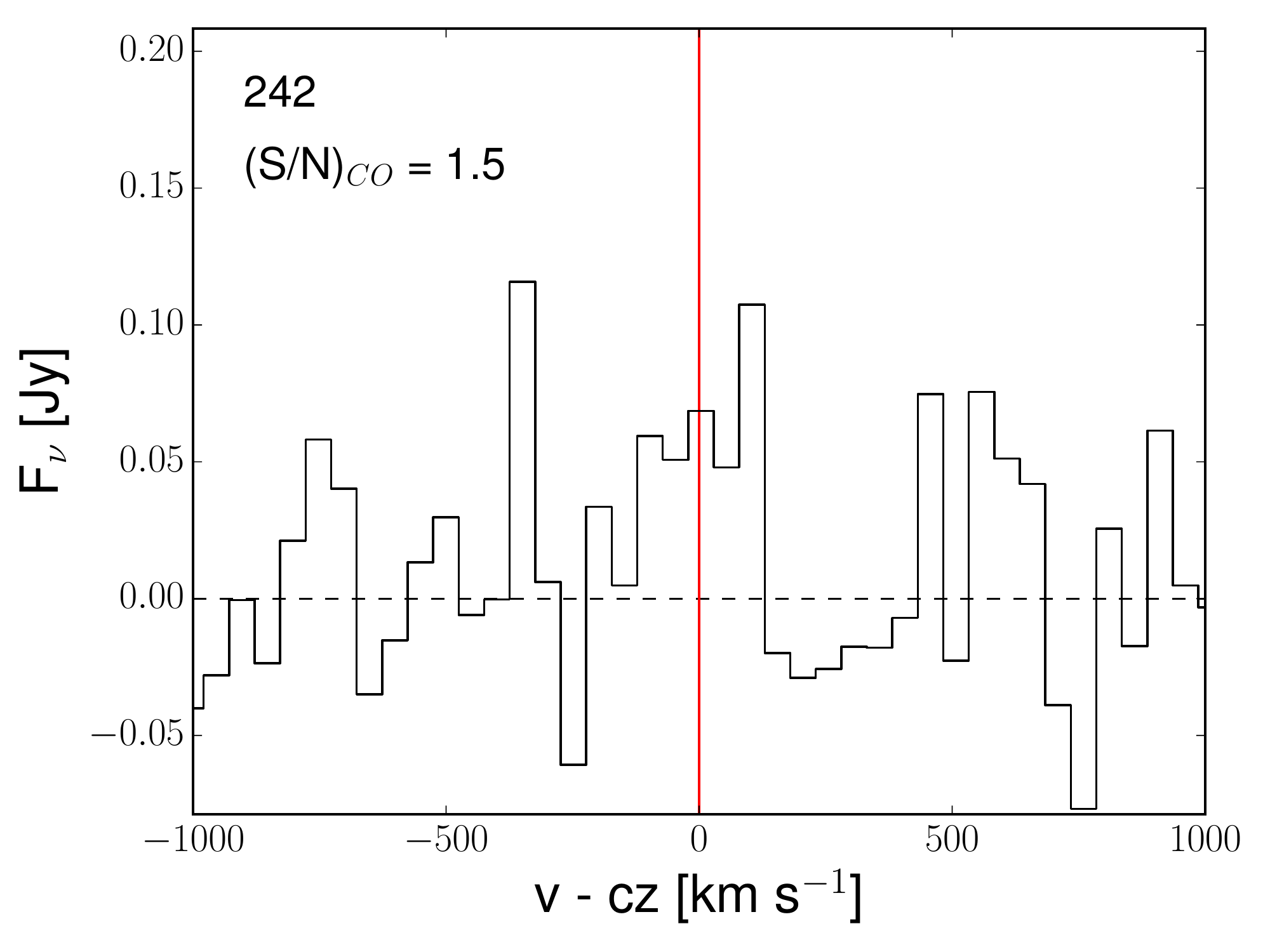}
\caption{Similar to Figure \ref{fig:CO21_spectra}, for the undetected sources not already shown in the text.
} 
\label{fig:CO21_spectra_all_undetect1}
\end{figure*}

\begin{figure*}
\centering
\raggedright
\includegraphics[width=0.18\textwidth]{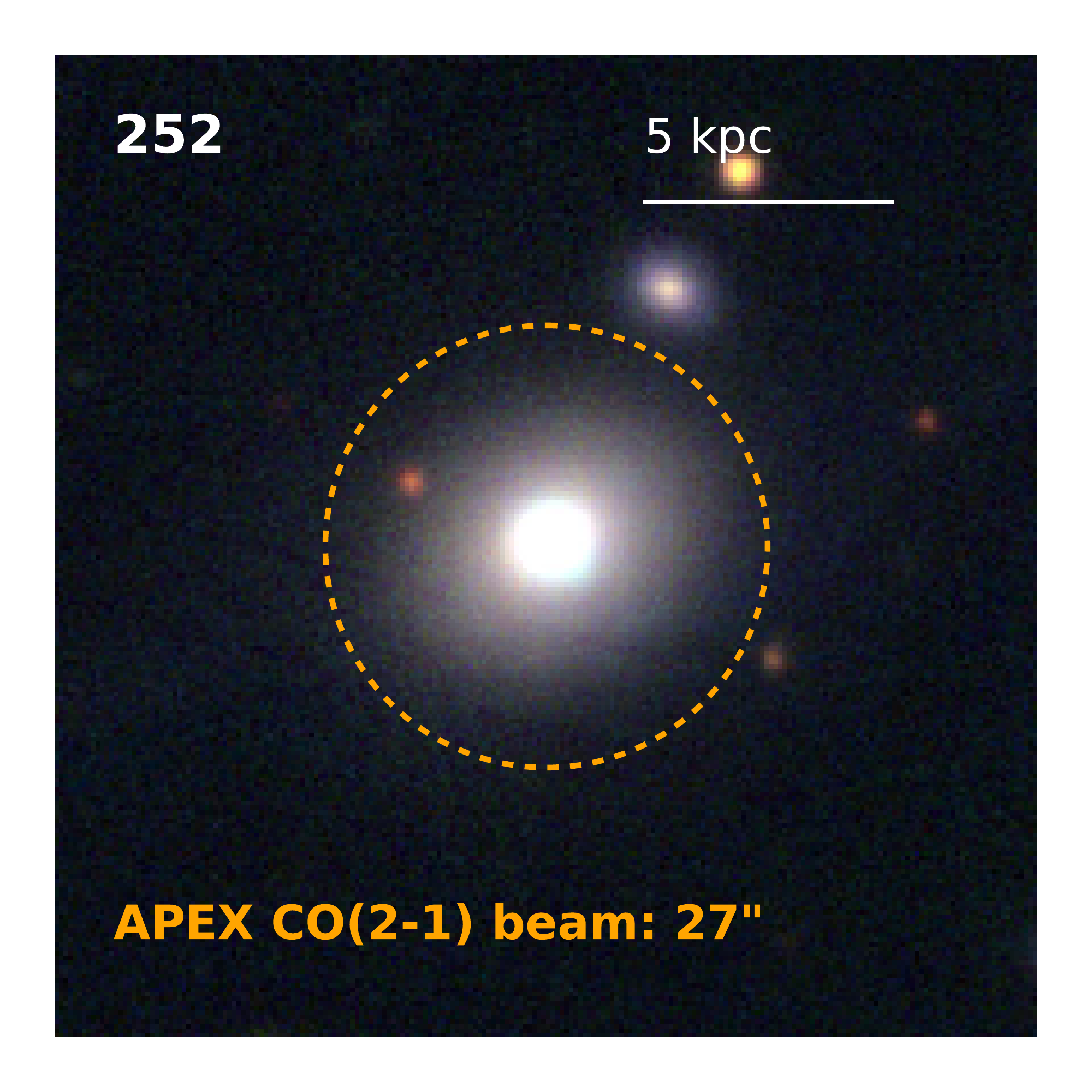}\includegraphics[width=0.26\textwidth]{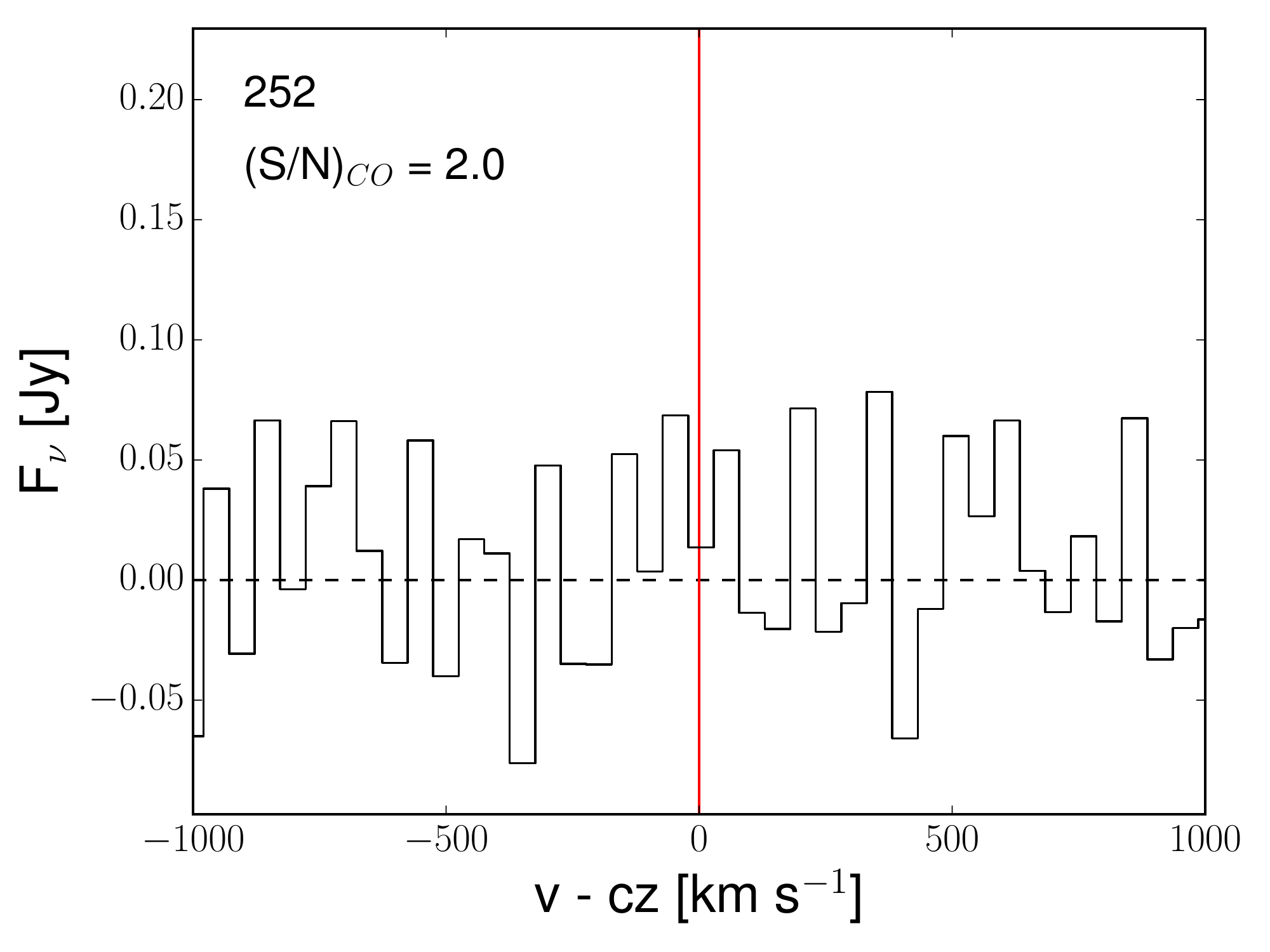}
\includegraphics[width=0.18\textwidth]{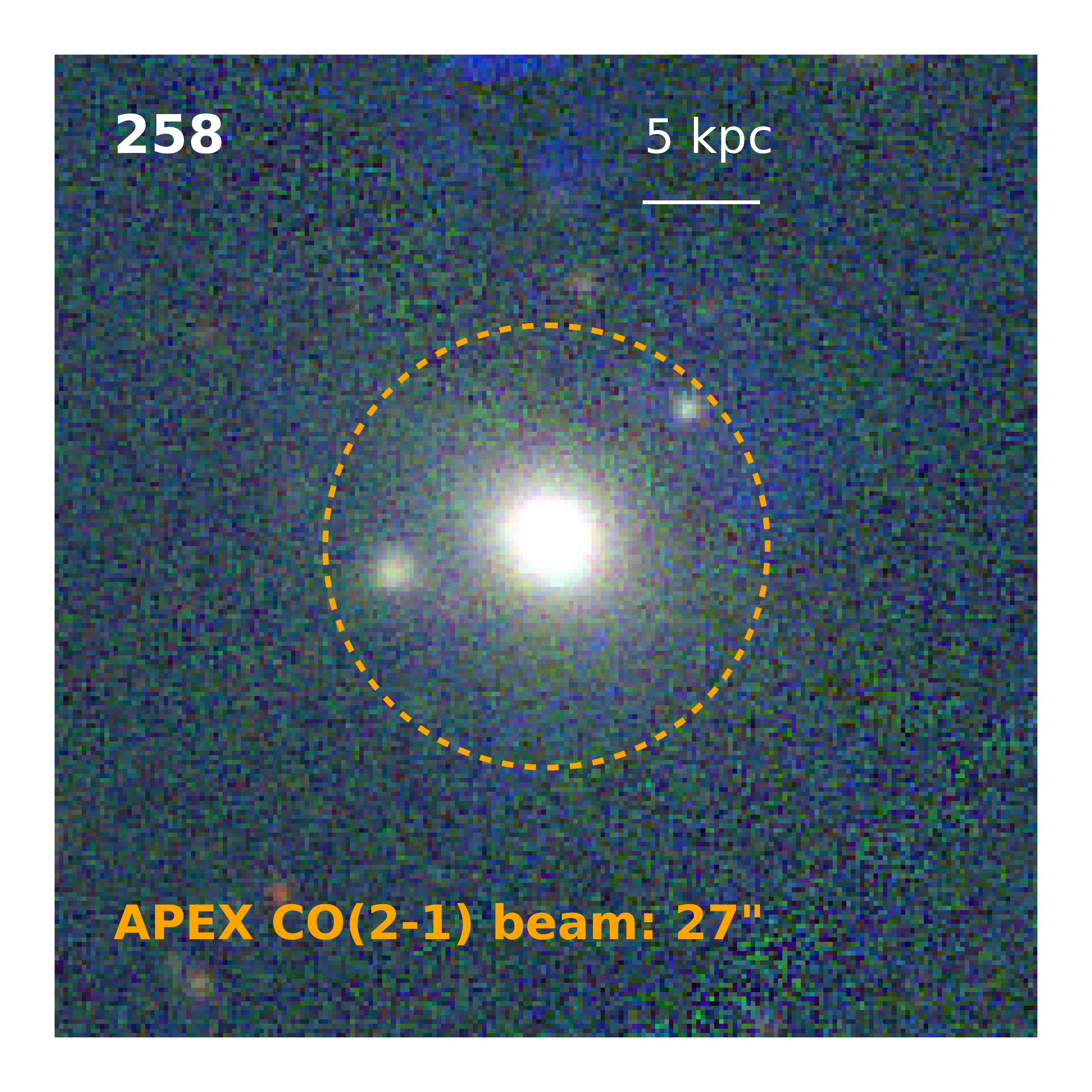}\includegraphics[width=0.26\textwidth]{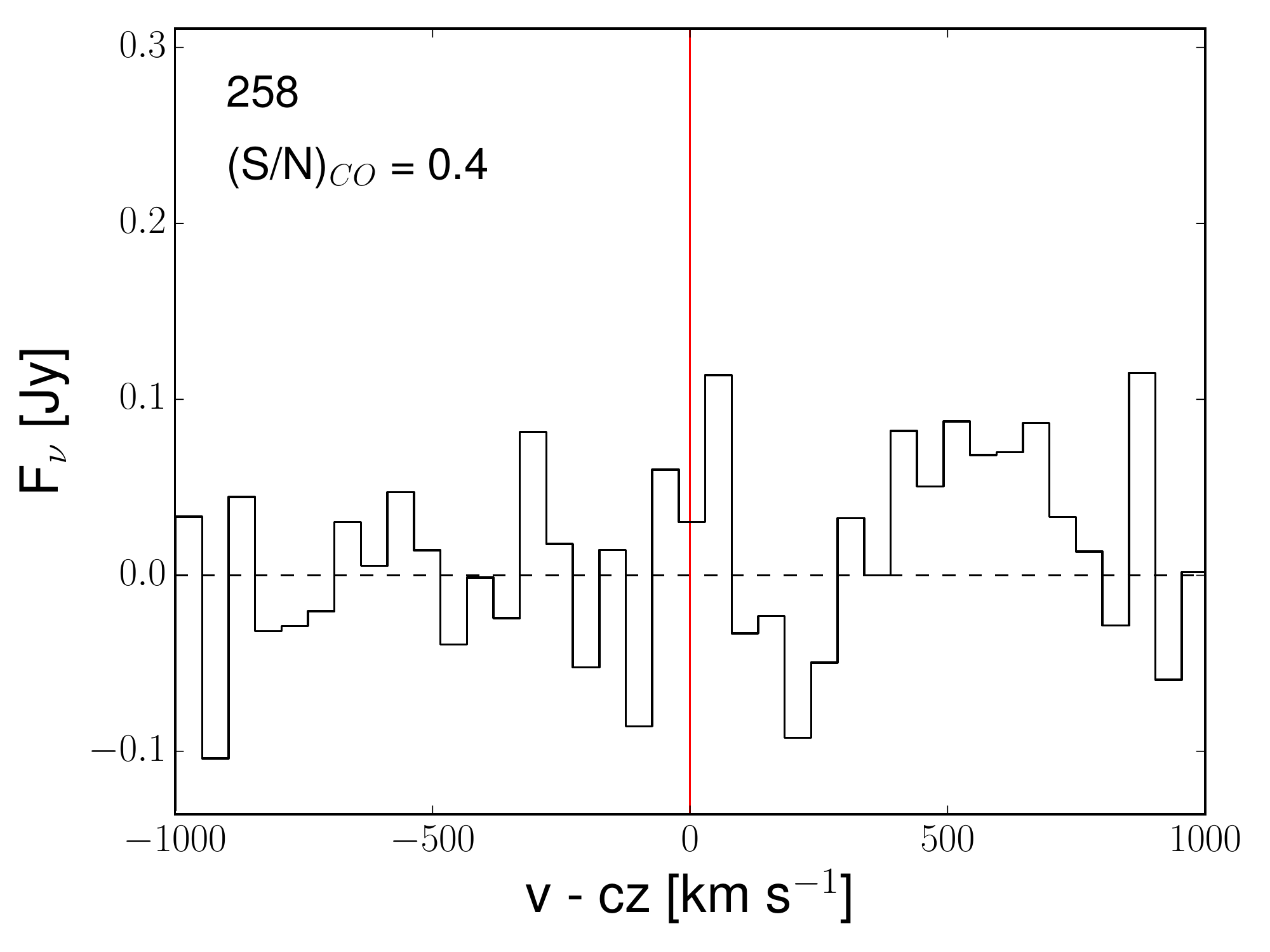}
\includegraphics[width=0.18\textwidth]{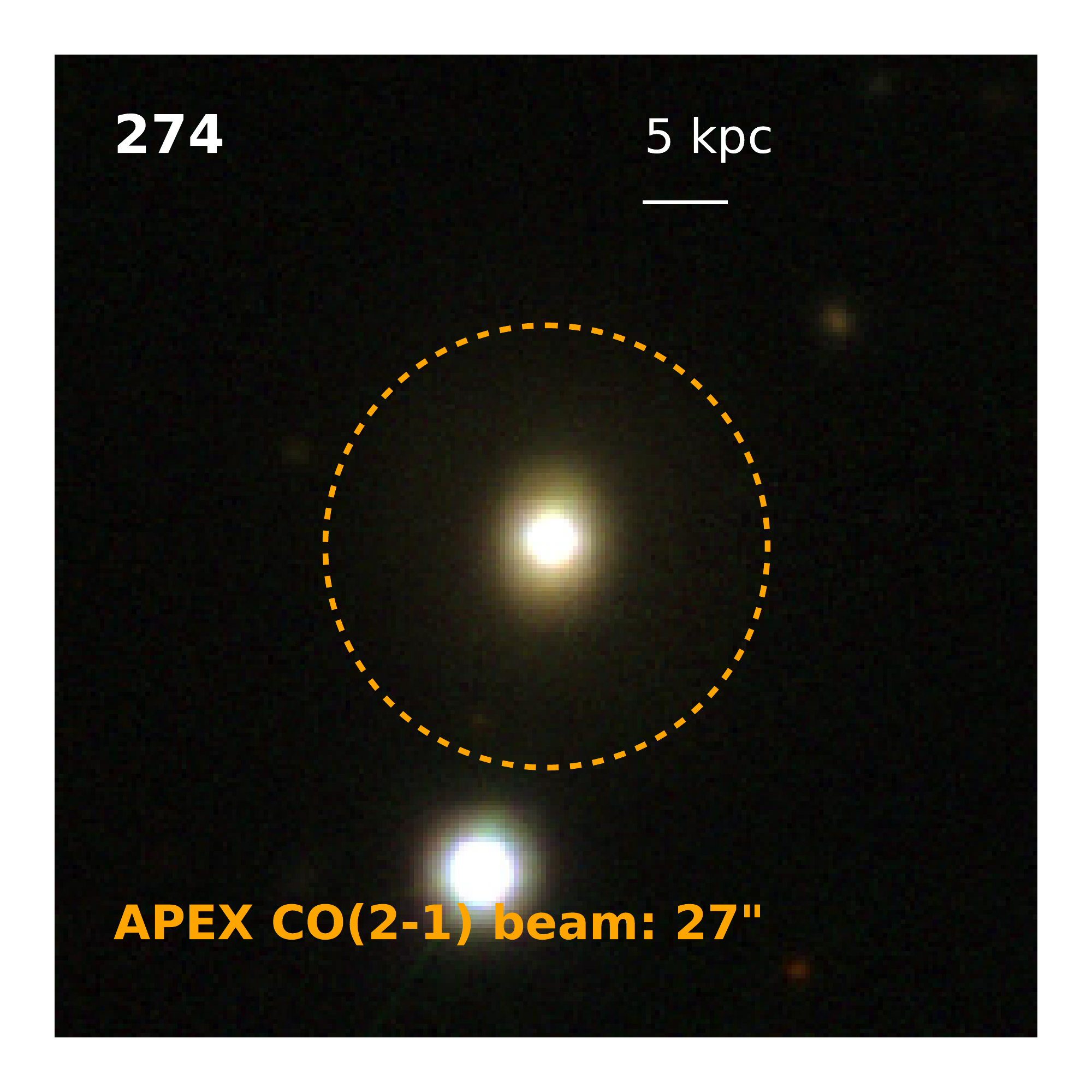}\includegraphics[width=0.26\textwidth]{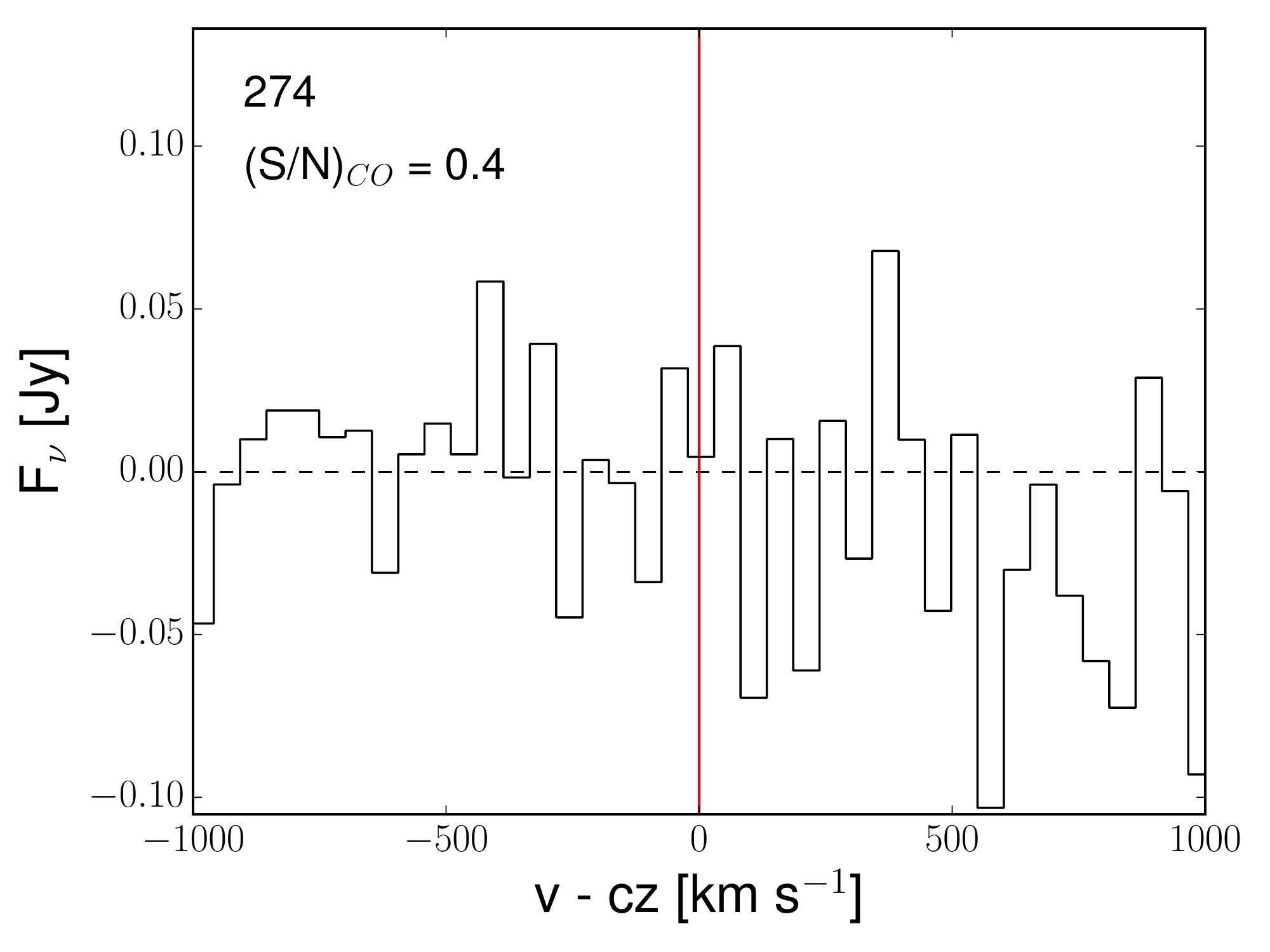}
\includegraphics[width=0.18\textwidth]{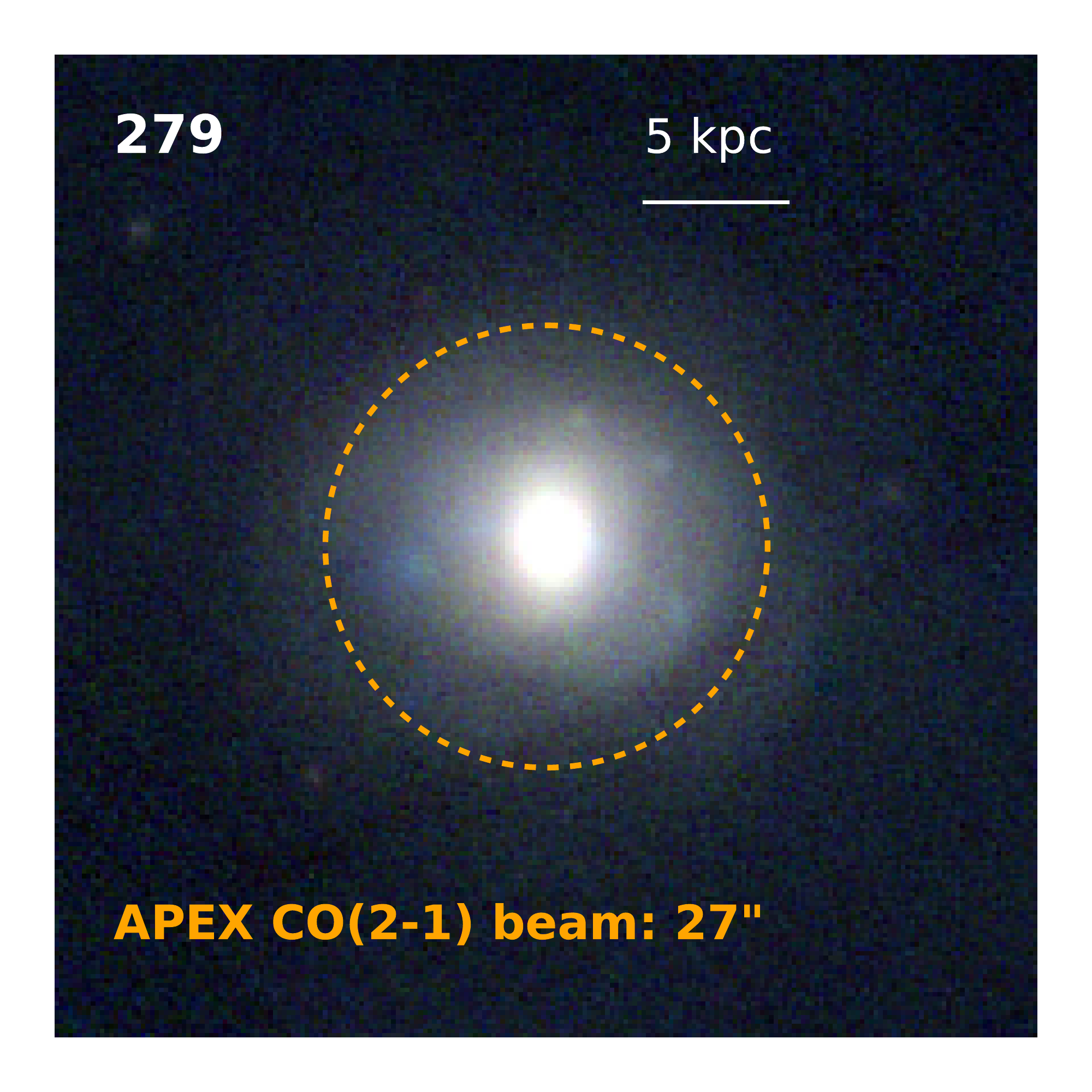}\includegraphics[width=0.26\textwidth]{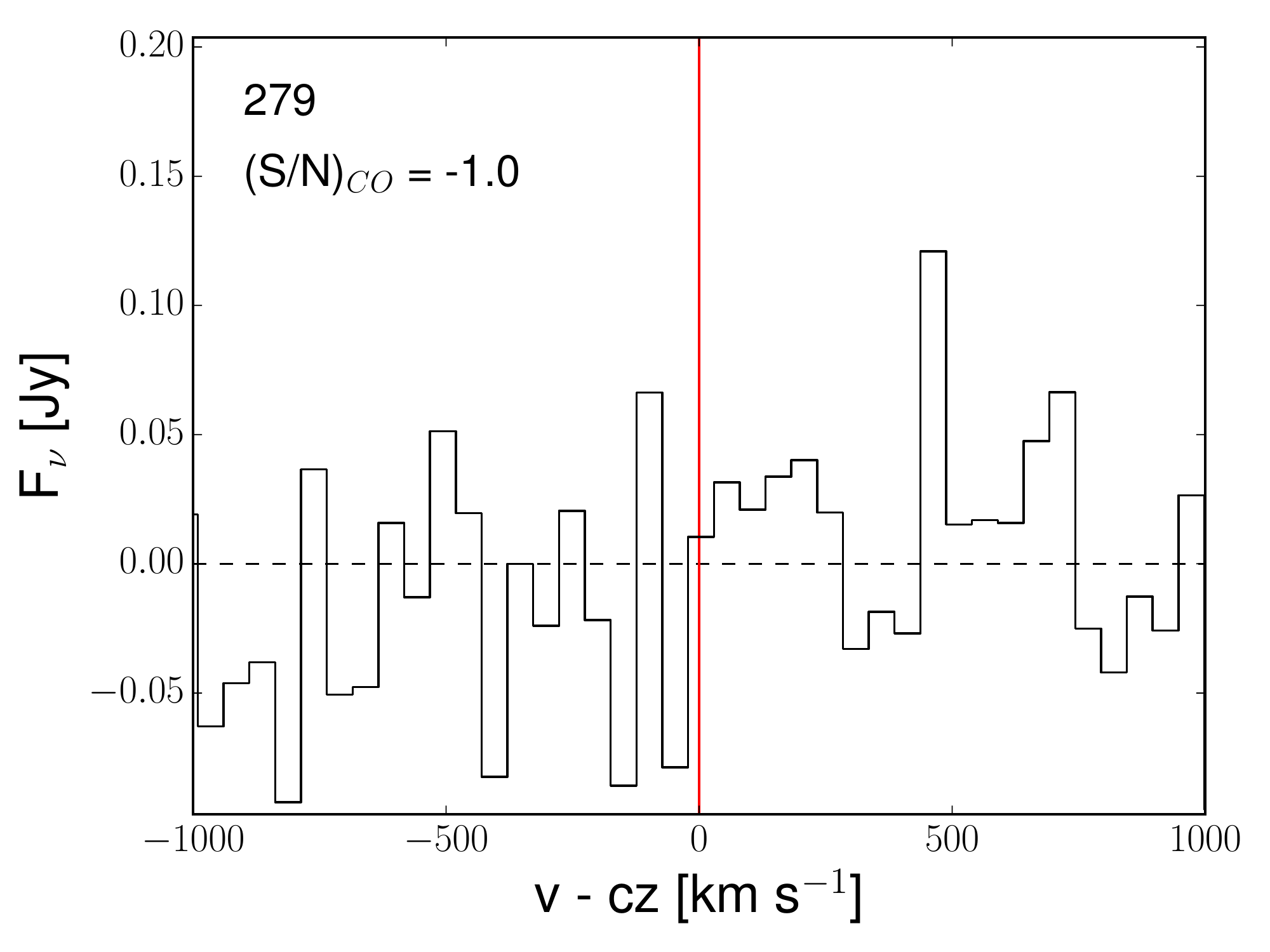}
\includegraphics[width=0.18\textwidth]{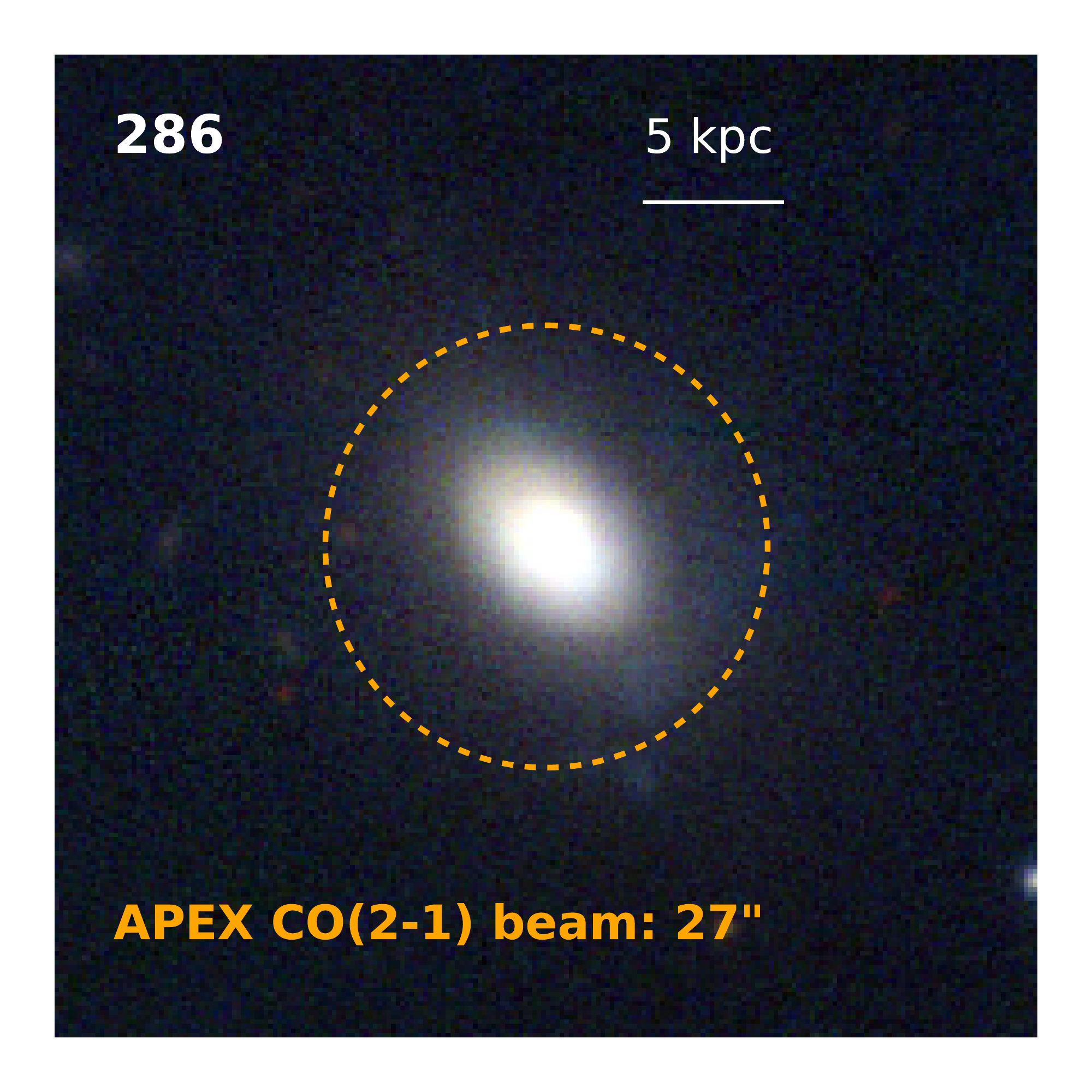}\includegraphics[width=0.26\textwidth]{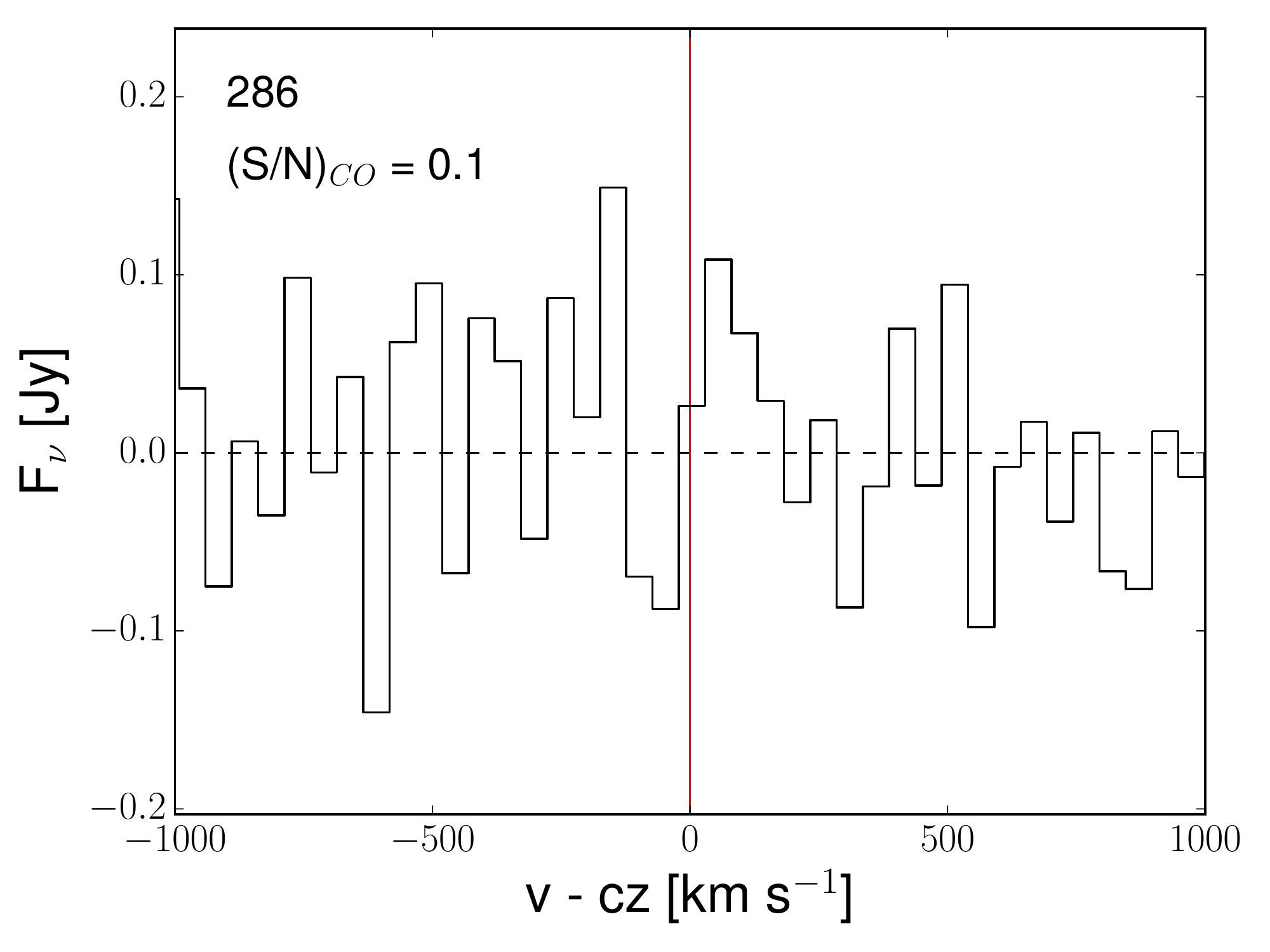}
\includegraphics[width=0.18\textwidth]{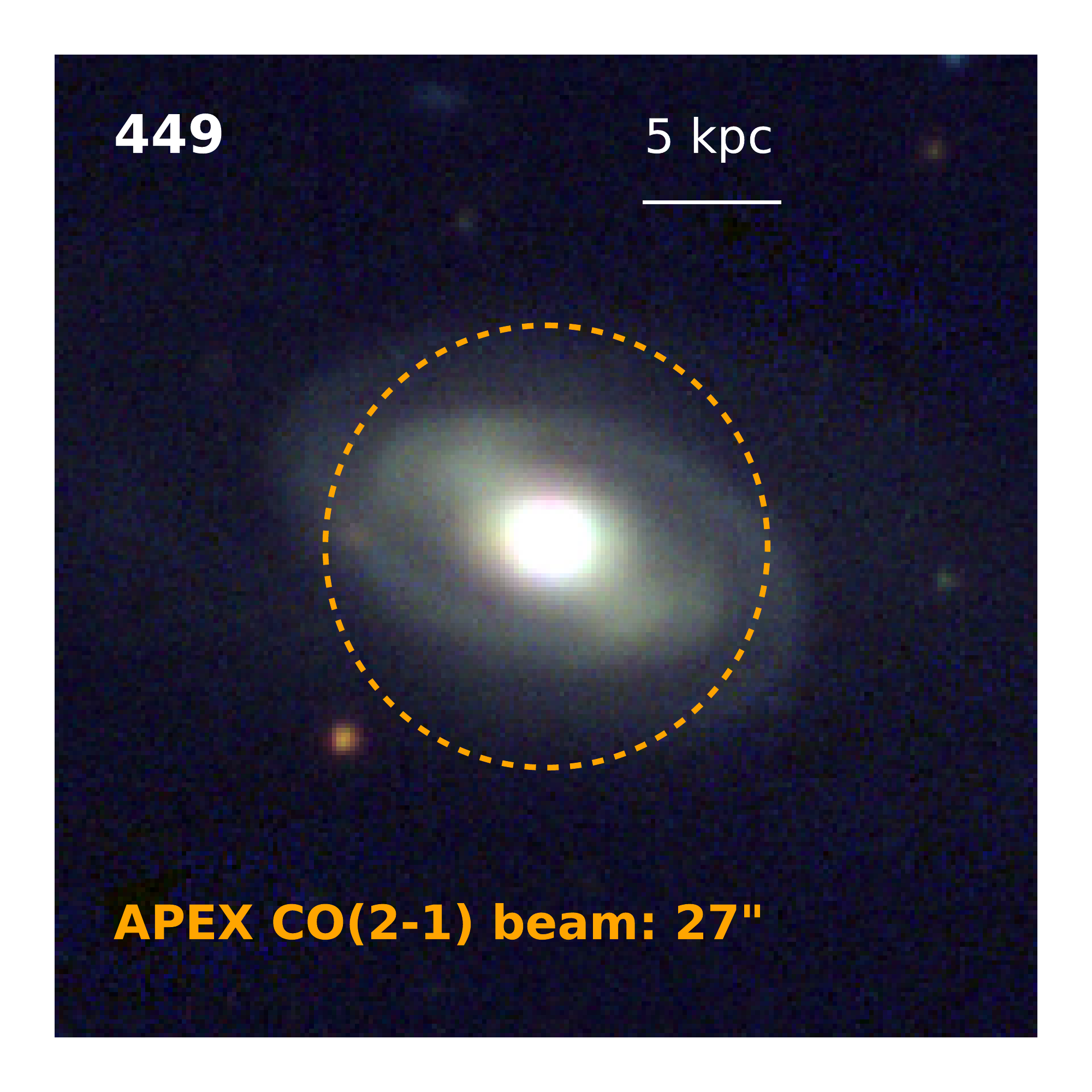}\includegraphics[width=0.26\textwidth]{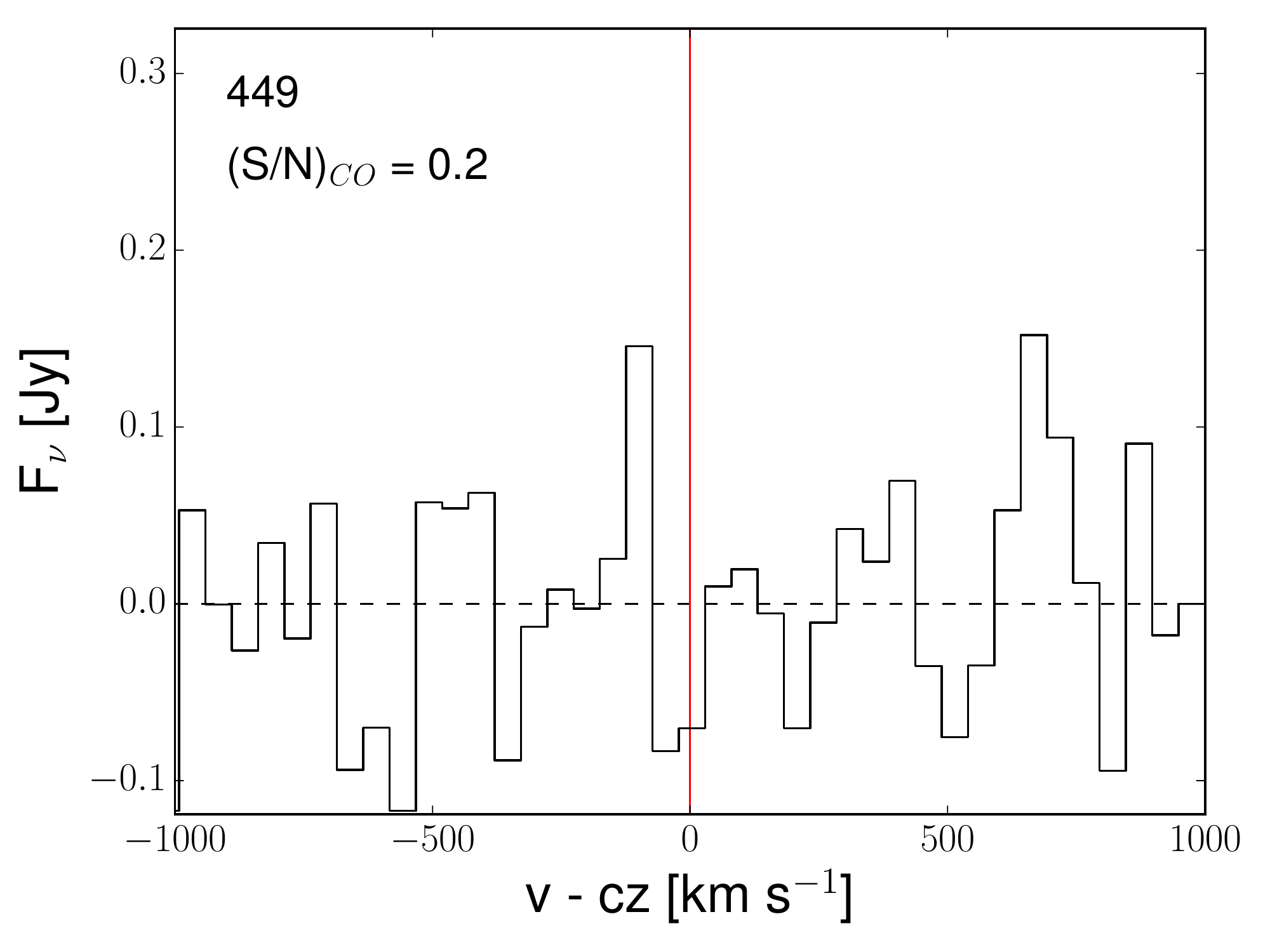}
\includegraphics[width=0.18\textwidth]{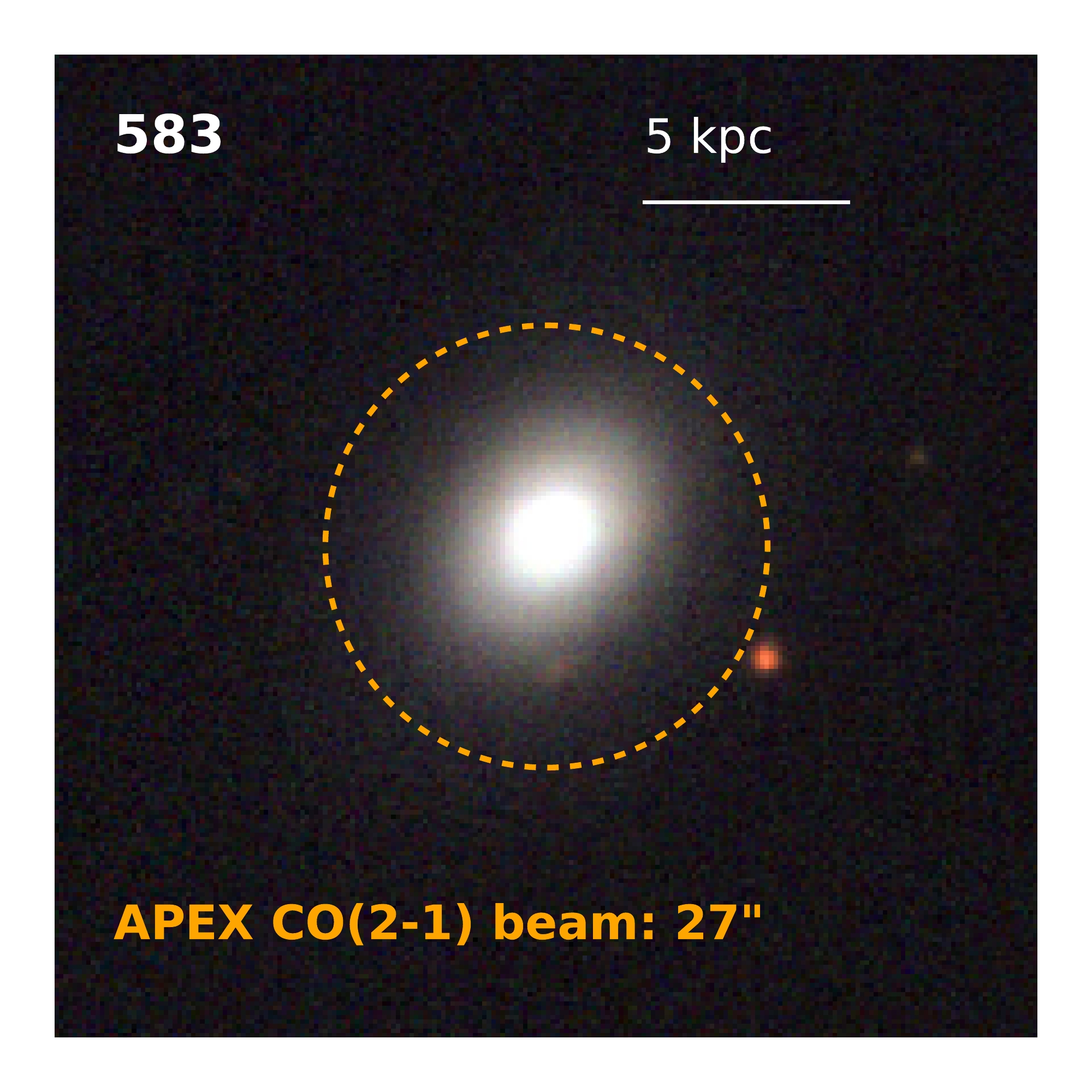}\includegraphics[width=0.26\textwidth]{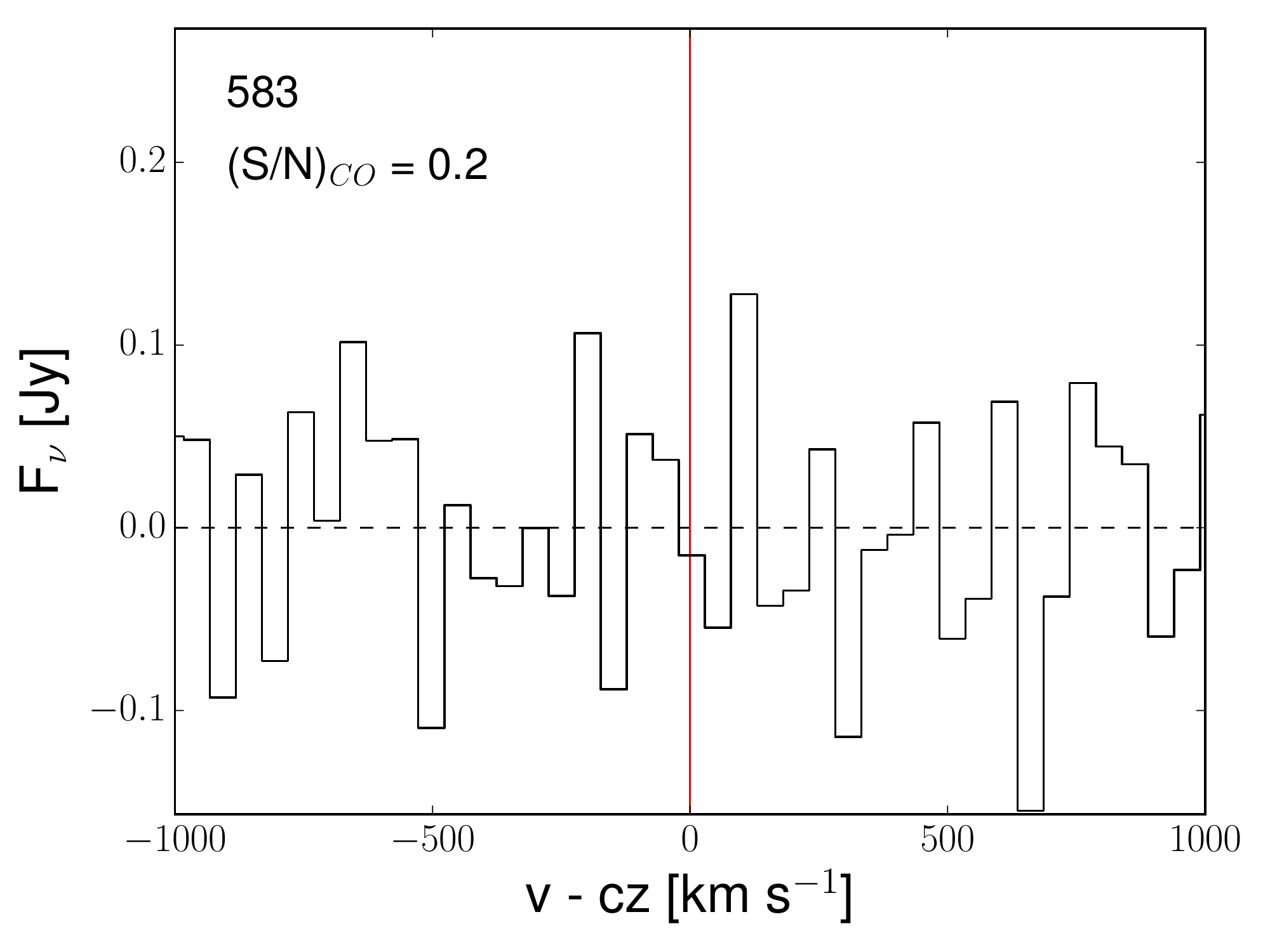}
\includegraphics[width=0.18\textwidth]{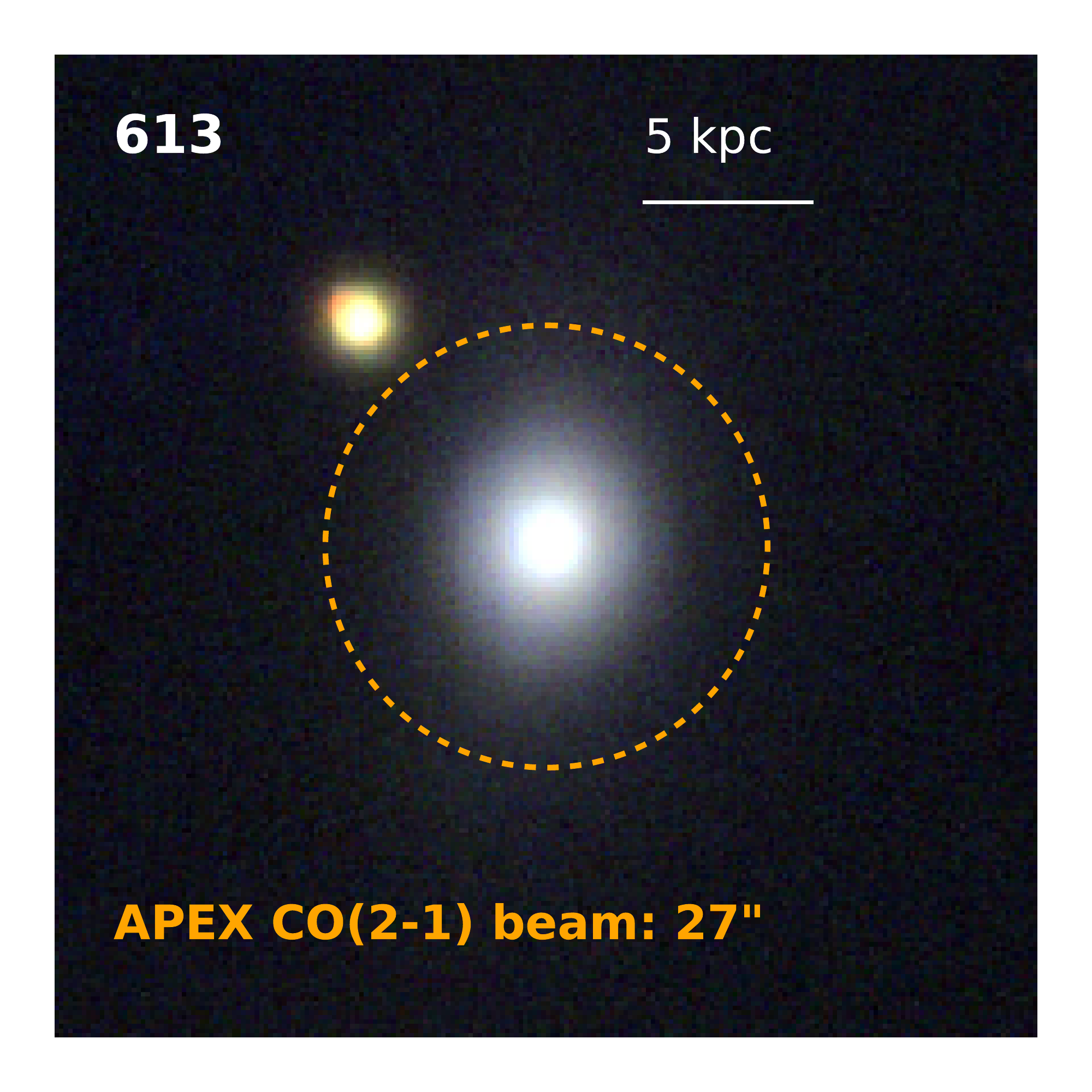}\includegraphics[width=0.26\textwidth]{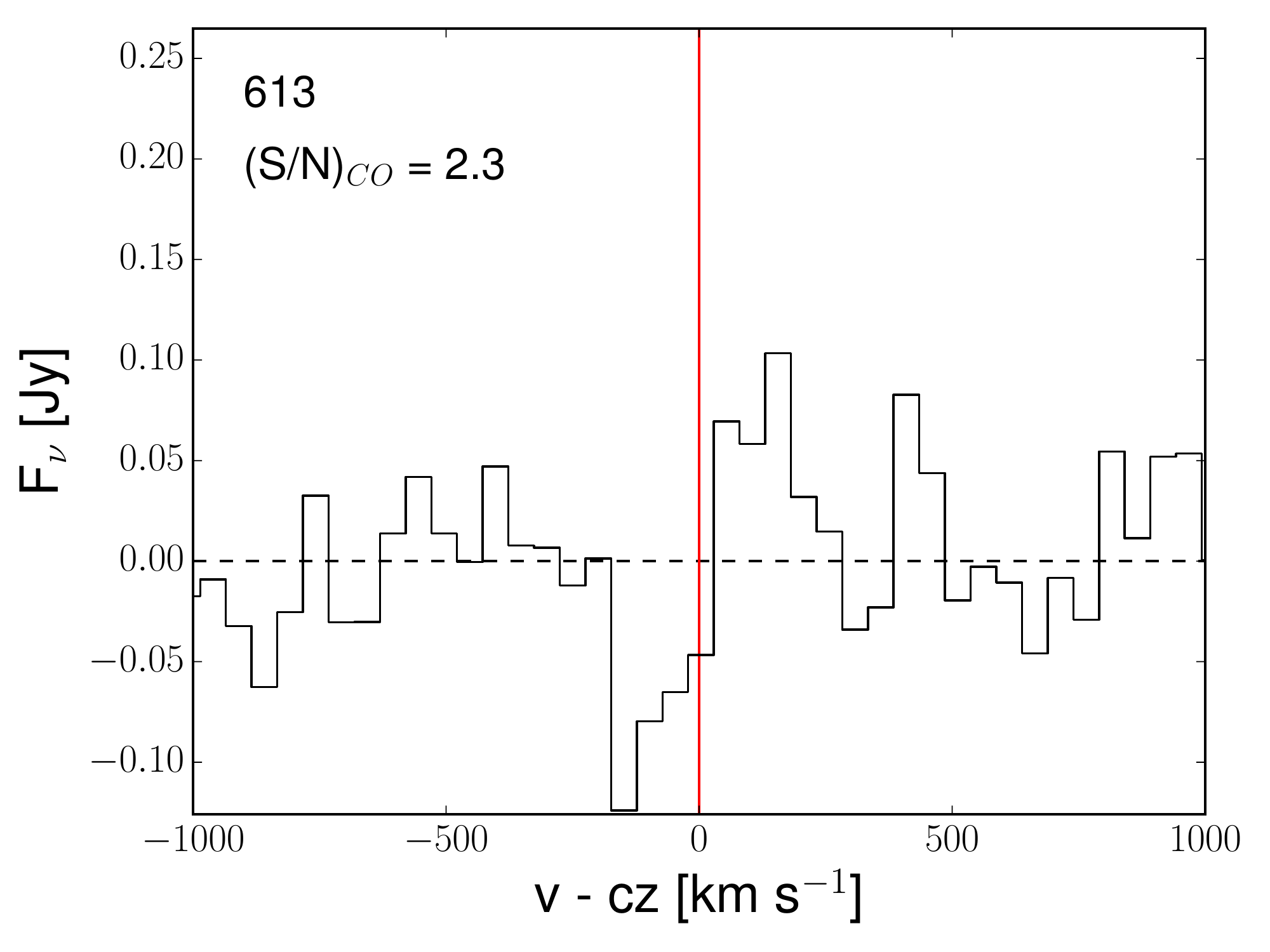}
\includegraphics[width=0.18\textwidth]{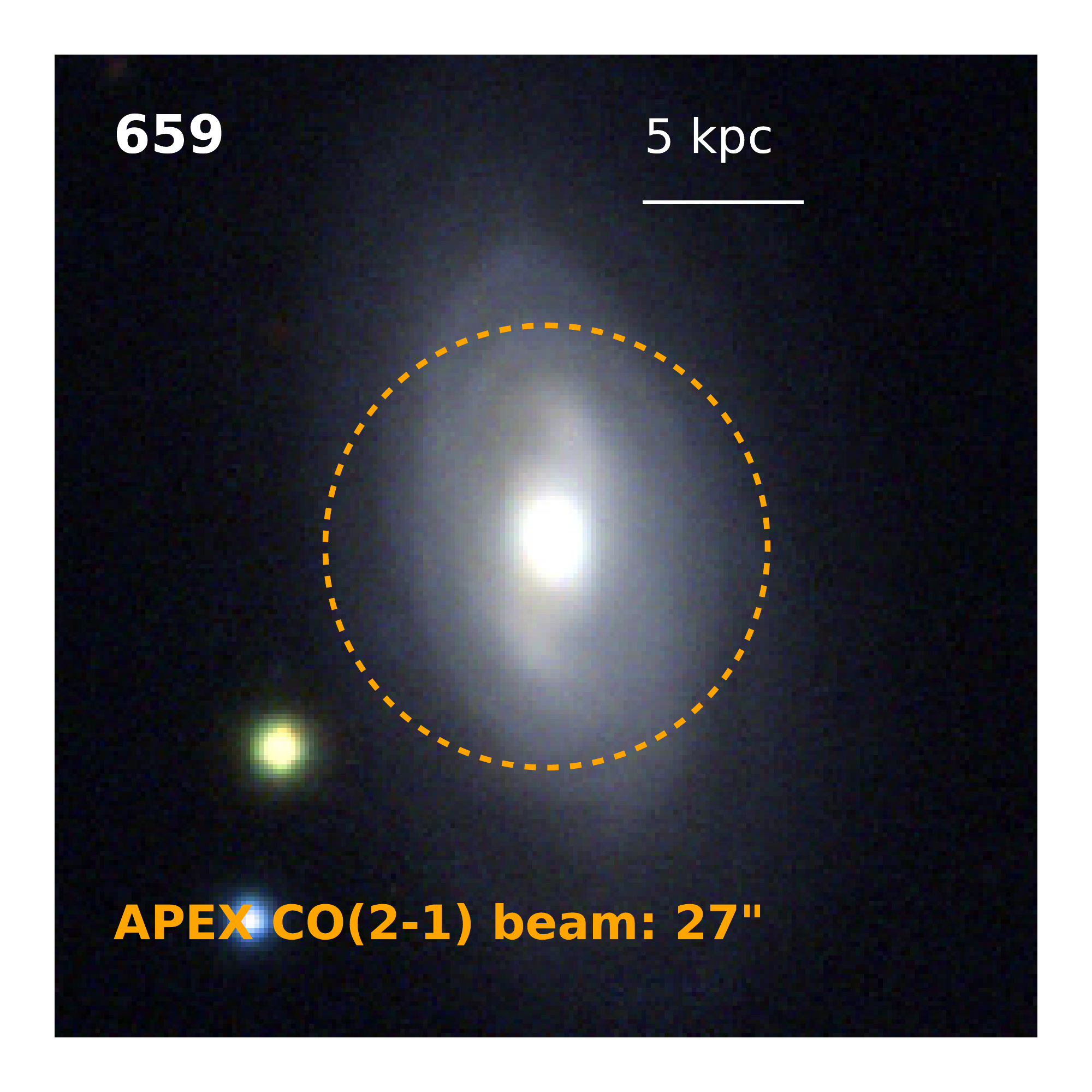}\includegraphics[width=0.26\textwidth]{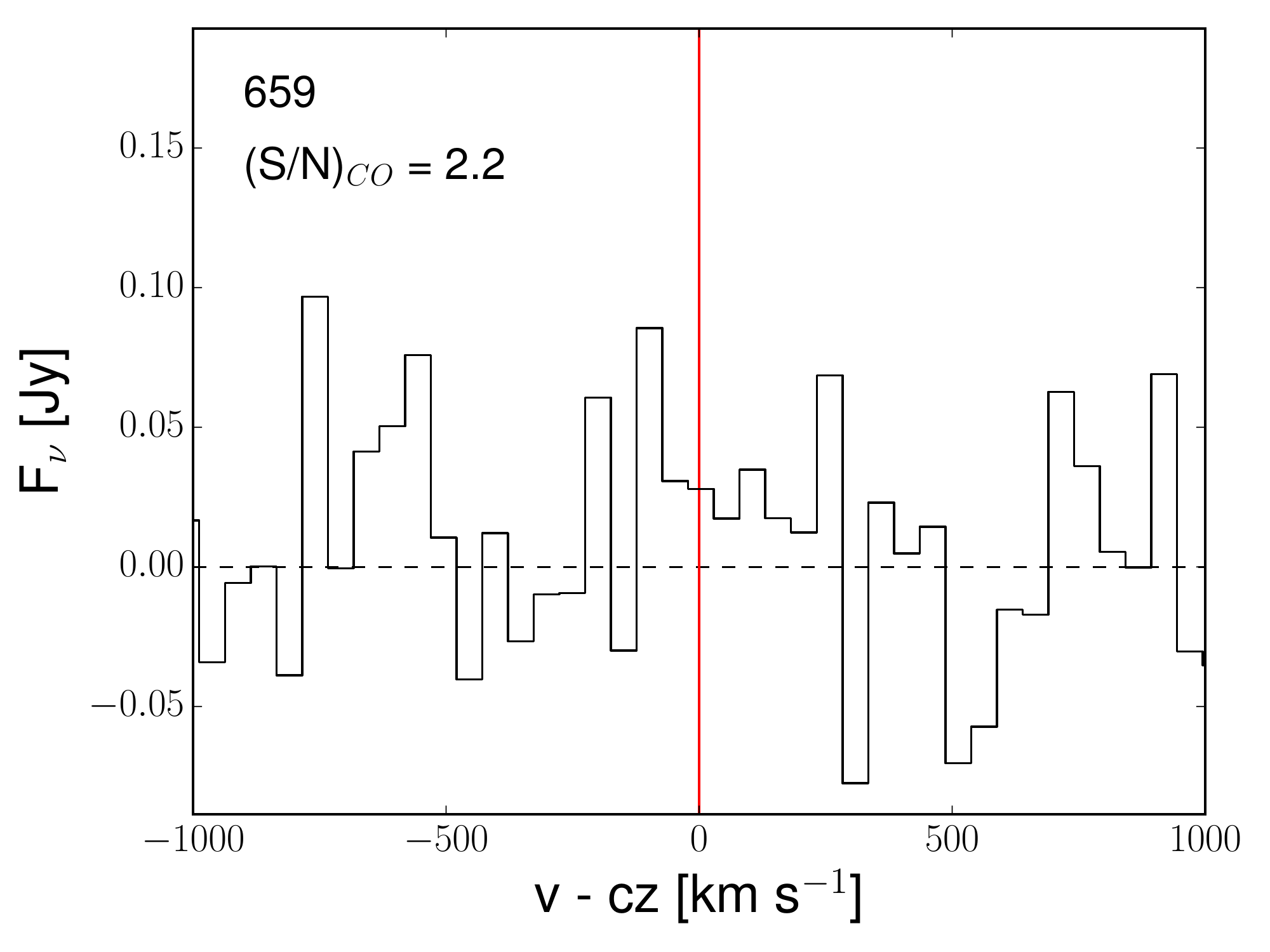}
\includegraphics[width=0.18\textwidth]{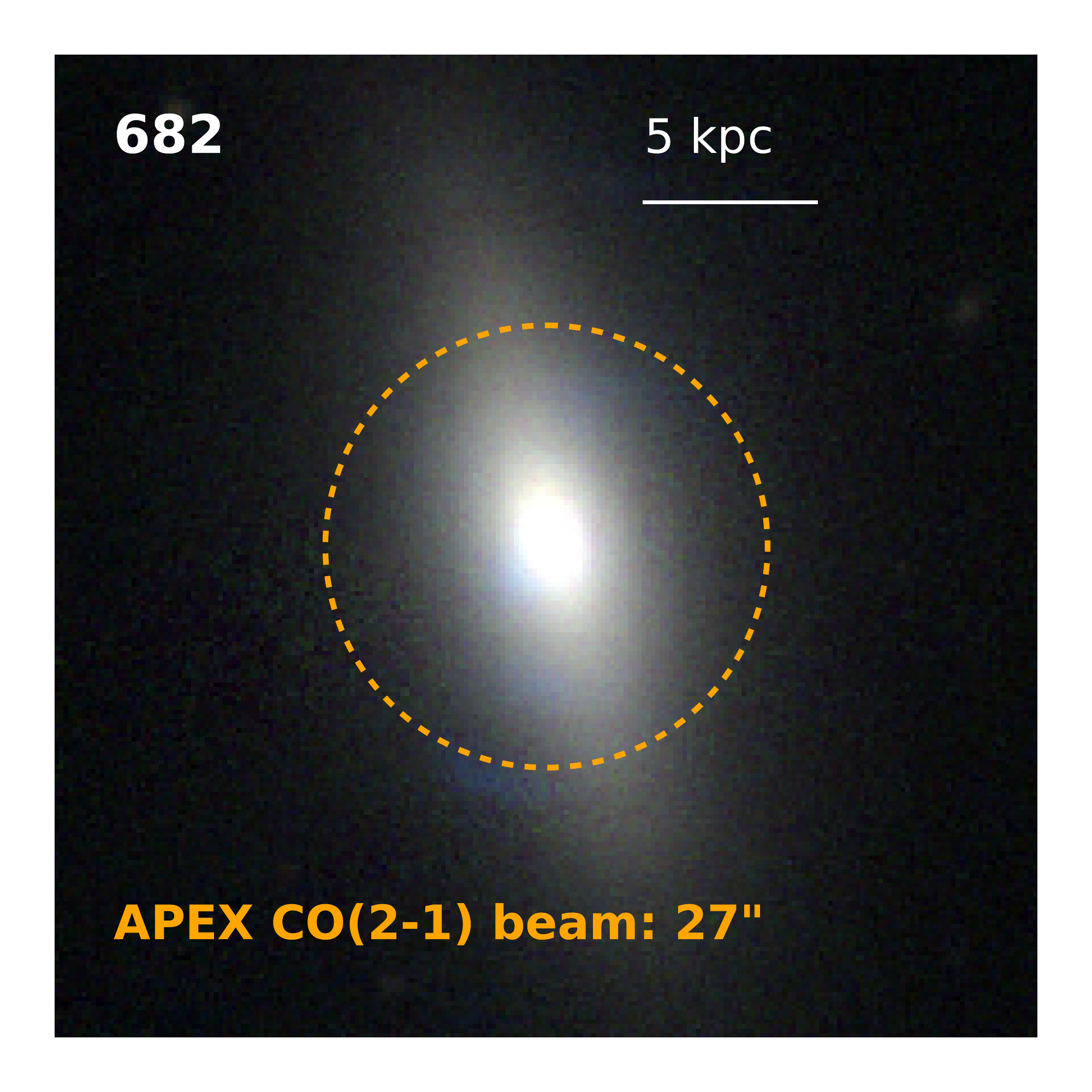}\includegraphics[width=0.26\textwidth]{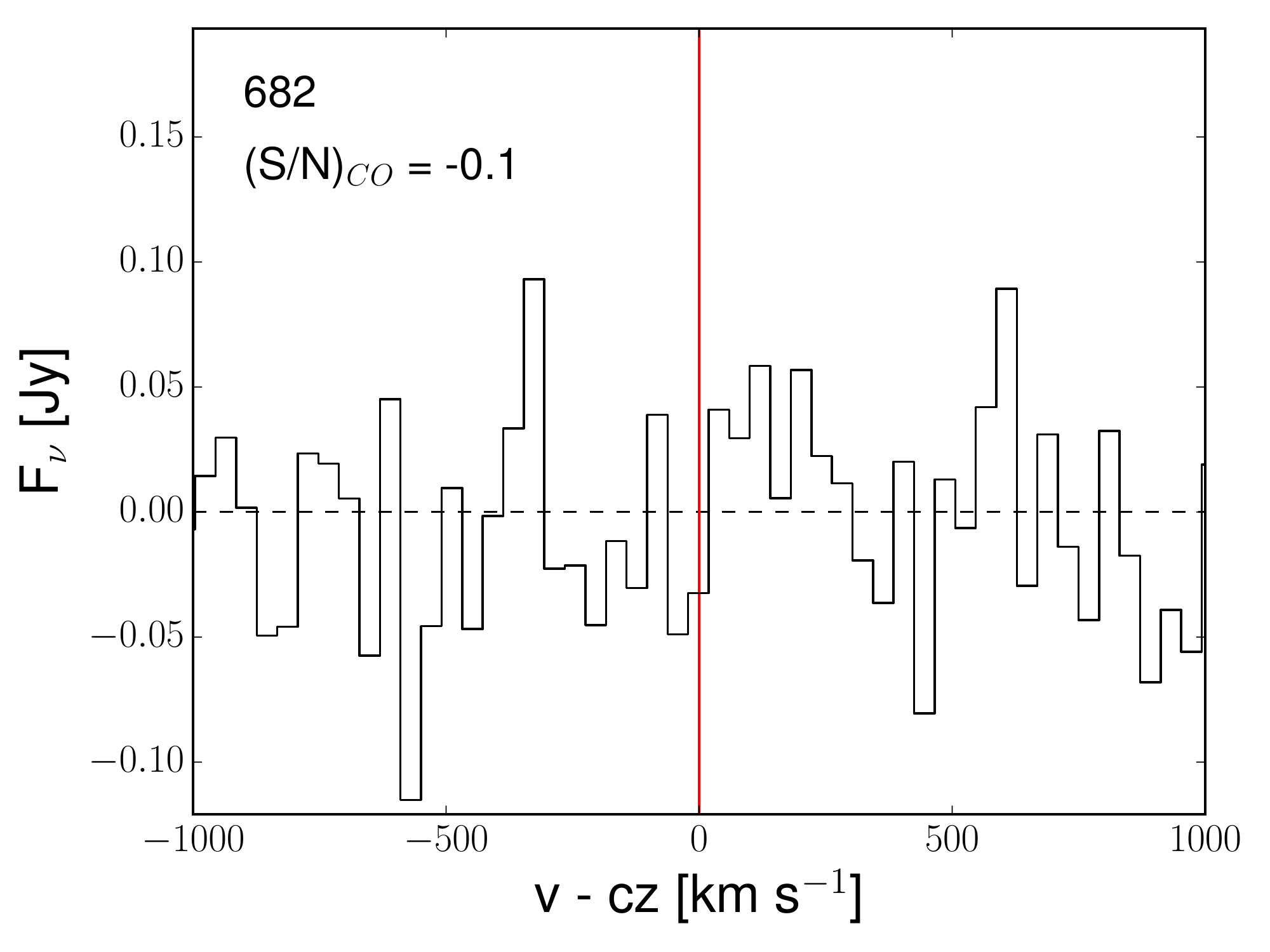}
\includegraphics[width=0.18\textwidth]{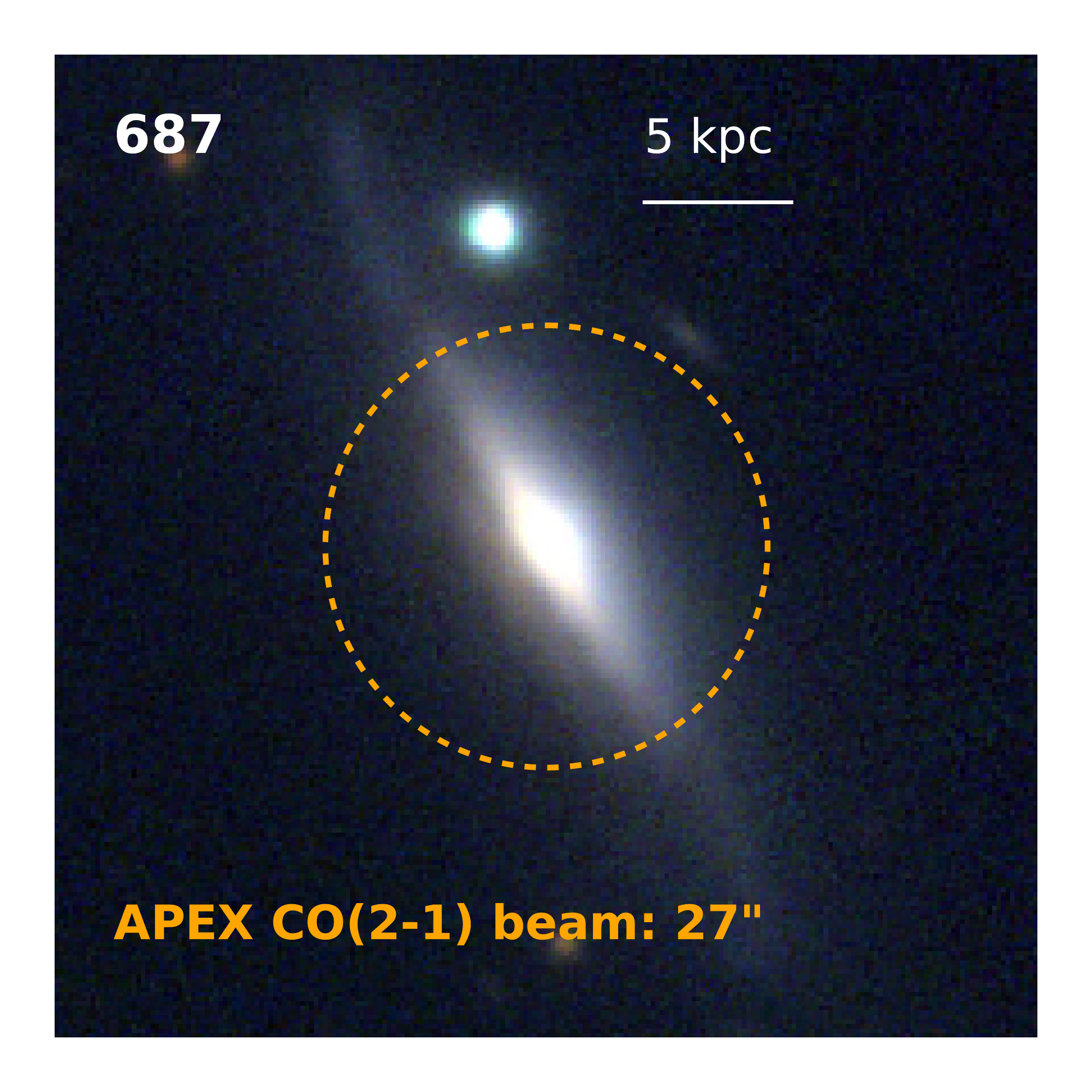}\includegraphics[width=0.26\textwidth]{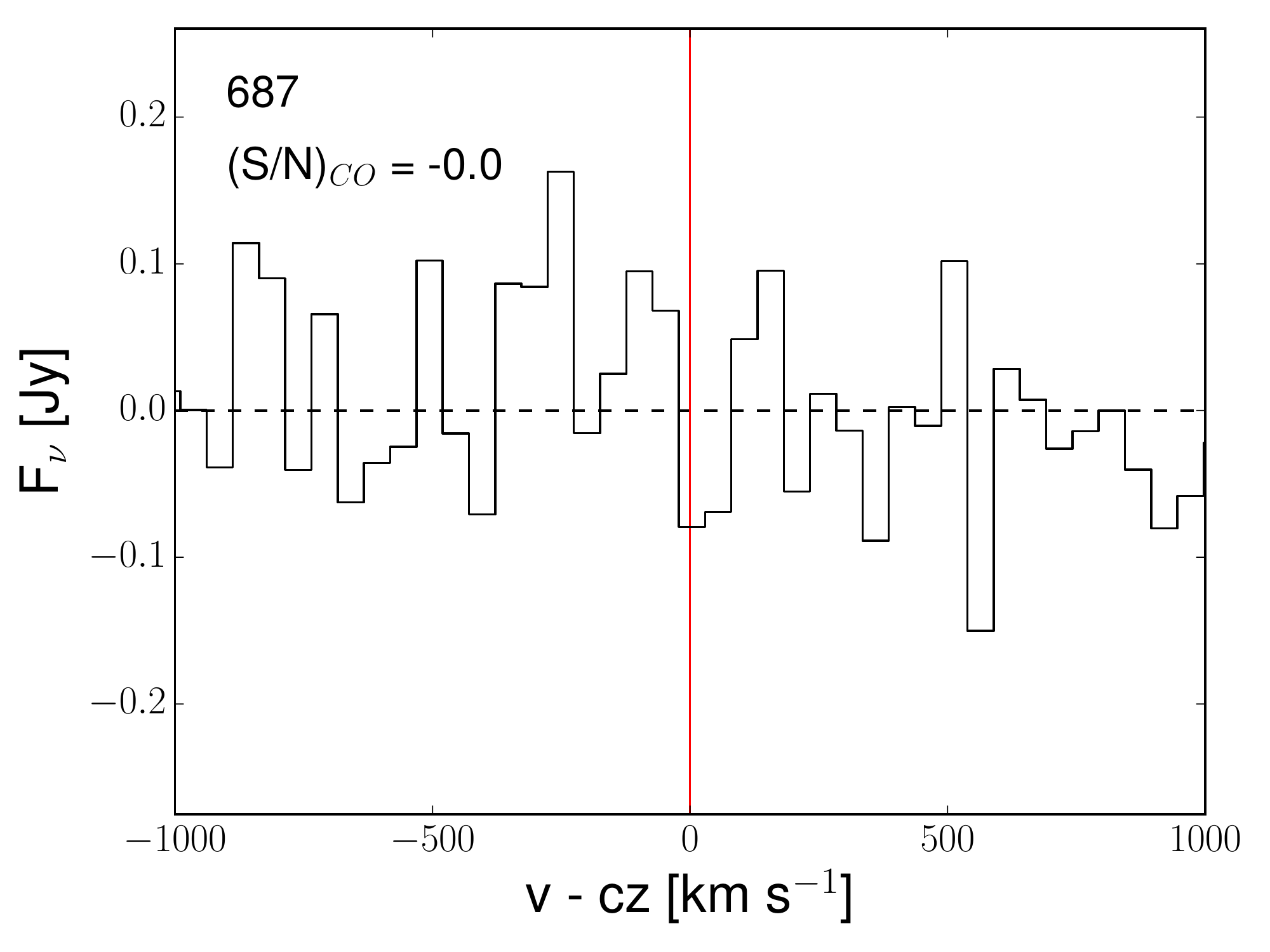}
\includegraphics[width=0.18\textwidth]{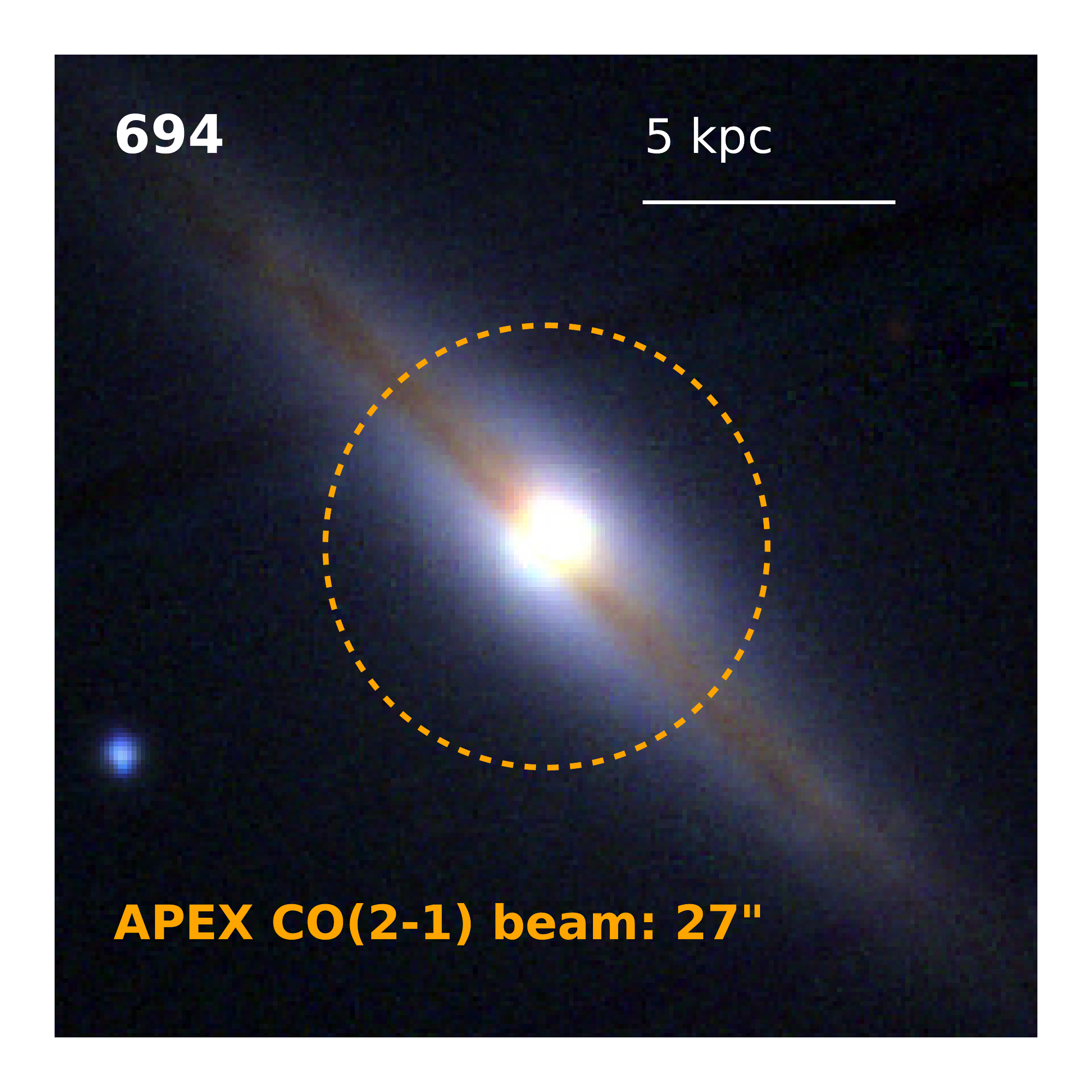}\includegraphics[width=0.26\textwidth]{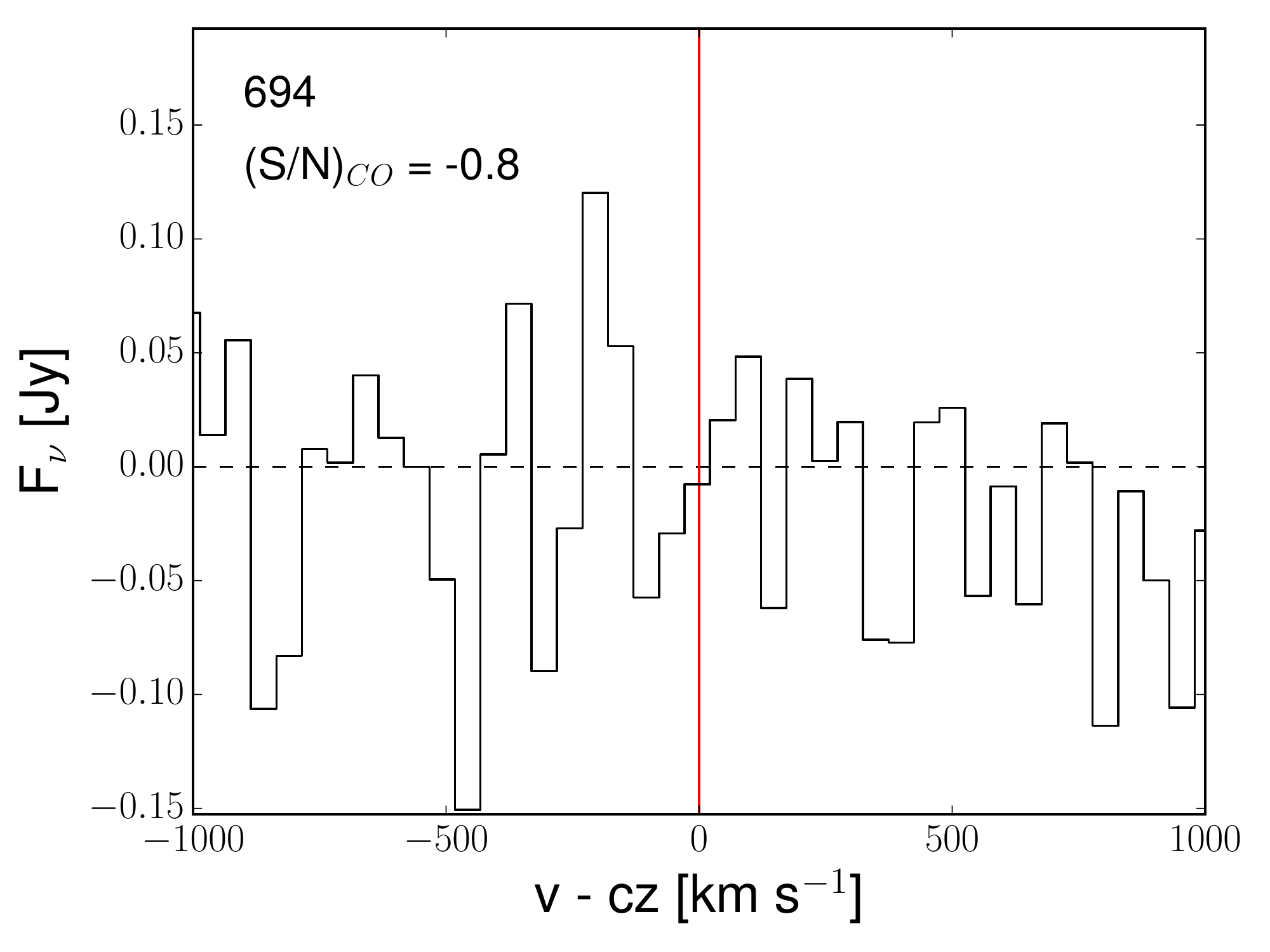}
\caption{continued from Fig.~\ref{fig:CO21_spectra_all_undetect1}
} 
\label{fig:CO21_spectra_undetected2}
\end{figure*}

\begin{figure*}
\centering
\raggedright
\includegraphics[width=0.18\textwidth]{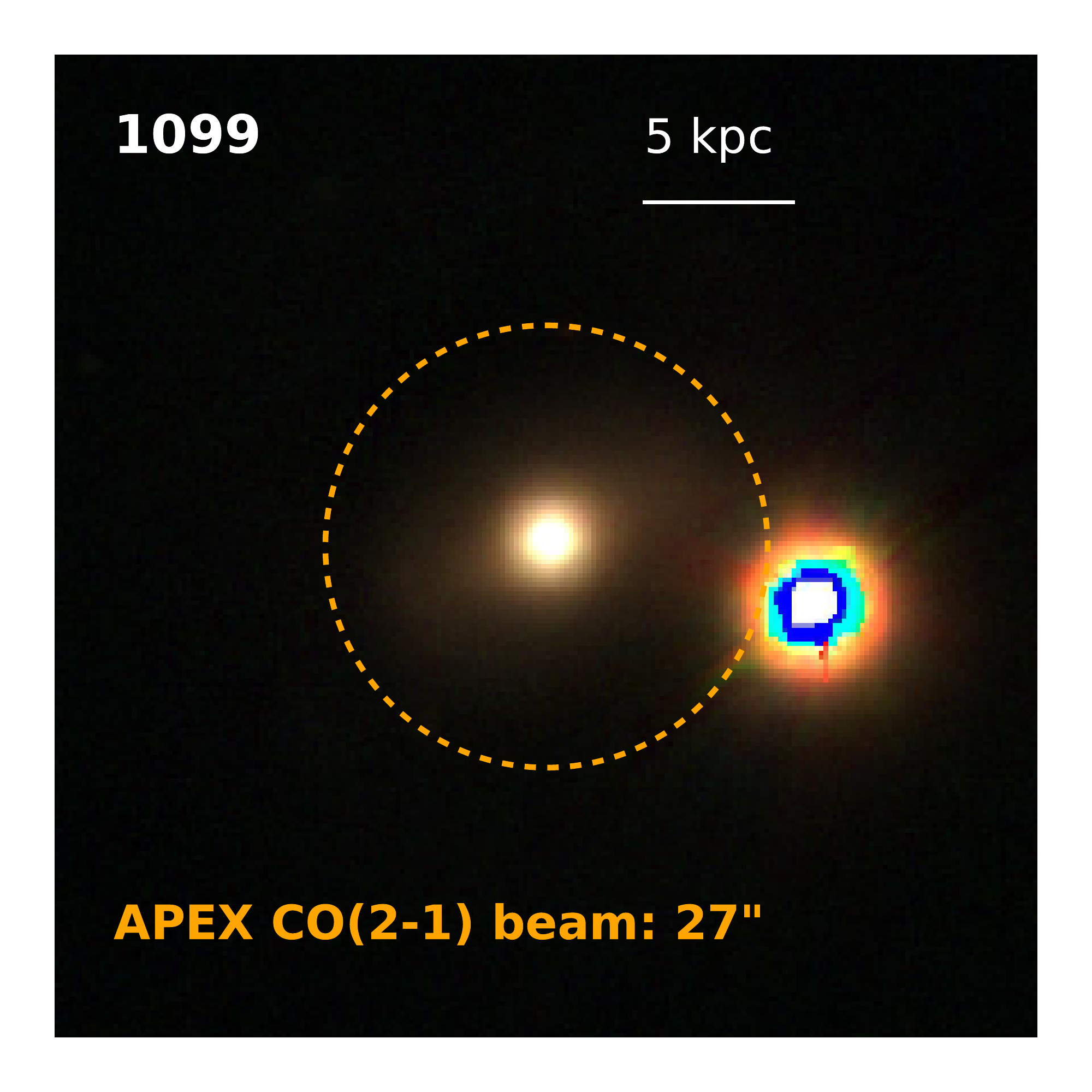}\includegraphics[width=0.26\textwidth]{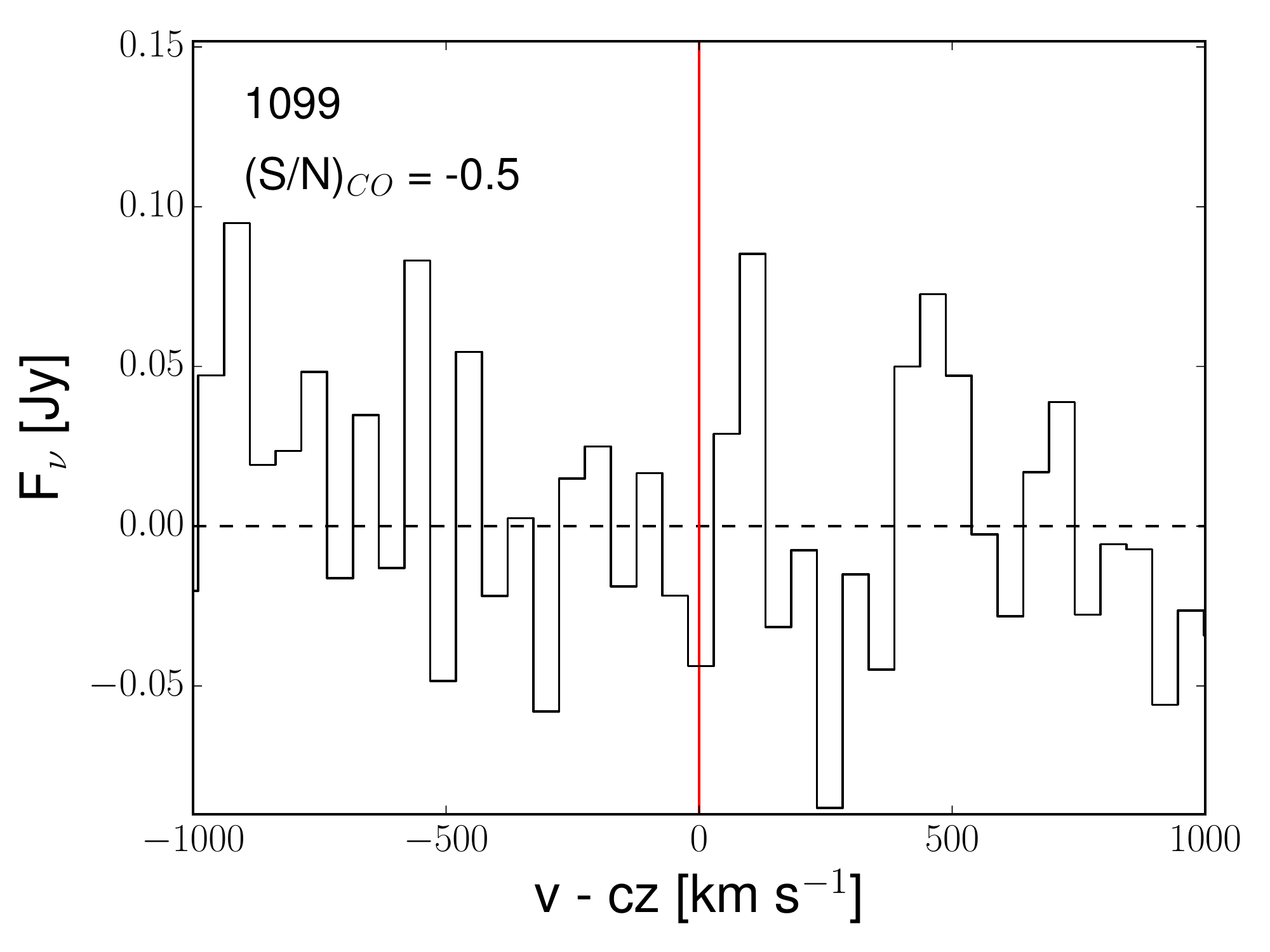}
\includegraphics[width=0.18\textwidth]{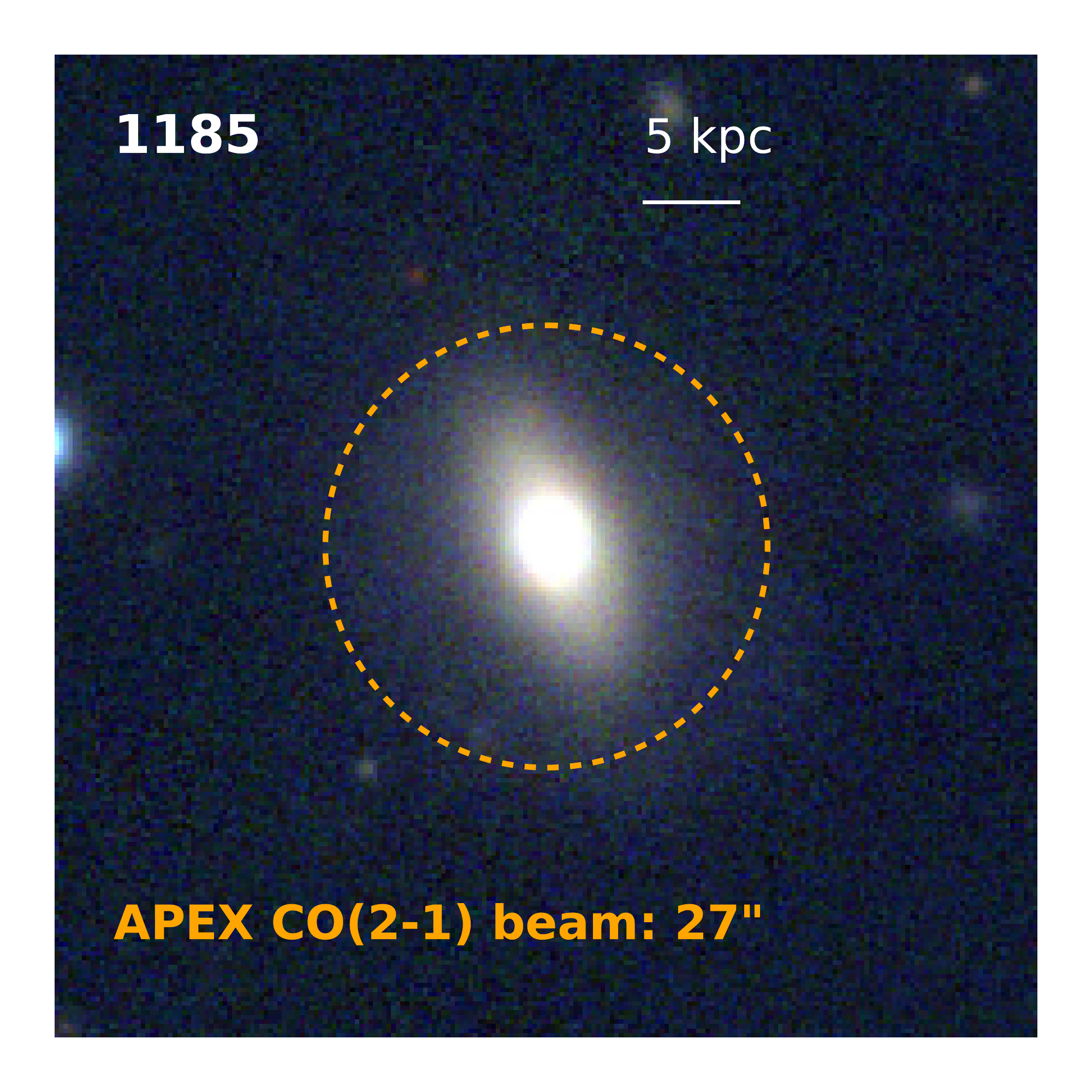}\includegraphics[width=0.26\textwidth]{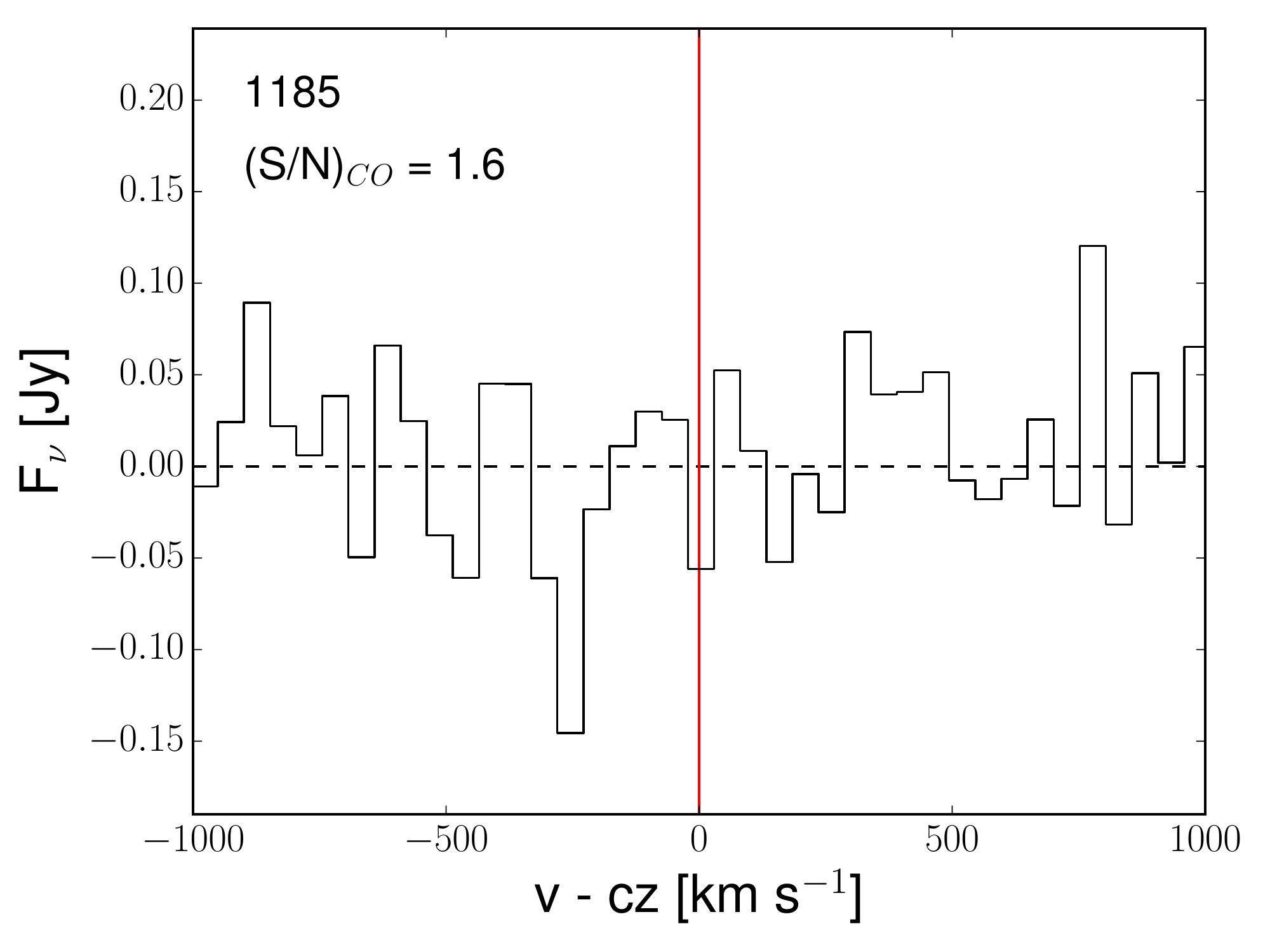}
\includegraphics[width=0.18\textwidth]{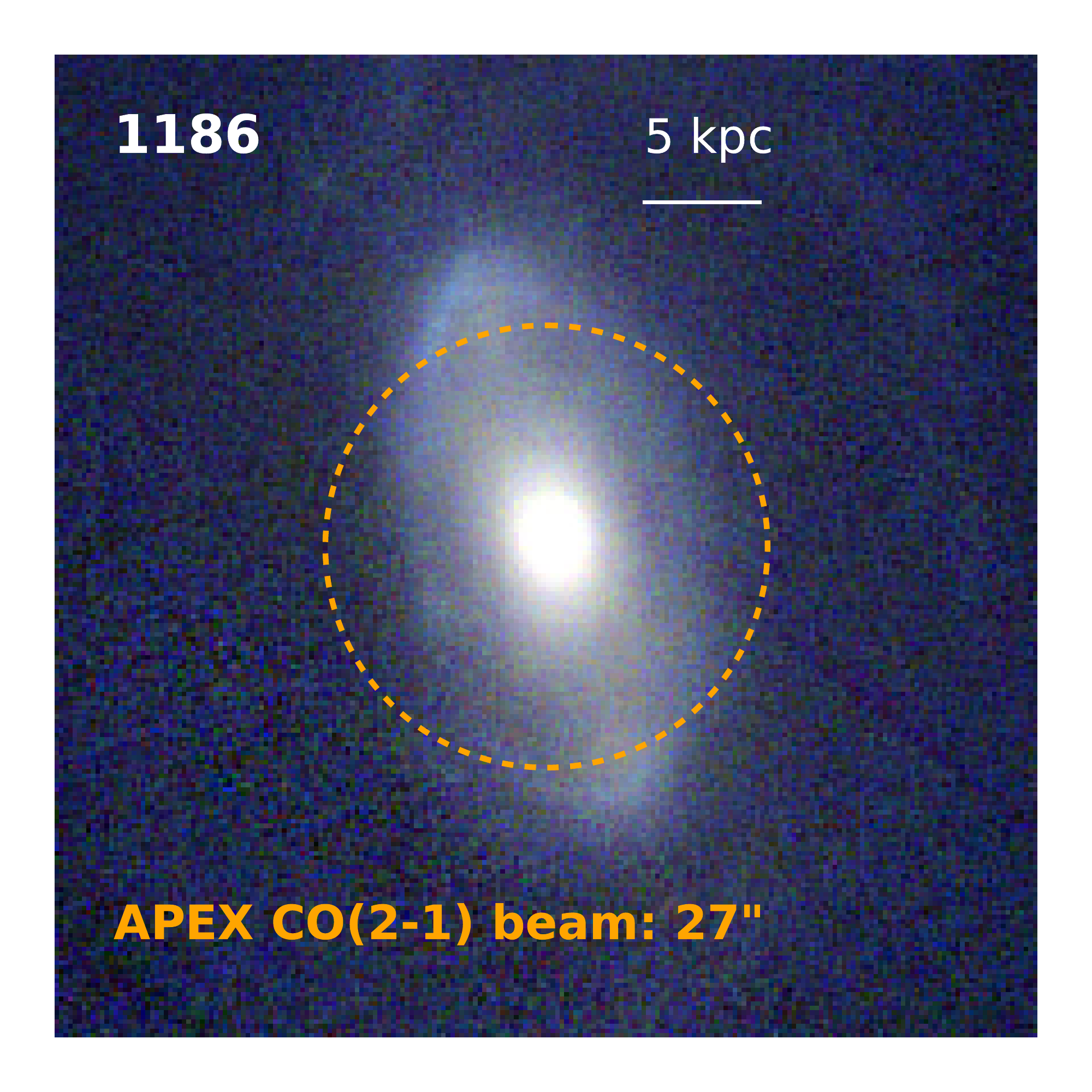}\includegraphics[width=0.26\textwidth]{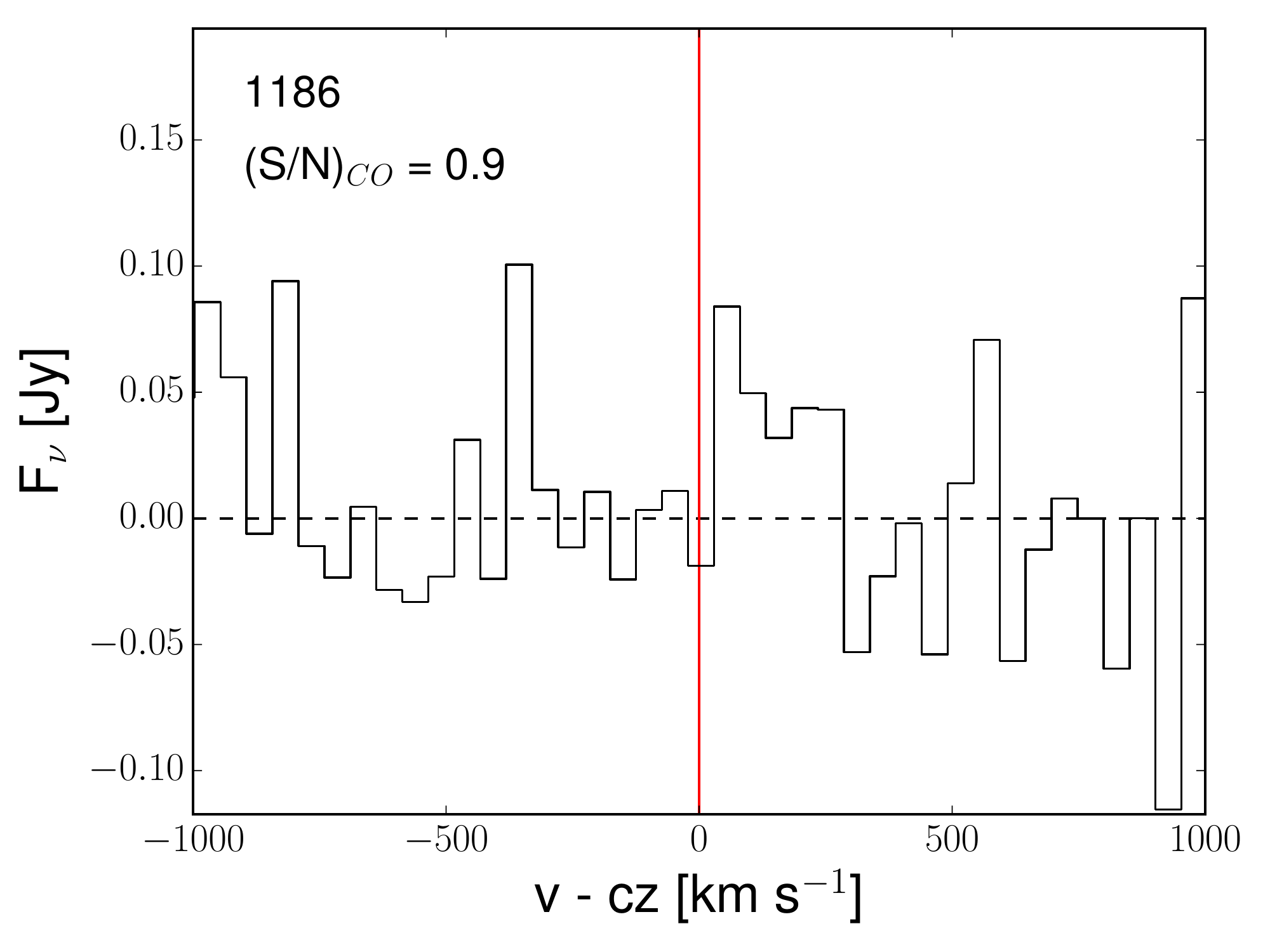}
\includegraphics[width=0.18\textwidth]{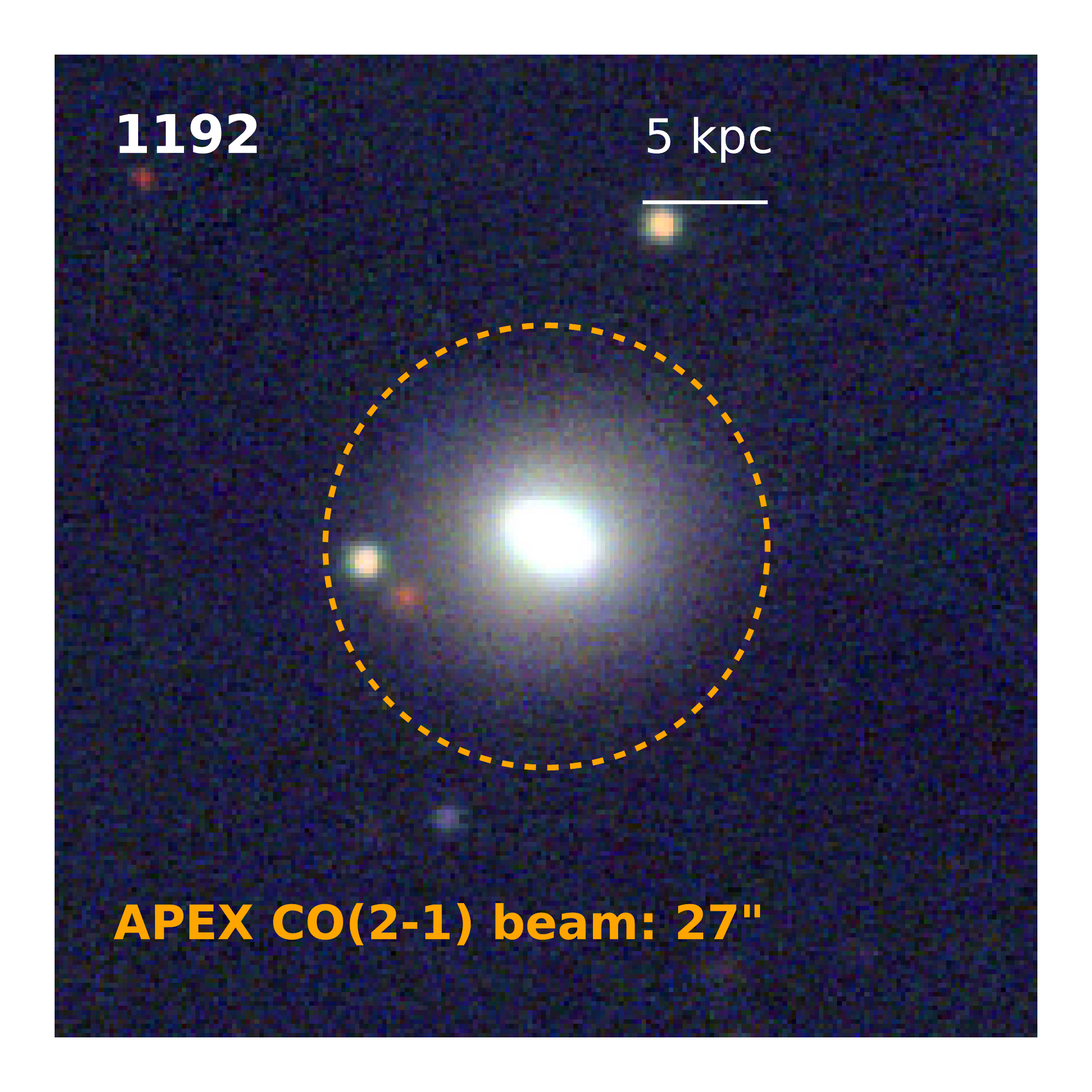}\includegraphics[width=0.26\textwidth]{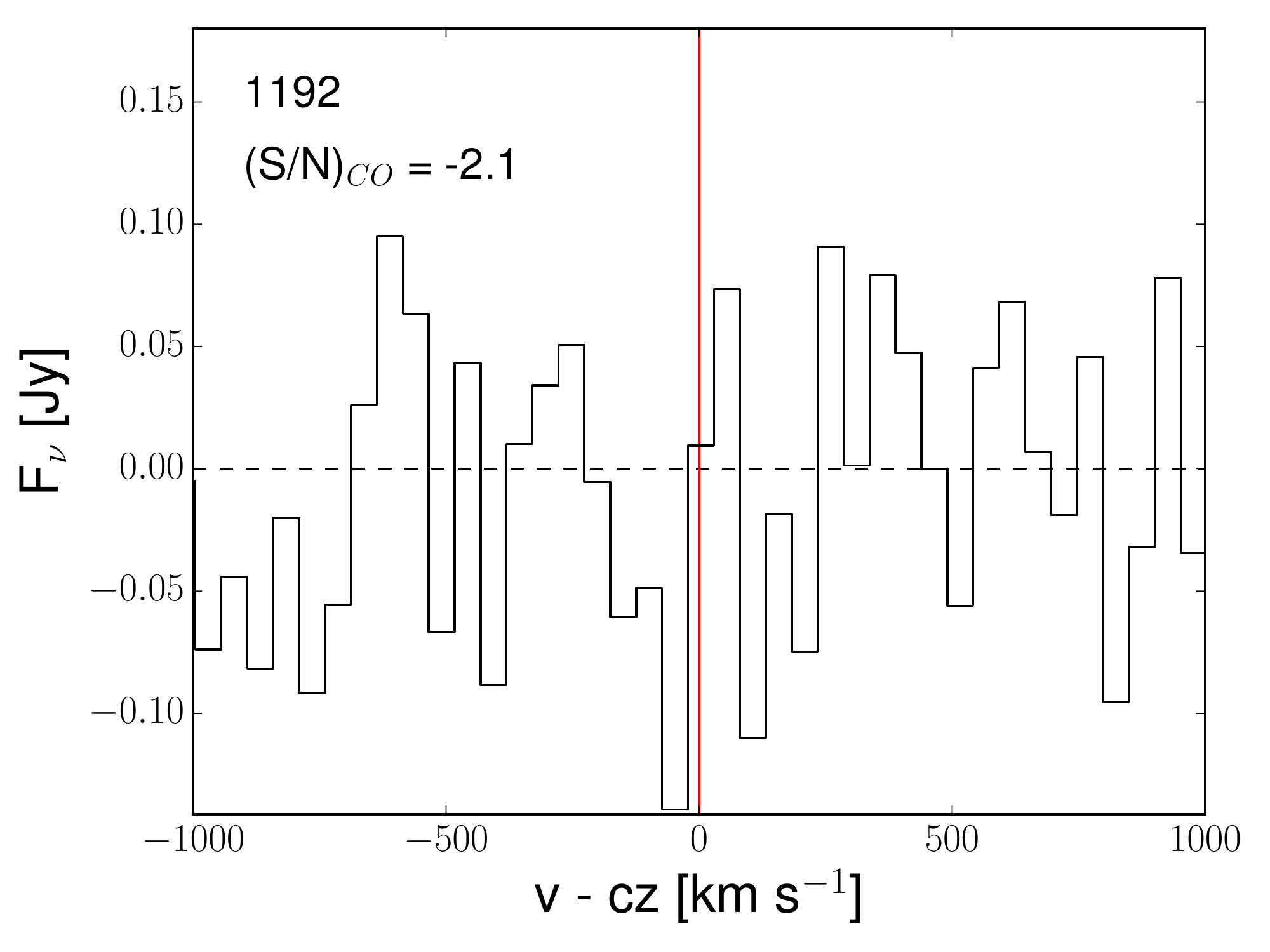}
\caption{continued from Fig.~\ref{fig:CO21_spectra_all_undetect1}
} 
\label{fig:CO21_spectra_undetected3}
\end{figure*}

\begin{figure*}
\centering
\raggedright
\includegraphics[width=0.18\textwidth]{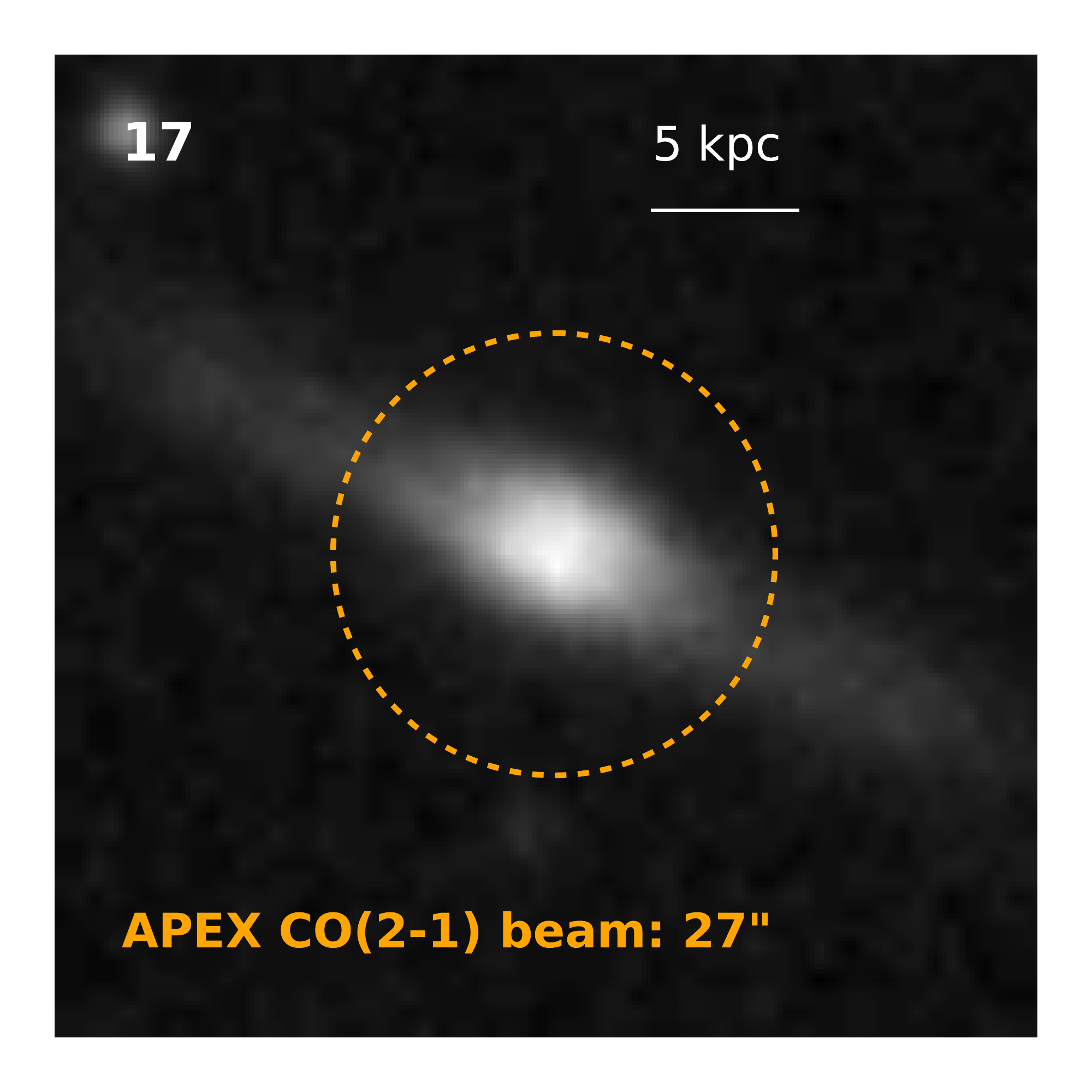}\includegraphics[width=0.26\textwidth]{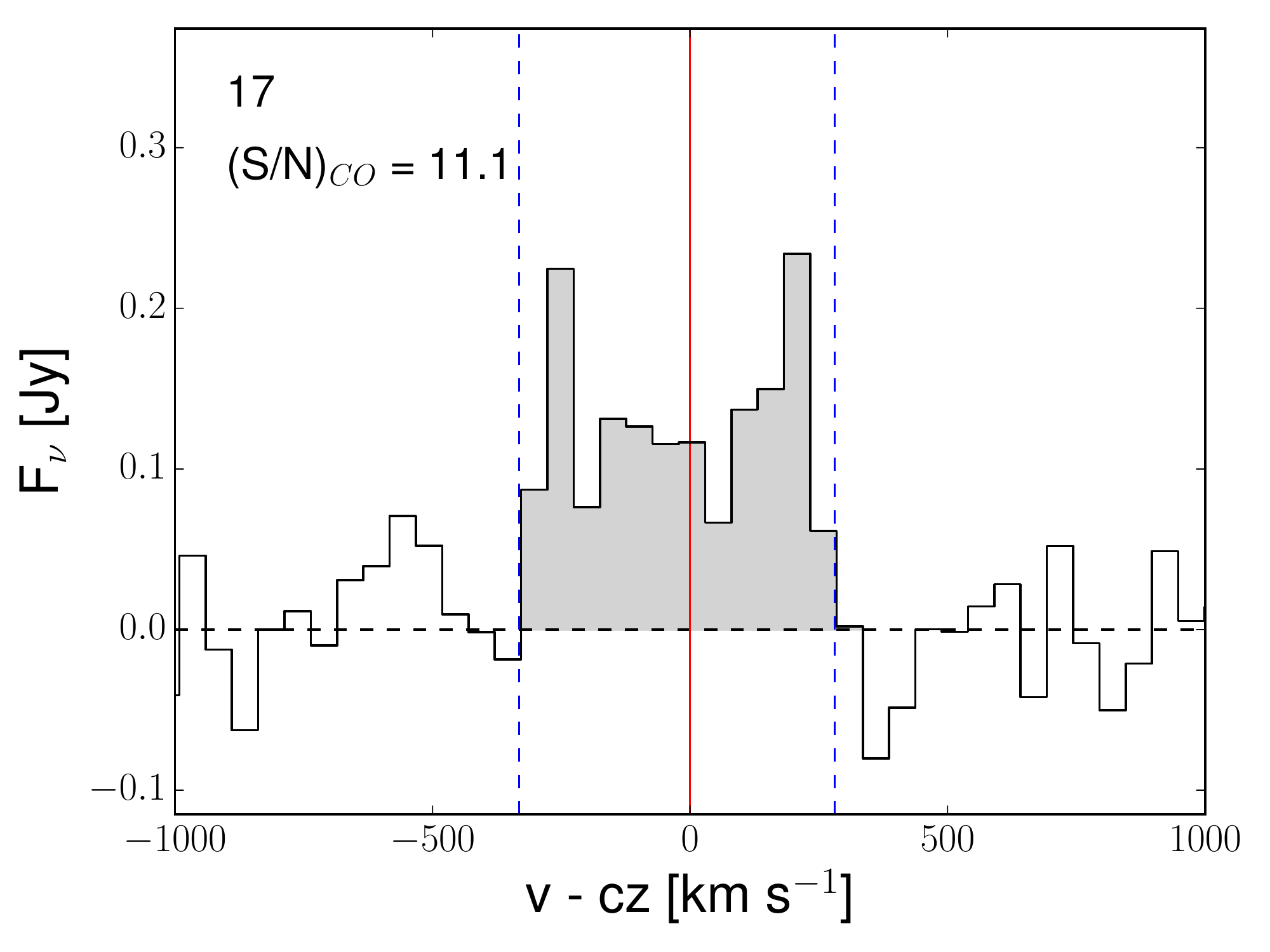}
\includegraphics[width=0.18\textwidth]{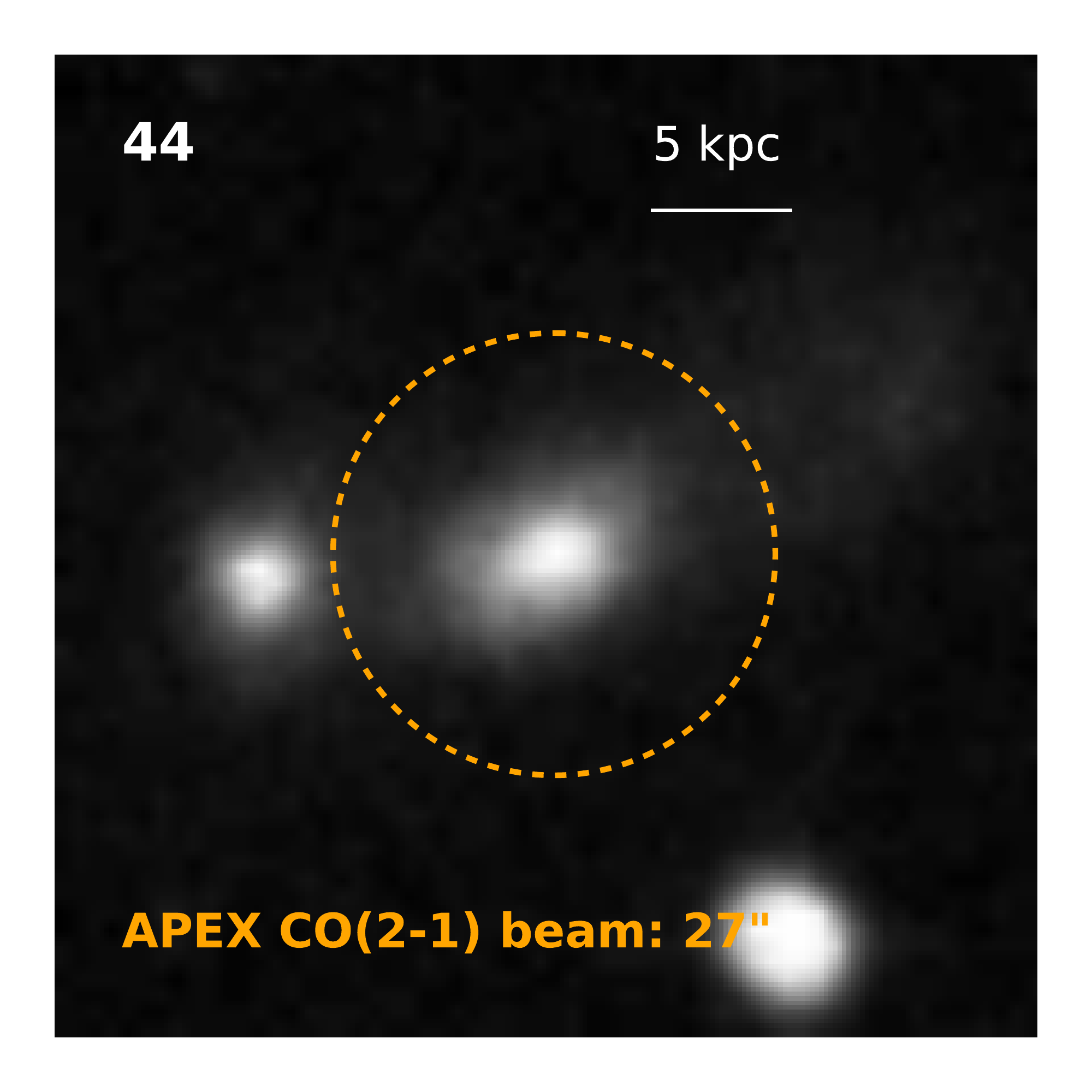}\includegraphics[width=0.26\textwidth]{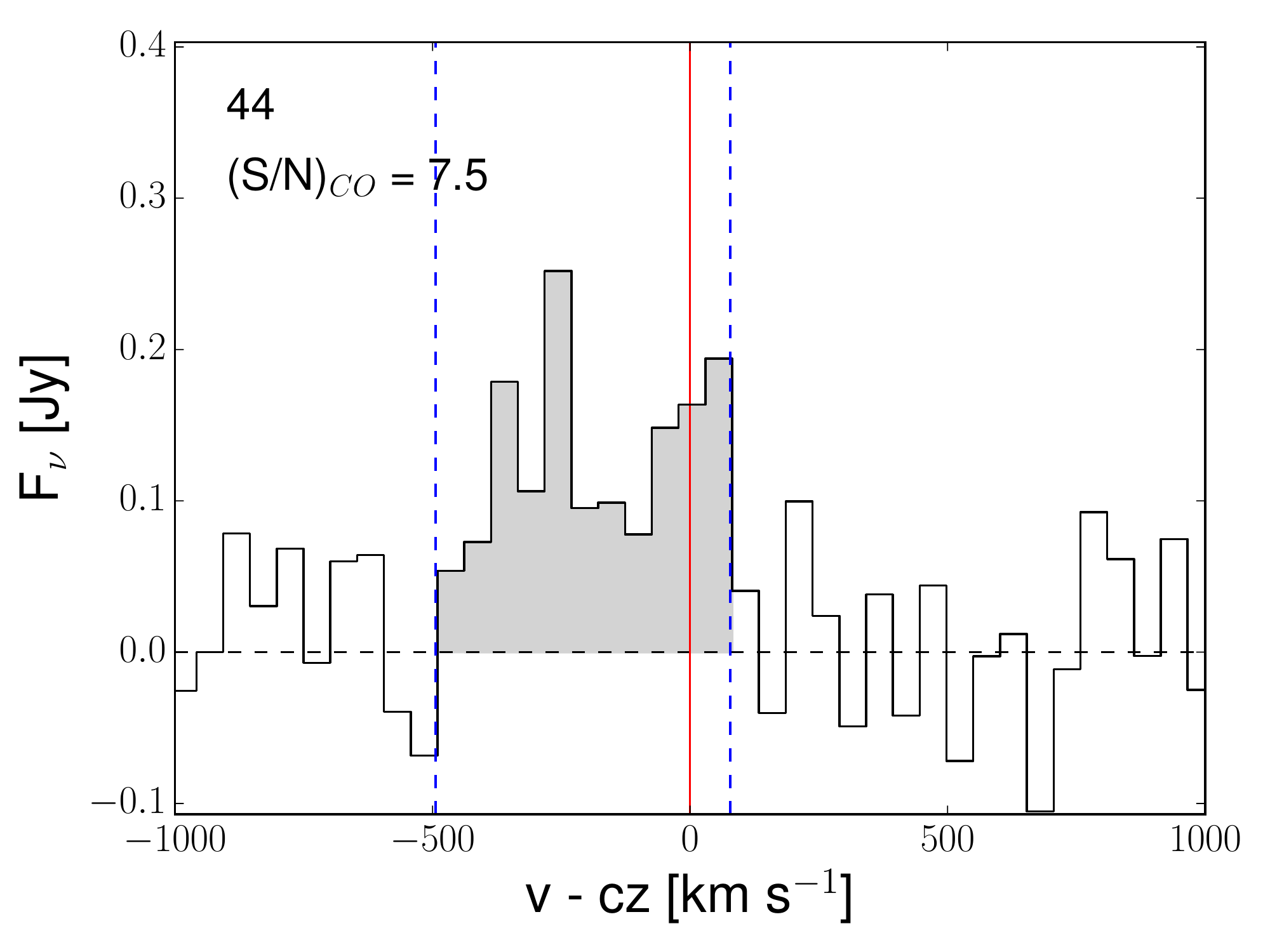}
\includegraphics[width=0.18\textwidth]{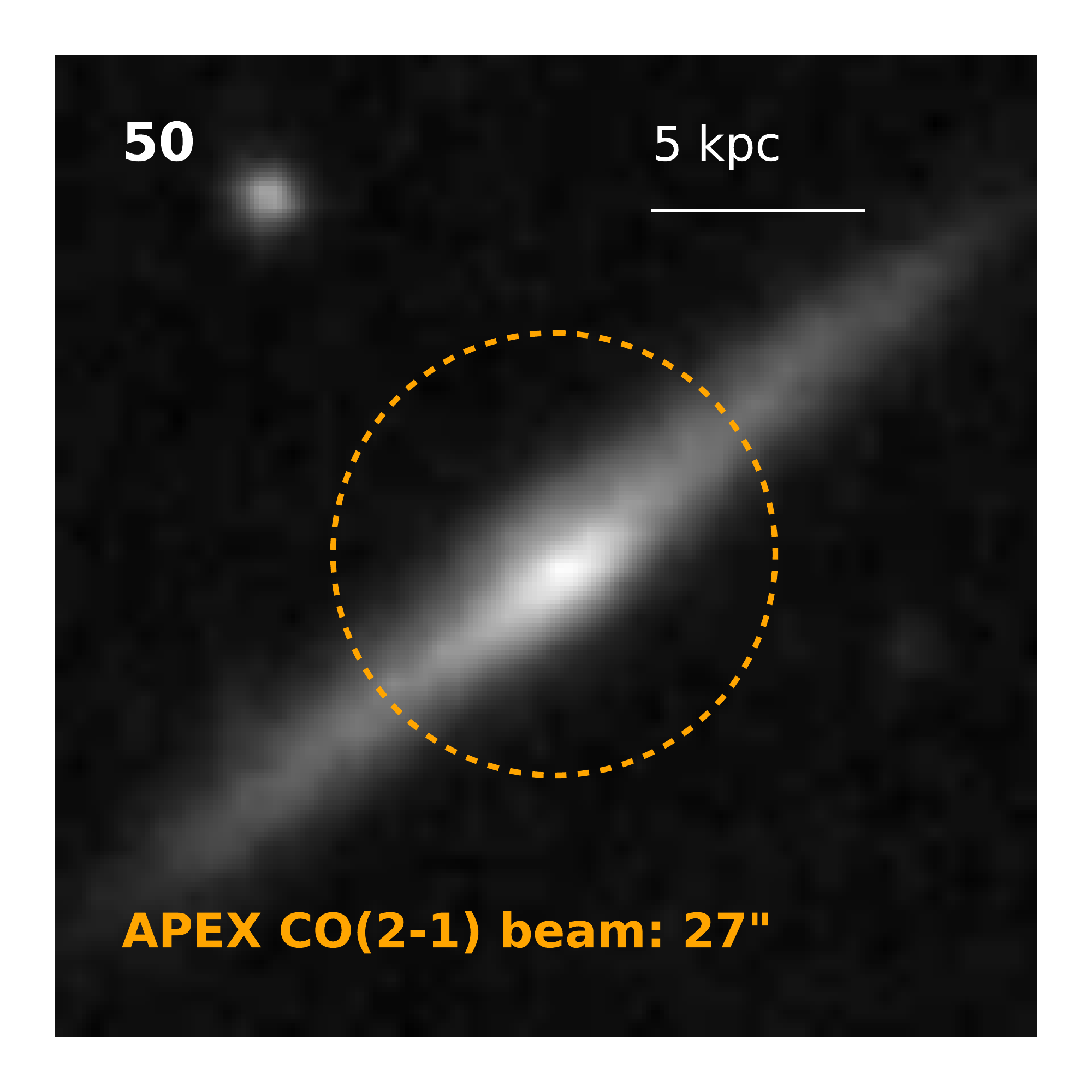}\includegraphics[width=0.26\textwidth]{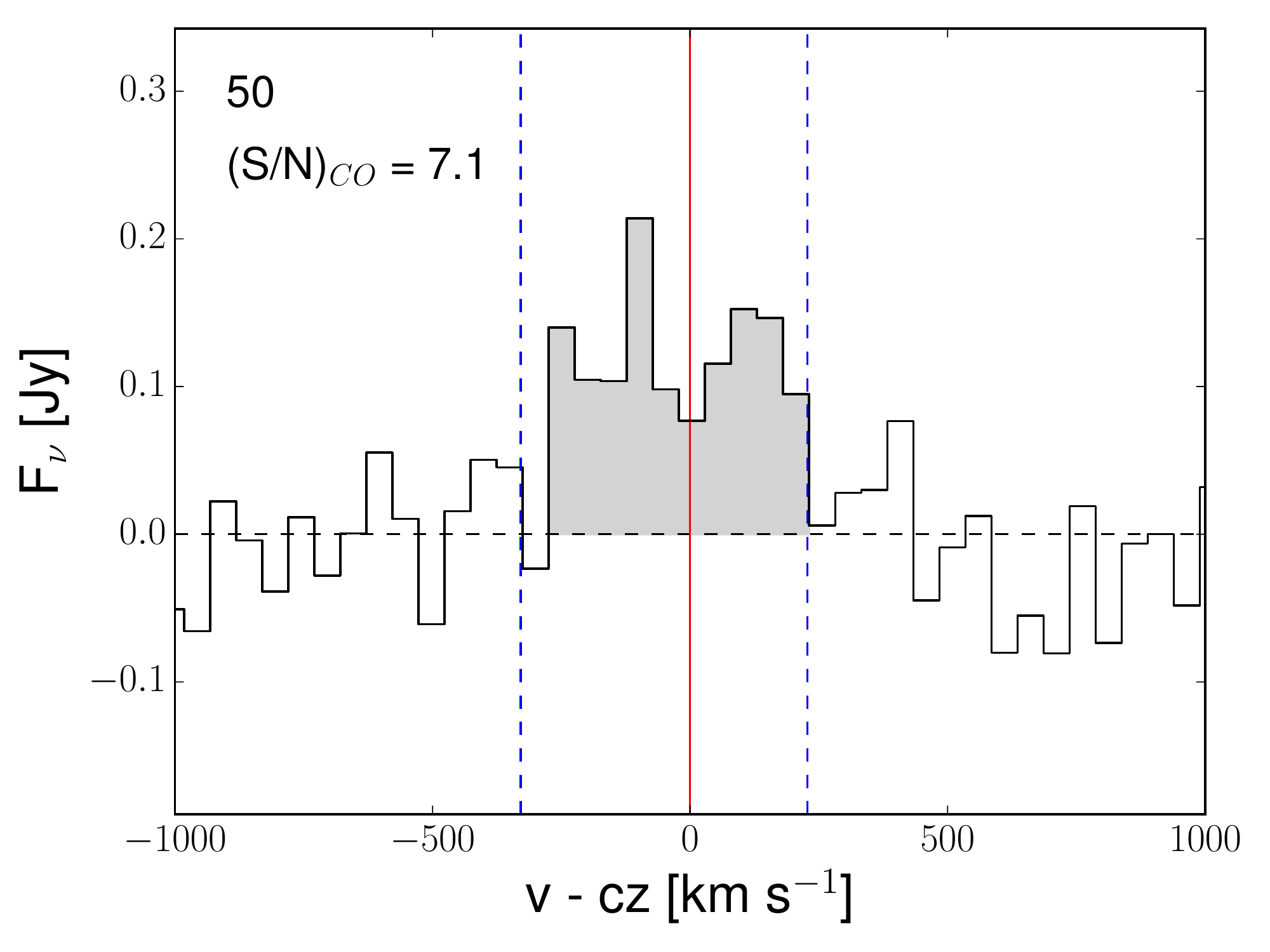}
\includegraphics[width=0.18\textwidth]{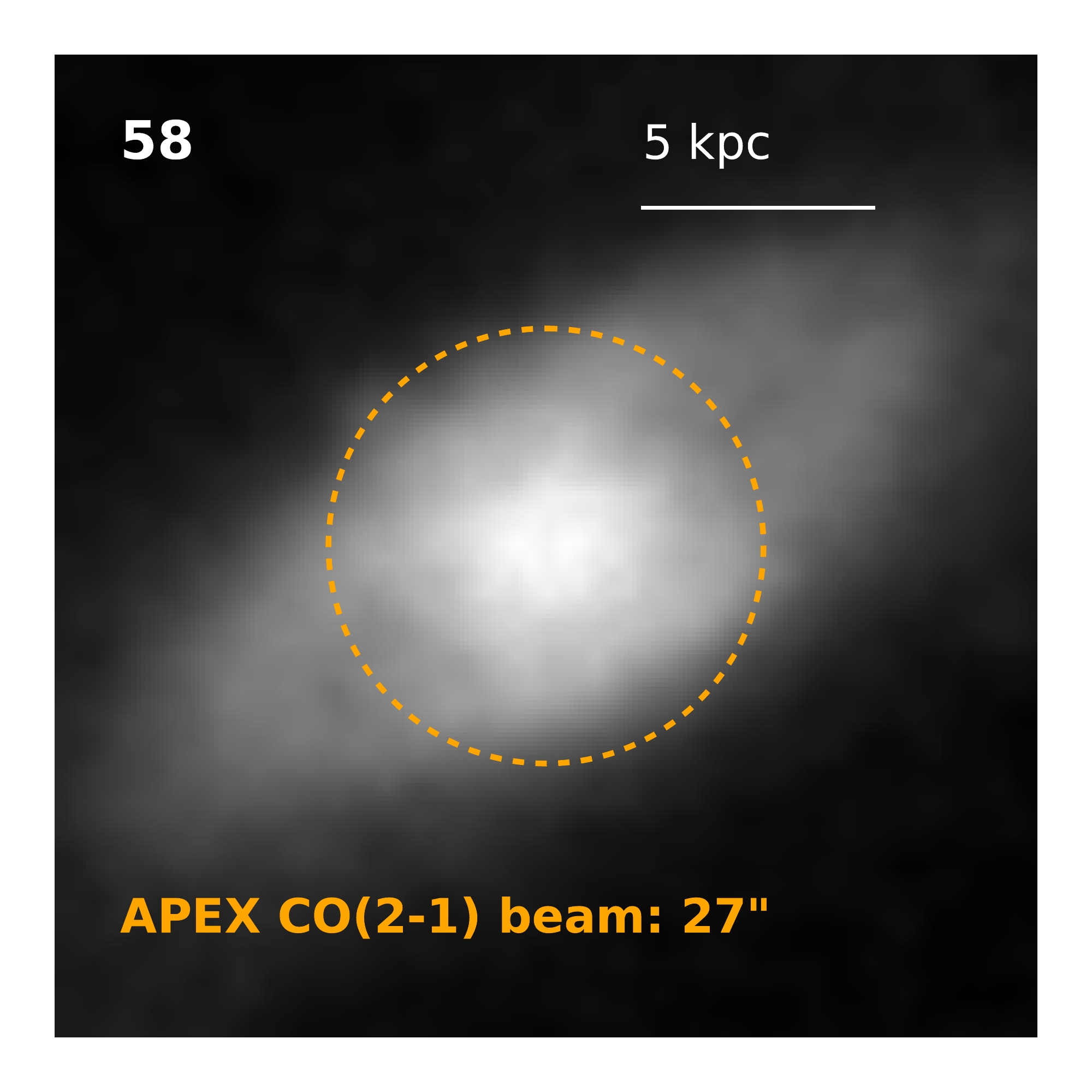}\includegraphics[width=0.26\textwidth]{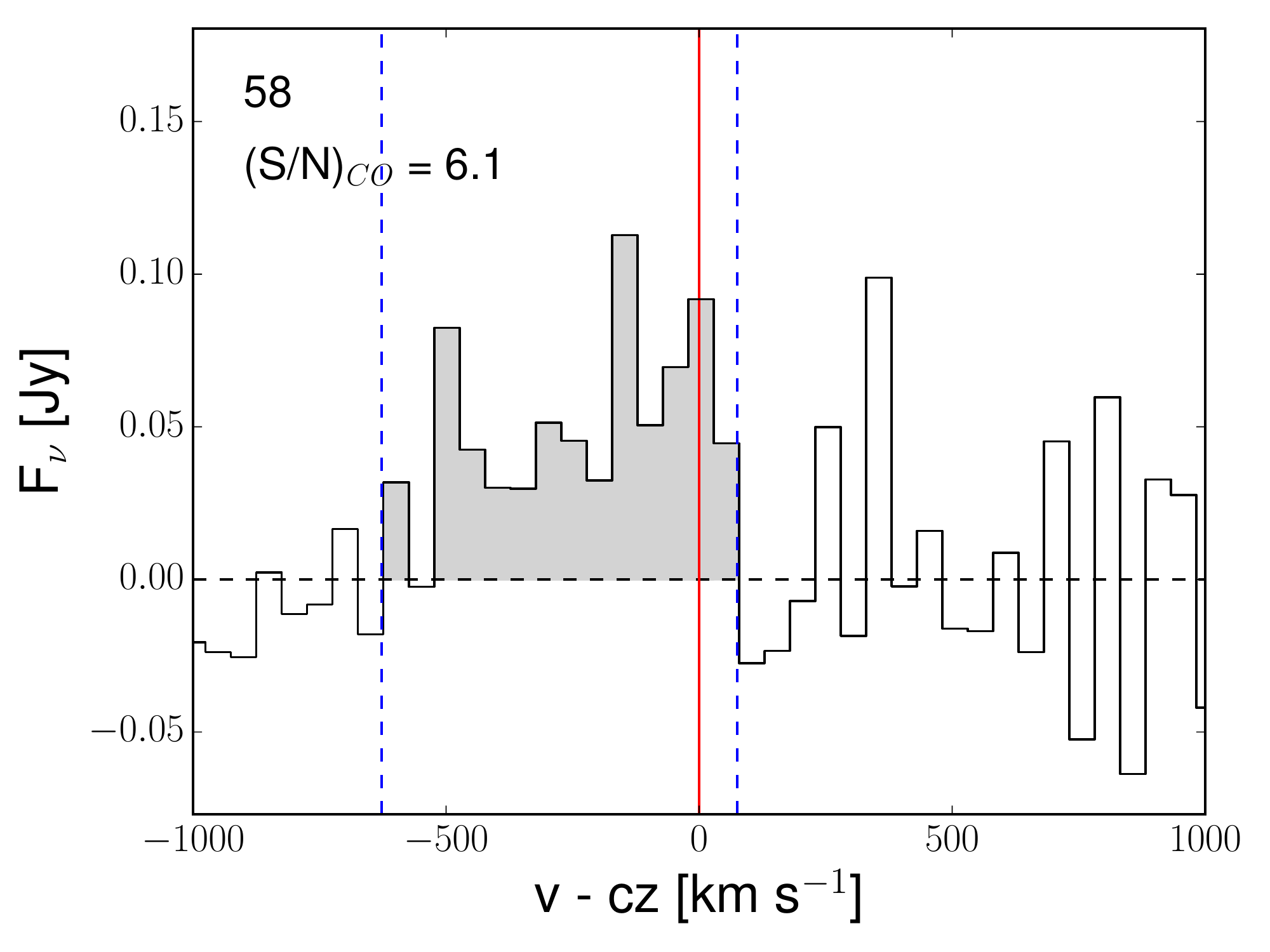}
\includegraphics[width=0.18\textwidth]{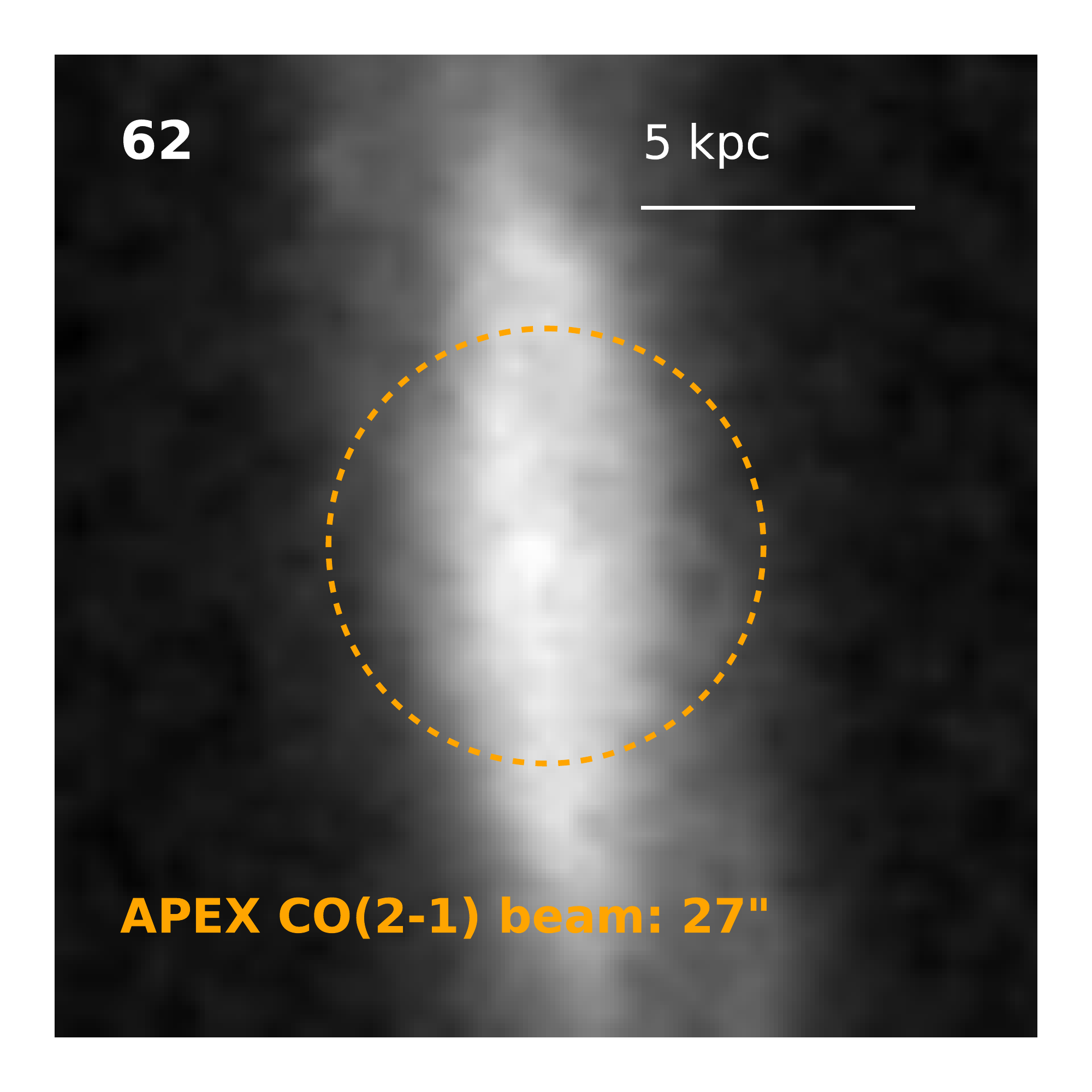}\includegraphics[width=0.26\textwidth]{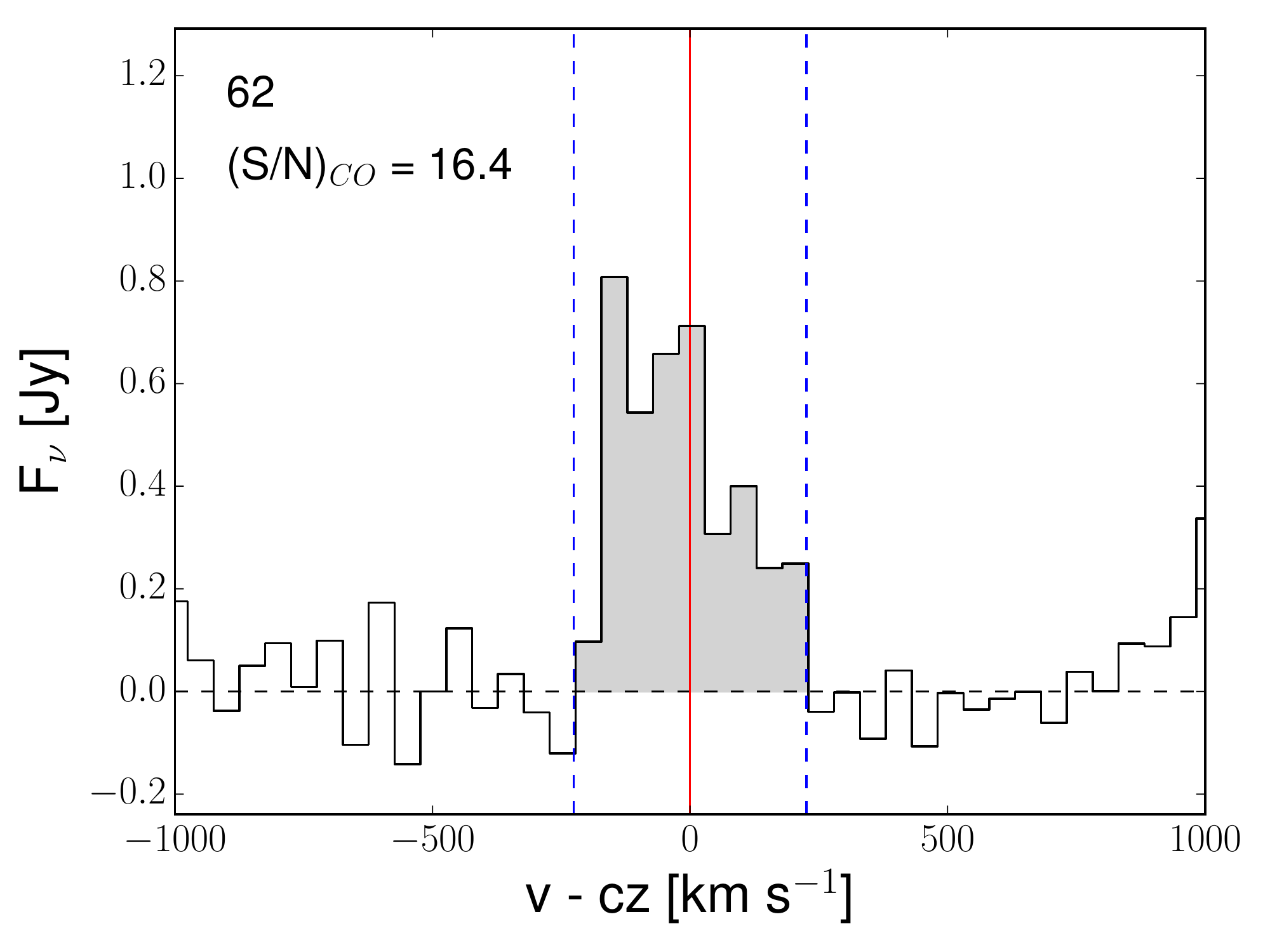}
\includegraphics[width=0.18\textwidth]{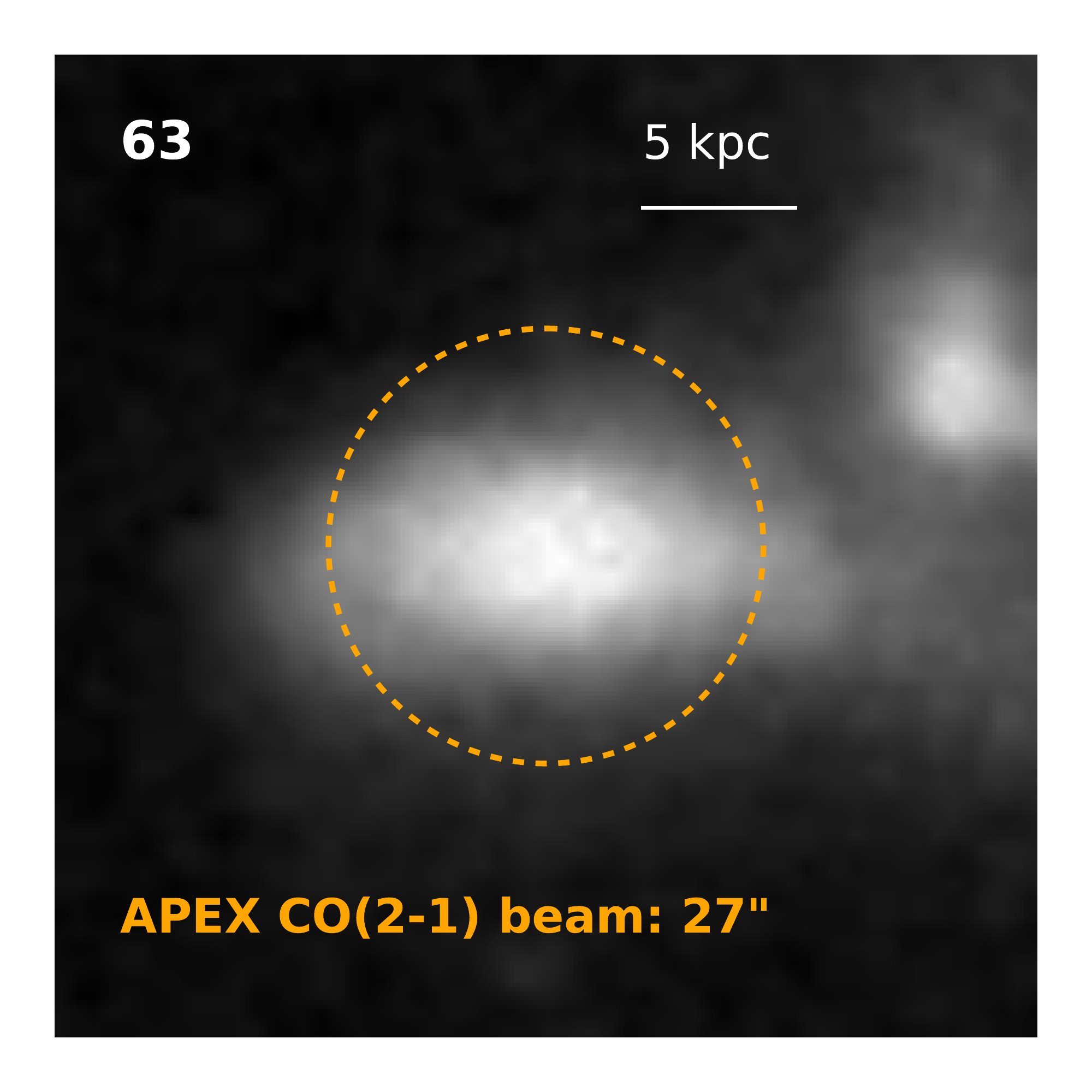}\includegraphics[width=0.26\textwidth]{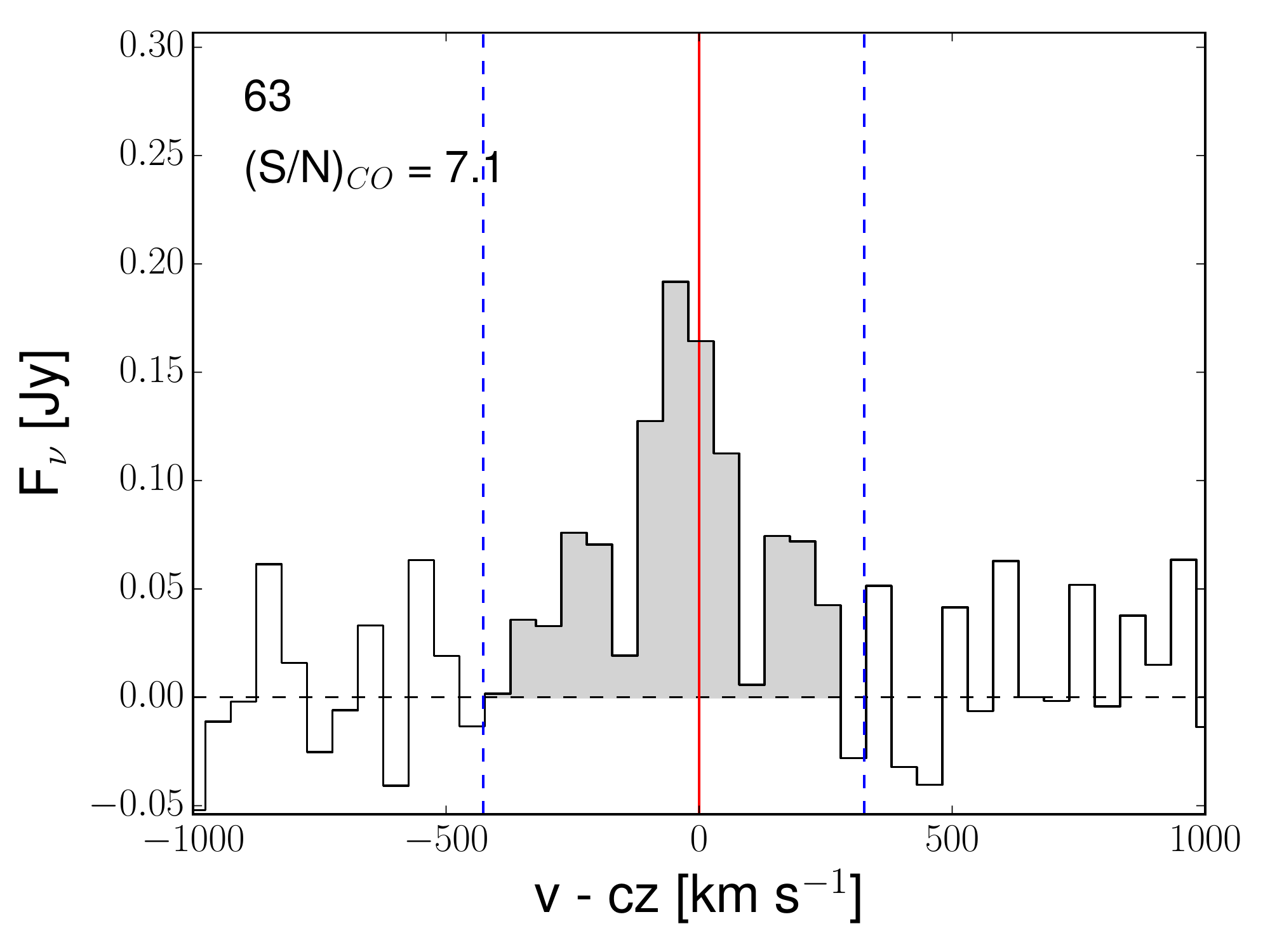}
\includegraphics[width=0.18\textwidth]{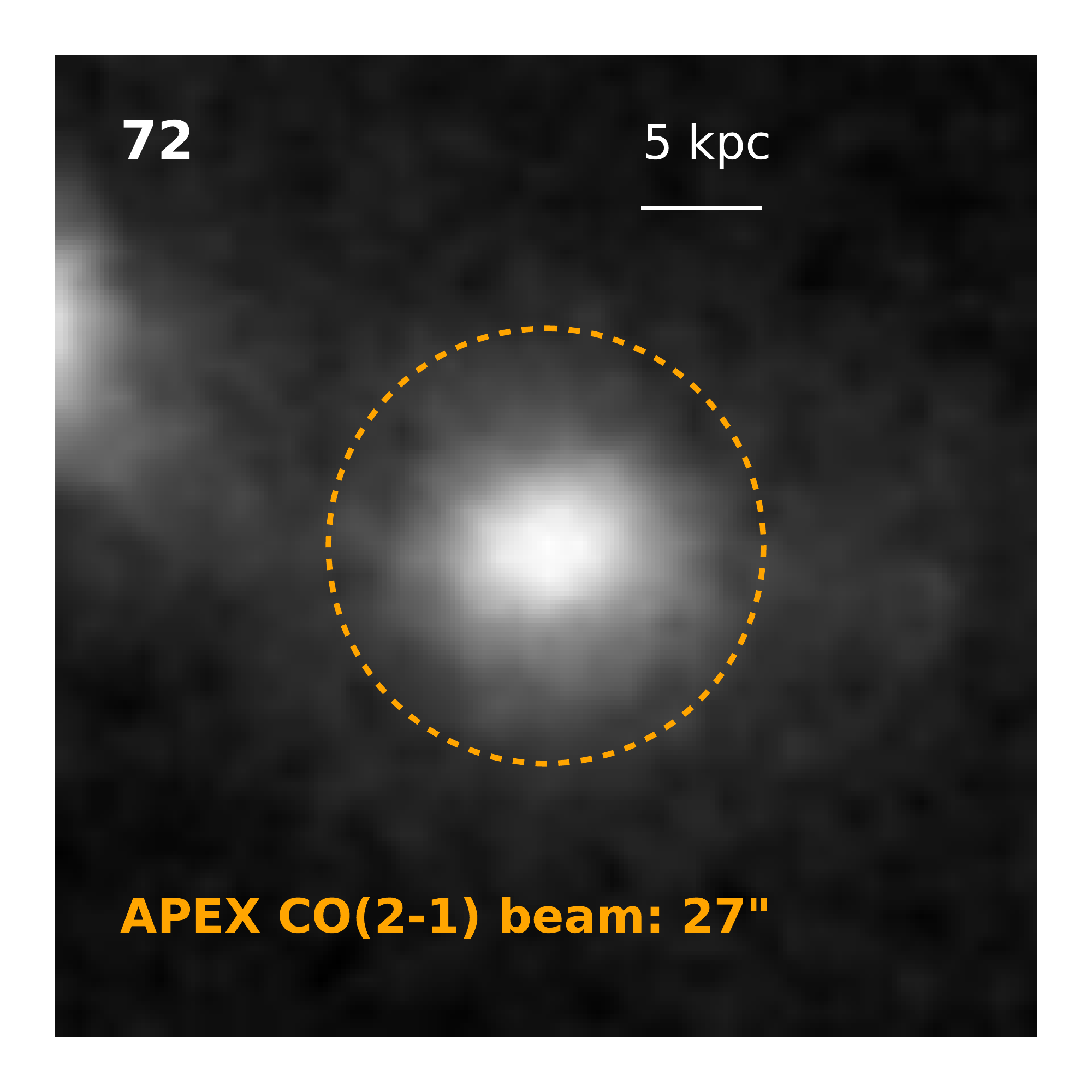}\includegraphics[width=0.26\textwidth]{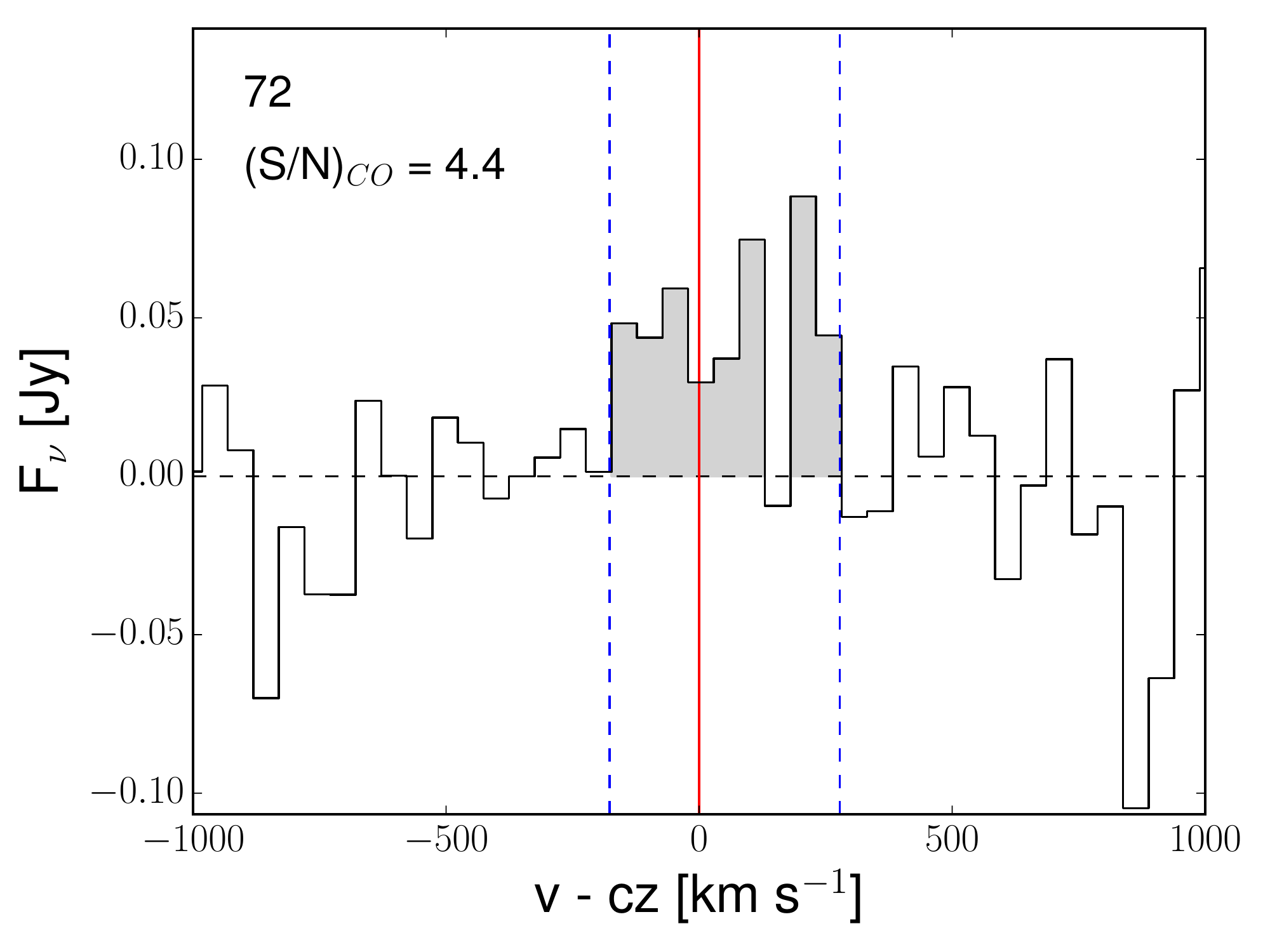}
\includegraphics[width=0.18\textwidth]{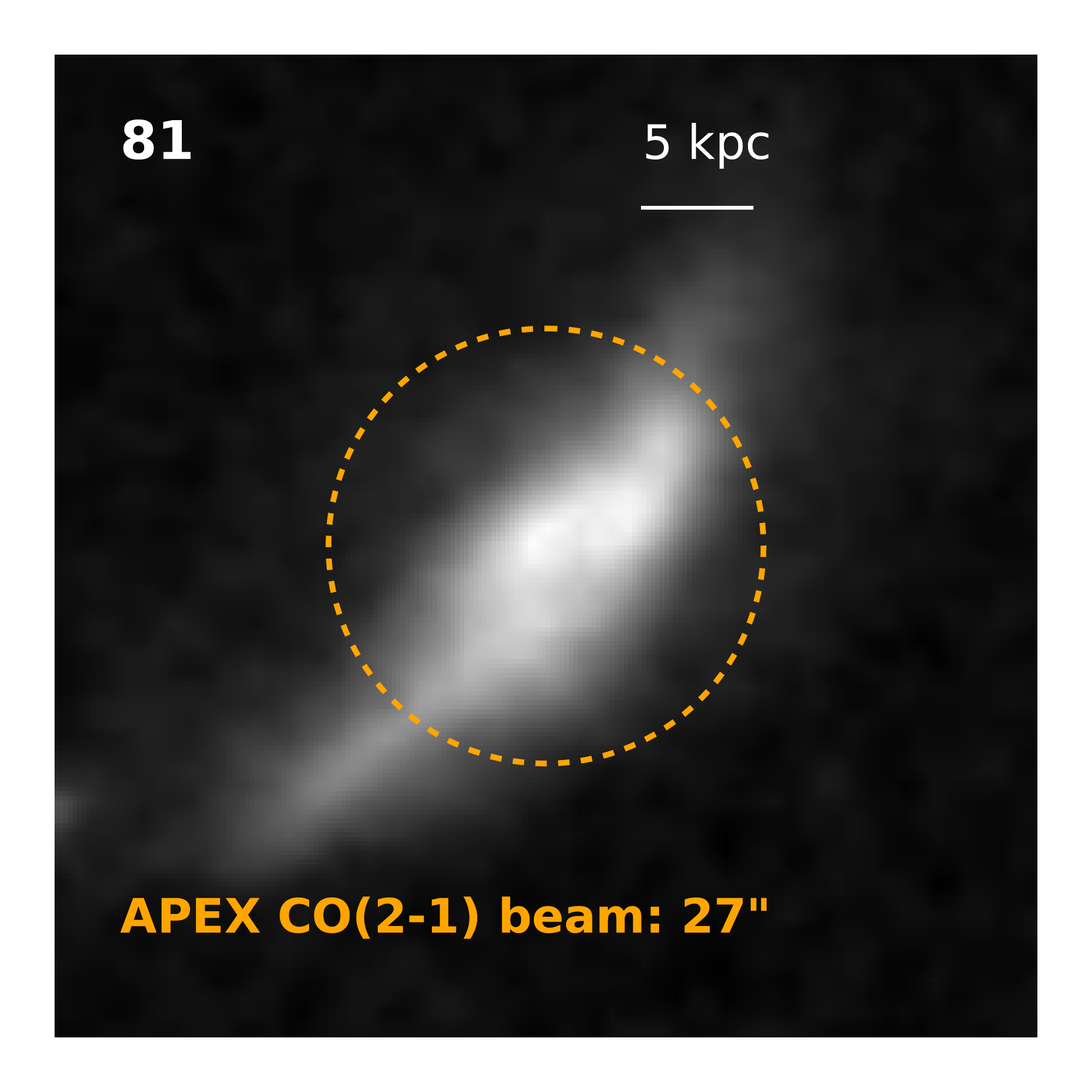}\includegraphics[width=0.26\textwidth]{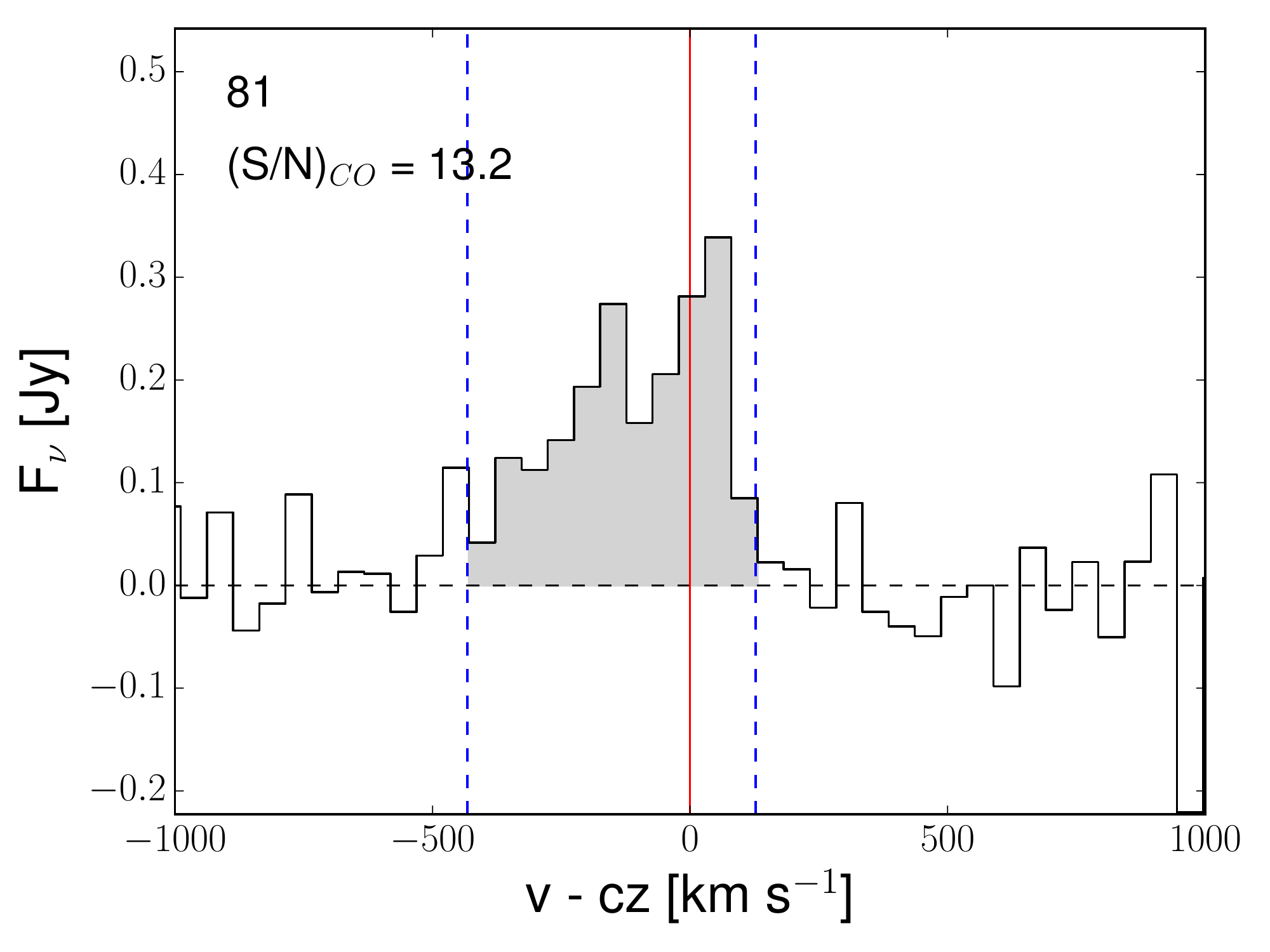}
\includegraphics[width=0.18\textwidth]{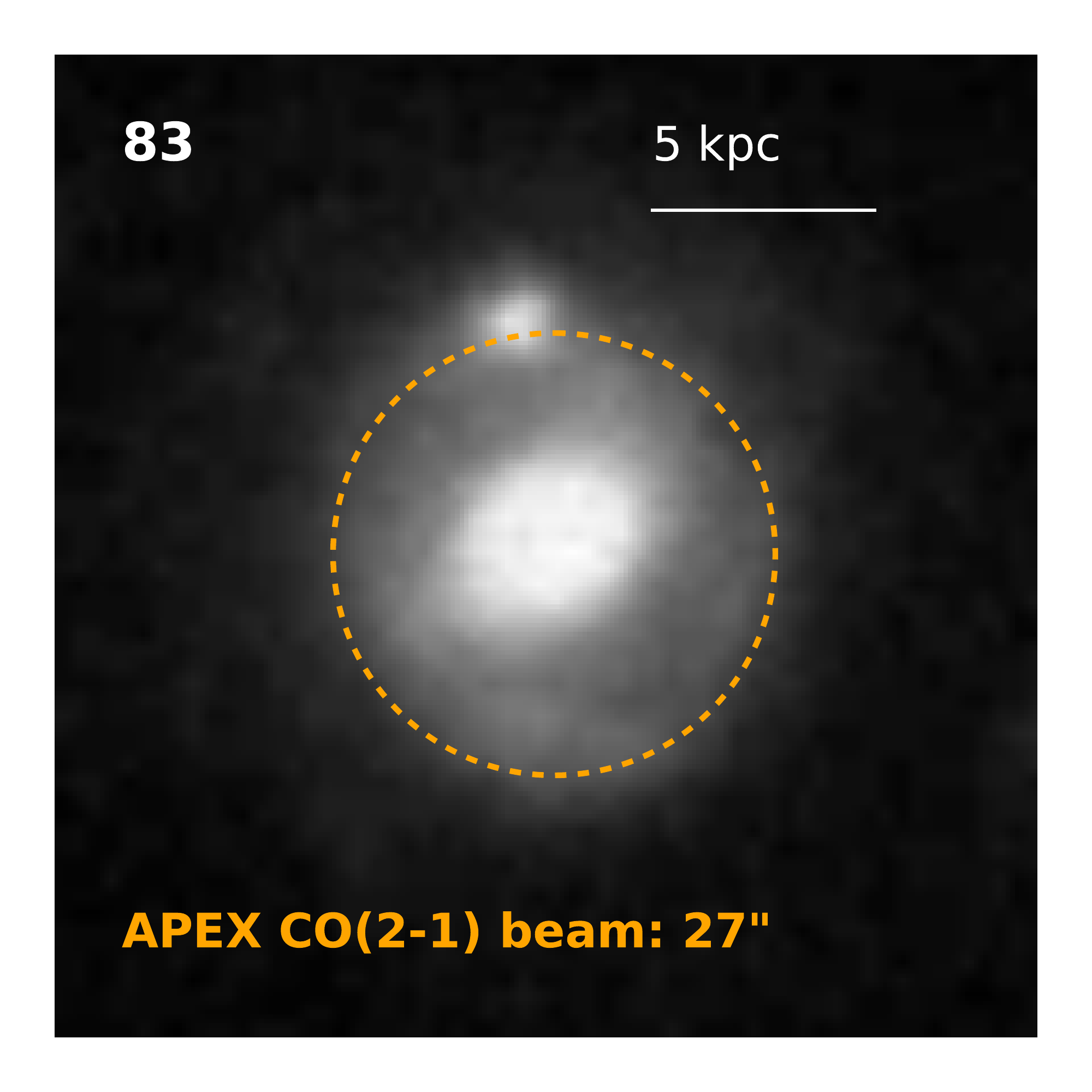}\includegraphics[width=0.26\textwidth]{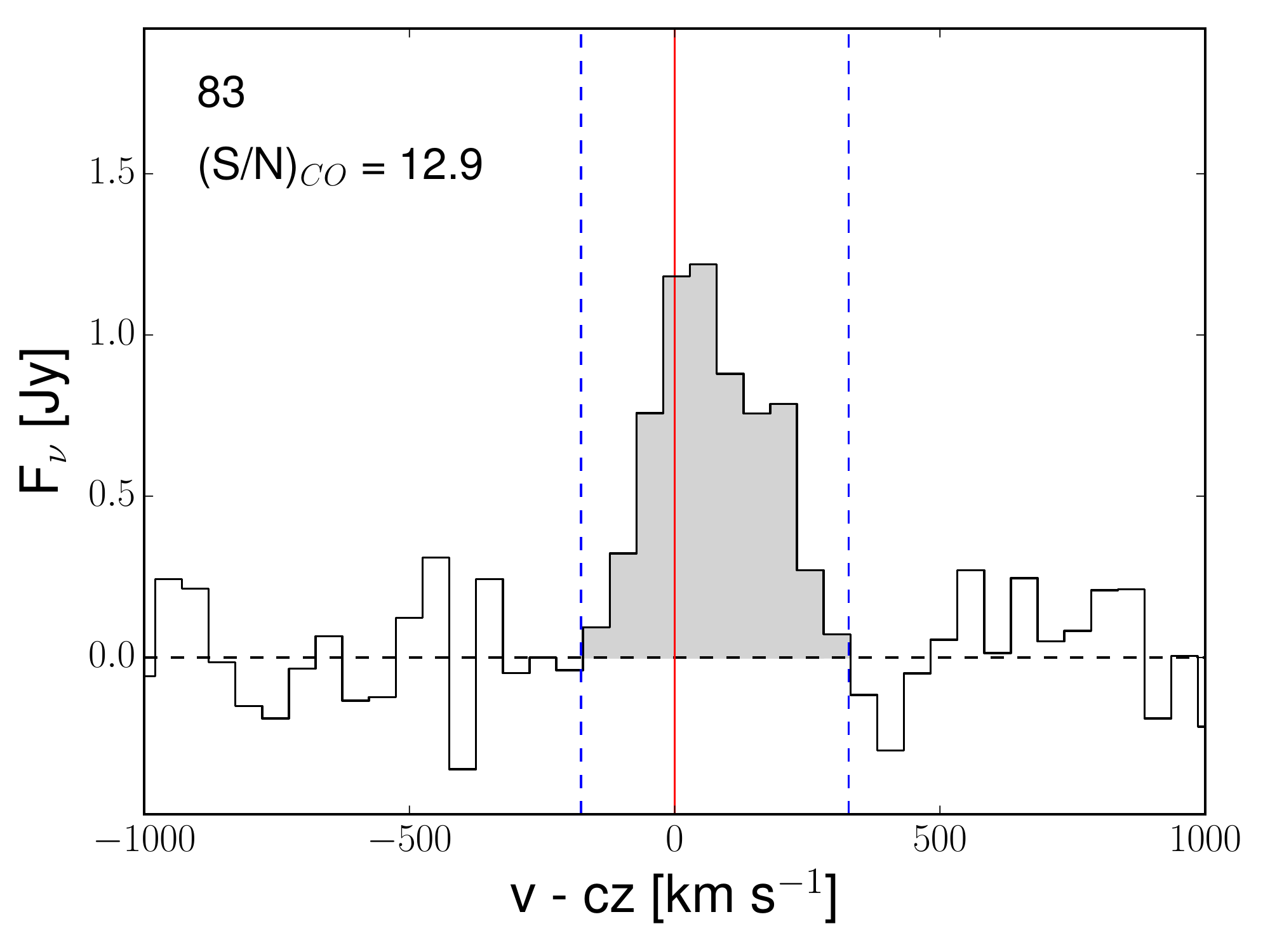}
\includegraphics[width=0.18\textwidth]{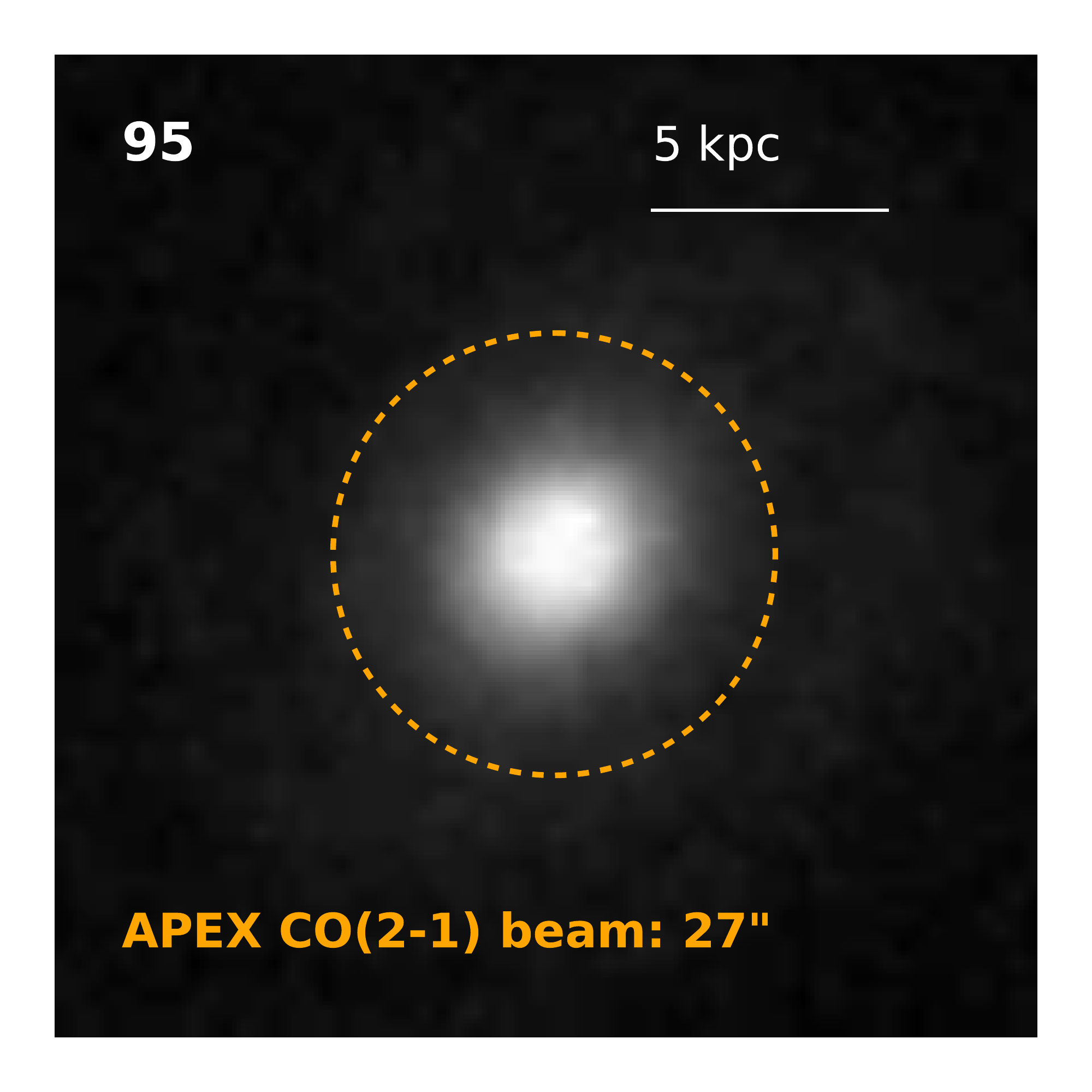}\includegraphics[width=0.26\textwidth]{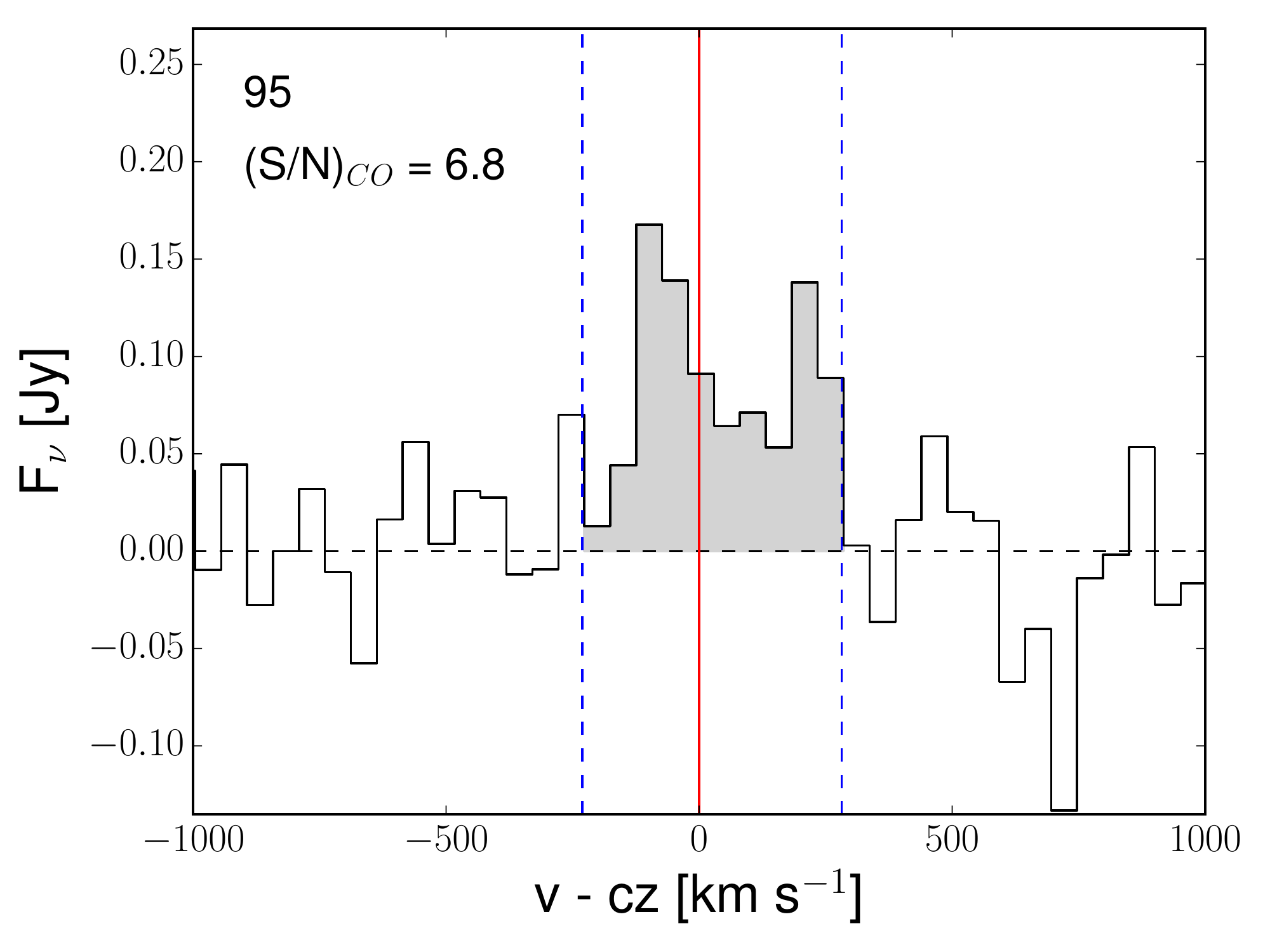}
\includegraphics[width=0.18\textwidth]{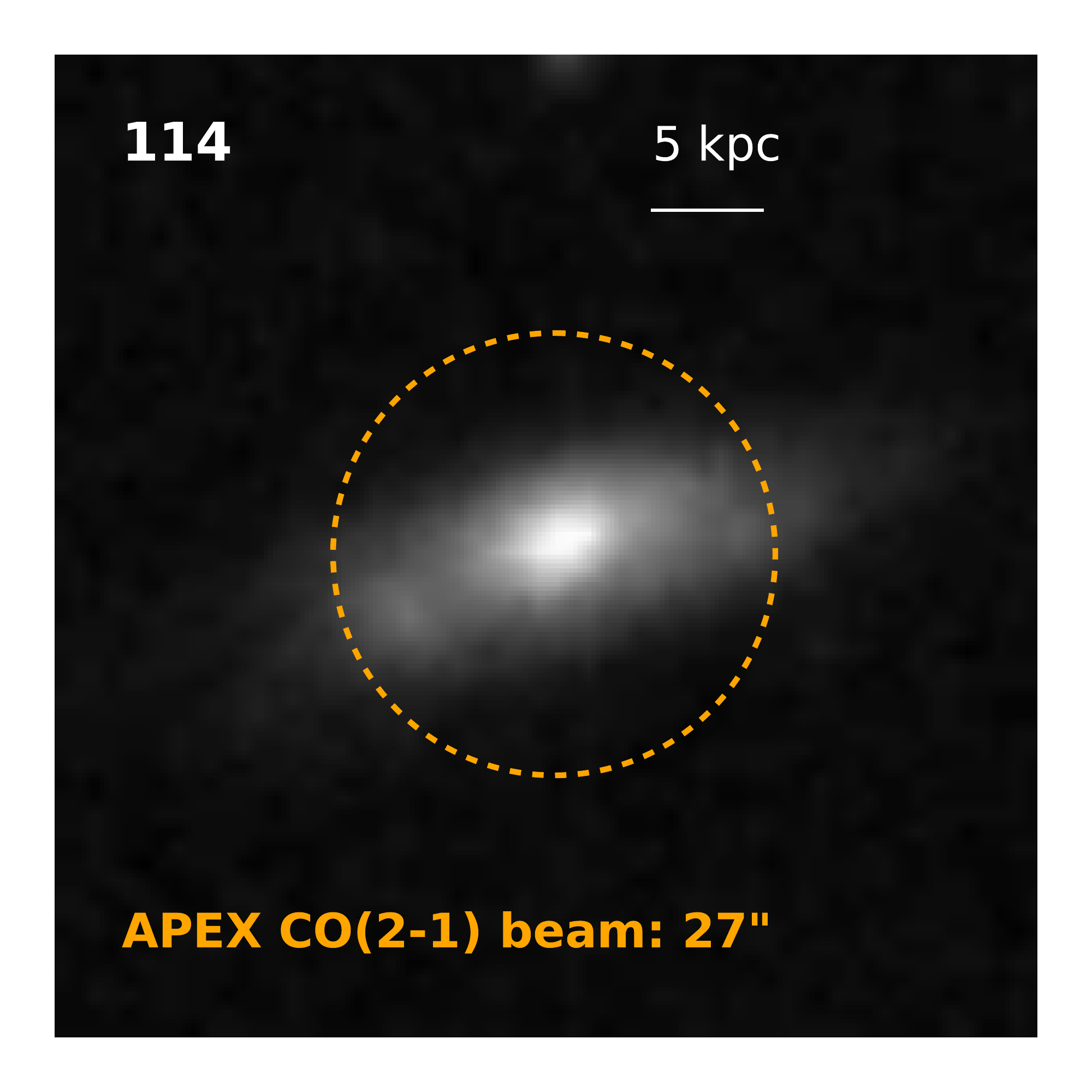}\includegraphics[width=0.26\textwidth]{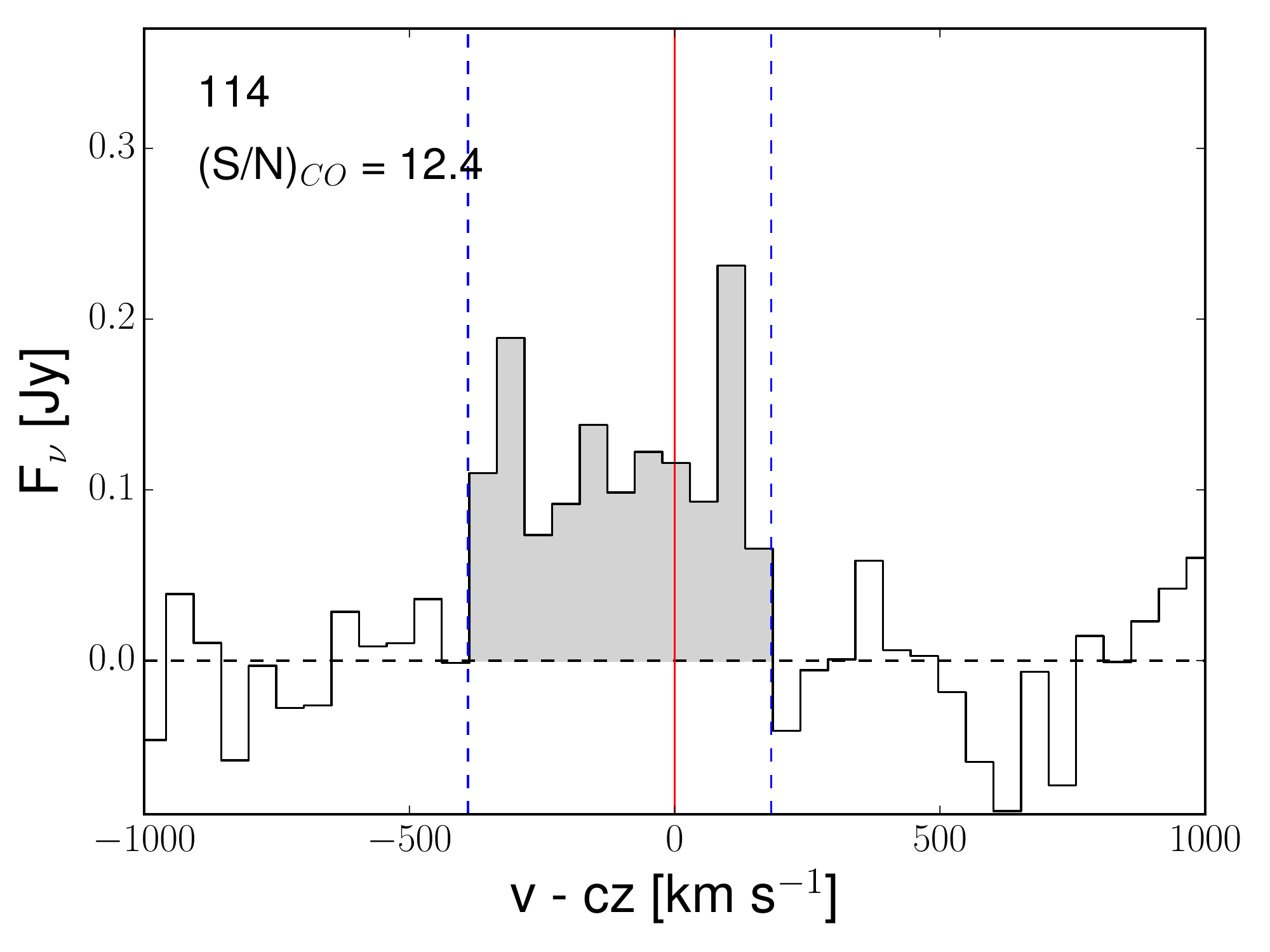}
\includegraphics[width=0.18\textwidth]{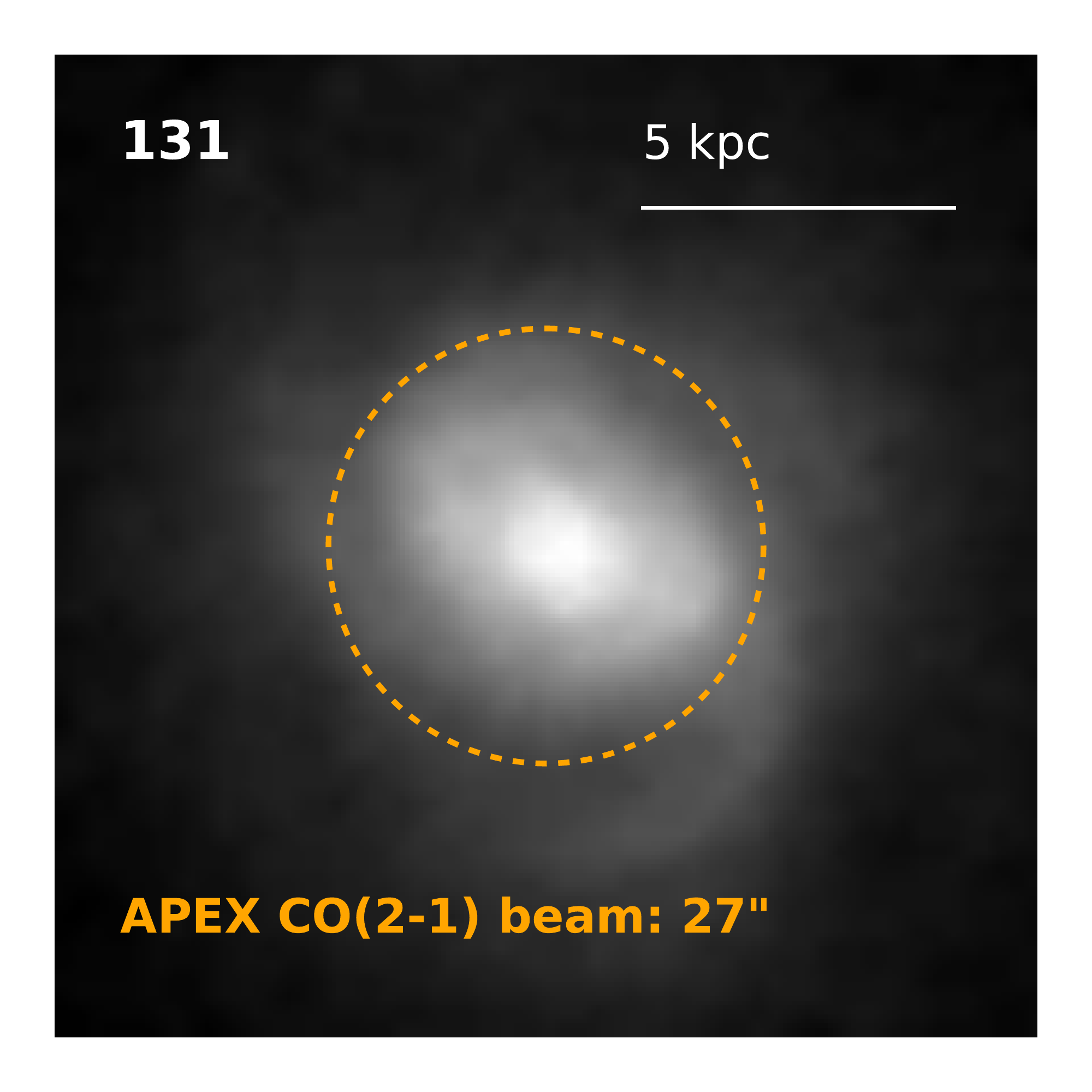}\includegraphics[width=0.26\textwidth]{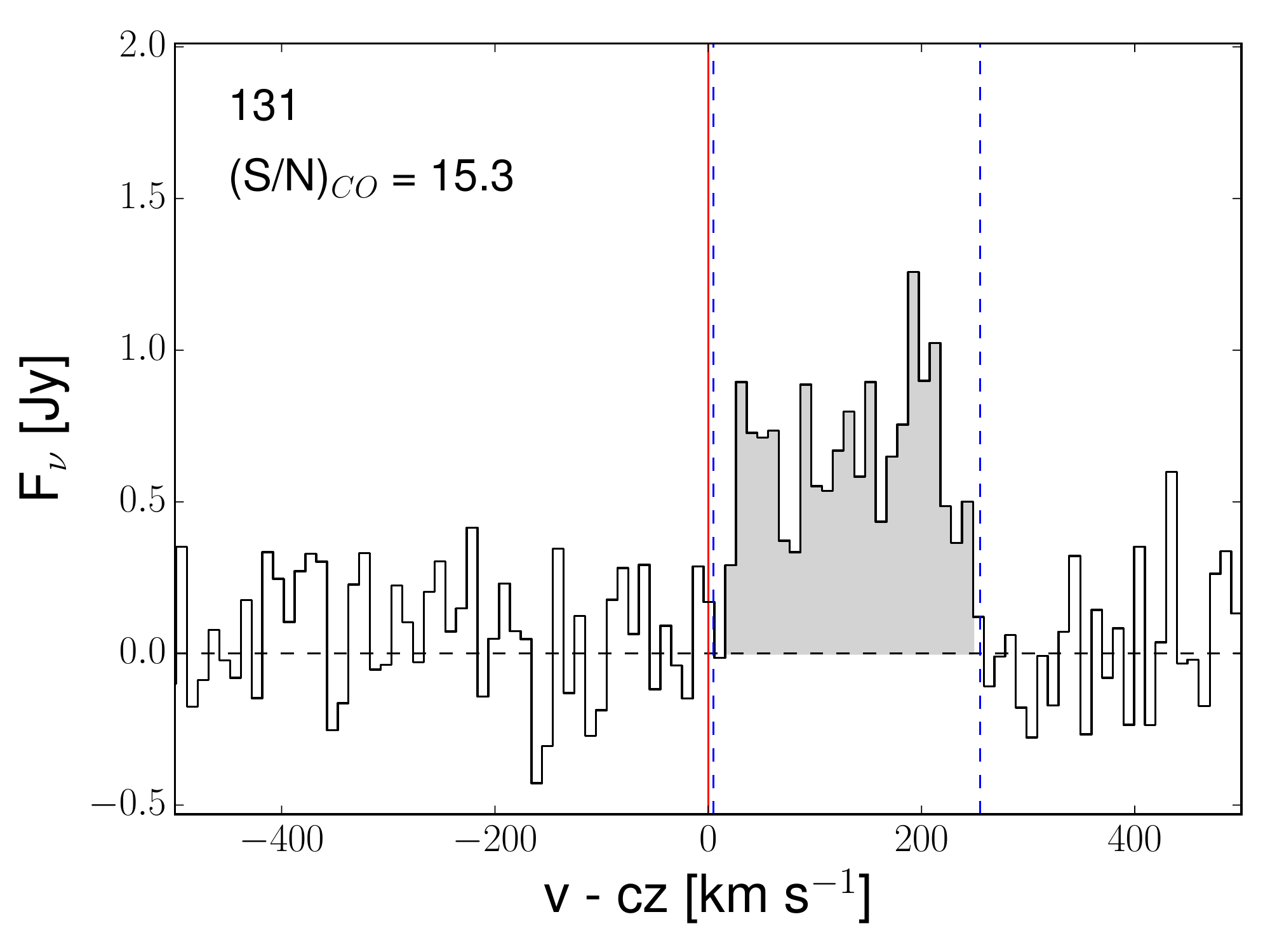}
\caption{Similar to Fig.~\ref{fig:CO21_spectra}, for detected sources with only DSS imaging.
} 
\label{fig:CO21_spectra_dss1}
\end{figure*}

\begin{figure*}
\centering
\raggedright
\includegraphics[width=0.18\textwidth]{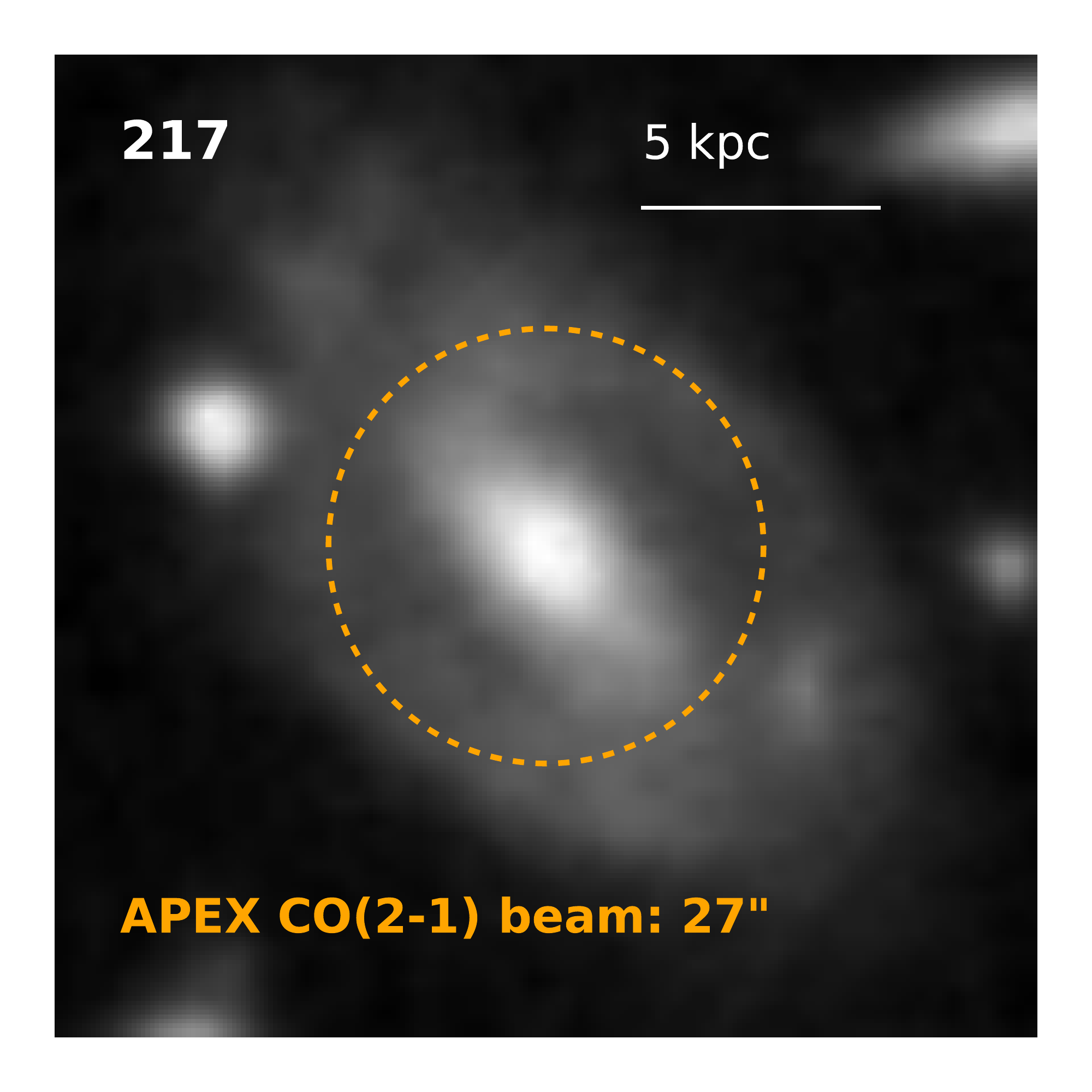}\includegraphics[width=0.26\textwidth]{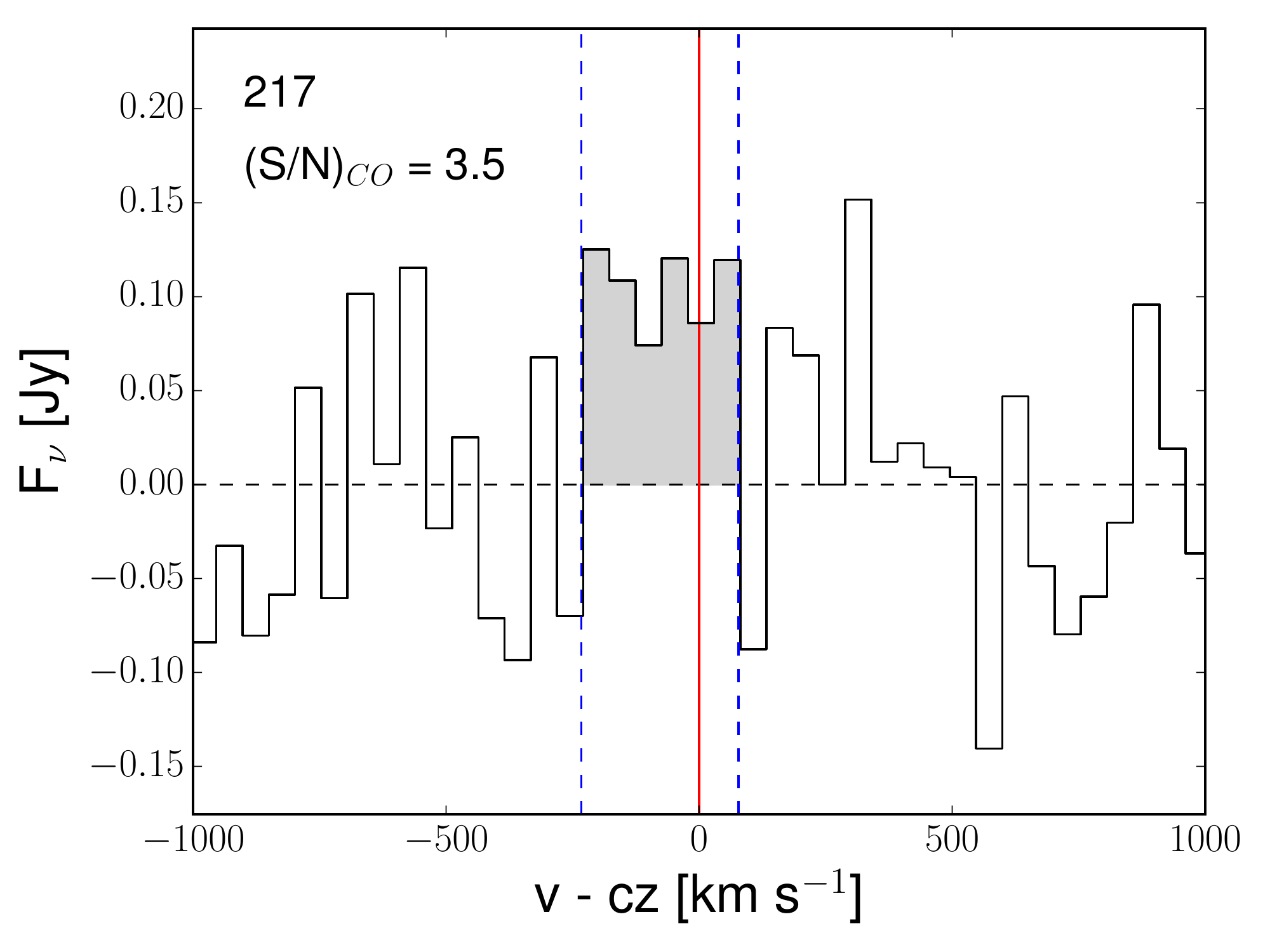}
\includegraphics[width=0.18\textwidth]{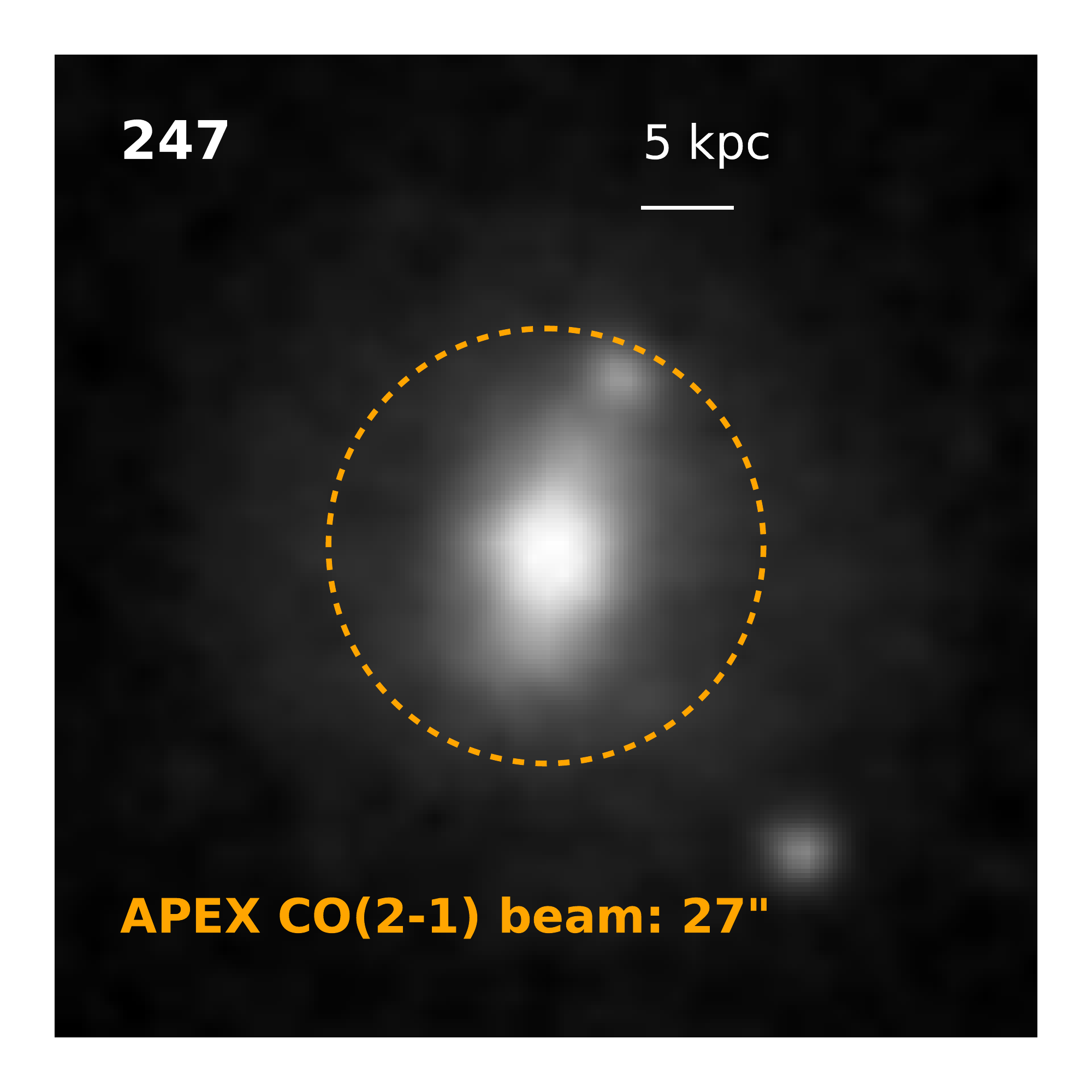}\includegraphics[width=0.26\textwidth]{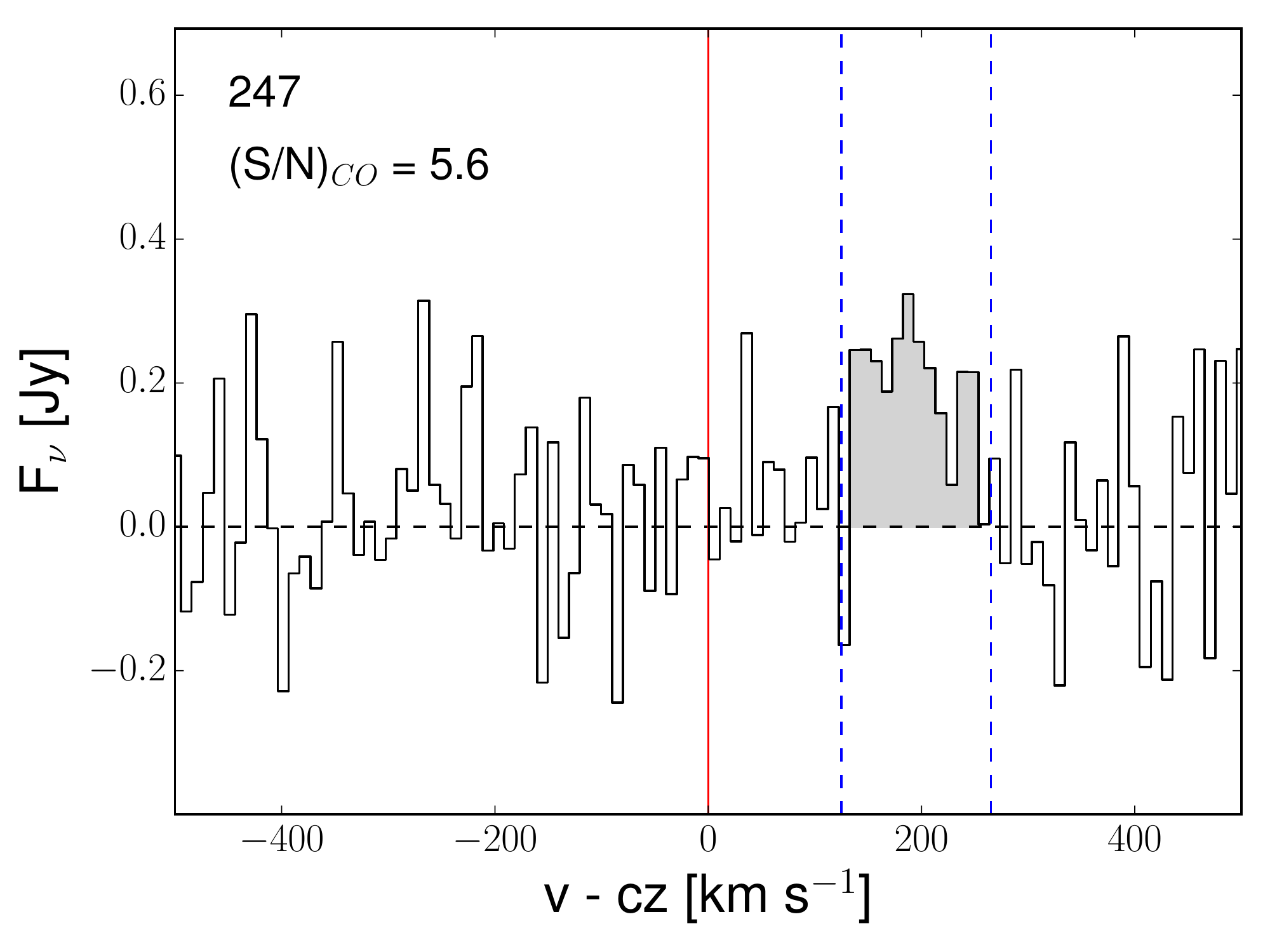}
\includegraphics[width=0.18\textwidth]{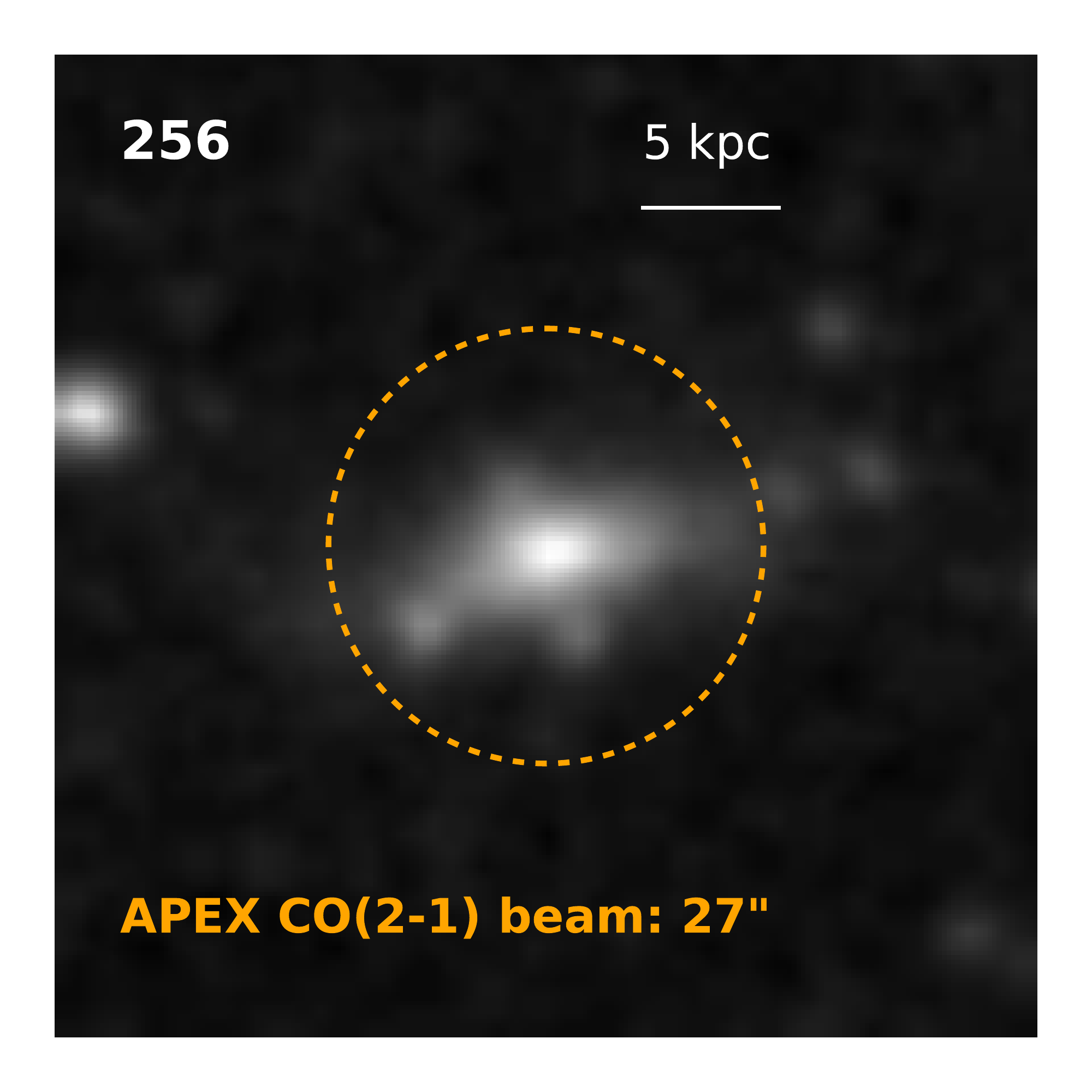}\includegraphics[width=0.26\textwidth]{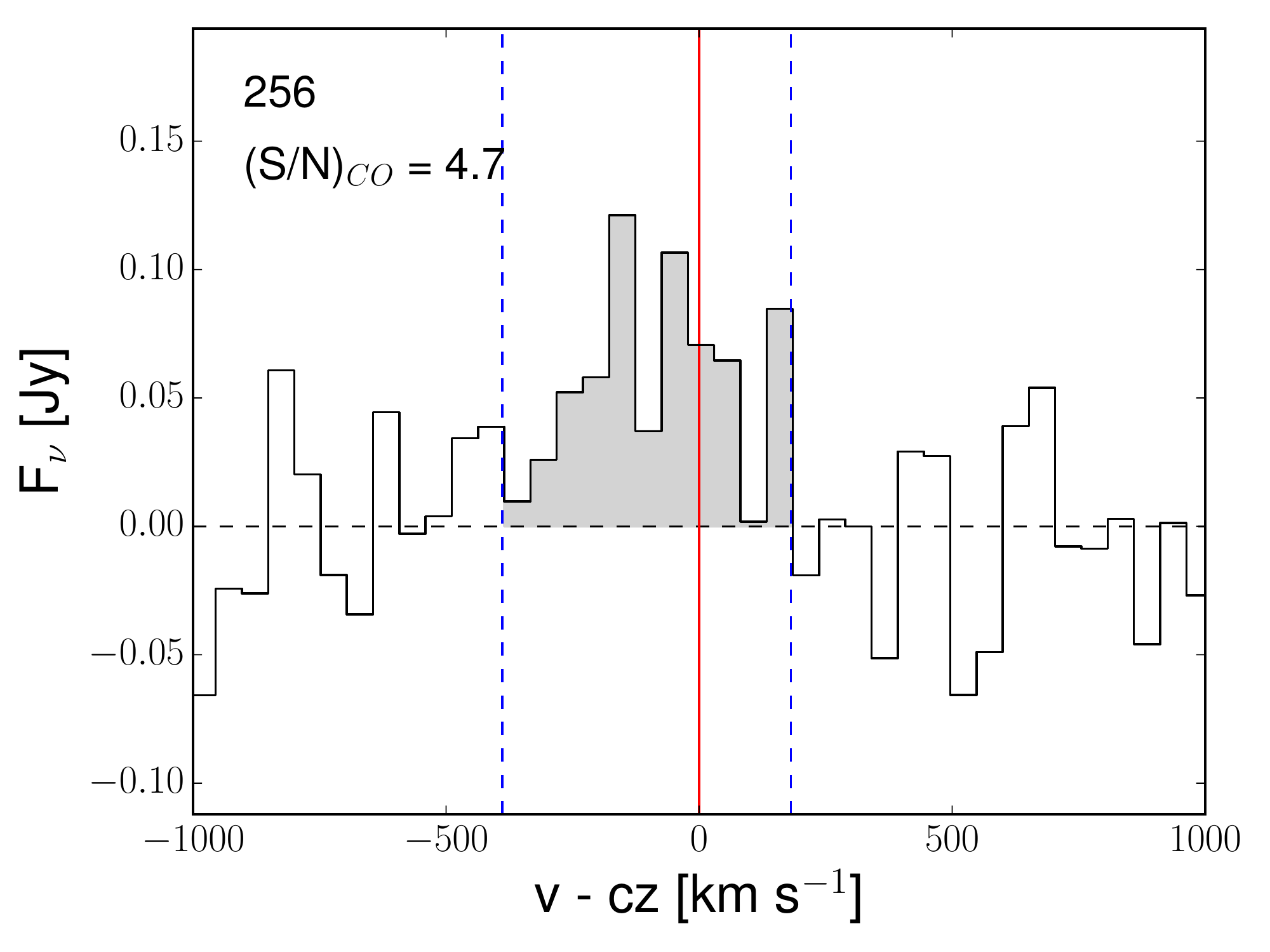}
\includegraphics[width=0.18\textwidth]{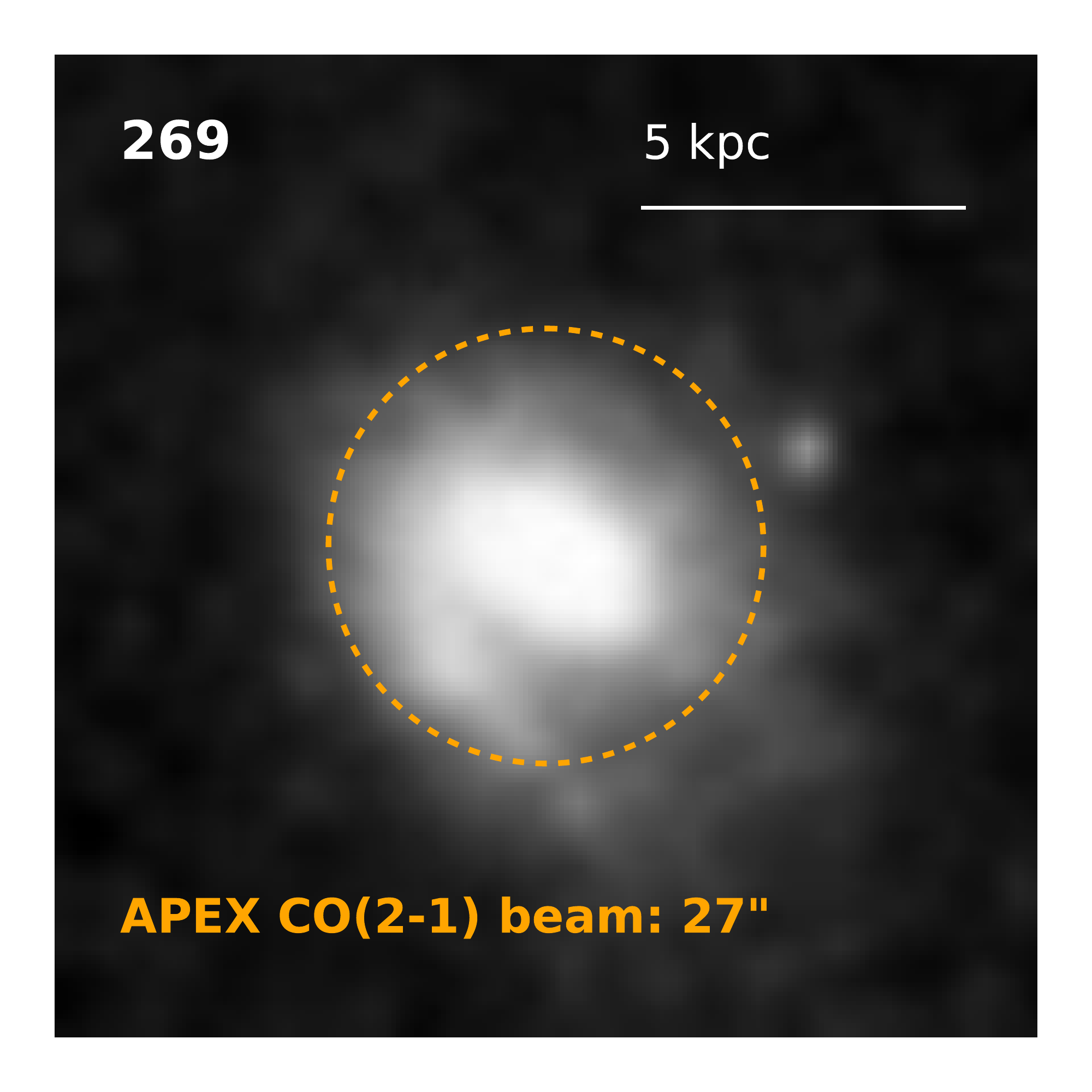}\includegraphics[width=0.26\textwidth]{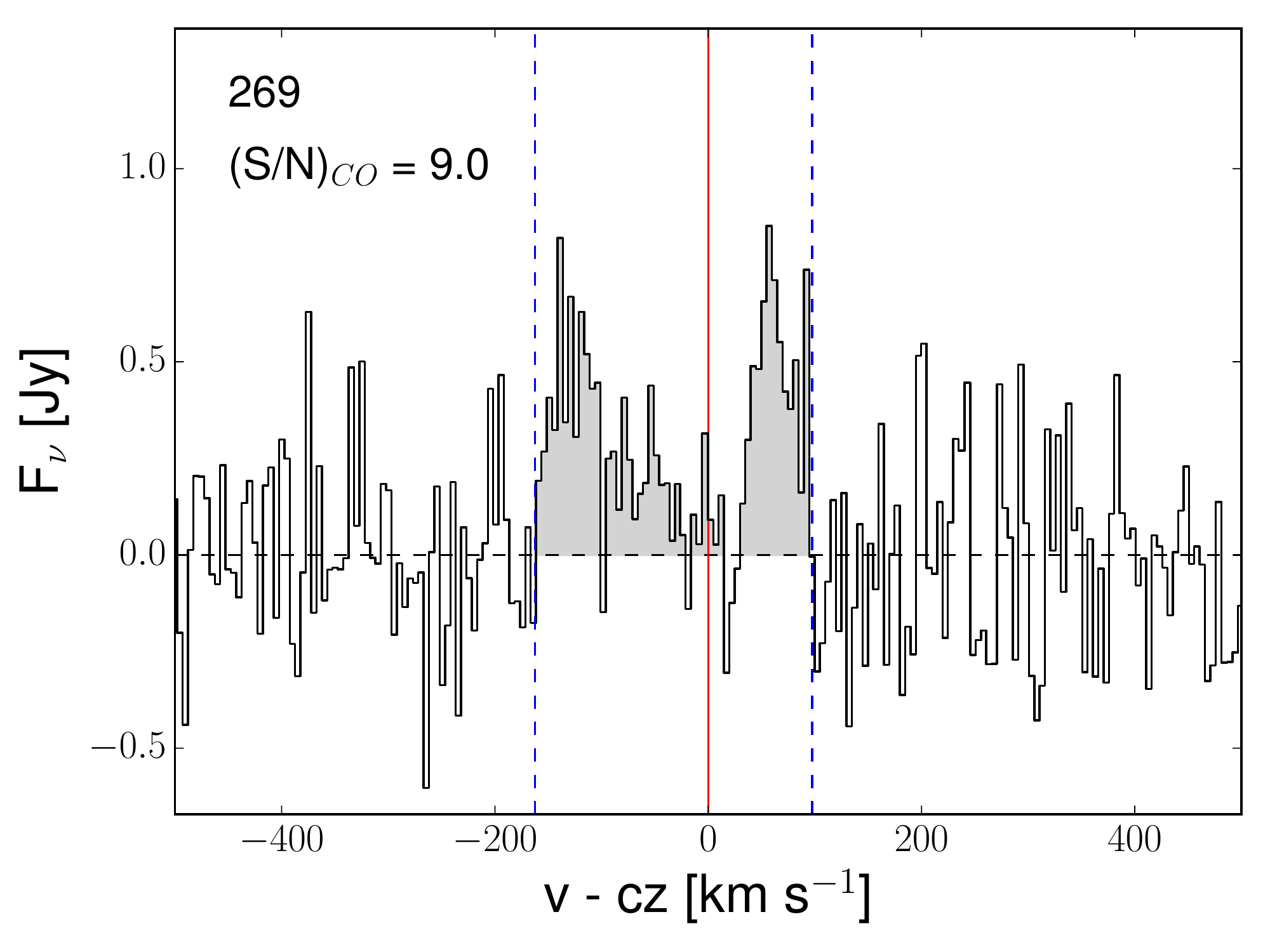}
\includegraphics[width=0.18\textwidth]{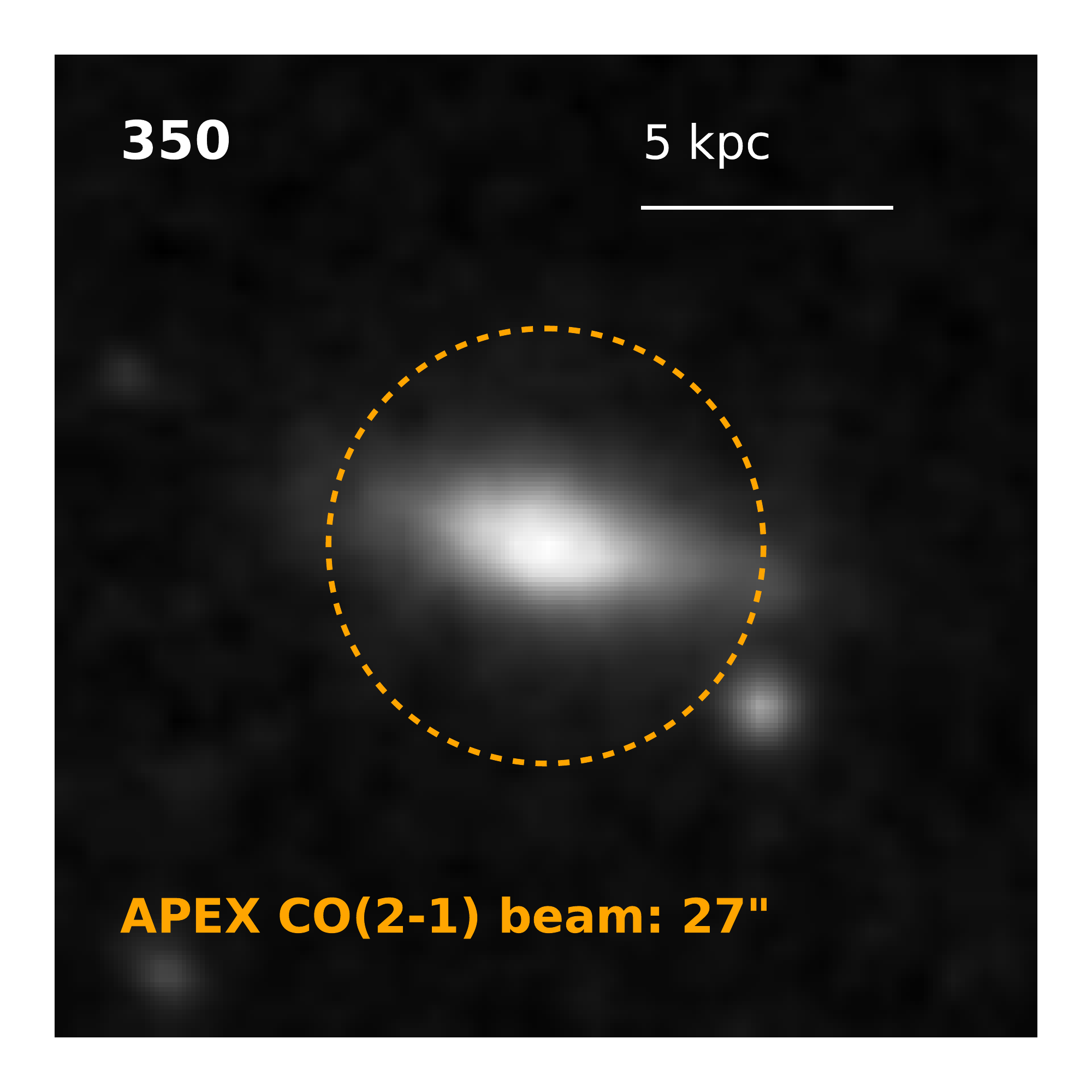}\includegraphics[width=0.26\textwidth]{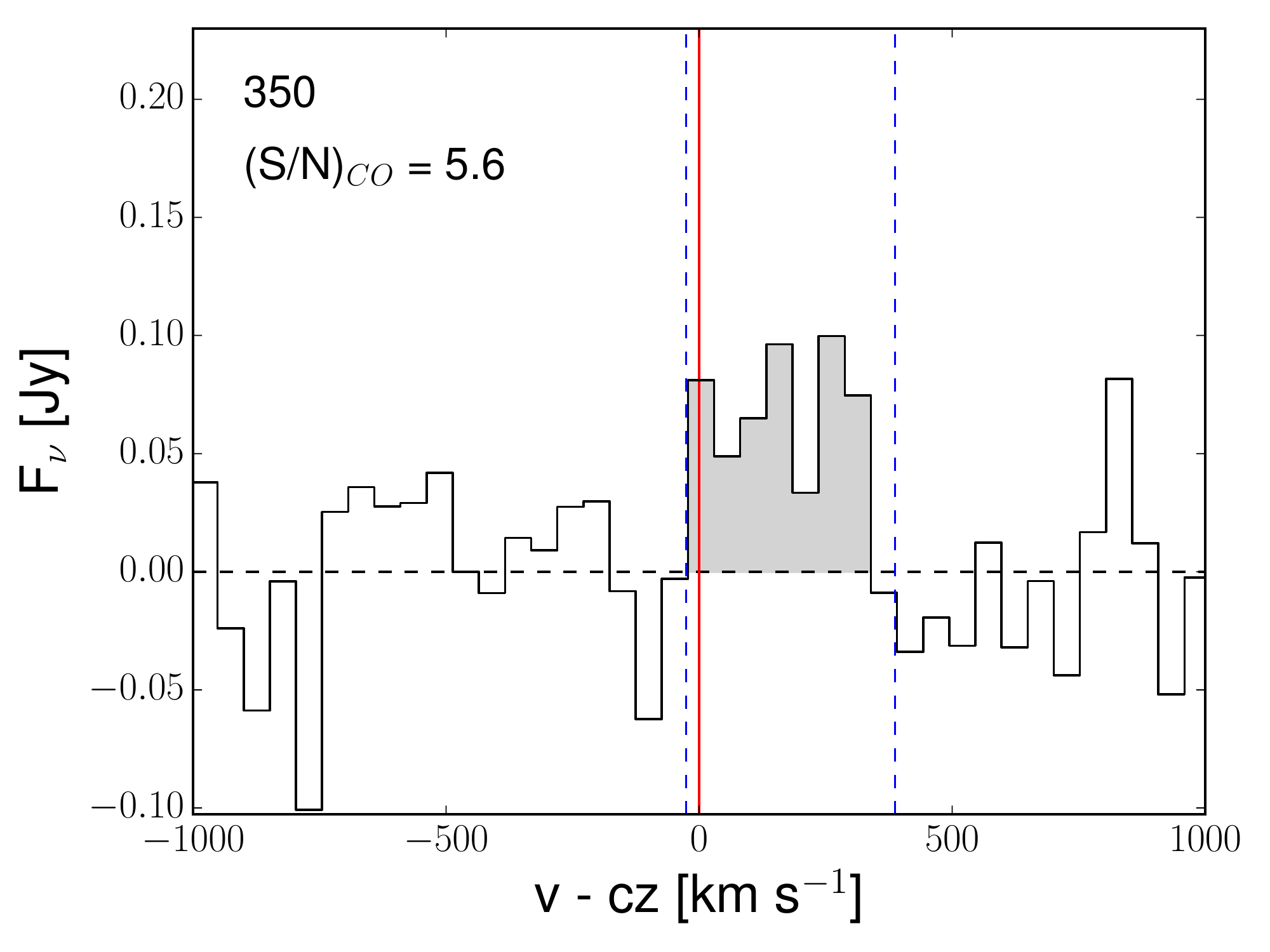}
\includegraphics[width=0.18\textwidth]{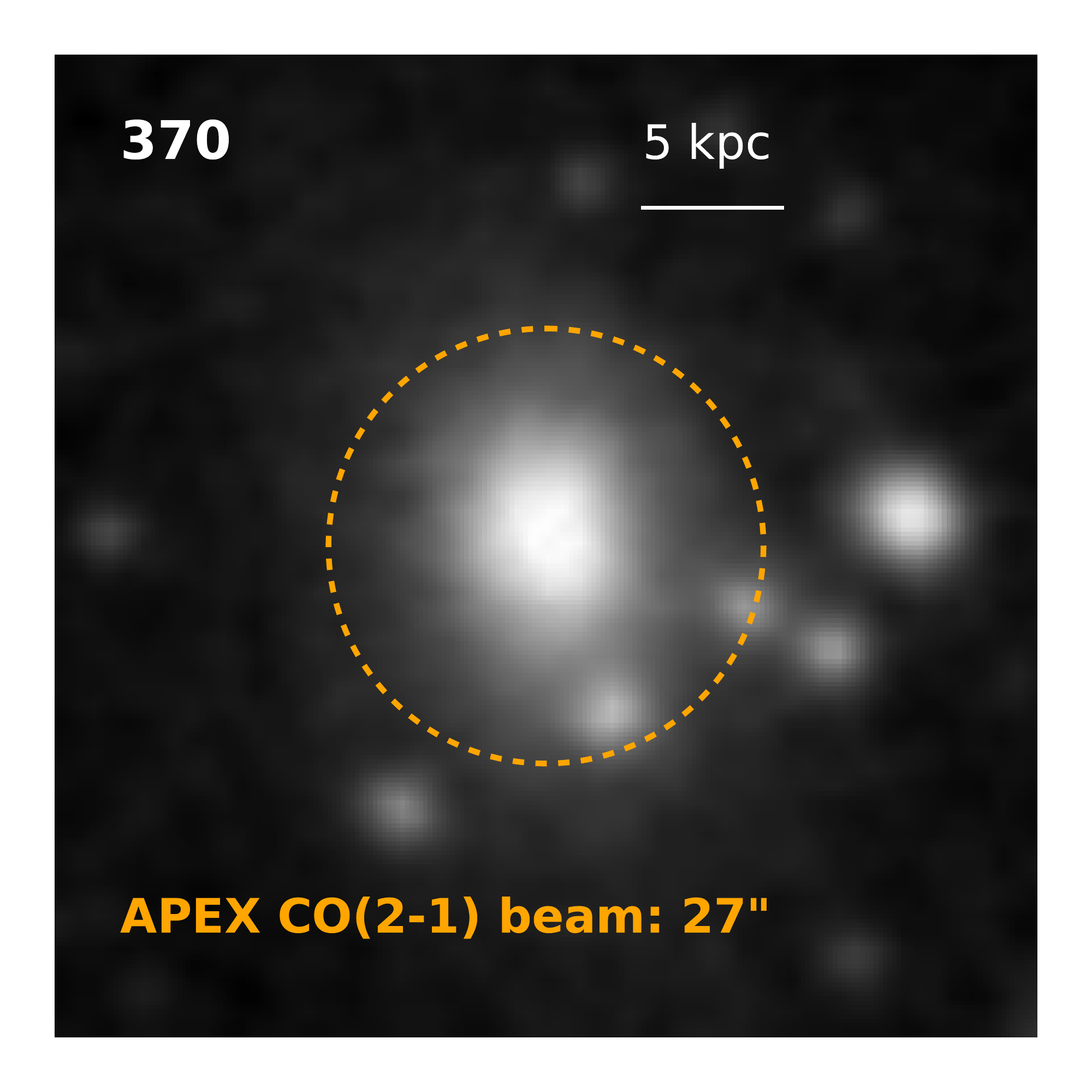}\includegraphics[width=0.26\textwidth]{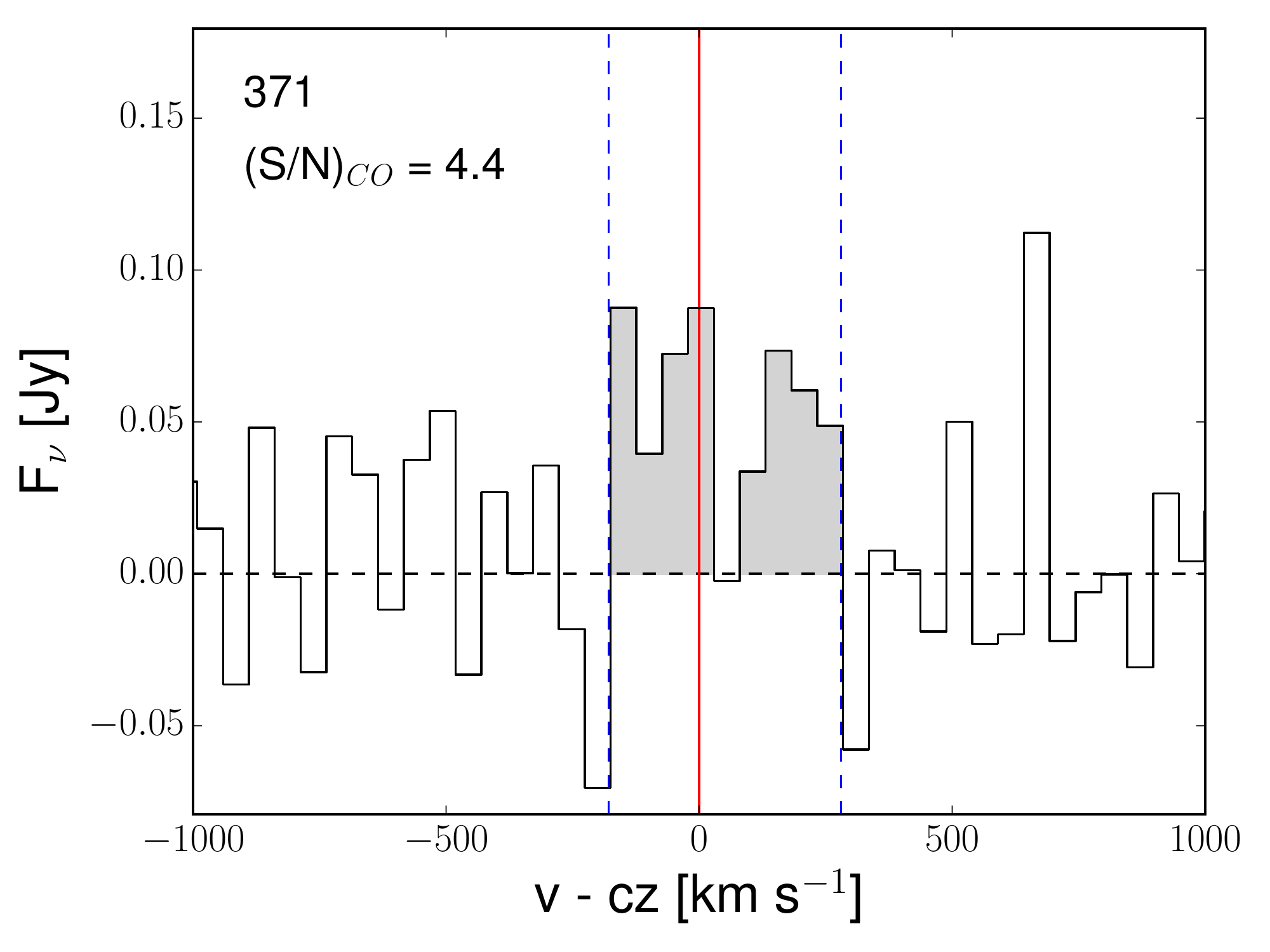}
\includegraphics[width=0.18\textwidth]{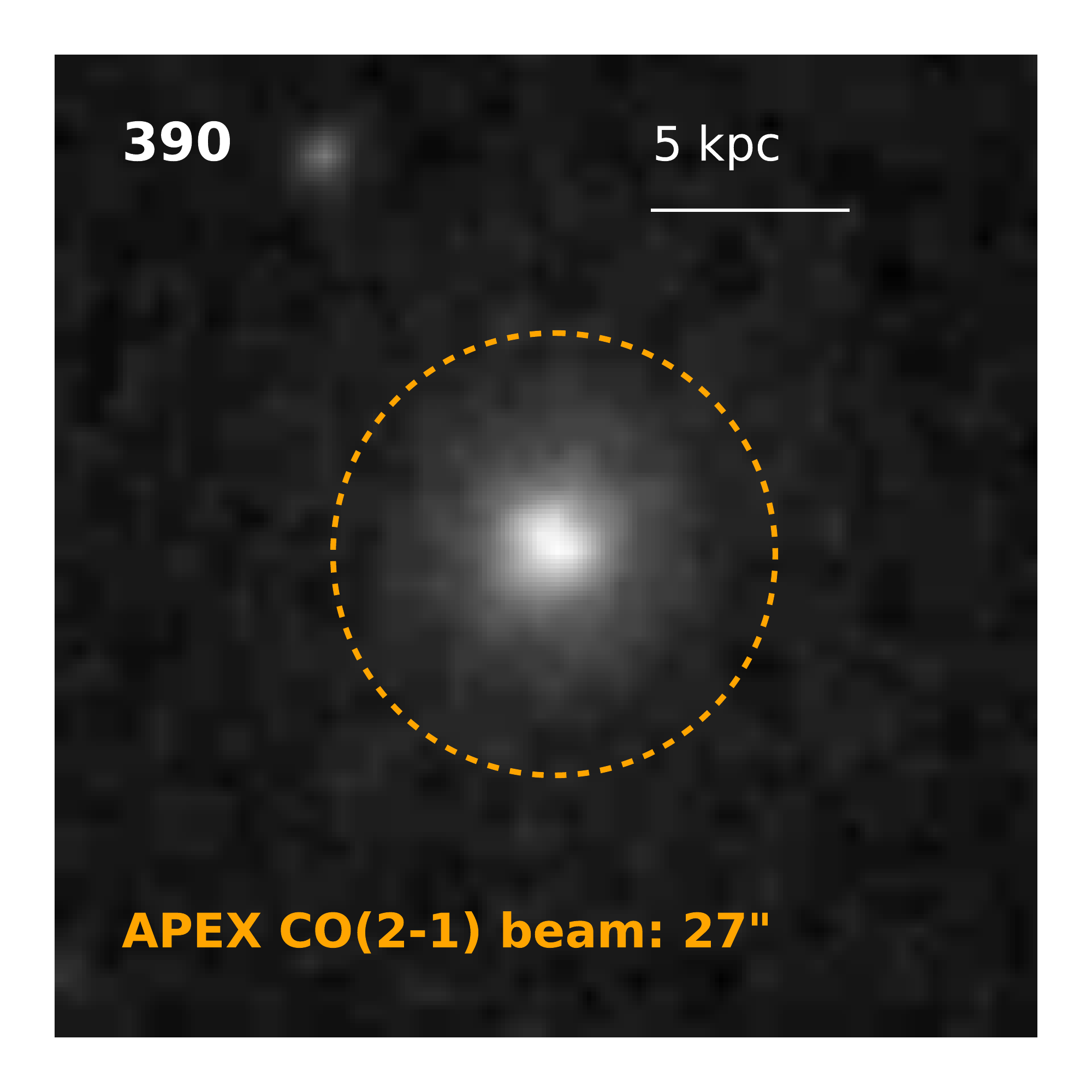}\includegraphics[width=0.26\textwidth]{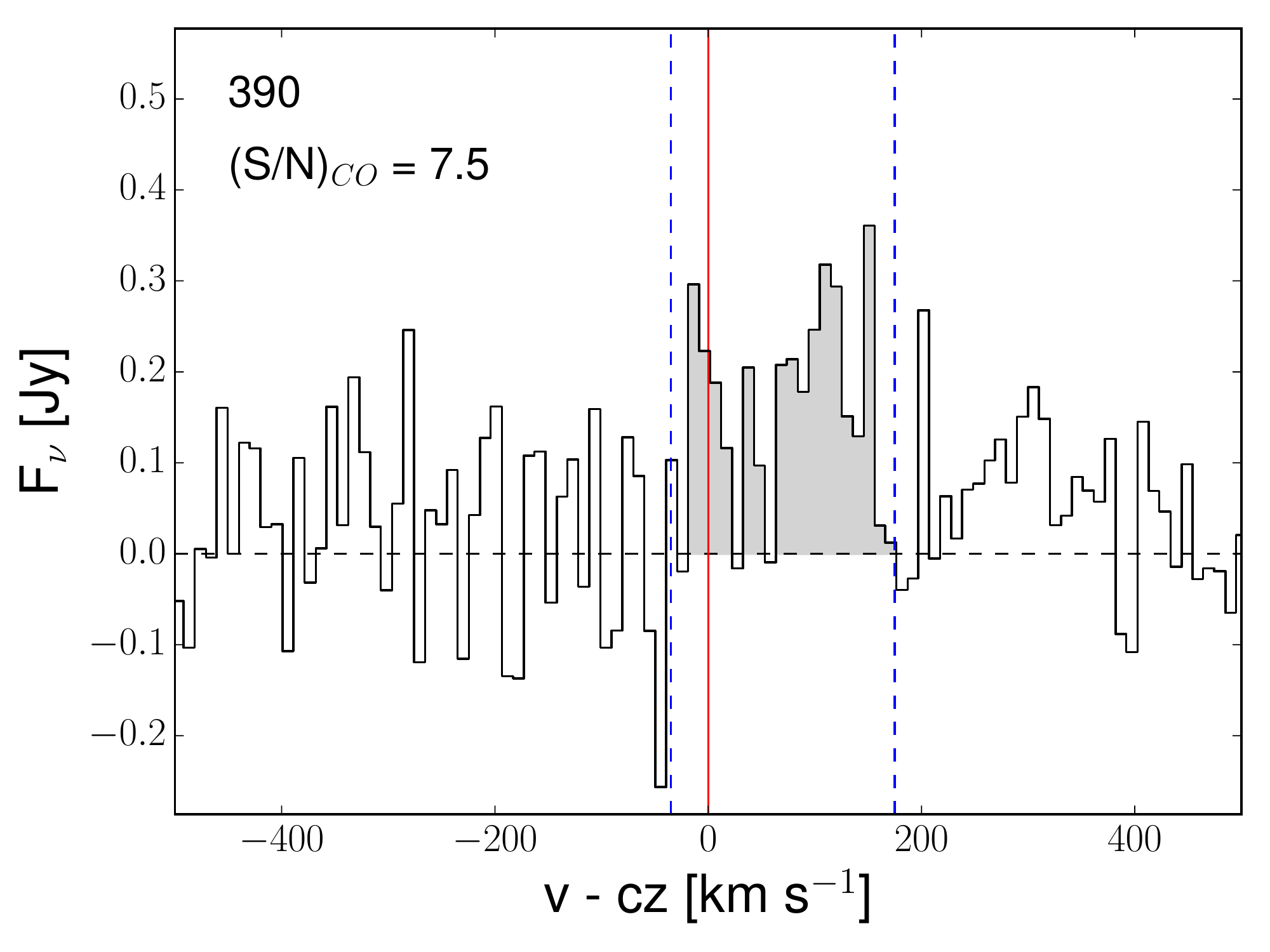}
\includegraphics[width=0.18\textwidth]{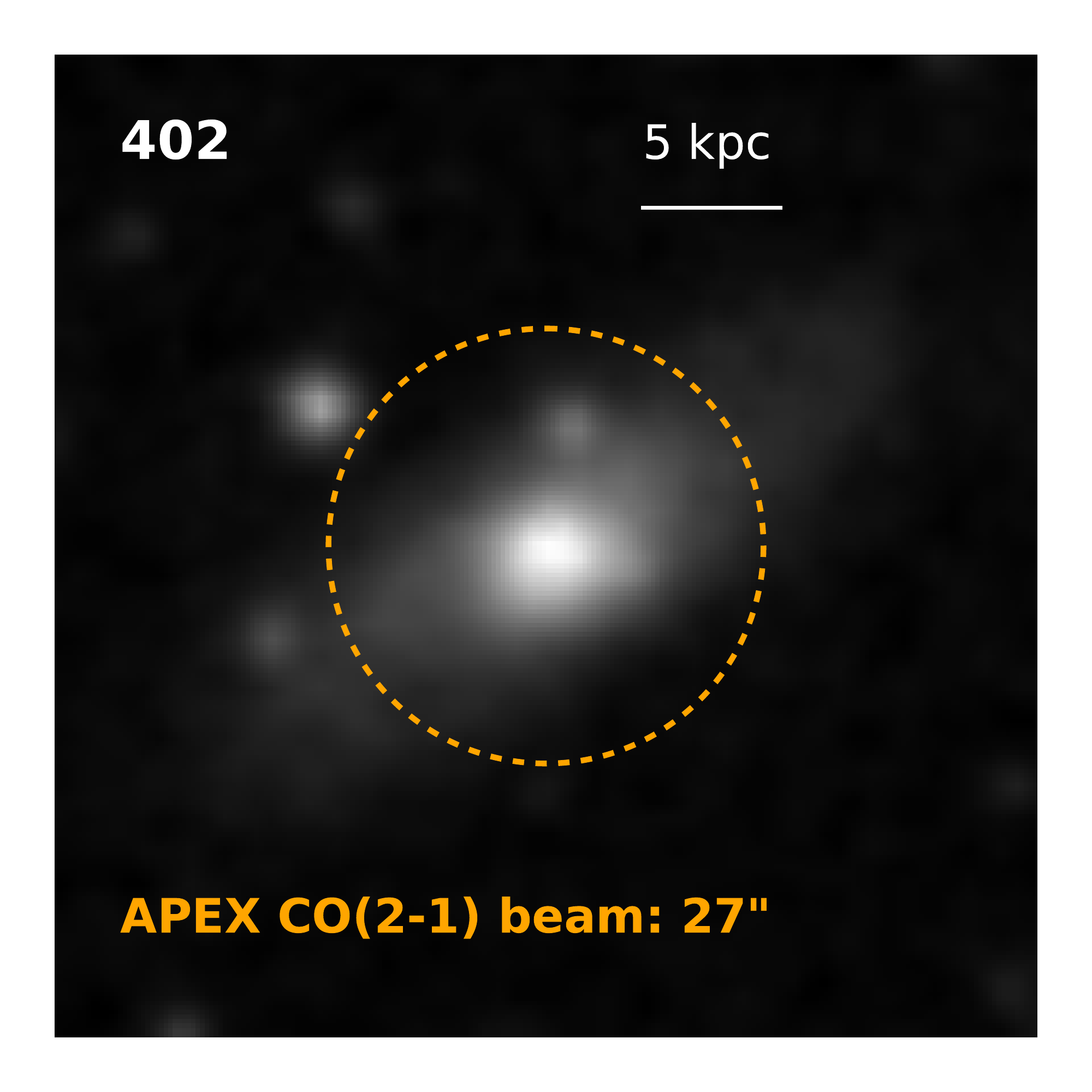}\includegraphics[width=0.26\textwidth]{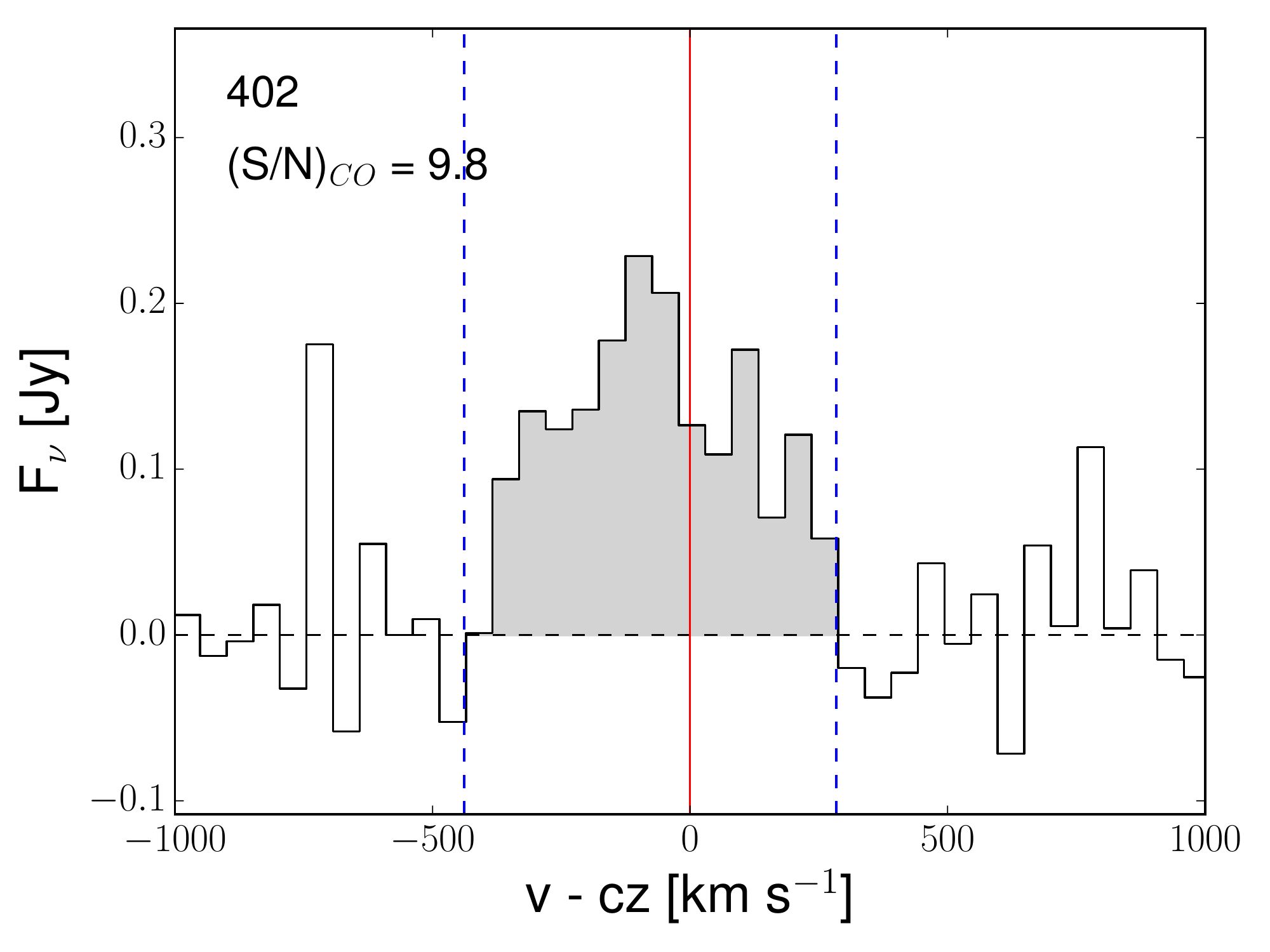}
\includegraphics[width=0.18\textwidth]{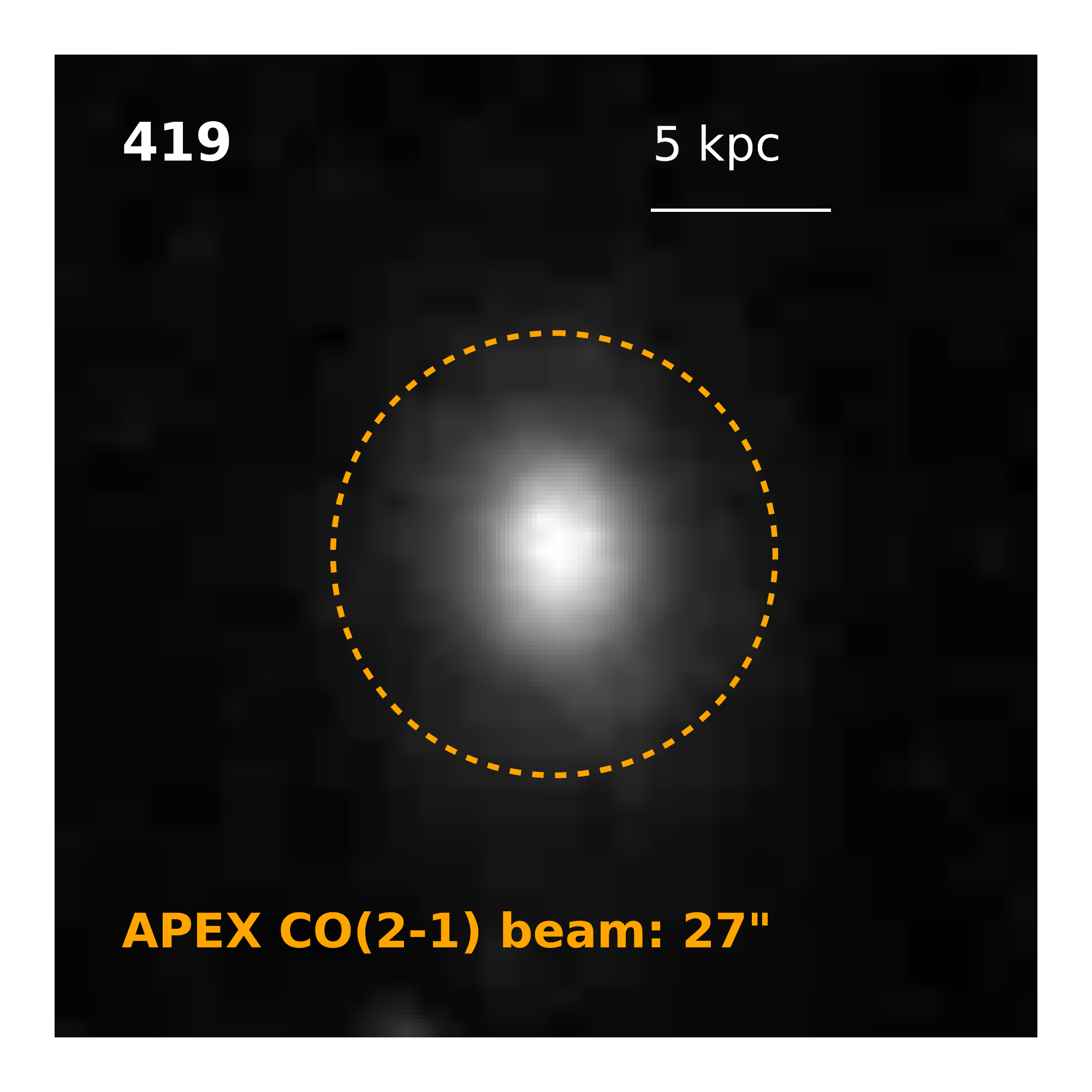}\includegraphics[width=0.26\textwidth]{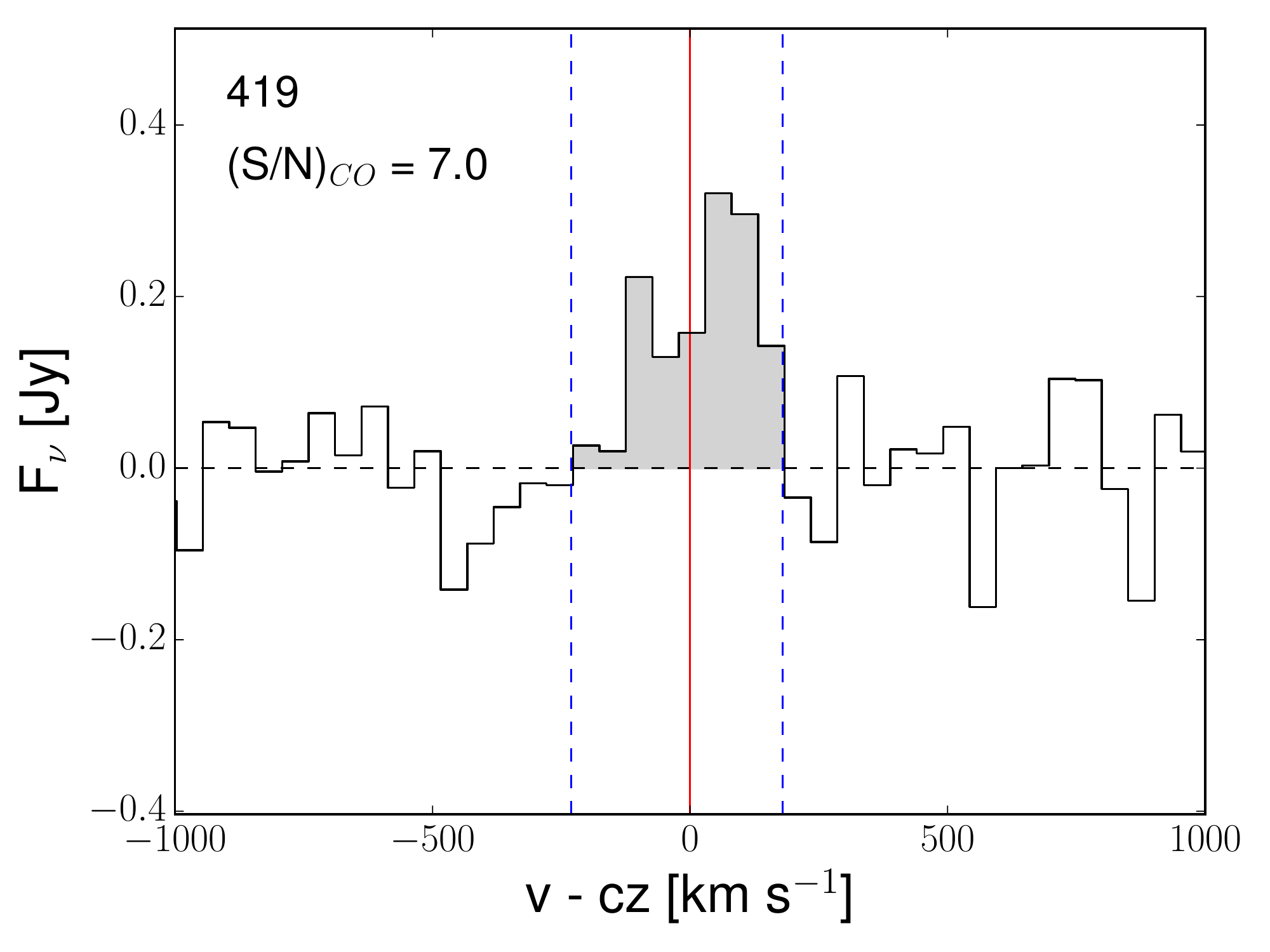}
\includegraphics[width=0.18\textwidth]{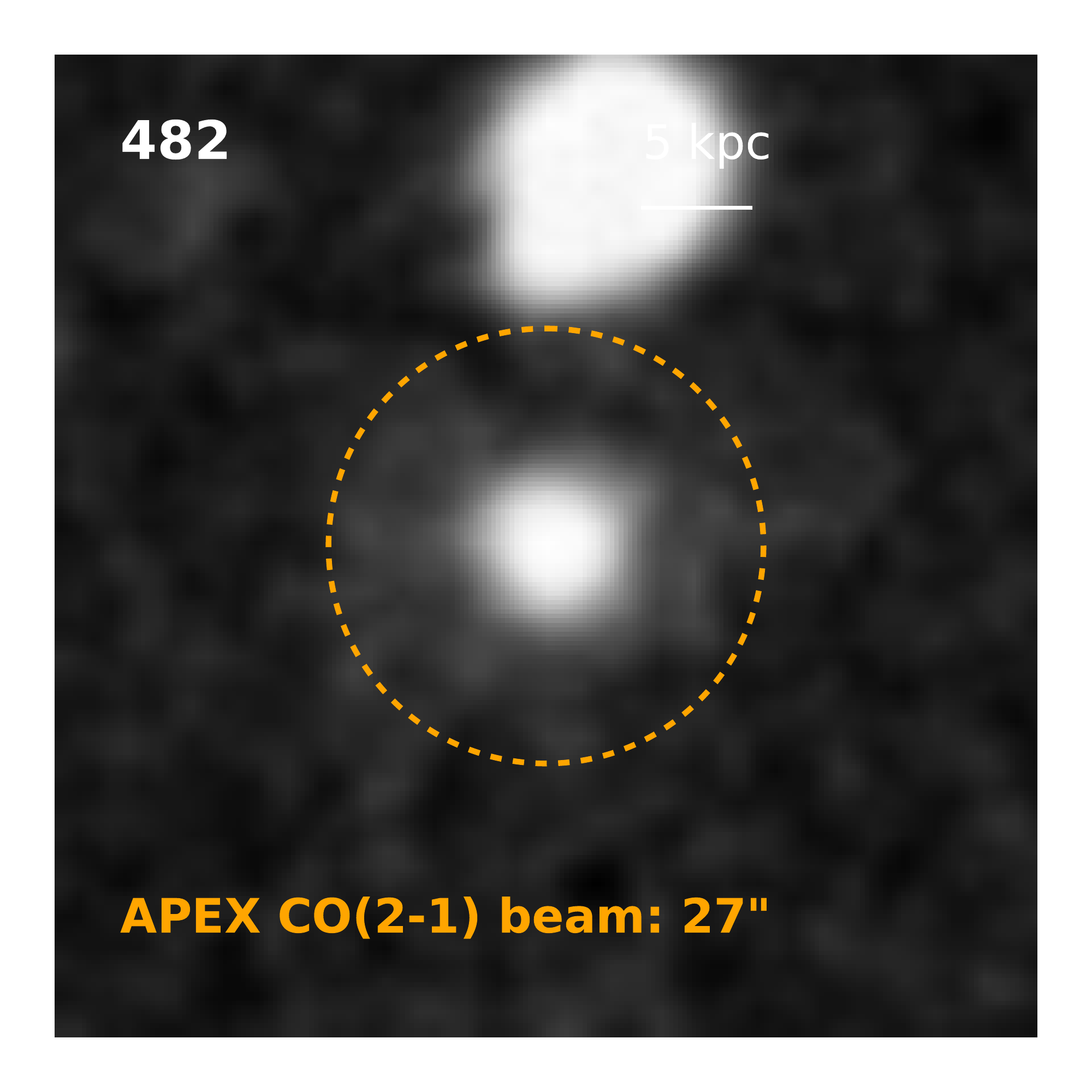}\includegraphics[width=0.26\textwidth]{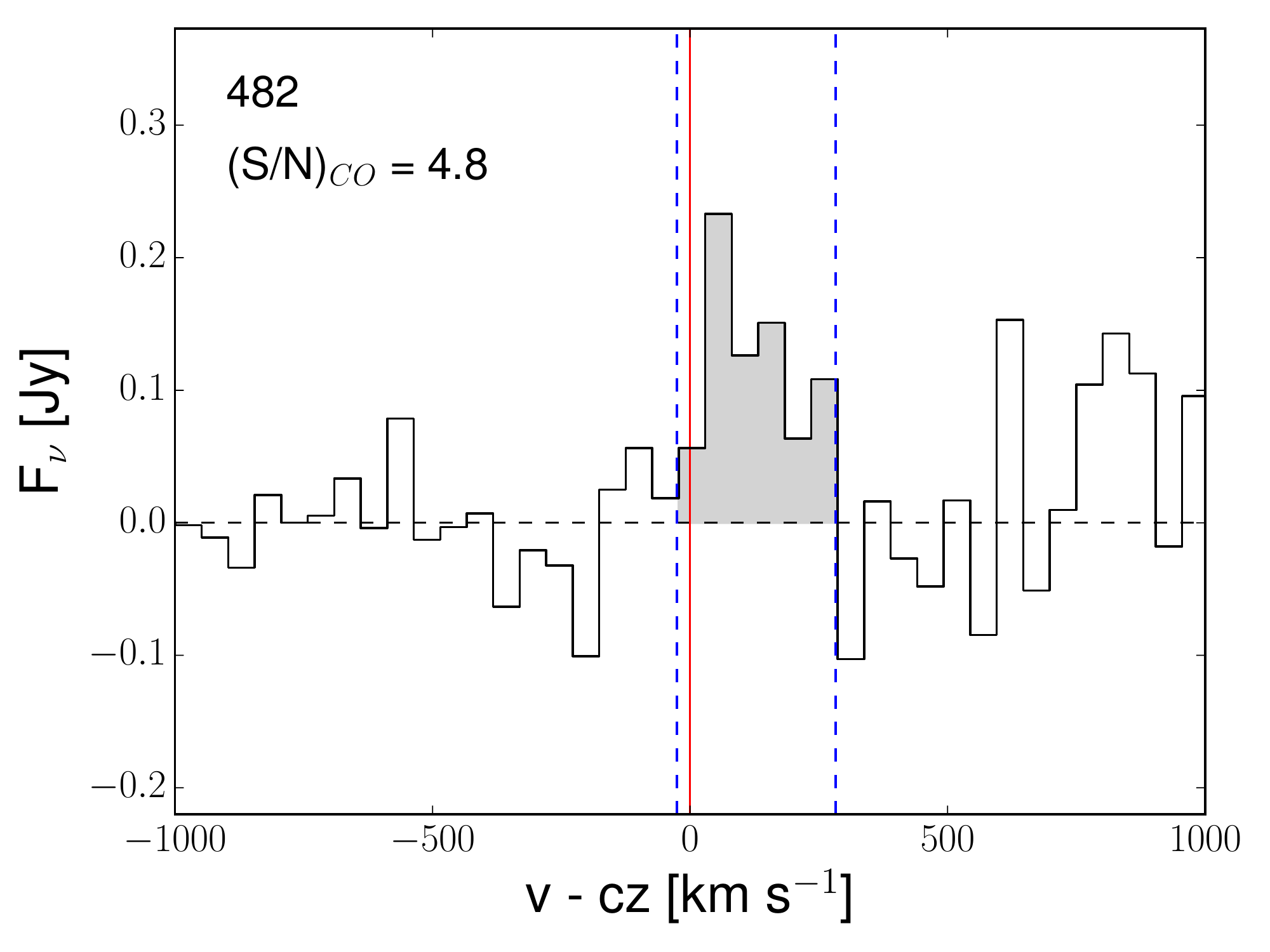}
\includegraphics[width=0.18\textwidth]{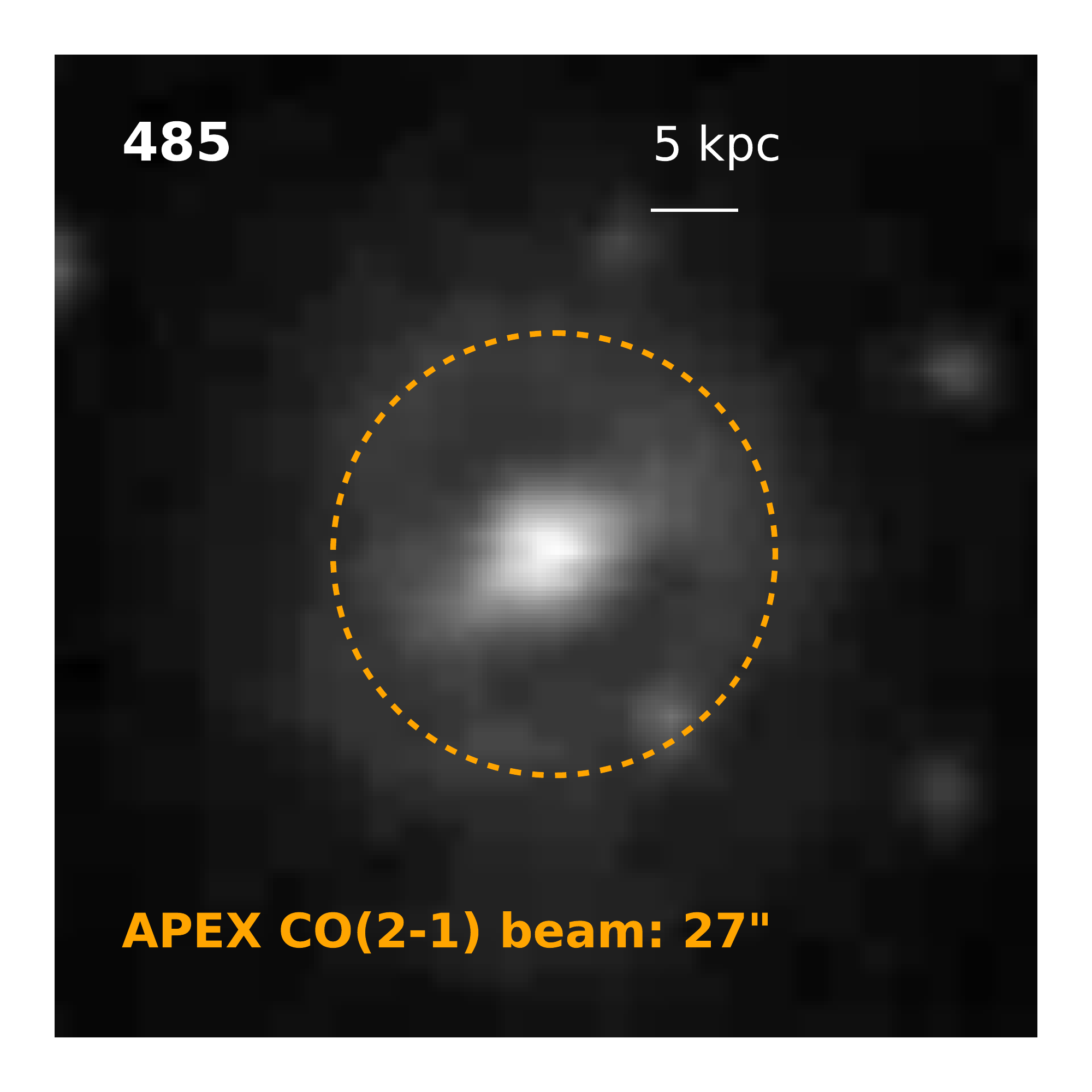}\includegraphics[width=0.26\textwidth]{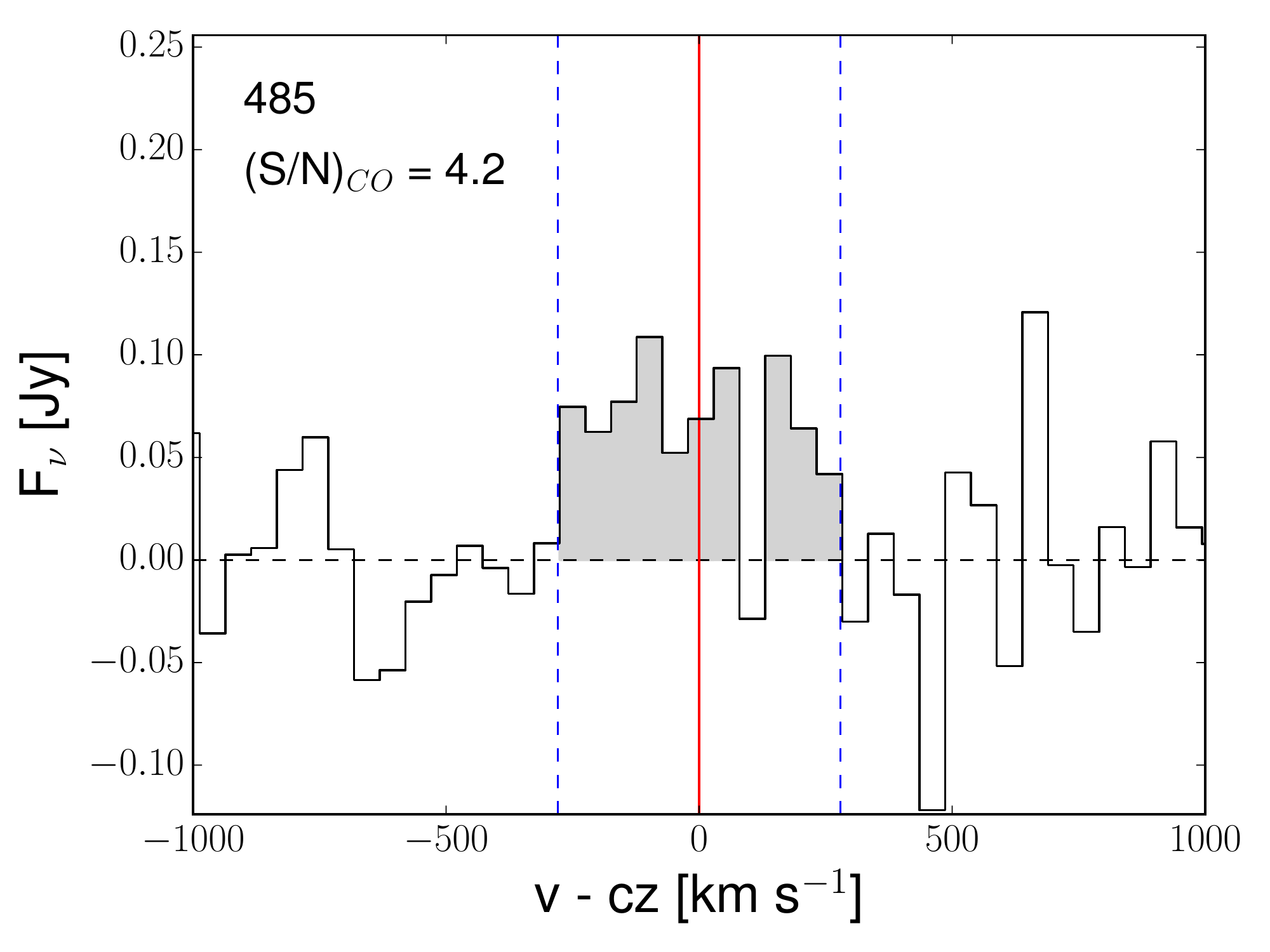}
\includegraphics[width=0.18\textwidth]{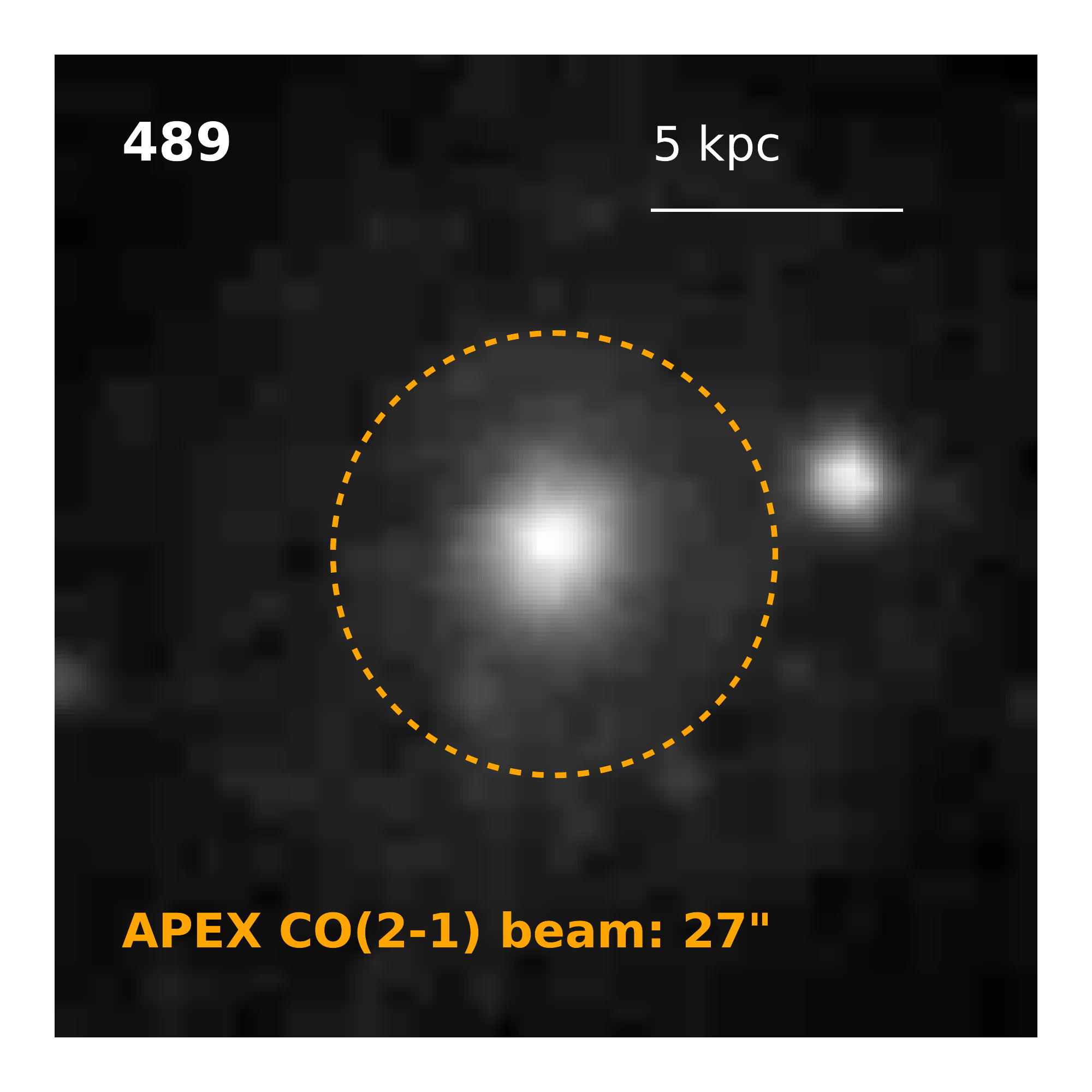}\includegraphics[width=0.26\textwidth]{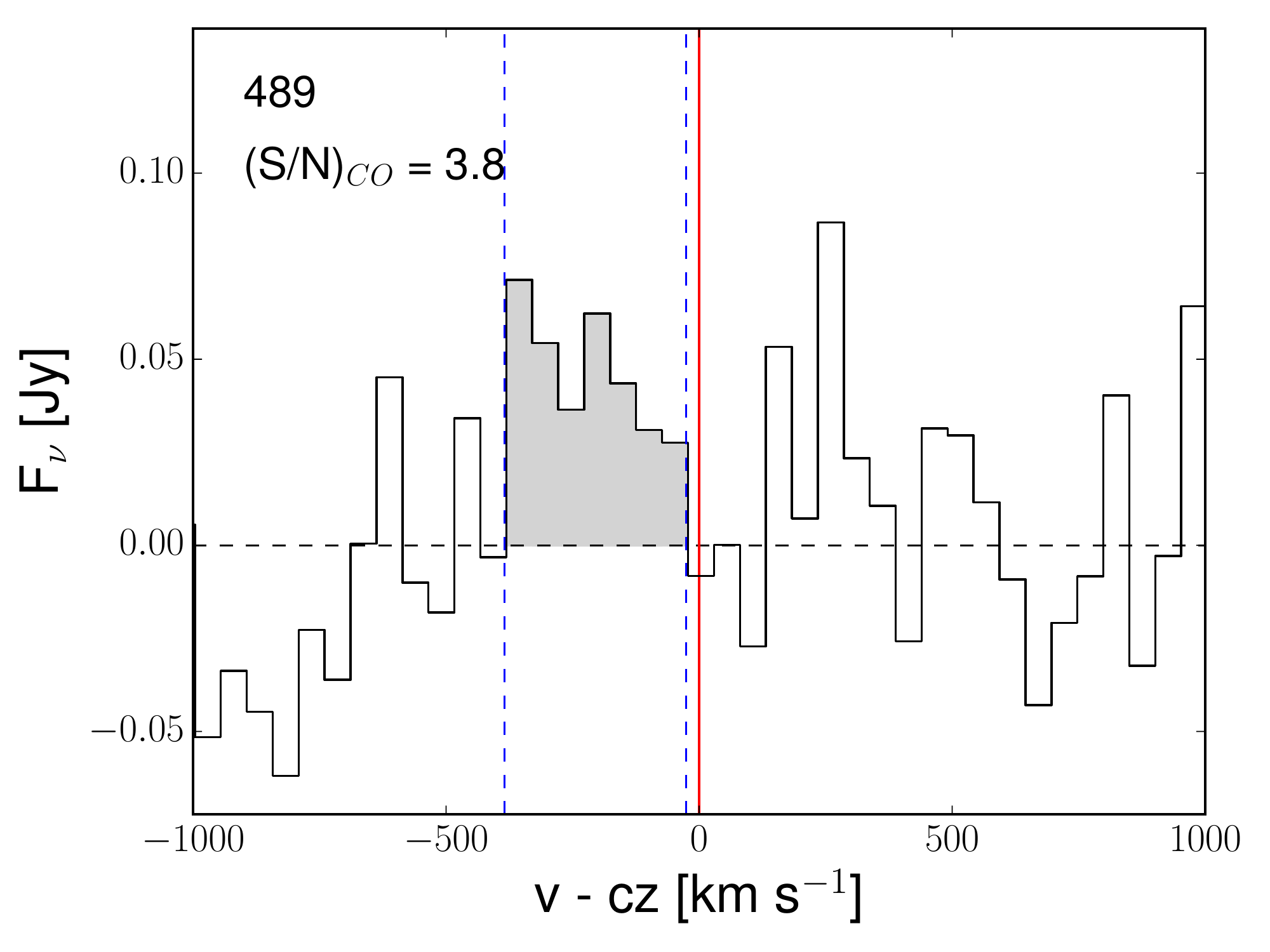}
\caption{continued from Fig.~\ref{fig:CO21_spectra_dss1}
} 
\label{fig:CO21_spectra_dss2}
\end{figure*}

\begin{figure*}
\centering
\raggedright
\includegraphics[width=0.18\textwidth]{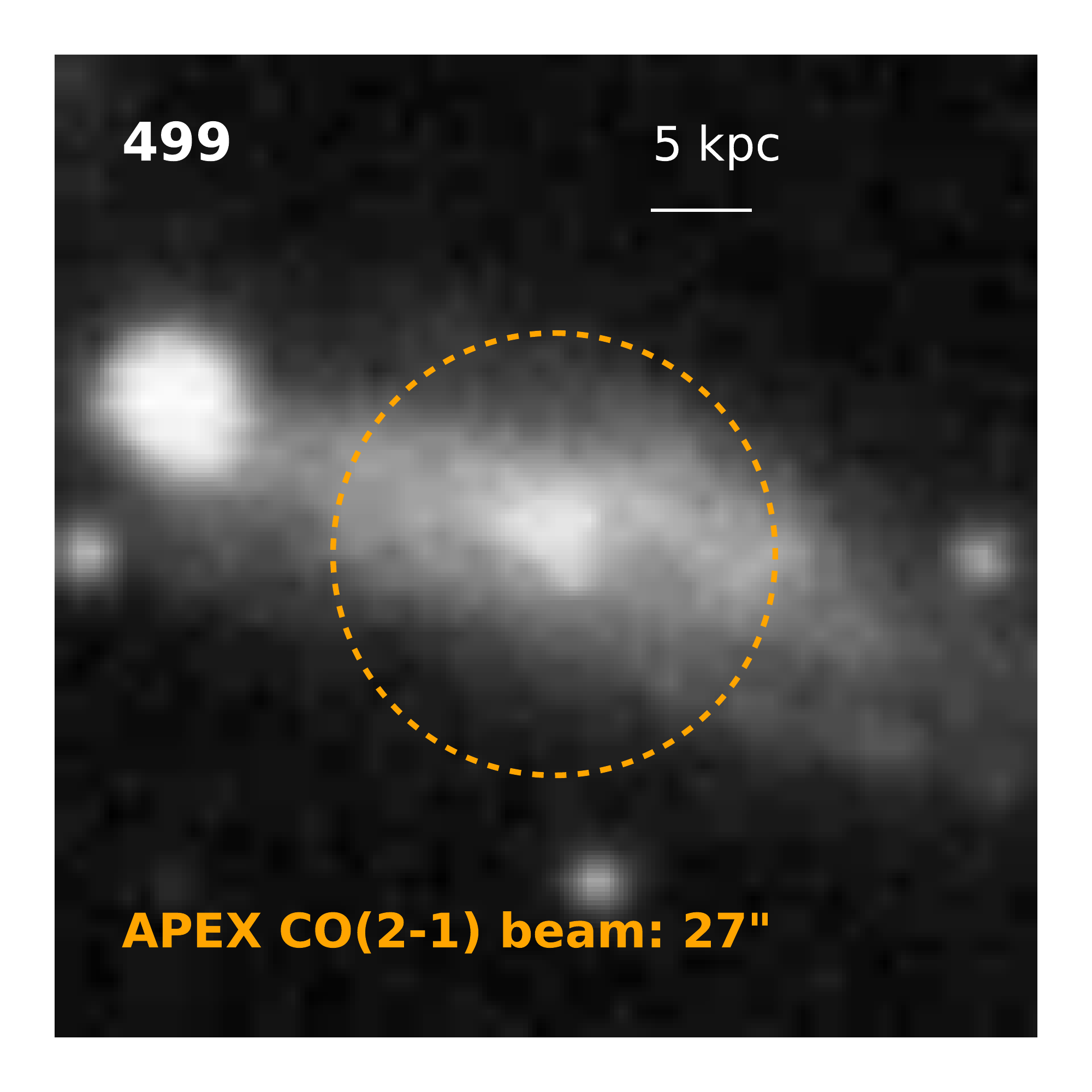}\includegraphics[width=0.26\textwidth]{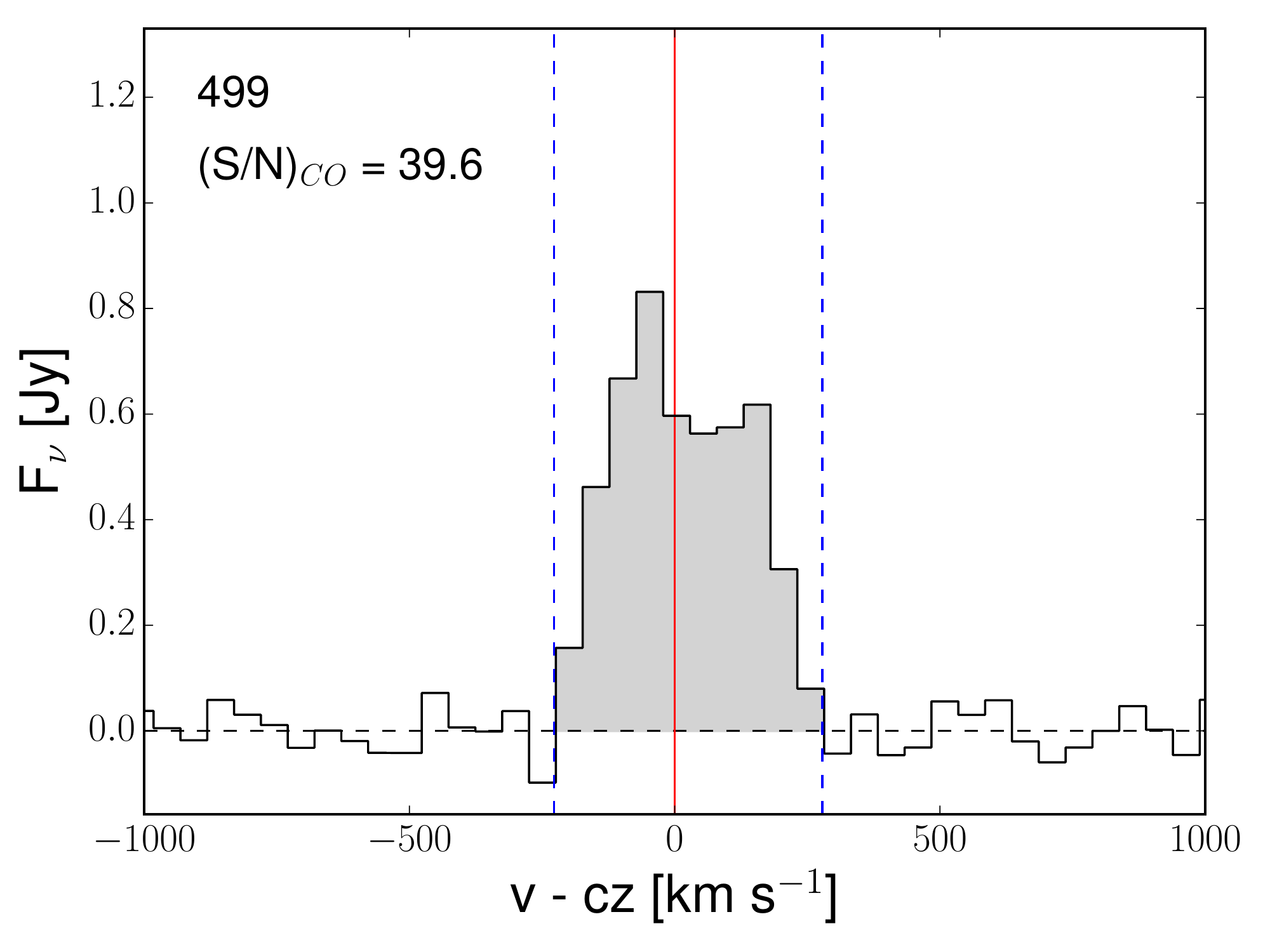}
\includegraphics[width=0.18\textwidth]{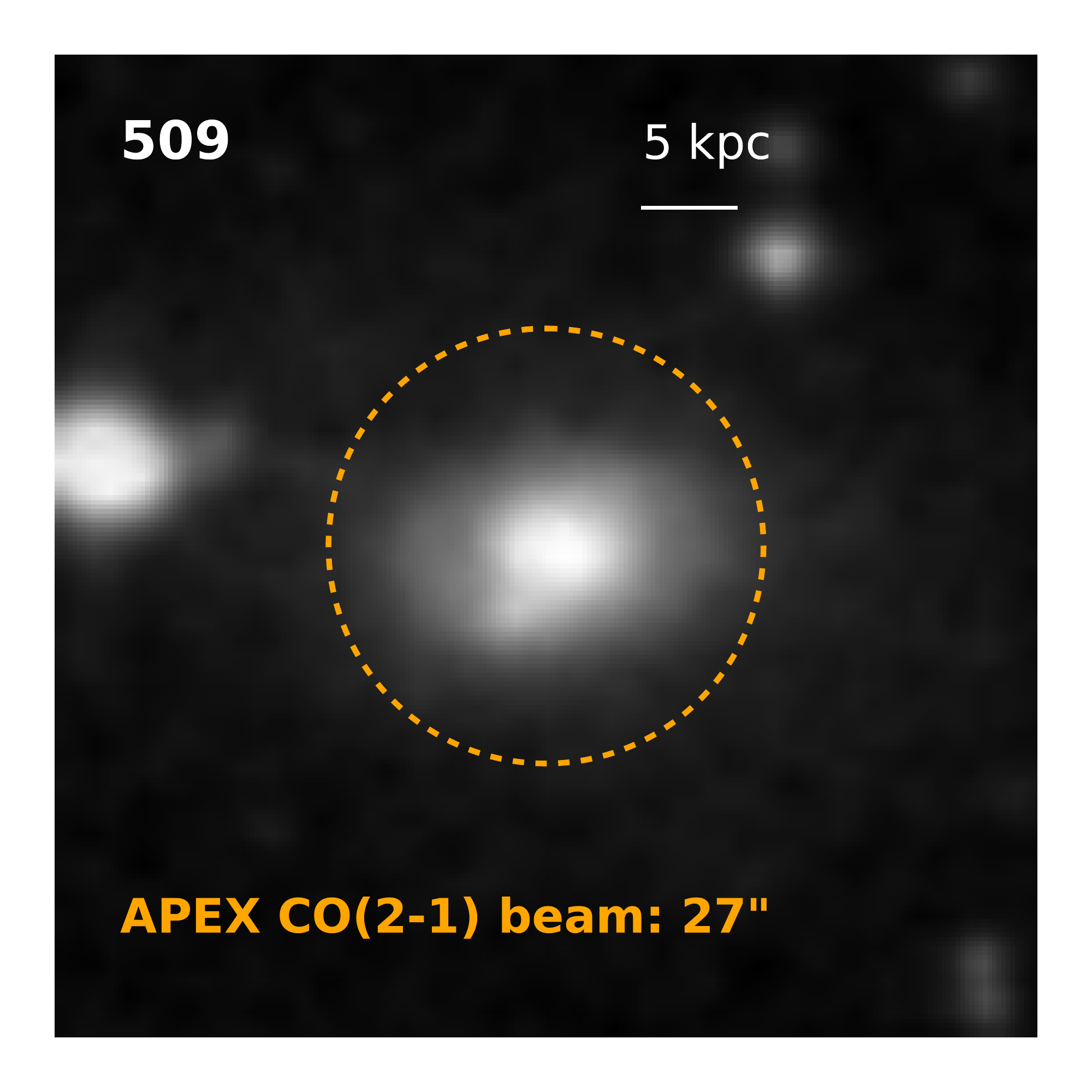}\includegraphics[width=0.26\textwidth]{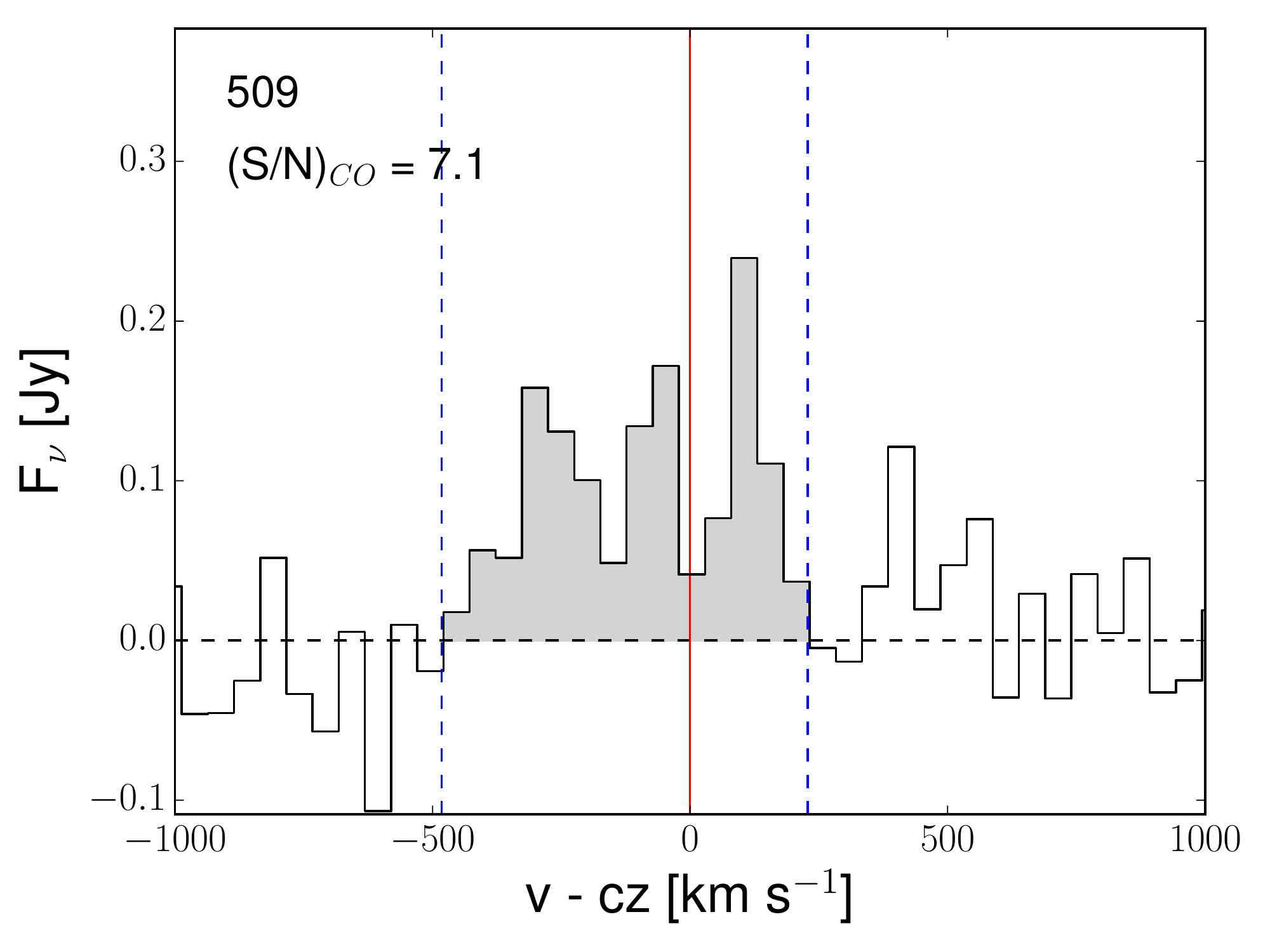}
\includegraphics[width=0.18\textwidth]{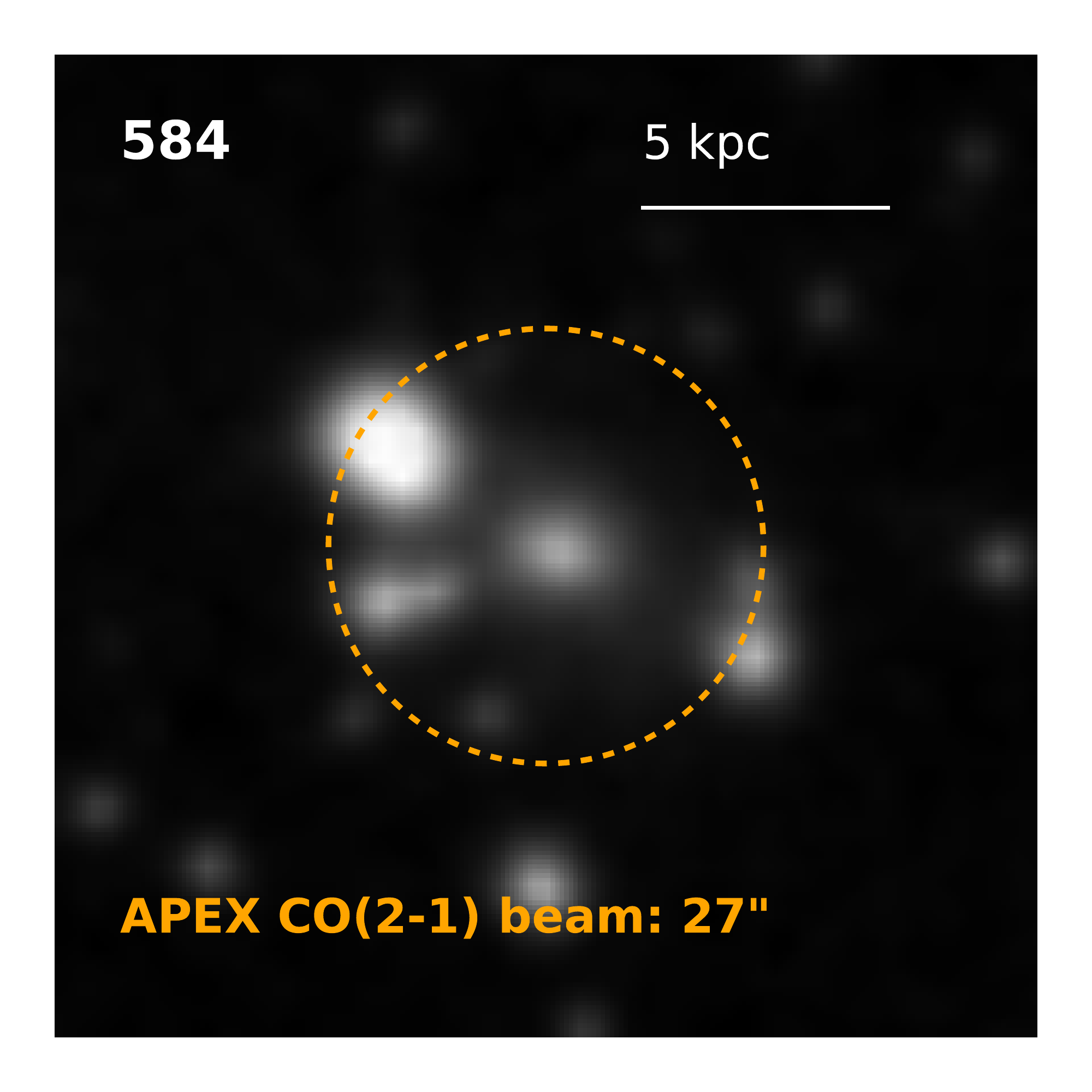}\includegraphics[width=0.26\textwidth]{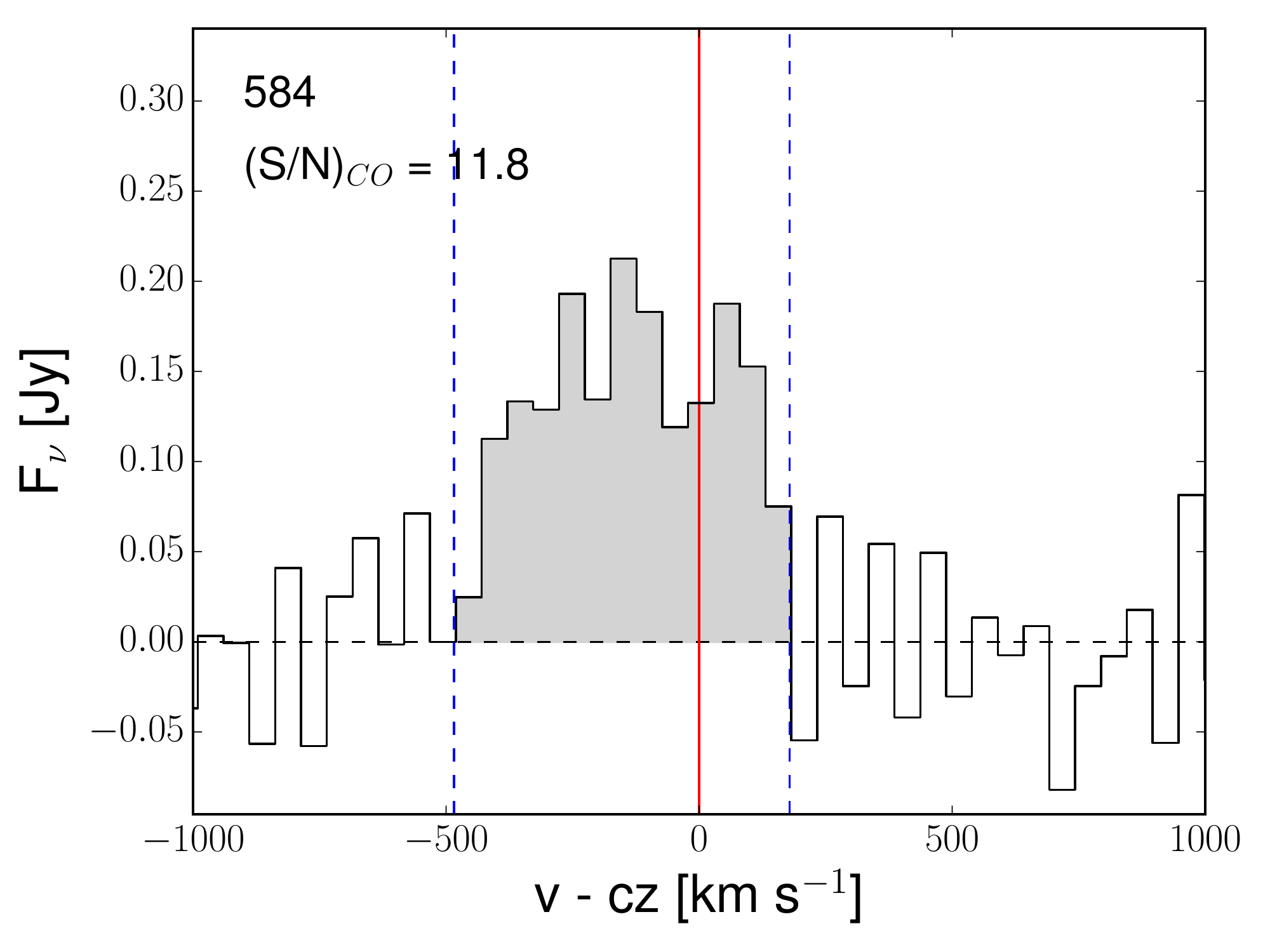}
\includegraphics[width=0.18\textwidth]{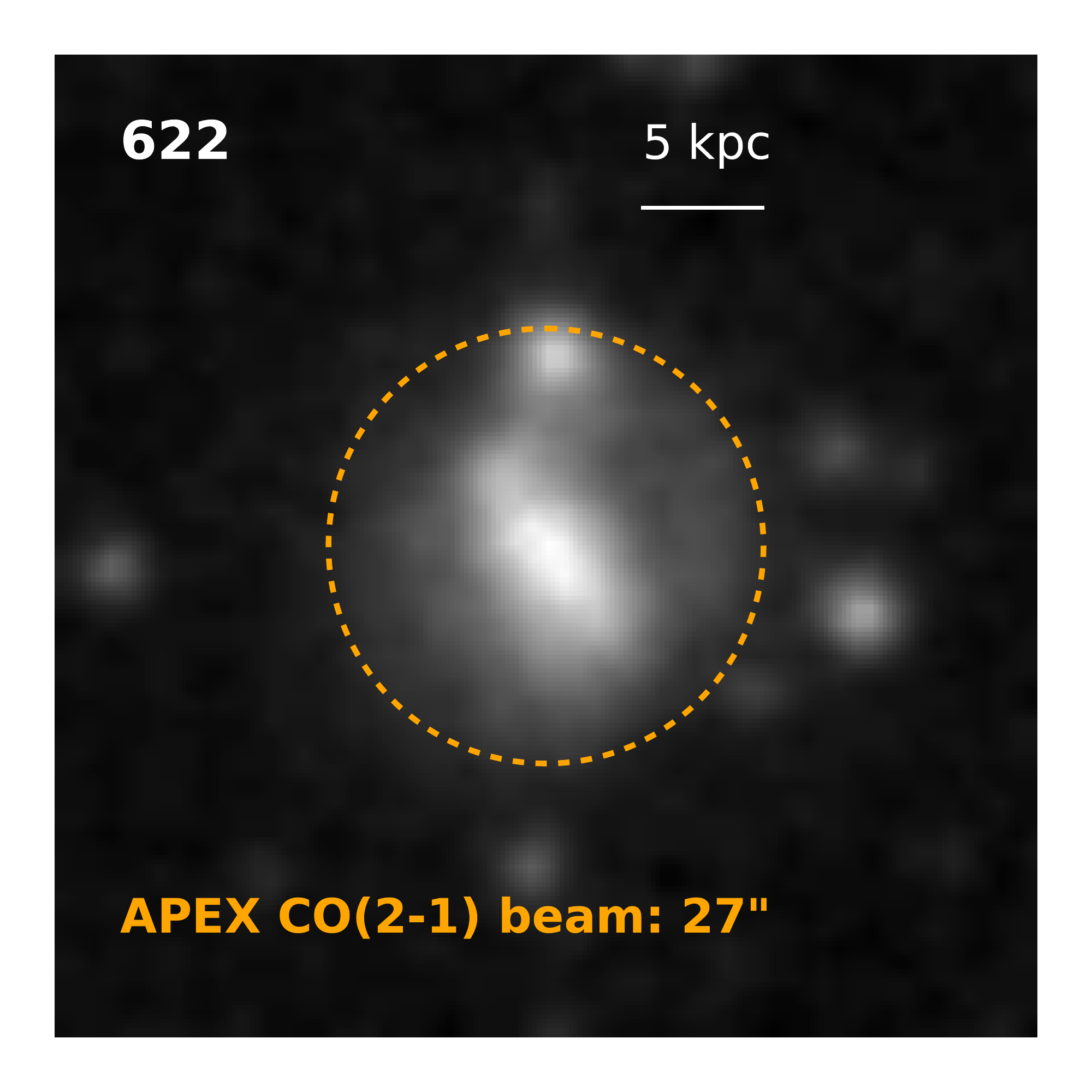}\includegraphics[width=0.26\textwidth]{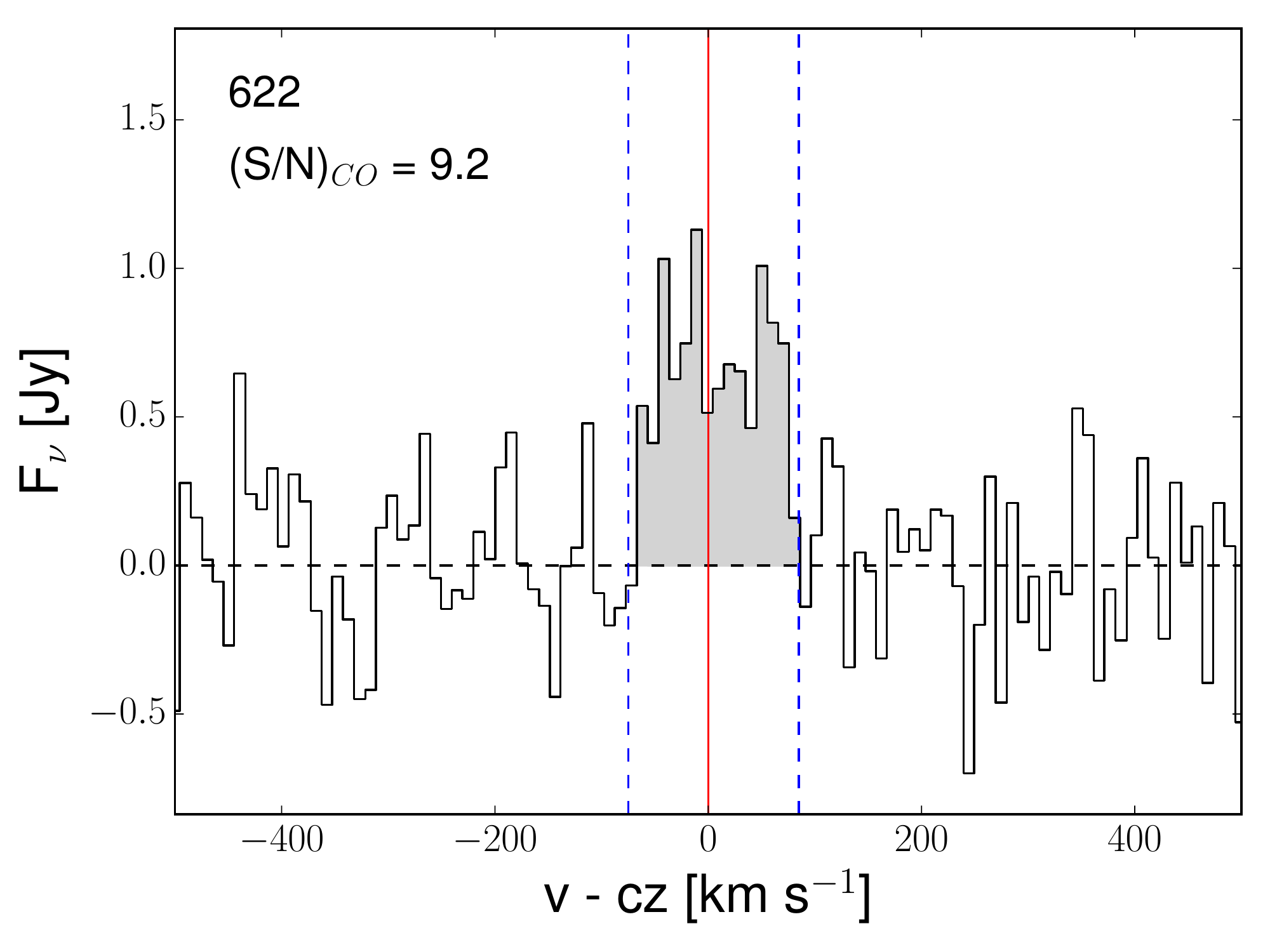}
\includegraphics[width=0.18\textwidth]{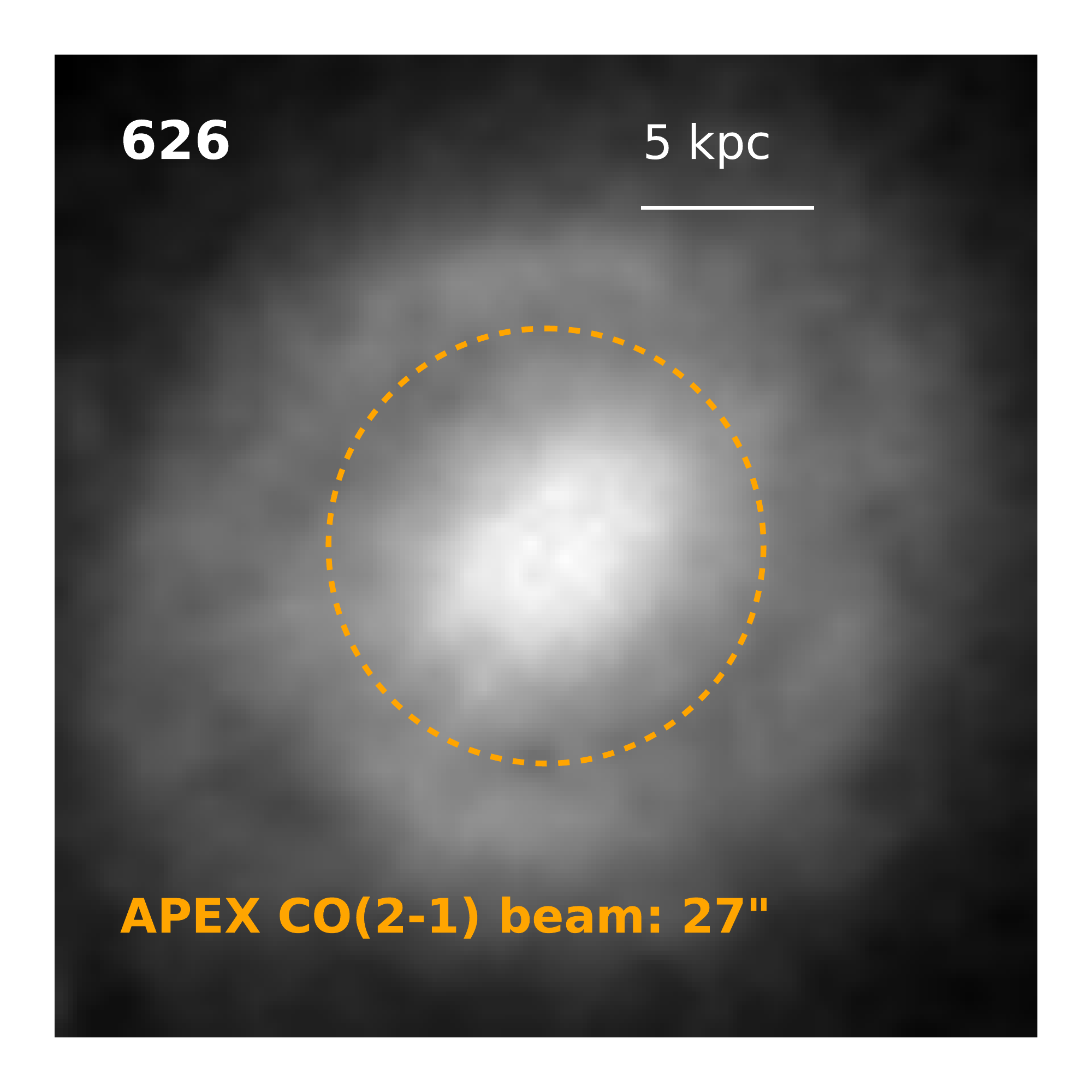}\includegraphics[width=0.26\textwidth]{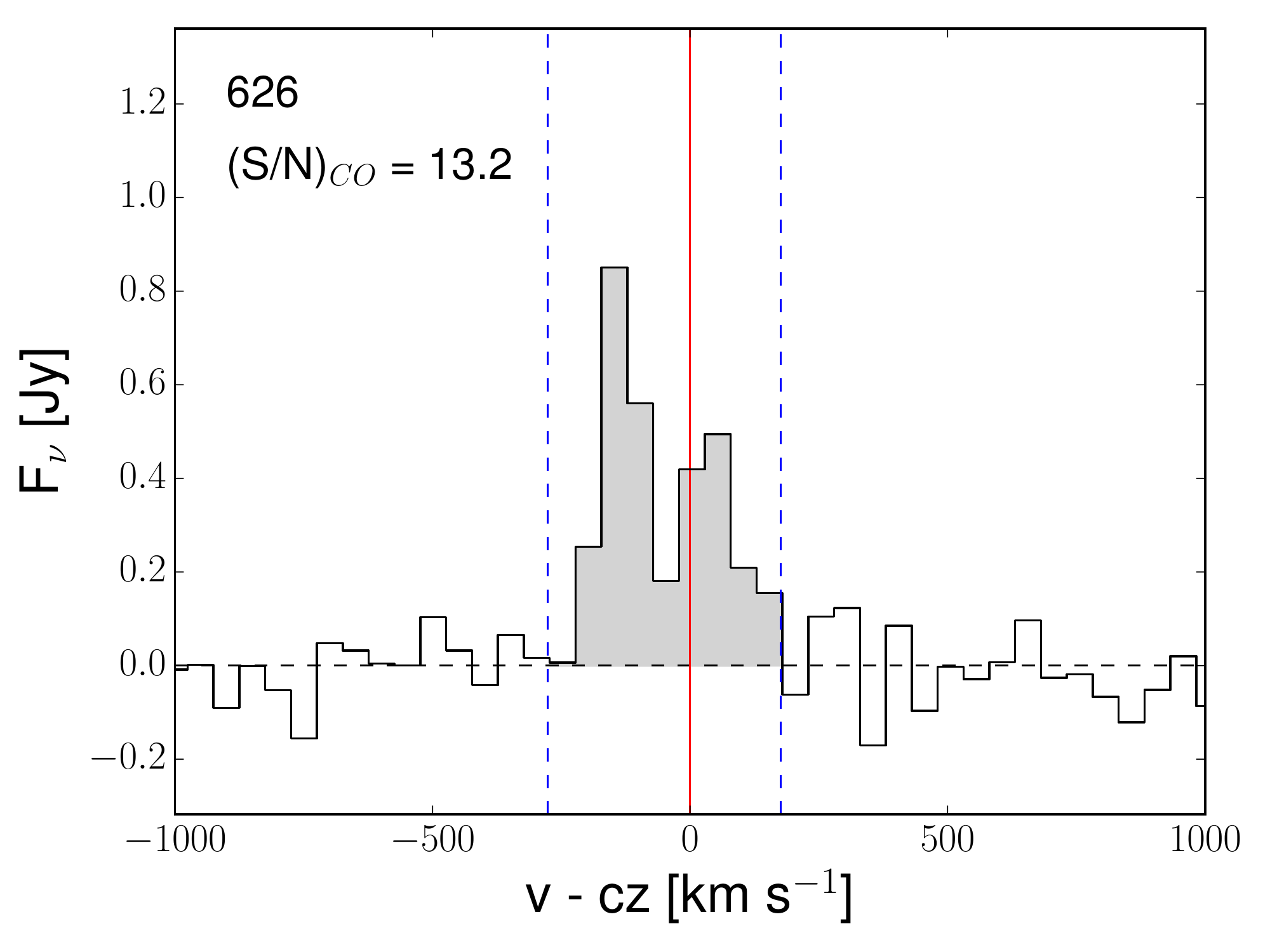}
\includegraphics[width=0.18\textwidth]{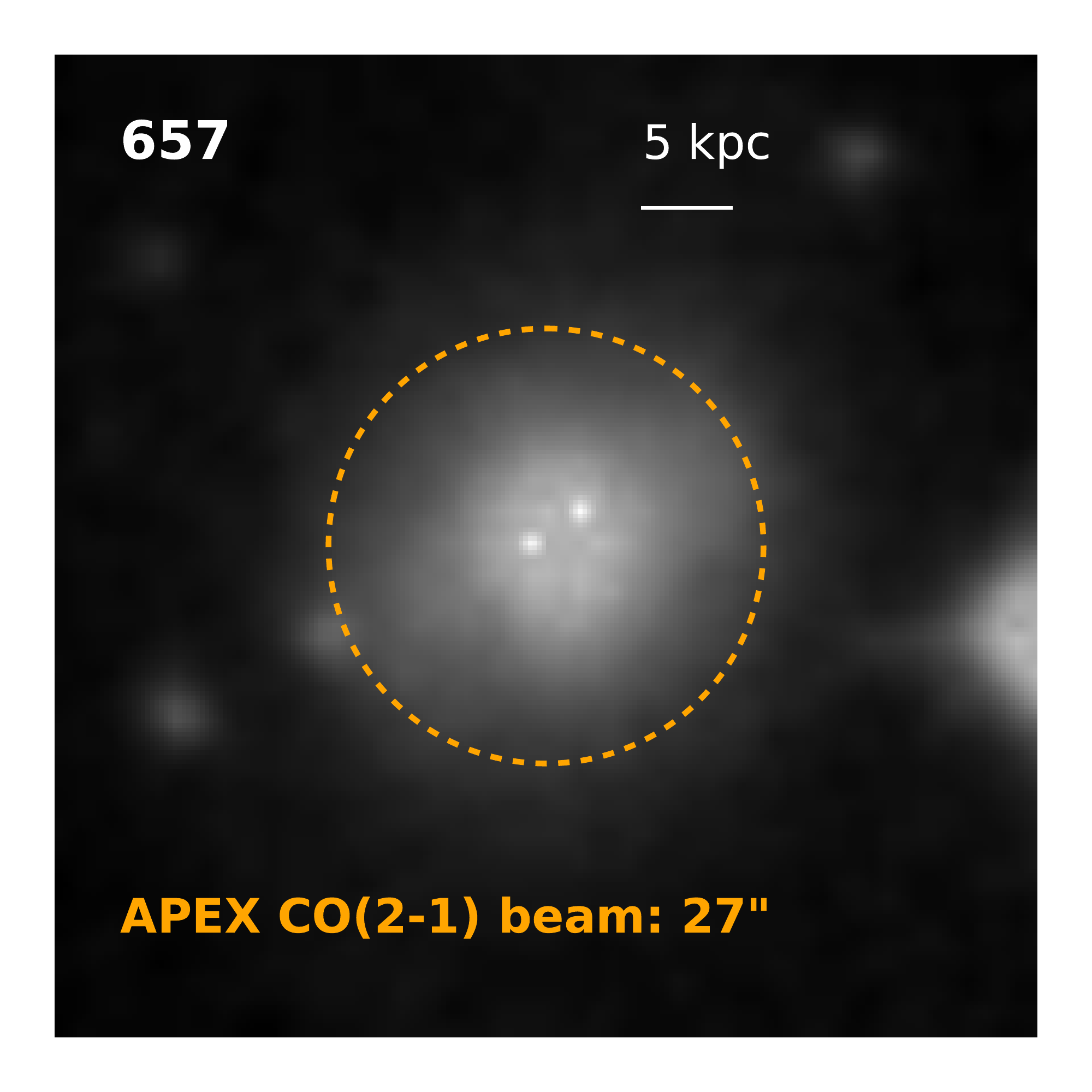}\includegraphics[width=0.26\textwidth]{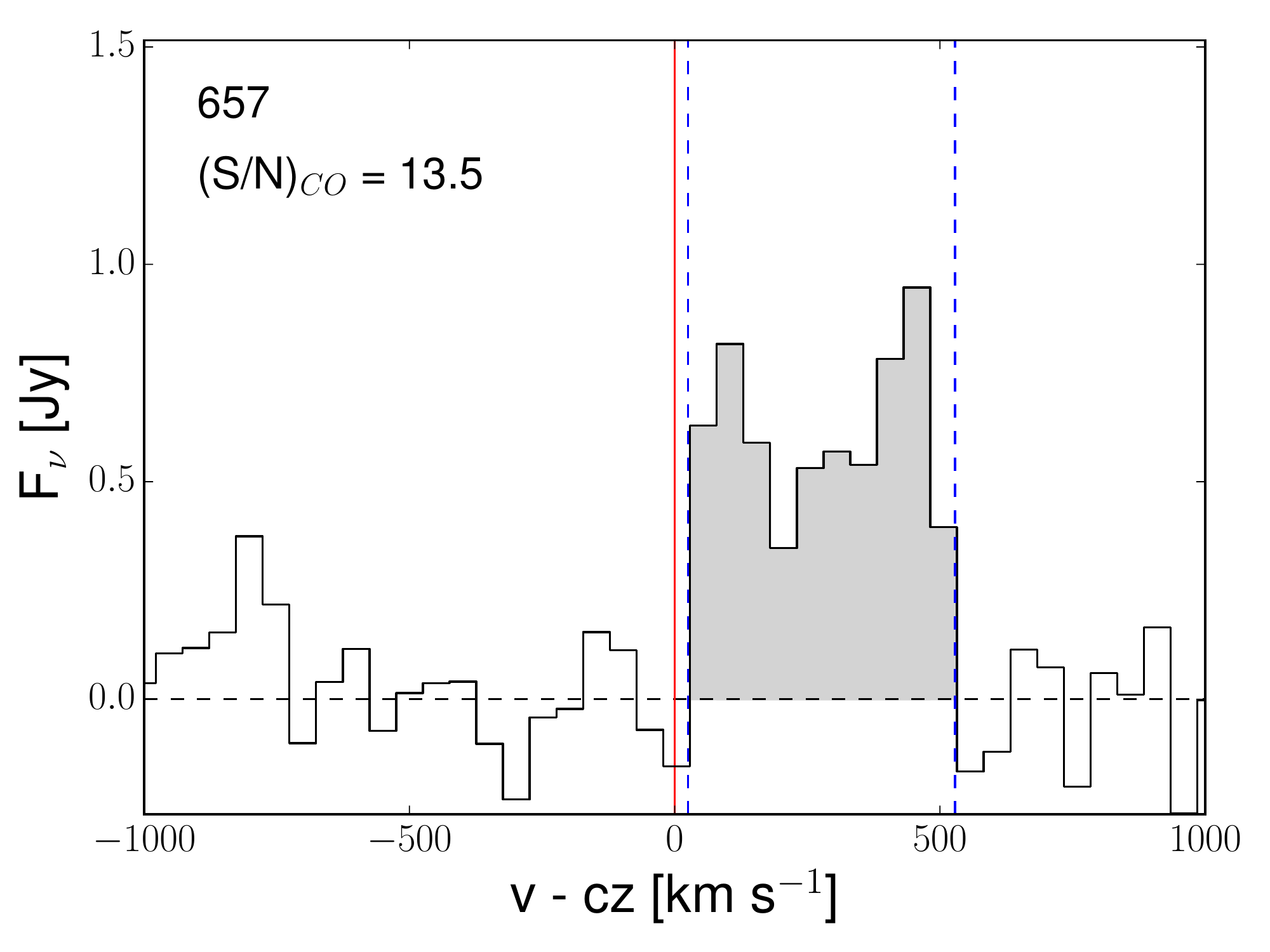}
\includegraphics[width=0.18\textwidth]{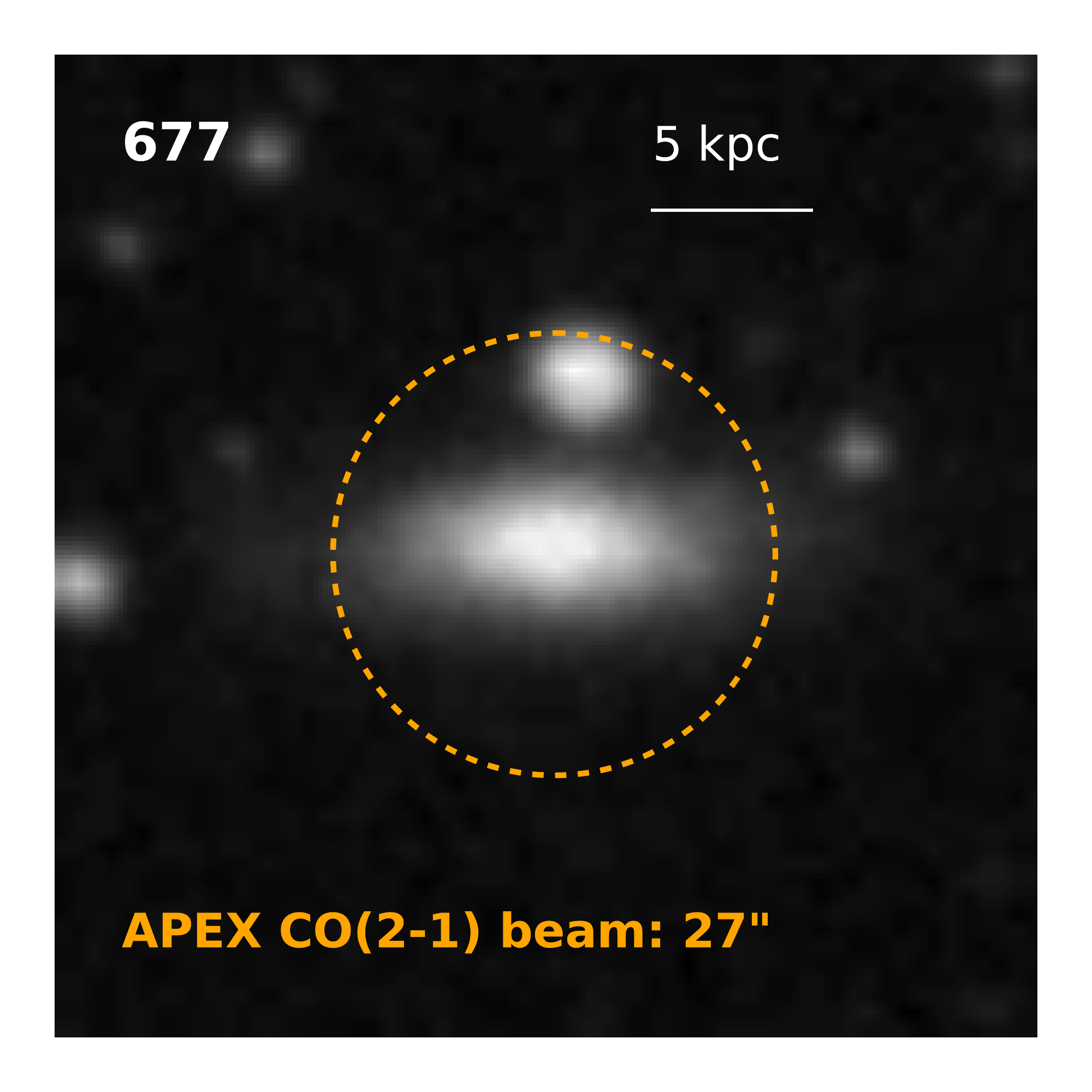}\includegraphics[width=0.26\textwidth]{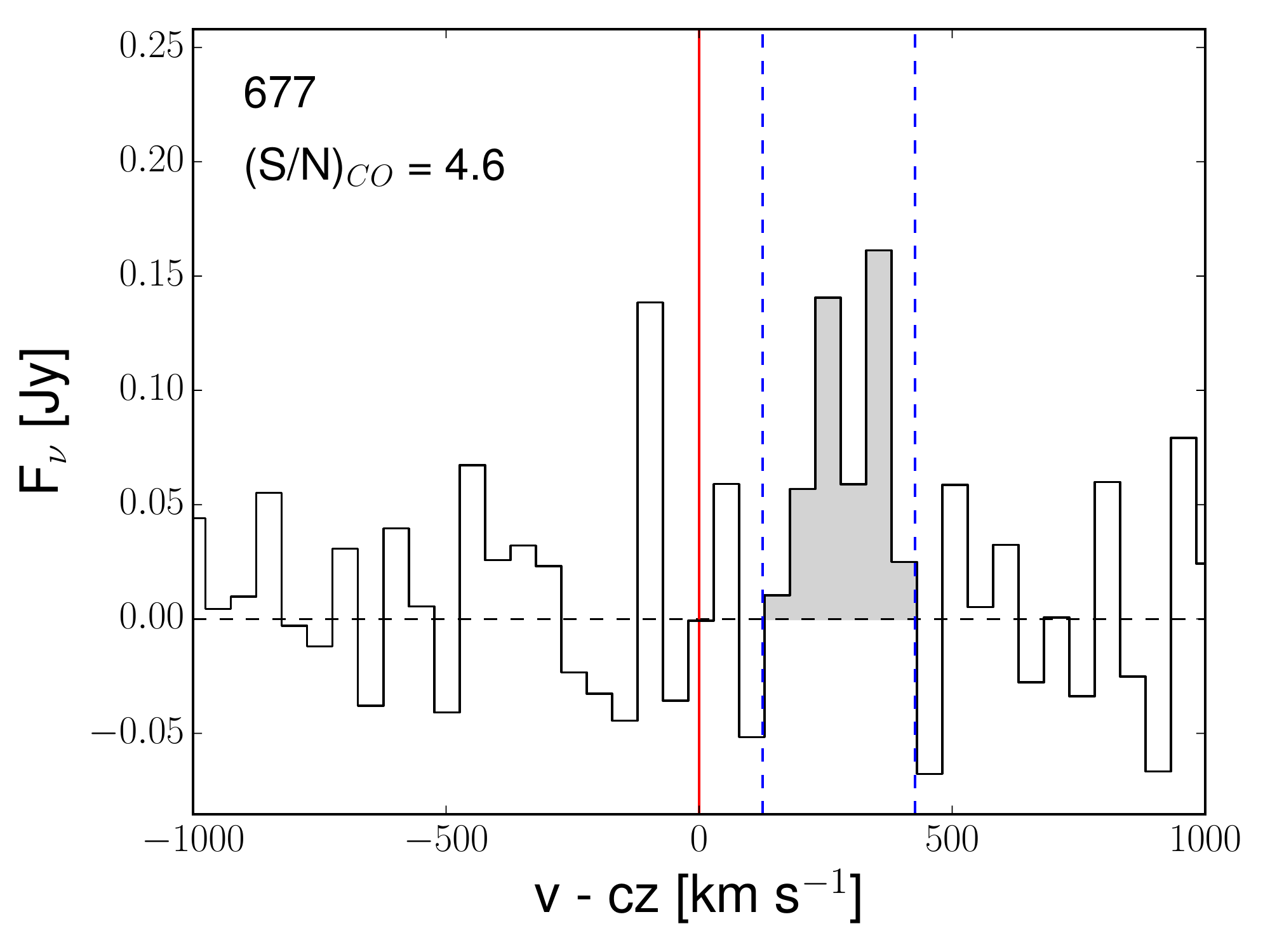}
\includegraphics[width=0.18\textwidth]{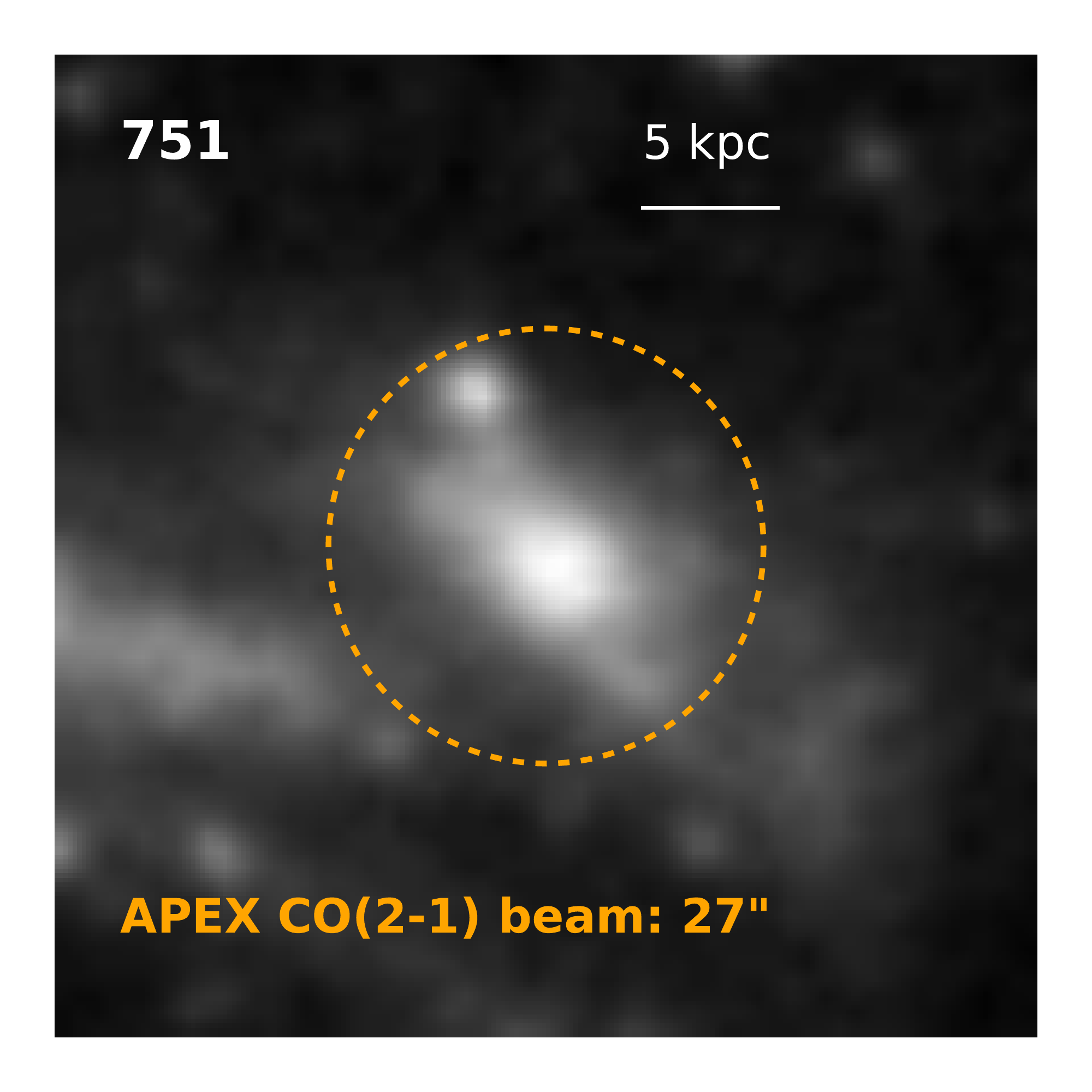}\includegraphics[width=0.26\textwidth]{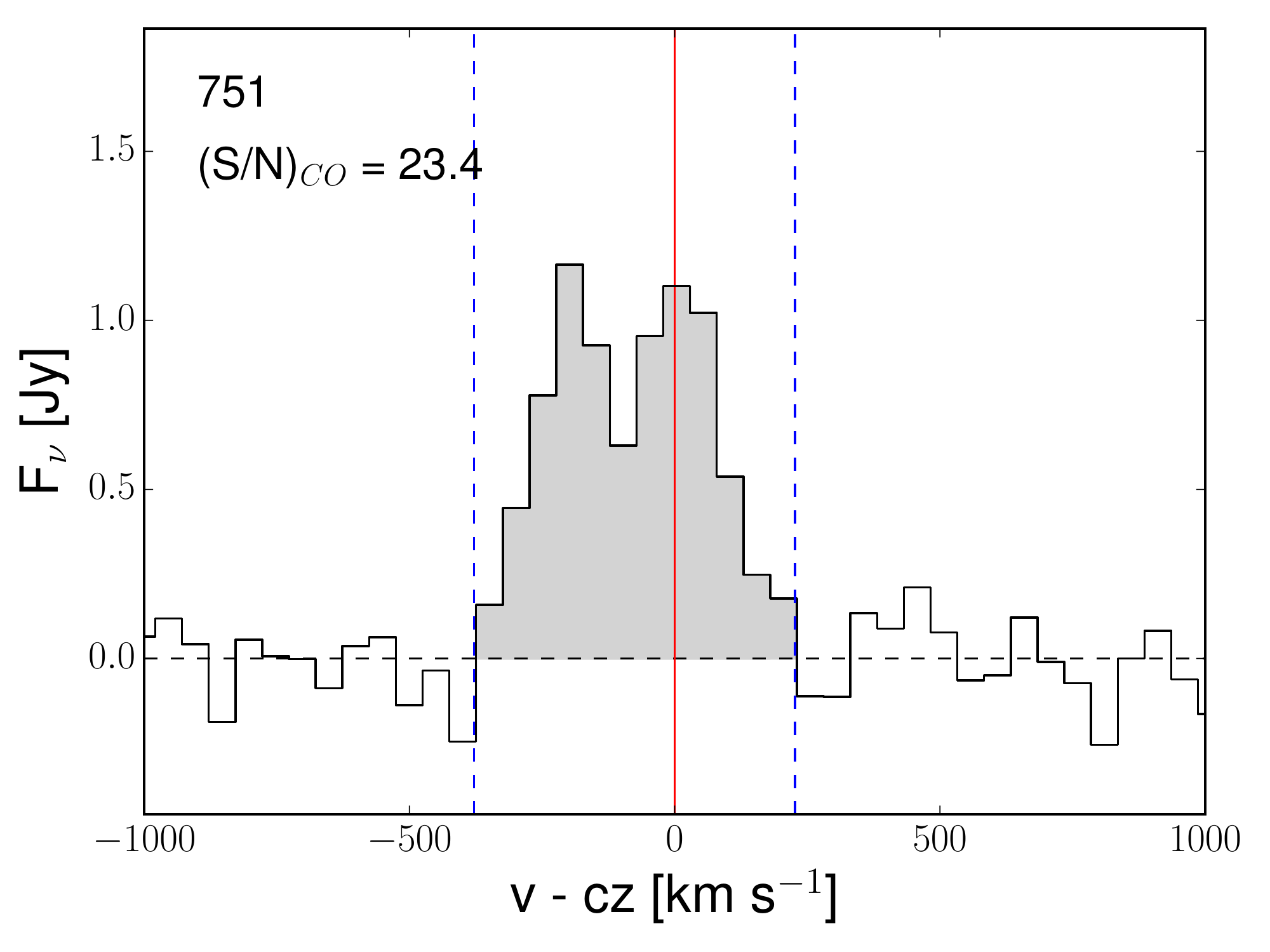}
\includegraphics[width=0.18\textwidth]{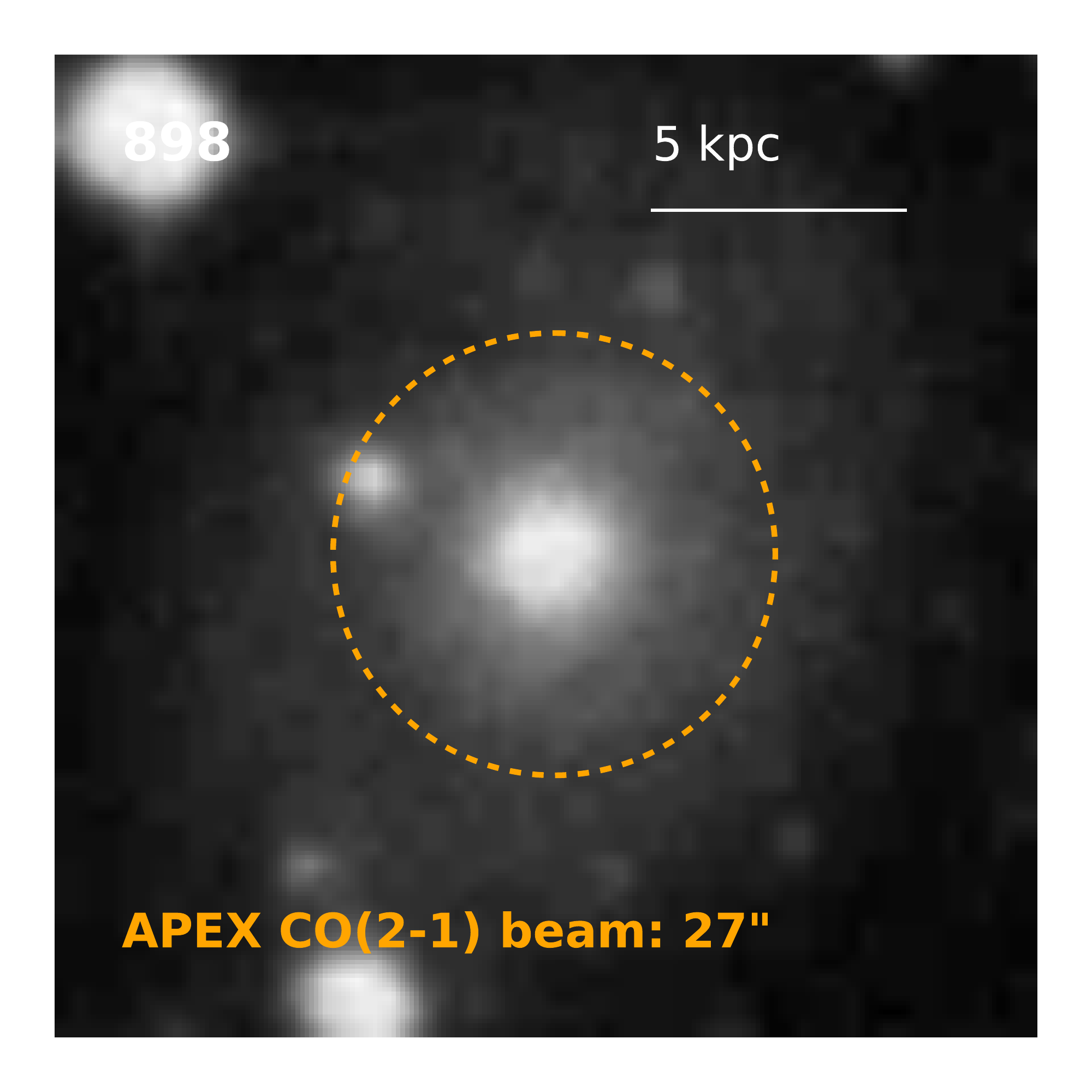}\includegraphics[width=0.26\textwidth]{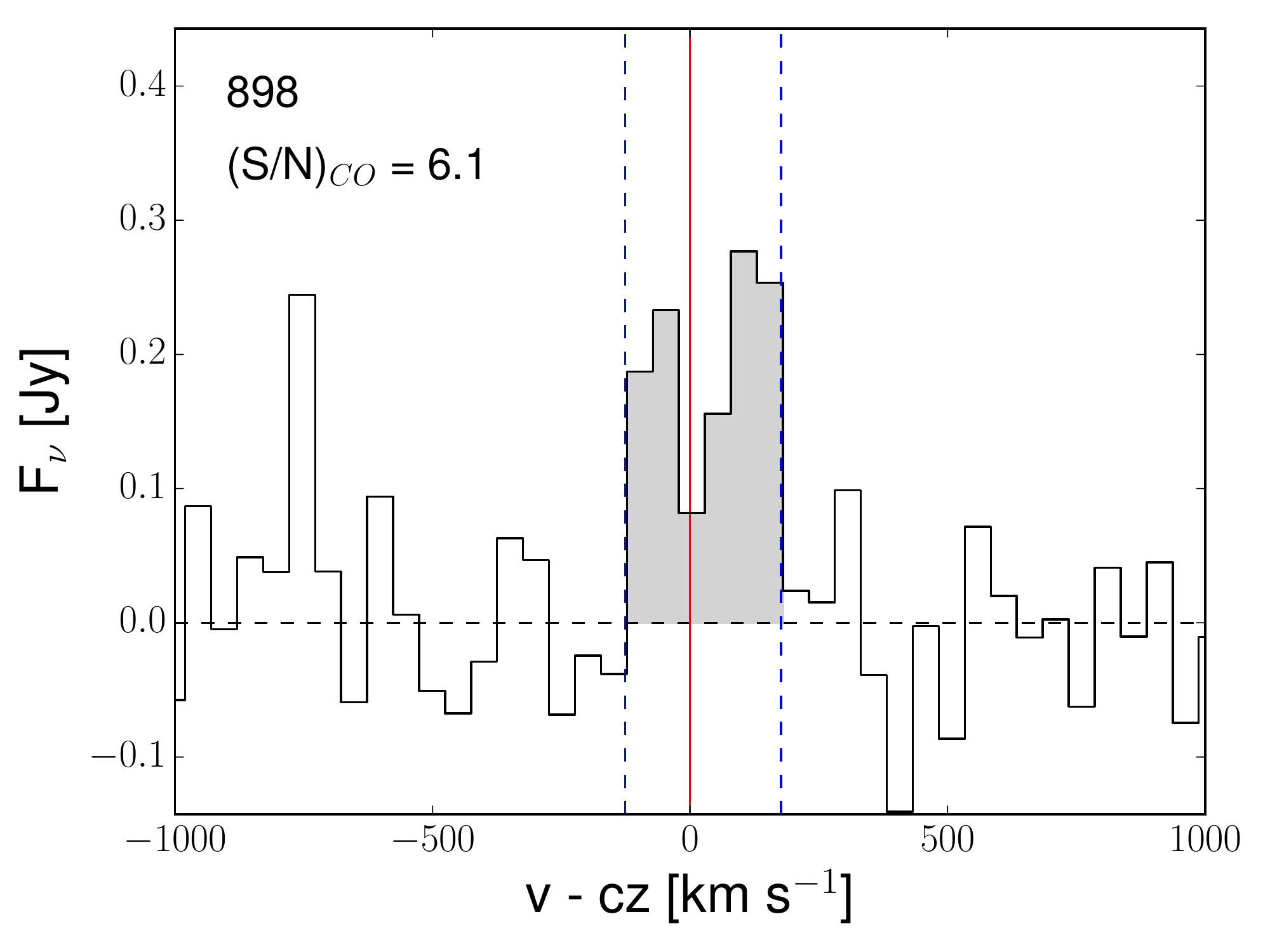}
\includegraphics[width=0.18\textwidth]{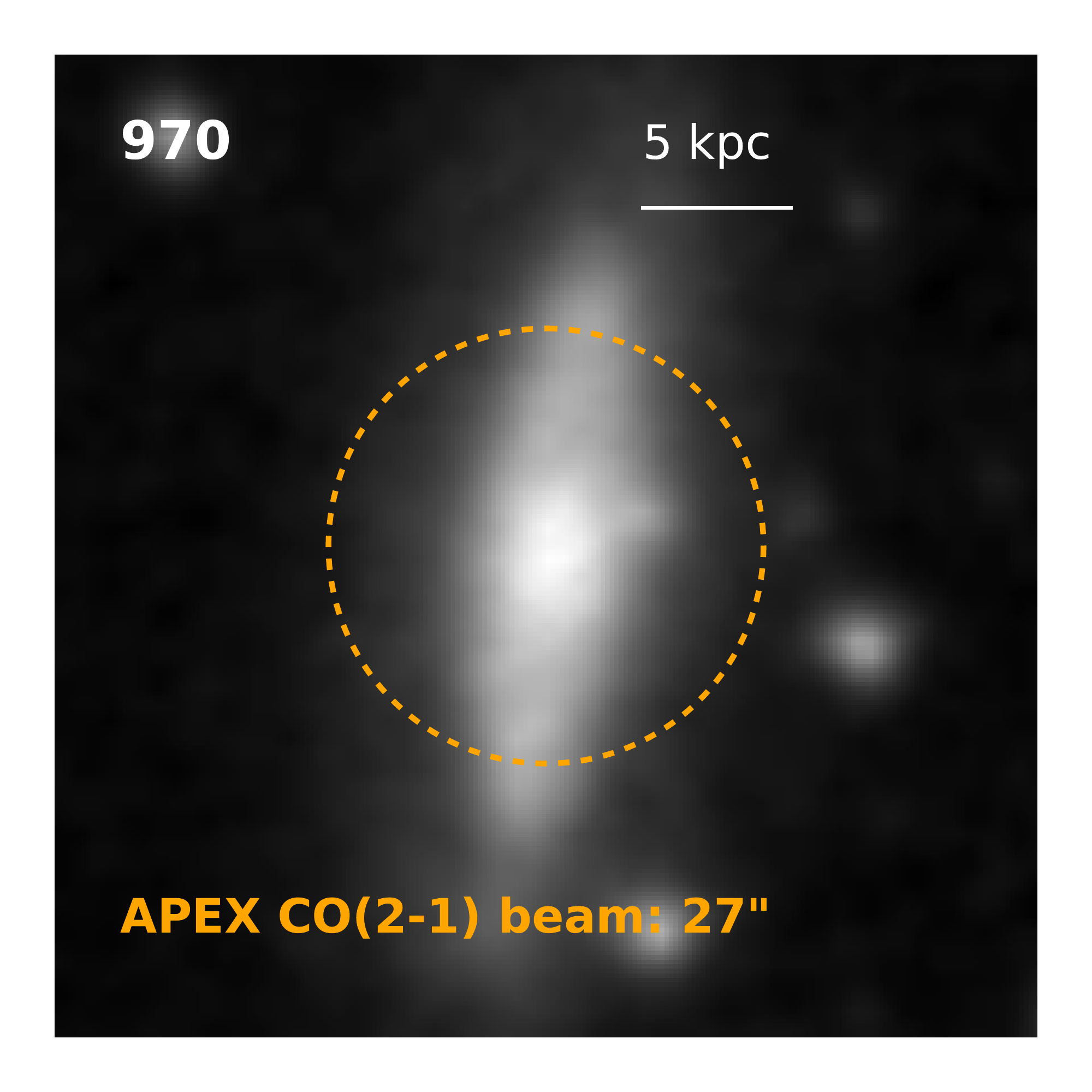}\includegraphics[width=0.26\textwidth]{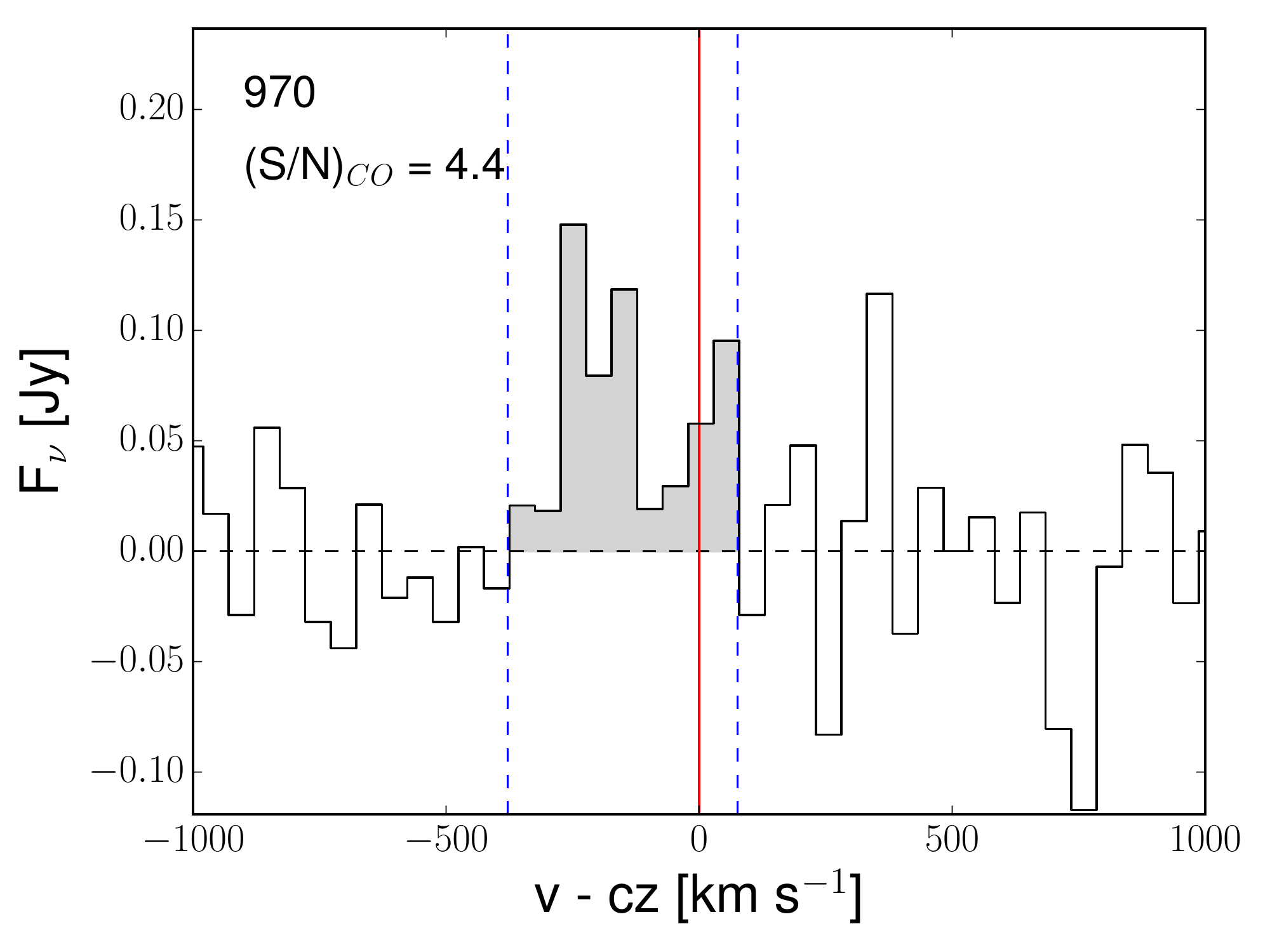}
\includegraphics[width=0.18\textwidth]{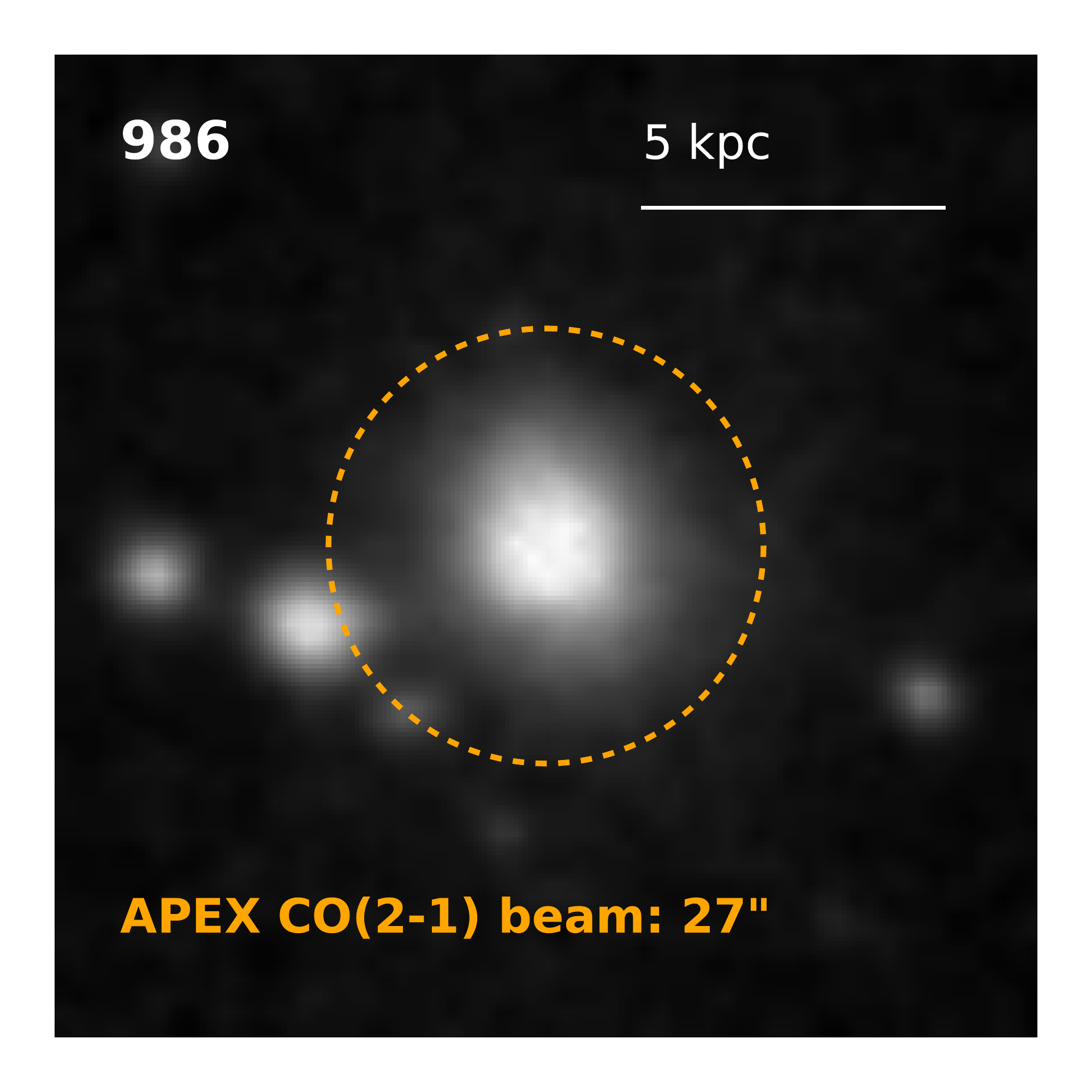}\includegraphics[width=0.26\textwidth]{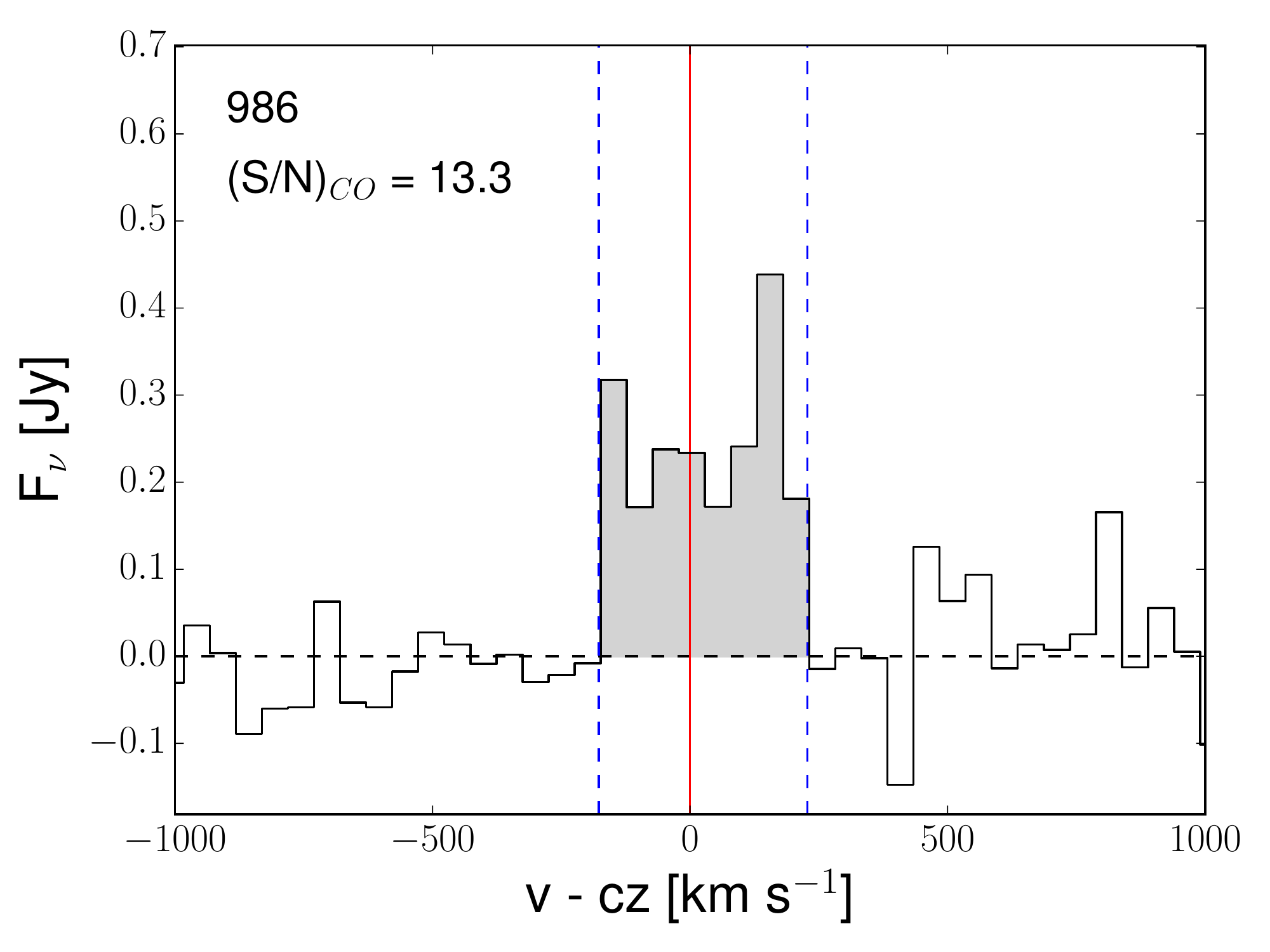}
\caption{continued from Fig.~\ref{fig:CO21_spectra_dss1}
} 
\label{fig:CO21_spectra_dss3}
\end{figure*}

\begin{figure*}
\centering
\raggedright
\includegraphics[width=0.18\textwidth]{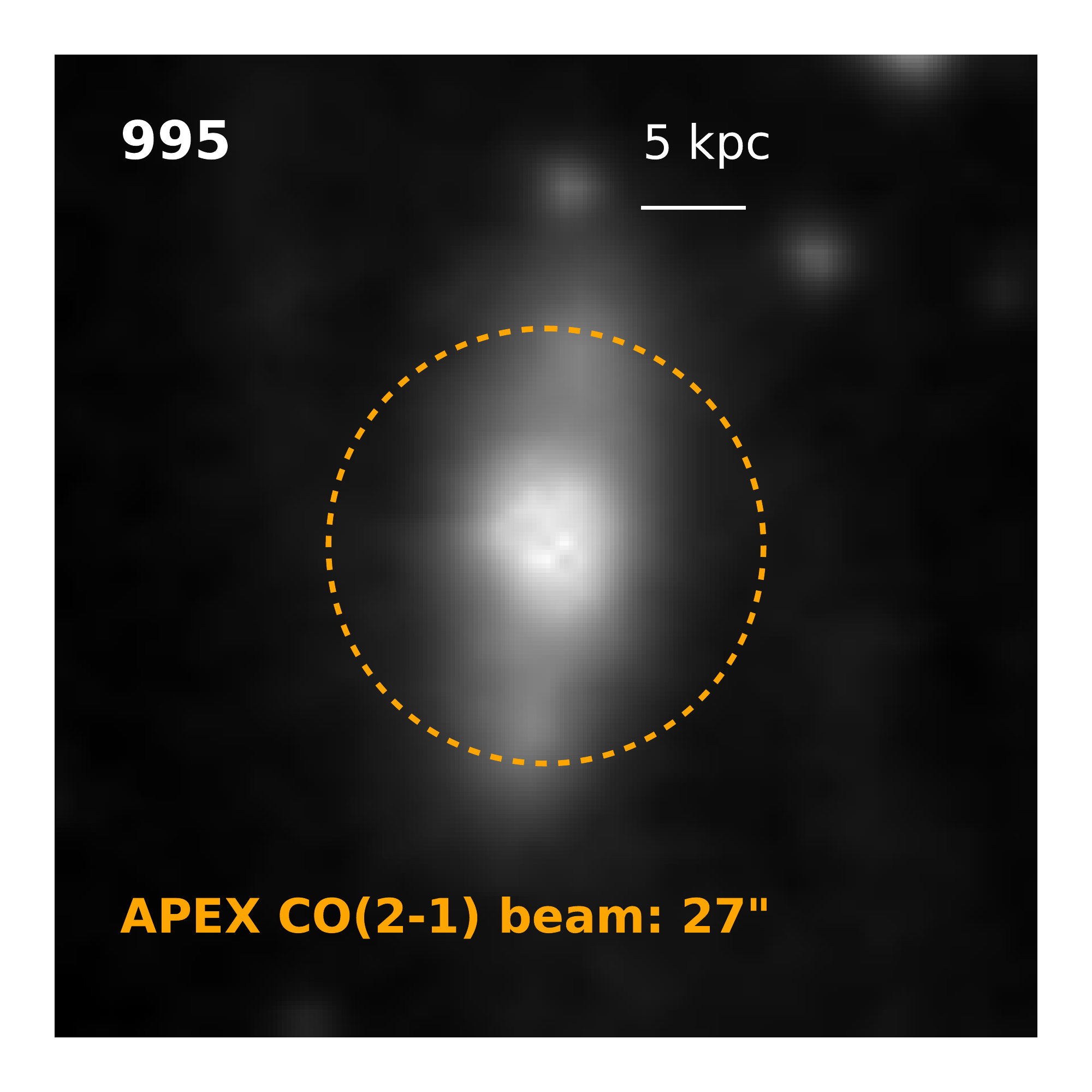}\includegraphics[width=0.26\textwidth]{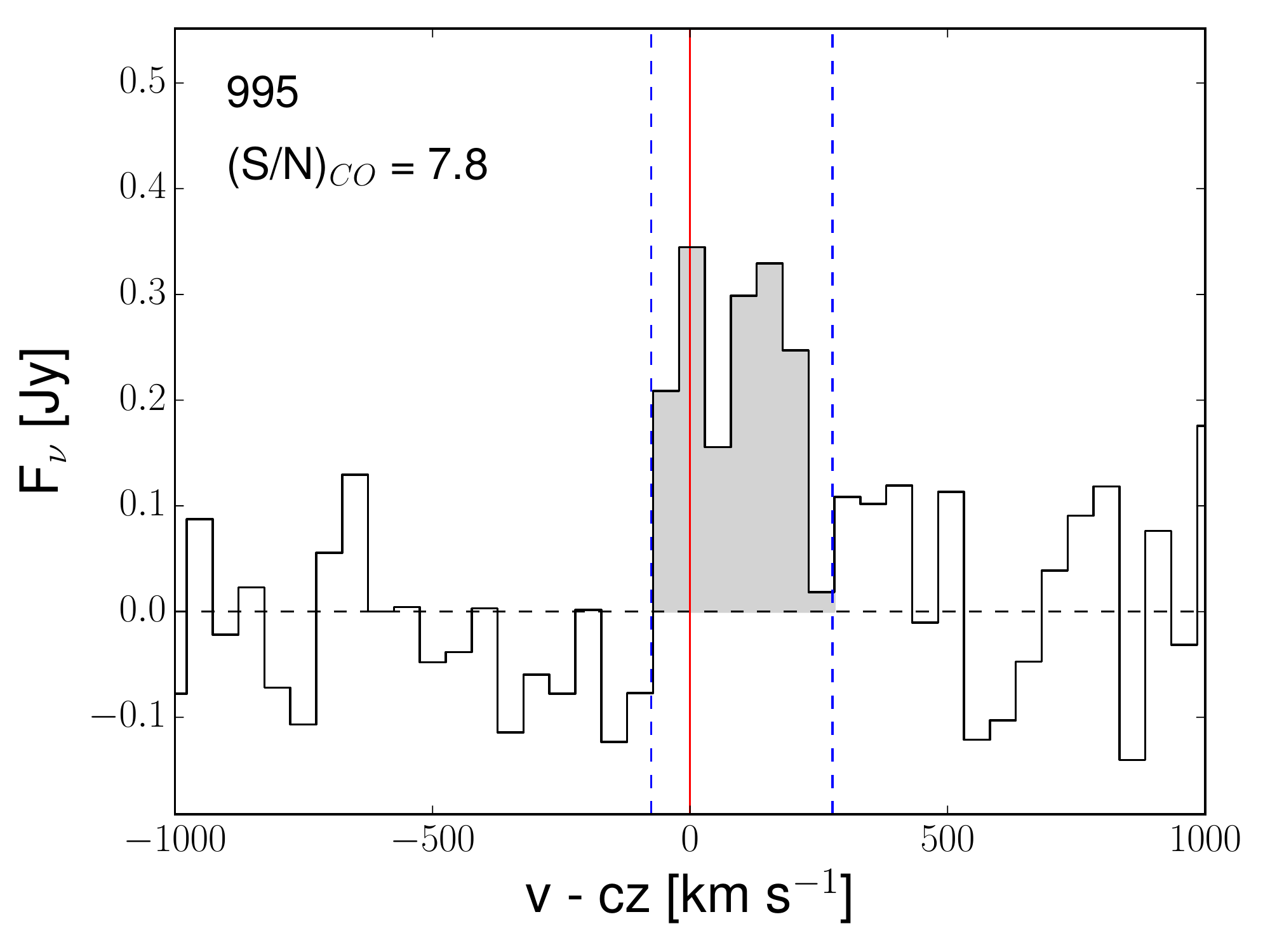}
\includegraphics[width=0.18\textwidth]{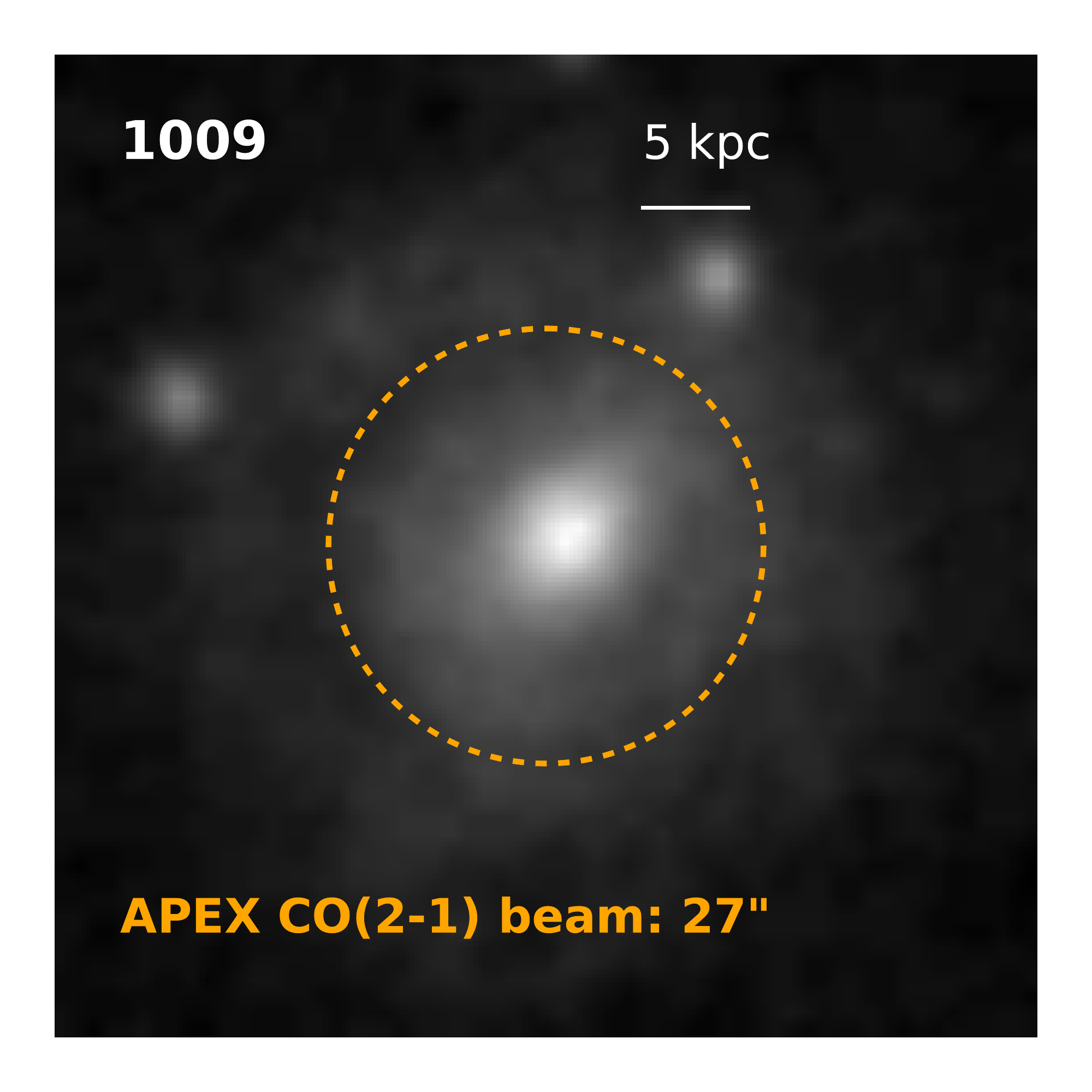}\includegraphics[width=0.26\textwidth]{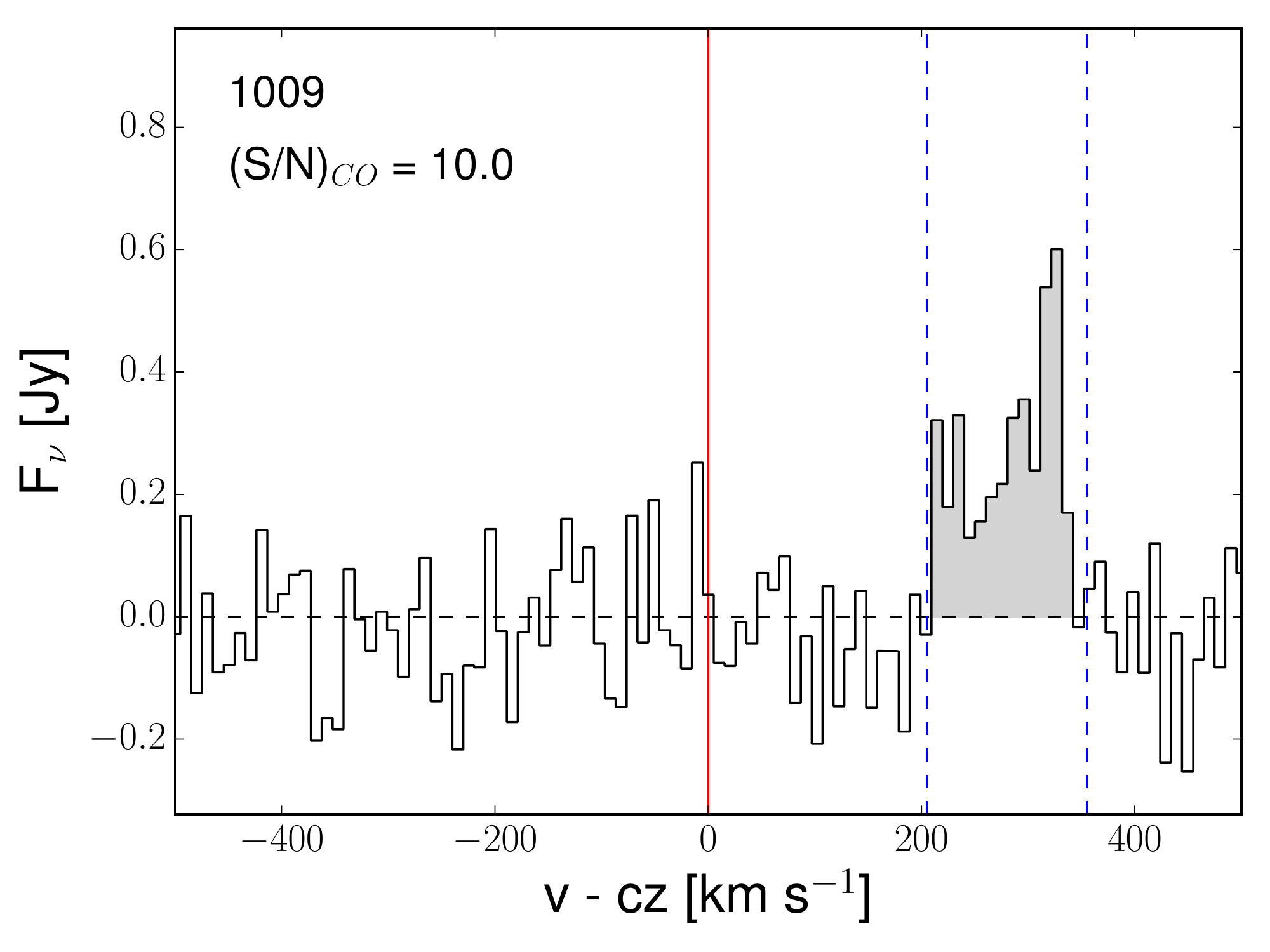}
\includegraphics[width=0.18\textwidth]{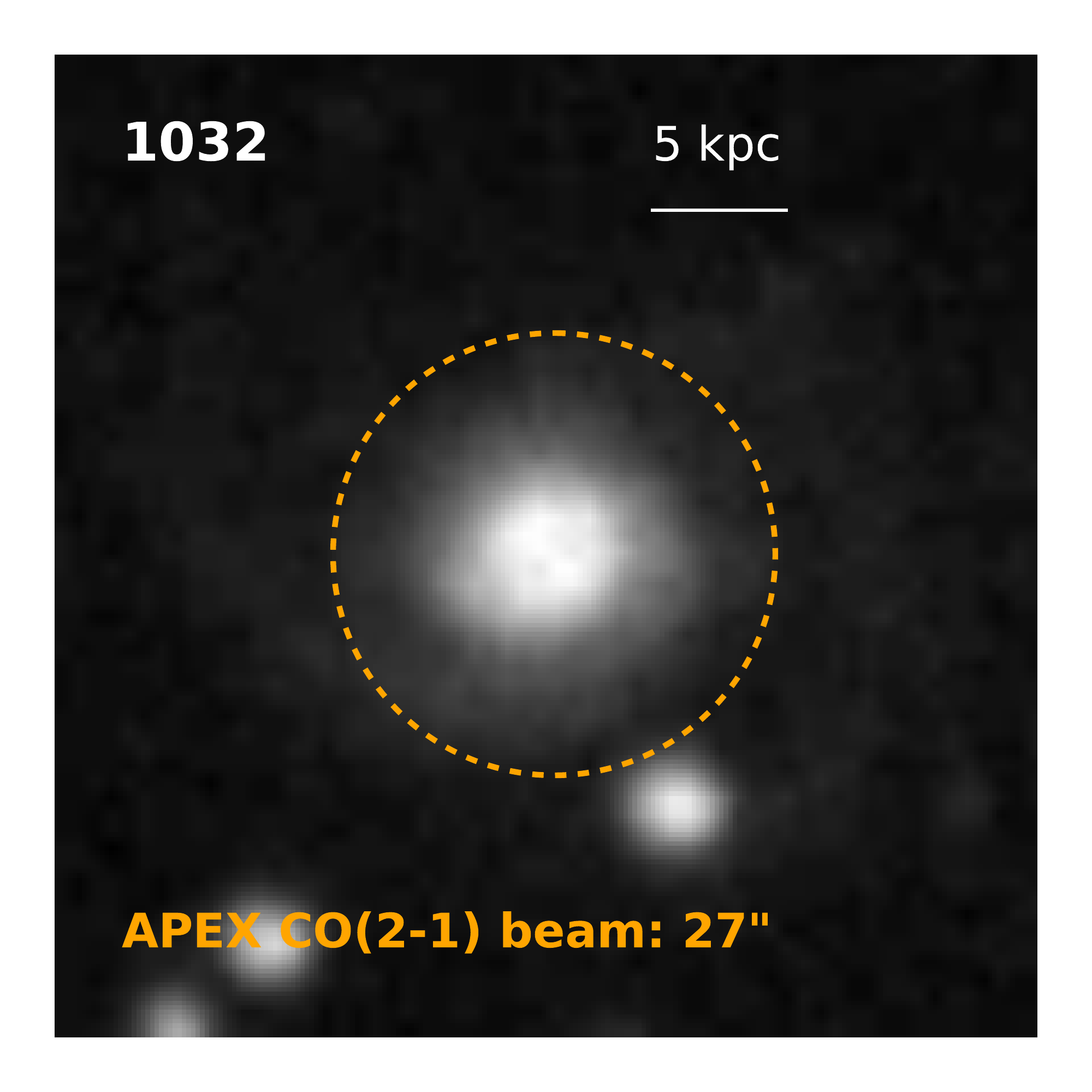}\includegraphics[width=0.26\textwidth]{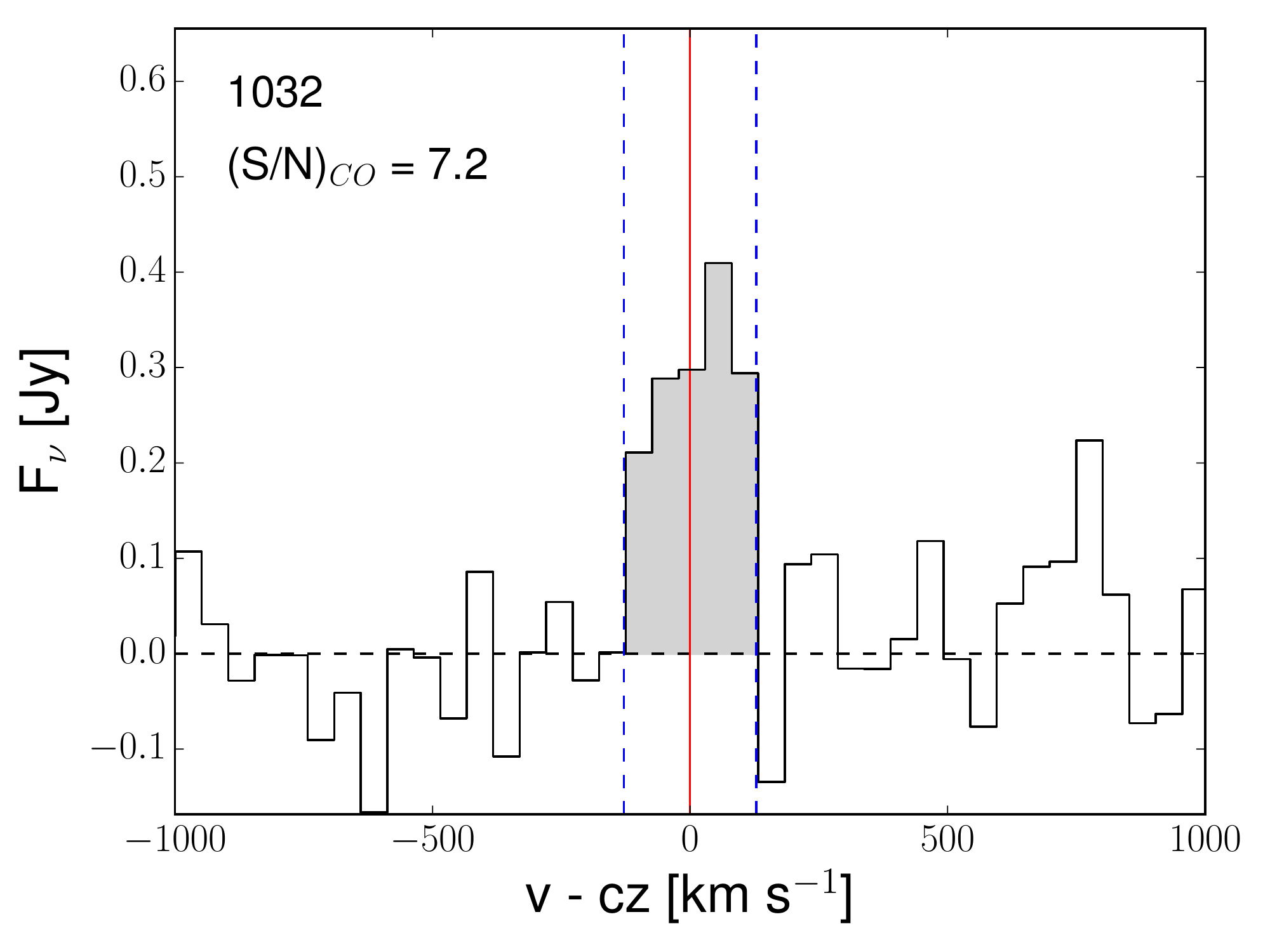}
\includegraphics[width=0.18\textwidth]{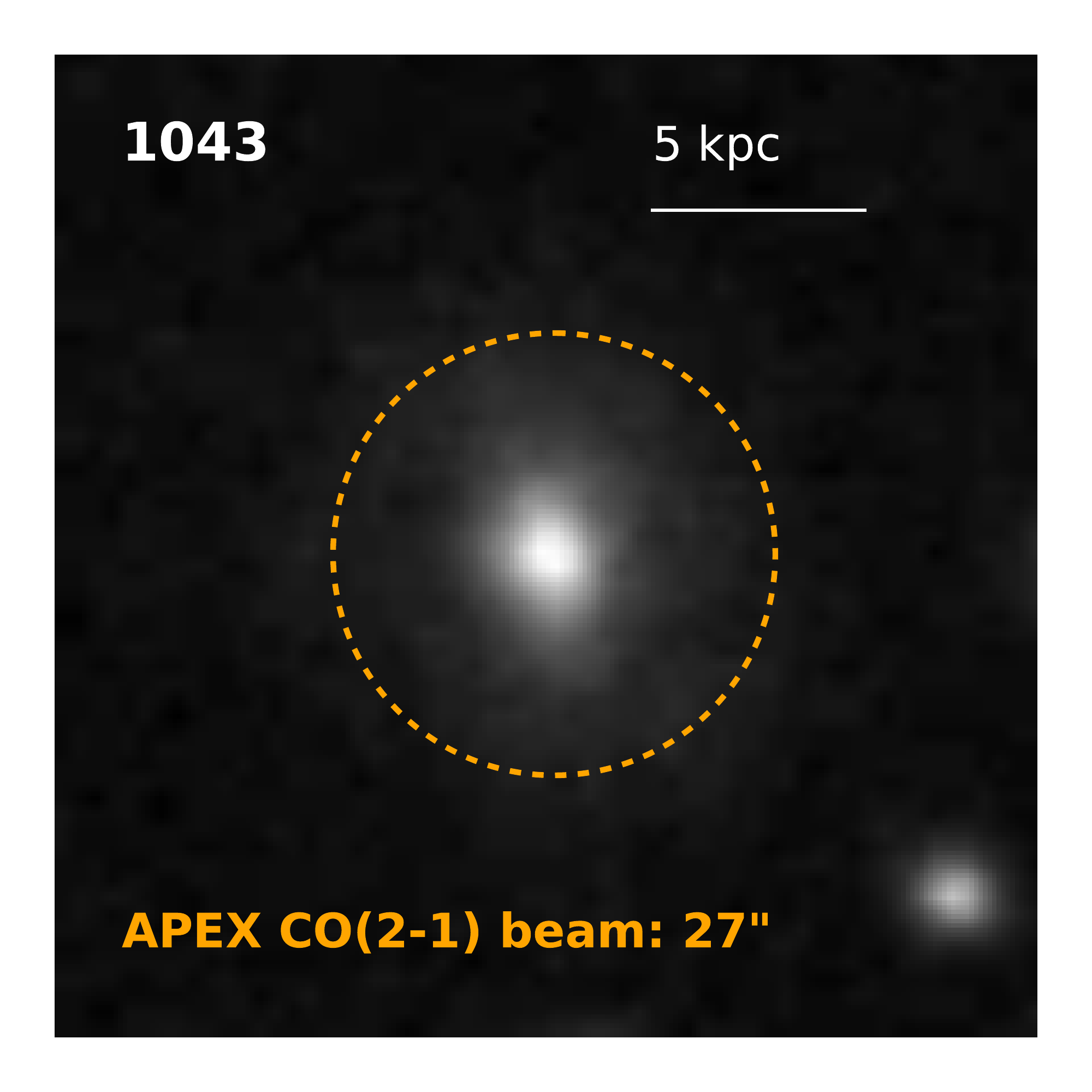}\includegraphics[width=0.26\textwidth]{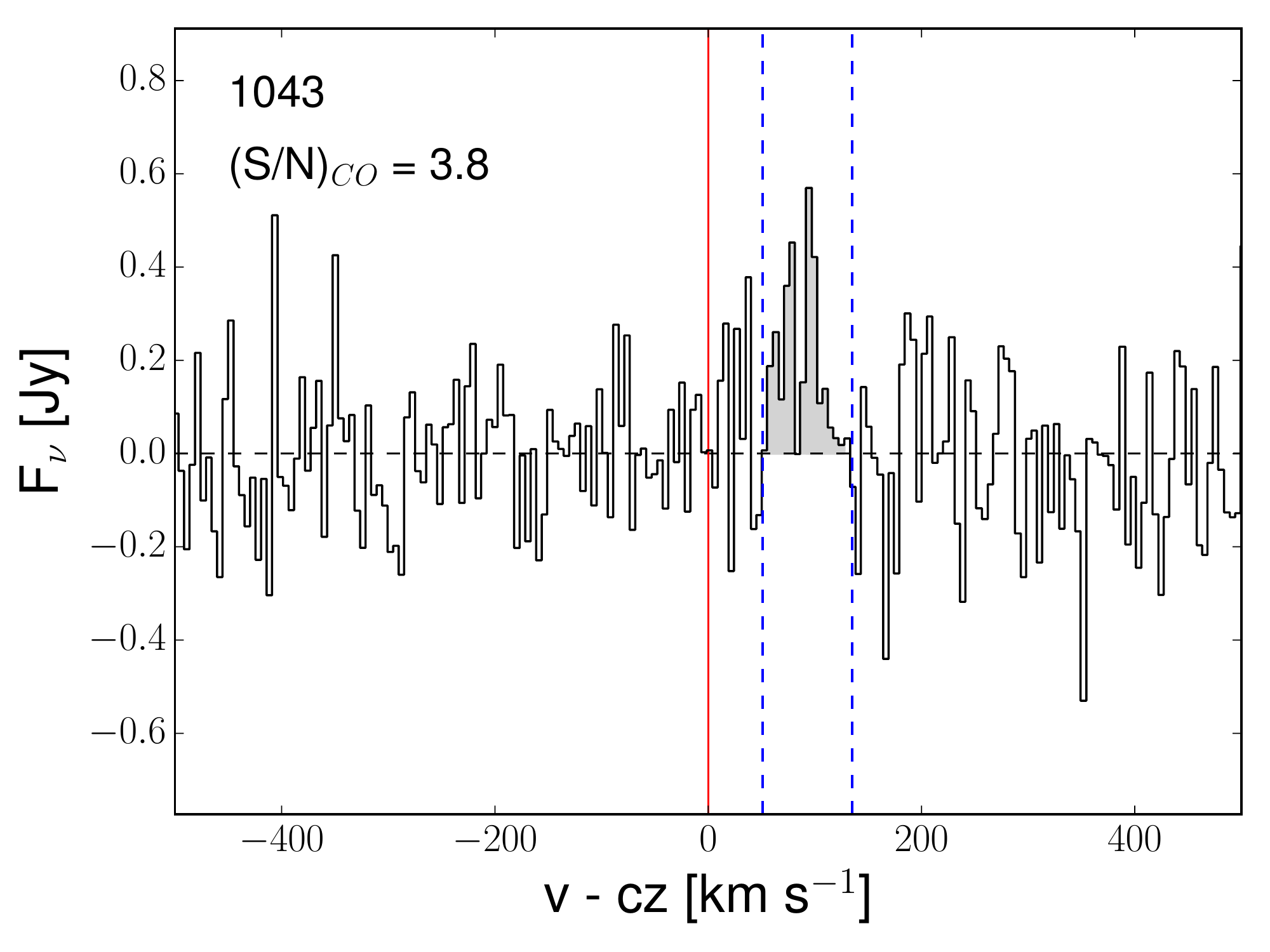}
\includegraphics[width=0.18\textwidth]{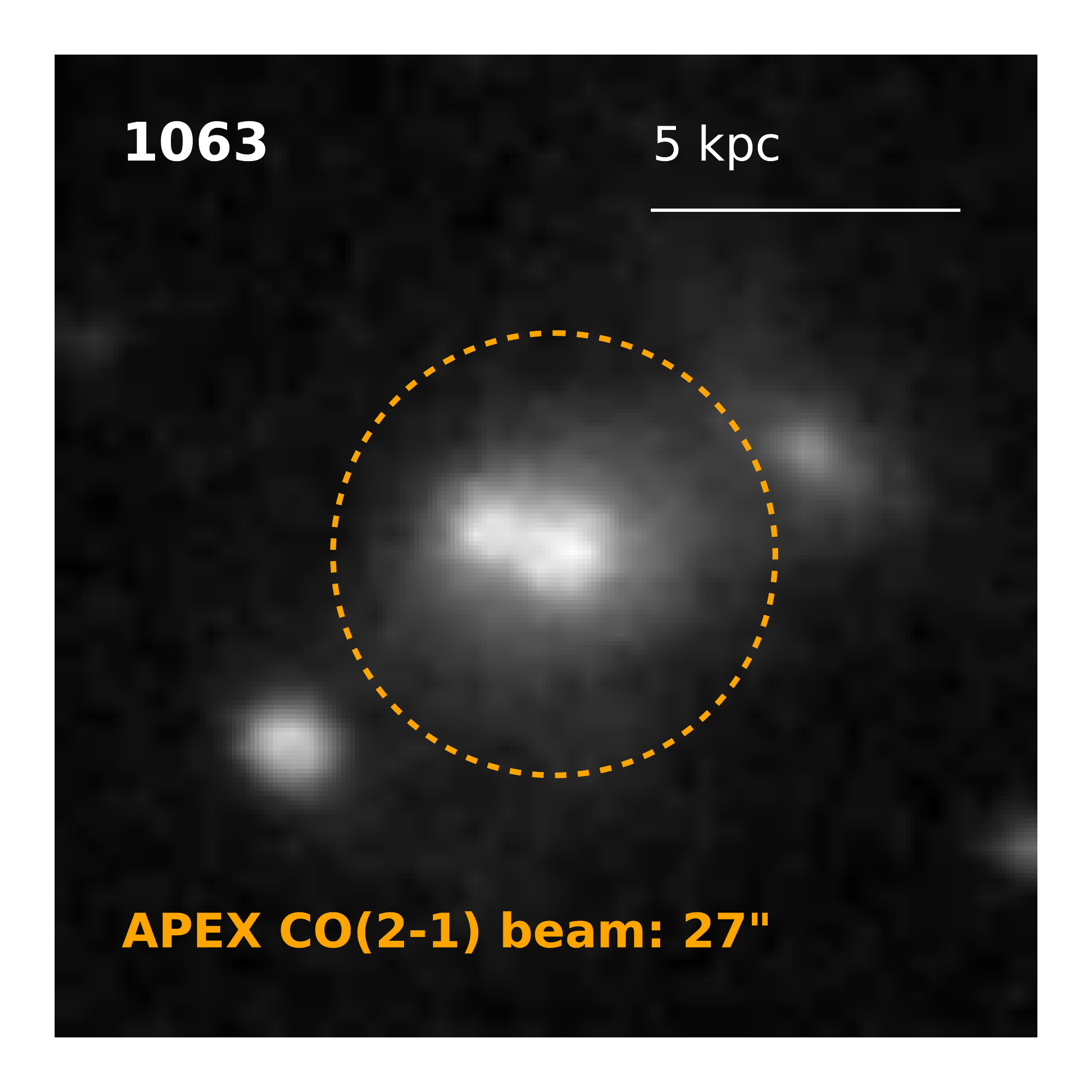}\includegraphics[width=0.26\textwidth]{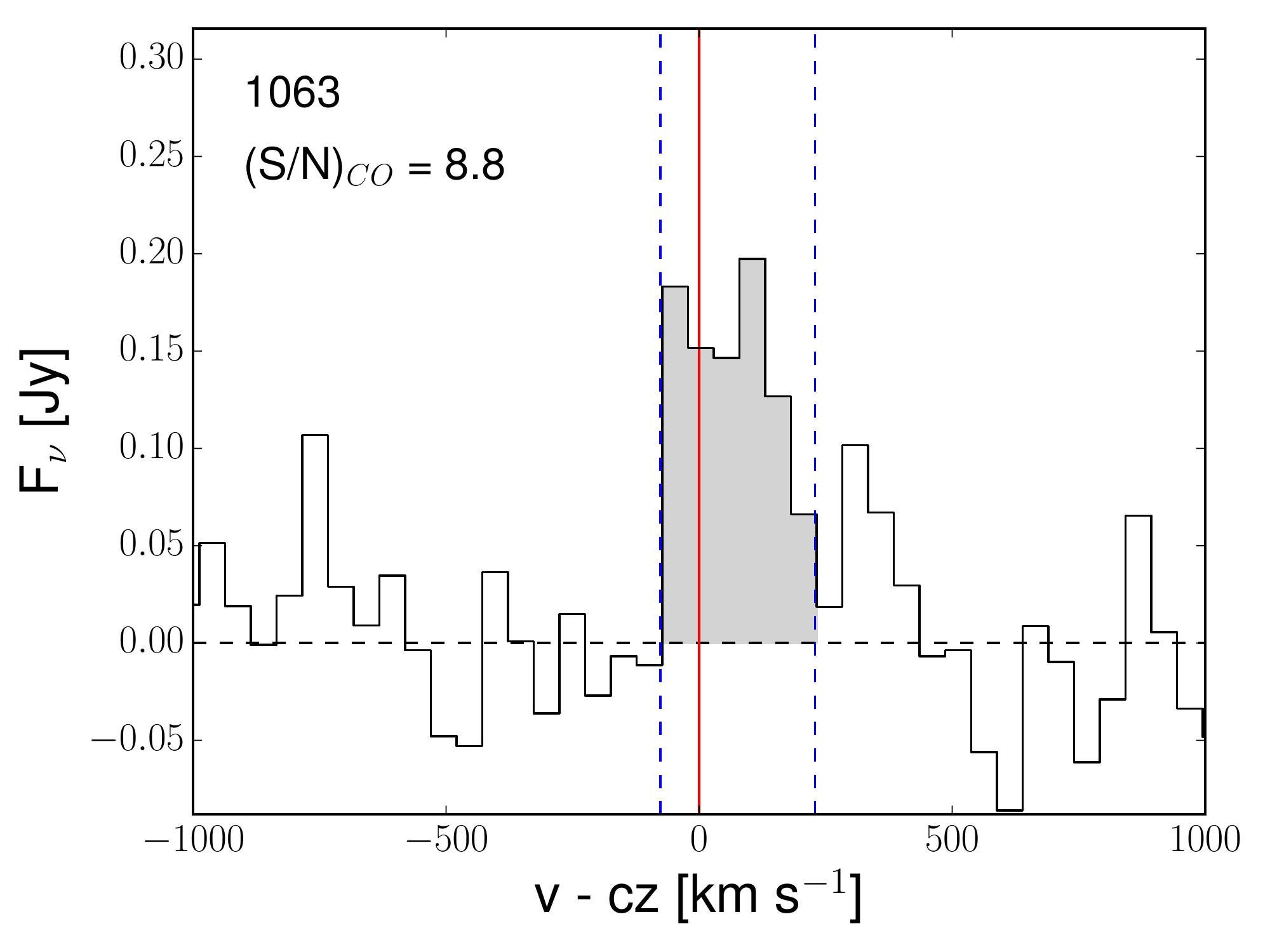}
\includegraphics[width=0.18\textwidth]{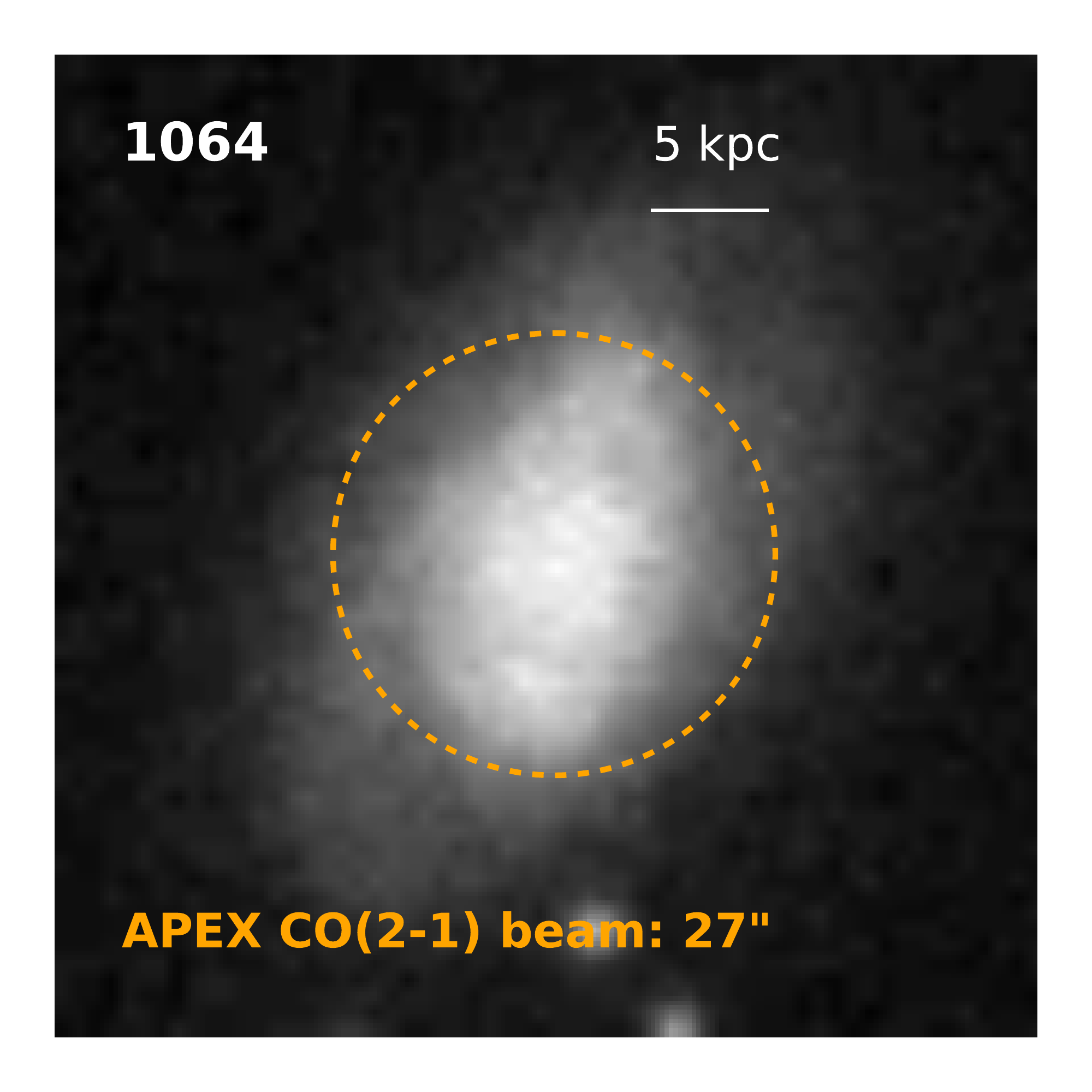}\includegraphics[width=0.26\textwidth]{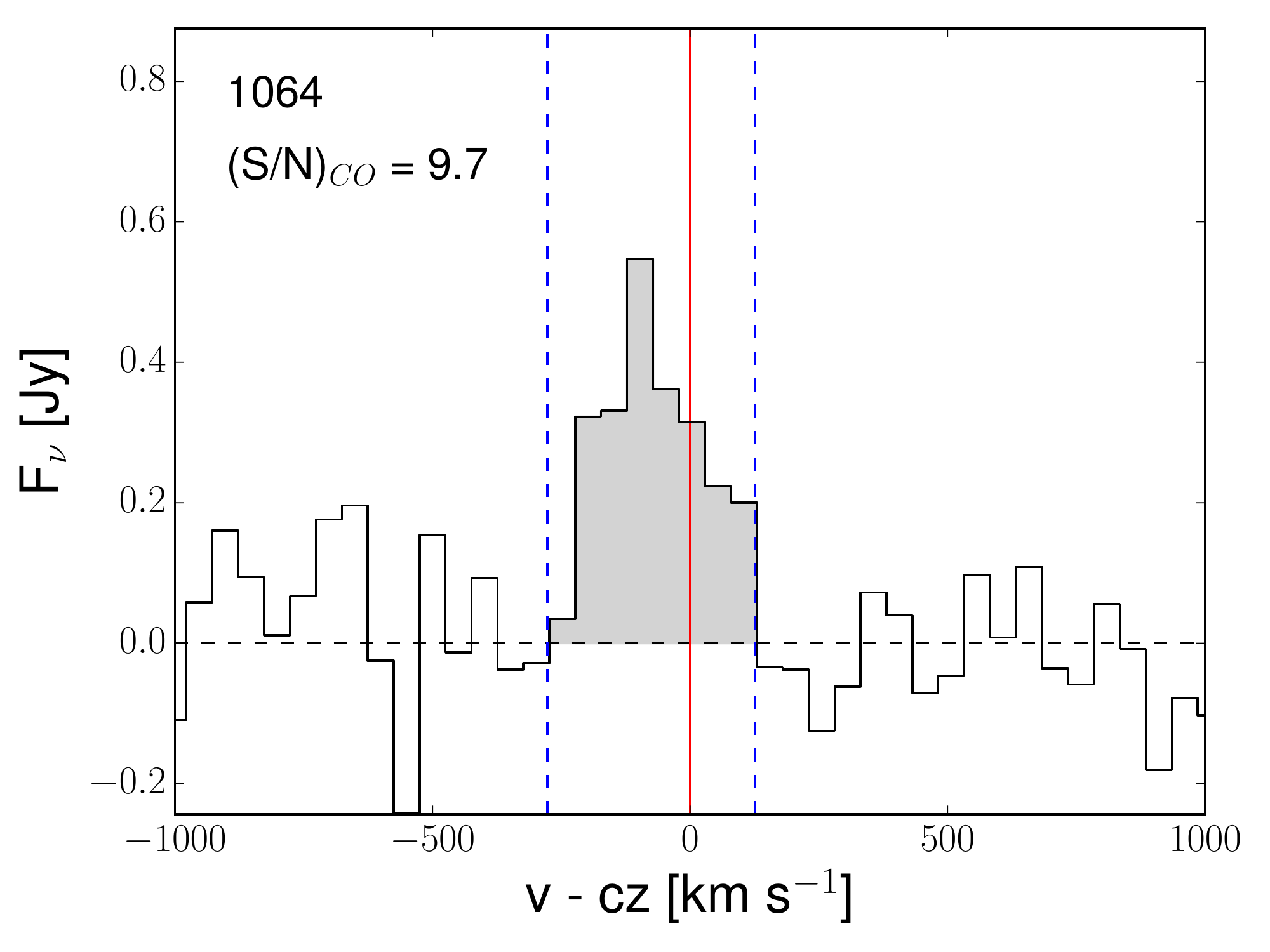}
\includegraphics[width=0.18\textwidth]{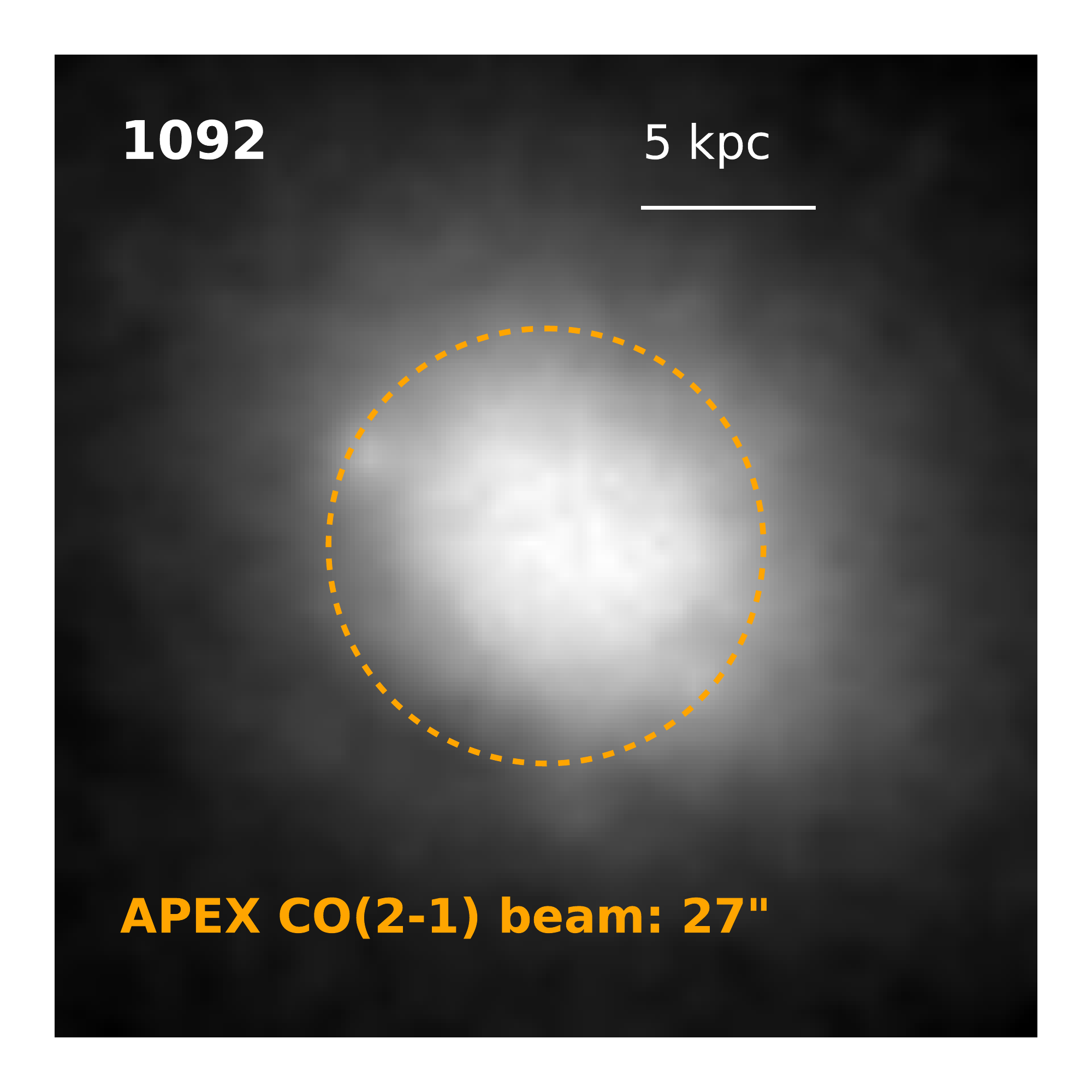}\includegraphics[width=0.26\textwidth]{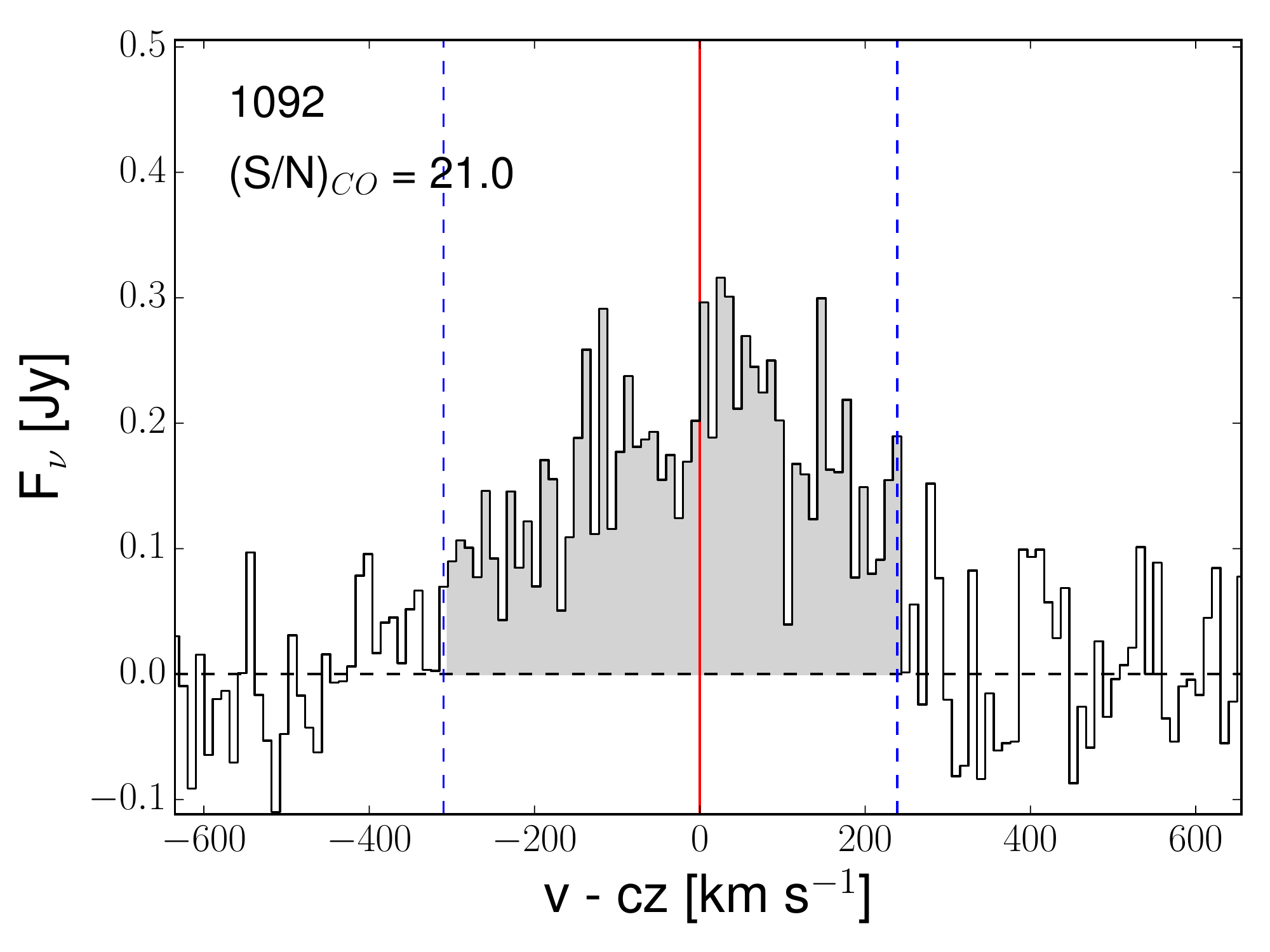}
\includegraphics[width=0.18\textwidth]{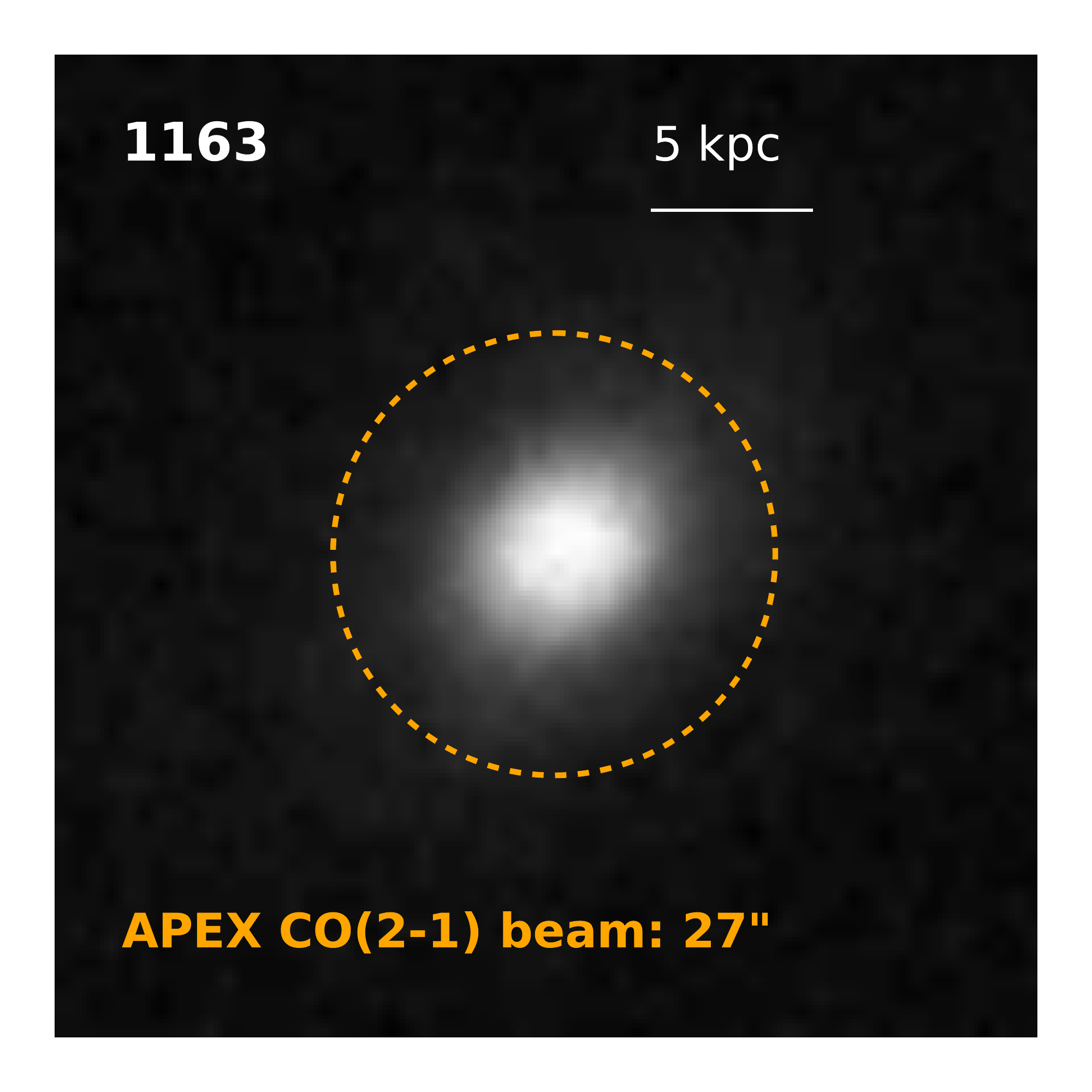}\includegraphics[width=0.26\textwidth]{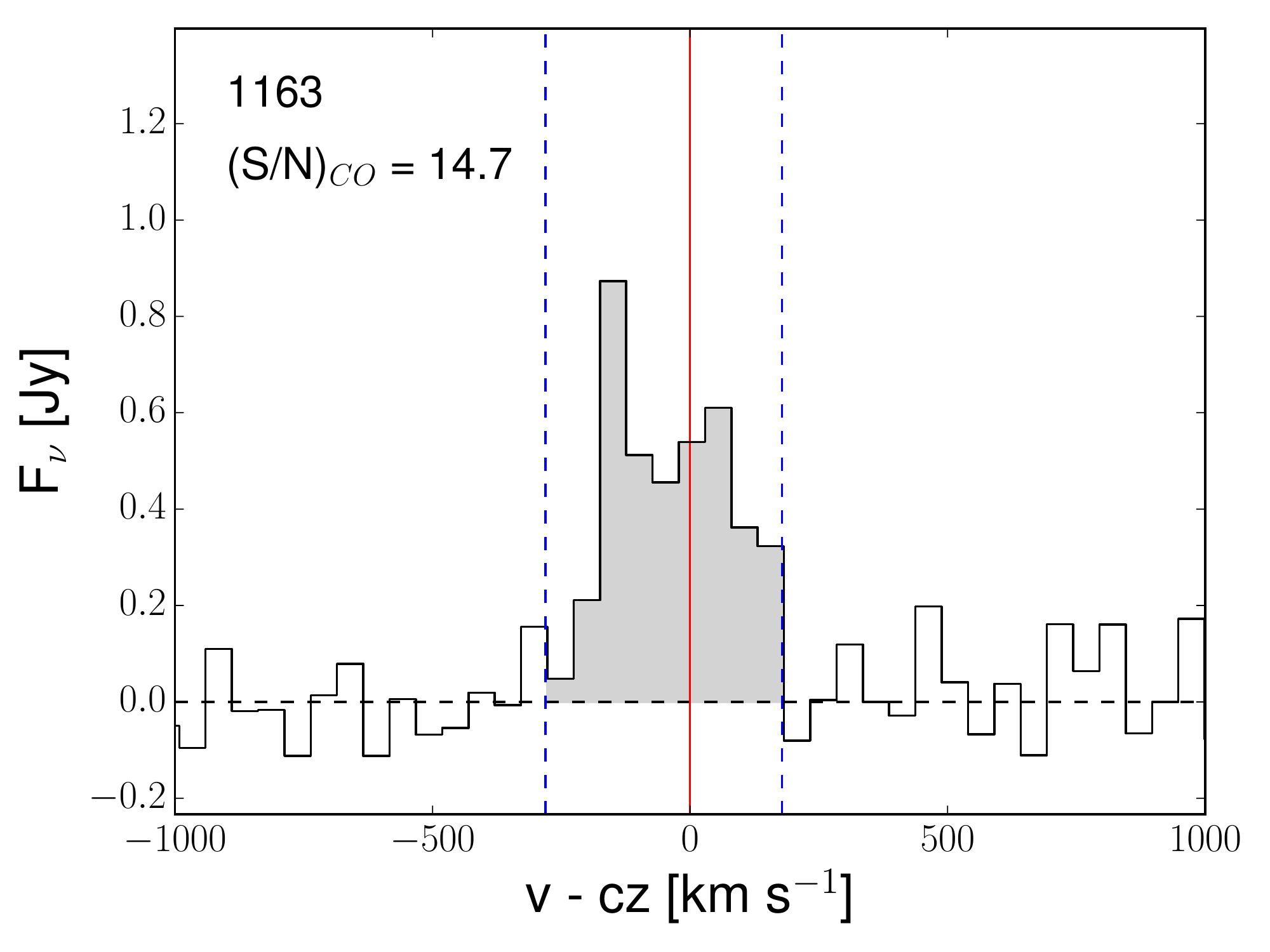}
\includegraphics[width=0.18\textwidth]{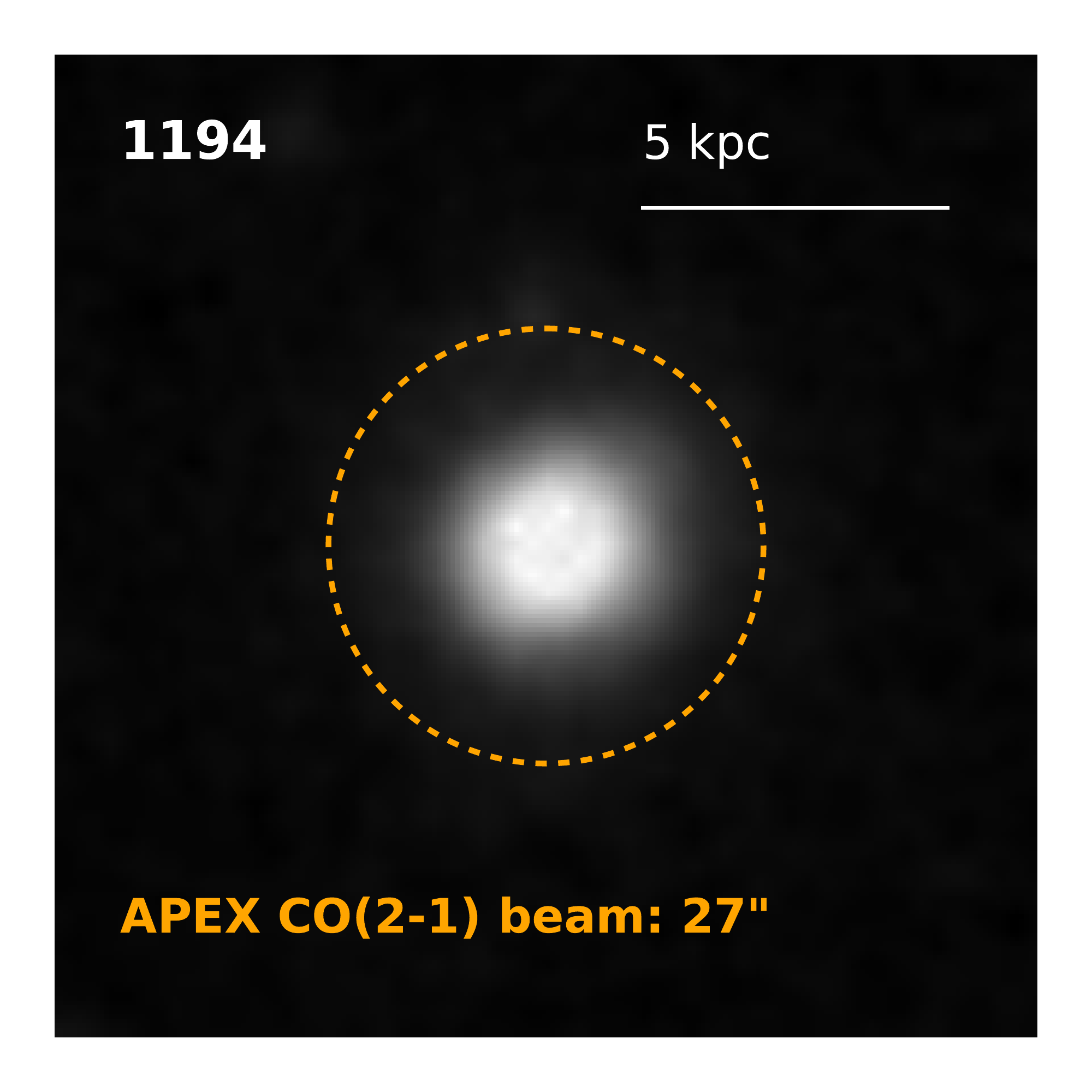}\includegraphics[width=0.26\textwidth]{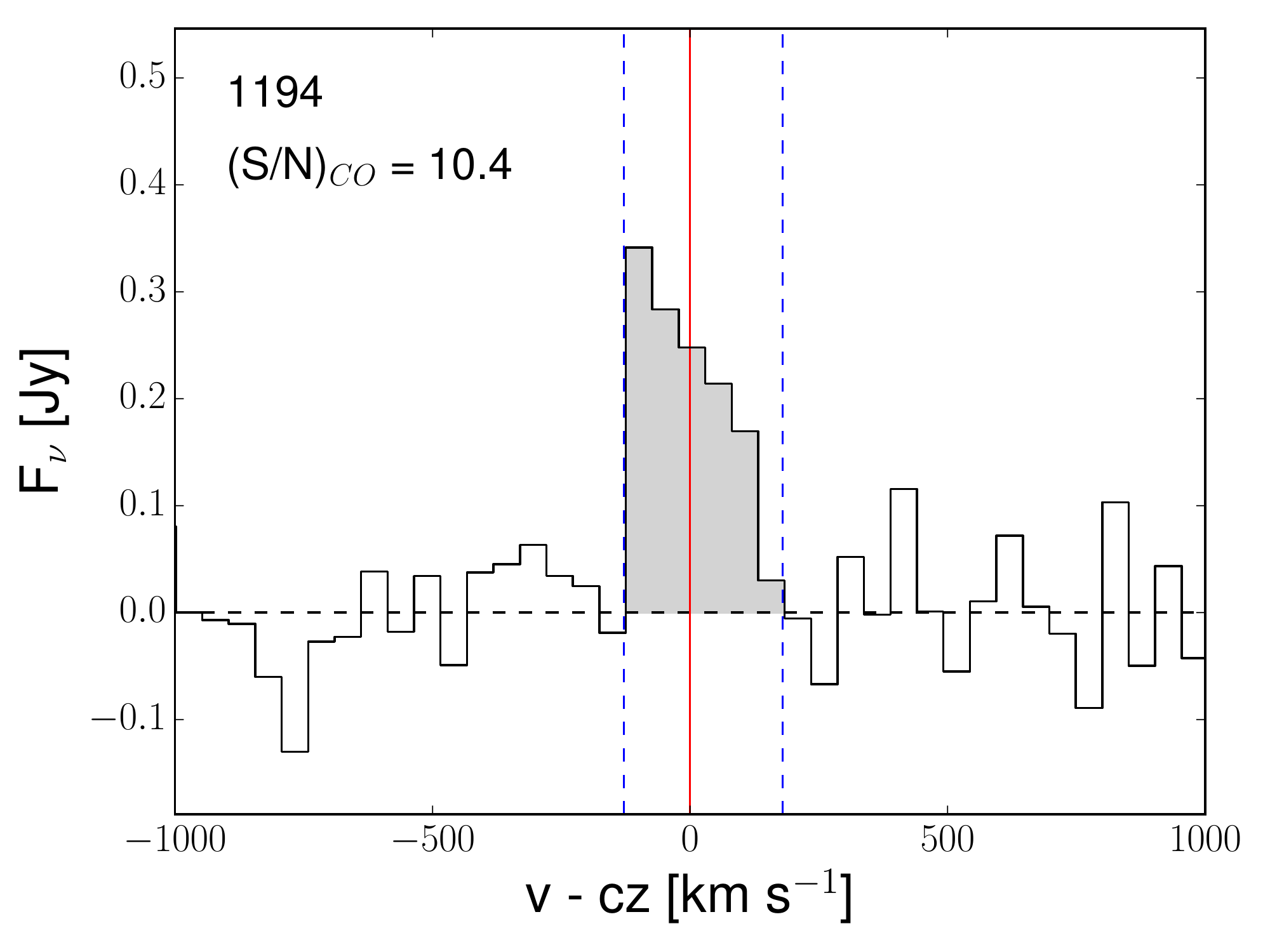}
\caption{continued from Fig.~\ref{fig:CO21_spectra_dss1}
} 
\label{fig:CO21_spectra_dss4}
\end{figure*}

\begin{figure*}
\centering
\raggedright
\includegraphics[width=0.18\textwidth]{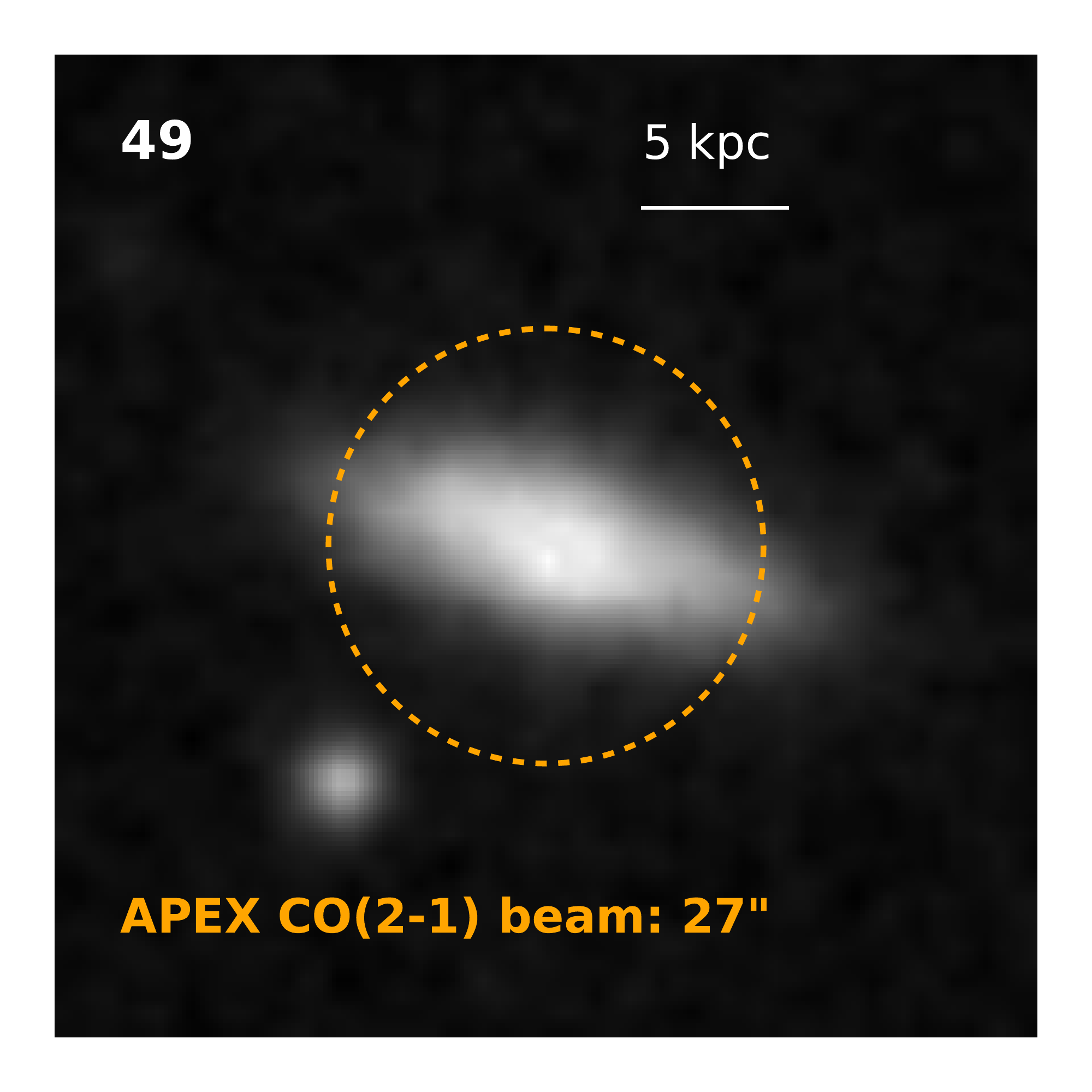}\includegraphics[width=0.26\textwidth]{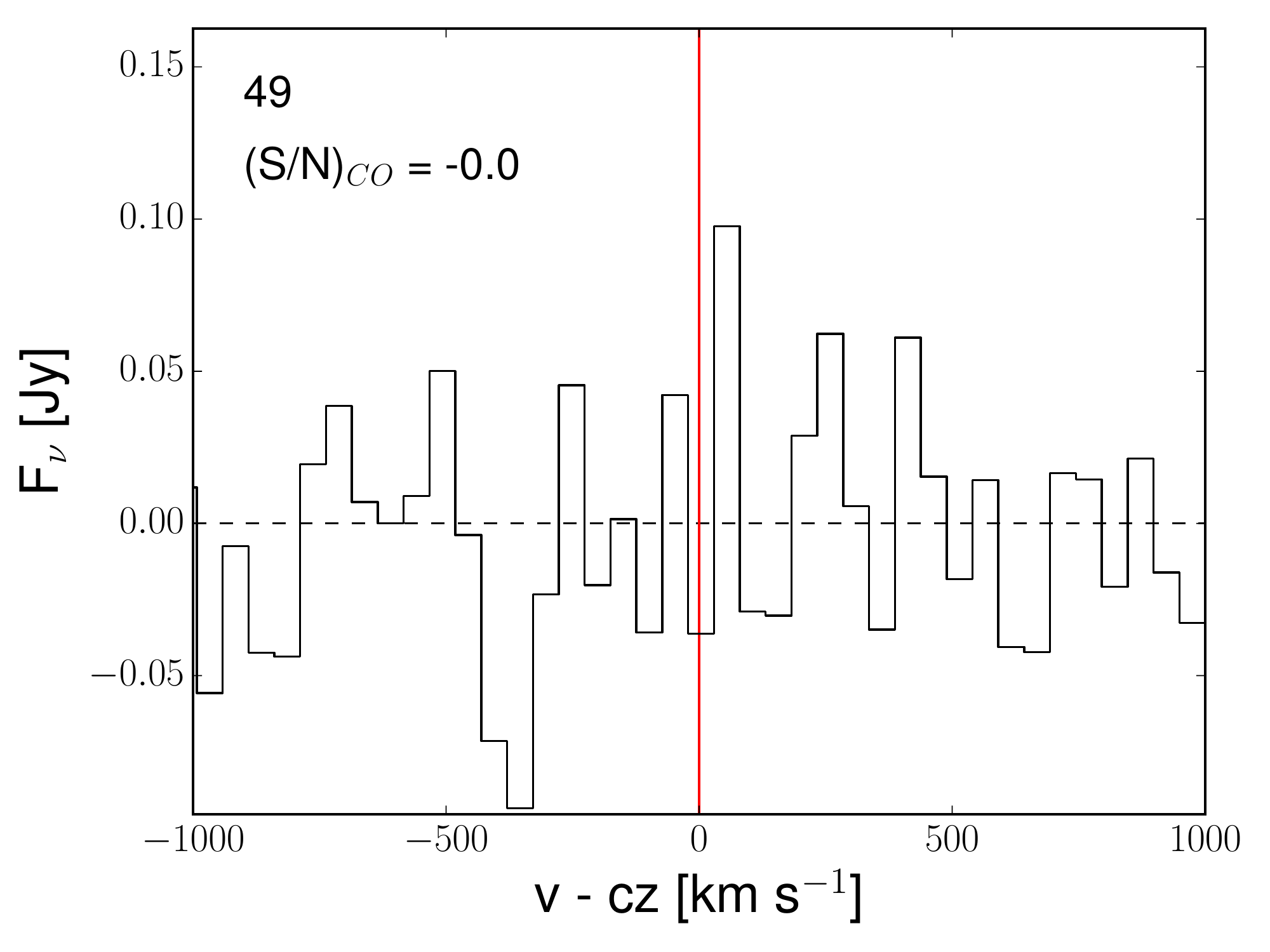}
\includegraphics[width=0.18\textwidth]{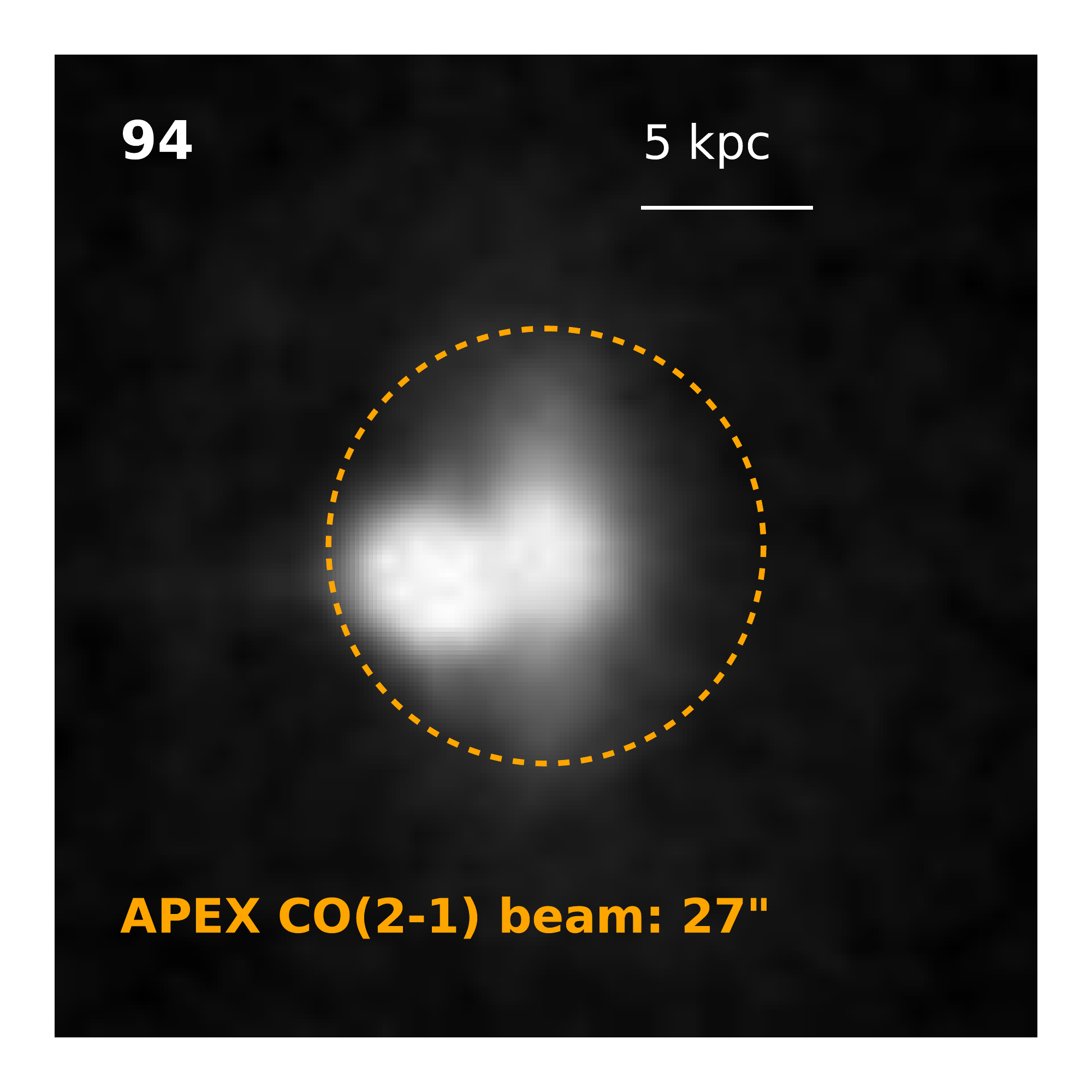}\includegraphics[width=0.26\textwidth]{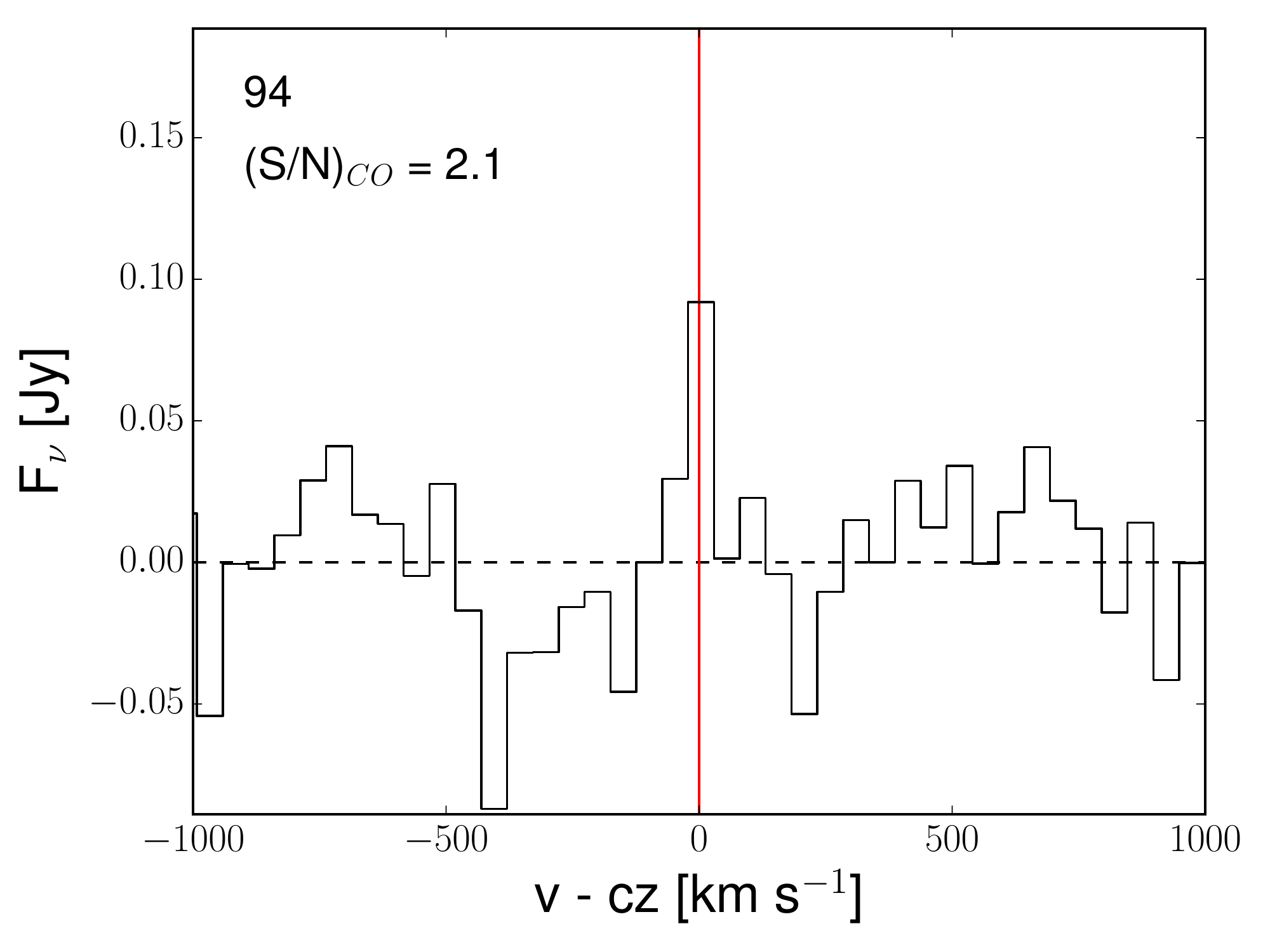}
\includegraphics[width=0.18\textwidth]{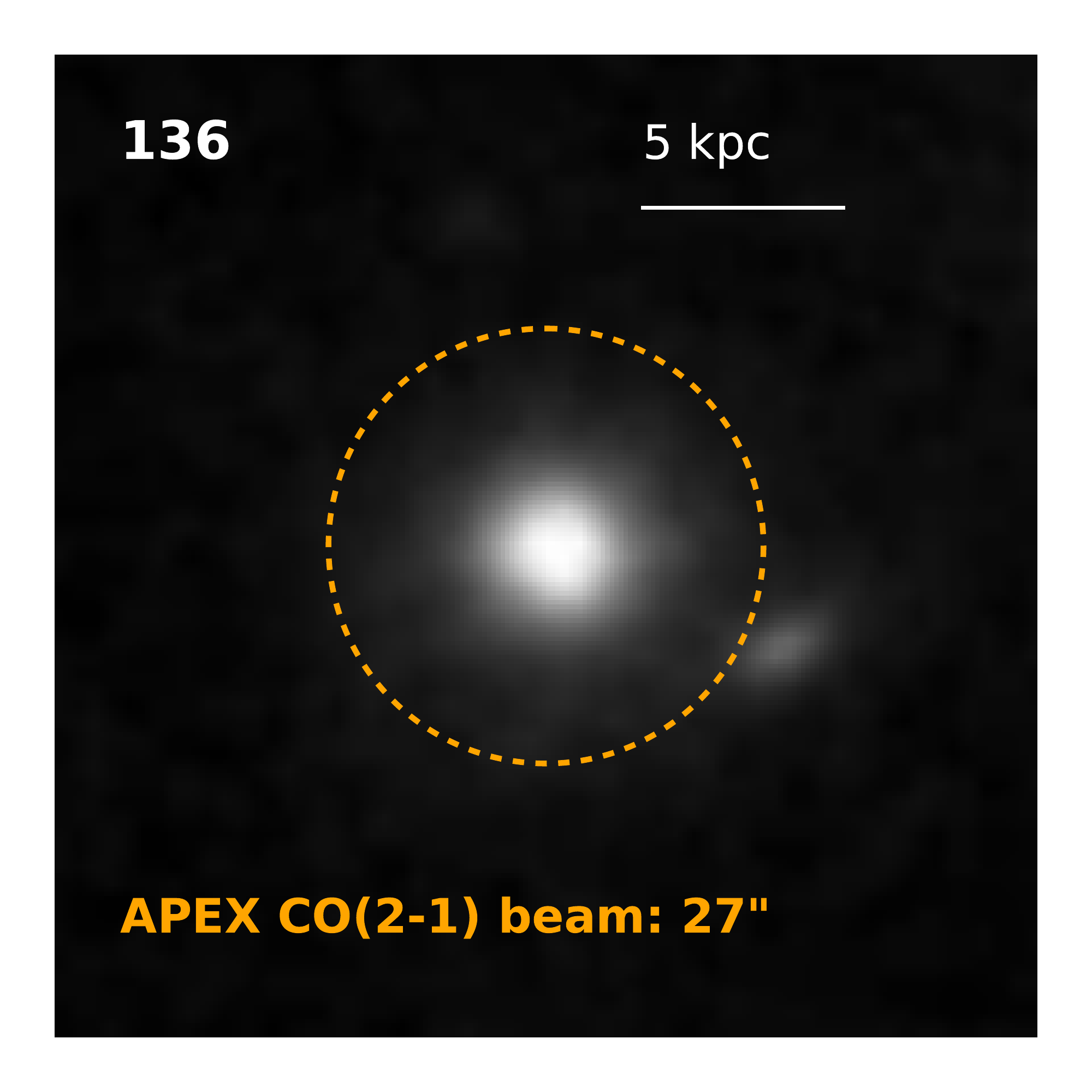}\includegraphics[width=0.26\textwidth]{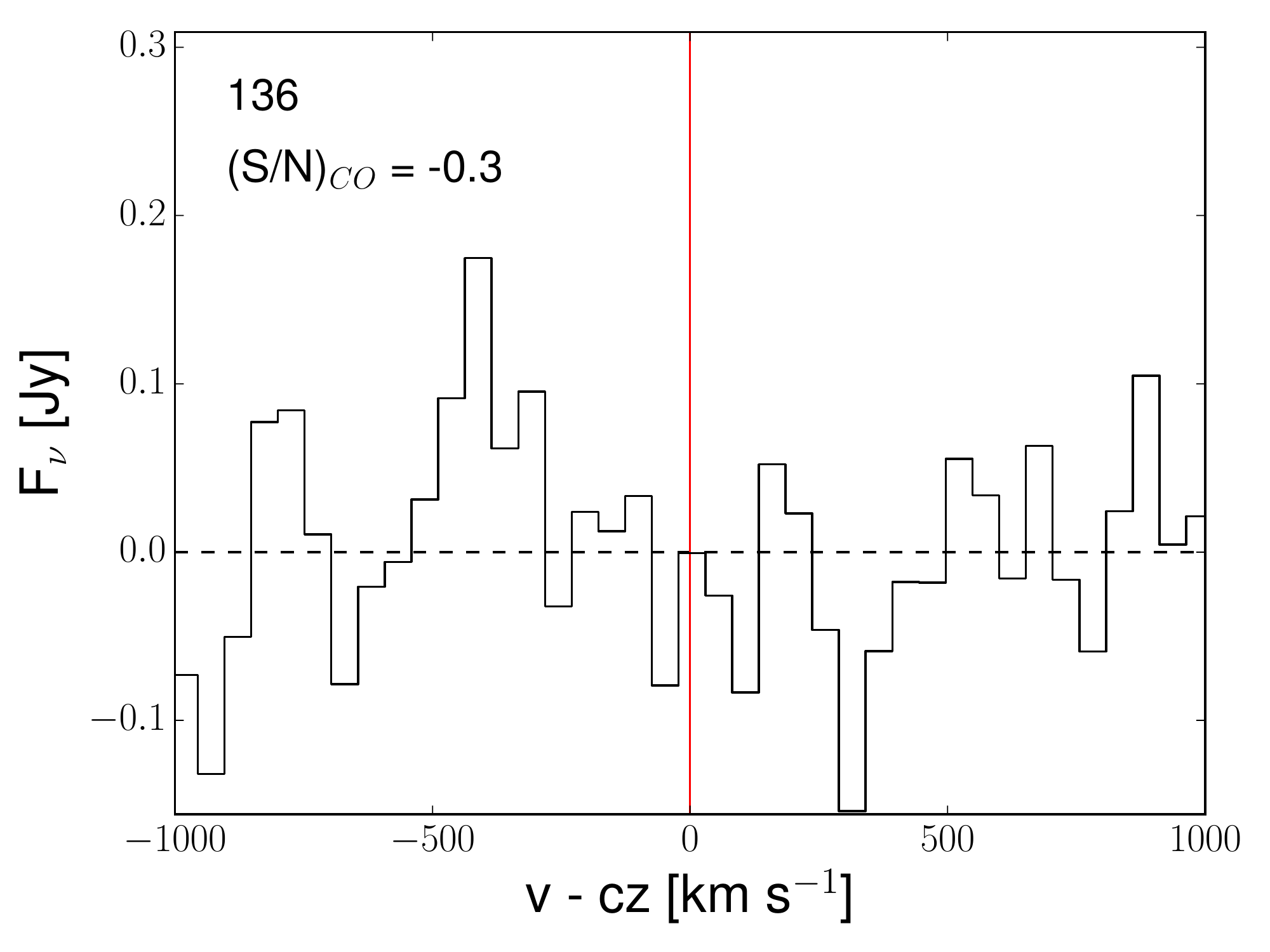}
\includegraphics[width=0.18\textwidth]{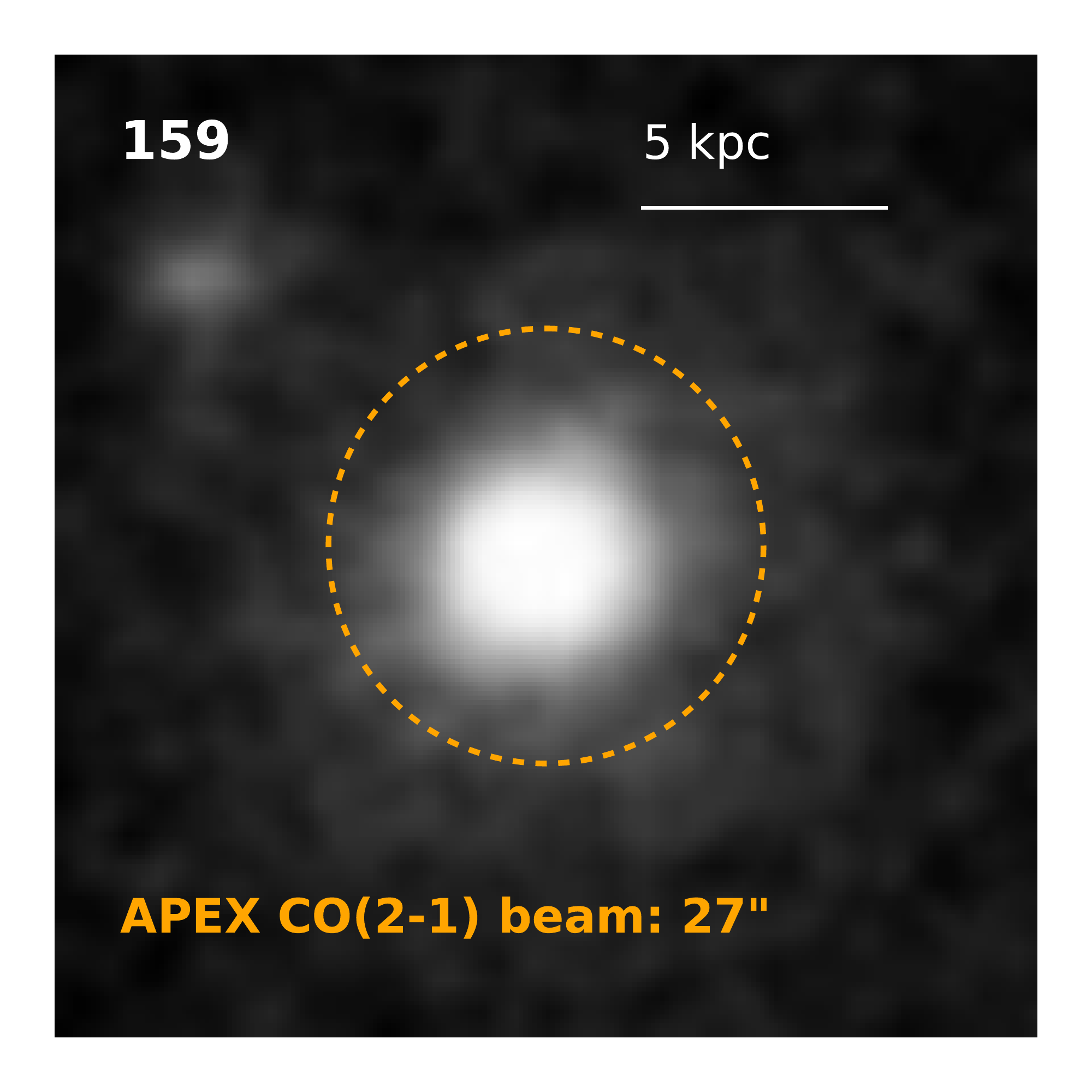}\includegraphics[width=0.26\textwidth]{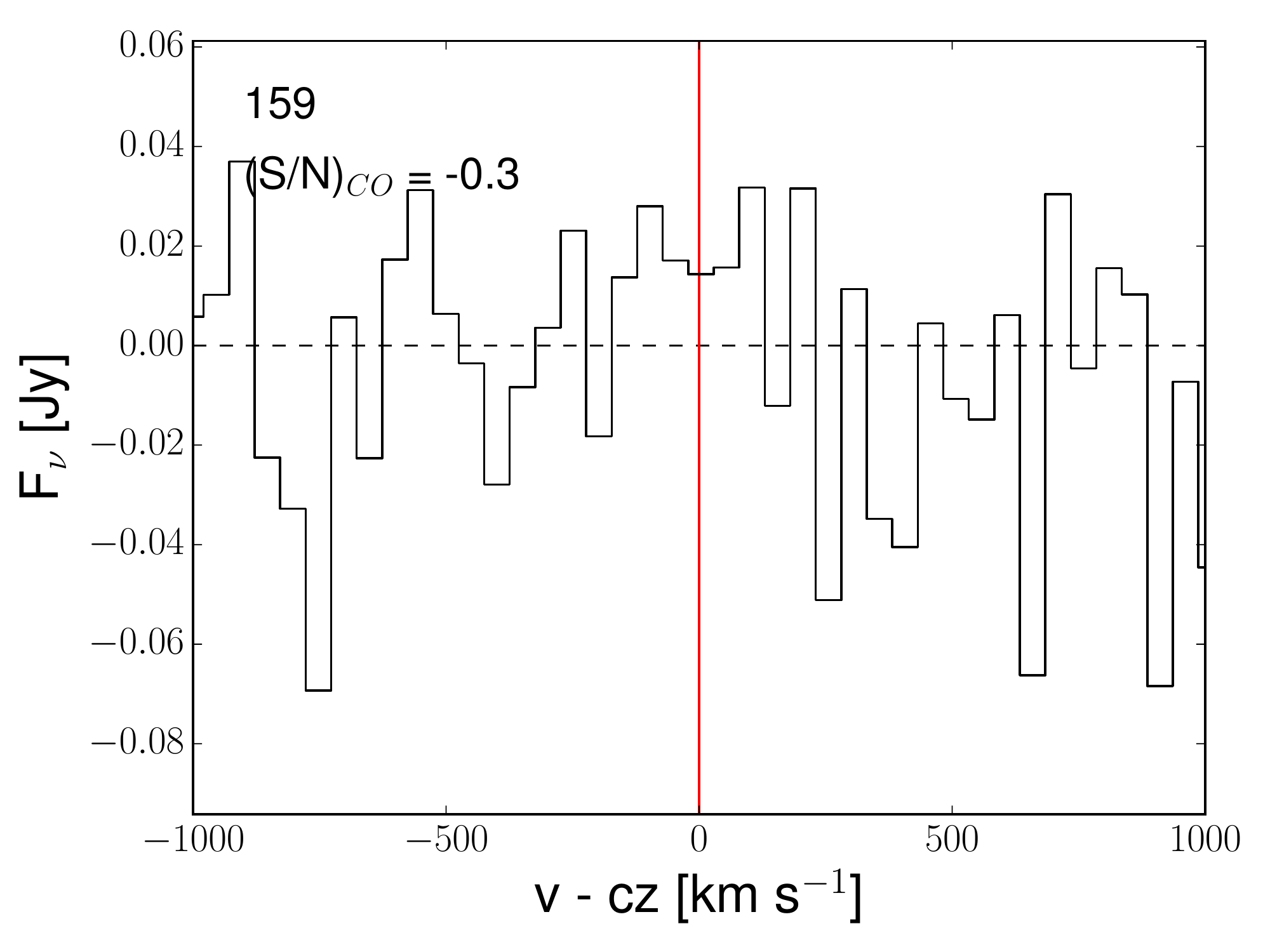}
\includegraphics[width=0.18\textwidth]{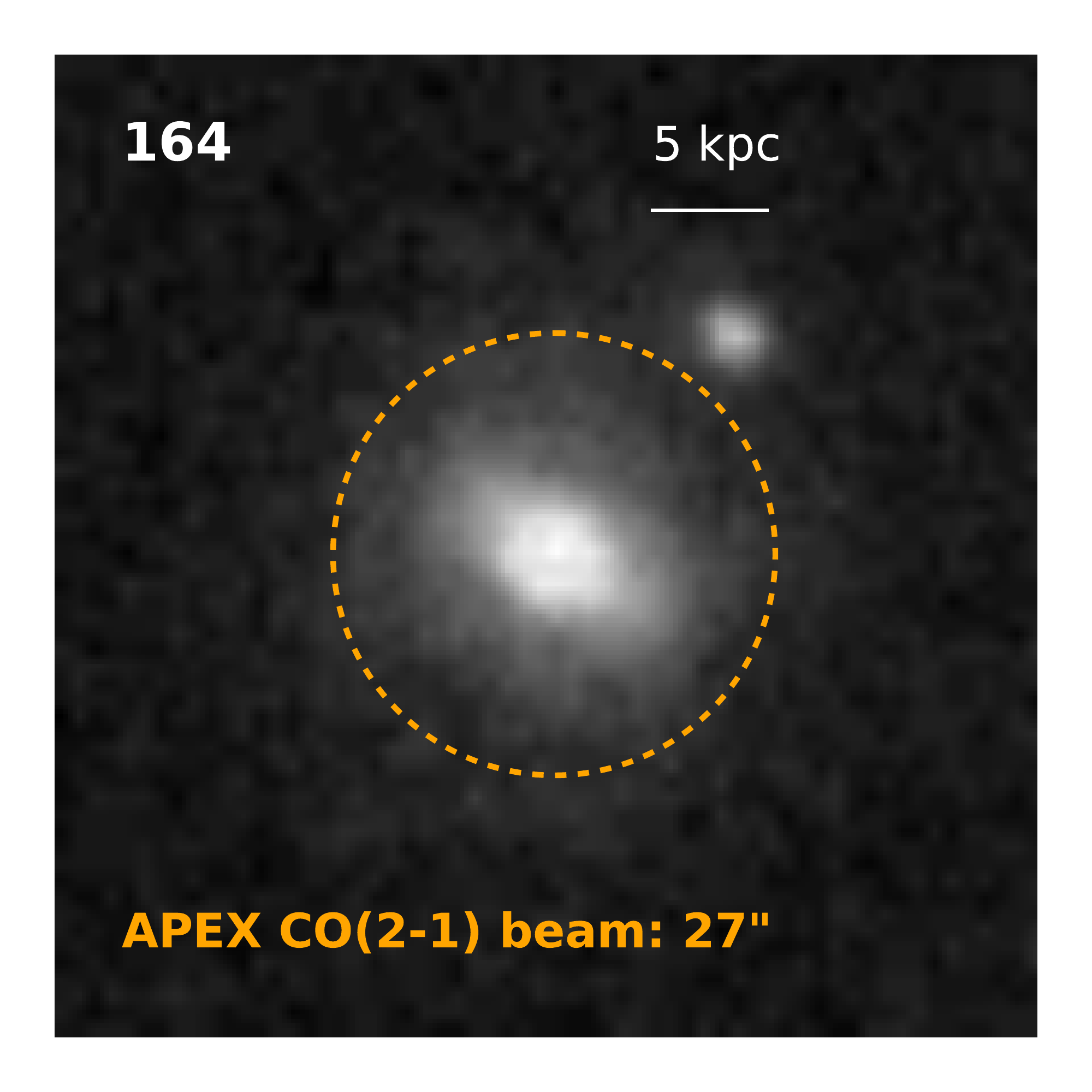}\includegraphics[width=0.26\textwidth]{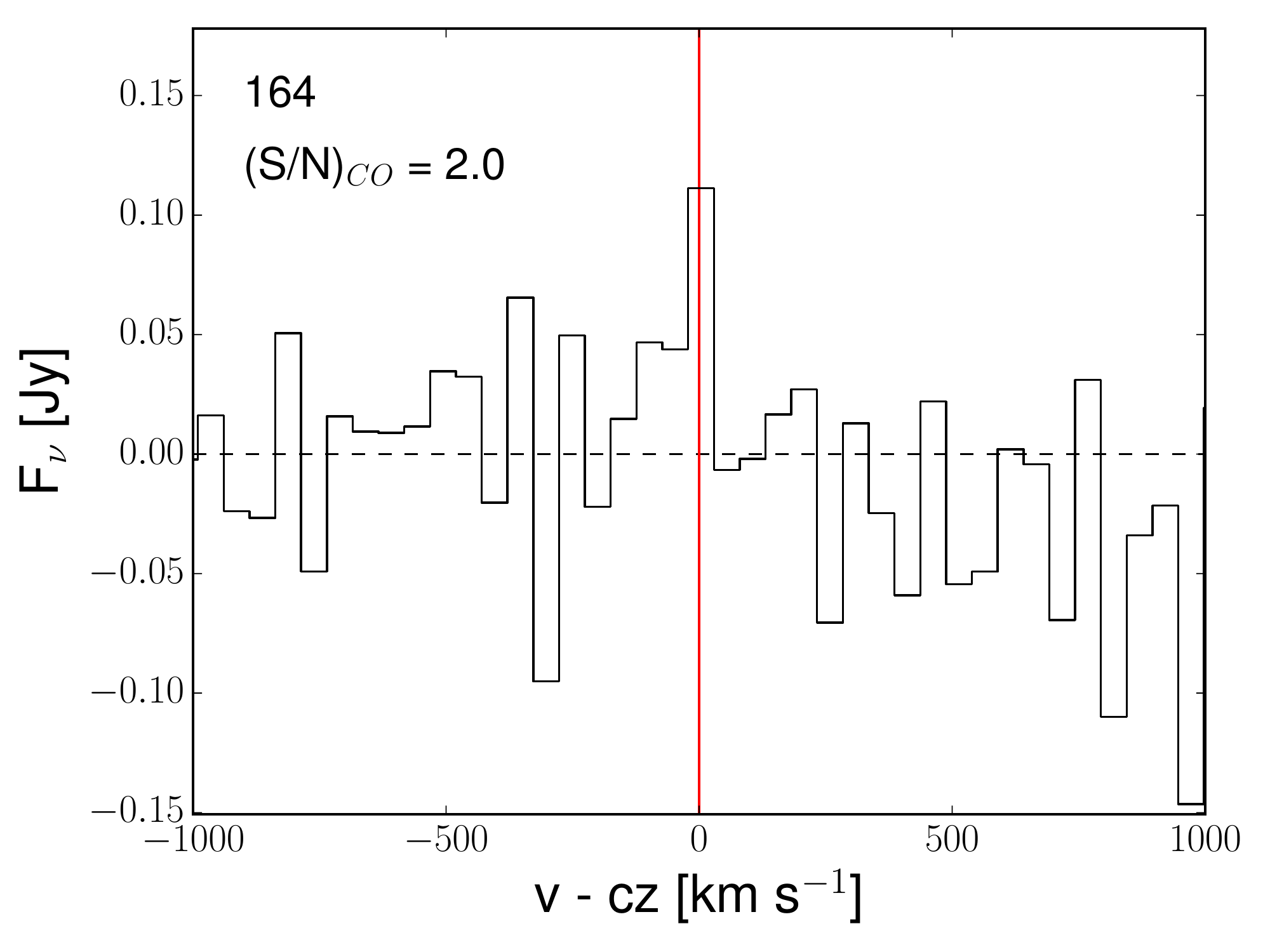}
\includegraphics[width=0.18\textwidth]{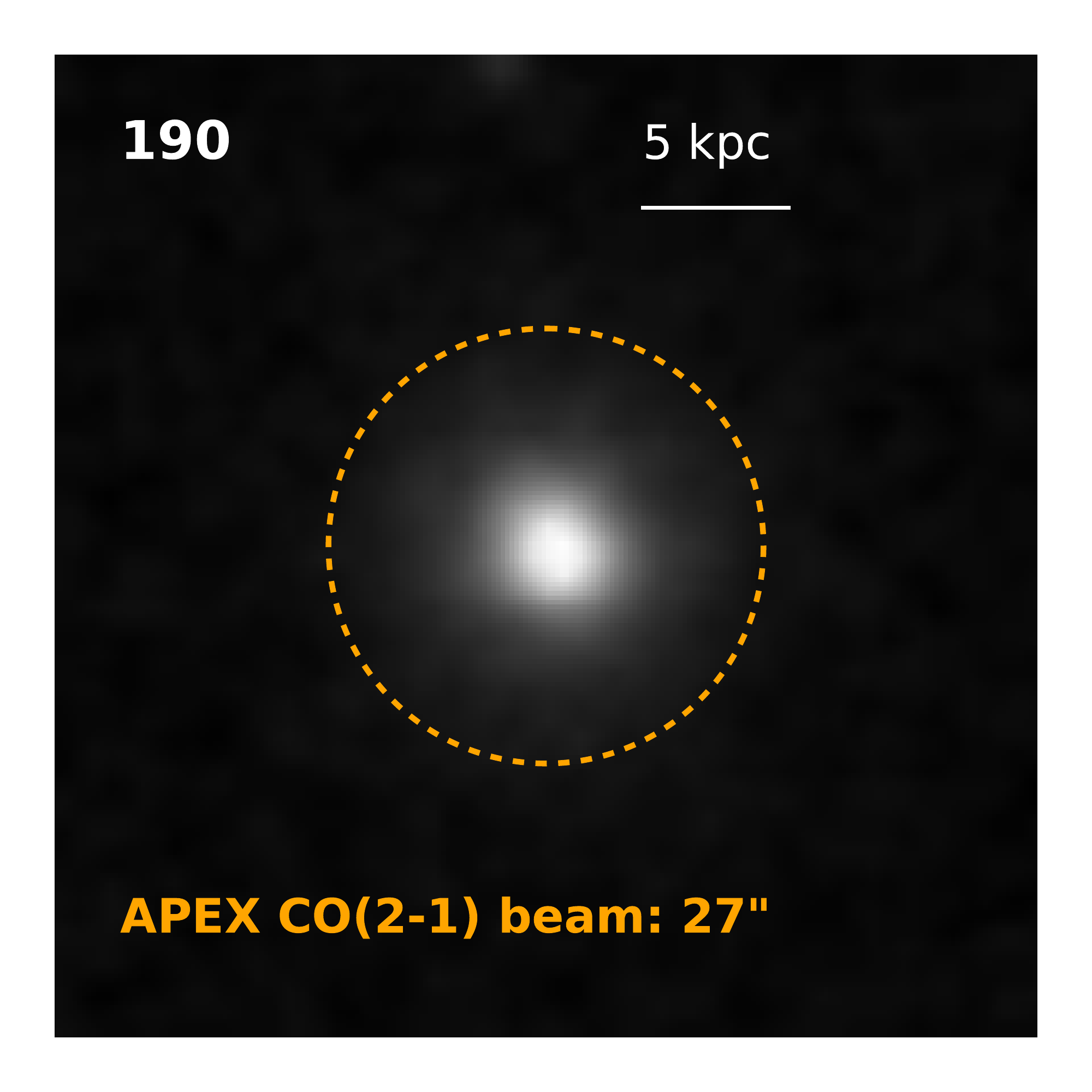}\includegraphics[width=0.26\textwidth]{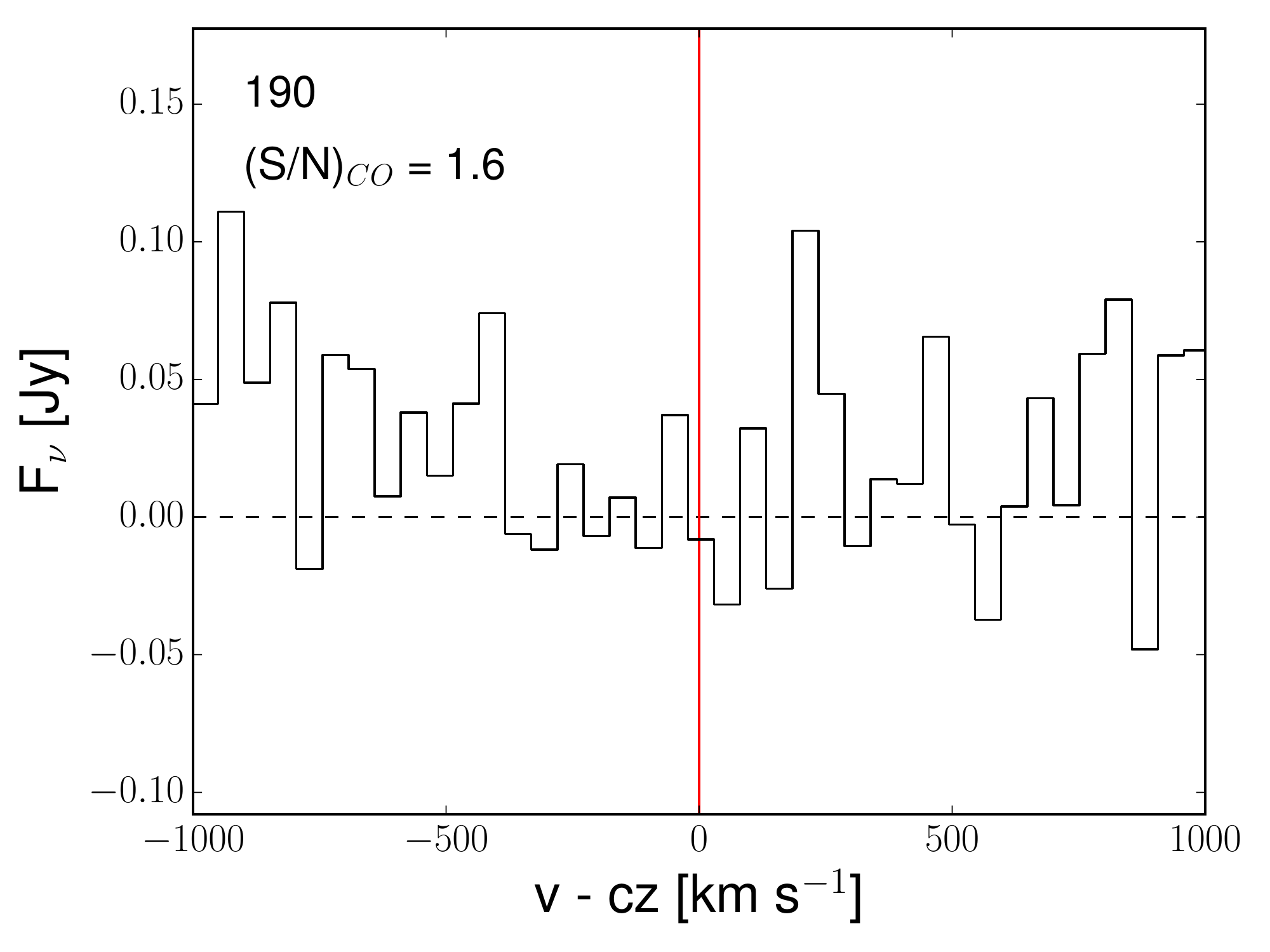}
\includegraphics[width=0.18\textwidth]{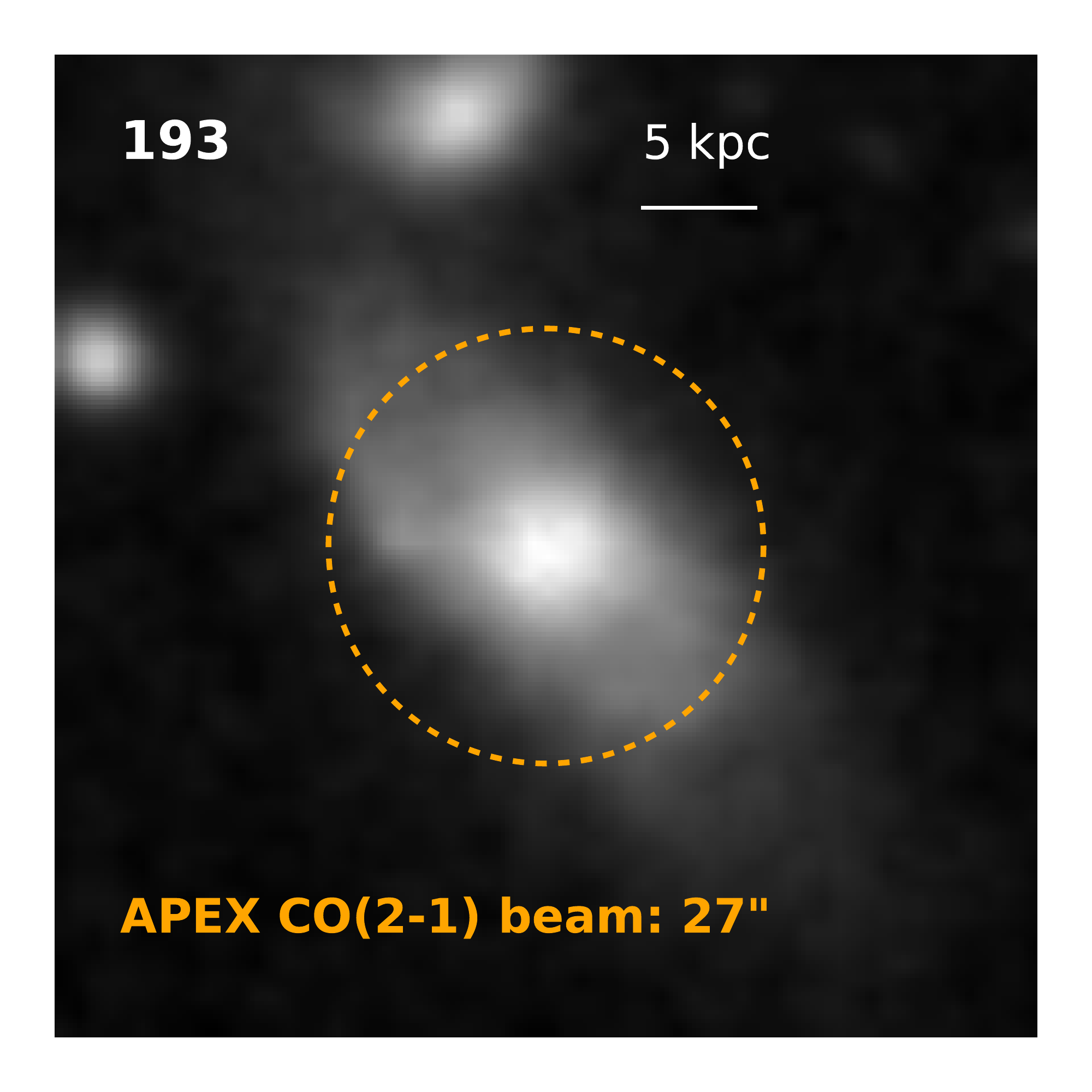}\includegraphics[width=0.26\textwidth]{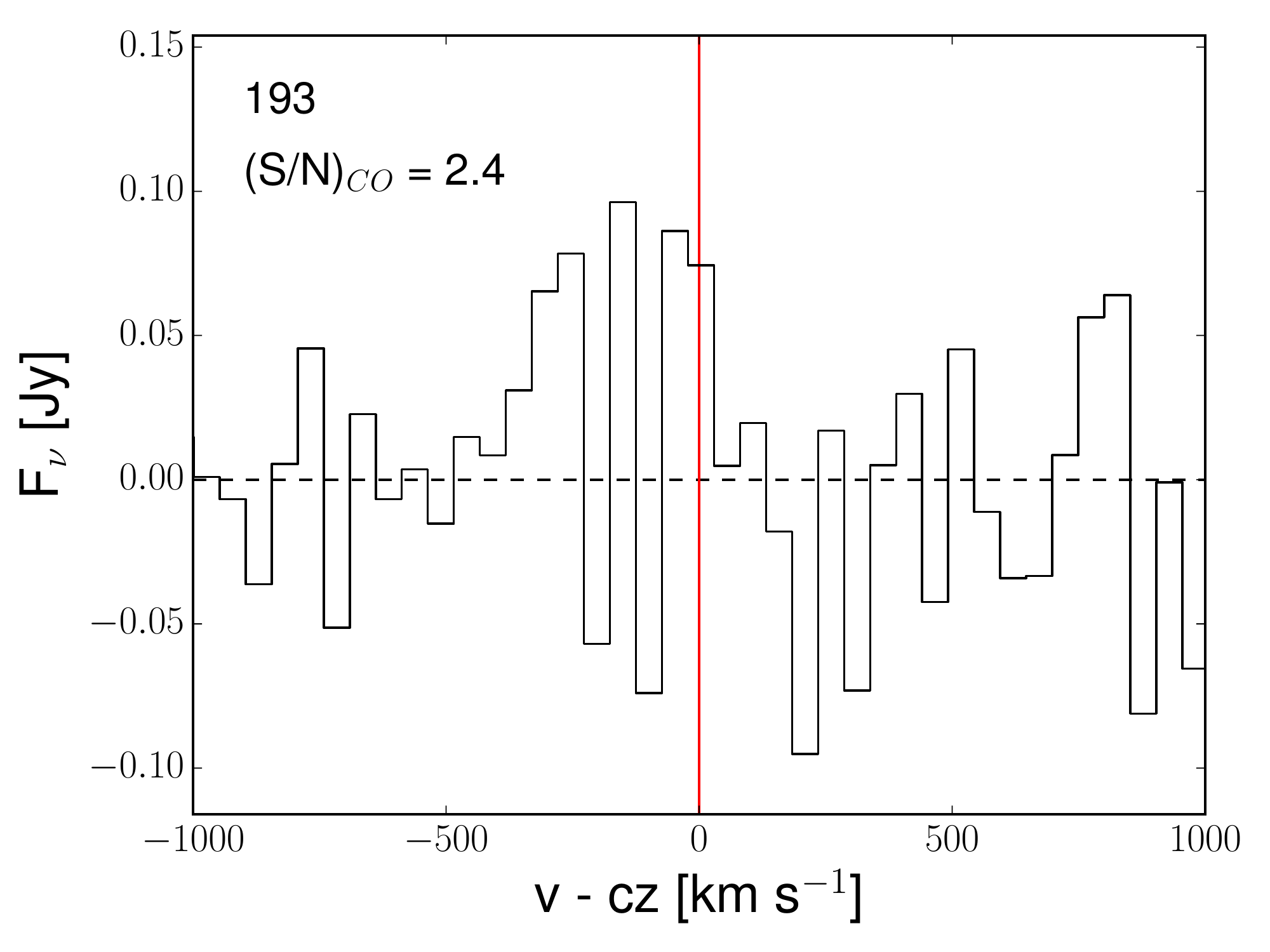}
\includegraphics[width=0.18\textwidth]{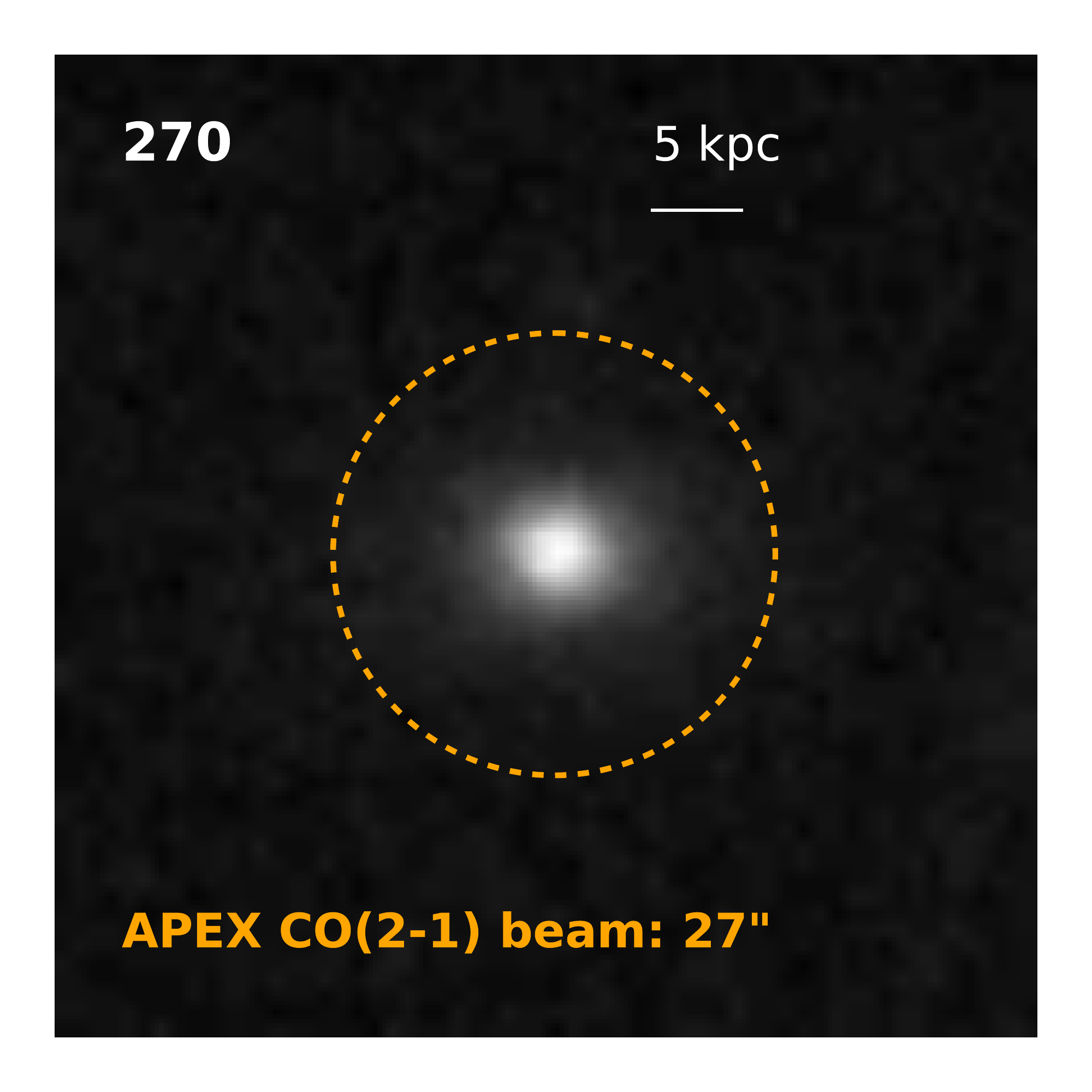}\includegraphics[width=0.26\textwidth]{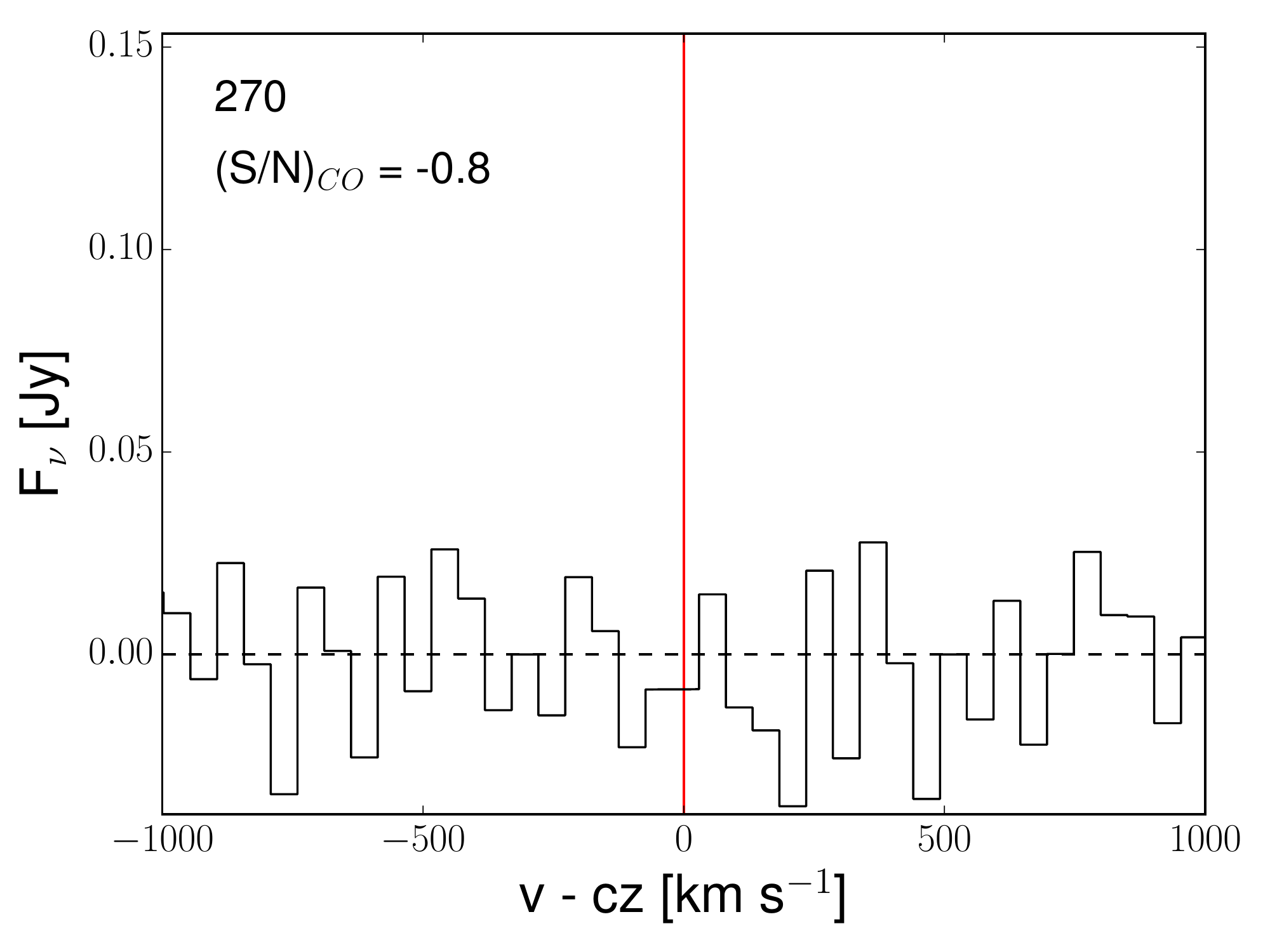}
\includegraphics[width=0.18\textwidth]{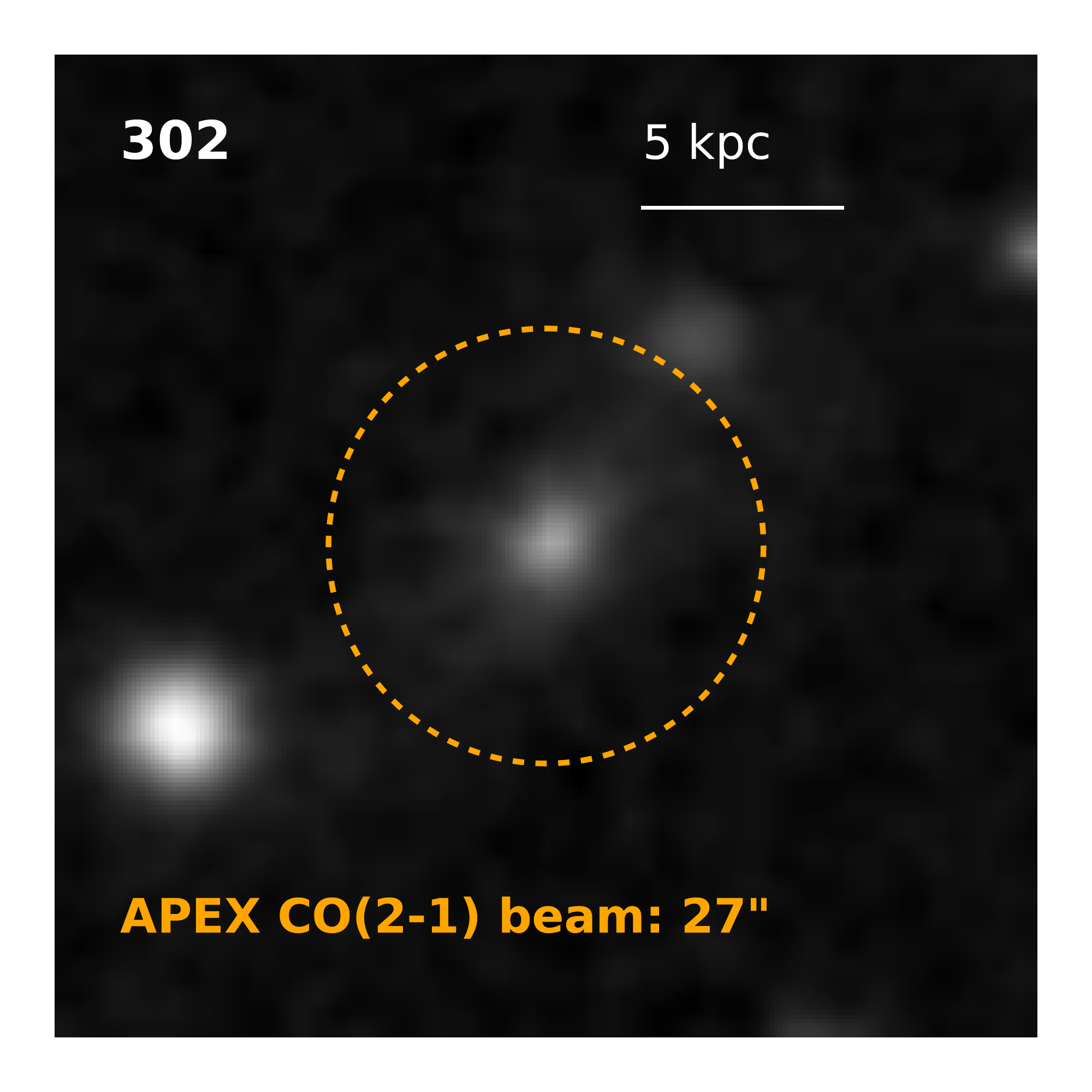}\includegraphics[width=0.26\textwidth]{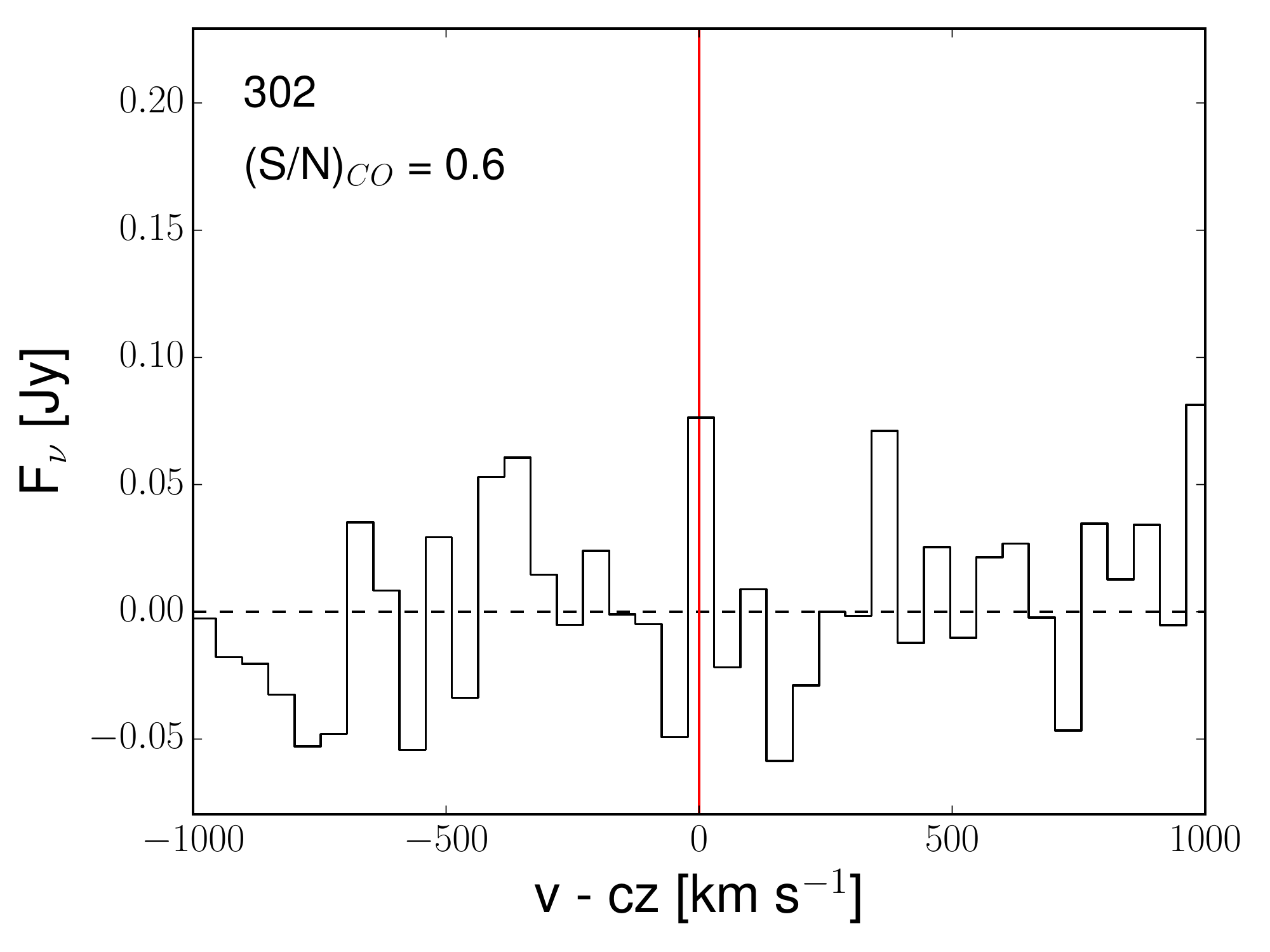}
\includegraphics[width=0.18\textwidth]{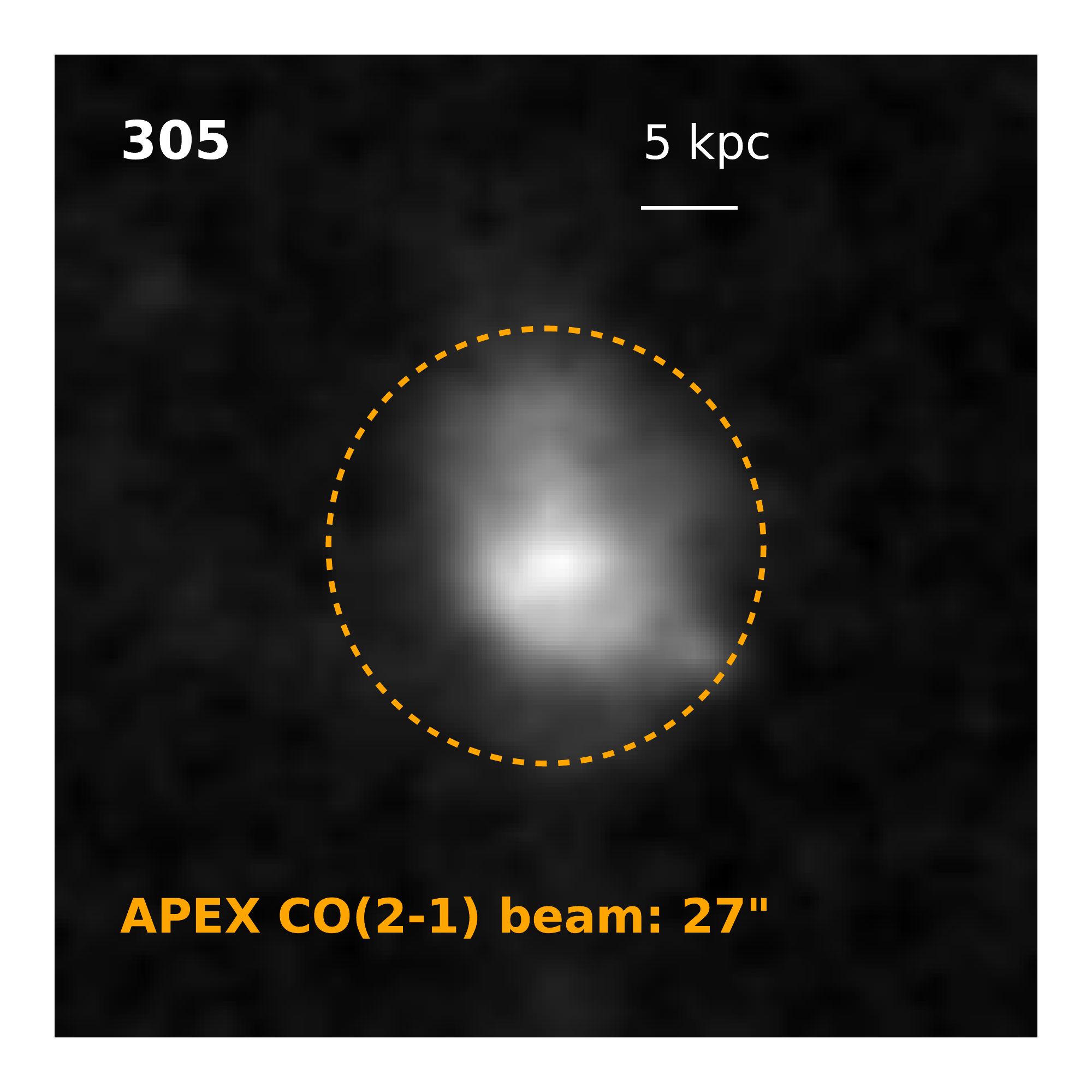}\includegraphics[width=0.26\textwidth]{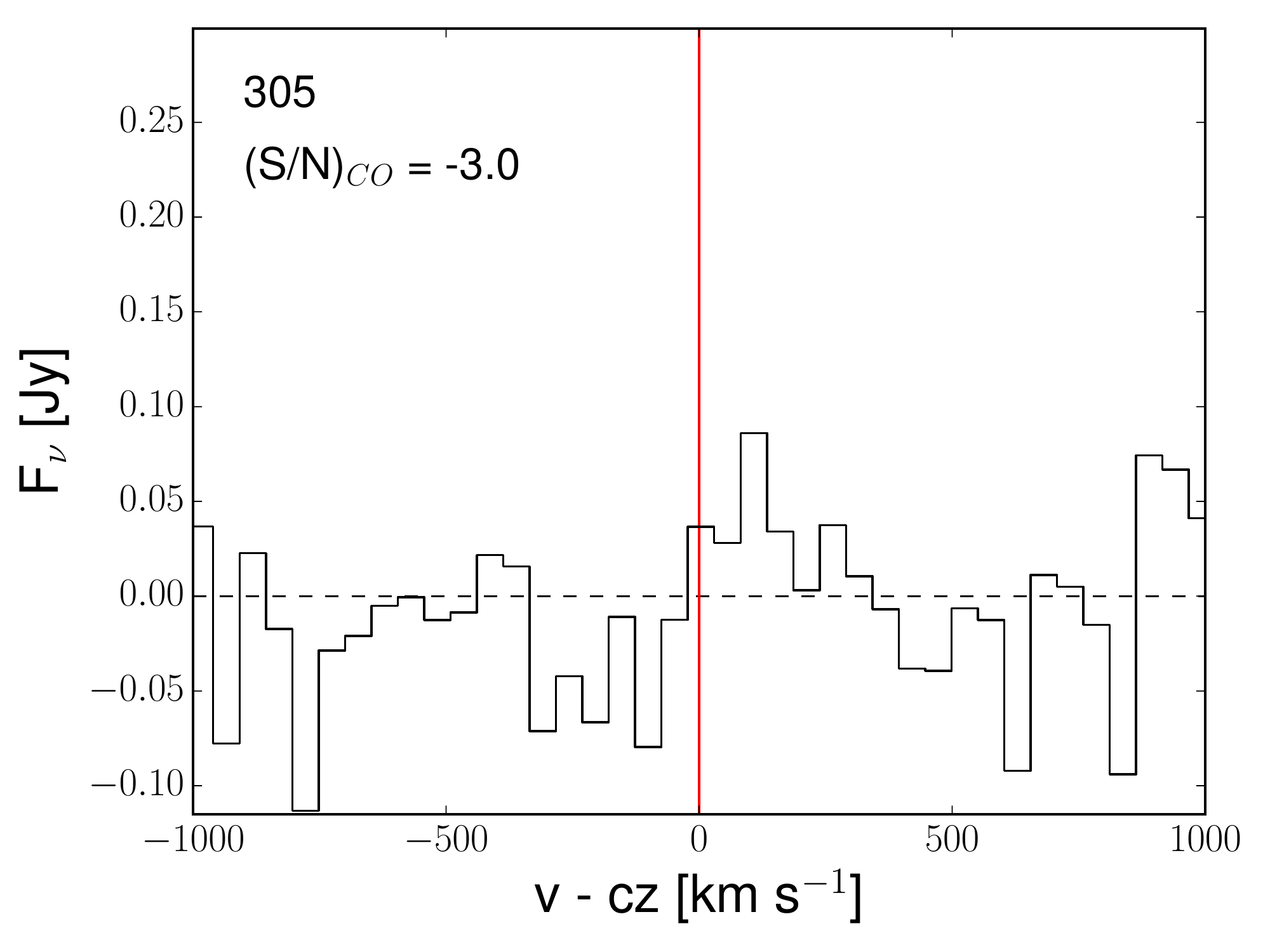}
\includegraphics[width=0.18\textwidth]{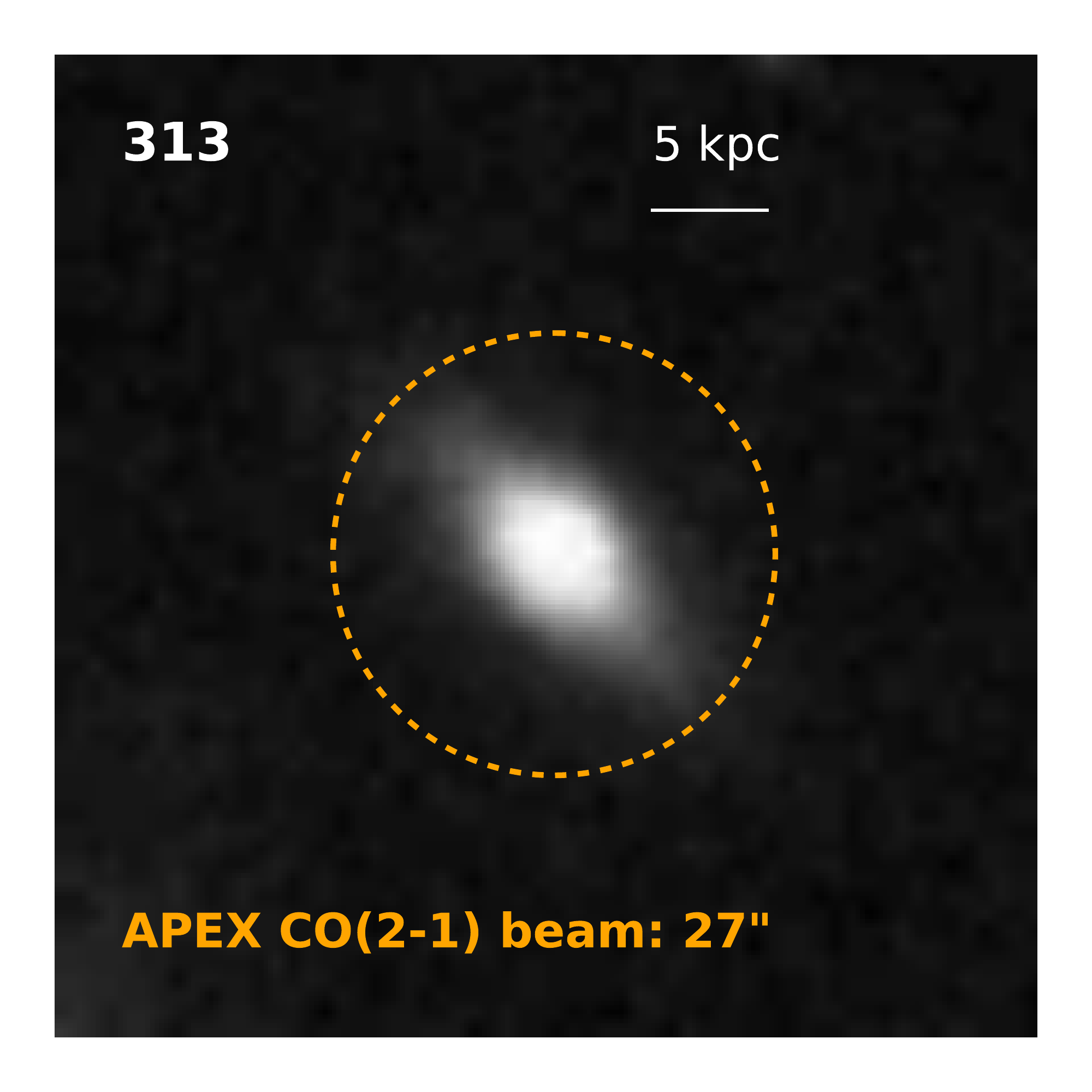}\includegraphics[width=0.26\textwidth]{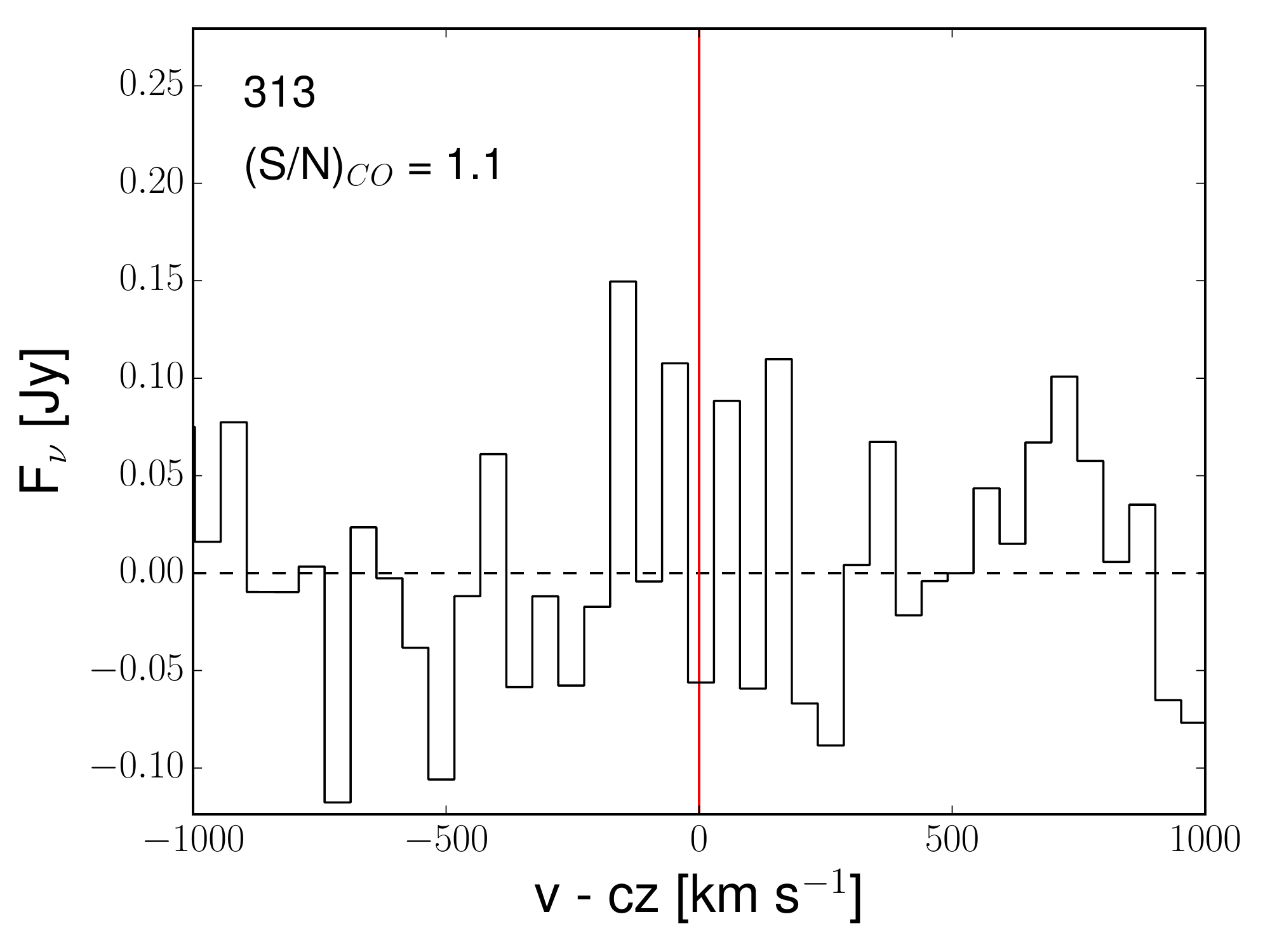}
\includegraphics[width=0.18\textwidth]{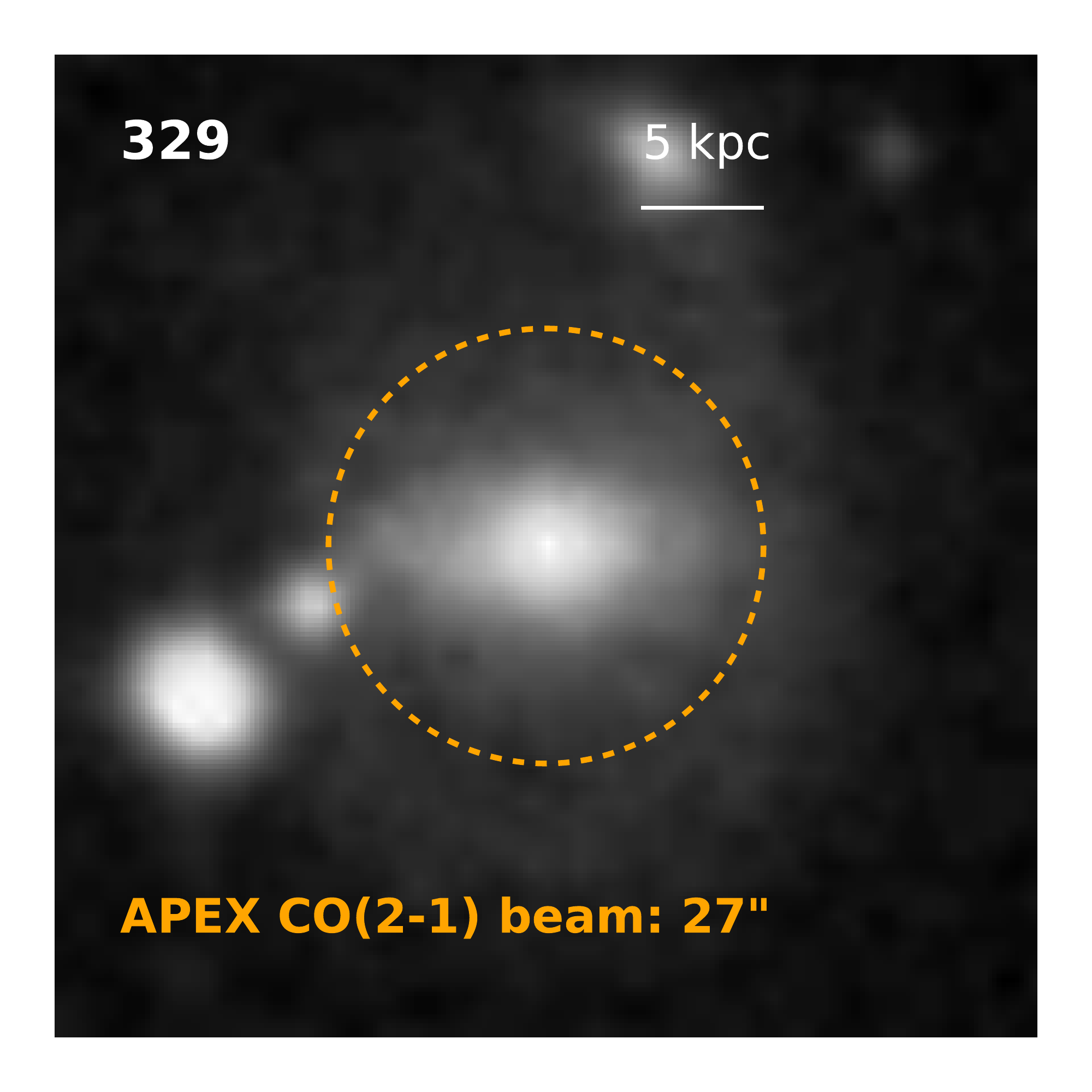}\includegraphics[width=0.26\textwidth]{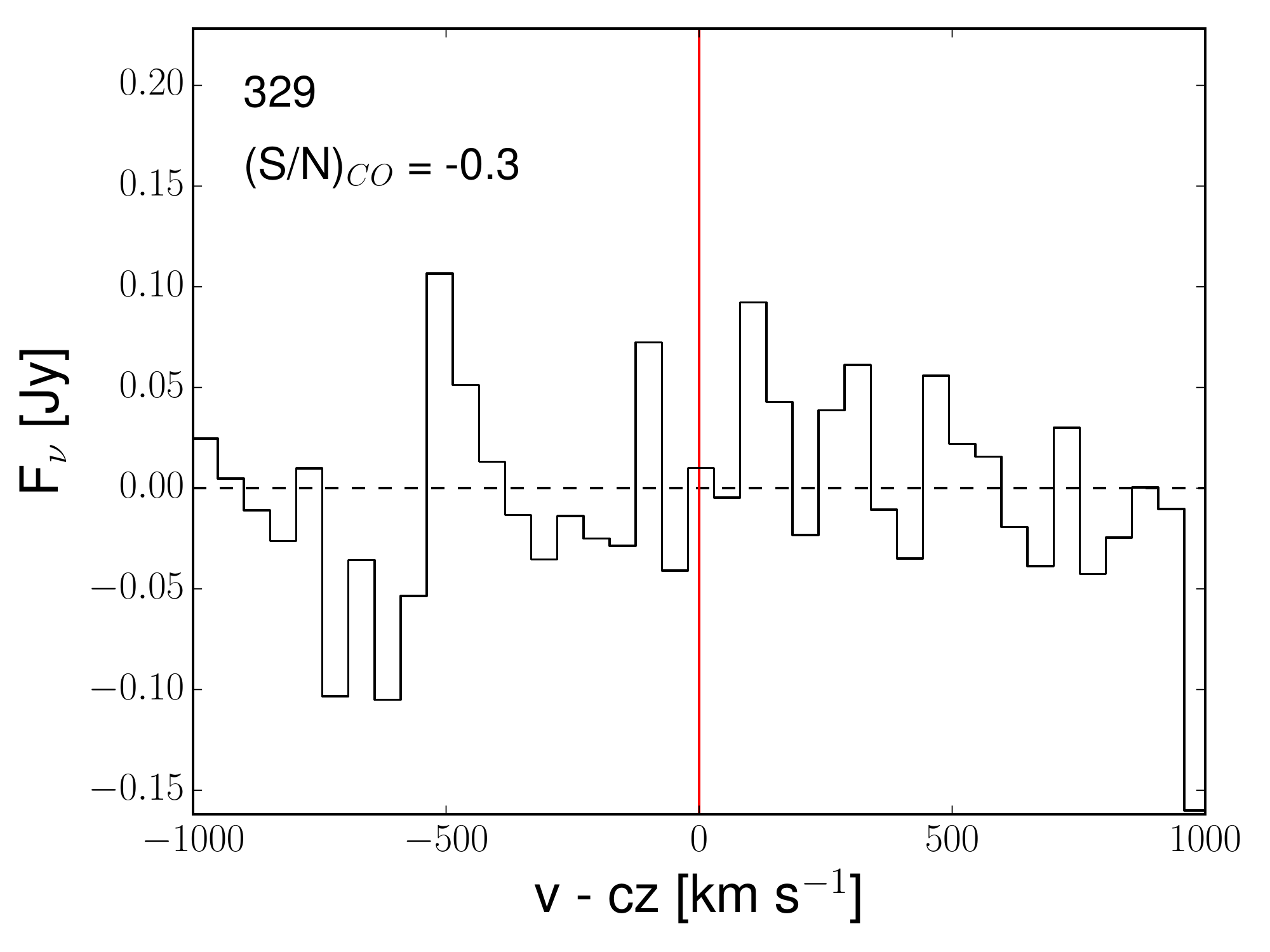}
\caption{Similar to Fig.~\ref{fig:CO21_spectra}, for undetected sources with only DSS imaging.
} 
\label{fig:CO21_spectra_dss5}
\end{figure*}

\begin{figure*}
\centering
\raggedright
\includegraphics[width=0.18\textwidth]{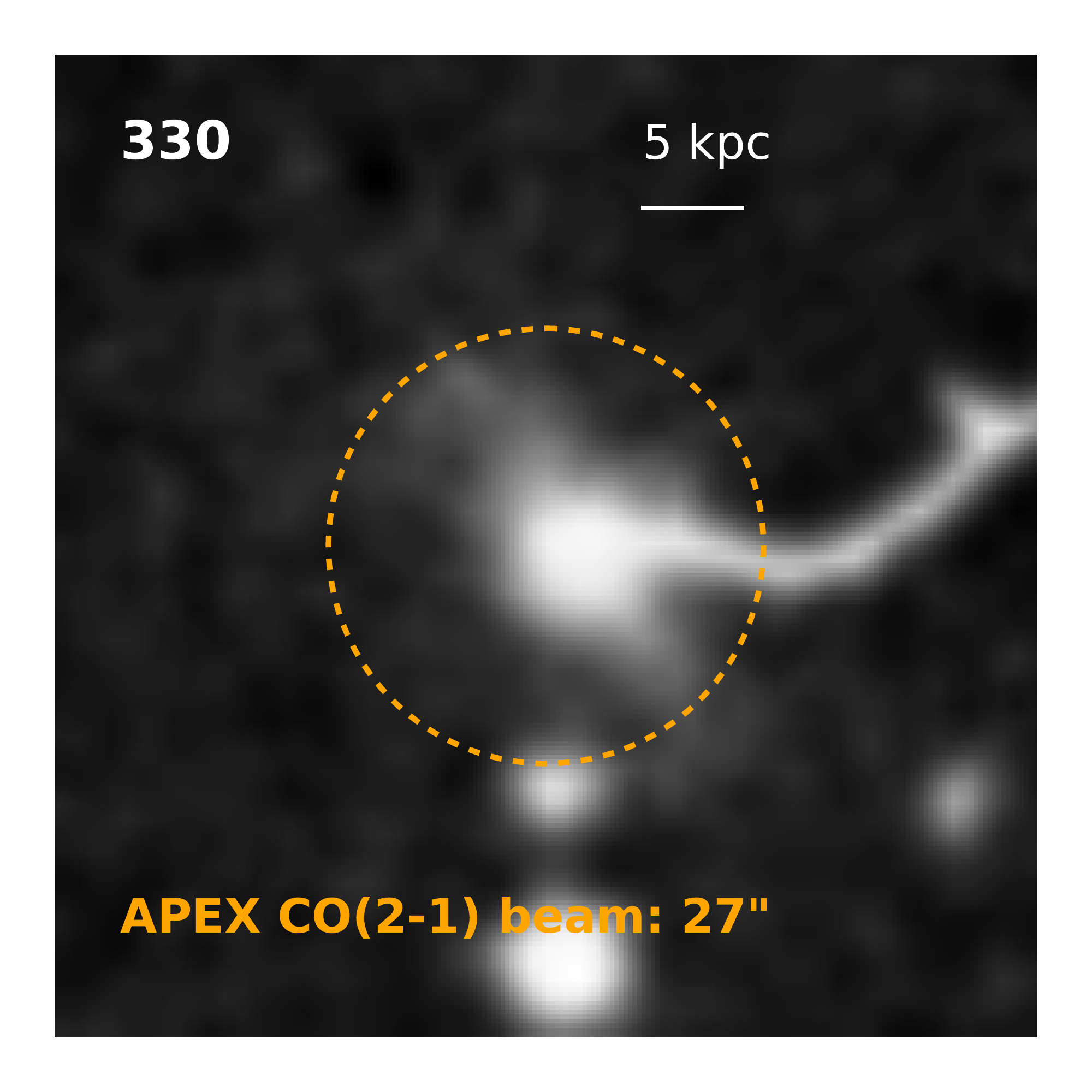}\includegraphics[width=0.26\textwidth]{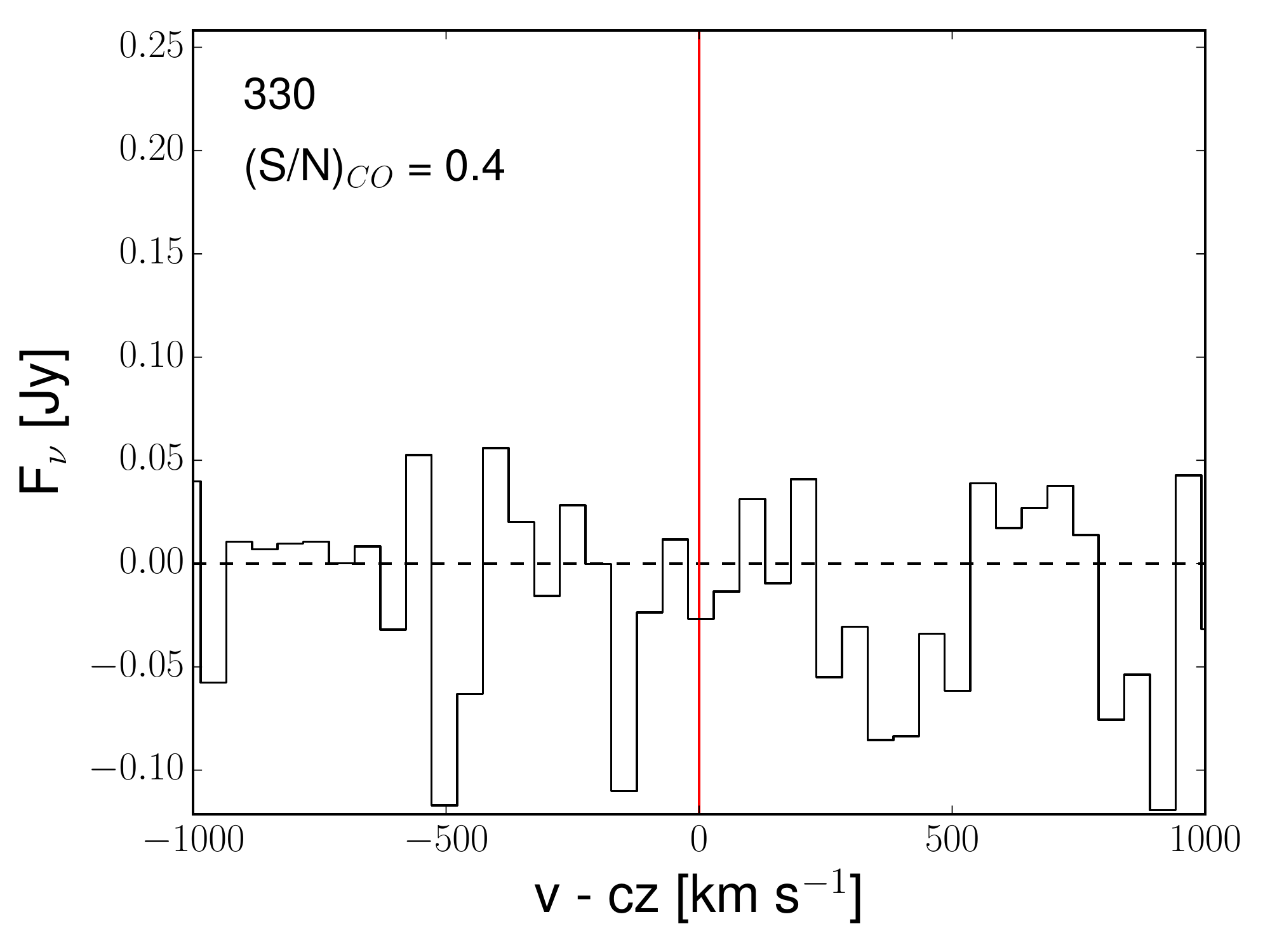}
\includegraphics[width=0.18\textwidth]{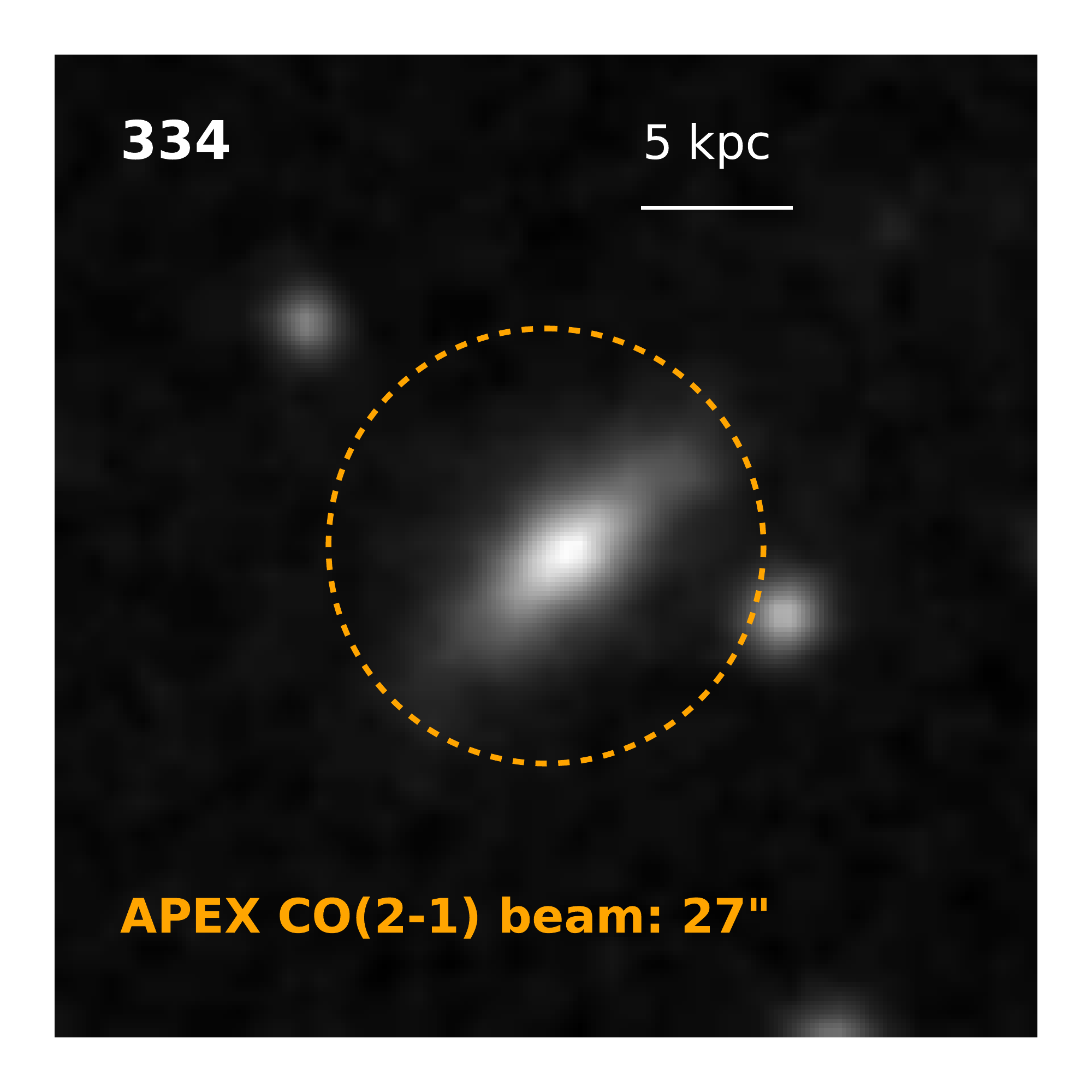}\includegraphics[width=0.26\textwidth]{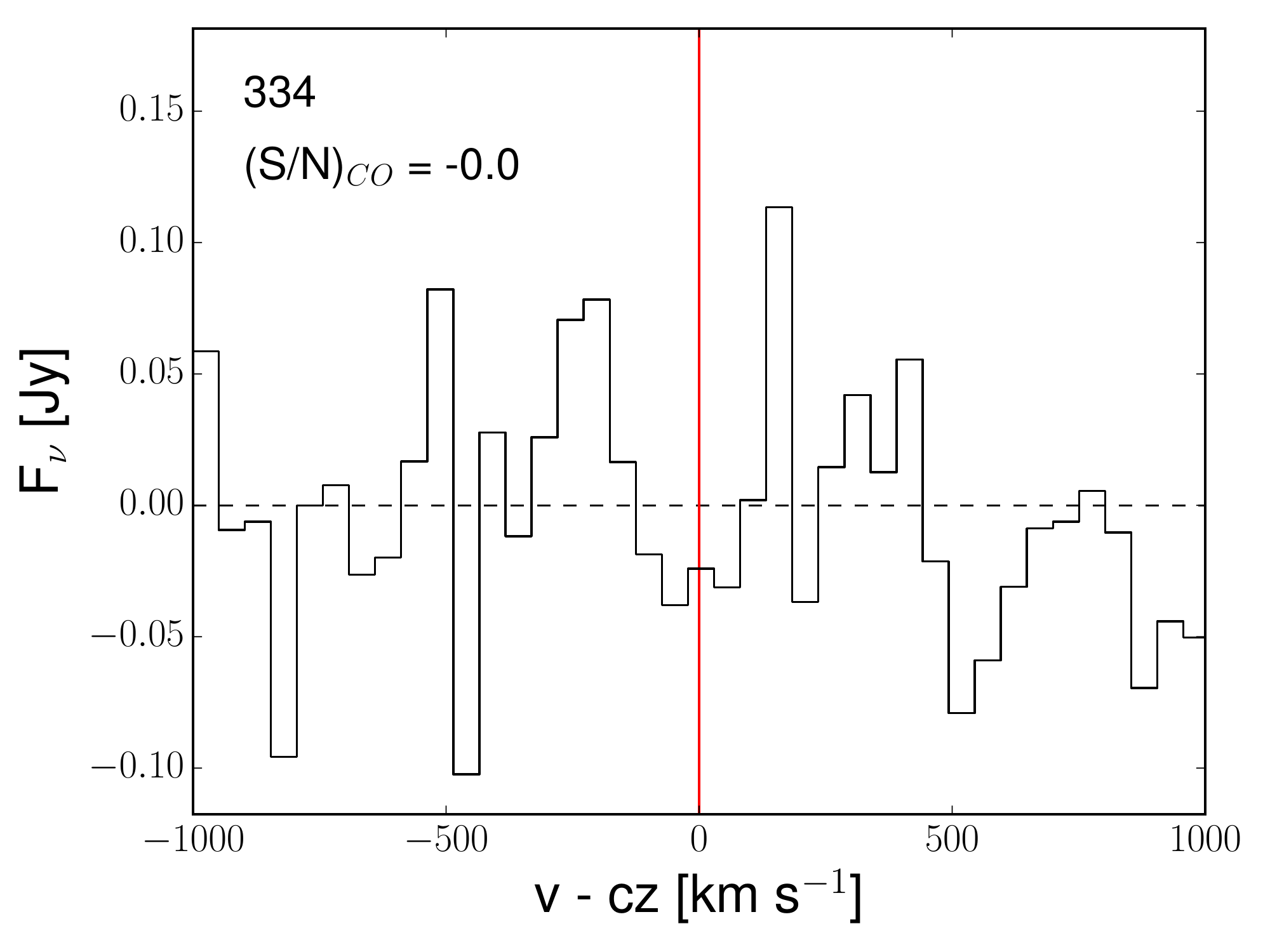}
\includegraphics[width=0.18\textwidth]{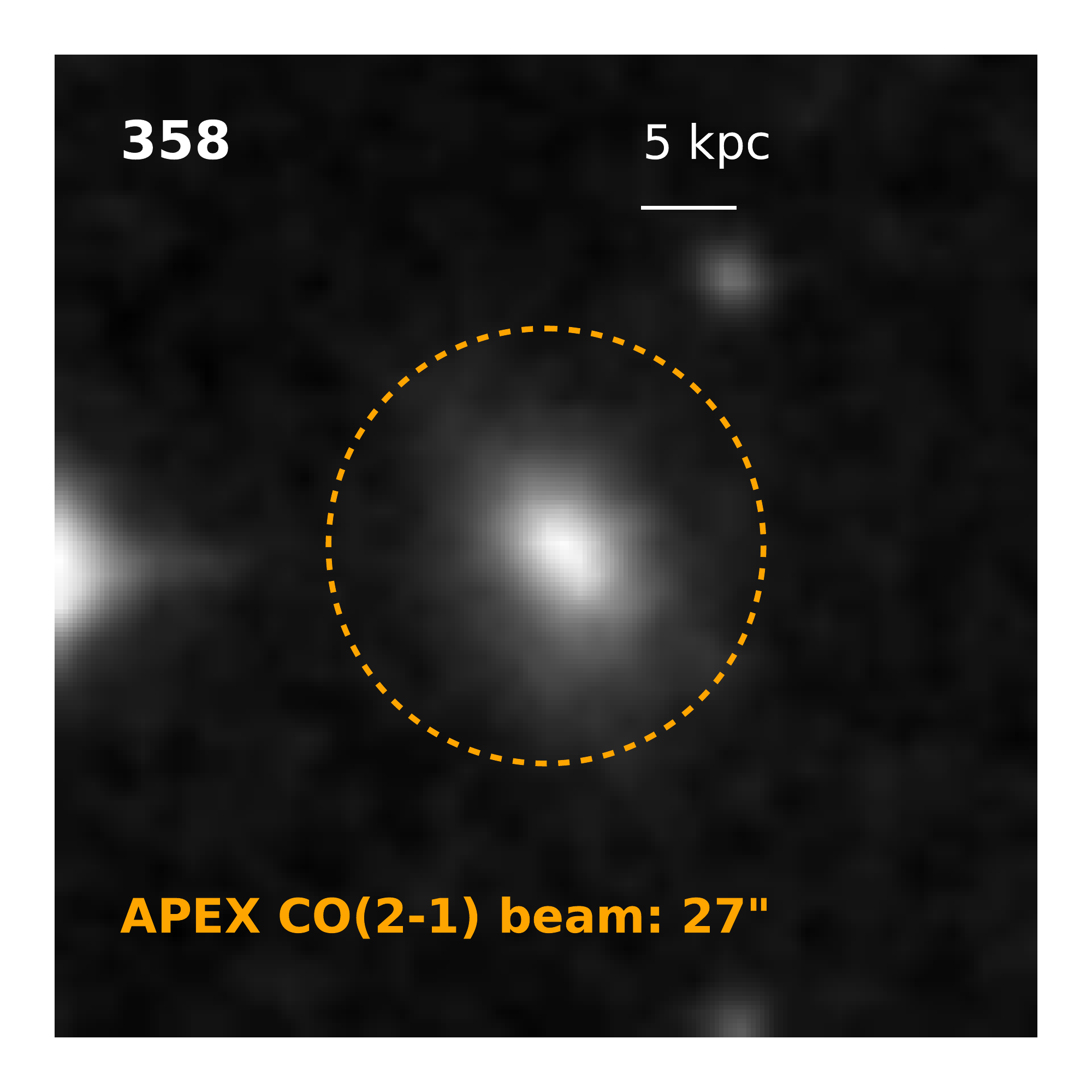}\includegraphics[width=0.26\textwidth]{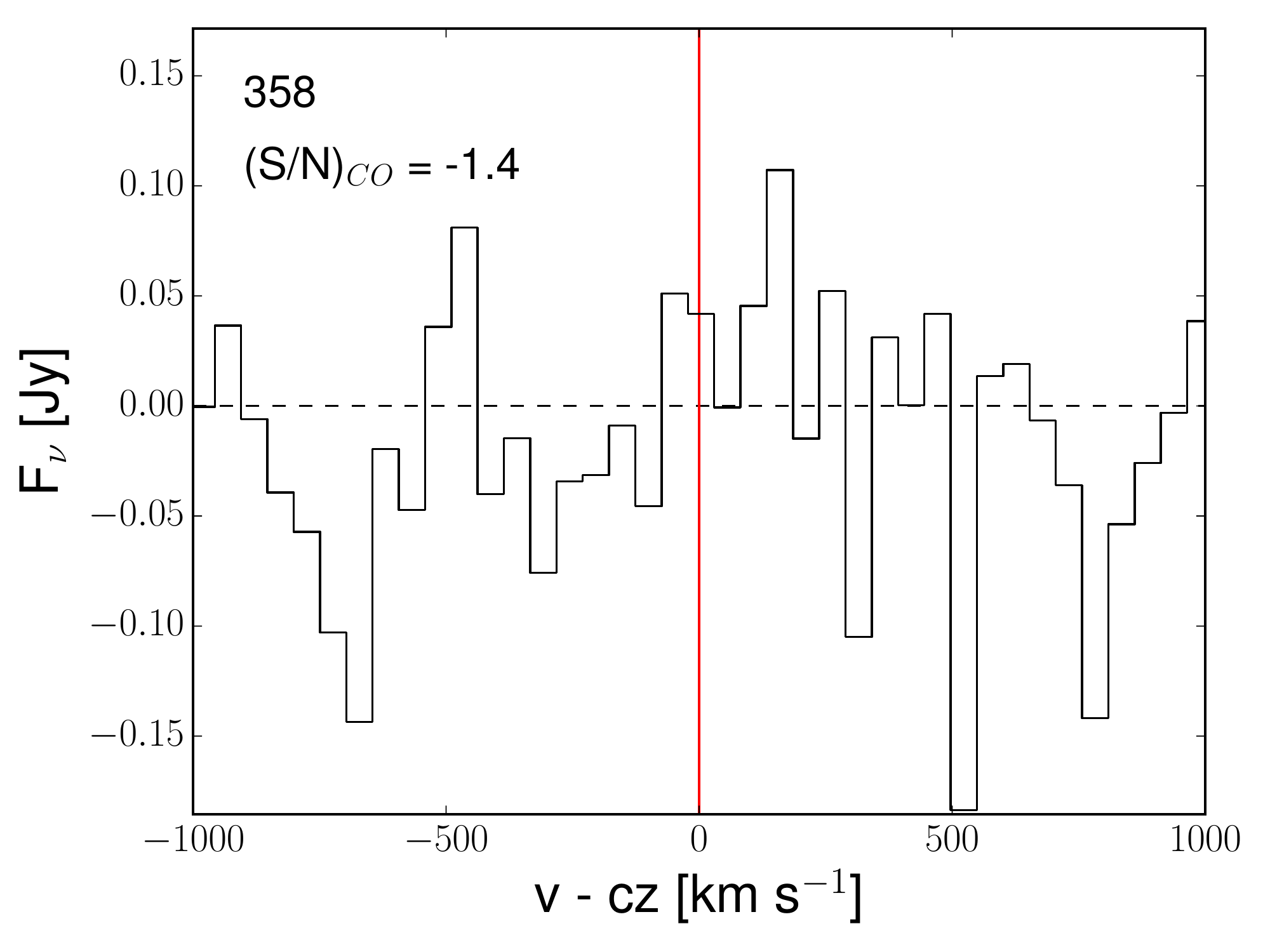}
\includegraphics[width=0.18\textwidth]{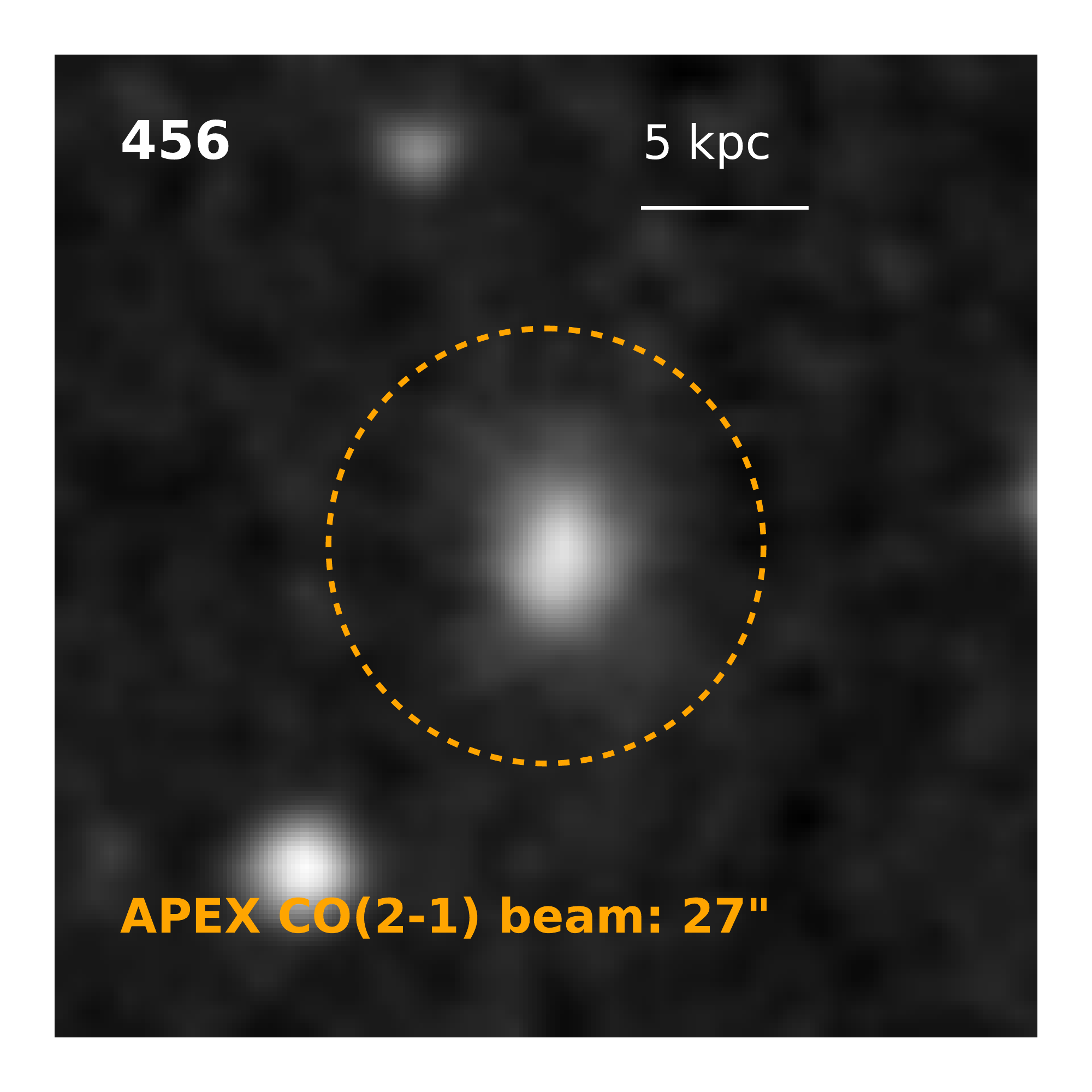}\includegraphics[width=0.26\textwidth]{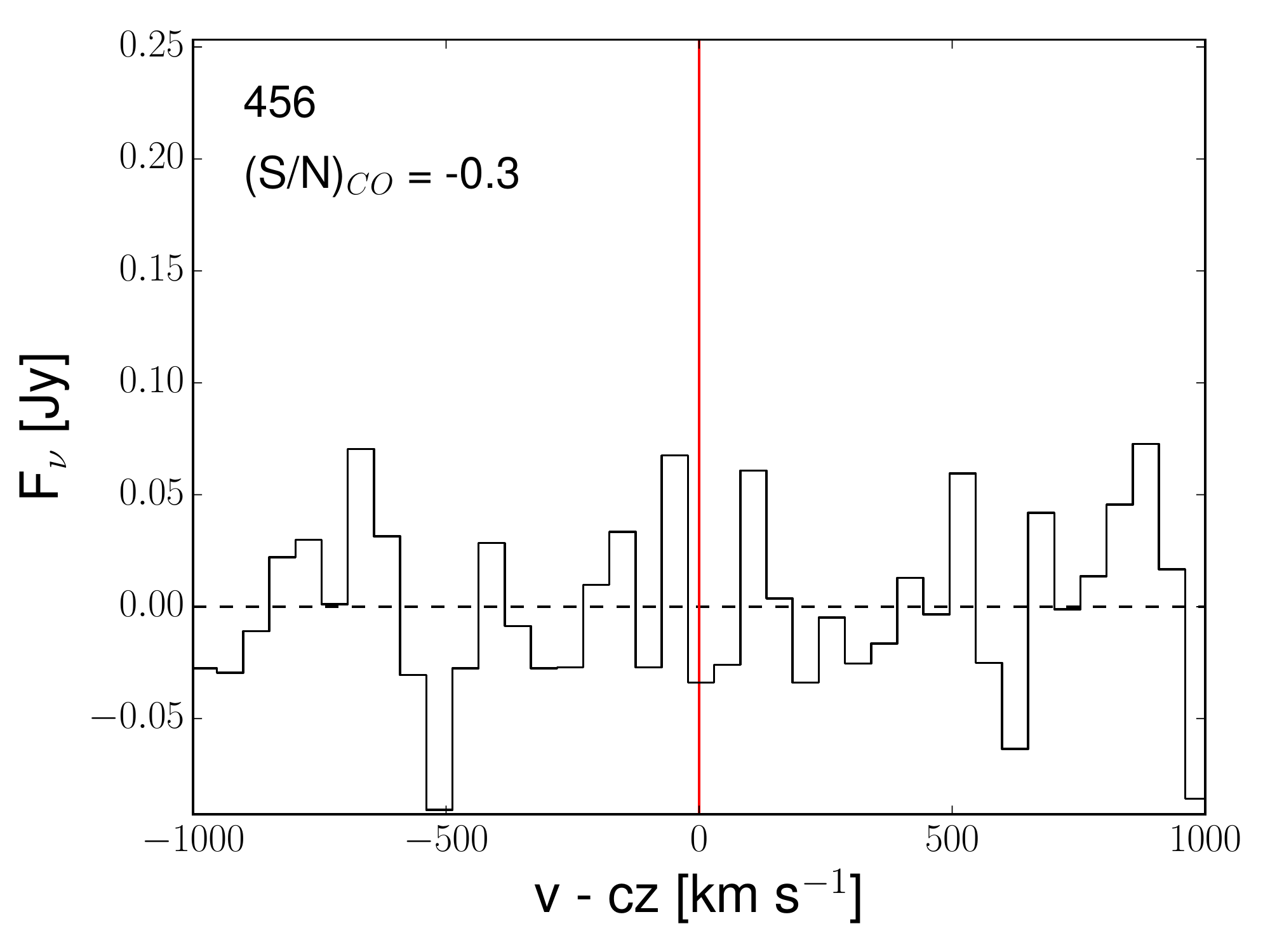}
\includegraphics[width=0.18\textwidth]{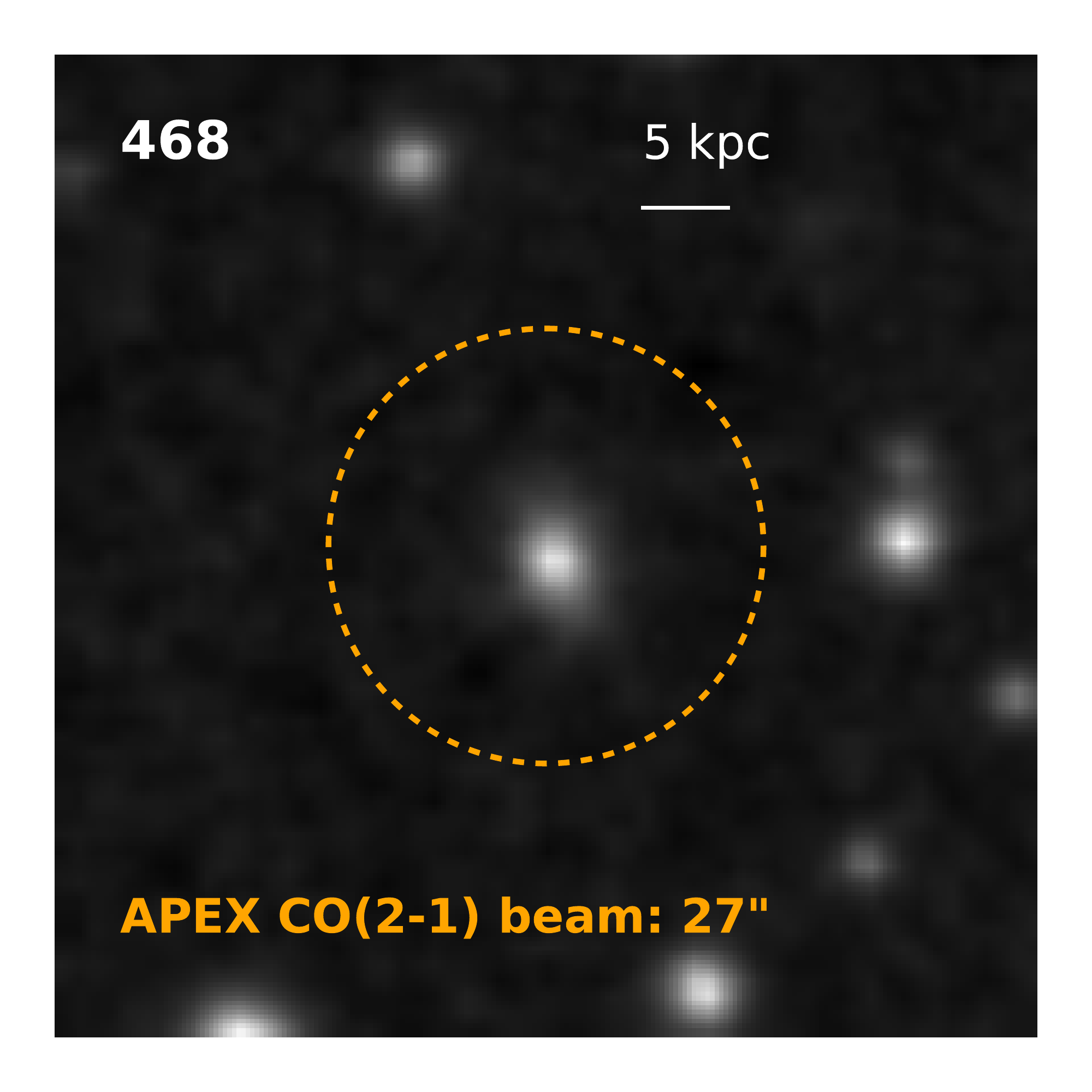}\includegraphics[width=0.26\textwidth]{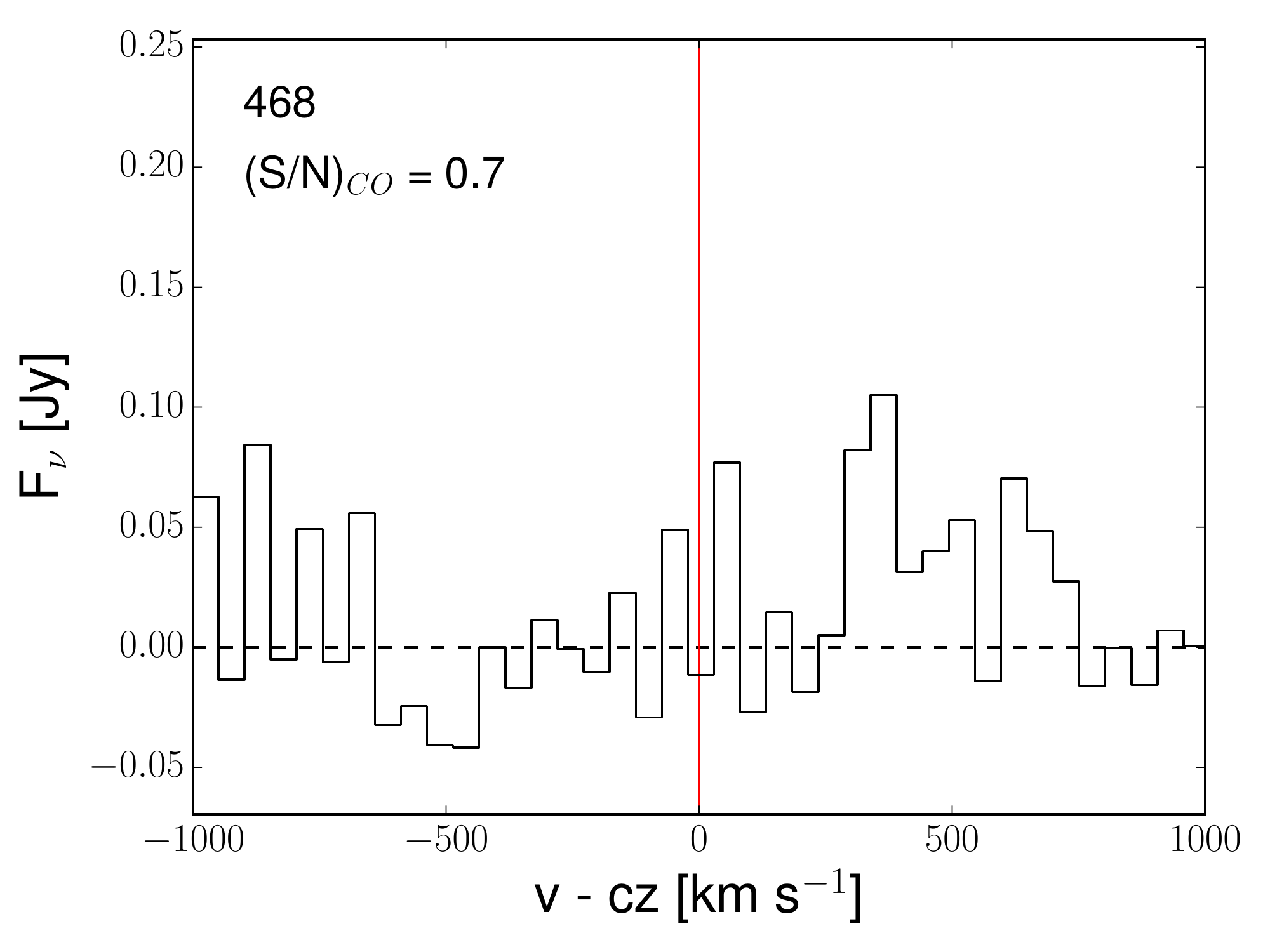}
\includegraphics[width=0.18\textwidth]{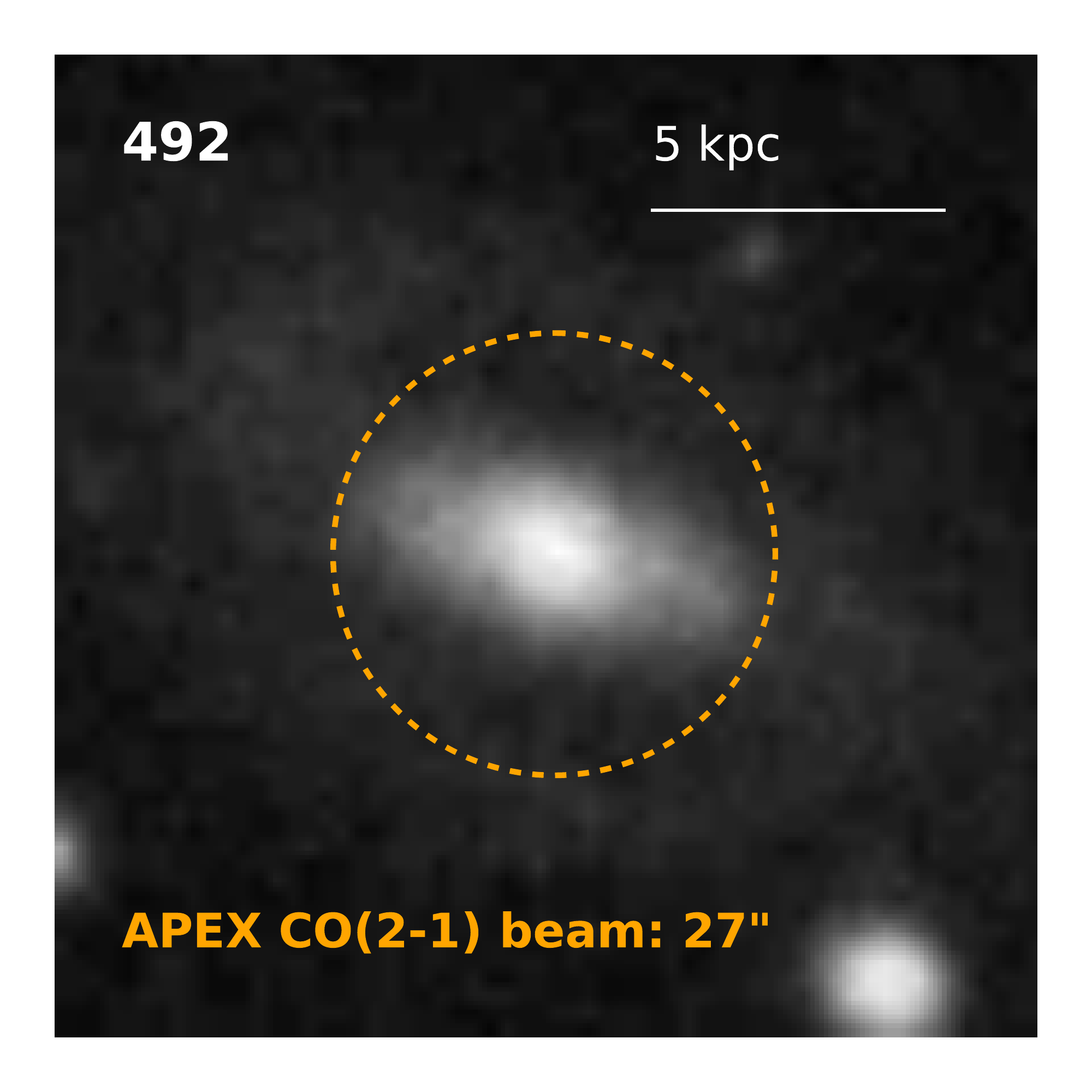}\includegraphics[width=0.26\textwidth]{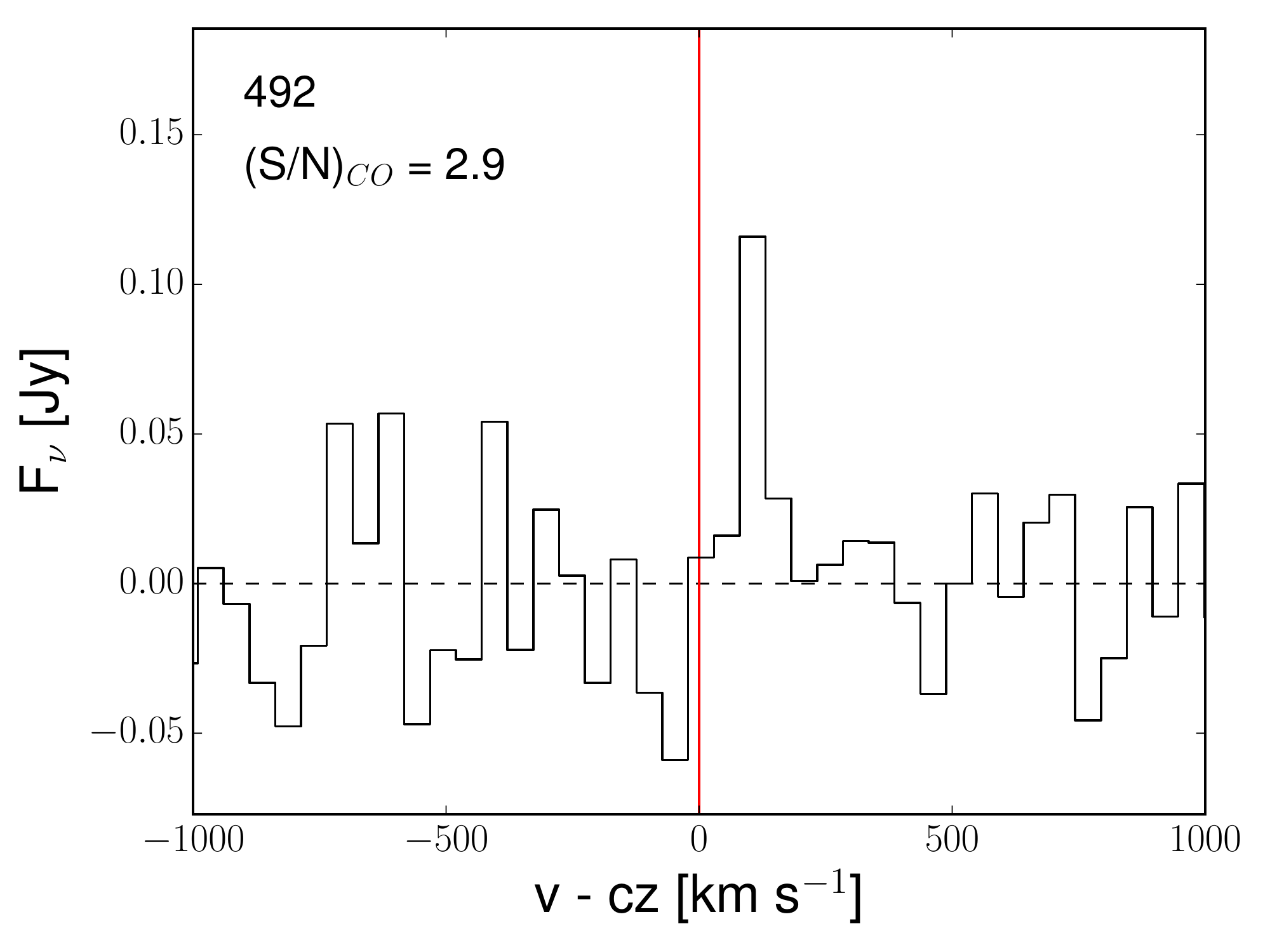}
\includegraphics[width=0.18\textwidth]{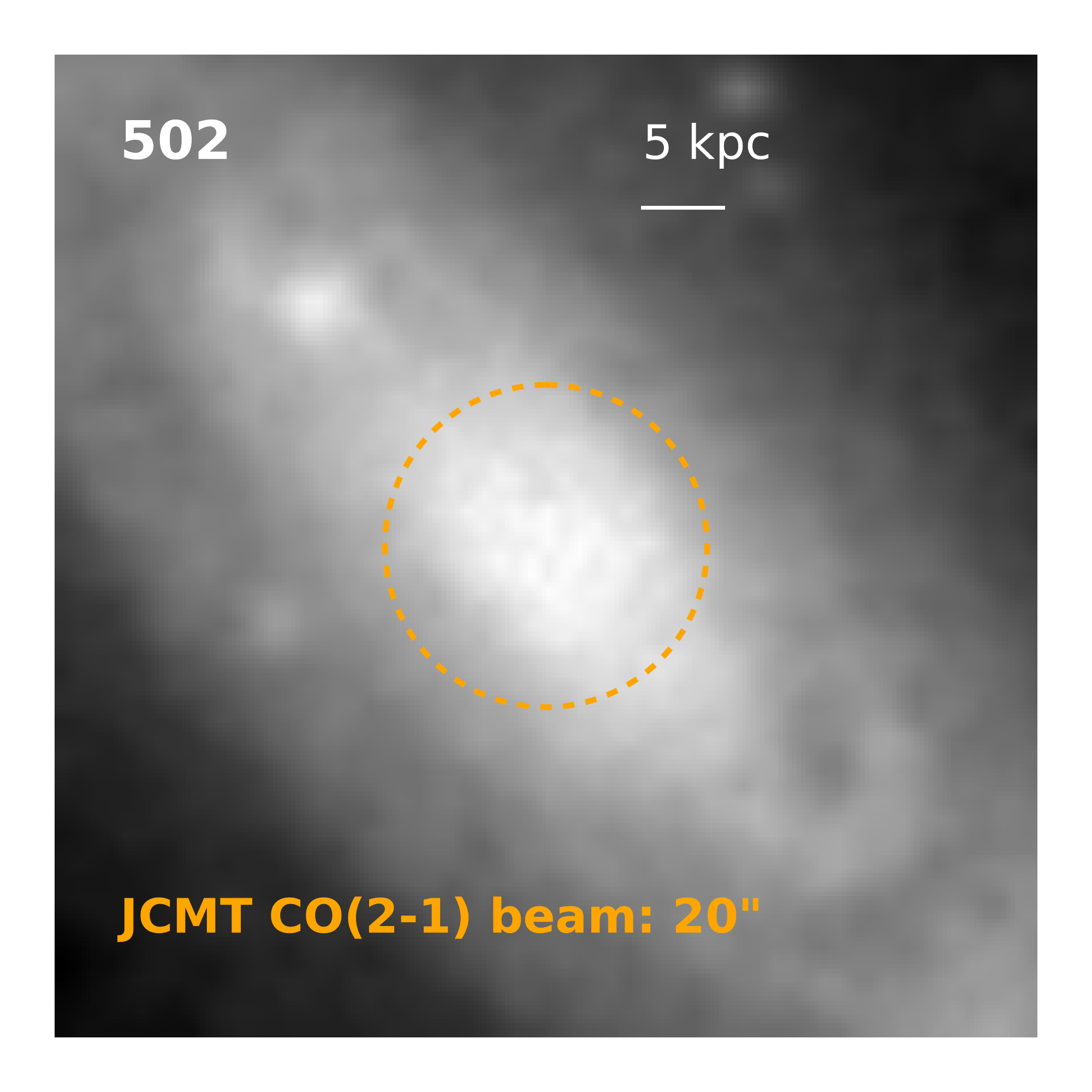}\includegraphics[width=0.26\textwidth]{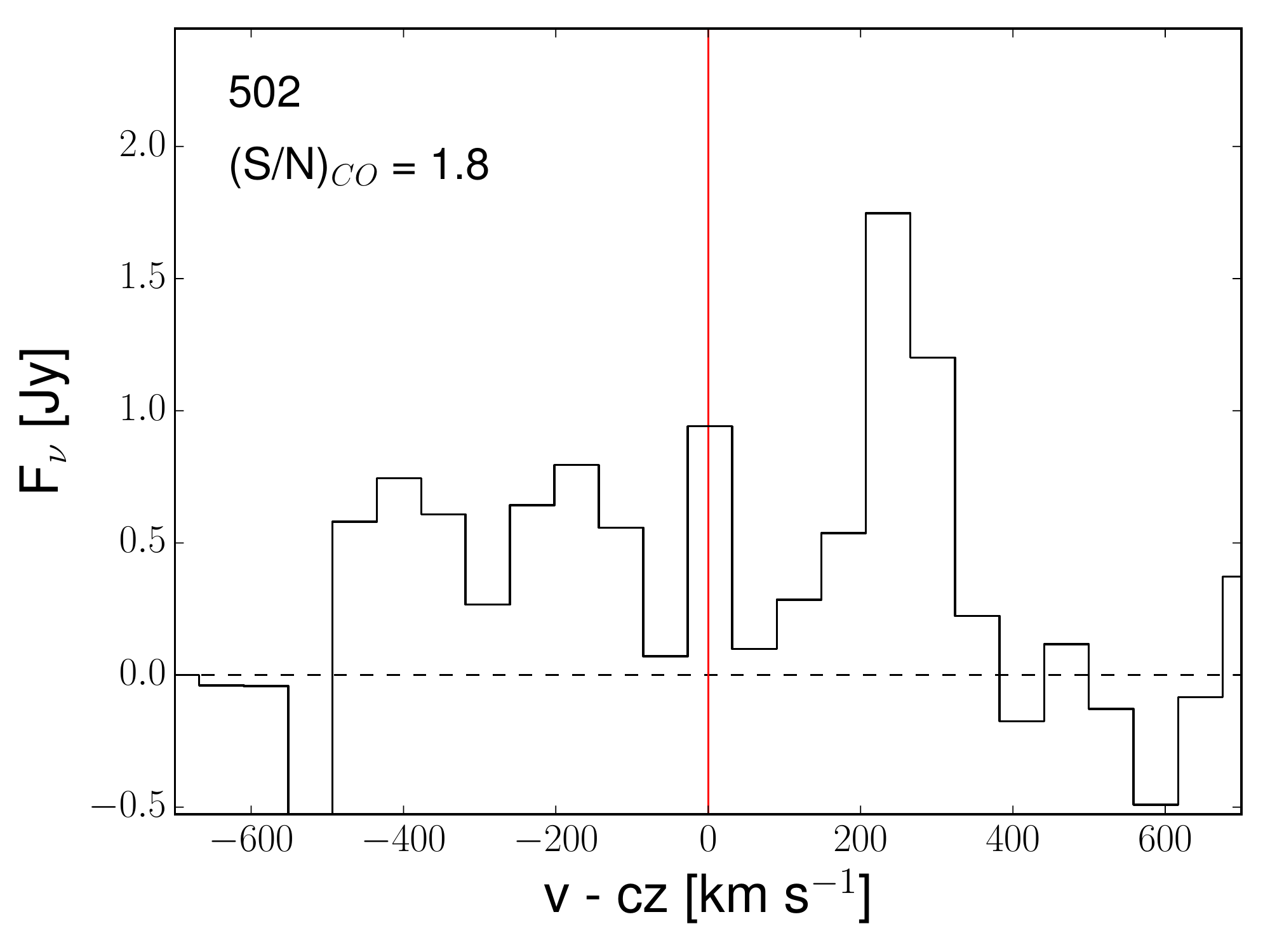}
\includegraphics[width=0.18\textwidth]{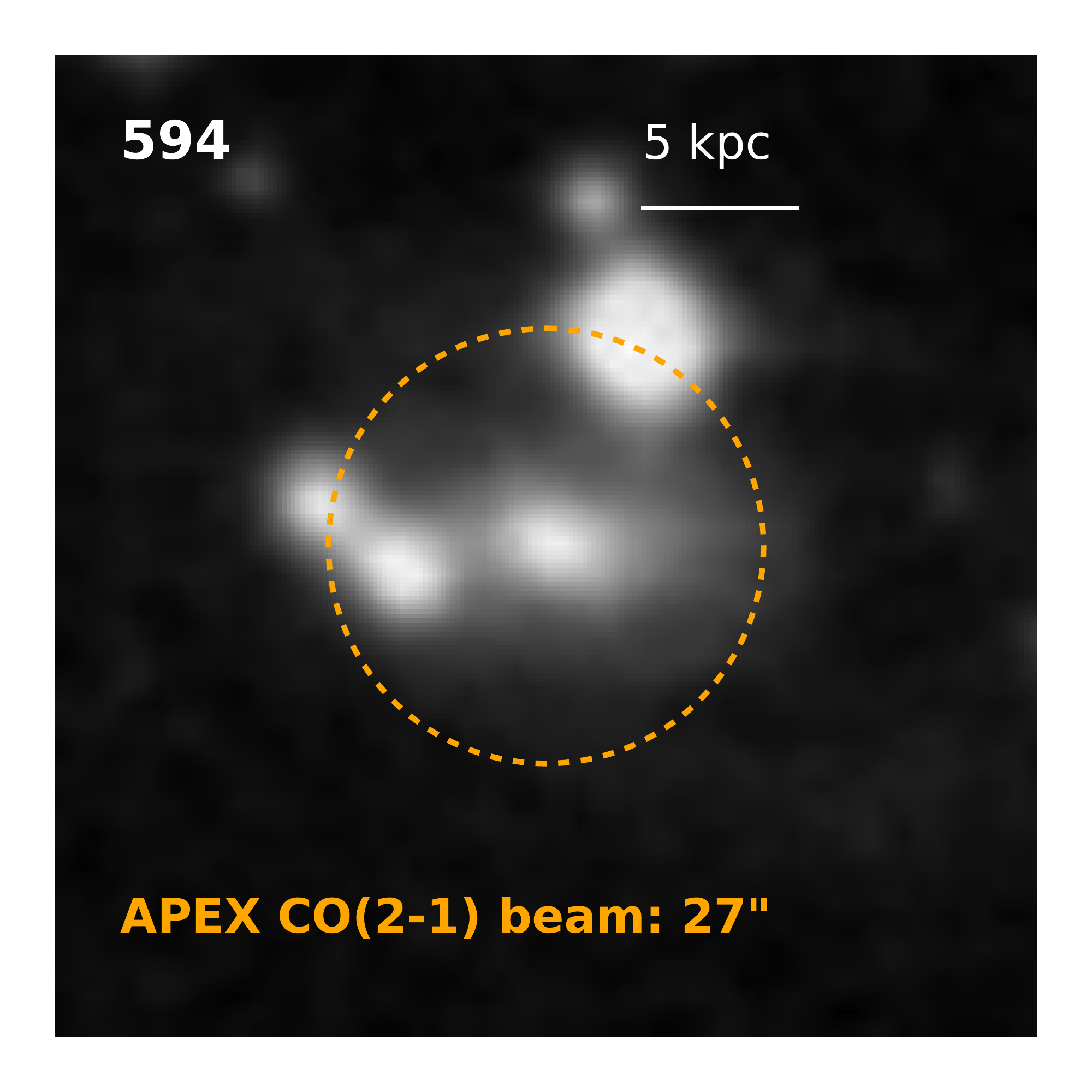}\includegraphics[width=0.26\textwidth]{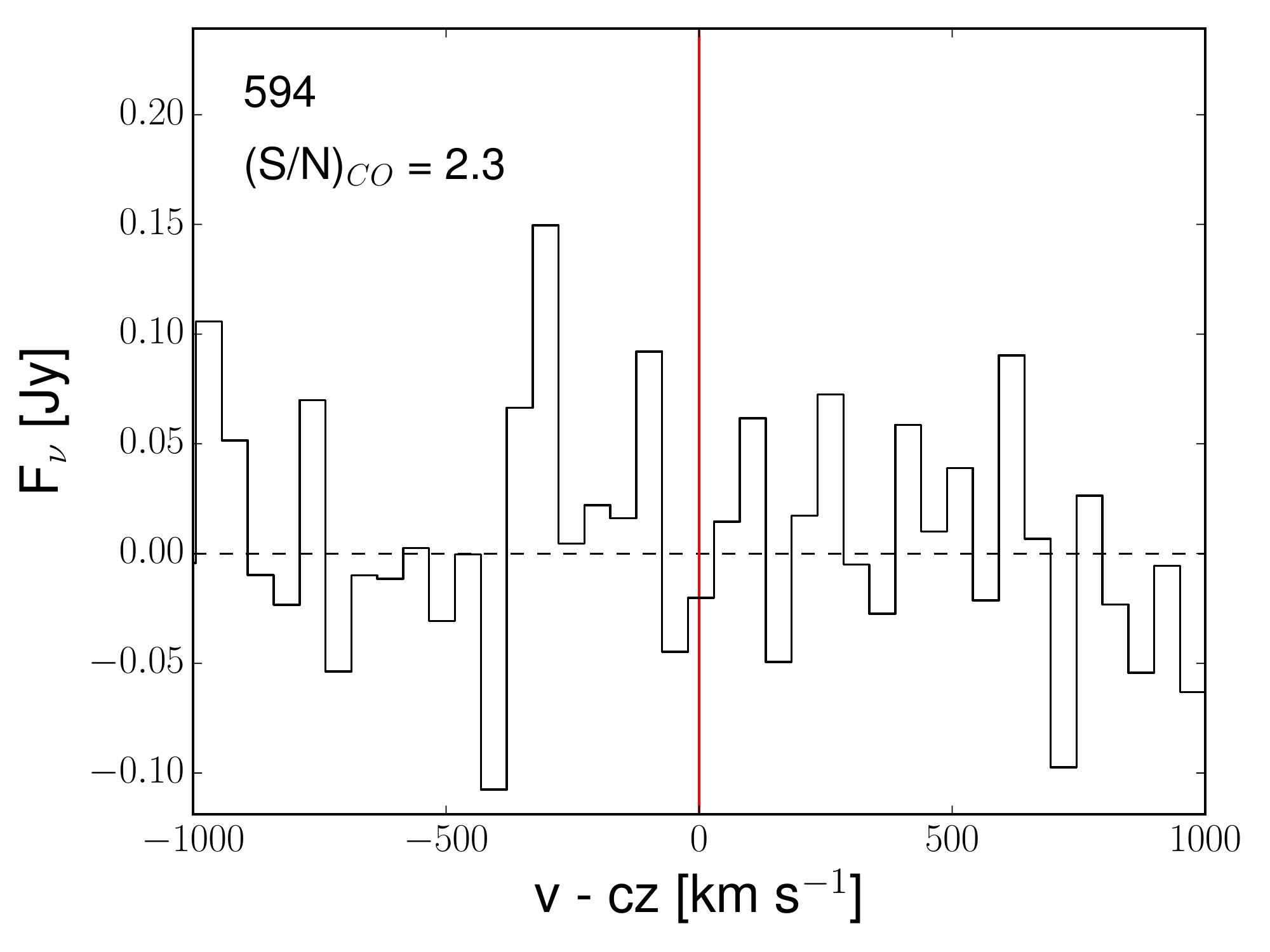}
\includegraphics[width=0.18\textwidth]{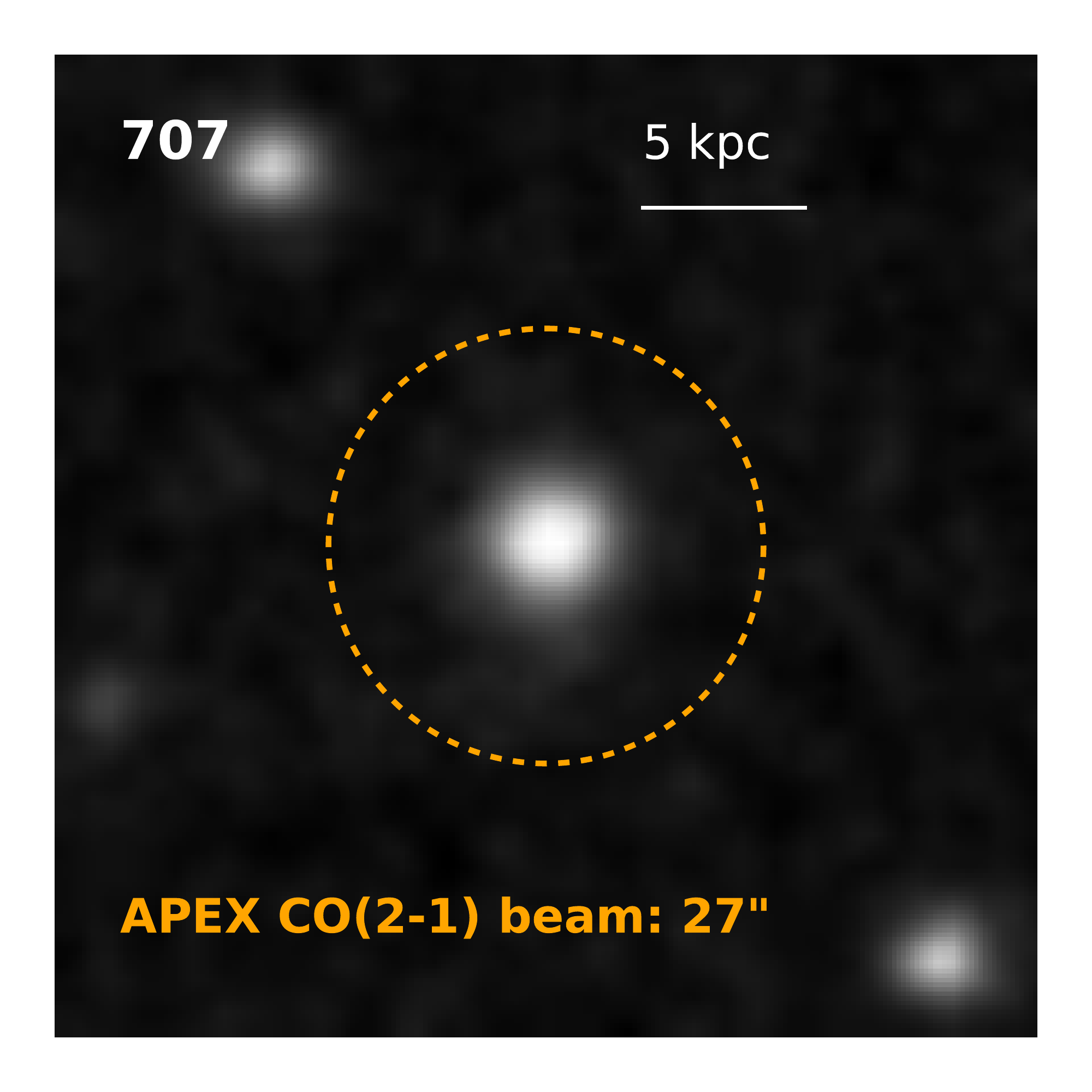}\includegraphics[width=0.26\textwidth]{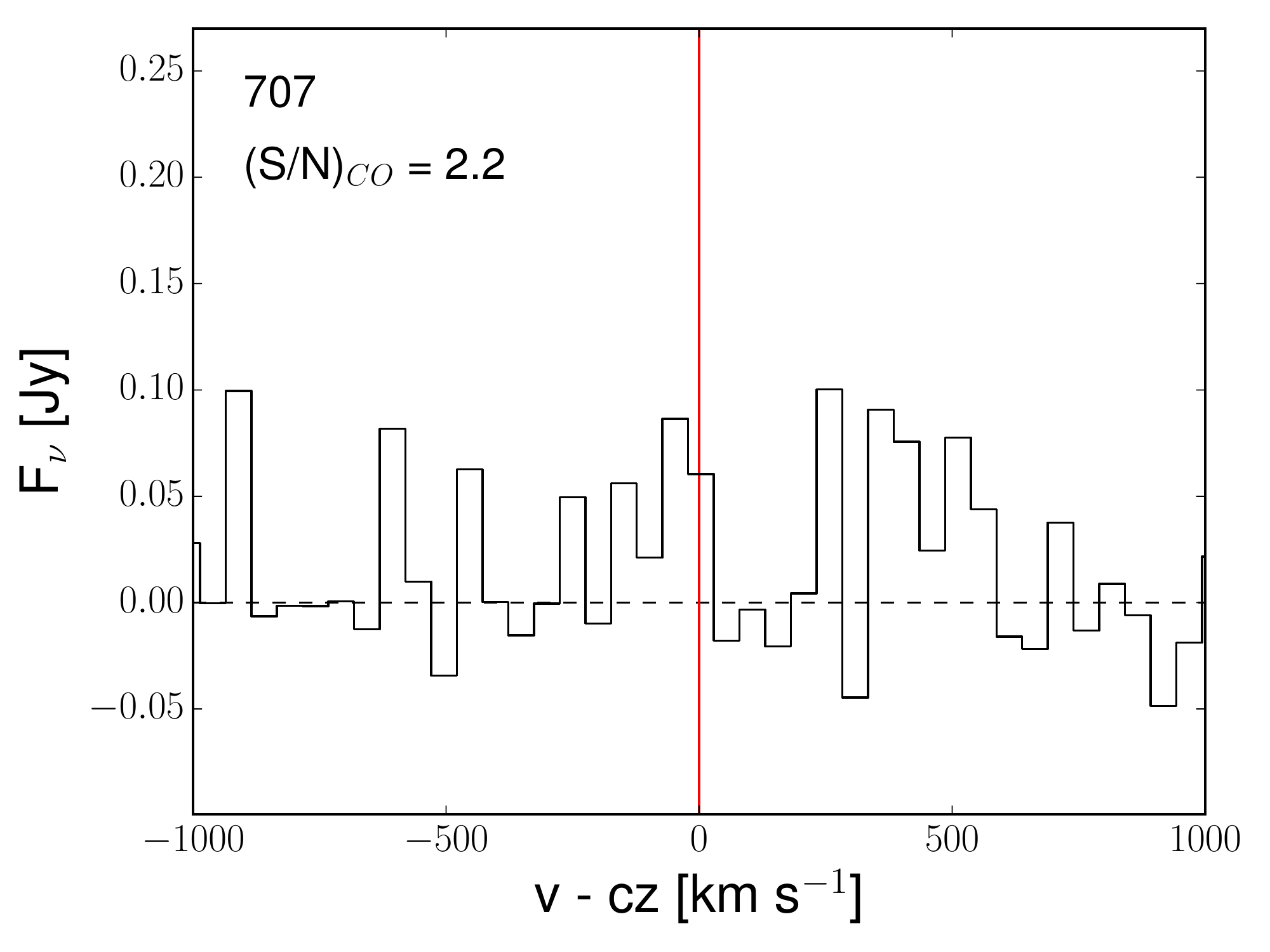}
\includegraphics[width=0.18\textwidth]{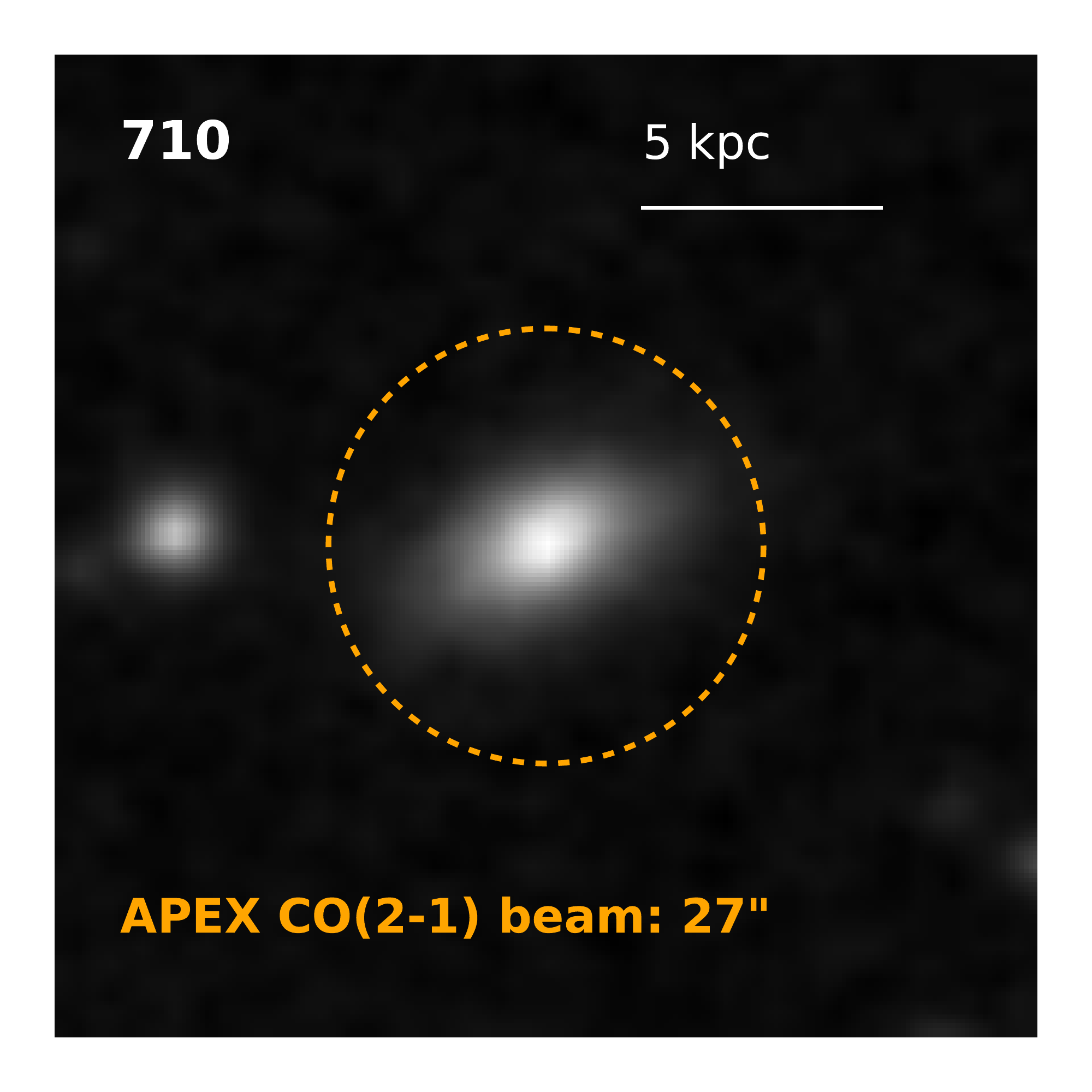}\includegraphics[width=0.26\textwidth]{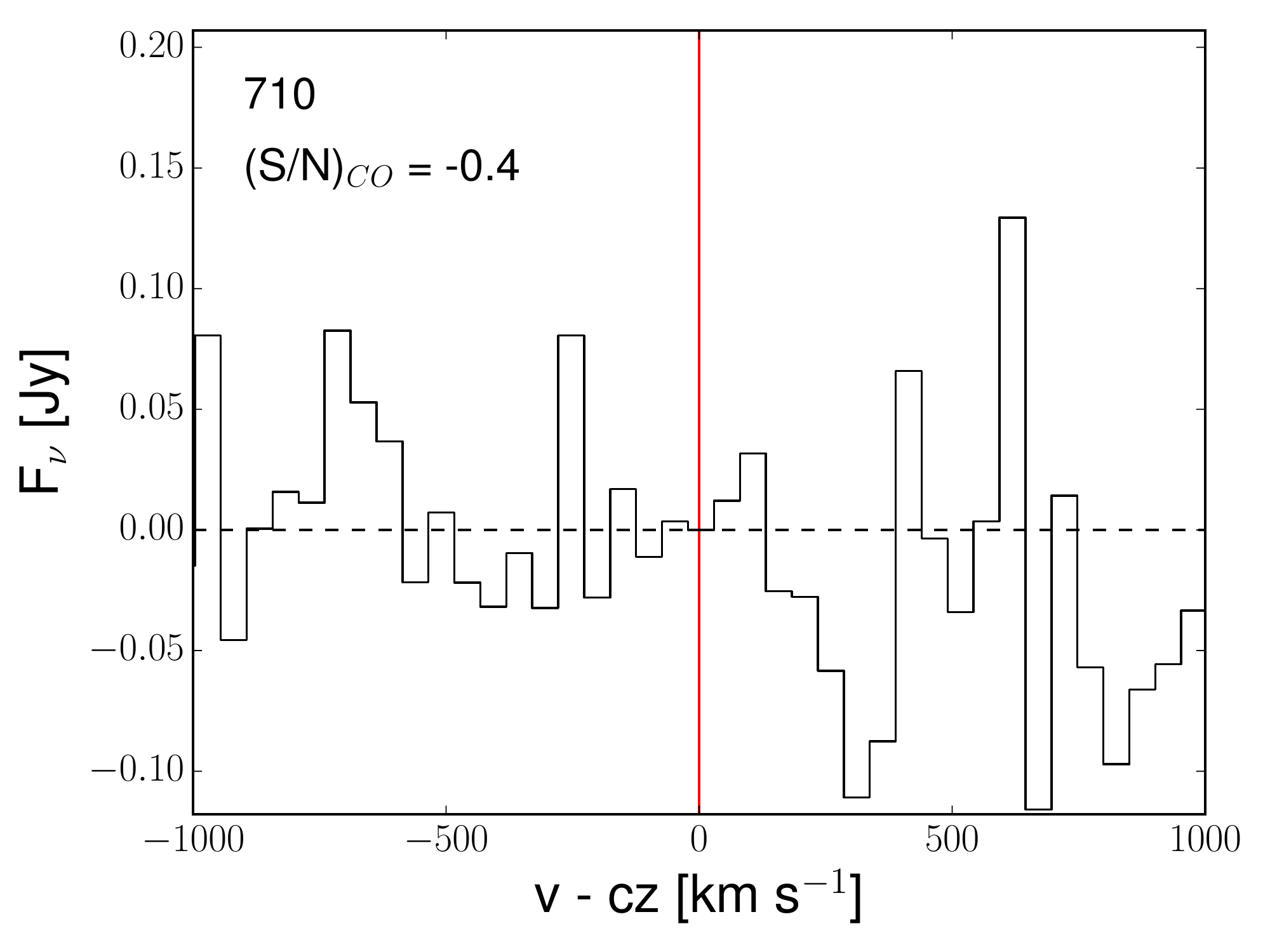}
\includegraphics[width=0.18\textwidth]{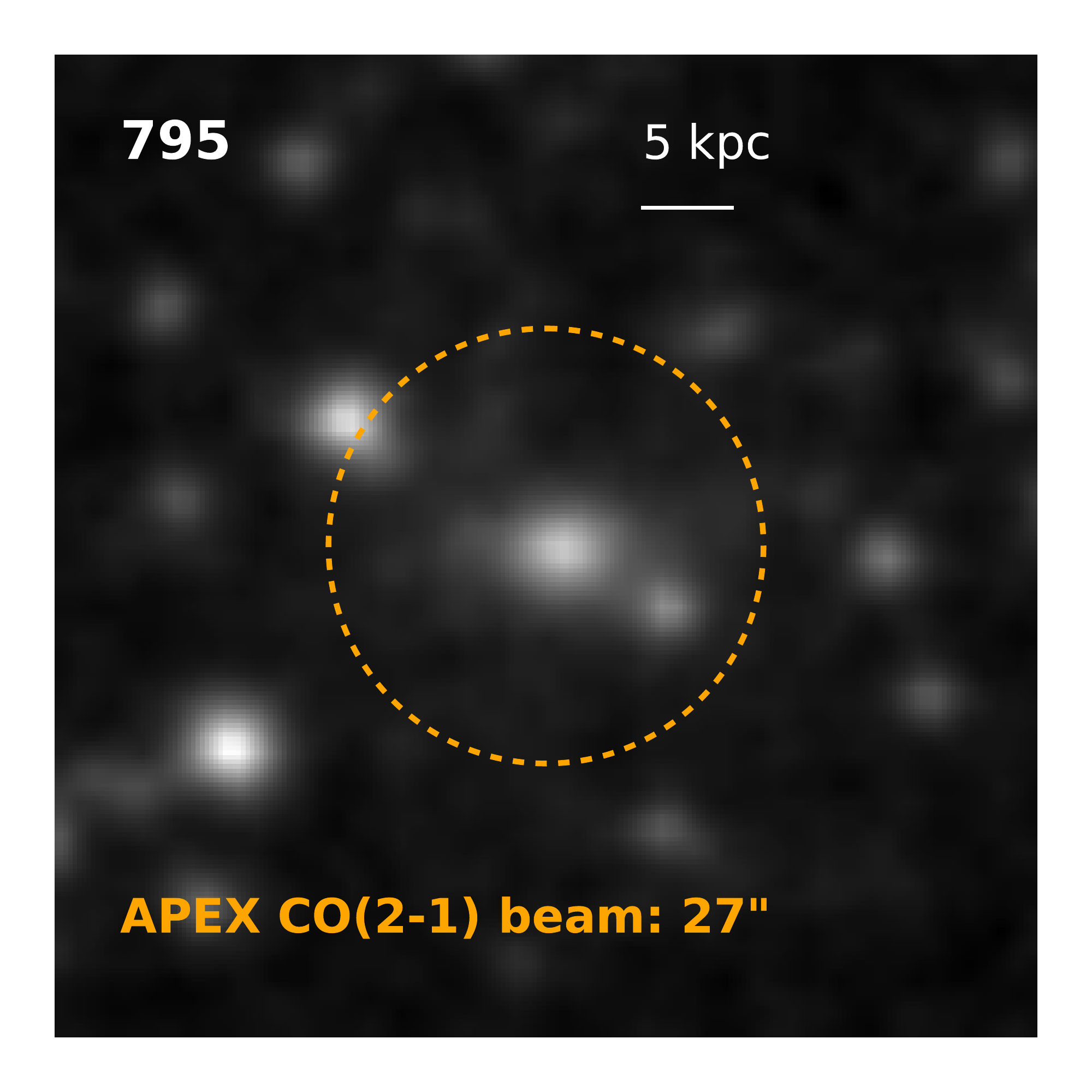}\includegraphics[width=0.26\textwidth]{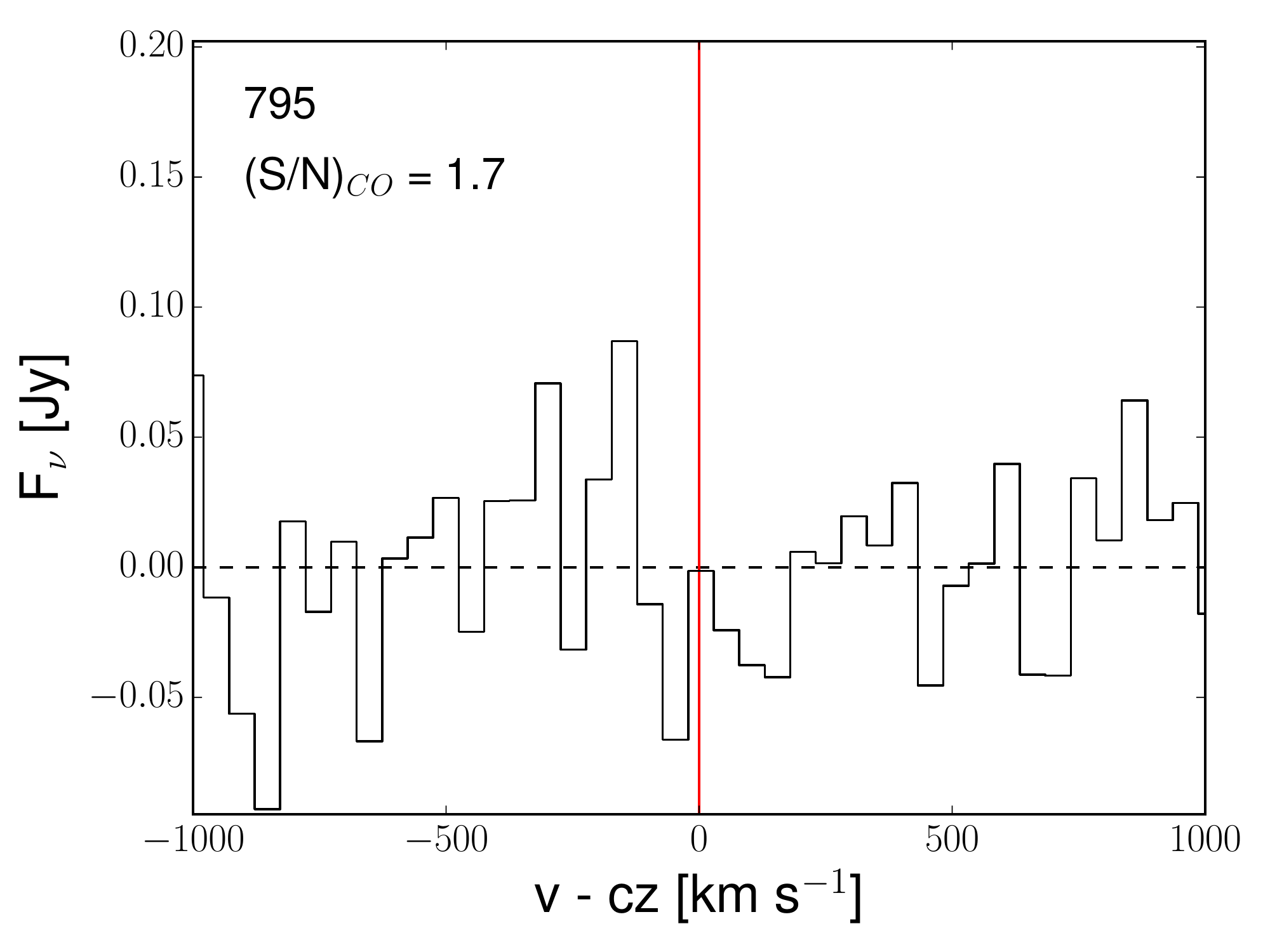}
\includegraphics[width=0.18\textwidth]{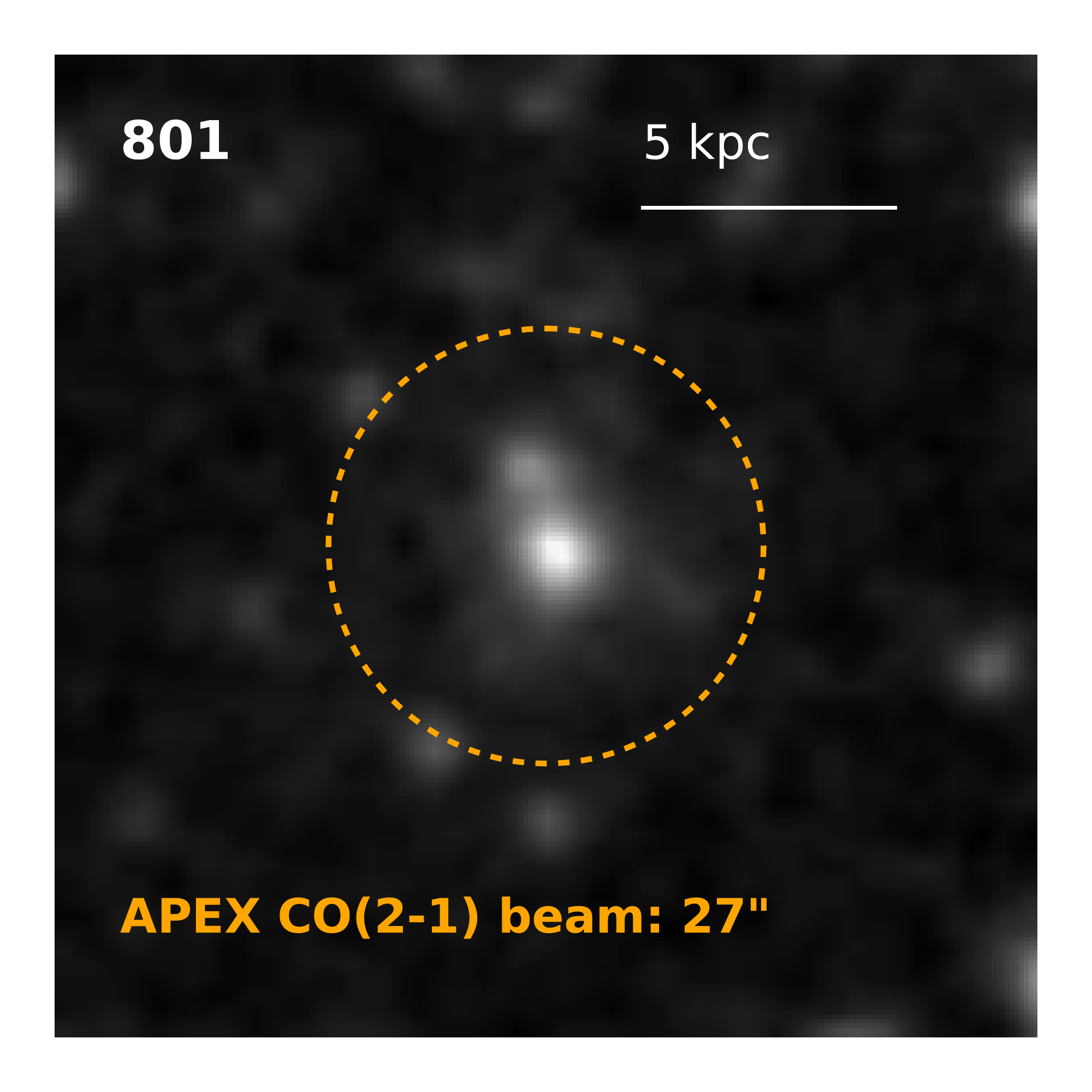}\includegraphics[width=0.26\textwidth]{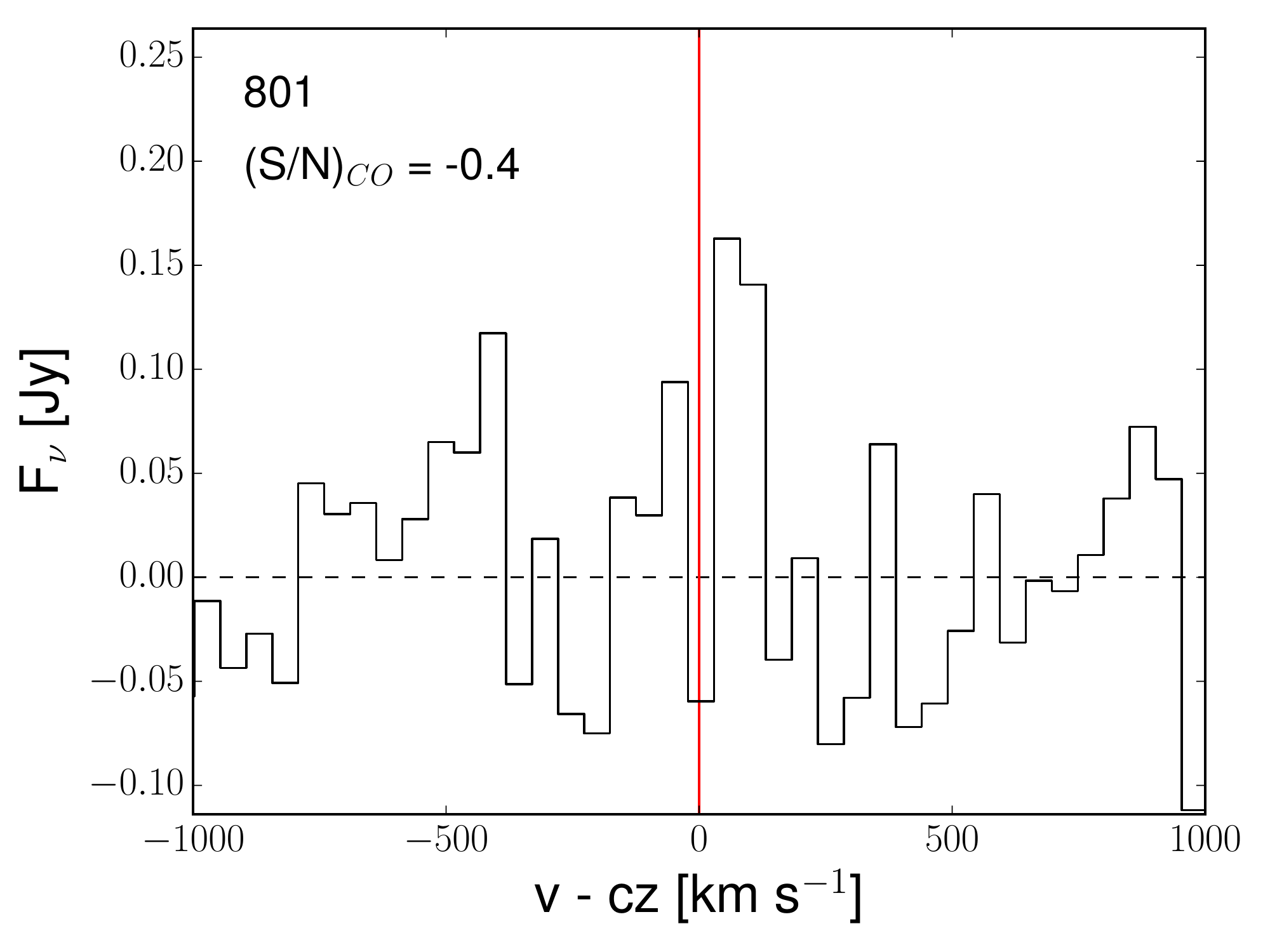}
\caption{continued from Fig.~\ref{fig:CO21_spectra_dss5}
} 
\label{fig:CO21_spectra_dss6}
\end{figure*}

\begin{figure*}
\centering
\raggedright
\includegraphics[width=0.18\textwidth]{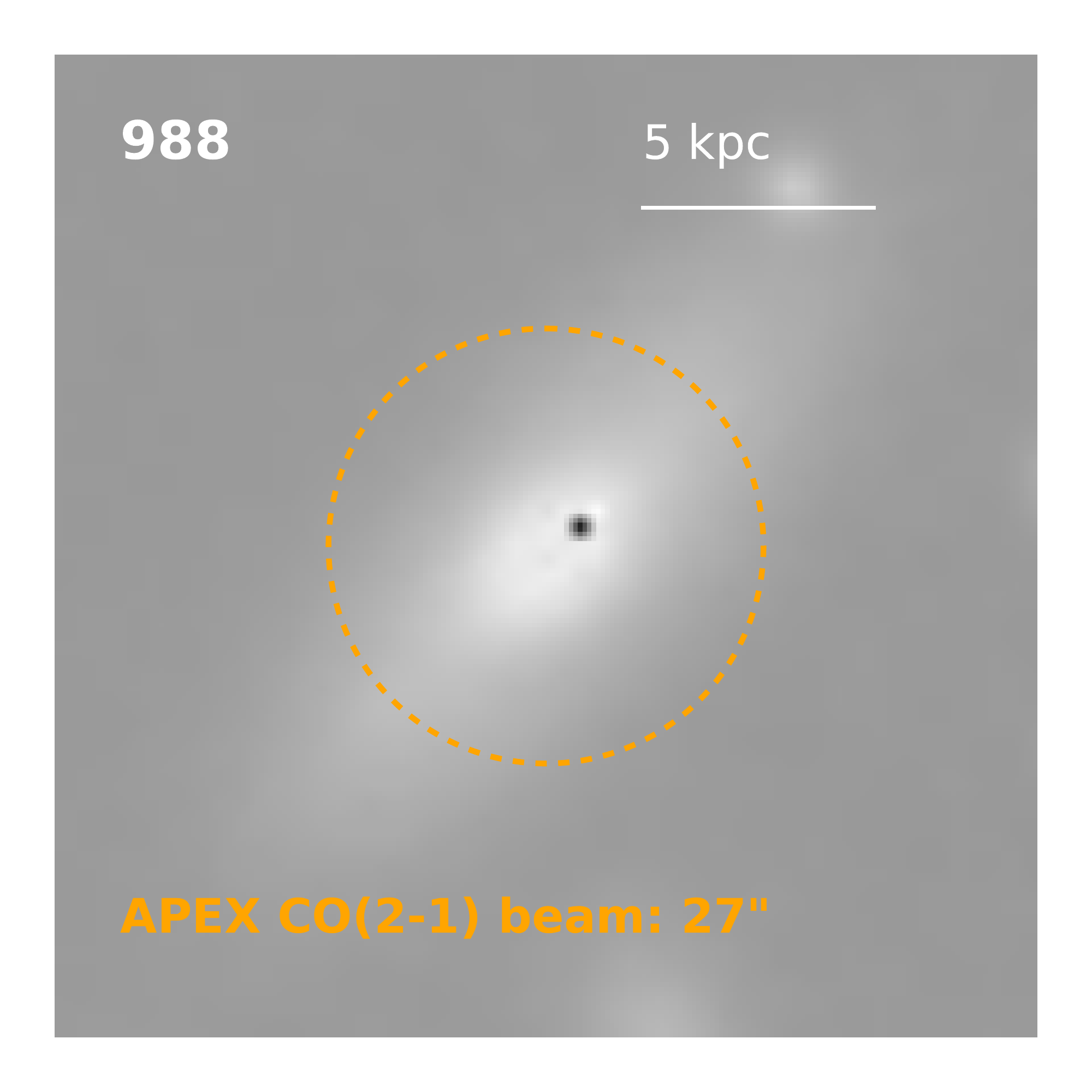}\includegraphics[width=0.26\textwidth]{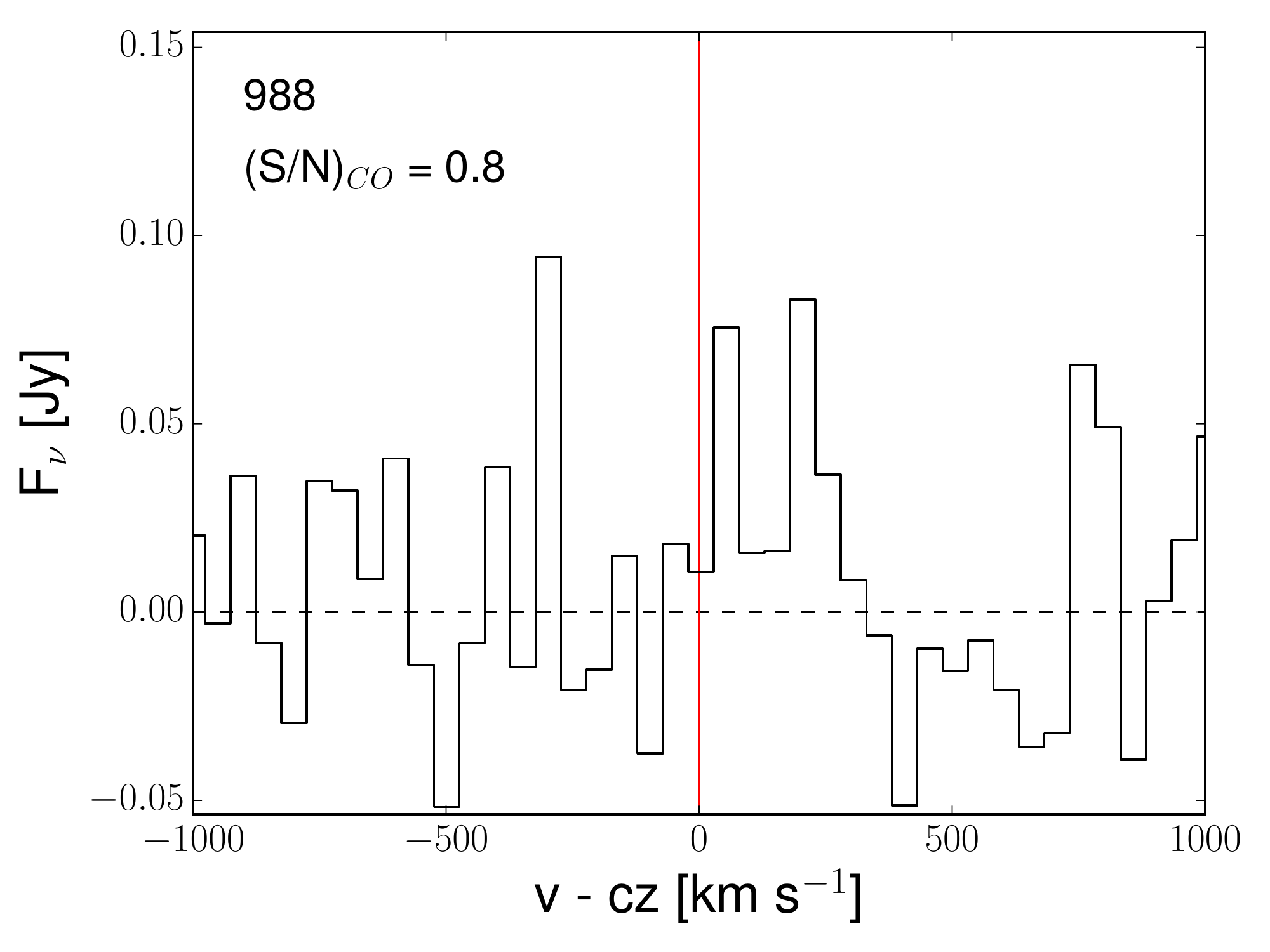}
\includegraphics[width=0.18\textwidth]{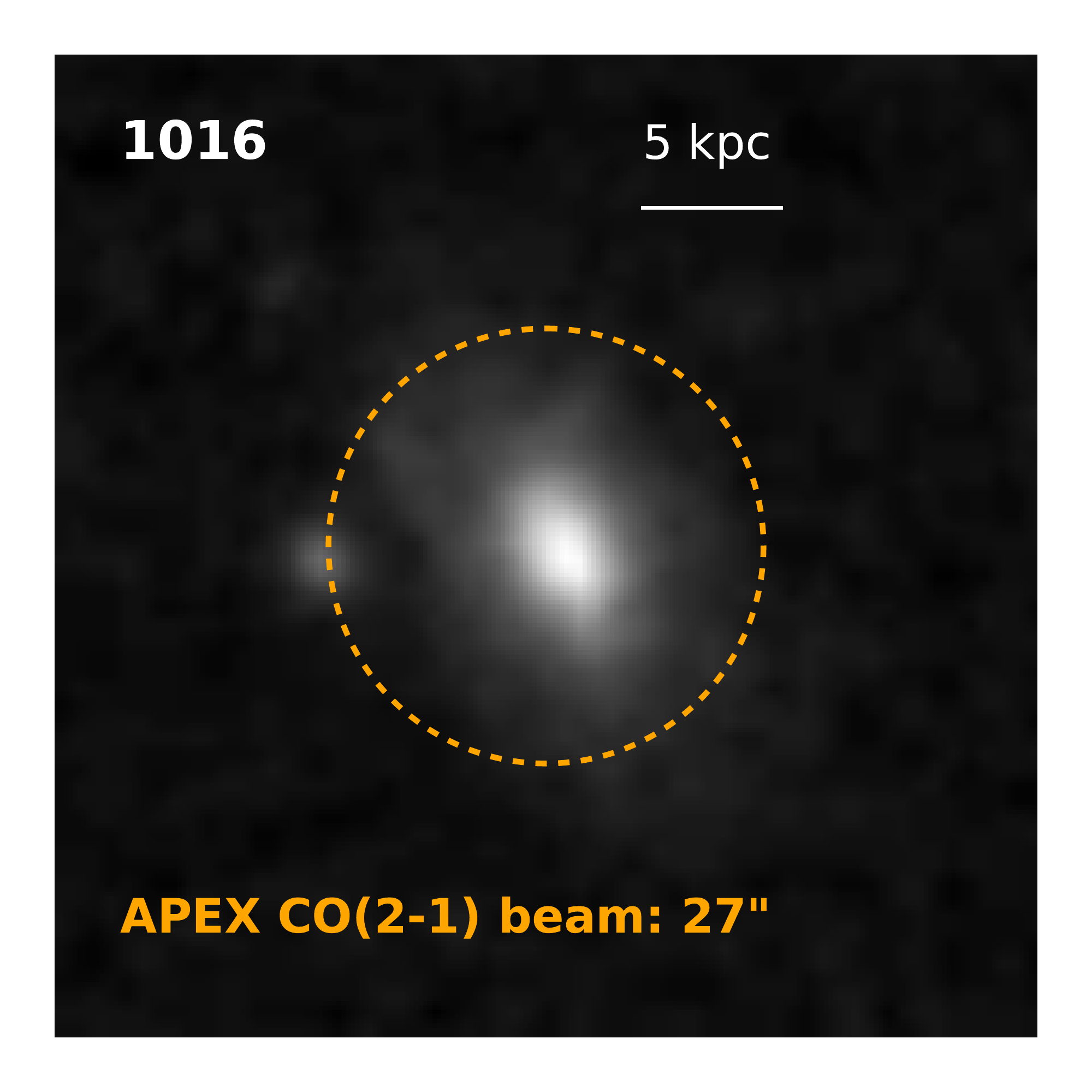}\includegraphics[width=0.26\textwidth]{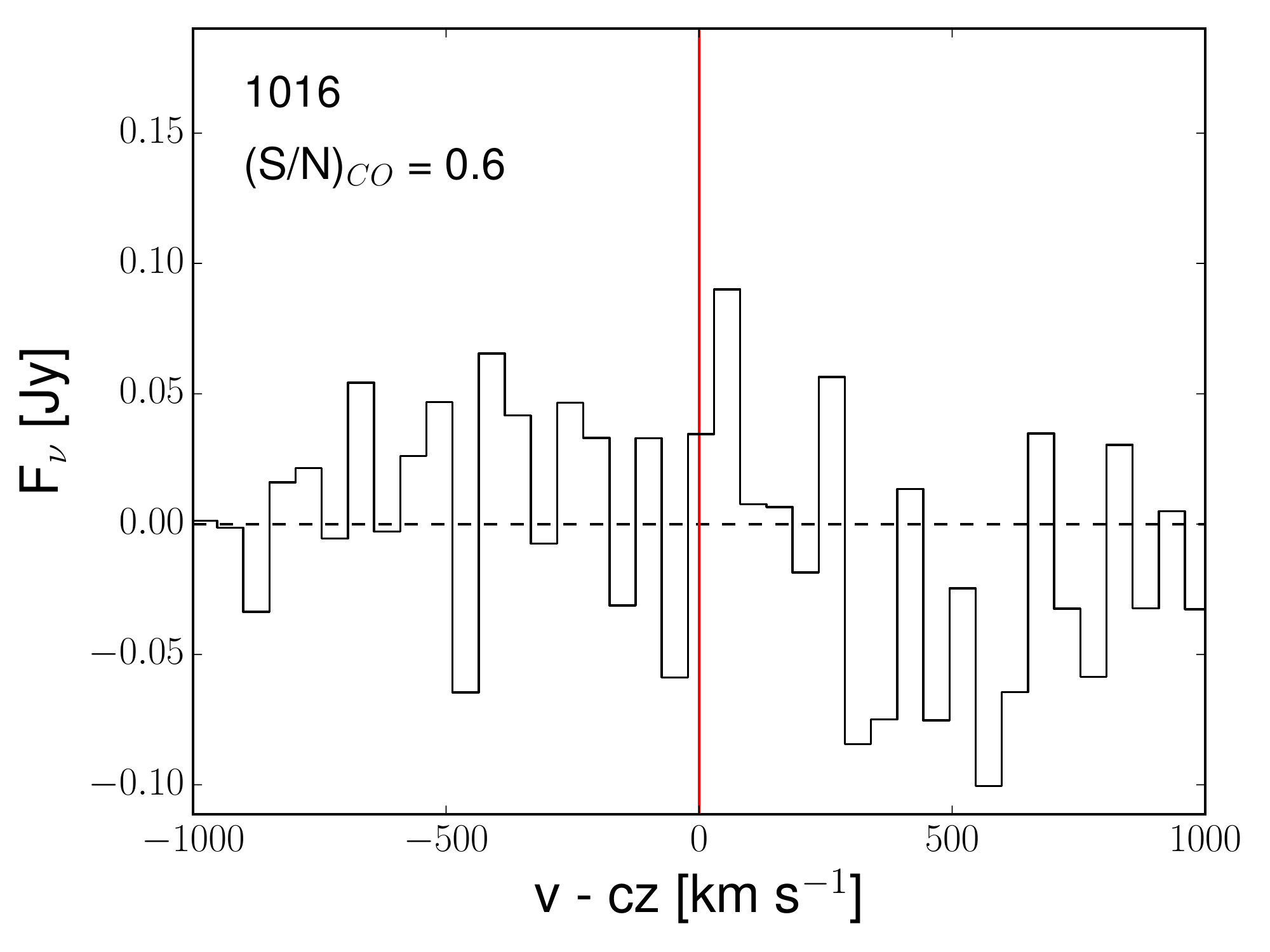}
\includegraphics[width=0.18\textwidth]{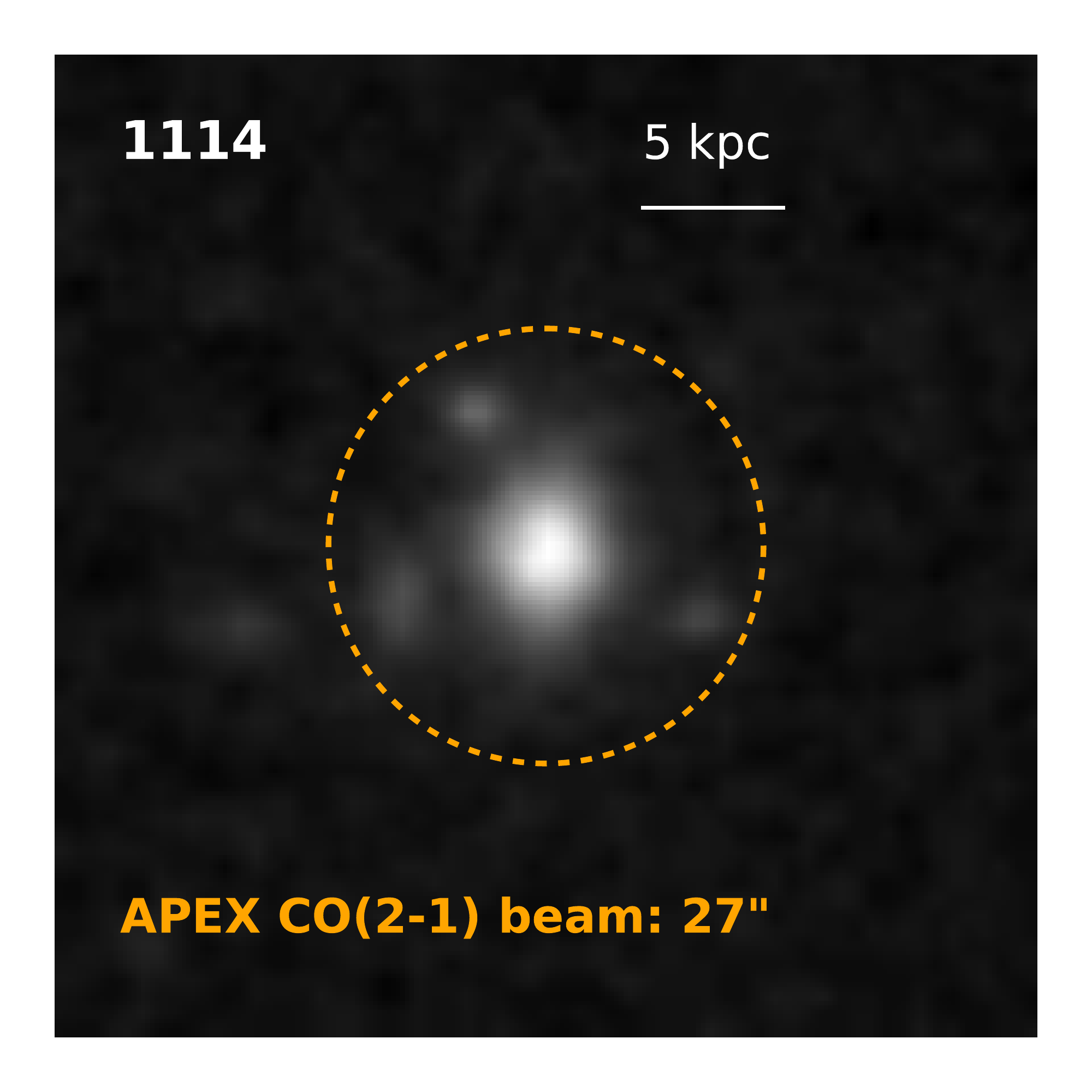}\includegraphics[width=0.26\textwidth]{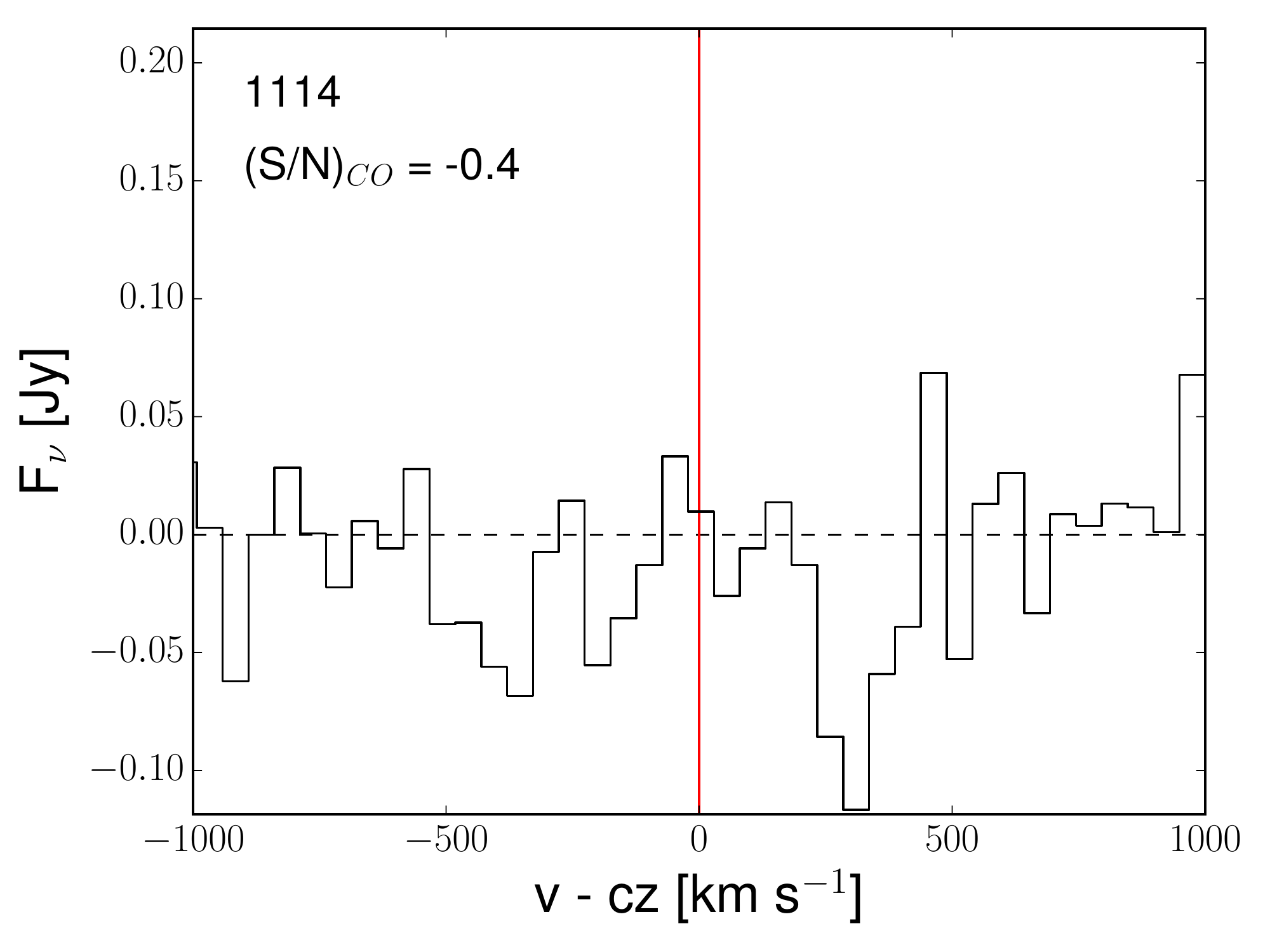}
\includegraphics[width=0.18\textwidth]{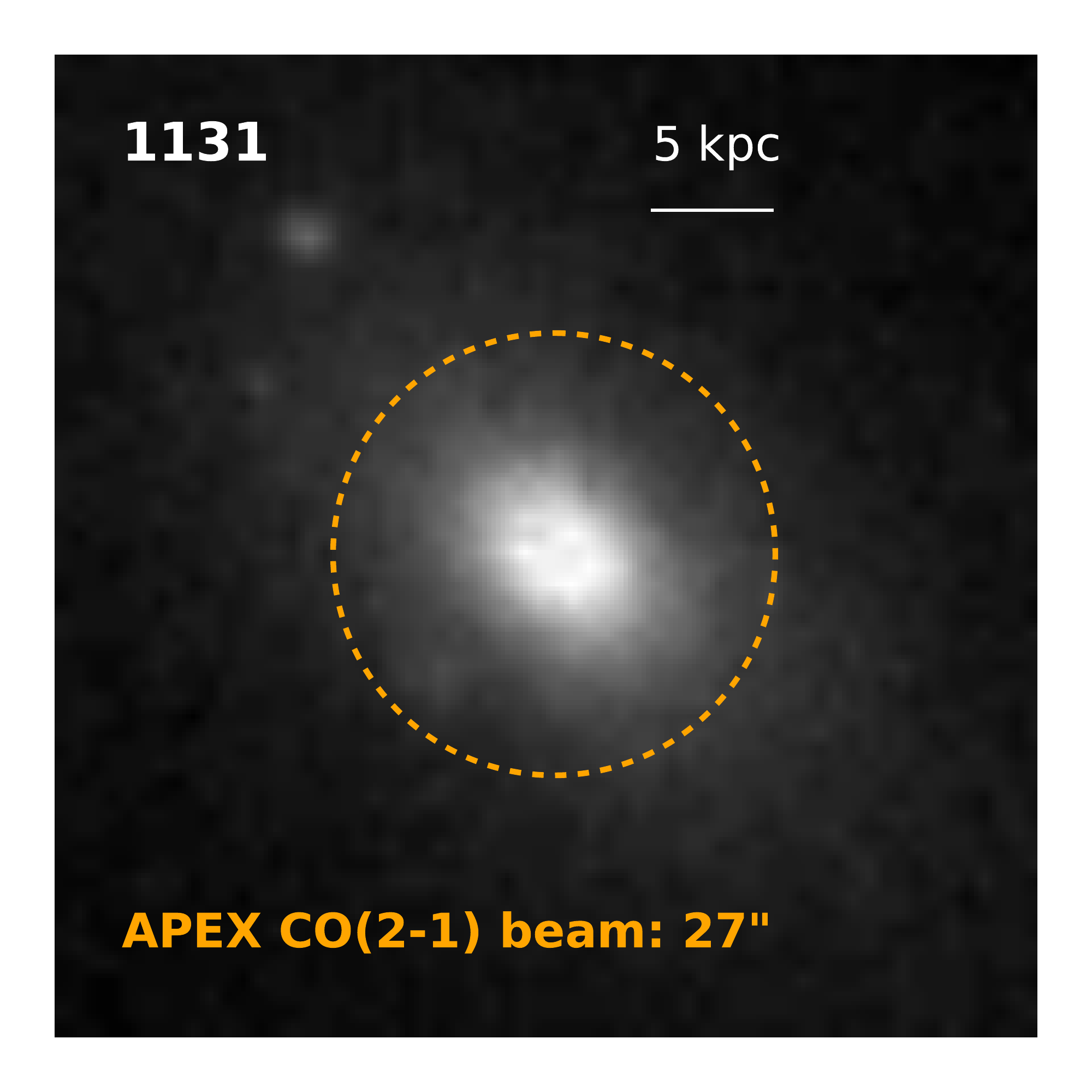}\includegraphics[width=0.26\textwidth]{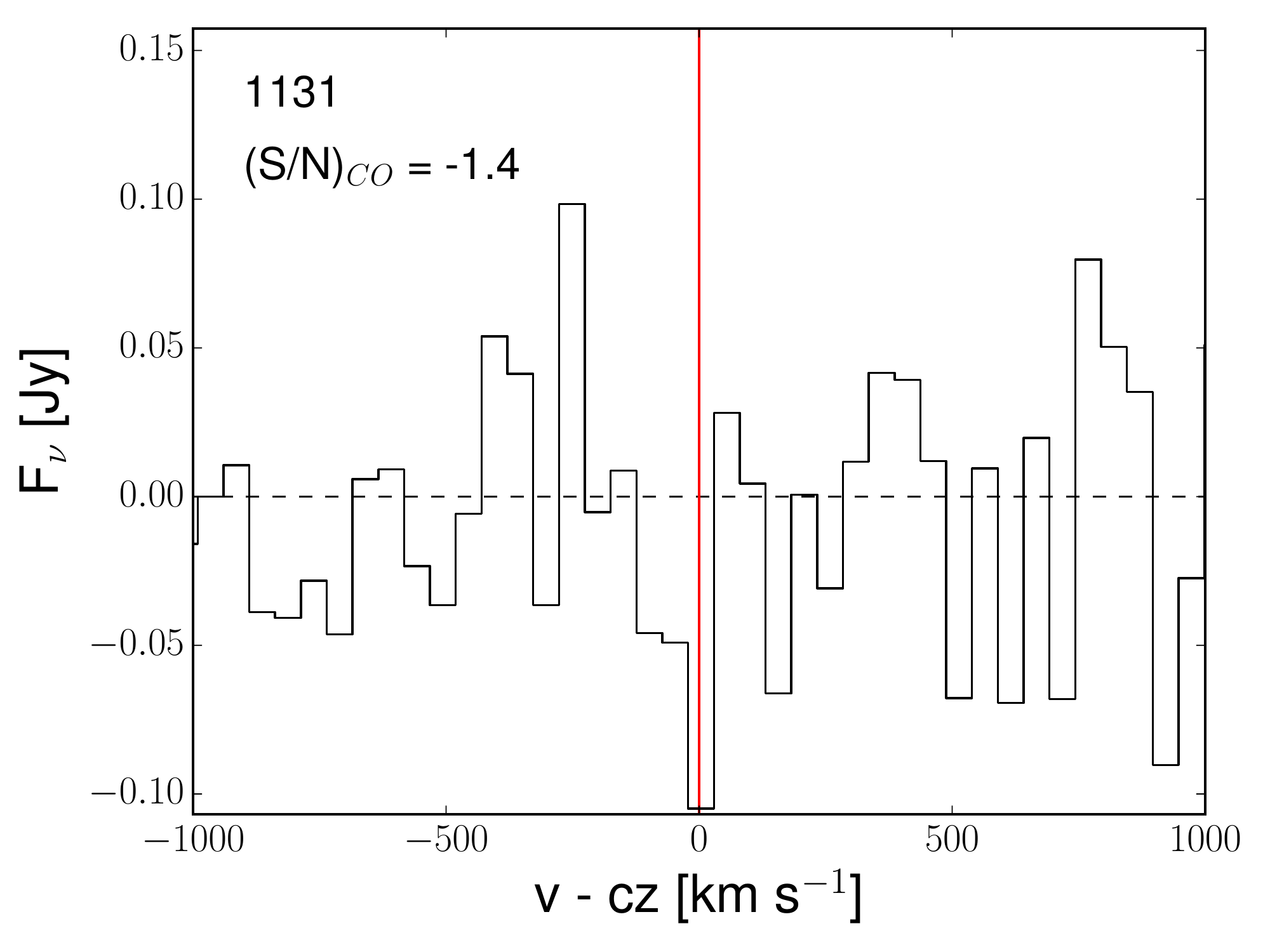}
\caption{continued from Fig.~\ref{fig:CO21_spectra_dss5}
} 
\label{fig:CO21_spectra_dss7}
\end{figure*}


\bibliographystyle{aasjournal}
\end{document}